\pdfoutput=1
\documentclass[11pt,headsepline,cleardoubleempty,twoside,openright]{scrbook}

\usepackage{SciBook}

\begin{document}


\begin{titlepage}
\begin{center}

\vspace*{\stretch{3}}

{\Huge\bfseries\scshape LSST Science Book} 

\vspace*{\stretch{2}}

\begin{center}
{\Large\bf Version 2.0\\

\vspace*{\stretch{0.2}}

November 2009}
\end{center}

\vspace*{\stretch{2.5}}

{\Large  Prepared by the LSST Science Collaborations,}\\
\vspace*{\stretch{0.15}}
{\Large with contributions from the LSST Project. }\\
\vspace*{\stretch{1}}

\end{center}
\end{titlepage}


This book is a living document. The most recent version can be found at \url{http://www.lsst.org}.

The front cover of the Science Book features an image of the LSST created from
mechanical drawings by Todd Mason, Mason Productions, Inc., shown against an image
created from Deep Lens Survey data. The back cover shows a simulated 15-second
LSST exposure from one of the $\rm 4\,K \times 4\,K$ CCDs in the focal plane. Graphic design by E.
Acosta (LSSTC).

For more information, contact:\\
\qquad J. Anthony Tyson, Director 530.752.3830 -- tyson@lsst.org\\
\qquad Donald W. Sweeney, Project Manager 925.487.2134 -- sweeney@lsst.org\\
\qquad Michael A. Strauss, Chair of Science Collaborations 609.258.3808 -- strauss@astro.princeton.edu\\

LSST is a public-private partnership. Design and development activity is supported in
part by the National Science Foundation. Additional funding comes from private
foundation gifts, grants to universities, and in-kind support of Department of Energy
laboratories and other LSST Member Institutions. The project is overseen by the LSST
Corporation, a non-profit 501(c)3 corporation formed in 2003, with headquarters in
Tucson, AZ.

\bigskip
\copyright\ 2009 by the LSST Corporation\\
No part of this book may be reproduced or utilized in any form or by any means without
the prior written permission from the LSST Corporation.

\bigskip

\begin{center}
LSST Corporation\\
933 North Cherry Avenue\\
Tucson, AZ 85721-0009\\
520.881.2626\\
contact@lsst.org\\
\url{http://www.lsst.org}
\end{center}

\begin{figure*}[h]
\begin{center}
\includegraphics[width=\textwidth]{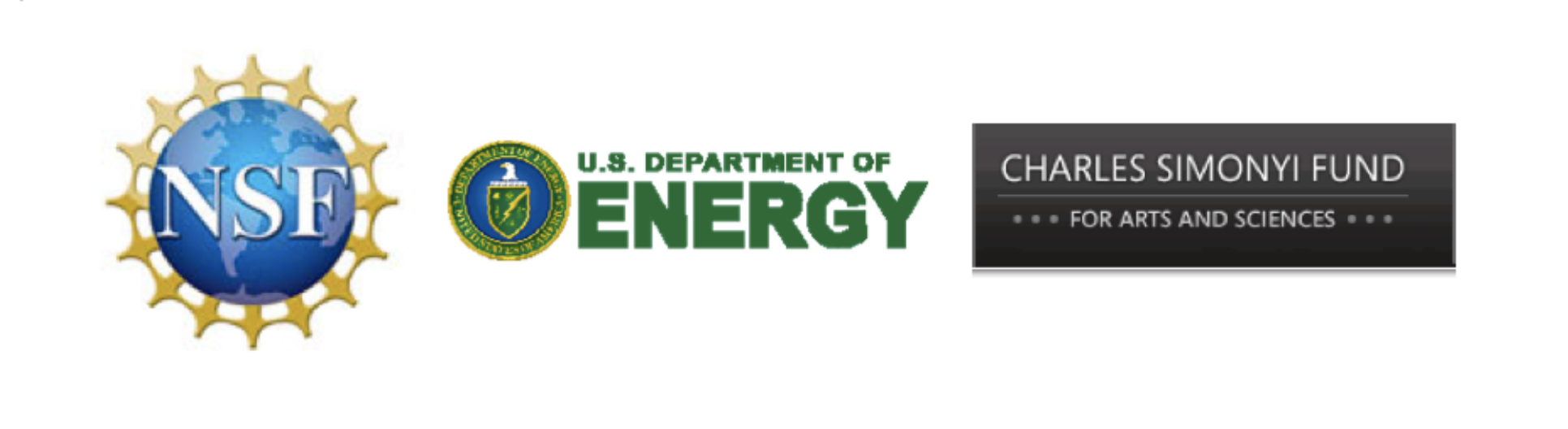}
\end{center}
\end{figure*}

\tableofcontents




\clearemptydoublepage

\ifthenelse{\boolean{draft}}
{     
}{    
     \typeout{ }
}

\chapter*{Preface}
\addcontentsline{toc}{section}{Preface}
\markboth{}{}

\noindent 

Major advances in our understanding of the Universe over the history
of astronomy have often arisen from dramatic improvements in our
ability to observe the sky to greater depth, in previously
unexplored wavebands, with higher precision, or with improved spatial,
spectral, or temporal resolution.  Aided by rapid progress in
information technology, current sky surveys are again changing the way
we view and study the Universe, and the next-generation instruments,
and the surveys that will be made with them, will maintain this
revolutionary progress.  Substantial progress in the important
scientific problems of the next decade (determining the nature of dark
energy and dark matter, studying the evolution of galaxies and the
structure of our own Milky Way, opening up the time domain to discover
faint variable objects, and mapping both the inner and outer Solar
System) all require wide-field repeated deep imaging of the sky in
optical bands.

The wide-fast-deep science requirement leads to a single wide-field
telescope and camera which can repeatedly survey the sky with deep
short exposures.  The Large Synoptic Survey Telescope (LSST), a
dedicated telecope with an effective aperture of 6.7 meters and a field
of view of 9.6 deg$^2$, will make major contributions to all these
scientific areas and more. It will carry out a survey of 20,000
deg$^2$ of the sky in six broad photometric bands, imaging each region
of sky roughly 2000 times (1000 pairs of back-to-back 15-sec exposures)
over a ten-year survey lifetime.

The LSST project
will deliver fully calibrated survey data to the United States
scientific community and the public with no proprietary period.  Near
real-time alerts for 
transients will also be provided worldwide. A goal is worldwide participation in all data products. The survey will
enable comprehensive exploration of the Solar System beyond the
Kuiper Belt, new understanding of the structure of our Galaxy and that
of the Local Group, and vast opportunities in cosmology and galaxy
evolution using data for billions of distant galaxies.  Since many of
these science programs will involve the use of the world's largest
non-proprietary database, a key goal is maximizing the usability of
the data.  Experience with previous surveys is that often their most
exciting scientific results were unanticipated at the time that the
survey was designed; we fully expect this to be the case for the LSST
as well.  

The purpose of this Science Book is to examine and document in detail
science goals, opportunities, and capabilities that will be provided
by the LSST. The book addresses key questions that will be confronted
by the LSST survey, and it poses new questions to be addressed by
future study. It contains previously available material (including a
number of White Papers submitted to the ASTRO2010 Decadal Survey) as
well as new 
results from a year-long campaign of study and evaluation.
This book does not attempt to be complete; there are many other
scientific projects one can imagine doing with LSST that are not
discussed here.  Rather, this book is intended as a first step in a
collaboration with the 
world scientific community to identify and prepare for the scientific
opportunities that LSST will enable.  It will also provide guidance to the
optimization and implementation of the LSST system and to the
management and processing of the data produced by the LSST survey.

The ten LSST Science Collaborations, together with others in the world
astronomy and physics community, have authored this Science Book; the
full list of over 200 contributors may be found in
\autoref{chp:ap:authors}.  These collaborations perform their work
as semi-autonomous organizations in conjunction with the LSST Project,
and provide access to the LSST and its support infrastructure for
large numbers of scientists.  These scientists are 
laying the groundwork necessary to carry out LSST science projects, 
defining the required data products, and developing optimal algorithms and
calibration strategies for photometry, astrometry, photometric
redshifts, and image analysis.  Membership in the science
collaborations is open to staff at the member institutions, and two US
community-wide open call for applications for membership have already
been issued.  There will be regular future opportunities to join the science
collaborations. 

This Science Book is a living document. Our understanding of the
scientific opportunities that LSST will enable
will surely grow, and the authors anticipate future updates of the material in this book as
LSST approaches first light.

\vspace*{5cm}
\begin{flushright} 
\emph{November 2009} 
\end{flushright}

\clearemptydoublepage



%
%
%
%
%
%
%
%
%
%
%
%
%
%
%
%
%
%
%
%
%
%
%
%
%
%
\chapter[Introduction]
{Introduction}
\label{chp:introduction}

{\it Anthony Tyson, Michael A. Strauss, \v{Z}eljko Ivezi\'{c}}



Wide-angle surveys have been an engine for new discoveries throughout
the modern history of astronomy, and have been among the most highly
cited and scientifically productive observing facilities in recent
years. Over the past decade, large scale sky surveys in many
wavebands, such as the Sloan Digital Sky Survey (SDSS), Two-Micron All
Sky Survey (2MASS), Galaxy Evolution Explorer (GALEX), Faint Images of
the Radio Sky at Twenty-centimeters (FIRST), and many others have
proven the power of large data sets for answering fundamental
astrophysical questions. This observational progress, based on
advances in telescope construction, detectors, and above all,
information technology, has had a dramatic impact on nearly all fields
of astronomy and many areas of fundamental physics. The hardware and
computational technical challenges and the exciting science
opportunities are attracting scientists from high-energy physics,
statistics, and computer science. These surveys are most productive
and have the greatest impact when the data from the surveys are made
public in a timely manner. The LSST builds on the experience of these
surveys and addresses the broad scientific goals of the coming decade.


\section{Astronomy-Physics Interaction}

The astronomical discovery that ordinary matter, i.e., that made of
familiar atoms, comprises only 4\% of the mass-energy density of the
Universe is the most dramatic in cosmology in the past several
decades, and it is clear that new physics will be needed to explain
the non-baryonic dark matter and dark energy. At the same time, data
from particle physics suggests a corresponding need for physics beyond
the Standard Model. Discovering and understanding the fundamental
constituents and interactions of the Universe is the common subject of
particle physics and cosmology. In recent years, the frontier
questions in both fields have become increasingly
intertwined; in addition to the dark matter and dark energy questions,
astronomical observations have provided the best evidence to date for
non-zero neutrino masses, have suggested phase transitions leading to
inflation in the early Universe, give the best constraints on
alternative theories of gravity on large scales, and allow us to test
for time variations in the fundamental physical constants. 

The emerging common themes that astrophysics and particle physics are
addressing have crystallized a new physics-astronomy community. The
number of particle physicists taking active roles in astrophysics has
increased significantly. 
Understanding the origin of dark matter and dark energy will require
simultaneous progress in both particle physics and cosmology, in both
theory and experiment. Discoveries with LSST and the Large Hadron
Collider will rely on scientists covering a broader intellectual
frontier, and require enhanced collaboration between theorists and
experimentalists in particle physics, cosmology, and astrophysics
generally. 


\section{What a Telescope with Enormous \'Etendue can Accomplish}
\label{sec:introduction:etendue}

A survey that can cover the sky in optical bands over wide fields to
faint magnitudes with a fast cadence is required in order to explore
many of the exciting science opportunities of the next decade. The
most important 
characteristic that determines the speed at which a system can survey
the sky to a given depth is its \'etendue (or grasp): the product of
its primary mirror area (in square meters) and the area of its
field-of-view (in square degrees). Imaging data from 
a large ground-based active optics telescope with sufficient \'etendue
can address many scientific missions
simultaneously rather than sequentially. By providing unprecedented
sky coverage, cadence, and depth, the LSST makes it possible to attack
multiple high-priority scientific questions that are far beyond the
reach of any existing facility.  

The effective \'etendue for LSST will be $319 ~ \rm m^2 deg^2$, more
than an order of magnitude larger than that of any existing
facility. Full simulations of LSST's capabilities have been carried
out, as described below. The range of scientific investigations that
will be enabled by such a dramatic improvement in survey capability is
extremely broad. These new investigations will rely on the statistical
precision obtainable with billions of objects. Thus hundreds of deep
exposures are required in each band to gain control of low-level
systematics. Hundreds of deep and short exposures are also required in
order to fully explore the faint time domain on short timescales.
This wide-fast-deep requirement led to the LSST design. The history of
astronomy has taught us that there are unanticipated surprises
whenever we view the sky in a new way. The wide-fast-deep survey
capability of LSST promises significant advances in virtually all
areas of astrophysics. 


\section{The History of the Idea}

The value of wide area imaging of the sky has long been recognized: motivated
by the opportunities of statistical astronomy, telescope and detector research and development (R\&D)
campaigns in the 1930s and 1940s at Caltech and Kodak gave rise to the
Palomar Observatory 
Sky Survey (POSS, 1948-1957).  While POSS enabled significant advances
in astronomy 
through follow-up observations, the next revolution -- very deep imaging --
had to wait 25 years for digital data from a new detector technology.  Early
Charge-Coupled Devices (CCDs) were ten thousand times smaller in area
than the POSS 
plates, but the promise of high quantum efficiency for astronomical
applications (including the Hubble Space Telescope (HST)) kept R\&D on scientific grade CCDs alive in
the 1970s and 1980s. With their higher sensitivity and linearity,
these early CCDs led to many astronomical advances.   Eventually
larger scientific CCDs were developed, leading to focal 
plane mosaics of these CCDs in the early 1990s.  The Big Throughput
Camera \citep[BTC]{BTC98} on the 4-meter Blanco telescope enabled the
surveys that discovered high-redshift supernovae and suggested the
existence of dark energy.  A mosaic of these same CCDs led to the
Sloan Digital Sky Survey \citep[SDSS]{York++00}, which has imaged over
10,000 deg$^2$ of sky in five broad bands. SDSS has been 
hugely successful because the high \'etendue of the
telescope/camera combination enabled a wide survey with
well-calibrated digital data. 

Wide surveys are very productive; the SDSS, for example, was cited as
the most productive telescope in recent years
\citep{Madrid+Macchetto09}. The discovery space could be made even
larger if the survey could be made deep and with good time resolution
(fast). LSST had its origin in the realization in the late 1990s --
extrapolating from the BTC on the 4-meter telescope -- that a
wide-fast-deep optical sky survey would be possible if the size and
field of view of the camera+telescope were scaled up. The challenge
was to to design a very wide field telescope with state-of-the-art
image quality. Originally, 4-meter designs with several square degrees 
field of view were studied. However, it was soon
realized that larger \'etendue and better image performance than
realizable in two-mirror+corrector designs would be required to address a
broad range of science opportunities simultaneously with the same
data. Indeed the three-mirror modified Paul-Baker design suggested by
Roger Angel in 1998 for the ``Dark Matter Telescope'' (DMT) had its origin
in two very different wide-fast-deep survey needs: mapping dark matter
via weak gravitational lensing and detecting faint Solar System
bodies \citep{Ang++00, Tys++01}.

Plans for the ``6-meter class'' DMT wide field telescope and camera
were presented at a workshop on gravity at SLAC National Accelerator
Laboratory in August 1998 \citep{Tys98}. The science case for such a
telescope was submitted to the 2000 Astronomy and Astrophysics Decadal
Survey in June 1999. That National Research Council (NRC) report
recommended it highly as a facility to discover near-Earth asteroids
as well as to study dark matter, and renamed it the Large Synoptic
Survey Telescope (LSST). In order to explore the science opportunities
and related instrument requirements, a Science Drivers Workshop was
held at the National Optical Astronomy Observatory (NOAO) in November
2000. A summer workshop on wide field astronomy was held at the Aspen
Center for Physics in July 2001, arguably the beginning of wide
involvement by the scientific community in this project. Many
alternative system designs were studied, but the need for short, deep,
and well sampled wide-field exposures led naturally to a single large
telescope and camera. At the behest of the National Science Foundation
(NSF) astronomy division, NOAO set up a national committee in
September 2002, with Michael Strauss as chair, to develop the LSST
design reference mission \citep{Str++04}. Plans for a Gigapixel focal
plane \citep{Sta++02}, as well as initial designs for the
telescope-camera-data system \citep{Tys02}, were presented in 2002. In
2002 the NSF funded development of the new imagers required for LSST,
supplementing an investment already made by Bell Labs. Lynn Seppala modified
Roger Angel's original three-mirror optical design \citep{Ang++00}, creating a wider, very low distortion
field. Also in 2002 the LSST Corporation was formed to manage the
project.  A construction proposal was submitted to the NSF in early
2007 and favorably reviewed later that year.  In 2008 the LSST 8.4-m
primary-tertiary mirror (\autoref{sec:design:optical}) was cast, and
in early 2009 the secondary mirror blank was cast as well.


\section{Overview of LSST Science}

Guided by community-wide input, the LSST is designed to achieve
multiple goals in four main science themes: Taking an Inventory of the
Solar System, Mapping the Milky Way, Exploring the Transient Optical
Sky, and Probing Dark Energy and Dark Matter. These are just four of
the many areas on which LSST will have enormous impact, but they span
the space of technical challenges in the design of the system and the
survey and have been used to focus the science requirements.  
  
The LSST survey data will be public with no proprietary period in the
United States, with a goal to make it world-public. As
was the case with SDSS, we expect the scientific community will
produce a rich harvest of discoveries.  Through the science
collaborations, the astronomical and physics communities are already
involved in the scientific planning for this telescope. 

Each patch of sky will be visited 1000 times (where a {\em visit} consists
of two 15-second exposures back to back in a given filter) in ten
years, producing a trillion line database with temporal astrometric
and photometric data on 20 billion objects. The 30 terabytes
of pipeline processed data (32 bit) obtained each night will open the
time domain window on the deep optical universe for variability and
motion. Rarely observed events will become commonplace, new and
unanticipated phenomena will be discovered, and the combination of
LSST with contemporary space-based near-infrared (NIR) missions will provide powerful
synergies in studies of dark energy, galaxy evolution, and many other
areas. The deep coverage of ten billion galaxies provides unique 
capabilities for cosmology. Astrometry, six-band photometry, and time
domain data on 10 billion stars will enable studies of Galactic
structure.  All LSST data and source code will be non-proprietary,
with public accessibility and usability a high priority.  A goal is to
have worldwide participation in all data products. 

This book describes in detail many of the scientific opportunities
that LSST will enable.  Here we outline some of the themes developed
in the chapters that follow: 

\begin{itemize}

\item {\bf A Comprehensive Survey of the Solar System (\autoref{chp:ss}):}

The small bodies of the Solar System offer a unique insight into its
early stages. Their orbital elements, sizes, and color distributions
encode the history of accretion, collisional grinding, and
perturbations by existing and vanished giant planets. Farther out,
runaway growth never occurred, and the Kuiper belt region still
contains a portion of the early planet population. Understanding these
distributions is a key element in testing various theories for the
formation and evolution of our planetary system. LSST, with its
unprecedented power for discovering moving objects, will make major
advances in Solar System studies. The baseline LSST cadence will
result in orbital parameters for several million moving objects; these
will be dominated by main belt asteroids (MBAs), with light curves and
colorimetry for a substantial fraction of detected objects. This
represents an increase of factors of ten to one hundred over the numbers of
objects with documented orbits, colors, and variability
information.


Our current understanding of objects beyond Neptune (trans-Neptunian
Objects, or TNOs) is limited by small sample sizes. Fewer than half of
the $\sim 1000$ TNOs discovered to date are drawn from surveys whose
discovery biases can be quantified, and only several hundred TNOs have
measured colors.  The LSST will survey over half the celestial sphere
for asteroids, get superb
orbits, go tremendously faint, and measure precise colors,
allowing measurement of light curves for thousands of TNOs, producing
rotation periods and phase curves, yielding shape and spin properties,
and providing clues to the early environment in the outer Solar
System.  Moreover, these objects fall into a wide variety of dynamical classes, which encode clues to the formation of the Solar System. 

  Many asteroids travel in Earth-crossing orbits, and Congress has
  mandated that National Aeronautics and Space Administration (NASA) catalog 90\% of all potentially hazardous
  asteroids larger than 140 meters in diameter.  The LSST is the 
  only ground-based survey that is capable of achieving this goal
  \citep{Ivezic++08}.

\item {\bf Structure and Stellar Content of the Milky Way (Chapters
  \ref{chp:stellarpops} and \ref{chp:mw}):}

Encoded in the structure, chemical composition and kinematics of stars
in our Milky Way is a history of its formation. Surveys such as 2MASS
and SDSS have demonstrated that the halo has grown by accretion and
cannibalization of companion galaxies, and it is clear that the next
steps require deep wide-field photometry, parallax, proper motions, and spectra
to put together the story of how our Galaxy formed. LSST will enable
studies of the distribution of numerous main sequence stars beyond the
presumed edge of the Galaxy's halo, their metallicity distribution
throughout most of the halo, and their kinematics beyond the thick
disk/halo boundary, and will obtain direct distance measurements below
the hydrogen-burning limit for a representative thin-disk sample. 
LSST is ideally suited to answering two basic questions about the
Milky Way Galaxy: What is the structure and accretion history of the
Milky Way?  What are the fundamental properties of all the stars
within 300 pc of the Sun?
     
LSST will produce a massive and exquisitely accurate photometric and
astrometric data set. Compared to SDSS, the best currently available
optical survey, LSST will cover an area more than twice as large,
using hundreds of observations of the same region in a given filter
instead of one or two, and each observation will be about two magnitudes
deeper. LSST will detect of the order 10$^{10}$ stars, with sufficient
signal-to-noise ratio to enable accurate light curves, geometric
parallax, and proper motion measurements for about a billion stars.
Accurate multi-color photometry can be used for source classification
(1\% colors are good enough to separate main sequence and giant stars,
\citealt{Hel++03}), 
and measurement of detailed stellar properties such as effective
temperatures to an rms accuracy of 100 K and metallicity to 0.3 dex
rms. 

To study the metallicity distribution of stars in the Sgr tidal stream
\citep{Maj++03} and other halo substructures at distances beyond the
presumed boundary between the inner and outer halo ($\sim30$ kpc,
\citealt{Car++07}), the coadded depth in the $u$ band must reach
$\sim24.5$. To detect RR Lyrae stars beyond the Galaxy's tidal radius
at $\sim300$ kpc, the single-visit depth must be $r \sim24.5$. In
order to measure the tangential velocity of stars to an accuracy of 10
\kms\ at a distance of 10 kpc, where the halo dominates over the disk,
proper motions must be measured to an accuracy of at least 0.2 mas
yr$^{-1}$. The same accuracy follows from the requirement to obtain
the same proper motion accuracy as Gaia \citep{Per++01} at its faint
limit ($r \sim20$). In order to produce a complete sample of solar
neighborhood stars out to a distance of 300 pc (the thin disk scale
height), with 3$\sigma$ or better geometric distances, trigonometric
parallax measurements accurate to 1 mas are required. To achieve the
required proper motion and parallax accuracy with an assumed
astrometric accuracy of 10 mas per observation per coordinate,
approximately 1,000 observations are required.  This requirement on
the number of observations is close to the independent constraint
implied by the difference between the total depth and the single visit
depth.

\item {\bf The Variable Universe (\autoref{chp:transients}):}

Characterization of the variable optical sky is one of the true
observational frontiers in astrophysics.  No optical telescope to date
has had the capability to search for transient phenomena at faint
levels over enough of the sky to fully characterize the
phenomena. Variable and transient phenomena have historically led to
fundamental insights into subjects ranging from the structure of stars
to the most energetic explosions in the Universe to
cosmology. Existing surveys leave large amounts of discovery parameter
space (in waveband, depth, and cadence) as yet unexplored, and LSST is
designed to start filling these gaps.

LSST will survey the sky on time scales from years down
to 15 seconds. Because LSST extends time-volume space a thousand times
over current surveys, the most interesting science may well be the
discovery of new phenomena.  With its repeated, wide-area coverage to
deep limiting magnitudes, LSST will enable the discovery and analysis
of rare and exotic objects, such as neutron star and black hole
binaries and high-energy transients, such as optical counterparts to
gamma-ray bursts and X-ray flashes (at least some of which apparently
mark the deaths of massive stars). LSST will also characterize in
detail active galactic nuclei (AGN) variability and new classes of transients, such as binary
mergers and stellar disruptions by black holes.  Perhaps even more
interesting are explosive events of types yet to be discovered, such
as predicted mergers among neutron stars and black holes. These may
have little or no high-energy emission, and hence may be discoverable
only at longer wavelengths or in coincidence with gravitational wave
events.

LSST will also provide a powerful new capability for monitoring
periodic variables such as RR Lyrae stars, 
which will be used to map the Galactic halo and
intergalactic space to distances exceeding 400 kpc.  The search for
transients in the nearby Universe (within 200 Mpc) is interesting and
urgent for two reasons. First, there exists a large gap in the
luminosity of the brightest novae ($-10$ mag) and that of sub-luminous
supernovae ($-16$ mag). However, theory and reasonable speculation point
to several potential classes of objects in this ``gap''. Such objects
are best found in the Local Universe. Next, the nascent field of
gravitational wave astronomy and the budding fields of ultra-high
energy cosmic rays, TeV photons, and astrophysical neutrinos are likewise
limited to the Local Universe due to physical effects (GZK effect,
photon pair production) or instrumental sensitivity (neutrinos and
gravitational waves). Unfortunately, the localization of these new
telescopes is poor, precluding identification of the host galaxy
(with corresponding loss of distance information and physical diagnostics). Both
goals can be met with a fast wide field optical imaging survey in concert with
follow-up telescopes.

\item {\bf The Evolution of Galaxies (Chapters \ref{chp:galaxies} and 
  \ref{chp:agn}):}

Surveys carried out with the current generation of ten-meter-class
telescopes in synergy with deep X-ray (Chandra X-ray Observatory,  X-ray Multi-mirror Mission) and infrared
(Spitzer Space Telescope) imaging have resulted in the outline of a picture of how
galaxies evolve from redshift 7 to the present. We now have a rough
estimate, for example, of the star formation history of the Universe,
and we are starting to develop a picture of how the growth of
supermassive black holes is coupled to, and influences, the growth of
galaxy bulges. But the development of galaxy morphologies and the
dependence on environment are poorly understood. In spite of the
success of the concordance cosmological model and the hierarchical
galaxy-formation paradigm, experts agree that our understanding of
galaxy formation and evolution is incomplete. We do not understand how
galaxies arrive at their present-day properties. We do not know if the
various discrepancies between theory and observations represent
fundamental flaws in our assumptions about dark matter, or problems in
our understanding of feedback on the interstellar medium due to star
formation or AGN activity. Because the process of galaxy
formation is inherently stochastic, large statistical samples are
important for making further progress.  

The key questions in galaxy evolution over cosmic time require a deep
wide-area survey to complement the more directed studies from HST,
James Webb Space Telescope (JWST), and Atacama Large Millimeter Array
(ALMA) and other narrow-field facilities. The essential correlation of
galaxy properties with dark matter --- both on small scales in the
local Universe and in gravitational lenses, and on the Gpc scales
required for large-scale structure --- requires a new generation
wide-area survey. LSST promises to yield insights into these problems.

It is likely that AGN spend most of their lives in low-luminosity
phases, outshone by their host galaxies, but recognizable by their
variability.  These will be revealed with great statistical accuracy by
LSST in synergy with other facilities.  The systematic evolution of
AGN optical variability is virtually unexplored in large samples and
would provide a new window into accretion physics.

\item {\bf Cosmological Models, and the Nature of Dark Energy and Dark
  Matter (Chapters \ref{chp:sne}-\ref{chapter-cosmology}):} 
 
Surveys of the Cosmic Microwave Background (CMB), the large-scale
distribution of galaxies, the redshift-distance relation for
supernovae, and other probes, have led us to the fascinating situation
of having a precise cosmological model for the geometry and expansion
history of the Universe, whose principal components we simply do not
understand. A major challenge for the next decade will be to gain a
physical understanding of dark energy and dark matter. Doing this will
require wide-field surveys of gravitational lensing, of the
large-scale distribution of galaxies, and of supernovae, as well as
next-generation surveys of the CMB (including polarization).

Using the CMB as normalization, the combination of these LSST deep
probes over wide area will yield the needed precision to distinguish
between models of dark energy, with cross checks to control systematic
error. LSST is unique in that its deep, wide-field, multi-color
imaging survey can undertake four cosmic probes of dark matter and
dark energy physics with a single data set and with much greater
precision than previously: 1) Weak lensing cosmic shear of galaxies
as a function of redshift; 2) Baryon acoustic oscillations (BAO) in
the power spectrum of the galaxy distribution; 3)
Evolution of the mass function of clusters of galaxies, as measured
via peaks in the weak lensing shear field; and 4) measurements of
redshifts and distances of type Ia
supernovae.  The synergy between these probes breaks degeneracies and
allows cosmological models to be consistently tested.  By
simultaneously measuring the redshift-distance 
relation and the growth of cosmic structure,
LSST data can test whether the recent acceleration is due to dark
energy or modified gravity. Because of its wide area coverage, LSST
will be uniquely capable of constraining more general models of dark
energy. LSST's redshift coverage will bracket the epoch at which dark
energy began to dominate the cosmic expansion. Much of the power of
the LSST will come from the fact that all the different measurements
will be obtained from the same basic set of observations, using a
facility that is optimized for this purpose. The wide-deep
LSST survey will allow a unique probe of the isotropy and
homogeneity of dark energy by mapping it over the sky, using weak
lensing, supernovae and BAO, especially when normalized by Planck observations.

Gravitational lensing provides the cleanest and farthest-reaching
probe of dark matter in the Universe, which can be combined with other
observations to answer the most challenging and exciting questions
that will drive the subject in the next decade: What is the
distribution of mass on sub-galactic scales? How do galaxy disks form
and bulges grow in dark matter halos? How accurate are CDM predictions
of halo structure? Can we distinguish between a need for a new
substance (dark matter) and a need for new gravitational physics? What
is the dark matter made of anyway?  LSST's wide-field, multi-filter,
multi-epoch optical imaging survey will probe the physics of dark
matter halos, based on the (stackable) weak lensing signals from all
halos, the strong lensing time domain effects due to some, and the
distribution of 3 billion galaxies with photometric redshifts.  LSST will
provide a comprehensive map of dark matter over a cosmological volume.
\end{itemize}


\section{The LSST Science Requirements}
\label{sec:introduction:reqs}

The superior survey capability enabled by LSST will open new windows
on the Universe and new avenues of research.  It is these scientific
opportunities that have driven the survey and system design. These
``Science Requirements'' are made in the context of what we forecast
for the scientific landscape in 2015, about the time the LSST survey
is planned to get underway.  
Indeed, LSST
represents such a large leap in throughput and survey capability that
in these key areas the LSST remains uniquely capable of addressing
these fundamental questions about our Universe. The long-lived
data archives of the LSST will have the astrometric and photometric
precision needed to support entirely new research directions which
will inevitably develop during the next several decades.

We have developed a detailed LSST Science Requirements
Document\footnote{\url{http://www.lsst.org/Science/docs.shtml}},
allowing the goals of all the 
science programs discussed above (and many more, of course) to be
accomplished.  The requirements are summarized as follows: 

\begin{enumerate}
\item  {\it The single visit depth} should reach $r\sim24.5$
  ($5\,\sigma$, point source). This limit is
   primarily driven by need to image faint, fast-moving potentially
   hazardous asteroids, as well as variable and transient sources
   (e.g., supernovae, RR Lyrae stars, gamma-ray burst afterglows), and
   by proper motion and trigonometric parallax measurements for
   stars. Indirectly, it is also driven by the requirements on the
   coadded survey depth and the minimum number of exposures required by
   weak lensing science (\autoref{chp:wl}) to average over systematics
   in the point-spread function.  
\item  {\it Image quality} should maintain the limit set by the 
     atmosphere (the median free-air seeing is 0.7 arcsec in the $r$ band 
     at the chosen site, see \autoref{fig:design:seeing}),
     and not be degraded appreciably by the hardware. In addition to stringent 
     constraints from weak lensing, the requirement for good image quality is driven by the 
     required survey depth for point sources and by image differencing
     techniques. 
\item  {\it Photometric repeatability} should achieve 5 millimag precision
     at the bright end, with zeropoint stability across the sky of 10 millimag
     and band-to-band calibration errors not larger than 5 millimag.
     These requirements are driven by the need for photometric redshift accuracy,
     the separation of stellar populations, detection of low-amplitude variable
     objects (such as eclipsing planetary systems), and the search for
     systematic effects in Type Ia supernova light curves.
\item  {\it Astrometric precision} should maintain the limit set by 
     the atmosphere of about 10 mas rms per coordinate per visit at
     the bright end 
     on scales below 20 arcmin. This precision is driven by the desire to 
     achieve a proper motion uncertainty of 0.2 mas yr$^{-1}$ and parallax
     uncertainty of  
     1.0 mas over the course of a 10-year survey (see \autoref{sec:apSize}).
\item  {\it The single visit exposure time} (including both exposures in a 
    visit) should be less than about a minute 
    to prevent trailing of fast moving objects and to aid control 
    of various systematic effects induced by the atmosphere. It should
    be longer than $\sim$20 seconds to avoid significant efficiency
    losses due to finite readout, slew time, and read noise
    (\autoref{sec:intro:exposure}).  
\item  {\it The filter complement} should include six filters
    in the wavelength range limited by atmospheric absorption and 
    silicon detection efficiency (320--1050 nm), with roughly
    rectangular filters and no large gaps in the coverage, in order
    to enable robust and accurate photometric redshifts and stellar typing. An 
    SDSS-like $u$ band is extremely important for separating 
    low-redshift quasars from hot stars and for estimating the metallicities of 
    F/G main sequence stars. A bandpass with an effective wavelength of
    about 1 micron  will enable studies of sub-stellar objects, high-redshift 
    quasars (to redshifts of $\sim$7.5), and regions of the Galaxy that are obscured 
    by interstellar dust.
\item  {\it The revisit time distribution} should enable determination of
   orbits of Solar System objects and sample SN light curves every few days, 
   while accommodating constraints set by proper motion and trigonometric 
   parallax measurements.
\item  {\it The total number of visits} of any given area of sky, when
  summed over all 
   filters, should be of the order of 1,000, as mandated by weak
   lensing science, the asteroid survey, and proper motion and 
   trigonometric parallax measurements. Studies of variable and
   transient sources of all sorts also benefit from a large number of
   visits. 
\item  {\it The coadded survey depth} should reach 
    $r\sim27.5$ ($5\,\sigma$, point source), with sufficient
  signal-to-noise ratio in other bands 
    to address both extragalactic and Galactic science drivers.
\item  {\it The distribution of visits per filter} should enable 
   accurate photometric redshifts, separation of stellar populations,
   and sufficient depth to enable detection of faint extremely red
   sources (e.g., brown dwarfs and high-redshift quasars). Detailed
   simulations of photometric redshift estimators (see
   \autoref{sec:common:photo-z}) 
   suggest an approximately flat distribution of visits among bandpasses
   (because the system throughput and atmospheric properties are 
    wavelength-dependent, the achieved depths are different in different
    bands). The adopted time allocation
   (see \autoref{tab:intro:syspar}) gives a slight preference to the
    $r$ and $i$ bands because of their 
   dominant role in star/galaxy separation and weak lensing measurements. 
\item  {\it The distribution of visits on the sky} should extend over
   at least $\sim 20,000$ deg$^2$ to obtain the required number of galaxies
   for weak lensing studies, to study the distribution of galaxies on
   the largest scales and to probe the structure of the Milky Way and
   the Solar System, with attention paid to include ``special''
   regions such as the ecliptic, the Galactic plane, and the Large and Small
   Magellanic Clouds. 
\item  {\it Data processing, data products, and data access} should enable 
   efficient science analysis. 
To enable a fast and efficient response to
   transient sources, the processing latency for objects that change
   should be less than a minute after the close of the shutter,
together with a robust and accurate preliminary classification 
   of reported transients.
\end{enumerate}

Remarkably, even with these joint requirements, none of the 
individual science programs is severely over-designed. That is, despite 
their significant scientific diversity, these programs are highly 
compatible in terms of desired data characteristics. Indeed, any one
of the four main science drivers: the Solar System inventory, mapping
the Milky Way, transients, and dark energy/dark matter,  could be removed, and the remaining 
three would still yield very similar requirements for most system 
parameters. As a result, the LSST system can adopt a highly 
efficient survey strategy where {\it a single data set serves most science
programs} (instead of science-specific surveys executed in series). 
One can think of this as {\it massively parallel astrophysics}.

About 90\% of the observing time will be devoted to a uniform
deep-wide-fast (main) survey mode.  All scientific investigations will
utilize a common database constructed from an optimized observing
program. The system is designed to yield high image quality as well as
superb astrometric and photometric accuracy. The survey area will
cover 30,000 deg$^2$ with $\delta < +34.5$ deg, and will be imaged
many times in six bands, $ugrizy$, spanning the wavelength range
320--1050 nm. Of this 30,000 deg$^2$, 20,000 deg$^2$ will be
covered with a deep-wide-fast survey mode, with each area of sky
covered with 1000 visits (summed over all six bands)
during the anticipated 10 years of operations.  This will result in 
measurements of 10 billion stars to a depth of 27.7 mag and photometry
for a roughly equal number of galaxies.  The remaining 10\% of the observing
time will be allocated to special programs such as a Very Deep + Fast
time domain survey, in which a given field is observed for an hour
every night.  
  
The uniform data quality, wavelength coverage, deep 0.7 arcsec imaging
over tens of thousands of square degrees together with the time-domain
coverage will be unmatched. LSST data will be used by a very large
fraction of the astronomical community -- this is a survey for
everyone. 

\begin{table}
\small
\begin{tabular}{p{2.5in}|p{3.5in}}
\hline
\hline
\multicolumn{2}{l}{\textbf{Main System and Survey Characteristics}}\\
\hline
\'Etendue                              & 319 m$^2$ deg$^2$\\
Area and diameter of field of view    & 9.6 deg$^2$\ \ (3.5 deg)\\
Effective clear aperture (on-axis)     & 6.7 m (accounting for obscuration) \\
Wavelength coverage (full response) & 320-1080 nm \\
Filter set                          & $u, g, r, i, z, y$ (five
concurrent in camera at a time) \\
Sky coverage & 20,000 deg$^2$ (Main Survey) \\
\hline
\multicolumn{2}{l}{\textbf{}}\\
\multicolumn{2}{l}{\textbf{Telescope and Site}}\\
\hline
Configuration                           & three-mirror, Alt-azimuth \\
Final f/ratio; plate scale              & f/1.23\ \ 50 microns/arcsec \\
Physical diameter of optics            & M1: 8.4m\ \ M2: 3.4m\ \   M3: 5.02 m \\
First camera lens; focal plane diameter & Lens: 1.55 m\ \ field of view: 63 cm \\
Diameter of 80\% encircled energy & $u$: $0.26''$\ \ $g$: $0.26''$\ \ $r$: $0.18''$\ \  $i$: $0.18''$\ \ $z$: $0.19''$\ \ $y$:  $0.20''$ \\
\ \ \ spot due to optics          & \\
\hline
\multicolumn{2}{l}{\textbf{}}\\
\multicolumn{2}{l}{\textbf{Camera}}\\
\hline 
Pixel size; pixel count           & 10 microns (0.2 arcsec);  3.2 Gpixels\\
Readout time                      & 2 sec\\
Dynamic range                    & 16 bits \\
Focal plane device configuration  & 4-side buttable, $>90$\% fill factor \\
Filter change time                & 120 seconds \\
\hline
\multicolumn{2}{l}{\textbf{}}\\
\multicolumn{2}{l}{\textbf{Data Management}}\\
\hline 
Real-time alert latency             & 60 seconds\\
Raw pixel data/night                & 15 TB     \\
Yearly archive rate (compressed)    & Images; 5.6 PB; Catalogs: 0.6 PB \\
Computational requirements          & Telescope: $<$1 Tflop; Base facility: 30 Tflop; \\
                                    & Archive Center: 250 Tflop by year 10 \\
Bandwidth:                          & Telescope to base: 40 Gbits/sec \\
                                    & Base to archive: 2.5 Gbits/sec avg \\
\hline
\multicolumn{2}{l}{\textbf{}}\\
\multicolumn{2}{l}{\textbf{System Capability}}\\
\hline 
Single-visit depths (point sources; $5\sigma$)    & $u$: 23.9\ \ $g$: 25.0\ \ $r$: 24.7\ \ $i$: 24.0  $z$: 23.3\ \ $y$: 22.1\ \ AB mag\\
Baseline number of visits over 10 years     & $u$: 70\ \  $g$: 100\ \  $r$: 230\ \  $i$: 230\ \ $z$: 200\ \  $y$: 200\\
Coadded depths (point sources; $5\sigma$) & $u$: 26.3\ \ $g$: 27.5\ \ $r$: 27.7\ \ $i$: 27.0 $z$: 26.2\ \ $y$: 24.9\ \ AB mag     \\
Photometry accuracy (rms mag)            & repeatability: 0.005;   zeropoints:  0.01 \\
Astrometric accuracy at $r=24$ (rms)     & parallax: 3 mas;  proper
motion: 1 mas yr$^{-1}$ \\
\hline
\hline
\end{tabular}
\caption{LSST System Parameters}
\label{tab:intro:syspar}
\end{table}

\section{Defining the Telescope Design Parameters}

Given the science requirements listed in the previous section, we now 
discuss how they are translated into constraints on the
main system design parameters: the aperture size, the
survey lifetime, and the optimal exposure time.  The basic parameters
of the system are outlined in \autoref{tab:intro:syspar}.  

\subsection{The Aperture Size}
\label{sec:apSize}

The product of the system's \'etendue and the survey
lifetime, for given observing conditions, determines the
sky area that can be surveyed to a given depth.
The LSST field-of-view area is set to the practical limit possible
with modern optical designs, 10 deg$^2$, determined by the
requirement that the delivered image quality be dominated
by atmospheric seeing at the chosen site (Cerro
Pach\'on in Northern Chile; \autoref{sec:design:site}). A larger
field-of-view would 
lead to unacceptable deterioration of the image quality.
This leaves the primary mirror diameter and survey lifetime
as free parameters. Our adopted survey lifetime is 
ten years.  Shorter than this would imply an excessively large and
expensive mirror (15 meters
for a three-year survey and 12 meters for a five-year survey),
while a much smaller telescope would require much more
time to complete the survey with the associated increase
in operations cost and evolution of the science goals.  

\begin{figure*}[!t]
\centering
\begin{minipage}{0.5\textwidth}
\centerline{\includegraphics[width=0.9\textwidth,angle=0]{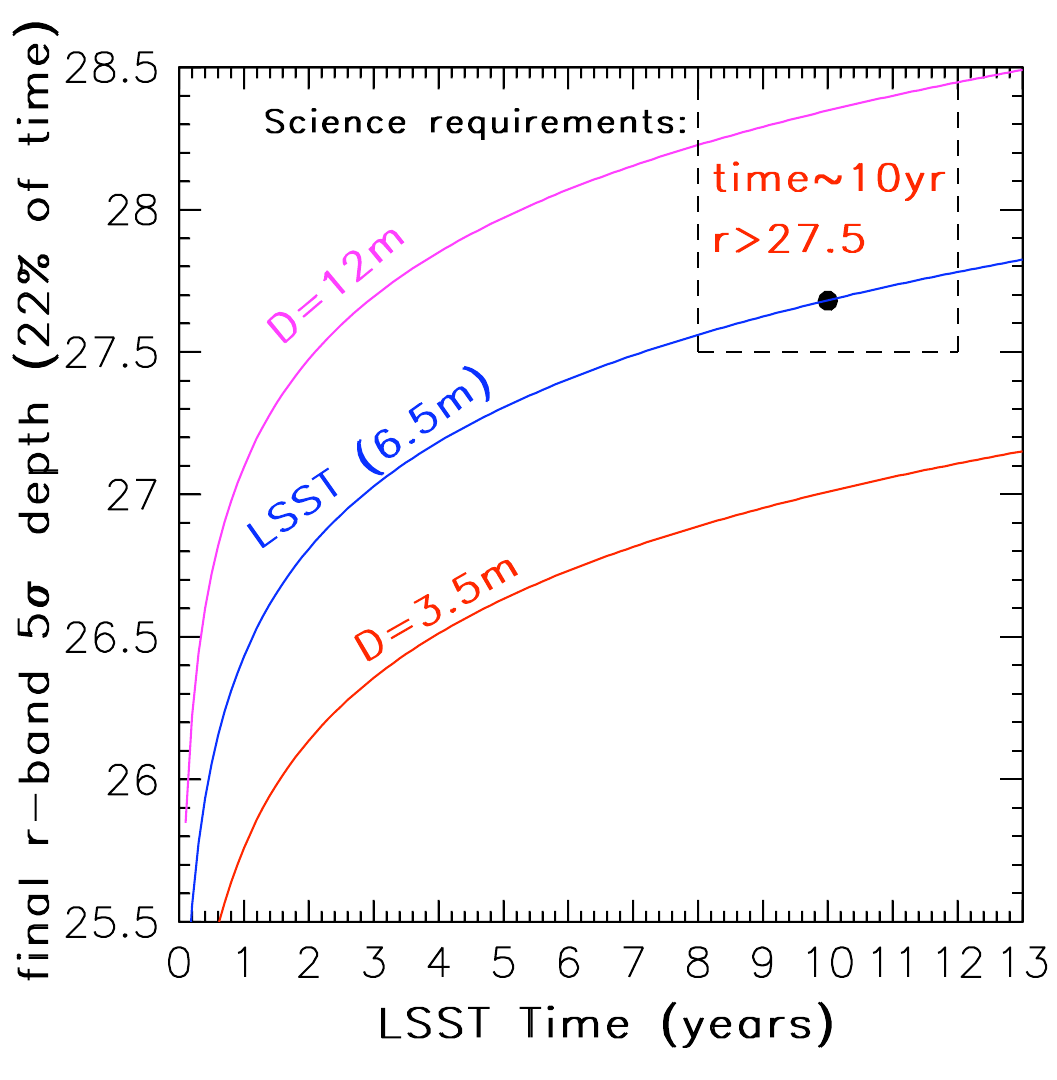}}
\end{minipage}\hfill
\begin{minipage}{0.5\textwidth}
\centerline{\includegraphics[width=0.9\textwidth]{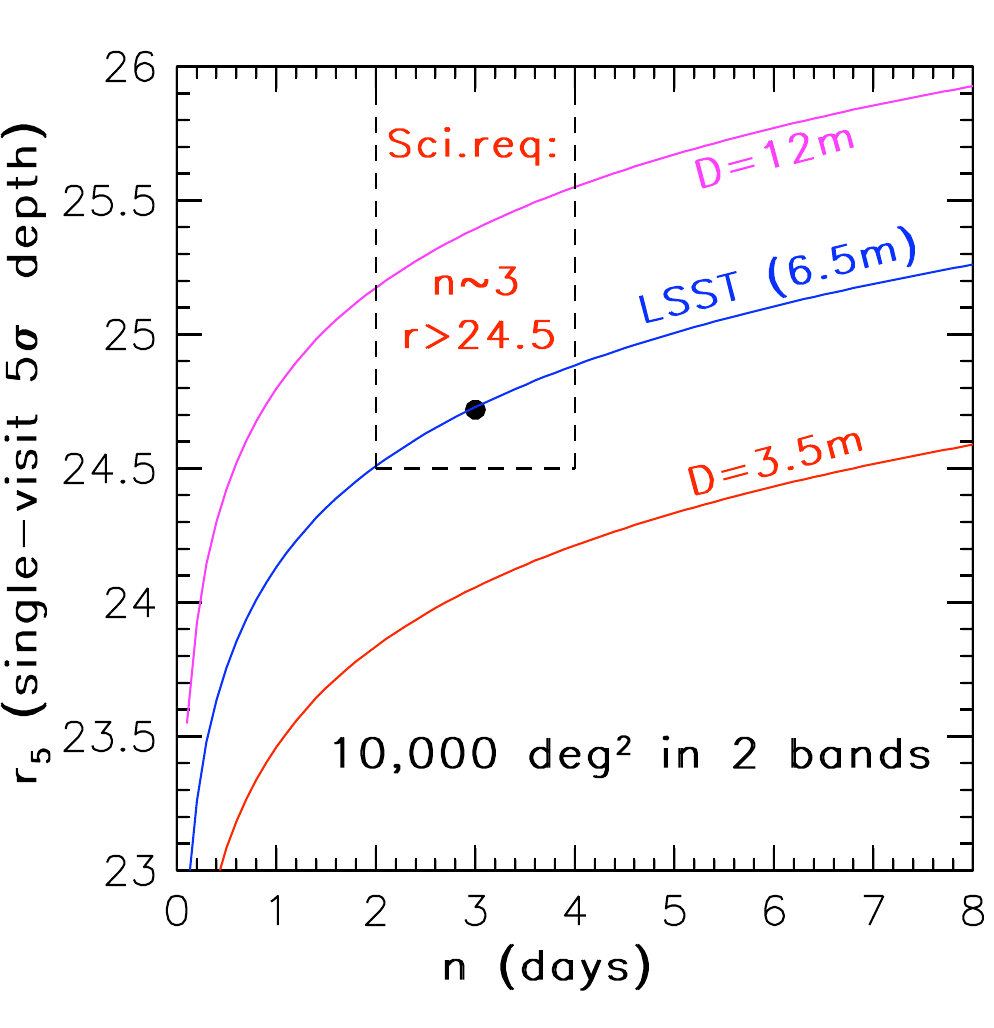}}
\end{minipage}
\caption{
(a) The coadded depth in the $r$ band (AB magnitudes)
vs. the effective aperture and the survey lifetime. It is assumed
that 22\% of the total observing time (corrected for weather and
other losses) is allocated for the $r$ band, and that the ratio of the
surveyed sky area to the field-of-view area is 2,000.
(b) The single-visit depth in the $r$ band ($5 \sigma$ detection
for point sources, AB magnitudes) vs. revisit time, $n$ (days), as a
function of the effective aperture size.  With a coverage of 10,000
deg$^2$ in two bands, the revisit time directly constrains the visit exposure
time, $t_{vis} = 10 n$ seconds; these numbers can be directly scaled
to the 20,000 deg$^2$ and six filters of LSST. In addition to direct
constraints 
on optimal exposure time, $t_{vis}$ is also driven by requirements on
the revisit time, $n$, the total number of visits per sky position over
the survey lifetime, $N_{visit}$, and the survey efficiency, $\epsilon$ (see \autoref{eq:introduction:epsilon}). 
Note that these constraints result in a fairly narrow range of
allowed $t_{vis}$ for the main deep-wide-fast survey.  From
\citet{Ivezic++08}.
}
\label{fig:introduction:coaddedDepth}
\end{figure*}

The primary mirror size is a function of the required
survey depth and the desired sky coverage. Roughly speaking, 
the anticipated science outcome scales with the number
of detected sources. For practically all astronomical
source populations, in order to maximize the number of
detected sources, it is more advantageous to maximize
first the area and then the detection 
depth.\footnote{The number of sources is proportional to area, but
  rises no faster than Euclidean with survey depth, which increases by
  0.4 magnitude for a doubling of exposure time in the sky-dominated
  regime; see \citet{Nemiroff03} for more details.}
For this reason, the sky area for the main survey is maximized
to its practical limit, 20,000 deg$^2$, determined by the
requirement to avoid large airmasses ($X \equiv \sec ({\rm zenith\ 
  distance})$), which would
substantially deteriorate the image quality and the survey
depth.
With the adopted field-of-view area, the sky coverage
and the survey lifetime fixed, the primary mirror diameter
is fully driven by the required survey depth. There
are two depth requirements: the final (coadded) survey
depth, $r \sim 27.5$, and the depth of a single visit, $r \sim 24.5$.
The two requirements are compatible if the number of
visits is several hundred per band, which is in good
agreement with independent science-driven requirements
on the latter.
The required coadded survey depth provides a direct
constraint, independent of the details of survey execution
such as the exposure time per visit, on the minimum
{\it effective} primary mirror diameter of 6.5 m, as illustrated
in \autoref{fig:introduction:coaddedDepth}. This is the effective diameter of the LSST taking into account the actual throughput of its entire optical system.

\subsection{The Optimal Exposure Time}
\label{sec:intro:exposure}

The single visit depth depends on both the primary
mirror diameter and the chosen exposure time, $t_{\rm vis}$. In
turn, the exposure time determines the time interval to
revisit a given sky position and the total number of visits,
and each of these quantities has its own science drivers.
We summarize these simultaneous constraints in terms
of the single-visit exposure time:
\begin{itemize}
	\item The single-visit exposure time should not be longer
	than about a minute to prevent trailing of fast solar
	system moving objects, and to enable efficient
	control of atmospheric systematics.
	\item The mean revisit time (assuming uniform cadence)
	for a given position on the sky, $n$, scales as
	\begin{equation}
	n = \left( \frac {t_{\rm vis}}  {10 {\rm ~sec}} \right) \left
	( \frac  {A_{\rm sky}} {{\rm 10,000 ~deg}^2} \right)  \left(
	\frac {{\rm 10 ~deg}^2}  {A_{\rm FOV} } \right) {\rm  ~days,}
	\end{equation}
	where two visits per night are assumed (this is needed to get
	velocity vectors for main belt and near-Earth asteroids), and the
	losses for realistic observing conditions have been	
	taken into account (with the aid of the Operations
	Simulator described in \autoref{sec:design:opsim}). Science
	drivers such as supernovae and moving objects in the Solar
	System require 
	that $n < 4$ days, or equivalently $t_{\rm vis} < 40$ seconds
	for the nominal values of $A_{\rm sky}$ and $A_{\rm FOV}$.
	\item The number of visits to a given position on the sky,
    	$N_{\rm visit}$, with losses for realistic observing conditions
    	taken into account, is given by
    	\begin{equation}
	N_{\rm visit} = \left( \frac{3000}{n}\right) \left(
	\frac{T}{10 {\rm  ~yr}}\right), 
	\end{equation}
	where $n$ is the mean time, in days, between visits to a given
	position.  The requirement $N_{\rm visit} > 800$ again implies that
	$n < 4$ and $t_{\rm vis} < 40$ seconds if the survey lifetime,
	$T$ is about 10 years.

	\item These three requirements place a firm upper limit
	on the optimal visit exposure time of $t_{\rm vis} < 40$
	seconds. Surveying efficiency (the ratio of open shutter
	time to the total time spent per visit) considerations
	place a lower limit on $t_{\rm vis}$ due to 
	finite read-out and slew time.  The read-out time of the
	camera is in fact two seconds for each exposure (\autoref{sec:design:camera}), and the slew and
	settle time is set to five seconds, including the readout
	time for the second exposure in a visit:
	\begin{equation}
	\epsilon =  \frac{t_{\rm vis}}  {t_{\rm vis} + 9 {\rm  ~sec}} .
        \label{eq:introduction:epsilon}
	\end{equation}
	To maintain efficiency losses below 30\% (i.e., at
	least below the limit set by weather),
	and to minimize the read noise impact, $t_{\rm vis}$ should be
	less than 20 seconds.
\end{itemize}

Taking these constraints simultaneously into account,
as summarized in \autoref{fig:introduction:coaddedDepth}, 
yields the following reference design:
\begin{itemize}
 \item A primary mirror effective diameter of $\sim 6.5$ m.
 With the adopted optical design, described below,
 this effective diameter corresponds to a geometrical
 diameter of $\sim 8$ m. Motivated by the characteristics of
 the existing equipment at the Steward Mirror Laboratory,
 which has cast the primary mirror, the
 adopted geometrical diameter is set to 8.4 m.
 \item A visit exposure time of 30 seconds (using two 15-second exposures to efficiently reject cosmic rays), 
 yielding $\epsilon = 77$\%. 
 \item  A revisit time of three days on average per 10,000 deg$^2$
 of sky (i.e., the area visible at any given time of the year), with two visits per night (particularly useful for
 establishing proper motion vectors for fast moving asteroids). 
\end{itemize}

To summarize, the chosen primary mirror diameter is
the minimum diameter that simultaneously satisfies the
depth ($r \sim 24.5$ for single visit and $r \sim 27.5$ for coadded
depth) and cadence (revisit time of 3-4 days, with 30
seconds per visit) constraints described above.

\bibliographystyle{SciBook}
\bibliography{introduction/introduction} 


%
%
%
%
%
%
%
%
%
%
%
%
%
%
%
%
%
%
%
%
%
%
%
%

\chapter[LSST System Design]
{LSST System Design}
\label{chp:design}

\noindent{\it 
John R. Andrew,
J. Roger P. Angel, 
Tim S. Axelrod, 
Jeffrey D. Barr, 
James G. Bartlett, 
Jacek Becla, 
James H. Burge,
David L. Burke, 
Srinivasan Chandrasekharan,
David Cinabro, 
Charles F. Claver,
Kem H. Cook,
Francisco Delgado,
Gregory Dubois-Felsmann, 
Eduardo E. Figueroa,
James S. Frank, 
John Geary, 
Kirk Gilmore, 
William J. Gressler,
J. S. Haggerty, 
Edward Hileman,
\v Zeljko Ivezi\'c, 
R. Lynne Jones, 
Steven M. Kahn, 
Jeff Kantor, 
Victor L. Krabbendam,
Ming Liang,
R. H. Lupton, 
Brian T. Meadows, 
Michelle Miller,
David Mills,
David Monet, 
Douglas R. Neill,
Martin Nordby, 
Paul O'Connor,
John Oliver, 
Scot S. Olivier, 
Philip A. Pinto, 
Bogdan Popescu, 
Veljko Radeka, 
Andrew Rasmussen, 
Abhijit Saha, 
Terry Schalk, 
Rafe Schindler,
German Schumacher,
Jacques Sebag,
Lynn G. Seppala,
M. Sivertz, 
J. Allyn Smith, 
Christopher W. Stubbs, 
Donald W. Sweeney, 
Anthony Tyson, 
Richard Van Berg,
Michael Warner,
Oliver Wiecha,
David Wittman
}

This chapter covers the basic elements of the LSST system design, with particular
emphasis on those elements that may affect the scientific analyses discussed in 
subsequent chapters.  We start with
a description of the planned observing strategy in
\autoref{sec:design:cadence}, and then go on to describe the key technical aspects of 
system, including the choice of site
(\autoref{sec:design:site}), the telescope and optical design
(\autoref{sec:design:optical}), and the camera including the characteristics of its
sensors and filters
(\autoref{sec:design:camera}).  The key elements of the data management system are
described in \autoref{sec:design:dm}, followed by overviews of the procedures
that will be invoked to achieve the desired photometric
(\autoref{sec:design:calsim}) and astrometric 
(\autoref{sec:design:calsimAstro}) calibration.  

\section{The LSST Observing Strategy}
\label{sec:design:cadence}

{\it \v Zeljko Ivezi\'c, Philip A. Pinto, Abhijit Saha, Kem H. Cook}

The fundamental basis of the LSST concept is to scan
the sky deep, wide, and fast with a single observing strategy, giving
rise to a data set 
that simultaneously satisfies the majority of the science
goals. This concept, the so-called ``universal cadence,''
will yield the main deep-wide-fast survey (typical single
visit depth of $r \sim 24.5$) and use about 90\% of the observing
time. The remaining 10\% of the observing time will
be used to obtain improved coverage of parameter space
such as very deep ($r \sim 26$) observations, observations
with very short revisit times ($ \sim 1$ minute), and observations
of ``special'' regions such as the ecliptic, Galactic
plane, and the Large and Small Magellanic Clouds. We
are also considering a third type of survey, micro-surveys,
that would use about 1\% of the time, or about 25 nights over ten
years. 

The observing strategy for the main survey will be optimized
for the homogeneity of depth and number of visits over 20,000 deg$^2$ of sky, 
where a ``visit" is defined as a
pair of 15-second exposures, performed back-to-back in a given filter, and separated
by a four-second interval for readout and opening and closing of the shutter.  
In times of good seeing and at low airmass, preference is
given to $r$-band and $i$-band observations, as these are the bands in
which the most seeing-sensitive measurements are planned. As often as
possible, each field will be observed twice, with visits
separated by 15-60 minutes. This strategy will provide
motion vectors to link detections of moving objects, 
and fine-time sampling for measuring
short-period variability. The ranking criteria also ensure
that the visits to each field are widely distributed in position
angle on the sky and rotation angle of the camera
in order to minimize systematics that could affect some sensitive analyses, such
as studies of cosmic shear.

The universal cadence will also provide the primary data set for the 
detection of near-Earth Objects (NEO), given that it naturally
incorporates the southern half of the ecliptic. NEO survey completeness
for the smallest bodies ($\sim 140~$m in diameter per the
Congressional NEO mandate\footnote{H.R. 1022: The George E. Brown, Jr. Near-Earth
  Object Survey Act;\\ \url{http://www.govtrack.us/congress/bill.xpd?bill=h109-1022}}) 
is greatly enhanced, however,
by the addition of a crescent on the sky within 10$^\circ$
of the northern ecliptic. Thus, the
``northern Ecliptic proposal'' extends the universal cadence
to this region using the $r$ and $i$ filters only, along
with more relaxed limits on airmass and seeing. Relaxed
limits on airmass and seeing are also adopted for $\sim 700$
deg$^2$ around the South Celestial pole, allowing coverage
of the Large and Small Magellanic Clouds.

\begin{figure}
\begin{center}
\includegraphics[width=12cm,angle=0]{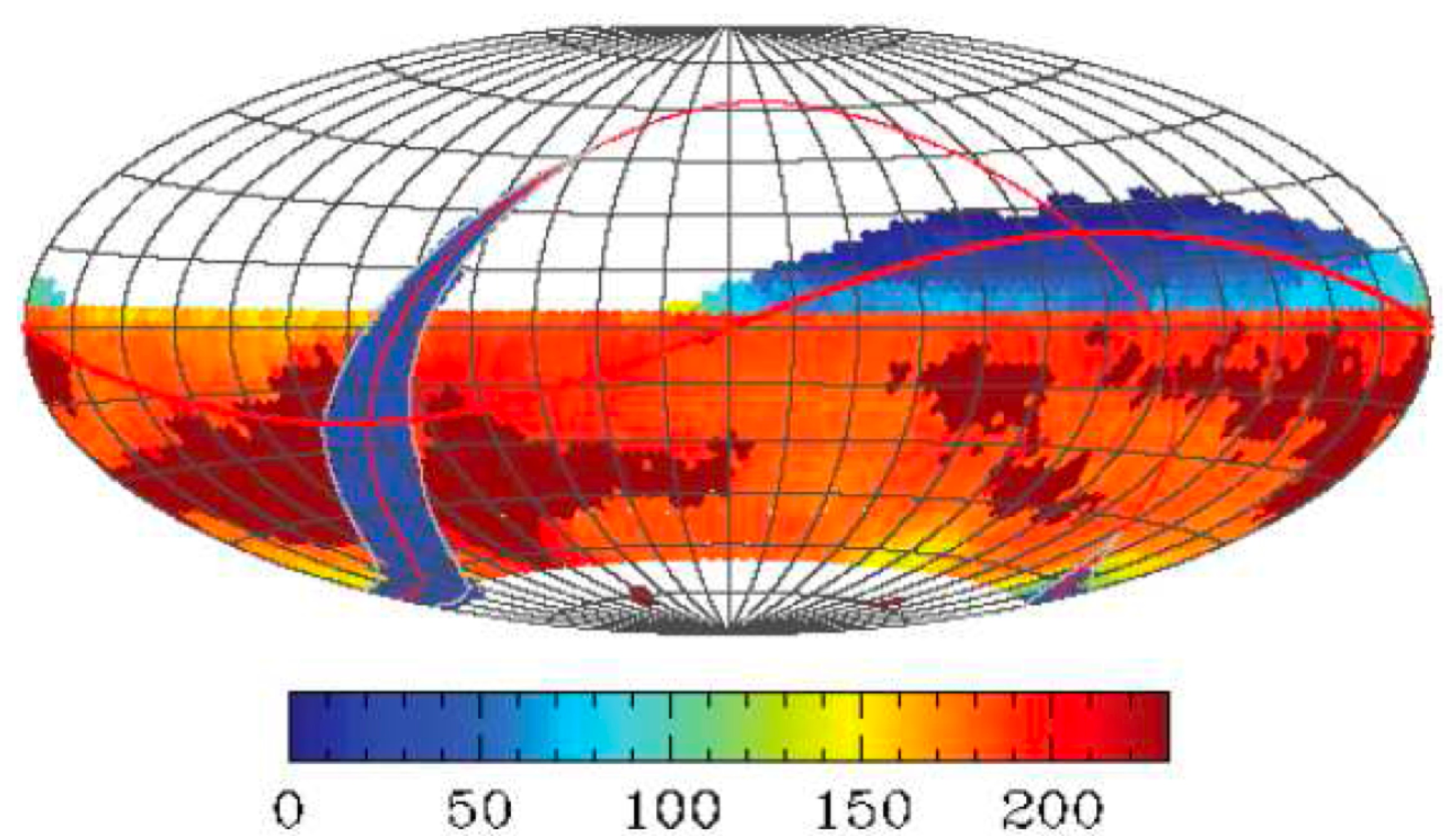}
\end{center}
\caption{
The distribution of the $r$ band visits on the sky
for one simulated realization of the baseline main survey. The
sky is shown in Aitoff projection in equatorial coordinates and the
number of visits for a 10-year survey is color-coded according to the
inset. The two regions with smaller number of visits than the main
survey (``mini-surveys'') are the Galactic plane (arc on the left)
and the so-called ``northern Ecliptic region'' (upper right). The
region around the South Celestial Pole will also receive substantial
coverage (not shown here).}
\label{fig:design:rband}
\end{figure}

Finally the universal cadence proposal excludes observations
in a region of 1,000 deg$^2$ around the
Galactic Center, where the high stellar density leads to a
confusion limit at much brighter magnitudes than those
attained in the rest of the survey. Within this region,
the Galactic Center proposal provides 30 observations in
each of the six filters, distributed roughly logarithmically
in time (it may not be necessary to use the bluest $u$ and
$g$ filters for this heavily extincted region).
The resulting sky coverage for the LSST baseline cadence,
based on detailed operations simulations described in
\autoref{sec:design:opsim}, is shown 
for the $r$ band in \autoref{fig:design:rband}. The anticipated total number
of visits for a ten-year LSST survey is about 2.8 million
($\sim5.6$ million 15-second long exposures). The per-band
allocation of these visits is shown in \autoref{tab:intro:syspar}.

Although the uniform treatment of the sky provided by
the universal cadence proposal can satisfy the majority
of LSST scientific goals, roughly 10\% of the time may be
allocated to other strategies that significantly enhance
the scientific return. These surveys aim to extend the
parameter space accessible to the main survey by going
deeper or by employing different time/filter sampling.

In particular, we plan to observe a set of ``deep drilling fields,''
whereby one hour of observing time per night is devoted to the
observation of a single field to substantially greater depth in
individual visits. Accounting for read-out time and filter changes,
about 50 consecutive 15-second exposures could be obtained in each of
four filters in an hour.  This would allow us to measure light curves
of objects on hour-long timescales, and detect faint supernovae and
asteroids that cannot be studied with deep stacks of data taken with a
more spread-out cadence.  The number, location, and cadence of these
deep drilling fields are the subject of active discussion amongst the
LSST Science Collaborations; see for example the plan suggested by the
Galaxies Science Collaboration at \autoref{sec:gal:deep_drilling}.
There are strong motivations, e.g., to study extremely
faint galaxies, to go roughly two magnitudes deeper in the final
stacked images of these fields than over the rest of the survey. 

These LSST deep fields will have widespread scientific value, both as
extensions on the main survey and as a constraint on systematics.
Having deeper data to treat as a model will reveal critical systematic
uncertainties in the wider LSST survey, including photometric
redshifts, that impact the measurements of weak lensing, clustering,
galaxy morphologies, and galaxy luminosity functions.
A vigorous and systematic research effort 
is underway to explore the enormously large parameter space of
possible survey cadences, using the Operations Simulator tool
described in \autoref{sec:design:opsim}. The commissioning period
will be used to test the usefulness of various observing modes and to
explore alternative strategies. Proposals from the community and the
Science Collaborations for specialized cadences (such as 
mini-surveys and micro-surveys) will also be considered.


\section{ Observatory Site}
\label{sec:design:site}

{\it Charles F. Claver,
Victor L. Krabbendam,
Jacques Sebag,
Jeffrey D. Barr,
Eduardo E. Figueroa,
Michael Warner
}

\begin{figure}
\begin{center}
\includegraphics[width=15cm,angle=0]{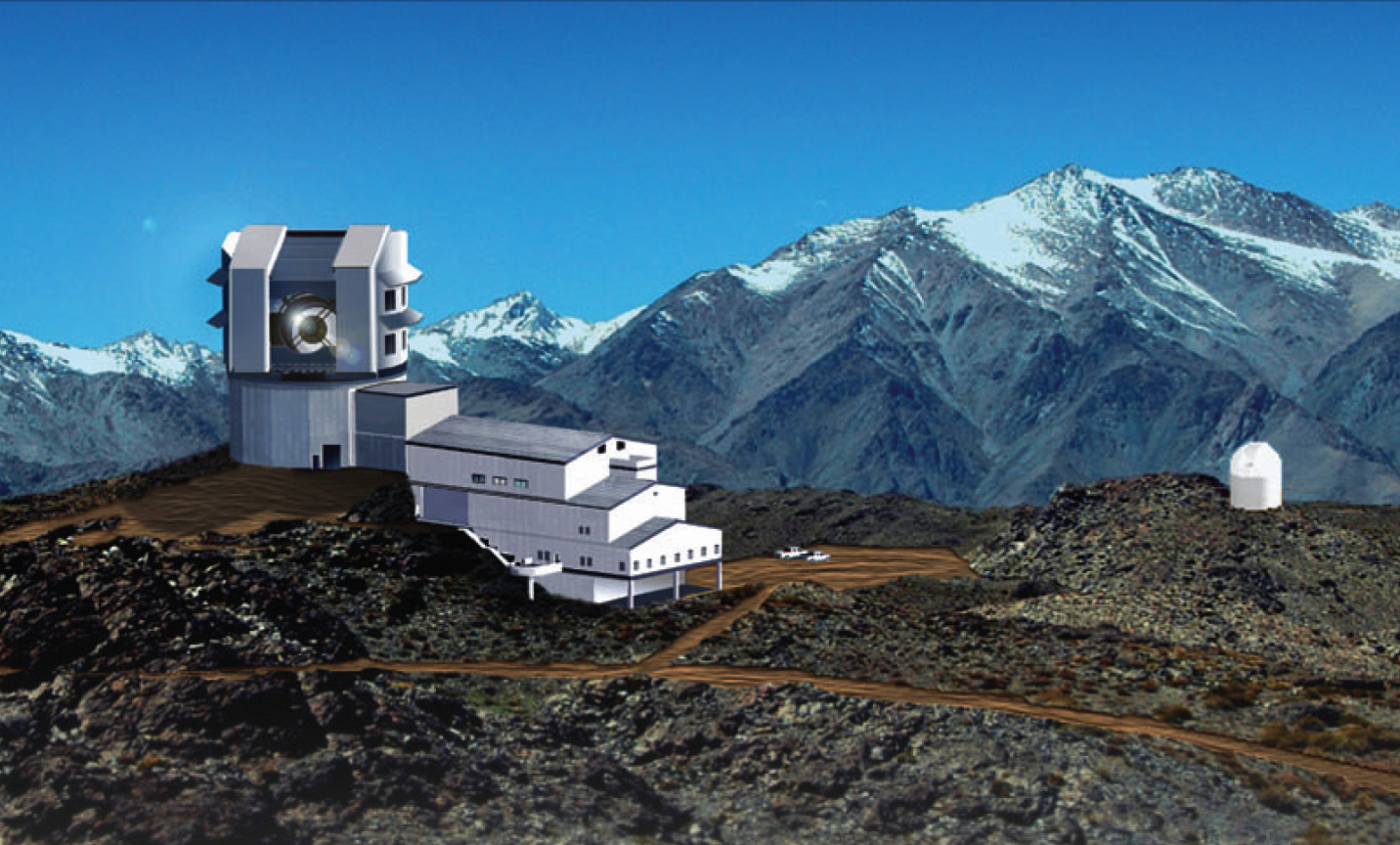}
\caption{
Artist's rendering of the LSST and dome enclosure on the summit of Cerro Pach\'on.  
The Auxiliary calibration telescope (\autoref{sec:design:calsim}) is
also illustrated on a neighboring peak. 
(Image Credit: Michael Mullen Design, LSST Corporation.)
}
\label{fig:design:telescope_artist}
\end{center}
\end{figure}

\begin{figure}
\begin{center}
\includegraphics[width=0.48\textwidth,angle=0]{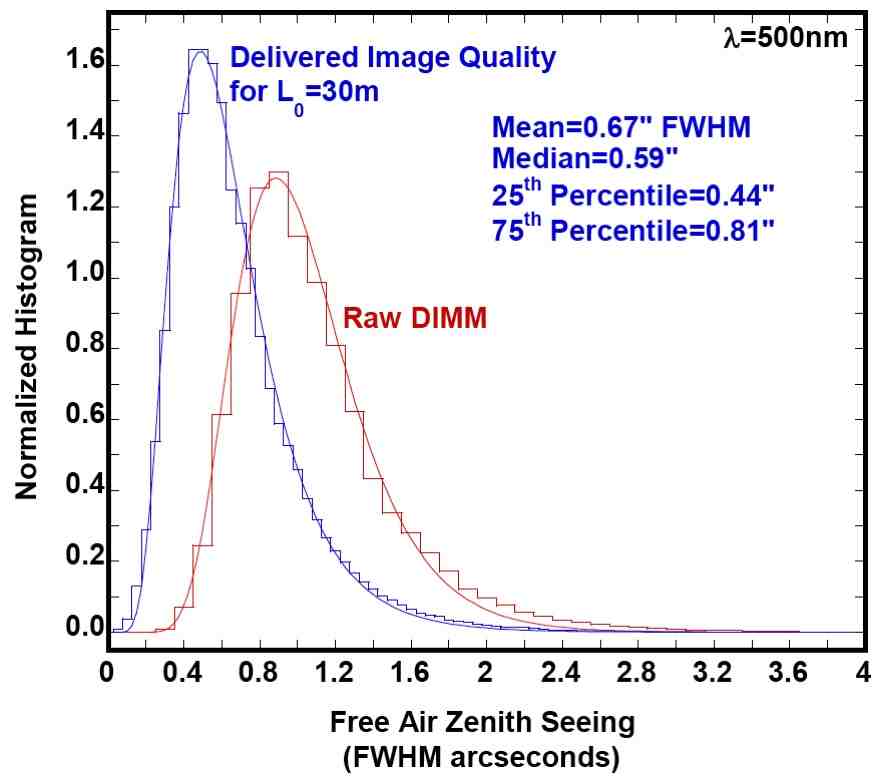}
\end{center}
\caption{
The distribution of ``seeing'' (FWHM of the image of a point source)
at 500 nm based on ten years of measurements from CTIO (10 km from the LSST
site).  The red curve shows results from a Differential Image Motion
Monitor (DIMM), while the blue curve shows the delivered image
quality.  The mean is $0.67''$, and the median is $0.59''$.  
}
\label{fig:design:seeing}
\end{figure}

The LSST will be constructed on El Pe\~n\'on Peak
(\autoref{fig:design:telescope_artist}) of Cerro Pach\'on in the
Northern Chilean Andes.  This choice was the result of a formal site selection
process following an extensive
study comparing seeing conditions, cloud cover and other weather patterns,
and infrastructure issues at a variety of potential candidate sites around the
world.  Cerro Pach\'on is located ten kilometers away from Cerro Tololo Inter-American
Observatory (CTIO) for which over ten years of detailed weather data have
been accumulated.
These data show that more than 80\% of the nights are usable, with excellent
atmospheric conditions.  Differential image motion monitoring (DIMM)
measurements made on Cerro Tololo show that the expected mean
delivered image quality is $0.67''$ in $g$ 
(\autoref{fig:design:seeing}).  

Cerro Pach\'on is also the home of the 8.2-m diameter Gemini-South and
4.3-m diameter 
Southern Astrophysical Research (SOAR)
telescopes.  Observations with those telescopes have confirmed the excellent image quality
that can be obtained from this site.  In addition, LSST will benefit from the extensive
infrastructure that has been created on Cerro Pach\'on and La Serena
to support these other facilities.  
The property is owned by the Association of Universities for
Research in Astronomy (AURA), which also supports operation of CTIO,
Gemini-South, 
and SOAR.

The LSST Observatory as a whole will be distributed over four sites: 
the Summit Facility on El Pe\~n\'on, the Base Facility, the Archive Center, and the Data Centers.
The Base Facility will be at the AURA compound in the town of La Serena, 57 km away
from the mountain.  The Archive Center will be at the National Center for Supercomputing
Applications (NCSA) on the campus of the University of Illinois at Urbana-Champaign.  There will 
be two Data Centers, one co-located with the Archive Center at NCSA, and one at the Base Facility
in La Serena.
Although the four facilities are distributed geographically, 
they are functionally connected via dedicated high-bandwidth fiber optic links.


\section{Optics and Telescope Design}
\label{sec:design:optical}
      
{\it Victor L. Krabbendam,
Charles F. Claver,
Jacques Sebag,
Jeffrey D. Barr, 
John R. Andrew,
Srinivasan Chandrasekharan,
Francisco Delgado,
William J. Gressler,
Edward Hileman,
Ming Liang,
Michelle Miller,
David Mills,
Douglas R. Neill,
German Schumacher,
Michael Warner,
Oliver Wiecha,
Lynn G. Seppala,
J. Roger P. Angel, 
James H. Burge
}

\begin{figure}
\begin{center}
\includegraphics[width=14cm,angle=0]{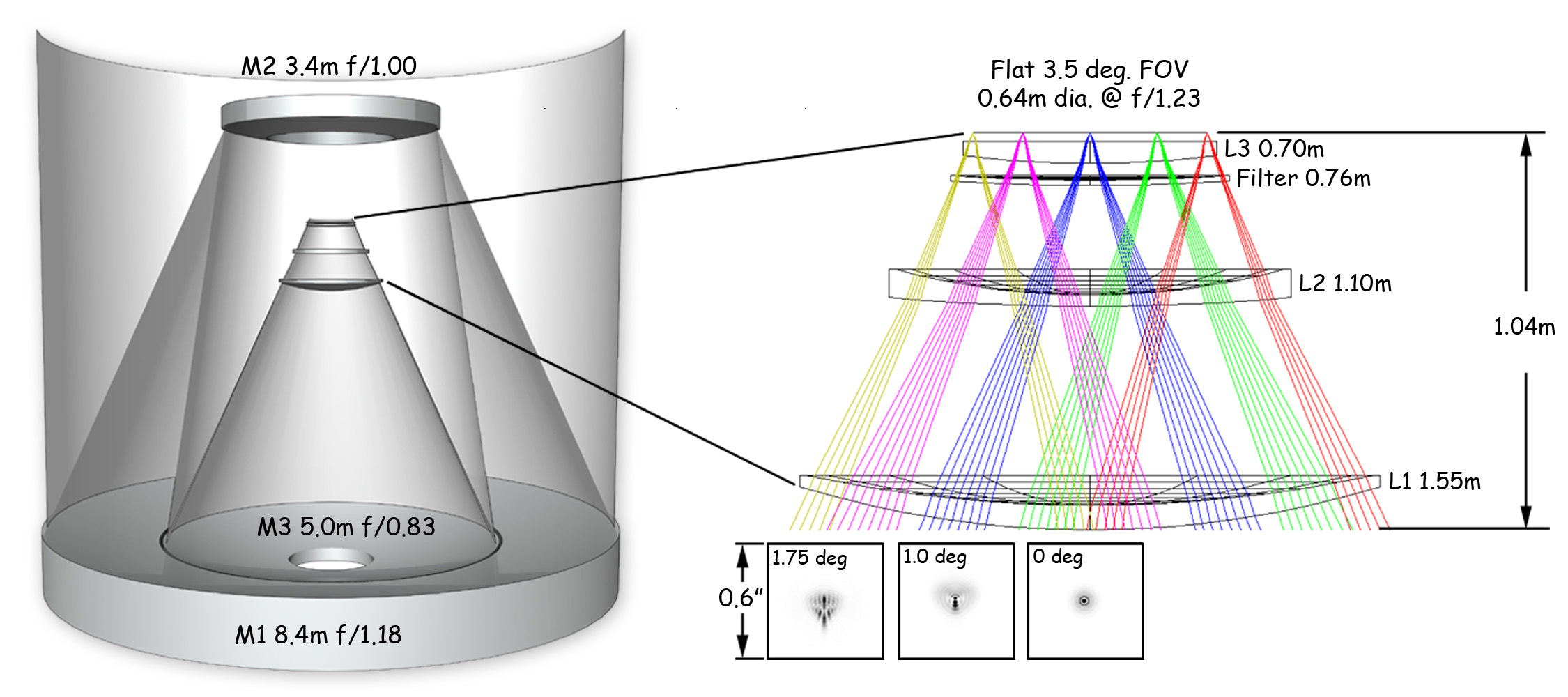}
\caption{
The optical design configuration showing the telescope (left) and camera (right) layouts. 
Diffraction images in $r$ for three field radii, 0, 1.0, and 1.75 degrees, are
shown in boxes 0.6 arcseconds square ($3\times 3$ pixels).
}
\label{fig:design:telescope_schematic}
\end{center}
\end{figure}

\begin{figure}[t]\centering
\includegraphics[width=9cm]{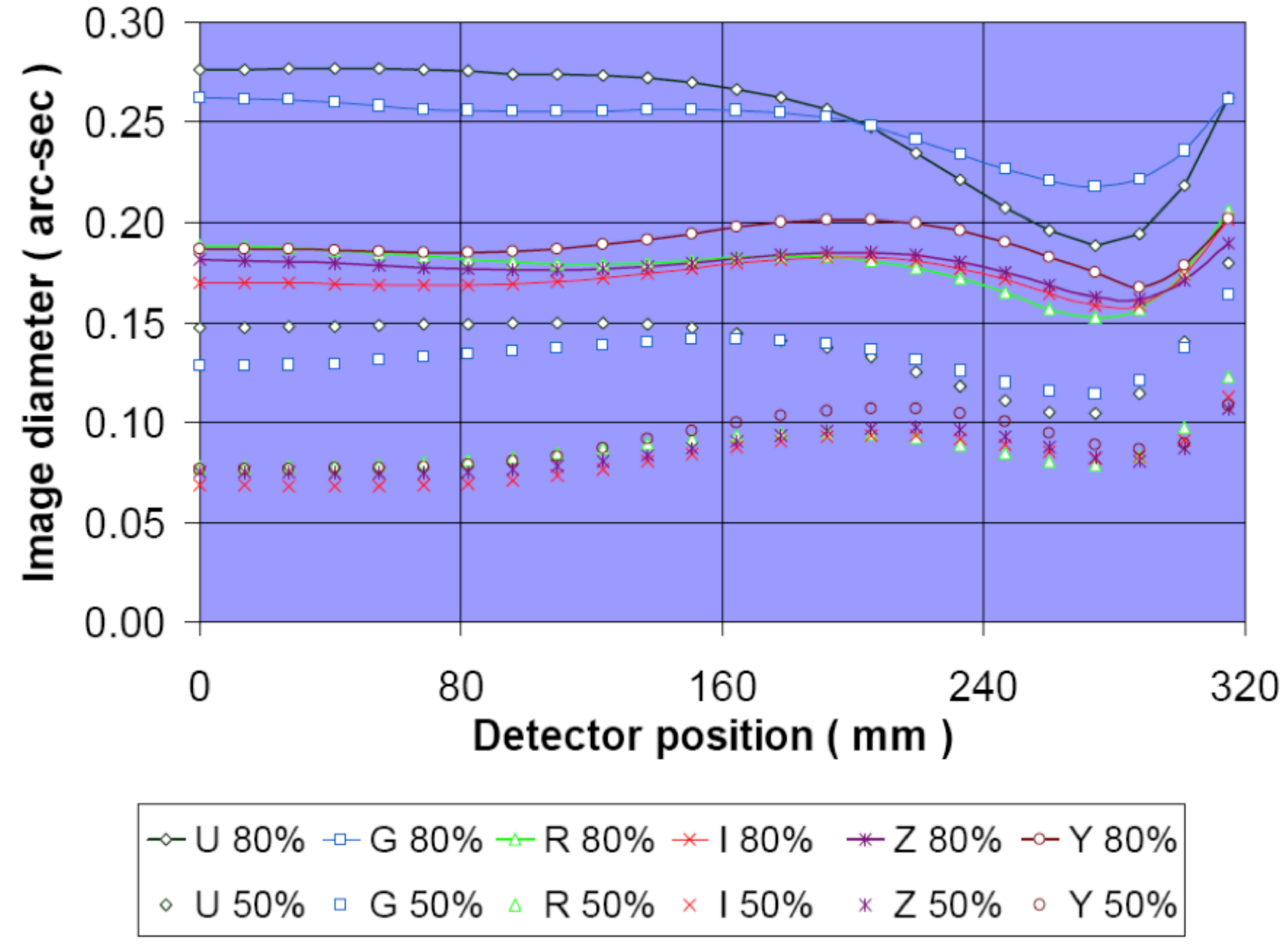}\caption{
The 50\% (plain symbols) and 80\% (symbols with lines) encircled energy 
diameter as a function of radius in the field of view for the LSST 
baseline optical design.  The image scale is 50 microns per
arcsec, or  180 mm per degree.
\label{fig:encircled_energy}}\end{figure}

The LSST optical design shown in \autoref{fig:design:telescope_schematic} 
is a modified Paul-Baker three-mirror system (M1, M2, M3) with three
refractive lenses (L1, L2, L3) and a color filter before the sensor at the focal plane.
Conceptually, it is a generalization of the well-known Mersenne-Schmidt family of 
designs and produces a large field of view with excellent image quality 
\citep{Wil84, Ang++00, Sep02}. Spot diagrams are shown in the figure
inset; these are made quantitative in \autoref{fig:encircled_energy},
which shows the encircled energy diameters at 50\% and 80\% in each
filter as delivered by the baseline optical design.  The uniformity
across the field is striking. 

The LSST \'{e}tendue (including the effects of camera vignetting) is 319 m$^2$deg$^2$.
The effective focal length of the optical system is 10.3 m, making the final f/number 1.23. 
The plate scale is 50 microns per arcsecond at the focal surface. 
This choice of effective focal length represents an optimum balance of image sampling, 
overall system throughput, and manufacturing feasibility. 
The on-axis collecting area is 35 m$^2$, equivalent to a 6.7-m diameter
unobscured clear aperture.  


\begin{figure}
\includegraphics[width=15cm,angle=0]{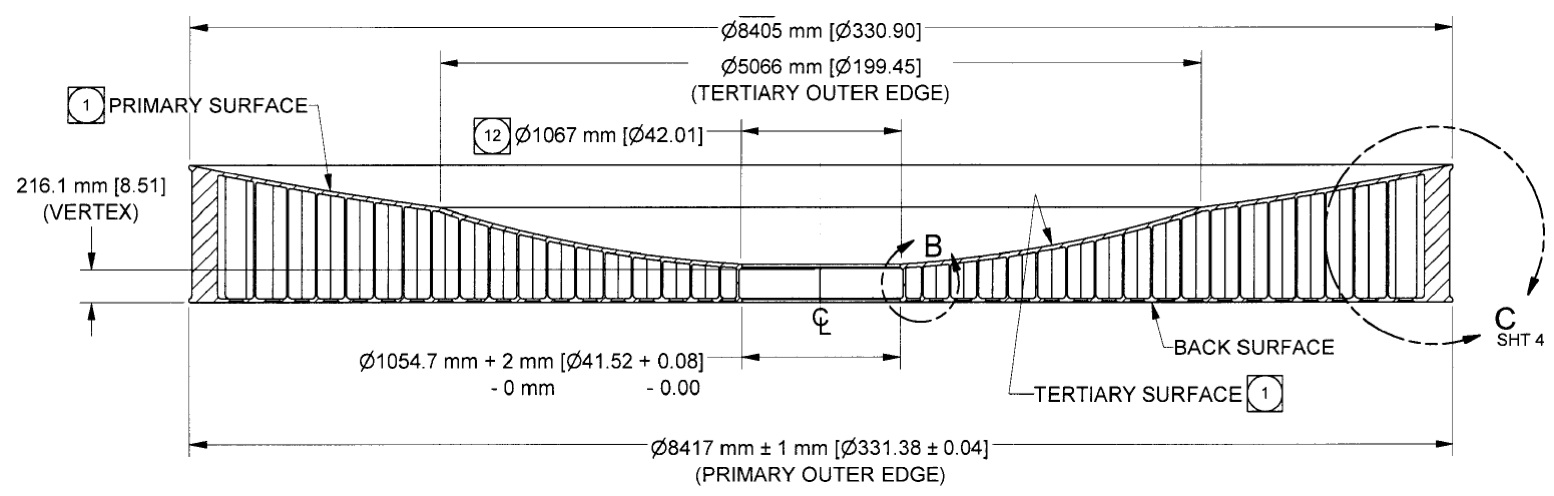}
\caption{
Design and dimensions of the primary and tertiary mirror, showing that
the two are built out of a single mirror blank.  
}
\label{fig:design:m1-3}
\end{figure}

The primary mirror (M1) is 8.4 m in diameter with a 5.1-m inner clear 
aperture.  The tertiary mirror (M3) is 5 m in diameter. 
The relative positions of M1 and M3 were adjusted during the design process 
so that their surfaces meet with no axial discontinuity at a cusp, 
allowing M1 and M3 to be fabricated from a single substrate (see
\autoref{fig:design:m1-3}). 
The 3.4-m convex secondary mirror (M2) has a 1.8-m inner opening. 
The LSST camera is inserted through this opening in order to access
the focal surface. 

The three reflecting mirrors are followed by a three-element
refractive system that corrects field flatness and chromatic
aberrations introduced by the filter and vacuum window.  The
$3.5^\circ$ field of view (FOV) covers a 64-cm diameter flat focal
surface.  Spectral filters reside between the second and third
refractive lens as shown on the right side of
\autoref{fig:design:telescope_schematic}.

The image brightness is constant to a field radius of 1.2 degrees and
gradually decreases afterward by about 10\% at the 1.75-degree field
edge.  The intrinsic image quality from this design is excellent.  The
design also has very low geometrical distortion, with the distortion
in scale $\Delta l / l <0.1$\% over the full FOV, making the LSST an
excellent system for positional astrometry.

There are five aspheric surfaces in the optical design: each of the
three mirror surfaces and one surface each on two of the camera
lenses.  The asphericity on the two concave surfaces of M1 and M3 are
well within standard fabrication methods used for astronomical
mirrors.  During the optimization process, the asphericity of M2 was
minimized to 18.9 microns of departure from the best-fit sphere in
order to reduce the technical challenge for this optic.  The three
fused-silica refractive elements, which have clear apertures of
1.55 m, 1.10 m, and 0.72 m, while large, do not present any particular
challenge in their fabrication.  The 0.75-m diameter spectral filter
is located just prior to L3.  The filter thickness varies from 13.5 to
26.2 mm depending on the choice of spectral band, and is used to
maintain the balance of lateral chromatic aberration.  The zero-power
meniscus shape of the filters keeps the filter surface perpendicular
to the chief ray over the full field of view.  This feature minimizes
shifting of the spectral band wavelength with field angle.  The last
refractive element, L3, is used as the vacuum barrier to the detector
cryostat.  The central thickness of L3 is 60 mm to ensure a
comfortable safety margin in supporting the vacuum stresses.

\begin{figure}
\begin{center}
\includegraphics[width=0.9\textwidth,angle=0]{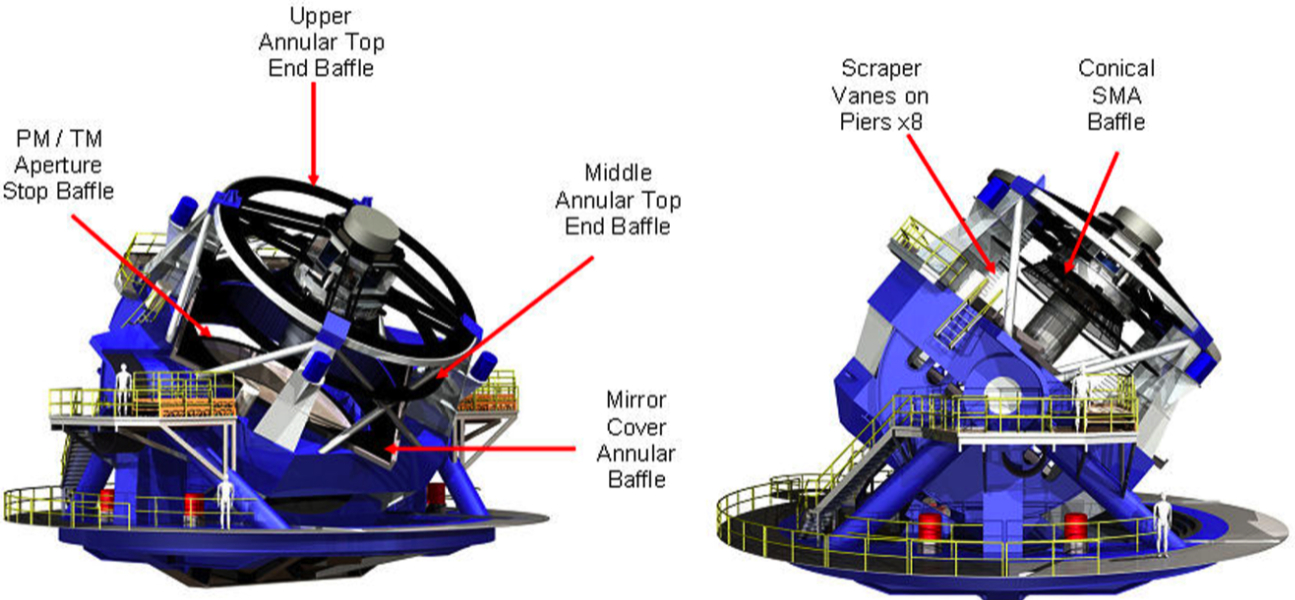}
\end{center}
\caption{
Rendering of the telescope, showing mirror support structures, top end
camera assembly, and integrated baffles. 
}
\label{fig:design:telescope_mount}
\end{figure}

The proposed LSST telescope is a compact, stiff structure with a
powerful set of drives, making it one of the most accurate and agile
large telescopes ever built. 
The mount is an
altitude over azimuth configuration
(\autoref{fig:design:telescope_mount}). The telescope structure is a
welded and bolted steel system designed to be a stiff metering
structure for the optics and a stable platform for observing
\citep{Neill06, Neill08}. The primary and tertiary mirrors are supported in a
single 
cell below the elevation ring; the camera and secondary mirror are
supported above it. The design accommodates some on-telescope
servicing as well as efficient removal of the mirrors and camera, as
complete assemblies, for periodic maintenance.

The stiffness of this innovative design is key to achieving a slew and
settle time that is beyond the capability of today's large
telescopes. The size and weight of the systems are a particular
challenge, but the fast optical system allows the mount to be short
and compact. Finite element analysis has been used to simulate the
vibrational modes of the telescope system, including the concrete
pier.  The frequencies of the four modes with largest amplitudes are
(in order): 
\begin{itemize}
\item 8.3 Hz: Transverse telescope displacement;
\item 8.7 Hz: Elevation axis rotation;
\item 11.9 Hz: Top end assembly optical axis pumping; and
\item 12.6 Hz: Camera pivot.
\end{itemize}


As described in \autoref{sec:design:cadence}, the standard visit time
in a given field is only 34 seconds, quite short for most telescopes.
The time required to reorient the telescope must also be short to keep
the fraction of time spent in motion below 20\%
(\autoref{sec:intro:exposure}).  The motion time for a nominal
$3.5^\circ$ elevation move and a $7^\circ$ azimuth move is five
seconds. In two seconds, a shaped control profile will move the
telescope, which will then settle down to less than $0.1''$
pointing error in three seconds. The stiffness of the support
structure and drive system has been designed to limit the amplitude
and damp out vibrations at these frequencies within this time.  The
mount uses 400 horsepower in the azimuth drive system and 50
horsepower in the elevation system. There are four motors per axis
configured in two sets of opposing pairs to eliminate hysteresis in
the system. Direct drive systems were judged overly complicated and
too excessive, so the LSST design has each motor working through a
multi-stage gear reduction, with power applied through helical gear
sets.  The 300-ton azimuth assembly and 151-ton elevation assembly are
supported on hydrostatic bearings. Each axis uses tape encoders with
$0.001''$ 
resolution. Encoder ripple from these tapes often dominates control
system noise, so LSST will include adaptive filtering of the
signal in the control loop. All-sky pointing performance will be
better than $2''$.  Pointing will directly impact trailing and
imaging systematics for LSST's wide field of view, so accurate
pointing is key to tracking performance.  Traditional closed loop
guiding will achieve the final level of tracking performance.

\section{Camera}  
\label{sec:design:camera}

 {\it Kirk Gilmore, Steven M. Kahn, John Geary, Martin Nordby, Paul O'Connor,  John Oliver,
Scot S. Olivier, Veljko Radeka, Andrew Rasmussen, Terry Schalk, Rafe Schindler, Anthony Tyson, Richard Van Berg}

\begin{figure}
\includegraphics[width=14cm,angle=0]{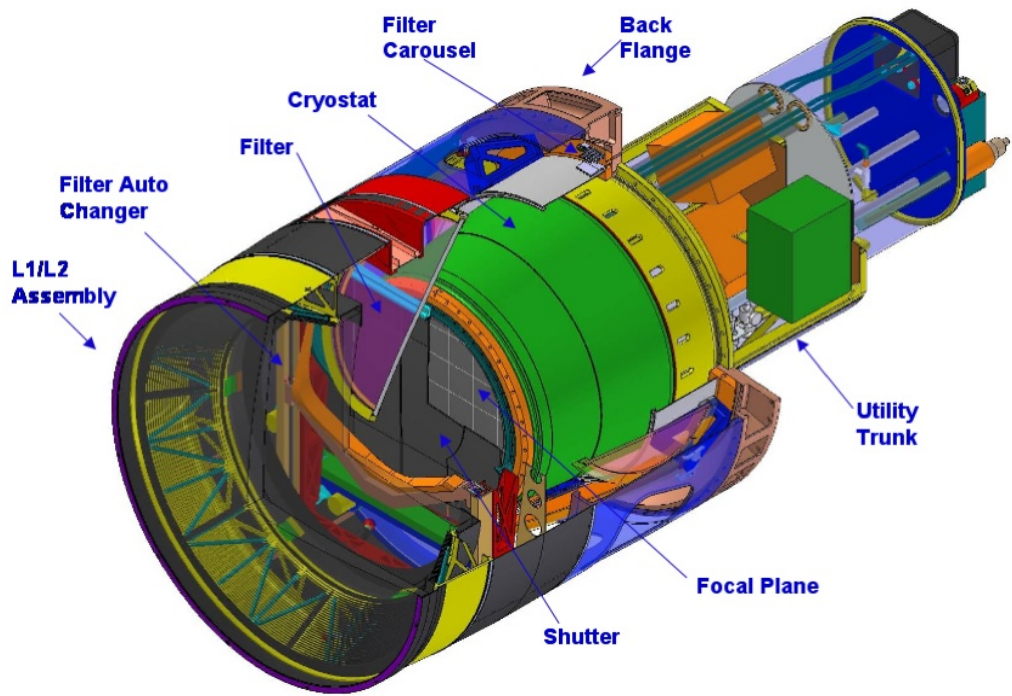}
\caption{
Cutaway drawing of the LSST camera.  The camera body is approximately 1.6 m
in diameter and 3.5 m in length.  The optic, L1, is 1.57 m in diameter.
}
\label{fig:design:camera}
\end{figure}

\begin{figure}
\begin{center}
\includegraphics[width=0.48\textwidth,angle=0]{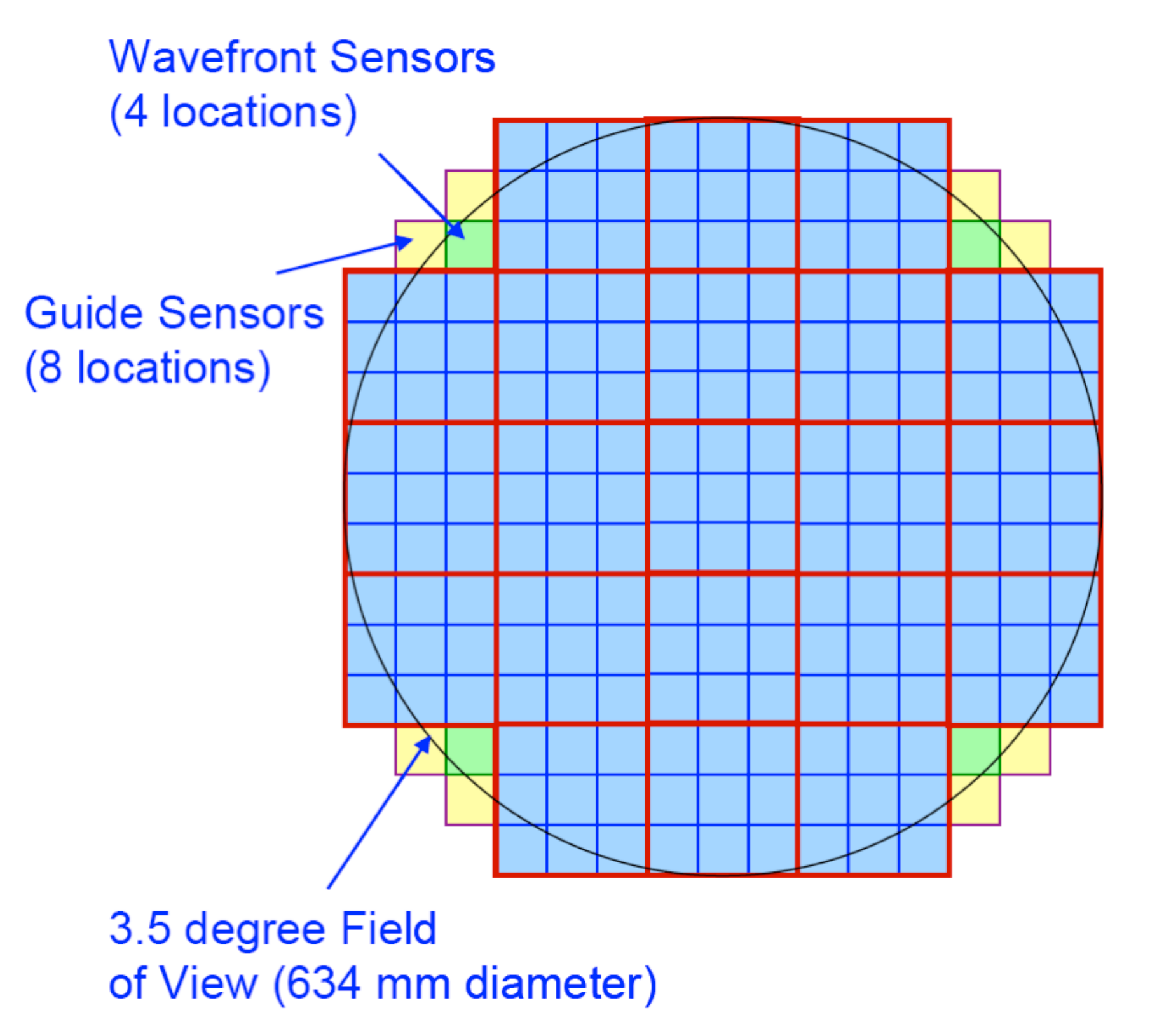}
\end{center}
\caption{
With its 189 sensors, each a $\rm 4\,K \times 4\,K$ charge-coupled device (CCD), the focal
plane of the camera images 9.6 deg$^2$ of the sky per exposure.  Note
the presence of wavefront sensors, which are fed back to the mirror
support/focus system, and the guide sensors, to keep the telescope
accurately tracking on a given field.  
}
\label{fig:design:focal_plane}
\end{figure}

\begin{figure}
\begin{center}
\includegraphics[width=9cm,angle=0]{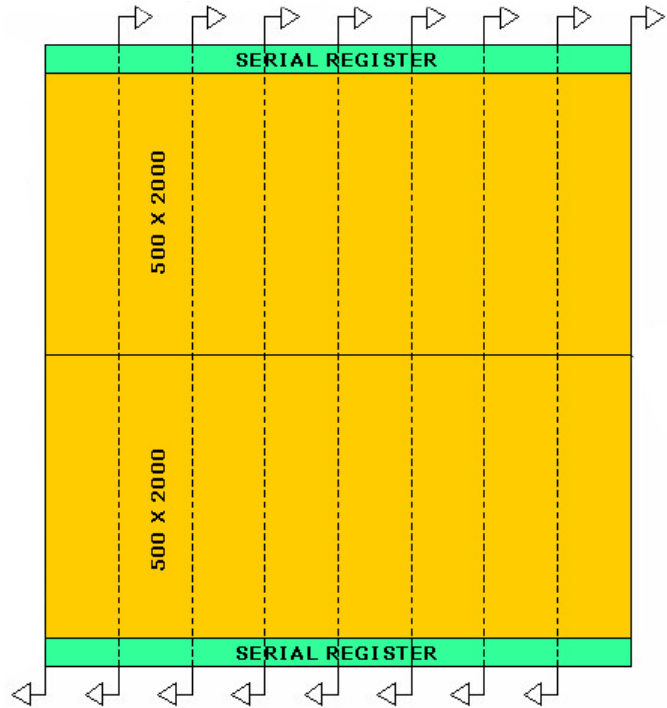}
\caption{
A schematic of the LSST sensor, showing the segmentation into 16
channels, each of which is read out in parallel. 
}
\label{fig:design:sensor}
\end{center}
\end{figure}

The LSST camera, shown in \autoref{fig:design:camera}, contains a 3.2-gigapixel focal plane array
(\autoref{fig:design:focal_plane})
comprised of  189 $4\,\rm K\times4\,K$ CCD sensors with 10 $\mu$m pixels.  The 
focal plane is 0.64 m in diameter, and covers 9.6 deg$^2$ field-of-view with a plate
scale of $0.2''\,\rm pixel^{-1}$. 
The CCD sensors are deep depletion, back-illuminated devices with a 
highly segmented architecture, 16 channels each, that enable the 
entire array to be read out in two seconds (\autoref{fig:design:sensor}). 

The detectors are grouped into $3 \times 3$ arrays called ``rafts.''
All the rafts are identical; each contains its own dedicated front-end
and back-end electronics boards, which fit within the footprint of its
sensors, thus serving as a 144-Megapixel camera on its own.  The rafts
and associated electronics are mounted on a silicon carbide grid
inside a vacuum cryostat, with an intricate thermal control system
that maintains the CCDs at an operating temperature of $-100^\circ$C.
The grid also contains two guide sensors and a wavefront sensor
positioned at each of the four corners at the edge of the field.  The
entrance window to the cryostat is the third of three refractive
lenses, L3 in \autoref{fig:design:camera}.  The other two lenses, L1
and L2, are mounted in an ``optics housing'' at the front of the
camera body.  The camera body also contains a mechanical shutter and a
filter exchange system holding five large optical filters, any of
which can be inserted into the camera field of view for a given
exposure.  The system will in fact have six filters; the sixth filter
can replace any of the five via an automated procedure accomplished
during daylight hours.  

\subsection{Filters}
\label{sec:design:filters}
\begin{figure}
\includegraphics[width=0.95\linewidth]{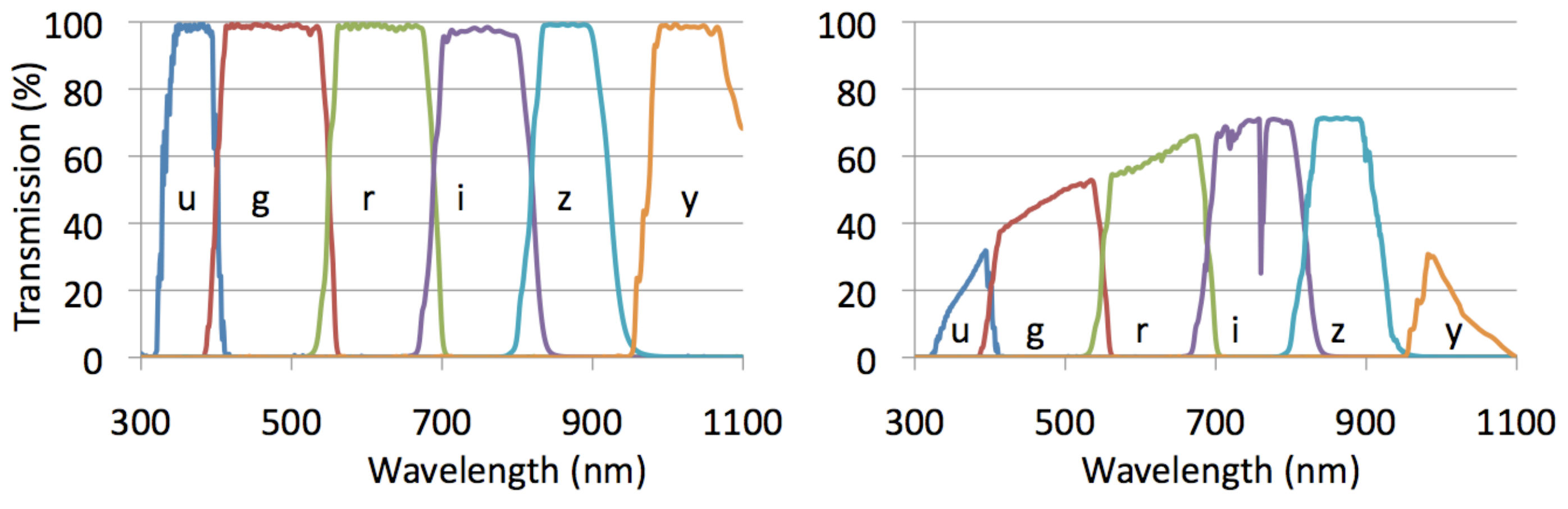}
\caption{
The left panel shows the transmission efficiency of the
$ugrizy$ filters by themselves 
as calculated from models of the filter performance.  
The total throughput, accounting for the transmission through the
atmosphere at the zenith, the reflectivity of the reflective optics,
the transmissivity of the refractive optics, and the quantum efficiency
of the sensors is displayed in the panel on the right.
}
\label{fig:design:filters}
\end{figure}

The LSST filter complement ($u$, $g$, $r$, $i$, $z$, $y$) is
modeled on the system used for the SDSS \citep{Fuk++96}, which covers
the available wavelength range with roughly logarithmic spacing while
avoiding the strongest telluric emission features and sampling the
Balmer break.  
Extension of the SDSS system to longer 
wavelengths ($y$-band) is possible because the deep-depletion CCDs
have high sensitivity to 1 $\mu$m (\autoref{fig:design:filters}).  

The current LSST baseline design has a goal of 1\% relative
photometric calibration (\autoref{sec:introduction:reqs}), which drives
the requirements on the filter set.   The filter set wavelength 
design parameters and the approximate FWHM transmission points for each filter 
are given in \autoref{tab:design:filters} and in \autoref{fig:design:filters}.

\begin{table}
\caption{Design of Filters: Transmission Points in nanometers}
\begin{center}
\begin{tabular}{|c|r|r|l|}
\hline  
Filter     &    Blue Side  &   Red Side   &    Comments  \\       
\hline  
  $u$      &       320     &      400     & Blue side cut-off depends on AR coating \\
  $g$      &       400     &      552     & Balmer break at 400 nm \\
  $r$      &       552     &      691     & Matches SDSS \\
  $i$      &       691     &      818     & Red side short of sky emission at 826 nm \\
  $z$      &       818     &      922     & Red side stop before H$_2$O bands \\
  $y$      &       950     &     1080     & Red cut-off before detector cut-off \\
\hline                         
\end{tabular}
\end{center}
\label{tab:design:filters}
\end{table}

The filters consist of multi-layer dielectric interference coatings
deposited on fused silica substrates. The baseline design has the
first surface of the filters concentric about the chief ray in order
to keep the angles of the light rays passing through the filters as
uniform as possible over the entire range of field positions. The
central thickness and the curvature of the second surface are
optimized for image quality.

\subsection{Sensors}

The heart of the
camera is the science sensor. Its key characteristics are as follows: 

{\it High quantum efficiency from 320 to 1080 nm.}  This is achieved using
a large depletion depth ($ 100~\mu$m) and implementation
of the sensor in a back-illuminated configuration with a thin entrance
window.

{\it Minimal detector contribution to the point spread function.} To
reduce charge diffusion, the sensor is fully depleted, and a high
internal field is maintained within the depletion region. This
is made possible by the use of high resistivity substrates, high applied
voltages, and back-side contacts. Light spreading prior to photo-conversion 
at longer wavelengths is a minor contributor at the $ 100~\mu$m depletion
depth.  

{\it Tight flatness tolerances.} The fast LSST beam ($f$/1.23) yields a short
depth of field, requiring $< 10$ $\mu$m
peak-to-valley focal plane flatness with piston, tip, and tilt
adjustable to $\sim1  \mu$m.  This is achieved through precision alignment
and mounting both within the rafts, and within the focal plane grid.

{\it High fill factor.} A total of 189 $4\,\rm K\times4\,K$ sensors are
required to cover the 3200 cm$^2$ focal plane. To maintain high
throughput, the sensors are mounted in four-side buttable packages and are
positioned in close proximity to one another with gaps of less than a few hundred
$\mu$m.  The resulting ``fill factor,'' i.e., the fraction of the focal plane
covered by pixels, is 93\%. 

{\it Fast readout.} The camera is read out in two
seconds.  To reduce the read noise associated with
higher readout speeds, the sensors are highly segmented.  The large
number of I/O connections then requires that the detector electronics
be implemented within the cryostat to maintain a manageable number of
vacuum penetrations.

Our reference sensor design is a CCD with a
high degree of segmentation, as illustrated in \autoref{fig:design:sensor}. 
A $\rm 4\,K\times 4\,K$ format was chosen because it is the largest
footprint consistent with good yield. Each amplifier will read out
1,000,000 pixels (a $2000\times500$ sub-array), allowing a pixel readout
rate of 500 kHz per amplifier.  
The sensors are
mounted on aluminum nitride (AlN) packages.  Traces are plated
directly to the AlN insulator to route signals from the CCD to the
connectors on the back of the package.  The AlN package provides a
stiff, stable structure that supports the sensor, keeps it flat, and
extracts heat via a cooling strap.

\subsection{Wavefront Sensing and Guiding}

Four special purpose rafts, mounted at the corners of the
science array, contain wavefront sensors and guide sensors
(\autoref{fig:design:focal_plane}). 
Wavefront measurements are accomplished using curvature sensing,
in which the spatial intensity distribution of stars is measured at equal 
distances on either side of focus. Each curvature sensor is composed 
of two CCD detectors, with one positioned slightly above the focal 
plane, the other positioned slightly below the focal plane. 
The CCD technology for the curvature sensors is identical to that used 
for the science detectors in the focal plane, except that the curvature 
sensor detectors are half-size so they can be mounted as an in-out defocus pair. 
Detailed analyses have verified that this configuration can reconstruct the wavefront to the required accuracy.
These four corner rafts also hold two guide sensors each. 
The guide sensors monitor the locations of bright stars at
a frequency of $\sim10$ Hz to provide feedback for a loop 
that controls and maintains the tracking of the telescope 
at an accurate level during an exposure. 
The baseline sensor for the guider is the Hybrid Visible Silicon hybrid-CMOS 
detector.  We have carried out extensive evaluation to 
validate that its characteristics (including wide spectral response, 
high fill factor, low noise, and wide dynamic range) are consistent
with guiding requirements. 

\section{Data Management System}
\label{sec:design:dm}

\noindent {\it Tim S. Axelrod, Jacek Becla, Gregory Dubois-Felsmann, \v Zeljko Ivezi\'c, 
R. Lynne Jones, Jeff Kantor, R. H. Lupton, David Wittman}


The LSST Data Management System
(``DMS'') is required to generate a set
of data products and to make them available to scientists and the
public. To carry out this mission the DMS performs the following major
functions:

\begin{itemize}
\item Continually processes the incoming stream of images generated by the camera
  system during observing to produce transient alerts and to archive
  the raw images.

\item Roughly once per year\footnote{In the first year of operations, we anticipate
  putting out data releases every few months.}, creates and archives a Data Release (``DR''),
  which is a static self-consistent collection of data products
  generated from all survey data taken from the date of survey
  initiation to the cutoff date for the Data Release. The data
  products include optimal measurements of the properties (shapes,
  positions, fluxes, motions) of all
  objects, including those below the single visit sensitivity limit,
  astrometric and photometric calibration of the full survey object
  catalog, and limited classification of objects based on both their
  static properties and time-dependent behavior.  Deep coadded images
  of the full survey area are produced as well.

\item Periodically creates new calibration data products, such as bias
  frames and flat fields, that will be used by the other processing
  functions.

\item Makes all LSST data available publicly through an interface and
  databases that
  utilize, to the maximum possible extent, community-based standards
  such as those being developed by the Virtual Observatory (``VO''), and
  facilitates user data analysis and the production of user-defined
  data products at Data Access Centers and at external sites.
\end{itemize}

\begin{figure}
\begin{center}
\includegraphics[width=9cm,angle=0]{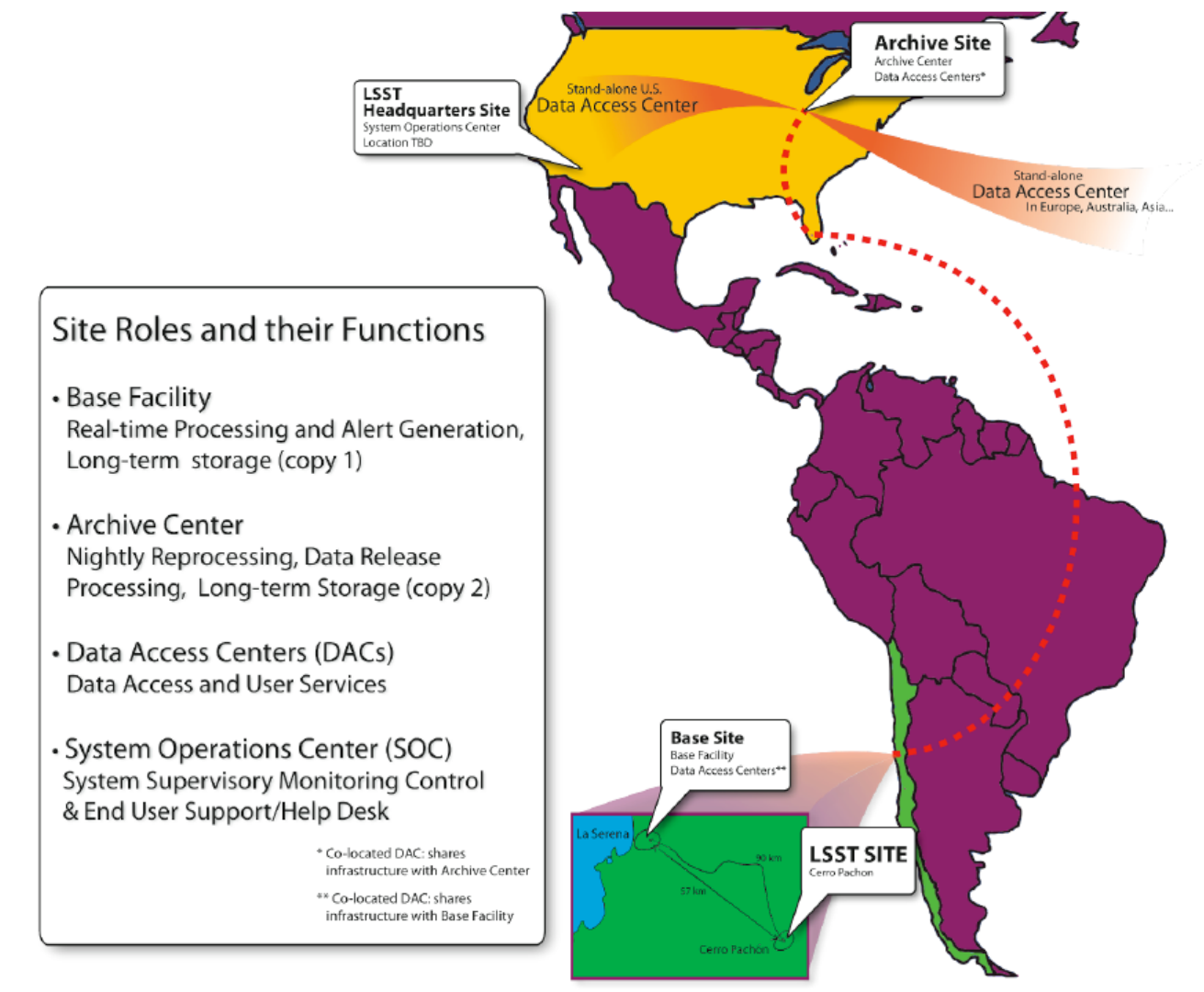}
\caption{A schematic map of the LSST DMS facilities. The LSST
  Telescope Site, located on Cerro Pach\'on, Chile, is connected to
  the Base Facility, located in La Serena, Chile by a dedicated fiber
  link. The Base Facility is connected to the Archive Center, located
  at the National Center for Supercomputing Applications in Illinois
  using commercial high-speed network links. The Archive Center, in
  turn, fans out data to Data Access Centers which serve the data to
  clients, and may be located anywhere in the world. The System
  Operations Center monitors and controls the overall operation of the
  DMS facilities, and provides end-user support facilities.}
\label{fig:design:DM_Geo.pdf}
\end{center}
\end{figure}

The geographical layout of the DMS facilities is shown in
\autoref{fig:design:DM_Geo.pdf}; the facilities include the Mountain
Summit/Base Facility at Cerro Pach\'on and La Serena, the central
Archive Center at NCSA,
the Data Access Centers at NCSA and La Serena, and a System Operations
Center.  The data management system begins at the data acquisition
interface between the camera and telescope subsystems and flows
through to the data products accessed by end users. On the way, it
moves through three types of managed facilities supporting data
management, as well as end user sites that may conduct science using
LSST data or pipeline resources on their own computing infrastructure.

\begin{itemize}
\item The data will be transported over existing high-speed optical
  fiber links from the Mountain Summit/Base Facility in Chile to the
  archive center in the U.S. Data will also flow from the Mountain
  Summit/Base Facility and the archive center to the data access
  centers over existing fiber optic links.  The Mountain Summit/Base
  Facility is composed of the mountaintop telescope site, where data
  acquisition must interface to the other LSST subsystems, and the
  Base Facility, where rapid-turnaround processing will occur for data
  quality assessment and near real-time alerts.

\item The Archive Center is a super-computing-class data center with
  high reliability and availability. This is where the data will
  undergo complete processing and re-processing and permanent
  storage. It is also the main repository feeding the distribution of
  LSST data to the community.

\item Data Access Centers for broad user access are envisioned,
  according to a tiered access model, where the tiers define the
  capacity and response available especially to computationally
  expensive queries and analyses. There are two project-funded Data
  Access Centers co-located with the Base Facility and the Archive
  Center. These centers provide replication of all of the LSST data to
  ensure that disaster recovery is possible. They provide Virtual
  Observatory interfaces to the LSST data products. LSST is
  encouraging non-US/non-Chilean funding for potential partner
  institutions around the world to host
  additional Data Access Centers, which could increase end user access bandwidth,
  provide local high-end computation,
  and help amortize observatory operations costs.  

\item The System Operations Center provides a control room and
  large-screen display for supervisory monitoring and control of the
  DM System. Network and facility status are available as well as the
  capability to ``drill down'' to individual facilities. The Center
  will also provide DM support to observatory science operations, as
  well as an end user help desk. 
\end{itemize}

\subsection{LSST Data Product Overview}
\label{sec:design:dataproduct}

\subsubsection{Level 1, 2, and 3 Data Products}

 The data products are organized into three groups, based largely on
 where and when they are produced.

\begin{itemize}
\item Level 1 products are generated by pipeline processing the stream
  of data from the camera system during normal observing.  Level 1
  data products are, therefore, continuously generated and/or
  updated every observing night. This process is of necessity highly
  automated, and must proceed with absolutely minimal human
  interaction.  Level 1 products include {\em alerts}, i.e.,
  announcements that the flux or position of a given object has
  changed significantly relative to the long-term average.  The
  alerts will be released within 60 seconds of the closing of the
  shutter at the end of a visit (\autoref{sec:introduction:reqs}).  
In addition to science data products, a number of
  Level 1 science data quality assessment (``SDQA'') data products are generated to
  assess quality and to provide feedback to the Observatory Control System.

\item Level 2 products are generated as part of a yearly Data
  Release. 
Level 2   products use Level
  1 products as input, and include data products for which extensive
  computation is required (such as variability information, detection,
  and measurement of the properties of faint objects, and so on),
  often because they combine information from the stack of
  many exposures.  Although the steps that generate Level 2 products
  will be automated, significant human interaction may be required at
  key points to ensure the quality of the data.


\item Level 3 data products are derived from Level 1 and/or Level 2
  data products to support particular science goals, often requiring
  the combination of LSST data across significant areas on the sky.
  The DMS will facilitate the creation of Level 3 data
  products, for example by providing suitable Applications Programming
  Interfaces (APIs) and computing infrastructure, but is not
  itself required to create any Level 3 data product. Instead these
  data products are created externally to the DMS, using software
  written by, for example, science collaborations. Once created, Level 3 data
  products may be associated with Level 1 and Level 2 data products
  through database federation\footnote{See Wikipedia's article on the
    subject at \url{http://en.wikipedia.org/wiki/Federated\_database}.}.  In rare
  cases, the LSST Project, with the agreement of the Level 3 creators,
  may decide to incorporate Level 3 data products into the DMS
  production flow, thereby promoting them to Level 2 data products.

\end{itemize}

Level 1 and Level 2 data products that have passed quality control
tests will be accessible to the public without restriction.
Additionally, the source code used to generate them will be made
available, and LSST will provide support for building the software
system on selected platforms. The access policies for Level 3 data
products will be product- and source-specific, and in some cases will
be proprietary.

\subsubsection{Overview of Pipeline Processing}

In the overall organization of the DMS pipelines and productions,
``production'' has a particular meaning: it is a
coordinated group of pipelines that together carry out a large-scale
DMS function.

\paragraph{Alert Production}
Astronomers interested in transient phenomena of many sorts
(\autoref{chp:transients}) need to know of objects whose flux has
changed significantly as soon as possible after the data are taken.
Therefore, the most visible aspect of Level 1 processing is the
production of {\em alerts}, i.e., announcements of such variability.  The
Alert Production is directly fed by the output data stream from the
camera Science Acquisition System (SDS) during observing. This
data stream contains both unprocessed (raw) camera images, and images
that have been corrected for crosstalk by the SDS on the mountain. At
the end of a visit, the Alert Production: 

\begin{itemize}
\item Acquires the raw science images from the camera, and moves them to
  the Archive Center for permanent storage.
\item Processes the crosstalk-corrected images from the camera to detect
  transient events within 60 seconds of shutter closure for the second
  exposure in a visit.  This will probably be done with a variant of
  the \citet{Alard+Lupton98} image-subtraction algorithm.  
\item Packages catalog information, together with postage-stamp images
  of detected transients as
  alerts and past history of the object, and distributes them to the community as VO events.
\item Continuously assesses the data quality of the data stream.
\end{itemize}

The major steps in the processing flow are:

\begin{itemize}
\item Image processing of the raw exposures to remove the instrumental
  signature, such as bias, flat-field, bad columns, and so on.  
\item Determination of the World Coordinate System (WCS), image
  Point-Spread Function (PSF), and rough photometric
  zeropoint. This produces processed exposures.
  \item Subtraction of a registered template exposure (a co-addition of
    previous high-quality images of a given field, created in an
    earlier data release) from the processed
  exposure, producing a difference exposure.  
\item Detection of sources (both positive and negative!) in the difference
  exposure, producing what we refer to hereafter as ``DIASources.''
  \item Visit processing logic, which compares the DIASources from the
  two exposures in the visit to discriminate against cosmic rays, and
  to flag very rapidly moving Solar System objects.
\item ``FaintSources,'' abbreviated measurements of low
  signal-to-noise ratio (S/N) detections,
  are produced for objects of particular interest, e.g., predicted
  positions for Solar System objects, or objects that have previously
  produced {\em alerts}.
\item Comparison of positive flux DIASources with predictions from the
  Moving Object Pipeline (MOPS; see \autoref{ss:mops}) for already known Solar
  System objects, as contained in the Moving Object table.
\item The Association Pipeline is run to match DIASources to already known astronomical objects,
  as contained in the Object table.
\item DIASources that are detected in both
  exposures of a visit, and are not matched to a known Solar System
  object,  produce an {\em alert}. 
\item Quality Assessment is performed at every pipeline stage, stored in database
  tables, and fed to the Observatory Control System as required.
\item The Moving Object Pipeline (\autoref{ss:mops}) is run during the day to
  interpret each new detection of a moving object as a new measurement of
  a Solar System object already in the Moving Object table, or as a
  previously unknown object, which will be added to the Moving Object
  table.  All orbits are refined based on the new measurements from
  the night.
\end{itemize}

The community has strongly expressed the
preference that {\em alerts} not be significantly filtered prior to
distribution so that science opportunities are not closed off.  We
have, therefore, adopted very simple criteria for issuing an {\em alert}: 
$5\,\sigma$ DIASources seen in both exposures of a visit which
are not consistent with cosmic ray events. 

Note that no explicit classification of an {\em alert} is provided, but
users can readily construct classifiers and filters based on information in the
Science Database; indeed, this is likely to be part of Level 3
software produced by the transient, stellar populations, and supernova
science collaborations.  The information that could be used for this
classification includes the light curve, colors, and
shape information for the associated object.  Additionally, database
queries can readily be formulated which will identify exposures that
have generated anomalously large numbers of {\em alerts}, presumably due to
image artifacts or processing problems.
  
As the raw images arrive at the Archive Center, the same processing
flow is performed there, with the consistency of the databases at the
Base and Archive Centers being periodically checked.  The duplication
of processing is carried out to reduce the data bandwidth required
between the Base and Archive Centers.

 \paragraph{Data Release Production}
At yearly intervals (more often during the first year of the survey) a
new Data Release (DR) is produced.  A DR includes all data taken by the
survey from day one to the cutoff date for the DR, and is a
self-contained set of data products, all produced with the same
pipeline software and processing parameters.  The major steps in the
processing flow are:

\begin{itemize}
\item As in the Alert Production, all raw exposures from the camera
  are processed to remove the instrumental signature, and to determine
  the WCS and PSF, producing processed exposures.  This is done with
  the best available calibration products, which in general will be
  superior to those available when the processing was initially done.
\item The survey region is tessellated into a set of sky patches of
  order the size of a CCD, and
  several co-added exposures are produced for each patch from the
  processed exposures.  These are a per-band template co-add used for image
  subtraction; a detection co-add used in the Deep Detection Pipeline
  (see next item),
  possibly per-band; and a RGB co-add used for visualization.
\item The Deep Detection Pipeline is run, populating the Object,
  Source, and FaintSource tables.  Rather than working from the co-add,
  Deep Detection will use the ``Multifit'' algorithm
  (\autoref{sec:design:algorithms}; \citealt{Multifit08}), whereby a model (e.g., a PSF for
  a stellar object or an exponential profile for a disk galaxy) is
  fit to the entire stack of exposures which contain the object.  Thus
  each exposure is fit using its own PSF; this results in a set of
  optimal measurements of the object attributes over the full time
  span of the survey, including astrometric parameters such as proper
  motion and parallax.
\item The Image Subtraction Pipeline is run, as in the Alert
  Production, yielding DIASources and FaintSources for transient
  objects.  
\item The Moving Object Pipeline is run on DIASources, to yield a
  complete set of orbits for Solar System Objects in the Moving Object
  table.
\item The Photometric Calibration Pipeline is run on the full set of
  measurements in the Source, DIASource, and FaintSource catalogs,
  incorporating measurements from the Auxiliary Telescope and other
  sources of data about the atmosphere to perform a global photometric
  calibration of the survey (\autoref{sec:design:calsim}).  In
  addition to accurate photometry for every measurement, this yields
  an atmosphere model for every exposure.
\end{itemize}

\subsection{Detection and Measurement of Objects}
\label{sec:design:algorithms}

Here we provide more detail on the specific algorithms used to define and measure object
properties that are issued with the Data Releases:


\subsubsection{Deep Detection Processing}


The survey region is organized into overlapping sky patches of order
the size of a CCD, and a deep co-added image is created for each patch.
The details of the co-add algorithm are still undecided, but the current
baseline is to use the \citet{Kaiser04} algorithm on the full stack of
survey images contained within the Data Release.  The Kaiser algorithm
convolves each image with the reflection of its PSF, and then
accumulates with weight inversely proportional to the sky variance.
Care will be taken to ensure that rapidly moving objects, such as
Solar System objects, do not appear in the co-add.  An object detection
algorithm is then run on the co-add, generating an initial Object
catalog.  An ``Object'' at this stage is nothing more than a pixel
footprint on the sky, possibly with links to related Objects in a
segmentation tree that has been created by segmenting (deblending)
overlapping Objects.  The tree will be organized so that the root node
is the largest object in the hierarchy, with the leaf nodes being the
smallest.  The segmentation/deblending algorithm to be employed is
still under investigation, with Sextractor \citep{1996A&AS..117..393B} or the SDSS
photometric pipeline \citep{2002AJ....123..485S} being examples of the
kind of processing involved.  The properties of the Objects that are
segmented in this way are then determined with Multifit as described
below. 

\subsubsection{Difference Exposure Processing}

A new object is created whenever a transient source that is
detected in both difference images from a visit does not match any
object already in the table.  The match will take account of
extendedness as well as position on the sky, so that a new point
source at the location of a galaxy already in the catalog (for
example, due to a supernova or variable AGN) will result in a new
object.  

Note that this process cannot be perfect, since measuring the
extendedness of objects near the PSF size will always be uncertain.
Consequently, there will be cases where flux from a supernova or AGN
point source will be incorrectly added to the underlying galaxy rather
than to a new point source.  Between successive Data Releases,
however, these errors will decrease in severity: As the survey goes
deeper, and accumulates images in better seeing, extendedness will be
better measured by the Multifit procedure, as discussed below. 

\subsubsection{Measuring the Properties of Objects}

The image pixels containing an object from all relevant exposures are
fit to one or more object models using Multifit, generating model
parameters and a covariance matrix.
Our choice of models is driven by astrophysics, by
characteristics of the LSST system, and by computing practicalities.  The 
initial model types are as follows:

{\bf Slowly Moving Point Source Model.}\ 
The Slowly Moving Point Source (SMPS) Model is intended to account for
the time varying fluxes and motion on the sky of point sources
(usually stars) with proper motions between zero and roughly
$10''\,\rm yr^{-1}$. 
The model accounts for motion with respect to the local
astrometric reference frame that is generated by proper motion,
parallax, and possibly orbital motion with respect to a binary
companion.  The object properties are measured in every exposure that
contains it.  If the S/N in the exposure is
above a predetermined threshold, perhaps 5, the measurement generates
a row in the Source table.  If the S/N 
is lower than the threshold, a FaintSource row is generated instead.
 \citet{Lang++09} have successfully used a similar modeling and measurement approach to detect very faint brown
dwarfs with high proper motion.  

The SMPS Model will be fit only to objects that are leaf nodes in the
segmentation tree.

{\bf Small Object Model.}\ 
The Small Object (SO) Model is intended to provide a robust
parametrization of small (diameter $< 1'$) galaxy images for weak
lensing shear measurement and determination of photometric redshifts.
The definition of the model flux profile is still undecided (Sersic profiles?
Superpositions of exponential and de Vaucouleurs profiles?), but should
be driven by the needs of photometric redshifts
(\autoref{sec:common:photo-z}).  The measurement of the elliptical 
shape parameters will be driven by the needs of weak lensing
(\autoref{chp:wl}).  

The SO Model will be fit only to objects that are leaf nodes in the
segmentation tree.

{\bf Large Object Model.}\ 
A ``large'' object is one for which the 20 mag/arcsec$^2$ isophotal
diameter is greater than $1'$, and less than 80\% of the patch
size. 
This includes, for
example, the majority of NGC galaxies.  The vast
majority of the LSST science 
will be accomplished with measurements made using the SMPS and SO
Models, but much valuable 
science and numerous EPO applications will be based on larger
objects found in LSST images.  To at least partially satisfy this
need, large objects will have entries in the Object table, but will
not have any model fitting performed by Multifit.  

{\bf Solar System Model.}\ 
The predicted ephemerides from the orbit for an object in the moving
object table 
constitutes an object model which is used to measure the object
properties in each exposure that contains the object.  
It is not yet decided whether the measurements of faint detections
should be at a position entirely fixed by the orbit prediction, or
should be allowed to compensate for prediction error by ``peaking up''
within some error bound around the prediction.

\subsubsection{The Multifit Algorithm}

Objects are detected on co-added images, but their models will be fit
to the full data set of exposures on which they appear ($n\sim 400$ at
the end of the survey in each filter).  The motivation for doing this
is two-fold \citep{Multifit08}.
First, the co-add will have a very complicated and discontinuous PSF
and depth patchiness due to detector gaps and masked moving objects.
Second, although the Kaiser co-add algorithm is a
sufficient statistic for the true sky under the assumptions that sky
noise dominates, and is Gaussian, those assumptions do not strictly
hold in real data.  

An initial model will be fit to the co-add, to provide a good starting
point for the fit to the full data set.  Multifit will then read in all
the pixels from the $n$ exposures and perform a maximum likelihood fit
for the model which, when convolved with the $n$ PSFs, best matches
the $n$ observations.  This naturally incorporates the effects of
varying seeing, as the contribution of the better-seeing images to the
likelihood will be sharper.  This approach also facilitates proper
accounting for masked areas, cosmic rays, and so on.  The best-fit model
parameters and their uncertainties will be recorded in an Object table
row.



\subsubsection{Model Residuals}

The measurement process will produce, in conjunction with every
source, a residual image that is the difference of the associated
image pixels and the pixels predicted from the model over the
footprint of the model.  Characterizing these residuals is important
for science such as strong lensing and merging galaxies, that will
identify interesting candidates for detailed analysis through their
residuals.  Selecting the most useful statistical measures of the
residuals will be the outcome of effort during the continuing design
and development phase of the project.

\subsection{The Moving Object Processing System (MOPS)}
\label{ss:mops}

Identifying moving objects and linking individual detections into
orbits, at all distances and solar elongations, would be a daunting
task for LSST without advanced software. Each observation from the
telescope is differenced against a ``template" image (built from many
previous observations), allowing detection of only transient,
variable, or moving objects in the result. These detections are fed
into the Moving Object Processing System (MOPS), which attempts to
link these individual detections into orbits.

MOPS uses a three-stage process to find new moving objects \citep{kubica2005thesis, kubica2005b, kubica2007}. 
In the first stage, intra-night associations are proposed by searching for detections
forming linear ``tracklets.''  By using loose bounds on the linear fit and 
the maximum rate of motion, many erroneous initial associations can be ruled
out. In the current model of operations, LSST will revisit observed fields twice each night, with approximately 20--45 minutes between these observations. These two detections are what are linked 
into tracklets. 
 In the second stage, inter-night associations are proposed by searching 
for sets of tracklets forming a quadratic trajectory.  Again, the algorithm 
can efficiently filter out many incorrect associations while retaining most 
of the true associations. However, the use of a quadratic approximation means 
that a significant number of spurious associations still
remains. Current LSST operations simulations
(\autoref{sec:design:opsim}) show that LSST will image
the entire visible night sky approximately every three nights - thus
these inter-night associations of ``tracklets'' into ``tracks'' are
likely to be separated by 3--4 nights.  

In the third stage, initial orbit determination and differential corrections
algorithms \citep{milani08} are used to further filter out erroneous associations by rejecting 
associations that do not correspond for a valid orbit.  Each stage of this
strategy thus significantly reduces the number of false candidate associations 
that the later and more expensive algorithms need to test. After orbit determination
has occurred, each orbit is checked against new or previously detected (but unlinked)
tracklets, to extend the orbit's observational arc. 

To implement this strategy, the LSST team has developed, in a collaboration with 
the Pan-STARRS project \citep{kaiser2002ps}, a pipeline based on multiple 
k-dimensional- (kd-) tree data structures \citep{kubica2007, barnard2006}.
These data structures provide an efficient way to
index and search large temporal data sets. Implementing a variable tree 
search we can link sources that move between a pair of observations, merge these 
tracklets into tracks spread out over tens of nights, accurately predict where a 
source will be in subsequent observations, and provide a set of candidate asteroids 
ordered by the likelihood that they have valid asteroid tracks. Tested on
simulated data, this pipeline recovers 99\% of correct tracks for
near-Earth and main belt asteroids, and requires less than a day of
CPU time to analyze a night's worth of  
data. This represents a several thousand fold increase in speed over a na\"{i}ve 
linear search. It is noteworthy that comparable amounts of CPU time are spent 
on the kd-tree based linking step (which is very hard to parallelize) and on 
posterior orbital calculations to weed out false linkages (which can be
trivially parallelized).

\subsection{Long-term Archive of LSST Data}
The LSST will archive all observatory-generated data products during its
entire 10-year survey.  A single copy of the resultant data set will be in
excess of 85 petabytes.  Additional scientific analyses of these data have
the potential to generate data sets that significantly exceed this amount.

The longer-term curation plan for the LSST data beyond the survey period is
not determined, but it is recognized as a serious concern.  This issue
is important for all large science archives and it is impractical
(perhaps impossible) for individual facilities or researchers to address
this problem unilaterally.

The NSF has recognized this issue and has begun soliciting input for
addressing long-term curation of scientific data sets via the DataNet and
other initiatives.  The LSST strongly endorses the need for this issue to be
addressed at the national level, hopefully via a partnership involving
government, academic, and industry leaders.


\section{Photometric Calibration}
\label{sec:design:calsim}

{\it David L. Burke, Tim S. Axelrod, James G. Bartlett, David Cinabro, Charles F. Claver, James S. Frank, J. S. Haggerty, \v Zeljko Ivezi\'c, R. Lynne Jones,
Brian T. Meadows, David Monet, Bogdan Popescu, Abhijit Saha, M. Sivertz, J. Allyn Smith, Christopher W. Stubbs, Anthony Tyson}

\subsection{ Natural LSST Photometric System  }

A ground-based telescope with a broad-band detector will observe the
integral of the source specific flux density {\it at the top} of the Earth's
atmosphere, $F_\nu(\lambda)$, weighted by the normalized 
response function (which includes the effects of the atmosphere and all 
optical elements), 
$\phi_b(\lambda)$,
\begin{equation}
\label{eqn:Fmeas}
  F_b = \int_0^\infty {F_\nu(\lambda) \phi_b(\lambda) d\lambda},
\end{equation}
where the index $b$ corresponds to a filter bandpass ($b=ugrizy$).
The chosen units for $F_b$ are Jansky (1 Jansky = 10$^{-26}$ W
Hz$^{-1}$ m$^{-2}$ = 10$^{-23}$ erg cm$^{-2}$ s$^{-1}$ Hz$^{-1}$), and by definition, 
$\int_0^\infty {\phi_b(\lambda) d\lambda}=1$. 
The corresponding astronomical magnitude is defined as
\begin{equation}
\label{eqn:stdmag}
     m_b \equiv -2.5\log_{10}\left({F_b \over F_{AB}}\right).
\end{equation}
The flux normalization $F_{AB} = 3631$ Jy 
follows the standard of \citet{Oke+Gunn83}. 

The normalized response function is defined as 
\begin{equation}
\label{PhiDef}
  \phi_b(\lambda) \equiv {\lambda^{-1} T_b(\lambda) \over \int_0^\infty 
                {\lambda^{-1} T_b(\lambda) d\lambda}}.
\end{equation}
The $\lambda^{-1}$ term reflects the fact that the CCDs used as
sensors in the camera are photon-counting devices rather than 
calorimeters. Here, $T_b(\lambda)$ is the system response function,
\begin{equation}
\label{eqn:TbDef}
    T_b(\lambda) = T_b^{instr}(\lambda) \times T^{atm}(\lambda), 
\end{equation}
where $T^{atm}$ is the optical transmittance from the top of the
atmosphere to the input pupil of the telescope, and $T_b^{instr}$ is 
the instrumental system response (``throughput'') from the input pupil 
to detector (including filter $b$). This function is proportional to 
the probability that a photon starting at the top of the atmosphere
will be recorded by the detector. Note that the overall normalization of both 
$T_b^{instr}$ and $T^{atm}$ cancels out in \autoref{PhiDef}. 

An unavoidable feature of ground-based broad-band photometry is 
that the normalized response function, $\phi_b(\lambda)$, varies with 
time and position on the sky and detector due to variations in 
shapes (spectral profiles) of $T^{atm}(\lambda)$ and
$T_b^{instr}(\lambda)$. Traditionally, these
effects are calibrated out using a set of standard stars. Existing 
data (e.g., from SDSS) demonstrate that this method is insufficient to deliver the
required photometric precision and accuracy in general observing 
conditions. 
Instead, the LSST system will measure $T^{atm}(\lambda)$ and
$T_b^{instr}(\lambda)$ (yielding {\em measured} quantities $S^{atm}$
and $S_b^{instr}$) on the 
relevant wavelength, temporal, and angular scales. 

In summary, the basic photometric products will be reported
on a {\it natural photometric system}, which means that for
each photometric measurement, $F_b^{meas}$, a corresponding
measured normalized response function, $\phi_b^{meas}(\lambda)$,
will also be available. Of course, error estimates for both 
$F_b^{meas}$ and $\phi_b^{meas}(\lambda)$ will also be reported. 
The survey will collect $\sim 10^{12}$ such ($F_b^{meas}$,$\phi_b^{meas}$) 
pairs over a ten year period -- one pair for each source detection.

\subsection{ Standardized Photometric System }

One of the fundamental limitations of broad-band photometry is 
that measurements of flux, $F_b^{meas}$, cannot be accurately 
related to $F_\nu(\lambda)$ unless $\phi_b(\lambda)$ is known.
An additional limitation is that $F_b^{meas}$ can vary
even when $F_\nu(\lambda)$ is constant because $\phi_b$ 
is generally a variable quantity. This variation needs to be 
accounted for, for example, when searching for low-amplitude
stellar variability, or construction of precise color-color and
color-magnitude diagrams of stars. 

Traditionally, this flux variation is calibrated out using
atmospheric extinction and color terms, which works for sources with 
relatively smooth spectral energy distributions. 
However, strictly 
speaking this effect cannot be calibrated out unless the 
shape of the source spectral energy distribution,
\begin{equation}
\label{eqn:fnu}
       f_\nu(\lambda) =  F_\nu(\lambda) / F_0,
\end{equation}
where $F_0$ is an arbitrary normalization constant, is known.
If $f_\nu(\lambda)$ is known, then for a pre-defined ``standard'' 
normalized response function, $\phi_b^{std}(\lambda)$ (obtained 
by appropriate averaging of an ensemble of $\phi_b^{meas}$ during
the commissioning period), 
the measurements expressed on the natural photometric system 
can be ``standardized'' as 
\begin{equation}
\label{Eq:dmStd}
  m_b^{std} - m_b^{meas} \equiv  \Delta m^{std} = 
  2.5\,\log\left({\int_0^\infty f_\nu(\lambda) \phi_b^{meas}(\lambda) d\lambda 
 \over \int_0^\infty f_\nu(\lambda) \phi_b^{std}(\lambda) d\lambda}
  \right), 
\end{equation}
where we have used magnitudes for convenience. 
While this transformation is in principle exact, $m_b^{std}$
inherits measurement error in $m_b^{meas}$, as well as an additional 
error due to the difference between the {\em true} $\phi_b(\lambda)$ and
the {\em measured} $\phi_b^{meas}$ which will be used in practice. 
Uncertainties in our knowledge of $f_\nu(\lambda)$ will contribute
an additional error term to $m_b^{std}$. Depending on the science
case, users will have a choice of correcting $m_b^{meas}$ 
using pre-computed $\Delta m^{std}$ for typical spectral
energy distributions (various types of galaxies, stars, and
solar system objects, average quasar spectral energy distribution, 
etc.), or computing their own $\Delta m^{std}$ for their particular
choice of $f_\nu(\lambda)$. 

\subsection{  Measurement of Instrumental System Response, $S_b^{sys}$  }
\label{sec:sys}

A monochromatic dome projector system will be used to provide a well-controlled source of light for 
measurement of the relative throughput of the full LSST instrumental system.
This includes the reflectivity of the mirrors, transmission of the refractive optics and filters,
the quantum efficiency of the sensors in the camera, and the gain and linearity of the sensor read-out electronics.

An array of projectors mounted in the dome of the LSST enclosure will be illuminated with both broadband
(e.g., quartz lamp) and tunable monochromatic light sources.
These ``flat-field'' projectors are designed to fill the LSST \'etendue with uniform illumination,
and also to limit stray light emitted outside the design acceptance of the system.
A set of precision diodes will be used to normalize the photon flux integrated during flat-field exposures.
These photodiodes, together with their read-out electronics, will be calibrated at the U.S. National Institute of Standards (NIST)
to $\sim 0.1\%$ relative accuracy across wavelength from 450 nm to 950 nm.
The response of these diodes varies smoothly across this range of wavelength and provides a well-behaved reference \citep{stubbs05}.
Adjustment of the wavelength of the light source can be as fine as one nanometer, and will
allow precise monitoring of the shape of the bandpasses of the instrumental system during the course of the survey \citep{S+T06}.

It is anticipated that the shapes of the bandpasses will vary only slowly, so detailed measurement will need be done only once per month or so.
But build-up of dust on the surfaces of the optics will occur more rapidly.  
The dimensions of these particles are generally large, and their shadows will be out of focus at the focal plane.
So the loss of throughput due to them will be independent of wavelength -- i.e., ``gray'', and the pixel-to-pixel gradients of their shadows will not be large. 
Daily broadband and ``spot-checks'' at selected wavelengths with the monochromatic source will be used to measure day-to-day changes in the system passbands.  

\subsection{  Measurement of Atmospheric Transmittance, $S^{atm}$  }
\label{sec:atmo}

Many studies have shown that atmospheric transmission can be factored into the product of
a frequency dependent (``non-gray'') part that varies only on spatial
scales larger than the telescope field-of-view and temporal scales long compared with the interval between LSST exposures;
and a frequency independent part (``gray'' cloud cover) that varies on moderately short spatial scales (larger than
the PSF) and temporal scales that may be shorter than the interval between exposures:
\begin{equation}
\label{Eq:satmo}
S^{atm}(alt,az,t,\lambda)~=~S^{atm}_{g}(alt,az,t) \times S^{atm}_{ng}(alt,az,t,\lambda).
\end{equation}

\noindent The measurement strategies to determine $S^{atm}_g$ and $S^{atm}_{ng}$ are quite different:
\begin{itemize}
\item $S^{atm}_{ng}$ is determined from repeated spectroscopic measurements of a small set of probe stars by a dedicated auxiliary telescope.
\item $S^{atm}_{g}$ is determined from the LSST science images themselves, first approximately as each image is processed, and later more
precisely as part of a global photometric self-calibration of the survey.
The precise measurement of $S^{atm}_{g}$ is based on the measured fluxes of
a very large set of reference stars that cover the survey area and are
observed over many epochs.  
Every exposure contains a large enough set of sufficiently stable
stars that a spatial map can be made of $S^{atm}_{g}$ across each image.
\end{itemize}

The LSST design includes a 1.2-m auxiliary calibration telescope
located on Cerro Pach\'on near the LSST that will  
be used to measure $S^{atm}_{ng}(alt,az,t,\lambda)$.
The strategy is to measure the full spatial and temporal variation in atmospheric extinction throughout each night
independently of operations of the main survey telescope. 
This will be done by repeatedly taking spectra of a small set of probe stars as they traverse the sky each night.
These stars are spaced across the sky to fully cover the area surveyed by the LSST main telescope.
The calibration will use state-of-the-art atmospheric models
\citep{Stubbs++07} and readily available codes (MODTRAN4) to
accurately compute  
the signatures of all significant atmospheric components in these spectra.
This will allow the atmospheric mix present along any line of sight at any time to be interpolated from the measured data.
The probe stars will be observed many times during the LSST survey, so the SED of each star can be bootstrapped from the data.
The instrumental response of the spectrograph can also be bootstrapped
from the data by including stars with a variety of SEDs over a broad
range of airmass. 

\subsection{Calibration Procedure}
\label{Sec:calproc}

Two levels of LSST calibration will be carried out at differing
cadences and with differing performance targets.  A nightly data
calibration based on the best available set of prior calibrated
observations will provide ``best-effort'' precision and accuracy.
This calibration will be used for quality assurance, generation of
alerts to transients, and other
quantities appropriate for Level 1 Data Products
(\autoref{sec:design:dataproduct}). 
A more complete analysis will
recalibrate the data accumulated by the survey at periodic ``Data
Release'' dates (Level 2 in the terminology of
\autoref{sec:design:dataproduct}).  It is this repeated calibration of
the accumulated 
survey that will be held to the survey requirements for photometric
repeatability, uniformity, and accuracy.  

LSST photometric calibration is then separated into three parts that
address different science requirement specifications:
\begin{itemize}
\item Relative calibration: normalization of internal (instrumental) measurements in a given bandpass relative to all
other measurements made in the same bandpass across the sky. 
\item Absolute calibration of colors: determination of the five unique differences between flux normalizations of the six bands (color zero points).
\item Absolute calibration of flux: definition of the overall physical
scale of the LSST magnitude system, i.e., normalization to $F_{AB}$ in
\autoref{eqn:stdmag}. 
\end{itemize}

\subsubsection{Relative Calibration}

Precision relative calibration of LSST photometry will be accomplished by analysis of the repeated
observations of order $10^8$ selected bright ($17 < r < 20$) isolated stars during science operations of the survey.
The LSST image processing pipelines will extract raw ADU counts for these stars from each image,
and the data release Calibration Pipeline will process data from the calibration auxiliary subsystems to determine
the optical bandpass appropriate for each image.
These measurements will be used to determine calibrations for all sources detected on each image.

After reduction of each image in the accumulated survey,
the Calibration Pipeline will execute a global self-calibration
procedure that will seek to minimize the dispersion of the errors
in all observations of all reference stars.  This process is based on
techniques used in previous imaging surveys \citep{Gla++94, Mac++04},
and the specific implementation used by LSST will be 
based on the ``\"Ubercal'' procedure developed for SDSS
\citep{Pad++08}.  ``Calibration patches'' of order the size of a
single CCD will be defined on the camera focal plane.  
The LSST survey will dither pointings from epoch to epoch to
control systematic errors, so stars will fall on different patches on
different epochs across the sky.  The measured magnitudes of reference
stars will be transformed (\autoref{Eq:dmStd}) to the LSST standard
bandpass using the accumulated estimates of the colors of each star
and the corresponding measured observational bandpasses.  The
Calibration Pipeline will minimize the relative error $\delta_b(p,j)$
in the photometric zero-point for each patch, $p$, on each image, $j$, of
the accumulated survey by minimizing,
\begin{equation}
\label{eqn:uberchisq}
\chi^{2} = \sum_{(i,j)} {\frac{\left( m_b^{std,meas}(i,j) - \left( m_b^{std,true}(i) + \delta_b(p,j) \right) \right)^2}
                                           {\left( \delta m_b^{std,meas}(i,j) \right)^2}}, 
\end{equation} 
where the magnitudes are in the standard system, and the summation is over all stars, $i$, in all images, $j$.
These $\delta_b(p,j)$ will be used to correct the photometry for all other sources in patch, $p$, on image, $j$.


\subsubsection {Absolute Calibration of Colors and Flux} 
\label{Sec:Deltam}

There are six numbers, $\Delta_b$, for the entire survey that set the zeropoints of
the standard bandpasses for the six filters.  
These six numbers can be expressed in terms of a single fiducial band,
which we take to be the $r$ band,
\begin{equation}
             \Delta_b = \Delta_r + \Delta_{br}.
\end{equation}

The LSST strategy to measure the observational bandpass for each
source is designed to reduce errors, $\Delta_{br}$, in the five color
zero points, to meet specifications in the survey requirements.  This
process will be validated with the measured flux from 
one or more celestial sources, 
most likely hot white dwarfs whose simple
atmospheres are reasonably well-understood.

At least one external flux standard will be required to determine
$\Delta_r$ (one number for the whole survey!).  While one celestial
standard would be formally sufficient, choosing a number of such
standards would provide a powerful test for $\Delta_r$.
Identification of such a standard, or set of standards, has not yet
been done.

\section{Astrometric Calibration}
\label{sec:design:calsimAstro}

{\it David Monet, David L. Burke, Tim S. Axelrod, R. Lynne Jones, \v Zeljko Ivezi\'c}

The astrometric calibration of LSST data is critical for many aspects
of LSST operations (pointing, assessment of camera stability, etc.)
and scientific results ranging from the measurement of stellar
parallaxes and proper motions to proper performance of difference image
analysis. 

The core of the astrometric algorithm is the simultaneous solution for
two types of unknowns, the coefficients that transform the coordinates
on the focal plane measured in a given exposure into some common
coordinate system (absolute astrometry), and the 
positions and motions of each star (relative astrometry).  Whereas a
direct solution exists, 
it involves the inversion of relatively large matrices and is rarely
used. Instead, the solution is based on an iterative improvement given
the prior knowledge of positions of a relatively small number of stars
(from a reference catalog or 
similar). All observations for all stars in a small area
of sky are extracted from the database. Using the catalog positions
for the stars as a first guess, the transformations from each
observation to the catalog system are computed, and then all measures
for each star are used to compute the new values for position and
motion. 


\subsection{ Absolute Astrometry} 

The current realization of the International Celestial References
Frame (ICRF) is defined by the stars in ESA's Hipparcos mission
catalog.  ESA's Gaia mission, set to launch in 2012, will improve the
ICRS and ICRF by another two orders of magnitude down to the level of
a few micro-arcseconds.

Absolute calibration consists of computing the positions of all the
detected sources and objects in the LSST imaging with respect to the
ICRF.  Were no improved catalogs available between now and LSST
commissioning, the reference catalogs would be the US Naval
Observatory's UCAC-3 catalog for bright optical stars (down to about
16th magnitude, uncomfortably close to LSST's saturation limit) or the
NASA 2MASS catalog whose near-IR positions for optical stars have an
accuracy of 70-100 milli-arcseconds (mas) for individual stars and
systematic errors in the range of 10-20 mas.  There are large numbers
of 2MASS stars in each and every LSST field of view, so the
astrometric calibration is little more than the computation of a
polynomial that maps position on the focal plane into the system of
right ascension and declination defined by the measured positions of
catalog objects.
The transformation is
encapsulated in the World Coordinate System (WCS) keywords in the Flexible Image Transport System (FITS) header
for each image.

One of the key
astrometric challenges in generating and using these WCS solutions is
the distinction between ``observed" and ``catalog" coordinates.  When
LSST takes an image, the stars and galaxies are at their observed
positions.  These positions include the astrometric effects of proper
motion, parallax, differential refraction, differential aberration,
and others.  Most applications work in catalog coordinates such as the
J2000 positions for objects or the equivalent for image
manipulation. The astrometric calibration will provide a rigorous
method for going between these coordinate systems.

\subsection{ Differential Astrometry }

Differential astrometry is for most science the more important job to
be done. The differential
solution, which provides measures for the stellar parallax, proper
motion, and perturbations (e.g., due to binary companions), can be
substantially more accurate than the 
knowledge of the absolute coordinates of an object. The task is to
measure centroids on images and to compute the transformation from the
current frame into the mean coordinate system of other LSST data, such
as the deep image stacks or the different images from the multi-epoch
data set.  The photon noise limit in determining the position of the
centroiding of a star is roughly half the FWHM of the seeing disk,
divided by the signal-to-noise ratio of the detection of the star. 
The expectation is that atmospheric seeing will be the
dominant source of astrometric error for sources not dominated by
photon statistics.  Experiments with wide-field imaging on the Subaru
Telescope (\autoref{sec:com:PMacc}) suggest that
accuracy will be better than 10 mas per exposure in the
baseline LSST cadence, although it may be worse with objects with
unusual SEDs such that simple differential color
refraction analysis fails, or for exposures taken at extreme zenith
angles. 

Perhaps the biggest unknown in discussion of differential astrometry
is the size of the ``patch" on the sky over which the astrometric
solution is taken. If the patch is small enough, the astrometric
impact of the unaveraged turbulence can be mapped with a simple
polynomial, and the differential astrometric accuracy approaches that
set by the photon statistics.  Our current understanding of
atmospheric turbulence suggests that we will be able to work with
patches between a few and 10 arcmin in size, small enough that the
geometry can fit with low-order spatial polynomials.  The current
approach is to use the 
JPL HEALPix tessellation strategy. 
For each solution HEALPix(el),
separate spatial transformations are computed for each CCD of each
observation.  These produce measures for each object in a mean coordinate
system, and these measures can be fit for position, proper motion, parallax,
refraction, perturbations from unseen companions, and other astrometric
signals.  Given
the very faint limiting magnitude of LSST, there should be 
be a sufficient number of astrometrically useful galaxies to deliver a
reasonable zero-point within each HEALpix\footnote{Quasars will be
  less useful; they are less numerous, and their very different SEDs
  cause different refraction from stars.}. The characterization of the
zero-point errors and the astrometric utility of galaxies will be the
major work area for the astrometric calibration team.

\bibliographystyle{SciBook}
\bibliography{design/design}

\chapter[System Performance]
{LSST System Performance}
\label{chap:common}

{\it Steven M. Kahn, 
Justin R. Bankert, 
Srinivasan Chandrasekharan, 
Charles F. Claver, 
A. J.  Connolly, 
Kem H. Cook,
Francisco Delgado, 
Perry Gee, 
Robert R. Gibson, 
Kirk Gilmore, 
Emily A. Grace,
William J. Gressler, 
\v Zeljko Ivezi\'c, 
M. James Jee,
J. Garrett Jernigan, 
R. Lynne Jones, 
Mario Juri\'{c}, 
Victor L. Krabbendam, 
K. Simon Krughoff,
Ming Liang,
Suzanne Lorenz, 
Alan Meert, 
Michelle Miller, 
David Monet,
Jeffrey A. Newman, 
John R. Peterson, 
Catherine Petry, 
Philip A. Pinto, 
James L. Pizagno, 
Andy Rasmussen,
Abhijit Saha, 
Samuel Schmidt, 
Alex Szalay, 
Paul Thorman, 
Anthony Tyson,
Jake VanderPlas, 
David Wittman}

In this chapter, we review the essential characteristics of the LSST system performance.  We begin with descriptions
of the tools that have been developed to evaluate that performance:  the 
Operations Simulator (\autoref{sec:design:opsim}), the Exposure Time Calculator 
(\autoref{sec:com:expos}), the Image Simulator (\autoref{sec:design:imsim}), and raytrace calculations used to
evaluate stray and scatter light (\autoref{sec:common:scattered}).
We then discuss the expected photometric accuracy that will be achieved
(\autoref{sec:photo-accuracy}), and the expected accuracy of trigonometric
parallax and proper motion measurements (\autoref{sec:com:PMacc}).   
Next, we provide
estimates of discrete source counts in the main LSST survey, both 
for stars in the Milky Way
(\autoref{Sec:stellarCounts}), and for galaxies as a function of redshift
(\autoref{sec:common:galcounts}).  We conclude with a discussion of the accuracy
with which redshifts of galaxies can be determined from LSST photometry (\autoref{sec:common:photo-z}).  

\section{Operations Simulator}
\label{sec:design:opsim}
{\it Philip A. Pinto, R. Lynne Jones, Kem H. Cook, Srinivasan
  Chandrasekharan, Francisco Delgado, \v Zeljko Ivezi\'c, Victor L. Krabbendam,
  K. Simon Krughoff, 
  Michelle Miller, Cathy Petry, Abhijit Saha}

During its ten-year survey, LSST will acquire $\sim5.6$~million
15-second images, 
spread over $\sim2.8$~million pointings.
Their distribution on the sky, over time, and among its six filters has a strong
impact on how useful these data are for almost any astronomical
investigation. The LSST Project has developed a detailed operations 
simulator (LSST OpSim : \url{http://www.noao.edu/lsst/opsim}) 
in order to develop algorithms for scheduling these exposures --
addressing the question ``what observation should be made next?'' -- and
to quantitatively evaluate the observing strategies discussed in 
\autoref{sec:design:cadence}. These algorithms will become a fundamental
component of the LSST design, as part of the scheduler driving the largely robotic observatory.
In addition, the simulator will remain an important tool allowing
LSST to adapt and evaluate its observing strategy in response to the changing 
scientific demands of the astronomical community.

The operations simulator incorporates detailed models of the site conditions and
hardware performance, as well as the algorithms for
scheduling observations.  It creates realizations of the set of visits
(back-to-back 15 second exposures in a given filter) that the LSST will make
during a ten-year survey, this being the primary output of the OpSim. 
These outputs include the position on the sky, time, and filter of each
visit, and the signal-to-noise ratio achieved. These outputs can be further
processed to generate estimates of the depth of the final stacked
images in each filter as a function of position on the sky (\autoref{fig:design:rband}),
histograms of the airmass distribution of visits (\autoref{fig:common:opsimairmass}),
or other figures of merit relevant to particular science goals. 

The simulation of observing conditions includes a model for seeing
drawn from observing records at Cerro Tololo (\autoref{fig:design:seeing}).
This model is consistent with the auto-correlation
spectrum of seeing with time over intervals from minutes to seasons as
measured at the site on Cerro Pach\'on. Weather data, including their
correlations, are taken from ten years of hourly measurements made at nearby
Cerro Tololo. 
The $5\,\sigma$ PSF depth of each observation is determined using a sky background
model which includes the dark sky brightness in each filter passband, the
effects of seeing and atmospheric transparency, and an explicit model for
scattered light from the Moon and/or twilight at each observation.

The time taken to slew the telescope from one observation to the next is given
by a detailed model of the camera, telescope, and dome. It includes such effects
as the acceleration/deceleration profiles employed in moving in altitude,
azimuth, camera rotator, dome azimuth, and wind/stray light screen altitude, the
time taken to damp vibrations excited by each slew, cable wrap, and the time
taken for active optics lock and correction as a function of slew distance,
filter changes, and focal plane readout. The detail of this model ensures an
accurate relation between system parameters and modeled performance, making the
operations simulator a valuable tool for optimizing design.

After each visit, all possible next
visits are assigned a score according to a set of scientific requirements, which
depend upon the current conditions and the past history of the survey. For
example, if a location in the ecliptic has been observed in the $r$-band, the
score assigned to another $r$-band visit to the same location will initially be
quite low, but it will rise with time to peak about an hour after the first
observation, and decline thereafter. This results in these observations being
acquired as pairs of visits roughly an hour apart, enabling efficient
association of near-Earth object (NEO) detections. To ensure uniform sky coverage, locations on the
sky with fewer previous visits will be scored more highly than those observed
more frequently. Observations with higher expected signal-to-noise
ratio are ranked
more highly, leading to a majority of visits being made near the local meridian,
and a tendency for visits in redder bands to be made closer to twilight and at
brighter phases of the Moon. Higher scores are given to observations in the $r$-
and $i$-bands during periods of better seeing to aid in shape determination for
weak lensing studies.

Once all possible next visits have been ranked for scientific priority, their
scores are modified according to the cost of making the observation. Visits to
locations which require more slew time are penalized, as are those which require
filter changes, unwrapping cables in the camera rotator, and so on. After this
modification according to cost, the highest-ranked observation is performed, and
the cycle repeats. The result of a simulator run is a detailed history of which
locations have been observed when, in what filter, and with what sky backgrounds,
airmass, seeing, and other observing conditions. A detailed history of all
telescope, dome, and camera motions is also produced for engineering studies.

Each of the two exposures in a visit requires 16 seconds to complete; while
every pixel is exposed for 15 seconds, the shutters require one second to
traverse the entire 63 cm of the active area in the focal plane. Two seconds are
required to read out the array between exposures. After the second exposure, a
minimum of five seconds is required to slew to an adjacent location on the sky,
settle, and acquire active optics lock and correction, during which time the
array is read out for the second time. Thus a complete visit to an adjacent
field, with no filter change, takes a minimum of 39 seconds to perform; this
amounts to spending 87\% of the time exposing the detector to the sky. This, of
course, does not take into account the time spent in changing filters (two
minutes per change) or any of the scientific requirements on
scheduling.  In one specific realization of the full ten-year survey,
80\% of the available time (i.e., when 
weather permitted) was spent exposing on the sky, which is about 92\%
of the na\"ive estimate above.  

\begin{figure}[t]
  \begin{center}
  \includegraphics[width=\textwidth]{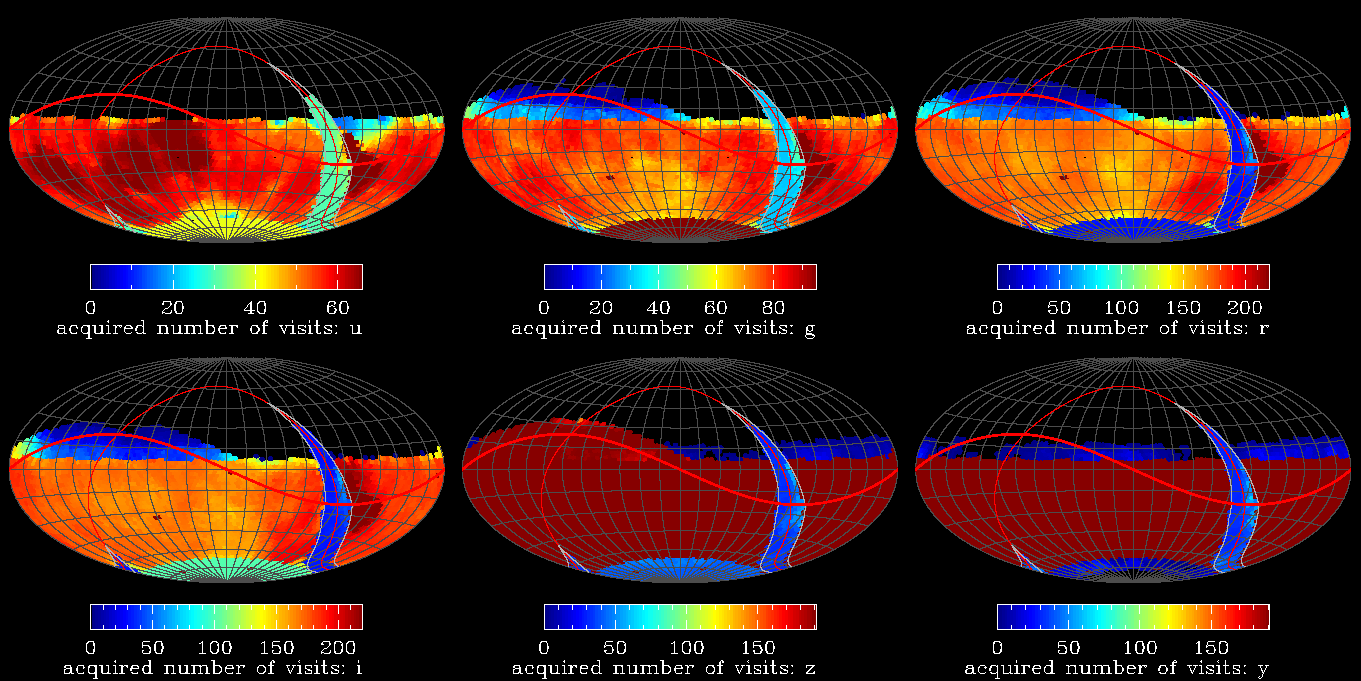}
  \caption{The number of visits in one realization of a simulated ten-year survey in all six LSST
    filters, shown in Equatorial coordinates.
    The project goals are to have 56, 80, 180, 180, 164, and 164 visits in the
    $u$, $g$, $r$, $i$, $z$, $y$ filters, respectively, over 20,000
    deg$^2$ of sky. One of the deep-drilling field is apparent at
    $\alpha=90^\circ,\delta=-32^\circ$.}
  \label{fig:common:opsimnvisit}
  \end{center}
\end{figure}

\begin{figure}[t]
  \begin{center}
  \includegraphics[width=\textwidth]{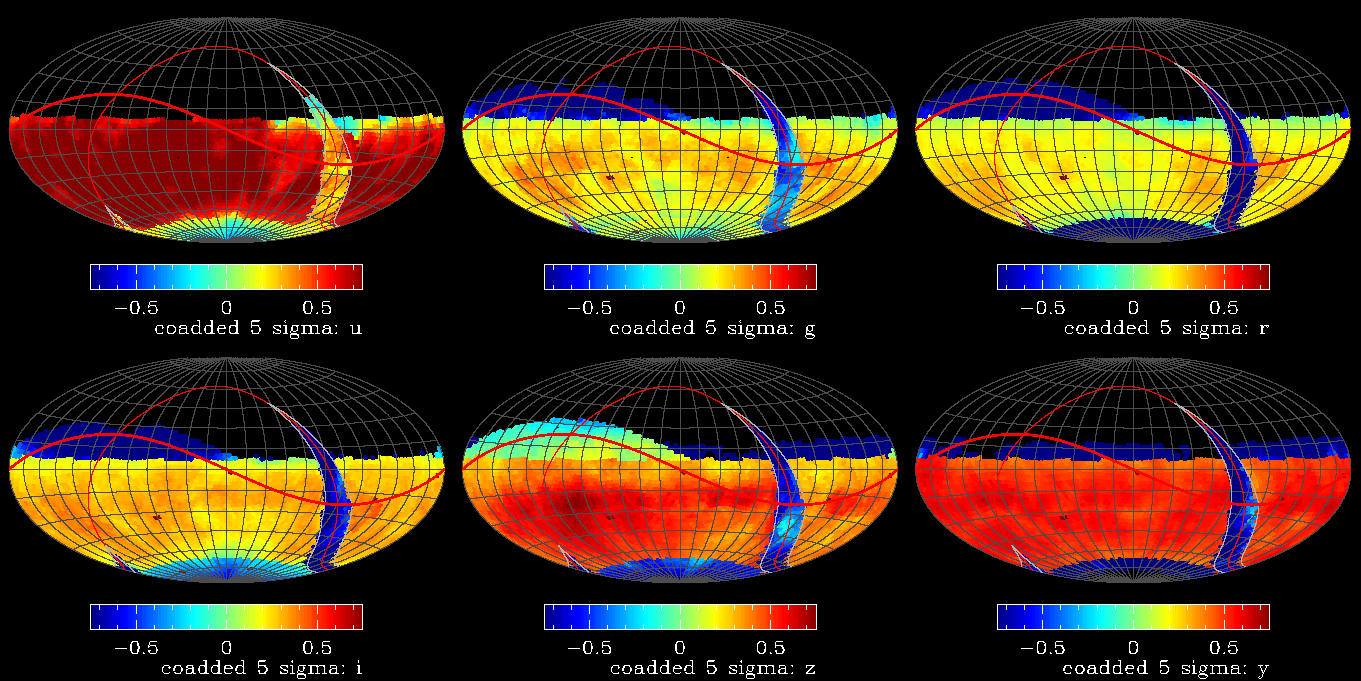}
  \caption{The $5\,\sigma$ stacked point-source depth of the simulated ten-year survey 
   shown in \autoref{fig:common:opsimnvisit}. The scale in each panel
   shows the depth of the stack relative to the fiducial values of
   25.8, 27.0, 27.2, 27.0, 25.7, and 24.4 in $u$, $g$, $r$, $i$, $z$,
   $y$ respectively.} 
 \label{fig:common:opsimcoaddeddepth}
 \end{center}
\end{figure}

\begin{figure}[t]
  \begin{center}
  \includegraphics[width=\textwidth]{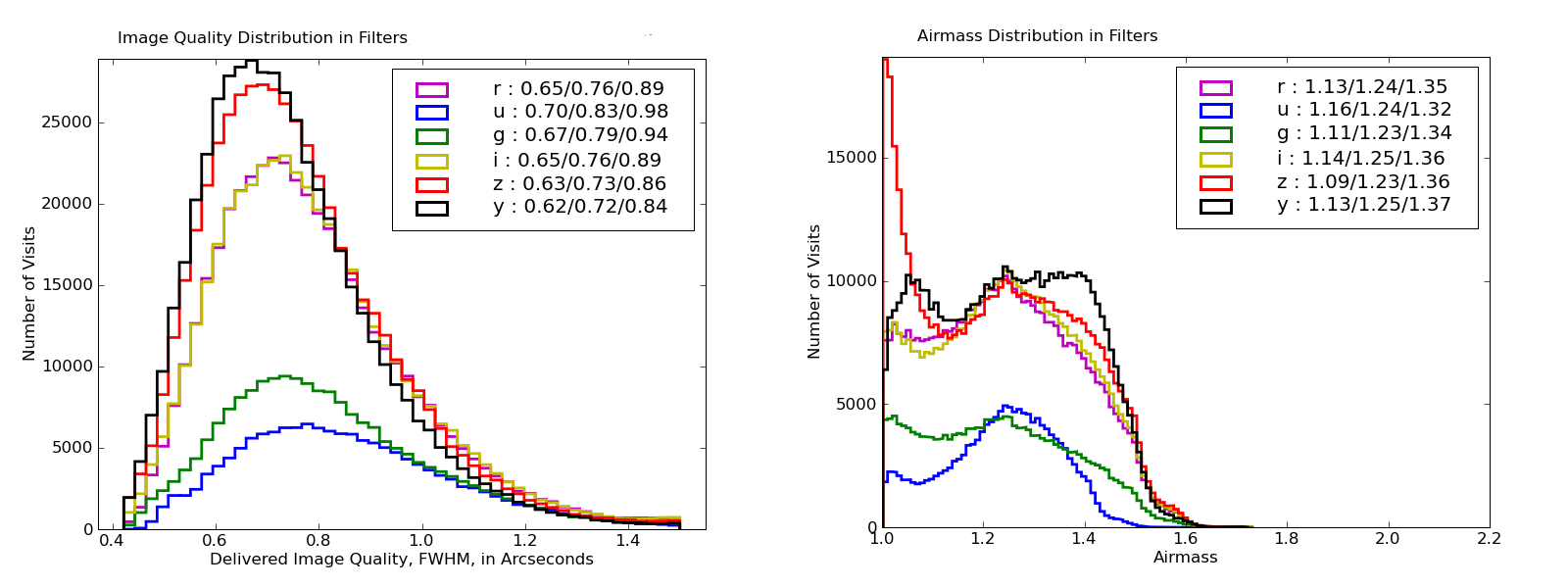}
  \caption{Histograms of the delivered seeing and airmass distributions for all
    visits in the simulated survey shown in
    \autoref{fig:common:opsimnvisit}. Also shown in the legend are the
    25-, 50-, and 75-percentile values in each filter.}
  \label{fig:common:opsimairmass}
 \end{center}
\end{figure}

\autoref{fig:common:opsimnvisit} shows the number of visits across the
sky in this simulation, while \autoref{fig:common:opsimcoaddeddepth} shows the 5 $\sigma$ limiting
magnitude for point sources achieved in the stacked images. \autoref{fig:common:opsimairmass} shows a 
histogram of the air-mass and seeing delivered during observations in each filter.

\begin{figure}[ht]
\begin{center}
\includegraphics[width=0.6\textwidth]{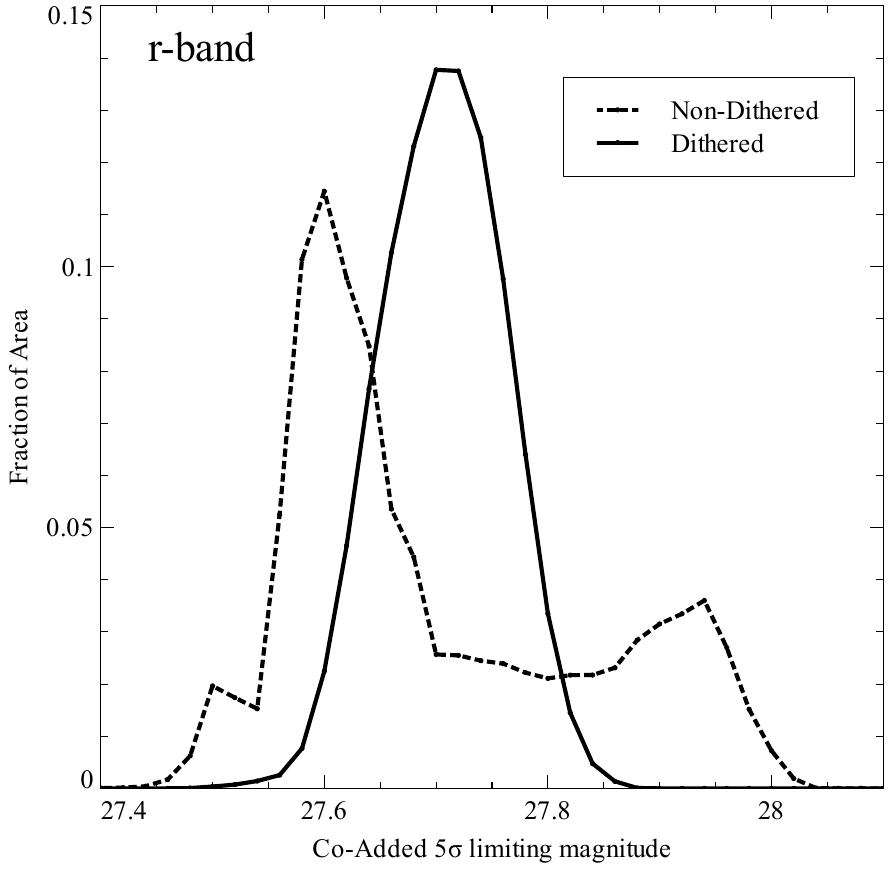}
 \caption{Histogram of the $r$ band $5\sigma$ limiting magnitude of the
   ten-year stacked image depth.  The
   histogram represents the stacked limiting magnitude evaluated over
   the full survey footprint on a 
   grid with resolution of $0.6'$, much finer than the grid of
   field centers. The dashed line indicates the $5\sigma$ stacked
   depth in the non-dithered simulation, with two discrete peaks where
   fields are not overlapped (peak near 27.6 mag) and where they do
   overlap (peak near 27.9). The solid line indicates the $5\sigma$
   stacked depth evaluated in the same simulation, with dithering
   added to each field's central position.  Dithering increases the
   median $5\sigma$ stacked depth by approximately 0.2 magnitudes. }
\label{fig:common:opsimdith}
\end{center}
\end{figure}

The current output from the OpSim assumes each visit is taken with the
field centers placed onto a fixed grid on an optimally packed
tessellation.  This gives a variation of the effective depth across
the sky, as is shown in the dashed line in
\autoref{fig:common:opsimdith}.  
To evaluate the effects of dithering on LSST performance, we
simply added a small ($<0.5$ times the field of view) dithering
pattern to the position of each pointing, leaving other aspects of the
simulation unchanged.  We added a
different offset in right ascension (RA) and declination (dec) for each night of the survey, following 
a pattern which stepped through a hexagonal grid on the field of
view. This dithering makes the coverage substantially more uniform, as is shown
by the solid line in \autoref{fig:common:opsimdith}.  

We are continuing to work on developing improved scheduling algorithms,
replacing the algorithm which simply observes the field with the
highest score at each step with one which looks ahead
for a few hours, using a path optimization algorithm to further reduce the slew
time required, and including more feedback from science metrics (on already acquired data) into the 
scheduling algorithms. We are also working with the LSST Science Collaborations to
refine our current cadences to enhance the utility of the LSST data set for
the widest possible applicability.

\section{Exposure Time Calculator}
\label{sec:com:expos}

{\it Anthony Tyson, Perry Gee, Paul Thorman}

In order to enable fast predictions of the signal-to-noise ratio for
the detection of both point and diffuse sources, we have developed 
an Exposure Time Calculator (ETC; \url{http://lsst.org/etc}). 
The ETC incorporates models of the extinction, telescope and camera optics, 
detector efficiency, and filter response to calculate the throughput of 
the system in each band.  It uses a sky brightness model 
based on data taken at CTIO, United Kingdom Infra-red Telescope (UKIRT), and SDSS.   

An input source model is shifted to 
the correct redshift and normalized to a selected brightness or surface 
brightness. The resulting flux density is multiplied by the system 
response as a function of wavelength for a given filter band to produce a predicted 
photon count-rate within a specified aperture. The integral sky 
brightness is also calculated for the same aperture, so that the signal-to-noise ratio for
detection can be calculated. The aperture is fully adjustable, and an option for PSF-weighted photometry 
is also provided.

The ETC allows 
the source spectral energy distribution, surface brightness profile, the extinction, and 
the redshift to be varied, and includes a library of stellar and 
extragalactic source spectra. 
For specified seeing, Moon and cloud conditions, 
and for multiple exposures of a specified time and fraction of exposure 
on source, the ETC estimates both the signal-to-noise ratio for a single 
exposure and the exposure time required to achieve a threshold 
signal-to-noise ratio. 

In \autoref{tab:common:etc} we provide the predicted signal-to-noise
ratios (S/N) for some canonical source types. For each object, we
quote S/N based on a single visit, and on the full ten-year survey.  The calculations in the table
are based on $z=0$ template spectra of galaxies, quasars, and stars,
without evolution (although the absolute magnitudes of galaxies at each redshift
are rough estimates of $M^*$).  The quoted S/N includes
sky subtraction and PSF-optimized filtering for galaxies of typical
angular size at the given redshift, but no provision for other
systematic errors (thus values of S/N more than several hundred should
be taken with a grain of salt).  The sky background was estimated assuming
three-day-old lunar phase and solar minimum sky levels. The seeing was
assumed to be $0.7''$ in $r$-band with clear skies.  For the
high-redshift quasars, no entries are given below the Lyman limit; the
flux is taken to be essentially zero at shorter wavelengths. 

The ETC also allows estimates of the saturation limits of the LSST
camera.  In $0.7''$ seeing under photometric skies, and for a 15 sec exposure, the detectors will
saturate with a star of $u,g,r,i,z,y = 14.7, 15.7, 15.8, 15.8, 15.3$
and 13.9, respectively.  

\begin{sidewaystable}
\caption{Typical apparent magnitudes and Signal-to-Noise ratios, S/N}
\label{tab:common:etc}
\begin{tabular}{|l|c|c|c|c|c|c|c|c|c|c|c|c|}
\hline
\multicolumn{ 1}{|c|}{Object} & \multicolumn{ 2}{|c|}{$u$} & \multicolumn{ 2}{|c|}{$g$} & \multicolumn{ 2}{|c|}{$r$} & \multicolumn{ 2}{|c|}{$i$} & \multicolumn{ 2}{|c|}{$z$} & \multicolumn{ 2}{|c|}{$y$} \\ \hline
\multicolumn{ 1}{|r|}{} &  mag &  S/N         & mag & S/N          & mag &    S/N       & mag &  S/N         & mag & S/N           & mag  & S/N      \\
\multicolumn{ 1}{|r|}{} &          &  visit/full &         & visit/full  &        &   visit/full &         &  visit/full &         &visit/full    &          &visit/full \\ \hline
Stars                            &         &                 &         &                 &        &                 &          &                &         &                  &           &               \\ 
        O5V, 100 kpc      &  18.6 &190/1600 & 19.0 & 250/3500&19.6 & 240/3600 & 20.0 &140/2200 & 20.2 & 65/930     & 20.5  &  17/250 \\ 
        A0V, 100 kpc      &  21.0 &    46/380 & 20.3 & 170/1700&20.5 & 140/2100 & 20.7 &86/1300   & 20.9 & 39/560     &  20.6 & 13/180 \\
       G2V, 100 kpc       &  25.9 &    0.8/6.6 & 24.7 & 5.6/56     &24.3 &  6.6/100   & 24.1 & 4.6/69     & 24.1 &  2.2/31     &  24.0 & 0.7/9.9 \\ 
      K4III, 100 kpc        & 23.5 &    6.6/55  & 20.6 &154/1500 &19.4 & 260/3900 & 19.0 & 270/4000& 18.8 & 200/2800 & 18.3  & 95/1300 \\ 
        M3V, 1 kpc          & 25.6 &    1.1/8.9 & 23.0 & 23/230    &21.8 & 55/830     & 20.6 & 92/1400  & 20.0 & 79/1100   & 19.3  & 42/590   \\ 
L1 Dwarf, 500 pc         &    -    &      -        &   -    &        -       & 27.1& 0.48/7.4  & 24.8 & 2.4/37     & 23.2 & 4.8/69      &  22.2 & 4.2/59    \\ \hline
Elliptical                      &         &                 &         &                 &        &                 &          &                &         &                  &           &               \\ 
$z=0.5,M_B=-20.8$    & 24.2 &    3.4/28  & 22.8 & 31/330    & 21.1 & 91/1400 &   20.2 &130/1900 &19.8 & 93/1300   & 19.5   & 47/670  \\ 
$z=1, M_B=-21.3$      & 25.3 &     1.3/11 & 25.0 &  4.2/45    & 23.7 & 11/170   &   22.6 & 17/260    &21.6 & 20/280     &   21.0 & 13/190   \\
$z=2, M_{B}=-21.9$     & 25.7 &  0.9/7.3  & 25.6 &  2.2/23     & 25.6 & 2.0/30    &  25.2 & 1.6/24     & 24.5 &1.4/20       &   23.6 & 1.2/17    \\ \hline
Spiral                           &          &               &          &                 &         &                &          &                &         &                  &           &                 \\ 
$z=0.5, M_B=-20.9$   & 22.6 &   14/120 &  22.0 & 59/630    &  20.8 &110/1700& 20.2 &120/1800 & 20.2 & 67/950    &   19.5 & 43/610    \\
$z=1, M_{B}=-21.2$    &  23.7 &    5.7/47 &  23.6 & 15/160    &  23.0 &19/290    & 22.1 & 26/390    & 21.7 & 18/250    &   21.0 & 12/170    \\
$z=2,M_B=-21.9$       & 23.6 &  6.9/57   &   24.0 & 9.9/110  &  23.9 & 8.8/130   & 23.7 & 6.3/95     & 23.5 & 3.7/52     &   23.1 & 1.9/27    \\ \hline
Quasar                        &         &                 &            &                &          &                &         &                 &         &                &            &                \\ 
$z=1, M_B=-24.4$     & 18.8 &170/1400 & 18.7 & 430/4600 & 18.5 & 430/6500 & 18.5 &360/5400 & 18.4 & 250/3500 &   18.2 &140/2000 \\
$z=3$, $M_{B}=-26.0$ & 22.8 & 14/120  & 21.3  & 93/930    & 20.8 & 110/1700 & 20.7 & 84/1300  & 20.7 & 44/620     & 20.5    & 19/270    \\
$z=5$, $M_{B}=-26.0$ &    -    &     -         & 26.9  & 0.7/7.2    & 24.1 & 7.3/110    & 21.6 & 41/630    & 21.3 & 26/370     & 21.2   & 10/140      \\
$z=7$, $M_{B}=-26.0$ &     -   &   -           &  -       &    -          &   -     &      -        &  -      &     -         &  -      &      -         &     21.4 & 9.1/130 \\
\hline
\end{tabular}
\end{sidewaystable}

\section{Image Simulator}
\label{sec:design:imsim}

{\it John R. Peterson, J. Garrett Jernigan, 
Justin R. Bankert, 
A. J. Connolly, 
Robert R. Gibson, 
Kirk Gilmore, 
Emily A. Grace,
M. James Jee,
R. Lynne Jones, 
Steven M. Kahn, 
K. Simon Krughoff,
Suzanne Lorenz, 
Alan Meert, 
James L. Pizagno, 
Andrew Rasmussen} 

\begin{figure}[ht!]
  \centerline{\resizebox{!}{5in}{\includegraphics{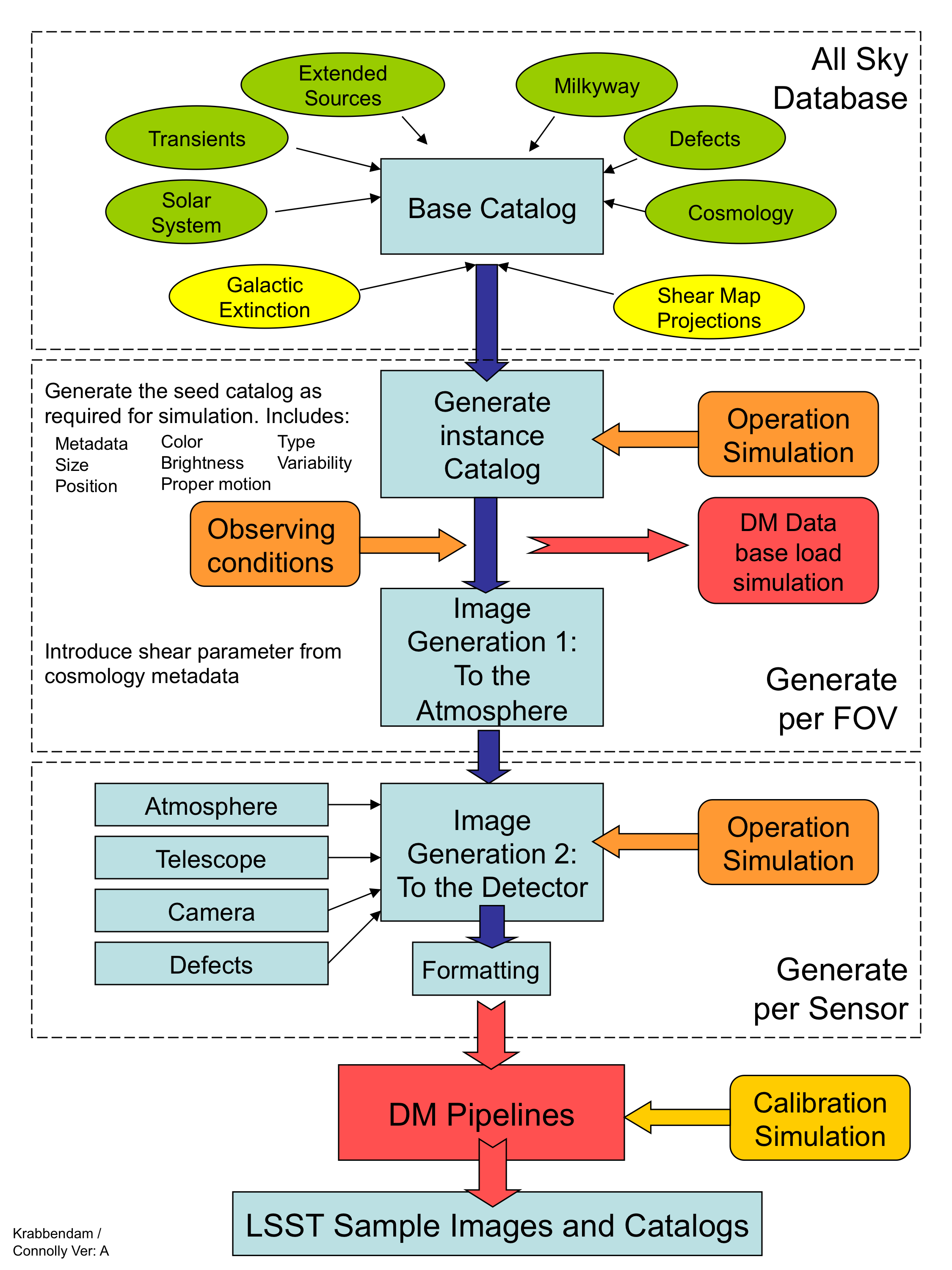}}}
  \caption{The flow of information through the image simulation. The
    top level describes a static view of the sky that is sampled to
    provide instance catalogs (based on the operations simulations, \autoref{sec:design:opsim}). These
    catalogs are then passed into the Image Simulator resulting in a
    set of FITS images and catalogs. \label{fig:flow}}
\end{figure}

The project team has developed a detailed Image Simulator (\url{http://lsst.astro.washington.edu}) to evaluate the system sensitivity for particular science
analyses, and to test 
data analysis pipeline performance on representative mock data sets.  The
simulated images and catalogs that it produces extend to $r = 28$ (deeper than the expected
ten year depth of the LSST stacked images).  These have proven useful in designing
and testing algorithms for image reduction,
evaluating the capabilities and scalability of the
analysis pipelines, testing and optimizing the scientific returns of
the LSST survey, and providing realistic LSST data to the science
collaborations.
\autoref{fig:flow} shows the flow of data through the LSST
simulation framework. 

\begin{figure}
\begin{center}
\includegraphics[width=1.0\linewidth]{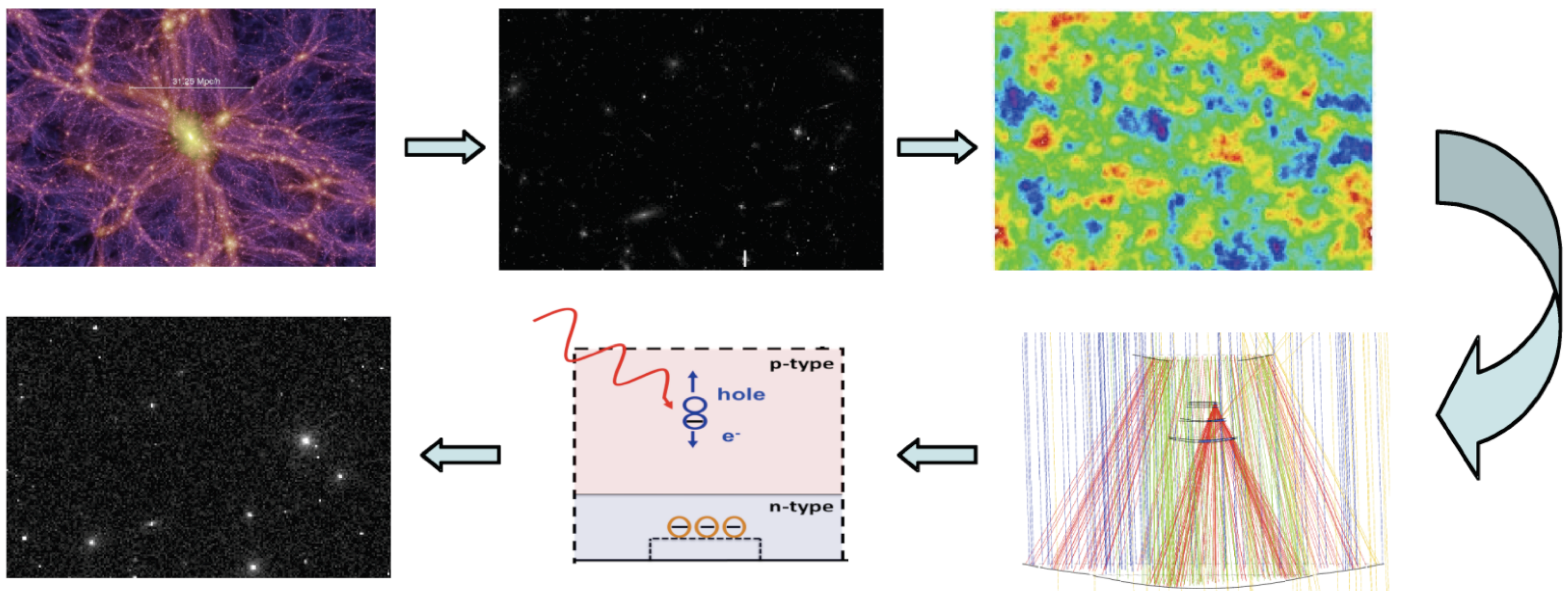}
\caption{
A schematic of the key steps leading to the production of a simulated
image.  First, a cosmological simulation is used to produce a three-dimensional dark
matter map of a limited region of sky (upper left).  This is then decorated with
galaxies, which, along with a set of stars generated from an
associated Milky Way model, are collected into a catalog of objects in
the field (upper middle).  This catalog is sampled to generate Monte
Carlo photons in angle and color, which are propagated through a set
of turbulent atmospheric screens (upper right) that move as a function
of time according to input wind velocity vectors. Photons are then
reflected and refracted through the mirrors and lenses of the LSST
optics with an assumed set of displacements and distortions (lower
right), and propagated into the detector (lower middle) where they
convert to photoelectrons detected in a pixel.  Background sky counts
are added to produce the final simulated image of a single 15-second
exposure at the lower left.
}
\label{fig:imsim5}
\end{center}
\end{figure}

The image simulator \citep{peterson05}
is a set of fast codes that begins with a catalog of objects (possibly lensed), and then traces photons 
through the atmosphere and the refractive and reflective optics, and into the detector where they
photo-convert into electrons. The
simulation can have extremely high fidelity in that all
wavelength-dependent optics, detector, and atmospheric effects can be readily
incorporated. The code is also sufficiently fast that a single 15-second
3.2 Gigapixel image from the LSST camera can be simulated in $\sim $ 6 or 7 hours 
using a large pool of parallel machines. 

The simulator constructs catalogs of objects drawn from
cosmological and Galactic structure models ({\em base catalogs}), 
which are then used to generate a view of the sky
above the atmosphere. These base catalogs
include the
spectral, photometric, astrometric, and morphological properties of the
astronomical sources. 
The base catalogs are queried based on simulated observation sequences
from the Operations Simulator (\autoref{sec:design:opsim}) creating a
series of {\em instance catalogs}. 
Given the time, RA and Dec of each
pointing, the appropriate airmass, sky background, and observing
conditions can be determined. 

Each object in the catalog has a sky position, magnitude at some wavelength, and spectral energy
distribution (SED) file to determine the probabilities with which wavelengths are assigned to photons. Each object
can either be represented by a set of parameters describing a spatial
model or a hyper-resolved image to determine the probability
distribution of where the photons are emitted. Additionally, objects
can vary in flux during the exposure, they can move during the
exposure (in the case of Solar System objects), or can be distorted 
due to gravitational lensing. Photons are drawn from this catalog in
proportion to their magnitude and both the SED and spatial model are
sampled. In this way, photons are chosen one at a time with two sky
positions, a wavelength, and a time.

Galaxy positions and properties in the simulations are taken from the
Millennium cosmological Simulation, with baryon physics included
following \citet{K+W07}.  Galaxy SEDs use \citet{bruzual03} models,
giving apparent magnitudes in all the LSST bands.  
Every galaxy is further assigned 
a realistic morphological profile via a disk-to-total flux ratio, position angle in the sky, inclination
along the line-of-sight, bulge radius, and disk radius. More accurate galaxy
profiles, including high-frequency spatial structure such as H~II regions and
spiral arms, can be simulated using FITS images as input into the Image
Simulator. 
The use of more detailed galaxy morphological profiles in the Image
Simulator will allow LSST to study how galaxy morphology varies with
environment and redshift.

Currently, stars are included in the Image Simulator with full SEDs, spatial velocities, and positions. 
The SEDs for stars are derived from Kurucz models. The model used to
generate main sequence stars is based on work done by Mario Juri\'c
and collaborators. The model includes user-specified amounts of thick-disk,
thin-disk, and halo stars. Each version of a catalog contains metadata
on metallicity, temperature, luminosity-type, and surface gravity,
allowing the user to search for correlations between observed LSST
photometry and physical information about stars using the simulated
data. The catalog will be updated to include dwarf and giant stars. 

After the photons are selected from the astronomical source list, they are propagated through the atmosphere and are 
refracted due to atmospheric turbulence.  The model of the
atmosphere is constructed by generating roughly half a dozen atmospheric screens as illustrated
in \autoref{fig:imsim5}. These model screens incorporate density fluctuations following a
Kolmogorov
spectrum, truncated both at an outer scale
(typically known to be between 20 m and 200 m) and at an inner scale
(representing the viscous limit). In practice the inner scale does not
affect the results. 
 The screens are moved during the
exposure according to wind velocity vectors, but, consistent with the well-established
``frozen-screen approximation,''  the nature of the
turbulence is assumed to stay approximately fixed during the relatively short time
it takes for a turbulent cell to pass over the aperture. With these screens, we
start the photons at the top of the atmosphere and then alter their
trajectory according to the refractions of the screen at each
layer. 
The part of the screen that a given photon will hit depends on the time that
photon is emitted in the simulation and the wind vector of the screen.

After passing through the atmosphere, photons enter the
telescope and are appropriately reflected and refracted as they hit the mirrors
and lenses. On the surface of the mirrors we introduce a spectrum of perturbations
that has been obtained by inverting wavefront data from existing
telescopes. We also allow the orientation of each optic to be perturbed in six
degrees of freedom within expected tolerances. The raytrace uses
extremely fast techniques to find the intercepts on the aspheric
surface and alter the trajectory by reflection or wavelength-dependent
refraction. Photons can be ``destroyed'' as they pass through 
the filter in a Monte Carlo sense with a probability related to the wavelength and
angle-dependent transmission function. 
The raytrace for the LSST
configuration is illustrated in \autoref{fig:imsim5}. 
The raytrace has been compared with
commercial raytrace codes and is found to be accurate to a fraction of a micron. 
We also incorporate diffraction spikes
appropriate for the design of the spider of the telescope.

In the last step, photons are ray-traced into the silicon in the
detector. Both the wavelength and temperature dependent conversion
probability and refraction at the interface of the silicon are included. The photons are
then converted into photoelectrons which drift to the readout electrodes
according to the modeled electric field profile. 
The
misalignments and surface roughness of the silicon can also be
included. The positions of the photoelectrons are pixelated and can
include blooming, charge saturation, cross-talk, and charge transfer
inefficiency to simulate the readout process. Finally, a
simulated image is built as the photoelectrons reach the readout. The read
noise and sky background are added in a post-processing step.  The sky background is
generated based on an SED for the full Moon and an SED for the dark sky, with an
empirically derived angular function for the Rayleigh scattering of the Moon's
light.  The background is vignetted according to the results of raytrace
simulations.

\begin{figure}[t]
\centerline{\resizebox{4in}{!}{\includegraphics{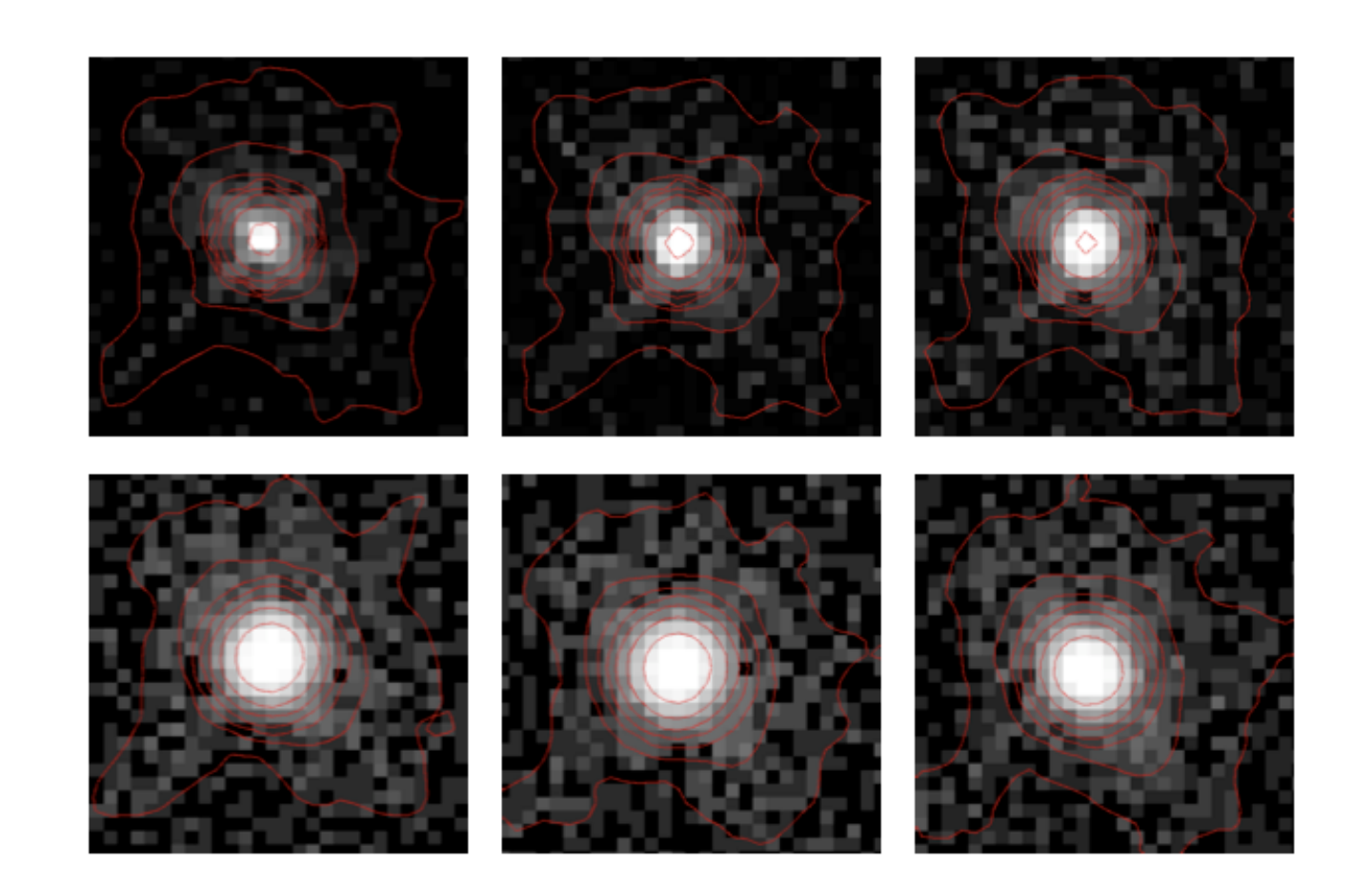}}}
\caption{
  The capabilities of the simulator are demonstrated by examples of
the point-spread function (PSF) for a single star $1.32^\circ$ off-axis
seen in the $r$ filter, in which various components of the simulator
are turned on or off. The images show a region $5.6''$ on a side, and
the stretch is logarithmic. The panels are from 
top left to bottom right: only optical aberrations of the system
design, adding mirror perturbation and misalignments, adding diffusion
of charge in the detector, adding a static atmosphere, adding an
atmosphere with wind, and a full atmosphere but a perfect
telescope. Both atmosphere and the optics contribute to the
ellipticity of the PSF. The FWHM of the PSF with telescope, atmosphere
and wind is about $0.6''$, with an ellipticity of 7\%. 
\label{fig:imsim3}}
\end{figure}

The simulator can
generate about 22,000 photons per second on a typical workstation. For
bright stars that saturate, it can simulate photons much faster since
tricks can be used during the simulation to figure out if a pixel will
saturate. Thus, we have a fast simulator with high fidelity.  \autoref{fig:imsim3}
shows images of stars with various components of the simulator turned
on or off.  \autoref{fig:imsim6} shows a simulated image from one LSST CCD.

\begin{figure}[th!]
\ifthenelse{\boolean{final}}
{\centerline{\resizebox{6in}{!}{\includegraphics{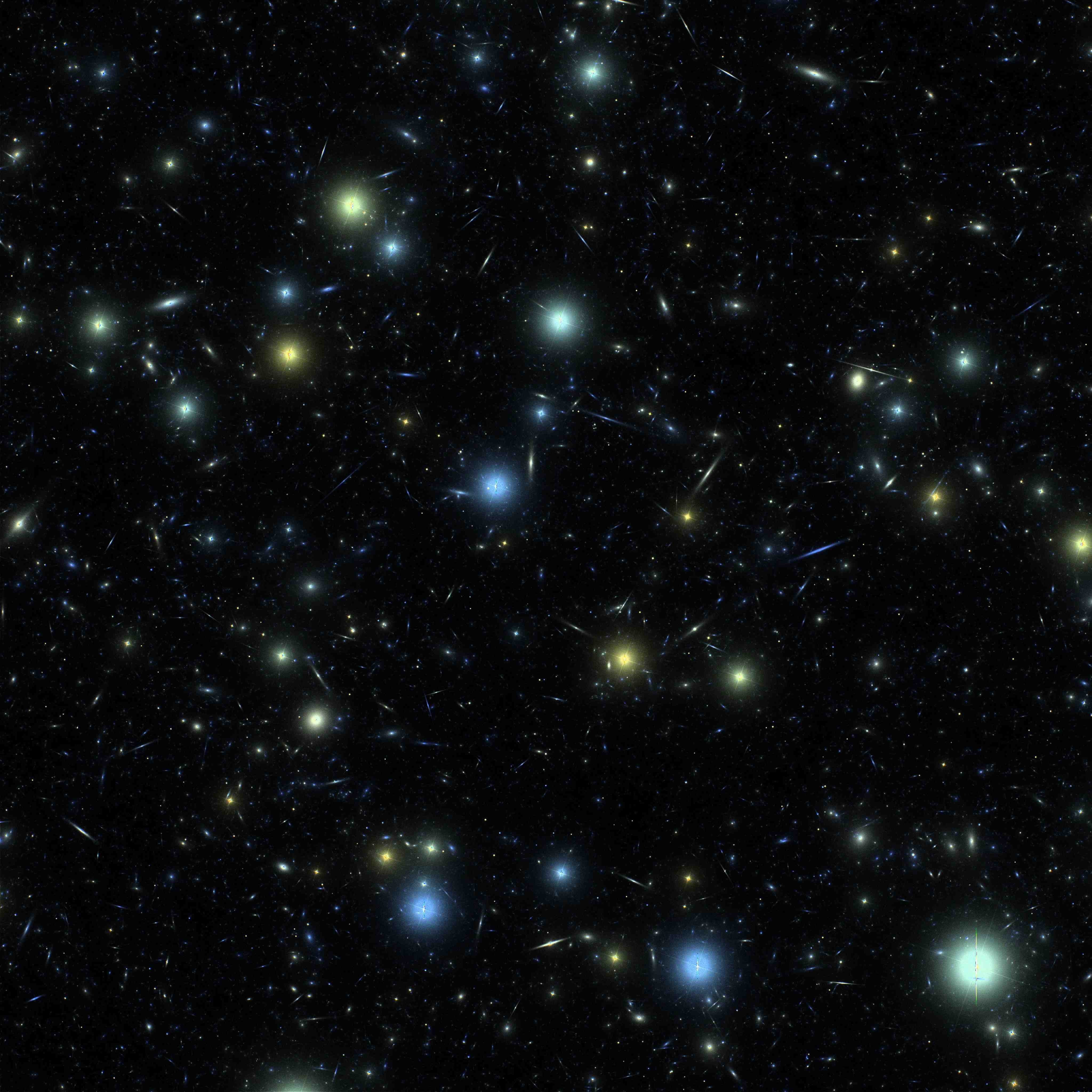}}}
\caption{A simulated image of a 15-second exposure of one LSST CCD
  (4K$\times$4K) with $0.2''$ pixels, $0.4''$ seeing  
and a field of view $13.7 ' \times 13.7'$, representing
roughly 0.5\% of the LSST focal plane.   
The brightest stars in 
the image are $\sim 12$ magnitude. An object of brightness $\sim 33$ magnitude would 
emit $\sim 1$ photon in a 15 second exposure. 
The image is a true color composite of three images, with the $g,  r$,
and $i$  
filters mapped into B, G, and R colors respectively. 
Each color channel is on a logarithmic intensity scale.
In its ten-year survey, LSST will produce $\sim2\times10^9$
single-band images of the same size. 
\label{fig:imsim6}}}
{\centerline{\resizebox{6in}{!}{\includegraphics{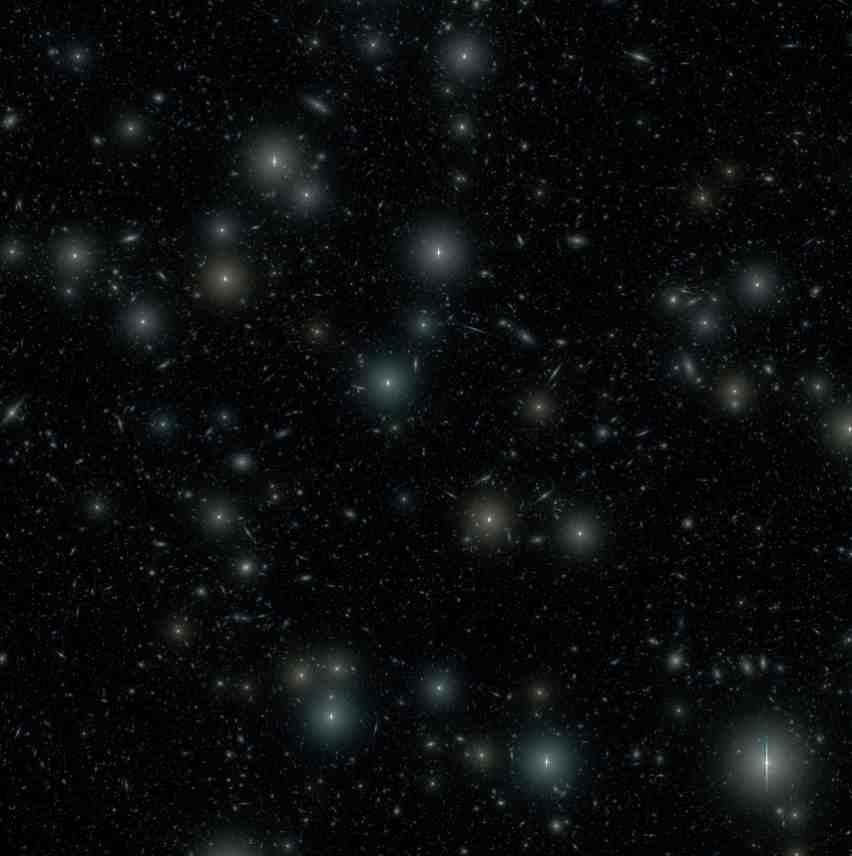}}}
\caption{The small simulated image above is one LSST CCD (4K$\times$4K) with 0.2" pixels, 0.4'' seeing 
and a field of view $\sim$13.7 ' $\times \sim13.7$'. This CCD is one of 145 fully illuminated 
CCDs that comprise the LSST focal plane. The image includes all photons ($\sim 3 \times 10^{10}$) 
for a 15 second exposure emitted by stars and galaxies in the base catalog. The brightest stars in 
the image are $\sim$12 magnitude. An object of brightness $\sim$33 magnitude would 
emit $\sim$1 photon in a 15 second exposure. The catalog includes objects as faint as
 $\sim$43 magnitude. The image is a true color composite of three images ({\it g, r} and $i$  
filters mapped into B, G and R colors respectively: display RGB color table). 
Each color channel is on a logarithmic intensity scale.
LSST will produce $\sim2\times10^9$ single-band images of this size.
\label{fig:imsim6}}}
\end{figure}

\section{Stray and Scattered Light}
{\it Charles F.  Claver, Steven M. Kahn, William J. Gressler, Ming Liang}
\label{sec:common:scattered}

Stray and scattered light is a major concern given the extremely large field of view of LSST.  
There are two major categories of stray
and scattered light:  structured and diffuse.
Structured stray light comes from diffraction, ghosts
from the refractive optics, and semi-focused scattering from surfaces
nearby the focal plane.  Diffuse scattered light is generated from the
micro-surface qualities of the mirrors, dust on the optical surfaces,
and scattering from non-optical surfaces in the telescope and dome. 

\subsection{Structured Stray Light}

\begin{figure}[t]\centering\includegraphics[width=9cm]{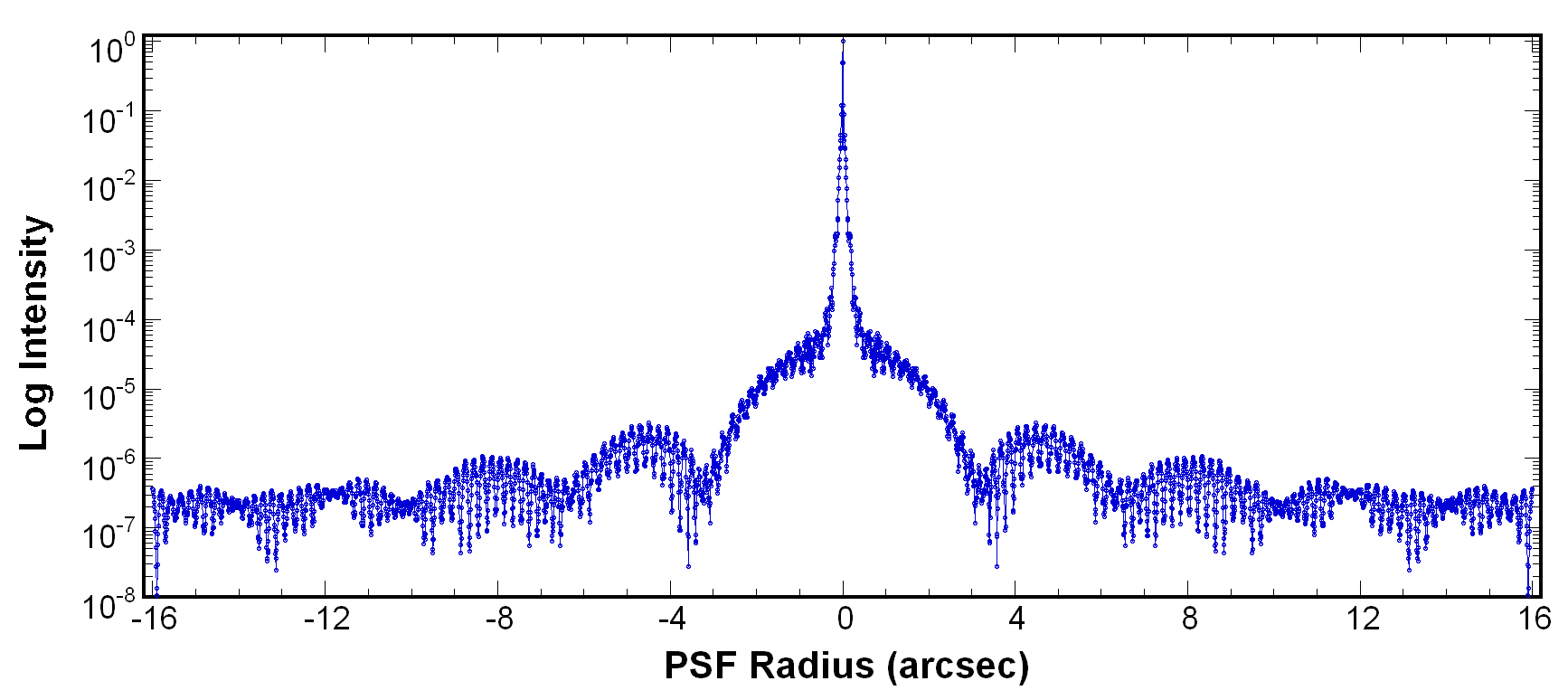}\caption{The
    log intensity of the r-band point-spread function along a
    diffraction spike.  The plot spans $32''$. 
\label{fig:common:psf}}\end{figure}

The fast optical beam and physical geometry of LSST help to minimize 
the impact of structured stray light at the focal
plane.  The relatively small cross-section ($\sim 0.7\,\rm m^2$) of the
support vanes holding the secondary and camera assemblies results in
very low intensity diffraction spikes.  The diffraction spike in the
$r$ band (see \autoref{fig:common:psf}) is down by six orders of
magnitude from the peak
at a radius of $4''$. 

\begin{figure}[t]\centering\includegraphics[width=12cm]{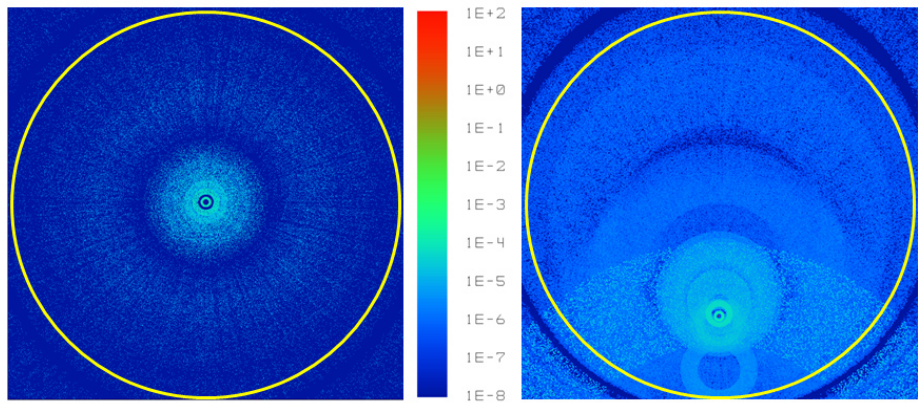}\caption{Calculated
    two-surface ghost images on-axis (left panel) and $1^\circ$ off
    axis (right panel).  The LSST's full field of view is 
represented by the yellow circle.  Note that this does not yet take
into account the reduction of detected surface brightness for the
designed anti-reflection coating 
performance, and thus somewhat overestimates the effect of ghosts. 
\label{fig:common:ghosts}}\end{figure}

Structured stray light from ghosting in the refractive elements is
further reduced by using state-of-the art broad band anti-reflection
coatings.  The relative surface brightness of the ghost images are
calculated using optical ray tracing with the lens surface treated
both transmissively (first pass) and reflectively (second pass); see
\autoref{fig:common:ghosts}.  The reflective properties of the
detector are assumed to be $\rm 1 - QE(\lambda)$. This overestimates
the ghost brightness at the extreme red wavelength since the QE
performance is dominated by the mean free path of the photon in
silicon rather than the reflection at the surface.  In any case, for
any reasonably bright source in the LSST's field of view, the ghost
image  
surface brightness will be well below that of the natural night sky.

\subsection{Diffuse Scattered Light}

\begin{figure}[t]\centering\includegraphics[width=12cm]{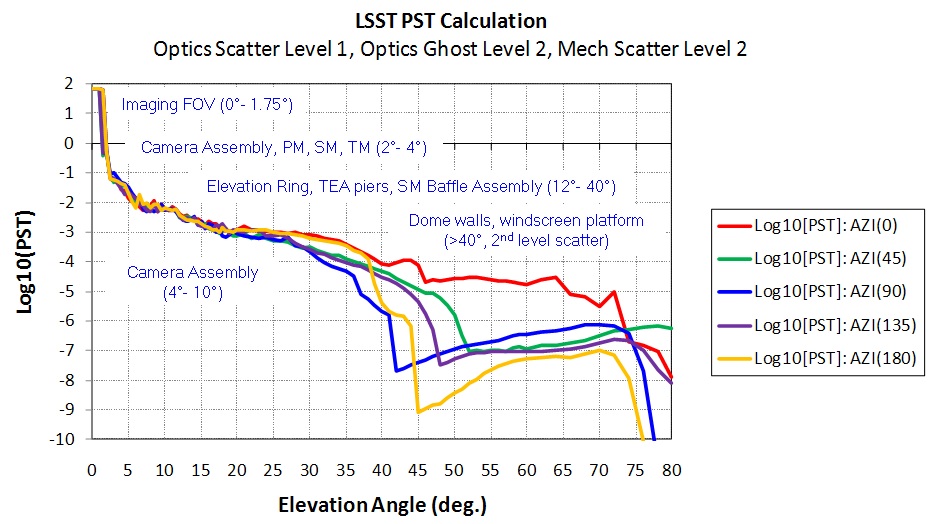}\caption{The
    LSST Point Source Transmittance (PST) as a function of source
    angle along five selected azimuth scans.  The primary sources of
    scattering are identified along the scans.  
\label{fig:common:PST}}\end{figure}

The first line of defense for unwanted diffuse scattered light is the
dome enclosure.  LSST's dome, like most modern domes, is well
ventilated to equalize the inside temperature with the exterior
ambient temperature, and is also well-baffled to reject external
sources of light.   A key feature in the LSST dome vent design is
light-restricting blackened louvers that have been aerodynamically
optimized 
to minimize restriction in air flow.  Light passing through the vents
must scatter from at least two louver surfaces before entering the
dome.  Using a specialized coating (Aeroglaze Z306) these dome vents
will allow $<3\%$ of the incident light through, while having $>95\%$
air
flow efficiency.  The wind screen within the dome slit will provide a
circular aperture to 
restrict unwanted light outside the LSST's field of view from entering
the dome.  Even with these measures some light will naturally enter
the dome and illuminate objects in a way that will create unwanted
light at the focal plane.  A detailed analysis using non-sequential
ray tracing and three-dimensional CAD models of the dome, telescope,
and camera has
been done to quantify the diffuse scattering contribution to the
overall natural sky background.   The initial analysis
(\autoref{fig:common:PST})
computes the Point Source Transmittance (PST) for a single point
source at various angles
with respect to the telescope's optical axis.  The PST is the
integrated flux over the entire focal plane from the point source
including the desired optical path and all first- and second-order
scattered light.  Each surface is specified with properties
anticipated for the realized design, including contamination on the
optical surfaces, micro-surface roughness, paint on non-optical
surfaces, and so on. 

The PST analysis shown in \autoref{fig:common:PST} indicates that the LSST
has excellent
rejection of diffuse scattered light from out-of-field objects, with the PST dropping nearly
three
orders of magnitude beyond the imaging field of view \citep{Elis09}.
Spreading this over the field of view of the LSST, the 
surface brightness contribution of a point source from diffuse
scattering is at least 11 orders of magnitude below that of the direct
image of the source. 

\section{The Expected Accuracy of Photometric Measurements } 
\label{sec:photo-accuracy}

{\it \v Zeljko Ivezi\'c}

The expected photometric error for a point source in magnitudes
(roughly the inverse of signal-to-noise ratio) 
for a single visit (consisting 
of two back-to-back 15-second exposures) can be written as 
\begin{equation}
\label{PHquad}
          \sigma_1^2 = \sigma_{sys}^2 + \sigma_{rand}^2,
\end{equation}
where $\sigma_{rand}$ is the random photometric error and
$\sigma_{sys}$ is the systematic photometric error (which includes errors
due to, for example, imperfect modeling of the point spread function, but
does not include errors in the absolute photometric zeropoint). For more
details, see Section 3.3.1 in the LSST overview paper,
\citet{lsst08}. For $N$ stacked observations, we assume
$\sigma_{rand}(N) = \sigma_{rand}/\sqrt{N}$. This theoretically
expected behavior has been demonstrated for repeated SDSS scans for
$N$ as large as 50 \citep {Ive++07,Ses++07,Bra++08}. The LSST
calibration system and procedures are designed to maintain
$\sigma_{sys}<0.005$ mag and this is the value we adopt for a single
LSST visit. Some effects that contribute to $\sigma_{sys}$ will be
uncorrelated between observations (e.g., errors due to imperfect
modeling of the point spread function) and their impact will decrease
with the number of stacked observations similarly to random
photometric errors. For the final irreducible errors in LSST stacked
photometry, we adopt $\sigma_{sys}$=0.003 mag (which will be probably
dominated by errors in the transfer of the photometric zeropoints
across the sky).
\begin{table}
\caption{The Parameters from \autoref{PHrand} and \autoref{m5}}
\begin{center}
\begin{tabular}{|r|r|r|r|r|r|r|}
\hline  
              &   $u$  &   $g$   & $r$   &  $i$  & $z$  & $y$  \\
\hline  
   $m_{\rm sky}^a$ &   21.8    & 22.0    & 21.3    & 20.0    & 19.1    &  17.5  \\
   $\theta^b$      &   0.77    & 0.73    & 0.70    & 0.67    &  0.65   &  0.63  \\
  $\gamma^c$       &   0.037   & 0.038   & 0.039   & 0.039   & 0.040   & 0.040 \\
    $C_m^d$        &   23.60   & 24.57   & 24.57   & 24.47   & 24.19   & 23.74 \\
     $k_m^e$       &    0.48   &  0.21   &  0.10   &  0.07   &  0.06   &  0.06 \\
    $m_5^f$        &   23.9    & 25.0    & 24.7    &  24.0   & 23.3    & 22.1  \\
    $\Delta m_5^g$ &   0.21    & 0.16    & 0.14    &  0.13   & 0.13    & 0.13  \\
\hline                         
\end{tabular}
\end{center}
\vskip 0.05in 
\small  $^a$ The expected median zenith sky brightness at Cerro Pach\'on, assuming
       mean solar cycle and three-day old Moon (mag/arcsec$^2$). \\
  $^b$ The expected delivered median zenith seeing (arcsec). For larger
       airmass, $X$, seeing is proportional to $X^{0.6}$. \\
  $^c$ The band-dependent parameter from \autoref{PHrand}. \\
  $^d$ The band-dependent parameter from \autoref{m5}. \\
  $^e$ Adopted atmospheric extinction. \\
  $^f$ The typical 5 $\sigma$ depth for point sources at zenith, assuming exposure time of 
       2$\times$15 sec, and observing conditions as listed. For larger
       airmass the 5 $\sigma$ depth is brighter; see the bottom row. \\
  $^g$ The loss of depth at the median airmass of $X=1.2$ due to seeing degradation 
       and increased atmospheric extinction. \\
\label{Tab:depths}
\end{table}
LSST's photometry will be limited by sky noise, and the random photometric 
error as a function of magnitude (per visit) can be described by
\begin{equation}
\label{PHrand}
  \sigma_{rand}^2 = (0.04-\gamma)\, x + \gamma \, x^2 \,\,\, {\rm (mag^2),}
\end{equation}
with $x = 10^{0.4\,(m-m_5)}$. Here $m_5$ is the 5 $\sigma$ depth (for
point sources) in a given band, and $\gamma$ depends on the sky 
brightness, readout noise, and other factors. Using the LSST exposure time 
calculator (\autoref{sec:com:expos}), we have obtained the values of $\gamma$ 
listed in \autoref{Tab:depths}. The 5 $\sigma$ depth for point sources is determined from 
\begin{eqnarray}
\label{m5}
  m_5 = C_m + 0.50\,(m_{sky}-21) + 2.5\,\log_{10}\frac{0.7}{\theta} +  \nonumber \\
        + 1.25\,\log_{10} \frac{t_{vis}}{30} - k_m(X-1) \phantom{xxxxx}
\end{eqnarray}
where $m_{sky}$ is the sky brightness (mag/arcsec$^2$), $\theta$ is
the seeing (FWHM, in arcsec), $t_{vis}$ is the exposure time
(seconds), $k$ is the atmospheric extinction coefficient, and $X$ is
airmass.  The constants, $C_m$, depend on the overall throughput of the
instrument and are determined using the LSST exposure time
calculator. The assumptions built into the calculator were tested
using SDSS observations and by comparing the predicted depths to the
published performance of the Subaru telescope \citep{Kas++03}. The
adopted values for $C_m$ and $k$ are listed in \autoref{Tab:depths},
as well as the expected $m_5$ in nominal observing conditions.  See
also \autoref{tab:com:T3} for the expected photometric accuracy at
higher S/N.  

\section{Accuracy of Trigonometric Parallax and Proper Motion Measurements } 
\label{sec:com:PMacc}

{\it \v Zeljko Ivezi\'c, David Monet}

Given the observing sequence for each sky position in the main survey
provided by the LSST Operations Simulator (\autoref{sec:design:opsim}), 
we generate a time sequence of mock
astrometric measurements. The assumed astrometric accuracy is a
function of $S/N$. Random astrometric errors
per visit are modeled as $\theta/(S/N)$, with $\theta=700$ mas and $S/N$
is determined using expected LSST 5 $\sigma$ depths for point sources.
The estimated proper motion and parallax accuracy at the bright end
($r<20$) is driven by systematic errors due to the
atmosphere. Systematic errors of 10 mas are added in quadrature, and
are assumed to be uncorrelated between different observations of a
given object. Systematic and random errors become similar at about
$r=22$, and there are about 100 stars per LSST sensor (0.05 deg$^2$)
to this depth (and fainter than the LSST saturation limit at
$r\sim16$) even at the Galactic poles.

\begin{figure}
\centering\includegraphics[width=10cm]{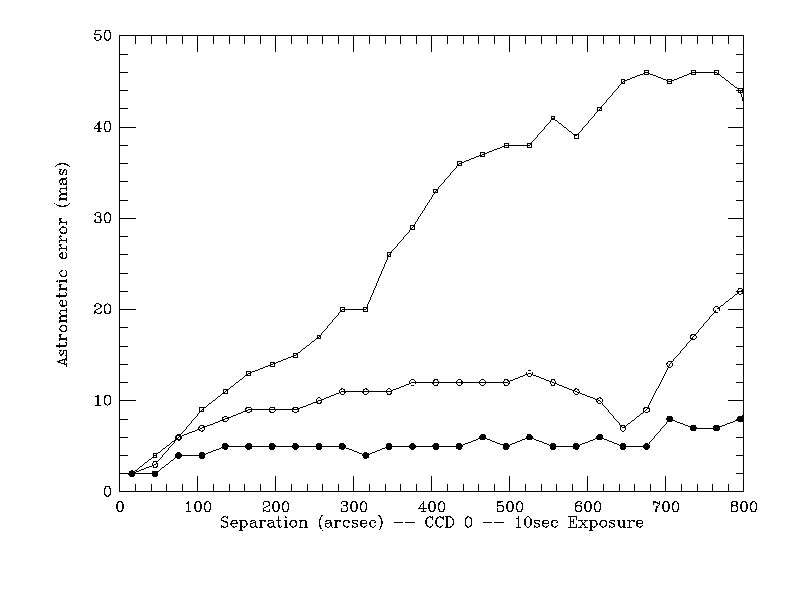}
\caption{Differential astrometric error as a function of angular separation
   derived from a sequence of 10-second Subaru Suprime-Cam observations.
   The upper curve is computed from transformation using only offsets between
   frames.  The middle curve includes linear transformation coefficients
   and the bottom curve includes cubic transformation coefficients.  The
   improvement in astrometric accuracy suggests that low-order polynomials
   are a reasonable model for the geometric impact of atmospheric turbulence
   over spatial scales of several arcminutes. From
   \citet{Saha+Monet2005}, with permission. 
\label{fig:common:subaru}}
\end{figure}

Precursor data from the Subaru telescope (\autoref{fig:common:subaru})
indicate that systematic
errors of 10 mas on spatial scales of several arc-minutes are
realistic. Even a drift-scanning survey such as SDSS delivers
uncorrelated systematic errors (dominated by seeing effects) at the
level of 20-30 mas rms per coordinate (measured from repeated scans, \citealt{PSK03}),
and the expected image quality for LSST will be twice as good as for
SDSS. Furthermore, there are close to 1000 galaxies per sensor with
$r<22$, which will provide exquisite control of systematic astrometric
errors as a function of magnitude, color, and other parameters, and
thus enable absolute proper motion measurements.

The astrometric transformations for a given CCD and exposure, and
proper motion and parallax for all the stars from a given CCD, are
simultaneously solved for using an iterative algorithm (\autoref{sec:design:calsimAstro}). The
astrometric transformations from pixel to sky coordinates are modeled
using low-order polynomials and standard techniques developed at the
U.S. Naval Observatory \citep{Mon++03}. The expected proper motion and
parallax errors for a ten-year long baseline survey, as a function of
apparent magnitude, are summarized in
\autoref{tab:com:T3}. Roughly speaking, trigonometric parallax errors
can be obtained by multiplying the astrometric errors by 0.039, and
proper motion errors (per coordinate) can be obtained by multiplying
the single-visit astrometric errors by 0.014 yr$^{-1}$.

Blue stars
(e.g., F and G stars) fainter than $r\sim23$ will have about 50\% larger
proper motion and parallax errors than given in the table due to
decreased S/N in $z$ and $y$. The impact on red
stars is smaller due to the relatively small number of observations in
the $u$ and $g$ bands, but extremely red objects, such as L and T
dwarfs, will definitely have larger errors, depending on details of
their spectral energy distribution.  After the first three years of
the survey, the proper motion errors are about five times as
  large, and parallax errors will be about twice as large as the values
given in \autoref{tab:com:T3}; the errors scale as $t^{-3/2}$ and
$t^{-1/2}$ respectively.

\begin{table}
\caption{The expected proper motion, parallax, and accuracy for a ten-year long baseline survey.}
\begin{center}
\begin{tabular}{|l|c|c|c|c|c|}
\hline  
    $r$   &  $\sigma^a_{xy} $  & $\sigma^b_\pi$  &   $\sigma^c_\mu$   &  $\sigma^d_1$  &  $\sigma^e_C$  \\
    mag &       mas            &      mas  & mas/yr &   mag   &    mag  \\
\hline  
       21 &  11  &  0.6  &  0.2   &   0.01  &   0.005 \\
       22 &  15  &  0.8  &  0.3   &   0.02  &   0.005 \\
       23 &  31  &  1.3  &  0.5   &   0.04  &   0.006 \\
       24 &  74  &  2.9  &  1.0   &   0.10  &   0.009 \\
\hline                         
\end{tabular}\\
\end{center}
\vskip 0.05in
  $^a$ Typical astrometric accuracy (rms per coordinate per visit). \\
  $^b$ Parallax accuracy for 10-year long survey. \\
  $^c$ Proper motion accuracy for 10-year long survey. \\
  $^d$ Photometric error for a single visit (two 15-second exposures). \\
  $^e$ Photometric error for stacked observations (see Table 1). \\
\label{tab:com:T3}
\end{table}

For comparison with \autoref{tab:com:T3}, the SDSS-POSS proper
motion measurements have an accuracy of $\sim$ 5 mas/yr per coordinate
at $r=20$ \citep{Mun++04}. Gaia is expected to deliver parallax
errors of 0.3 mas and proper motion errors of 0.2 mas/yr at its faint
end at $r\sim20$. Hence, LSST will smoothly extend Gaia's error
vs. magnitude curve four magnitudes fainter, as discussed in detail in
\autoref{sec:Gaia}.  

\section{Expected Source Counts and Luminosity and Redshift Distributions\label{sec:com:counts}}

{\it \v Zeljko Ivezi\'c, A. J.  Connolly, Mario Juri\'{c}, Jeffrey A. Newman, Anthony Tyson, Jake VanderPlas, David Wittman}

The final stacked image of LSST will include about ten billion galaxies and ten
billion stars, mostly on the main sequence. The data sources and assumptions
used to derive these estimates are described here.  Of course, 
LSST will also detect very large samples of many other types of objects such as asteroids, white dwarfs,
and quasars (roughly ten million in each category).  We defer discussion of those more specific 
topics to the relevant
science chapters that follow.

\subsection{Stellar Counts }
\label{Sec:stellarCounts}

\begin{figure}
\centerline{\resizebox{6in}{!}{\includegraphics{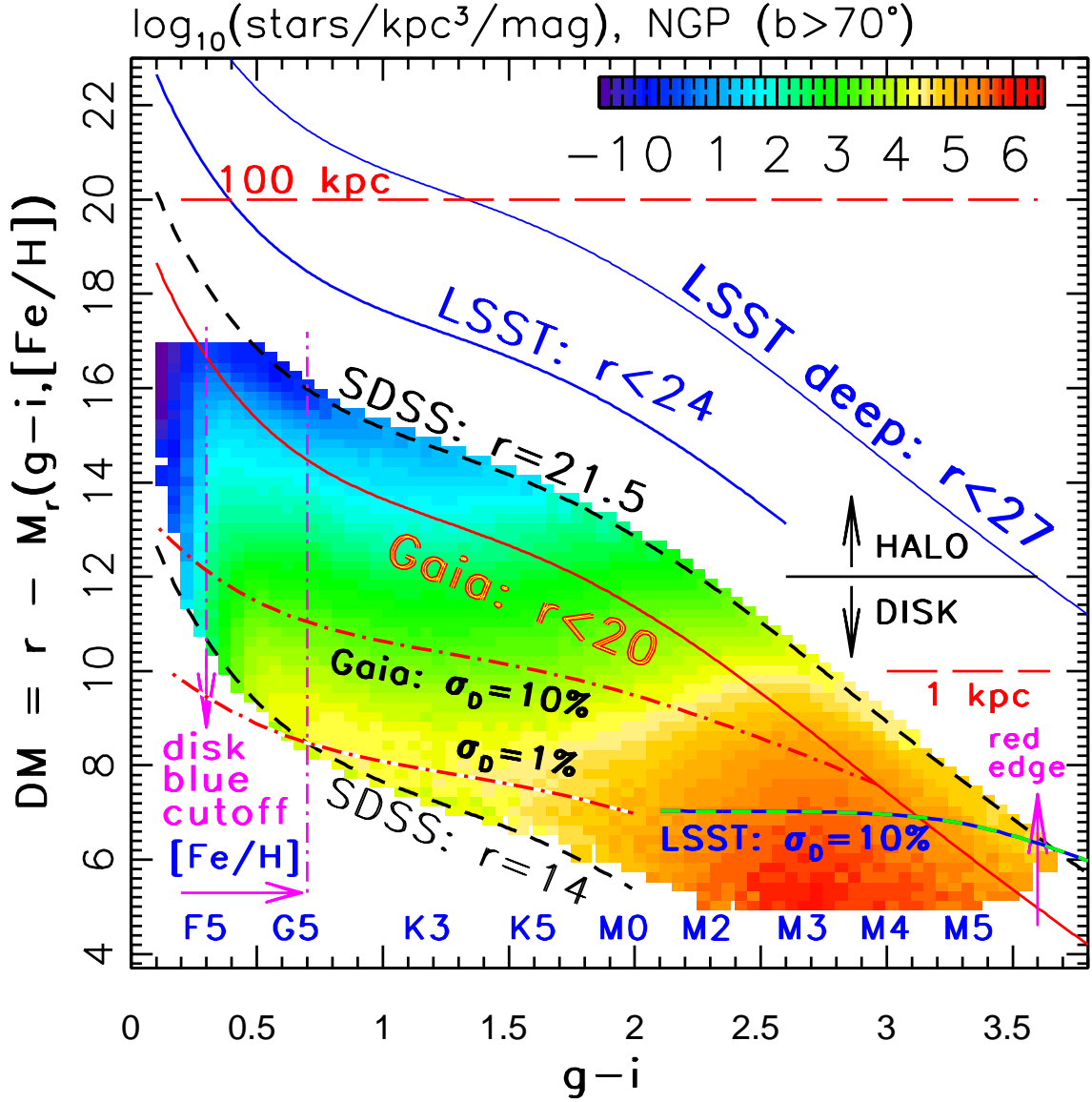}}}
\caption{The volume number density (stars/kpc$^3$/mag, log scale
according to legend) of $\sim $2.8 million SDSS stars with $14<r<21.5$ 
and $b>70^\circ$, as a function of their distance modulus 
(distances range from 100 pc to 25 kpc) and their $g-i$ color. 
The sample is dominated by color-selected main sequence stars. 
The absolute magnitudes are determined using the photometric parallax 
relation from I08. The metallicity correction is applied using
photometric metallicity for stars with $g-i<0.7$, and by assuming 
$[Fe/H]=-0.6$ for redder stars. The relationship between the MK spectral type and $g-i$ color 
from \citet{Cov++07} is indicated above the $g-i$ axis; $g-i=0.7$ roughly corresponds to G5. The two 
vertical arrows mark the turn-off color for disk stars and the red edge of 
the M dwarf color distribution. The $[Fe/H]$ label shows the color range 
($g-i<0.7$) where the photometric metallicity estimator from I08 performs
best. The two diagonal dashed lines, marked $r=14$ and $r=21.5$, show 
the apparent magnitude limits for SDSS data. At a distance of $\sim$ 2-3
kpc ($DM=12$), halo stars begin to 
dominate the counts. The diagonal solid lines mark the apparent magnitude
limits for Gaia ($r<20$), LSST's single epoch data ($r<24$,
10 $\sigma$), and LSST's stacked data ($r<27$, 10 $\sigma$). The dashed
line in the lower right corner marks 
the distance limits for obtaining 10\% accurate trigonometric distances 
using LSST data. The two dot-dashed lines mark analogous limits for obtaining
1\% and 10\% accurate trigonometric distances using Gaia's data
(\autoref{sec:Gaia}).} 
\label{Fig:CountsStars1}
\end{figure}

In order to accurately predict stellar source counts for the LSST survey, both a
Galactic structure model and a detailed estimate of the stellar luminosity
function are required.  SDSS data can be used to guide these choices.
\autoref{Fig:CountsStars1} shows the stellar
counts, as a function of distance and color, for stars observed with SDSS
towards the North Galactic Pole. Stars are selected to have colors
consistent with main sequence stars following criteria from
\citet[hereafter I08]{Ive++08}. This color selection is sufficiently
complete to represent true stellar counts, and sufficiently efficient
that contamination by giants, white dwarfs, quasars, and other
non-main sequence objects is negligible. Distances are computed using
the photometric parallax relation and its dependence on metallicity
derived by I08. The displayed density variation in the horizontal
direction represents the luminosity function, and the variation in the
vertical direction reflects the spatial volume density profiles of
disk and halo stars. Both effects need to be taken into account in
order to produce reliable counts for the LSST survey.

To extrapolate
stellar counts from the SDSS faint limit at $r=22.5$ to the faint limit of
the stacked LSST map ($r=27.5$), we use the Milky Way
model by \citet[hereafter, J08]{2008ApJ...673..864J}. This model reproduces the SDSS
count data to within 10\% (except in regions with significant
substructure) as shown in \autoref{Fig:CountsStars1}, as well as the
count variation as a function of position on the sky.
Using photometric data for 50 million stars from SDSS Data Release 4, 
sampled over a distance range from 100 pc to 15 kpc, J08 showed that the stellar number density distribution, 
$\rho(R,Z,\phi)$ can be well described (apart from local 
overdensities; the J08 best-fit was refined using residual 
minimization algorithms) as a sum of two cylindrically symmetric components,
\begin{equation} 
      \rho(R,Z,\phi) = \rho_D(R,Z) + \rho_H(R,Z).
\end{equation}
The disk component can be modeled as a sum of two exponential disks
\begin{eqnarray} 
\rho_D(R,Z)=\rho_D(R_\odot) \times \phantom{xxxxxxxxxxxxx}
\nonumber \\ 
     \left[{\rm e}^{-|Z+Z_\odot|/H_1-(R-R_\odot)/L_1} 
   + \epsilon_D {\rm e}^{-|Z+Z_\odot|/H_2-(R-R_\odot)/L_2} \right],
\end{eqnarray} 
and the halo component requires an oblate power-law model
\begin{equation} 
 \rho_H(R,Z)= \rho_D(R_\odot)\,\epsilon_H\, \left({R_\odot^2 \over R^2 
               + (Z/q_H)^2}\right)^{n_H/2}.
\end{equation} 
The best-fit parameters are discussed in detail by J08. For LSST
simulations, we have adopted parameters listed in the second column 
of their Table 10.

This Galaxy model gives star counts accurate only to about a factor of
two, due to our incomplete 
knowledge of the three-dimensional dust distribution in the Galactic
plane, and the uncertain location of the edge of the stellar halo. 
As illustrated in \autoref{Fig:CountsStars1}, if this
edge is at 100 kpc or closer to the Galactic center, it will be detected 
as a sudden drop in counts of blue faint stars beyond some 
color-dependent flux limit. For example, blue turn-off stars 
with $M_r<5$ should display a sharp decrease in their differential 
counts for $r>25$, if there is a well-defined end to the distribution 
of halo stars at 100 kpc. We obtain approximate estimates
by extrapolating counts for $r<21$ from USNO-B all-sky catalog 
to fainter magnitudes using models described above. 
There are $10^9$ stars with $r<21$ in the USNO-B catalog, and 
this count is probably accurate to better than 20\%, which 
is a smaller uncertainty than extrapolations described below.

The ratio of stellar counts to $r<24.5$ and $r<27.8$ (LSST's single 
visit and stacked depths) to those with $r<21$ varies significantly
across the sky due to Galactic structure effects and the interstellar
dust distribution. For the dust distribution, we assume an exponential
dependence in radial and vertical directions with a scale height 
of 100 pc and a scale length of 5 kpc. We assume a dust opacity of 
1 mag/kpc (in the $r$ band) which produces extinction of 0.1 mag towards
the North Galactic pole, 20 mag towards the Galactic center, and
5 mag towards the anticenter, in agreement with ``common wisdom.''
Using the stellar counts model described above, and this dust model, 
we evaluate the counts' ratios as a function of location on the sky
and integrate over the sky to be covered by LSST's main survey. 
In the regions observed by SDSS, the predicted counts agree to 
better than 20\% (the models were tuned to SDSS observations, but
note that the normalization comes from USNO-B). The counts' ratios for 
several special directions are listed in \autoref{Tab:counts}.
The predicted total number of stars is 4 billion for $r<24.5$
with an uncertainty of $\sim$50\%, and 10 billion for $r<27.8$,
with an uncertainty of at most a factor of 2.

\begin{table}
\begin{center}
\caption{ Stellar counts based on USNO-B and model-based extrapolations}
\begin{tabular}{|r|r|r|r|r|}
\hline  
         &  N($r<24.5$)$^a$ &  ratio(24.5/21) & ratio(27.8/24.5) & ratio(27.8/21) \\
\hline  
Galactic center     &  172  &       6.4       &       3.8        &	24   \\
anticenter          &  120  &       4.5       &       2.4        &	11   \\
South Galactic Pole &    4  &       2.6       &       2.0        &      5    \\
\hline                    
\end{tabular}
\end{center}
\vskip 0.05in
  $^a$ The number of stars with $r<24.5$ in thousands per deg$^2$. The
       entries are computed using counts based on the USNO-B catalog 
       and extrapolated from its $r=21$ limit using
       model-based count ratios, listed in the second column.
       LSST will detect $\sim 4$ billion stars with $r<24.5$ and
       10 billion stars with $r<27.8$. 
  \\
\label{Tab:counts}
\end{table}

\subsection{Galaxy Counts }
\label{sec:common:galcounts}

\begin{figure}
\begin{center}
\includegraphics[scale=0.25]{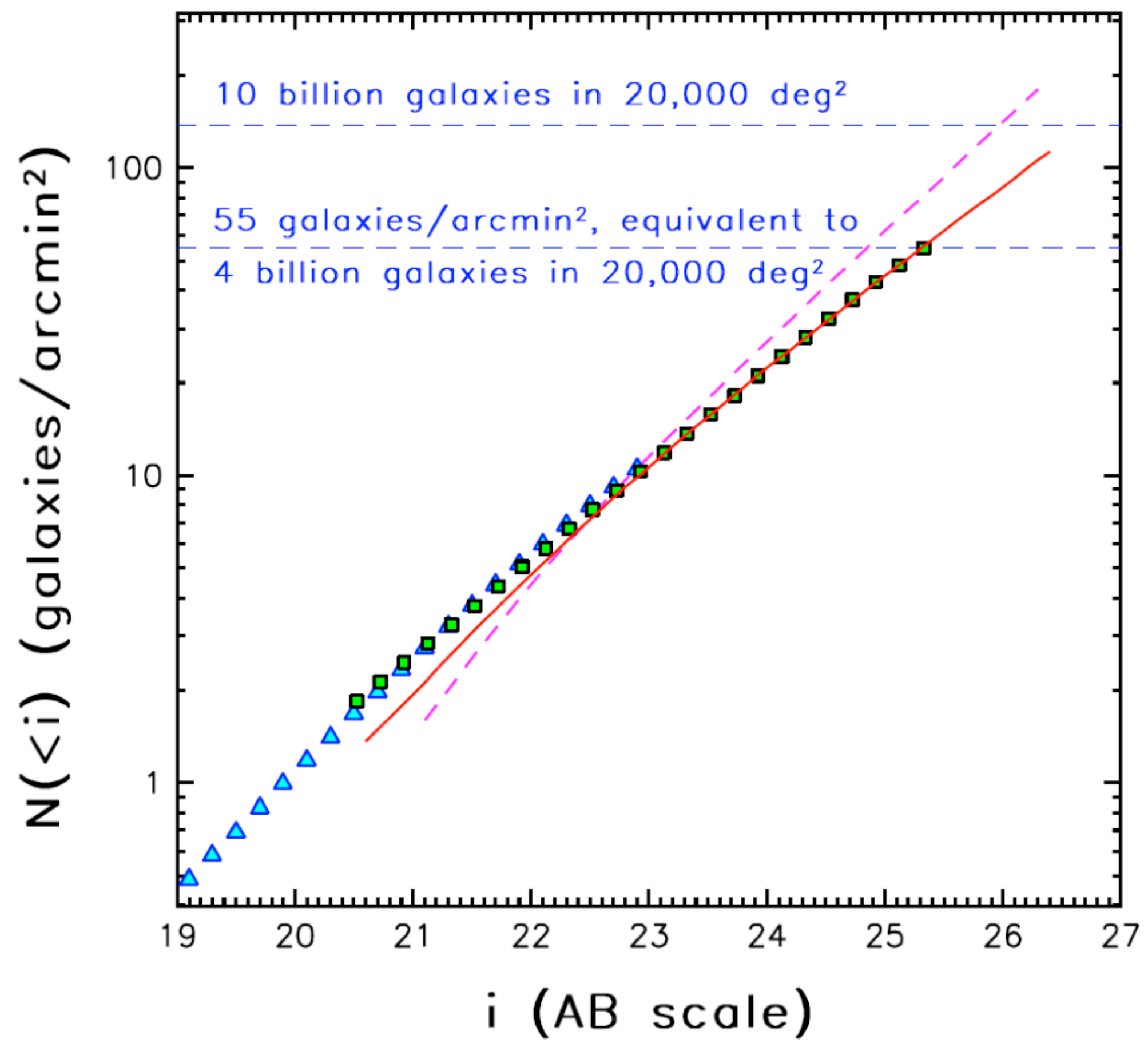}
\caption{Cumulative galaxy counts in the SDSS $i$ band. The triangles 
show SDSS counts from the so-called ``stripe 82" region \citep{A+S08} and 
the squares show counts from the CFHTLS Deep survey
\citep{Gwyn08}. The dashed diagonal line is based on the Millennium  
Simulation \citep{2005Natur.435..629S} and the solid line is a simulation 
based on a model with evolving luminosity function from the DEEP2 and VVDS
surveys (measured at redshifts up to unity) and non-evolving SEDs. 
The two dashed horizontal lines are added 
to guide the eye. LSST will detect 4 billion galaxies with $i<25.3$, which 
corresponds to an S/N of at least 20 for point sources in
median observing conditions. The full LSST sample may include as many as 
10 billion galaxies.}
\label{Fig:CountsGals1}
\end{center}
\end{figure}

Model-independent, empirical estimates of galaxy counts with LSST can be
gleaned from a number of deep multicolor photometric surveys that have
been performed over the last decade.
These are sufficient to predict the counts for the
LSST galaxies \citep[e.g.,][]{Ilb++06} with an uncertainty of about
20\% (most of this uncertainty comes from photometric systematics
and large-scale structure). Based on the CFHTLS Deep survey 
\citep{Hoekstra++06, Gwyn08}, 
the cumulative galaxy counts for $20.5<i<25.5$ are well described by 
\begin{equation}
       N_{gal} = 46 \times 10^{0.31(i-25)} \,\, {\rm galaxies\,arcmin^{-2}}. 
\end{equation}
The so-called ``gold'' sample of LSST galaxies with a high S/N defined by $i<25.3$ (corresponding to S/N $>$ 20 for point sources
assuming median observing conditions), 
will include four billion galaxies (or 55 arcmin$^{-2}$) over 
20,000 deg$^2$ of sky (see \autoref{Fig:CountsGals1}). The effective surface
density of galaxies useful for weak lensing analysis in the ``gold'' sample
will be about 40 arcmin$^{-2}$ with an uncertainty of 20\%. The total number 
of galaxies detected above the faint limit of the stacked map ($r<27.5$, 
corresponding to $i \sim 26$ given the typical colors of galaxies) will be 
close to 10 billion over 20,000 deg$^2$. 

\begin{figure}
\begin{center}
\includegraphics[scale=0.28]{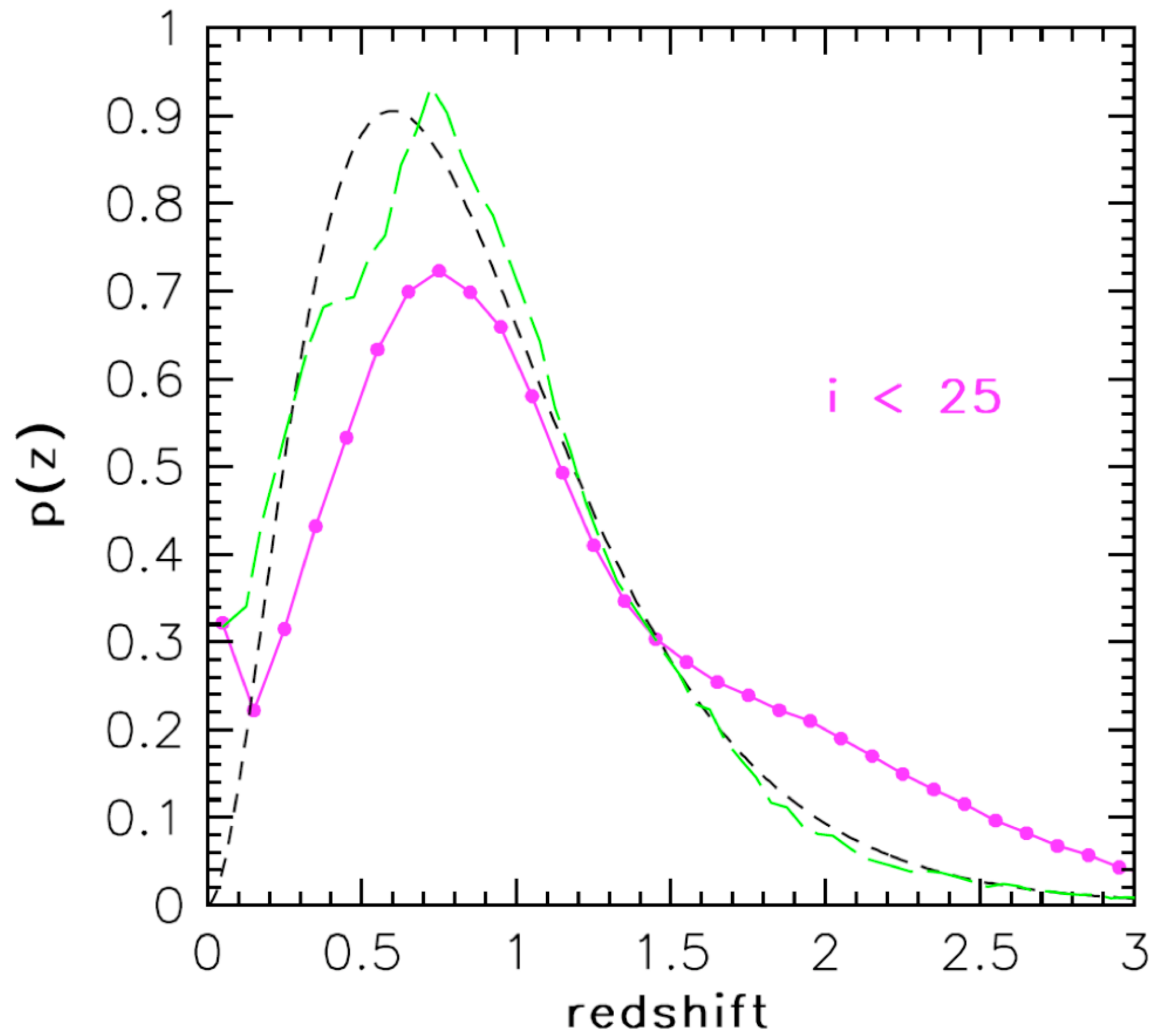}
\caption{The redshift probability distributions for faint galaxies. 
The dashed curve shows a best fit to the measured and debiased DEEP2 redshift 
distribution for galaxies with $i<23$, and extrapolated to $i<25$ (see text). 
The other two curves show model predictions for galaxies with $i<25$ 
(magenta: the Millennium Simulation; green: an evolving luminosity 
function; see \autoref{Fig:CountsGals1}).}
\label{Fig:CountsGals2}
\end{center}
\end{figure}

The redshift and rest-frame color distributions of these sources are
much less well understood due to the lack of any complete spectroscopic sample 
to the depth of the LSST. To estimate the redshift
distributions for the LSST we, therefore, use both simple extrapolation
of observations and more sophisticated simulations that have been designed 
to match available observational data sets. In \autoref{Fig:CountsGals2},
we show a prediction for the redshift distribution of galaxies of the form
\begin{equation}
      p(z) = {1 \over 2 z_0} \left({z \over z_0}\right)^2 \, \exp{(-z/z_0)}.  
\end{equation}
This functional form fits DEEP2 data well (after completeness corrections) 
for $i<23$. We estimate $z_0$ for the $i<25$ sample by extrapolating the tight 
linear relationship between $z_0$ and limiting $i$ magnitude in the DEEP2 
survey, $z_0=0.0417\,i-0.744$ (measured for $21.5<i<23$). The mean redshift 
of a sample is $3z_0$ and the median redshift is $2.67z_0$; for the $i<25$ sample, 
the mean redshift is 0.9 and the median is 0.8.

This prediction is compared to a model based on an empirical evolving luminosity
function, and to the simulations of
\citet[hereafter KW07]{K+W07}. Based on the Millennium simulations 
\citep{2005Natur.435..629S}, the baryonic
physics in KW07 models includes gas cooling, heating from star
formation, supernovae, and radio mode feedback. Comparisons with
existing imaging surveys has shown that the model for the dust used in
KW07 provides a good match to the color-luminosity relation seen in
deep surveys to $z\sim 1.4$ (although the simulations predict more than the 
$K$-band number counts).

\newcommand{\Cs}{{\bf C}}
\newcommand{\Os}{{\bf O}}
\newcommand{\SB}{{\rm SB}}
\newcommand{\obs}{{\rm obs}}

\section{Photometric Redshifts}
\label{sec:common:photo-z}

{\it A. J. Connolly, Jeffrey A. Newman, Samuel Schmidt, Alex Szalay, Anthony Tyson}

The estimation of galaxy redshifts from broad band photometry, i.e.,\
photometric redshifts \citep{baum62,koo85,loh++86,con++95}, has become
a widely used tool in observational cosmology \citep{col++04, wada05,
  carl++07,gwyn++96, lan++96,saw++97,bud++00,ilb++06b}. These
probabilistic redshift (and galaxy property) estimates are derived from
characteristic changes in the observed colors of galaxies due to the
redshifting of features in galaxy spectral energy distributions
through a series of broad band filters. At optical and ultraviolet
wavelengths, the Lyman and Balmer breaks (at 1000\AA\ and 4000\AA\
respectively) are the primary source of the redshift information. To
first order, the accuracy to which we can determine the position of
these breaks from the observed colors determines the scatter within
the photometric redshift relation, and our ability to correctly
distinguish between the breaks determines the amount of systematic
error (or catastrophic outliers) in the relation.

The LSST reference filter system, covering the $u, g, r, i, z, $ and
$y$ passbands, provides leverage for redshift estimation from $z=0$ to
$z>6$ (although, as we will describe later, the redshift interval,
$1.4<z<2.5$, will be less well constrained as the Balmer break has
transitioned out of the $y$ band and the Lyman break has yet to enter
the $u$ band). We describe here the expected photometric redshift
performance for LSST based on empirical studies and simulations
(including scatter, bias, and the fraction of sources that are
catastrophic outliers) and describe ongoing work to characterize and
minimize these uncertainties.

\subsection{Photometric Redshifts for the LSST}

Photometric redshifts for LSST will be applied and calibrated over
the redshift range $0<z<4$ for galaxies to $r\sim$ 27.5. For the
majority of science cases, such as weak lensing and BAO, a subset of
galaxies with $i<25.3$ will be used. For this high S/N gold standard subset (\autoref{sec:common:galcounts}) over the
redshift interval, $0<z<3$, the photometric redshift requirements are:
\begin{itemize}
\item The root-mean-square scatter in photometric redshifts,
  $\sigma_z/(1+z)$, must be smaller than 0.05, with a goal of 0.02.
\item The fraction of 3$\sigma$ outliers at all redshifts must be
  below 10\%.
\item The bias in $e_z=(z_{photo}- z_{spec})/(1+z_{spec})$ must
  be below 0.003 (\autoref{sec:wl:zphot}, or 0.01 for combined
  analyses of weak lensing and baryon acoustic oscillations);  
the uncertainty in $\sigma_z/(1+z)$ must also be known to similar accuracy.
\end{itemize}

\begin{figure}
\centerline{\resizebox{6in}{!}{\includegraphics{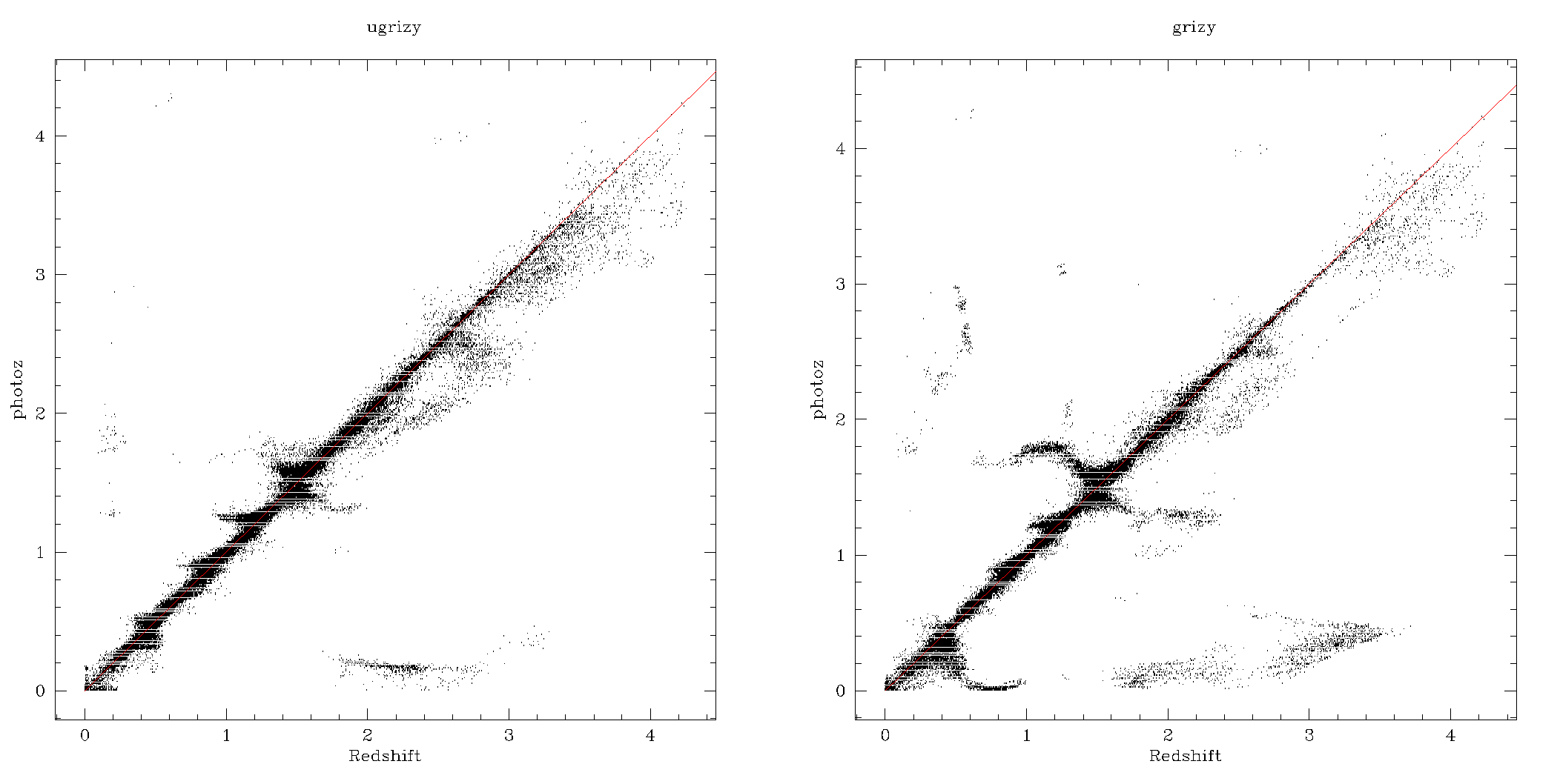}}}
\caption{
Impact of using the $u$ filter to improve measurement and  resolve 
degeneracies in photometrically determined redshifts.  On  the left is 
the correlation between the photometric redshifts and  spectroscopic 
redshifts with the full complement of LSST  filters. The right panel 
shows the photometric redshift relation for  data excluding the $u$ filter. 
The addition of $u$ data reduces the  scatter substantially for $z<0.5$ and 
removes degeneracies over the full redshift range.
\label{fig:photozUZband}}
\end{figure}

\begin{figure}
\begin{center}
\vskip -2in
\includegraphics[scale=0.2]{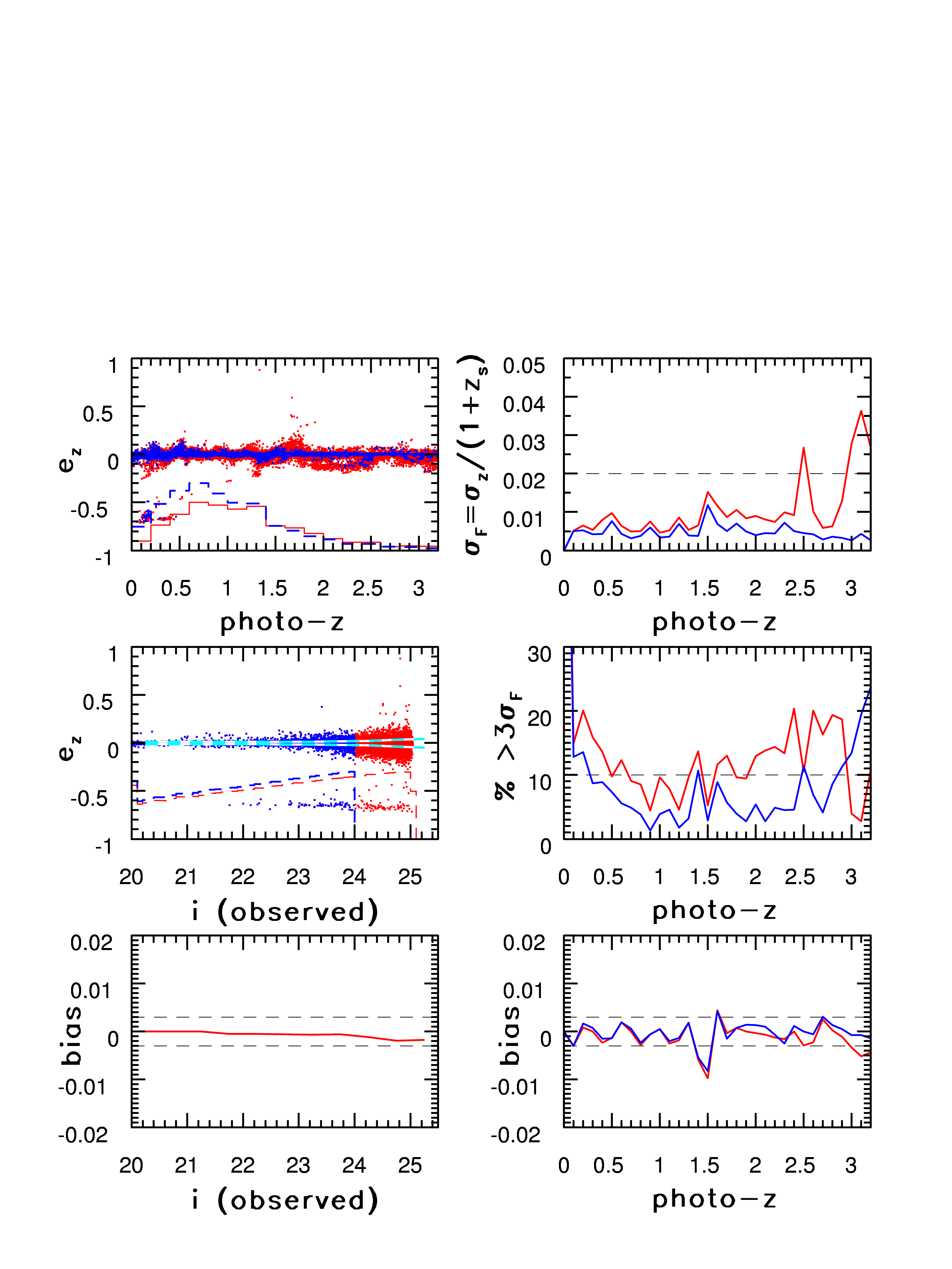}
\caption{Illustration of the photometric redshift performance as a
  function of apparent magnitude and redshift, for a simulation based
  on the LSST filter set ($ugrizy$). Red points and curves correspond to
  the gold sample defined by $i<25$, and blue points and curves to a
  subsample with $i<24$. The photometric redshift error is defined as
  $e_z =(z_{photo}-z_{spec})/(1+z_{spec})$. {\it Top left}: $e_z$
  vs. photometric redshift. The two histograms show redshift distributions
  of the simulated galaxies. Top right: the root-mean-square scatter
  (rms, determined from the interquartile range) of $e_z$ as a
  function of photometric redshift. The horizontal dashed line shows
  the science driven design goal. {\it Middle left}: $e_z$ vs. observed $i$ band
  magnitude. Two histograms show the logarithmic differential counts
  (arbitrary normalization) of simulated galaxies. The two horizontal
  cyan lines show the 3$\sigma$ envelope around the median $e_z$
  (where $\sigma$ is the rms from the top right panel). {\it Middle right}:
  the fraction of 3$\sigma$ outliers as a function of redshift. The
  horizontal dashed line shows the design goal. {\it Bottom left}: the
  median value of $e_z$ (bias) as a function of apparent
  magnitude. The two dashed lines show the design goal for limits on
  bias. {\it Bottom right}: the median value of $e_z$ (i.e., the bias in estimated redshifts) as a function
  of redshift. The two dashed lines show the design goal for this quantity.
\label{fig:photoz}}
\end{center}
\end{figure}

\autoref{fig:photozUZband} and \autoref{fig:photoz} show the expected
performance for the LSST gold sample on completion of the survey.
These results are derived from simulated photometry that reproduces
the distribution of galaxy colors, luminosities, and colors as a
function of redshift as observed by the COSMOS \citep{Lil++09}, DEEP2 (Newman et al. 2010, in preparation), and
VVDS  \citep{Gar++08} surveys. The simulations include the effects of evolution
in the stellar populations, redshift, and type dependent luminosity
functions, type dependent reddening, and of course photometric errors.
The photometric redshifts are determined using a likelihood technique
as outlined below.  

\autoref{fig:photoz} shows the residuals, fraction of outliers,
dispersion, and bias associated with the photometric redshifts as a
function of $i$ band magnitude and redshift. For this case, magnitude
and surface brightness priors have been applied to the data and all
sources with broad or multiply peaked redshift probability functions
have been excluded (see \S\ref{photoz:priors}). For the brighter
sample, ($i<24$), the photometric redshifts meet or exceed our
performance goals for all except the highest redshift bin. 
For the gold sample, the photometric redshifts meet the science
requirements on dispersion and bias at all redshifts. At redshifts
$z>2$, the fraction of outliers is a factor of two larger than the goal
for LSST.   These outliers reduce the 
size of the samples with usable photometric redshifts by
approximately 10\%. Other cuts and priors will reduce the outlier
fraction further.  This demonstrates that highly accurate 
photometric redshifts should be attainable with LSST photometry, assuming perfect 
knowledge of SED templates (or equivalently the span of galaxy properties).  
For selected subsets of objects (e.g., bright red sequence galaxies),  we may be able to do much better attaining $\sigma_z/(1+z)$ of 0.01 or less.



\subsection{Dependence on Filter Specification and Signal-to-Noise Ratio}

The accuracy of LSST photometric redshift depends on both the
characteristics of the filter system and our ability to
photometrically calibrate the data. For the LSST reference filters the
scatter in the photometric redshifts in simulated data scales approximately linearly
with S/N, with a floor of $\sigma_z \sim 0.02$ (when
including all galaxy types). This is consistent with the $\sigma_z$ obtained for 
photometric redshifts obtained for $r<17.77$ galaxies in SDSS (e.g., \citealt{2008ApJ...683...12B, 2009MNRAS.tmp.1053F}), several magnitudes 
brighter than the depths of the photometry (and hence in some ways analogous to the gold sample).  

However, it is significantly better than has been achieved to date for
photometric redshift algorithms for fainter samples and to higher
redshifts, likely due to the fact that such samples must handle a
broader range of galaxy types over a broader redshift range, and are
increasingly dominated by strongly star forming galaxies (which
possess only weak 4000\AA\ breaks) as they extend fainter and to
higher redshifts.  With CFHT Legacy Survey deep $ugriz$ imaging,
\citet{ilb++06b} achieve $\sigma_z \sim 0.03$ for $i<21.5$, degrading to
$0.04-0.05$ for $22.5<i<24$; while with deep 16-band photometry, and
restricting to a subset of galaxies with $z < 1.2$ and $K < 21.6$
(AB), \citet{mob++07} attain $\sigma_z \sim 0.03$ for a sample with
$20<i<24$.  Unfortunately, these numbers are difficult to compare due
to the larger number of bands and the $K$ band limit (which will favor
massive, lower star formation rate galaxies at higher redshifts) used
by \citet{mob++07}. The fundamental limitation which puts a floor on
$\sigma$ in these empirical tests is unclear (likely depending on
poorly known template spectra, errors in photometric measurements due
to blended galaxies, and variations in the emission line properties of
galaxies with redshift and type). The number of catastrophic failures
also depends on S/N, but the exact scaling remains
unclear \citep{mob++07}; in \citet{ilb++06b}, the catastrophic failure
rate is $<1\%$ for $i<21.5$, $\sim 2\%$ for $21.5<i<22.5$, $\sim 4\%$
for $22.5<i<23.5$, and $\sim 9\%$ for $23.5<i<24$.  Regardless, based
upon the SDSS experience, we can expect that with greater zero point
uniformity, better bandpass characterization, and improved calibration
LSST should yield significantly better photometric redshift results
than previous optical broad-band surveys.


In \autoref{fig:photozUZband} we show the impact of the $u$ band
filter for redshift estimation assuming a survey to the nominal depth
of the LSST and including magnitude and surface brightness priors.  At
low redshift the redshifting of the Balmer break through the $u$
filter enables the estimation of the photometric redshifts for $z<0.5$
(where the break moves into the $g$ and $r$ bands). At higher
redshift, the transition of the Lyman break into the $u$ band filter
increases the accuracy of the photometric redshifts for $z>2.5$. The
result of this is two-fold; the scatter in the redshift estimation is
decreased at low redshift, improving studies of the properties of
galaxies in the local Universe and reducing the number of catastrophic outliers
(mistaking the Lyman break for the Balmer break results in a
degeneracy between $z=0.2$ and $z=3$ galaxies) by a factor of two.  
Removal of the $u$ band results in a
deterioration of the photometric redshifts for $z<0.6$ to such an extent
that they fail to meet the required performance metrics as described
above.

At redshifts $1.3<z<1.6$, the photometric redshift constraints are
most dependent upon the $y$ filter. 
For $z>1.6$ the Balmer break transitions out of the $y$ band and hence the
photometric redshifts are only poorly constrained until the Lyman break 
enters the $u$ band at
$z>2.5$. 

Addition of near-infrared passbands from, for example, a space-based
imager yielding S/N=10 photometry in both the $J$ and $H$ bands at an AB magnitude of 25
results in a reduction in $\sigma_z$, the fraction of outliers, and the
bias by approximately a factor of two for $z>1.5$. At redshifts $z<1.5$
there is no significant improvement in photometric redshift
performance from near-infrared data, in contrast to the $u$
band data which impacts photometric redshifts below $z=1$ (even when the
$J$ and $H$ bands are included already).

\subsection{Priors in Redshift Estimation}
\label{photoz:priors}

In order to mitigate catastrophic failures in photometric redshifts,
Bayesian approaches for redshift estimation have been developed
\citep{ben00}. In this case we search for the two-dimensional
posterior distribution $P(z,T|\Cs,\Os)$, where $z$ is the redshift of
the galaxy, $T$ is the ``template'' or galaxy type, $\Cs$ is the
vector of fluxes from the data, and $\Os$ is a vector of galaxy observables
independent of the fluxes, such as size, brightness, morphology, or 
environment. If we make the assumption that $\Os$ and $\Cs$ are
independent then,
\begin{equation}
  \label{eq:pz2}
  P(z,T|\Cs,\Os) = \frac{P(\Cs|z,T)P(z,T|\Os)}{P(\Cs)}.
\end{equation}
The posterior distribution $P(z,T|\Cs,\Os)$ is given in terms of the
likelihood function $P(\Cs|z,T)$ and the prior distribution
$P(z,T|\Os)$; the prior encompasses all knowledge about galaxy
morphology, evolution, environment, brightness, or other quantities. 

The most common prior used in photometric redshifts has been 
magnitude \citep{ben00}; e.g. a prediction for the overall redshift distribution of
galaxies given an apparent magnitude and type. Other priors that
have been considered include morphological type and surface brightness \citep{stab++08}. Surface brightness for a given galaxy scales as 
$(1+z)^4$, suggesting it should be a powerful constraint, but it
evolves strongly with redshift, depends on spectra type, and depends
on accurate measurements of the size of galaxies close to the seeing
size, making it less useful. 
However, \citet{stab++08} show that if a ground-based survey can precisely
measure the angular area of galaxies, achieve a seeing of $1''$ or
less, and attain a surface brightness sensitivity below 26.5 mag/arcsec$^2$, then 
it should be able to do almost as well as one from
space in constraining photometric redshifts via
surface brightness (resulting in a decrease in the photometric
redshift bias of up to a factor of six).

In general even with the use of priors, photometric redshifts are
typically taken to be simply the redshift corresponding to the maximum likelihood point of the redshift probability
function. Template-based photometric redshift algorithms, however, can provide
a full probability distribution over the entire redshift and
spectral type range.  Using the full redshift probability distribution
function for each galaxy can significantly reduce the number of catastrophic
outliers by, for example, excluding galaxies with broad or multiply
peaked probability distributions.

By identifying and pre-filtering problematic regions in photometric redshift
space, we can exclude the galaxies most likely to produce outliers
while retaining the galaxies that have redshifts that are well-constrained. For most statistical studies, it is far more important to
eliminate outliers than to maximize the total number of galaxies in
the sample. 
Application of a simple photometric redshift space filter (for example, excluding 
galaxies classified as blue galaxies at $1.5<z<1.8$, which are
particularly susceptible to catastrophic failure) gives an outlier
fraction for an $i<25.3$
sample  a factor of two smaller than those we've described in this
section. 
Other priors can further reduce outliers.

Both the specific photometric redshift technique used, the appropriate
selection methods, priors, and their weights will be science case
specific. For high redshift galaxies or for searches for unusual
objects, heavily weighting priors based on galaxy
properties may suppress those sources. For science cases
requiring galaxies of specific types (e.g.,\ baryon acoustic
oscillation measurements, \autoref{sec:lss:bao}) or for galaxies with
particular observed attributes (e.g.,\ resolved galaxies for weak
lensing studies), methods for optimizing priors must be defined.



\subsection{Photometric Redshift Calibration: Training Sets}
\label{sec:common:phz:cal}


Calibration of photometric zero points, SEDs, 
and priors will be critical for developing photometric redshifts for the LSST.
If the range of spectral types is only coarsely sampled, the
uncertainty in predicted redshift will increase, as the exact SED for
an individual galaxy may not be present in the training set.  For
example, if we use only 50\% of the model galaxy templates used to generate spectra when
computing photometric redshifts with the methods used for Figures \ref{fig:photozUZband} 
and \ref{fig:photoz}, the scatter ($\sigma_z$)
increases by 40\% and the bias by 50\% overusing all of
the templates.  This outcome highlights the need for significant numbers of
spectroscopic galaxies to train our template SEDs, and also illustrates the 
need for training sets to span the properties of galaxies in the samples 
to which we apply photometric redshifts.

It remains unclear both how small a subset of the complete data is
sufficient to determine the 
overall redshift structure, and how we might select that subset. If
the objects we seek reside only in certain regions of color 
space or have some specific properties, then simple sampling
strategies can be used to pick an appropriate subset for spectroscopy 
(e.g., \citealt{2009arXiv0902.2782B}). We often
cannot, however, isolate a problematic population a priori.  We could rely
on the ``standard'' technique of either applying a sharp selection
threshold in a particular attribute (e.g.,\ galaxy size or magnitude) or picking
a suitable random fraction of the underlying sample and then studying
this population in detail. Neither of these na\"\i ve techniques is
optimal in any statistical sense; their only appeal is their apparent
simplicity.


For certain well-defined parameter estimation problems there are
classical stratification techniques \citep{ney38}
if we want the optimum variance estimator over a sample consisting of
discrete ``classes,'' each with its own variance. These stratified
sampling strategies lie between the limiting cases of totally randomly
sampling from the full ensemble or randomly sampling each category. In
astronomy, however, it is quite rare that a single estimator will
suffice. More likely we seek a distribution of a derived quantity;
that is, we seek the distribution of an intrinsic quantity but have
only the observed quantity available (consider measurements of the luminosity function:
we seek to determine the probability distribution of the true physical brightnesses of a
population when only apparent magnitude can be measured).

Sampling strategies using Local Linear Embedding (LLE,
\citealt{row00}) can preserve the distribution of
spectral types of galaxies in local spectroscopic surveys with $\sim
10,000$ galaxies (compared to an initial sample of 170,000 
spectra). This is done by considering how much new information is
added as we increase the number of galaxies within a sample
\citep{vPlas++09}. This reduction in sample size required to
encapsulate the full range of galaxy types is also consistent with the
sample sizes used for Principal Component Analyses of SDSS spectra
\citep{yip++04}.


Based on this fact, one approach would be to generate a series of
selected fields distributed across the sky with galaxies to $r \sim
26$ calibrated via selected deep spectroscopy.  We would need
sufficient numbers of galaxies per redshift bin to beat down the
statistical errors to at least the level of the systematic errors.  If
we take the previously stated dark energy systematic targets as a goal
($\delta_{z}=0.003(1+z),\ \Delta\,\sigma_{z}=0.004(1+z)$), then we
need $\sim\ 6000$ galaxies per bin.  
In fact, given that we need to
characterize the full distribution function, as it is non-Gaussian, it
is more likely that we would need $\sim 100,000$ galaxies total if the sample were
split up into ten redshift bins.  The number needed can, however, be
reduced by almost a factor of two by sampling the redshift distribution in an optimized manner
\citep{ma++08}.  This number is comparable to that needed
for calibration of the templates and zero points.
For the gold sample, $i<25$, obtaining redshifts for 50,000
galaxies over several calibration fields is not an unreasonable goal
by 2015; there are existing samples of comparable size already down to 
somewhat brighter magnitude limits.  For instance, the DEEP2 Galaxy Redshift 
Survey has obtained spectra of $>50,000$ galaxies down to $R_{AB}=24.1$ 
(Newman et al. 2010, in preparation), while VVDS \citep{Gar++08} and 
zCOSMOS \citep{Lil++09} have both obtained spectra of $\gtsim 20,000$ 
galaxies down to $i=22.5$, and smaller samples extending to $i=24$.

\subsection{Photometric Redshift Calibration: Cross-correlation}  

\label{sec:photoz:cross}

An alternative method that can get around any incompleteness issues in
determining redshift distributions is to employ cross-correlation
information \citep{Newman08}.  Past experience suggests we may not be
successful in obtaining redshifts for all of the galaxies selected for
spectroscopy; recent relatively deep ($i<22.5$ or $R<24.1$) surveys have obtained high-confidence
($>95\%$ certainty) redshifts for from 42\% (VVDS; \citealt{Gar++08}) 
to 61\% (zCOSMOS; \citealt{Lil++09}) to 75\%
(DEEP2; Newman et al. 2010, in preparation) of targeted galaxies, and extremely
high-confidence ($>99.5\%$) redshifts for 21\% (VVDS) -- 61\% (DEEP2).  
Surveys of fainter galaxies have even higher rates of failure \citep{2004AJ....127.2455A}.  
Redshift success rate in these surveys is a strong function of both
galaxy properties and redshift; i.e., the objects missed are not a fair sample.  

Deep infrared spectroscopy from space has problems of its own.  The
field of view of JWST is quite small, resulting in large cosmic
variance and small sample size, and Joint Dark Energy Mission (JDEM) or Euclid spectroscopy will
be limited to emission-line objects.  Even with a spectroscopic
completeness as high as that of SDSS ($\sim
99\%$; \citealt{Strauss2002}), the missed objects are 
not a random subsample, enough to bias redshift
distributions beyond the tolerances of dark energy
experiments \citep{2008MNRAS.386.1219B}. 

Even if spectroscopic follow-up systematically misses some populations,
however, any well-designed spectroscopic campaign will have a large
set of faint galaxies with well-determined redshifts.  These 
can then be used to determine the actual redshift
distribution for any set of galaxies selected photometrically, such as
objects in some photometric redshift bin, via angular
cross-correlation methods.

\begin{figure}
\begin{center}
\includegraphics[scale=.25]{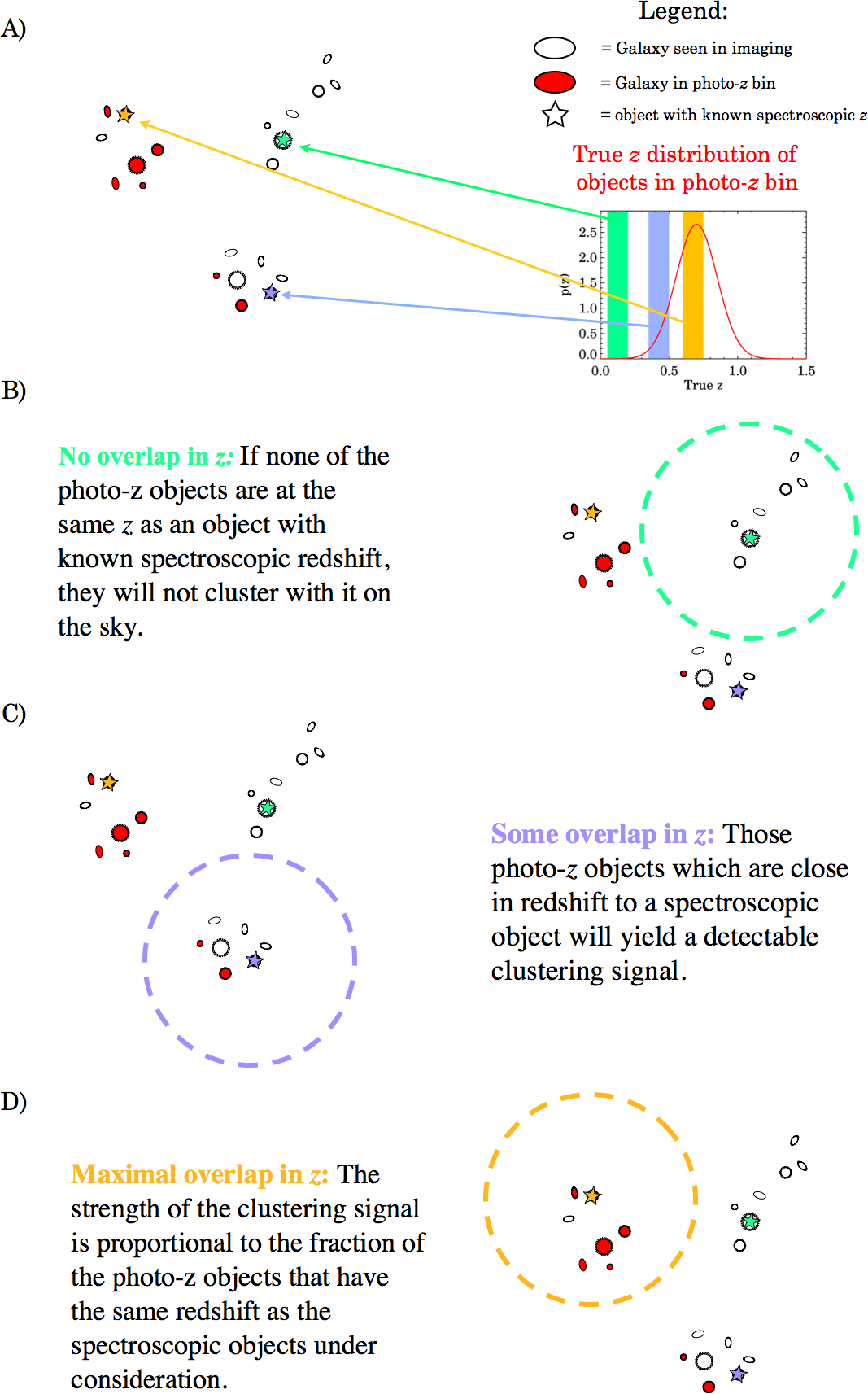}
\end{center}
\caption{\footnotesize Cartoon depiction of cross-correlation
photometric redshift calibration \citep{Newman08}.   Panel A) shows the
basic situation: we have imaging for many galaxies (circles/ellipses),
some of which fall in a photometric redshift bin of interest (red).
Galaxies that are near each other in three dimensions cluster together
on the sky.  We also know the spectroscopic redshifts of a smaller
sample of objects (stars).  The true redshift distribution for the
objects in the photometric redshift bin is here assumed to be a
Gaussian with mean 0.7 (plot); the stars are color-coded according to
the redshift range the galaxy in question was determined to lie in
with the color-coded ranges shown on the plot.  B) For
spectroscopic redshift objects that do not overlap in $z$ with the
photometric redshift objects, there will be no excess of neighbors
that lie in the photometric redshift sample.  C) If there is some
overlap with the true redshift range of the photometric redshift
sample, there will be some excess of neighbors around the
spectroscopic object that lie in the photometric redshift bin.  D) The
strength of this clustering signal will be stronger the greater the
fraction of the photometric redshift sample lies at the same $z$ as
the spectroscopic object in question.  Because of this, we can
reconstruct the true redshift distribution of the photometric redshift
sample by measuring its clustering with objects of known redshift as a
function of the spectroscopic $z$.  \label{fig:cartoon}} 
\end{figure}

Because galaxies cluster together over only relatively small
distances, any observed clustering between a photometric sample and
galaxies at some fixed redshift, $z_s$, can only arise from galaxies in
the photometric sample that have redshifts near $z_s$
(\autoref{fig:cartoon}).  Therefore, by 
measuring the angular cross-correlation function (the excess number of
objects in one class near an object of another class on the sky, as a
function of separation) between a photometric sample and a
spectroscopic sample as a function of the known spectroscopic $z$, we
can recover information about the redshift distribution of the
photometric sample (hereafter denoted by $n_p(z)$; \citealt{Newman08}).  If we only
measure this cross-correlation, the redshift distribution would be
degenerate with the strength of the intrinsic three-dimensional
clustering between the two samples; however, the two-point
autocorrelation functions
of the photometric and spectroscopic samples 
provide sufficient information to break that degeneracy.  
Other cross-correlation techniques for testing photometric redshifts have been 
developed \citep{2006ApJ...644..663Z, 2006ApJ...651...14S}, but they do not break the 
clustering-redshift distribution degeneracy.


%


In the limit where sample cosmic variance is negligible
(e.g., because many statistically independent fields on the sky have
been observed spectroscopically), and spectroscopic surveys
cover $\gtsim$ 10 deg$^2$ on the sky, Monte Carlo simulations
\citep{Newman08} find that the errors in determining either $\langle z
\rangle$ or $\sigma_z$ for a Gaussian $n_p(z)$ for a single
photometric redshift bin are nearly identical, and are fit within 1\%
by: 
\begin{equation}
	\sigma = 9.1\times 10^{-4} \left({\sigma_z \over 0.1}\right)^{1.5} \left({\Sigma_p \over 10}\right)^{-1/2} \left( {dN_s/dz \over 25,000 }\right)^{-1/2} \left({4\,h^{-1}\, {\rm Mpc} \over r_{0,sp}}\right)^{\gamma} \left({10\, h^{-1} \,{\rm Mpc} \over r_{max}}\right)^{2-\gamma}  \,,
	\label{eq:scaling0cv}
\end{equation}
where $\sigma_z$ is the Gaussian sigma of the true redshift
distribution, $\Sigma_p$ is the surface density of objects in the
given photometric redshift bin in galaxies arcmin$^{-2}$,
$dN_s/dz$ is the number of objects with spectroscopic redshifts per
unit $z$, $r_{0,sp}$ is the true scale length of the two-point
cross-correlation function between spectroscopic and photometric
galaxies (the method provides a measurement of this quantity as a
free byproduct); and $r_{max}$ is the maximum radius over which
cross-correlations are measured (larger radii will reduce the impact
of nonlinearities, at the cost of slightly lower S/N).  Typical values of $r_0$ and $\gamma$ for both local and $z\sim 1$ galaxy samples are 3-5 $h^{-1}$ Mpc and 1.7--1.8, respectively \citep{2005ApJ...630....1Z,2006ApJ...644..671C}.  

\begin{figure}
\centerline{\resizebox{3in}{!}{\includegraphics{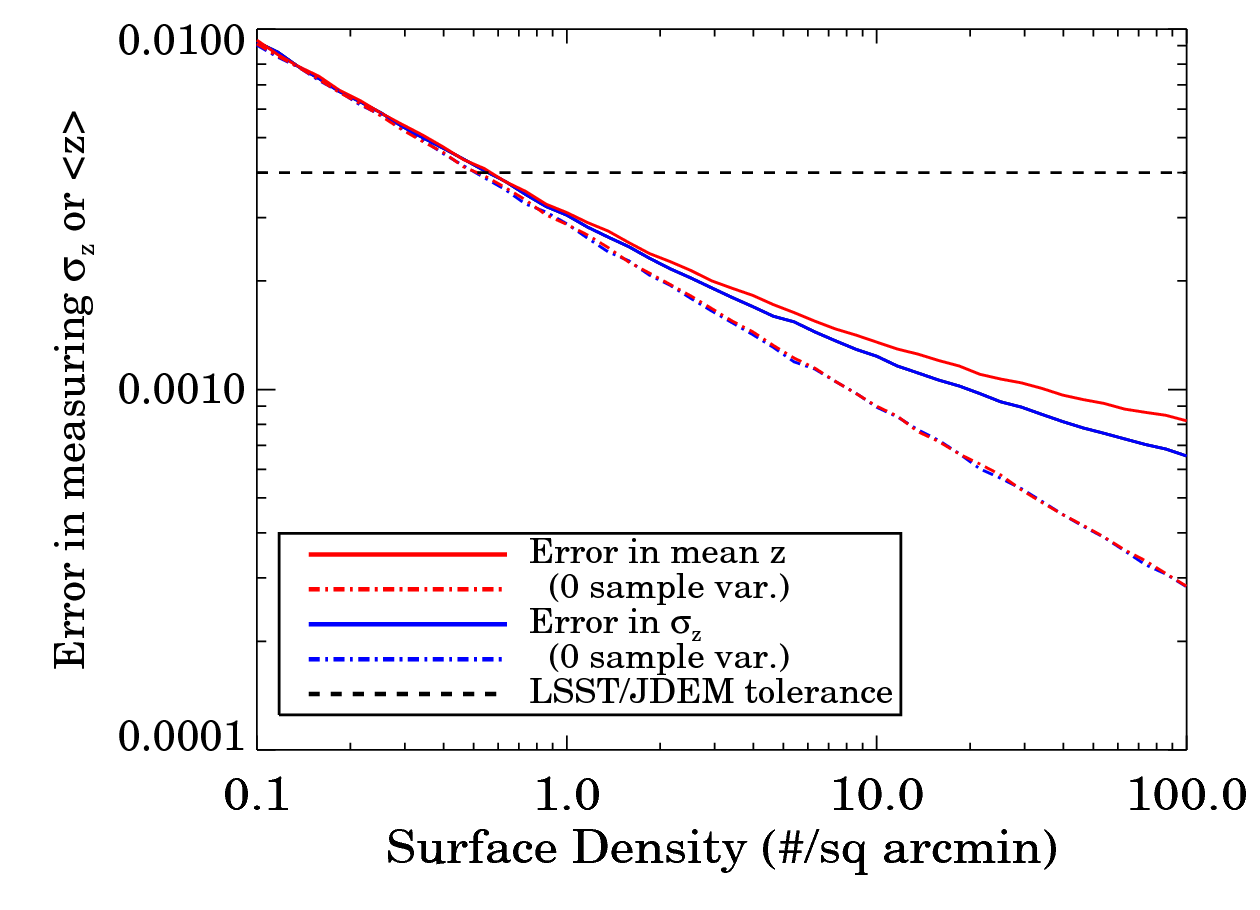}}}
\caption{Results from Monte Carlo simulations of uncertainties in
  cross-correlation measurements of redshift distributions.  Plotted
  are the rms errors in the recovery of the mean and sigma of a
  photometric sample distributed as a Gaussian in $z$ with
  $\sigma_z=0.1$, as a function of the surface density of that sample
  (representing objects in a single photometric redshift bin) on the
  sky.  We assume a fiducial spectroscopic survey of 25,000 galaxies
  per unit redshift.  Current and planned spectroscopic samples are
  sufficient to reach the required LSST photometric redshift calibration tolerances
  at $z<2.5$, but larger sets of redshifts than currently available at
  $z>2.5$ may be required.
\label{fig:xcorr}}
\end{figure}

Errors are roughly 50\% worse in typical scenarios if sample variance
is significant (i.e., a small number of fields, covering relatively
area, are sampled); see
\autoref{fig:xcorr} for an example of these scalings.  Detecting
non-Gaussianities such as tails in the photometric redshift 
distributions is straightforward in this method.  The
number of spectroscopic galaxies required to meet LSST photometric redshift bias and error characterization requirements is similar to the
number in current and funded redshift samples for $z<2.5$.  More
details on cross-correlation photometric redshift calibration and on potential systematics are given in \citet{Newman08}.


\begin{figure}
\centerline{\resizebox{2.5in}{!}{\includegraphics{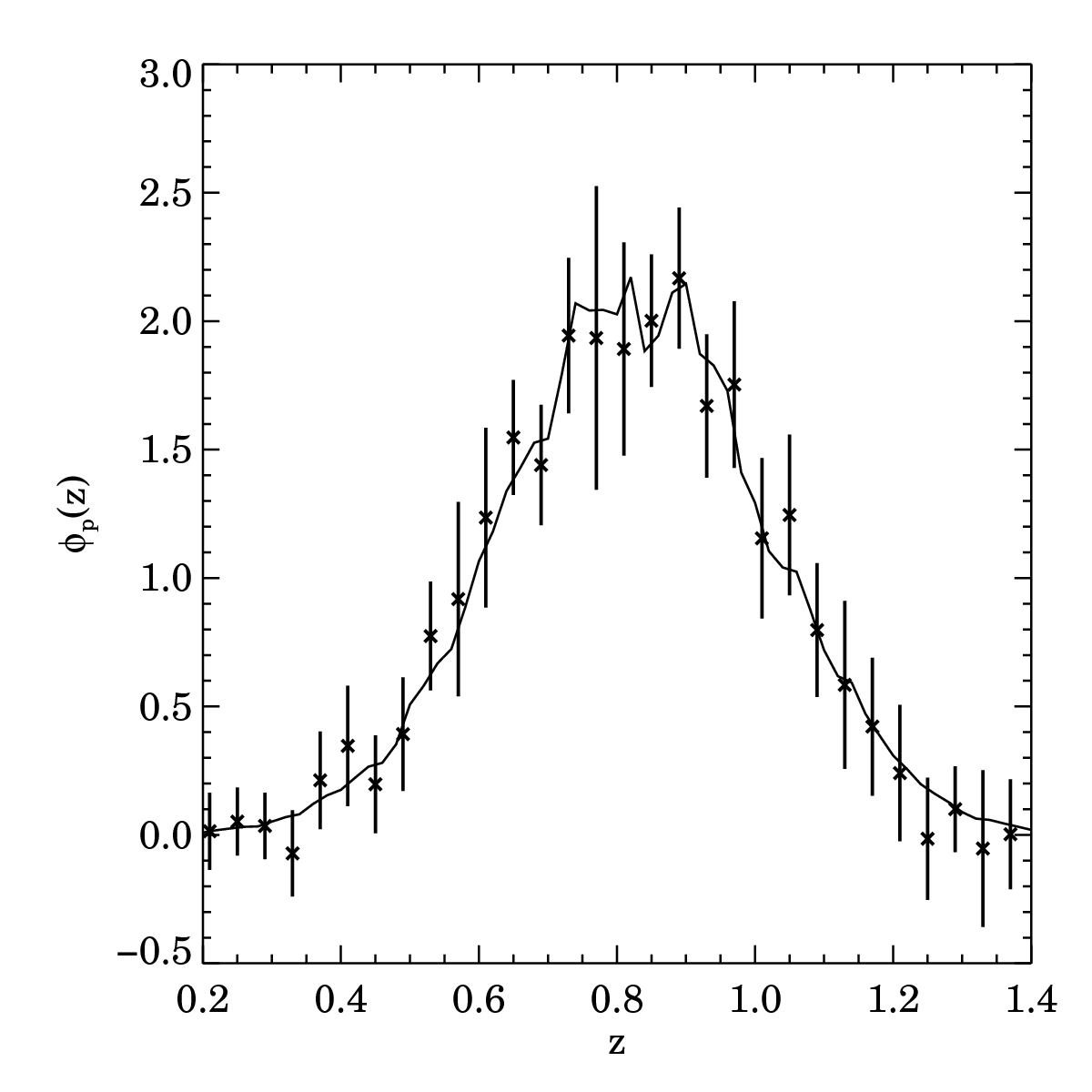}}}
\caption{A demonstration of the recovery of redshift distributions
  with cross-correlation techniques. Results for a single redshift bin
  are shown.  The solid line is the true
  redshift distribution of a subset of those galaxies with $M_B-5 {\rm log}
  h<-17$ in 24 0.5$\times$2 degree light-cone mock catalogs
  constructed from the Millennium Simulation semi-analytic models of
  \citet{Cro++06}, with the probability of being included in the
 set given by a Gaussian in redshift, centered at $z=0.8$ and with
 dispersion $\sigma_z=0.2$.  Deviations from a Gaussian curve are due
  to sample (or ``cosmic'') variance in the redshift distribution.
  Points and error bars show the median and rms variation in the
  cross-correlation reconstruction of this true distribution using a
  spectroscopic sample consisting of 60\% of all galaxies down to
  $R_{AB}=24.1$ in only 4 of the 24 fields.  The true distribution may
  be reconstructed to the accuracy required by LSST using spectroscopic samples of realistic size.
\label{fig:xcorr_demo}}
\end{figure}

We have tested these Monte Carlo simulations by applying
cross-correlation techniques to mock catalogs produced by
incorporating semi-analytic galaxy evolution prescriptions into the
Millennium Run simulation \citep{Cro++06, K+W07}.  Although these
simulations do not perfectly match reality, they do present the same
sorts of obstacles (e.g., clustering evolution) as we will encounter
with LSST samples.  As seen in \autoref{fig:xcorr_demo},
cross-correlation techniques can accurately reconstruct the true
redshift distribution of a sample of faint galaxies using only
spectroscopy of a subset of bright ($R<24.1$) objects over 4 deg$^2$
of sky.  The dominant uncertainty in the Millennium Run
reconstructions is due to the variance in the integral constraint
\citep{1994ApJ...424..569B}, which was not included in the error model
of \citet{Newman08}.  This variance can be suppressed, however, by use
of an optimized correlation estimator (e.g.,
\citealt{2007MNRAS.376.1702P}), and is negligible if spectroscopic
surveys cover $\gtsim 10$ deg$^2$ fields. 

Cross-correlation methods can accurately determine photometric error distributions 
for faint galaxies even if we obtain spectra of only the brightest objects at a given 
redshift (there are many $z=2$ galaxies with $R<24$, for instance, or $z=3$ galaxies with $R<25$).  
This is in contrast to methods which calibrate photometric redshifts via spectroscopic samples, 
due to the differences in SEDs between bright and faint galaxies and the substantial impact 
of confusion/blending effects on samples of faint, high-redshift galaxies (Newman et al. 2010, in preparation).  
To apply cross-correlation techniques, we do not
need an excessively deep sample, nor must it be uniformly complete,
only well-defined.  The proposed BigBOSS survey
\citep{2009arXiv0904.0468S} or a proposed wide-field spectrograph on Subaru would be ideal for these purposes, providing samples of millions of galaxies, Lyman $\alpha$ absorbers, 
and QSOs with spectroscopic redshifts to $z=2.5$, each with different clustering characteristics facilitating cross-checks.  Even if BigBOSS only overlaps with LSST
around the Celestial Equator, it should  
provide large enough numbers of redshifts to meet LSST calibration goals.  

It would, of course, be preferable to obtain statistically complete spectroscopy down to a 
limit approaching the LSST photometric depth rather than relying on these less direct techniques.  
Even if spectroscopy does not prove to be sufficiently complete to test calibrations, the closer 
we can come to that goal, the better the photometric redshift algorithms we will be able to develop.  
It would be extremely difficult to tune those algorithms using cross-correlation techniques alone, 
without also using a set of objects with well-known redshifts and SEDs.    
Furthermore, as seen in \autoref{eq:scaling0cv}, the better our photometric redshifts are 
(i.e., the smaller $\sigma_z$ is), the more precisely we can calibrate them.  
Making sure we are achieving the tight calibration requirements for LSST dark energy studies 
will require cross-checks.  Cross-correlation techniques will allow us to do this by repeating 
the analysis with very different spectroscopic samples; if all is working properly, the redshift 
distributions from each spectroscopic sample should agree.  As an example, one could use one set 
of spectroscopy going faint in the deep LSST calibration fields, and another, shallower or sparser 
set, covering the main wide field LSST survey.  Alternatively, one could use spectroscopic 
subsamples with very different clustering properties (e.g., star forming galaxies versus luminous 
red galaxies) to do the test. The recovered redshift distribution for a photometric redshift bin must be 
consistent when applying any variety of~type, redshift, and magnitude cuts to the spectroscopic 
sample if the reconstruction is accurate.



\bibliographystyle{SciBook}
\bibliography{common/common}

\chapter{Education and Public Outreach}
\label{chp:epo}

{\it Suzanne H. Jacoby, Kirk D. Borne, Julia K. Olsen, M. Jordan Raddick, Sidney C. Wolff}

\section{Introduction}

Goals of the Education and Public Outreach (EPO) program include
engaging a broad audience in LSST's science mission, increasing public
awareness of scientific research, contributing to science, technology,
engineering, and math (STEM) education, and enhancing 21st century
workforce skills.  LSST will contribute to the national goals of
improving scientific literacy and increasing the global
competitiveness of the US science and technology workforce.  The open
data access policy and survey operations mode of LSST facilitates the
active engagement of a broad audience in many venues: in the
classroom, through science centers, and in our homes, anywhere with
access to the Internet.  The LSST project will provide value-added
products to enable both student and public participation in the
process of scientific discovery.  The LSST EPO program is well
planned, tuned to our audience needs, aligned with national education
standards, and integrated with the science mission of LSST.  The
program is organized around three main threads: 1) inquiry-based,
scientifically authentic exploration in the classroom;
2) visualization of LSST data in science centers and on computer
screens of all sizes; and 3) support of public involvement in
activities that may be as simple as browsing through the data or as
sophisticated as contributing to active research projects through
Citizen Science opportunities.

\section{National Perspective on Education Reform}

Scientific literacy, defined as the ``knowledge and understanding of
scientific concepts and processes required for personal decision
making, participating in civic and cultural affairs, and economic
productivity,'' is a requirement in today's complex society
\citep{NRC96}. Yet, only 28\% of American adults currently qualify as
scientifically literate; nearly 70\% of Americans adults cannot read
and understand the Science Times section of the New York Times
\citep{Miller07}.  Policy makers, scientists and educators have
expressed growing concern over the fact that most people in this
country lack the basic understanding of science that they need to make
informed decisions about many complex issues affecting their lives
\citep{Singer05}.

The influential report ``A Nation at Risk'' investigated the declining
state of the educational system in the US, identified specific
problem areas, and offered various recommendations for improvement
\citep{NCEE83}. The report specifically documented the need for
greatly improved science education in this country and galvanized the
inclusion of the quality of education as a prominent element of the
national political agenda.  A succession of education reform efforts
set forth to remedy the situation: standards-based reform, the
establishment in 1989 of National Education Goals, National Science
Education Standards put forth by the National Research Council in
1996, and most recently the No Child Left Behind legislation.  All
have sought to standardize classroom learning goals, improve
instructional methods, and enforce accountability.

Today, science education standards exist for content, teaching, and
assessment, such as the National Science Education Standards
\citep{NRC96} or Project 2061 \citep{AAAS94}.  Consistent with the ``A
Nation at Risk'' report, expectations are defined for high school
graduates, whether or not they plan additional education.  These
expectations include the ability to know, use, and interpret specific
mathematical and scientific concepts, but also the ability to evaluate
scientific evidence, understand scientific development, and
participate in scientific practices and discourse.  Beyond mere facts,
it is these ``habits of mind'' that result in a scientifically
literate populace, capable of participating in an increasingly complex
global society.

In a changing world, new kinds of knowledge and skills are as valuable
as core subjects.  The 21st century worker must have strengths and
attitudes dramatically different from typical workers of today, who were trained in the 20th century.  In
particular, three areas of proficiency must be addressed in preparing
the 21st century workforce: core knowledge in science, mathematics,
and other content areas; learning and thinking skills; and information
and communications technology. Critical for workers of the future is
the ability to incorporate high-level cognitive abilities with
inventive thinking skills such as flexibility, creativity, problem
solving, effective communication, and collaboration. The use of
technology as a tool for research, organization, evaluation and
communication of information is an integral aspect for the future
workforce. It is skillfulness in both proactive learning and response
to innovation that will separate students who are prepared for the
work environment of the 21st century from those who are not.
Scientific literacy, education reform, and workforce preparedness are
all elements of the educational environment in which LSST is poised to
contribute.

\section{Teaching and Learning in the Classroom}

LSST data can become a key part of projects emphasizing
student-centered research in middle school, high school, and
undergraduate settings.  Taught in an exemplary fashion, using
technology to its best advantage, students can participate in
cutting-edge discovery with authentic classroom research opportunities
developed through the LSST EPO effort.  The LSST education program
will design and develop a number of student research projects in
conjunction with a teacher professional development program.

As described in ``How People Learn'' \citep{Bra++00}, the goal of
education is to help students develop needed intellectual tools and
learning strategies, including how to frame and ask meaningful
questions about various subject areas.  This ability will help 
individuals to become self-sustaining, lifelong learners.

Engaging students by using real data to address scientific questions
in formal education settings is known to be an effective instructional
approach \citep{Man++02}.  The National Science Education Standards
\citep{NRC96} emphasize that students should learn science through
inquiry (Science Content Standard A: Science as Inquiry) and should
understand the concepts and processes that shape our natural world
(Science Content Standard D: Earth and Space Science).  Students learn
best if they are not passive recipients of factual information but
rather are engaged in the learning process \citep{Wan++94, Hake98,
  Pra++04, Duncan06}.

Professional development, including the preparation and retention of
highly qualified teachers, plays a critical role. The importance of
teachers cannot be underestimated.  The most direct route to improving
mathematics and science achievement for all students is better
mathematics and science teaching.  In fact, ``\ldots teacher
effectiveness is the single biggest factor influencing gains in
achievement, an influence bigger than race, poverty, parent's
education, or any of the other factors that are often thought to doom
children to failure'' \citep{Carey04}.

One goal of having teachers and their students engage in data analysis
and data mining, is to help them develop a sense of the methods
scientists employ, as well as a familiarity with the tools they use to
``do science.''  The common lecture-textbook-recitation method of
teaching, still prevalent in today's high schools, prevents students
from applying important scientific, mathematical, and technological
skills in a meaningful context.  This model of teaching science is
akin to teaching all the rules of a sport, like softball, to a child
-- how to bat, catch, throw, slide, and wear the uniform -- but never
letting the child actually play in a game \citep{Yager82}.

In order to support implementation of scientific inquiry in classrooms
using public databases, the LSST EPO team is exploring the technological and
pedagogical barriers to educational use of data mining and integrating
that knowledge into planned professional development and classroom
implementation modules.  We refer to this effort as CSI: The Cosmos,
capitalizing on public appeal of crime scene investigation television
shows.  We model a research question as a crime scene, with a mystery
to be solved, and answers are found through clues mined from the
database.  Our goal is to develop a feasible plan promoting data
mining as an instructional approach and successful classroom
implementation, facilitating authentic research experiences using the
LSST database.  This approach provides an authentic experience of
astronomy as a forensic (evidence-based) science.  What is learned and
what is known about our Universe comes entirely from evidence that is
presented to us for observation through telescopes and preserved by us
for exploration in databases.  The CSI model of learning science
resonates with the inquisitiveness of the human mind --- everyone
loves a good detective story.

The LSST EPO group has adopted the formal process of Understanding by
Design \citep{W+M05} to facilitate the cohesive planning and
implementation of LSST education for specific audiences.  Experience
shows that the most successful classroom research projects fall into
two categories, both of which are natural outcomes of the LSST
database:
\begin{enumerate}
\item projects that use the same analysis techniques with a changing
  data set, e.g., measuring lightcurves of a series of novae or supernovae, and
\item the classification or organization of large samples of a
  particular object type, such as galaxies.
\end{enumerate}

Sample Learning Experiences being explored for formal settings are all
aligned with NRC content standards for Earth \& Space Science,
Technology, and Physical Science.  Those involving large number
statistics and classification are aligned with mathematical content
standards.  All can be taught in an inquiry-based approach and
supported with appropriate professional teacher development.  These
Learning Experiences include:

\begin{enumerate}

\item Wilderness of Rocks: Students classify asteroids (by rotation
  curve, light curve, and colors), make maps of their interplanetary
  distribution and orbital paths, and use colors to determine
  composition.  Students also deduce the shape, orientation, and
  family membership (and possible binarity of the system) from LSST
  asteroid observations. Learning goal: to understand scientific
  classification and inference through synthesis of information; to
  understand the scientific measurement process, data calibration, and
  reduction; and to understand properties of primordial Solar System
  bodies. This broad area of investigation could be implemented at
  middle school, high school, or undergraduate levels.

\item Killer Asteroids: Students measure the locations of small Solar
  System bodies in multiple LSST images to calculate their orbital
  parameters and to see if a planetary impact is possible.  If an
  asteroid will pass near a planet, the odds of an impact are also
  determined.  Learning goal: to understand orbits, hazards from
  space, detection methods, and mitigation strategies.

\item Type Ia Supernovae in the Accelerating Universe: An analysis of
  Type Ia Supernovae light curves could be developed in partnership
  with the SDSS-II survey during the LSST construction phase.
  Students would monitor the images of $\sim$ several hundred nearby
  galaxies as measured by LSST, and try to find supernovae. This
  project is most appropriate for physics classes and astronomy
  research classes at the high school and undergraduate levels.
  Learning goal: to understand scientific data collection, and to
  understand fundamental physics as it applies to cosmology and stars.

\item Photometric Redshifts: Using optical colors from the
  LSST database, students apply the photometric redshift technique to
  measure the distance to high-redshift galaxies and to estimate their
  star formation history.  Learning goal: to understand the concepts
  of photometric redshift, star and galaxy evolution, and
  model-fitting.

\item Galaxy Crash (Train Wreck): Using deep, wide surveys at many
  wavelengths, students track the rate of galaxy collisions as a
  function of redshift.  While we can't watch individual galaxies
  collide and merge, we can use a wide survey to catch an ensemble of
  colliding galaxies in all stages of interaction in order to
  understand the processes of environment-driven galaxy-building and
  cosmological mass assembly.  Learning goal: to understand galaxy
  evolution timescales and the concepts of dynamical evolution,
  hierarchical galaxy formation, and the development of the Hubble
  sequence of galaxies.

\item Star Cluster Search: Students search for overdensities of stars,
  to determine if a star cluster or star stream may be contained
  within the data.  Students plot a simple H-R diagram and estimate
  the age of the star cluster or star stream (from the H-R diagram).
  If the overdensity looks promising, students can check lists of
  known clusters (e.g., WEBDA\footnote{\url{http://www.univie.ac.at/webda/}}) to determine other properties of
  the star system and to verify their age estimate.  Learning goal: to
  understand the HR diagram, star formation in groups, stellar
  evolution, the difference between apparent and absolute magnitudes,
  gravitational clustering in astrophysical settings, and how to check
  online databases.

\end{enumerate}

\section{Outside the Classroom --- Engaging the Public}

The formal education system does not exist in a vacuum; students,
teachers, and families are all part of the broader context in which we
learn.  Opportunities that exist for learning outside the classroom
include Informal Science Education (ISE), Out-of-School Time (OST),
and the world of Citizen Science, where non-specialist volunteers
assist scientists in their research efforts by collecting, organizing,
or analyzing data.  More than a decade of research now shows that
sustained participation in well-executed OST experiences can lead to
increases in academic achievement and positive impact on a range of
social and developmental outcomes \citep{Harvard08}.

Adults play a critical role in promoting children's curiosity, and
persistence studies show that that one of the best indicators of
likely success in the educational system (i.e., matriculation all of
the way to graduation) is a home environment that is supportive of
education \citep{NIU09}.  Engagement of parents in informal education,
visits to museums and planetaria, and now Citizen Science can all help
to create an environment that encourages young people to pursue
challenging courses in math and science. As then-candidate Barack
Obama said in his speech, ``What is Possible for our Children" in May
2008, ``There is no program and no policy that can substitute for a
parent who is involved in their child's education from day one''
\citep{Den08}.

``Experiences in informal settings can significantly improve science
learning outcomes for individuals from groups, which are historically
underrepresented in science, such as women and minorities.
Evaluations of museum-based and after-school programs suggest that
these programs may also support academic gains for children and youth
in these groups'' \citep{Bell++09}.

Two concepts are under development to engage the interested public in
LSST through the Internet outside of the classroom.  It is expected
that these public interfaces can provide a gateway to more formal
activities in the classroom as described above, once interest is
established.

\begin{enumerate}
\item Cosmic News Network (CN$^{2}$): A web-based news report on
  ``changes'' in the world of physics and astronomy; that is, a News,
  Weather, and Traffic Report of the Universe.  Presented in the
  format of an online popular news source like cnn.com or msnbc.com,
  we will collect, organize, and present information on everything
  that could be reported as news in the Universe: phases of the Moon,
  eclipses, planet positions, satellite locations, the discovery of
  new asteroids, new Kuiper Belt objects, extra-solar planet transits,
  supernovae, gamma ray bursts, gravitational microlensing events,
  unusual optical transients, particle physics experiments, solar
  weather data, launches, comets, hot stories, and more.  New media
  technologies will be used on the site, including an LSST blog and
  links to existing podcasts and video casts, RSS feeds and widgets of
  interest.  Just as someone checks the morning on-line or on-paper
  news source to learn what happened overnight in the world, they
  would access the CN$^{2}$ web portal to learn about recent
  happenings in the Universe, including daily reports of the most
  significant LSST alerts and transient events.
\item LSST@HOME: A way for the general public or classrooms to adopt a
  piece of the celestial highway and call it their own.  As in the
  public ``Adopt-A-Highway'' service along our nation's highways,
  individuals and organizations would register at no cost to be
  identified with a patch of the Universe.  ``Owners'' of the patch
  can contribute their own inputs: images, links to other data and
  information resources for sources in the region, news events based
  in that region, tracks of asteroids that have passed or that will
  pass through the area, new measurements (astrometry, photometry,
  redshifts), links to related published papers, etc. These celestial
  patches may provide the starting point for robotic telescope
  observation requests for ancillary data on objects and/or LSST
  events within the region. We will develop a mechanism to collect,
  distribute, and archive all metadata about the adoptable, small
  parcels of the ``LSST sky'' (e.g., one square degree), including a
  table of historical VOEvents within the region.  A user will be able
  to click anywhere on the LSST sky to learn about objects and
  discoveries within the selected stamp. Some users will be interested
  only in monitoring their patch of sky, while active users will be
  able to explore events and return their findings to the professional
  scientific community, for follow-up observations or publication.
  The gateway to the data can be provided through the World-Wide
  Telescope (WWT) or Google Sky interfaces.  The LSST EPO Database
  would serve the cutouts.  The VOEvent database would serve the
  alerts.
\end{enumerate}


With survey projects like LSST (and its predecessors) on the sky, the
role of amateur astronomers will shift away from discovery space into
opportunities for follow-up and data mining.  LSST saturates at
magnitude 16, well within the reach of many well-equipped amateurs.
Thousands of alerts per night will point to objects to be understood
and monitored.  Two windows of opportunity are particularly well
suited to amateur observations: 1) following an object's
brightness as its light curve rises above what LSST can observe and
2) filling in observations between LSST visits to increase time
coverage of suitable objects.  Working with the American Association
of Variable Star Observers (AAVSO), pro-am collaborations and Citizen
Science venues will be developed into partnerships that extend the
scientific productivity of LSST.

\section{Citizen Involvement in the Scientific Enterprise}
\label{sec:epo:Citizens}

Citizen Science is emerging as a popular approach to engaging the
general public and students in authentic research experiences that
contribute to the mission of a scientific research project \citep{Rad++09}.  
Citizen Science specifically refers to projects in which volunteers, many of
whom have little or no specific scientific training, perform or manage
research-related tasks such as classification, observation,
measurement, or computation.  As reported at the Citizen Science
Toolkit Conference held in Ithaca, NY, June 20th-23rd, 2007,
successful Citizen Science projects are known to include authentic
contributions to the field, not just ``busy work,'' as well as
validation for volunteer's effort.  LSST recognizes the importance of
Citizen Scientists in the astronomical endeavor and the vital
contributions to research activities made by volunteers from the
American Association of Variable Star Observers (AAVSO), NASA's Lunar
Impact Monitoring project, and others.\\
\begin{figure} [hbt]
\begin{center}
\includegraphics[width=0.4\textwidth,angle=0]{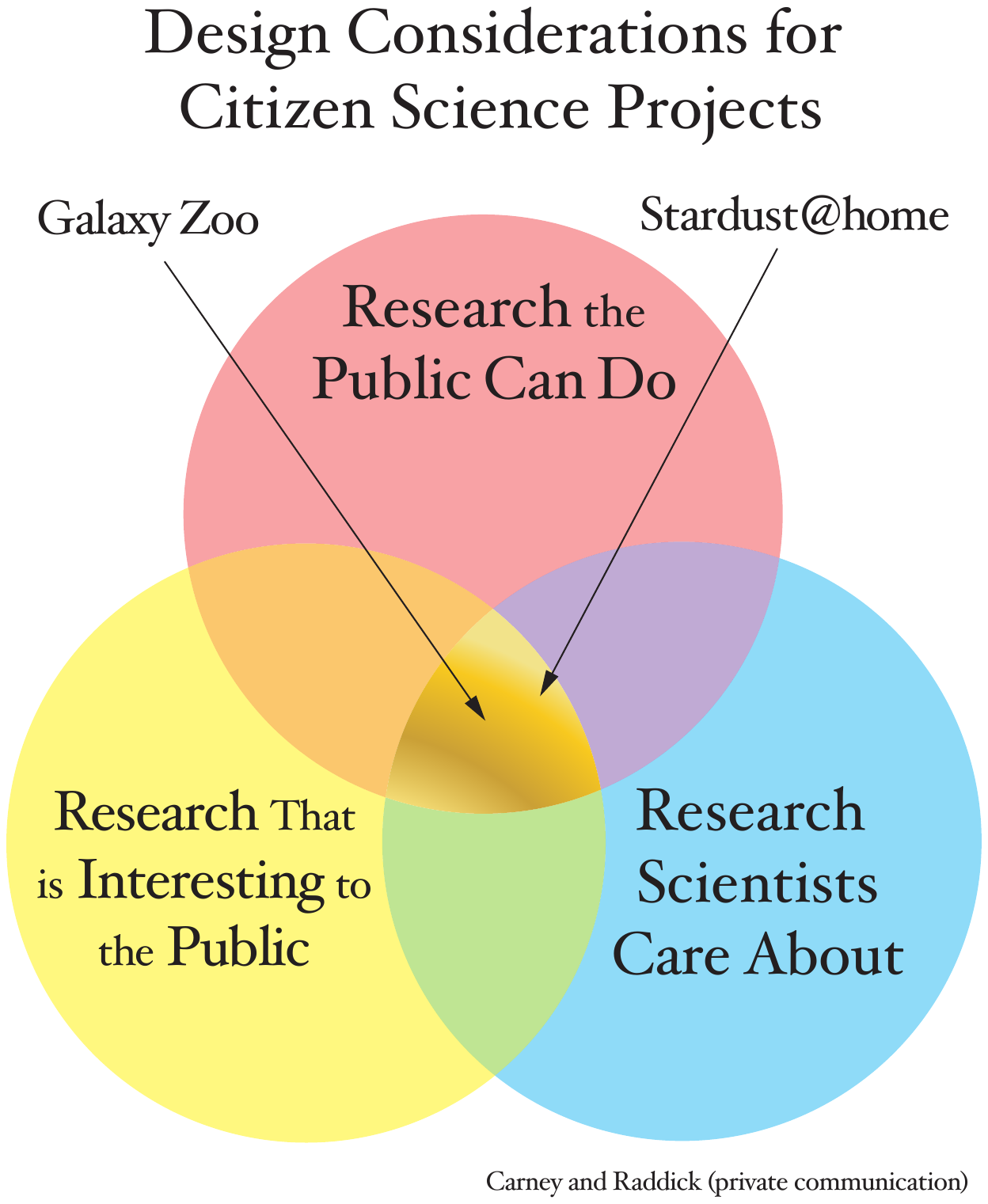}
\end{center}
\caption{ \scriptsize Citizen Science offers volunteers a fun and
  meaningful way to contribute to science. It offers scientific
  researchers the means to complete projects that are otherwise
  impossible to do on a reasonable time scale. The most successful
  projects maximize volunteer contributions and scientific
  value. Three overlapping circles symbolize design considerations for
  Citizen Science projects: Research the Public Can Do, Research that
  is Interesting to the Public, and Research that Scientists Care
  About.  Finding out exactly why a particular project occupies one
  portion of the diagram over another is a key part of the research
  agenda for Citizen Science.  In all cases, Citizen scientists work
  with real data and perform duties of value to the advancement of
  science.  }
\label{fig:epo:Citizen_Science}
\end{figure}

Citizen Science is one approach to informal science education,
engaging the public in authentic scientific
research. \autoref{fig:epo:Citizen_Science} illustrates design
considerations for Citizen Science Projects, showing three overlapping
circles: projects that people want to do, projects that people can do,
and projects that scientists want done.  A recent and highly
successful astronomy Citizen Science project, Galaxy Zoo, sits in the
sweet spot of the intersection of these three circles.  Galaxy Zoo has
involved more than 200,000 armchair astronomers from all over the
world in classifying the morphology of galaxies from the SDSS,
resulting in four papers published in peer-reviewed journals already
\citep{Lan++08, Lin++08, Slo++09, Bam++09}.  In the 18 months prior to
February 2009, 80 million classifications of galaxies were submitted
on 900,000 galaxies at galaxyzoo.org.

The Stardust@home project \citep{Mendez08} where volunteers pass a test to qualify to
participate in the search for grains of dust in aerosol gels from the
NASA Stardust Mission, has attracted smaller numbers of Citizen
Scientists (24,000), perhaps because of the more sophisticated
training and analysis required by participants, or perhaps because
images of galaxies are inherently more interesting to the larger
public than cracks in an aerosol gel.  In all cases, Citizen
Scientists work with real data and perform authentic research tasks of
value to the advancement of science.  The human is better at pattern
recognition (Galaxy Zoo) and novelty (outlier) detection
(Stardust@home) tasks than a computer, making Galaxy Zoo's galaxy
classification activity and others like it good candidates for
successful Citizen Science projects.

Within the realm of LSST, many Citizen Science projects are possible,
including these proposed by the Science Collaboration Teams:
\begin{enumerate}
\item Galaxy Zoo Extension: Continue the Galaxy Zoo classification
  project with LSST data, adding billions of candidates to the sample.
  Extend classification categories to include low surface brightness
  galaxies and mergers.  Put interacting galaxies in a sequence, and
  understand the timescales for the collision to produce detectable
  distortions in the galaxies and for the eventual merger of the two
  galaxies.  Explore how to use a large sample to probe changes over a
  Hubble time (\autoref{chp:galaxies}).
\item Light Curve Zoo: Classify light curves generated from the
  automatically provided photometry of variable objects.  Use trained
  human volunteers for the initial classifications
  (\autoref{chp:transients}).  The AGN group offers several
  suggestions (\autoref{sec:agn:epo}): Once a gravitationally lensed AGN has been identified
  via the presence of multiple images, one of the key projects would
  be to measure the brightness of the lensed images as the
  magnification caustics sweep across the accretion disk.  These light
  curves will then be used in a sophisticated statistical analysis to
  infer AGN accretion disk sizes.  One important and attainable
  project would be a study of the light curves of different classes of
  Active Galactic Nuclei, which are then used to model the differences
  due to obscuration and luminosity.
\item Lens or Not: Find new gravitational lens candidates via the
  Galaxy Zoo model.  Human recognition of arcs, rings, and
  multiply-imaged sources can supplement the pattern recognition tools
  within the LSST image processing pipeline and aid in the discovery
  of rare unique objects.  This investigation could be extended to the
  classroom by having students interactively model variables of mass,
  light, and placement to recreate the observed lensed
  candidates. (\autoref{chp:sl}).
\item Human Computing: Label and annotate LSST images, along the lines
  of the Google image labeler
  (\url{http://images.google.com/imagelabeler/}) or the ESP Guessing
  Game\\ 
  (\url{http://www.gwap.com}) in which participants select words to
  describe and annotate each image; the most popular descriptors
  become part of the image header.
\end{enumerate}

\section{Diversity}

A negative trend over the past 25 years is the increasing numbers of
students -- now nearly 1/3 -- who do not graduate from high school
\citep{G+W05} and who therefore do not posses the minimum education
required to be functioning Citizens and workers in a global
environment.  A disproportionate number of these students are from
groups of ethnic and racial minorities, students from low-income
families, and recent immigrants, all of whom have been ill-served by
our educational system.  The Greene and Winters study said: ``the
national graduation rate\footnote{Graduation rate is defined by the
  Manhattan Institute study to be: graduation rate = regular diplomas
  from 1998 divided by adjusted 8th grade enrollment from 1993.} for
the class of 1998 was 71\%.  For white students the rate was 78\%,
while it was 56\% for African-American students and 54\% for Latino
students.'' Sixteen of the 50 largest school districts in the US failed to
graduate more than half of their African-American students.  All but
15 of the districts for which rates can be computed have Latino
graduation rates below 50\%.  Minorities comprise the fastest growing
segment of the US workforce, yet these are the same individuals most
likely to be undereducated and consequently unqualified for positions
in the science and technology fields.  The statistics underscore the
importance of diversity and inclusion, as aging baby boomers leave the
workforce to an increasingly diverse pool of replacement workers.

LSST is well positioned to broaden participation of underrepresented
groups in astronomy and physics with its open access policy and EPO
plan integrating science and education.  The data-intensive aspects of
LSST includes research and education opportunities specifically in the
contexts of computer science, instrumentation, and the data sciences
\citep{B+J09}.  Thinking beyond the traditional types of students will
open up a vastly larger pool of talent encompassing a diversity of
disciplinary backgrounds and educational levels. LSST scientists and
engineers throughout the project will partner with Faculty and Student
Teams (FaST) from minority-serving institutions to develop long-term
research and educational opportunities.  This work builds on two years
experience with NSF/DOE sponsored FaST teams at three LSST
institutions: BNL (focal plane sensor development), SLAC (system
calibration), and UW (variability detection sensitivity).

\section{Summary}

The challenge of today is not only to build excellence in students and
teachers, but also to provide access to this excellence -- quality
education for all.  To do this, we engage the entire community --
students, teachers, parents, and the public -- with pathways to
lifelong learning.  With its open data policy and data products that
offer vast potential for discovery, congruence with educational
standards, and relevance to problems that are inherently interesting
to students, LSST offers a unique opportunity to blend research and
education and to achieve the national goal of quality education for
all students and enhanced scientific literacy for all citizens.

This engagement of the public in LSST-enabled formal and informal
education is not entirely altruistic on our part.  Full exploration of
the LSST databases (to maximize specific scientific goals) is likely
to require the engagement of large numbers of people outside the
formal LSST project structure, and even beyond the traditional
professional astronomy research community.  By welcoming educators,
students, and amateur astronomers to the LSST database, the doors will
be opened wide to all.  ``And why not open the doors wide? It's hard
to imagine that this data will ever get used up -- that all the good
discoveries will one day be wrung out of it -- so the more minds
working away at it, the better''  \citep{Bec09}.

LSST is uniquely positioned to have high impact with the interested
public and K-16 educational programs. Engaging the public in LSST
activities has, therefore, been part of the project design from the
beginning.  This involvement and active participation will allow LSST
to fulfill its public responsibility and extend its scientific
potential -- a truly transformative idea for the 21st century
telescope system.

\bibliographystyle{SciBook}
\bibliography{epo/epo}

%
%
%
%
%
%
%
%
%
%
%
%
%
%
%
%
%
%
%
%
%
%
%
%
%
%
%

\newcommand\x         {\hbox{$\times$}}
\newcommand\othername {\hbox{$\dots$}}
\def\eq#1{\begin{equation} #1 \end{equation}}
\def\eqarray#1{\begin{eqnarray} #1 \end{eqnarray}}
\def\eqarraylet#1{\begin{mathletters}\begin{eqnarray} #1 %
                  \end{eqnarray}\end{mathletters}}
\def\mic              {\hbox{$\mu{\rm m}$}}
\def\about            {\hbox{$\sim$}}
\def\Mo               {\hbox{$M_{\odot}$}}
\def\Lo               {\hbox{$L_{\odot}$}}
\def\comm#1           {{\tt (COMMENT: #1)}}
\def\kms   {\hbox{km s$^{-1}$}}

\chapter[The Solar System]
{The Solar System}
\label{chp:ss}

{\it R. Lynne Jones, Steven R. Chesley, Paul A. Abell, Michael E. Brown,  Josef {\v D}urech, Yanga R. Fern{\'a}ndez,
Alan W. Harris, Matt J. Holman, \v Zeljko Ivezi\'c, R. Jedicke, Mikko Kaasalainen, Nathan A. Kaib, Zoran Kne\v zevi\'c,
Andrea Milani, Alex Parker, Stephen T. Ridgway, David E. Trilling, Bojan Vr\v snak} 

LSST will provide huge advances in our knowledge of
millions of astronomical objects ``close to home'"-- the small bodies
in our Solar System. Previous studies of these small bodies have led to
dramatic changes in our understanding of the process of planet
formation and evolution, and the relationship between our Solar System
and other systems. Beyond providing asteroid targets for space missions or
igniting popular interest in observing a new comet or learning about a
new distant icy dwarf planet, these small bodies also serve as large
populations of ``test particles," recording the dynamical history of the
giant planets, revealing the nature of the Solar System impactor population 
over time, and illustrating the size distributions of planetesimals, which
were the building blocks of planets.

In this chapter, a brief introduction to the different populations of 
small bodies in the Solar System
(\autoref{ss:overview}) is followed by a summary of the number of objects
of each population that LSST is expected to find
(\autoref{ss:counts}).  Some of the Solar System science that LSST
will address is presented through the rest of the chapter, starting with the insights into 
planetary formation and evolution gained through the small body 
population orbital distributions (\autoref{ss:orbits}). 
The effects of collisional evolution in the Main Belt and Kuiper Belt 
are discussed in the next two sections, along with the implications for the
determination of the size distribution in the Main Belt (\autoref{ss:mainbelt_familsize})
and possibilities for identifying wide binaries and understanding the 
environment in the early outer Solar System in \autoref{ss:binaries}. 
Utilizing a ``shift and stack" method for delving deeper into the faint
end of the luminosity function (and thus to the smallest sizes) is discussed
in \autoref{ss:faint}, and the likelihood of deriving physical properties of individual objects 
from light curves is discussed in the next section (\autoref{ss:lightcurves}). 
The newly evolving understanding of the overlaps between different populations
(such as the relationships between Centaurs and Oort Cloud objects) and LSST's 
potential contribution is discussed in the next section (\autoref{ss:overlappingpops}).
Investigations into the properties of comets are described in \autoref{sec:ss:comets}, and
using them to map the solar wind is discussed in \autoref{ss:coronalmass}.  
The impact hazard from Near-Earth Asteroids (\autoref{ss:neo_hazard}) and 
potential of spacecraft missions to LSST-discovered Near-Earth Asteroids 
(\autoref{ss:spacecraft}) concludes the chapter.

\section{A Brief Overview of Solar System Small Body Populations}
\label{ss:overview}
{\it Steven R. Chesley, Alan W. Harris, R. Lynne Jones}

A quick overview of the different populations of small objects of our Solar System, which
are generally divided on the basis of their current dynamics, is: 
\begin{itemize}

{\item \textbf{Near-Earth Asteroids (NEAs)} are defined as any asteroid
  in an orbit that comes within 1.3 astronomical unit (AU) of the Sun (well
  inside the orbit of Mars). Within this group, a subset in orbits
  that pass within 0.05 AU of the Earth's orbit are termed
  \textbf{Potentially Hazardous Asteroids (PHAs)}. Objects in more
  distant orbits pose no hazard of Earth impact over the next century or so,
  thus it suffices for impact monitoring to pay special attention to
  this subset of all NEAs. Most NEAs have evolved into planet-crossing
  orbits from the Main Asteroid Belt, although some are
  believed to be extinct comets and some are still active comets. }

{\item Most of the inner Solar System small bodies are \textbf{Main
    Belt Asteroids (MBAs)}, lying between the orbits of Mars and
  Jupiter. Much of the orbital space in this range is stable for
  billions of years.  Thus objects larger than 200~km found there are probably primordial,
  left over from the formation of the Solar System. However, the zone
  is crossed by a number of resonances with the major planets, which
  can destabilize an orbit in that zone. The major resonances are
  clearly seen in the distribution of orbital semi-major axes in the Asteroid
  Belt: the resonances lead to clearing out of asteroids in such zones, called  Kirkwood gaps.
  As the Main Belt contains most of the
  stable orbital space in the inner Solar System and the visual
  brightness of objects falls as a function of distance to the fourth power (due
  to reflected sunlight), the MBAs also compose
  the majority of observed small moving objects in the Solar System. }

{\item \textbf{Trojans} are asteroids in 1:1 mean-motion resonance
  with any planet. Jupiter has the largest group of Trojans, thus
  ``Trojan'' with no clarification generally means Jovian Trojan
  (``TR5'' is also used below as an abbreviation for these).
Jovian Trojan asteroids are found in two swarms around the L4 and L5 Lagrangian 
points of Jupiter's orbit, librating around these resonance points with periods
on the order of a hundred years. Their orbital 
eccentricity is typically smaller ($<$0.2) than those of Main Belt asteroids, but 
the inclinations are comparable, with a few known Trojans having inclinations 
larger than 30 degrees. It seems likely that each planet captured planetesimals into its
Trojan resonance regions, although it is not clear at what point in the history of the 
Solar System this occurred or how long objects remain in Trojan orbits, as not all Trojan orbits
are stable over the lifetime of the Solar System. } 


{\item Beyond Neptune, the \textbf{Trans-Neptunian Objects (TNOs)} occupy a large area of stable orbital space. When these objects were first discovered, it was thought that they were truly primordial remnants of the solar nebula, both dynamically and chemically primordial. Further discoveries proved that this was not the case and that the TNOs have undergone significant dynamical processing over the age of the Solar System. Recent models also indicate that they are likely to have been formed much closer to the Sun than their current location, as well as being in high relative velocity, collisionally erosive orbits. Thus, they are likely to also have undergone chemical processing. TNOs can be further broken down into \textbf{Scattered Disk Objects (SDOs)}, in orbits which are gravitationally interacting with Neptune (typically $e>0.3$, $q<38$~AU); \textbf{Detached Objects}, with perihelia beyond the gravitational perturbations of the giant planets; \textbf{Resonant Objects}, in mean-motion resonance (MMR) with Neptune (notably the ``Plutinos," which orbit in the 3:2 MMR like Pluto); and the \textbf{Classical Kuiper Belt Objects (cKBOs)}, which consist of the objects with $32<a<48$~AU on stable orbits not strongly interacting with Neptune (see \citealt{nomenclature} for more details on classification within TNO populations). The \textbf{Centaurs} are dynamically similar in many ways to the SDOs, but the Centaurs cross the orbit of Neptune.}
 
\item{ \textbf{Jupiter-family comets (JFCs)} are inner Solar System
comets whose orbits are dominantly perturbed by Jupiter. They are
presumed to have derived from the Kuiper Belt in much the same manner
as the Centaur population. These objects are perturbed by the giant
planets into orbits penetrating the inner Solar System and even evolve
into Earth-crossing orbits. The Centaurs may be a key step in the
transition from TNO to JFC.  The JFCs tend to have orbital
inclinations that are generally nearly ecliptic in nature. A second
class of comets, so-called \textbf{Long Period comets (LPCs)}, come
from the \textbf{Oort Cloud (OC)} 10,000 or more AU distant, where
they have been in ``deep freeze'' since the early formation of the
planetary system. Related to this population are the \textbf{Halley
Family comets (HFCs)}, which may also originate from the Oort Cloud,
but have shorter orbital periods (traditionally under 200 years).
Evidence suggests that some of these HFCs may be connected to the
\textbf{Damocloids}, a group of asteroids that have dynamical
similarities to the HFCs, and may be inactive or extinct comets. A
more or less constant flux of objects in the Oort Cloud is perturbed
into the inner Solar System by the Galactic tide, passing stars, or
other nearby massive bodies to become the LPCs and eventually
HFCs. These comets are distinct from JFCs by having very nearly
parabolic orbits and a nearly isotropic distribution of inclinations.
Somewhat confusingly, HFCs and JFCs are both considered ``short-period
comets'' (SPCs) despite the fact that they likely have different
source regions. } 
\end{itemize}

\section{Expected Counts for Solar System Populations} 
\label{ss:counts}
{\it \v Zeljko Ivezi\a'c, Steven R. Chesley, R. Lynne Jones}

In order to estimate expected LSST counts for populations of small solar 
system bodies, three sets of quantities are required:
\begin{enumerate}
\item the LSST sky coverage and flux sensitivity;
\item the distribution of orbital elements for each population; and 
\item the absolute magnitude (size) distribution for each population.
\end{enumerate}

Discovery rates as a function of absolute magnitude can be 
computed from a known cadence and system sensitivity without
knowing the actual size distribution (the relevant parameter is
the difference between the limiting magnitude and absolute
magnitude). For an assumed value of
absolute magnitude, or a grid of magnitudes, the detection 
efficiency is evaluated for each modeled population. 
We consider only observing nights when an object
was observed at least twice, and consider an object
detected if there are three such pairs of detections
during a single lunation. The same criterion was used in 
recent NASA NEA studies. 

\begin{figure}[htb]
\begin{center}
\includegraphics[width=4in]{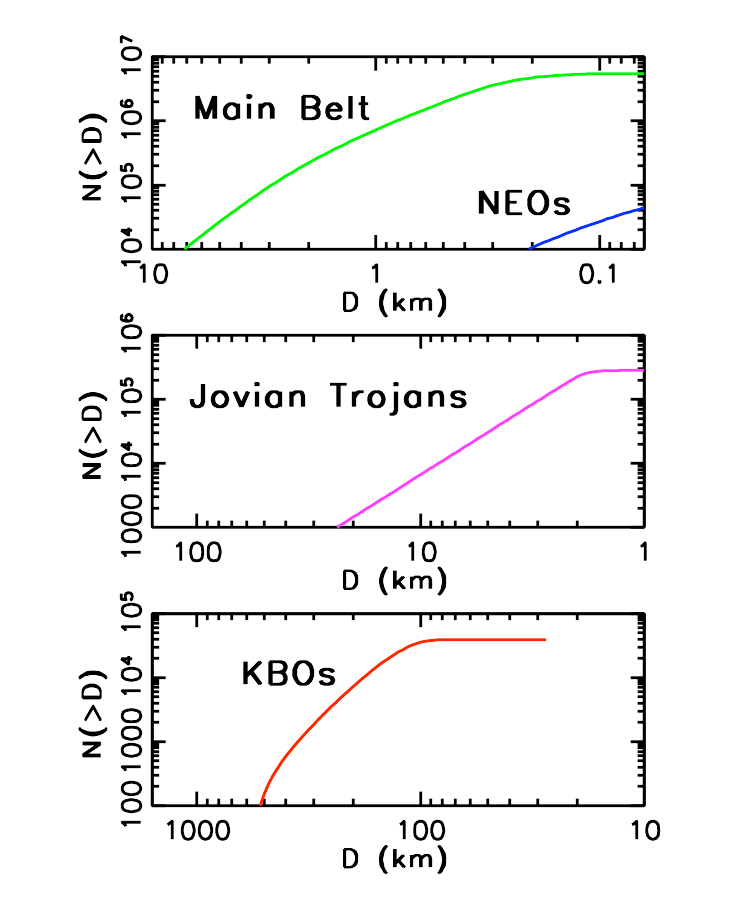}
\end{center}
\caption{Cumulative counts of asteroids detected by LSST vs. size for dominant populations of Solar System bodies,
as marked. The total expected numbers of objects detected by LSST
are 5.5 million Main Belt asteroids, 100,000 NEAs, 280,000 Jovian
Trojans, and 40,000 TNOs (marked KBO).}
\label{Fig:ss:obsCounts}
\end{figure}

\autoref{Fig:ss:obsCounts} summarizes our results, and 
\autoref{Tab:ss:selEff} provides differential completeness (10\%, 50\%, 90\%)
values at various $H$ magnitudes\footnote{The absolute magnitude $H$
of an asteroid is the apparent magnitude it would have 1 AU from both
the Sun and the Earth with a phase angle of $0^\circ$.}.
The results essentially reflect the geocentric (and for
NEAs, heliocentric), distance distribution of a given
population. The details in orbital element distribution 
are not very important, as indicated by similar 
completeness curves for NEAs and PHAs, and for
classical and scattered disk TNOs. 

The next subsections provide detailed descriptions of the adopted quantities. 

\begin{table*}[htb]
\begin{center}
\caption{Absolute magnitude at which a given detection completeness is reached$^a$}
\begin{tabular}{rrrrc}
\hline
 Population  & H(90\%) & H(50\%) & H(10\%) & N$_{LSST}^b$ \\
\hline 
      PHA  &   18.8  &  22.7  &  25.6  &   ---        \\  
       NEA  &   18.9  &  22.4  &  24.9  & 100,000      \\ 
       MBA  &   20.0  &  20.7  &  21.9  &  5.5 million \\
       TR5  &   17.5  &  17.8  &  18.1  &  280,000     \\
       TNO  &    7.5  &   8.6  &   9.2  &   40,000     \\
       SDO  &    6.8  &   8.3  &   9.1  &   ---        \\
\hline 
\end{tabular}
\end{center} 
\medskip
$^a${Table lists absolute magnitude $H$ values at which a differential completeness of 90\%, 50\% or 10\% is reached. 
This is not a cumulative detection efficiency (i.e. completeness for $H>X$), but a differential efficiency (i.e. completeness at $H=X$).}
$^b${Approximate total number of objects detected with LSST, in various populations. PHAs and SDOs are included in the counts of NEAs and TNOs.}
\label{Tab:ss:selEff}
\end{table*}

\subsection{LSST Sky Coverage and Flux Sensitivity} 

A detailed discussion of the LSST flux limits for moving objects and
impact of trailing losses is presented in \citet{lsstoverview},
\S 3.2.2. Here we follow an identical procedure, except that we
extend it to other Solar System populations: Near-Earth Asteroids, Main
Belt asteroids, Jovian Trojans, and TNOs.

The sky coverage considered for the cumulative number of objects in
each population includes the universal cadence fields and the northern ecliptic
spur, as well as the ``best" pairs of exposures from the deep drilling fields. However,
the increased depth in the deep drilling fields which is possible from co-adding
the exposures using shift-and-stack methods is not considered here. Instead, 
the results of deep drilling are examined in \autoref{ss:faint}. 

\subsection{Assumed Orbital Elements Distributions} 
\label{ss:assumedorbits}

We utilize orbital elements distributed with the MOPS code described in \autoref{ss:mops}. The MOPS code
incorporates state-of-the-art knowledge about various 
Solar System populations \citep{ssmodel}.  The availability of MOPS synthetic 
orbital elements made this analysis fairly straightforward. 
In order to estimate the efficiency of LSST cadence for
discovering various populations, we extract 1000 sets of orbital
elements from MOPS for each of the model populations of
NEAs, PHAs, MBAs, Jovian Trojans, TNOs and SDOs. 

Using these orbital elements, we compute the positions
of all objects at the time of all LSST observations 
listed in the default cadence simulation 
(see \autoref{sec:design:opsim}). We use the JPL 
ephemeris code implemented as described in \citet{juric2002asteroids}.
We positionally match the two lists and retain all
instances when a synthetic object was within the field of 
view. Whether an object was actually detected or not depends 
on its assumed absolute magnitude, drawn from the adopted
absolute magnitude distribution (see \autoref{ss:assumedabsmag}).  

\begin{figure}[htb]
\begin{center}
\includegraphics[width=4in]{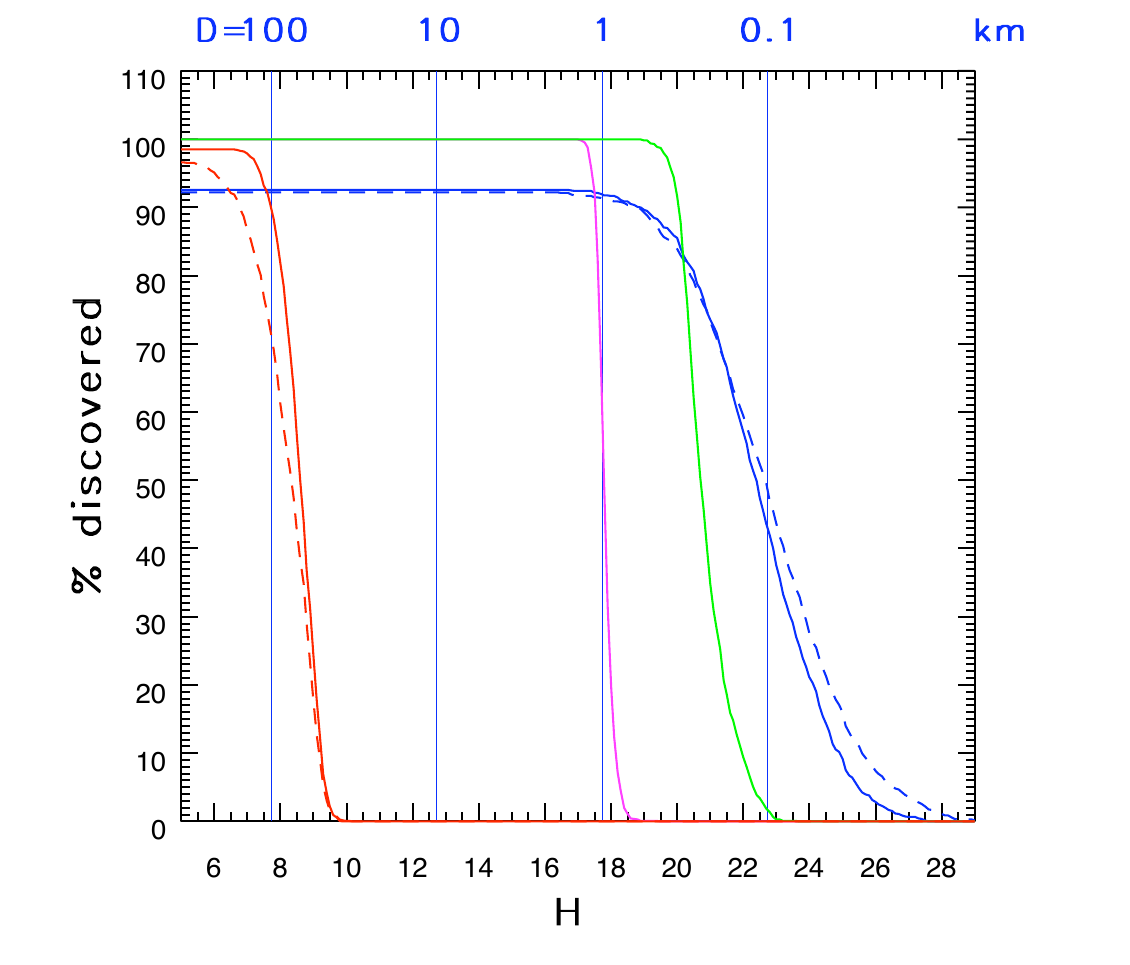}
\end{center}
\caption{A comparison of LSST discovery efficiency for dominant 
populations of Solar System bodies. Solid lines correspond to 
classical TNOs (red), Jovian Trojans (magenta), Main Belt Asteroids
(green), and NEAs (blue). The red dashed line corresponds to 
scattered disk objects, and the blue dashed line to PHAs. Note that
the completeness for NEAs and PHAs does not reach 100\% even for
exceedingly large objects (due to finite survey lifetime).}
\label{Fig:ss:eff}
\end{figure}

These orbital element distributions are, of course, only approximate. 
However, they represent the best current estimates of these populations, 
and are originated from a mixture of observations and theoretical modeling. 
This technique provides an estimate of the fraction of detectable objects in each population, 
at each absolute magnitude. The results of this analysis are shown in \autoref{Fig:ss:eff}.

\subsection{The Absolute Magnitude Distributions} 
\label{ss:assumedabsmag}

\begin{figure}[htb]
\begin{center}
\includegraphics[width=4in]{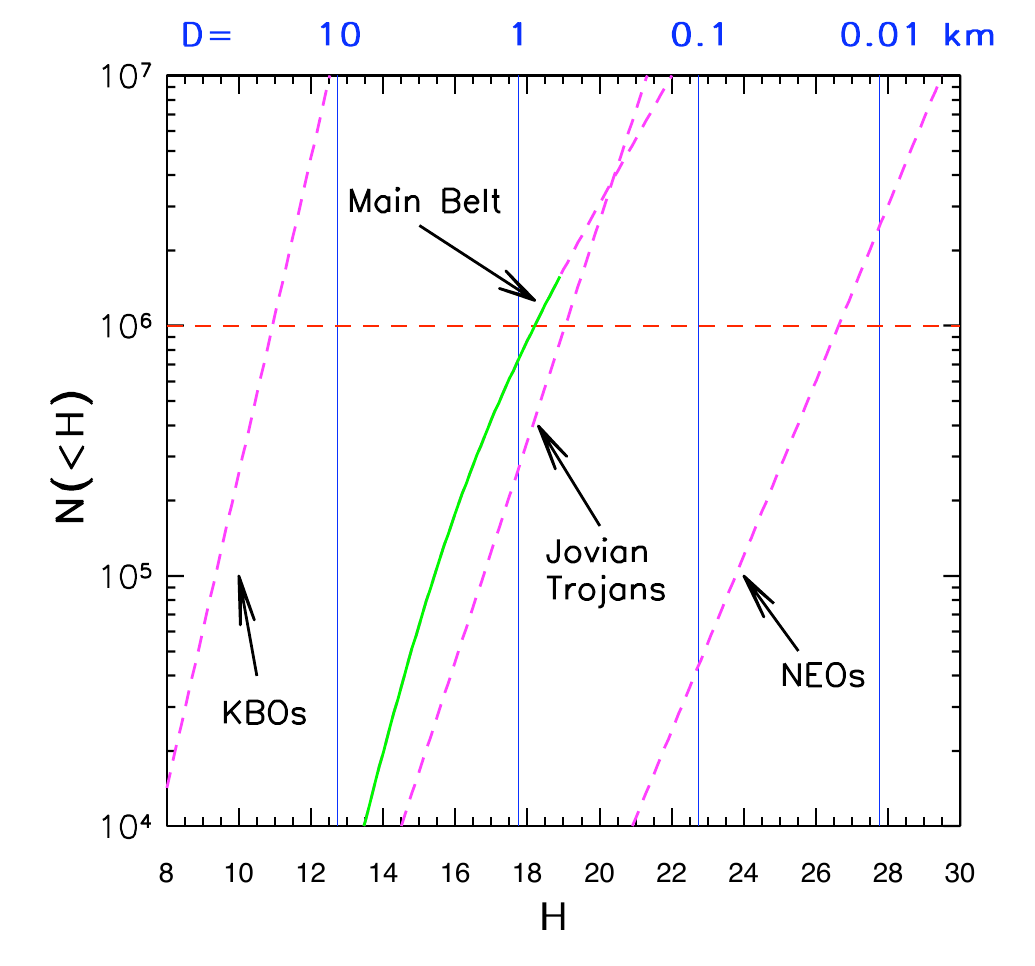}
\end{center}
\caption{A comparison of cumulative count vs. absolute magnitude
curves
for dominant populations of Solar System bodies. The solid portion
of the line for Main Belt Asteroids signifies directly constrained
counts; all dashed lines are extrapolations from brighter $H$. The
horizontal line at $N=10^6$ is added to guide the eye. The object
diameters marked on top correspond to an albedo of 0.14. Populations
with low median albedo, such as Jovian Trojans and TNOs, have
2-3 times larger $D$ for a given $H$. In particular, there are
comparable numbers of Main Belt Asteroids and Jovian Trojans
down to the same {\it size} limit.}
\label{Fig:ss:popCounts}
\end{figure}

LSST's flux limit will be about five magnitudes 
fainter that the current completeness of various
Solar System catalogs. Hence, to estimate expected 
counts requires substantial extrapolation of known
absolute magnitude distributions. We adopt the 
following cumulative distributions, which are 
illustrated in \autoref{Fig:ss:popCounts}. 

For MBAs, we adopt the shape of the cumulative
counts curve based on SDSS data and given by Equation~12 of
\citet{ivezic2001_sdssasteroids}, including their
normalization of 774,000 objects larger than $D=1$ km
\begin{equation}
 N_{cum}^{MBA} = 267,000 \, {10^{0.43x} \over 10^{0.18x} + 10^{-0.18x}},
\end{equation}
where $x = H - 15.7$ 
and a fiducial albedo of 0.14 is assumed
(so that $H=22$ corresponds to a size of 140 m, as discussed
in the NEA context, see \autoref{ss:neo_hazard}).  
This normalization agrees within 10\% with the \citep{durda1997} 
result that there are 67,000 objects with $H<15.5$ 
(assuming a mean albedo for MBAs of 0.10), and is consistent
at the same level with the latest SDSS results \citep{parker2008_sdssasteroids}.
This approach is accurate to only several tens 
of a percent, because the shape of the count vs.\ $H$ curve 
varies across the belt and between families and background, 
as well as among individual families. At this level
of accuracy, there are about a million Main Belt 
Asteroids larger than 1 km. We note that the MOPS normalization 
implies twice as many objects as given by this 
normalization. About half of this discrepancy could 
be due to faulty $H$ values in contemporary asteroid 
catalogs (for more details, see \citealt{parker2008_sdssasteroids}).
For other populations, we adopt the cumulative
counts implemented in MOPS.

For NEAs, we adopt the \citet{bottke2002_neo} result
\begin{equation}
      N_{cum}^{NEA} = 960 \times 10^{0.35(H-18)}.
\end{equation}

For Jovian Trojans, we adopt the \citet{szabo2007_trojans}
result
\begin{equation}
      N_{cum}^{Tr5} = 794 \times 10^{0.44(H-12)}.
\end{equation}
This expression was constrained using SDSS data to $H=14$, 
and implies similar counts of Jovian Trojans and Main Belt 
Asteroids down to the same size limit, for sizes larger 
than $\sim$10 km. Note that this does not imply similar
observed number counts of Jovian Trojans and MBAs, since the Main Belt 
is much closer. The extrapolation of this expression to
$H>14$ may be unreliable. In particular, the Jovian Trojan
counts become much larger than the cumulative counts of MBAs for $H>20$, because the counts slope at the faint end 
becomes smaller for the latter. A recent study
based on SDSS data by \citet{szabo2008a}
demonstrated that existing moving object catalogs are complete 
to $r\sim19.5$, or to a size limit of about 20 km,
giving a total count of the order a thousand known Jovian Trojans.

For TNOs, we adopt results obtained by
\citet{trujillo2000, trujillo2001}
\begin{equation}
      N_{cum}^{TNO} = 71,400 \times 10^{0.63(H-9.1)},
\end{equation}
where we assumed a normalization of 71,400 objects 
larger than 100 km, and an albedo of 0.04. This 
normalization includes classical, scattered disk
and resonant TNOs, with equal numbers of classical 
and resonant objects and 0.8 Scattered Disk Objects per classical TNO. 


\subsection{ Expected Cumulative Counts}
\label{ss:cumcounts}

Given the adopted cumulative counts (\autoref{ss:assumedabsmag}) and completeness
curves (\autoref{ss:assumedorbits}), it is straightforward to generate the expected observed
counts. 
\autoref{Tab:ss:selEff} provides the expected LSST
sample size for each population. 

Unsurprisingly, the largest observed sample will be MBAs, which will be probed to a size
limit as small as $\sim 100$ m. It is remarkable that the Jovian 
Trojan sample will include $\sim280,000$ objects, on the order of the number
of currently known MBAs -- currently
there are only a few thousand known Jovian Trojans. 
In addition, the expected detection of 40,000 objects in the TNO sample, with accurate 
color and variability measurements for a substantial
fraction of these objects, will enable investigation of 
these distant worlds with a thoroughness 
that is currently only possible for MBAs.

\begin{figure}[htb]
\begin{center}
\includegraphics[width=4in]{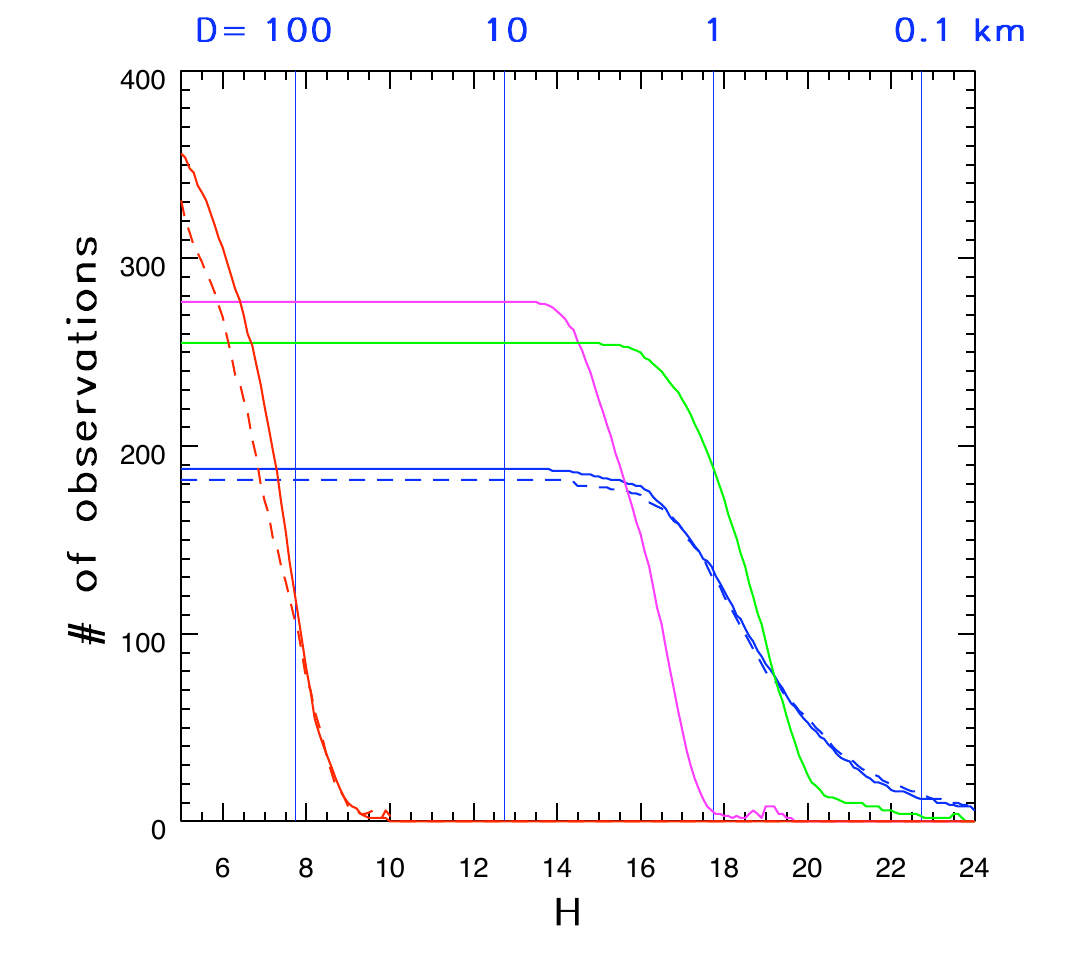}
\end{center}
\caption{The median number of expected LSST detections of a given
object as a function of $H$ for dominant populations of Solar System bodies. 
Solid lines correspond to 
classical TNOs (red), Jovian Trojans (magenta), MBAs (green), and NEAs (blue). The red dashed line corresponds to 
Scattered Disk Objects, and the blue dashed line to PHAs. Nights 
with only one detection are not counted.}
\label{Fig:ss:nObs}
\end{figure}

\autoref{Fig:ss:nObs} shows the median number of expected LSST 
observations (based on the Operations Simulator;
\autoref{sec:design:opsim}) for dominant populations of Solar System bodies. We
do not include nights with only one detection. A significant fraction
of discovered objects will have several hundred detections. For
example, more than 150 observations will be available for about 500
NEAs, one million MBAs, 50,000 Jovian Trojans and 7,000 TNOs.  The corresponding counts for objects with more than 100
observations are 1,400 NEAs, 1.6 million MBAs, 80,000 Jovian Trojans, and 11,000 TNOs. These large numbers of multi-color light curves will
enable numerous novel research directions in studies such as
light-curve inversion for a significant fraction of these Solar System
populations.

\section{The Orbital Distributions of Small Body Populations}
\label{ss:orbits}
{\it R. Lynne Jones, Michael E.  Brown}

LSST will produce large catalogs of well-measured orbits for
moving objects throughout the Solar System from NEAs to TNOs. These
orbital catalogs are important for many reasons, the most obvious of which is
the necessity of predicting highly accurate ephemerides (positions and magnitudes)
for the study of individual objects in greater detail. Just as (or more) important,
however, is the study of the ensemble of orbits (as the distribution of orbital parameters) 
in order to understand the current state and previous
evolution of each population of small bodies, as this is inextricably
linked to the evolution of the giant planets. Information about this
evolution is preserved in the orbital parameters of the small bodies.

The importance of this record was first clearly realized when the discovery of large numbers of
TNOs in mean-motion resonance with Neptune, together with the
discovery of giant extrasolar planets at small distances from their
stars, created a new vision of our Solar System. Instead of a static
place, where the giant planets formed in their current locations,
\citet{malhotra95} proposed that a gradual outward migration in
Neptune's orbit could have gathered TNOs into 2:3 mean-motion resonance (MMR) with Neptune.
This migration trapped TNOs (Plutinos in this resonance)
into the 2:3 MMR resonance at a density higher than in the rest of the Kuiper Belt.  In
this new vision of a more dynamic Solar System, the orbital
distributions of large populations of small bodies serve as ``test particles" and preserve an
invaluable fossil record of the orbital evolution of the giant
planets.

In recent years, the Nice model \citep{niceGiants}  has proposed that
all giant planets formed at less than 14~AU from the Sun and the solar
nebula was truncated near 30~AU. The giant planets and small bodies in
the Solar System subsequently evolved to their current state through
planetary migration due to angular momentum exchange with
planetesimals. The Nice model presents an intriguing theory which
could account for many previously unexplained problems in various
small body populations: the mass depletion observed in the Kuiper Belt
\citep{niceKuiper} and the Asteroid Belt \citep{niceAsteroids},
the 
orbital distribution of Trojans \citep{niceTrojans}, and the late
heavy bombardment \citep{niceLHB}. However, the Nice model has 
no obvious way to produce the detached TNOs with perihelion beyond 50~AU
(such as 2004~XR$_{190}$) and also has problems reproducing the orbital distribution
(particularly the inclinations) of the cold classical Kuiper Belt. 

There are other older but still competitive theories: models related to 
the slow planetary migration first detected in the Plutino fraction \citep{gomes03, gomes04, hahnmalhotra05}, 
models where a rogue planetary embryo or large planetesimal pass through or orbit
briefly in the outer Solar System \citep{petitrogue_kb, gladmanrogue}, or models of nearby 
stellar passages early in the history of the Solar System
\citep{ida00, kenyonbromley04, morbihal_sedna, brasserIII, kaibquinn08}. 
Each of these theories has particular strengths. The stellar flyby model is able to produce
objects with large semi-major axes, high perihelions, and high eccentricities such as Sedna. 
The rogue planetary embryo model is able to produce objects like 2004~XR$_{190}$ with 
perihelion beyond the reach of Neptune's perturbation, high inclination or eccentricity,
but semi-major axis just outside the classical belt, without
perturbing the classical belt as strongly as the stellar flyby model would. The slow migration
model can drop objects into low eccentricity orbits, perhaps even to the level of
creating the dynamically cold classical belt, while creating a distribution of inclinations.
A major problem with all of these theories beyond various problems recreating specifics of 
the inclination and eccentricity distributions is that the mass required to build
the largest objects we see in the Kuiper Belt today is much larger than the total
mass detected \citep{stern97}; therefore the mass must have been depleted somehow.
The amount of mass depletion required would likely have left its trace in
the orbit of Neptune \citep{gomes04}, resulting in a different orbit than observed
today (circular at 30~AU). However, none of these models has been conclusively ruled
out, and it seems likely that one or more of these mechanisms has contributed to
the current distribution of TNOs, in particular since migration is known to have occurred in some form, and passing
stars in the solar birth environment is likely \citep{ladalada03}. 

It becomes clear from this range of models that can potentially fit the available data that the current statistical sample of TNOs ($<2,000$ objects) is unable to make strong distinctions
among the theories. With a vastly increased sample size, LSST will provide much stronger statistical
tests. In particular, the inclination and eccentricity distributions of the classical belt will be well measured,
along with obtaining $griz$ color measurements for further understanding of the ``cold" and ``hot"
classical belt members -- this alone should provide strong constraints on the Nice model and determine whether
a rogue embryo or planetesimal must have passed through the primordial Kuiper Belt. By measuring the perihelion
distribution of Scattered Disk Objects to greater distances (LSST can detect objects down to 400~km in diameter 
as far as 100~AU assuming an albedo of 0.1) and larger amounts of sky than currently possible, LSST will provide 
direct tests of the stellar flyby models.

In addition, the detection of ``rare" objects can provide strong leverage to distinguish among
models, or even rule out theories which are unable to create such objects. As an example of a
currently known rare population, there are a handful of TNOs, called ``detached" TNOs
\citep{nomenclature}, which generally show the signature of some strong dynamical perturbation
in the past through a current high eccentricity or inclination but without a strong indication
of the cause of this perturbation. As the detached TNOs have perihelia beyond $\sim45$~AU, the
perturbations cannot be due to gravitational interaction with the giant planets.
For some of these detached objects, such as 2000~CR$_{105}$ 
\citep{2000CR105} or Sedna \citep{brownSedna} (whose orbit is entirely contained beyond
the outer edge of the classical Kuiper Belt, $\sim50$~AU, and inside the inner edge of the Oort Cloud, $\sim20,000$~AU), interaction with a passing star seems the most likely cause \citep{morbihal_sedna}. 
For others, such as 2004~XR$_{190}$ \citep{buffy} or 2008~KV$_{42}$ \citep{gladman_drac} (the first known
retrograde TNO, having an inclination of $102^{\circ}$), the source of the perturbation is much less clear. 
A complication in the interpretation of these unusual objects is knowing if the newly discovered TNO 
is just an unlikely outlier of an underlying distribution, or if it truly is the 
``first discovery of its kind." Many of these problems in interpretation are due to observational
selection biases in flux, inclination, and observational followup
\citep{jjbias} or miscalculated orbits \citep{biaspaper}. For example, retrograde TNOs are not only difficult
to detect due to their apparent rarity, but in a short series of observations (a few days), the orbit can
appear to be that of a much more common nearby high-eccentricity asteroid instead of a distant 
retrograde or high-inclination TNO.  The frequent observing schedule and well-characterized
in limiting magnitude and sky coverage of LSST will minimize the effect of these biases.
With the total sample size of $\sim40,000$ TNOs expected by LSST, it will also be possible to 
characterize these rare objects, which likely compose at most a few percent of the observed population.

In general, a large sample of TNOs with well-measured
orbits, detected in a well-characterized survey, will provide strong statistical tests for the current theories
of Solar System evolution and strong pointers to where the models need to go for the next generation
of theories. 

These tests can be carried on into the inner Solar System, although in these
regions the populations have been much more strongly affected by perturbations
from the planets. For example, resonances in the Main Asteroid Belt were
long ago cleared of primordial objects, since these resonances are
unstable to gravitational perturbations from Jupiter. Asteroids which
chance to drift into these zones (i.e., by Yarkovsky drift), will be
promptly removed by resonant perturbations to become planet-crossing,
and from there suffer collision or ejection by close encounters with
major planets. Interestingly, the Main Belt itself seems to have been severely
depleted of mass, beyond the expected losses due to ejection by
gravitational perturbation from the planets in their current locations.
The Main Belt inclination distribution also has been dynamically excited, 
in a manner similar to the classical Kuiper Belt. 
Theories to explain this mass depletion and/or dynamical
excitation include similar models as used to explain the mass
depletion or dynamical excitation of the Kuiper Belt -- a planetary
embryo (or large planetesimal) passing through the region \citep{petitrogue_asteroid}, 
secular resonances sweeping through the Main Belt \citep{nagasawa00}, or gravitational
perturbations resulting from the large-scale rearrangement of the
Solar System occurring during the rapid evolution phase of the Nice
model \citep{niceAsteroids, minton09_niceasteroid}. In the Asteroid Belt, 
the colors of objects are strongly correlated with the history of the object's
formation and dynamical evolution, suggesting that obtaining $griz$ colors
as well as orbital parameters will provide further strong constraints for
these models. 

\begin{figure}[htb!]
\begin{center}
\includegraphics[width=6in]{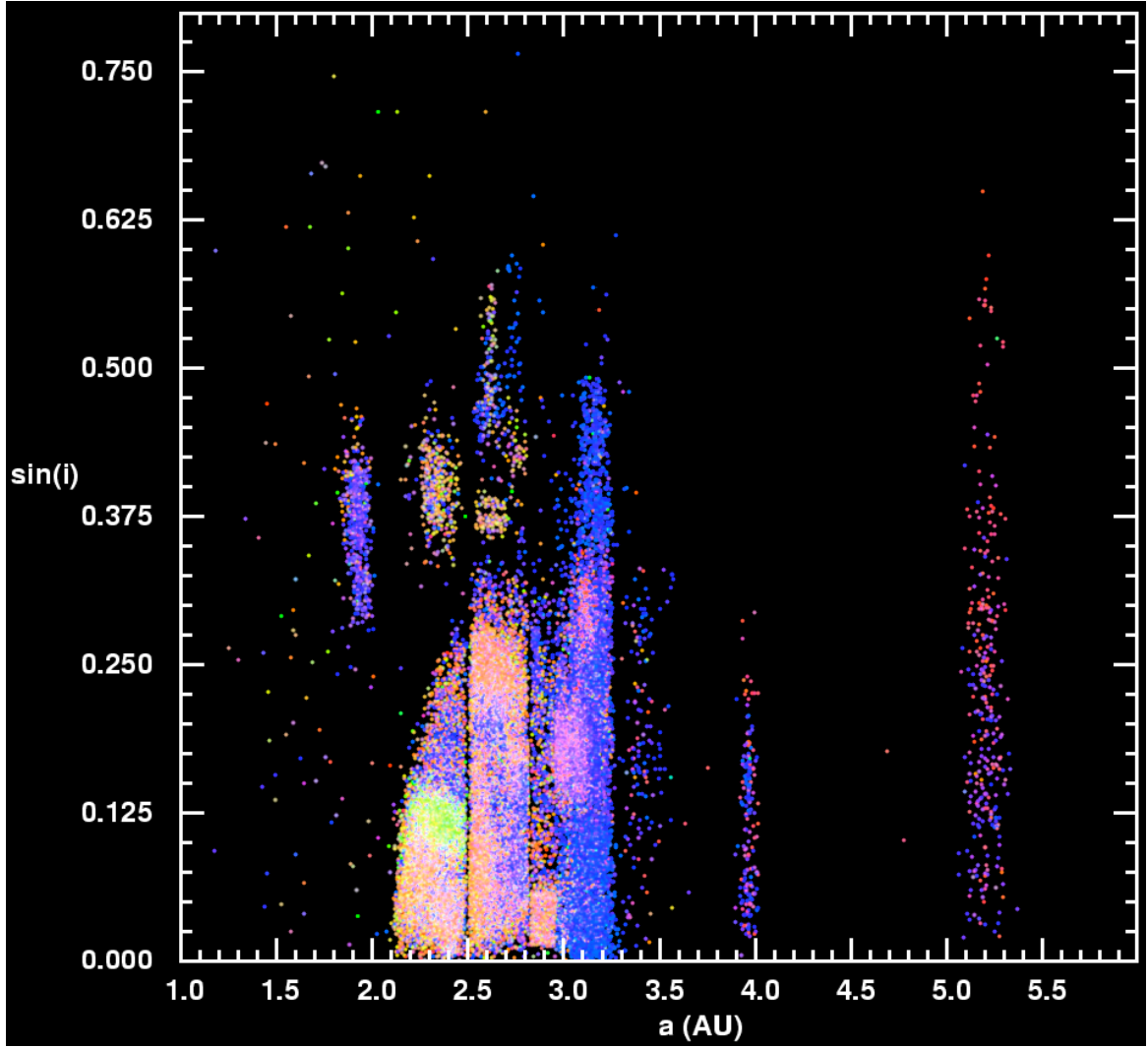}
\caption{The dots show the sine of the osculating orbital inclination 
vs. orbital semi-major axis ($a$) distribution of $\sim$ 43,000 unique 
moving objects detected by the SDSS, and matched to objects with known 
orbital parameters. The dots are color-coded according to their colors
measured by SDSS. About 1,000 Jovian Trojans are seen at $a\sim$ 5.2 AU, 
and display a correlation between the color and orbital inclination
\citep{szabo2007_trojans}. LSST will enable the construction of 
such a diagram with several million objects, including about
300,000 Jovian Trojans (50,000 with more than 150 detections).}
\label{Fig:ss:Trojans1}
\end{center}
\end{figure}

The orbital distribution of Jovian Trojans also provides useful constraints on
the environment of the early Solar System. 
One hypothesis for the origin of the Trojans is that they were formed simultaneously 
with Jupiter and then captured and stabilized near the growing
Jupiter's L4 and L5 points \citep{peale1993}. An alternative hypothesis suggests they
were captured over a much longer
period after forming elsewhere in the Solar System \citep{jewitt1996}. 
The colors of many known Trojans are similar to SDOs from the outer Solar System
and others appear similar to the colors of outer MBAs, as in \autoref{Fig:ss:Trojans1}, 
lending support to the second hypothesis, with
implications for the importance of gas drag in the early Solar System. 
The Nice model suggests a more complex
picture, where the present permanent Trojan population is built up by planetesimals
trapped after the 1:2 mean-motion resonance crossing of Saturn and Jupiter \citep{niceTrojans}. 

A clear picture of the orbital distribution of small bodies throughout
the entire Solar System would provide the means to test each of these
models and provide constraints for further model development. In
particular, these orbital distributions need to be accompanied by a
clear understanding of the selection biases present in the observed
distributions.

\subsection{Adding Colors: $ugrizy$ Photometry}

Combining the orbits with color information accurate to $\sim0.01-0.02$
magnitudes for a significant fraction of the
objects allows for additional exploration of
sub-populations and investigation of similarities among the different groups. 
This is complicated
by the fact that LSST will not take simultaneous color measurements; observations
in different filters will often be separated by at least 30 minutes. For slow rotators
this will not be a significant problem, especially when combined with many repeat
measurements over the lifetime of the survey (however, it may increase the
effective error in LSST color measurements).

This color information is useful in providing insights beyond the orbital distributions as shown in studying differences between the ``hot" and ``cold" classical Kuiper Belt. 
These two populations are
just barely distinguishable by looking at the statistical distribution of
inclinations of classical belt objects. However, the statistical color
differences between the two groups
are clear \citep{doressoundiramcolors, meudon, des}, indicating a
strong likelihood of significantly different 
dynamical histories, rather than just a bimodal distribution of
inclinations. 
The colors of ``cold" (low inclination, low eccentricity)
classical belt members tend to be only red, while the
colors of ``hot" (wider range of inclination and eccentricity)
classical belt members range from red to gray. These differences
are hard to explain with any of the current models of the outer Solar
System, thus providing an important challenge for testing these and future
models of the evolution of the Solar System.

\begin{figure}[htb!]
\begin{center}
\includegraphics[width=5.5in]{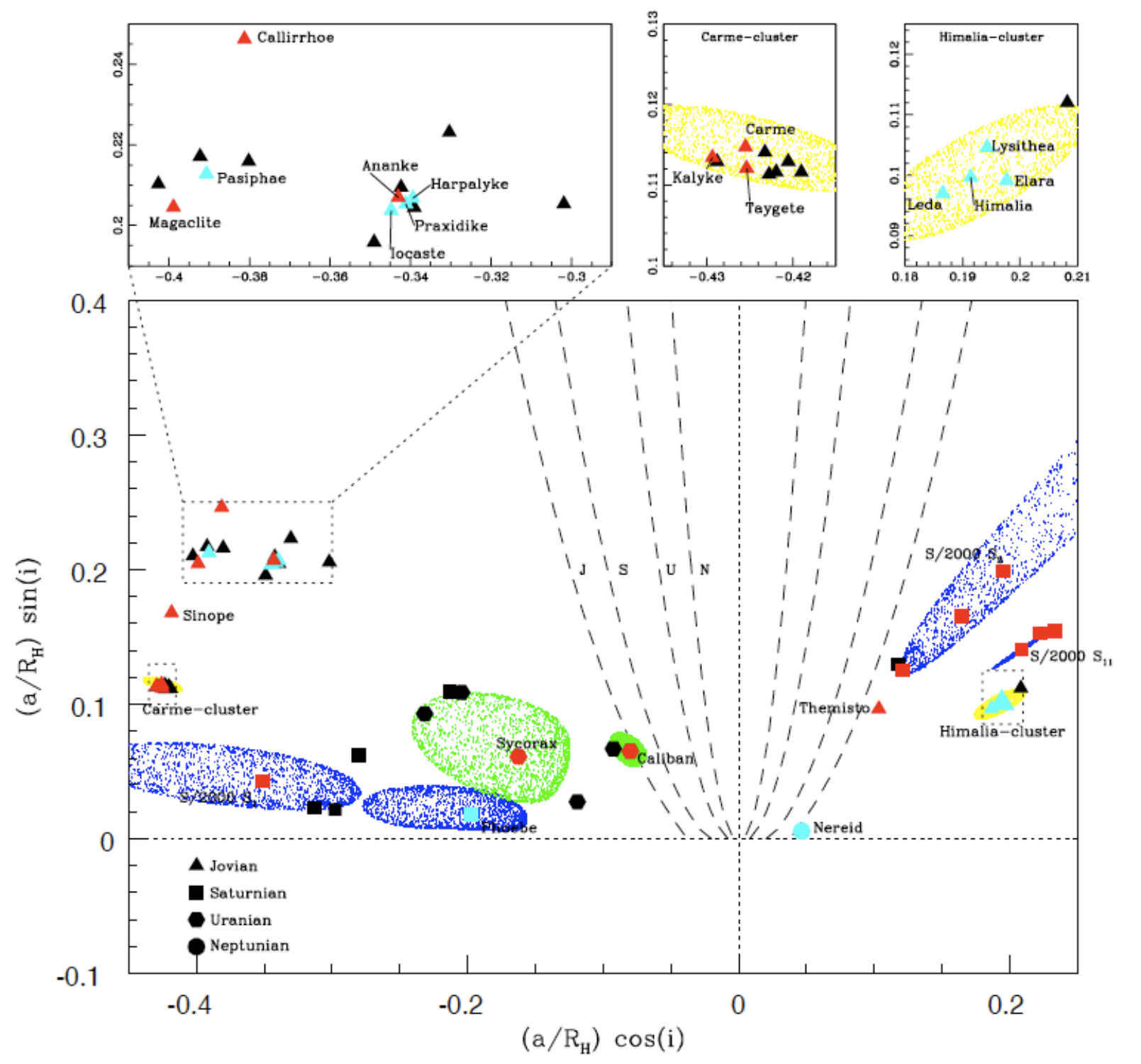}
\caption{Distribution of irregular satellites around the giant
planets.  The x-axis is the component of the distance of the
semi-major axis of each satellite along the axis of rotation of the
planet, normalized by the planet's Hill-sphere radius; the y-axis is
the component perpendicular to the axis.  Irregular satellites with
measured colors have been binned into ``gray" or ``red" color bins and
are plotted according to blue for ``gray" objects and red for
``red." The colored ellipses indicate the area of $a-i$ space where
each cluster could disperse, given a catastrophic fragmentation
event. From \citet{gravcolors}, with permission.}
\label{Fig:ss:grav}
\end{center}
\end{figure}

As another example of the application of color data to understanding
the history of small bodies, giant planet irregular satellites with a
variety of inclinations show clear ``families" when their orbital
parameters are combined with color information (\citealt{gravcolors}; see
\autoref{Fig:ss:grav}).  With the addition of this information, the
likelihood of different methods of capture mechanisms --- gas drag
capture of a series of small bodies versus capture of one parent body
which was then broken apart through tidal stresses or collisions ---
can be evaluated.


\section{The Main Belt: Collisional Families and Size Distributions}
\label{ss:mainbelt_familsize}

Due to its large intrinsic and nearby (thus bright) population, the Main
Asteroid Belt has historically provided the largest observational samples
of small body populations. The size and color distributions of the Main Belt 
signal the history of individual bodies and their place within the larger
population, providing clues to their history of accretion and collisional
disruption. It is necessary to understand the role collisions and accretion
play within each population if we are to understand planetary formation in detail.

\subsection{Identifying Collisional Families in the Main Belt}
\label{ss:families}
{\it Zoran Kne\v zevi\'c, Andrea Milani}


Collisional families provide constraints on all parameters appearing
in evolutionary theories of small body populations: collision
frequency and mean lifetime between disruptions, material strength of
the bodies, timescales of dynamical diffusion due to chaos,
secular resonances and non-gravitational perturbations, space
weathering of surfaces, evolution of multiple systems, and rotation
states. LSST is more likely to discover statistically significant numbers
of collisional families (and their members) in the Main Belt rather than
in the Kuiper Belt, due to the lower ratio of velocity dispersion (among 
the family members) to relative velocities (compared to non-family 
members). 

The primary requirement for identifying collisional families in the Main Belt is
obtaining accurate proper orbital elements for all objects which are
``regular" or at least in stable chaos. Orbital elements calculated from
observations are ``osculating elements" -- most reported orbital elements
are osculating elements. ``Proper elements" can be
computed, starting from the osculating elements, in different ways: a 
typical method is to integrate the osculating orbital elements forward over
a long time scale, averaging the osculating elements to calculate proper elements.
The distinctive property of proper elements is that they are nearly constant over very
long time scales, thus a similarity of the orbits is preserved for the
same time span.

The algorithms to compute proper orbital elements depend on the
orbital region of the object. In the
Main Asteroid Belt, proper elements are stable for time spans between
a few times $10^6$ and a few times $10^8$ years. If a catastrophic
disruption event occurred even a very long time ago, the proper
elements show clustering. These clusterings can be identified because
the ejection velocities of the fragments, which are of the order of
the escape velocity from the parent body, are smaller than the orbital
velocities by two orders of
magnitude. The ratio increases as catalogs 
reach smaller sizes of bodies. 

The processing load for the computation of proper elements is
expected to be quite significant for LSST's expected rate of discovery. Sophisticated tools of parallel computing are
being developed to calculate proper elements. Development is ongoing
in identifying clustered groups of objects within a denser background.
The main families within each orbital region can be identified, using
only comparatively large objects to avoid the chaining effects which
prevent the use of currently known mathematical taxonomy methods for
overly dense samples. Thus ``core families'' with larger objects can be
formed with well tested methods, such as hierarchical clustering with
the nearest neighbor metric. Given these defined families, the smaller
objects can be tested for classification into potential families
within their same orbital region. There is unavoidably some potential
for objects being classified in more than one family, the removal of
this ambiguity can be obtained, at least partially, by using
multicolor photometry. 

Adding information about the photometric colors can aid in identifying families, as objects 
coming from the same parent body might be expected to have similar 
colors. The most striking example of this
is the Vesta family, which is too large (also as a result of
non-gravitational perturbations) to be discriminated by proper
elements only, but is characterized by a very distinctive spectral
signature. It should be noted that this is only an aid in 
identifying some potential families; if the parent object 
was differentiated and then completely disrupted, the family
members could have very different spectral signatures (and thus colors)
depending on their point of origin in the parent body.

The families act as a probe of the orbital stability of their members,
taking into account both conservative chaotic diffusion and
non-gravitational perturbations such as the Yarkovsky effect. The
instability gaps and leaks detected in the families should be
investigated for their dynamical mechanism and long-term evolution.
They allow one to estimate the age of the families, as with the Veritas
family, and to constrain physical properties such as thermal
conductivity.  Combined with the sparse light curve inversion, which
should allow the determination of the rotation axis, the
family member leakage could be used to validate and constrain
Yarkovsky effect models. Another method to estimate the family age
uses the distribution of proper semi-major axis as a function of
absolute magnitude and thus size.

Individual objects break up due to collisions, tidal and rotational
instabilities, and possibly other causes. A goal for future work is to identify
recent and small events, as opposed to the large and ancient (millions
of years) disruptions documented by the families. It is necessary to
use very accurate proper elements in combination with direct numeric
and semi-analytic computations to find and analyze such cases. Very
recent breakups could belong to two categories: disruption of a binary
into a two-component family or collisional catastrophic disruption of
small bodies. Very recent collisional breakups with ages of the order of a million years are already known, and
their number should increase very significantly by increasing the
inventory of small objects.

There is an excess of pairs of asteroids on very similar orbits that
indicates a common origin between the paired objects. Given the
extremely low relative velocities (down to $< 1 \rm \,m\,s^{-1}$), these cases
appear most likely to be generated by fission of a
solitary body or separation of binary components. Mapping the
frequency, size distribution, and other properties of these pairs will
provide constraints on the rate and nature of the fissions induced by
tides and/or non-gravitational perturbations. With the single-visit limiting magnitude
$r =24.7$, LSST will produce a more 
complete catalog down to a given size 
range, which should increase the number of identified asteroid pairs enormously.
This applies in particular to the Hungaria
region, which is the subset of asteroids best observable from Earth in
the context of a very large field of view survey such as LSST. Given
the expected limiting magnitude of LSST, Hungaria family members with absolute
magnitudes $H$ up to 23 should be very well observable, and their number is
expected to be comparable to the total number of MBAs
presently known. Pairs with a primary of less than 500~m
diameter, and a secondary around 200~m diameter should be found. This
in turn will constrain the rate of formation and the stability of
binary asteroids although most of them will not be directly observable
with LSST.

\subsection{The Size Distribution of Main Belt Asteroids}
{\it \v Zeljko Ivezi\a'c, Alex Parker, R. Lynne Jones}
\label{ss:mbsizes}

The size distribution of asteroids is one of most significant 
observational constraints on their history and is considered to 
be the ``planetary holy grail'' (\citealt{jedicke1998}, and references 
therein).  It is also one of the hardest quantities 
to determine observationally because of strong selection effects
in the extant catalogs. Based on a comparison of recent known object 
catalogs (the ASTORB compilation of asteroid orbits from January 2008,
\citealt{ASTORB}) and the SDSS 
Moving Object Catalog 4 \citep{ivezic2001_sdssasteroids}, 
\citet{parker2008_sdssasteroids} concluded that the former is complete
to $r=19.5$. LSST will produce a moving object catalog complete 
to a limit 5 magnitudes fainter. 

Determining the size distribution of Main Belt Asteroids requires
unraveling a complex combination of the background size distribution and varying
size distributions of asteroid families. Asteroid dynamical 
families are identified as groups of asteroids in orbital element space 
\citep{gradie1979, gradie1989, valsecchi1989}. 
This clustering was first 
discovered by Hirayama (\citealt{hirayama1918}, for a review
see \citealt{binzel1994}),
who also proposed that families may be the remnants of parent 
bodies that broke into fragments. About half of all known 
asteroids are believed to belong to families; for example,
\citet{zappala1995} applied a hierarchical clustering 
method to a sample of 12,487 asteroids and found over 30 families.

Asteroid families are traditionally defined as clusters of 
objects in orbital parameter space, but SDSS data shows that
they often have distinctive optical colors \citep{ivezic2002_astcolors}.
Recently, \citet{parker2008_sdssasteroids} studied the asteroid size distribution 
to a sub-km limit for Main Belt families using multi-color 
photometry obtained by SDSS. They
showed that the separation of family members from background 
interlopers can be significantly improved with the aid of 
colors as a qualifier for family membership, although this 
method is not generally applicable for families resulting 
from the breakup of a differentiated parent body whose members
could have significantly different colors. 

\begin{figure}[htb!]
\begin{center}
\includegraphics[width=6.3in]{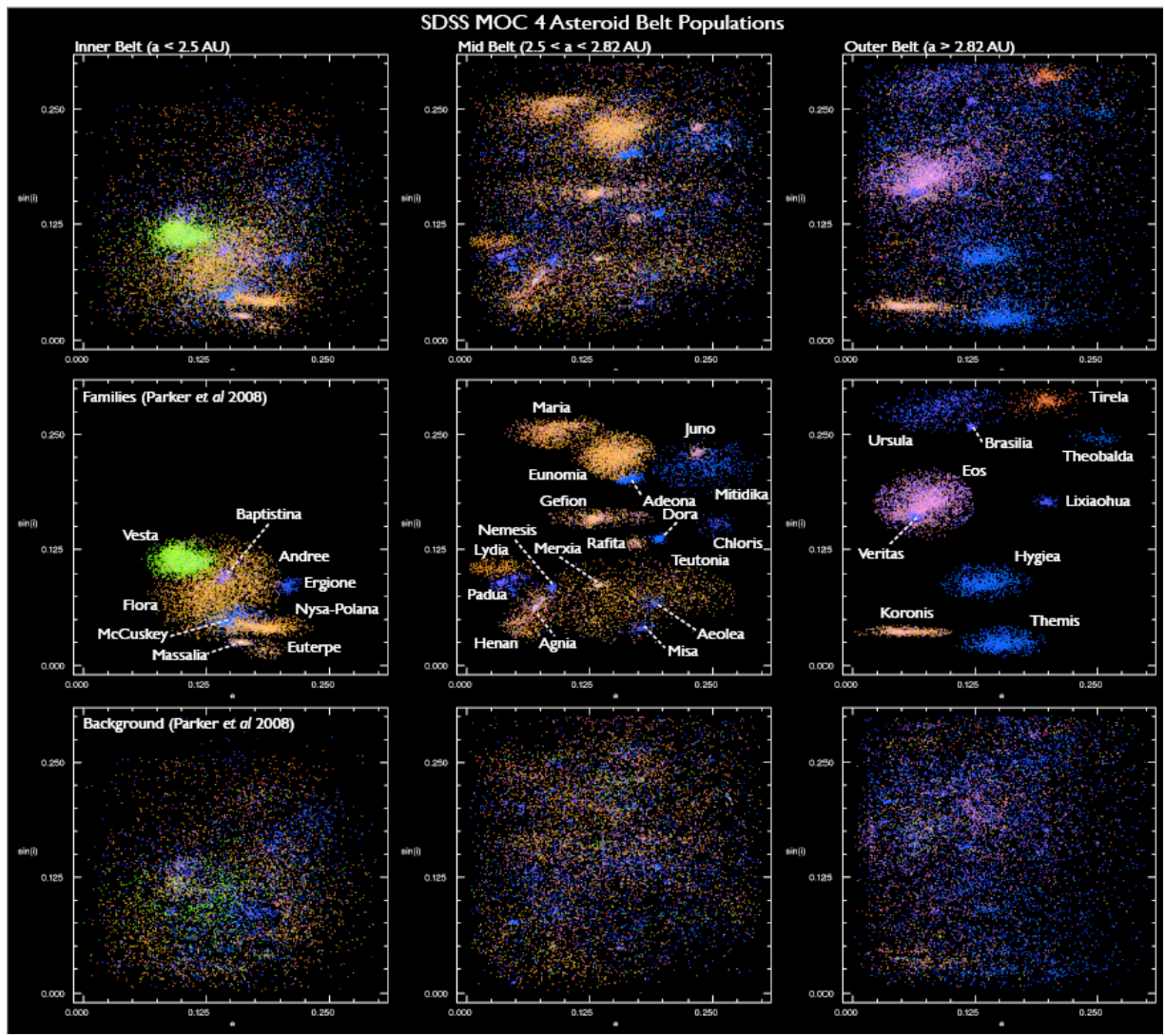}
\caption{Illustration of the decomposition of the Main Belt asteroid population
into families and background objects using proper orbital elements and color 
(adapted from \citealt{parker2008_sdssasteroids}). The top three panels show the sine 
of the orbital inclination vs. orbital eccentricity diagrams for three 
regions of the Main Asteroid Belt defined by semi-major axis range 
(see the top labels). Each dot represents one object, and is color-coded
according to its color measured by SDSS (see also \autoref{Fig:ss:Trojans1} for a ``zoomed-out'' view). 
The three middle panels show objects from 37 identified families, and 
the bottom three panels show the background population. Examples of
size distributions for several families are shown in 
\autoref{Fig:ss:Parker2}. These results
are based on about 88,000 objects. The LSST data set will 
include several million objects and will also provide exquisite time 
domain information.}
\label{Fig:ss:Parker1}
\end{center}
\end{figure}

Using a data set 
with $\sim$ 88,000 objects, they defined 37 statistically robust 
asteroid families with at least 100 members (see 
\autoref{Fig:ss:Parker1}). About 50\% of objects in this data 
set belong to families, with the fraction increasing from 
about 35\% to 60\% as asteroid size drops below $\sim$ 25 km. 

\begin{figure}[htb!]
\begin{center}
\includegraphics[width=6in]{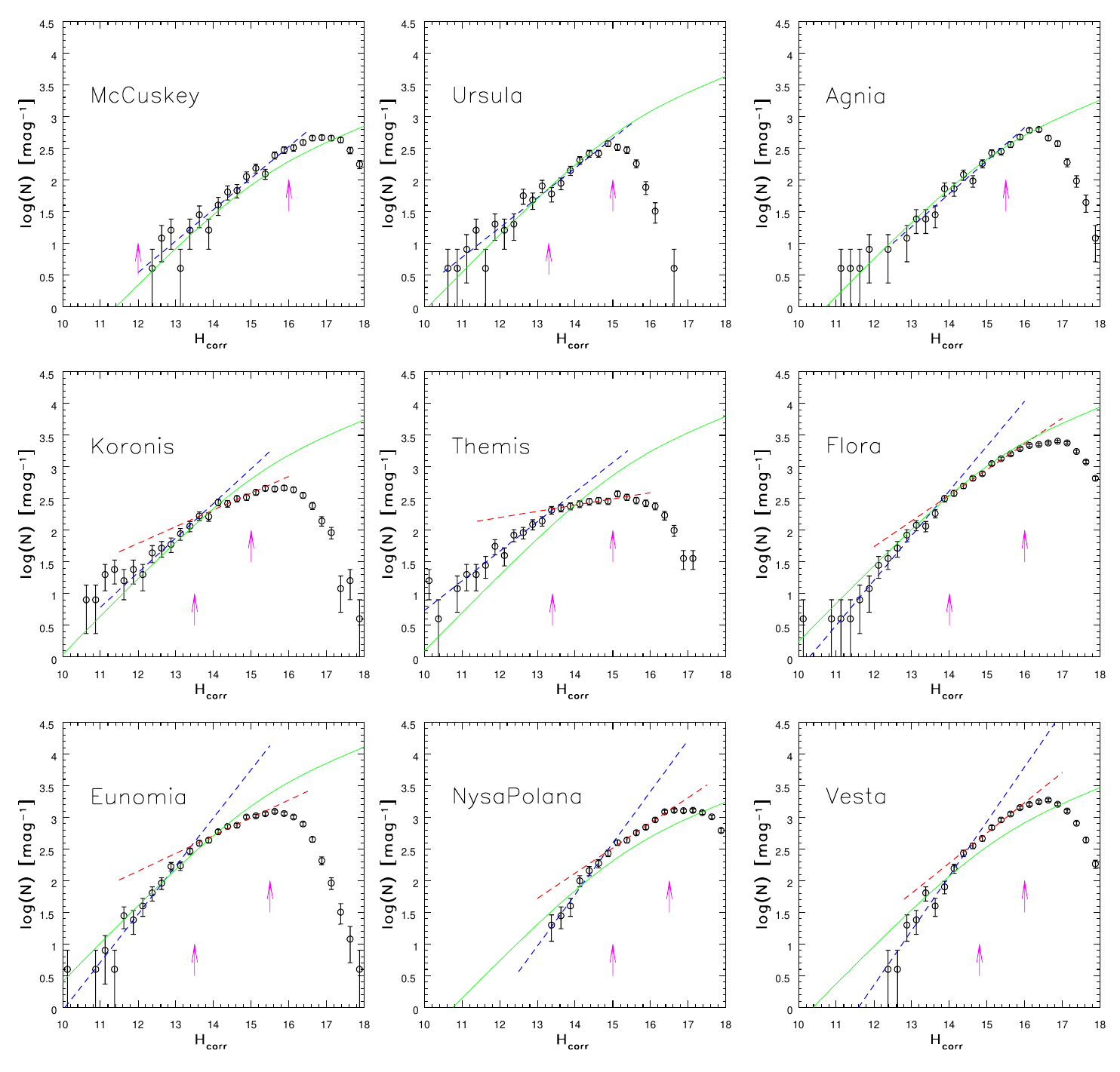}
\caption{The differential absolute magnitude distributions from SDSS
data 
for selected asteroid families (see \autoref{Fig:ss:Parker1}), 
shown as symbols with Poisson error bars
(adapted from \citealt{parker2008_sdssasteroids}). 
The green solid line in each panel shows the distribution for
the whole Main Belt with amplitude fit to the data. The two dashed
lines show the best-fit power-law fits for the bright (blue) and faint
(red) ends separately. The two  
arrows show the best-fit break magnitude (left) and the adopted completeness 
limit (right). The current catalogs are limited to $H<15$; the LSST data set 
will extend these studies to  $H\sim20$.
\label{Fig:ss:Parker2}
}
\end{center}
\end{figure}

According to \citet{parker2008_sdssasteroids}, the size distribution 
varies significantly 
among  families, and is typically different from the size distributions 
for background populations. The size distributions for 15 families 
display a well-defined change of slope and can be modeled as a 
``broken'' double power-law (see \autoref{Fig:ss:Parker2}). These 
complex differences between size distributions probably depend 
on the collisional history of individual families and offer an
observational tool to study the evolution of the Solar System. 

The currently available data set is limited to $H\sim15$, and includes
several hundred thousand objects. The LSST data set will include several 
million objects, and will extend these studies to $H\sim20$ (a
limit ten  times smaller, corresponding to about several hundred meters). In addition, 
over 150 detections will be available for about million objects
(see \autoref{ss:cumcounts}) enabling studies of asteroid rotation via light curve inversion,
(see \autoref{ss:lightcurves}) and providing exquisitely accurate colors for taxonomy. 
While taxonomy is not representative of composition, it can provide a first 
set of guidelines if spectra is not available. 

Previous surveys have shown that the albedo distribution of asteroids
is bimodal, with one peak having a mean albedo of 0.06 while the other
peak has a mean of 0.20 in $g$ or about $0.25$ in $r$ or $i$.  These
two different albedo peaks are correlated with asteroid color,
representing their taxonomic types.  Low albedo MBAs are C-, D-, and P-types
asteroids, while those MBAs with higher albedos are S-, R-, V-, E-, and M-type
asteroids.



LSST data can be used to measure MBA taxonomies, which may be used to constrain the albedos of the MBA population.  However, it is important to note that asteroids of the same taxonomic type can have a wide range of compositions and albedos.  In addition, asteroids of disparate compositions may appear to belong to the same taxonomic group, but have completely different albedo values.  Hence any broad generalizations about the MBA population albedo distribution with respect to taxonomy should be made with the utmost of caution.  Even with this caveat, the real power of the LSST photometry will be in its
large number statistics, which may help in improving the size estimates of a large portion of the MBA population, perhaps improving the uncertainty on the size estimate from 30\% -- 50\%. 



\subsection{Determining the Masses of Large Main Belt Asteroids}
{\it Steven R. Chesley, Zoran Kne\v zevi\'c, Andrea Milani}

While the size distribution is estimated from the photometric observations
of color and absolute magnitude, one can also attempt 
to measure the masses of larger asteroids directly from the perturbation of
other, typically smaller, ``test particle'' asteroids that pass near the
perturber. At present only a few dozen asteroids have mass estimates
based on perturbations, but LSST will produce astrometry that is both
prolific and precise, at the same time that it dramatically expands
the pool of potential test particles. LSST data should allow the
estimation of the mass of several hundred or so main belt
asteroids with an uncertainty of $\sim30\%$ or less. These estimates will provide many more mass bulk density
estimates than are currently known, constraining the internal
structure and/or mineralogy of many asteroids. Moreover, asteroid mass
uncertainty remains the largest source of error for precise asteroid
(and planetary) ephemerides. Driving this uncertainty lower will
afford more precise predictions of asteroid and planetary
trajectories.

The main problem of this technique is the complexity of explicit, simultaneous computation of a large number of asteroid orbits; while the target objects for which the mass may be computed are few, the list of objects potentially having a close approach is on the order of millions. To avoid intractable computational complexity, the candidate couples need to be selected through a sequence of filters. After an elementary selection based on absolute magnitude, perihelia and aphelia, one of the filters is based on the computation of the Minimum Orbital Intersection Distance (MOID) between two asteroids; this computation can be refined by also taking into account the orbital uncertainties. If the MOID is small, the maximum amount of deflection can be computed from a two-body hyperbolic formula. Only when the result of these preliminary computations indicate the possibility of a measurable deflection, then an accurate orbit propagation for the smaller asteroid, including the larger asteroid in the dynamic model, needs to be performed. If the close approach actually occurs with an observable signal, for the given (or expected) set of observations, then actual orbit determination with mass as an additional fit parameter takes place (this both in simulated/predicted cases and in actual data processing).

\section{Trans-Neptunian Families and Wide Binaries}
\label{ss:binaries}
{\it Michael E. Brown, R. Lynne Jones, Alex Parker}

Only one collisional family of objects is currently known in the outer
Solar System. Haumea, the fourth largest object known beyond Neptune, orbits within a
dynamical cloud of debris left over from a giant impact with a
comparably-sized object \citep{brownHaumea}. Such a giant impact is
exceedingly improbable in the current environment, and even difficult
to explain in a more dense earlier environment. \citet{levisonSat08}
realized that collisions between objects being scattered by Neptune
could potentially explain this family. This suggests that many
collisional families should exist in the outer Solar System and their
orbital distributions could trace the scattering history of the early
Kuiper Belt. 

\begin{figure}[htb!]
\begin{center}
\includegraphics[width=4in]{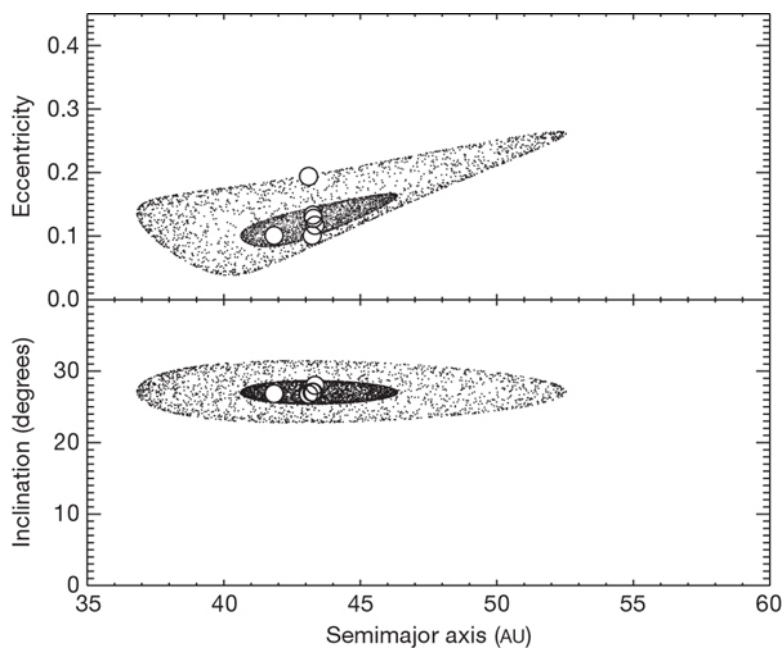}
\caption{
Figure from \citet{brownHaumea}. The open circles give the proper
orbital elements of the KBOs thought to be part of a collisional
family with Haumea. The widely dispersed
small dots show the orbital elements possible from a collision
centered on the average position of the fragments and with a
dispersive velocity of 400 $\rm m\,s^{-1}$.  The more tightly concentrated dots
show orbital elements expected if the collision had a dispersive
velocity of 140 $\rm m\,s^{-1}$. The orbital dispersion from these collisions
indicates that identifying collisional families in the Kuiper Belt
will require accurate orbital elements for a large number of objects
and may be strongly aided by color or other physical measurements. }
\label{Fig:ss:brown}
\end{center}
\end{figure}

The Haumea family was recognized only because each of its members
shares the same distinct infrared spectrum: a surface dominated by
almost pure water ice. Without the spectra, the family could not have
been recognized as no statistically significant concentration could be
identified by dynamics alone (\autoref{Fig:ss:brown}). The icy
surface of the family members is likely the result of the
differentiation of proto-Haumea before impact, where the family
members are pieces of the pure ice mantle. As there are 
strongly identifiable spectral features associated with only a few TNOs, other collisional families
in the Kuiper Belt cannot currently be identified by their spectra,
but rather will have to be identified as significant concentrations in
dynamical space, as the asteroid families are identified. Such
identification  may be possible with LSST due to the large
number of TNOs discovered with well-measured orbits, and will be
aided by information on colors and perhaps other physical properties (such as 
rotation rate). 

Along with collisional families, Kuiper Belt binaries offer a unique window into 
understanding the physical structure and composition of TNOs. Accurate 
mutual orbits allow determination of component masses and, if coupled with size measurements 
derived from thermal observations or direct detection, densities. The ice-to-rock fraction of objects in 
the Kuiper Belt is not constrained other than in the Pluto-Charon system, but is a strong indicator of 
the chemical environment at the time of formation \citep{Lunine93}. Density measurements are therefore 
essential in establishing the composition in the early solar nebula, similar in importance to the 
compositional gradient observed in the Main Belt of asteroids. 

Binarity in the Kuiper Belt looks distinctly different than that in the Main Belt: known TNO binaries are likely to be widely separated and roughly equal mass. Among NEAs and small MBAs, binaries tend to be closely bound with the primary rapidly spinning, suggesting that they have formed by fission, perhaps due to
over-spinning of a single body by the YORP radiation torque as
described below in \autoref{ss:sparselightcurve}. Widely
separated binaries of nearly equal-sized bodies suggest completely
different formation mechanisms, and as a result of the different evolutionary history in the Main Belt compared to the Kuiper Belt,
most resolved binary systems detected by LSST will be wide TNO binaries. 

A number of theories describe the formation of TNO binaries, and to some degree offer testable predictions. In the early dense environment, satellites can be captured by the effects of dynamical friction \citep{goldreichBin}, through two-body collisions, or exchange reactions in the presence of a third planetesimal \citep{weidenschillingBin, funatoBin}. Large Kuiper Belt objects appear to have tiny satellites formed as a result of giant impacts \citep{brownSatellites}, which may be related to yet-unidentified collisional families. Each of these processes preserves traces of the environments of the regions where the objects formed, which are likely dramatically different from the current Kuiper Belt environment, where low interaction rates among TNOs make forming binaries extremely unlikely. 

Work with the Hubble Space Telescope (HST) has shown that the cold classical Kuiper Belt has a significantly higher fraction of resolved satellites than any other TNO population: 22\% rather than 5.5\% \citep{nollBinSSB}. However, the sample of known binaries is small. LSST, in the course of detecting $>20,000$ TNOs, will also find many satellites ($\sim50-100$) separated by arcseconds, allowing detailed study of these systems. Measuring the statistical properties of the large-separation binary orbit distribution, which are most 
sensitive to disruption and formation mechanisms, will tell us which mechanism(s) were at work, 
provide constraints on the dynamical history and space densities of the Kuiper Belt, and help us 
understand how those objects survived until present time in the disruptive dynamical environment of 
the Kuiper Belt.

\section{The Size Distribution for Faint Objects---``Shift and Stack''}
{\it Steven R. Chesley, R. Lynne Jones, David E. Trilling}
\label{ss:faint}

In addition to measuring the size distribution through a near complete
inventory of larger objects, LSST can extend the size distribution
estimate to much smaller sizes through a special program of deep
fields (\autoref{sec:design:cadence}), capitalizing on the large LSST
aperture and quick CCD read-out 
times to search for very faint TNOs, Trojans, MBAs, and potentially even NEAs.

The strategy for such an LSST ``deep drilling'' project is to maintain
a given pointing for successive exposures until the desired depth can
be obtained in a sum, or ``stack,'' of all images. For routine
follow up and recovery work the individual images are stacked with the
known rate of motion of the target body, but for initial discovery
with LSST deep fields, a family of stacks is necessary to cover the
range of motion vectors for each of the target populations. The large
number of stacks, in the thousands for MBAs, leads to a non-trivial
computational problem, with the challenge proportional to the time
duration of the stack, since more stacking rates are required to avoid
trailing of a given target in at least one of the stacks. Thus single night
stacks are significantly more attractive than multi-night stacks.

As an example, LSST will be able to track a single opposition field
for up to eight hours in winter, during which time it could obtain
about 850 ``single-visit'' exposures of 30 seconds each. Since the signal-to-noise
ratio for $N$ exposures follows $\sqrt{N}$, the stack from these
images will reveal detections about 3.7 magnitudes fainter than the
single visit $5\,\sigma$ limit of $r=24.7$. However, $7\,\sigma$ limits are
generally more appropriate for detections in deep stacks, and so we
estimate that a single night stack will reach $r \approx 28.0$. This
translates to diameters more than 5 times smaller than the single
visit limit. To reach this limiting magnitude, stacking will have to
be done with a few thousand different assumed rates, the vast majority being at main belt
rates. All observations would be carried out in the same
filter---probably $r$. Potentially, exposure times other than 30~seconds
would be explored for this mode of operation to reduce the number of
images required to shift and stack. 

The first epoch is repeated at a later time, ideally on the next night
so that more than 90\% of MBAs will remain in the field, which, for opposition fields, is generally
sufficient to obtain a reliable topocentric distance and hence
absolute magnitude. For TNOs, the field should be repeated a few
months later when it is at a significantly lower solar elongation. In that case, it may take two partial nights of staring to reach the
desired limiting magnitude, since the field is not observable the
entire night. However, the stacking requirements for TNOs are much
less demanding (they move much more slowly!), and so multi-night stacking appears tractable. At a
cost of ~1-2\% of survey time, this three-night deep-drilling cadence
process could be repeated annually on the same field for a few years,
building up a large set of MBA detections and
solidifying the orbits of the TNOs in the field.
 
The deep TNO survey should have several unique pointings overall. In
consultation with other science drivers, these should be divided
between ecliptic and off-ecliptic pointings. The ecliptic
pointings---at various ecliptic longitudes---would allow a
longitudinal probe of the outer Solar System small body
population. This is particularly important since the sky density
membership of resonant objects -- a key probe of outer Solar System
evolution -- varies as a function of ecliptic latitude and
longitude. The off-ecliptic pointings would provide a
three-dimensional map of the outer Solar System down to very small
sizes. It is worth noting that other science drivers would
profit from the same deep stack data sets.

Deep drilling fields targeting the outer Solar System could profitably
also include the Trojan clouds of Jupiter and Neptune (as well as
hypothesized Trojan clouds of Saturn and Uranus, though no Trojan
asteroids for these planets are currently known). Jupiter and Neptune
are in conjunction in mid-2022, and so their leading and trailing
Trojan clouds will be respectively aligned at this time, making it a
good opportunity to probe the Trojan populations of both planets---in
addition to MBAs and TNOs---with a minimum of telescope time. About
four years earlier, in 2018, the leading Jupiter cloud coincides with
the trailing Neptune cloud, forming another good opportunity.

The deepest search for TNOs to date reached $r\sim29$ over $0.02$
deg$^2$, obtained with HST/ACS \citep{bernstein_hst}. Thus, a 
deep drilling experiment with even a single LSST field will increase
the areal coverage by a factor of $\sim500$. Using the
\citet{bernstein_hst} result to predict LSST results at $r\sim28$, we
expect something like 1000 TNOs per deep drilling field on the
ecliptic; off-ecliptic fields may have densities one tenth this value. For
MBAs, each deep field should yields upwards of 20,000
detections.

Successfully detecting faint ($r \sim28$) MBAs enables science in a different
size regime than the projects described in \autoref{ss:mbsizes}. For
example, the size distribution of MBAs is known to have significant
structure that records the intrinsic strength of asteroids (e.g.,
\citealt{Obrien2005}), 
and probing to this size regime will allow studies of the global
internal properties of asteroids. Additionally, small MBAs are 
the direct predecessors of NEAs (that is, the sizes of typical NEAs 
are comparable to those of ``very small" MBAs that are only
available through an LSST-type deep drilling project as described in
\autoref{sec:design:cadence}). Therefore, 
by measuring the properties of very small MBAs (i.e., size distribution,
orbital distribution), we can probe the links and processes by
which MBAs become NEAs. A study of the MBA-NEA connection is only
possible with both an NEA survey {\em and} a very deep MBA survey
such as described here. This link between the two is described further in 
\autoref{ss:neos_mbas}.


\subsection{Detection of Extremely Faint Objects through Real-Time Collisions}
{\it R.  Jedicke}

We will measure or set a limit on the collision rate of MBAs too
small to detect directly with LSST. We will do this by searching for
signatures of the transient dust clouds produced in the catastrophic
collision of two objects that are otherwise too small to detect, or by
detecting transient increases in the brightness of asteroids. This
will allow us to
\begin{itemize}
{\item test whether the size--frequency distribution (SFD) measured for the larger Main Belt objects can be
  extrapolated to smaller sizes,} 
{\item test and refine collisional models, and} 
{\item understand the physical structure of asteroids.}
\end{itemize}

There is expected to be roughly one catastrophic disruption of a 10 m
diameter main belt object every day and, given LSST's sky coverage, we
expect to image about one of these disruptions every week. As the dust cloud
from a catastrophic disruption expands, its apparent
brightness increases as long as the optical depth $\tau > 1$ after
which the clouds brightness will decrease. A 10 m diameter asteroid's
disruption could create a dust cloud 1 km in diameter which would have
the apparent brightness of a 1 km diameter asteroid (easily detected
by LSST).

The difficulty lies in knowing the expansion rate of the dust cloud
and therefore determining how long the cloud is visible. If the cloud
is visible for many days to a week we might detect the expanding dust
cloud on each of three nights during a lunation. The brightness of the
cloud could vary dramatically from night to night, and it will be
impossible to recover the object or assign a detection to a previously
detected object. If the dust cloud does
not last that long it is possible that we will detect bright but
`orphaned' tracklets that are impossible to link to other tracklets.

It may also be possible to detect the collision of small objects into
larger objects that are easily detected by LSST. By continuously
monitoring many objects over the LSST operational lifetime we can
search for unusual and unrepeated brightening of asteroids as a
signature of a recent collision.

With a sufficient number of collisions we may determine the collision
rate of these objects. The rate at which the dust clouds brighten and
fade will provide details on the physical structure of the
asteroids. Color measurements or detailed spectroscopic followup of
the dust clouds will provide information on the dust properties.
If the collisions produce enough large grains, the clouds may be observable
in the infrared for much longer if followup could be obtained from space.


\section{Lightcurves: Time Variability} 
\label{ss:lightcurves}
{\it Stephen T.  Ridgway, R. Lynne Jones}

The variation in the apparent brightness of solid Solar System bodies
can be a valuable source of information about their history, their
surfaces and even their interiors.  Cyclic variations can show the
rotational period and rotational axis orientation, the shape, compositional clues, the
density, and information about the surface roughness.  Many objects
have brightness variations on the order of only 0.2 magnitudes, and
require accurate, well sampled light curves for unambiguous
interpretation.  LSST will provide outstanding period coverage through
the method of sparse light-curve inversion. 

Asteroidal rotation and the direction of its spin axis are an obvious
consequence of the accretion and collision process.  Photometry can
provide periods, and in some cases the spin axis can be estimated by
the timing of brightness extrema \citep{taylor1983}. On the order of a
few thousand asteroids have reliably measured rotation rates --
\citet{harris2006} provide a brief overview on asteroid rotational
periods, which range from 2 hours up to about a day, reflecting
tensile strengths and rubble pile or monolithic
structures. \citet{kryszczy2007} point to an online catalog of
asteroid spin states and pole positions, illustrating a non-random
distribution of pole axis positions likely due to radiation pressure
torques. Some fraction of the asteroids will have detectable
rotational lightcurves, which will allow determination of their
rotational periods. 

The amplitude of a rotational light curve can give a measure of the
object shape, commonly modeled as a triaxial ellipsoidal. Contact or
small separation (unresolved) binaries can be inferred from
characteristic brightness variations, or in some cases,
eclipses. Observed brightness variations may not be due entirely to
object shape, but may also depend on varying albedo associated with
compositional variations across the surface. Multicolor measurements
can support separation of these effects, thanks to the known colors of
a number of surface compositions -- mineral or carbonaceous materials,
or in the outer Solar System, ices. The albedo and apparent brightness
then support a reliable estimate of object size. For rapidly rotating
objects, the size gives a lower limit to the mass consistent with a
rubble structure. Even for unresolved binary objects, the orbital
period gives a dynamical measurement of the masses. Mass and size
provide a measure of the densities, which constrain the ratio of
minerals to ices and the porosity of the object. \citet{kaasalainen01}
and
\citet{kaasalainen01b} have shown that several hundred accurate phase data are sufficient to support
optimal inversion of lightcurves to determine shape and albedo distributions (see \autoref{ss:sparselightcurve} for
more information). 

To date, even after painstaking work, little is known about rotations of objects in the outer Solar System
\citep{sheppardRot}. At present, to measure a rotation, each object must be individually tracked and monitored with a large telescope for hours or days. Some rotations show up easily on these time scales, 
some are heavily aliased or too subtle for detection and the current
sample of objects with known rotation periods is small. Nonetheless, a
few interesting objects stand out. The large objects Varuna and Haumea
have extremely rapid rotations (six and four hours respectively),
which cause them to elongate into triaxial ellipsoids
\citep{lacerda2007}. Haumea is suspected to have suffered a
family-producing collision, which likely imparted the spin. No such
family has yet been dynamically linked to Varuna. Observations of
rotations have suggested with poor statistics that a large fraction
of objects could be contact binaries \citep{shepjewitt04}. Such
contact binaries could be a natural consequence of the
dynamical-friction induced capture in the early Solar System
\citep{goldreichBin} if the dense-early environment persisted for long
periods of time allowing orbits of captured satellites to decay.

Small bodies do not normally reflect as Lambertian surfaces owing to
shadowing in the surface microstructure.  Thus the asteroid magnitude
system employs two numbers to represent the brightness: a mean
(normalized) magnitude, $H$, and also a phase factor, $G$, that describes
the observed brightness variation as a function of the scattering
angle.  More detailed models attempt to relate the phase effect to the
surface microstructure.  LSST photometry will provide a massive body of
homogeneously obtained phase data for on the order of a million asteroids
(see \autoref{ss:cumcounts}). Measurements at very small phase angles
($<2^{\circ}$) are particularly valuable \citep{domingue1989}, and while LSST will
observe most of these asteroids near opposition as a matter of regular operations,
additional follow-up targeted measurements could be scheduled at other facilities. 

\subsection{Sparse lightcurve inversion}
{\it Josef {\v D}urech, Mikko Kaasalainen}
\label{ss:sparselightcurve}

LSST will provide us with accurate photometry of a large number of
asteroids. As has been suggested by many simulations
\citep{kaasalainen2004, durech2005, durech2007},  
this so-called
``sparse photometry" can be used the same way as standard dense
lightcurves to derive basic physical parameters of observed asteroids:
the global shape, the spin axis direction, and the rotation
period. Simulations that have been done so far showed that, roughly
speaking, once we have at least $\sim100$ sparse brightness 
measurements of an asteroid over $\sim5$ years calibrated with a
photometric accuracy of $\sim5$\% or better, a coarse model can be
derived. This approach is much more time-efficient than the usual
lightcurve photometry. The sparse data inversion gives correct results
also for fast ($0.2-2$~h) and slow ($>24$~h) rotators, although it may 
give best results with large amplitude variations and moderate periods. 

As can be seen in \autoref{Fig:ss:nObs} the number of observations of
individual asteroids is generally sufficient for lightcurve inversion. The median number of
expected LSST detections over 10~years is $\sim190$ for NEAs with $H\le15$ mag 
and $\sim260$ for MBAs with $H\le16$ mag.

An important issue is to use all available data, so we will combine
LSST sparse photometry with sparse and dense data from other sources
(e.g., Pan-STARRS, follow-up observations, existing databases,
etc.). Photometry can be also combined with adaptive optics images \citep{marchis06} and
occultation profiles to obtain more detailed models with accurate
dimensions.

We expect to derive about $10^4$ to $10^5$ Main Belt and Near-Earth
Asteroid shape models from LSST photometry, which means that we will be able
to map a substantial part of the asteroid population. This will bring
new insights into its structure, history, and evolution. We will be
able to detect Yarkovsky and Yarkovsky-Radzievskii-O'Keefe-Paddock
(YORP) effects that can secularly change orbits and spins of
asteroids. Both effects are caused by the anisotropic thermal emission
of the heated surface.  While the Yarkovsky effect describes the
change of the orbit caused by the net thermal force, the YORP effect
describes the influence of the thermal net torque on the spin
state (see \citealt{bottkeyorp06} for a review). The distribution of spin rates and obliquities will allow us to
quantify the YORP evolution. We also expect to
reveal new populations in spin-orbit resonances
\citep{vokrouhlicky03}.  In addition, by constraining the Yarkovsky
effect, this would be potentially very important in discerning the
history of genetic pairs.

For TNOs, the viewing/illumination geometry changes very slowly and
the full solution of the inverse problem is not possible. However,
accurate sparse photometry can be used for period determination.

Due to the stability and uniqueness properties of the inverse problem
solution derived from the disk-integrated photometry, asteroids are
mostly modeled as convex bodies.  LSST sparse photometry can be also
used for detecting (but not modeling) ``non-standard" cases such as binary
and tumbling asteroids. A fully synchronous binary system behaves like
a single body from the photometric point of view \citep{durech03}. Its
binary nature can be revealed by the rectangular pole-on silhouette
and/or large planar areas of the convex model. In some cases -- when
mutual events are deep enough -- asynchronous binaries can be detected
from sparse photometry. Interesting objects can then be targeted for
follow-up observations.

\section{Overlapping Populations} 
\label{ss:overlappingpops}

As we discover and characterize more small bodies throughout the Solar
System, more surprises are uncovered. One such area is the discovery
of linkages and overlaps between different populations of objects. The
discovery of asteroids showing cometary activity is an example of the
overlap of physical properties between different
populations. Simulations demonstrating that objects can have orbits
which slowly cycle between the inner Oort Cloud and the Scattered Disk
or even Centaur regions, or from the MBAs into NEA
orbits, imply that to fully understand each of these groups requires
understanding the Solar System as a whole. 

\subsection{The Relationship between NEAs and MBAs}
{\it Alan W. Harris, Steven R. Chesley, Yanga R. Fern{\'a}ndez, R. Lynne Jones}
\label{ss:neos_mbas}

Orbits crossing the orbits of the giant planets have
lifetimes of only thousands of years; those crossing the terrestrial
planets have lifetimes of millions of years, which is still short enough
that none of the current population of NEAs is ``primordial'' in their
current orbits. Their dynamical lifetimes are only on the order of $10^6$ to $10^8$ years due to interactions with other objects in the inner Solar System that cause them to either impact one of the inner planets or the Sun, or be ejected from the Solar System altogether \citep{morbiglad98}.  Hence the continued presence of these objects within near-Earth space requires a mechanism(s) and source region(s) to replenish and maintain the NEA population over time.  
 
Current dynamical models and orbit integrations \citep{bottke2002_neo}
suggest that NEAs are delivered primarily from specific regions
within the Main Belt that are particularly affected by certain
secular and mean-motion resonances. However the Yarkovsky effect
can push objects from different parts of the
Main Belt into orbits that make them more likely to be thrown
inward. Therefore it is crucial to study the migration
within the Main Belt if we are to learn where NEA material comes from.

A key to understanding the transfer of MBAs into near-Earth orbital
space is to determine the population of both classes, especially in
the same size range. Presently, we only know the size frequency distribution (SFD) of MBAs down to a size of several km
diameter. Unfortunately, only the largest hundred or so NEAs are that
large, so there is very little overlap of our measured SFD of NEAs
with that of MBAs. LSST will extend that overlap down to sizes of
$\sim 100$ meters diameter in the Main Belt, providing enough overlap to examine
the differences of the SFDs. This will shed
light on the efficiency of migration into Earth-crossing orbits versus
size, or whether close planetary encounters modify the distribution,
say by tidal disruptions, and the effect that Yarkovsky and YORP have 
in these transfer mechanisms.

By the time LSST begins operations in 2014, nearly all of the NEAs
with diameters greater than 1 km will have been cataloged by surveys
such as Pan-STARRS. At smaller
sizes, down to perhaps 150m, LSST, over its lifetime, will discover
and catalog nearly all ($\sim90\%$) of the NEAs. In the size ranges where nearly
all of the NEAs have been discovered, the orbits of each asteroid can
be propagated forward to determine the probability of future impacts
with the Earth and the Moon. At sizes smaller than that at which the
catalog is complete, characterizing the future impact hazard will
remain a statistical problem of estimating size frequency
distributions and orbital distributions from a limited sample of
objects. At these smaller sizes, a statistical description of the size
frequency distribution and orbital distribution along with taxonomic
identifications can yield insight into the source regions that
resupply the NEAs and whether the resupply processes differ by
size. There is also utility in characterizing the past impact flux on
the Earth, the Moon, and other bodies, in comparison with the
cratering record, to understand whether and how impact fluxes have
changed over the history of the Solar System. 

\subsection{Damocloids and Main Belt Comets: Asteroids on Cometary Orbits and Comets on Asteroidal Orbits}
{\it Paul A. Abell, Yanga R. Fern{\'a}ndez}

The Main Belt asteroids have been recognized as one of the primary sources of material for the NEA population \citep{mcfadden85}, but several investigators have suggested that a non-negligible portion of the NEA population could also be replenished by cometary nuclei that have evolved dynamically into the inner Solar System from such reservoirs as the Edgeworth-Kuiper Belt and the Oort Cloud \citep{weissmanbottke02}. Evidence used to support the hypothesis of a cometary component to the NEA population has been based on: observations of asteroid orbits and associated meteor showers (e.g., 3200 Phaethon and the Geminid meteor shower); low activity of short-period comet nuclei, which implied nonvolatile surface crusts (e.g., 28P/Neujmin 1, 49P/Arend-Rigaux); lack of recent cometary activity in NEAs observed to have apparent transient cometary activity in the past (e.g., 4015 Wilson-Harrington); and a similarity of albedos among cometary nuclei and asteroids in comet-like orbits.  Recent studies have estimated that approximately 5 -- 10\% of the entire NEA population may be extinct comets \citep{fernandezjewitt05, demeobinzel08}.

Thus several observational investigations have focused on examining low-activity short period comets or asteroids in apparent comet-like orbits.  A population that has been thought to have probable connections to the Oort Cloud and the isotropic comets are the Damocloid asteroids.  The Damocloid-class objects are thought to be possible dormant or extinct comets because these asteroids have high-inclinations and large semi-major axes 
just like those of Halley-family and long-period comets \citep{asherbailey94, bailey96}.
About 50 such objects are known (as of Sept 2009), although all
of the objects so far seem to have evolved orbits. That is, none of
the objects is new in the Oort sense.
Most observations of these objects suggest that they have similar spectral characteristics to those of 
Jupiter-family comets and
outer Main Belt asteroids, but show no evidence of coma
\citep{jewitt2005}.  
However, at least one Damocloid object (C/2001 OG$_{108}$)
demonstrated intense coma 
during its perihelion passage 1 AU
from the Sun after showing no coma for several months beforehand, 
which supports the notion that Damocloids
in general could be dynamically evolved objects 
from the Oort Cloud \citep{abell05_og108}. 

In addition, it seems that the conventional dynamical and physical
demarcation between asteroids and comets is becoming even less clear.
Observations of a few objects located within the Main Belt asteroid
population show degrees of activity that are normally a characteristic
of cometary objects \citep{hseih06}.  Dynamical modeling of
the dust generated from these Main Belt objects suggests that this
level of activity requires a sustained source, and is not the result
of impulsive collisions.  Thus it is plausible that an additional
cometary reservoir exists within the Solar System among the main belt asteroids \citep{hseih06}. 
If these objects were formed in-situ, they would suggest that
condensed water ice survived to the present-day much closer than
traditionally believed. However there could be dynamical
mechanisms that can place outer Solar System objects
into low-eccentricity, outer Main Belt orbits \citep{levison08DDA}, 
so the origin
of these objects is an important science question.
Only four such main belt comets (MBCs) have been discovered to date, but given the low level of activity in these objects, many more could be present below the current detection limits of existing ground-based sensors.

During survey operations, the LSST will discover many more low albedo Damocloid objects, and have the capability to detect faint/transient activity from MBC candidates.  
A large statistical database of several hundred
Damocloids and MBCs would be an invaluable resource for understanding
volatile distribution in the Solar System and 
thermal evolution of small bodies.
In addition, objects originating in the different
cometary reservoirs (Oort Cloud, Edgeworth-Kuiper Belt, and 
potentially the Main Belt)
may have distinct physical characteristics.  
LSST will not only be the optimal system for discovering a majority of these objects, but will let us use gross physical 
properties (e.g., lightcurve, colors, taxonomy, etc.) to 
make comparisons across many Solar System populations at 
different stages of their evolution.
This will enable investigators to get a much clearer picture of these enigmatic Damocloid and MBC populations as a whole, which in turn will aid in the refinement of Solar System formation models.

\subsection{The Source(s) of Centaurs}
{\it Nathan A. Kaib}

Identifying the source population for Centaurs, which are similar in dynamical properties to Scattered Disk Objects but have orbits which cross interior to Neptune and are unstable over the lifetime of the Solar System, has proven
difficult. The generally accepted source region for Centaurs is the
Scattered Disk.  As SDOs chaotically diffuse into Neptune-crossing
orbits on Gyr timescales, they naturally produce a population of
unstable planet-crossers qualitatively similar to observed Centaurs.
However, due to perturbations from passing stars and the Galactic
tide, the Oort Cloud also steadily injects bodies into planet-crossing
orbits.  Because the Oort Cloud has a much higher typical semi-major
axis than the Scattered Disk, objects with an Oort Cloud origin will
dominate the high-$a$ range of Centaurs, whereas objects from the
Scattered Disk will dominate the low-$a$ population of Centaurs
\citep{kaib09}.  However, energy kicks from planetary encounters will
act to smear these two $a$-distributions leading to an Oort Cloud
contribution even for Centaurs with semi-major axes less than that of
the actual Oort Cloud.  

With a semi-major axis of 796 AU, 2006 SQ$_{372}$ was recently shown
to have the highest probability of an Oort Cloud origin for any known
Centaur \citep{kaib09}.  Even using a conservative estimate for the
total population of Oort Cloud objects, it was shown that this body is
16 times more likely to originate from the Oort Cloud compared to the
Scattered Disk.  Furthermore, the same analysis showed another known
centaur, 2000 OO$_{67}$ is 14 times as likely to come from the Oort
Cloud as from the Scattered Disk.  Even more intriguingly, dynamical
modeling of these objects' production shows that they almost
exclusively come from the inner 10$^4$ AU of the Oort Cloud.  Known
LPCs only provide an upper limit on the population of objects in this
region and provide no constraints on the actual radial distribution of
material in the Oort Cloud \citep{kaibquinn09}, which is intimately
linked to the Sun's formation environment \citep{fern97}.  Any
additional constraints on this reservoir would be highly valuable.  
Although little information can be gleaned from only the two currently
known objects of 2000 OO$_{67}$ and 2006 SQ$_{372}$, LSST will have
nearly 100 times the sky coverage of the survey that detected
2006$_{372}$.  LSST will also be able to detect objects 4 magnitudes
fainter as well.  As a result, it is reasonable to expect LSST to
discover a hundreds to thousands of objects analogous to 2006
SQ$_{372}$.  Studying the orbital distributions of a large sample of
these types of bodies will be able to further constrain the population
size and provide the first constraints on the radial distribution of
objects in the Oort Cloud. 

\subsection{The Source(s) of Comet Families}
{\it Yanga R. Fern{\'a}ndez}

The conventional idea is that Halley Family comets (HFC) and Long Period comets (LPC) originate from the Oort Cloud. However, dynamical modeling finds this very challenging to reconcile
with current theories about the state of the Oort Cloud (see
e.g., \citealt{duncan2008}).  
The difficulty lies in determining what structural differences there
are (if any) between the inner and outer Oort Clouds, and how the physical
aging and fading of HFCs and LPCs changes the population over time from what
is injected into the inner Solar System to what we observe today. 
There is also a hypothesis that the Scattered Disk is
responsible for some of the HFCs \citep{levison2006}, which is
interesting in light of recent compositional studies showing that
there is more overlap in parent-molecule abundance between Jupiter Family Comets (JFCs) and
LPCs than previously thought \citep{disanti2008}. LSST will
be able to address this situation by dramatically improving the
number of HFCs and LPCs that are known. In particular, astrometry
of LPCs while they are far from the Sun will make it easier to
identify those that are new in the Oort sense (i.e., on their first
trip in from the Oort Cloud) more quickly. The orbital elements of
the HFCs and LPCs will give us a less biased view of the current
distribution of these comets in our Solar System, thereby constraining
the dynamical models.

\section{Physical Properties of Comets}
\label{sec:ss:comets}
{\it Yanga R. Fern{\'a}ndez}

Comets are the most pristine observable remnants left over from the
era of planet formation in our Solar System. As such, their composition
and structure can in principle tell us much about the chemical and
thermophysical conditions of our protoplanetary disk. This can then be used
to understand the place of our Solar System in the wider context of planetary
disks throughout our Galaxy. 

Achieving this understanding of the protoplanetary disk using comets
requires determining the evolutionary processes that have affected the comets we see
today. In the 4.5 Gyr since formation, and even before the comets
felt significant insolation by traveling into the inner Solar System,
they suffered various processes -- e.g., collisions, cosmic-ray
bombardment, flash heating by nearby supernovae -- that changed
their physical and chemical properties from the primordial. It is
crucial to understand evolutionary processes of small bodies in
order to interpret what they may tell us about planetary formation. While this applies
to all small bodies throughout the Solar System, it is particularly interesting in the 
case of comets (and especially comets inbound from the Oort Cloud for the first time)
because they may be closer to the primordial state.

Currently only about 350 JFCs and 50 HFCs 
are known. LSST will discover on the order of 10,000 comets, with 50
observations or more of each of them \citep{Solontoi++09}.  This will dwarf the 
current roster, providing answers to many questions regarding the physical properties
of today's cometary population.

The size distribution will tell us about the
competing evolutionary processes that affect a comet's radius, e.g., its creation as a collisional fragment, its self-erosion from
activity, and its stochastic ejection of significant fragments.
The shape of the JFC size distribution is starting to be understood
(for example, \citealt{meech2004}), although there are still strong discovery
biases in the known population, as evinced by the fact that many
large JFCs (~3-4 km radius) with perihelia beyond 2 AU have only
been discovered in the last few years \citep{fernandez2008}. LSST
will provide us with a much more complete survey of the JFC population,
since it will see ~400-m radius inactive nuclei at 3 to 4 AU and
even 1-km radius nuclei at 6 AU (the typical JFC aphelion). Perhaps
even more important will be LSST's discoveries of HFCs and LPCs. The size distributions of these groups are completely
unknown, suffering from low-number statistics and the fact that
these comets are discovered or recovered inbound only after they
have become active.

While adequately explaining the
measured color distributions of TNOs and Centaurs
has proved challenging, the dichotomy between TNO/Centaur colors
and cometary colors is striking \citep{jewitt2002, grundy2009}.
Cometary nuclei seem to be on average less red than their
outer Solar System counterparts.  In the case of the JFCs, the
nuclei are presumably direct descendants of Centaurs and TNOs, so
understanding how a comet's surface changes as it migrates deeper
into the center of the Solar System is an important question. Perhaps
cometary activity rapidly changes surface properties, but if so,
then there should be a correlation between colors of comets and
active Centaurs.  In addition to finding TNOs and Centaurs that are
closer in size to cometary nuclei, LSST will provide us with a large
number of cometary colors with which to make statistically strong
comparisons.  In particular, LSST will let us measure the colors
of HFCs and LPCs, a field that is right now almost totally unexplored.
A very exciting possibility is that LSST will discover some
``new-in-the-Oort-sense'' LPCs that have not yet turned on, giving us
an opportunity to study a cometary surface unchanged from its time
in the deep freeze of the Oort Cloud.

Traditionally, comets were thought to ``turn off" beyond
3 AU, but in recent years that paradigm has started to change as
we observe low but definitely non-zero mass loss from comets even
all the way to aphelion in the case of JFCs (e.g., \citealt{snodgrass2008, 
mazzotta2008}) and out beyond 25 AU in the
case of Hale-Bopp \citep{szabo2008}.  LSST's 10-year lifespan and
deep magnitude limit will allow us to monitor many comets for
outgassing activity over a significant interval of time (and for JFCs, over
all or nearly all their orbits). The excellent spatial resolution
will let us monitor even low levels of activity using point spread function  comparisons,
where the comet shows some coma that extends just slightly beyond
the seeing disk. LSST will also be able to address how long comets stay
active after perihelion and for what fraction of comets is crystallization of water
ice and/or supervolatile sublimation a source of energy at high heliocentric distances.

Understanding the gas-to-dust ratio of comets and how this varies among
comets of different dynamical classes and ages 
could let us understand the nature of the cometary activity process itself.
The LSST $u$-band peaks near the CN violet (0-0) band at 387 nm.  While CN
is not the most abundant dissociation product from cometary volatiles,
its violet band is second in intrinsic brightness only to the OH
(0-0) band at 309 nm, which is much harder to observe. Thus CN
emission can be used as a proxy for the overall gas production rate.
This $u$-band throughput peak occurs at the longward edge of the
bandpass; the rest of the bandpass will detect shorter wavelength
continuum, and since a comet's continuum is reflected sunlight, it
gets weaker toward the violet and near-UV.  So the $u$-band will be
particularly sensitive to a comet's gas coma.  In combination with
the $r$, $i$, $z$, and $y$ bandpasses, which will be mostly sensitive to
the continuum, a comet's colors should yield a rough estimate of
the CN band strength and hence an approximate 
CN production rate. Thus LSST
provides the very exciting opportunity to produce a large database
of CN production rates for the known comets and for many of the
new comets that it will discover. Existing databases 
\citep{ahearn1995, schleicher2008} will not be able to
match the size of an LSST-produced catalog.  Trends of
the gas-to-dust ratio as a function of other parameters -- perihelion
distance, heliocentric distance, active fraction, statistical age,
dynamical group -- will give clues about how pulsed insolation
affects the evolution of a comet's surface.


\section {Mapping of Interplanetary Coronal Mass Ejections} 
\label{ss:coronalmass}
{\it Bojan Vr\v snak, \v Zeljko Ivezi\a'c}

Large-scale solar eruptions, called coronal mass ejections (CMEs), are 
the most powerful explosive events in the Solar System, where the total 
released energy can be as high as 10$^{26}$ J. During the eruption, a
magnetic flux of the order 10$^{23}$ Weber is launched into interplanetary 
space at velocities of the order of 1000 \kms, carrying along
$10^{11}-10^{14}$ kg of coronal plasma. The Earth-directed CMEs, and 
the shocks they drive, are the main source of major geomagnetic storms 
\citep{gosling1990}, so understanding their propagation through 
interplanetary space is one of central issues of Space Weather 
research.
	
The propagation of CMEs in the high corona can be traced by space-borne 
coronagraphs onboard spacecraft missions such as the Solar and Heliospheric Observatory (SoHO) and
the Solar Terrestrial Relations Observatory (STEREO). At larger heliocentric distances, the interplanetary
counterparts of CMEs (hereafter ICMEs) can be followed with very high 
sensitivity coronagraphs onboard STEREO and Solar Mass Ejection Imager (SMEI) missions, by mapping 
the interplanetary scintillation of distant radio sources \citep{manoharan2006}, 
or by employing the long-wavelength radio type II bursts excited 
at shocks that are driven by ICMEs \citep{reiner2007}. The 
physical characteristics of ICMEs can also be directly determined by  
in situ measurements of various space probes that register 
solar wind characteristics.

LSST will offer a novel method for three dimensional mapping of ICME
propagation,  when combined with in situ solar wind measurements. 
This method has already been applied, although in a very limited form, 
in the 1970s \citep{dryer1975}. The cometary plasma is affected by the
passage of an ICME due to the enhanced ram and magnetic-field pressure 
associated with the ICME. This causes sudden changes of the cometary 
brightness and morphological changes of the coma and the
tail\footnote{For an impressive demonstration, please see 
\url{http://smei.nso.edu/images/CometHolmes.mpg}.} \citep{dryer1975, dryer1976}. 
The comprehensive spatial and temporal LSST sky coverage will locate 
a sufficient number of comets that could be used as probes to detect 
passages of ICMEs. The three-day time resolution of the LSST deep-wide-fast
survey is sufficient to track ICME-forced changes at distances larger 
than a few AU \citep{dryer1975}. At closer distances the changes 
could be monitored by a network of large amateur telescopes, which
will be provided by the comet positions from LSST, as well as by 
monitoring comet activity by STEREO and SMEI. 

The unprecedented capabilities of LSST, in combination with 
comet observations by STEREO and SMEI, as well as by follow-up 
observations by networks of telescopes such as those anticipated for
the Las Cumbres Observatory, will provide a high-quality monitoring of 
a large number of comets, and enable exquisite three dimensional mapping of the ICME 
activity in interplanetary space. The detected passages of ICMEs 
and their shocks will be used to:
\begin{itemize}
\item measure kinematic properties of the ICME propagation 
(position and velocity as functions of time), which will provide 
valuable information about forces acting on ICMEs;
\item determine the angular extent of ICMEs and their shocks;
\item estimate the distance range up to which ICMEs preserve their 
identity; and
\item study interaction of cometary plasma with solar wind.
\end{itemize}


\section{The NEA Impact Hazard}
\label{ss:neo_hazard}
{\it Alan W. Harris, R. Lynne Jones}

Although the possibility of a catastrophic impact of an asteroid or
comet with the Earth has been recognized for decades and even
centuries (Edmund Halley articulated the possibility in his
publication of the orbit of the comet that now bears his name), only in
the past few decades have surveys targeted Near Earth Asteroids (NEAs)
with the specific intent of cataloging all or as many objects as possible
in order to understand this risk. 

In 2005, Congress issued a mandate calling for the detection and tracking of 90\% 
of all NEAs larger than 140~m in diameter by 2020. This has typically been interpreted as applying
to 90\% of all Potentially Hazardous Asteroids (PHAs), which are NEAs with a perihelion distance of less
than 1.3~AU. The date deadline was chosen to be 15 years after signing the mandate, which at the time seemed a reasonable period to build a system (either space or Earth-based) to catalog these PHAs. The size limit (140~m in diameter) and completeness level (90\%) were chosen through a careful calculation of potential risks from impactors, weighed against increasing costs to detect smaller and smaller objects, as well as a consideration for previous cataloging efforts.


Previous and on-going surveys such as Spacewatch and the Catalina Sky
Survey have already come close to identifying 90\% of all PHAs larger
than 1~km in diameter (NASA's so-called ``Spaceguard'' goal), using modest sized ($<2$m) telescopes with
limiting magnitudes in the range of $V\sim21$.  These 1~km PHAs would
be capable of causing global catastrophe if one impacted the Earth.
To date, over 800 PHAs
have been detected above this size limit and while tracking must be
ongoing (particularly for objects which pass particularly close to
gravitational perturbation sources such as Earth), none is currently
known to be on an impacting orbit.  

\begin{figure}[htb!]
\begin{center}
\includegraphics[width=6in]{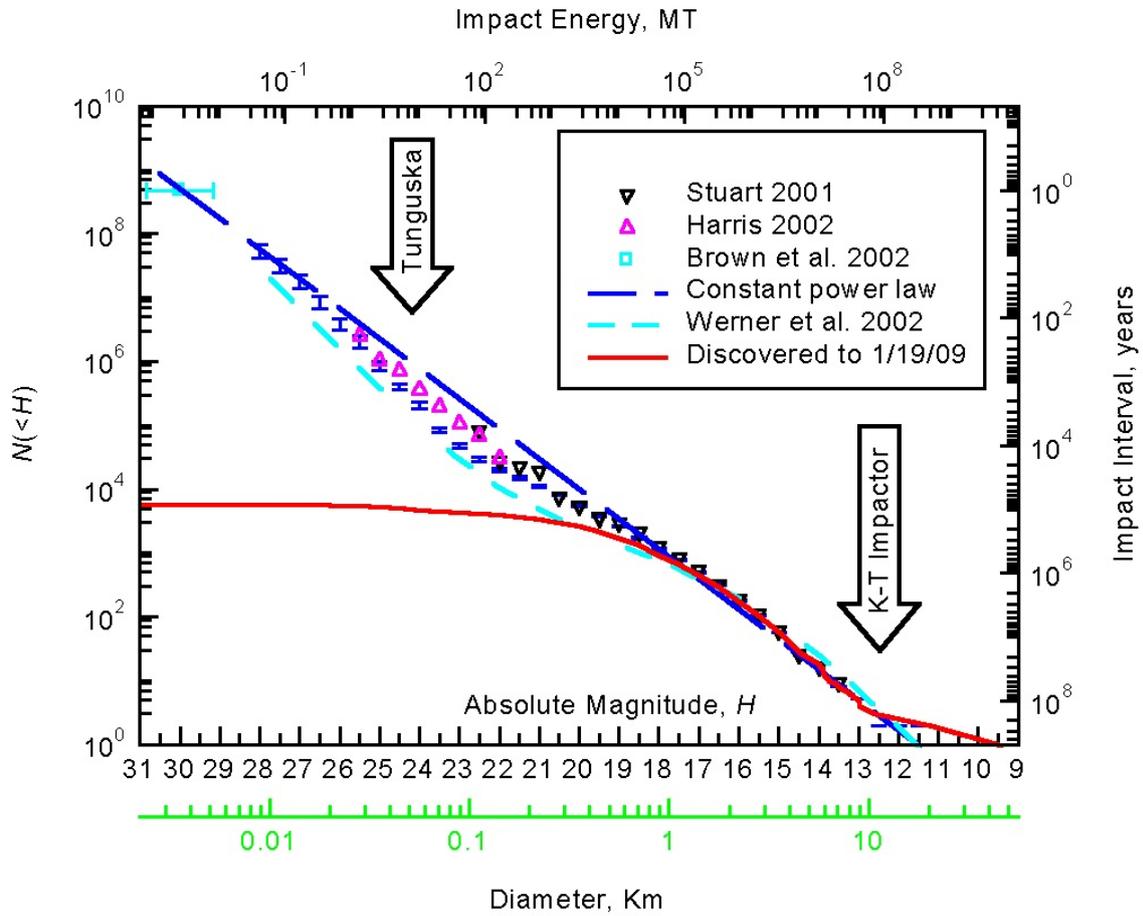}
\caption{Various estimates of the size vs. impact frequency
of NEAs (dashed lines). Equivalent astronomical absolute magnitude and impact
in megatons are shown. The potential damage from a cosmic impact can be divided roughly 
into four categories.  Below a diameter of $\sim30$~m, incoming bodies explode high enough in the atmosphere
that no ground damage occurs in the form of a blast wave.  In the next size
range extending up to 100-150 m or so, most of the impact
energy is released in the atmosphere resulting in ground damage more
or less similar to a large nuclear blast. Over land this has the
potential to create major devastation as can be seen by the scar of
the Tunguska event of a century ago. Even larger events
in which the incoming body would reach the ground still traveling at
cosmic velocity would cause even greater damage over land,
but it is expected that the larger risk in this size range is from
tsunami from impacts occurring into the ocean.  At some size,
variously estimated between 1 and 2 km diameter, it is expected that
the impact event would lead to a global climatic catastrophe (for either land or sea impact) 
due to dust lofted into the stratosphere, with the possibility of ending
civilization, perhaps killing a quarter or more of the human
population from famine, disease, and general failure of social order. An example of
this mass-extinction level event is the K-T Impactor. (Alan W. Harris,
modified from \url{http://neo.jpl.nasa.gov/neo/report2007.html}). 
\label{Fig:ss:popcomb3}}
\end{center}
\end{figure}

However, smaller PHAs certainly could be on impact trajectories. This
was recently brought home by the asteroid 2008~TC$_3$, detected less than
24 hours before it entered the Earth's atmosphere, ultimately
impacting in a remote part of Sudan \citep{jenniskens2009, mcgaha2008,
chesley08_tc3}. While 2008~TC$_3$ was a small PHA and impacts of this
size are actually fairly common, it does illustrate that the possibility 
exists for larger PHAs to hit the Earth. By cataloging all PHAs above 140~m in diameter, 
the congressional mandate is intended to increase our awareness of potential risk in
terms of death and property damage by approximately an order of magnitude 
beyond that which had been posed by 1~km objects. \autoref{Fig:ss:popcomb3} and its caption 
describes more of the hazards posed by various sizes of PHAs. 

Technology has improved beyond that available when the 2005
Congressional mandate was issued, although the funding available to
fulfill this mandate has not materialized. A 140~m PHA has an absolute magnitude of
approximately $H=22$. Integrating models of the orbital distribution
of PHAs to determine their positions and distances indicate that 10\%
of PHAs larger than 140~m never become brighter than $V=23.5$ over a
10 year period.  In addition, PHAs can move up to a few degrees per day,
thus requiring detection during short exposure times. This short exposure time,
coupled with this required limiting magnitude and the necessary sky coverage, 
requires a system with a large field of view and sensitive detection limit. 
LSST has the potential to reach the goal of detecting 90\% of all PHAs 
larger than 140~m by 2028, as described in \autoref{ss:NEOc}.

\subsection{The NEA Completeness Analysis} 
\label{ss:NEOc}
{\it \v Zeljko Ivezi\a'c}

To assess the LSST completeness for PHAs, 
the PHA 
population is represented by a size-limited complete sample of 800 true
PHAs whose orbital elements are taken from the Minor Planet Center.
The simulated baseline survey is used to determine which PHAs are present in
each exposure and at what signal-to-noise ratio they were observed. In 
addition to seeing, atmospheric transparency, and sky background
effects, the signal-to-noise computation takes into account losses 
due to non-optimal filters and object trailing. Using SDSS observations
of asteroids \citep{ivezic2001_sdssasteroids}, we adopt the following mean colors to 
transform limiting (AB) magnitudes in LSST bandpasses
to an `effective' limiting magnitude in the standard $V$ band: 
$V-m=(-2.1, -0.5, 0.2, 0.4, 0.6, 0.6)$ for $m=(u,g,r,i,z,y)$. Due to 
very red $V-u$ colors, and the relatively bright limiting magnitude in the $y$ 
band, the smallest objects are preferentially detected in the $griz$ bands.
The correction for trailing is implemented by subtracting from the 
$5\,\sigma$ limiting magnitude for point sources
\begin{equation}
 \Delta m_5^{\rm trailing} = 1.25\,\log_{10}\left(1+0.0267{v \,t_{vis}
 \over \theta}\right), 
\end{equation}
where the object's velocity, $v$, is expressed in deg/day. For the nominal
exposure time ($t_{vis}$) of 30 seconds and seeing $\theta=0.7''$, the loss of limiting 
magnitude is 0.16 mag for $v=0.25$ deg day$^{-1}$, typical for objects in the main 
asteroid belt, and 0.50 mag for $v=1.0$ deg day$^{-1}$, typical of NEAs passing
near Earth.

\begin{figure}
\begin{center}
\includegraphics[width=3.75in]{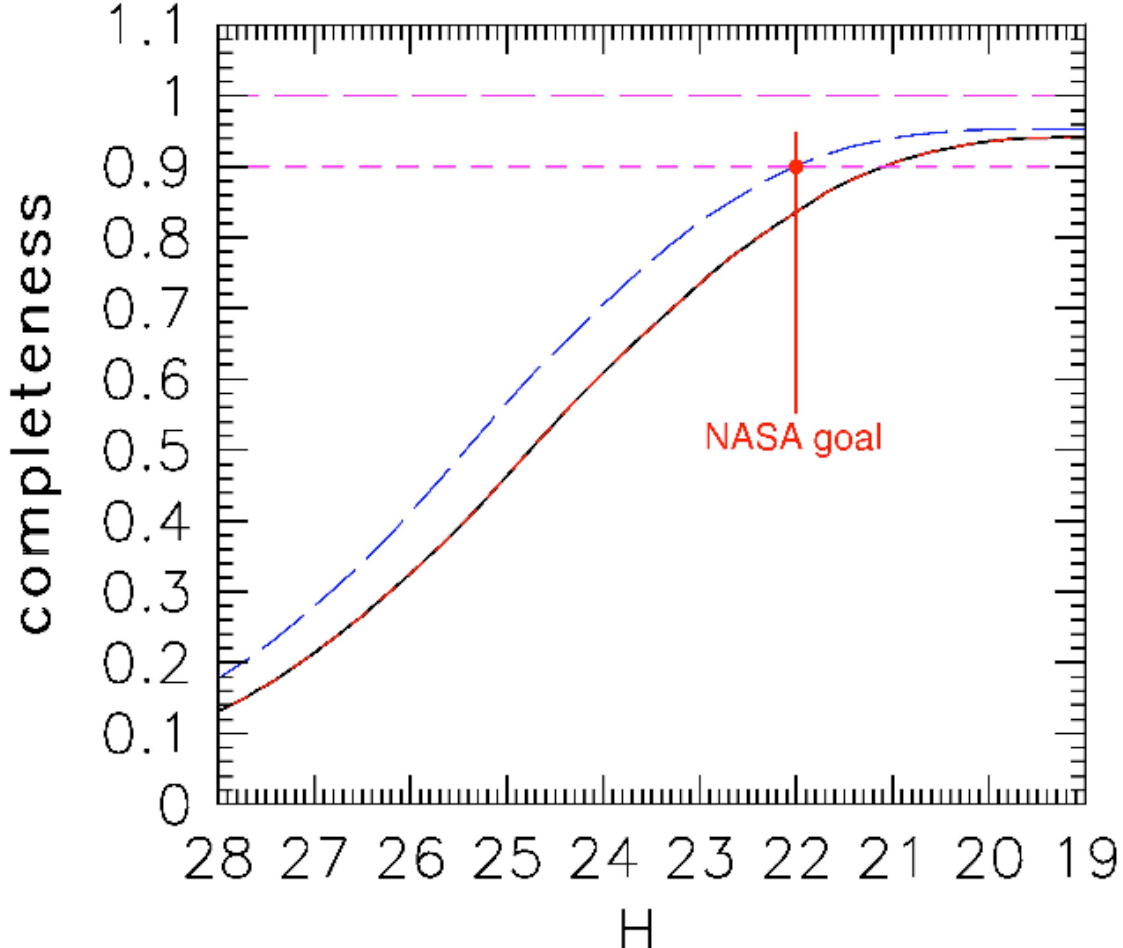}
\caption{Completeness of the LSST survey for PHAs brighter than a given absolute
magnitude (related to the size of the object and albedo; 
$H$=22 mag is equivalent to a typical 140 m asteroid and $H$=24 mag is
equivalent to a 50 m asteroid). Two scenarios are shown: the lower curve is the 
10-year long baseline survey where 5\% of the total observing time is spent 
on NEA-optimized observations in the Northern Ecliptic (NE) region, and it reaches a 
completeness of 84\% after 10 years. The upper dashed curve results from spending 15\% of the 
observing time in an NEA-optimized mode, and running the survey for 12 years.  
It meets the 90\% completeness level for 140 m objects mandated by the U.S. Congress.} 
\label{Fig:ss:Cneo}
\end{center}
\end{figure}

The completeness of LSST in cataloging NEAs was calculated
by propagating a model NEA source population (taken from the MOPS
Solar System model, as in \autoref{ss:mops}), over the lifetime of the 
LSST survey mission, and simply counting the number of times LSST would
be expected to detect the object under a variety of methods of operation (more on these 
observing cadences below).  An object's orbit is considered to be 
cataloged if the object was detected on at least three nights during a single 
lunation, with a minimum of two visits per night. The same criterion
was used in NASA studies\footnote{The NASA 2007 NEA study is available from
\url{http://neo.jpl.nasa.gov/neo/report2007.html}.}, and is confirmed as
reliable by a detailed analysis of orbital linking and determination using
the MOPS code (\autoref{ss:mops}). The MOPS software
system and its algorithms are significantly more advanced than anything
fielded for this purpose to date. Realistic MOPS simulations show 
$>99$\% linking efficiency across all classes of Solar System objects. 

For the LSST baseline cadence (\autoref{sec:design:cadence}), objects
counted as cataloged are observed on 20 different nights on average over ten
years. A more stringent requirement that an object must be detected on at least 
five nights decreases the completeness by typically 3\%. The completeness is also a
function of the assumed size distribution of NEAs: the flatter the distribution, the 
higher the completeness. If the latest results for the NEA size
distribution by Alan W. Harris (personal communication) are taken into account, the
completeness increases by 1-2\%. Due to these issues, the completeness
estimates have a systematic uncertainty of at least 2\%. 
Once the completeness rises above 60\%, an increase in 10\% in completeness
corresponds to roughly a decrease of one magnitude in $H$.

The LSST baseline cadence provides orbits for 82\% of PHAs larger than 140
m after 10 years of operations. With a minor change of this cadence, such as
requiring that all observations in the so-called North Ecliptic (NE) region, 
defined by $\delta >5^\circ$) are obtained in the $r$ band, the completeness 
for 140 m and larger PHAs is 84\%, with 90\% completeness reached for 200 m and 
larger objects. The completeness curve as a function of an object's size is shown 
in \autoref{Fig:ss:Cneo} (lower curve). The observing cadence described here spends only 5\% of the
total observing time on NEA-optimized observations in the NE region.

Various adjustments to the baseline cadence can boost the completeness for
140 m and larger PHAs to 90\%. We find that such variations can have an unacceptably 
large impact on other science programs, if the 90\% completeness is to be reached 
within the 10 year survey lifetime. However, with a minor adjustment of the 
baseline cadence, such that 15\% of the time is spent in the NE region to reach
fainter limiting magnitudes, this completeness level can be reached with a 12-year 
long survey, and with a negligible effect on other science goals. The completeness 
curve as a function of an object's size for such a modified cadence is shown in 
\autoref{Fig:ss:Cneo} (upper curve).

Our analysis assumes that no NEAs are known prior to LSST. Currently known
NEAs do not have a significant impact on this calculation. However, if a precursor 
survey, such as Pan-STARRS 4, operated for three years prior to LSST, the time to
fulfill the Congressional mandate by LSST could be shortened by about a year.


\section{NEAs as Possible Spacecraft Mission Targets}
{\it Paul A. Abell}
\label{ss:spacecraft}
%
%

LSST has the capability of detecting and characterizing more than
90\% of the NEAs equal to, or larger than 140 m in diameter in just 12
years of operation.  This is not only important for characterizing the
potential impact threat from these objects, but these observations
will also provide a wealth of information on possible spacecraft
targets for future investigation.  NEAs are objects of interest from a
hazard perspective given that their orbits can bring them into close
proximity with the Earth.  However, this makes them prime candidates
for in situ investigation given that they are also some of the easiest
objects to reach in the Solar System.  These objects have relatively
low velocities relative to Earth (5 to 7 \kms) and are good targets
for possible future science and sample return missions.  NASA's NEAR
Shoemaker spacecraft to (433) Eros, JAXA's Hayabusa probe to (25143)
Itokawa, and ESA's Rosetta mission to comet 67P/Churyumov-Gerasimenko
are examples of the types of missions that can be sent to NEAs.  Given
that a subset of the total NEA population has orbital parameters
similar to that of the Earth (i.e., low inclination and low
eccentricity), new discoveries made by LSST will expand the
currently known target list for future robotic and human-led
spacecraft missions.   

NASA's Constellation Program is developing the next generation of
vehicles for human exploration, as mandated by the United States Space
Exploration Policy. These vehicles are currently under development for
missions to the International Space Station (ISS) and the Moon.
However, these missions are not the only ones currently under
consideration at NASA.  Crewed voyages to NEAs are also being analyzed
as possible alternative missions for NASA.  The 2009 Augustine
Committee review of U.S. human spaceflight plans has included NEAs as
high-profile astronaut destinations in several of its exploration
options.  In addition, an agency-sponsored internal study has
determined that the new Constellation vehicles have the capability to
reach several NEAs, conduct detailed scientific and exploration
operations of these objects, and return to Earth after 180 days.
Using the existing NEA database, currently only about ten known targets are reachable using NASA's Constellation
systems within the desired 2020 to 2035
time-frame.  
New data from LSST would expand this list of dynamically viable targets by more
than an order of magnitude and help to refine target selection based
on the observed physical characteristics (taxonomy, rotation state,
etc.) of the objects discovered.  LSST is uniquely qualified for
this type of effort given its sensitivity for detecting and
characterizing NEAs.

The next stages in the human exploration and exploitation of space
will be highly dependent on the feasibility of extracting materials
(primarily water and minerals) from in situ sources.  In addition, to
their accessibility from Earth, NEAs are potentially the most
cost-efficient sources for providing propulsion and life support, and
for building structures in space.   It is highly probable that the
success and viability of human expansion into space beyond low-Earth
orbit depends on the ability to exploit these potential resources.
Therefore, a detailed physical and compositional assessment of the NEA
population will be required before any human missions are sent to
these objects. LSST will be a key asset in NEA discovery and play
a significant role in the initial reconnaissance of potential NEA
resources necessary for future human exploration of the Solar System. 


\bibliographystyle{SciBook}
\bibliography{solarsystem/solarsystem}

%
%
%
%
%
%
%
%
%
%
%
%
%
%
%
%
%
%
%
%
%
%
%
%
%
%
\chapter[Stellar Populations]
{Stellar Populations in the Milky Way and Nearby Galaxies}
\label{chp:stellarpops}

\noindent{\it  Abhijit Saha, Kevin R. Covey, Timothy C. Beers, John J. Bochanski, Pat Boeshaar, Adam J. Burgasser,  Phillip A. Cargile, You-Hua Chu, Charles F. Claver, Kem H. Cook, Saurav Dhital, Laurent Eyer, Suzanne L. Hawley, Leslie Hebb, Eric J. Hilton, 
  J. B. Holberg, \v{Z}eljko Ivezi\'{c}, Mario Juri\'c, Jason Kalirai,
  S\'ebastien L\'epine,  
  Lucas M. Macri, Peregrine M. McGehee, David Monet, Knut Olsen, Edward W. Olszewski, 
  Joshua Pepper, Andrej Pr\v{s}a, Ata Sarajedini, Sarah Schmidt, Keivan G. Stassun, Paul
  Thorman, Andrew A. West, Benjamin F. Williams}  

\section{Introduction}

Stellar populations, consisting of individual stars 
that share coherent spatial, kinematic, chemical, or age 
distributions, are powerful probes of a wide range of 
astrophysical phenomena.  The coherent properties of
stellar populations allow us to use measurements
of an individual member to inform our understanding of the larger system and vice versa.  As examples, 
globular cluster metallicities are often derived from
measurements of the brightest few members, while the
overall shape of the cluster color magnitude diagram (CMD) enables us to assign
ages to an individual star within the system.  Leveraging
the wealth of information available from such analyses
enables us to develop a remarkably detailed and nuanced
understanding of these complex stellar systems.  

By providing deep, homogeneous photometry for billions of
stars in our own Galaxy and throughout the Local Group, LSST 
will produce major advances in our understanding of
stellar populations.  In the sections that follow, we
describe how LSST will improve our understanding
of stellar populations in external galaxies
(\autoref{sec:sp:magellanic} and \autoref{sec:sp:sfh}) and in our own Milky Way (\S\S~\ref{sec:sp:distances}--\ref{sec:sp:ages}),
and will allow us to study the properties of rare stellar systems
(\S\S~\ref{sec:sp:metalpoor}--\ref{sec:sp:wds}). 

Many of the science cases in this chapter are based on the rich
characterization LSST will provide for stars in the solar
neighborhood.  This scientific landscape will be irrevocably altered
by the Gaia space mission, however, which will provide an exquisitely
detailed catalog of millions of solar neighborhood stars shortly after
its launch (expected in 2011).  To illuminate the scientific areas
where LSST provides a strong complement to Gaia's superb capabilities,
\autoref{sec:Gaia} develops a quantitative comparison of the astrometric and
photometric precision of the two missions; this comparison highlights
LSSTs ability to smoothly extend Gaia's solar neighborhood catalog to
redder targets and fainter magnitudes.


\section{The Magellanic Clouds and their Environs}
\label{sec:sp:magellanic}
\noindent {\it Abhijit Saha, Edward W. Olszewski, Knut Olsen, Kem H. Cook}

The Large and Small Magellanic Clouds (LMC and SMC respectively;
collectively referred to hereafter as ``the Clouds'') are laboratories
for studying a large assortment of topics, ranging from stellar
astrophysics to cosmology. Their proximity allows the study of their
individual constituent stars: LSST will permit broad band photometric
``static" analysis to $M_{V} \sim +8$ mag, probing well into the M
dwarfs; and variable phenomena to $M_{V} \sim +6$, which will track
 main sequence stars 2 magnitudes fainter than the turn-offs for the 
oldest known systems. 

A sense of scale on the sky is given by the estimate of ``tidal
debris" extending to 14 kpc from the LMC center \citep{Weinberg2001}
based on 2MASS survey data. Newer empirical data (discussed below) confirm such spatial scales.  The stellar bridge between the
LMC and SMC is also well established. 
Studies of the full spatial extent of the clouds thus require a
wide area investigation of the order  of 1000 deg$^2$.  We 
show below that several science applications call for reaching
``static" magnitudes to $V ~ {\rm or} ~  g \approx 27$ mag, and time domain
data and proper motions reaching 24 mag or fainter.  The relevance of
LSST for these investigations is unquestionable.

Nominal LSST exposures will saturate on stars about a magnitude brighter 
than the horizontal branch luminosities of these objects, and work on such 
stars is not considered to be an LSST forte. Variability 
surveys like MACHO and SuperMACHO have already covered much ground on 
time-domain studies, with SkyMapper to come between now and LSST. We do not 
consider topics dealing with stars bright enough to saturate in nominal 
LSST exposures.

\subsection{ Stellar Astrophysics in the Magellanic Clouds }
\label{MCastroph}

For stellar astrophysics studies, the Clouds present a sample of stars
that are, to first order, at a common distance, but contain the
complexity of differing ages and metallicities, and hence an
assortment of objects that star clusters within the Galaxy do not
have. In addition, the age-metallicity correlation in the Clouds is
known to be markedly different from that in the Galaxy. This allows
some of the degeneracies in stellar parameters that are present in
Galactic stellar samples to be broken. 

LSST's  extended time sampling will reveal, among other things, eclipsing binaries on the main
sequence through the turn-off.  We plan to use them to calibrate the
masses of stars near the old main sequence turnoff.  Only a small
number of such objects are known in our own Galaxy, but wide area coverage of the
Clouds (and their extended structures) promises a sample of $ \sim
80,000 $ such objects with $22 <g < 23$ mag, based on projections
from MACHO and SuperMACHO.  Binaries in this brightness range track
evolutionary phases from the main sequence through turn-off. The direct determination
of stellar masses (using follow-up spectroscopy of eclipsing binaries
identified by LSST) of a select sub-sample of eclipsing binaries in
this range of evolutionary phase will confront stellar evolution
models, and especially examine and refine the stellar age ``clock,"
which has cosmological implications.

For this question, we need to determine the number of eclipsing
binaries (EB) that LSST can detect within 0.5 magnitudes of the old turn-off
($r \sim 22.5$).  Additionally, because the binary mass depends on
$(\sin i)^3$, we need to restrict the EB sample to those
with $ i = 90^\circ \pm 10^\circ $ in order to determine masses to 5\%.
Such accuracy in mass is required for age sensitivity of $\sim 2$ Gyr at
ages of 10-12 Gyr.  We determined the number of LMC EBs
meeting these restrictions that could be discovered by LSST by
projecting from the 4631 eclipsing binaries discovered by the MACHO
project (e.g. \citealt{Alcocketal95}). Selecting only those MACHO
EBs with colors placing them on the LMC main sequence, we calculated the
minimum periods which these binaries would need to have in order for
them to have inclinations constrained to be $90^\circ \pm 10^\circ$, given
their masses and radii on the main sequence. Of the MACHO EB sample,
551 systems (12\%) had periods longer than the minimum, with the
majority of the EBs being short-period binaries with possibly large
inclinations.  Next, we constructed a deep empirical LMC stellar
luminosity function (LF) by combining the V-band luminosity function
from the Magellanic Clouds Photometric Survey \citep[][MCPS]{Zaritsky++04}
with the HST-based LMC LF measured by \citet{Dolphin02}, using \citet{Smecker++02} observations of the LMC bar, where we scaled the HST LF to match the MCPS LF over the
magnitude range where the LFs overlapped.  We then compared this
combined LF to the LF of the MACHO EBs, finding that EBs comprise
$\sim 2\%$
of LMC stellar sources.  Finally, we used our deep combined LF to
measure the number of LMC stars with $22 < V < 23$, multiplied this
number by 0.0024 to account for the fraction of sufficiently
long-period EBs, and found that LSST should be able to detect $\sim 80,000$
EBs near the old main sequence turnoff.  Based on the MACHO sample,
these EBs will have periods between $\sim 3$ and $\sim 90$ days, with an average of $\sim 8$ days.

LSST is expected to find $\sim 10^{5}$ RR~Lyrae stars over the
full face of both Clouds (the specific ratio of RR Lyraes can vary by a
factor of 100, and the above estimate, which is based on 1 RR~Lyrae per $\sim
10^{4} L_{\odot}$, represents the geometric mean of that range and
holds for HST discoveries of RR~Lyrae stars in M31 and for
SuperMACHO results in the Clouds).  The physics behind the
range of subtler properties of RR Lyrae stars is still being
pondered: trends in their period distributions as well as possible
variations in absolute magnitude with period, age, and metallicity.
Our empirical knowledge of these comes from studying their properties
in globular clusters, where the distance determinations may not
be precise enough (at the 20\% level).  The range of distances within
the LMC is smaller than the uncertainty in relative distances between
globular clusters in our Galaxy.  Ages and metallicities of the
oldest stars (the parent population of the RR Lyraes) in any given
location in the Clouds may be gleaned from an analysis of the local
color-magnitude diagrams, as we now discuss, and trends in RR Lyrae
properties with parent population will be directly mapped for the
first time.

\subsection{ The Magellanic Clouds as ``Two-off'' Case Studies of Galaxy Evolution}
\label{MCevol}

The Clouds are the only systems larger and more complex than dwarf
spheroidals outside our own Galaxy where we can reach the main sequence stars with LSST. Not only are these the most numerous, and therefore the most
sensitive tracers of structure, but they {\it proportionally represent
stars of all ages and metallicities}.  Analyzing the ages,
metallicities, and motions of these stars is the most effective and
least biased way of parsing the stellar sub-systems within any
galaxy, and the route to understanding the history of star formation,
accretion, and chemical evolution of the galaxy as a whole. Decades of
work toward this end have been carried out to define these elements within
our Galaxy, but the continuing task is made difficult not only because
of the vastness on the sky, but also because determining distances to
individual stars is not straightforward. The Clouds are the {\it only}
sufficiently complex systems (for the purpose of understanding galaxy
assembly) where the spatial perspective allows us to know where in the
galaxy the stars we are examining lie, while at the same time being
close enough for us to examine and parse its component stellar
populations in an unbiased way through the main sequence stars.  LSST will
provide proper motions of individual stars to an accuracy of $ \sim
50  {\rm km\,s}^{-1} $ in the LMC, but local ensembles of thousands of
stars on spatial scales from 0.1 to several degrees will be able to
separate disk rotation from a ``stationary'' halo.  Internal motions
have been seen using proper motions measured with only 20 positional
pointings with the HST's Advanced Camera for Surveys (ACS) with only a few arc-min field of view and a 2-3 year time baseline \citep{Piatek2008}.

Color-magnitude and Hess diagrams from a composite stellar census
can be decomposed effectively using stellar evolution models
(e.g., \citealt{Tolstoy1996,Dolphin2002}). While the halo of our Galaxy bears
its oldest {\it known} stars, models of galaxy formation lead us to
expect the oldest stars to live in the central halo and bulge.  Age
dating the oldest stars toward the center of the Galaxy is thwarted
by distance uncertainties, complicated further by reddening and
extinction. The Clouds present objects at a known distance, where
color-magnitude diagrams are a sensitive tool for evaluating ages.  A
panoramic unbiased age distribution map from the CMD turn-off is not
possible at distances larger than 100 kpc.  The Clouds are a gift in
this regard.

\subsection{The Extended Structure of the Magellanic Clouds}
\label{MCstruct}

Knowledge of the distribution and population characteristics in
outlying regions of the LMC/SMC complex is essential for understanding
the early history of these objects and their place in the $\Lambda$CDM
hierarchy.  In our Galaxy the most metal poor, and (plausibly) the
oldest observed  stars are distributed in a halo that extends beyond 25 kpc.
Their spatial distribution, chemical composition and kinematics
provide clues about the Milky Way's early history, as well as its
continued interaction with neighboring galaxies.  If the Clouds also
have similar halos, the history of their formation and interactions
must also be written in their stars. In general how old are the
stars in the extremities of the Clouds? How are they distributed (disk
or halo dominated)? How far do such stellar distributions extend? What
tidal structure is revealed? Is there a continuity in the stellar
distribution between the LMC and SMC?  Do they share a common halo
with the Galaxy?  What do the kinematics of stars in outlying regions
tell us about the dark matter distribution? Is there a smooth change from 
disk to non-disk near the extremities?

\begin{figure}
\centering\includegraphics[width=0.6\linewidth]{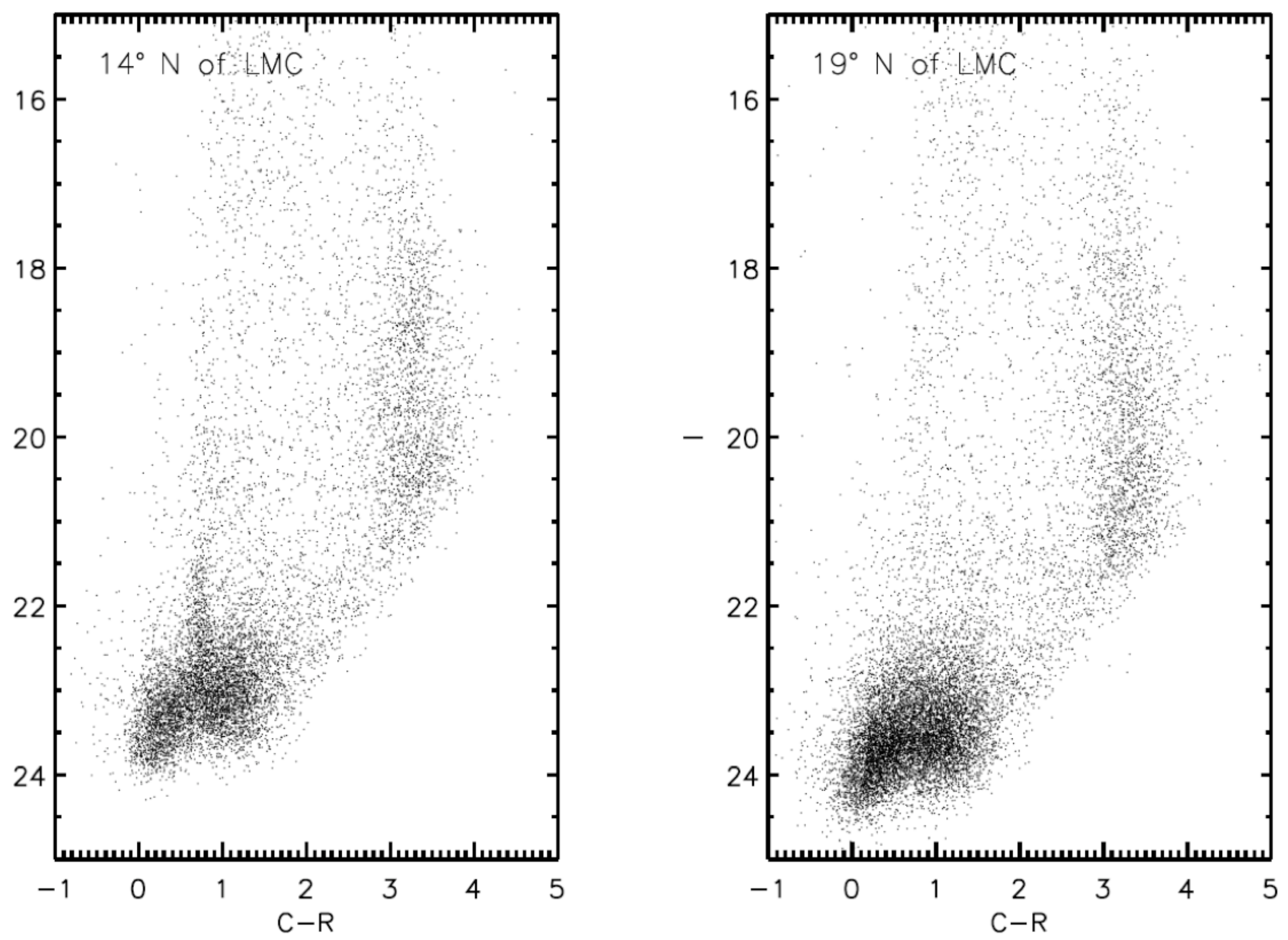}
\caption{The color-magnitude diagrams in $C-R$ vs.~$R$ for two fields, $14^{\circ}$ (left) and  
$19^{\circ}$ (right) due north of the LMC center.
The stub-like locus of stars near $C-R \sim 0.7$ and $I > 21.0$ that 
can be seen on the left panel for the $14^{\circ}$
field corresponds to the locus of old main sequence stars from the LMC, 
which have a turn-off at $I \sim 21.0$. 
This shows that stars associated with LMC extend past 10 disk scale lengths. 
The feature is absent in the $19^{\circ}$ field,
which is farther out. 
Mapping the full extent of the region surrounding the Clouds 
on these angular scales is only feasible with LSST.
}
\label{fig:sp:lmcn14}
\end{figure}

Past panoramic studies such as with 2MASS and DENIS have taught us
about the LMC disk {\it interior} to 9 kpc (10$^\circ$,
e.g., \citealt{vanderMarel2001}). Structure beyond that had not been 
systematically probed in an {\it unbiased} way (studies using HII
regions, carbon stars, and even RR Lyrae exist, but they are heavily
biased in age and metallicity) until a recent pilot study (NOAO
Magellanic Outer Limits Survey) with the MOSAIC
imager on the Blanco 4-m telescope at Cerro Tololo Inter-American Observatory, which uses main sequence stars as
tracers of structure. Even with their very selective spatial sampling
of a total of only $\sim 15$ deg$^2$ spread out over a region
of interest covering over $\sim 1000$ deg$^2$, the LMC disk is
seen to continue out to 10 disk scale lengths, beyond which there are
signs either of a spheroidal halo that finally overtakes the disk (a
simple scaled model of how our own Galaxy must look when viewed
face on), or a tidal pile-up. Main sequence stars clearly associated with the
LMC are seen out to 15 degrees along the plane of the disk
(\autoref{fig:sp:lmcn14}). This exceeds the tidal radius estimate of
11 kpc (12.6$^{\circ}$) by \citet{Weinberg2000}, already a
challenge to existing models of how the LMC has interacted with the
Galaxy.  (This extended LMC structure has a surface brightness density
of $\sim 35$ mag per square arc-sec, which underscores the
importance of the Clouds and the opportunity they present, because
this technique will not work for objects beyond 100 kpc from us.)  In
contrast, the structure of the SMC appears to be very truncated, at
least as projected on the sky.  Age and metallicity of these tracer
stars are also derived in straightforward manner.

Not only will LSST map the complete extended stellar distribution
(where currently less than 1\% of the sky region of interest has been
mapped) of the Clouds using main sequence stars as tracers that are unbiased in
age and metallicity, but it will also furnish proper motions. The
accuracy of ensemble average values for mapping streaming motions, such
as disk rotation and tidal streams, depends eventually on the
availability of background quasars and galaxies, which do not move on the
sky.  The HST study of proper motions \citep{Kallivayalil2006a,
Kallivayalil2006b, Piatek2008} was able to use quasars with a surface density
of 0.7 deg$^{-2}$. We expect that LSST, using the hugely more
numerous  
background galaxies as the ``zero proper motion reference," and a longer time 
baseline should do even better.  Individual proper motions of stars 
at these distances can be measured to no better than $\sim 50~ \kms$, 
but the group motions of stars will be determined to much higher
accuracy, depending ultimately on the positional accuracy 
attainable with background galaxies. Over scales of $0.1^\circ$, statistical analysis of group motions can be expected to yield systemic 
motions with accuracies better than $10\,  \kms$. This would not only discriminate between
disk and halo components of the Clouds in their outer regions but
also identify any tidally induced structures at their extremities.

\subsection{ The Magellanic Clouds as Interacting Systems }

In addition to interacting gravitationally with each other, both
Clouds are in the gravitational proximity of the Galaxy. Until
recently, it was held that the Clouds are captive satellites of the
Galaxy and have made several passages through the Galaxy disk. The
extended stream of $ \rm { H I }$, called the Magellanic Stream, which
emanates from near the SMC and wraps around much of the sky, has been
believed to be either a tidal stream or stripped by ram pressure from
passages through the Galaxy disk.  This picture has been challenged
recently by new proper motion measurements in the Clouds from HST data
analyzed by two independent sets of investigators \citep{Kallivayalil2006a,
Kallivayalil2006b, Piatek2008}. Their results indicate significantly higher
proper motions for both systems, which in turn imply higher space
velocities. Specifically, the LMC and SMC may not have begun bound
to one another, and both may be on their first approach to the Milky
Way, not already bound to it.  Attempts to model the motion of the
Clouds together with a formation model for the Magellanic Stream in
light of the new data (e.g., \citealt{Besla2009}) are very much works in
progress.  Even if a higher mass sufficient to bind the Clouds is
assumed for the Galaxy, the orbits of the Clouds are changed radically
from prior models: specifically, the last peri-galacticon could not
have occurred within the last 5 Gyrs with the high eccentricity orbits
that are now necessary \citep{Besla2009}, indicating that the
Magellanic Stream cannot be tidal. The proper motion analyses have also
determined the rotation speed of the LMC disk \citep{Piatek2008}. The new result
of $120 \pm 15\,  {\rm kms}^{-1}$ is more reliable than older
radial velocity-based estimates for this nearly face-on galaxy, and as
much as 
twice as large as some of the older estimates.  This new scenario
changes the expected tidal structures for the Clouds and argues
against a tidal origin for the Magellanic Stream.  These expectations
are empirically testable with LSST. For instance, a tidal origin requires 
a corresponding stream of stars, even though the stellar stream can be 
spatially displaced with respect to the gas stream: to date, such a star 
stream, if its exists, has escaped detection.  A definitive conclusion about whether such a stellar stream exists or not, awaits a deep multi-band 
wide area search to detect and track main sequence stars, which are 
the most sensitive tracers of such a stellar stream.

Aside from the specific issue of a stellar stream corresponding to the Magellanic gas stream, the full area 
mapping of extremities via the main sequence stars described in
\autoref{MCstruct} will reveal any tidally induced asymmetries in
the stellar distributions, e.g., in the shape of the LMC disk as result
of the Galactic potential as well as from interaction with the SMC.
Proper motions of any tidal debris (see \autoref{MCstruct}) will
contribute to determining the gravitational field, and eventually to a
modeling of the halo mass of the Galaxy. How far out organized
structure in the LMC persists, using kinematic measures from
proper motions, will yield the mass of the LMC, and thus the size of
its dark matter halo. 

\subsection{Recent and On-going Star Formation in the LMC}

\noindent{\it You-Hua Chu}

Studies of recent star formation rate and history are complicated
by the mass dependence of the contraction timescale.  For example,
at $t \sim 10^5$ yr, even O stars are still enshrouded by 
circumstellar dust; at $t \sim 10^6$ yr, massive stars have formed
but intermediate-mass stars have not reached the main sequence;
at $t \sim 10^7$ yr, the massive stars have already exploded as 
supernovae, but the low-mass stars are still on their way to the
main sequence.

The current star formation rate in the LMC has been determined
by assuming a Salpeter initial mass function (IMF) and scaling 
it to provide the ionizing flux required by the observed H$\alpha$ 
luminosities of HII regions \citep{Kennicutt1986}.
Individual massive stars in OB associations and in the field
have been studied photometrically and spectroscopically to
determine the IMF, and it has been shown that the massive end of the IMF
is flatter in OB associations than in the field \citep{Massey1995}.

\begin{figure}
\centering\includegraphics[width=0.5\textwidth] {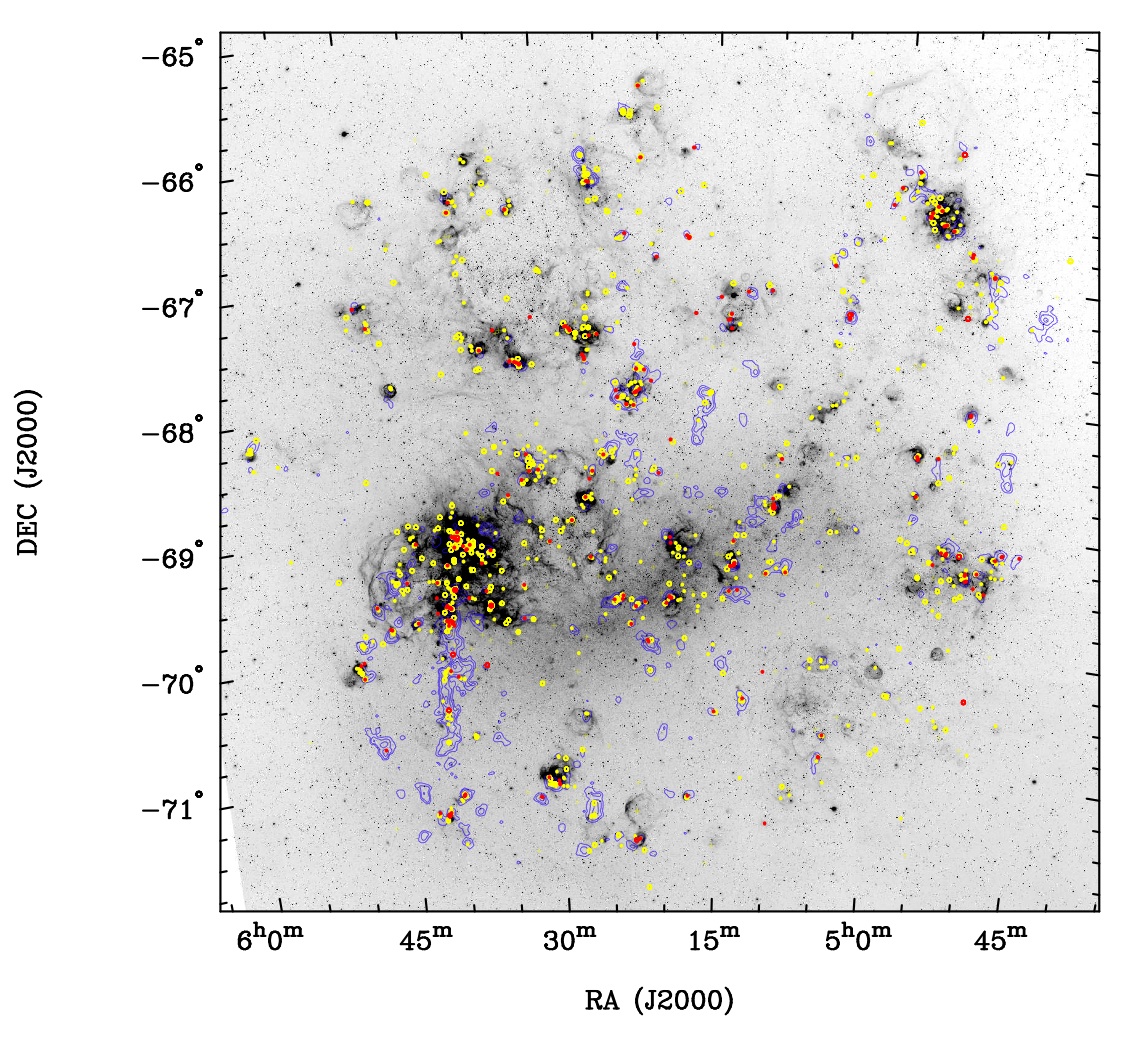}
\caption{H$\alpha$ image of the LMC.  CO contours extracted from
the NANTEN survey are plotted in blue to show the molecular clouds.
Young stellar objects with 8.0 $\mu$m magnitude brighter than 8.0 
are plotted in red, and the fainter ones in yellow.  Roughly, the
brighter objects are of high masses and the fainter ones of 
intermediate masses.  Adapted from \citet{Gruendl2009}.}
\label{sp:fig:lmc_sf}
\end{figure}

The Spitzer Space Telescope has allowed the identification
of high- and intermediate-mass young stellar objects (YSOs),
representing ongoing (within 10$^5$ yr) star formation,
in the LMC \citep{Caulet2008, Chen2009}.
Using the Spitzer Legacy program SAGE survey of the central
$7^\circ \times 7^\circ$ area of the LMC with both IRAC and MIPS,
YSOs with masses greater than $\sim 4\, M_\odot$ have been identified
independently by \citet{Whitney2008} and \citet{Gruendl2009}. 
\autoref{sp:fig:lmc_sf} shows the distribution of YSOs, HII regions, and molecular 
clouds, which represent sites of on-going, recent, and future star 
formation respectively.  It is now possible to fully specify the 
formation of massive stars in the LMC.

\begin{figure}
\begin{center}$
\begin{array}{ccc}
\includegraphics[width=0.3\textwidth]{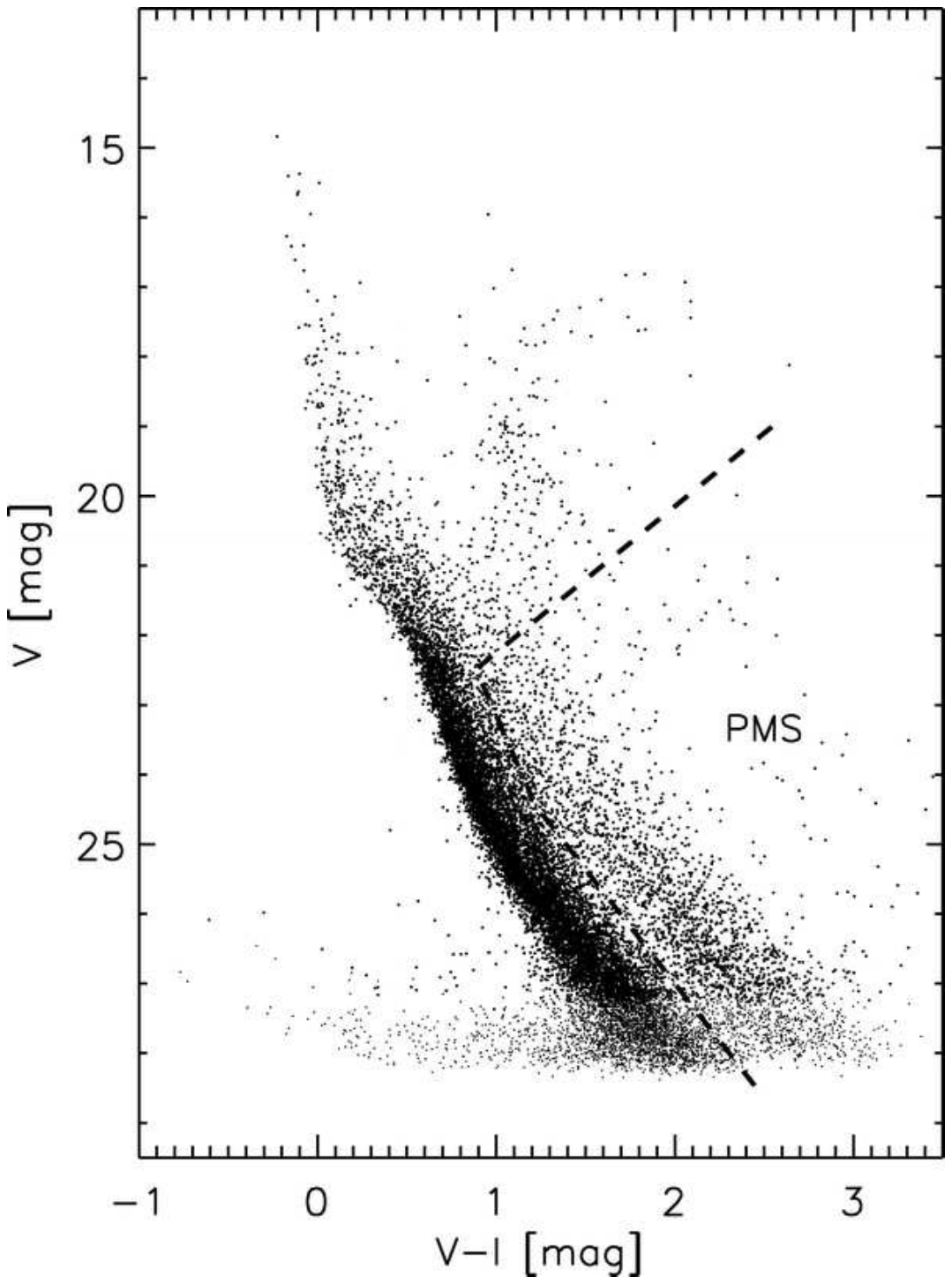}
\includegraphics[width=0.3\textwidth]{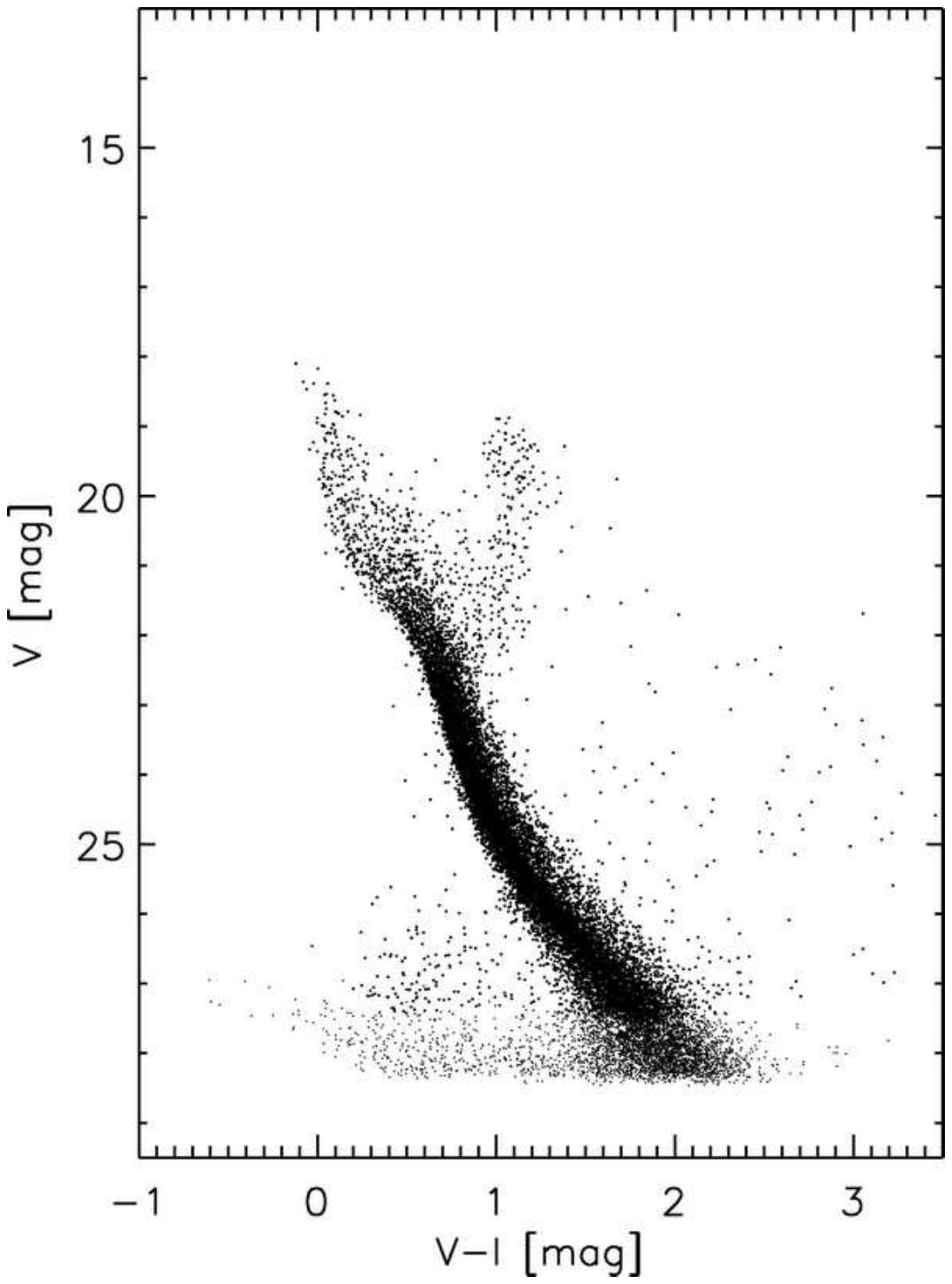}
\includegraphics[width=0.3\textwidth]{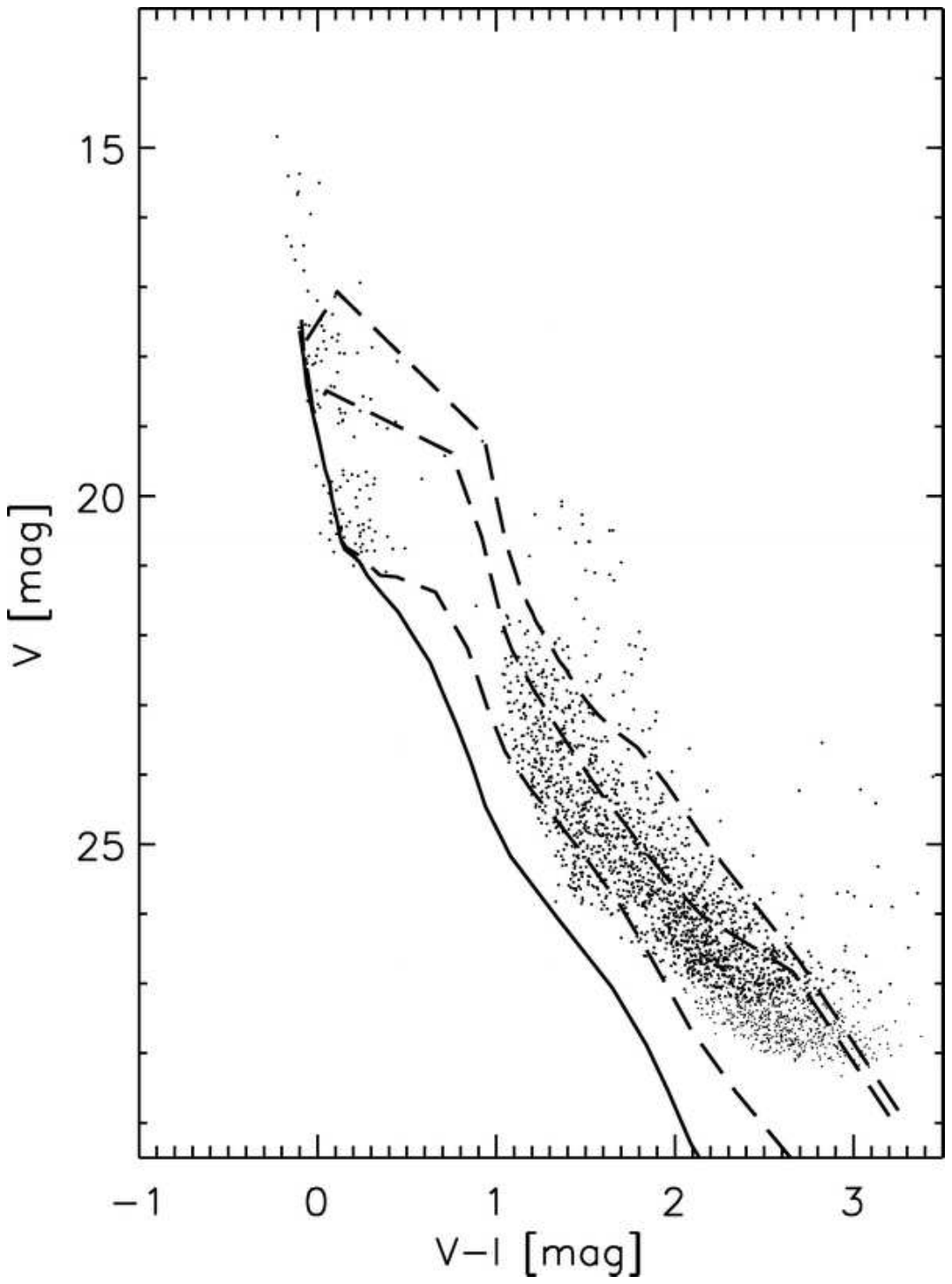}
\end{array}$
\end{center}
\caption{V-I vs. V color-magnitude diagram of stars detected in the
OB association LH95 (left), surrounding background region (middle),
and the difference between the two (right).  The zero-age main sequence
is plotted as a solid line, and PMS isochrones for ages 0.5, 1.5, and
10 Myr are plotted in dashed lines in the right panel. Adapted from
\citet{Gouliermis2007} with permission.}
\label{sp:fig:lmc_cmd}
\end{figure}

The formation of intermediate- and low-mass stars in the LMC
has begun to be studied only recently by identifying pre-main
sequence (PMS) stars in $(V-I)$ vs $V$ color-magnitude diagrams
(CMDs), as illustrated in \autoref{sp:fig:lmc_cmd}.  Using HST WFPC2 observations,
low-mass main sequence stars in two OB associations and in the field have 
been analyzed to construct IMFs, and different slopes are also 
seen \citep{Gouliermis2006a, Gouliermis2006b, Gouliermis2007}.

Using existing HST image data in LMC molecular clouds to estimate how
crowding will limit photometry from LSST images, we estimate that PMS
stars can be detected down to 0.7-0.8 $M_\odot$ ($g \sim 24$ mag: see
right hand panel of \autoref{sp:fig:lmc_cmd}).  LSST will provide a
mapping of intermediate- to low-mass PMS stars in the entire LMC except the bar, where crowding will prevent reliable photometry at
these magnitudes. This young lower-mass stellar population, combined
with the known information on massive star formation and
distribution/conditions of the interstellar medium (ISM), will allow
us to fully characterize the star formation process and provide
critical tests to different theories of star formation.

Conventionally, star formation is thought to start with the 
collapse of a molecular cloud that is gravitationally unstable.
Recent models of turbulent ISM predict that colliding HI clouds 
can also be compressed and cooled to form stars. Thus, both the neutral
atomic and molecular components of the ISM need to be considered 
in star formation. The neutral atomic and molecular gas in the 
LMC have been well surveyed: the ATCA+Parkes map of HI \citep{Kim2003}, the NANTEN survey of CO \citep{Fukui2008}, and the MAGNA 
survey of CO \citep[][Hughes et al., in prep.]{Ott2008}.
\autoref{sp:fig:lmc_sf} 
shows that not all molecular clouds are forming massive stars:
How about intermediate- and low-mass stars?  Do some molecular
clouds form only low-mass stars?  Do stars form in regions with
high HI column density but no molecular clouds?  These questions
cannot be answered until LSST has made a complete mapping of intermediate- and
low-mass PMS stars in the LMC.

\section{ Stars in Nearby Galaxies}
\label{sec:sp:sfh}
\noindent{\it Benjamin F. Williams, Knut Olsen, Abhijit Saha}   

\subsection{Star Formation Histories}
\subsubsection{General Concepts}

Bright individual stars can be distinguished in nearby galaxies with
ground-based observations.  In galaxies with no recent star formation
(within $\sim$1 Gyr), the brightest stars are those on the asymptotic
giant branch (AGB) and/or the red giant branch (RGB).  The color of
the RGB depends mostly on metallicity, and only weakly on age, whereas
the relative presence and luminosity distribution of the AGB stars is
sensitive to the star formation history in the range $2 < t < 8$ Gyrs.
The brightest RGB stars are at $I \sim -4$, and in principle are
thus visible to LSST out to distances $\sim 10$ Mpc. The presence of
significant numbers of RR Lyrae stars indicates an
ancient population of stars, 10 Gyrs or older. RR~Lyrae, as well the
brightest RGB stars, are standard candles that measure the distance of
the host galaxy.

In practice, object crowding at such distances is severe for galaxies
of any significant size, and resolution of individual RGB stars in
galaxies with $M_{V} \sim -10$ and higher will be limited to distances
of $\sim 4$ Mpc, but that includes the Sculptor Group and the
Centaurus and M83 groups. Within the Local Group, the ``stacked"
photometry of individual stars with LSST will reach below the
Horizontal Branch, and certainly allow the detection of RR Lyraes in
addition to the RGB.

Galaxies that have made stars within the last 1 Gyrs or so contain luminous supergiant stars (both blue and red). The luminosity
distributions of these stars reflect the history of star formation
within the last 1 Gyr. The brightest stars (in the youngest systems)
can reach $M_{V} \sim -8$, but even stars at $M_{V} \sim -6$
(including Cepheids) will stand out above the crowding in LSST images
of galaxies at distances of $\sim 7$ Mpc.

A great deal of work along these lines is already being done, both from 
space and the ground.  LSST's role here will be to (1)
cover extended structures, and compare, for example, how populations
change with location in the galaxies -- important clues to how
galaxies were formed, and (2)
identify the brighter variables, such as RR~Lyraes, Cepheids, and the brighter 
eclipsing binaries wherever they are reachable. 

\subsubsection{Methods and Techniques}

Methods of deriving star formation histories (the distribution
of star formation rate as a function of time and chemical composition)
from Hess diagrams given photometry and star counts in two or more bands
(and comparing with synthetic models) are adequately developed,
e.g., \citet{Dolphin2002}. For extragalactic systems and in the solar
neighborhood, where distances are known independently, the six-band LSST data
can be used to self-consistently solve for extinction and star formation history. This is
more complicated if distances are not known independently,
such as within the Galaxy, where other methods must be brought to bear.
For nearby galaxies, distances are known at least from the bright termination 
of the RGB. 

\begin{figure}
\centering\includegraphics[width=0.85\linewidth]{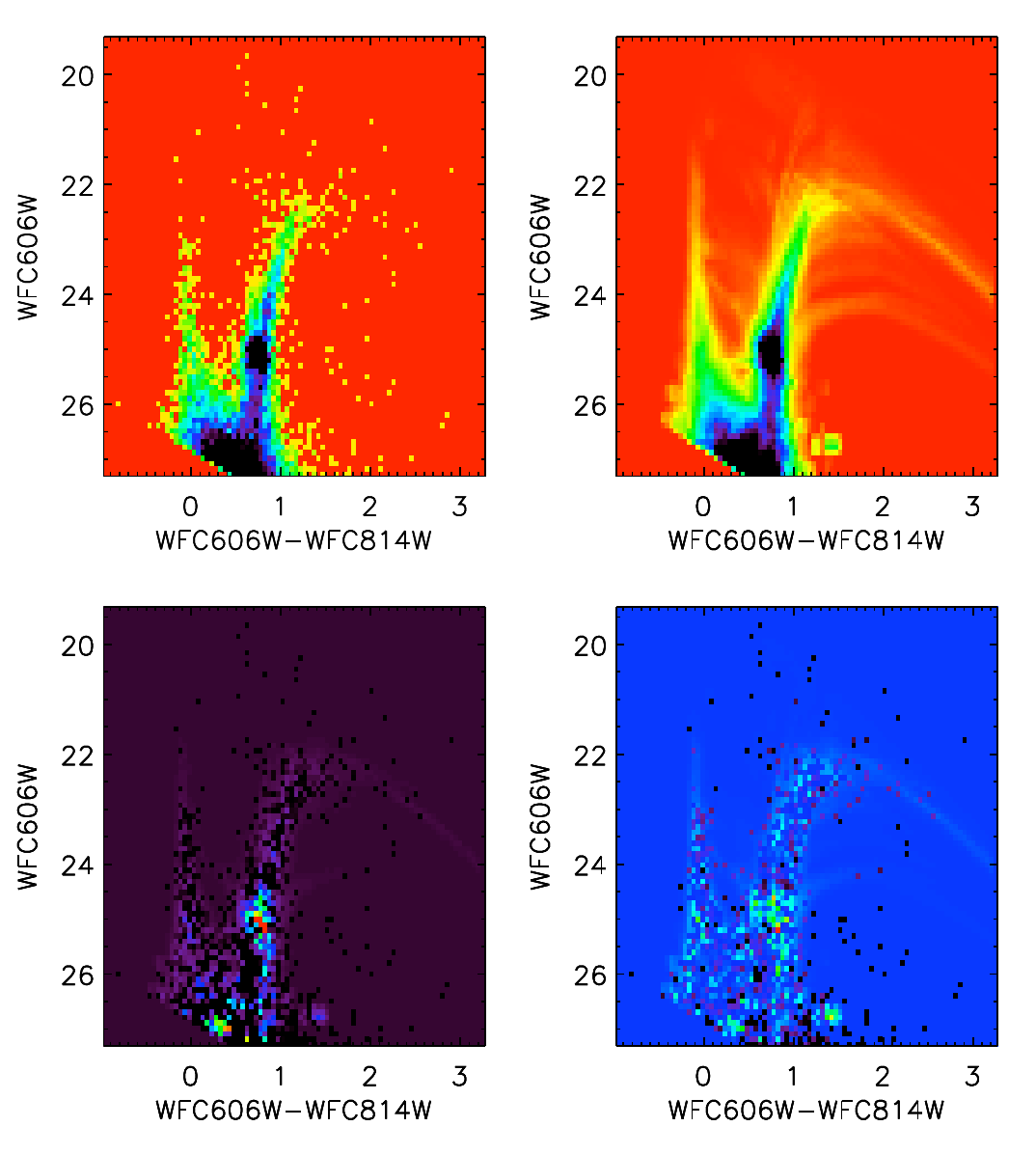}
\caption{Best fit to a CMD from an archival Hubble Space Telescope Advanced Camera for Surveys field in M33.
{\it Upper~left:} The observed CMD. {\it Upper~right:} The
best-fitting model CMD using stellar evolution models. {\it
  Lower left:} The residual CMD.  Redder 
colors denote an overproduction of model stars.  Bluer colors denote
an underproduction of model stars. {\it Lower~right:} The deviations
shown in {\it lower left} normalized by the Poisson error in each CMD
bin, i.e., the statistical significance of the residuals.  
Only the red clump shows statistically significant residuals.}
\label{sfhbw1}
\end{figure}

\begin{figure}[ht]
\begin{center}
\includegraphics[width=0.5\linewidth]{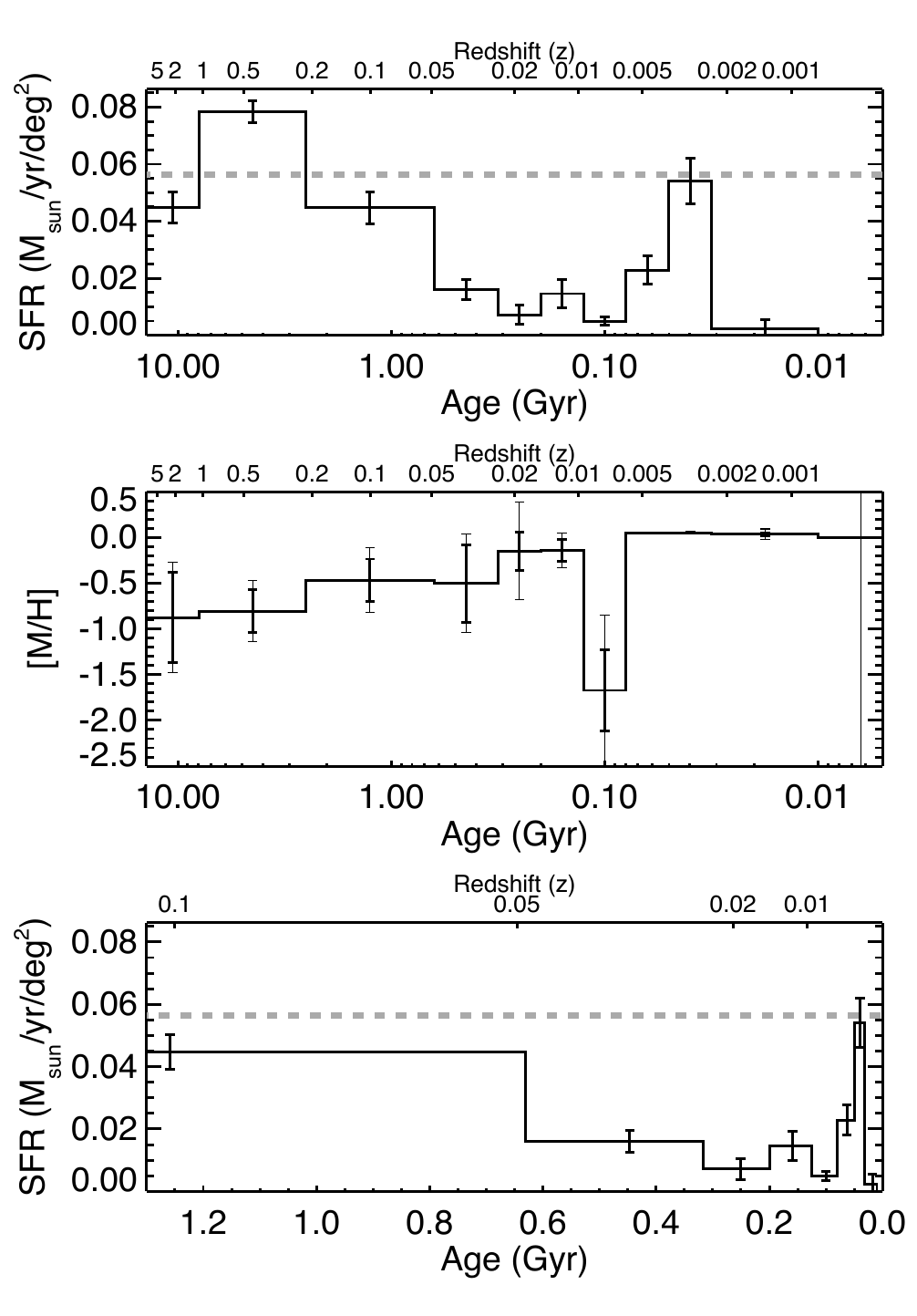}
\caption{The star formation history from the CMD shown in \autoref{sfhbw1}. {\it Top:}
The solid histogram marks the star formation rate (normalized by sky
area) as a function of time for the past 14 Gyr. The dashed line marks
the best-fitting constant star formation rate model. {\it Middle:} The
mean metallicity and metallicity range of the population as a function
of time.  Heavy error bars mark the measured metallicity range, and
lighter error bars mark how that range can slide because of errors in
the mean metallicity. {\it Bottom:} Same as {\it top}, but showing
only the results for the past 1.3 Gyr.}
\label{sfhbw2}
\end{center}
\end{figure}

Analysis of a composite population, as observed in a nearby galaxy, 
is performed through detailed fitting of stellar evolution
models to observed CMDs.  An example CMD of
approximately LSST depth is shown in \autoref{sfhbw1}, along with an
example model fit and residuals using the stellar evolution models of
\citet{Gir++02} and \citet{Mar++08}.  The age and metallicity distribution
from this fit are shown in \autoref{sfhbw2}.  These kinds of
measurements can show how star formation has progressed within a
galaxy over the past Gyr
\citep[e.g.][]{Dohm-Palmer++02,williams2003}, and provides the
possibility of looking for radial trends that provide clues about galaxy
formation.

This work requires obtaining as much information as possible about the
completeness and photometric errors as a function of position, color,
and magnitude.  The most reliable way to determine these
values is through artificial star tests in which a point spread
function typical of the LSST seeing at the time of the each
observation is added to the LSST data, and then the photometry of the
region is remeasured to determine 1) if the fake star was recovered
and 2) the difference between the input and output magnitude.  This
action must be performed millions of times to get a good sampling of
the completeness and errors over the full range in color and magnitude
over reasonably small spatial scales.  Furthermore, extinction in the
field as a function of position must be well-characterized, which
requires filters that are separated by the Balmer break.  The LSST $u$
filter fits the bill nicely.  

In order to be able to perform detailed studies of the age and
metallicity distribution of stars in the LSST data, we will need
to add artificial stars to the data to test our completeness.  We will
also need a reliable model for foreground
Galactic contamination, because the halos of nearby galaxies may be sparsely
populated and contain stars with colors and apparent
magnitudes similar to those of the Galactic disk and halo.

\section{Improving the Variable Star Distance Ladder}
\label{sec:sp:distances}
\noindent{\it Lucas M. Macri,  Kem H. Cook,  Abhijit Saha, Ata Sarajedini}   

Pulsating variable stars such as Cepheids and RR Lyraes have been 
indispensable in the quest to understand the scale of the Universe. 
The Cepheid Period-Luminosity-Color relation has been long established
and used initially to determine the distance to the Large Magellanic Cloud, 
and then to our nearest spiral neighbor M31. Their shorter period and 
fainter cousins, the RR Lyraes, are ubiquitous in globular clusters and among
the field star population; they can also be used in a relatively 
straightforward manner to measure distances. 

\subsection{Cepheids and Long Period Variables}
There is a major scientific interest in the use of Cepheid variables
to calibrate the absolute luminosity of type Ia supernovae (SNe) and other
cosmological distance indicators like the Tully-Fisher relation,
leading to improved determination of the Hubble constant ($H_0$). The
discovery of dark energy a decade ago brought new attention to this
topic because an increase in the precision of the measurement of $H_0$
results in a significant reduction of the uncertainty in $w$, the
parameter that describes the equation of state of dark energy
(\autoref{sec:com:cos}). 

Current efforts are aimed at measuring $H_0$ with a precision of 5\% or
better, through a robust and compact distance ladder that starts with
a maser distance to NGC 4258. Next comes the discovery of Cepheids in
that galaxy using optical data (acquired with HST, Gemini, and LBT),
which is followed up in the near-infrared with HST to establish a NIR
period-luminosity relation that is accurately calibrated in terms of absolute
luminosity and exhibits small scatter. Lastly, Cepheids are
discovered in galaxies that were hosts to modern type Ia SNe to
calibrate the absolute luminosity of these events and determine $H_0$
from observations of SNe in the Hubble flow. 

In the next few years before LSST becomes operational, we anticipate
using HST and the ladder described above to improve the precision in
the measurement of $H_0$ to perhaps 3\%. Any further progress will
require significant improvement in several areas, and LSST will be
able to contribute significantly to these goals as described below. 

\begin{itemize} 
   \item  We need to address the intrinsic variation of Cepheid
     properties from galaxy to galaxy. This can only be addressed by
     obtaining large, homogeneous samples of variables in many
     galaxies. LSST will be able to do this for all southern spirals
     within 8 Mpc.  

   \item  We need to calibrate the absolute luminosity of type Ia SNe
     more robustly, by increasing the number of host galaxies that
     have reliable distances. Unfortunately HST cannot discover
     Cepheids (with an economical use of orbits) much further out
     than 40 Mpc, and its days are numbered. Here LSST can play a
     unique role by accurately characterizing long-period variables
     (LPVs), a primary
     distance indicator that can be extended to much greater
     distances. 

   \item  LPVs are hard to characterize because of the long time
     scales involved (100-1000 days). Major breakthroughs were
     enabled by the multi-year microlensing surveys of the LMC (MACHO
     and OGLE) in combination with NIR data from 2MASS, DENIS and the
     South African/Japanese IRSF. An extension to Local Group spirals
     (M31, M33) is possible with existing data. LSST would be the
     first facility that could carry out similar surveys at greater
     distances and help answer the question of intrinsic variation
     in the absolute luminosities of the different LPV period-luminosity relations.  

   \item  The LSST observations of these nearby ($D<8$ Mpc) spirals would
   result in accurate Cepheid distances and the discovery of large
   LPV populations. This would enable us to accurately calibrate the
   LPV period-luminosity relations for later application to galaxies that hosted
   type Ia SNe or even to galaxies in the Hubble flow. This would
   result in further improvement in the measurement of $H_0$. 
\end{itemize}

\subsection{RR Lyrae Stars}
\label{sec:sp:rrlyrae}

\begin{figure}[ht]
\begin{center}
\includegraphics[width=0.5\linewidth]{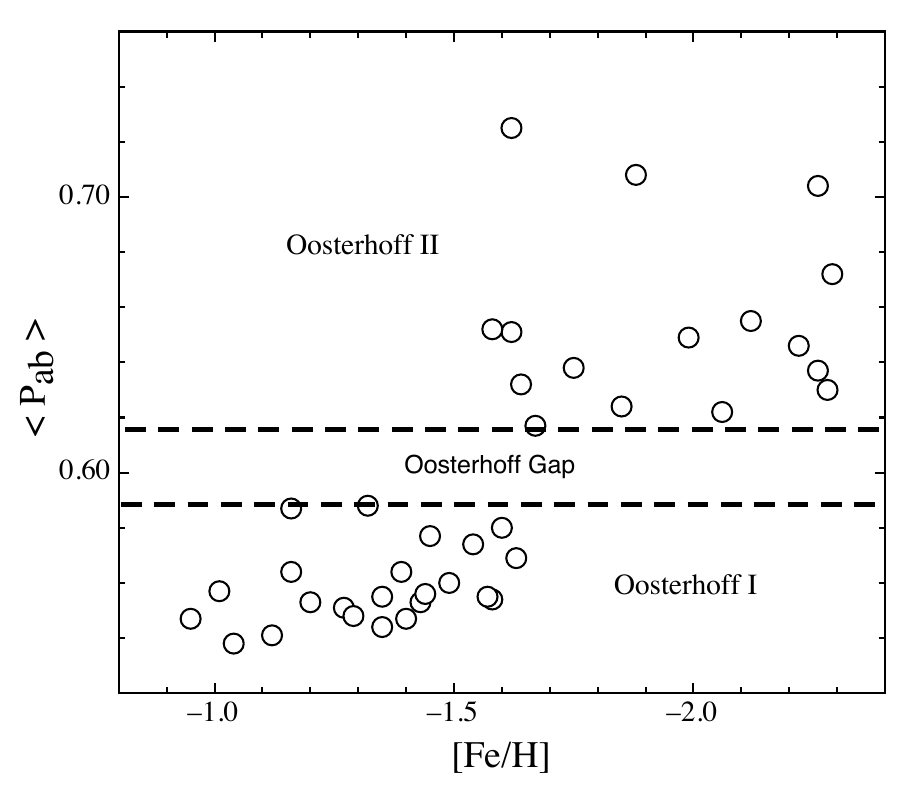}
\caption{Plot of the mean fundamental period for RR Lyraes in 
Galactic globular clusters, as a function of the cluster metal 
abundance \citep[from data compiled by ][]{Catelan++05}. The two 
types of Oosterhoff clusters divide naturally on either side of 
the Oosterhoff gap at 0.6 days.}
\label{oosterfig}
\end{center}
\end{figure}

While the empirical properties of RR Lyrae stars have been well
studied due to their utility as standard candles, theoretical models
that help us understand the physics responsible for these
properties are not as advanced. For example, it has been known for a
long time that Galactic globular clusters divide into two groups
\citep{oosterhoff1939} based on the mean periods of their ab-type RR
Lyrae variables - those that pulsate in the fundamental mode.  As
shown in \autoref{oosterfig}, Oosterhoff Type I clusters have ab-type
RR Lyraes with mean periods close to $\sim$0.56 days while type II
clusters, which are more metal-poor, harbor RR Lyraes with mean
periods closer to $\sim$0.66 days \citep{clement2001}. There have been
numerous studies focusing on the Oosterhoff dichotomy trying to
understand its origin (e.g. \citealt{LDZ1990, Sandage1993}). There is
evidence to suggest that globular clusters of different Oosterhoff
types have different spatial and kinematic properties, perhaps from
distinct accretion events in the Galactic halo
\citep{kinman1959, vdb1993}. There is also evidence favoring the notion
that the Oosterhoff Effect is the result of stellar evolution on the
horizontal branch (HB). In this scenario, RR Lyraes in type I clusters
are evolving from the red HB blueward through the instability strip
while those in type II clusters are evolving from the blue HB becoming
redward through the instability strip \citep{LeeCarney1999}. Yet
another explanation proposes that the Oosterhoff gap is based on the
structure of the envelope in these pulsating stars.
\citet{KanburFernando2005} have suggested that understanding the
physics behind the Oosterhoff Effect requires a detailed investigation
of the interplay between the photosphere and the hydrogen ionization
front in an RR Lyrae variable. Because these features are not co-moving
in a pulsating atmosphere, their interaction with each other can
affect the period-color relation of RR Lyraes, possibly accounting for
the behavior of their mean periods as a function of metallicity and,
therefore, helping to explain the Oosterhoff Effect. Clearly the
Oosterhoff Effect is one example of a mystery in need of attention
from both observers and theoreticians.

One reason there are so many open questions in our theoretical
understanding of RR Lyraes and other pulsating variables is that
progress requires observations that not only cover the time domain in
exquisite detail but also the parameter space of possible pulsation
properties in all of their diversity. This is where the LSST will make
a significant contribution. We expect to have a substantial number of
complete light curves for RR Lyraes in Galactic and Large Magellanic
Cloud globular clusters (see the estimate of the RR Lyrae recovery rate in 
\autoref{tr:opsim}). In addition, the data set will contain field
RR Lyraes in the Milky Way, the LMC, and the SMC, as well as a number
of dwarf spheroidal galaxies in the vicinity of the Milky Way. Some of
these RR Lyraes may turn out to be members of eclipsing binary systems,
further adding to their utility, as described in detail in \autoref{sec:sp:eb}.  The depth and breadth of this
variability data set will be unprecedented thus facilitating
theoretical investigations that have been heretofore impossible.

\section{A Systematic Survey of Star Clusters in the Southern Hemisphere}
\label{sp:sec:clusters}
\noindent{\it Jason Kalirai, Peregrine M. McGehee}   

\subsection{Introduction -- Open and Globular Star Clusters}

Nearby star clusters in the Milky Way 
are important laboratories for 
understanding 
stellar processes. There are two distinct classes of clusters in the
Milky Way, population I open clusters, which are lower mass (tens to
thousands of stars) and mostly confined to the Galactic disk, and
population II globular clusters (tens of thousands to hundreds of
thousands of stars), which
are very massive and make frequent excursions into the Galactic
halo.  
The systems are co-eval, co-spatial, and iso-metallic, and, therefore,
represent controlled testbeds with well-established
properties. 
The knowledge 
we have gained from studying these 
clusters grounds basic understanding of how stars evolve, and
enables us to interpret light from unresolved galaxies in the Universe.

\begin{figure}
\centering\includegraphics[width=0.95\linewidth]{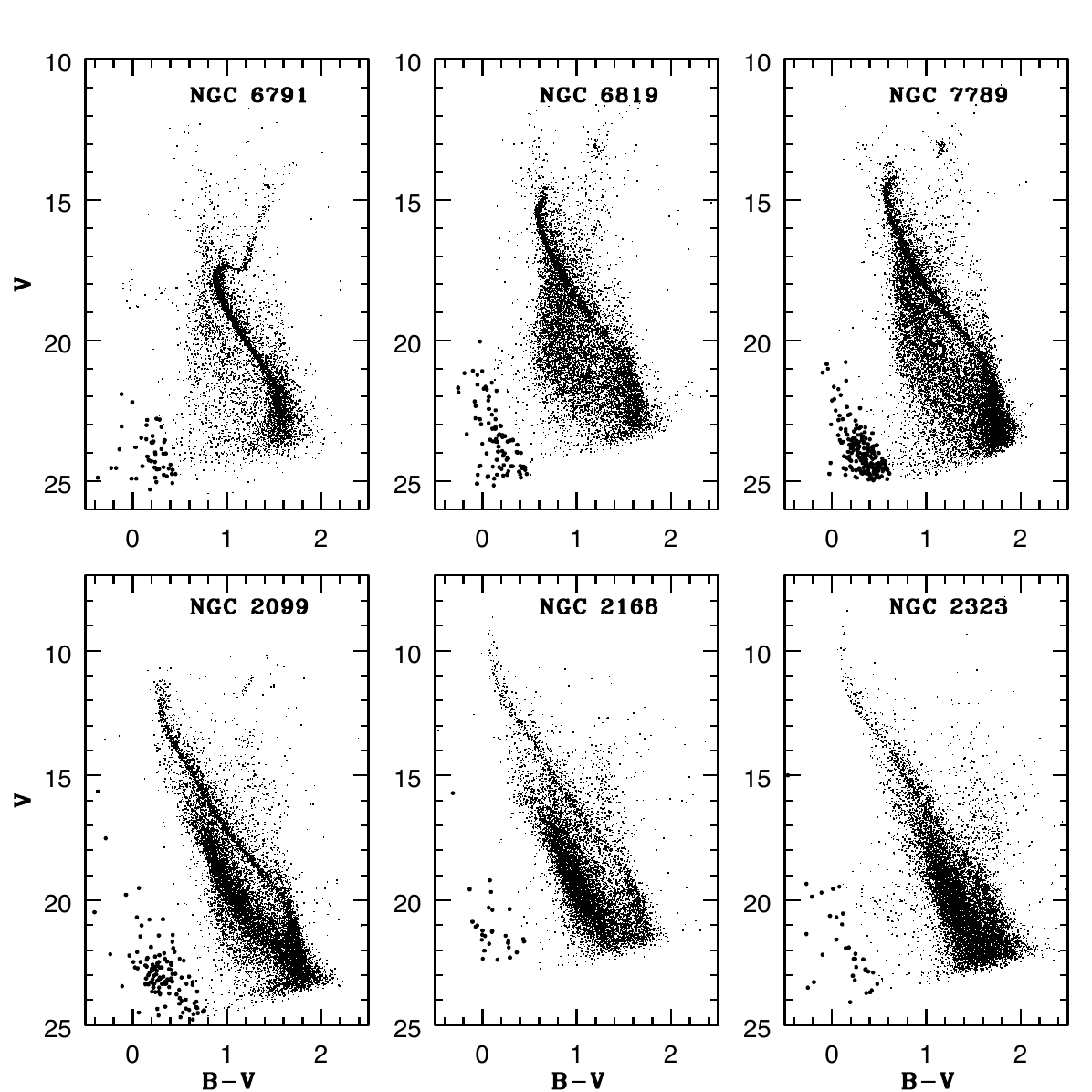}
\caption{Color-magnitude diagrams of six rich open star clusters observed 
as a part of the Canada-France-Hawaii Telescope Open Star Cluster Survey 
\citep{kalirai01b}.  The clusters are arranged from oldest in the 
top-left corner (8~Gyr) to the youngest in the bottom-right corner (100~Myr).  
Each color-magnitude diagram presents a rich, long main sequence stretching 
from low mass stars with $M \lesssim$ 0.5~$M_\odot$ up through the turn-off, 
including post-main sequence evolutionary phases.  The faint blue parts of 
each color-magnitude diagram illustrate a rich white dwarf cooling sequence 
(candidates shown with larger points).}
\label{fig:6cmds}
\end{figure}

Despite their importance to stellar astrophysics, most rich star
clusters have been relatively poorly surveyed, a testament to the
difficulty of observing targets at large distances or with large
angular sizes.  The advent of wide-field CCD cameras on 4-meter class
telescopes has recently provided us with a wealth of new data on these
systems.  Both the CFHT Open Star Cluster Survey \cite[][see
\autoref{fig:6cmds}]{kalirai01b} and the WIYN Open Star Cluster Survey
\citep{mathieu00} have systematically imaged nearby northern
hemisphere clusters in multiple filters, making possible new global
studies.  For example, these surveys have refined our understanding of
the fundamental properties (e.g., distance, age, metallicity,
reddening, binary fraction, and mass) of a large set of clusters and begun
to shed light on the detailed evolution of stars in post main sequence
phases (e.g., total integrated stellar mass loss) right down to the
white dwarf cooling sequences \citep{kalirai08}.

Even the CFHT and WIYN Open Star Cluster Surveys represent pencil beam
studies in comparison to LSST.  The main LSST survey will provide
homogeneous photometry of stars in all nearby star clusters in the
southern hemisphere (where no survey of star clusters has ever been
undertaken).  The LSST footprint contains 419 currently known
clusters; of these, 179 are within 1 kpc, and several are key
benchmark clusters for testing stellar evolution models.  Only 15 of
the clusters in the LSST footprint, however, have more than 100 known
members in the WEBDA database, demonstrating the relative paucity of
information known about these objects. LSST's deep,
  homogeneous, wide-field photometry will greatly expand this census,
  discovering new, previously unknown clusters and providing a more
  complete characterization of the properties and membership of
  clusters already known to exist.  Analysis of this homogeneous,
complete cluster sample will enable groundbreaking advances in several
fields, which we describe below.

\subsection{New Insights on Stellar Evolution Theory}

A century ago, Ejnar Hertzsprung and Henry Norris Russell found that
stars of the same temperature and the same parallax and, therefore, at the same distance, could have very different
luminosities \citep{hertzsprung05,russell13,russell14}.  They coined
the terms ``giants'' and ``dwarfs'' to describe these stars, and the
initial work quickly evolved into the first Hertzsprung-Russell (H-R)
Diagram in 1911.

The H-R diagram has since become one of the most widely used plots in
astrophysics, and understanding stellar evolution has been one of the
most important pursuits of observational astronomy.  Much of our
knowledge in this field, and on the ages of stars, is based on our
ability to understand and model observables in this plane, often for
nearby stellar populations.  This knowledge represents
fundamental input into our understanding of many important
astrophysical processes.  For example, stellar evolution aids in our
understanding of the formation of the Milky Way \cite[e.g., through
age dating old stellar populations,][]{krauss03}, the history of star
formation in other galaxies (e.g., by interpreting the light from
these systems with population synthesis models), and chemical
evolution and feedback processes in galaxies (e.g., by measuring the
rate and timing of mass loss in evolved stars).

With the construction of sensitive wide-field imagers on 4-m and 8-m
telescopes, as well as the launch of the HST, astronomers have
recently been able to probe the H-R diagram to unprecedented depths
and accuracy for the nearest systems \cite[e.g.,][]{richer08}.  These
studies have made possible detailed comparisons of not only the
positions of stars in the H-R diagram with respect to the predictions
of theoretical models, but also a measurement of the distribution of
stars along various evolutionary phases \cite[e.g.,][]{kalirai04b}.
Such comparisons provide for a more accurate measurement of the
properties of each system (e.g., the age), and also yield important
insight into the binary fraction, initial mass function, and initial
mass of the clusters.  Unfortunately, these comparisons have thus far
been limited to those clusters that are nearby and for which we have
such photometry, thus only sampling a small fraction of
age/metallicity space.

LSST will yield homogeneous photometry of star clusters in multiple
bands down to well below the main sequence turn-off, out to
unprecedented distances, and, therefore, will provide a wealth of
observational data to test stellar evolution models. With a detection
limit of 24 -- 25th magnitude in the optical bandpasses in a single
visit, and a co-added 5-$\sigma$ depth in the $r$-band of 27.8, LSST
will yield accurate turn-off photometry of all star clusters in its
survey volume out to beyond the edge of the Galaxy. For a 12~Gyr
globular cluster, this photometry will extend over three magnitudes
below the main sequence turn-off.

The H-R diagrams LSST will produce for thousands of star clusters will
completely fill the metallicity/age distribution from [Fe/H] = $-$2,
12~Gyr globular clusters to super-solar open clusters with ages of a
few tens of millions of years (including those in the LMC and SMC).
The multi-band photometry will constrain the reddenings to each cluster
independently and, therefore, allow for detailed tests of the physics
involved in the construction of common sets of models, as well as
atmospheric effects.  For example, slope changes and kinks along the
main sequence can yield valuable insights into the treatment of
convection and core-overshooting, the importance of atomic diffusion
and gravitational settling \citep{vandenberg96,chaboyer00}, and the
onset of rotational mixing in massive stars (e.g., younger clusters).  Examples
of these effects on the color-magnitude diagrams can be seen in 
\autoref{fig:6cmds}, for example, from the morphology of the hook at the
main sequence turn-off in NGC~6819, NGC~7789, and NGC~2099, and the
slope of the main sequence in NGC~2099 at $V$ = 14 -- 16.  For the
first time, these comparisons can be carried out in sets of clusters
with different ages but similar metallicity, or vice versa, thus
fixing a key input of the models.  Taken further, the data may allow
for new probes into the uncertainties in opacities, nuclear reaction
rates, and the equation of state, and, therefore, lead to new
understandings on both the micro- and macrophysics that guide stellar
evolution theory.

\subsection{The Stellar Mass Function}

An important goal of stellar astrophysics in our local neighborhood is
to characterize the properties of low luminosity stars on the lower
main sequence; such studies will be greatly advanced by the LSST data, as described in more detail in 
\autoref{sec:MW:stellarcensus}. 
Such
studies feed into our knowledge of the color-magnitude relation and
the initial mass function of stars, which themselves relate to the
physics governing the internal and atmospheric structure of stars.  In
fact, knowledge of possible variations in the initial mass function
has widespread consequences for many Galactic and extragalactic
applications (e.g., measuring the star formation mechanisms and mass
of distant galaxies).  Measuring these distributions in nearby star
clusters, as opposed to the field, offers key advantages as the stars
are all at the same distance and of the same nature (e.g., age and
metallicity).

Previous surveys such as the SDSS and 2MASS have yielded accurate
photometry of faint M dwarfs out to distances of $\sim$2~kpc. LSST,
with a depth that is two and five magnitudes deeper than Pan-STARRS
and Gaia respectively, will enable the first detection of such stars to beyond
10~kpc.  At this distance, the color-magnitude relation of hundreds of
star clusters will be established and permit the first systematic
investigation of variations in the relation with age and metallicity.
The present day mass functions of the youngest clusters will be
dynamically unevolved and, therefore, provide for new tests of the
variation in the initial mass function as a function of environment.
Even for the older clusters, the present day mass function can be
related back to the initial mass function through dynamical
simulations \cite[e.g.,][]{hurley08}, enabling a comparison between
these cluster mass functions and that derived from LSST detections of
Milky Way field stars.

\subsection{A Complete Mass Function of Stars: Linking White Dwarfs to Main Sequence Stars}

The bulk of the mass in old stellar populations is now tied up in the
faint remnant stars of more massive evolved progenitors.  In star
clusters, these white dwarfs can be uniquely mapped to their
progenitors to probe the properties of the now evolved stars (see
\autoref{sec:sp:wds} below).  The tip of the sequence, formed from
the brightest white dwarfs, is located at $M_V \sim$ 11 and will be
detected by LSST in thousands of clusters out to 20~kpc.  For a 1~Gyr
(10~Gyr) cluster, the faintest white dwarfs have cooled to $M_V$ = 13
(17), and will be detected in clusters out to 8~kpc (1~kpc).  These
white dwarf cooling sequences not only provide direct age measurements
\cite[e.g.,][]{hansen07} for the clusters and, therefore, fix the
primary leverage in theoretical isochrone fitting, allowing secondary
effects to be measured, but also can be followed up with current
Keck, Gemini, Subaru, and future (e.g., TMT and/or GMT) multi-object
spectroscopic instruments to yield the mass distribution along the
cooling sequence. These mass measurements represent the critical
input to yield an initial-final mass relation \citep{kalirai08} and,
therefore, provide the progenitor mass function above the present day
turn-off.  The relations, as a function of metallicity, will also yield
valuable insight into mass loss mechanisms in post-main sequence
evolution and test for mass loss-metallicity correlations.  The
detection of these white dwarfs can, therefore, constrain difficult-to-model phases such as the asymtotic giant branch (AGB) and planetary nebula (PN) stages.

\subsection{The Utility of Proper Motions}

The temporal coverage of LSST will permit the science discussed above
to be completed on a proper motion cleaned data set. To date, only a
few star clusters have such data down to the limits that LSST will
explore. Those large HST data sets of specific, nearby systems that we
do currently possess \citep[e.g.,][]{richer08} demonstrate the power
of proper motion cleaning to produce exquisitely clean H-R diagrams.
Tying the relative motions of these cluster members to an
extragalactic reference frame provides a means to measure the space
velocities of these systems and, therefore, constrain their orbits in
the Galaxy.  As open and globular clusters are largely confined to two
different components of the Milky Way, these observations will enable
each of these types of clusters to serve as a dynamical tracer of the potential
of the Milky Way and help us understand the formation processes of
the disk and halo (e.g., combining the three-dimensional distance, metallicity, age,
and star cluster orbit).

\subsection{Transient Events and Variability in the H-R Diagram}

The finer cadence of LSST's observations will also yield the first
homogeneous survey of transient and variable events in a well studied
sample of clusters (cataclysmic variables, chromospherically active stars, dwarf novae,
etc.). For each of these systems, knowledge of their cluster
environment yields important insight into the progenitors of the
transients, information that is typically missing for field stars.
Virtually all of the Galactic transient and variable studies outlined
in this chapter and in \autoref{chp:transients} will be possible
within these star clusters. 

\section{Decoding the Star Formation History of the Milky Way}
\label{sec:sp:ages}
\noindent{\it Kevin R. Covey, Phillip A. Cargile, Saurav Dhital}   

Star formation histories (SFHs) are powerful tools for understanding
galaxy formation.  Theoretical simulations show that galaxy mergers
and interactions produce sub--structures of stars sharing a single age
and coherent spatial, kinematic, and chemical properties
\citep{Helmi1999,Loebman2008}.  The nature of these sub--structures
places strong constraints on models of structure formation in a
$\Lambda$CDM universe \citep{Freeman2002}.

The Milky Way is a unique laboratory for studying these
Galactic sub-structures.  Detailed catalogs of stars in the Milky Way
provide access to low contrast substructures that cannot be detected in
more distant galaxies.  Photometric and spectroscopic surveys have
identified numerous spatial--kinematic--chemical substructures: the
Sagittarius dwarf, Palomar 5's tidal tails, the Monoceros Ring, etc.
\citep{Ibata1994,Odenkirchen2001,Yanny2003,Grillmair2006,Belokurov2006}.
LSST and ESO's upcoming Gaia mission will produce an order of
magnitude increase in our ability to identify such spatial--kinematic
substructures (see Sections \ref{sec:MW:intro}, \ref{sec:MW:tomog},
and \ref{sec:Gaia}).

Our ability to probe the Galactic star formation history has severely lagged these rapid
advances in the identification of spatial--kinematic--chemical
sub--structures.  Age distributions have been constructed for halo
globular clusters and open clusters in the Galactic disk
\citep{de-la-Fuente-Marcos2004}, but the vast majority of clusters
dissipate soon after their formation \citep{Lada2003}, so those that
persist for more than 1 Gyr are a biased sub--sample of even the clustered
component of the Galaxy's star formation history.  The star formation histories of distributed populations
are even more difficult to derive: in a seminal work, \citet{Twarog1980}
used theoretical isochrones and an age--metallicity relation to
estimate ages for Southern F dwarfs and infer the star formation history of the Galactic
disk. The star formation history of the Galactic disk has since been
inferred from measurements of several secondary stellar age
indicators: chromospheric activity--age relations
\citep{Barry1988,Soderblom1991,Rocha-Pinto2000a,Gizis2002,Fuchs2009};
isochronal ages \citep{Vergely2002,Cignoni2006,Reid2007}; and white
dwarf luminosity functions \citep{Oswalt1996,Harris2006}.  Despite
these significant efforts, no clear consensus has emerged as to the
star formaiton history of the thin disk of the Galaxy: most derivations contain episodes
of elevated or depressed star formation, but these episodes rarely
coincide from one study to the next, and their statistical
significance is typically marginal ($\sim 2 \sigma$).

Two questions at the next frontier in stellar and Galactic
archeology are: How well can we understand and calibrate stellar age
indicators?  What is the star formation history of the Milky Way, and
what does it tell us about galaxy formation and evolution?  Answering
these questions requires LSST's wide-field, high-precision photometry
and astrometry to measure proper motions, parallaxes, and
time--variable age indicators (rotation, flares, and so on) inaccessible to
Gaia.  Aspects of LSST's promise in this area are described elsewhere
this science book; see, for example, the discussions of LSST's promise
for  measuring the age distribution of Southern Galactic Star Clusters
(\autoref{sp:sec:clusters}), identifying the lowest metallicity stars
(\autoref{sec:sp:metalpoor}), and deriving stellar ages from white
dwarf cooling curves (\autoref{sec:sp:wds}).  Here, we describe three
techniques (gyrochronology, age--activity relations, and binary star
isochronal ages) that will allow LSST to provide reliable ages for
individual field stars, unlocking fundamentally new approaches for
understanding the SFH of the Milky Way.

\subsection{Stellar Ages via Gyrochronology} 
Since the seminal
observations by \citet{Skumanich1972}, we have known that rotation,
age, and magnetic field strength are tightly coupled for solar--type
stars.  This relationship reflects a feedback loop related to the
solar--type dynamo's sensitivity to inner rotational shear: fast
rotators generate strong magnetic fields, launching stellar winds that
carry away angular momentum, reducing the star's interior rotational
shear and weakening the star's magnetic field.  This strongly
self--regulating process ultimately drives stars with the same age and
mass toward a common rotation period.

Over the past decade, the mass--dependent relationship between stellar
rotation and age has been calibrated for the first time
\citep{Barnes2003, Meibom2008, Mamajek2008}.  These calibrations are
based on rotation periods measured for members of young clusters ($t
< 700$ Myrs) and the Sun, our singular example of an old ($t \sim
4.5$ Gyrs), solar--type star with a precise age estimate. The Kepler
satellite is now acquiring exquisite photometry for solar--type stars
in NGC 6819 and NGC 6791, providing rotation periods for stars with ages
of 2.5 and 8 Gyrs, respectively, and placing these gyrochronology
relations on a firm footing for ages greater than 1 Gyr
\citep{Meibom2008a}.

\begin{figure}
\begin{center}
\includegraphics[width=0.5\linewidth,angle=0]{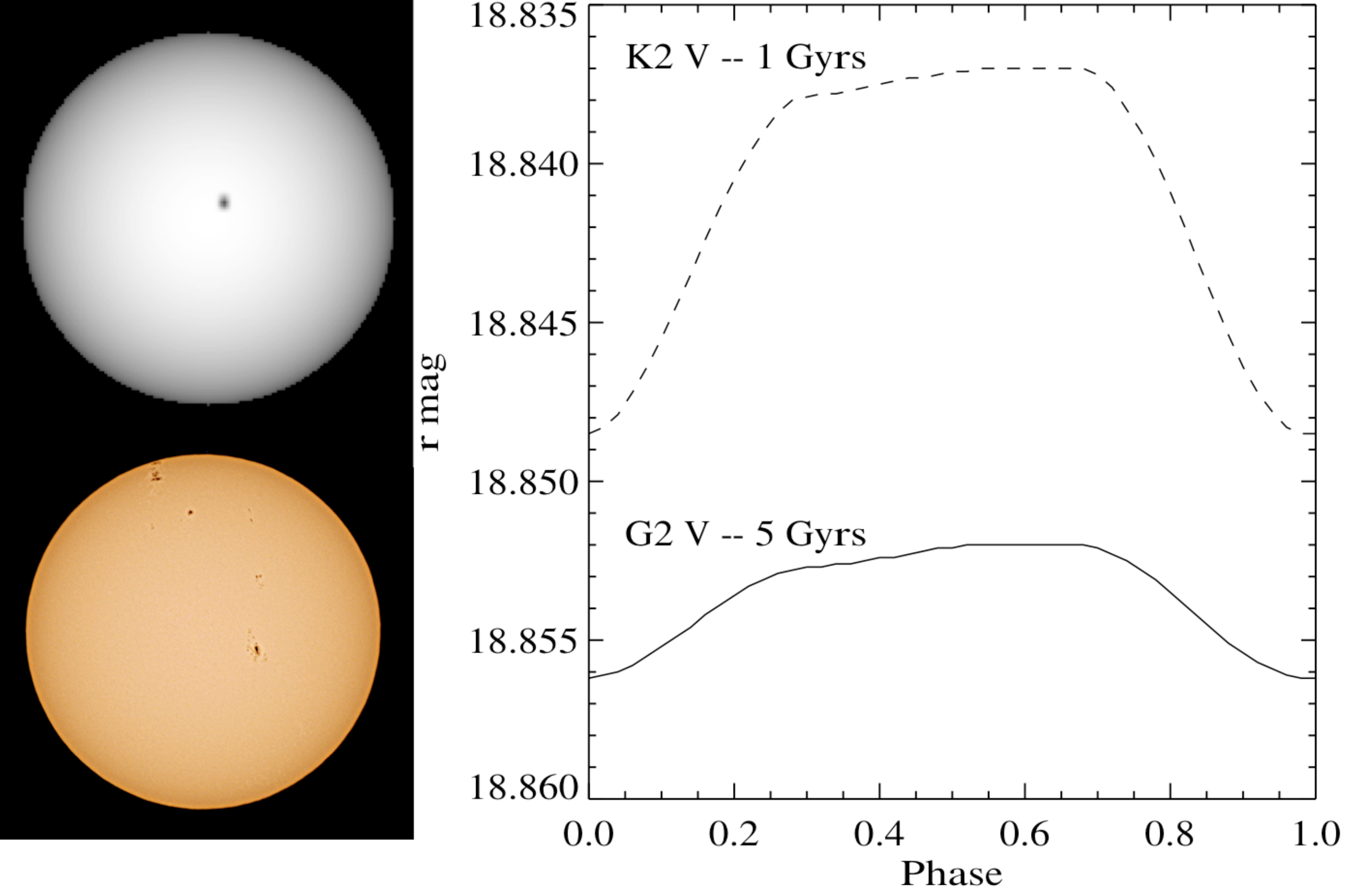}
\caption{\textit{Left:} Comparison of our synthetic model of a 5 Gyr
solar analog (top panel; image produced by A. Frasca's star spot light
curve modeling code, Macula.pro) with an actual image of the Sun
(bottom panel) from Loyd Overcash, with permission.  
\textit{Right:} Synthetic LSST light curves for the 5
Gyr Solar analog model shown above (solid line), as well as for a 1
Gyr K2 dwarf. \label{sp:fig:SunSpotModel}}
\end{center}
\end{figure}

We have performed a detailed simulation to identify the domain in
age--distance--stellar mass space where LSST will reliably measure
stellar rotation periods, and thus apply gyrochronology relations to
derive ages for individual field stars.  We begin with a detailed
model of a rotating, spotted star, kindly provided by Frasca et
al. (private communication).  Adopting appropriate synthetic spectra for
the spotted and unspotted photosphere, the disk-averaged spectrum is
calculated as a function of stellar rotational phase; convolving the
emergent flux with the LSST bandpasses produces synthetic light curves
for rotating spotted stars (see \autoref{sp:fig:SunSpotModel}).
Using this model, we produced a grid of synthetic $r$ band light
curves for G2, K2, and M2 dwarf stars with ages of 0.25, 0.5, 1.0,
2.5, and 5.0 Gyrs.  The rotation period and spot size were set for
each model to reproduce the age-period-amplitude relations defined by
\citet{Mamajek2008} and \citet{Hartman2009}.  An official LSST tool
({\tt Interpolator0.9}, S. Krughoff, private communication) then sampled this
grid of synthetic light curves with the cadence and observational
uncertainties appropriate for the main LSST survey.

\begin{figure}
\begin{center}
\includegraphics[width=0.9\linewidth,angle=0]{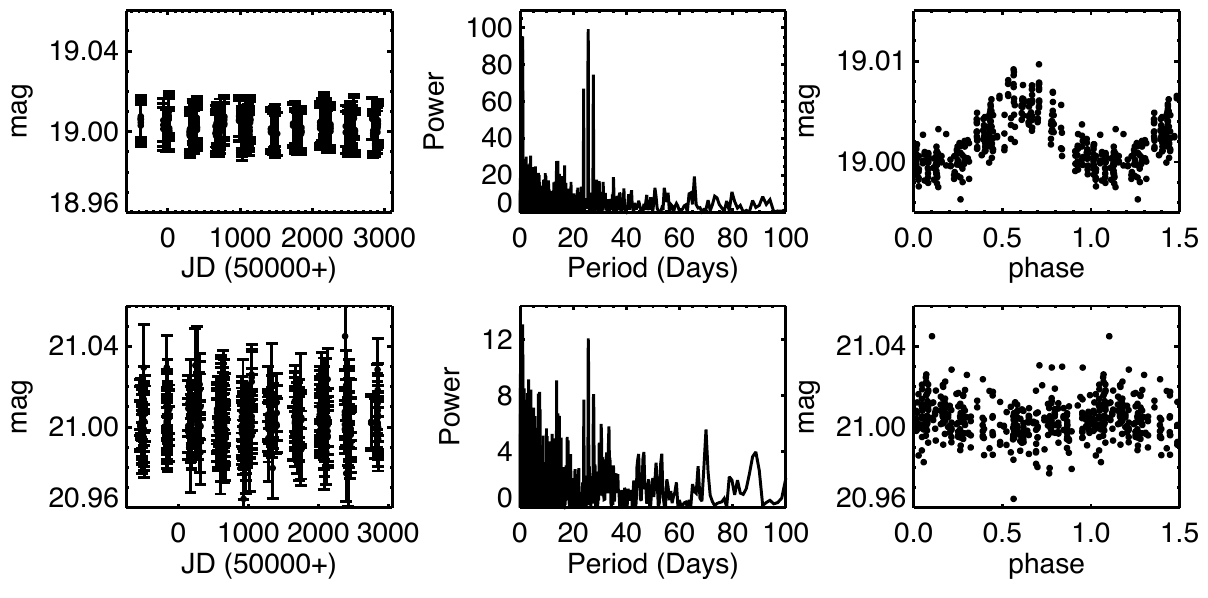}
\caption{The unfolded light curve, Lomb-Scargle periodogram, and the
folded light curve for a K2 dwarf of age 2.5 Gyr with $r=$ 19 and 21
magnitudes. We were able to easily recover the period at the bright
end, with the efficiency decreasing at the faint end, especially for
older stars, as noise starts to dominate. Our search through
parameter space shows that rotation periods can be recovered for G to
early--M spectral types, for ages up to a few Gyrs, and up to $r=21$
(see text for a detailed description).\label{sp:fig:K2}}
\end{center}
\end{figure}

We identify rotation periods from these simulated LSST light curves
using a Lomb-Scargle periodogram \citep{Scargle1982, Horne1986}, where
we identify the most significant frequency in the Fourier transform of
the simulated light curve.  Folding the data at the most significant
frequency then allows visual confirmation of the rotation
period. \autoref{sp:fig:K2} shows the unfolded light curve, the
periodogram, and the folded light curve for a K2 star of age 2.5 Gyr
``observed'' at $r=$ 19 and 21.  As the first panel of each row shows,
the noise starts to swamp the signal at fainter magnitudes, making it
harder to measure the period. This problem is most important for the
oldest stars: with diminished stellar activity producing small
starspots, these stars have light curves with small amplitudes.
However, with LSST's accuracy, we will still be able to measure
periods efficiently for G, K, and early-M dwarfs with $r \leq 20$ and
ages $\lesssim$2 Gyr.  All periods in these regimes were recovered,
without prior knowledge of the rotation period.  At older ages and
fainter magnitudes, the periodogram still finds peaks at the expected
values, but the power is low and the folded light curves are not
convincing.  Periods could potentially be recovered from lower
amplitude and/or noisier light curves by searching for common periods
across LSST's multiple bandpasses; with coverage in the $ugrizy$
bands, at least four of the bands are expected to exhibit the
periodicity. This will allow us to confirm rotation periods using
light curves with low amplitudes in a single band by combining the
results at the various bands.

Our simulations indicate LSST will be able to measure rotation periods
of 250 Myr solar analogs between 1 and 20 kpc; the inner distance
limit is imposed by LSST's $r \sim 16$ saturation limit, and the outer
distance limit identifies where LSST's photometric errors are
sufficiently large to prevent detection of photometric variations at
the expected level.  Older solar analogs will have smaller photometric
variations, reducing the distance to which periods can be measured:
LSST will measure periods for 5 Gyr solar analogs over a distance
range from 1 to 8 kpc.  Lower mass M dwarfs, which are significantly
fainter but also much more numerous, will have reliable rotation
period measurements out to 500 pc for stars as old as 5 Gyrs.
Measuring photometric rotation periods for thousands of field
  stars in a variety of Galactic environments, LSST will enable
  gyrochronology relations to map out the SFH of the Galactic disk
  over the past 1-2.5 Gyrs, and as far back as 5 Gyrs for brighter
  stars within the extended solar neighborhood.

LSST will also significantly improve our understanding of the 
gyrochronology relations that form the foundation of this analysis.
One fundamental requirement of any stellar dating technique, including 
gyrochronology, is that it should be able to accurately predict the age of
an object (or collection of objects) whose age(s) we know very well from an
independent measure. Open clusters (\autoref{sp:sec:clusters}) with precise age determinations
are essential to this calibration process.

\begin{table}
\begin{center}
\caption{Young LSST Benchmark Open Clusters}
\begin{tabular}{l c c c c c c}
\hline\hline
Cluster & Age   & Distance & [Fe/H] & Known & Spectral Type & M$_z$\\
        & [Myr] &  [pcs]   &        & Members & at LSST Limit & limit \\
\hline
ONC (NGC~1976)   &  1   & 414     &  0.00 & 733  & L3 & 15.92  \\
NGC~2547         &  35  & 474     & -0.16 &  69 & L1 & 15.65   \\
IC~2602/IC~2391  &  50  & 145/149 & -0.09 & 196/94 & T5 & 17.93\\
Blanco~1         &  80  & 207     &  0.04 & 128  & L6 & 16.84  \\
NGC~2516         &  120 & 344     &  0.06 & 130  & L2 & 15.79  \\
NGC~3532         &  355 & 411     & -0.02 & 357 & L1 & 15.51   \\
\end{tabular}
\label{sp:tab:youngclusters}
\end{center}
\end{table}

The LSST footprint contains several open clusters that are critical 
testbeds for testing of stellar evolution theory over the first 0.5 Gyrs (see 
\autoref{sp:tab:youngclusters}). These clusters have precise 
age estimates from robust dating techniques (e.g., lithium depletion boundary 
ages) and, therefore, will provide the necessary calibration to accurately 
determine how stellar rotation evolves with age over the initial portion
of each star's lifetime. 

In addition, the WEBDA open cluster database lists over 400 known open 
clusters in the LSST footprint; many of these have poorly constrained 
cluster memberships (e.g., fewer than 20 known members), especially for 
the oldest clusters (for example, see \autoref{sp:tab:oldclusters}). 
LSST's deep, homogeneous photometry and proper motions will significantly 
improve the census of each of these cluster's membership, providing new 
test cases for gyrochronology in age domains not yet investigated with 
this dating technique.

\begin{table}
\begin{center}
\caption{Selection of Old LSST Open Clusters}
\begin{tabular}{l c c c c c c}
\hline\hline
Cluster & Age   & Distance & [Fe/H] & Known & Spectral Type & M$_z$\\
        & [Myr] &  [pcs]   &        & Members & at LSST Limit & limit \\
\hline
        IC~4651 & 1140 &  888 &  0.09 & 16 & L0 & 14.08 \\
    Ruprecht~99 & 1949 &  660 &   ... &  7 & L1 & 14.83\\
       NGC~1252 & 3019 &  640 &   ... & 22 & L2 & 14.94\\
       NGC~2243 & 4497 & 4458 & -0.44 &  8 & M5 & 10.68\\
    Berkeley~39 & 7943 & 4780 & -0.17 & 12 & M5 & 10.43\\
  Collinder~261 & 8912 & 2190 & -0.14 & 43 & M6 & 11.90\\
\end{tabular}
\label{sp:tab:oldclusters}
\end{center}
\end{table}

\subsection{Stellar Ages via Age--Activity Relations} 
Age--activity relations tap the same physics underlying the gyrochronology
relations \citep{West2008, Mamajek2008}, and provide an opportunity
to sample the star formation history of the Galactic disk at ages inaccessible to gyrochronology.
Although inherently intermittent and aperiodic, stellar flares, which trace
the strength of the star's magnetic field, are one
photometric proxy for stellar age that will be accessible to LSST. The same
cluster observations that calibrate
gyrochronology relations will
indicate how the frequency and intensity of
stellar flares vary with stellar age and mass (see
\autoref{sec:tr:mdwarf_flare}), 
allowing the star formation history of the Galactic disk to be inferred
from flares detected by LSST in field dwarfs.  The primary limit on
the lookback time of a star formation history derived from stellar flare rates
relations is the timescale when flares become
too rare or weak to serve as a useful proxy for stellar age. We do not
yet have a calibration of what this lifetime is, but early
explorations suggest even the latest
M dwarfs become inactive after $\sim$5 Gyrs \citep{Hawley2000,West2008}.

\subsection{Isochronal Ages for Eclipsing Binaries in the Milky Way Halo}
Halo objects are $\sim$0.5\% of the stars in
the local solar neighborhood, so the ages
of nearby high velocity stars provide a first glimpse
of the halo's star formation history.  The highly substructured nature of the Galactic halo,
however, argues strongly for sampling its star formation history
in situ to understand the early Milky Way's full accretion history.  The
stellar age indicators described
in the previous sections are not useful for probing the distant halo, as
stellar activity indicators (rotation, flares) will
be undetectable for typical halo ages.

Eclipsing binary stars (EBs; \autoref{sec:sp:eb}), however,
provide a new opportunity for measuring the
SFH of the distributed halo population.
Combined analysis of multi-band light curves
and radial velocity measurements of detached,
double--lined EBs yield direct and accurate
measures of the masses, radii, surface gravities,
temperatures, and luminosities of the two stars
\citep{Wilson1971,Prsa2005}.
This wealth of information enables the derivation of \textit{distance
independent} isochronal ages for EBs by comparing to stellar evolution models in
different parameter spaces, such as the mass--radius
plane.
Binary components with $M>1.2 M_{\odot}$
typically appear
co--eval to within 5\%, suggesting that the age estimates of
the individual components are reliable at that level
\citep{Stassun2009}.  Lower mass binary components have larger errors,
likely
due to the suppression of convection by strong magnetic
fields \citep{Lopez-Morales2007}; efforts to include these
effects in theoretical models are ongoing, and should
allow for accurate ages to be derived for lower--mass binaries as
well. By identifying a large sample of EBs in the
Milky Way halo, LSST will enable us to begin mapping out the star formation history of the
distributed halo population.

\section{Discovery and Analysis of the Most Metal Poor Stars in the Galaxy}
\label{sec:sp:metalpoor}
\noindent{\it Timothy C. Beers}   

Metal-poor stars are of fundamental importance to modern astronomy and astrophysics
for a variety of reasons.  This long and expanding list includes:

\begin{itemize}

\item {\bf The Nature of the Big Bang:} Standard Big Bang cosmologies predict, with
  increasing precision, the amount of the light element lithium that was present
  in the Universe after the first minutes of creation.  The measured abundance
  of Li in very metal-poor stars is thought to provide a direct estimate of the
  single parameter in these models, the baryon-to-photon ratio.

\item {\bf The Nature of the First Stars:} Contemporary models and observational
  constraints suggest that star formation began no more than a few hundred
  million years after the Big Bang, and was likely to have been responsible for
  the production of the first elements heavier than Li. The site of this first
  element production has been argued to be associated with the explosions of
  stars with characteristic masses up to several hundred solar masses. These short-lived
  objects may have provided the first ``seeds'' of the heavy elements, thereby
  strongly influencing the formation of subsequent generations of stars.

\item {\bf The First Mass Function:} The distribution of masses with which stars have
  formed throughout the history of the Universe is of fundamental importance to
  the evolution of galaxies. Although the Inital Mass Function (IMF) today
  appears to be described well by simple power laws, it is almost certainly
  different from the First Mass Function (FMF), associated with the
  earliest star formation in the Universe. Detailed studies of elemental
  abundance patterns in low-metallicity stars provide one of the few means by
  which astronomers might peer back and obtain knowledge of the FMF.

\item {\bf Predictions of Element Production by Supernovae:} Modern computers
enable increasingly sophisticated models for the production
  of light and heavy elements by supernovae explosions. Direct insight into the
  relevant physics of these models can be obtained from inspection of the
  abundances of elements in the most metal-deficient stars, which presumably
  have not suffered pollution from numerous previous generations of stars.

\item {\bf The Nature of the Metallicity Distribution Function (MDF) of the Galactic
  Halo:} Large samples of metal-poor stars are now making it possible 
to confront detailed Galactic chemical evolution models with the observed
  distributions of stellar metallicities. Tests for structure in the MDF at low
  metallicity, the constancy of the MDF as a function of distance throughout the
  Galactic halo, and the important question of whether we are approaching, or
  have already reached, the limit of low metallicity in the Galaxy can all be
  addressed with sufficiently large samples of very metal-poor stars.

\item {\bf The Astrophysical Site(s) of Neutron-Capture Element Production:} Elements
  beyond the iron peak are formed primarily by captures of neutrons, in a
  variety of astrophysical sites.  The two principal mechanisms are
  referred to as the slow (s)-process, in which the time 
  scales for neutron capture by iron-peak seeds are longer than the time
  required for beta decay, and the rapid (r)-process, where the
  associated neutron capture occurs faster than beta decay.  These are
  best explored at low metallicity, where one is examining the
  production of heavy elements from a limited number of sites, perhaps even a single
  site. 

\end{itemize}

Owing to their rarity, the road to obtaining elemental abundances for metal-poor
stars in the Galaxy is long and arduous. The process usually involves three
major observational steps: 1) A wide-angle survey must be carried out, and
candidate metal-poor stars selected; 2) Moderate-resolution spectroscopic
follow-up of candidates is required to validate the genuine metal-poor stars
among them; and finally, 3) High-resolution spectroscopy of the most
interesting candidates emerging from step 2) must be obtained. 

The accurate $ugriz$ photometry obtained by LSST will provide for the
photometric selection of metal-poor candidates from the local neighborhood out
to over 100 kpc from the Galactic center. Similar techniques have been (and are
being) employed during the course of SDSS-II and SDSS-III in order to identify
candidate very metal-poor ([Fe/H] $< -2.0$) stars for subsequent follow-up with
medium-resolution ($R = 2000$) spectroscopic study with the SDSS spectrographs.
This approach has been quite successful, as indicated by the statistics shown in
\autoref{sp:tab:metalpoor}, based on work reported by \citet{Beers2009}. See
\citet{Beers+Christlieb05} for more discussion of the classes of metal-poor
stars. 

\begin{table}[htbp]
  \label{sp:tab:metalpoor}
  \caption{Impact of SDSS on Numbers of Metal-Poor Stars}
  \begin{center}
    \begin{tabular}{lrr}\hline\hline
      [Fe/H]      &  Pre SDSS-II & Post SDSS-II \rule{0ex}{2.3ex} \\ \hline
      $< -1.0$    & $\sim$ 15000 &	150000+ \\
      $< -2.0$    & $\sim$ 3000  &  30000+ \\
      $< -3.0$    & $\sim$ 400   &    1000+ \\ 
      $< -4.0$    &   5	      &   5     \\
      $< -5.0$    &   2         &   2     \\
      $< -6.0$    &   0         &   0     \\\hline
    \end{tabular}
  \end{center}
\end{table}

LSST photometric measurements will be more accurate than those SDSS obtains \citep{Ivezic2008} (\autoref{tab:intro:syspar}). This has three
immediate consequences: 1) Candidate 
metal-poor stars will be far more confidently identified, translating to much
more efficient spectroscopic follow-up; 2) Accurate photometric metallicity
estimates will be practical to obtain down to substantially lower metallicity
(perhaps [Fe/H] $< -2.5$) than is feasible for SDSS photometric selection
([Fe/H] $\simeq -2.0$); and 3) The much deeper LSST photometry means that
low-metallicity stars will be identifiable to 100 kpc, covering a thousand times
the volume that SDSS surveyed. The photometrically determined metallicities from
LSST will be of great scientific interest, as they will enable studies of the
changes in stellar populations as a function of distance based on a sample that
includes over 99\% of main sequence stars in the LSST footprint. This sample
will also enable studies of the correlations between metallicity and stellar
kinematics based on measured proper motions \citep[for an SDSS-based example,
see ][]{Ivezic2008}. Detailed metallicity measurements will of course require
high S/N, high-resolution spectroscopic follow-up of the best candidates.  

Proper motions obtained by LSST will also enable spectroscopic targeting of what
are likely to be some of the most metal-poor stars known, those belonging to the
so-called outer-halo population. \citet{Carollo2007} used a sample of some
10,000 ``calibration stars'' with available SDSS spectroscopy, and located
within 4 kpc of the Sun, to argue that the halo of the Galaxy comprises (at
least) two distinct populations: a slightly prograde inner halo (which dominates
within 10 kpc) with an MDF that peaks around [Fe/H] $= -1.6$ and an outer halo
(which dominates beyond 15-20 kpc) in net retrograde rotation with an MDF that
peaks around [Fe/H] $= -2.2$. The expectation is that the tail of the outer-halo
MDF will be populated by stars of the lowest metallicities known. Indeed, all
three stars recognized at present with [Fe/H] $< -4.5$, including two stars with
[Fe/H] $< -5.0$, exhibit characteristics of membership in the outer-halo
population. Stars can be selected from LSST with proper motions that increase
their likelihood of being members of this population either based on large
motions consistent with the high-energy outer-halo kinematics, or with proper
motion components suggesting highly retrograde orbits. 

%






\section{ Cool Subdwarfs and the Local Galactic Halo Population}
\noindent{\it S\'ebastien L\'epine, Pat Boeshaar, Adam J. Burgasser}   

Cool subdwarfs are main sequence stars, which have both a low mass and
a low abundance of metals. Locally they form the low-mass end of the
stellar Population II. Cool subdwarfs have historically been
identified from catalogs of stars with large proper motion, where they
show up as high velocity stars. Kinematically they are associated
with the local thick disk and halo populations. Because they are the
surviving members of the earliest generations of stars in the Galaxy
with evolutionary timescales well exceeding a Hubble
time, cool subdwarfs are true fossils of the early history of star
formation in the Galaxy, and hold important clues to
the formation of the Galactic system. While these stars have already
traveled dozens of orbits around the Galaxy and undergone some
dynamical mixing, a study of their orbital characteristics and
metallicity distribution can still shed light on the formation and
dynamical evolution of our Galaxy. In particular, cool subdwarfs do
not undergo any significant enrichment of their atmospheres, but
largely retain their original elemental composition from the time of their
birth. This makes them perfect tracers of the early chemical
composition of the gas that formed these first generations of low-mass
stars.

\begin{figure}
\centering
\includegraphics[width=0.5\linewidth]{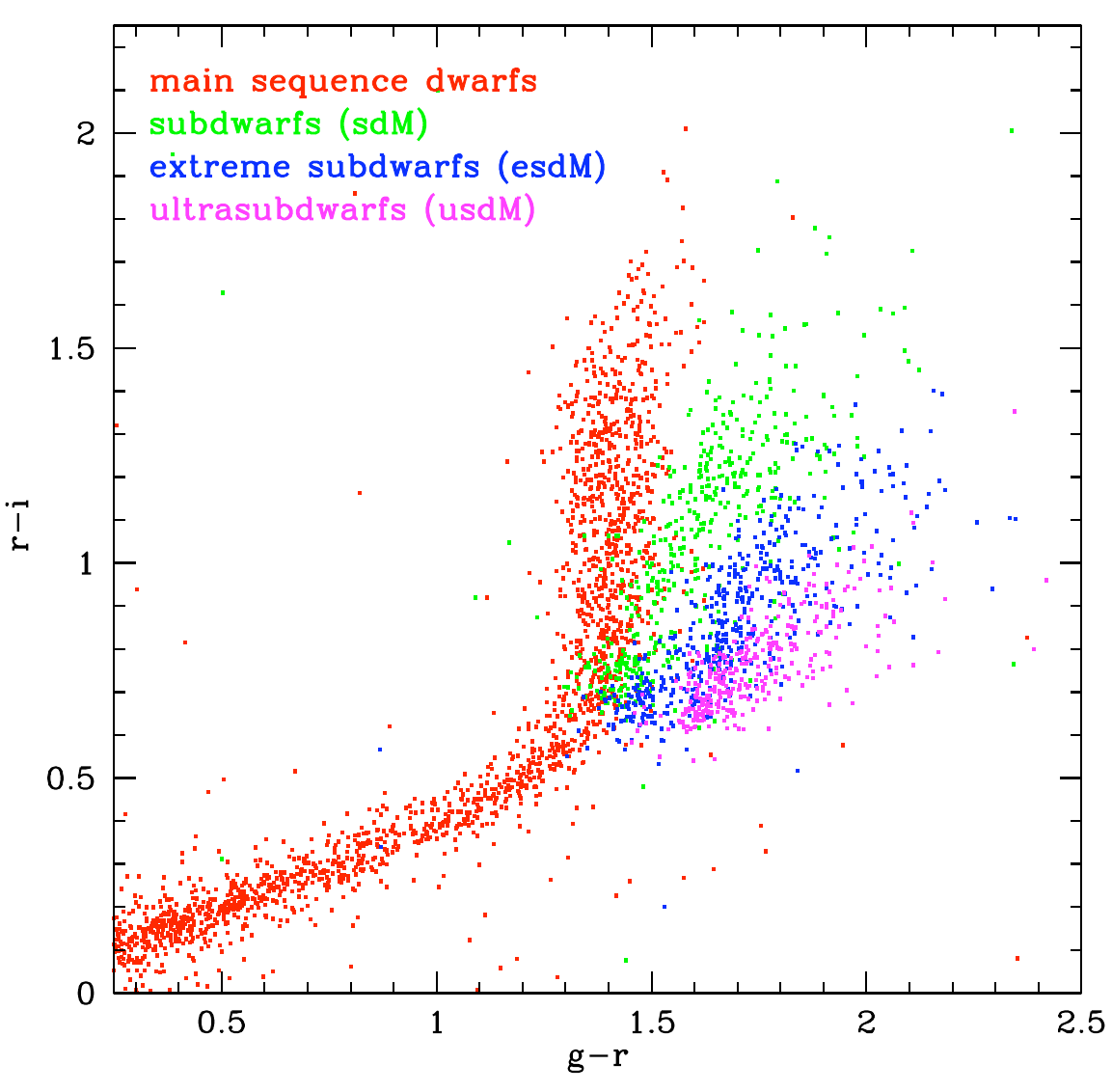}
\caption{Distribution of cool subdwarfs in the $g-r / r-i$ color-color
  diagram. The four metallicity classes (dwarfs, subdwarfs, extreme
  subdwarfs, and ultrasubdwarfs) are represented in different
  colors. The segregation according to metallicity class allows one to
  identify the halo subdwarfs and estimate their metallicities (and
  temperatures) based on photometry alone. Photometric data and
  spectroscopic confirmation of the stars have been obtained from SDSS.}
\label{sp:fig:subdwarf_colors}
\end{figure}

Cool stars of spectral type M have atmospheres that are dominated by
molecular bands from metal hydrides and oxides, most notably CaH, FeH,
TiO, and VO. Metallicity variations result in marked differences in
the absolute and relative strengths of these bands. As a result, M
dwarfs and subdwarfs also display significant variations in their broadband
colors depending on their metal abundances. The significantly
metal-poor subdwarfs from the Galactic halo populate a distinct
locus in the $g-r/r-i$ color-color diagram. The strong color-dependence
makes the M subdwarfs easy to identify
(\autoref{sp:fig:subdwarf_colors}), and also 
potentially allows one to determine the metallicity from broadband
photometry alone. The only caveat is that this part of color-color
space is also populated by extragalactic sources, which may be
distinguishable by their extent and their zero proper motions.

\begin{figure}
\centering
\includegraphics[width=0.95\linewidth]{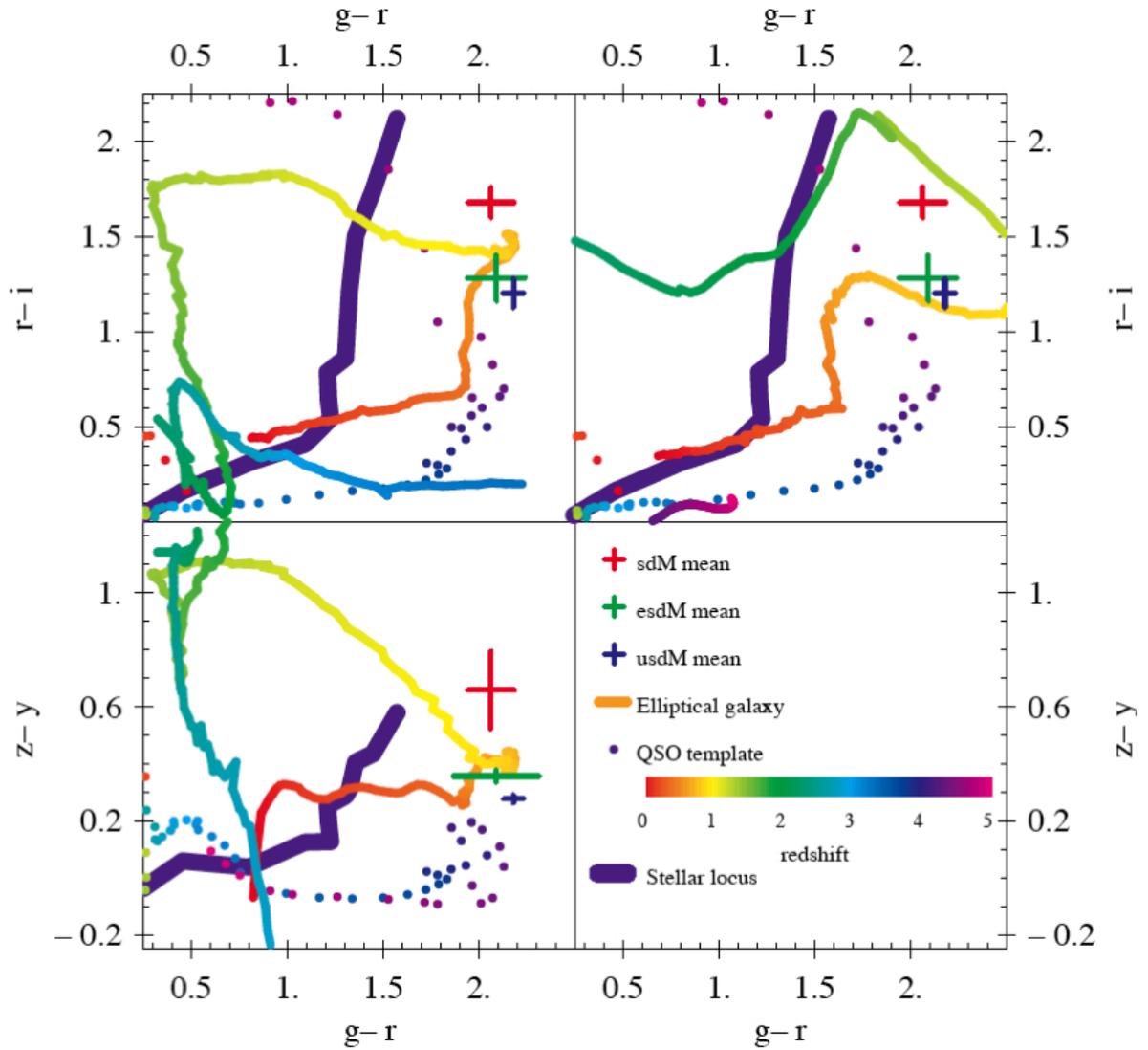}
\caption{{\it Upper left:} $g-r/r-i$ color-color diagram similar to 
\autoref{sp:fig:subdwarf_colors} showing the location of the main sequence 
stars (purple line), mean position of  extreme-, ultra-, and M subdwarfs (crosses), 
quasars (dots) and unevolved elliptical galaxies (thin colored line) as a function of redshift.
{\it Upper right:}  Same as previous figure with the \citet{B+C03} single
stellar population 5 Gyr 
evolutionary model for the elliptical galaxies.  {\it Lower left:}  g-r/z-y color-color 
diagram for objects in first figure. 
See also \autoref{fig:agn:color_select} in the AGN chapter.}
\label{sp:fig:extragalactic}
\end{figure}

In \autoref{sp:fig:extragalactic} panels, the blue line shows the
mean positions of \citet{Pic98} main sequence stars, the dots refer
to quasars from $z = 0$ to 5, and the mean position of the coolest ultra-,
extreme-, and M subdwarfs are noted by crosses in $grizy$ color
space.  Redshifting the locally observed elliptical galaxy template
\citep{CWW80} over the range $z = 0 - 2$ clearly
results in colors that occupy the same color space as the subdwarfs
of all classes, and even extend into the region of the coolest
subdwarfs. Estimates from the Deep Lens Survey \citep{Boeshaar2003} indicate that at high
Galactic latitudes, up to 30\% of the ``stellar" objects with M
subdwarf colors detected at 23-25 mag in $\geq 1''$ seeing will actually be unresolved
ellipticals at $z = 0.25 - 1$. The $g-z$ vs. $z-y$ plot clearly
separates the high redshift quasars from the subdwarf region, but quasars should
be only a minor contaminant due to their low spatial density.  The net
effect of including evolution into stellar population models 
is to shift the elliptical tracks by
several tenths of a magnitude in $g-r$ and $r-i$. Thus unresolved ellipticals may
still fall within the overlap envelope.  Additional synthetic $z-y$
colors for stars, brown dwarfs plus quasars and unevolved galaxies as a
function of redshift with color equations between the UKIRT
Wide-Field Camera and SDSS can be found in \citet{Hew++06}.  A proper motion detection is
thus required for formal identification.

Very large uncertainties in the luminosity function and number density
of such objects exist \citep{Digby2003}. It is not known whether the subdwarfs 
have a mass function similar to that of the disk stars. Their metallicity
distribution is also poorly constrained. The main limitation in using
the low-mass subdwarfs to study the Galactic halo resides in their
relatively low luminosities. M subdwarfs have absolute magnitudes in
the range $10<M_r<15$. With the SDSS magnitude limit of $r=22$ and
proper motion data to only $r=20$, M subdwarfs can thus only be
detected out to a few hundred parsecs. With a local density yielding
$\sim$ 1,000 objects within 100 parsecs of the Sun (all-sky), SDSS
can only formally identify a few thousand M subdwarfs.

LSST will open the way for a study of the low-mass halo stars on a
much grander scale. With photometry to $r\simeq 27$ and 
proper motion data available to
$r=24.5$, the LSST survey will detect all stellar subdwarfs
to 1 kiloparsec. In the Sun's vicinity, halo stars have large
transverse velocities ($v_T>100$ \kms), which yield proper motions
$\mu>20$ mas yr$^{-1}$ up to 1 kpc. With the required proper motion
accuracy of 0.2 mas yr$^{-1}$ for LSST, virtually all the subdwarfs
will be confirmed through proper motion detection. The ability to
estimate metallicity classes for the halo subdwarfs based on the LSST $gri$
magnitudes alone will make it possible to determine the approximate metallicity
distribution of the halo stars from an unprecedented sample of
$>$500,000 objects.

Relatively accurate photometric distances can also be determined for
low-mass stars, yielding distances generally accurate to better than
50\%. These, combined with the proper motion data, will make it
possible to plot large numbers of stars in tangential velocity space and search
for possible substructure in the tangential velocity space distribution.

Besides determining the subdwarf number density and distribution in
metallicity, mass, and luminosity, the exploration of the time
domain by LSST will identify eclipsing doubles, monitor rotational
modulation, and search for unexpected flaring activity. 
The multiplicity fraction of the halo population is only weakly constrained
due to the paucity of subdwarfs and their greater 
distances relative to their main sequence counterparts. 
A direct comparison with
the number of eclipsing binaries also expected to be discovered among
the disk stars will determine whether close double stars are more or
less common in the  halo population.  More critically, no eclipsing
system comprised of cool subdwarfs has ever been identified, but LSST's
systematic monitoring has excellent prospects for
finding at least several sdM+sdM eclipsing systems. Such systems would
be immensely useful in determining the mass and radii of low-mass,
metal-poor stars, which is now poorly constrained due to a paucity of
known binary systems.

Beyond building up very large samples of M-type subdwarfs, LSST will also 
uncover the first substantial samples of cooler L-type subdwarfs, 
metal-deficient analogs to the L dwarf population of very low mass stars 
and brown dwarfs \citep{burgasser2008}.  L subdwarfs have masses 
spanning the metallicity-dependent, hydrogen-burning limit, making them 
critical probes of both low-mass star formation processes in the halo and 
thermal transport in partially degenerate stellar interiors.  The subsolar 
metallicities of L subdwarfs are also important for testing chemistry models 
of low-temperature stellar and brown dwarf photospheres, in particular 
condensate grain and cloud formation, a process that largely defines the 
properties of L dwarfs but may be inhibited or absent in L subdwarfs
 \citep{burgasser2003b,reiners2006}.  Only a few L subdwarfs have been 
identified to date, largely serendipitously with 2MASS and SDSS.  But with 
$15 \lesssim M_i \lesssim 18$, they will be detected in substantial 
numbers with LSST, with volume-complete samples
out to at least 200 pc.   LSST should also discover specimens of the even
cooler T-type subdwarfs, whose spectral properties are as yet unknown but 
likely to be substantially modified by metallicity effects.
Collectively these low-temperature subdwarfs will facilitate the first 
measurement of the hydrogen-burning gap in the halo luminosity function, 
a population age indicator that can constrain the formation history of the 
low-mass halo and its subpopulations \citep[e.g.][]{burrows1993}.

\section{Very Low-Mass Stars and Brown Dwarfs in the Solar Neighborhood}
\noindent{\it Kevin R. Covey, John J. Bochanski, Paul Thorman, Pat Boeshaar, Sarah
  Schmidt, Eric J. Hilton, Mario Juri\'c, \v Zeljko Ivezi\'c, Keivan G. Stassun, Phillip A. Cargile, Saurav Dhital, Leslie Hebb, Andrew A. West, Suzanne L. Hawley}  

\subsection{The Solar Neighborhood in the Next Decade}

The least--massive constituents of the Galactic population are brown
dwarfs, objects too small to sustain hydrogen fusion in their
cores. These are divided into L dwarfs \citep{Kirkpatrick1999}, which
have $T_{eff}$ between 1500K and 2200K, whose spectra show weakening
of the TiO absorption bands, and T dwarfs with $T_{eff}$ between 800K
and 1500K,  which show the presence of CH$_4$ in their atmospheres. An
additional spectral type, the Y dwarf \citep[after][]{Kirkpatrick1999},
 has been reserved to describe possible dwarfs
with even cooler temperatures, which are expected to show NH${}_3$
absorption, a weakening in the optical alkali lines, and a reversal of
the blueward $J-K$ trend caused by CH${}_4$ absorption in T dwarfs
\citep{Kirkpatrick2005}. Although their masses are low, these brown
dwarfs are relatively common in the solar neighborhood, with 600 L and
150 T dwarfs now confirmed. Most of these have been discovered by
combining near--infrared imaging \citep[e.g. 2MASS;][]{Skrutskie2006}
with optical surveys \citep[e.g. SDSS;][]{York2000}; the faintness of
these objects at optical wavelengths ($M_z>13$) and the depth of
existing NIR imaging are such that these brown dwarfs are
overwhelmingly located within the immediate solar neighborhood
($d<65$ pc).  Informing theoretical models of these objects requires
measurements of precise physical parameters such as radius and mass. 

A number of ongoing or near--term surveys will expand the census of
brown dwarfs by 2017, when LSST will begin standard survey operations.
The largest single epoch catalog of warm (i.e., L and early T) 
brown dwarfs will be compiled by
the UKIDSS NIR survey \citep{Hew++06}, which is currently in
progress and capable of detecting L0 dwarfs in the J-band within $\sim
250$ pc.  At the cool end, the upcoming WISE mid-IR space telescope
\citep{Mainzer2005} will provide [3.3]-[4.7] $\mu$m colors for
early T dwarfs within 200 pc  \citep[assuming colors and magnitudes from][]{Patten2006} and all T9 dwarfs within 20 pc, and exquisite sensitivity to cooler Y dwarfs. 
Multi-epoch photometric surveys enable initial measurements of
a source's trigonometric parallax and potential binarity, such that
brown dwarfs detected by such surveys have significantly more value
for constraining theoretical models. The SkyMapper survey
\citep[SSSS][]{Keller2007}, a large-area imaging survey covering the southern
sky to depths similar to SDSS, will begin within 
the next few years. SkyMapper's six-year campaign will produce parallaxes 
for brown dwarfs within 20 pc. The Pan-STARRS PS1 survey will detect L0 dwarfs
in the $i$ band out to 400 pc and measure parallaxes for those within
100 pc; the volume sampled shrinks for fainter, cooler brown dwarfs.

\subsection{Simulating LSST's Yield of Solar Neighborhood MLTY Dwarfs}

Unlike previous surveys, LSST will not depend on separate
NIR surveys in order to distinguish L and T dwarfs from possible
contaminants such as transient detections, high-redshift quasars, and red
galaxies (see \autoref{fig:agn:z7quasars}). A narrow range of late L
dwarfs may overlap with $z\simeq 6.25$ quasars in color; shortly after
the start of the survey, even these brown dwarfs will be identifiable by their
proper motions. The addition of the $y$
band allows color identification based on detection in only the three
reddest LSST bands, allowing LSST to detect (5 $\sigma$ in full 10-year
co-adds) L0 dwarfs out to 2100 pc, and T0 dwarfs to 100 pc.  
This L dwarf detection limit extends well into the Thick disk, enabling 
LSST to probe the physics of old, metal--poor substellar objects 
(see \autoref{sec:sp:metalpoor}), and potentially
decode the star formation history of the thick and thin disks of the 
Milky Way from the age distribution of field brown dwarfs. 

We have constructed a detailed simulation of the very-low mass (VLM)
stars and brown dwarfs in the solar neighborhood; using the baseline
specifications for the LSST system, we have identified the subsets of
this population that LSST will characterize with varying degrees of
precision.  To model the stellar population in the stellar
neighborhood, we have adopted $ugriz$ absolute magnitudes for
VLM stars as tabulated by \citet{Kraus2007}, synthetic $z-y$ colors
calculated from optical and infrared template spectra provided by
\citet{Bochanski2007} and \citet{Cushing2005}, and the mass function
and space densities of low-mass stars measured by Bochanski et
al.\ (in preparation).

The lack of a hydrogen-burning main sequence in the brown dwarf regime
introduces strong degeneracies into the relationships between the
masses, ages, and luminosities of substellar objects.  These
degeneracies are an important consideration for studies of the
properties of field brown dwarfs, as the properties of brown dwarfs in
the solar neighborhood are sensitive to the star formation history of
the Milky Way and the shape of the stellar/substellar mass function.
We are currently developing simulations that explore the brown dwarf
samples LSST would observe assuming different star formation histories
and mass functions; for simplicity, however, the simulations described
below assume a single population of 3 Gyr brown dwarfs. In detail,
this substellar population is described by:

\begin{itemize}
\item For $2100  > T_{eff} > 1200$ (L dwarfs and the earliest Ts) 
we adopt empirical SDSS $riz$ magnitudes (Schmidt et al. in preparation), 
supplemented by synthetic $ugy$ magnitudes from cloudy 
\citet{Burrows2006} models.
\item For $1200 > T_{eff} > 600$ (mid-late Ts), we adopt synthetic 
$ugrizy$ magnitudes calculated from cloud-free \citet{Burrows2006} models.
\item For $T_{eff} > 600$ (as yet undiscovered Y dwarfs), we adopt 
synthetic $ugrizy$ magnitudes calculated from the ultra-cool 
\citet{Burrows2003} models.
\item 
We adopt the \citet{Cruz2007} luminosity function for L dwarfs; for 
T and Y dwarfs, we define a luminosity function with a linear extrapolation anchored 
by the coolest bin of the Cruz et al. luminosity and the empirical T dwarf 
space density measured by \citet{Metchev07}.
\end{itemize}

Using the above parameters as inputs, we simulated the properties of
field stars and brown dwarfs within 200 pc of the Sun.
\autoref{sp:tab:MLTY} summarizes the number of stars and brown dwarfs
LSST will likely detect in various filter combinations as a function of spectral
type with reliable parallaxes, to the single-visit depth.  
To make this estimate, we assume the
relationship between parallax uncertainty vs. $r$ band magnitude
reported in \autoref{tab:com:T3}, but we increase these uncertainties 
by a factor
of 1.25 to reflect that many stars and brown dwarfs will lack $u$ and
$g$ band detections.   

These results indicate that LSST will greatly expand the sample of
ultra--cool objects with reliable parallax measurements.  These
extremely red objects are ill-suited for parallax measurements with
Gaia's blue filterset, while LSST's greater depth and astrometric
precision will enable it to measure parallaxes for brown dwarfs
significantly beyond the parallax limit of Pan-STARRS.  This sample will, therefore, provide a key set of well-characterized brown dwarfs which
can confront the predictions of theoretical models of brown dwarf and
planetary atmospheres.

\begin{table}
\begin{center}
\caption{LSST's MLTY Dwarf Sample}
\begin{tabular}{l r r r r}
\hline\hline
Spectral Class & N$_{rizy}$  & N$_{izy}$ & N$_{zy}$ & N$_{\pi}$ \\
\hline
M & $>$347,000 & $>$347,000 & $>$347,000 & $>$347,000 \\
L & 18,500 & 27,500 & 35,600 & 6,550 \\
T & $\sim$3-4 & 50 & 2,300 & $\sim$260 \\
Y & 0 & 0 & $\sim$18 & $\sim$5 \\
\end{tabular}
\label{sp:tab:MLTY}
\end{center}
\end{table}

While we focus here primarily on brown dwarfs in the Galactic field,
 we also note that LSST's deep photometric limits and its ability to 
select cluster members via high-precision proper motions will provide a 
high-fidelity census of the very-low-mass populations of many Southern 
open clusters.  LSST will be able to easily identify VLM 
objects near or below the hydrogen-burning limit in most of the 
clusters listed in Tables \ref{sp:tab:youngclusters} and 
\ref{sp:tab:oldclusters}, with ages ranging from 1 Myr to $\sim$10 Gyr 
and to distances as far as $\sim$1.5 kpc.  LSST will provide 
colors and magnitudes for a large sample of L and T dwarfs with ages 
and metallicities derived from the morphology of each cluster's 
upper main sequence. This sample will define empirical brown dwarf 
cooling curves over a wide range of ages, providing a key calibration for 
understanding the properties of nearby field brown dwarfs whose ages are almost entirely undetermined.

\subsection{Science Results}

\subsubsection{Measuring Fundamental Physical Parameters of VLM stars
  and Brown Dwarfs}

The wide areal coverage, depth, precision, and temporal coverage of
LSST photometry make it an ideal instrument for the detection and
characterization of low--mass ($M < 0.8 M_{\odot}$) eclipsing binaries
within the Milky Way.  Currently, only $\sim15$ low--mass eclipsing
binaries are known \citep{Demory2009}, and even fewer VLM binaries,
presumably due to their intrinsic faintness ($L < 0.05\, L_{\odot}$
for a $0.4 M_{\odot}$ early M dwarf) and a stellar binary
fraction that decreases with mass \citep{Duquennoy1991,Burgasser2007}.
The large volume probed by LSST, with typical low-mass pairs being
detected at $d< 200$ pc, will discover a slew of these rare systems.

To quantify the expected yield of VLM star and brown dwarf EBs,
we start with the expected yield of MLTY dwarfs from the $N_{zy}$
column of \autoref{sp:tab:MLTY}. Such objects will already have
LSST light curves in the $z$ and $y$ bands, which can be complemented
by follow-up light curves in the $JHK$ bands. A five-band light curve
analysis is sufficient for modeling the EB parameters to
$\sim 1$\% \citep[e.g.][]{Stassun2004}.  Assuming that approximately
1/10 of the M dwarfs are M9 brown dwarfs, then the total brown dwarf yield from
\autoref{sp:tab:MLTY} is $\sim 70,000$. 
Recent surveys of binarity
among brown dwarfs yield fractions of 10--15\% for visual binaries with
separations of $> 1$~AU \citep[e.g.,][]{Martin2003,Bouy2003}. Thus
a very conservative estimate for the overall binary fraction of brown dwarfs
is 10\%.  Next, assuming a distribution of binary separations similar
to that for M dwarfs \citep{Fischer1992} implies that $\sim 10$\% of
these will be tight, spectroscopic binaries with physical separations $<
0.1$~AU. Finally, among these, the probability of an eclipse is of order
$(R_1 + R_2)/a$, where $R_1$ and $R_2$ are the component radii and $a$
is the semi-major axis, so for two brown dwarfs each with $R \sim 0.1$~R$_\odot$
and $a \sim 0.1$ AU, we have an eclipse probability of $\sim 1$\%. Thus
the overall expected brown dwarf EB yield will be $0.1 \times 0.1
\times 0.01 \times 70,000 \approx 7$.  With only one brown dwarf EB
currently known \citep{Stassun2006,Stassun2007}, the LSST yield represents
a critical forward advance for substellar science. The calculation above 
implies an overall yield of $\sim 40$ VLM eclipsing binary systems, 
a factor of several increase over the number currently known.

These fundamental astrophysical laboratories will redefine the
empirical mass--radius relations, for which current data are sparse and
derived from heterogeneous sources. For nearby eclipsing systems,
native LSST parallaxes will result in model--free luminosity
estimates, and can help constrain the effective temperature
distribution of low--mass stars.  This will be especially important at
smaller masses ($M < 0.1 M_{\odot}$), where only one eclipsing binary
is known 
\citep{Stassun2006}\footnote{This system exhibits
interesting behavior:  the hotter component (primary) is actually
fainter than its companion.}.  This regime will be well suited to LSST's
survey specifications and complement the Pan-STARRS survey as LSST will be probing a different area of the sky (the Southern Hemisphere).  At large distances, the currently elusive 
halo binaries, identified kinematically through proper motions,
may serve as probes for changes in low--mass stellar structure due to
metallicity.  Low--mass stars inhabit an interesting regime in stellar
structure.  At masses $\sim 0.4 M_{\odot}$, the interiors of low--mass
stars transition from a convective core surrounded by a radiative
shell to a fully convective interior.  Observations do not currently constrain how metallicity may affect this transition, and only the deep photometry that LSST will provide will enable an empirical investigation of this phenomenon. The
eclipsing binaries discovered in LSST will enable new science and
redefine the empirical understanding of stellar structure and binary
properties.
With the ``brown dwarf desert" significantly limiting the existence of
F/G/K+brown dwarf binaries\footnote{For example, \citet{Grether2006} find that
approximately 16\% of solar-type stars have companions with $P< 5$~yr, $M>
1$~M$_{Jup}$. Of these, $4.3\pm 1.0$\% have companions of planetary mass,
0.1\% have brown dwarf companions, and $11.2 \pm 1.6$\% have companions
of stellar mass.}, VLM+brown dwarf eclipsing binaries are the only systems in
which masses and radii of brown dwarfs can be measured---along with their
temperatures.

High-resolution near-infrared spectroscopic follow-up on 10-m class
telescopes will be critical for determining the radial velocity orbit
solutions for the discovered eclipsing binaries. For example, the one known
brown dwarf eclipsing binary \citep{Stassun2006,Stassun2007} is in the Orion Nebula
cluster at a distance of $\sim 500$~pc, the outer limit for the systems
included in the estimated yields calculated above. Its radial velocity
curve was obtained with the {\it Phoenix} spectrograph on the Gemini South
8-m telescope operating in the $H$ band (1.5~$\mu$m).

\subsubsection{Variability of VLM Stars and Brown Dwarfs}

The temporal coverage of LSST opens a window on the time variability
of VLM stars, including flares, spot modulation, and rotation periods.
 Additionally, by discovering new eclipsing binaries, LSST will
provide new laboratories for measuring fundamental stellar parameters like mass and radius.

Stellar magnetic activity has been observed and studied on M dwarfs for several decades
(see \autoref{sec:tr:mdwarf_flare} for a review of this subject), but
much less is known about activity on brown dwarfs. 
The fraction of stars with H$\alpha$ in emission, an indicator of magnetic 
activity, peaks around M7 and decreases through mid-L 
\citep{Gizis2000,West2004,Schmidt2007,West2008} although the changing continuum 
level with spectral type makes this an imperfect tracer of magnetic field
strength. 
\citet{Burgasser2003} report three T dwarfs with H$\alpha$ activity, 
one of which has strong, sustained emission \citep{Burgasser2002}.
 Flares on brown
dwarfs appear to be less frequent than on M dwarfs, but have been seen
in both X-ray \citep{Rutledge2000} and radio \citep{Berger2001}.
Optical spectra have shown variable H$\alpha$ emission in L dwarfs
\citep{Hall2002,Liebert2003, Schmidt2007,Reiners2008} 
that may be the result of flares.
LSST's new observations of such a large number of brown
dwarfs over dozens of epochs will provide much--needed empirical
determinations of flare rates.

LSST's temporal coverage will permit precise, dense coverage
of most main sequence stars with spots.  This subject is modeled in
detail in \autoref{sec:sp:ages}, and discussed below in the context of
low--mass stars and brown dwarfs.  Starspots, analogous to their solar
counterparts, provide a measure of the relative magnetic field
strength for stars of a given spectral type (and mass), assuming that the spot coverage is not so uniform as to prevent rotational modulation of the star's observed flux.  If spot variations can be detected, the $ugrizy$
light curves of these stars can be used to estimate temperature, from 
relative depths due to spot modulation, and filling factors, from the 
absolute deviations from a pristine stellar photosphere.  

Furthermore, the photometric signatures imprinted by these cooler
regions provide the opportunity to measure rotation periods that are
shorter than the lifetime of a typical starspot
($\sim$weeks to months). As demonstrated in \autoref{sec:sp:ages}, LSST
will be adept at measuring the rotation periods of coherently spotted,
magnetically active, low--mass stars.  Combining the measured rotation
periods with other proxies of stellar magnetic activity will provide a fundamental test of magnetic dynamo
generation theory.   This is
particularly interesting within the low--mass regime. At masses $\sim
0.4 M_{\odot}$, the interior of low--mass stars transition from an
convective core surrounded by a radiative shell to a fully convective
interior.  The transition region between convective core and radiative
exterior is thought to drive magnetic activity in earlier type
low--mass stars \citep{West2008}.


\subsubsection{Substellar Subdwarfs}

The deep
LSST survey imaging will photometrically identify statistically significant 
numbers of L dwarfs at large distances from the Galactic
plane. These dwarfs will allow the spatial distribution of dwarfs in
the thin and thick disk populations to be determined, and allow a
search for additional members of a halo population of metal--poor
subdwarf brown dwarfs to be discovered
\citep[e.g., ][]{Cushing2009}. The kinematics of the L dwarf and
subdwarf populations will also provide an empirical test of the
metallicity dependence of the hydrogen burning limit, based on the
model cooling curves. The existence of a population of substellar
subdwarfs may also indicate star formation due to infalling primordial
gas, or be a relic of the Milky Way's recent merger history.

\section{ Eclipsing Variables}
\label{sec:sp:eb}
%
%
%
%
%
%
%
%
%
%
%
%
%
%
%
%
%
%
%
%
%
\noindent{\it Andrej Pr\v{s}a, Keivan G. Stassun, Joshua Pepper}   

The importance of eclipsing binary stars (EBs) can hardly be overstated. Their analysis provides:

\begin{itemize}
\item {\bf Calibration-free physical properties of stars (i.e., masses, radii, surface temperatures, luminosities).} Masses are measured dynamically via radial velocities with no $\sin i$ ambiguity because the eclipses provide an accurate measure of $\sin i$. Radii are measured directly from the eclipse durations, the temperature ratio from the eclipse depths. The radii and  temperatures together yield the luminosities.
\item {\bf Accurate stellar distances.} With luminosities measured directly from the component radii and temperatures, the distance to the EB follows directly from the observed fluxes.
\item {\bf Precise stellar ages.} By comparing the measured mass-radius relationship with stellar evolution models, precise stellar ages can be determined. The accuracy of the age determination is of course model dependent and also mass dependent, with typical accuracy of $\sim 5$\% for $M_\star > 1.2\; {\rm M}_\odot$ and $\sim 50$\% for $M_\star < 0.8\; {\rm M}_\odot$ \citep{Stassun2009}.
\item {\bf Stringent tests of stellar evolution models.} With accurate, directly determined properties of the two stars in the EB, the basic predictions of stellar evolution models can be tested. For example, the two stars should lie on a single model isochrone under the assumption that they formed together as a binary, or the observed parameter relationships (mass-radius, temperature-luminosity, and so on)
can be compared against the model predictions.
\end{itemize}
 
The products of state-of-the-art EB modeling are seminal to many areas of astrophysics:

\begin{itemize}
    \item calibrating the cosmic distance scale;
    \item mapping clusters and other stellar populations (e.g., star-forming regions, streams, tidal tails, etc) in the Milky Way;
    \item determining initial mass functions and studying stellar population theory;
    \item understanding stellar energy transfer mechanisms (including activity) as a function of temperature, metallicity, and evolutionary stage;
    \item calibrating stellar color-temperature transformations, mass-radius-luminosity relationships, and other relations basic to a broad array of stellar astrophysics; and
    \item studying stellar dynamics, tidal interactions, mass transfer, accretion, chromospheric activity, etc.
\end{itemize}

LSST will be ideally suited for extensive mapping of EBs. As the
simulations described below demonstrate, LSST will detect essentially
all EBs with orbital periods less than 0.3 days, and 50\% of those
with periods up to $\sim$10 days (see
\autoref{EB_fig_detection_rate}). This completeness estimate is based
on analysis of a single passband; simultaneous analysis of all six
LSST bands will in reality improve this completeness. With a nominal
detection limit of $r=24.5$, a magnitude of $r=22.0$ should allow
detection of targets with a $S/N$ of 3.5, $r=19.5$ will have $S/N$ of
10 per data point. \autoref{EB_table_interesting_targets} shows the distance out to
which certain fiducial EB types can be detected. For example, a pair
of eclipsing M2 dwarfs will be detected out to 1 kpc with $S/N$ of
10.  

We can estimate the number of EBs that LSST will be able to fully
characterize (our experience modeling EB light curves shows that $S/N
\sim 3.5$ per data point typically suffices for the determination of physical and geometric parameters to a few percent). Gaia will observe $\sim$1 billion stars down to $r\sim20.5$ over the whole sky. We can expect that LSST will observe $\sim$0.5 billion stars to this same depth in the southern hemisphere. Extrapolating the results from Hipparcos (917 EBs in the sample of 118,218 observed stars; or 0.8\%), the LSST sample will contain $\sim$16 million EBs down to $r \sim 22.0$. The average detection rate for EBs over all periods will be around 40\% ($\sim 100$\% for $P < 0.3$ days, $\sim$50\% for $P \sim 10$ days, $\sim 20$\% for $P<30$ days; \autoref{EB_fig_detection_rate}), bringing the total number to $\sim$6.4 million EBs. Roughly 25\% of those will have components of similar luminosities (double-lined systems), yielding $\sim$1.6 million EBs with S/N$\geq 10$ for ready detailed modeling.

\begin{table}[t]
\begin{center}
\caption{Distance Limits for LSST Detection of Sample EBs.}
\scriptsize
\begin{tabular}{c|ccc}
Sample Binary${}^a$ Type & Binary Absolute Magnitude & Distance${}^b$ for $r=22.0$ [kpc] & Distance${}^b$ for $r=19.5$ [kpc] \\
\hline
M5V + M5V    & 12.9 &    0.7 &   0.2 \\
M2V + M2V    &  9.0 &    4.0 &   1.3 \\
K0V + K0V    &  5.0 &   25.1 &   7.9 \\
G2V + MxV    &  4.6 &   30.2 &   9.5 \\
G5III + GxV  &  2.9 &   66.1 &  20.1 \\
\hline
\multicolumn{4}{p{0.9\textwidth}}{${}^a$Scientifically interesting EB systems. EBs with M-dwarf components are rare in the literature. Their discovery will permit detailed testing of stellar models in this important mass regime. G-dwarf/M-dwarf pairs will be particularly valuable for pinning down the properties of M-dwarfs, since the temperature scale of G-dwarfs is relatively well established.  A particularly exciting prospect are Cepheids (G giants) in EB systems.} \\
\multicolumn{4}{p{0.9\textwidth}}{${}^b$Assuming no extinction.}
\end{tabular}
\label{EB_table_interesting_targets}
\end{center}
\end{table}

\subsection{Simulating LSST's Harvest of Eclipsing Binary Stars}

With LSST's six-band photometry and a cut-off magnitude of $r \sim 24.5$, the limiting factor for the detection of EB stars will be the cadence of observations. To estimate LSST's EB detection efficiency, we set up a test-bed by employing PHOEBE \citep{Prsa2005}, a \citet{Wilson1971} based eclipsing binary modeling suite. We first partitioned the sky into 1558 fields, covering all right ascensions and declinations between $-90^\circ$ and $10^\circ$.  The cadence of observations of these fields was then determined from the Simulated Survey Technical Analysis Report (SSTAR) for the operations simulations described in \autoref{sec:design:opsim}. 

\begin{figure}
\begin{center}
\includegraphics[width=7cm]{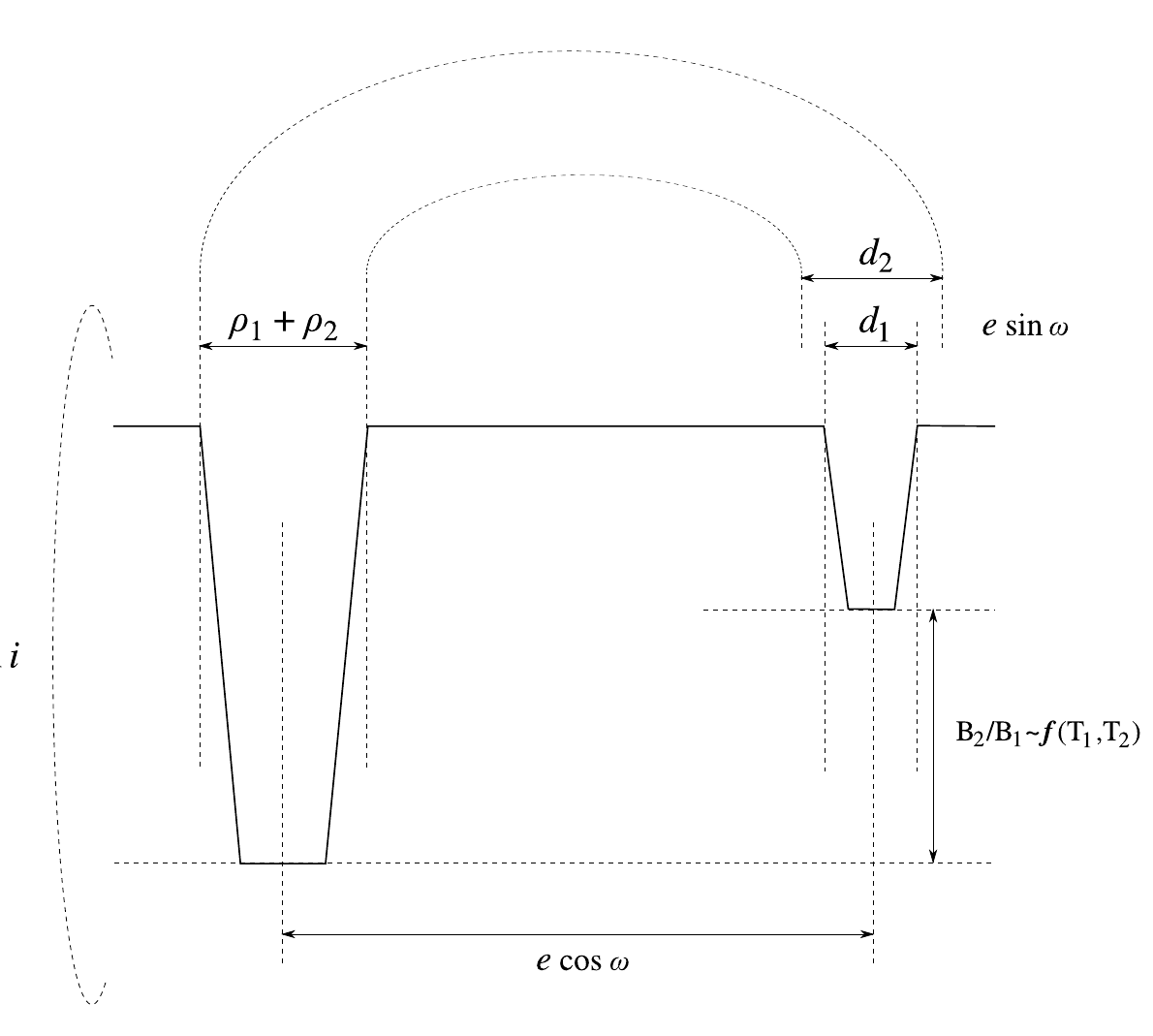} \\
\end{center}
\caption{Schematic view of an EB light curve. Surface brightness ratio $B_2/B_1$ directly determines the ratio of depths of both eclipses, and can be roughly approximated by the temperature ratio $T_2/T_1$. The sum of fractional radii $\rho_1+\rho_2$ determines the baseline width of the eclipses, while $e \sin \omega$ determines the ratio between the widths of each eclipse, $d_1$ and $d_2$. The phase separation of the eclipses is governed by $e \cos \omega$, and the overall amplitude of the light curve, as well as the shape of eclipses, are determined by $\sin i$.  \label{EB_fig_schematic_view}}
\end{figure}

To estimate LSST detection effectiveness, we synthesized light curves for five EBs that are representative of the given morphology type: well detached, intermediate detached, close detached, close, and contact. These most notably differ in fractional radii and orbital periods, hence in the number of observed data points in eclipses. Each EB light curve is described by its ephemeris (HJD${}_0$ and period $P_0$) and five principal parameters: $T_2/T_1$, $\rho_1+\rho_2$, $e \sin \omega$, $e \cos \omega$ and $\sin i$ (cf.~\autoref{EB_fig_schematic_view}; for a thorough discussion about the choice of principal parameters please refer to \citealt{prsa2008}). 

Let $N_1$ and $N_2$ be the numbers of data points observed in each eclipse. To detect and correctly classify light curves, we need as many points in \emph{both} eclipses as possible. We thus selected the
product $\mathcal C = N_1 N_2$ for the cost function. This way, if all data points are observed during one eclipse but not the other, this quantity will be zero. Consecutive observations of long period EBs
present another complication: although they contribute equally to the in-eclipse count, they cover essentially the same point in phase space because of the prolonged duration of eclipses. To account for that, all adjacent data points in phase space that are separated by less than some threshold value -- in our simulation we used 1/1000 of the period -- are counted as a single data point. 

The cost function $\mathcal C$, shaped according to these insights,
was computed for all five synthesized EBs (the details of the study
are presented in \citealt{Prsa2009}). The light curves are computed in
phase space, assuming that periodicity can be found correctly by a
period search algorithm if the S/N of a single data point exceeds 3.5
(or, in terms of LSST, $r < 22.0$). Simulation steps are as follows: 

\begin{enumerate}
\item given the $P_0$, pick a random phase shift between 0.0 and 1.0 and convert the time array to the phase array; 
\item sort the array and eliminate all data points with adjacent phases closer than the threshold value required to resolve them;
\item given the $\rho_1+\rho_2$, count the number of data points in eclipses ($N_1$, $N_2$);
\item compute the cost function value $\mathcal C = N_1 N_2$;
\item repeat steps 1-4 for a predefined number of times (say, 100), and find the average value of $\mathcal C$;
\item repeat steps 1-5 for all 1588 fields $(\alpha_i, \delta_i)$; and
\item repeat steps 1-6 for a range of periods sampled from a uniform distribution in $\log (P_0) \in [-1, 3]$.
\end{enumerate}

\begin{figure}
\begin{center}
\includegraphics[width=0.7\linewidth]{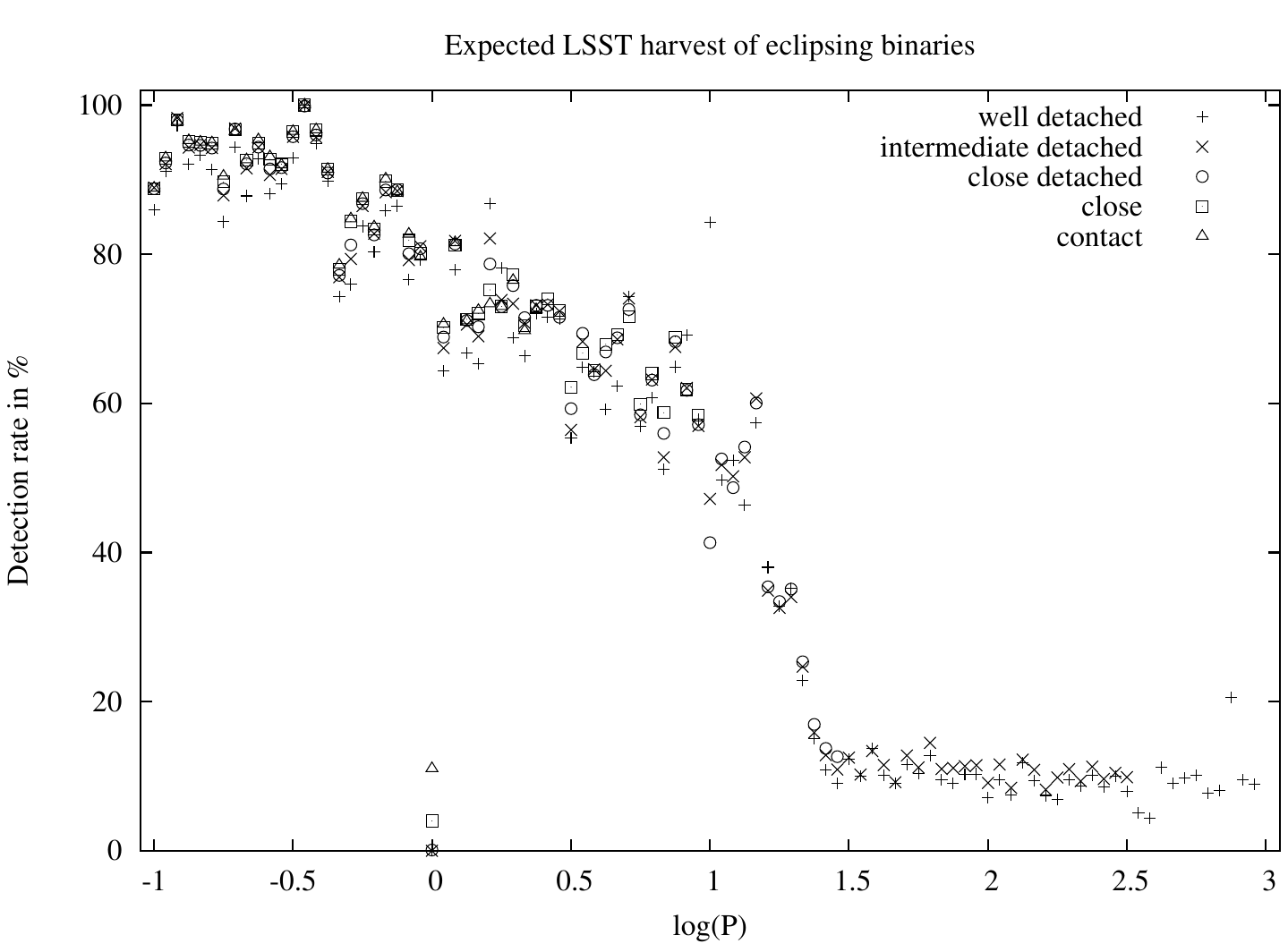}
\caption{The detection rate of eclipsing binary stars based on their
morphology and their orbital period. Assuming that S/N$\geq 3.5$ per
data point 
suffices for reliable recovery of orbital periods, LSST will detect 
almost all short period EBs, around 50\% of intermediate EBs, and around 10\% of long period EBs down to $r \sim 22.0$. The simulation is based on a single passband, implying that the values quoted here correspond to the worst case scenario. Since short and intermediate period EBs are most interesting for stellar population studies, it is clear that the expected LSST harvest of EB stars will be unprecedented. \label{EB_fig_detection_rate}} 
\end{center}
\end{figure}

\autoref{EB_fig_detection_rate} depicts the results of our simulation. Under the assumption that the variability analysis provides correct periods, the LSST sample of short period eclipsing binary stars will be essentially complete to $r\sim24.5$; these stars have the best characteristics to serve as calibrators -- both because of their physical properties and because of the feasibility for the follow-up studies.

\subsection{Effectiveness of EB Parameter Determination from LSST Data}

To further qualify LSST's harvest of EBs, we generated a sample of 10,000 light curves across the southern sky, using the cadence coming out of the Operations Simulations (\autoref{sec:design:opsim}). The values of principal parameters were sampled randomly, according to the following probability distribution functions: 

\begin{itemize}
\item $T_2/T_1$ is sampled from a normal distribution $\mathcal G(1.0, 0.18)$;
\item $P_0$ is sampled from a log-uniform distribution $[-1,4]$;
\item $\rho_1+\rho_2$ is sampled from a uniform distribution $[0.05, \delta_\mathrm{max}-0.05]$, where $\delta_\mathrm{max}$ is the morphology constraint parameter that depends exponentially on the value of $\log P_0$:
$$ \delta_\mathrm{max} (\log P_0) = 0.7 \exp \left( -\frac{1+\log P_0}{4} \right); $$
\item The eccentricity $e$ is sampled from an exponential distribution $\mathcal E(0.0, \epsilon_\mathrm{max}/2)$, where $\epsilon_\mathrm{max}$ is the attenuation parameter that depends exponentially on the value of $\rho_1+\rho_2$:
$$ \epsilon_\mathrm{max} = 0.7 \exp \left( -\frac{\rho_1+\rho_2-0.05}{1/6} \right); $$
\item The argument of periastron $\omega$ is sampled from a uniform
distribution $[0, 2\pi]$; the combination of the $e$ and $\omega$
distributions produces a sharp, normal-like distribution in $e \sin
\omega$ and $e \cos \omega$; 
\item The inclination $i$ is sampled from a uniform distribution $[i_\mathrm{grazing}, 90^\circ]$, where $i_\mathrm{grazing}$ is the inclination of a grazing eclipse.
\end{itemize}

Once the light curve sample was created, we added random Gaussian errors with $\sigma$ ranging from 0.001 to 0.2 (simulating different distances and, hence, different S/N), and we measured best fit parameters with {\tt ebai} (Eclipsing Binaries via Artificial Intelligence; \citealt{prsa2008}), an efficient artificial intelligence based engine for EB classification via trained neural networks.  
%
%
Backpropagation network training, the only computationally intensive part of {\tt ebai}, needs to be performed only once for a given passband; this is done on a 24-node Beowulf cluster using OpenMPI. Once trained, the network works very fast; 10,000 light curves used in this simulation were processed in 0.5\,s on a 2.0GHz laptop, where most of this time was spent on I/O operations.

\begin{figure}
\begin{center}
\includegraphics[width=0.48\linewidth]{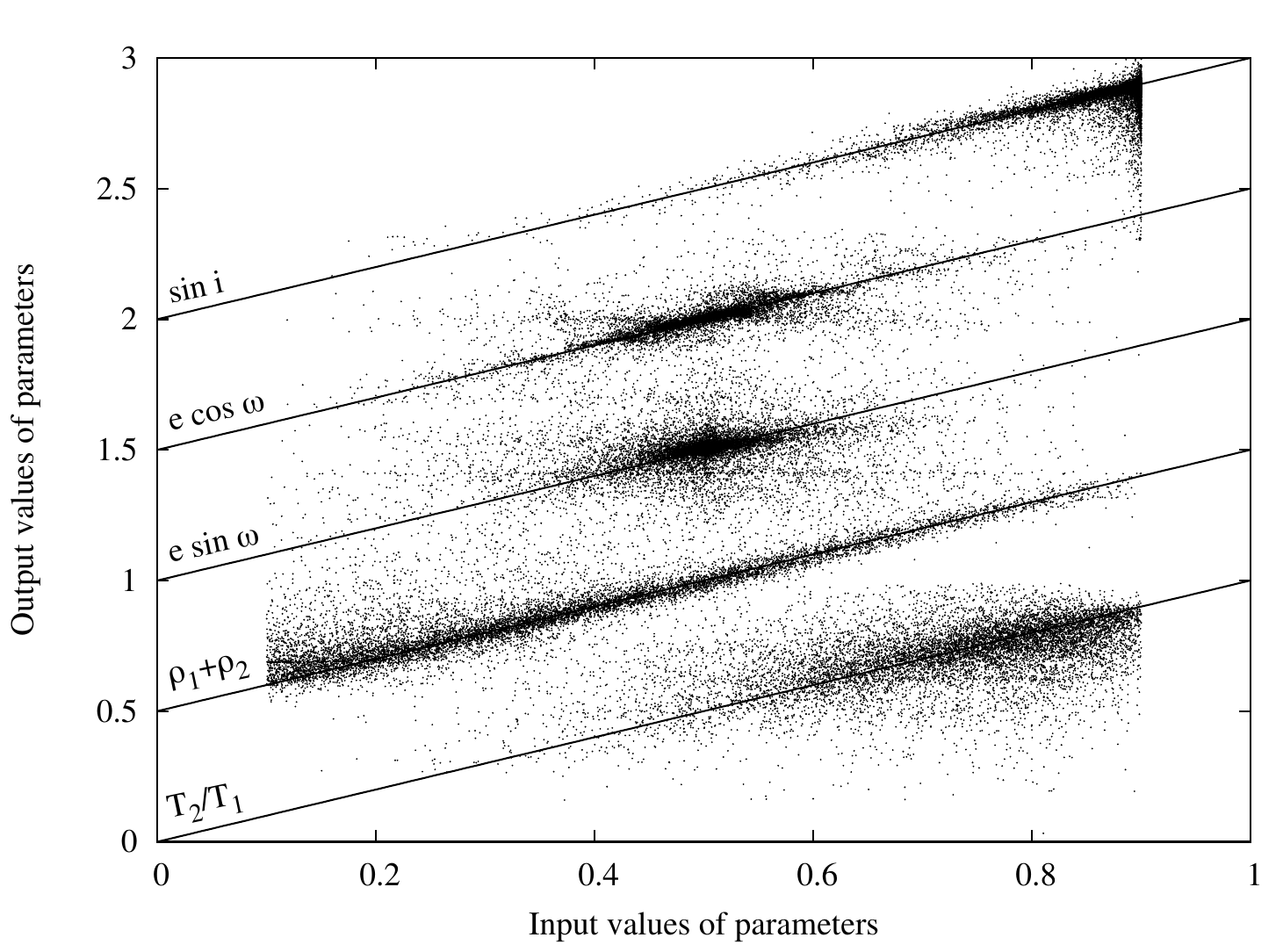}
\includegraphics[width=0.48\linewidth]{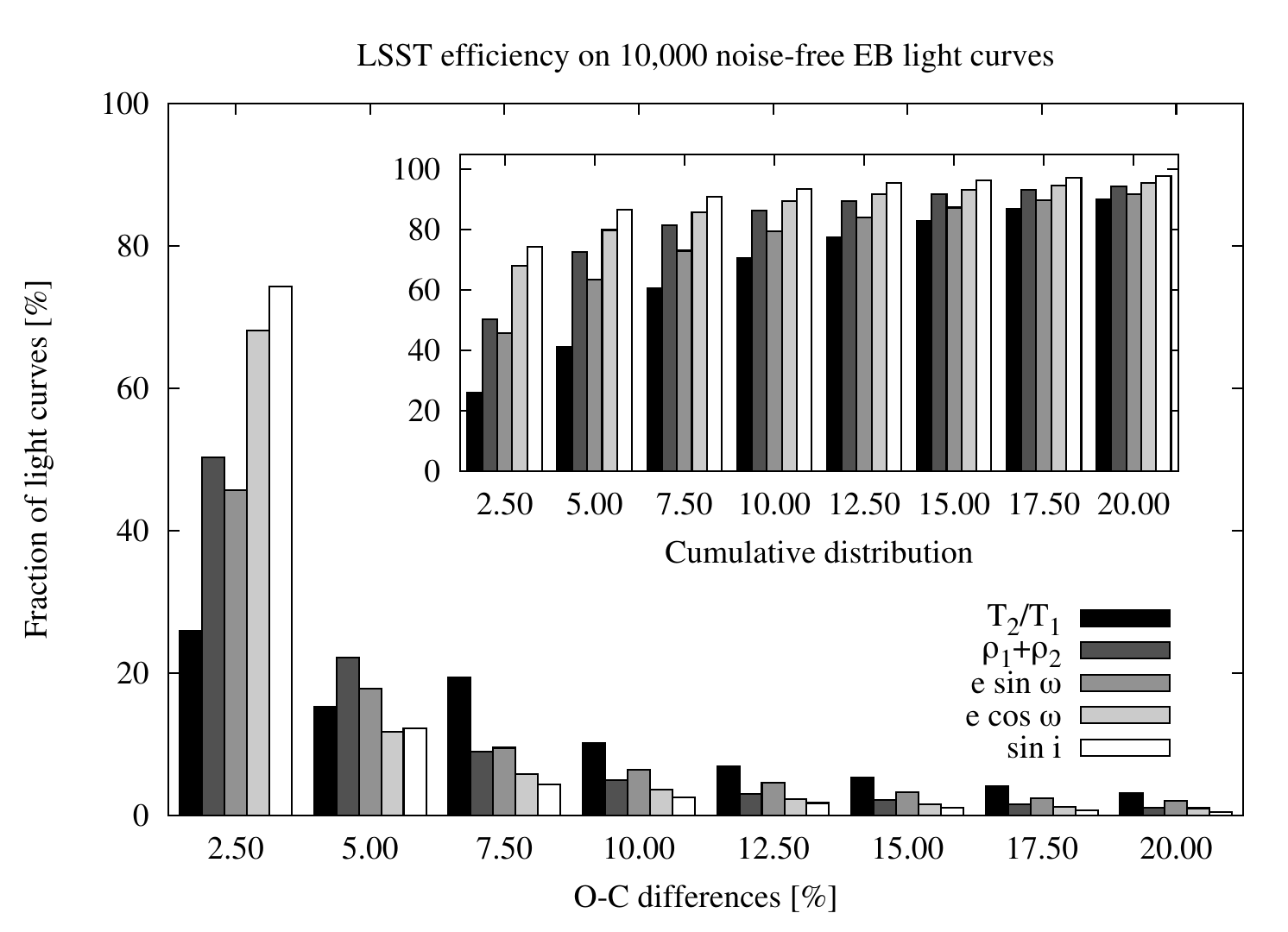} \\
\end{center}
\caption{{\it Left:} comparison between input and {\tt ebai}-computed
values of orbital parameters in a simulation of LSST-observed
eclipsing binaries. For neural network optimization purposes,
parameter values were rescaled to the interval $[0.1,
0.9]$. Successive parameters are vertically offset by 0.5 and
correlation guidelines are provided to facilitate comparison. {\it
Right:} histogram of the residuals computed by {\tt ebai}. Parameter
$T_2/T_1$ is most weakly determined (26\% of all light curves have a
corresponding error less than 2.5\%) -- this is due to the only
approximate relationship between $T_2/T_1$ and the surface brightness
ratio $B_2/B_1$. Parameters $\sin i$ and $e \cos \omega$ have the
highest success rate (75\% and 68\% of all light curves have $\sin i$
and $e \cos \omega$, respectively, determined to better than 2.5\%),
meaning that the cadence suffices for accurate determination of
orbital properties. The inset depicts the cumulative distribution of
the residuals: over 80\% of the sample has errors in \emph{all}
parameters less than 15\%. \label{EB_fig_ebai_results}} 
\end{figure}

\autoref{EB_fig_ebai_results} depicts the results of {\tt ebai}: 80\% of all stars passed through the engine have less than 15\% error in \emph{all five parameters}. A 15\% error might seem large at first
(typical error estimates of state-of-the-art EB modeling are close to 2-3\%), but bear in mind that {\tt ebai} serves to provide an \emph{initial} estimate for parameter values that would subsequently be improved by model-based methods such as Differential Corrections or Nelder \& Mead's Simplex, as implemented in {\tt PHOEBE}. 

These two simulations indicate LSST will provide a sample of short period EBs ($<1$\,day) essentially complete to $r\sim22.0$; a sample of EBs with periods of tens of days will be $\sim$50\% complete; a sample of long-period EBs will be $\sim$10\% complete. Since short period EBs carry the most astrophysical significance, and since parameter determination is most accurate for those stars because of the large number of data points in eclipses, LSST's high detection efficiency and accurate parameter measurements promise to revolutionize EB science and the many fields that EBs influence.

\section{White Dwarfs}
\label{sec:sp:wds}
\noindent{\it Jason Kalirai, Charles F. Claver, David Monet, \v{Z}eljko Ivezi\'{c}, J.B. Holberg}   

\subsection{The Milky Way White Dwarf Population}
\label{sec:sp:wdpop}

Over 97\% of all stars end their lives passively, shedding 
their outer layers and forming low mass white dwarfs.  These stellar 
cinders are the burnt out cores of low and intermediate mass hydrogen burning 
stars and contain no more nuclear fuel. As time passes, white dwarfs 
will slowly cool and release stored thermal energy into space becoming dimmer and dimmer. Although they are difficult to study given their intrinsic 
faintness, successful observations of white dwarfs can shed light on a very 
diverse range of astrophysical problems.

The largest sample of white dwarfs studied to date comes from SDSS, 
which has increased the known population of 
these stars by over an order of magnitude to more than 10,000 stars 
\citep{eisenstein06}.  This has enhanced our 
knowledge of stellar chemical evolution beyond 
the main sequence, uncovered new species of degenerate 
stars such as highly magnetized white dwarfs and accretion 
disk objects that may harbor planets, and provided a more accurate white 
dwarf luminosity function for the Galactic disk \citep{Harris2006}. 


\begin{figure}
\centering\includegraphics[width=0.95\linewidth]{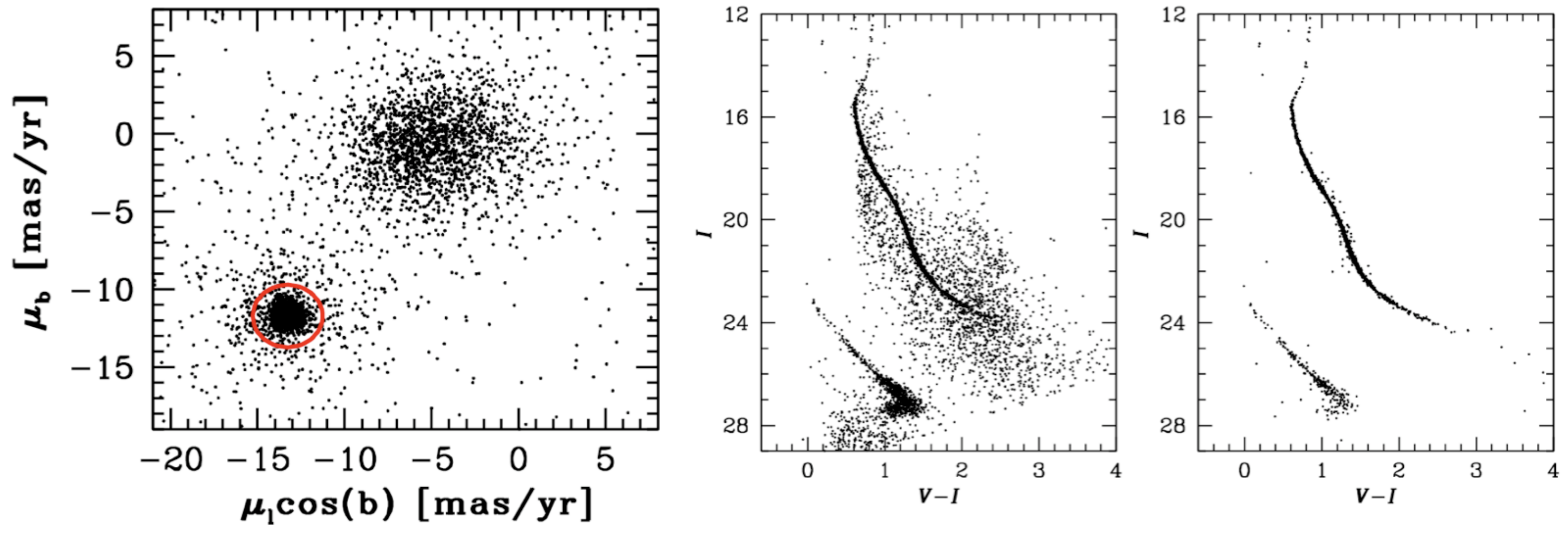}
\caption{The color-magnitude diagram of NGC~6397 from HST observations 
with ACS \citep{richer08} based on all stars (center) and proper 
motion members over a fraction of the field (right).  
The proper motion diagram (left) illustrates the motion of the 
cluster with respect to the field, over a 10 year baseline.  The rich white 
dwarf cooling sequence of the star cluster (faint-blue end) is modeled 
in \cite{hansen07} to yield an independent age for the cluster 
of $t$ = 11.5 $\pm$ 0.5 Gyr.}
\label{fig:hstwds}
\end{figure}


The luminosity function of white dwarfs rises with increasing 
photometric depth, such that LSST's sensitivity and wide areal 
coverage is expected to yield over 13 million white dwarfs with $r$ $<$ 24.5 
and over 50~million to the final co-added depth (model luminosity functions 
are presented later).  LSST will completely sample the brightest white 
dwarfs in our Galaxy (with $M_V \sim$ 11) to 20~kpc and beyond.  This 
broadband study of white dwarfs will yield important leverage on 
the overall baryon mass budget of the Milky Way and provide an 
unprecedented sample of white dwarfs, of all spectral types, to improve 
our understanding of a variety of astrophysical problems.  For example, 
based on MACHO predictions, LSST will be sensitive to thousands of dark 
halo white dwarfs and can therefore verify or rule out whether an 
appreciable fraction of the Galactic dark matter is tied up in these 
stars \citep{alcock00}.  LSST's photometry of 
white dwarfs will also be more than three times as precise as SDSS photometry, 
particularly in the $u$ band, which is often definitive for these stars.  
This greatly facilitates the matching of observed colors with 
predicted colors at the 1\% level making it possible to estimate white 
dwarf temperatures, gravities, and spectral types over a much wider 
range of parameter space then is now practical.  Below we outline 
several of the key science cases that LSST will address, followed by 
a specific discussion of the simulated distribution of white dwarfs that 
LSST will be sensitive to.

\subsection{White Dwarfs as Chronometers -- Dating Stellar Components of the Milky Way}

Although SDSS found abundant white dwarfs in the Galactic disk, the 
survey was too shallow to uncover large numbers of halo white dwarfs.  
These distant objects, which LSST will detect, will allow for the 
first time the construction of a luminosity function for field 
{\it halo} white dwarfs (see \autoref{sec:sp:wd_sample}
for the expected white dwarf spatial 
distribution). The structure in this luminosity function 
(and in particular, the turnover at the faint end) holds important clues  about the formation 
time of each specific Galactic component because the white dwarfs cool 
predictably with time. Therefore, an older population of white dwarfs 
is expected to show a fainter turnover as the stars have had more time 
to cool.  As we show in \autoref{sec:sp:wd_sample}, a simulation of 
the expected LSST white dwarf number counts indicates that over 
400,000 halo white dwarfs will be measured to $r$ $<$ 24.5.

The faintest white dwarfs in the nearest globular star clusters have 
now been detected with HST (see \autoref{fig:hstwds}); at 
$M_{\rm V} \gtrsim$ 16 \citep{hansen07}, they are a full 
magnitude fainter than their counterparts in the Galactic disk 
\citep{Harris2006}.  This work 
provides independent age measurements for nearby globular clusters 
and suggests that these objects formed several Gyr before the Galactic 
disk.  By extending these studies to the remnants in the Milky Way 
field halo, LSST will provide us with a direct measurement of the 
age of the Galactic halo, a vital input into the construction of 
Galactic formation models.  These measurements can not only help 
answer when our Galaxy formed, but also constrain the formation timescales 
of different populations within the same component.  
For example, the age distribution of Milky Way globulars can be 
contrasted with the field halo population to shed light 
on the formation processes of the clusters themselves (e.g., 
in situ formation vs. accretion).

LSST will also improve, by several orders of magnitude, the statistics 
of the field Milky Way disk white dwarf luminosity function.  \cite{Harris2006} 
comment on the lack of a precise ($\sigma$ $\sim$ 2~Gyr) age measurement 
\citep{leggett98,hansen02} given the low numbers of low-luminosity white dwarfs in the SDSS 
sample.  LSST will not only constrain the age of the oldest stars in the 
Galactic disk to a much higher accuracy than currently possible, but also 
map out the complete star formation history of the disk.  Epochs of 
enhanced star formation in the Galactic disk's history will leave imprints 
on the white dwarf luminosity function in the form of brighter peaks.  The 
luminosity and width of these peaks can be inverted to shed light on the 
formation time and timescale of the star forming events. In
\autoref{sec:sp:wd_sample} we simulate  the expected LSST white dwarf
disk luminosity function. 

A key component of LSST's study of Milky Way white dwarfs will be a 
kinematic analysis.  With LSST we will look for dependencies of the white 
dwarf luminosity function in 
the disk with the population's velocity, and therefore verify the age difference 
between the thin and thick disks.  The velocity may also be correlated 
with other Galactic parameters, such as metallicity, to give indirect 
age-metallicity estimates. Alternatively, dependencies of the luminosity 
function (and, therefore, age) may exist with scale height above/below the Galactic 
plane, improving our understanding of Galactic structure. The expected kinematic separation of these populations, based on 
LSST statistics, is also discussed near the end of this chapter. 

\subsection{White Dwarfs in Stellar Populations}


\begin{figure}
\begin{center}
\leavevmode 
\includegraphics[height=10.0cm,angle=270]{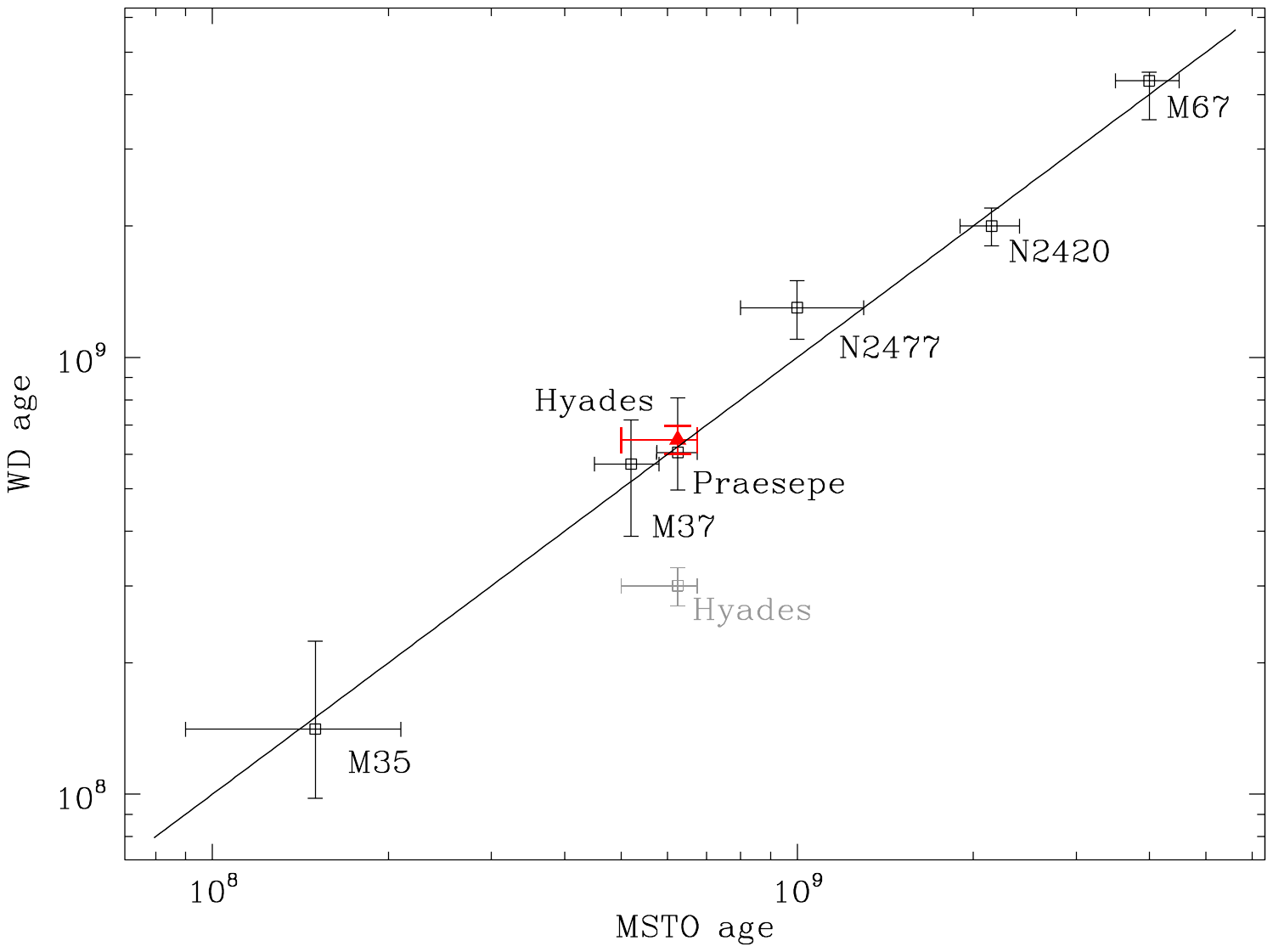}
\end{center}
\caption{The white dwarf cooling ages of several nearby solar metallicity 
open star clusters are compared with their corresponding main sequence 
turn-off ages based on a set of theoretical isochrones.  The two independent 
age measurements are in good agreement with one another for the 
assumed models and would not agree if for example, core-overshooting was 
not allowed.  Taken from \citet{Deg++09}.}
\label{fig:WDcoolages}
\end{figure}


The comparison of theoretical isochrones to observational color-magnitude 
diagrams has historically been used to infer the age of a nearby stellar population, 
provided that the distance is known through independent methods (e.g., 
main sequence fitting with the Hyades cluster, for which individual parallax 
measurements exist).  In practice this comparison is often limited by our 
lack of knowledge of fundamental quantities (e.g., the distance and metallicity) 
and so the isochrones are used to estimate multiple parameters at once.  
When combined with the uncertainties in the microphysics of the models (e.g., the 
role of gravitational settling or the treatment of convective core overshooting), 
the absolute uncertainty on the age of any stellar population using the main sequence 
turn-off method is $\sim$2~Gyr for old stellar populations \citep{dantona02}.  At 
higher redshifts, the theoretical isochrones are used to interpret 
light from distant galaxies in terms of the properties of the systems (e.g., age and 
metallicity).  These age and metallicity measurements form a major component of 
our understanding of galaxy formation and evolution.

The study of white dwarfs with LSST will naturally extend to stellar populations 
such as nearby star clusters (\autoref{sp:sec:clusters}).  LSST 
will detect the tip of the white dwarf cooling sequence in star clusters located over 20~kpc 
from the Sun.  It will also completely map the entire white dwarf cooling 
sequence in nearby globular and open clusters.  For example, the faintest white 
dwarfs in a cluster with $t$ = 1~Gyr have $M_V$ = 13, and will be seen out to 
8~kpc.  The white dwarf cooling sequences of these clusters provide 
age and distance measurements \citep{hansen07}.  
This age measurement is not affected by our knowledge of rotation, 
diffusion, overshooting, even metallicity, and is, therefore, independent of 
the main sequence turn-off approach.  By fixing the age and distance of the stellar 
population using white dwarf cooling theory, we will be able to test stellar 
evolution models in exquisite detail and constrain many of the microphysics.  
These improved models will directly impact our ability to deconvolve the colors 
of distant galaxies using population synthesis methods.

In \autoref{fig:WDcoolages}, we compare the main sequence turn-off age with 
the white dwarf cooling age for the handful of open star clusters where both 
measurements exist (DeGennaro et~al.\ 2009, in preparation).  This work has already 
shown that synthetic color-magnitude diagrams, based on various sets of 
theoretical isochrones, that do not adopt convective core-overshooting yield 
ages are too low to fit the white dwarf cooling measurements 
\citep{kalirai01a,kalirai04b}.  LSST will increase the sample of 
clusters in which these measurements exist by over an order of magnitude, and thus allow these comparisons to be made over a substantial range in age 
and metallicity to test a broad region of parameter space in the models. 

\subsection{White Dwarfs as Probes of Stellar Evolution}

As an intermediate or 
low mass star evolves off the main sequence and onto the red giant and asymptotic giant branch, it quickly sheds its 
outer layers into space.  
The mass loss mechanisms (e.g., helium flash and thermal pulses on the 
asymptotic giant branch) are poorly understood theoretically \citep{habing96} 
and observational constraints are rare given the very short lifetimes 
of stars in these phases ($\sim  $10$^5$ years) and heavy obscuration by dusty shells.  The end products of this stellar evolution 
are white dwarfs, and studying these stars in detail beyond the initial 
imaging observations can directly constrain the total integrated stellar 
mass loss.

\begin{figure}
\centering\includegraphics[width=0.40\linewidth]{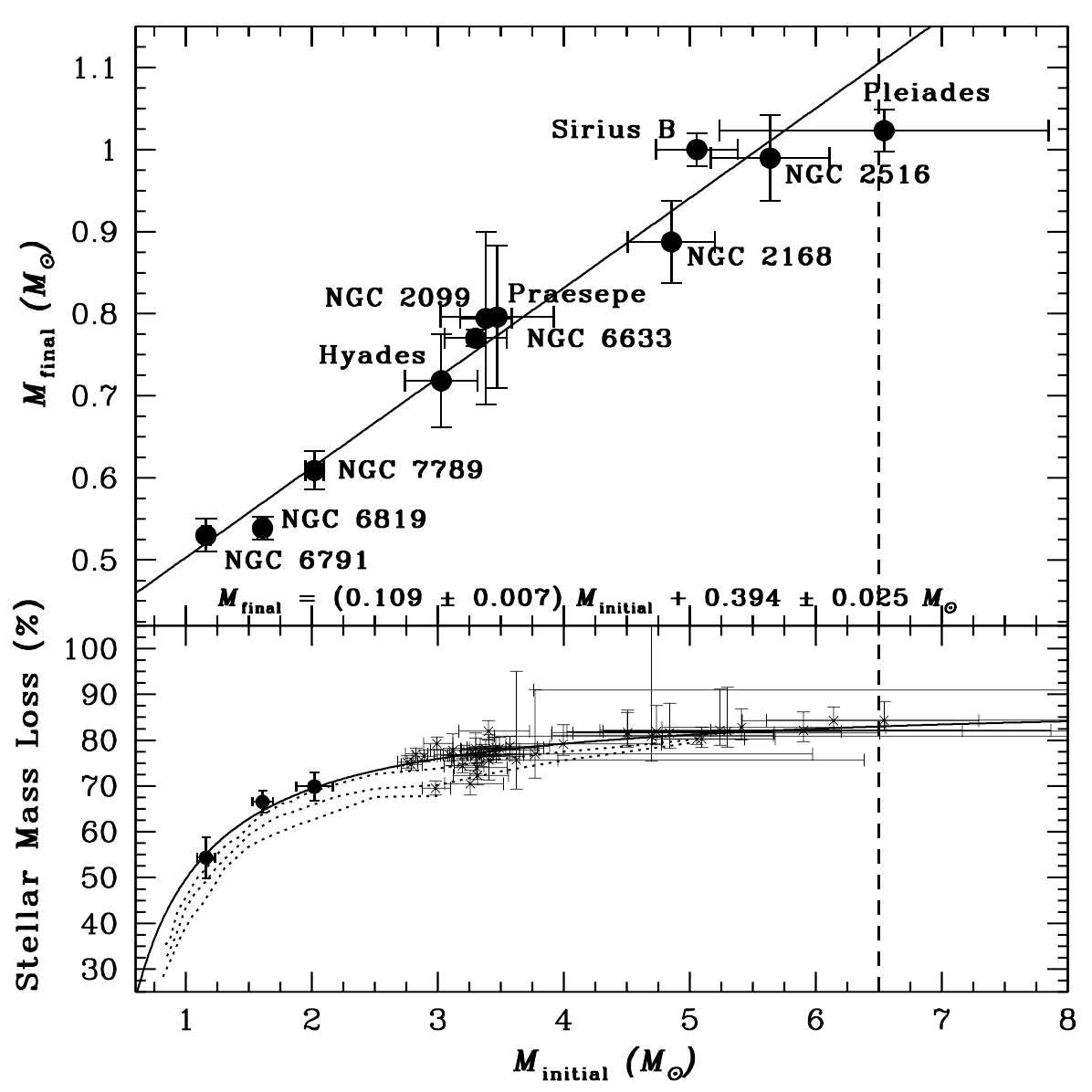}
\caption{{\it Top}: The relation between the masses of white dwarf progenitors and
  their final masses from \cite{kalirai08}, 
with best linear fit.  The entire white dwarf population of a given
cluster is represented by a single data point.  
For the older clusters (e.g., lower initial masses), the white dwarf cooling 
lifetimes are negligible relative to the age of the cluster, and, therefore, all of 
the stars at the top of the cooling sequence came from progenitors with the same 
approximate mass.  For the younger clusters, this method averages over small ranges 
in initial and final mass within each star cluster.
The relation shows a roughly linear rise in the remnant mass as a function of the 
initial mass (see empirical relation on the plot).  {\it Bottom}: The lower panel illustrates 
the total integrated stellar mass lost through standard 
evolution, directly constrained from the initial-final mass relation.  The individual 
data points, except at the low mass end, correspond to individual progenitor-white dwarf 
measurements.}
\label{fig:ifmr}
\end{figure}

As a follow up study to the initial imaging observations that LSST 
will undertake, the brightest (i.e., youngest) white dwarfs in nearby stellar 
populations can be spectroscopically measured with multi-object 
technology on 8 -- 10-m ground-based telescopes (and possibly with TMT or GMT).  The spectra of the DA white dwarfs are remarkably simple, 
showing pressure broadened Balmer lines caused by the thin hydrogen 
envelope in the atmosphere of the stars.  These Balmer lines can be 
easily modeled to yield the temperature and gravity of the stars, 
and, therefore, the individual stellar masses \citep{bergeron95}.  
These mass measurements can be uniquely connected to the initial 
mass of the progenitor star for each white dwarf (e.g., the total 
cluster age is the sum of the white dwarf cooling age and the 
main sequence lifetime of the progenitor), and, therefore, an 
initial-to-final mass relation can be constructed as shown in 
\autoref{fig:ifmr} \citep{kalirai08}.  

LSST will revolutionize our study of the initial-to-final mass relation.  The new relation, consisting of hundreds 
of data points over the full range in initial mass of stars that will 
form white dwarfs, will directly constrain the amount of mass loss that 
occurs through stellar evolution. This forms a powerful input to chemical 
evolution models of galaxies (including enrichment in the interstellar 
medium) and, therefore, enhances our understanding of star formation 
efficiencies in these systems \citep{somerville99}.  Moreover, LSST will 
provide new insights into how stellar evolution and mass loss rates are 
affected by metallicity variations. Theoretically it is expected 
that mass loss rates in post main sequence evolution depend on 
metallicity \cite[e.g., see ][and references therein]{kalirai07}.  
These dependencies can be directly tested by constructing relations 
specifically for clusters of different metallicities that LSST will observe.

LSST's detection of white dwarfs in the youngest stellar systems will also 
provide new insights into the threshold mass that separates white 
dwarf production from type II SNe formation.  For example, the most 
massive singly evolved white dwarf that can be connected to a progenitor 
mass in \autoref{fig:ifmr} is currently the Pleiades star, which has 
$M_{\rm initial}$ = 6.5~$M_\odot$.  However, the remnant star of this 
progenitor is 1.0~$M_\odot$, much smaller than the Chandrasekhar limit, 
suggesting that more massive {\it singly} evolving white dwarfs remain to be found 
in star clusters.  Theoretically, this threshold mass is difficult to constrain 
as models do not include rotation and are very sensitive to overshooting and 
rotationally induced mixing.  A shift in the critical mass from 9~$M_\odot$ (as 
suggested by an extrapolation of the present initial-final mass relation above) to 
6~$M_\odot$ \cite[as suggested by several models for the ignition of 
carbon in the core of the star, e.g.,][]{girardi00} results 
in a 80\% increase in the numbers of type II SNe based on a Salpeter mass 
function.  This changes the amount of kinetic energy imparted into the 
inter-galactic medium (IGM) and would, in fact, be in better agreement with some observations 
of the IGM \citep{binney01} as well as the mass function of stars in 
the solar neighborhood \citep{vandenbergh91}.  Such an effect should be seen as a steepening 
of the initial-final mass relation at higher masses, which LSST will probe by sampling 
white dwarf populations in successively younger systems.  For example, 
LSST's detection of white dwarfs in a cluster of age 50~Myr, where 8~$M_\odot$ 
stars are still burning hydrogen on the main sequence, would suggest that 
the critical mass is above 8~$M_\odot$.  Such young open clusters do exist 
in the southern hemisphere but lack current deep imaging data (e.g., NGC~2451 and 
NGC~2516).


\subsection{Rare White Dwarf Species and the Physics of Condensed Matter}

The temporal coverage of LSST observations in multiple filters will lead to exciting 
discoveries of exotic stellar species that are astrophysically important.  These 
will include eclipsing short period double degenerate systems, transits of 
white dwarfs by planetary bodies and other accretion disk objects down to 
asteroidal dimensions (see also the discussion in
\autoref{sec_transits}), and a very large number of pre-cataclysmic
variable/post-common envelope  
systems.  The synoptic nature of LSST will be critical in identifying these 
systems.  For some classes, such as eclipses by planets, it may that
the LSST cadence will be adequate only for identifying candidates requiring follow-up on smaller telescopes with a much faster cadence.  

Eclipsing short period double-degenerate systems are of great interest 
for several reasons. Follow-up studies of such systems will yield direct 
determinations of white dwarf radii as well as astrometric masses, which 
can be used to accurately populate the degenerate mass-radius relation.  
The catastrophic merger of double degenerate systems is believed to be 
one potential source of type Ia supernova events.  Identifying such systems 
through their eclipse signals with LSST is a real possibility.  Continued 
monitoring of such systems could reveal the gravitational decay rate of 
the mutual orbit.  It is even conceivable that particular systems found 
with LSST could be linked to specific gravitational wave signals detected 
by Laser Interferometer Space Antenna (LISA), since merging white dwarf systems are thought to constitute a major source 
of the Galactic noise background for LISA.

Some white dwarfs are now known to be orbited by dusty disks and 
even more show spectral features of heavy elements (Si, Mg, Ca, 
Fe, and so on), which quickly settle out of the atmosphere indicating
on-going accretion.  In both cases the source of the dust is believed to be 
collisions of asteroidal bodies in tight orbits around the white dwarf.  
Because white dwarfs have small diameters, it is quite possible that 
favorable orbital plane orientations will reveal transits of substantial 
bodies from the size of massive Jupiters to asteroids having diameters 
of tens of km.  The gravitational perturbations of such massive bodies 
are thought to play a role in promoting asteroidal collisions and in 
maintaining any dusty ring structures that result. 

Finally, it should be possible to identify a large number of
pre-cataclysmic variable and 
post-common envelope systems.  In general, eclipses (although helpful) 
are not even necessary since reflection effects produced by the hot 
white dwarf on the low mass secondary are a frequent signature of 
these sources.  Having a large number of such systems to study will 
help map out the spectrum of stellar masses and 
orbital separations that constitute the end states of post-common 
envelope evolution. 

With large numbers of detected white dwarfs, LSST can select those that are 
variable to the limit of LSST's photometric precision ($\sim1$\%), and therefore 
identify new candidate pulsating white dwarfs.  Follow-up time-series photometry 
of these candidates on other telescopes will lead to a substantial number of new white dwarf 
pulsators, and, therefore, provide a more accurate mapping of the boundaries of 
the known white dwarf instability strips for pulsation (DAV - H, DBV - He, PG1159
- C) in the HR diagram and in $\log g$ vs $T_{\rm eff}$, and also 
allow exploration to search for previously unknown instability strips along 
the white dwarf cooling sequence.  A more detailed discussion of LSST's connection to 
pulsating white dwarfs is provided in \autoref{anjum:pulsWDs}.

LSST will provide a new test for the internal physics of white dwarf stars.  
The low luminosity end of the white dwarf luminosity functions (WDLF), 
$\log(L/L_\odot)<-3$, contains information about the equation of state
of condensed (degenerate) matter.  The shape of the disk WDLF at  
the turnover (discussed in \autoref{sec:sp:wdpop}) due to the disk's finite age is affected 
by the release of latent heat of crystallization of the carbon-oxygen 
white dwarf core. The release of latent heat provides an energy source 
in an otherwise dead star and slows the white dwarf cooling 
process. This slowdown manifests itself as an increase in number density 
of white dwarfs over the luminosity range corresponding to the crystallization
event. Even more intriguing is the possibility of having a halo WDLF
that is sufficiently populated that we can fully
resolve the crystallization bump. LSST's large white dwarf sample will determine
if the crystallization bump is indeed present, and if so, at what luminosity 
(i.e., age), providing new constraints on the equation of state for 
carbon-oxygen white dwarfs.

\subsection{The LSST White Dwarf Model Sample } 
\label{sec:sp:wd_sample}

In the following sections we calculate the expected distributions of 
white dwarfs that LSST will see. The main purpose of these simulations
is to estimate the accuracy LSST will obtain in calibrating the white dwarf
photometric parallax relation, kinematically separating the 
disk and halo populations, and measuring their 
luminosity functions. 

In order to generate a simulated sample of disk and halo
white dwarfs, five sets of quantities need to be adopted: 
\begin{enumerate}
\item The expected astrometric and photometric measurement errors.
\item The spatial distribution for each Galaxy component.
\item The distributions of three velocity components.
\item The bolometric luminosity functions.
\item The mapping from bolometric luminosity to 
      broad-band luminosity in each LSST bandpass.
\end{enumerate}
The astrometric and photometric measurement errors are computed
as described in \autoref{chap:common}. We proceed with detailed 
descriptions of the remaining quantities.

\subsubsection{The Spatial Distribution } 

LSST will detect white dwarfs to distances much
larger than the scale heights and lengths of the Galactic disk. Hence the spatial variation of volume density in the Galaxy must be taken
into account. We assume that the spatial distribution of 
white dwarfs traces the distribution of main sequence stars,
both for halo and disk populations (the impact of their
different ages is handled through adopted luminosity functions).  We
ignore bulge white dwarfs in the simulations as they represent only a
small fraction of the population. 
The adopted spatial distribution of main sequence stars,
based on recent SDSS-based work by \citet{Juric2008} 
is described in \autoref{Sec:stellarCounts}. 


\subsubsection{The Kinematic Distributions} 
 
We assume that the kinematics (distributions of three velocity components)
of white dwarfs are the same as the corresponding distribution of
main sequence stars, both for halo and disk populations.
The adopted kinematic distribution of main sequence stars is 
based on recent SDSS-based work by \citet{Ivezic2008}. 

\subsubsection{The White Dwarf Luminosity Function }

For disk stars, we adopt the measured luminosity function based on 
SDSS data \citep{Harris2006}. Using their Figure 4, we obtained
the following parameters for a power-law approximation to the
measured bolometric $\Phi$ (the number of white dwarfs per cubic 
parsec and magnitude),
\begin{eqnarray}
\log\Phi = -2.65 + 0.26\,(M_{\rm bol} -15.3) \,\,\,\, {\rm for} \,\, 7<M_{\rm bol}<15.3 \nonumber \\
\log\Phi = -2.65 - 1.70\,(M_{\rm bol} -15.3) \,\,\,\, {\rm for} \,\, 15.3<M_{\rm bol}<17.0,       
\end{eqnarray}
which agrees with the data to within 10\% at the faint end. 
The observational knowledge of the halo 
white dwarf luminosity function is much poorer.  Theoretical predictions 
\citep[][and references therein]{Torres2005} indicate an overall shift
of the halo luminosity distribution toward fainter absolute
magnitudes due to its larger age compared to the disk. Motivated by 
these predictions and the desire to 
test the ability to distinguish different luminosity functions when
analyzing the simulated sample, we simply shift the \citet{Harris2006}
luminosity function by 0.7 mag toward the faint end. 

\begin{figure}
\begin{center}
\includegraphics[width=0.5\linewidth]{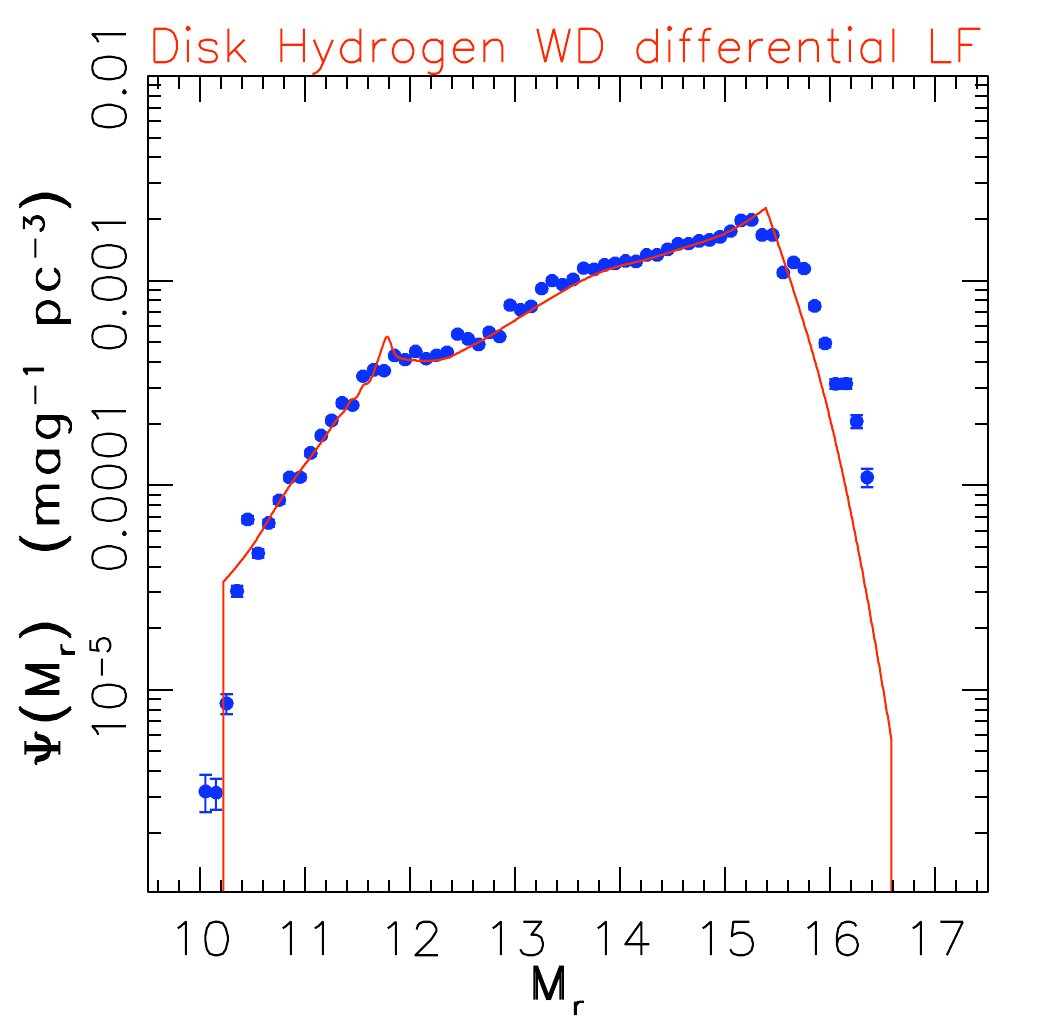}
\caption{Simulated differential luminosity function for candidate hydrogen 
white dwarfs in the disk sample (normalized to solar neighborhood). 
The dots with error bars show the result obtained by binning the
cumulative luminosity function computed using Lynden-Bell's 
$C^-$ method in 0.1 mag wide $M_r$ bins (based on $\sim$ 200,000 stars). 
The red line shows the input luminosity function in the simulation. Note the 
``feature'' in the input luminosity function at $M_r=11.8$.}
\label{Fig:LFdisk}
\end{center}
\end{figure}

\begin{figure}
\begin{center}
\includegraphics[width=0.5\linewidth]{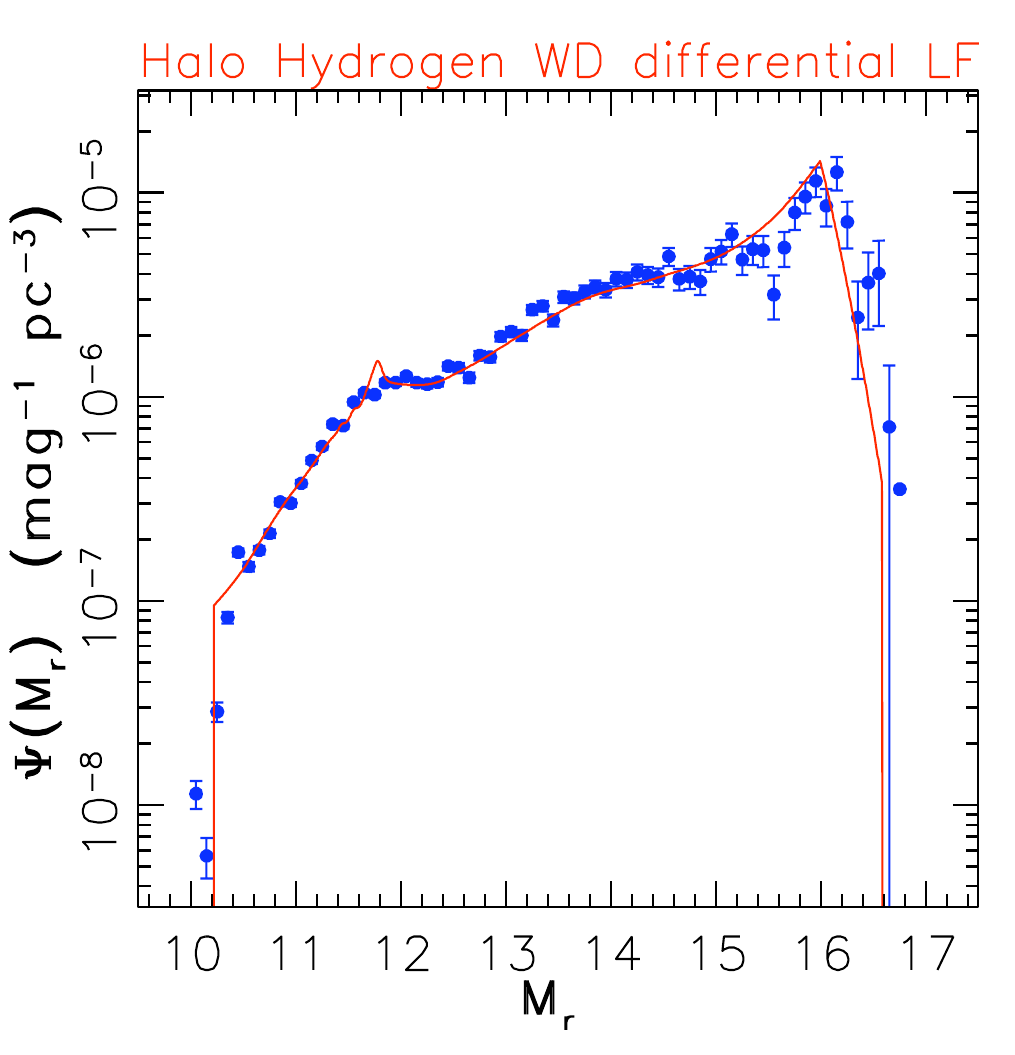}
\caption{Similar to \autoref{Fig:LFdisk}, now showing the 
  luminosity function for the candidate halo sample. 
}
\label{Fig:LFhalo}
\end{center}
\end{figure}

We re-express the luminosity function per unit $M_r$ magnitude,
$\Phi_r$ by multiplying by 
$dM_{bol}/dM_r$, determined from the spectral energy distribution, described below. The
resulting luminosity functions for disk and halo 
white dwarfs are shown in Figures~\ref{Fig:LFdisk} and
\ref{Fig:LFhalo}, respectively. The integral of the adopted disk luminosity 
function is 0.0043 stars pc$^{-3}$ (about 1/10 of the integrated
luminosity function for main sequence stars). The disk luminosity 
function reaches its maximum around $M_r=15.4$, and the halo 
luminosity function at $M_r=16$. Both luminosity functions show 
a $\sim$ 0.2 mag wide and 20-30\% strong ``feature'' at 
$M_r\sim11.8$ which is due to the behavior of $dM_{bol}/dM_r$. 


We assume that 90\% of all white dwarfs are hydrogen (DA) 
white dwarfs and the rest are helium (DB) white dwarfs, but assumed
the same luminosity function for both. 


\subsubsection{The White Dwarf Spectral Energy Distribution}

We use models by \citet{bergeron95}, which produce color
tracks that agree with SDSS measurements at the $\sim$0.02 mag level
\citep{eisenstein06}. Using a sample of $\sim$ 10,000 white 
dwarfs with SDSS spectroscopic data, \citet{eisenstein06} found a 
very narrow distribution (0.1 dex) of $\log g$ centered on 
$\log g=7.9$. Motivated by this result and the desire to
simplify analysis of the simulated sample, we adopt a fixed
value of $\log g=8.0$ (Bergeron's models are computed with a $\log g$
step of 0.5 dex). Hence, for a given type of white dwarf
atmosphere (hydrogen vs. helium), the models provide unique relationships 
between $M_r$ and all relevant colors (including bolometric 
corrections). For hydrogen white dwarfs with $\log g=8.0$,
$M_r=15.4$ corresponds to an effective temperature of 4500 K,
mass of 0.58 M$_\odot$ and age of 7.6 Gyr. For $M_r=16$, the
effective temperature is 3900 K, the mass is unchanged and
the age is 9.3 Gyr. A 13 Gyr old hydrogen white dwarf, according 
to Bergeron's models, would have $M_r=17.4$ and an effective
temperature of 2250 K. 

\subsubsection{Preliminary Analysis of the Simulated White Dwarf Sample } 

The simulated sample includes $\sim$ 35 million objects with $r<24.5$
over the whole sky. Here we briefly describe the expected counts
of white dwarfs in the main (deep-wide-fast, DWF; see \autoref{sec:design:cadence}) LSST survey,
discuss how objects with good trigonometric parallax measurements
can be used to derive an empirical photometric parallax relation,
and how this relation can be used with proper motion measurements
to separate disk and halo candidates. We conclude with preliminary
estimates of the accuracy of disk and halo white dwarf luminosity 
function measurements.

\subsubsection{Counts of Simulated White Dwarfs }


\begin{figure}
\begin{center}
\includegraphics[width=0.45\linewidth]{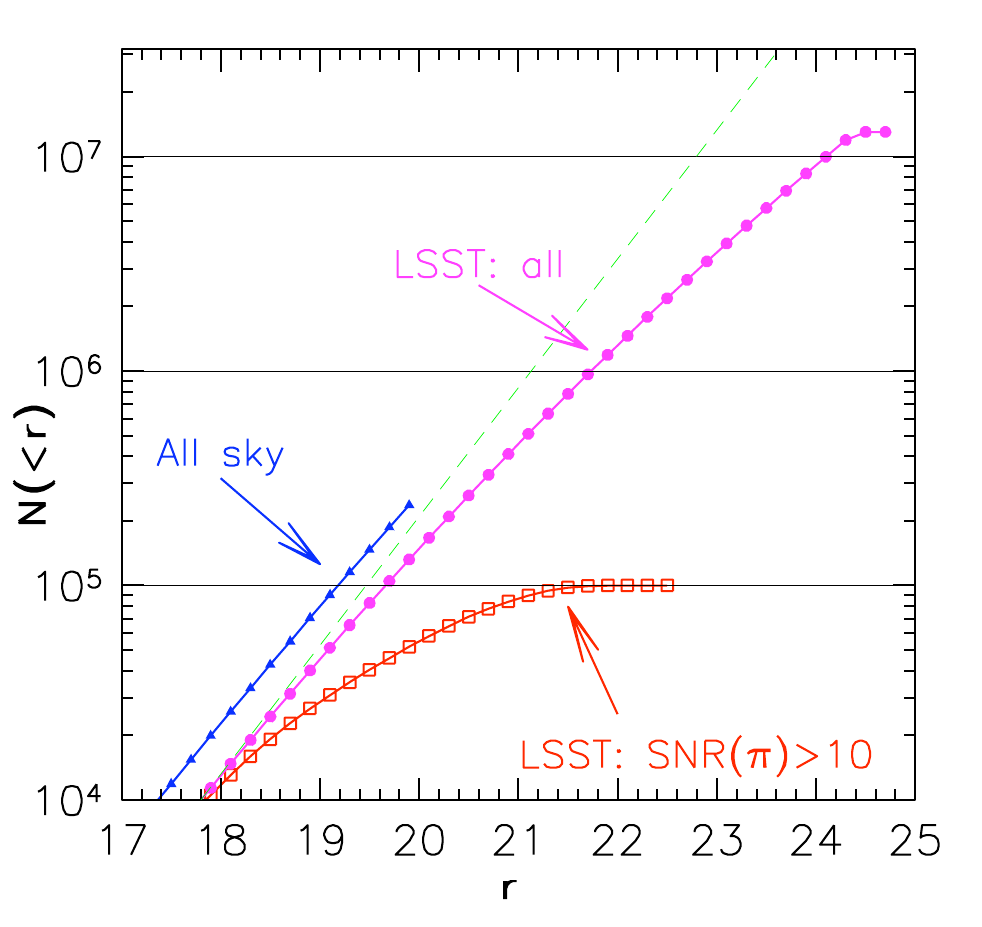}
\caption{A comparison of cumulative white dwarf counts for 
several samples. The triangles (blue curve) show the counts
over the full sky in the magnitude range corresponding to Gaia
survey ($r<20$). The squares (red curve) show the counts
of white dwarfs from the main LSST survey (about 1/2 of the
sky) that have anticipated signal-to-noise ratio for trigonometric 
parallax measurements greater than 10. The circles (magenta
curve) show the counts of all white dwarfs from the main 
LSST survey that will have proper motion measurements 
($r<24.5$). The predicted magnitudes are not corrected for
the interstellar dust extinction. The dashed line shows the
behavior expected for a spatially uniform distribution of sources
($\log[N(<r)] \propto 0.6\,r$) - the impact of Galactic structure
is evident in the much shallower slope for simulated counts around 
$r=24$.}
\label{Fig:counts}
\end{center}
\end{figure}


The main DWF LSST survey is expected to deliver about 1000
visits (summed over all bands) over a $\sim$ 20,000 deg$^2$
area, and without including the Galactic plane. \autoref{Fig:counts} compares cumulative white dwarf counts 
for several samples. The simulations predict that Gaia's all-sky 
survey will
detect about 240,000 white dwarfs with $r<20$. Of those, 
about 1,200 will be halo white dwarfs. These counts are
in fair agreement with the results of \citet{Torres2005}
who simulated Gaia's performance on white dwarfs. We have also
compared the simulated counts to photometrically
selected white dwarf candidates from SDSS (see bottom
left panel in Fig.~24 of \citealt{Ivezic2007}).
We selected 355 white dwarf candidates over 203 deg$^2$ defined by 
$330^\circ < \alpha < 50^\circ$ and $|\delta|<1.267^\circ$; we
required that the objects be non-variable (rms scatter less than 0.07 mag
in $g$) and have $16 < g < 20$, $-0.3 < u-g < 0.5$, $-0.4 < g-r < -0.2$.
With the same color-magnitude criteria, the simulated sample
includes 340 objects in the same sky region. Given that 
the observed color-selected sample might include some
contamination, this is a robust verification of the model
count normalization.  The simulations do not include the effects of
interstellar extinction, but the extinction over this area is small,
and most white dwarfs are close enough to be in front of the majority
of the dust. 

As illustrated in \autoref{Fig:counts}, there will
be about 13 million white dwarfs with $r<24.5$ in the DWF
survey. While the number of all detected white dwarfs in 
LSST will be much larger (about 50 million for the $r<27.5$
limit of co-added data),
here we focus only on objects with $r<24.5$ because they 
will have, in addition to highly accurate photometry, 
trigonometric parallax and proper motion measurements. 
In particular, about 375,000 simulated objects have 
anticipated signal-to-noise ratio for trigonometric 
parallax measurements greater than 5 and 104,000
greater than 10. This latter subsample (whose cumulative counts are
shown in \autoref{Fig:counts}) can be used 
to empirically constrain photometric parallax relations 
for hydrogen and helium white dwarfs and to train 
color-based classification algorithms, as described
next. In the remainder of this analysis, we assume
no knowledge of the input model parameters except 
when estimating the performance parameters such as
sample completeness and contamination.

\subsubsection{White Dwarf Photometric Parallax Relations }

The distribution of 
the difference between trigonometric and true distance moduli for the 104,000
white dwarfs with parallax S/N$>10$
is close to Gaussian, with a median value of 
$-0.03$ mag and an rms scatter of 0.15 mag. For the subset of 10,000
objects with $r<18$, the rms scatter is 0.10 mag and the bias is below
0.01 mag. 

\begin{figure}
\begin{center}
\includegraphics[width=0.45\linewidth]{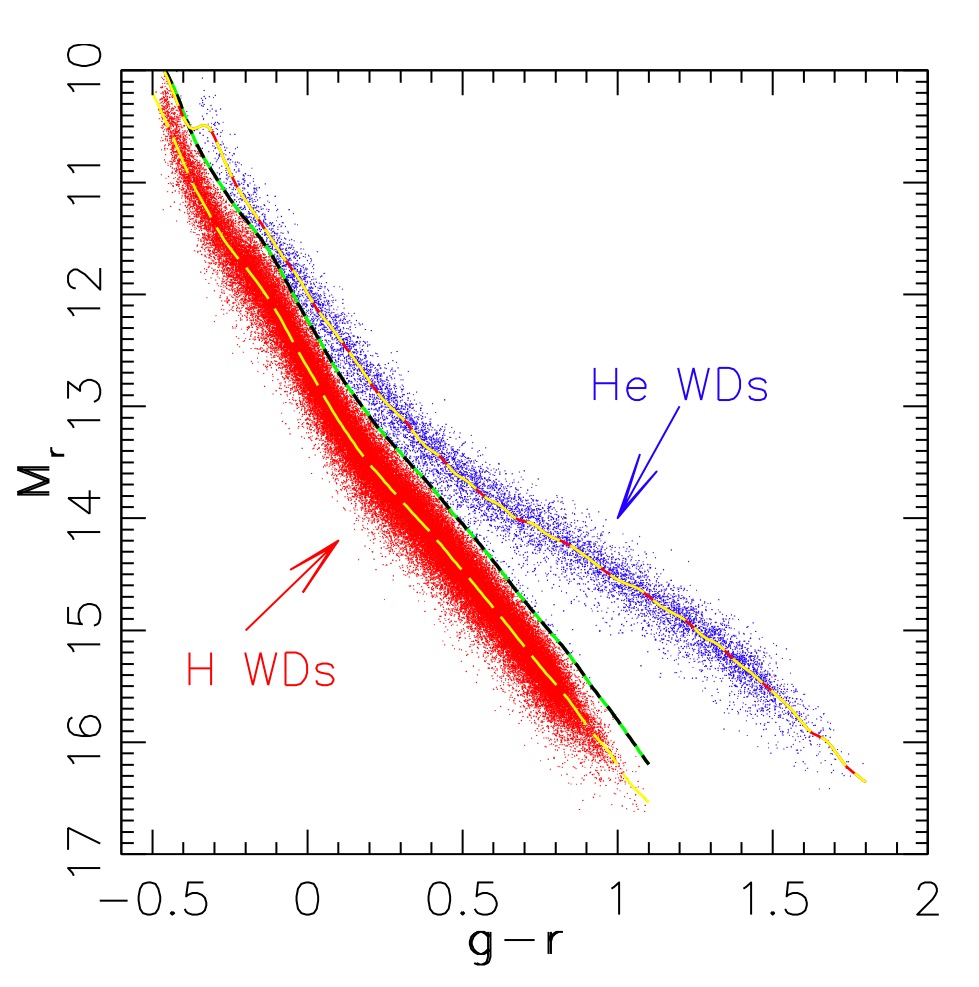}
\caption{The calibration of the photometric parallax relation, 
$M_r(g-r)$, for white dwarfs. The $M_r$ values are based 
on trigonometric parallax and ``measured'' $r$ band magnitudes.
The dots represent $\sim$ 100,000 simulated objects with the 
signal-to-noise ratio for trigonometric parallax measurements 
greater than 10. The middle dashed line is the color-magnitude 
separator described in the text. The other two lines are the median 
$M_r$ vs. $g-r$ photometric parallax sequences. The true relations
used to generate the simulated sample are indistinguishable
(rms $\sim$ 0.01 mag) from these empirically determined median 
values.}
\label{Fig:pp}
\end{center}
\end{figure}

The absolute magnitude based on ``measured'' trigonometric 
parallax as a function of ``measured'' $g-r$ color is 
shown in \autoref{Fig:pp}. The two sequences that correspond
to hydrogen and helium white dwarfs are easily discernible.  
A simple separator of hydrogen and helium color-magnitude
sequences is obtained by shifting the median $M_r$ vs. $g-r$ 
curve for hydrogen white dwarfs by 0.4 mag towards the bright
end. A slightly better choice would be to account for the
shape of the helium sequence as well. 
The application of this separator results in
correct classification for 99.6\% of the objects in the candidate 
hydrogen sample and for 96.3\% of the objects in the candidate
helium sample. 

\subsubsection{Photometric Separation of Hydrogen and Helium White Dwarfs}

The separation of hydrogen and helium white dwarfs 
based on the $M_r$ vs. $g-r$ diagram is possible only
for objects with high S/N trigonometric 
parallax measurements. Since such objects represent 
only about 1\% of the full $r<24.5$ LSST white dwarf
sample, a color separation method is required to
classify the latter sample. Although helium white
dwarfs represent only 10\% of all objects, the
differences in $M_r$ vs. $g-r$ relations between
helium and hydrogen white dwarfs might significantly 
bias the luminosity function determination. 

\begin{figure}
\begin{center}
\includegraphics[width=0.5\linewidth]{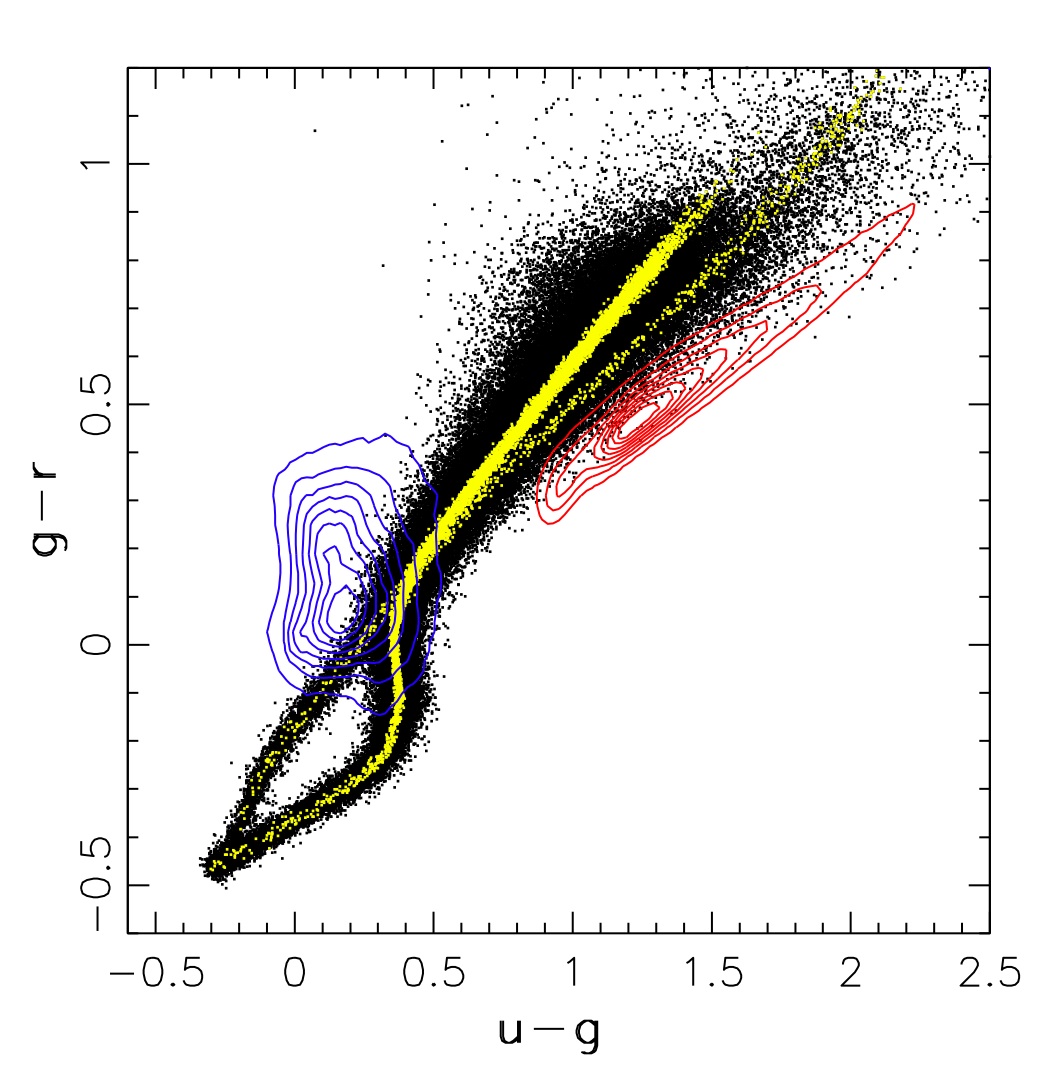}
\caption{The distribution of the simulated white dwarfs in the
$g-r$ vs. $u-g$ color-color diagram. Black points show objects
with $r<24.5$ and $b>60^\circ$. Yellow points show a subsample
of predominantly brighter sources that have 10$\sigma$ or 
better ``measurement'' of the trigonometric parallax. The two
sequences correspond to He and H white dwarfs. The distribution 
of low-redshift ($z<2.2$) quasars observed by SDSS is shown by 
blue contours. The blue part of the stellar locus (dominated by
F and G stars), as observed by SDSS, is shown by the red contours.
LSST photometry will be sufficiently accurate not only to
separate white dwarfs from quasars and main sequence stars, 
but also to separate hydrogen from helium white dwarfs (the sequences
do not overlap in the multi-dimensional color space, see text).}
\label{Fig:uggr}
\end{center}
\end{figure}

We use the two candidate samples with good trigonometric 
parallax measurements to quantify their multi-dimensional 
color tracks. \autoref{Fig:uggr} shows the two-dimensional 
projection of these tracks. At the hot end, the tracks 
for hydrogen and helium objects are well separated. 
Although they appear to cross around $g-r=0.2$, they
are still separated in the four-dimensional color space
spanned by the $u-g$, $g-r$, $r-i$ and $i-z$ colors\footnote{Reliable
  colors are not yet available for the $y$ band 
so we do not consider it here.}.


For each sample, we compute the median $u-g$, $r-i$ and 
$i-z$ color for each 0.01 mag wide bin of $g-r$ color.
Using these tracks, for each star we compute the 
shortest distance to each locus, denoted here $D_{He}$
and $D_H$. The difference between these two four-dimensional
color distances (4DCD) can be used as a simple color-based 
classifier. For the training sample, which has small 
photometric errors due to the relatively bright flux limit
imposed by requiring high trigonometric parallax signal-to-noise
ratio, the separation is essentially perfect (mis-classification
rate, or sample contamination, is less than 1\%). 

\begin{figure}
\begin{center}
\includegraphics[width=0.5\linewidth]{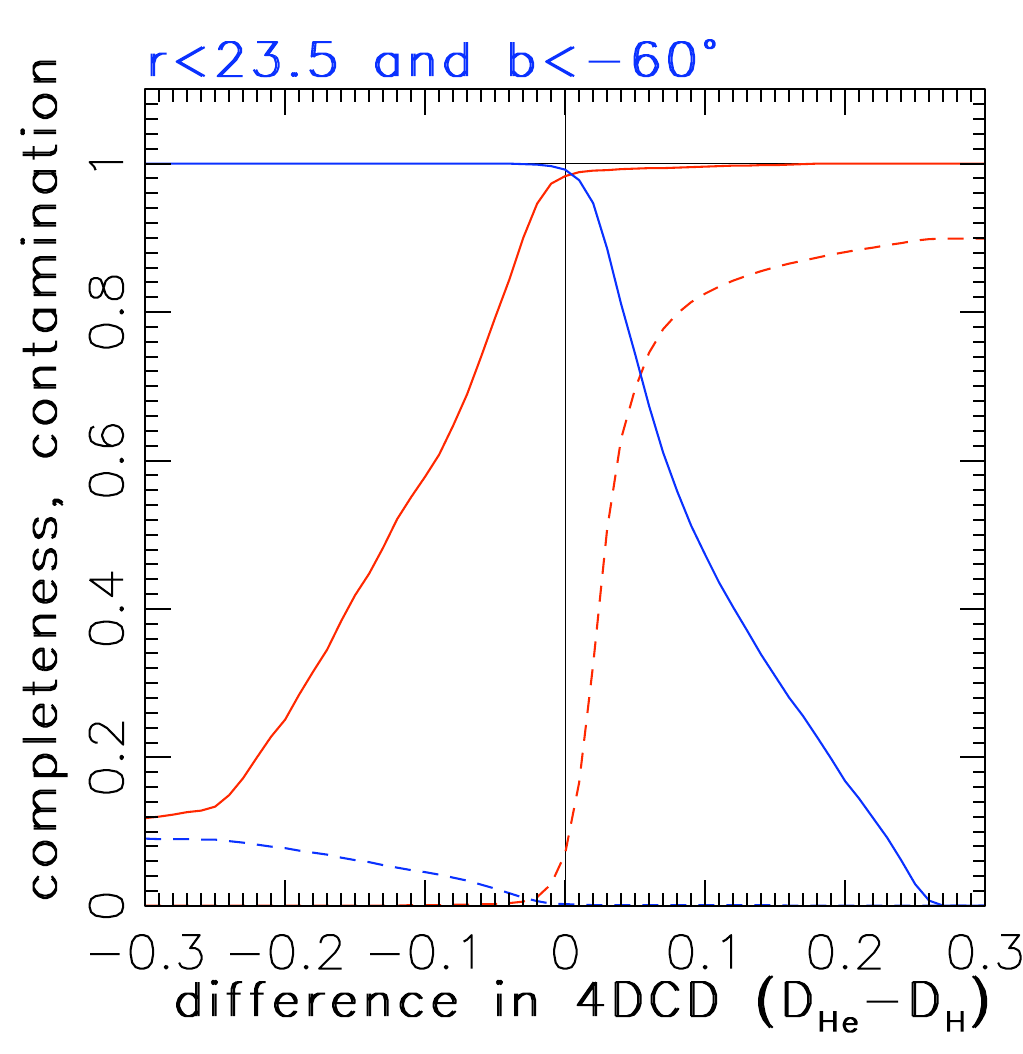}
\caption{The completeness and contamination for color-selected 
subsamples of hydrogen and helium white dwarfs, as a function
of the difference between distances to each four-dimensional color
sequence ($\delta_{4DCD}=D_{He}-D_H$). The solid 
lines show completeness and dashed lines show contamination. 
The blue lines correspond to 
hydrogen subsample, and red lines to helium subsample.  Objects 
are classified as helium white dwarfs if their $\delta_{4DCD}$ is 
smaller than the adopted cut-off value. The panel shows a 
flux-limited sample with $r<23.5$.}
\label{Fig:4DCDsep}
\end{center}
\end{figure}

We assess the performance of color separation at the faint 
end by resorting to true input class, and study the
completeness and contamination of  candidate samples as
a function of $\delta_{4DCD}=D_{He}-D_H$ (see 
\autoref{Fig:4DCDsep}). The optimal value of $\delta_{4DCD}$
for separating two object types is a trade-off and depends on 
whether a particular science case requires high completeness
or low contamination. Typically the best $\delta_{4DCD}$ value
is not zero because hydrogen white dwarfs are ten times as numerous 
as helium white dwarfs. These effects can be elegantly 
treated using the Bayesian formalism developed by \citet{Mortlock2008},
hereafter MPI08. 
Here we follow a simpler approach and, informed by the results 
shown in \autoref{Fig:4DCDsep}, adopt $\delta_{4DCD}=-0.05$
for the rest of the analysis presented here. For $r<23.5$,
the candidate helium
sample completeness and contamination are 79\% and 0.2\%, 
respectively (see the right panel in \autoref{Fig:4DCDsep}). Where $r<24.5$, the completeness of 99\% with a contamination of 3\% for the candidate
hydrogen sample, and 73\% and 14\%, respectively, for the candidate
helium sample, the degraded but still remarkable performance being attributed to larger photometric errors.


We note that despite high completeness for the helium subsample,
there are ranges of $M_r$, such as $M_r\sim12.5$, where it 
is sufficiently small to induce large systematic errors in 
luminosity function. To properly treat the helium subsample,
a more sophisticated method, such as that described by MPI08,
is required. Nevertheless, the simplistic $\delta_{4DCD}$ 
method used here produces sufficiently clean samples of 
candidate hydrogen white dwarfs for further analysis.

\subsubsection{Kinematic Separation of Disk and Halo White Dwarfs }

The measured proper motion and distance estimate can be used to 
probabilistically assign disk or halo membership, if suitable 
kinematic models exist, for an arbitrary direction on the sky.
In the general case, the observed proper motion depends on a linear combination
of all three velocity components, and the probabilistic class 
assignment can be computed following the approach outlined in MPI08
(the standard method for separating disk and halo stars based on 
reduced proper motion diagram will fail at kpc distances probed 
by LSST, see Appendix B in \citealt{Sesar2008}). In this preliminary
analysis, we limit our sample  
to the region with $b<-60^\circ$, where proper motion primarily 
depends on radial, $v_R$, and azimuthal (rotational), $v_\phi$, 
components, while the vertical velocity component, $v_Z$, is by
and large absorbed into the radial (along the line of sight) 
velocity component. 

\begin{figure}
\begin{center}
\includegraphics[width=0.45\linewidth]{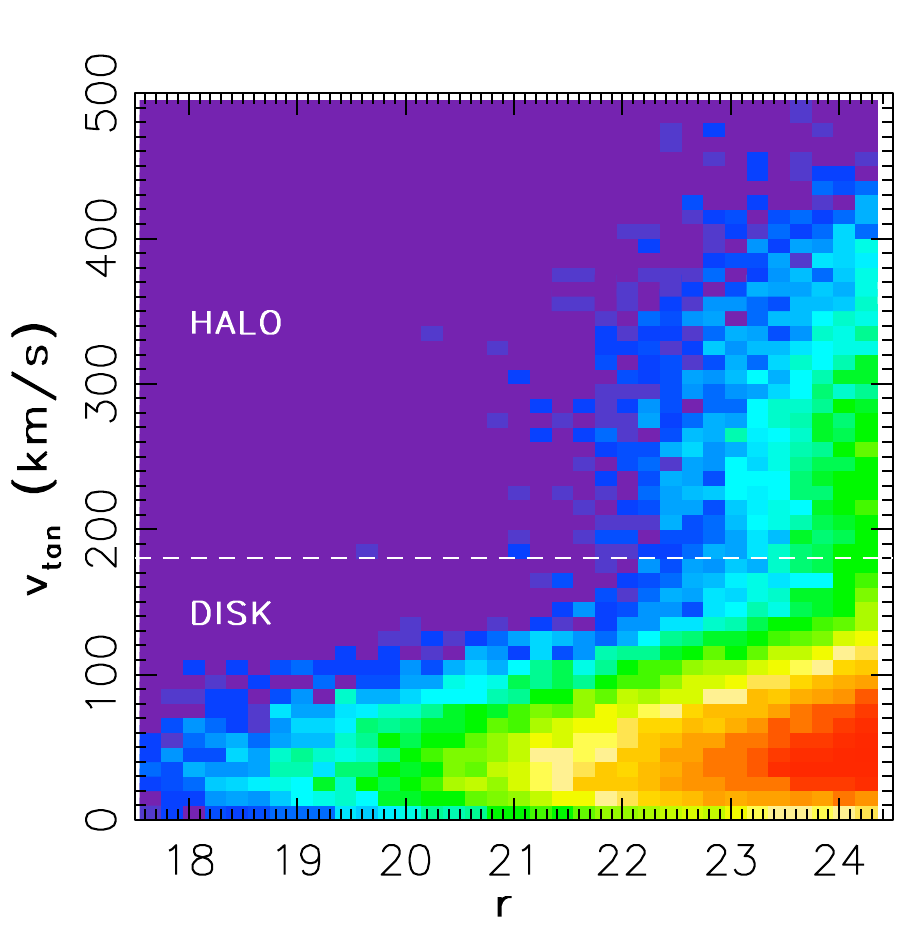}
\caption{The dependence of tangential velocity on apparent magnitude for
white dwarfs with $b>60^\circ$. The map shows counts of stars in each bin
on logarithmic scale, increasing from blue to red. The tangential 
velocity is computed from each star's measured proper motion and 
distance estimate from the photometric parallax relation shown in \autoref{Fig:pp}. At faint
magnitudes ($r>22$), the sample contains a large fraction of halo white 
dwarfs. The horizontal line at 180 \kms\ separates disk and halo stars
with sample completeness and contamination of 99\% and 3\%, 
respectively, for disk stars, and 78\% and 6\%, respectively, 
for halo stars.}
\label{Fig:vtan2D}
\end{center}
\end{figure}

From $\sim$ 273,000 simulated objects with $r<24.5$ and $b<-60^\circ$
(2,680 deg$^2$), we select $\sim$ 250,000 candidate hydrogen white 
dwarfs using the color-based classification described above. We determine 
their distances using a photometric parallax relation, and compute the
absolute value of their tangential velocity, $v_{tan}$. The distribution 
of $v_{tan}$ as a function of measured apparent $r$ band magnitude
for this sample is shown in \autoref{Fig:vtan2D}. The median 
difference between the ``measured'' and true $v_{tan}$ is 3 \kms,
and ranges from 1 \kms\ at distances smaller than 400 pc, to 
30 \kms\ at a distance of 5 kpc.


\begin{figure}
\begin{center}
\includegraphics[width=0.5\linewidth]{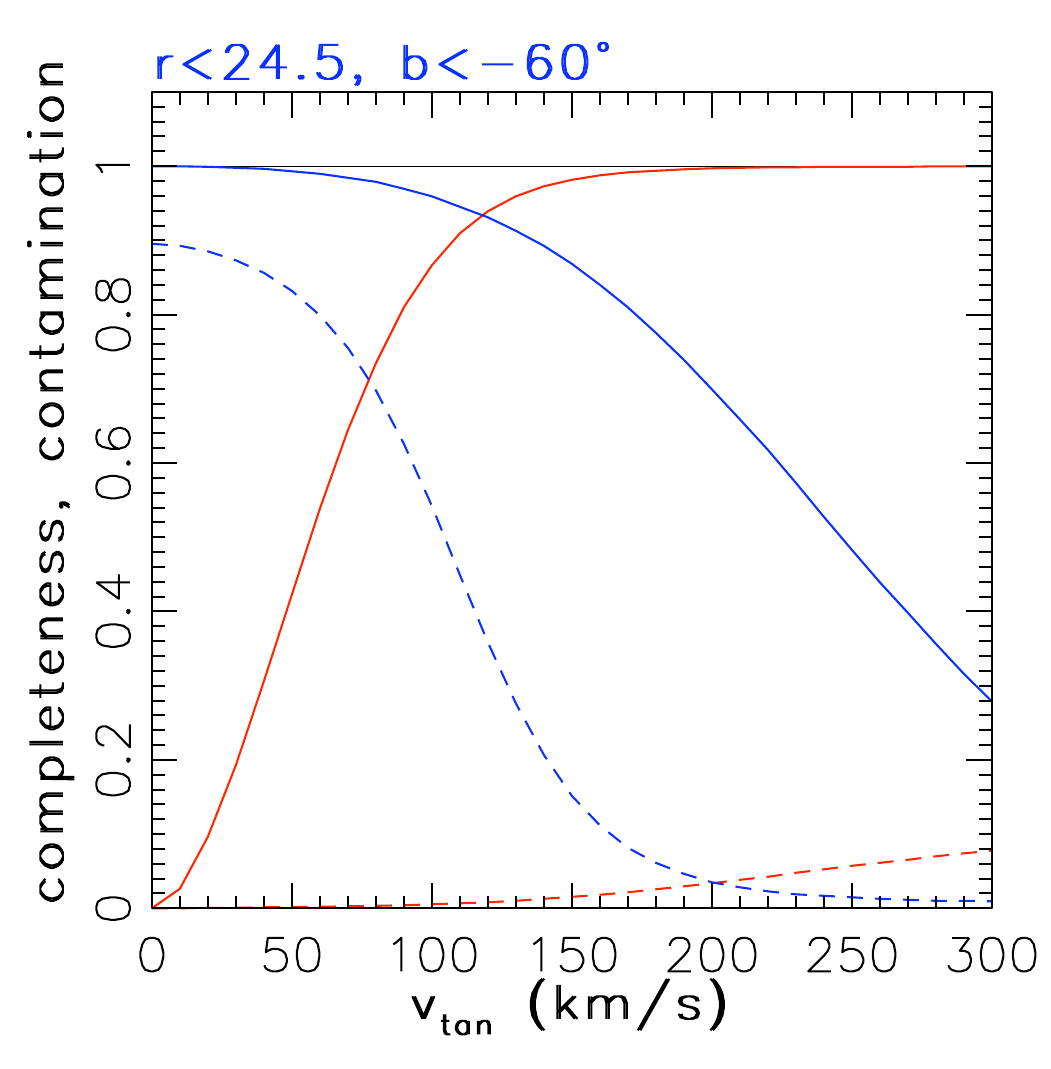}
\caption{The completeness and contamination of candidate 
disk and halo subsamples selected by tangential velocity.
The solid lines show completeness and dashed lines show 
contamination. The blue lines correspond to halo subsample,
and red lines to disk subsample. Objects are classified as 
disk candidates if their tangential velocity is 
smaller than the adopted cut-off value.}
\label{Fig:vtanSep}
\end{center}
\end{figure}

The $v_{tan}$ distribution is clearly bimodal, with high 
$v_{tan}$ stars corresponding to the halo sample. Notably a 
significant fraction of halo white dwarfs is seen only 
at  {$r>22$}. 
Just as in the case of color separation of the hydrogen and helium
sequences, the optimal separation of disk and halo candidates
by $v_{tan}$ includes a trade-off between completeness and
contamination, as illustrated in \autoref{Fig:vtanSep}. 

Informed by \autoref{Fig:vtanSep}, we select $\sim$ 195,000
candidate disk members by requiring $v_{tan}<100$ \kms,
and $\sim$ 19,000 candidate halo members by requiring $v_{tan}>200$ 
\kms. These samples are optimized for low contamination:
the sample contamination for halo candidates is 3.5\% and 
0.5\% for disk candidates. The sample completeness is
70\% for the halo sample and 87\% for the disk sample. We proceed
to determine the luminosity function for these two samples. 


\subsubsection{Determination of Disk and Halo White Dwarf Luminosity Functions }

There are many different methods for estimating a luminosity
function from data \citep[e.g., ][ and references therein]{Kelly2008}. In the case of uncorrelated
variables (the luminosity function is independent of position
once disk and halo candidates are separated.  With real data
this assumption can be tested, e.g., \citealt{Fan2001}). One of the best methods is the $C^-$ method \citep{Lynden-Bell1971}, because it requires binning in only
one coordinate. We used the $C^-$ method to determine the
luminosity functions shown in Figures~\ref{Fig:LFdisk} and 
\ref{Fig:LFhalo}.

Although the sample analyzed here ($b<-60^\circ$) includes
only about 10\% of the total DWF area (and $\sim$2\% of the white
dwarf counts for the entire LSST sample), the random (statistical) 
errors for both disk and halo luminosity functions are negligible. 
The dominant systematic errors (with an rms scatter of about 10\%)
are due to errors in the photometric parallax relation: when the true 
$M_r$ values are used, the $C^-$ method reproduces the input 
luminosity function essentially perfectly. This nearly perfect 
agreement also demonstrates
that the hydrogen vs. helium separation, and disk vs. halo 
separation algorithms have satisfactory performance.
Most importantly, the faint end of the luminosity functions
for both disk and halo samples is correctly reproduced to
within 0.1-0.2 mag.

\section{A Comparison of Gaia and LSST Surveys}  
\label{sec:Gaia}

{\it Laurent Eyer, \v{Z}eljko Ivezi\'{c}, David Monet}

In this section, we compare the design predictions
for the astrometric and photometric performance of the Gaia mission 
and the LSST system. For Gaia's performance, we have collected and 
parametrized predictions from various technical documents, and 
discussed the adopted model with Gaia key technical personnel. 
For LSST performance, we adopted parameters listed in
\autoref{chp:introduction}.  While the various adopted 
errors are probably accurate to much better than a factor of two for 
both Gaia and LSST, their ultimate values cannot be more precisely 
known before their data products are delivered. 

\subsection{Photometric Errors }

To determine photometric errors for Gaia and LSST, we follow the
discussion in \autoref{sec:photo-accuracy}. 
For Gaia, we adopt $\sigma_{sys}=0.001$ mag and $\sigma_{sys}=0.0005$
mag for single transit and the end-of-mission values in the $G$ band, 
respectively. For LSST, we adopt $\sigma_{sys}=0.003$ mag. 
We model random photometric errors (per transit) for Gaia as
\begin{equation}
\label{Gphrand}
         \sigma_{rand} = 0.02 \times 10^{0.2 (G-20)} \,\,\, {\rm (mag)},
\end{equation}
where $G$ is the Gaia's broad-band magnitude\footnote{More elaborate
models have been produced, for example by C. Jordi; however, for our 
purpose this simplified model is a sufficiently accurate
approximation.}. We described the model
for LSST's photometric errors in \autoref{PHrand}. 


The behavior of photometric errors as a function of $r$ band magnitude
for Gaia, LSST and SDSS is illustrated in the top panel in
\autoref{Fig:GLerrors} (for SDSS, we used \autoref{PHrand} and 
$m_5=22.1$ in the $r$ band).

\subsection{Trigonometric Parallax and Proper Motion Errors }

Similarly to our treatment of photometric errors, we add 
systematic and random astrometric errors in quadrature(see \autoref{PHquad}). For Gaia, we
set a systematic trigonometric parallax error of 0.007 mas, and model
the random errors as
\begin{equation}
\label{Gphrand_astrom}
         \sigma_{rand}^{\pi} = 0.30 \times 10^{0.22 (G-20)} \,\,\, {\rm (mas)}. 
\end{equation}
We obtain proper motion errors (per coordinate) by multiplying 
trigonometric parallax errors by 0.66 yr$^{-1}$. 
We compute LSST trigonometric parallax and proper motion errors
using identical expressions with performance parameters listed in 
\autoref{tab:com:T3}.

\begin{figure}
\centering\includegraphics[width=0.75\linewidth]{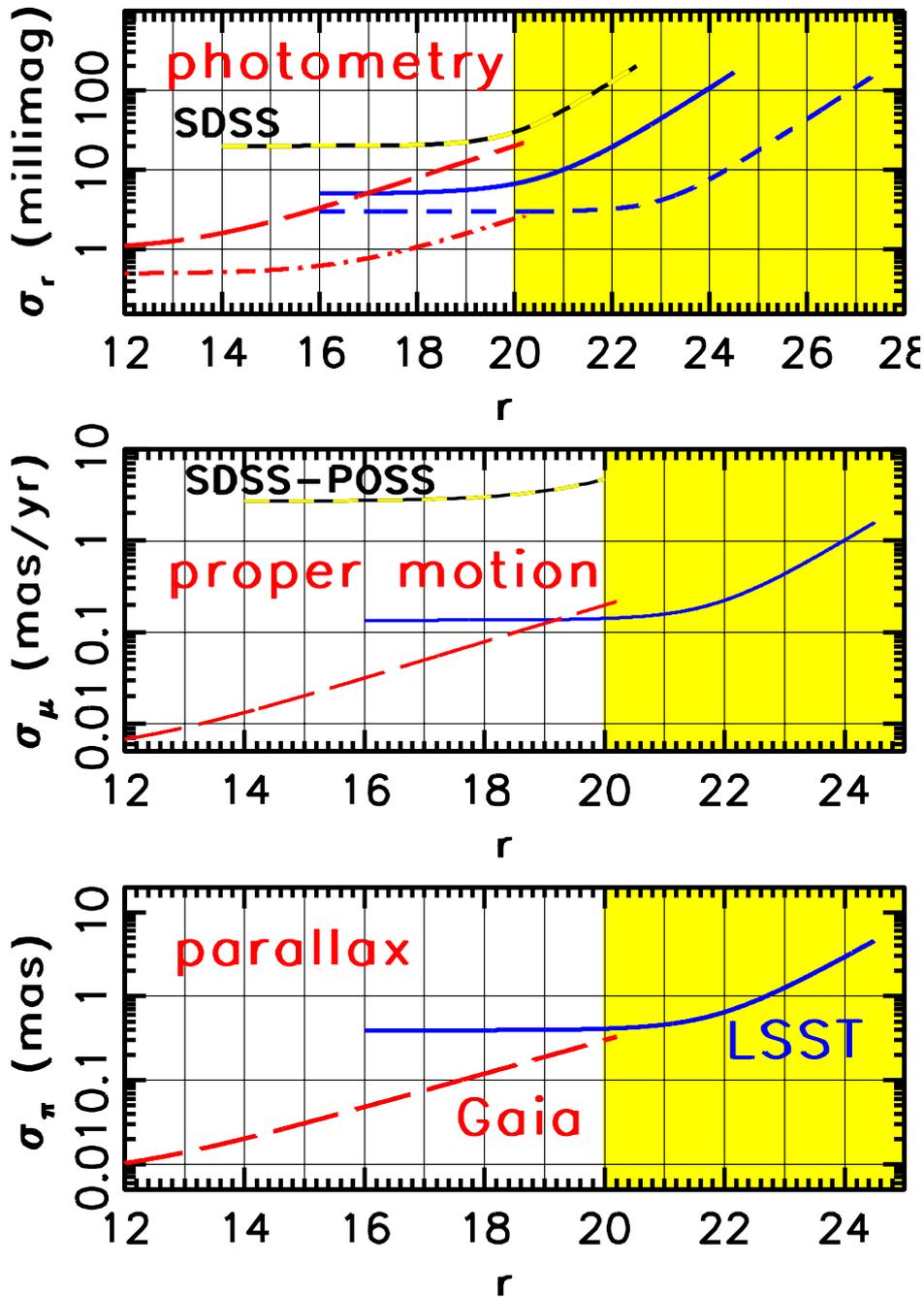}
\caption{A comparison of photometric, proper motion and parallax errors
for SDSS, Gaia and LSST, as a function of apparent magnitude $r$, for
a G2V star (we assumed $r=G$, where $G$ is the Gaia's broad-band
magnitude). In the top panel, the curve marked ``SDSS'' corresponds 
to a single SDSS observation. The red curves correspond to Gaia; the 
long-dashed curve shows a single {\it transit} accuracy, and the 
dot-dashed curve the end of mission accuracy (assuming 70 transits). 
The blue curves correspond to LSST; the solid curve shows a single 
{\it visit} accuracy, and the short-dashed curve shows accuracy for 
co-added data (assuming 230 visits in the $r$ band). The curve marked 
``SDSS-POSS'' in the middle panel shows accuracy delivered by the
proper motion catalog of \citet{Mun++04}. In the middle and bottom 
panels, the long-dashed curves correspond to Gaia, and the solid curves 
to LSST. Note that LSST will smoothly extend Gaia's error vs.~magnitude
curves four magnitudes fainter. The assumptions used in 
these computations are described in the text.}
\label{Fig:GLerrors}
\end{figure}

The behavior of trigonometric parallax and proper motion errors as 
function of $r$ band magnitude for Gaia and LSST is illustrated in 
the bottom two panels in \autoref{Fig:GLerrors}. For comparison,
we also show proper motion error behavior for the current
state-of-the-art large-area database constructed by
\citet{Mun++04} using SDSS and Palomar Observatory Sky Survey
data (a baseline of 50 years). Following \citet{Bond++09},
the SDSS-POSS proper motion errors (per coordinate) are modeled as 
\begin{equation}
\label{SDSSPOSS}
  \sigma_{\rm SDSS-POSS}^{\mu} = 2.7 + 2.0 \times 10^{0.4(r-20)} \,\,\, {\rm (mas/yr)}. 
\end{equation}

(Compare with \autoref{tab:com:T3} to see how much better LSST will do.)
All adopted performance parameters for LSST and Gaia are summarized in
\autoref{GLtable}.

\subsection{Implications for Science Projects }

Gaia will provide an all-sky map with exquisite trigonometric parallax,
proper motion and photometric measurements to $r\sim20$ for about
billion stars. LSST will extend this map to $r\sim27$ over half of 
the sky, and detect about 10 billion stars. Due to Gaia's superb
astrometric and photometric measurements, and LSST's significantly 
deeper data, the two surveys are highly complementary: Gaia will map 
the Milky Way's disk with unprecedented detail, and LSST will extend 
this map all the way to the halo edge.

For example, stars just below the main sequence turn-off with $M_r=4.5$
will be detected by Gaia to a distance limit of $\sim$10 kpc ($r<20$), 
and to $\sim$100 kpc with LSST's single-epoch data ($r<24.5$). 
\citet{lsst08} estimated that LSST will obtain metallicity 
measurements accurate to 0.2 dex or better, with proper motion 
measurements accurate to $\sim$0.2 mas/yr or better, for about 
200 million F/G dwarf stars within 100 kpc.
For intrinsically faint stars, such as late M dwarfs, L/T 
dwarfs, and white dwarfs, the deeper limit of LSST will enable
detection and characterization of halo populations. 
A star with $M_r=15$ will be detectable to a distance limit of 
100 pc with Gaia and $\sim$800 pc with LSST, and hence LSST samples 
will be about 100 times larger. For a substantial fraction of red
stars with $r>20$, LSST will provide trigonometric parallax
measurements accurate to better than 10\% (see \autoref{Fig:CountsStars1}). 
In summary, LSST will represent a deep complement to Gaia.

\begin{table}[t]
\caption{Adopted Gaia and LSST Performance}
\begin{center}
\begin{tabular}{|l|l|l|}
\hline  
   Quantity                              &  Gaia             &    LSST                  \\
\hline  
Sky Coverage                             & whole sky         & half sky                 \\
Mean number of epochs                    & 70 over 5 yrs     & 1000 over 10 yrs         \\
Mean number of observations              &320$^a$ over 5 yrs & 1000$^b$ over 10 yrs     \\
Wavelength Coverage                      &  320--1050 nm     & $ugrizy$                 \\
Depth per visit (5$\sigma$, $r$ band)    &   20              & 24.5; 27.5$^c$           \\
Bright limit ($r$ band)                  &   6               & 16-17                    \\
Point Spread Function (arcsec)           & 0.14$\times$0.4   & 0.70 FWHM                \\
Pixel count (Gigapix)                    &  1.0              &  3.2                     \\
Syst. Photometric Err. (mag)             & 0.001, 0.0005$^d$ &  0.005, 0.003$^e$        \\
Syst. Parallax Err. (mas)                &  0.007$^f$        &  0.40$^f$	        \\     
Syst. Prop. Mot. Err. (mas/yr)           &  0.004            &  0.14                    \\
\hline                         
\end{tabular}
\end{center}
\vskip 0.01in
$^a$ One transit includes the $G$-band photometry (data collected
over 9 CCDs), BP and RP spectrophotometry, and measurements by
the SkyMapper and RVS instruments.  \\
$^b$ Summed over all six bands (taken at different times). \\  
$^c$ For co-added data, assuming 230 visits. \\
$^d$ Single transit and the end-of-mission values for the 
$G$ band (from SkyMapper; integrated BP and RP photometry
will be more than about 3 times less precise). \\
$^e$ For single visit and co-added observations, respectively. \\  
$^f$ Astrometric errors depend on source color. The listed values
     correspond to a G2V star.  
\label{GLtable}      
\end{table}

\clearpage
\bibliographystyle{SciBook}
\bibliography{stellarpops/stellarpops}

%
%
%
%
%
%
%
%
%
%
%
%
%
%
%
%
%
%
%
%
%
%
%
%
%
%

\chapter[Milky Way and Local Volume Structure]
{Milky Way and Local Volume Structure}
\label{chp:mw}

{\it \noindent Beth Willman, John J. Bochanski, James S. Bullock, Roelof de
  Jong, Victor P. Debattista, Douglas Finkbeiner, Carl J. Grillmair, Todd J. Henry, Kathryn V.
  Johnston, Mario Juri\a'c, Jason Kalirai, Peregrine M. McGehee, Rok
  Ro\v{s}kar, Ata Sarajedini, Joshua D. Simon, Jay Strader, Michael
  A. Strauss}

\section{Introduction}
\label{sec:MW:intro}

{\it \noindent Kathryn V. Johnston, James S. Bullock, Michael A. Strauss}

The last decade has seen a renaissance in the study of our own and
other galaxies in the Local Volume (LV). The multi-dimensional,
contiguous maps of the Milky Way (MW) provided by star-by-star surveys
(e.g. HIPPARCOS, 2MASS, and SDSS) have demonstrated that the
smooth fitting functions developed to describe the properties of
galaxies and popularized by integrated light studies are neither
accurate nor complete descriptions of galaxy structure
\citep[e.g.][]{belokurov2006,juric08,ivezic08,bell08}. The tomographic studies
facilitated by the wide-field, depth, and uniformity of the SDSS
data set have revolutionized the way that the structure of the Milky
Way can be mapped. With only the photometric catalog of the SDSS,
photometric abundances were determined for millions of
Milky Way stars and proper motions were derived by comparison with
earlier observations.

Vast numbers of resolved stars and the addition of new dimensions have
revealed: structures in the disk due to dynamical resonances
\citep{dehnen00}; lumps in the halo from hierarchical structure
formation \citep[e.g.][]{newberg2002,majewski03,belokurov2006}; the shapes
of tails in abundance and velocity distributions
\citep{helmi06,kollmeier08}; and a new population of satellite
galaxies that have challenged previous conceptions about the faint
threshold of galaxy formation \citep{willman05a,belokurov07b}.  At the
same time, simulations of structure formation in the cosmological
context have for the first time resolved dark matter structure within
Galactic-scale halos \citep{moore99,klypin99} and made predictions for
the contribution of substructure to the stellar halo distribution
\citep{bullock2005,johnston08}.  These observational and theoretical
advances have combined to launch a new discipline of ``near-field
cosmology.''  The LSST will generate an unprecedentedly large data set
of photometric measurements of use for Galactic structure studies. It
will continue and dramatically accelerate this shift towards {\em
  mapping} studies of the Galaxy started by recent surveys such as
2MASS and SDSS.

Another triumph of the last decade was to demonstrate the broad
consistency of our expectations from hierarchical models of structure
formation with the discovery of substructure (both bound and unbound)
in the stellar halo \citep{bell08,tollerud08,koposov09}.  The
challenge of the next decade is to move beyond ``consistency'' checks
to fully exploit the potential of upcoming LV data sets as probes of
galaxy formation more generally.  The Dark Energy + Cold Dark Matter
hierarchical paradigm provides the necessary theoretical framework
that allows the interpretation of local data within a larger context:
the stars that make galaxies are expected to form within dark matter
halos that are themselves growing through gravitational collapse and
mergers.  In fact, we are very fortunate to live in a {\it
  hierarchical} Universe where the LV galaxies contain the signatures
not only of their own formation, but also of the hundreds of galaxies
that they accreted and merged with.  Assuming that every galaxy in the
Universe is shaped by the same underlying physics, the LV can then be
thought of as a laboratory for testing how stars form over a range of
timescales, within a variety of masses of dark matter halos, in
different environments in the early Universe, and with different
interaction histories.

LSST will contribute the vital framework for this endeavor, producing
the first maps of the stellar distribution in space reaching
throughout the LV --- maps that will define the limits of volume
probed and surface brightness sensitivity feasible in this field for
the next decade.  In simplest terms, these maps will provide a census
of LV structures. But this global view will tell us not only numbers
--- it will also tell us how the properties of structures (morphology,
density, and extent) vary as a function of location, allowing us to
make connections both to the local environment today, and to
early-Universe influences.  Combining this understanding with stellar
populations studies to make chemo-dynamical-spatial maps of local
galaxies will provide insight into their assembly histories and star
formation trajectories unrivaled by any studies that rely on
integrated light from the far field.  Only LSST will have the volume
sensitivity necessary to generalize the results from high-resolution
spectroscopic studies, which will be feasible for smaller, nearby
samples, to a statistical set of objects on larger scales.

This chapter outlines in more detail the maps attainable using
various tracers within the Milky Way and beyond, as well as raising
specific science questions that can be addressed by these data.

\section{Mapping the Galaxy -- A Rosetta Stone for Galaxy Formation}
\label{sec:MW:tomog}

{\it \noindent Mario Juri\a'c, James S. Bullock}

Historically Milky Way surveys have suffered from lack of data, and
instead relied on sparse samples and analytic density laws (fitting
functions often inspired by extra-galactic observations) to
characterize results.  But large, deep, and uniform data sets,
exemplified by the SDSS, have shifted the emphasis from model fitting
toward multidimensional mapping.  Such model-free maps were
instrumental in correctly characterizing the overall smooth
distribution of stars in the Galaxy \citep{juric08}, as well as
revealing a number of coherent, localized substructures
\citep{newberg2002,rochapinto03,juric08} that would have been missed or
misinterpreted by pencil-beam surveys.  Interestingly, some of these
structures have been found in the disk, suggesting a more complex
assembly history for the disk than previously suspected
\citep{kazantzidis08}.

Moreover, only recently has the distribution of Milky Way stars in
metallicity space revealed a more complete view of the Milky Way and
its formation than possible with number counts alone.
\citet{ivezic08} calibrated the relation between the position on the
$u-g,g-r$ diagram to  [Fe/H] using
SDSS imaging (for colors) and SEGUE spectra (for Fe/H) estimates. This
calibration provides photometric metallicity estimates good to $\sim 0.1$
dex.  The per-star estimate uncertainty is almost entirely determined
by the photometric precision in the $u$ band. One caveat is that this
calibration assumes that $[\alpha/Fe]$ does not have a large influence
on the $u-g,g-r$--[Fe/H] relation.

Using photometric metallicity indicators, one of the discovered
substructures (the Monoceros stream) was revealed as having a distinct
signature in metallicity space \citep{ivezic08}, thereby providing an
important constraint on its origin.  Finally, the SDSS has also mapped
the distribution of metallicities of near turn-off stars to distances
of $D=8$~kpc, and found an intriguing {\em lack} of radial metallicity
gradients at $Z > 500$~pc as well as a tantalizing lack of correlation
between metallicity and kinematics throughout the observed disk volume
\citep{ivezic08}. The latter discovery questions the physical meaning
of traditional decomposition of the Galactic disk into two distinct
and simple components (thin vs. thick) and hints at a kinematic and
chemical continuum that arises from a more complex formation process.

Despite these substantial benchmarks, studies of the Milky Way based
on SDSS are limited in distance and in coverage.  Except for a limited
number of imaging stripes, the SDSS nearly avoided the Galactic disk,
where most of the stellar mass, and all of the star formation,
actually occur. Thus all inferences about the disk drawn from the SDSS
come from stars a few scale heights above the midplane, or a sample
limited to a few hundred parsecs around the position of the Sun.  LSST
will have none of these limitations.  Between now and 2014, several
other ground-based, wide-field, multi-filter imaging surveys will take
place, such as Pan-STARRS1, the Southern Sky Survey, and the Dark
Energy Survey.  However, none of these has the depth, width, and
temporal coverage, as well as the simultaneous chemical
characterization capability, needed to obtain a complete map of our
Galaxy.

LSST will provide a uniform, multidimensional, star-by-star phase
space map of all Milky Way components, including two orders of
magnitude more stars than visible with SDSS.  It will for the first
time open the window to a complete picture of the spatial, kinematic,
and chemical makeup of Galactic components. LSST will uniformly cover
the Galactic plane, as well as provide up to one thousand epochs of
time-domain information. This information holds the promise of
becoming a true Rosetta Stone for galactic disk formation and
structure. It will provide a powerful complement to large scale galaxy
surveys, and may well be a linchpin in our efforts to build a
consensus theory of cosmology and galaxy formation.

\subsection{Mapping the Milky Way with LSST}

Specifically, LSST's data set will enable:
\begin{itemize}
\item The mapping of stellar number density with observations
      of $\sim 10$~billion main sequence stars to (unextincted) distances 
      of 100~kpc over 20,000~deg$^2$ of sky.
\item The mapping of stellar metallicity over the same volume, using
      observations of photometric metallicity indicators in 
      $\sim200$~million near turn-off main sequence (F/G) stars.
    \item Construction of maps of other more luminous tracers, such as
      RR Lyrae variables, to as far as 400 kpc -- the approximate
      virial radius of the Milky Way.
\item High fidelity maps of tangential velocity field to at least $10$~kpc
      (at 10 \kms\ precision) and as far as as $25$~kpc (at 60 \kms\ precision).
\end{itemize}
LSST can achieve such a complete map of the Milky Way only because
it has combined a series of unique enabling capabilities: 
 \begin{itemize}
 \item The existence of the $u$ band, allowing the measurement
       of stellar metallicities of near turn-off stars and its mapping
       throughout the observed disk and halo volume.
 \item The near-IR $y$ band, allowing the mapping of stellar number densities
       and proper motions even in regions of high extinction.
 \item Well sampled time domain information, allowing for the unambiguous 
       identification and characterization of variable stars (e.g., RR Lyrae),
       facilitating their use as density and kinematic tracers to large distances.
\item Proper motion measurements for stars 4 magnitudes fainter
       than will be obtained by Gaia (see \autoref{sec:com:PMacc}).
 \item The depth and wide-area nature of the survey, which combined with the
       characteristics listed above, permits a uniquely uniform,
       comprehensive, and global view of all luminous Galactic components.
 \end{itemize}

 With these characteristics, LSST will achieve a two orders of magnitude
 increase in the amount of data that will be available for Milky Way
 science \citep{ivezic08}.  The typical resolution of LSST Galactic
 maps will be on order of $\sim10-15\%$ in distance and $0.2-0.3\, {\rm
   dex}$ in metallicity. The former is fundamentally limited by
 unresolved multiple systems \citep{sesar08}, while the latter is
 limited by calibration and accuracy of $u$ band photometry.

\subsection{The Science Enabled by LSST Maps}

\begin{figure}
\centering\includegraphics[width=0.9\linewidth]{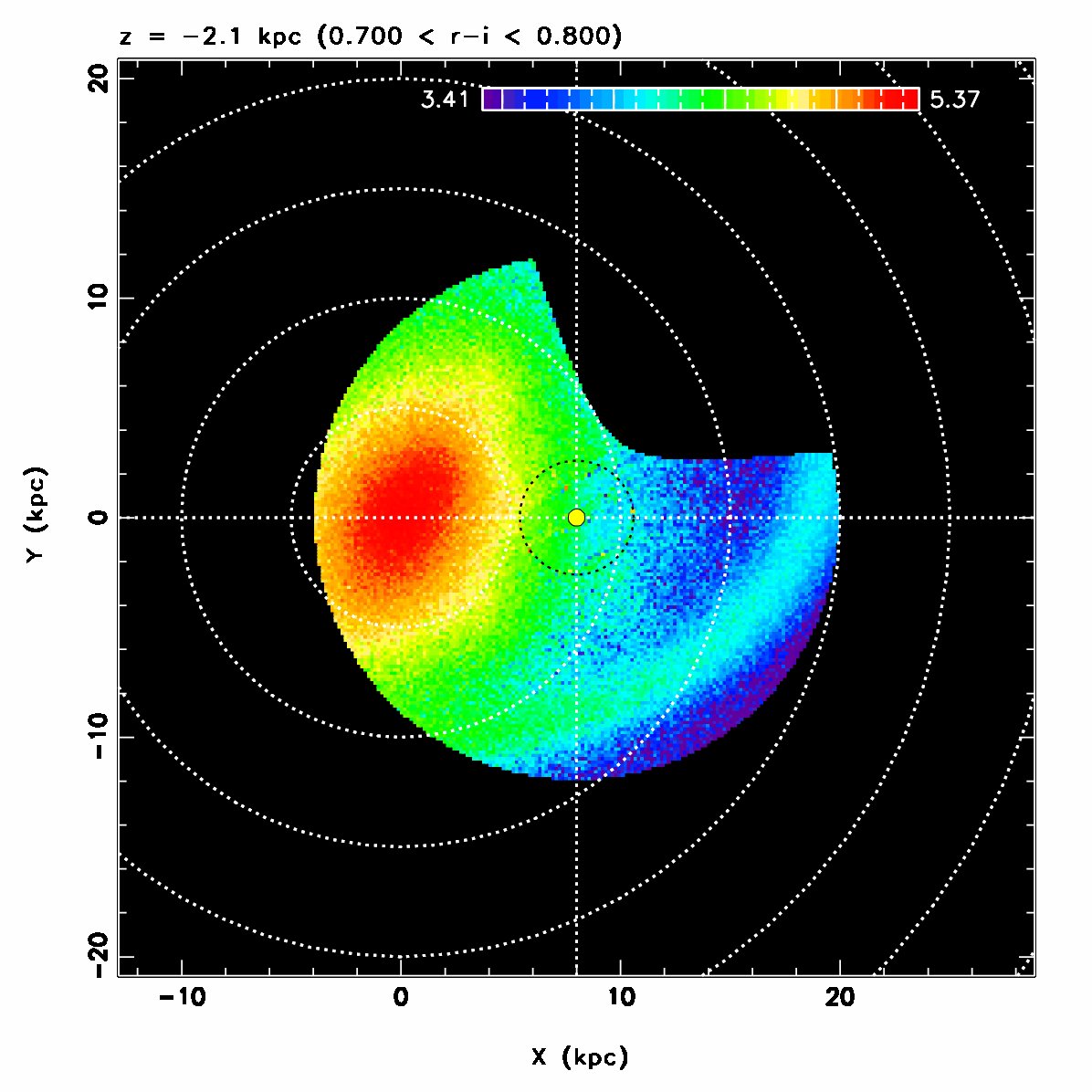}
\caption{ LSST view of the inner Galaxy. A plane-parallel slice
  through a simulated three-dimensional map of stellar number density (stars
  kpc$^{-3}$, log scale) taken at $Z=-2.1$ kpc (south of the Galactic
  plane). The simulation includes a full SDSS-like model of realistic
  instrumental and methodological uncertainties, and is directly
  comparable to Figs. 12-14 of \citet[hereafter J08]{juric08}. 
  The projected positions of the Galactic center and the Sun are at
  $X=Y=0$ and $X=8\,$kpc, $Y=0$, respectively. The stars were distributed
  according to the J08 density law, with the addition of an inner
  triaxial halo/bulge/bar component, and a nearly plane-parallel
  Monoceros-like tidal stream in the outer regions. Only data at
  Galactic latitudes $|b|>10$ are shown. The missing piece in the first
  quadrant is due to the $\delta < 34.5^\circ$ limit of the survey.
The small dotted circle centered at the position of the Sun denotes the reach
of the J08 SDSS study, and plotted within it are the actual J08 SDSS data
from the $Z=+2.1$~kpc slice. Neither the outer stream nor the triaxiality of the
inner halo/bulge were detected by the SDSS. LSST will easily
detect and characterize both.
} \label{fig:mwxyslice}
\end{figure}

\begin{figure}
\centering\includegraphics[width=0.9\linewidth]{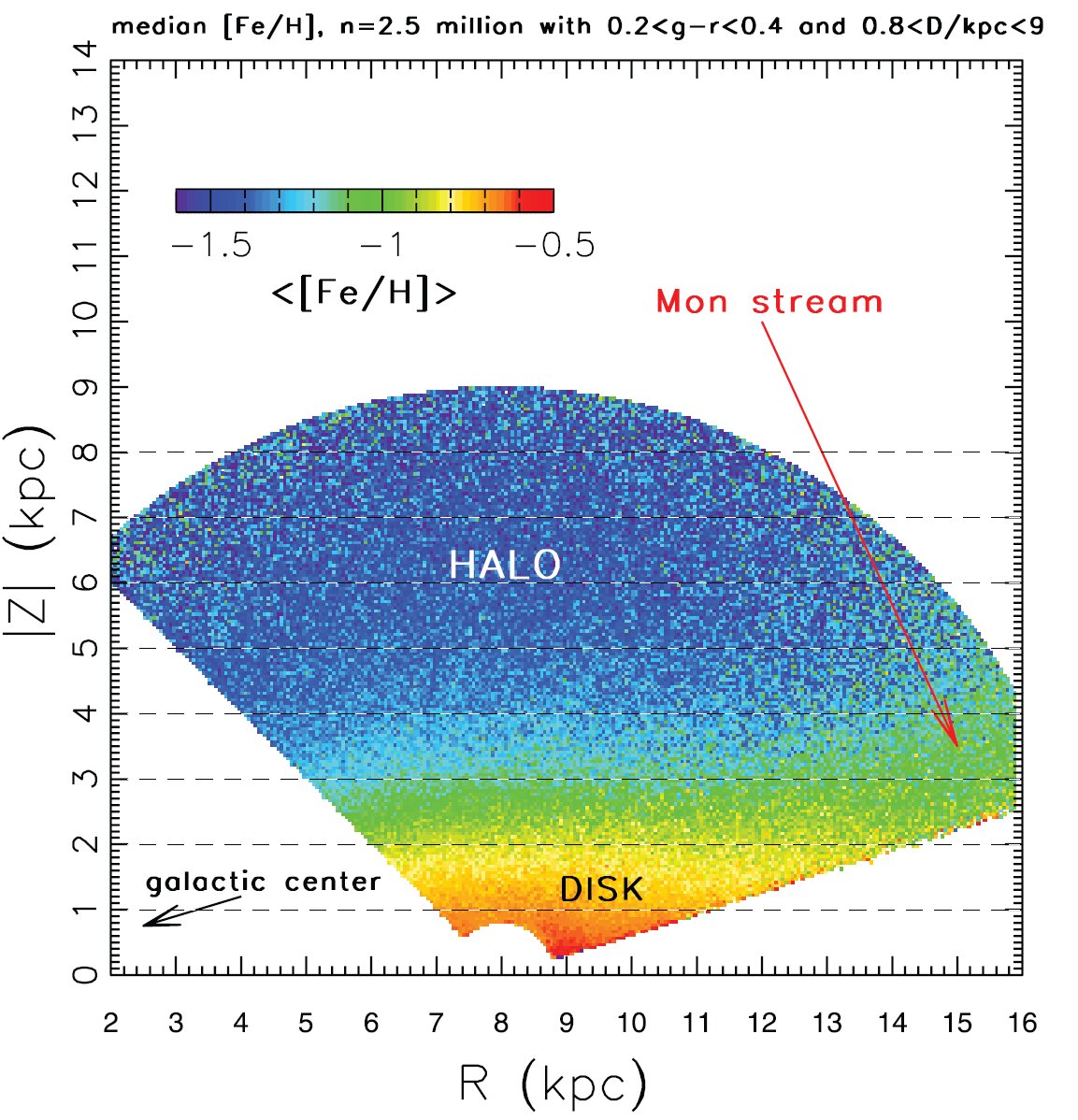}
\caption{ The map of median photometric metallicity for
  $\sim2.5$~million main sequence turn-off stars from SDSS DR6 in
  cylindrical Galactic coordinates $R$ and $|Z|$ \citep[adapted
  from][]{ivezic08}. There are $\sim 40,000$~pixels (50 pc x 50 pc)
  contained in this map, with a minimum of 5 stars per pixel and a
  median of 33 stars. Note the strong vertical metallicity gradient,
  and a marked difference of metallicity of the region coincident with
  the Monoceros stream (as marked). LSST will produce equivalent three
  dimensional maps with $\sim200$~million stars, that will extend to
  $100$~kpc in the halo and provide coverage of the Galactic plane (as
  allowed by extinction).  } \label{fig:FeHmap}
\end{figure}

The science immediately enabled by LSST maps of the stellar
distribution (Figures~\ref{fig:mwxyslice} and \ref{fig:FeHmap}) can be divided into a
number of headings:

 \begin{itemize}
 \item Characterization and understanding the overall smooth
   distribution of stars in the Milky Way (this section,
   \autoref{sec:MW:stellarcensus}) and other nearby galaxies
   (\autoref{sec:MW:LV})
 \item Characterization and understanding large-scale chemical
   gradients in the Milky Way (this section)
 \item Discovery and characterization of localized features, such as
   clumps and streams, in metallicity and phase space (Milky Way disk
   - this section; MW bulge - \autoref{sec:MW:diskbulge}; MW halo -
   \autoref{sec:MW:streams}, \autoref{sec:MW:UFs}; \autoref{sec:MW:LV})
 \item Inferring the distribution of mass and the potential of the
   Milky Way (\autoref{sec:MW:haloPMs})
 \end{itemize}

The stellar number density and proper motion maps will allow
measurements of structural parameters of all Galactic components
(bulge, disk, halo) including the hitherto poorly observed ones (e.g.,
the disk scale length). Together with kinematic information, these
will facilitate the construction of global dynamical models of the
Milky Way and may break the disk/halo degeneracies still present in
today's models \citep{binney87}. This would put observational
constraints on the distribution of matter in the Galactic disk and
halo, and most interestingly, the distribution of dark matter in the
inner Galaxy.

In addition to mapping the overall smooth distribution, the maps will
facilitate the discovery and characterization of disk ($|Z| \lesssim
2$~kpc) substructure to at least $D=12$~kpc heliocentric distance
(\autoref{fig:mwxyslice}) and at density contrasts of $\Delta \rho
/ \rho \gtrsim 20\%$.  In regions of small or well-measured
extinction, detection of substructure will be possible to
significantly deeper levels\footnote{With a single-epoch limiting
  magnitude of $r=24.7$, near turn-off stars can be observed to
  $\sim80\,{\rm kpc}$ distance on clear
  sightlines.}. Furthermore, the uniform coverage of the Galactic
plane will yield data for numerous star-forming regions, and the $y$
band data will penetrate through the interstellar dust layer.

These data will provide constraints on the merger history of the Milky
Way and shed light on how the thin disk formed and survived since
$z=1$.  Structure formation in the concordance cosmology model (and the vast
majority of suggested variants) is fundamentally hierarchical:
galaxies and their halos are assembled from the continuous merging of
smaller systems (e.g., \citealt{Purcell++08, Kazantzidis++09}).
Meanwhile, the majority of Milky-Way size galaxies in the Universe are
dominated by thin, cold disks of stars, which seem to be relatively
unmolested by violent mergers. This is one of the most pressing
problems of galaxy formation today, and any formulation of galaxy
formation must account for this tension. Indeed there are a number of
competing suggestions aimed at explaining how thin disks may survive
and/or emerge from the expected bombardment.  To first order, the
competing theories are {\em designed} to reproduce the broad-brush
statistics obtained from large galaxy surveys (e.g., the fraction of
galaxies that are disks).  In contrast, the rich kinematic, spatial,
and chemical data set offered by the Milky Way disk itself provides an
entirely disjoint testing ground for models aimed at explaining disk
formation in a cosmological context. The LSST maps described here will
provide such a data set.

Some of the detected disk substructure may be of secular (dynamical) and 
not merger origin (e.g., due to spiral arms; \autoref{sec:MW:diskbulge}).
Their detection and identification as such can constrain the distribution
of matter as well as the pattern speeds of non-axisymmetric features in 
the Galactic disk. Furthermore, recent simulations of galaxy formation in a fully 
cosmological context  \citep{Read++08} have reopened the discussion 
about the existence and distribution of disk dark matter.
While its dynamical influence is (theoretically) expected to be small, 
it is highly uncertain and may still be detectable in global disk kinematics,
or in local kinematics and morphology of phase-space substructure.

Photometric metallicity measurements will be available for about 200
million main sequence F/G stars. These will sample the disk to the
extent allowable by extinction, provide three-dimensional maps of the metallicity
distribution, and reveal large-scale metallicity gradients both in the
disk and the halo. As well as being crucial for differentiating
between various models of chemical evolution and disk assembly, this
metallicity information will aid in determining the nature of detected
substructures. Both have been powerfully demonstrated on a smaller
sample by the SDSS (\autoref{fig:FeHmap}). LSST will be capable of
producing analogous maps that are fully three dimensional, extend up
to 5 magnitudes deeper, and cover the Galactic plane.

The metallicity of the halo will be mapped to distances of 100 kpc. No other 
existing or planned survey will provide such a comprehensive data set to study 
the outer halo (including Gaia, which is flux limited at $r=20$, and Pan-STARRS, which does not have the $u$ band). Maps of RR Lyrae and classical novae 
will extend the observable distances to $\sim 400$kpc and enable the
exploration of the extent and structure of Galactic halo out to beyond the
presumed virial radius. Thus, the LSST will enable studies of the
distribution of main sequence stars beyond the presumed edge of Galaxy's
halo, of their metallicity distribution throughout most of the halo, and of
their kinematics beyond the thick disk/halo boundary. It will also obtain
direct distance measurements via trigonometric parallax below the
hydrogen-burning limit for a representative thin-disk sample.

Taken together, these six dimensional phase-space (two angular
positions, one photometric distance, 
two proper motions, and metallicity) maps of the Galaxy will provide a
detailed accounting of the Galaxy's true makeup and have the potential
to spawn a revolution in our understanding of galaxy formation in
general. They will facilitate comprehensive dynamical and chemical
modeling of the structure and evolution of all Galactic components,
including mergers in the full cosmological context, and provide a rich
data set with detailed features whose explanation will present a
challenge for the decades to come.




\section{Unravelling the Secular Evolution of the Bulge and Disk}
\label{sec:MW:diskbulge}


{\it \noindent Victor P. Debattista, Rok Ro\v{s}kar, Mario Juri\a'c, Jay Strader}

\subsection{The Bulge}

The Milky Way is a barred spiral galaxy of Hubble type SBbc with a triaxial bulge
\citep{ger_vie_86, bin_etal_91, nak_etal_91, wei_etal_94, dwe_etal_95, zhao_96,
bin_etal_97, sta_etal_97}. The bar and spiral arms break the axisymmetry of the disk
and lead to secular evolution as gas is transported to the central regions where it
forms stars. Heating of stars in the center can also occur as disk stars scatter off
a bar \citep{kor_ken_04}.

Bulges formed secularly in this manner are termed pseudobulges, as
distinct from the merger-built bulges that inhabit early-type
spirals. Pseudobulges have shallow, exponential light profiles
(corresponding to $n \lesssim 2$ in Sersic fits) and may be
flattened. The kinematics of a pseudobulge are dominated by rotation.

The Milky Way presents one of the largest challenges to the
pseudobulge hypothesis. Its bulge is boxy and flattened, with
cylindrical kinematics \citep{howard09}---all pseudobulge
characteristics. Yet the stars in the bulge are old, metal-rich, and
enhanced in $\alpha$-elements \citep{zoc_etal_06}; such properties are
inconsistent with gradual secular enrichment.

Observationally, it is clear that LSST will provide a unique map of
the kinematic properties and metallicity distribution of the
bulge. However, more theoretical work is needed to determine the most
informative way to constrain bulge formation in detail. It is worth
recalling that LSST bulge studies will take place in the context of
other large upcoming surveys, such as SDSS-III/APOGEE, which will
obtain high-resolution near-IR spectra of $10^5$ bulge giants to
determine precise radial velocities and chemical abundances for many
elements.

Let us consider the kinematic constraints available with LSST. Old
main sequence turn-off stars have unextinguished magnitudes of $r \sim
19$ in the bulge. Recalling the proper motion limits of \autoref{sec:com:PMacc}, single
stars with $r=21$ will have proper motion accuracies of $\sim 8$ \kms,
increasing to $\sim 40$ \kms\ at $r=24$. These apparent magnitudes
correspond to extinctions of $A_r \sim 2$ and $\sim 5$ mag respectively.  Using the extinction map of \citet{popowski03}, these
mean extinctions are reached at $b=4^{\circ}$ (550 pc---this is the
latitude of Baade's Window) and $b=1.6^{\circ}$ (220 pc) moving toward
the Galactic Center. Thus the detailed kinematics of the bulge, well
into the central parts, can be studied quite readily with proper
motions of turn-off stars.

Estimating stellar metallicities will be more challenging, since there
is a degeneracy between metallicity and extinction for main sequence
stars. Red clump giants can be used as standard candles to give
reddening-independent magnitudes and estimate the local extinction;
these values can then be applied to turn-off stars to yield intrinsic
colors and thus metallicities.

\subsection{Spiral Structure}

Surprisingly little is known about the spiral arms of the Milky Way, from
their vertical structure to even whether there are two or four arms
\citep{bis_ger_02}. Spiral structure drives large-scale radial mixing
of stars without heating the disk. In models of inside-out disk formation,
such mixing tends to erase correlations between age and metallicity that
would otherwise be present at a given radius \citep{Sellwood:2002,
Roskar:2008b, Roskar:2008}. It should be possible to use LSST data to trace
the evolution of the stellar populations of the disk toward the $l=270$
edge.

While there is strong theoretical motivation for LSST to study the
spiral structure of the disk, more work remains to be done to make
predictions specific to LSST. This work should include proper image
simulations to estimate the effects of crowding, saturated bright
stars, and extinction on studies of the disk.

\section{A Complete Stellar Census}
\label{sec:MW:stellarcensus}

{\it \noindent John J. Bochanski, Jason Kalirai, Todd J. Henry}


Hydrogen burning low--mass stars ($M < 0.8 M_{\odot}$) and evolved
white dwarfs are the dominant stellar constituents of the Milky Way
and comprise nearly 70\% of all stars.  Because they dominate the Galaxy
in both mass and numbers and have endured since the Galaxy's formation, these
samples hold unique information about the entire chemical enrichment
and dynamical history of the Galaxy. Yet until recently, their low
intrinsic luminosities ($L \lessim$ $0.4 L_{\odot}$) have limited
observational studies of these stars to distances $\sim 500$ pc.
Surveys such as 2MASS and SDSS have ameliorated this situation,
providing accurate, precise photometry that is sensitive to M dwarfs
at distances up to $\sim 2$ kpc.  The upcoming Gaia mission will provide
parallaxes out to only 10 pc for the latest M dwarfs.
LSST is poised to revolutionize this
field, with precise photometry of M dwarfs to distances $\sim 30$ kpc
and trigonometric parallaxes of stars within 300 pc (see
\autoref{sec:com:PMacc} and \autoref{tab:com:T3}).  The photometric sample
will contain $\sim$ 7 billion stars, providing a database of
unprecedented magnitude.  The parallactic sample will be a critical 
component of
future investigations, including the luminosity function and
corresponding mass function.  Studies of white dwarfs (WDs), the most
common stellar remnant, have also been limited by their diminutive
luminosities.  The sensitivity of LSST photometry will extend the
white dwarf luminosity function by several magnitudes as discussed in
detail in \autoref{sec:sp:wds}.  The structure and cutoff of the WD
luminosity function are sensitive to the star formation history,
progenitor initial mass function, and the initial epoch of star
formation.  Combining the initial mass functions measured by M dwarfs
and white dwarfs, along with estimates of the star formation history,
will provide a unique glimpse into the evolution of the Galactic disk
and halo, and provide a complete census of nearby Galactic stellar
populations.

Accurate distances are essential to a complete stellar census.
Distance estimates from LSST data will come in two forms:  direct,
trigonometric parallaxes and photometric parallaxes from
color--magnitude relations (CMRs). The accuracy of LSST trigonometric
parallaxes is described in \autoref{sec:com:PMacc}.  LSST will measure
accurate parallaxes for millions of low--mass hydrogen burning
dwarfs, with spectral types M4 and later (compare with SDSS, for which
only 10-20 stars have measured trigonometric parallax and native 2.5-m
photometry).  These distances will be used to construct CMRs in the
native LSST system, augmented with Gaia parallaxes. These CMRs will be
used to map the distribution of stars within $\sim 2$ kpc with
unprecedented resolution and place new constraints on the initial mass
function above and below the hydrogen burning limit.


These vastly improved CMRs and trigonometric parallaxes will make
possible a volume--complete sample of low--mass dwarfs within 300 pc.
In 2009, the largest volume--complete sample extends to $\sim 25$ pc,
containing roughly 500 systems \citep{reid2002}.  With LSST parallaxes, this volume
limit will grow by three orders of magnitude and contain millions of
stars.  Furthermore, a volume--complete, trigonometric parallax sample
will obviate any systematics introduced by the assumed CMR.  In order
to correctly account for unresolved binaries, follow--up radial
velocity studies will be necessary, although statistical corrections
can be made from existing data sets.  This project will yield a precise
measurement of the low--luminosity LF ($M_r > 16$), with a data set of
unprecedented size.

\begin{figure}
\begin{center}$
\begin{array}{cc}
\includegraphics[width=0.4\linewidth]{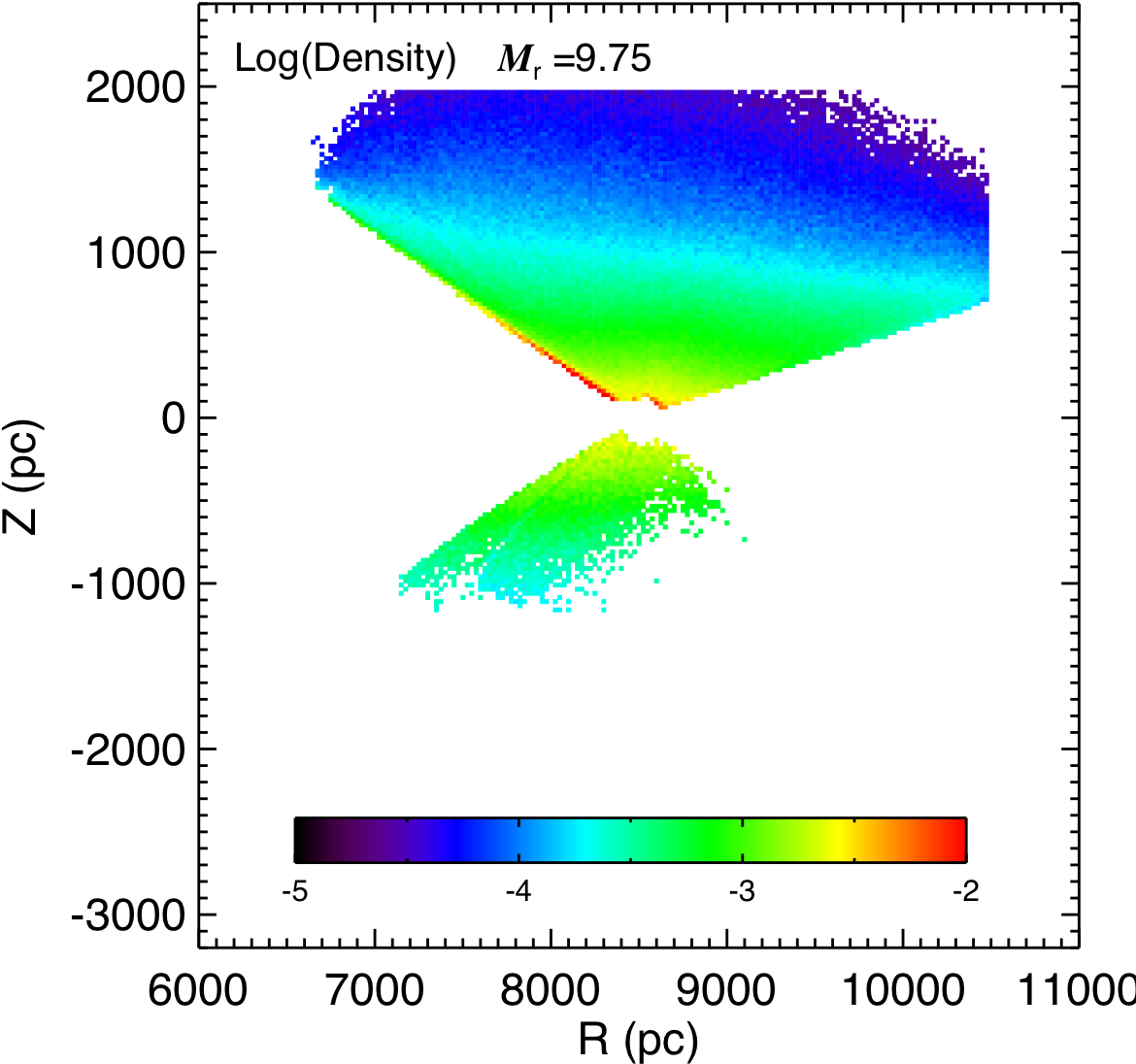}%
\includegraphics[width=0.4\linewidth]{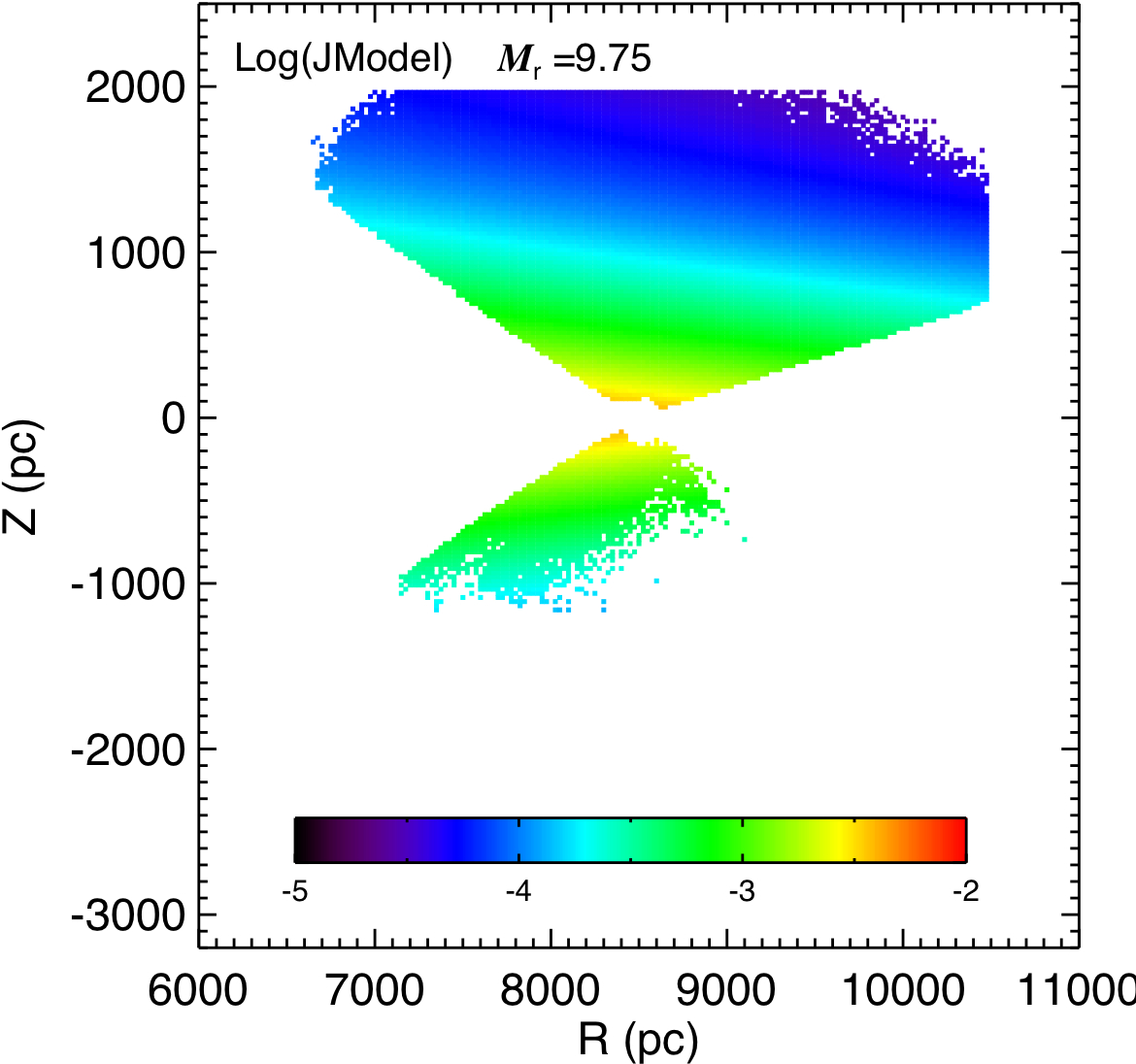}
\end{array}$
\end{center}
\caption{Left Panel: The stellar density profile of stars in a small
  (0.5 mag) slice in absolute magnitude, centered on $M_r = 9.75$.
  Right Panel: The corresponding Galactic density model.  The
  luminosity function and corresponding mass function is constructed
  by iterating this analysis over absolute magnitude.  Figures adapted
  from \cite{2008arXiv0810.2343B}.}
\label{fig:density_model}
\end{figure}

The CMRs and parallaxes from the LSST data set will also facilitate a
simultaneous mapping of the local Galaxy structure and the stellar
luminosity function.  This map will be made based on the stellar
luminosity function technique introduced by
\cite{2008arXiv0810.2343B}.  For this technique, distances are first
assigned to each star using a CMR (in this case, measured directly by
LSST).  Stellar density maps \citep[similar to][]{juric08} are
constructed for small slices in absolute magnitude.  A Galactic
density profile is fit to the maps, and the local density is recorded
for each slice in absolute magnitude.  An example of the stellar
density profile and corresponding model for one slice in absolute
magnitude is shown in \autoref{fig:density_model}.  These local
densities plotted as a function of absolute magnitude form the
luminosity function.  Applying this technique, the local Galactic
structure \textbf{and} luminosity function are thus measured
simultaneously. LSST observations would extend the distance limits to
$\sim 30$ kpc for the brightest M dwarfs, mapping out the thin and
thick disk with unprecedented precision.  This stellar census will
provide an estimate of Galactic structure and the total stellar mass
of the thin and thick disks.  It will also be sensitive to changes in
the LF and IMF as a function of position in the Galaxy.
The vast numbers of low-mass and low-luminosity stars to be revealed
by LSST will yield important constraints on the overall stellar mass
content of the Galaxy, the stellar initial mass function and the star
formation history of the Milky Way.  White dwarfs trace the
distribution of previous stellar generations, and their cooling curves
provide a rough age estimate.  Since 97\% of all stars exhaust their
fuel and cool to become white dwarfs, these stars become powerful
tracers of the Milky Way's star formation history and evolution.
Given the age of the Galactic halo, most of the mass in this component
is now tied up in these remnant stars, which LSST will uncover.
 See \autoref{sec:sp:wds} for a more detailed and nuanced
discussion of the white dwarf science that will be uniquely possible
with LSST.

\section{Three-Dimensional Dust Map of the Milky Way}
\label{sec:MW:dust}


%

{\it \noindent Peregrine M. McGehee, Douglas Finkbeiner}

Interstellar dust is a significant constituent of the Galaxy.
Its composition and associated extinction properties tell us
about the material and environments in which stars and their planets
are formed. Dust also presents an obstacle for a wide-range of
astronomical observations, causing light from stars in the plane of
the Milky Way to be severely dimmed and causing the apparent colors of
objects observed in any direction to be shifted from their intrinsic
values. These color shifts are dependent upon the dust column density
along the line of sight and the radiative transport properties
of the dust grains. 

The wavelength dependence of the absorption due to dust is parametrized 
in the widely used model of \citet{cardelli89}
by the ratio of general to selection extinction in the Johnson
$B$ and $V$ bands, defined as $R_V = A_V/E(B-V)$.
The value of $R_V$ depends on the dust composition and grain size
along the line of sight. In the low-density diffuse ISM, $R_V$ has
a value $\sim$ 3.1, while in dense molecular clouds, $R_V$ can be higher 
with values $4 < R_V < 6$. 

The fundamental importance of a well-characterized dust map
to astronomy is underscored by the $>5,000$ citations to the dust and
extinction maps by \citet{schlegel98}, henceforth SFD98.  The SFD98
maps are based on far-infrared observations and predict reddening in specific
bands by assuming a dust model and $R_V = 3.1$ as appropriate for
sky areas away from the Galactic plane.

Despite the great contribution that the SFD98 extinction map has made to
the field, these maps suffer from several issues that limit their
utility in some regimes of study.  1) While the SFD98 map seems to be
well calibrated at low column density, various tests using galaxy
counts, star counts and colors, and stellar spectrophotometry indicate
that SFD98 overpredicts dust by $\sim30\%$ above $E(B-V) \sim 1$ mag.
Because this overcorrection appears especially in cold clouds, it is
likely related to the temperature correction adopted in the SFD98
model. 2) In some cases, especially at low Galactic latitudes, $R_V$
variation is important and is not tracked by SFD98.  3) For study of
low-redshift, large-scale structure, contamination by unresolved point sources
can be important \citep[see][]{yahata07}.  4) Finally, the resolution
of the SFD98 map is $\sim6'$, which is larger than the angular scales
subtended by nearby, resolved, galaxies for which a carefully
characterized foreground dust distribution is particularly important.
For all these reasons, LSST stellar photometry, which can constrain the
temperature correction, overall calibration, and point source
contamination of SFD98, is valuable.

For the study of stellar populations and objects within the Galactic
disk it is also important to determine both the line of sight
extinction and the value of $R_V$ at a specific distance, neither of
which is dealt with by SFD98.  By analysis of the observed reddening
of stellar colors, we will verify both the dust column density and
$R_V$ values predicted by these maps and can also determine the local
spatial distribution of the dust.  We will do this utilizing two
specific stellar populations - the M dwarfs and the F turn-off stars.

The reddening of stellar colors due to the presence of interstellar
dust along the line of sight can, in principle, be used to map the
three-dimensional distribution of that dust. This requires that two important parameters
are determined - the amount the observed stellar color is reddened and
the distance to the star. By comparison of the color excess 
measured in stars at varying distances we can infer the location of the 
extincting medium.
However, given lack of an a priori knowledge
of the light of sight extinction, which is the very quantity we wish to
measure, it can be difficult to accurately assign 
intrinsic stellar colors and luminosities in order to determine the 
amount of color excess and the distance. This difficulty can be surmounted,
however, if we utilize reddening-invariant combinations of colors whose
values can be used to infer location on the stellar locus and hence 
intrinsic colors and luminosities. This technique is viable if we use
LSST photometry of M dwarfs as the stellar locus in $ugriz$ 
colors is nearly parallel to the reddening vector for all but coolest stars.

\subsection{Spatial Distribution of Dust} 

The use of stellar samples to create three-dimensional extinction maps has an
established history beginning with the work of \citet{neckel80};
however these, including studies based on SDSS photometry, are
typically limited to heliocentric distances of $1-2$ kpc.  In the
full co-added survey, LSST will be able to map dust structures out to
distances exceeding 15 kpc, thus revealing a detailed picture of this
component of the Milky Way Galaxy.

Mapping of the dust component of the Galactic ISM requires detection
of the reddening in the colors of stars at known distances.  The
reddening is determined from the color excess deduced by comparison of
the observed colors with those expected based on the stellar spectral
type.  In the absence of identifying spectra, the spectral type can be
inferred by dereddening the observed colors (assuming a specific
extinction law, i.e., a particular value of $R_V$) back to the
unreddened stellar locus in a color-color diagram. This dereddening is
equivalent to assignment of reddening-free colors along the stellar
locus, which measure the location in the color-color diagram along the
direction perpendicular to the reddening vector. Once the effective
line of sight reddening has been computed, the distance to each star
can be determined using dereddened photometry and well-calibrated
color-absolute magnitude relations.

\subsubsection{Reddening-invariant Indices}

Reddening-free colors were defined in the SDSS $ugriz$ system by
\citet{mcgehee05} for characterization of embedded pre-main sequence
stars and were subsequently used as part of the SDSS photometric
quality analysis system \citep{abazajian08}.  The general definition
is 
\begin{equation}
\begin{split}
Q_{xyz} &= (x-y) - (y-z) \times \frac{E(x-y)}{E(y-z)}
\end{split}
\end{equation}
where $(x-y)$ and $(y-z)$ are the colors used to construct the
color-color diagram. This extends the definition by 
\citet{johnson53} whose original $Q$ would be defined here as $Q_{UBV}$.
The reddening coefficients adopted by the SDSS
\citep{stoughton02} follow SFD98 and assume the ``standard'' dust law
of $R_V = 3.1$ and a $z=0$ elliptical galaxy spectral energy
distribution.

\begin{figure}
\begin{center}
\includegraphics[width=0.5\linewidth,angle=0]{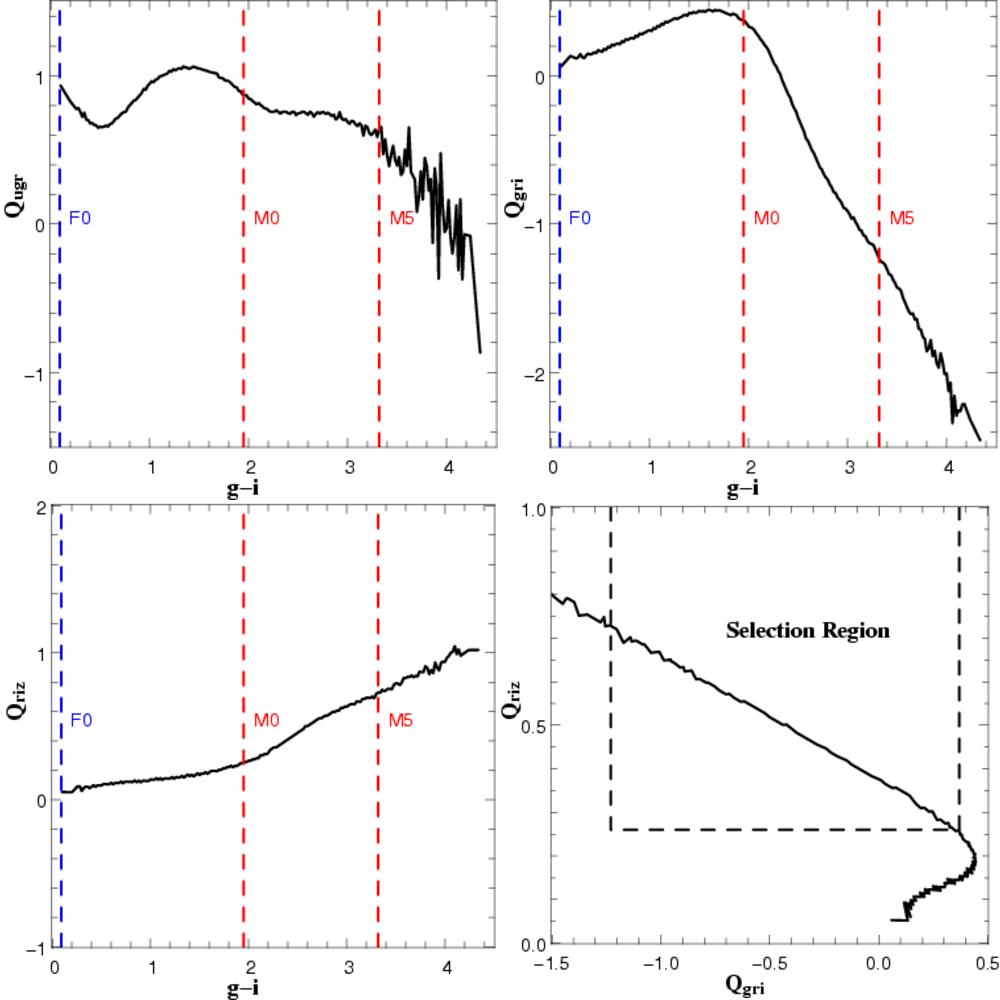}

\caption{The relation between reddening-invariant colors $Q_{ugr}$
({\it upper left}),
$Q_{gri}$ ({\it upper right}), 
and $Q_{riz}$ ({\it lower left}) and intrinsic $g-i$ color
is shown here for the SDSS median stellar locus \citep{covey07}. 
These 
indices, as a whole, show little variation for stars earlier than M0. The
vertical axis in these three plots spans three magnitudes in order to 
facilitate
comparison of the index ranges.
The lower right panel shows the selection of
M0 to M5 stars based
on $Q_{gri}$ and $Q_{riz}$, where the latter cut is primarily to discard 
earlier and more luminous background stars. 
\label{fig:mw:dust3d_locusq}}
\end{center}
\end{figure}

In \autoref{fig:mw:dust3d_locusq} we compare the variation of the
three reddening-invariant indices formed from the $ugriz$ passbands
($Q_{ugr}$, $Q_{gri}$, and $Q_{riz}$) with $g-i$, a proxy for stellar
spectral type \citep{covey07}. For $g-i < 1.9$ (spectral type earlier
than M0) there is little variation in any of these indices, indicating
that the stellar locus is approximately parallel to the reddening
vector in the corresponding color-color diagrams. For the M dwarfs we
see that the $Q_{gri}$ has the largest range between M0 and M5, and
thus is of the greatest utility for determination of spectral type.

\subsubsection{Selection of Reddening Probes}

For determination of spectral type and intrinsic stellar colors to be
accurate, the stars used as reddening probes must reside on the
portion of the stellar locus that is not aligned with the reddening
vector in a color-color diagram. As we have seen, this condition is
fulfilled by stars of spectral types M0 and later.  In the final
panel of
\autoref{fig:mw:dust3d_locusq} we show the criteria used to select for
early and mid M dwarfs based on the $Q_{gri}$ and $Q_{riz}$ indices,
where the latter is used to filter out earlier and more luminous
background stars whose $Q_{gri}$ colors are similar to M0 dwarfs.  The
threshold at M5 is chosen to remove the later spectral type stars, which
are too intrinsically faint to serve as probes for all but the nearest
dust structures. 

Analysis of the LSST imaging data will adapt the
following procedure as used in the SDSS High Latitude Cloud Survey 
\citep{mcgehee09}:
\begin{itemize}
\item The intrinsic $g-i$ color ($(g-i)_0$) is determined from the observed 
$Q_{gri}$ color based on a fifth-order polynomial fit using the median 
stellar locus \citep{covey07} and assuming $R_V  = 3.1$. 
\item The total reddening to each star is computed from the $g-i$ color excess.
\item Distances are assigned based on the
color-absolute magnitude relations of \citet{ivezic08} using 
the dereddened photometry.
\item $E(B-V)$ maps are created at specific distance ranges using the
adaptive technique of \citet{cambresy05} in which the reddening at each
pixel is the median of that computed for the $N$ nearest extinction probes.
\end{itemize}

\begin{figure}
\begin{center}
\includegraphics[width=0.9\linewidth,angle=0]{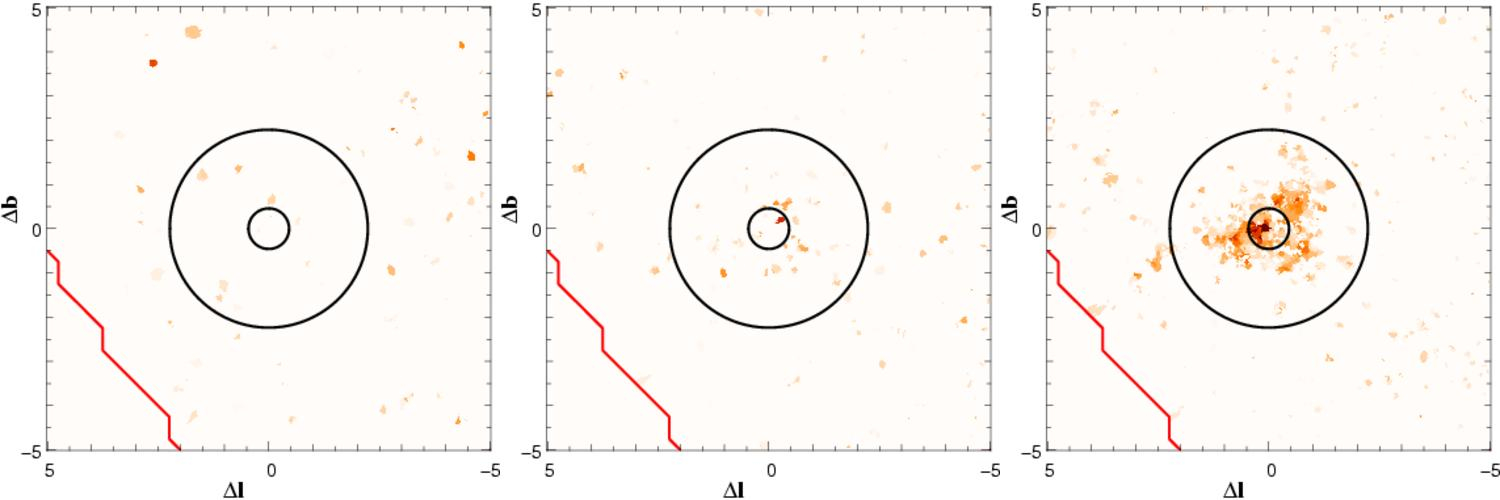}

\caption{Reddening maps from SDSS data for a $10^{\circ} \times 10^{\circ}$ field
containing the high latitude molecular cloud HRK 236+39. The 
computed reddening is shown for M dwarfs having distance moduli spanning 
7.0-8.0 ({\it left}),
8.0-9.0 ({\it center}), and 9.0-10.0 ({\it right}). The two circles
centered on the cloud position are based on the core and envelope
size as tabulated by \citet{dutra02}.
\label{fig:mw:dust3d_hrk236}}
\end{center}
\end{figure}

Example maps from the SDSS project are depicted in 
\autoref{fig:mw:dust3d_hrk236}
for a 10$^{\circ}$ by 10$^{\circ}$
field containing the high latitude molecular cloud HRK 236+39.
These maps are based on the reddening computed for stars having distance
moduli of $7.0 < m-M < 8.0$, $8.0 < m-M < 9.0$, and $9.0 < m-M < 10.0$.
The reddening shown at each pixel is computed as the median of the 
$E(B-V)$ values obtained for the $N=5$ nearest stars. The reddening
associated with the HRK 236+39 cloud is discernible at $m-M > 7.0$ 
($d > 250$ pc) and is obvious at $m-M > 8.0$ ($d > 400$ pc).

\subsubsection{Distance and $A_V$ Limits}

It has been demonstrated that
accurate three-dimensional mapping of the local ISM within a few kpc is possible 
using
SDSS photometry of M dwarfs \citep{mcgehee09}. Analysis of the $g-i$ color
excess in regions effectively free of interstellar reddening 
shows that distance modulus limits of 7.0 (at M5) to 11.2 (at M0)  
result in a volume-limited survey nearly free of the systematic color biases 
inherent in this
$g$-band limited data set. 

\begin{figure}
\begin{center}
\includegraphics[width=0.9\linewidth,angle=0]{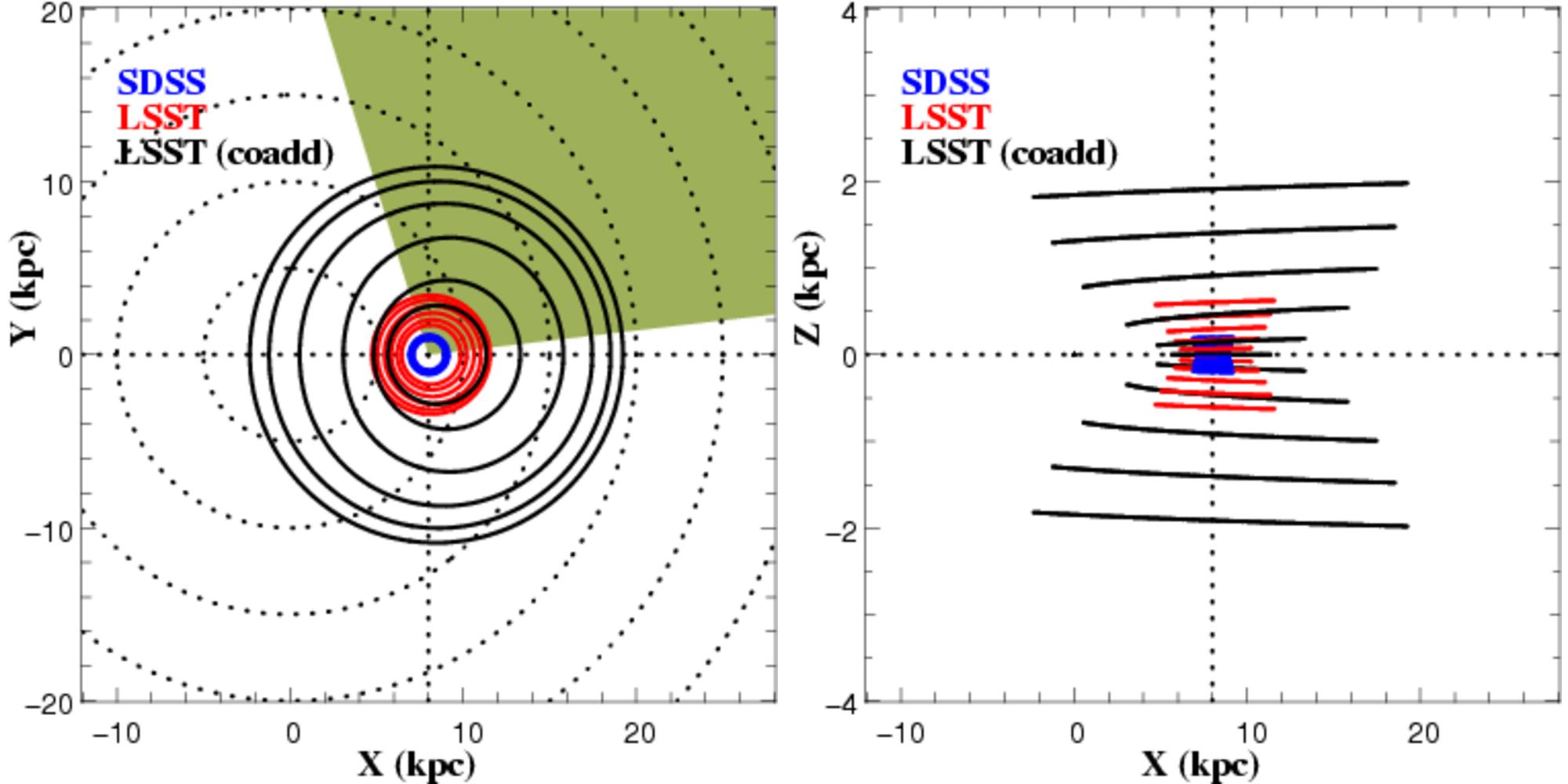}

\caption{
This plane-parallel view of the Galaxy ({\it left}) taken at Z=0.0 kpc
(the Galactic plane) is used to illustrate the dust mapping limit at
specific Galactic latitudes
for the SDSS ({\it blue}), single epoch LSST observations ({\it red}),
and the full LSST survey ({\it black}, assuming 100 visits in $g$). The survey
limits at $|b|=0$, 2, 4, 6, 8, and 10$^{\circ}$ are computed using
the vertical and radial exponential scale Galactic dust model described in 
\autoref{Sec:stellarCounts}.  
The projected positions of the Galactic center and the Sun are at
$X=Y=0$ and $X=8\,kpc, Y=0$, respectively. The shaded region indicates
the portion of the Galactic plane north of $\delta = 34.5^{\circ}$
limit of the survey.
The survey limits are shown on the right in projection onto the $X-Z$ 
plane to illustrate the ability of LSST to probe structures several kpc
above the Galactic disk at significant distances within the plane.
\label{fig:mw:dust3d_avdist}}
\end{center}
\end{figure}

These limits correspond to $g \sim 20.6$ and 
$\sigma_g \sim 0.02 - 0.03$ for single-epoch SDSS 
observations. Given the relative $g$-band $5\,\sigma$ limits of SDSS 
and the LSST single epoch and final co-added surveys, we estimate that
the the co-added LSST data will reach 5 magnitudes deeper in m-M,
allowing the LSST to probe dust structures across a significant portion of 
the Galaxy. In \autoref{fig:mw:dust3d_avdist} we depict the portion
of the Galactic disk accessible by the LSST single and co-added surveys as well
as the SDSS assuming the vertical and radial scale height dust model
outlined in \autoref{Sec:stellarCounts}.  

\subsection{Variation in Extinction Laws}

Changes in the absorption properties of dust grains,
as parametrized by $R_V$, result in a shift in both the direction
and length (for a specific dust column density) of the reddening
vector in a color-color diagram. This is reflected in the reddening-free
colors by variations in the scaling factor used when defining the linear
combination of colors, e.g., in the $E(g-r)/E(r-i)$ term for
$Q_{gri}$.
By analysis of the observed 
color shifts due to reddening it is possible to constrain the value
of $R_V$ along the line of sight and gain insight into the
nature and composition of the interstellar dust in that region of the
Galaxy.

The LSST will be in a unique position to measure the changes in the 
observed reddening vector due to $R_V$ variations due to its superb
photometric accuracy (see \autoref{sec:design:calsim}). The specifications
for LSST are a factor of two more stringent than typically achieved 
in previous surveys, including the SDSS (except for limited photometric 
conditions).

F turn-off stars ($g_{abs} \sim 4$) reside on the blue tip of the
stellar locus in $ugriz$ color space and for $g > 19$ trace the total
Galactic extinction along high-latitude lines of sight. This method
will provide a verification of the far-infrared-based SFD98 extinction model
and allow study of the variations in dust grain sizes as inferred from
$R_V$. The value of $R_V$ provides a general indicator of 
grain size, with the $R_V \sim 4.5-5$ values seen in star formation
regions suggestive of grain growth in cold molecular clouds.

\begin{figure}
\begin{center}
\includegraphics[width=0.9\linewidth,angle=0]{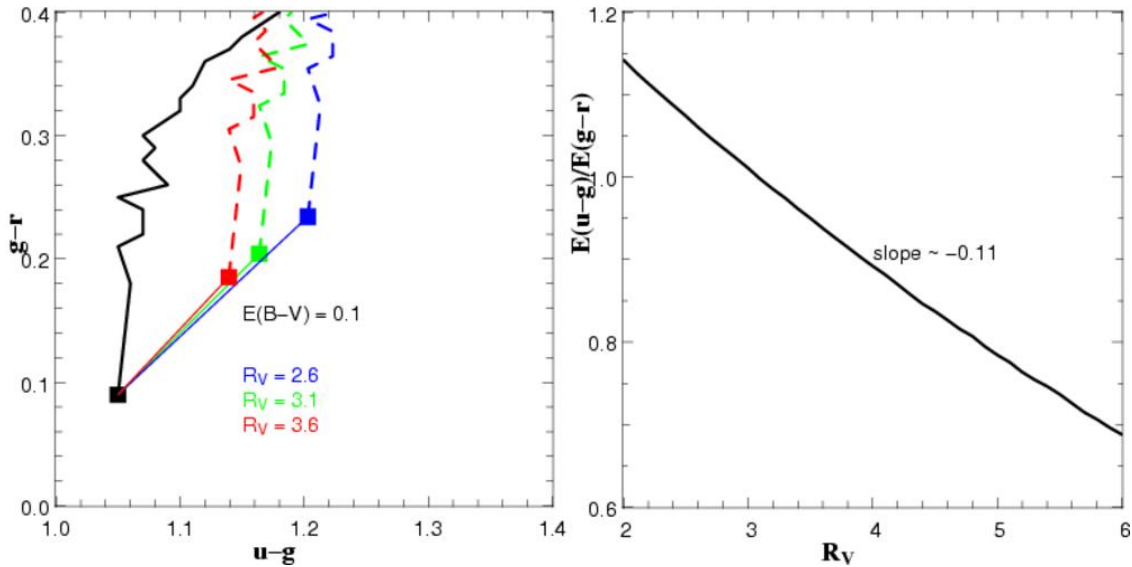}

\caption{The position of the blue tip of the stellar locus, 
populated by F turn-off stars,
can be used to constrain $R_V$, the ratio of general to selective
extinction. On the left panel is shown reddening vectors
of length $E(B-V) = 0.1$ for $R_V$ = 2.6, 3.1, and 3.6. The 
slope of the reddening vector ($E(u-g)/E(g-r)$) is a monotonic
function of $R_V$, having a mean derivative of $\sim -0.11$ in the domain 
$2 < R_V < 6$
({\it right}).
\label{fig:mw:dust3d_bluetip}}
\end{center}
\end{figure}

The slope of the reddening vector is sensitive to the value of
$R_V$ as shown in \autoref{fig:mw:dust3d_bluetip}. For the SDSS passbands
and an assumed F star source SED, the value of $E(u-g)/E(g-r)$ is larger
for small $R_V$ and decreases with a slope of approximately $-0.11$
with increasing $R_V$. This analysis mandates precise and well-calibrated
photometry. For example, determination of $R_V$ to within $\sigma_{RV} = 0.5$
requires the slope of the reddening vector to be measured to $\sigma_m = 0.06$.
If $E(B-V) = 1$ along the line of sight, then the required photometric
accuracy is 2\%. The photometric accuracy requirement becomes proportionally 
more stringent as the dust column density decreases due to the reduced 
movement of the blue tip in the color-color diagram. LSST, with better than
1\% photometric accuracy in the final co-added survey, will be able to
study $R_V$ variations in both Galactic plane and high latitude environments.

\section{Streams and Structure in the Stellar Halo}
\label{sec:MW:streams}

{\it \noindent Carl J. Grillmair, Ata Sarajedini}

\begin{figure}
\centering\includegraphics[width=0.9\linewidth,angle=0]{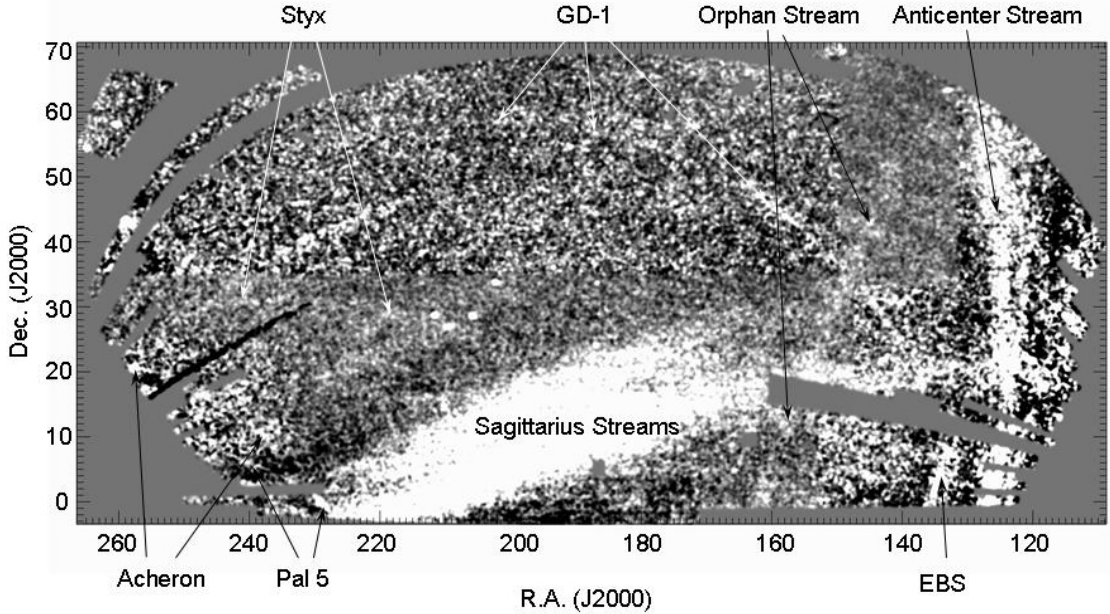}
\caption{A composite, filtered surface density map of stars in the
SDSS Data Release 5. Stars in DR5 have been filtered to select stellar
populations at different distances with color-magnitude sequences
similar to that of the globular cluster M 13
\citep{grillmair2009}. Lighter shades indicate areas of enhanced
surface density, and different portions of the field have been
filtered for stars at different distances. Varying noise levels are a
consequence of the very different levels of foreground contamination
using these different filters. The distances of the streams range from
4 kpc for Acheron, to 9 kpc for GD-1 and the Anticenter Stream, to 50
kpc for Sagittarius and Styx.} \label{fig:streams_composite}
\end{figure}

Cosmological simulations predict that the halo of our Galaxy should be
composed at least partly of tidal debris streams from disrupted dwarf
galaxies \citep{bullock2005}. Some fraction of the halo is also
believed to be made up of debris streams from both existing and
disrupted globular clusters \citep{grillmair1995, gnedin1997}.  At
least 11 substantial streams have now been detected in the SDSS and
2MASS \citep{newberg2002, yanny2003, majewski03, odenkirchen2003,
  rocha2004, grillmairdionatos2006a, grillmairjohnson2006,
  belokurov2006, grillmairdionatos2006b, belokurov2007,
  grillmair2006a, grillmair2006b, grillmair2009}. The more prominent
of these are shown in \autoref{fig:streams_composite}.  In this
section, we focus on identifying the stellar streams around the Milky
Way that can be studied with individual stars.  The detection and
study of very low surface brightness stellar streams based on diffuse
light is discussed in \autoref{sec:gal:lsb}.  Using the proper motions
of tidal stream stars to derive their orbits is discussed in
\autoref{sec:MW:haloPMs}.

Tidal streams provide powerful and sensitive new probes for studies of
Galactic structure and formation.  For example, the mapping of the
positions and motions of stars in tidal streams is the most accurate
method known for determining the mass distribution of the Galactic
halo \citep{johnston1999, odenkirchen2000}.  For dwarf galaxies and
globular clusters, tidal stripping is a relatively weak process, and
the stripped stars are left with very small random velocities ($\sigma
\approx 1-10$ km s$^{-1}$). These stars therefore travel in orbits
almost identical to those of their progenitors. By sampling the
motions of stream stars at various points along the orbit, it becomes
possible to accurately measure the exchange of potential and kinetic
energies, and thus the potential field of the Galaxy
\citep[e.g.][]{grillmair1998, johnston1999}. With a sample of many tidal
streams, both their orbits and the shape of the Galactic potential can
be determined in a self-consistent manner. Globular cluster streams
are particularly useful in this respect as they will be both numerous
and dynamically cold \citep{combes1999}. They will not only help to
constrain the overall shape of Galactic potential, but also to probe
its lumpiness and perhaps reveal the existence of pure dark matter
subhalos \citep{murali1999, johnston2002}.

Tidal streams also provide a new window on the formation process of
the Galaxy. The streams discovered to date appear to be very
long-lived structures, and simply counting streams will greatly
improve estimates of the number and distribution of dwarf galaxies and
star clusters which, through disruption, contributed to the buildup of
the Galactic halo \citep{bullock2005}. Cosmological models suggest
that there may be considerably more substructure at larger radii ($R >
50\, \rm kpc$), with orbits becoming predominantly radial for the more
remote objects.  As photometric and kinematic surveys reach ever
further and wider, we can look forward to a day when we will be able
to lay out a precise, chronological sequence of the major events that
led to the Galaxy as we see it today.

The detection of tidal streams is now reaching the limit of what is
possible with the SDSS; the most recently discovered streams having
been detected at the $\sim 7\,\sigma$ level
\citep{grillmair2009}. However, by virtue of its areal coverage and
much fainter limiting magnitude, the LSST survey will be able to
detect many more streams, both locally and throughout the Local
Group. Current simulations predict that at least 20\% of detectable
dwarf galaxy debris streams reside at $R > 50$ kpc \citep{johnston08}.
Due to both the limiting magnitude of the SDSS and a selection bias
that strongly favors long features in the plane of the sky
(e.g. \citealt{grillmair2009}), the seven known globular cluster streams
all lie within 10 degrees of being perpendicular to our line of sight.
Assuming that the orbits should be oriented more or less
isotropically, and that this selection bias can be overcome with
deeper photometry (to reach the populous turn-off and main sequence)
and improved search techniques, then scaling to all possible
orientations one would expect another $\sim 80 - 170$ globular cluster
streams within 50 kpc waiting to be discovered in the LSST survey
area.  Some fraction of these will be found by SkyMapper and
Pan-STARRS, but the more tenuous, inclined, and distant streams will
require the extended reach of LSST. If globular cluster progenitors
and their debris fall off as $R^{-3}$, then LSST could find another 60
to 130 debris streams beyond 50 kpc. The actual number will presumably
depend on the supply of relatively loosely bound clusters at these
distances, and/or whether the orbits are sufficiently radial that tidal
stresses can remove large numbers of stars.

The use of matched filters in color-magnitude space
\citep{rockosi2002, grillmair2009} is currently the most efficient way
to detect dwarf galaxies, tidal streams, and other low surface
density structures (\autoref{sec:mw:search}). This technique is particularly well suited to
LSST-like data.  By its nature, the matched filter makes optimal use
of every star in a structure of interest based on its color and
magnitude and how these relate to the color-magnitude distribution of
contaminating foreground stars and the unresolved background
galaxies. To first order, the signal-to-noise ratio of a stream
detection goes as $N_{s}/\sqrt{N_f}$, where $N_s$ refers to the number
of stars in the stream and $N_f$ to the number of foreground stars in
the same color-magnitude space.  By going deeper and improving the
photometry at all magnitudes, LSST will both greatly increase $N_s$,
and significantly reduce the relative contribution of foreground
stars.  For example, the globular cluster stream Lethe at 13 kpc is
detected at the 7$\,\sigma$ level in the SDSS
\citep{grillmair2009}. Using the luminosity function of $\Omega-Cen$
\citep{demarchi1999} and the Besancon model of the Galaxy
\citep{robin2003} to estimate the stream and field star populations
down to $g = 25$, we find that a single LSST pass would detect this
stream at the $\approx 20\,\sigma$ significance level.

The end-of-survey photometric depth that will be achieved by LSST is
important for two reasons: 1) a larger portion of the main sequence
will be accessible, where the stellar luminosity function provides
many more stars that can contribute directly to the signal and 2) the
useful range of a main sequence matched filter can be extended much
further out into the local volume. While matched filters have been
used to find dwarf galaxies and tidal streams in the SDSS out to $\sim
50$ kpc, the same techniques applied to end-of-survey LSST data will
enable similar detections out to nearly 0.5  Mpc (where the main
sequence turn-off for old populations falls below the detection
limit). The volume sampled by LSST will thus be nearly three orders of
magnitude larger than that of SDSS.  

With a magnitude limit similar to SDSS, SkyMapper \citep{keller07}
is expected to find the strongest substructures within 50 kpc in the
southern hemisphere. Working to a limit of $g \sim 24$, Pan-STARRS
(PS-4) is expected to find such structures out to 100 kpc in a single
pass of the $3\pi$ survey, and perhaps 250 kpc at end-of-survey. Gaia
is not expected to find $new$ structures at distances greater than 20
kpc.  End-of-survey LSST data will therefore
sample a volume almost an order of magnitude larger than any other
existing or planned survey.

At least three factors will tend to limit the value of increasing
depth: 1) The lower main sequence of even very old and metal poor
stars will have colors very similar to the bulk of the foreground
population, and a properly constructed filter will unweight these
stars to a degree where it becomes pointless to include them. Where
this happens will depend critically on the photometric precision - a
very narrow matched filter can be carried much further through the sea
of foreground stars than a broad one. 2) For nearby streams ($r < 40$
kpc), LSST photometry will ultimately push well beyond the peak of the
stream's stellar luminosity function, to where the increase in the
number of stream stars (the signal) is vastly exceeded by the increase
in the number of intervening foreground stars and of unresolved
galaxies (the noise).  The matched filter will naturally compensate
for this by unweighting the faintest stars, but it sets an upper limit
on the signal-to-noise ratio that can be achieved. 3) If the number of
dwarf galaxies and tidal structures surrounding the Galaxy falls off
faster than $R^{-3}$, then fewer of them will be found at the faintest
magnitudes. (This of course would be an important finding in itself).

Using several colors can yield significant improvement in
signal-to-noise ratio, provided that the photometric precision is
similar among the wave bands. Multiple colors can help to remove some
fraction of unresolved non-stellar sources (i.e., those with power-law
spectra), which at the faintest LSST magnitudes will vastly outnumber
stars. More importantly, since each color represents an independent
measurement, using all available colors can improve the placement of a
given star within the matched filter by the square root of 
the number of colors used, and reduce the noise accordingly.

The end-of-survey proper motions from LSST will also be useful, both
for detecting streams and substructures and for constraining their
orbits (see \autoref{sec:MW:haloPMs}). Indeed, as a completely
independent measurement, proper motions will enable the identification
of much fainter or diffuse remnants than would be possible with
color-magnitude filtering alone.  Unequivocally demonstrating a
physical association of stars in very large, sparse, amorphous, broken
up, or widely separated structures almost certainly will require
measuring similar (or at least consistent) mean proper motions among
all components.  While proper motion measurements for individual stars
in the halo will be uncertain ($\sigma \sim 100 \rm \, km \,s^{-1}$ at
100 kpc), the uncertainties are expected to be dominated by random
measurement errors and thus be amenable to averaging. Combining proper
motion measurements for many hundreds of stars selected by
color-magnitude filtering will reduce the error in the mean to a level
($< 10 \rm \, km \,s^{-1}$) where widely spaced or fragmentary
detections can be confidently related to one another, or significant
constraints can be placed on the orbits of structures or on the
Galactic potential \citep[e.g.][]{grillmair2009}. Since the measured
dispersion in the tangential velocities will be a convolution of the
intrinsic tangential velocity dispersion of stars in the structure
with the measurement errors, simply demonstrating that the intrinsic
velocity dispersion must be nearly zero (as opposed to $\approx 100
\rm \, km \,s^{-1}$ for random halo stars) will enhance the
significance of otherwise marginal photometric detections. Finally,
for a prescribed Galactic potential, proper motions can be used to put strong constraints on the orbit of the progenitor, even in the
absence of radial velocity measurements \citep{eyrebinney2009}.

Distances to streams will be estimated using both main sequence
fitting techniques \citep{grillmair2009} and (depending on the natures
of the progenitors) RR Lyrae stars. LSST data will be particularly
important in both respects, as the faint, end-of-survey magnitude
limit will enable robust, age-independent, main sequence comparisons,
and any RR Lyrae stars in these streams will most likely have been
discovered by LSST as well.  For streams detected in the SDSS,
relative distances estimated via matched-filtered, main sequence
fitting to two or three magnitudes below the turn-off are precise to
$\sim 5-10\%$ \citep{grillmairdionatos2006b, grillmair2009}, with
absolute accuracies limited both by age and metallicity mismatches
between the stellar populations in the streams and those in the
globular clusters used as templates, and by the RR Lyrae distance
estimates to these same globular clusters. Similar methods using LSST
photometry are expected to improve precision by at least factor of
two simply by virtue of the greater extent of the main sequence
available for fitting. Accuracy will continue to be limited by
template mismatches and RR Lyrae distances to template globular
clusters.

RR Lyrae stars in streams and substructures are useful for a number of
other reasons. First, their presence usually suggests that the stellar
population is older than $\sim$10 Gyr. The intrinsic color of the
ab-type RR Lyraes, those pulsating in the fundamental mode, is
constant with very little dependence on metal abundance
\citep{sturch66,guldenschuh05}, suggesting that reddenings to these RR
Lyraes can be determined with an error of $\pm$0.02 mag in $E(B-V)$. In
addition, a number of investigators \citep{sandage93,alcock00} have
found a correlation between period and metal abundance for ab-type and
c-type RR Lyraes, those pulsating in the first overtone. This relation
can yield individual abundances to $\pm$0.3 dex but will do much
better for populations of these stars in establishing their relative
metallicity scales. This in turn will aid in the selection of an
appropriate template population with which to map the stream. RR Lyrae
stars, and the science that they will facilitate, are discussed in
more detail in \autoref{sec:sp:rrlyrae}.

Using debris streams as precision mass tracers will require
considerable follow-up work with wide-field, multiplexing
spectrographs to obtain radial velocities of individual
stars. However, the single greatest hurdle in unraveling the halo
remains the detection, unique identification, and tracing of these
streams. With its unprecedented combination of depth, areal coverage,
and wavelength sampling, LSST will provide the most extensive and
detailed map of the structure of the Galactic halo yet conceived.


%
%

\section{Hypervelocity Stars: The Black Hole--Dark Halo Link?}
\label{sec:MW:HVSs}

{\it \noindent Jay Strader, James S. Bullock, Beth Willman}


Hypervelocity stars (HVS) were discovered as stars in the Milky Way's
halo with anomalously high velocities \citep{brown05,brown06a}. Their
large Galactic rest frame velocities ($v > 500\,\kms$) suggest
ejection from the center of the Galaxy in a three-body interaction
with the supermassive black hole. Velocities of 1000 \kms, coupled
with an estimated production rate of one HVS every $\sim 10^5$ years
(for binary-black hole interactions; \citealt{yu03}), give an estimated
total population of 1000 HVS within 100 kpc. However, there will be a
range of ejection velocities, and all HVS will decelerate on their way out
of the Galaxy, so the actual number of observable HVSs may be higher.

The study of HVS can: 1) provide important constraints on the dynamics
near the center of the Galaxy, including limits on multiple black
halos or black hole binaries; 2) distinguish among triaxial models for
the halo with an estimate accurate to several \kms\ for the
three-dimensional motions for two or three HVS \citep{gned05}; and 3)
provide an estimate of the initial mass function in the Galactic
Center, based on the relative numbers of low- and high-mass HVS
\citep{koll07}.

Three-body ejection of stars by supermassive black holes is not a
unique interpretation for the observed population of hypervelocity
stars. 
Runaway ejections of stars from binaries can account for a fraction of
the low-velocity tail of HVS. \citet{abad09} have suggested that many
HVS could be stars tidally stripped from accreted dwarf galaxies,
though this proposal requires a high virial mass for the Galaxy ($2.5
\times 10^{12} M_{\odot}$).

Most known HVS were discovered by radial velocity surveys of late B
stars (mass $\sim 2.5-4 M_{\odot}$) in the outer halo
\citep{brow06}. Such stars are uncommon in the halo but have long
enough main sequence lifetimes to have traveled from the Galactic
Center at their high velocities (this is generally not the case for
early B or O stars, while stars of spectral type A and later are
common in the halo).

Because the Sun is close to the Galactic Center, the transverse velocities 
of distant HVS will be small compared to their radial velocities, 
independent of the ejection vector. The magnitude of the effect is $v_{tr} 
\sim$ (8 kpc/$d$) $v_{tot}$. Known HVS lie at $\sim 50-100$ kpc and so 
will have small proper motions. Beyond $\sim 20$ kpc, it will be difficult 
to separate candidate HVS from normal halo stars on the basis of 
kinematics alone, although other stellar properties (such as 
metallicity, since HVS should be relatively metal-rich) may be used to 
distinguish HVS from halo dwarfs.

For this reason, an LSST search for HVS would focus on a volume within
$\sim 10-20$ kpc of the Sun. Gaia will obtain more accurate proper
motions than LSST for stars with $r < 20$; this
magnitude limit corresponds roughly to the old main sequence turn-off
at a distance of 10 kpc. For $V > 20$, LSST will dominate, with
estimated proper motion accuracy of $\sim 10$ \kms\ at a distance of
10 kpc. It follows that the HVS niche for LSST is in finding
\emph{low-mass} HVS. As an example, an HVS 10 kpc from the Sun with a
transverse velocity of 500 \kms\ will have a proper motion of 10
mas$\,\rm yr^{-1}$. The proper motion error at the single-visit limit
of the survey ($r=24$) is $\rm \sim 1\, mas\,yr^{-1}$, so HVS with
absolute magnitudes as faint as $M_r = 9$ (mass $\sim 0.4 M_{\odot}$)
will be identified.  Such stars are too faint to be studied by Gaia,
and are so rare in the solar neighborhood that they are unlikely to be
selected by any radial velocity survey.

If we assume a total of $10^3$ HVS emitted at uniform angles, 
LSST alone should discover $\sim 10$ HVS within 20 kpc of the Sun. It will be 
the only proposed survey sensitive to low-mass HVS over a significant 
volume. If properties such as metallicity can be used to efficiently
separate HVS from halo stars, then the yield could be higher by a factor
of several, especially if coupled with a follow-up radial velocity survey.

\section{Proper Motions in the Galactic Halo} 
\label{sec:MW:haloPMs}

{\it \noindent Joshua D. Simon, James S. Bullock}

LSST will provide a major step forward in our understanding of the
Milky Way in a cosmological context by enabling a new set of precise
constraints on the total mass, shape, and density profile of its dark
matter halo.  The key LSST deliverable that will allow these advances
is unprecedentedly accurate proper motion measurements for millions
of main sequence stars and hundreds of tracer objects in the outer
halo (see \autoref{sec:com:PMacc} and \autoref{sec:MW:tomog} for
details on proper motion measurements).  The orbits of the tracer
populations (e.g., dwarf galaxies, globular clusters, and
high-velocity stars) will also provide an important means for testing
models of the formation and evolution of the tracers themselves.

Constraints on the dark matter halo of the Milky Way are motivated by
at least three distinct scientific goals.  First, the total dark
matter mass of the Milky Way halo is an important zero-point for
models of galaxy formation \citep[e.g.][]{somerville08,mb04}.  Second,
the global shape of the Milky Way dark halo can be compared directly
to $\Lambda$CDM predictions for the shapes of dark matter halos
\citep{allgood06}.  Finally, the overall mass of the Milky Way halo is
a critical normalizing constraint for the local velocity dispersion of
dark matter particles, which is an important input for dark matter
direct detection experiments (see review by \citealt{gaitskell04}).

Because the Milky Way is the galaxy that we can study in the most
detail, it necessarily provides the benchmark normalization for
semi-analytic modeling of galaxies, which is a valuable tool for
comparing a wide variety of observations of galaxy evolution to
theoretical predictions.  The stellar mass and cold gas mass of the
Galaxy are already well-known; what remains uncertain at the factor of
$\sim2-3$ level is the mass of the dark matter halo
\citep[e.g.,][]{klypin_etal_02,battaglia05}.  Improved measurements of the
Milky Way halo mass will also offer new possibilities for solving the
missing baryon problem: the observation that the observed baryons in
galaxies account for half or less of the Big Bang nucleosynthesis
value of $\Omega_{b}$ \citep{fukugita98,mb04}.  Similarly, with
current halo mass estimates the observed baryon fraction of the Milky
Way is $f_{b} \sim 0.05$, well below the cosmic value of $f_{b}= 0.20$
\citep{komatsu09}.  Since the difference between the Milky Way baryon
fraction and the universal one is similar in magnitude to the
uncertainty of the halo mass, nailing down the total mass of the Milky
Way's dark matter halo could have important implications for the
severity of the discrepancy and the location of the missing baryons.

In addition to mapping out the Galactic potential, the orbits of Milky
Way satellite galaxies are a critical input to models of the formation
of dwarf galaxies. HST observations of the star formation
histories of Local Group dwarf galaxies reveal that each galaxy has a
unique history \citep{orban08}.  This result suggests that the
individual epochs of star formation and quiescence experienced by each
dwarf could be related to tidal effects from the Milky Way.  If so,
detailed knowledge of their orbits will allow predictions for star
formation histories that can be compared directly with observations.
Such measurements are particularly important for the Magellanic
Clouds, where recent HST proper motions have suggested that the
LMC and SMC are on their first passage around the Milky Way
\citep{besla08}, rendering preferred explanations for the origin of
the Magellanic Stream extremely problematic.  More fundamentally, the
discovery of the ultra-faint dwarf galaxies (\autoref{sec:MW:UFs}) has
raised a number of burning questions: how did such incredibly tiny
galaxies manage to form?  Are they merely the remnants of much more
luminous objects --- similar to the classical dSphs --- that have lost
most of their mass from tidal stripping?  Or did they never contain
more than the $10^{3} - 10^{5}$ stars that they host today?
Theoretical modeling has argued against the tidal stripping hypothesis
\citep{penarrubia08}, but the only conclusive test will be to derive
orbits for the ultra-faint dwarfs and determine whether they have been
subject to strong enough tidal forces to remove nearly all of their
stars.

Finally, full orbits from proper motions will offer the potential to
match dwarf galaxies and globular clusters to the tidal streams they
leave behind as they are assimilated into the halo.  Most of the
streams identified in SDSS data (see \autoref{sec:MW:streams}) lack an
obvious progenitor object, compromising their utility as tracers of
the Galaxy's accretion history.  Likewise, identifying the signatures
of tidal stripping in dwarf galaxies other than Sagittarius has proven
to be controversial \citep{munoz06,munoz08,sohn07,mateo08,lokas08}.
Confirming kinematic associations between stripped stars and their
parent objects will both clarify the impact of tidal interactions on
dark matter-dominated systems and provide new insight into the buildup
of the Milky Way's stellar halo by the destruction of dwarf galaxies
and star clusters.


LSST observations will produce resolved proper motion measurements for
individual stars in nearby dwarf galaxies and globular clusters.
Coupled with pre-existing line-of-sight velocities, these data will
yield three-dimensional orbital velocities for virtually all bright
satellites over more than half of the sky.  The orbits will
substantially strengthen constraints on the mass distribution of the
Galaxy, particularly at large radii.  Radial velocities alone do not
improve upon existing halo mass constraints, but three-dimensional velocities will
enable new mass measurements out to radii beyond 200~kpc, approaching
the expected virial radius of the Milky Way's dark matter halo.

As summarized in \autoref{tab:com:T3} and discussed in
\autoref{sec:com:PMacc}, LSST will provide
0.2~mas~yr$^{-1}$ (1~mas~yr$^{-1}$) proper motion accuracy for objects
as bright as $r=21$ (24) over its 10-year baseline.  For main sequence
stars at a distance of $\sim15$~kpc (60~kpc), this proper motion
accuracy corresponds to approximately $\sim15$~km~s$^{-1}$
(300~km~s$^{-1}$) velocity accuracy per star.  Measurements at these levels for
more than 200 million stars will enable high-precision mass models of
the Milky Way halo.  By the end of the survey, tangential velocities
with accuracies better than 100~km~s$^{-1}$ will be available for
every red giant star within 100~kpc.

While some of the measurements described in this section, particularly
orbital motions for the closest dwarf galaxies, may be obtained by HST by the time LSST is operational, the completely independent
observations obtained by LSST will be extremely valuable.  With its
large field of view and deep multicolor photometry, LSST is better
suited to astrometric measurements of very low surface density and
very distant (but spatially extended) objects like many of the Milky
Way dwarfs.  Moreover, the recent controversy over and revision of the
proper motions of such nearby and well-studied objects as the
Magellanic Clouds demonstrate the importance of multiple independent
measurements for deriving reliable proper motions.

\section{The Darkest Galaxies} 
\label{sec:MW:UFs}


{\it \noindent Beth Willman,  James S. Bullock}

The 11 dwarf galaxy companions to the Milky Way known prior to 2000
have been extensively studied.  The properties of these shallow
potential well objects are highly susceptible to ionizing radiation
(reionization), tides, and supernova feedback - poorly understood
processes of fundamental importance to galaxy formation on all scales.
Moreover, their resolved stellar populations provide a unique picture
of star formation and chemical enrichment in the early Universe, and
may provide direct information on the sources that reionized the
Universe. \autoref{sec:sp:sfh} describes the method that can be used
to derive the detailed formation histories of resolved galaxies using
LSST.  To use these detailed histories to disentangle the competing
effects of reionization, tides, and supernovae requires a
statistically significant sample of nearby resolved dwarf galaxies,
over the largest possible dynamic ranges of mass and environment.

In 1999, simulations of structure formation in a cold dark matter
(CDM) dominated Universe highlighted the discrepancy between the
number of dark matter halos observed to be lit up by these 11 Milky
Way dwarf galaxies and the number of dark matter halos predicted to
orbit around the Milky
Way 
\citep{klypin99,moore99}.  As simulation resolution improved, the
magnitude of this apparent discrepancy increased, with the most recent
simulation containing 300,000 gravitationally bound dark matter halos
within the virial radius of a galaxy with the mass of the Milky Way
\citep{springel08}.

There are three potential explanations for this observed discrepancy:
1) The present CDM dominated cosmological model is wrong; 2)
Astrophysical processes prevent the vast majority of low mass, dark
matter halos from forming stars; or 3) The dwarf galaxies are there,
but have not yet been found.  These explanations are not mutually
exclusive.  The least luminous dwarf galaxies thus bear great
potential to simultaneously reveal the micro- and macroscopic
properties of dark matter and the effects of environment and feedback
on galaxy suppression.  To fully exploit this potential {\it requires}
an unbiased and carefully characterized census of dwarf galaxies to
the faintest possible limits.  LSST is the only planned survey with
the depth, filter set, and wide-field necessary to search for the very
least luminous dwarf galaxies out to the virial radius of the Milky
Way.

Since 2004, 25 dwarf galaxy companions to the Milky Way and M31 have
been discovered that are less luminous than any galaxy known before
\citep[e.g.,][]{willman05a, zucker06a, belokurov07b, mcconnachie08}.
These new discoveries underscored the role of incompleteness in past
attempts to use nearby dwarf galaxies to pursue the answers to
cosmological questions.  These ``ultra-faint'' dwarfs have absolute
magnitudes of only $-2$ mag $< M_V < -8$ mag (L$_{V} \simeq 10^3 -
10^5$ L$_\odot$, a range extending below the luminosity of the
average globular cluster) and can be detected only as slight
overdensities of resolved stars in deep, uniform imaging
surveys. Follow-up spectroscopy reveals that these, the least luminous
galaxies known, are also the most dark matter dominated
\citep{martin07,simon07,strigari08a} and most metal poor
\citep{kirby08} galaxies known.

\subsection{The Edge of a Vast Discovery Space}
\begin{figure}
\begin{center}
\includegraphics[width=0.9\linewidth,angle=0]{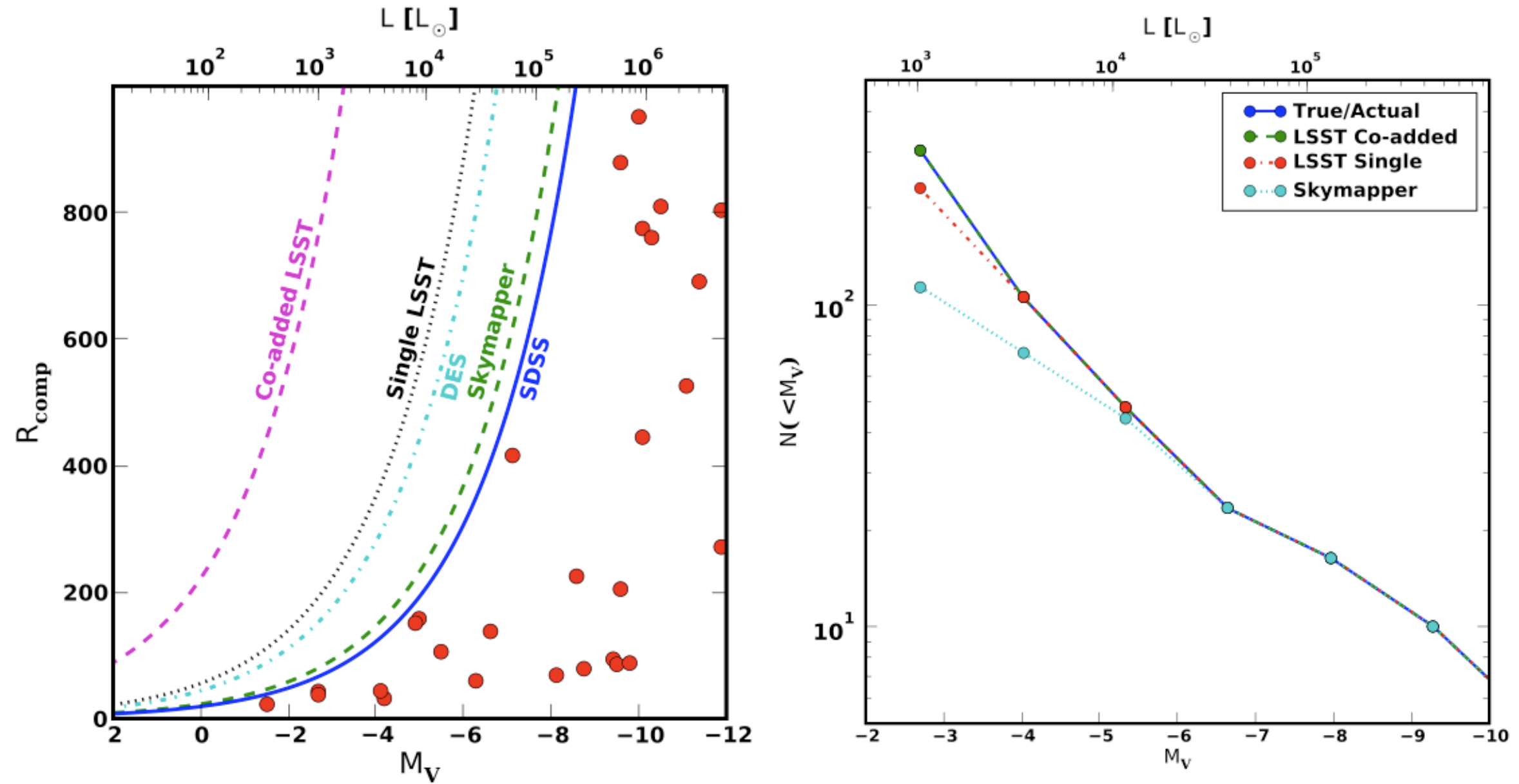}
\caption{Left panel: Maximum detection distance of dwarf galaxies in
  the SDSS Data Release 5 stellar catalogs, and projected for future
  surveys.  The dwarf galaxies known within 1 Mpc are overplotted in
  red.  Right panel: The predicted luminosity function of dwarf
  galaxies within 400 kpc of the Milky Way (blue line) over 4$\pi$
  steradians.  Overplotted are the expected number of these dwarfs
  that may be discovered over the entire sky with survey data similar
  to the upcoming SkyMapper Southern Sky Survey \citep{keller07} and
  LSST.  The green line for LSST is hiding behind the blue line.  Both
  figures are from \citet{tollerud08}, with permission.} 
\label{fig:mw:tollerud_fig}
\end{center}
\end{figure}

Both empirical and theoretical evidence suggest that the LSST data set
is likely to reveal hundreds of new galaxies with luminosities
comparable to those of this new class of ultra-faint dwarfs.  The
recent discovery of Leo V by \citet{belokurov08}, and the 30
statistically significant - yet previously unknown - stellar
overdensities identified by \citet{walsh09} highlight the possibility
that more ultra-faint dwarfs may yet be found in the relatively
shallow SDSS data set, and that many more may be hiding beyond the edge
of detectability.  Unfortunately, there will still be an unavoidable
luminosity bias in searches for dwarf galaxies possible with SDSS, the
Southern Sky Survey, and Pan-STARRS-1.  The distances of the dwarf galaxies known within 1 Mpc are plotted versus their $M_V$ in the left
panel of \autoref{fig:mw:tollerud_fig} (Figure 9 of
\citealt{tollerud08}).  The overplotted blue line shows that the known
dwarfs fill the volume accessible by SDSS.  The depth of the co-added
LSST survey (purple dashed line) could reveal objects, like the very
least luminous now known ($M_V \sim -2$) to distances of 600 kpc --
several thousand times the volume searchable by SDSS. A
straightforward luminosity bias correction suggests there may be as
many as 500 ultra-faint dwarf galaxies within the virial radius of the
Milky Way \citep{tollerud08}.  A survey of the depth of LSST should
detect all known Milky Way satellite galaxies within 420 kpc, assuming
a population of satellites similar to those known.  The right panel of
\autoref{fig:mw:tollerud_fig} shows the predicted luminosity function
of Milky Way dwarf satellites, overplotted with the expected number
that could be discovered in an LSST-like survey.

\subsection{New and Longstanding Questions}
The depth and wide field of LSST will
facilitate a complete census (within LSST's footprint) of the Milky
Way's satellite galaxies and will reveal ultra-faint dwarf galaxies
beyond the edge of the Local Group.  These improvements will
revolutionize knowledge of the ultra-faints in several ways that will
only be possible with an LSST-like survey:

{\noindent $\bullet$ {Taking the Temperature of Dark Matter:}} A
census that reveals the total number of dwarf galaxies without
assumptions may yield a large enough number of dwarfs to rule out dark
matter models with reduced power on small scales, although numerical
effects presently inhibit concrete predictions of such models
\citep{wang07}. Moreover, dwarf galaxy kinematic studies will be
useful in placing limits on (or measuring the existence of) a
phase-space limited core in their dark matter halos.  This will
provide an important constraint on the nature of dark matter. The
ability for dark matter to pack in phase space is limited by its
intrinsic properties such as mass and formation mechanism. CDM
particles have negligible velocity dispersion and very large central
phase-space density, resulting in cuspy density profiles over
observable scales.  Warm Dark Matter (WDM), in contrast, has smaller
central phase-space density, so that density profiles saturate to form
constant central cores.

{\noindent $\bullet$ {Low-luminosity Threshold of Galaxy Formation:}}
The discovery of dark-matter dominated galaxies that are less luminous
than a star cluster raises several basic questions, including the
possibility of discovering a threshold luminosity for galaxy
formation.  LSST will enable discovering, enumerating, and
characterizing these objects, and in doing so provide a testing ground
for the extreme limits of galaxy formation.


{\noindent $\bullet$ {The Underlying Spatial Distribution of the Milky
    Way's Dwarf Galaxy Population:}} The epoch of reionization and its
effect on the formation of stars in low mass dark matter halos also
leaves an imprint on both the spatial distribution
\citep{willman04,busha09} and mass function of MW satellites
\citep{strigari07b,simon07}.  Other studies have claimed that the
spatial distribution of MW satellites is inconsistent with that
expected in a Cold Dark Matter-dominated model
\citep{kroupa05a,metz08a}.  The interpretation of such results hinges
critically on the uniformity of the MW census with direction and with
distance.

{\noindent $\bullet$ {Indirect Detection of Dark Matter: Detecting
    dark matter through the products of its decay or self-annihilation
    in an astrophysical system is an exciting prospect. It is possibly
    the only way we can infer or confirm the physical nature of the dark
    matter in the Universe. Dark matter models from theories with new
    physics at the weak scale generically predict high-energy
    annihilation products such as gamma-rays.  The closest and densest
    dwarf galaxies are expected to be the brightest sources
    \citep{strigari08a} after the Galactic Center.}}

\subsection{General Search Technique}
\label{sec:mw:search}

\begin{figure}
\centering\includegraphics[width=1.0\linewidth,angle=0]{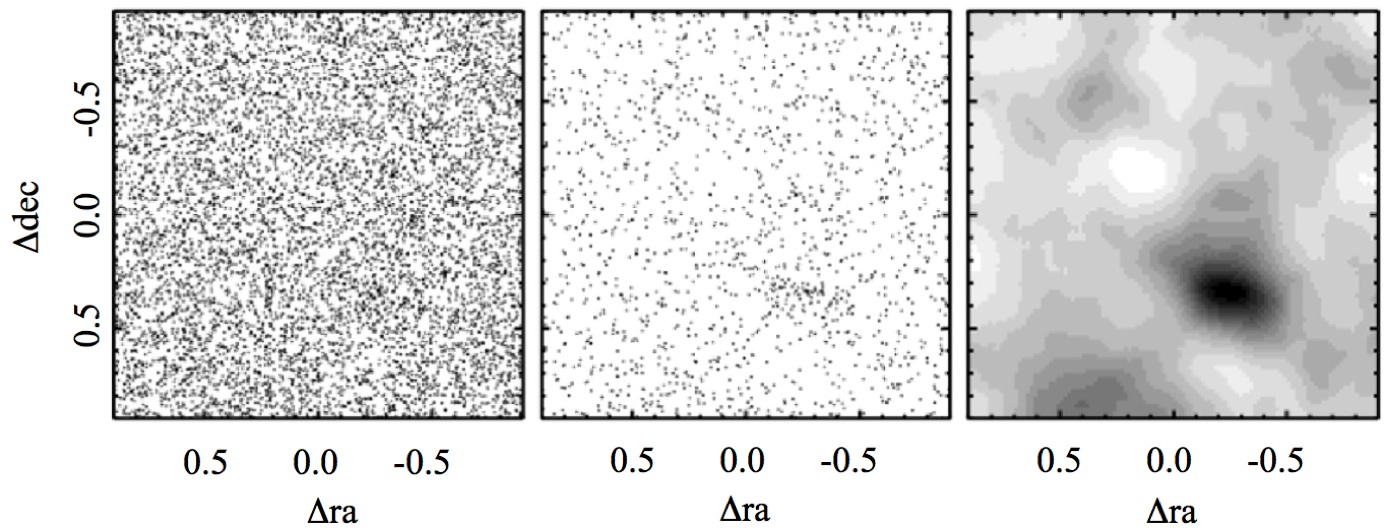}

 \caption{Far left: Map of all stars with $r < 21.5$ in the field around the Ursa Major
  I dwarf satellite, $M_V = -5.5$, $d = $100 kpc. Middle: Map of stars
  passing a color-magnitude filter projected $m-M = 20.0$
  Far right: Spatially smoothed number density map of the stars in the
  middle panel.  The galaxy has a central $V$-band surface
  brightness of only 27.5 mag arcsec$^{-2}$ \citep{martin08}. Figure and 
  caption from \citet{willman09a} with permission.  Data from SDSS Data Release 7.}

\label{fig:mw:search_technique}
\end{figure}

The known ultra-faint dwarfs are up to ten million times less luminous
than the Milky Way, and are invisible (except for Willman 1 and Leo T)
in the SDSS images that led to their discoveries, even in hindsight.
How can these invisible galaxies be discovered?  They are found
as statistically significant fluctuations in the number densities of
cataloged stellar objects, not from analysis of the images themselves.
\autoref{fig:mw:search_technique} illustrates the general technique
that has been used to search for ultra-faint dwarfs by \citet{koposov08}
and \citet{walsh09}, among others.  This general procedure, which can
also be used to find streams in the halo (\autoref{sec:MW:streams}), will be one
way to find ultra-faint dwarfs in the LSST era, although more
sophisticated algorithms will also be utilized then.

To filter out as much noise from the Milky Way stars and unresolved
galaxies as possible, a color-magnitude filter is applied to cataloged
stars.  The middle panel of \autoref{fig:mw:search_technique} shows
the distribution of stellar sources brighter than $r$ = 21.5 that
remain after a color-magnitude filter designed to select old,
metal-poor stars at 100 kpc has been applied to the SDSS star counts in a region
around the Ursa Major I ultra-faint dwarf.  A spatial smoothing filter
is applied to the stars passing the color-magnitude filter to enhance
the signal from stellar associations with the angular size expected
for nearby dwarf galaxies.  The right panel of
\autoref{fig:mw:search_technique} shows the strong enhancement of the
Ursa Major I dwarf that results from this spatial smoothing.

A well-defined and systematic search for ultra-faint dwarf galaxies
with LSST will differ in several important ways from analogous
searches performed on shallower data sets.  Some of these are discussed
in more detail in \citet{willman09a}.  The first difference is that
unresolved galaxies will be the primary source of noise diluting the
signal from dwarf galaxy stars in the LSST stellar catalog. In
SDSS-based searches, Milky Way stars have been the primary
noise. Another difference will be that the final LSST co-added stellar
catalog will provide point source photometry as deep as can be
obtained from the ground over a wide field-of-view.  The strategy used
to identify and study new objects thus necessarily will be different
from that used with with shallower surveys, where deep, wide-field
follow-up is often used to confirm the veracity of a tentative dwarf
galaxy detection and then to study its detailed properties.
Spectroscopic follow-up of ultra-faint dwarfs discoverable in LSST,
but not in the shallower SkyMapper Southern Sky Survey, will largely be
impossible until the advent of 30-m telescopes.


\section{Stellar Tracers of Low-Surface Brightness Structure in the
  Local Volume}
\label{sec:MW:LV}

%
%
%
%

{\it \noindent Beth Willman, Jay Strader, Roelof de Jong, Rok Ro\v{s}kar}

The unprecedented sensitivity to point sources and large sky coverage
of LSST will for the first time enable the use of resolved stellar
populations to uniformly trace structure in and around a complete
sample of galaxies within the Local Volume.  This section focuses on
the science that can be done by using resolved stars observed by LSST
to discover and study low-surface brightness stellar structures beyond
the virial radius of the Milky Way. \autoref{chp:stellarpops} of this
Science Book focuses on the science that can be done by using resolved
stars observed by LSST to study stellar
populations. \autoref{sec:gal:demo} and \autoref{sec:gal:lsb} of this
Science Book focus on the study of low surface brightness structures
using diffuse light.

\subsection{The Landscape of the Local Volume}
\label{sec:mw:LV:landscape}

The groups, galaxies, and voids that compose the Local Volume are the
landscape that resolved stars in LSST will be used to map.  In this
subsection, we provide an overview of this landscape.  We then
detail specific studies in the remainder of this Section.  

\citet{karachentsev04} (KK04) cataloged 451 galaxies within the Local
Volume (d$\lesssim$10 Mpc), hereafter referred to as the KK04
catalog. In a conference proceeding, \citet{karachentsev07} report
that the KK04 catalog has been updated to include 550 galaxies, half
of which have been imaged with HST and thus have distances measured
with an accuracy of $\sim 8\%$.  The searchable volume reachable by
LSST is expected to include over an order of magnitude more galaxies
than currently known in that volume (see also \autoref{sec:MW:UFs} and
\autoref{sec:gal:lsb}).

The KK04 catalog includes 13 galaxies to be LSST's footprint brighter
than $M_V = -17.5$ mag, that are beyond the Local Group and within 5
Mpc of the Milky Way.  These galaxies are luminous and nearby enough
for LSST to facilitate detailed studies of their stellar halos and
outer stellar disks with individual stars.  

The distribution of galaxies in the Local Volume is highly
inhomogeneous, and includes a number of groups (e.g. M81, IC 342 - aka
Maffei, Cen A/M83, Leo I, M101, NGC 6946, Sculptor filament;
\citealt{karachentsev05}) and voids - where void is defined to be a
volume of space that contains no currently known galaxy.  LSST will be
used to trace the structure and assembly histories of these galaxy
groups with individual stars.  
LSST will
also provide the first opportunity to search these voids for the
faintest dwarf galaxies.


\subsubsection{LSST Feasibility Limits within the Local Volume}

The kind of stars that can be used to trace and investigate low
surface brightness systems and features within the Local Volume has
been described in \autoref{sec:sp:sfh}. The magnitude limits of the LSST
are such that within the Local Group one can use RR Lyraes and blue
horizontal branch stars to trace structures. However, beyond the Local
Group these stars become too faint even for the 10-year LSST data
stack, and we have to rely on old ($\gtrsim1$ Gyr) RGB stars,
intermediate age ($\gtrsim0.5$ Gyr) AGB stars and young
($\lesssim0.5$ Gyr) red and blue super giants. For populations younger
than 50 Myr we can also detect main sequence stars.

Of these populations, the RGB stars have been most widely used to detect
faint structures. They are very numerous, and have a well-defined
upper luminosity for populations older than 3 Gyr \citep{SalaGira05}, 
providing Tip-of-the-RGB (TRGB) distance estimates.
If we obtain photometry to at least 1.5 mag below
the TRGB, we can measure accurate colors for the brightest (and most
metallicity-sensitive) RGB stars, and can detect all the brighter
stellar types mentioned above. 

The TRGB is almost independent of metallicity at $M_i\sim-3.6$ mag
and $r-i$ colors of about 0.5--1.0 depending on metallicity. With
10-year LSST survey limits of $r=27.7$, $i=27.0$ (\autoref{tab:intro:syspar}), the pure
detection distance limit for tracing faint structures (with stars to
1.5 mag below the TRGB) is about $m-M\sim29$, or about 6
Mpc. However, the surface brightnesses that can be reached are not
primarily limited by the point-source detection limit, but by low
number statistics and contamination (mainly unresolved background
galaxies) at the low surface brightness end, and by image crowding at
the high surface brightness end.

The faint limit of the equivalent surface brightness LSST can reach is
mainly determined by its ability to perform star-galaxy separation.
About half the galaxies brighter than the detection limit of $r =27.7$
have half light radii $<0.2''$, giving a
potential contamination of about 100 unresolved galaxies arcmin$^{-2}$ at
this depth. Careful multi-color selection may reduce the contamination
of background galaxies, but many background galaxies have
colors quite similar to RGB stars at low S/N. For a target galaxy at
$m-M=28$ (where most galaxies can be found in the Local Volume)
and local surface brightness of $\mu=29$ r-mag arcsec$^{-2}$ we expect
about 40 stars arcmin$^{-2}$ brighter than $r=27.7$. This results in
a S/N of $40/\sqrt{40+100}=3.3$ arcmin$^{-2}$. Many nearby targets
have extended halo structures of at least 30 arcmin diameter, so we
have many arcmin$^2$ to average to push detections toward
$\mu\sim30$ r-mag arcsec$^{-2}$.  One might be able to improve the
contamination by weighting the deep stack of LSST photometry
(\autoref{sec:design:algorithms}) more to
the best-seeing images, reducing depth a bit but improving star-galaxy
separation.  

The highest surface brightness magnitude limit we can reach is
determined by image crowding, which scales directly with distance
modulus and image resolution $a_{\rm res}$ (i.e., $\mu_{\rm lim}$ [mag
arcsec$^{-2}$]$ +(m-M)$[mag]$+5 \log(a_{\rm res}[\rm arcsec])$ is
constant). The theory of image crowding or confusion limited
photometry has been extensively studied \citep[e.g.,][and references
therein]{Hogg01,OlseBlum03}, but a general rule of thumb states that
photometry becomes confusion-limited when the background surface
brightness equals that which would be produced if the light from the
star were spread over about 30 resolution elements, depending somewhat
on the steepness of the luminosity function of sources. Figure\,1 of
\citet{OlseBlum03} shows that confusion sets in for a star 1 mag below
the TRGB ($M_V\sim-2$) at $\mu\sim27$ V mag arcsec$^{-2}$ for a galaxy at
4\,Mpc or $m-M=28$ and with a $0.7''$ PSF. Brighter AGB and
super giant stars can be resolved to $\mu\sim 25$ V mag arcsec$^{-2}$
at those distances. For nearer systems LSST can trace old populations
to higher surface brightnesses; for instance, for galaxies at 1.5\,Mpc
TRGB stars can be resolved at $\mu\sim25$ V mag arcsec$^{-2}$, in the
Local Group even much brighter.

\subsection{Stellar Halos in External Galaxies}
\label{sec:mw:LV:halos}

Stellar halos around other galaxies can be studied to very low surface
brightnesses with star counts of red giants.  The outstanding examples
of such work are that of \citet{ferg02} and \citet{ibat07}, who used
star counts to study the outer disk and halo of M31. They discovered
an astounding array of features that indicate an active accretion
history for M31, including a giant stream and warps or tidal features
in the outer disk.  Their star counts utilized only the brightest
$\sim 1$ mag of the red giant branch, with color cuts to isolate giant
stars of different metallicities.

In hierarchical structure formation, stellar halos are largely built
from the disruption of accreted satellites. Much of
this hierarchical accretion is happening in small units at high
redshift, for the foreseeable future not measurable by direct
observations, but information about this process is stored in the
fossil record of galactic stellar halos.  And while much information
about stellar halos can be gleaned from studying the Milky Way and M31
halo in detail, 
observations of other halos are needed to place the local,
detailed information in the overall context.
 

RGB stars in stellar halos can be detected to about 6\,Mpc using the
10-year co-added LSST stack as described in \autoref{sec:mw:LV:landscape} .
Within this distance, the two main accessible galaxy groups are the
Sculptor Filament and the NGC\,5128/M83 Group. The former has four galaxies
with $M_B < -18$ (NGC\,55, NGC\,247, NGC\,253, and NGC\,7793); the latter group
also has four galaxies above a similar luminosity limit
(NGC\,5128, M83, NGC\,4945, and NGC\,5102). NGC\,5128 itself is the
nearest  massive early-type galaxy.  There are three other galaxies of similar
luminosity outside of bound groups: NGC\,1313, E274-01, and the
Circinus galaxy (the latter is behind the Galactic Plane and highly
extinguished). In addition to these luminous galaxies, there are
hundreds of fainter dwarfs, many of them still to be discovered. 

\begin{figure}
  \centering\includegraphics[width=0.5\linewidth,angle=0]{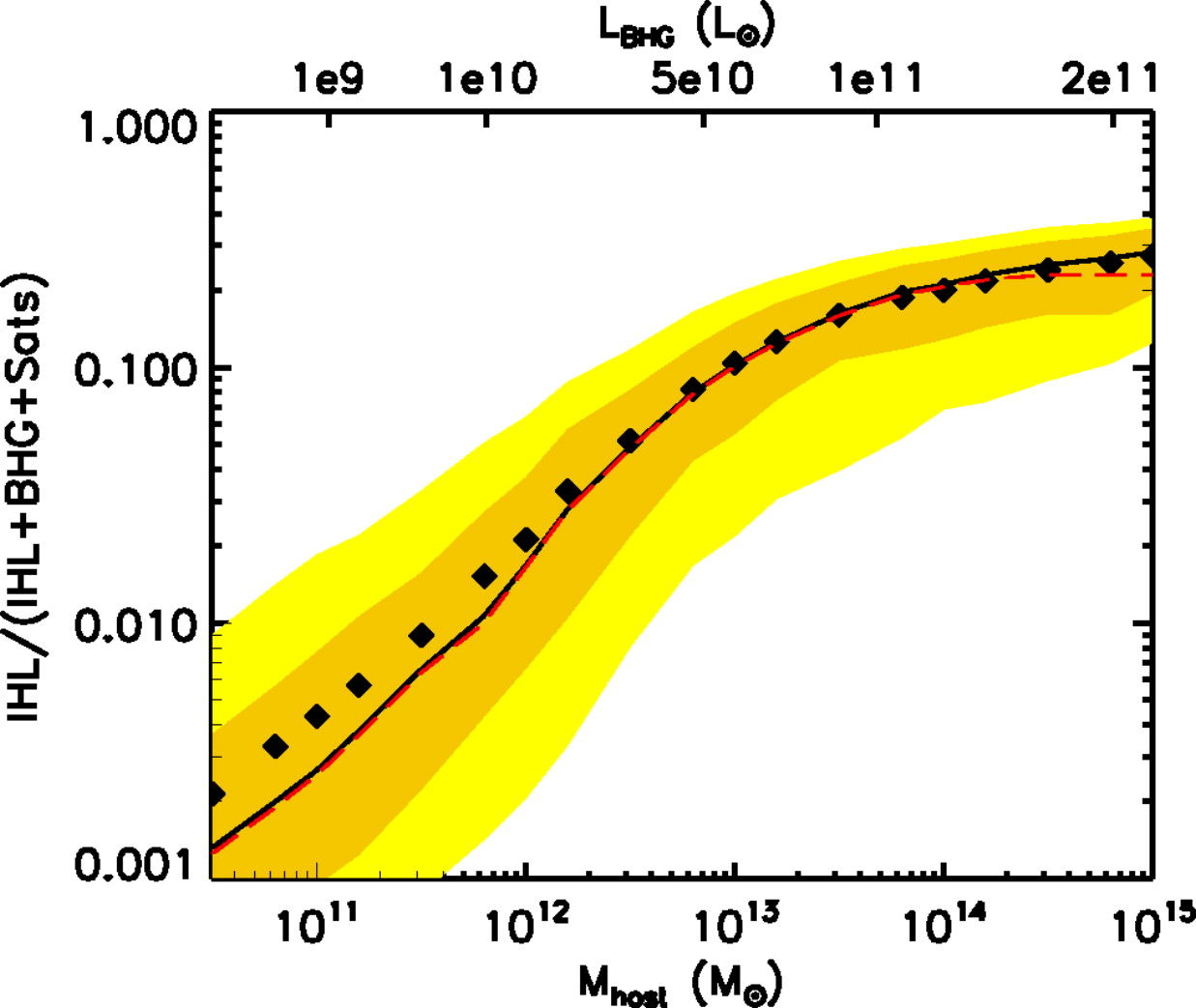}
\caption{The diffuse light fraction as a function of host halo mass,
  for systems with virial mass between $10^{10.5} M_{\odot}$ and
  $10^{15} M_{\odot}$.  The {\em diamonds} denote the mean of the
  distribution of intra-halo light (IHL) fractions at fixed mass based
  on 1000 realizations of the analytic model.  The {\em light} shaded
  region shows the 95\% range of the distribution of IHL fractions at
  fixed mass and the {\em dark} shaded region contains 68\% of the
  distribution.  The {\em solid} lines show the median of the
  distribution.  Note that the median differs markedly from the mean
  at small host masses, illustrating the skewness of the IHL
  distribution in that range.  The {\em dashed} line represents the
  preparatory IHL fraction, without the addition of pre-processed
  diffuse material already in subhalos at the time of accretion. The
  upper axes show the corresponding central galaxy (BHG)
  luminosity. Figure and caption from \citet{purc07}, with permission.
}
\label{fig:mw:halo_light_frac}
\end{figure}

The primary quantity to derive is the luminosity (or even better
mass) of stellar halos as function of total galaxy
luminosity/mass. \citet{purc07} predict that the stellar mass fraction
in diffuse, intrahalo light should rise {\em on average} from $\sim
0.5\%$ to $\sim 20\%$ from small galaxy halos ($\sim 10^{11}
M_{\odot}$) to poor groups ($\sim 10^{13} M_{\odot}$), and increase
only slowly to roughly $\sim 30\%$ on massive clusters scales ($\sim
10^{15} M_{\odot}$) (see \autoref{fig:mw:halo_light_frac}).  The
mass-dependent diffuse light fraction is governed primarily by the
empirical fact that the mass-to-light ratio in galaxy halos must vary
as a function of halo mass. Galaxy halos have little diffuse light
because they accrete most of their mass in small subhalos that
themselves have high mass-to-light ratios; stellar halos around
galaxies are built primarily from disrupted dwarf-irregular-type
galaxies with $M_*\sim{}10^{8.5} M_{\odot}$. While measurements of the
diffuse, accreted component at the massive end of the distribution can
be measured with integrated light measurements, only LSST can provide
enough statistics to nail down the halo light fraction at the low mass
end.

Beyond simple luminosities, it should be possible to derive the halo
density profiles with LSST---both radially and azimuthally. The halos
are predicted to have S\'ersic like density profiles \citep{AbaNav06},
but their scale size will depend critically on the star formation
history of the satellites before they are being accreted, which will
depend in turn on the epoch of reionization and the ability of supernovae to
remove gas from small systems \citep[e.g.,][]{BekChi05}. While halo
measurements for a few massive systems can be made from the ground
with targeted observations, only LSST will detect and fully map enough
smaller systems ($V_{\rm rot}<100$\,\kms) to quantify halo shapes for
smaller galaxies.

In M31, substantial spectroscopic followup is necessary to study the
halo density distribution (e.g., \citealp{kali06, chap06}) because of
confusion of M31 giants with Galactic dwarf stars. The other issue is
that the angular size of M31 is huge, enabling the spectroscopic study
of only a small portion of the halo at a time.  For more distant
galaxies these issues will be minimized: a single spectroscopic
pointing can cover a large fraction of the galaxy, and the main
contaminants will be distant unresolved galaxies that can be
efficiently rejected through multi-band imaging. The 30-m class
telescopes with adaptive optics will enable this kind of followup
enabling kinematic and abundance pattern analysis out to at least
3\,Mpc.

The next step up in complexity in parametrizing stellar halos will be
quantifying streams, minor mergers, and other such events in the halos
of nearby galaxies. In hierarchical structure formation, stellar halos
are largely built from the disruption of accreted satellites, and most
of the mass is donated by a relatively small number of massive
satellites \citep{bullock2005}. This scenario gives specific
predictions about (a) the typical frequency and amplitude of
accretion features in stellar halos, (b) the typical orbits of
satellites currently being accreted, and (c) the expected variation in
these features among galaxies of a range of halo masses. Current
predictions indicate that more massive accreted satellites have sunk
to the center of the potential well of the main galaxy and have been
completely disrupted to make a fairly smooth halo. 
At the present
time, it is mostly 
smaller satellites that are being accreted and disrupted, resulting in a
radial increase of the amount of substructure relative to the smooth
light profile.  
These
predictions can be directly tested with LSST observations of nearby
galaxy halos. The width of the streams is partly determined by how deep
the baryons have sunk into the potential well dominated by dark
matter. This is still a poorly determined parameter in the galaxy
models and LSST measurements of streams may help constrain this
parameter. 

For the nearest galaxies (within a few Mpc), we can go significantly
deeper than 1 mag below the RGB tip, increasing the contrast between
halo stars and background contaminants. In such galaxies, photometric
metallicity estimates will be available for individual giant stars,
enabling the study of abundance gradients and measurements of
abundance variations due to substructure. With more massive, and hence
more self-enriched, satellites sinking deeper to the center, we expect
the smooth underlying stellar halo of totally disrupted satellites to
have a metallicity gradient decreasing radially outward. However,
satellites currently being accreted and disrupted had more time to
chemically enrich themselves, and hence the substructure of streams
and loops is expected to have higher abundances than the smooth
underlying stellar halo component \citep{FonJoh08}. LSST can be expected to test these
predictions for the roughly ten nearest galaxies that are massive enough to
have sizable accretion features.

This is a field in which there is 
clear synergy between LSST and a 30-m class telescope. The combination
of spatial information and rough photometric metallicities (from LSST)
with kinematics and more detailed abundances (from 30-m spectroscopy) would
allow a comprehensive test of models of halo formation.

\subsection{Exploring Outer Disks} 
\label{sec:mw:LV:disks}

In addition to testing hierarchical
merging predictions with detailed anatomical studies of galactic
halos, observations of resolved stars can also shed light on the faint
outer disks of spiral galaxies. Outer disks of spiral galaxies offer a
unique window into the process of galaxy growth and, while significant
strides have been made in recent years toward their understanding, new
puzzles have also arisen with new data. A large number of disks, both
in the local Universe \citep{van-der-kruit:1979, Pohlen:2002,
  Trujillo:2005} and at higher redshifts \citep{Perez:2004}, deviate
from single-exponential surface brightness distribution in their
outskirts. The largest fraction have down-bending profiles, only a
small fraction are pure-exponential disks (e.g. NGC300
\citealt{Bland-Hawthorn:2005}), and a few have up-bending profiles.  A
slew of mechanisms have been proposed for down-bending profiles,
including disk response to bar formation \citep{Debattista:2006}, some
variant of star formation suppression \citep{Kennicutt:1989,
  Elmegreen:1994, Elmegreen:2006}, and the interplay of secular
evolution processes with the finite extent of star forming disks
\citep{Roskar:2008}. Up-bending profiles are believed to be relics of
recent interactions \citep{Younger:2007}.

Resolved star data for galaxies such as NGC4244 \citep{de-jong:2007}, 
NGC300 \citep{Vlajic:2009}, and M33 \citep{Barker:2007, Williams:2009}
are beginning to reveal how complex these tenuous outer regions
of galaxies may be,  thereby providing important constraints on possible 
formation models. Current models are reaching sufficient complexity to 
formulate detailed predictions regarding stellar populations in the outer 
disks (e.g., \citealt{Roskar:2008, Roskar:2008b}), therefore enabling the 
interpretation of the wealth of information contained in resolved star 
disk studies. Several studies have found very old stars in these outermost 
regions, defying the usual assumptions about inside-out disk growth 
(e.g., \citealt{Barker:2007, Vlajic:2009}), which stipulate that the outermost 
part of the disk should also be the youngest. One enticing explanation 
is that these old stars originated in the inner disk and migrated outwards 
via spiral arm scattering \citep{Sellwood:2002, Roskar:2008, Haywood:2008}. 
Based on truncated H$\alpha$ radial profiles, outer disks were thought
to be mainly devoid of star formation \citep{Kennicutt:1989}. However,
recent observations appear to defy our definitions of where star formation 
should take place within a galaxy, as evidenced by isolated H{\sc ii} regions 
\citep{Ferguson:1998} and UV emission \citep{Gil-de-paz:2005, Thilker:2005, 
Thilker:2007} well beyond the H$\alpha$ star forming disk. 

The LSST imaging of the Local Volume will allow us 
to create a complete census of the
outer disks of LV galaxies. While these galaxies have certainly been
studied extensively in the past, the observational expense required to
reliably detect individual stars at Mpc distances has limited the
exploration to a few localized pointings. The LSST will instead yield
an unbiased view of entire disks and combined with other recent and
upcoming nearby galaxy surveys (e.g., THINGS, SINGS), enable for the
first time a detailed multi-wavelength study of outer disks. The outer
disks may, therefore, at the same time provide us with a view of disk
assembly in progress as well as a glimpse of our Galactic
neighborhood's history. As our theoretical understanding of disk
formation and evolution within the $\Lambda$CDM paradigm develops in the
coming decade, the LSST view of the LV will become an invaluable
testbed for these models.

\subsection{Discovering New Galaxies}
\label{sec:mw:gals}


In \autoref{sec:MW:UFs}, we described the search for new dwarf
galaxies in the Milky Way by identifying 
subtle statistical overdensities of stars.  LSST will also
allow an unbiased search for new low surface brightness galaxies
throughout accessible regions of the Local Volume (see also
\autoref{sec:gal:lsb}).   

The census of dwarf galaxies within 2 Mpc is certainly not complete,
and the completeness of the dwarf galaxy census in this volume
substantially exceeds the completeness of the entire Local Volume
galaxy census as of 2004. Assuming that the census of galaxies in
their catalog is complete within 2 Mpc, KK04 estimate their
catalog to be 70 - 80\% complete within 8 Mpc.  This estimate only
accounts for the apparent magnitude limit of the Karachentsevs' galaxy
surveys, but does not account for surface brightness limitations of
past surveys or the correlation between surface brightness and
absolute magnitude observed in galaxies.  \autoref{sec:gal:lsb}
demonstrates that a simple extrapolation of the galaxy luminosity
function observed at brighter luminosities predicts $8 \times 10^3$
galaxies brighter than $M_V = -10$ within 10 Mpc, compared with the
550 galaxies now known within the Local Volume.

Most current and planned future strategies to find new nearby galaxies
have focused on deep imaging of known galaxy groups or clusters, with
a variety of subsequent detection algorithms. However, identifying low
surface brightness galaxies in low-density environments is more
challenging.  A substantial fraction of the Local Volume is occupied
by underdense regions.  \citet{tikhonov07} used the KK04 catalog to
show that a sphere of 7.5 Mpc radius is home to sixvoids of more
than 30 Mpc$^3$ each. 
 The six detected voids occupy 58\% of the volume within 7.5 Mpc.  Four
of these voids lie at, or overlap with, negative declination.

LSST will be able to identify LSB galaxies through their resolved
giant stars, in the same manner that stellar halos of nearby galaxies
will be studied. This project will be the first large-volume survey
for low surface brightness galaxies that is not strongly biased (e.g.,
to the presence of active star formation).

\subsection{Intragroup Stars - The Local Group, NGC 5128, and Sculptor}
\label{sec:mw:icstars}

Stars outside of individual virialized halos are a natural consequence
of structure formation in a hierarchical model
\citep{purc07}. Simulations suggest that the fraction of ``intrahalo''
light ranges from below 1\% for sub-$L^*$ spiral galaxies, to 20\%-30\%
on the mass scale of groups and clusters. Observationally, searches
for planetary nebulae in massive clusters like Virgo have found
estimated intrahalo light fractions of 10-20\% (e.g., \citealp{feld04})
and similar values from the direct detection of diffuse
light \citep{Mihos++05}.  However, little or no intergalactic light has been detected in
nearby groups such as Leo and the M81 Group \citep{cast03, feld01}, in
apparent conflict with the theoretical expectations.

LSST will offer several routes to studying intergalactic
stars. Imaging of the nearby ($\sim 4$ Mpc) Sculptor and NGC 5128
groups will allow a direct search for intragroup light through the
most luminous red giants. The co-added 10-year imaging will reach $\sim
1.5$ mag below the tip of the red giant branch for an old metal-poor
population; the sensitivity limits are somewhat worse by $\sim
0.5-1.0$ mag for more metal-rich stars.

Red giants are poor choices to study intergalactic light in the Local
Group, since their distances cannot be accurately estimated. However,
RR Lyraes and blue horizontal branch stars are both viable
alternatives, and can be identified to the edge of the Local Group
(1--1.5 Mpc) in the 10-year coadd. Both tracers carry the unfortunate
bias that they are most populous in metal-poor populations, and will
be less effective if the intragroup stars are largely
metal-rich. Nonetheless, the specific frequency of RR Lyraes in
metal-poor stellar populations is very high (1 per $\sim 1.5 \times
10^{4} L_{\odot}$; \citealp{brow04}), and so they should be numerous.

An intriguing alternative to these tracers is planetary nebulae. Work
with SDSS has indicated that planetary nebulae
can be selected solely from $ugr$ imaging with a surprising efficiency
($> 80$\%; \citealp{knia05}); they have very unusual colors due to
strong emission in the $g$ and $r$ bands. The specific frequency of
planetary nebulae 2.5 mag fainter than the most luminous objects is
one per $\sim$$2\times 10^7$ $L_{\odot}$ of integrated stellar luminosity; if
we go only 1.0 mag fainter than the most luminous planetary nebulae,
there is still one object per $\sim$$2 \times 10^8$ $L_{\odot}$
\citep{feld04}. The $u$-band is the limiting factor for this work, and
scaling from the M31 results of \citet{knia05}, the single epoch
distance limits for these two depths are $\sim$$800$ kpc and $\sim$$1.7$
Mpc; corresponding 10-year limits are $\sim$$2.1$ and $\sim$$4.2$ Mpc.

If we assume that the total stellar luminosity of the Local Group is
$5 \times 10^{10} L_{\odot}$ \citep{vand99} and only 1\% of the
stars are in intergalactic space, then there should be 50--100
intragroup planetary nebulae to LSST depth. Some fraction of these
will be in the Northern sky and so unobservable by LSST. Of course, this
estimate scales directly with the fraction of intergalactic light; if
this is 10\%, then there should be more than 500 nebulae, so a solid
null result will put an important upper limit to the fraction of stars
outside of virialized galaxies in the Local Group. Follow-up
spectroscopy will probably be required for the success of this
project, since a portion of the candidates will be emission line
galaxies at moderate to high redshifts.


\section{Globular Clusters throughout the Supralocal Volume}
\label{sec:MW:GCs}

{\it \noindent Jay Strader}

Globular star clusters (GCs) are powerful probes of the formation
epochs, assembly mechanisms, and subsequent evolution of galaxies
\citep{brod06}. This potency springs from the general
association of GC formation with the major star-forming episodes in a
galaxy's history, and from the survival of GCs through the long course
of galaxy assembly as largely unaltered bright beacons---particularly
in galaxy halos. As simple stellar populations, GCs are far more
easily analyzed and understood than a galaxy's diffuse field
starlight, which is a complicated mix of stars of different ages and
abundances.

All galaxies but the faintest dwarfs have GCs. Two flavors of GCs
dominate most systems: old metal-poor halo clusters and old metal-rich
bulge clusters. However, the formation of GCs continues to the present
day, and there are substantial numbers of young and intermediate-age
GCs in star-forming disk galaxies and in recent merger remnants.


LSST will offer a complete photometric characterization of the GC systems of
essentially every galaxy within $\sim 30$ Mpc in the LSST footprint, with partial
coverage extending to much larger distances (see below).

\subsection{Properties of Globular Cluster Systems}

The fundamental properties of GC systems that can be estimated using
broadband photometry are: 1) total numbers of GCs, usually normalized
to galaxy mass (``specific frequency'' or $T_N$), 2) two-dimensional spatial
densities, 3) mass functions (estimated from luminosity functions
with a knowledge of distance), and 4) color distributions, used to
infer GC ages and metallicities under certain assumptions. The
following subsections discuss a subset of science questions that can be
answered with such data, including ancillary topics such as
intergalactic stellar populations.

\subsubsection{Total GC Populations}

Specific frequencies vary in a systematic manner with galaxy mass. Very
massive ellipticals and dwarf galaxies have more GCs per unit stellar
mass than do $L^*$ galaxies. The characteristic $U$-shape of specific
frequency with stellar mass is similar to that of the
mass-to-light ratios for galaxies \citep{peng08}. The dispersion in
specific frequency is highest for dwarf galaxies---some, like the
Fornax dSph have surprisingly large GC populations, while other dwarfs
of similar stellar mass have no GCs.


GCs can offer important conclusions from simple observations
of total populations. Spirals have fewer metal-poor GCs per unit
mass than ellipticals \citep{rhod07}; since relatively recent
mergers should primarily involve enriched gas (and thus produce
metal-rich GCs), the immediate conclusion is that the progenitor disk
galaxies that built current ellipticals are a fundamentally different
population than nearby disk galaxies. 



It is not clear what physical parameters control these variations.
In the Milky Way, at least, we know that the metal-poor and
metal-rich GCs form in very different proportions to their associated
field star populations (halo and bulge stars, respectively). The ratio
of stellar to GC mass is $\sim 50$ for the halo and $\sim 1000$ for
the bulge \citep{stra05}. In nearby elliptical galaxies there is
also an offset in the efficiency of the formation of the two GC
populations, though the exact ratio is difficult to estimate because
stellar halo masses cannot be accurately determined. Upper limits on
the mass of the metal-poor stellar halo can be set by (for
example) optical spectroscopy at several effective radii.

If we make the assumption that the efficiency of metal-poor GC
formation does not vary strongly among galaxies, the number density of
GCs can be used to estimate masses of stellar halos. Despite the
uncertainties, these estimates are the \emph{best available}
by any method that can be expected for the foreseeable future.

\subsubsection{Spatial Distributions}

All available data suggest that metal-poor GCs are broadly accurate
tracers of metal-poor stellar halos. In the Milky Way and M31, the
radial distribution and mean metallicity of the metal-poor GCs matches
that of the stellar halo.

Such observations inspire the use of metal-poor GCs as general tracers
of stellar halos. Except in the nearest galaxies, it is not possible
to study the halo on a star-by-star basis. Radial distributions of
metal-poor GCs will be derived with LSST for literally thousands of galaxies,
allowing statistical estimates of such distributions as a function of
galaxy mass and environment.

To first order, radial distributions give collapse times. Less massive
halos that collapsed earlier are more centrally concentrated. This
simple prediction is consistent with existing observations:
metal-poor GCs in $\sim 10^{12} M_{\odot}$ halos like the Milky Way
and M31 have three-dimensional radial distributions that go as $\sim r^{-3.5}$,
while those in giant elliptical galaxies are more typically $\sim
r^{-2.5}$ \citep{bass06}. Assumptions about the typical
redshifts and halo masses of metal-poor GC formation then yield
predictions of radial distributions as a function of halo mass.

Very little is known about the azimuthal distribution of GCs. There is
some evidence that the halo of the Milky Way is flattened (see the
review of \citealt{helm08}), and if non-spherical stellar halos are typical,
then this should be detected with high significance by stacking the GC
systems of many galaxies together. To do this project independent of LSST
would require a significant investment of large telescope time
for pointed observations of many nearby galaxies.

%
%
%

\subsubsection{Stellar Populations with Globular Clusters}

Broadband photometry can provide the metallicity distributions of old
GCs, and with the $u$-band there is some hope of discriminating
between old and intermediate-age (1--5 Gyr) metal-rich GCs. For
systems with younger star clusters---for example, in ongoing galaxy
mergers or in disk galaxies, broadband photometry gives little
metallicity information, but is useful in age-dating clusters.

It is well-established that the mean metallicity of the metal-poor GCs
is correlated with the stellar mass of their parent galaxies 
\citep{peng06, stra04}. In simple terms, dwarf spheroidals
have metal-poor GCs of lower metallicity than does the Milky Way or
M87. To first order this indicates that the accreted dwarfs that built
the stellar halo of the Milky Way are a different population from the
surviving dwarfs observed today. This same point is frequently noted
in discussions of the chemical abundance of the stellar halo: typical
halo stars are enhanced in $\alpha$-elements, while stars in dwarfs
have solar enhancement. The resolution of this dilemma is generally
considered to be star formation timescales. Extant dwarfs have had a
Hubble time to enrich their stars to solar abundances, while those
that formed the halo were probably accreted early before such
enrichment had taken place.

The constraints from metal-poor GCs are actually much
stronger. Current populations of Milky Way dwarfs cannot be
representative of those that were accreted to form the halo in terms
of their \emph{old} stellar populations. Subsequent enrichment is
irrelevant.

The science case for age-dating metal-rich GCs rests on the star
formation history of bulges. Ellipticals are made by mergers, but the
formation and assembly histories of the stars in ellipticals are
partially decoupled; stars may have been formed in mergers at early
times and only been assembled into their final galaxy in more recent
``dry'' mergers.

Violent star formation---the kind found in gas-rich mergers---produces
prodigious populations of star clusters. If the mass functions of old
and young star clusters are similar, and the dynamical destruction of
massive clusters a minor factor, then one can set upper limits to the
amount of star formation in the last few Gyr by searching for
intermediate-age GCs down to a certain luminosity limit (younger GCs
are brighter than older GCs). The distance limits are more stringent
for this sort of work because $u$ is required to identify
intermediate-age GCs due to the age-metallicity degeneracy in standard
optical colors.

Studying the recent GC formation history of actively star-forming
galaxies will be possible given the $ugrizy$ filter set. Here the $u$
is also crucial because it breaks the degeneracy between extinction
and age in dusty galaxies. Example science projects here would be
deriving the star formation histories of galaxies like the Antennae
through their star clusters (see below).


\subsubsection{Intergalactic Globular Clusters}

A substantial fraction of the stellar mass in galaxy groups and clusters 
exists outside of the virial radii of individual galaxies
(\autoref{sec:mw:icstars}).  This 
``intergalactic'' or ``intracluster'' light makes up as much as 15-20\% of 
the total stellar mass in the Virgo Cluster (e.g, \citealp{vill05}). The 
prevailing formation model for this light is that in the galaxy mergers 
common in groups and proto-clusters, stars in the outer parts of galaxies 
are frequently sent on high-energy, radial orbits that escape the local 
potential. The total amount and distribution of intergalactic stars 
constrain models of galaxy assembly. Intergalactic GCs can be isolated with 
LSST by broadband imaging alone. Since stars are preferentially stripped 
from the outer parts of galaxies, most intergalactic GCs should be 
metal-poor.

The expected density of intergalactic GCs is poorly constrained, but
some estimates are available. In the Virgo cluster estimates range
from 0.2--0.3 GCs/arcmin$^2$ \citep{tamu06, will07}. 
Of course, this density will not be constant across the
entire cluster, but it is reasonable to expect well over $10^{4}$
intergalactic GCs in relatively low-mass galaxy clusters like Virgo
and Fornax. More distant, richer clusters will have more such GCs but
the detection limits will be at a higher luminosity. The theoretical
expectation is 
that galaxy groups will have a small fraction of their light in
intergalactic space; this can be directly tested by looking for 
intragroup GCs.

\subsection{Detection Limits}

The mass function of old GCs can be represented as a broken power law
or an evolved Schechter function \citep{jord07}. Thus, in
magnitudes the GC luminosity function has a characteristic ``turnover''
above which $\sim 50$\% of the clusters (and 90\% of the mass in the
GC system) lie; this turnover is analogous to the $L^*$ galaxy
luminosity in a traditional Schechter function fit.

In an old GC system, the turnover is located at $M_r \sim -7.7$,
equivalent to $\sim 2 \times 10^5 M_{\odot}$. Reaching $\sim 3$ mag
beyond this value finds $> 90$\% of the GC system and gives
essentially complete coverage of all of the GCs in a galaxy.
Conversely, the brightest GCs in a galaxy are typically $\sim 3$ mag
brighter than the turnover ($\sim 3\times 10^6 M_\odot$, although some
objects are even more massive). We can then define three distance
limits for LSST: galaxies with full GC system coverage, those with
data to the turnover, and those ``stretch'' galaxies for which we barely hit
the bright end of the GC sequence.

We can use the single-visit and 10-year co-added 5$\,\sigma$ depths 
to estimate the $r$-band distance limits in 
these three regimes. These are complete (7 and 30 Mpc), 50\% (30 and 
115 Mpc) and stretch (120 and 450 Mpc). Of course, these distances 
are upper limits to the true distance limits, since for nearly all 
goals GCs need to be detected in multiple filters, some fraction of 
the GC system is projected onto its brighter host galaxy, and for 
the most distant GC systems, cosmological, crowding, and projection 
effects become important.

If we impose the requirement that GCs must be detected in all of 
$griz$, then the principal limits come from the $z$-band for 
metal-poor GCs and from $g$ for the metal-rich GCs, especially the 
former. The joint constraints for GCs in the outer regions of 
galaxies are then: complete (5 and 17 Mpc), 50\% (18 and 65 Mpc), 
and stretch (70 and 260 Mpc). In terms of touchstone objects, over 
the 10-year mission we will have complete (or nearly so) GC samples 
to the Virgo and Fornax clusters and better than 50\% GC coverage to 
the Centaurus-Hydra supercluster. Interpreted as a luminosity 
distance, the stretch distance of 260 Mpc corresponds roughly to $z 
\sim 0.06$. However, the color information available will be limited 
for these objects, and only massive ellipticals will have 
substantial numbers of sufficiently bright clusters. As a 
comparison, the most distant GC system detected with HST imaging is 
in a massive galaxy cluster at $z \sim 0.18$ \citep{mies04}.

For studying young star cluster populations in star-forming and 
interacting galaxies, the limiting filter is $u$. A solar 
metallicity, 100 Myr old $10^5 M_{\odot}$ star cluster has $M_u 
\sim -9.2$. This gives single-visit and 10-year coadd $5\,\sigma$ 
distance limits of $\sim 40$ and 125 Mpc. These distances are 
similar to those of the $r$-band alone for old globular cluster 
systems, even given the lower mass assumed, owing to the much 
larger luminosities of young clusters.

At a given distance, the actual detection limits for each galaxy 
will, on average, be brighter than that for isolated clusters as 
outlined above. This is because many of the GCs will be superimposed 
on their host galaxy, and even in the case of an extremely smooth 
background (for example, an elliptical galaxy), shot noise over the 
scale of a seeing disk will swamp the light from the faintest 
clusters.

For a quantitative example, let us consider the single-visit limits 
for an $L^*$ elliptical at 20 Mpc. $r = 24.7$ corresponds to $M_r = 
-6.8$, about 1 mag fainter than the turnover of the globular cluster 
luminosity function, encompassing $> 95$\% of the mass of the cluster 
system. Such a galaxy might have a typical effective radius of $r_e 
\sim 50\arcsec$ = 5 kpc, with a galaxy surface brightness of 19.6 $r$ 
mag/arcsec$^2$ at 1 $r_e$. At this isophote, the 5$\,\sigma$ sensitivity 
is about 0.9 mag brighter ($M_r = -7.7$), but by mass nearly 90\% of
the globular cluster system is still detected.

\subsubsection{Photometry in the LSST Pipeline}

Extragalactic GCs present special issues for data processing in
LSST. Those clusters in the outermost regions of galaxies (beyond an
approximate isophote of 26 mag arcsec$^{-2}$) can be treated as normal
sources. However, accurate photometry for those GCs superimposed on
the inner regions of galaxies must be properly deblended.  This
problem is tractable for the case of a slowly changing background (for
example, elliptical galaxies), but more challenging for GCs in spirals
or star-forming dwarfs, in which there is structure on many
scales. This is an example of a Level 3 science application in the language of
\autoref{sec:design:dm}, which the Milky Way science collaboration
plans to develop. 






\bibliographystyle{SciBook}
\bibliography{milkyway/milkyway}

%
%
%
%
%
%
%
%
%
%
%
%
%
%
%
%
%
%
%
%
%
%
%
%
%
%

\newcommand{\teff}{${\mathrm{T}}_{\mathrm{eff}}$}
\newcommand{\zzc}{ZZ~Ceti}
\newcommand{\msun}{${\mathrm{M}}_{\odot}$}
\newcommand{\mjup}{${\mathrm{M}}_{\mathrm{J}}$}
\newcommand{\mstar}{${\mathrm{M}}_{\star}$}
\newcommand{\porb}{${\mathrm{P}}_{\mathrm{orb}}$}
\newcommand{\pspin}{${\mathrm{P}}_{\mathrm{spin}}$}
\newcommand{\ppuls}{${\mathrm{P}}_{\mathrm{puls}}$}

\chapter[The Transient and Variable Universe]{Transients and Variable Stars}
\label{chp:transients}
{\it R. Lynne Jones, Lucianne M. Walkowicz, Julius Allison, Scott F. Anderson, Andrew
C. Becker, Joshua S. Bloom, John J. Bochanski, W. N. Brandt, Mark W. Claire,
Kem H. Cook, Christopher S. Culliton, Rosanne Di Stefano, S.G. Djorgovski, Ciro Donalek, Derek
B. Fox, Muhammad Furqan, A. Gal-Yam, Robert R. Gibson, Suzanne L. Hawley, Eric J. Hilton, Keri Hoadley, Steve B. Howell, {\v
Z}eljko Ivezi{\'c}, Stylani (Stella) Kafka, Mansi M. Kasliwal, Adam Kowalski,
K. Simon Krughoff, Shrinivas Kulkarni, Knox S. Long, Julie Lutz, Ashish A. Mahabal, Peregrine M. McGehee, Dante Minniti, Anjum S. Mukadam, Ehud Nakar, Hakeem Oluseyi,
Joshua Pepper, Arne Rau, James E. Rhoads, Stephen T. Ridgway, Nathan Smith,
Michael A. Strauss, Paula Szkody, Virginia Trimble, Andrew A. West, Przemek Wozniak}




  
\section{Introduction}
{\it Rosanne Di Stefano, Knox S. Long, Virginia Trimble, Lucianne M. Walkowicz}  

Transient and variable objects have played a major role in astronomy
since the Chinese began to observe them more than two millennia ago.
The term nova, for example, traces back to Pliny.  Tycho's
determination that the parallax of the supernova of 1576 was small (compared
to a comet) was important in showing the Universe beyond the Solar
System was not static. In 1912, Henrietta Leavitt reported that a
class of pulsating stars (now known as Cepheid variables) had a
regular relation of brightness to period \citep{1912HarCi.173....1L}.
Edwin Hubble's subsequent discovery, in 1929, of Cepheids in the
Andromeda nebula conclusively showed that it was a separate galaxy,
and not a component of the Milky Way \citep{1929ApJ....69..103H}.

Fritz Zwicky's 18-inch Palomar Schmidt program was the first
systematic study of the transient sky. He undertook a vigorous search
for supernovae, and with Walter Baade promoted them as distance
indicators and the source of cosmic rays
\citep{1934PNAS...20..259B,1934PNAS...20..254B}.  Just 10 years ago,
supernovae came back into the mainstream. The first indication of a
new constituent of the Universe, dark energy, was deduced from the
dimming of type Ia supernovae located at cosmological distances
\citep{1999ApJ...517..565P, 1998AJ....116.1009R}. The last decade has
seen a flowering of the field of gamma ray bursts, the most
relativistic explosions in nature.  Meanwhile, the most accurate
metrology systems ever built await the first burst of gravitational
radiation to surpass their sensitivity threshold, revealing the
signature of highly relativistic interactions between two massive,
compact bodies.

Astronomical progress has been closely linked to technological
progress. Digital sensors (CCDs and IR detectors) were invented and
funded by military and commercial sectors, but their impact on
astronomy has been profound.  Thanks to Moore's law\footnote{The
number of transistors in commodity integrated circuits has been
approximately doubling every two years for the past five decades.},
astronomers are assured of exponentially more powerful sensors,
computing cycles, bandwidth, and storage. Over time, such evolution
becomes a revolution.  This windfall is the basis of the new era of
wide field optical and near-infrared (NIR) imaging. Wide-field imaging has become a
main stream tool as can be witnessed by the success of the Sloan
Digital Sky Survey (SDSS). The renaissance of wide field telescopes,
especially telescopes with very large \'etendue (the product of
the field of view and the light collecting area of the telescope,
\autoref{sec:introduction:etendue}), 
opens new opportunities to explore the variable and transient
sky. LSST will add to this legacy by exploring new sky and reaching
greater depth.

The types of variability LSST will observe depends on both intrinsic variability and limitations
of sensitivity.  From an observational perspective, transients are objects that fall
below our detection threshold when they are faint and for which individual events are
worthy of study, whereas by variables, we generally mean objects are always detectable, 
but change in brightness on various timescales.  From a physical perspective, transients are objects 
whose character is changed by the event, usually as the result of some kind of explosion or collision,
whereas variables are objects whose nature is not altered significantly by the event.  Furthermore,
some objects vary not because they are intrinsically variable, but because some aspect of their geometry causes them to vary.  Examples of this kind 
of variability are objects whose light is amplified by gravitational lenses, or simply
binary systems containing multiple objects, including planets, which occult other system
components.  

\newbox\grsign \setbox\grsign=\hbox{$>$}
\newdimen\grdimen \grdimen=\ht\grsign
\newbox\laxbox \newbox\gaxbox
\setbox\gaxbox=\hbox{\raise.5ex\hbox{$>$}\llap
        {\lower.5ex\hbox{$\sim$}}}\ht1=\grdimen\dp1=0pt
\setbox\laxbox=\hbox{\raise.5ex\hbox{$<$}\llap
        {\lower.5ex\hbox{$\sim$}}}\ht2=\grdimen\dp2=0pt
\def\simlt{\mathrel{\copy\laxbox}}
\def\simgt{\mathrel{\copy\gaxbox}}

In this chapter, we discuss some of the science associated with ``The
Transient and Variable Universe'' that will be carried out with LSST:
transients, or objects that explode (\autoref{sec_transients}-\autoref{sec_followup});
objects whose brightness changes due to gravitational lensing
(\autoref{sec_microlensing}); variable stars
(\autoref{sec_variable_stars}-\autoref{sec_eruptive}); and planetary
transits (\autoref{sec_transits}). In this chapter we focus on what the
variability tells us about the objects themselves; using such objects
to map the structure of galaxies, characterize the intracluster
medium, or study cosmology is discussed elsewhere in this book.

As discussed in \autoref{sec_transients}, LSST has a fundamental
role in extending our knowledge of transient phenomena. Its cadence is 
well-suited to the evolution of certain objects in particular, such as novae and supernovae.
The combination of all-sky coverage, consistent long-term 
monitoring, and flexible criteria for event identification will allow LSST to probe a large unexplored
region of parameter space and discover new types of transients. Many types of transient events are expected on theoretical grounds to inhabit this space, but have not yet 
been observed. For example, depending on the initial mass of a white
dwarf when it begins accreting matter, it may collapse upon achieving
the Chandrasekhar mass instead of exploding. This accretion-induced
collapse is expected to generate an event whose characteristics are difficult to
predict, and for which we have no good candidates drawn from nature.
LSST should be sensitive to accretion-induced collapse - just one of a wider range of transient phenomena than we have not yet been able to observe. 

As described in \autoref{sec_microlensing}, geometrical effects can cause the amount of light we
receive from a star to increase dramatically, even when the star itself has a
constant luminosity. Such a transient brightening occurs when starlight is
focused by an intervening mass, or gravitational lens.  LSST will either
discover MAssive Compact Halo Objects (MACHOs) by their microlensing signatures,
or preclude them. LSST will also detect tens of thousands of lensing events
generated by members of ordinary stellar populations, from brown dwarfs to black
holes, including nearby sources that have not been revealed by other measurement
techniques.  Microlensing can detect exoplanets in parameter ranges that are
difficult or impossible to study with other methods. LSST identification of
lensing events will, therefore, allow it to probe a large range of distant stellar
populations at the same time as it teaches us about the nature of dark and dim
objects, including black holes, neutron stars, and planets in the solar
neighborhood.

LSST will make fundamental contributions to our understanding of
variability in stars of many types as is described in
\autoref{sec_variable_stars}.  It will identify large numbers of known
variable types, needed both for population studies (as in the case of
cataclysmic variables) and for studies of Galactic structure (as in the
case of RR Lyrae stars).  Photometric light curves over the ten-year
lifetime of LSST of various source populations will establish patterns
of variability, such as the frequency of dwarf nova outbursts in
globular clusters and the time history of accretion in magnetic
cataclysmic variable 
and VY Sculptoris stars, differing behaviors of the various types of
symbiotic stars, and activity cycles across the main sequence.  Huge
numbers of eclipsing systems and close binary systems will be
revealed, allowing detailed studies of binary frequency in various
populations. The automatic generation of light curves will effectively
support all detailed studies of objects in the LSST field of view
during the period of LSST operations.

Finally in \autoref{sec_transits}, we describe another form of geometric variability: the
dimming of stars as they are occulted by transiting planets. The cadence of the
survey makes LSST most sensitive to large planets with short orbital periods. 
Much will be known about planets and planetary transits by the time LSST is
operational, both from ongoing studies from the ground and from space-based
missions.  However, LSST has the distinct advantages of its brightness and
distance limits, which will extend the extrasolar planet census to larger
distances within the Galaxy. Thousands of ``hot Jupiters'' will be discovered,
enabling detailed studies of planet frequency as a function of, for example,
stellar metallicity or parent population. 


\section{Explosive Transients in the Local Universe}
\label{sec_transients}
{\it
Mansi M. Kasliwal,
Shrinivas Kulkarni}

The types of objects that dominate the Local Universe differ from those typically found
at cosmological distances, and so does the corresponding science. The following discussion
of explosive transient searches with LSST reflects this distinction: we first discuss transients 
in the Local Universe, followed by more distant, cosmological transients. 


Two different reasons make the search for transients in the nearby
Universe ($d \simlt200\,$Mpc) interesting and urgent. First, there
exists a large gap in the luminosity of  the brightest novae
($M_{v}\sim-10\,$mag) and  that of sub-luminous supernovae ($M_{v}\sim-16$\,mag).
However, theory and reasonable speculation point to several potential
classes of objects in this ``gap.''  Such objects are best found
in the Local Universe.  Second, the nascent field of Gravitational
Wave (GW) astronomy and the budding fields of ultra-high energy
cosmic rays, TeV photons, and astrophysical neutrinos
 are likewise limited to the Local Universe by physical effects
(Greisen-Zatsepin-Kuzmin (GZK) effect, photon pair production), or instrumental sensitivity
(in the case of neutrinos and GWs). Unfortunately, the positional 
information provided by the telescopes dedicated to these new fields 
is poor, and precludes identification of the
host galaxy (with attendant loss of distance and physical
diagnostics). Both goals can be met with wide field imaging
telescopes acting in concert with follow-up telescopes. 

\subsection{Events in the Gap}
\label{sec:TheGap}

\begin{figure}[hbt]
\centerline{
\includegraphics[width=0.6\textwidth,angle=0]{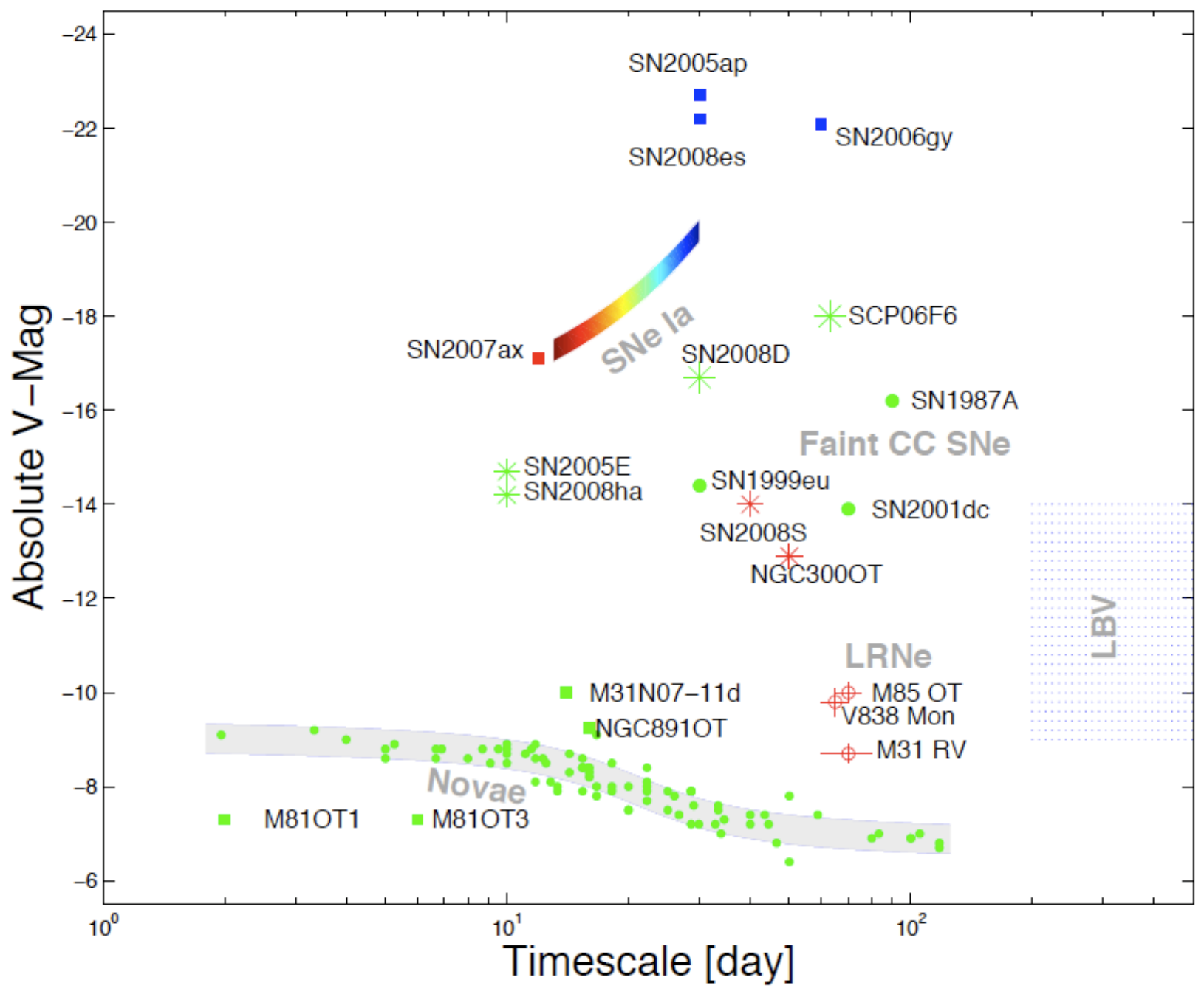}}
\caption[]{
The phase space of cosmic transients : peak $V$-band luminosity as a function
of duration, with color a measure of the true color at maximum.
Shown are the known explosive (supernovae) and eruptive (novae,
luminous blue variables (LBV) transients.  Also shown are new types
of transients (all found over the last two years):
the peculiar transients M\,85\,OT2006-1, M31-RV, and V838\,Mon, which
possibly form a new class of ``luminous red novae,'' for which a
variety of models have been suggested -- core collapse, common envelope event,
planet plunging into star, a peculiar nova, and a peculiar AGB phase;
the baffling transient with a spectrum of a red-shifted carbon star, SCP\,06F6
\citep[see][]{2009ApJ...690.1358B,Soker++08};
a possible accretion induced collapse event SN\,2005E \citep{2009arXiv0906.2003P};
the extremely faint, possibly fallback, SN\,2008ha \citep{2009arXiv0901.2074V};
and peculiar eruptive events with extremely red progenitors
SN\,2008S and NGC300-OT (\citealt{2008arXiv0809.0510T,2008arXiv0811.3929S,2009arXiv0901.0198B})
Figure adapted from \citet{2007Natur.447..458K}.
}
\label{fig:taumv}
\end{figure}

A plot of the peak luminosity versus characteristic duration 
(based on physics or convention) is a convenient way to summarize
explosive events.  We first focus on novae and supernovae of
type Ia (SN\,Ia). As can be seen from \autoref{fig:taumv}, novae and
SNe\,Ia form distinctly different loci. Brighter
supernovae take a longer time to evolve (the Phillips relation;
\citealt{1993ApJ...413L.105P}) whereas  the opposite is true of novae:
the faster the nova decays the higher the luminosity \citep[the ``Maximum
Magnitude Rate of Decline'', MMRD relation; see, for example,][]{1995ApJ...452..704D,2000AJ....120.2007D}.

The primary physical parameter determining the optical light curve in
SN\,Ia is the amount of 
nickel synthesized. There is almost a factor of 10 variation
between the brightest (``1991T-like'') and the dimmest
(``1991bg-like'') SN\,Ia. The Phillips relation has been quantified with
high precision, and the theory is well understood. In contrast, the
MMRD does not enjoy the same quantity or quality of light curves
as those of type Ia supernovae. Fortunately, dedicated ongoing nova searches in
M31 and the P60-FasTING project 
have vastly increased the number of well-sampled light curves.

A discussion of potential new classes of events in the gap would
benefit from a review of the basic physics of explosions.
An important
factor is the potential heat source at the center: a hot white
dwarf (novae) or gradual release of radioactive energy (supernovae).

The primary physical parameters are:  the mass of the ejecta ($M_{\rm
ej}$), the velocity of the ejecta ($v_s$), the radius of the
progenitor star ($R_0$), and the total energy of the explosion
(${\mathcal{E}}_0$).  Two distinct sources of energy contribute to
the explosive energy: the kinetic energy of the ejecta, ${\mathcal{E}_{\rm
k}}\equiv (1/2) M_{\rm ej} v_{s}^2$, and the energy in the photons
(at the time of the explosion), $\mathcal{E}_{\rm ph}$.

Assuming spherical symmetry and homogeneous density, the
following equation describes the gains and losses suffered
by the store of heat ($E$):
\begin{equation}
\dot E = \varepsilon(t)M_{\mathrm{ej}} - L(t) - 4\pi R(t)^2 P v(t).
\label{eq:dotE}
\end{equation}
Here, $L(t)$ is the luminosity radiated at the surface and
$\varepsilon(t)$ is the heating rate (energy per unit time) per gram from any source of energy
(e.g.,\ radioactivity or a long-lived central source).  $P$ is the
total pressure and is given by the sum of gas and photon pressure.

Next,  we resort to the so-called ``diffusion'' approximation
\citep[see][]{1996snih.book.....A,2000taap.book.....P},
\begin{equation}
L = E_{\rm ph}/t_{\rm d},
\label{eq:L}
\end{equation}
where  $E_{\rm ph}=aT^4V$ is the energy in
photons  ($V$ is the volume, $(4\pi/3)R^3$), and
\begin{equation}
t_d=B\kappa M_{\rm ej}/cR
\label{eq:td}
\end{equation}
is the timescale for a photon to diffuse from the center
to the surface. The pre-factor $B$ in \autoref{eq:td} depends
on the geometry and, following Padmanabhan, we set
$B=0.07$.  $\kappa$ is the mass opacity.

We will make one simplifying assumption: most of the acceleration
of the ejecta takes place on the initial hydrodynamic timescale, $\tau_h=R_0/v_s$,
and subsequently coasts at $R(t) = R_0 + v_{\rm s} t$.

First, let us consider a ``pure'' explosion i.e., no subsequent
heating ($\varepsilon(t)=0$).  If photon pressure dominates then
$P=1/3 (E/V)$  and an analytical formula for $L(t)$ can be obtained
\citep{1996snih.book.....A}:
\begin{equation}
L(t) = L_0 \exp\bigg(-\frac{t\tau_h + t^2/2}{\tau_h\tau_d}\bigg);
\label{eq:Ltphot}
\end{equation}
here,  $\tau_d=
B(\kappa M_{\rm ej}/cR_0)$ is the initial diffusion timescale and
$L_0={\mathcal{E}}_{\rm ph}/\tau_d$.

From \autoref{eq:Ltphot} one can see that the
light curve is divided into 1) a plateau phase which
lasts until about $\tau=\sqrt{\tau_d\tau_h}$ after which 2) the
luminosity undergoes a (faster than) exponential decay.  The duration
of the plateau phase is
        \begin{equation}
        \tau = \sqrt{\frac{B\kappa M_{\rm ej}}{cv_s}}
        \label{eq:tau}
        \end{equation}
and is independent of $R_0$. The plateau luminosity is
        \begin{eqnarray}
                L_p &=& {\mathcal{E}}_{\rm ph}/\tau_d =
                \frac{cv_s^2 R_0}{2B\kappa}\frac{\mathcal{E}_{\rm
                ph}}{\mathcal{E}_{\rm k}}.  \label{eq:Lp}
        \end{eqnarray}
As can be seen from \autoref{eq:Lp} the peak luminosity is
independent of the mass of the ejecta but directly proportional to
$R_0$. To the extent that there is rough equipartition\footnote{This
is a critical assumption and must be checked for every potential
scenario under consideration. In a relativistic fireball most of the energy
is transferred to matter. For novae, this assumption is violated (Shara, personal communication).} between the kinetic energy and the
energy in photons, the luminosity is proportional to the square of
the final coasting speed, $v_s^2$.

Pure explosions satisfactorily account for supernovae of type IIp.
Note that since $L_p \propto R_0$ the larger the star the higher
the peak luminosity.  SN~2006gy, one of the brightest supernovae,
can be explained by invoking an explosion in a ``star'' which is
much larger (160\,AU) than any star (likely the material shed by a
massive star prior to its death; see \citealt{2007ApJ...671L..17S}).

{\it Conversely}, pure explosions resulting from the deaths of compact
stars (e.g.,  neutron stars, white dwarfs, or even stars with radius
similar to that of the Sun) will be very faint.
For
such progenitors, visibility in the sky would require some sort of
additional subsequent heat input, which is discussed next.

First we will consider ``supernova''-like events, i.e., events in
which the resulting debris is heated by radioactivity. One can
easily imagine a continuation of the type Ia supernova sequence.  We
consider three possible examples for which we expect a smaller
amount of radioactive yield and a rapid decay (timescales of days):
coalescence of compact objects, accreting white dwarfs (O-Ne-Mg), and
final He shell flash in AM CVn systems.

Following \citet{1998ApJ...507L..59L}, \citet{2005astro.ph.10256K}
considers the
possibility of the debris of neutron star coalescence being heated
by decaying neutrons. Amazingly (despite the 10-min decay time of
free neutrons)  such events (dubbed as ``macronovae'') are detectable
in the nearby Universe over a period as long as a day, provided
even a small amount ($\gtrsim 10^{-3}\,M_\odot$) of free neutrons
is released in such explosions. \citet{2007ApJ...662L..95B}
consider a helium nova (which arise in AM
CVn systems). For these events (dubbed ``Ia'' supernovae), not
only radioactive nickel but also radioactive iron is expected.
Intermediate mass stars present two possible paths to sub-luminous
supernovae. The O-Ne-Mg cores could either lead to a disruption (bright
SN but no remnant) or a sub-luminous explosion \citep{2006A&A...450..345K}.
Separately, the issue of O-Ne-Mg white dwarfs accreting matter from a companion
continues to fascinate astronomers.
The likely possibility is a neutron star, but the outcome
depends severely on the unknown effects of rotation and magnetic
fields. One possibility is an explosion with low nickel yield
(see \citealt{2008arXiv0812.3656M}
for a recent discussion and review of the literature).

An entirely different class of explosive events is expected to
arise in massive or large stars: birth of black holes (which can
range from very silent events to gamma-ray bursts (GRBs) and everything in between),
strong shocks in supergiants \citep{2008Natur.451..775V}
and common
envelope mergers.  Equations~\ref{eq:tau} and \ref{eq:Lp} provide
guidance to the expected appearance of such objects.
\citet{2007ApJ...662L..55F}
developed a detailed model for faint, fast supernovae due to nickel
``fallback'' into the black  hole. For the case of the birth of a black
hole with no resulting radioactive
yield (the newly synthesized material could be advected into the
black hole), the star will slowly fade away on a timescale of
min($\tau_d,\tau$). Modern surveys are capable of finding such wimpy events
\citep{2008ApJ...684.1336K}.

In the spirit of this open-ended discussion
of new transients, we also consider the case where the gas pressure
could dominate over photon pressure. This is the regime of weak
explosions.  If so, $P=2/3 (E/V)$ and \autoref{eq:dotE} can
be integrated to yield:
        \begin{equation} L(t) = \frac{L_0}{(t/\tau_h+1)}
        \exp\bigg(-\frac{\tau_h t + t^2/2}{\tau_h \tau_d}\bigg).
        \label{eq:Ltgas}
        \end{equation}
In this case the relevant timescale is the hydrodynamic timescale.
This regime is populated by luminous blue variables and hypergiants.
Some of these stars are barely bound and suffer from bouts of
unstable mass loss and photometric instabilities.

As can be gathered from \autoref{fig:taumv}
the pace of discoveries over the past two years gives
great confidence to our expectation of filling in the phase
space of explosions.

\subsection{New Astronomy: Localizing LIGO Events}
\label{sec:NewAstronomy}

LSST's new window into the local transient Universe will complement four 
new fields in astronomy: the study of cosmic rays, very high energy
(TeV and PeV) photons, neutrinos, and gravitational waves. 
Cosmic rays with energies exceeding $10^{20}\,$eV are strongly
attenuated owing to the production of pions through interaction
with the cosmic microwave background (CMB) photons (the famous GZK effect). Recently, the Pierre Auger
Observatory \citep{PAO07} has found evidence showing that such cosmic
rays with energies above $6\times 10^{19}\,$eV are correlated with
the distribution of galaxies in the local 75-Mpc sphere. Similarly, very 
high energy (VHE) photons (TeV and PeV) have a highly restricted horizon. 
The TeV photons interact with CMB photons and
produce electron-positron pairs. A number of facilities are now
routinely detecting extra-galactic TeV photons from objects in the
nearby Universe (VERITAS, MAGIC, HESS, CANGAROO). Neutrino astronomy is another 
budding field with an expected vast increase in sensitivity. The horizon here is 
primarily limited by sensitivity of the telescopes (ICECUBE). GW astronomy suffers from both poor localization (small interferometer baselines) and sensitivity.
The horizon radius is 50 Mpc for  enhanced LIGO (e-LIGO) and about 200\,Mpc
for advanced LIGO (a-LIGO) to observe neutron star coalescence. The greatest gains 
in these areas, especially GW astronomy,  {\it require} arc-second electromagnetic localization of the event.


\begin{table*}[!hbt]
\begin{center}
\begin{footnotesize}
\caption{Galaxy Characteristics in LIGO Localizations \label{tab:galchar}}
\begin{tabular}{lcccccc}

\hline
 & & E-LIGO  &  & &  A-LIGO  & \\
\hline
 & 10\% & 50\% & 90\% & 10\% & 50\% & 90\% \\
\hline
GW Localization (deg$^2$) & 3 & 41 & 713 & 0.2 & 12  & 319 \\
Galaxy Area (arcmin$^2$) & 4.4 & 26 & 487 & 0.15 & 20.1 & 185 \\
Galaxy Number & 1 & 31 & 231 & 1 & 76 & 676 \\
Log Galaxy Luminosity ($M_\odot$) & 10.3 & 11.2 & 12.1 & 10.9 & 12.0 & 13.0  \\
\hline
%
\end{tabular}
\end{footnotesize}
\end{center}
\end{table*}

\begin{figure}[hbt]
\centerline{
\includegraphics[angle=0, width=0.6\textwidth]{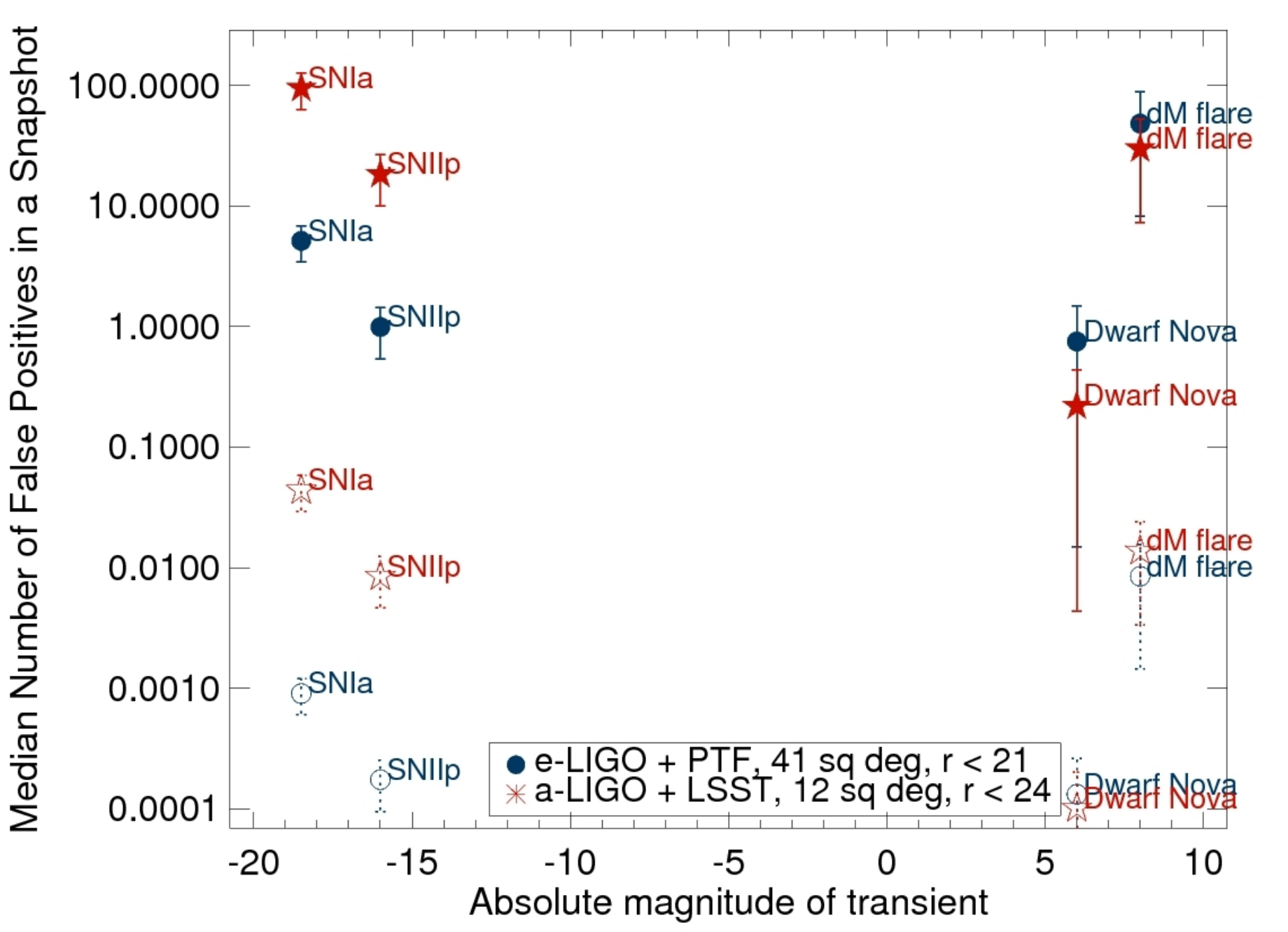}}
\caption{Number of false positives in a single LSST image in searching
for gravitational wave events. 
For e-LIGO (blue circles), we assume the median localization of
41 deg$^2$ and follow-up depth of  $r<21$. For a-LIGO
(red stars), we use the median localization of 12 deg$^2$ and follow-up depth
of $r<24$. Filled symbols denote false positives in the entire
error circle and open symbols show false positives that are spatially
coincident with nearby galaxies. Dwarf novae and
M-dwarf (dM) flares constitute the foreground fog and the error bars on
numbers represent the  dependence on galactic latitude. Supernovae (Ia,IIp)
constitute background haze.}
\label{fig:false-positive}
\end{figure}

We simulated a hundred GW events (Kasliwal et al. 2009a, in preparation) and computed the
exact localization on the sky (assuming a neutron-star neutron-star merger
waveform and triple coincidence data from LIGO-Hanford, LIGO-Louisiana and Virgo).
The localizations range between 3--700 deg$^2$ for e-LIGO and 0.2--300 deg$^2$
for a-LIGO (range quoted between 10th and 90th percentile). The Universe is
very dynamic and the number of false positives in a single LSST image
is several tens for a median localization  (see \autoref{fig:false-positive}).
Fortunately, the sensitivity-limited, $<200\,$Mpc horizon of GW astronomy
is a blessing in disguise. The opportunity cost can be substantially reduced by
restricting follow-up to those transients that are spatially coincident with
galaxies within $200\,$Mpc. Limiting the search to the area covered by galaxies
within a LIGO localization reduces a square degree problem to a square arc-minute problem
 --- a reduction in false positives by three orders of magnitude!

Given the total galaxy light in the localization,
we also find that the number of false positives due to unrelated supernovae 
or novae within the galaxy is negligible.
To be sensitive to transients with peak absolute magnitude as faint as
$-$13 (fainter than the faintest observed
short hard gamma ray burst optical afterglow), e-LIGO needs at least a 1-m class telescope for
follow-up (going to $m < 21$, or 50\,Mpc) and a-LIGO an 8-m class ($m<24$, 200\,Mpc).
Given the large numbers of galaxies within the localization (\autoref{tab:galchar}),
a large field of view camera ($>5$ deg$^2$) will help maximize depth and cadence as compared
to individual pointings. Thus in the present, the Palomar Transient
Factory (PTF; \citealt{2009arXiv0906.5350L}; \autoref{sec:PTF}) is well-positioned
to follow up e-LIGO events, and in the years to come, LSST to follow up a-LIGO events.

\subsection{Foreground Fog and Background Haze}
\label{sec:PilotPrograms}

\begin{figure}[hbt]
\centerline{
\includegraphics[width=0.6\textwidth,angle=0]{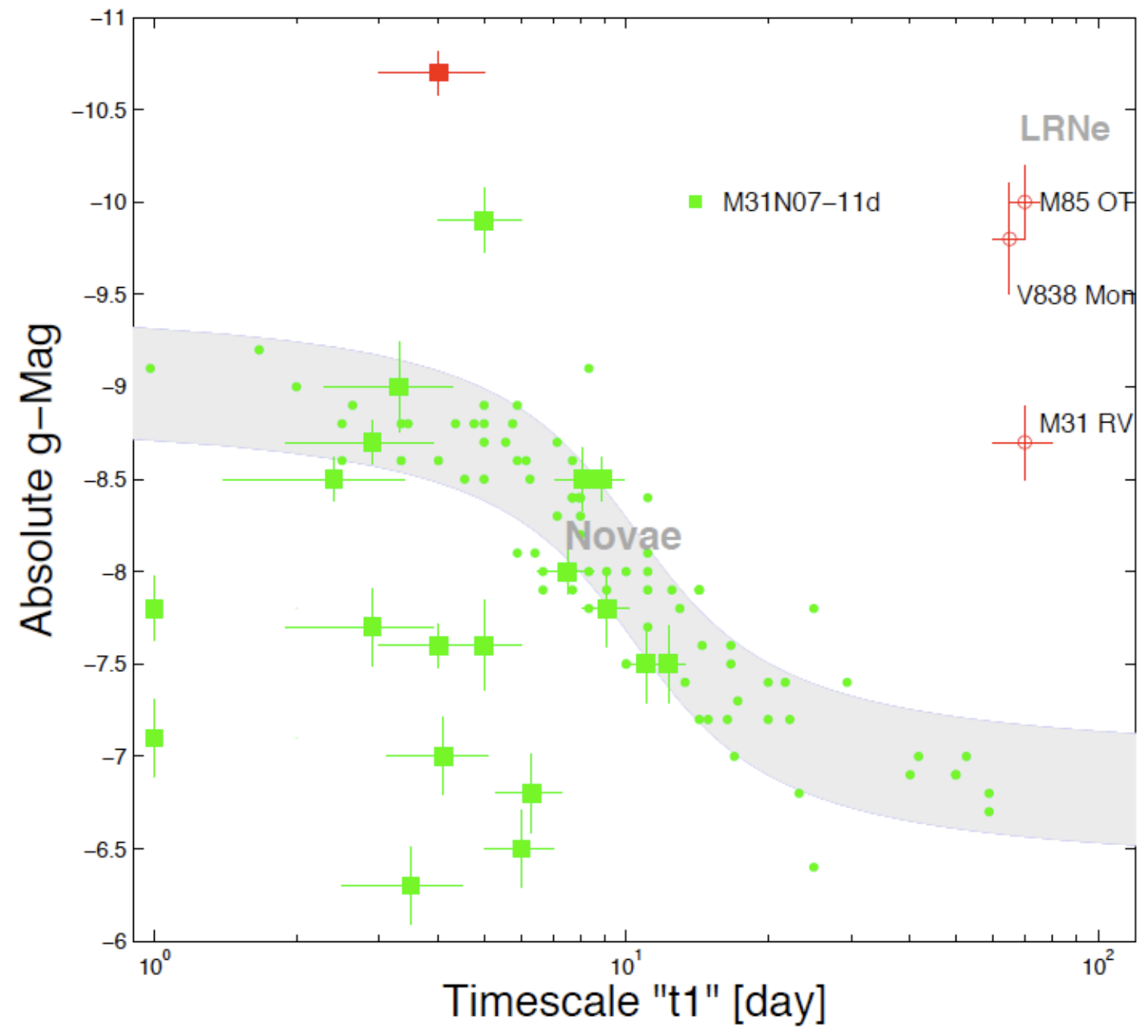}}
\caption[]{
A plot of the peak absolute magnitudes versus decay timescale of
novae discovered by the Palomar P60-FasTING project
(low luminosity region of \autoref{fig:taumv}). The shaded gray region represents
the Maximum Magnitude Rate of Decline (MMRD) relationship bounded by $\pm
3\,\sigma$ \citep{1995ApJ...452..704D}.
The data that defined this MMRD are shown by green circles. Squares indicate 
novae discovered by P60-FasTING in 2007-2008. 
(Preliminary results from Kasliwal et al. 2009b, in preparation.)
}
\label{fig:novae}
\end{figure}

Unfortunately, all sorts of foreground and background
transients {\it will} be found within the several to tens of deg$^2$
of expected localizations.  Studying
each of these transients will result in significant ``opportunity
cost.''  Ongoing projects of modest scope offer a glimpse of
the pitfalls on the road to understanding local transients. Nightly monitoring of M31 for novae (several groups)
and a Palomar 60-inch program of nearby galaxies (dubbed ``P60-FasTING'') designed to be sensitive
to faint and fast transients already show high
variance in the MMRD relation (\autoref{fig:novae}).
The large scatter of the new novae suggests
that in addition to the mass of the white dwarf, other physical
parameters play a role (such as accretion rate, white dwarf luminosity, for example, \citealt{1981ApJ...243..926S}).

\begin{figure}
\begin{center}
\includegraphics[width=0.8\textwidth]{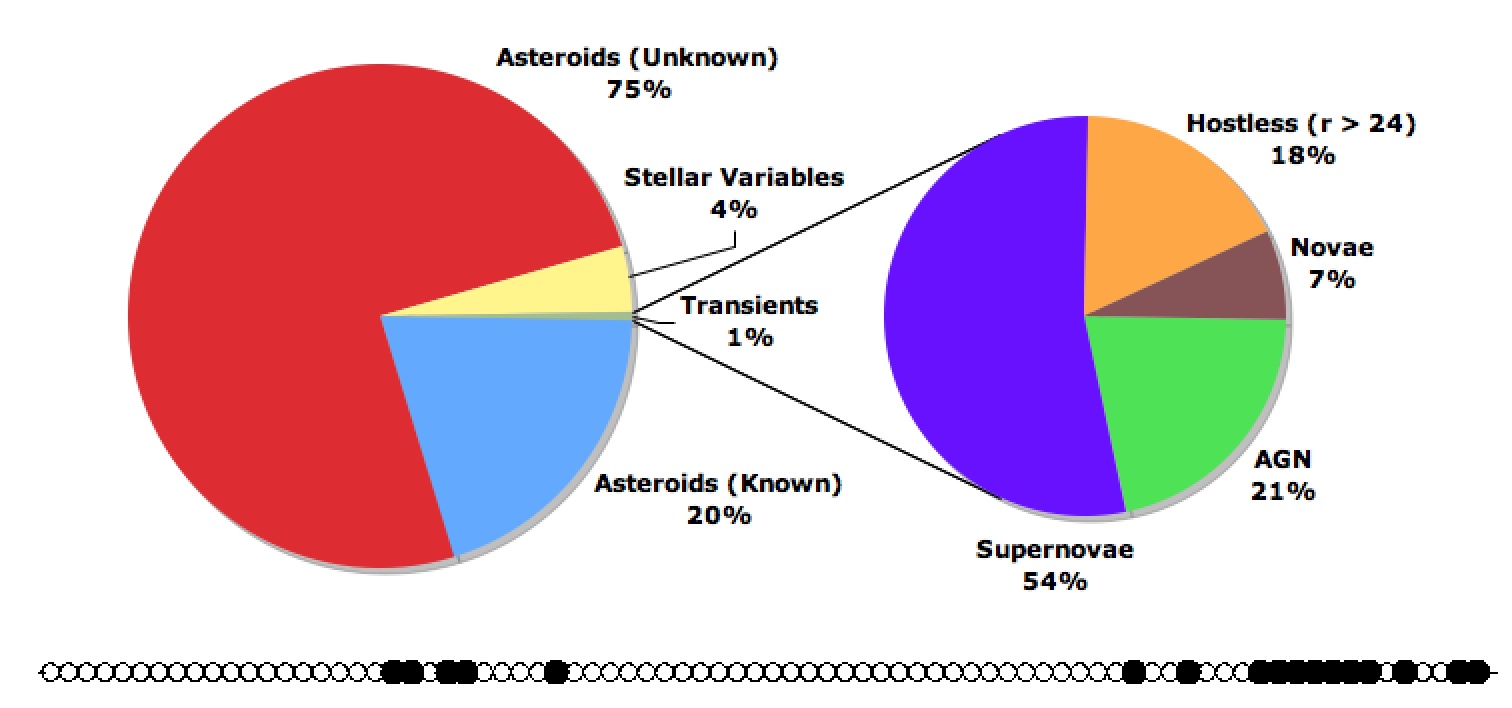}
\caption{
28 COVET transients were discovered during a
pilot run in 2008A (7 hours) -- two novae and
the remainder background supernovae and AGN. Transients with no point source or galaxy
host to a limiting magnitude of $r>24$ are classified as hostless.
Of the 2,800 candidates, the COVET pipeline automatically
rejected 99\% as asteroids or Galactic objects.
(Preliminary version from Kasliwal et al.  2009c, in preparation.)}
\label{fig:covetpie}
\end{center}
\end{figure}

A nightly targeted search of nearby rich clusters (Virgo, Coma,
and Fornax) using the CFHT (dubbed ``COVET'')
and the 100-inch du Pont \citep{2008ApJ...682.1205R}
telescopes has revealed the extensive foreground fog
(asteroids, M dwarf flares, dwarf novae) and the background
haze (distant, unrelated SN). However, even faint Galactic foreground
objects will likely be detected in 3-4 of LSST bands. If they
masquerade as transients in one band during outburst, basic
classification data may be used to identify these sources and
thus remove them as a source of ``true" transient pollution.
 The pie-chart in \autoref{fig:covetpie}
dramatically illustrates that {\it new discoveries require efficient
elimination of foreground and background events.}


\subsection{The Era of Synoptic Imaging Facilities}
\label{sec:PTF}

There is widespread agreement that we are now on the
threshold of the era of synoptic and
wide field imaging at optical wavelengths. This is
best illustrated by the profusion of operational
(Palomar Transient Factory, Pan-STARRS1), imminent
(SkyMapper, VST, ODI), and future facilities (LSST).

\begin{figure}[hbt]
\centerline{
\includegraphics[width=0.6\textwidth,angle=0]{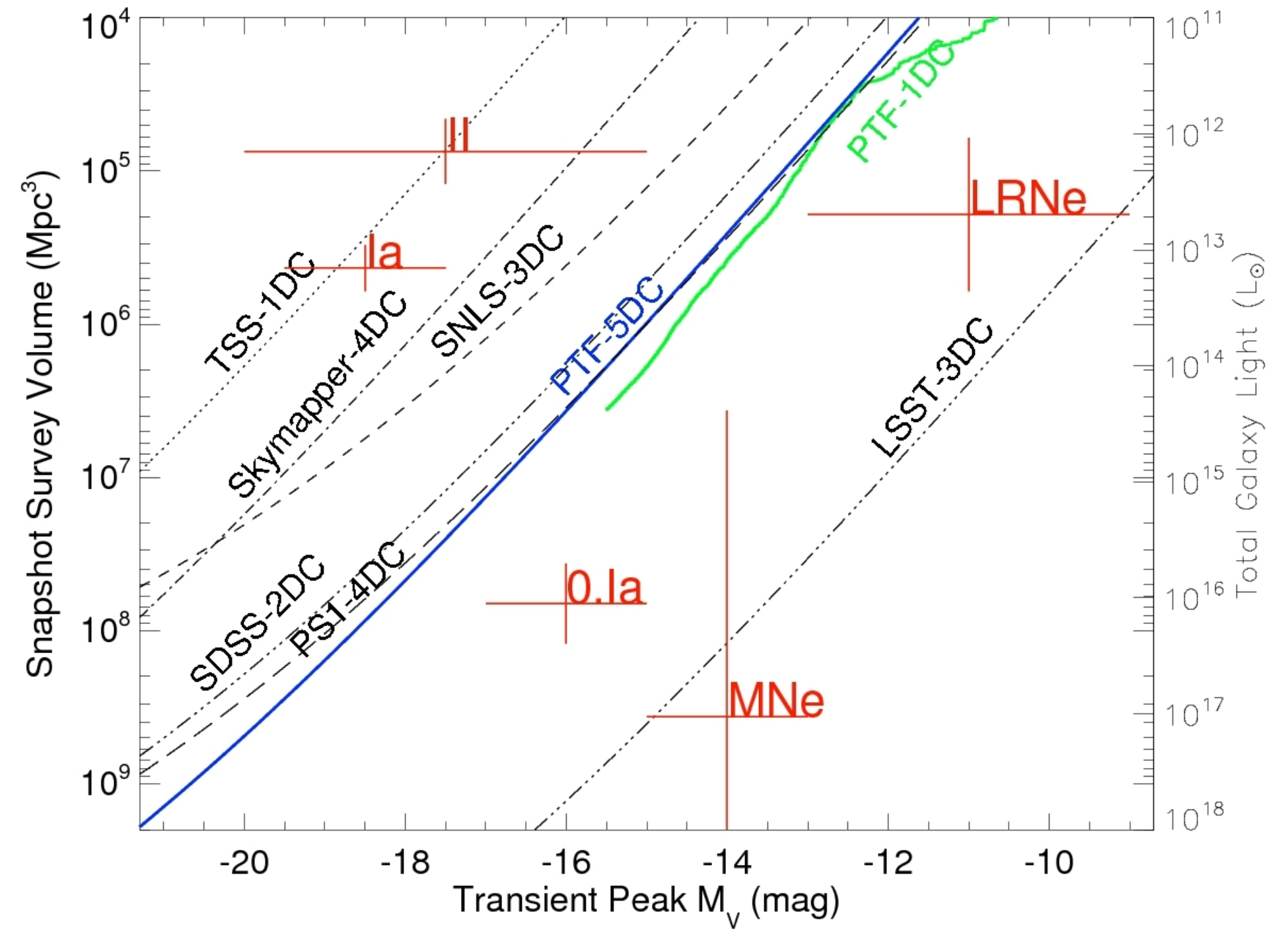}}
\caption[]{
Volume probed by various surveys as a function of transient absolute magnitude.
The cadence period to cover the volume is shown in days: e.g., 5DC
for a five-day cadence. Red crosses represent
the minimum survey volume needed to detect a single transient event
(the uncertainty in the y-axis is due to uncertainty in rates).
Palomar Transient Factory (PTF-5DC, blue-solid) is more sensitive than
Texas Supernova Search (TSS, dotted),
SkyMapper (dot-dashed), Supernova Legacy Survey (SNLS-3DC, dashed), and
SDSS Supernova Search (SDSS-2DC, double-dot dashed), and
is competitive with PanSTARRS-1 (PS1-4DC, long dashed).  Lines for each
survey represent one transient event in the specified cadence period.
PTF-1D (green solid line) represents a targeted 800 deg$^2$ survey
probing luminosity concentrations in
the local Universe, with a factor of three larger effective survey
volume than a blind survey with same solid angle. PTF will discover
hundreds of supernovae and possibly several rare events such as
``0.Ia'',  Luminous Red Novae (LRNe) and Macronovae (MNe) per year.
The LSST (Deep Wide Fast Survey) will discover hundreds of rare events
in the Local Universe. The corresponding plot for distant cosmological
transients is shown in \autoref{fig:survey_rates_distant}.
(Adapted from figure by Bildsten et al. 2009, in preparation.)
}
\label{fig:survey_rates_local}
\end{figure}

In \autoref{tab:rates_local} and \autoref{fig:survey_rates_local}, we present
current best estimates for the rates of various events and the
``grasp'' of different surveys.

\begin{table*}[ht]
\begin{center}
\caption{Properties and Rates for Optical Transients$^a$}
\begin{footnotesize}
\begin{tabular}{lcccccl}
\hline
\hline
Class           &  M$_v$ & $\tau$$^b$ & Universal Rate (UR) & PTF Rate & LSST Rate \\
& [mag] & [days] &  & [yr$^{-1}$] &  [yr$^{-1}$] \\
\hline
Luminous red novae & $-9..-13$ & 20..60 & $(1..10)\times10^{-13}$\,yr$^{-1}$ L$_{\odot,K}^{-1}$ & 0.5..8 & 80..3400 \\
Fallback SNe    & $-4..-21$ & $0.5..2$ &  $<5\times10^{-6}$\,Mpc$^{-3}$ yr$^{-1}$  &  $<$3 & $<$800  \\
Macronovae      & $-13..-15$ & 0.3..3 & $10^{-4..-8}$\,Mpc$^{-3}$ yr$^{-1}$ &  0.3..3 & 120..1200 \\
SNe~.Ia       & $-15..-17$ & 2..5 & $(0.6..2)\times10^{-6}$\,Mpc$^{-3}$ yr$^{-1}$ &  4..25 & 1400..8000 \\
SNe~Ia          & $-17..-19.5$ & 30..70 & $^c$ $3\times10^{-5}$\,Mpc$^{-3}$ yr$^{-1}$ & 700 & 200000$^d$ \\
SNe~II & $-15..-20$ & 20..300 & $(3..8)\times10^{-5}$\,Mpc$^{-3}$ yr$^{-1}$  & 300 & 100000$^d$ \\
\hline
\end{tabular}
\begin{itemize}
$^a${Table from \citet{2009arXiv0906.5355R}; see references therein.}
$^b${Time to decay by 2 magnitudes from peak.}
$^c${Universal rate at $z<0.12$.}
$^d${From M. Wood-Vasey, personal communication.}
\end{itemize}
\label{tab:rates_local}
\end{footnotesize}
\end{center}
\end{table*}

The reader should be cautioned that many of these rates are very rough.
Indeed, the principal goal of the Palomar Transient
Factory is to accurately establish the rates of foreground and
background events. Finding a handful of rare events with PTF
will help LSST to define the metrics needed to identify
these intriguing needles in the haystack.   
It is clear from \autoref{fig:survey_rates_local} that {\it the impressive
grasp of LSST is essential to uncovering and understanding the
population of these rare transient events in the Local Universe.}

\section{Explosive Transients in the Distant Universe}
{\it Przemek Wozniak,
Shrinivas Kulkarni,
W. N. Brandt,
Ehud Nakar,
Arne Rau,
A. Gal-Yam,
Mansi Kasliwal,
Derek B. Fox,
Joshua S. Bloom,
Michael A. Strauss,
James E. Rhoads
}

\begin{figure}[hbt]
\centerline{
\includegraphics[width=0.6\textwidth]{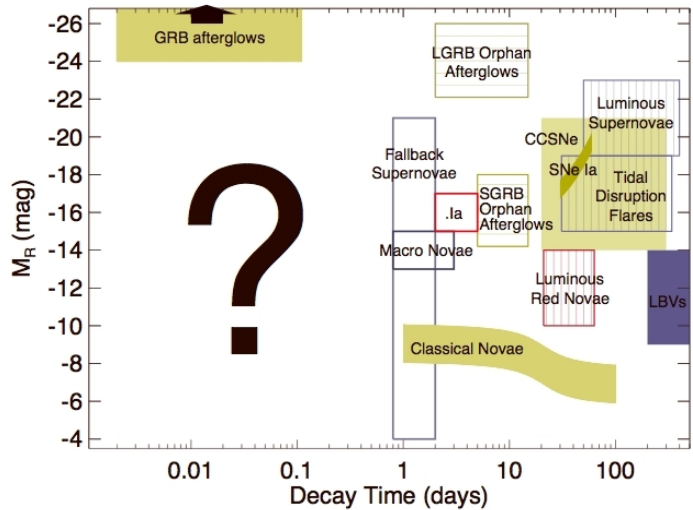}
}
\caption{
Discovery space for cosmic transients. Peak absolute $r$-band magnitude is plotted vs. decay
timescale (typically the time to fade from peak by $\sim2$ mag)
for luminous  optical transients and variables. Filled boxes
mark well-studied classes with a large number of known members
(classical novae, SNe Ia, core-collapse supernovae,
luminous blue variables (LBVs)). Vertically hatched boxes show
classes for which only a few candidate members have been
suggested so far (luminous red novae, tidal disruption flares,
luminous supernovae). Horizontally hatched boxes are  classes
which are believed to exist, but have not yet been detected
(orphan afterglows of short and long GRBs). The positions of
theoretically predicted events (fall back supernovae, macronovae,
0.Ia supernovae (.Ia)) are indicated by empty boxes.
The brightest transients (on-axis afterglows of GRBs) extend
to $M_R \sim -37.0$. The color of each box corresponds to the mean $g-r$ color at
peak (blue, $g-r<0$\,mag; green, $0 < g-r < 1$\,mag; red, $g-r>1$\,mag).
LSST will be sensitive to transients with a wide range of time scales
and will open for exploration new parts of the parameter space (question mark).
Figure adapted from \citet{rau08}.
}
\label{fig:phase_space}
\end{figure}

We now discuss the role of LSST in discovering and
understanding cosmological transients. The phase space of transients
(known and anticipated) is shown in \autoref{fig:phase_space}.  The
region marked by a big question mark is at present poorly explored and
in some sense represents the greatest possible rewards from a deep
wide field survey such as LSST. Here, we discuss a few example areas
in which LSST will provide exciting new discoveries and insights.  We
leave the discussion of transient fueling events in active galactic
nuclei due to tidal disruption of stars by the central black hole to
\autoref{sec:agn:transient}. 

\subsection{Orphan GRB Afterglows}

\begin{figure}[hbt]
\centerline{
\includegraphics[width=0.6\textwidth]{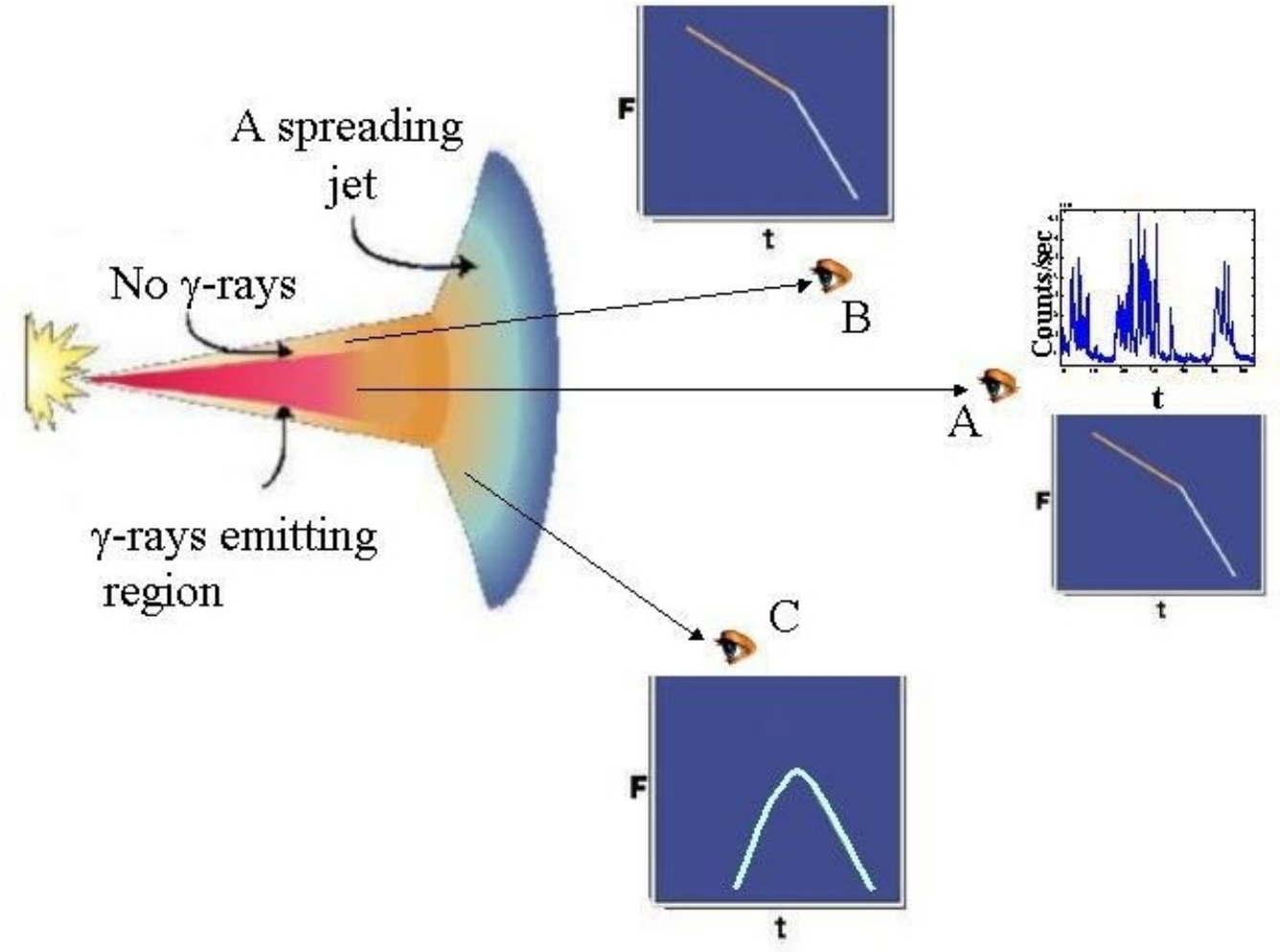}
}
\caption{
Geometry of orphan GRB afterglows.
Observer A detects both the GRB and an afterglow. Observer B does
not detect the GRB due to a low Lorentz factor of material in the line of sight,
but detects an on-axis orphan afterglow that
is similar to the one observed by A. Observer C detects an off-axis
orphan afterglow with the flux rise and fall that differs from the
afterglow detected by observers A and B \citep[from][]{NP03}.
}
\label{fig:orphans}
\end{figure}

Gamma-ray bursts (GRBs) are now established to be the most relativistic
(known) explosions in the Universe and as such are associated with the birth of
rapidly spinning stellar black holes. We believe that long duration GRBs
result from the deaths of certain types of massive stars
\citep{2006ARA&A..44..507W}. The explosion is deduced to be conical
(``jetted'') with opening angles ranging from less than a degree to
a steradian. The appearance of the explosion depends on the location
of the observer (\autoref{fig:orphans}). An on-axis observer sees the fastest material
and thus a highly beamed emission of gamma rays. The optical afterglow emission arises from the interaction
of the relativistic debris and the circumstellar medium. Due to decreasing relativistic
beaming in the decelerating flow, the light curve will show a characteristic break to a steeper decline
at $t_{\rm jet} \sim 1$--10 days after the burst \citep{1999ApJ...525..737R,1999ApJ...519L..17S}.
An observer outside the cone of the jet misses the burst of gamma-ray emission,
but can still detect the subsequent afterglow emission \citep{1997ApJ...487L...1R}.
The light curve will first rise steeply and then fade by $\sim$1 mag over a timescale of
roughly $\Delta t \sim 1.5 t_{\rm jet}$ (days to weeks).
We will refer to these objects as ``off-axis'' orphan afterglows.
The ``beaming fraction'' (the fraction of sky lit by gamma-ray bursts) is estimated to be
between 0.01 and 0.001, i.e., the true rate of GRBs is 100 to 1000
times the observed rate. Since a supernova is not relativistic and is spherical,
all observers can see the supernovae that accompany GRBs.
Finally, there may exist entire classes of explosive events which are
not as relativistic as GRBs (e.g., the so-called ``X-ray Flashes'' are argued
to be one such category; one can imagine ``UV Flashes,'' and so on).
Provided the events have sufficient explosive yield, their afterglows will
also exhibit the behavior shown in \autoref{fig:orphans} (case B).
We will call these ``on-axis'' afterglows with unknown parentage.

Pending SKA\footnote{
Square Kilometer Array, planned for the next decade, is designed to cover an instantaneous field of view
of 200 deg$^2$ at radio frequencies below 1 GHz.},
the most efficient way to detect all three types of events
discussed above is via synoptic imaging of the optical sky.
Statistics of off-axis afterglows, when compared to GRBs, will yield the
so-called ``beaming fraction,'' and more importantly, the true rate of GRBs.
The total number of afterglows brighter than $R \sim 24$ mag
visible per sky at any given instant is predicted to be $\sim$ 1,000,
and rapidly decreases for less sensitive surveys \citep{2002ApJ...576..120T}.
With an average afterglow spending 1--2 months above that
threshold, we find that monitoring 10,000 deg$^2$  every $\sim3$
days with LSST will discover 1,000 such events per year. LSST will also detect
``on-axis'' afterglows.  The depth and cadence of LSST observations will, in
many cases, allow the on- or off-axis nature of a fading afterglow to be
determined by careful light curve fitting \citep{2003ApJ...591.1097R}.
Continuous cross-correlation of optical light curves
with detections by future all-sky high energy missions (e.g., EXIST) will
help establish the broad-band properties of transients, including the orphan
status of afterglows.

\begin{figure}[hbt]
\centerline{
\includegraphics[width=0.6\textwidth]{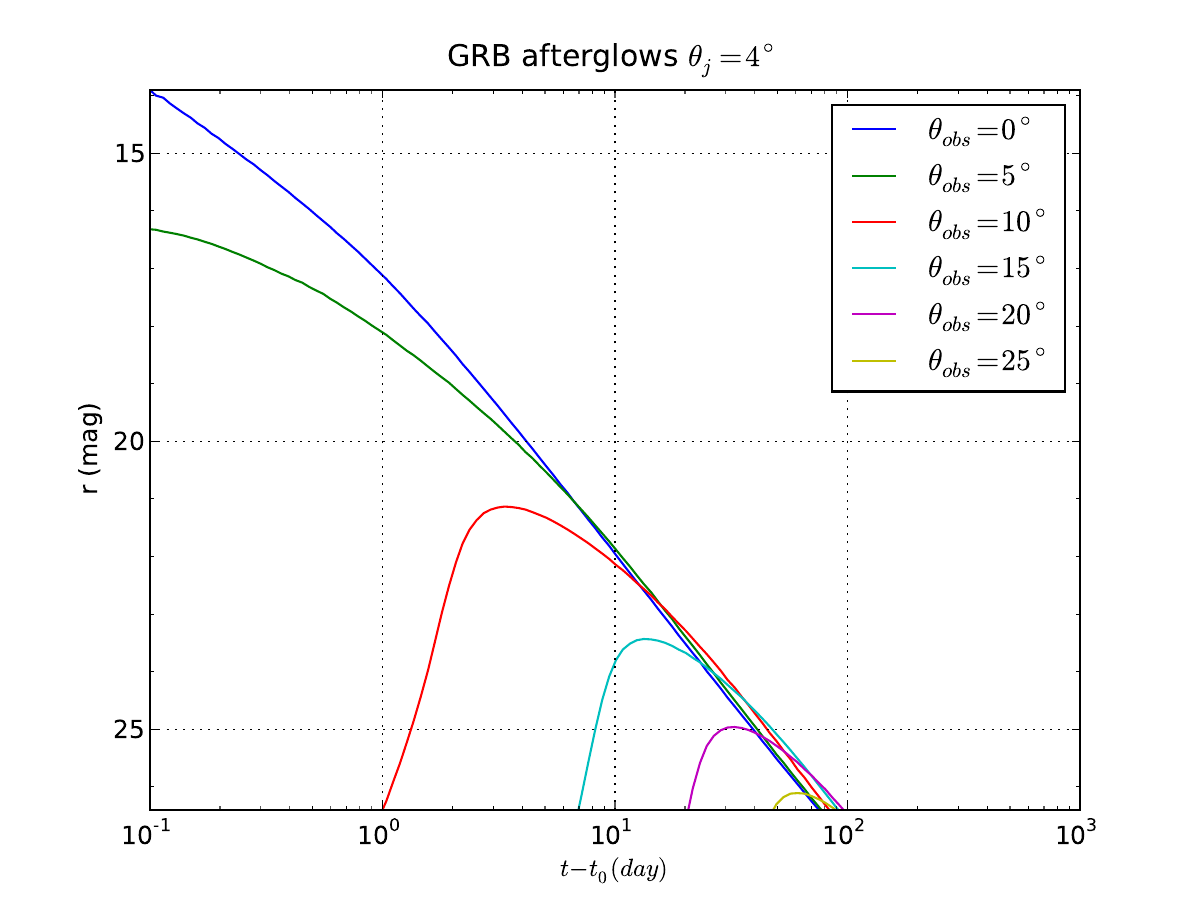}
}
\caption{
Predicted light curves of GRB afterglows. The model of the forward shock emission is from \citealt{2002ApJ...576..120T}
(code courtesy of Alin Panaitescu). The adopted global and microphysical parameters reproduce the properties
of well observed GRBs:
jet half opening angle $\theta_j = 4^\circ$, the isotropic equivalent energy of $E_{\rm iso} = 5\times10^{53} \rm erg$,
ambient medium density $n = 1$ g cm$^{-3}$, and the slope of the electron energy distribution $\rm p = 2.1$.
The apparent $r$-band magnitudes are on the AB scale assuming a source redshift $z = 1$ and a number
of observer locations with respect to the jet axis $\theta_{\rm obs}$.
}
\label{fig:afterglow_lcs}
\end{figure}

\begin{figure}[hbt]
\centerline{
\includegraphics[width=0.9\textwidth]{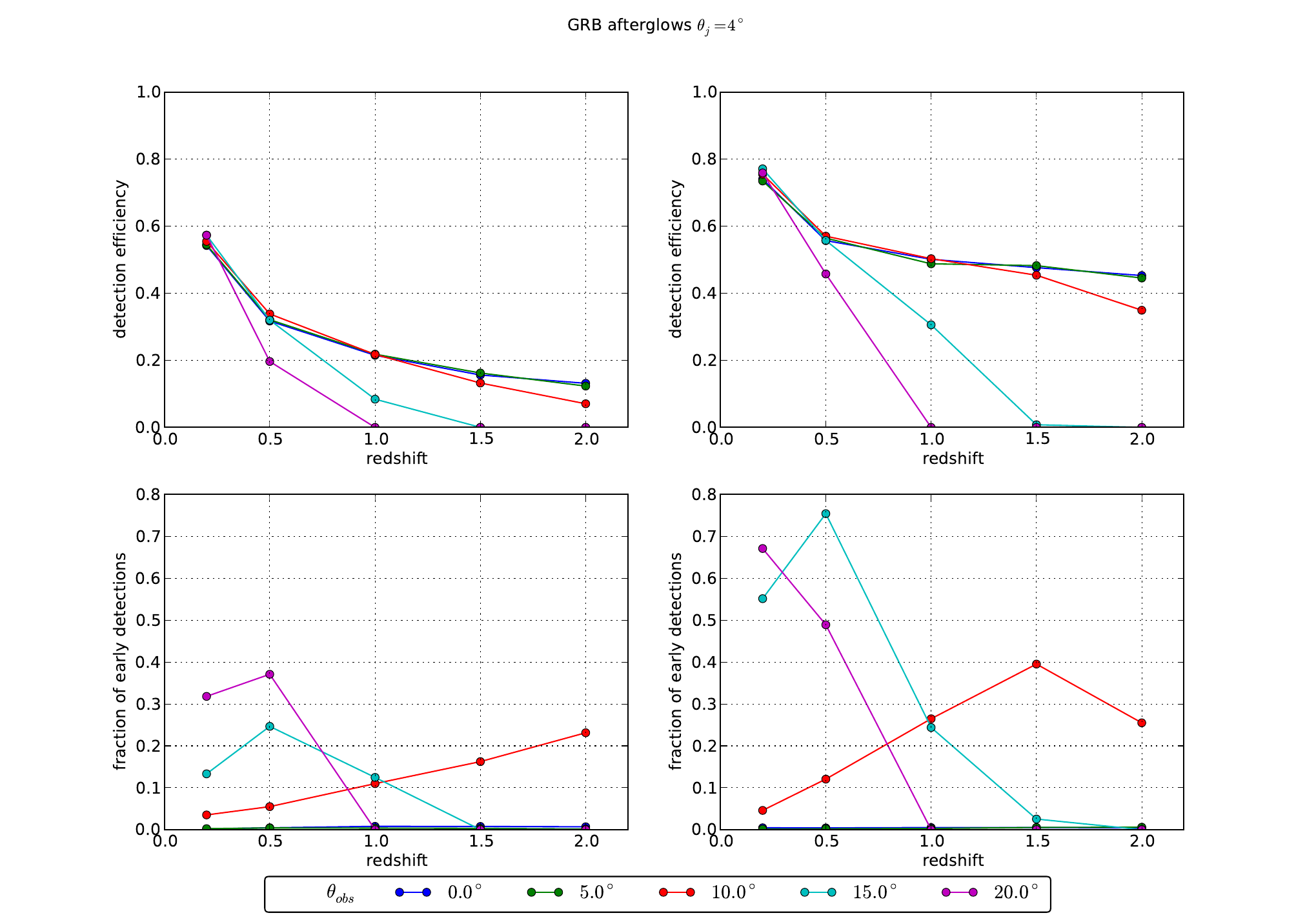}
}
\caption{
Predicted efficiency of detecting GRB afterglows in LSST (upper) and the fraction of early detections (lower)
using models from \autoref{fig:afterglow_lcs}. The main survey area (left) is compared to the seven deep drilling fields
(right). The efficiency calculation assumes that a transient is detected as soon as variability by 0.1 mag with $S/N > 10$
from at least two $5\,\sigma$ detections can be established. The early detections are those that occur before the maximum light
or within 1 mag of the peak on the fading branch. 
}
\label{fig:afterglow_detection}
\end{figure}

In \autoref{fig:afterglow_lcs} we show model predictions of the forward shock
emission from a GRB jet propagating into the circumstellar medium. The ability of LSST to detect
GRB afterglows, and the off-axis orphan afterglows in particular, is summarized in \autoref{fig:afterglow_detection}.
Time dilation significantly increases the probability of detecting off-axis orphans at redshifts $z > 1$
and catching them before or near the peak light. The peak optical flux of the afterglow rapidly decreases 
as the observer moves away from the jet. At $\theta_{\rm obs} \simeq 20^\circ$ only the closest events
($z < 0.5$) are still accessible to LSST and even fewer will have well-sampled light curves.
However, the true rate of GRBs and the corresponding rate of the off-axis orphans are highly uncertain.
Indeed, the discovery of orphan GRB afterglows will greatly reduce that uncertainty.

It is widely agreed that the detailed study of the associated supernovae is
the next critical step in GRB astrophysics, and synoptic surveys will speed
up the discovery rate by at least a factor of 10 relative to GRB missions.
Finally, the discovery of afterglows with unknown parentage will open
up entirely new vistas in studies of stellar deaths, as we now
discuss. 

\subsection{Hybrid Gamma-Ray Bursts}

The most popular explanation
for the bimodal distribution of GRB durations invokes the existence
of two distinct physical classes. Long GRBs typically last 2--100
seconds and tend to have softer $\gamma$-ray spectra, while short
GRBs are typically harder and have durations below $\sim2$ seconds,
sometimes in the millisecond range \citep[see review in][]{NP03}.
Short GRBs are expected to result from
compact binary mergers (NS-NS or NS-BH), and the available limits
rule out any significant supernova component in optical emission
\citep{2006ApJ...638..354B,2005Natur.437..845F}.

Recent developments suggest a richer picture.
Deep imaging of GRB\,060614
\citep{2006Natur.444.1053G,2006Natur.444.1050D,2006Natur.444.1047F}
and GRB\,060505
\citep{2007ApJ...662.1129O,2006Natur.444.1047F}
exclude a supernova brighter than $M_V\sim -11$.
The data for GRB\,060614 rule
out the presence of a supernova bump in the afterglow light curve
up to a few hundred times fainter than bumps seen in other bursts.
The host galaxy of this burst shows a smooth morphology and a low
star formation rate that are atypical for long GRB hosts
\citep{2006Natur.444.1053G}.
A very faint (undetected) event could have been powered with a
small amount of $^{56}$Ni \citep[e.g.,][]{2006Natur.444.1047F},
as in the original collapsar model with a relativistic jet, but without
a non-relativistic explosion of the star \citep{1993ApJ...405..273W}.
Such events would fall in the luminosity gap between novae and supernovae
discussed in \autoref{sec_transients}. Alternatively,
a new explosion mechanism could be at play.

\subsection{Pair-Instability and Anomalous Supernovae}
\label{sec:tr:pair_sn}

The first stars to have formed in the Universe were likely very
massive ($M > 100 M_{\odot}$) and died  as a result of thermonuclear
runaway explosions triggered by e$^+$e$^-$ pair production instability
and the resulting initial collapse.
The predicted light curve of a pair-instability supernova
is quite sensitive to the initial mass and radius of the progenitor with the brightest events exceeding $M_V \sim -22$ at maximum, lasting
hundreds of days and sometimes showing more than one peak
\citep{2008AIPC..990..263K}.
The pair instability should not take place
in metal-enriched stars, so the best place to look for the first
stellar explosions is the distant Universe at $z \geq 5$, where
events would appear most luminous in the $K$ band and take up to
1,000 days to fade away due to cosmological time dilation. Short of
having an all-sky survey sensitive down to $K_{AB} = 25$, the best
search strategy is a deep survey in red filters on a cadence of a few days
and using monthly co-added images to boost the sensitivity.

Recently, there have been random
discoveries of anomalously bright \citep[e.g., SN 2005ap;][]{2007ApJ...668L..99Q} and in one case
also long-lived  \citep[SN 2006gy;][]{2007ApJ...659L..13O} supernovae in the Local Universe. While
there is no compelling evidence that these objects are related to
explosive pair instability, there is also no conclusive case that
they are not. In fact, star formation and metal enrichment are very
localized processes and proceed throughout the history of the Universe
in a very non-uniform fashion. Pockets of very low metallicity material
are likely to exist at moderate redshifts ($z \sim 1-2$), and some
of those are expected to survive to present times \citep{2005ApJ...633.1031S}.
The anticipated discoveries of pair-instability SN and the characterization
of their environments can potentially transform our understanding of the interplay between
the chemical evolution and structure formation in the Universe.

\subsection{The Mysterious Transient SCP 06F6}

The serendipitous discovery of the peculiar transient SCP 06F6
\citep{2009ApJ...690.1358B}
has baffled astronomers, and its unique
characteristics have inspired many wild explanations. It had a
nearly symmetric light curve with an amplitude $>$6.5\,mag over a
lifetime of about 200 days with no evidence of a quiescent host
galaxy or star at that position down to $i > 27.5$ mag. Its spectrum was
dissimilar to any transient or star ever seen before, and its broad
absorption features have been identified tentatively as redshifted
Swan bands of molecular carbon.
One of the suggested explanations \citep{2008arXiv0809.2562G}
postulates an entirely new class
of supernovae -- a core collapse of a carbon star at redshift $z =
0.143$. However, the X-ray flux being a factor of ten more than the
optical flux and the very faint host ($M > -13.2$) appear inconsistent
with this idea.
\citet{Soker++08}
proposed that the emission comes from a CO white
dwarf being tidally ripped by an intermediate mass black hole in
the presence of a strong disk wind. Another extragalactic hypothesis
is that the transient originated in a thermonuclear supernova explosion with an AGB carbon
star companion in a dense medium. A Galactic scenario involves an asteroid
at a distance of 1.5\,kpc ($\sim$ 300\,km across; mass $\sim$ 10$^{19}$ kg)
colliding with a white dwarf in the presence of very strong magnetic fields.
The nature of this transient remains unknown.

\subsection{ Very Fast Transients and Unknown Unknowns}

As can be seen from \autoref{fig:phase_space} the discovery space of fast
transients lasting from seconds to minutes is quite empty.

On general grounds there are two distinct families
of fast transients: incoherent radiators (e.g., $\gamma$-ray bursts and afterglows)
and coherent radiators (e.g., pulsars, magnetar flares). It is a well-known result
that incoherent synchrotron radiation is limited to a brightness temperature
of $T_b \sim 10^{12}\,$K. For such radiators to be detectable from any reasonable distance
(kpc to Gpc) there must be a relativistic expansion toward the observer, so that the source appears
brighter due to the Lorentz boost. Coherent radiators do not have any such limitation
and can achieve very high brightness temperature (e.g., $T_b\sim10^{37}\,$K in pulsars).

Scanning a large fraction of the full sky on a time scale of $\sim$ 1 minute is still
outside the reach of large optical telescopes.
However, large telescopes with high \'etendue operating on a fast cadence will be the first to probe
a large volume of space for low luminosity transients on very short time-scales.
One of the LSST mini-surveys, for example, will cover a small number of 10 deg$^2$ fields
every $\sim$ 15 seconds for about an hour out of every night \citep{2008arXiv0805.2366I}.
Fast transients can also be detected by differencing the standard pair of 15-second exposures
taken at each LSST visit. Given the exceptional instantaneous sensitivity of LSST
and a scanning rate of 3,300 deg$^2$ per night, we can expect to find 
contemporaneous optical counterparts to GRBs, early afterglows, giant pulses from pulsars,
and flares from anomalous X-ray pulsars. But perhaps the most exciting findings
will be those that cannot be named before we look. The vast unexplored space
in \autoref{fig:phase_space} suggests new discoveries lie in wait.


\begin{figure}[hbt]
\centerline{
\includegraphics[width=0.6\textwidth]{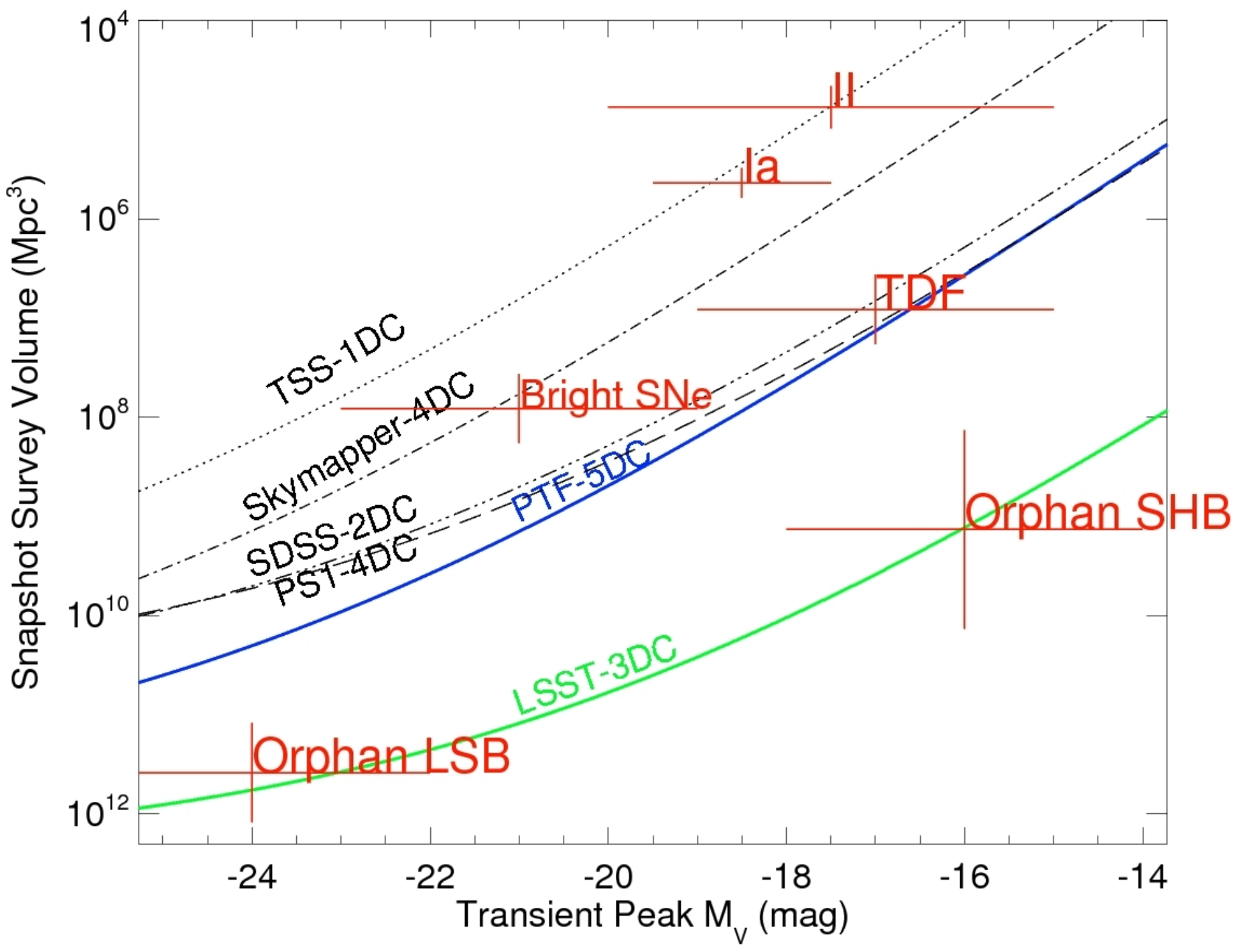}
}
\caption{
Volume probed by various surveys as a function of transient absolute magnitude.
The cadence period to cover the volume is shown in days: e.g., 5DC
for a five-day cadence.
Red crosses represent the minimum survey volume
needed to detect a single transient event.  The uncertainties in the rates
and luminosities translate to the displayed ``error.''  LSST will cover
10,000 deg$^2$ every three days down to the limiting magnitude r = 24.7,
and will have the grasp to detect rare and faint events such as
orphan afterglows of Long Soft Bursts (LSB) and Short Hard Bursts (SHB)
out to large distances in a single snapshot. The main LSST survey will also
discover a large number of Tidal Disruption Flares (TDF).
Palomar Transient Factory (PTF-5DC, blue-solid) is more sensitive
than Texas Supernova Search (TSS, dotted),
SkyMapper (dot-dashed), Supernova Legacy Survey (SNLS-3DC, dashed), and
SDSS Supernova Search (SDSS-2DC, double-dot dashed) and
competitive with PanSTARRS-1 (PS1-4DC, long dashed). Lines for each
survey represent one transient event in specified cadence period.
For example, TSS discovers one type Ia supernova every day - however,
since type Ia supernovae have a lifetime of one month, TSS discovers the
same type Ia supernova for a month. The corresponding plot for transients in the Local Universe
is shown in \autoref{fig:survey_rates_local}.
(Original figure provided by L. Bildsten, UCSB.)
}
\label{fig:survey_rates_distant}
\end{figure}

\section{Transients and Variable Stars in the Era of Synoptic Imaging}
\label{sec_followup}
{\it Ashish A. Mahabal, Przemek Wozniak, Ciro Donalek, S.G. Djorgovski}

The way we learn about the world was revolutionized when computers---a
technology which had been around for more than 40 years---were
linked together into a global network called the World Wide Web and
real-time search engines such as Google, were first deployed. Similarly,
the next generation of wide field surveys is positioned to revolutionize
the study of astrophysical transients by linking
heterogeneous surveys with a wide array of follow-up instruments as well as
rapid dissemination of the transient events using various mechanisms on 
the Internet.

\begin{table*}[ht]
\begin{center}
\caption{Properties and Rates for Optical Cosmological Transients$^a$}
\begin{footnotesize}
\begin{tabular}{lcccc}
\hline
\hline
Class           &  M$_v$ & $\tau$$^b$ & Universal Rate (UR) & LSST Rate \\
& [mag] & [days] &  & [yr$^{-1}$] \\
\hline
Tidal disruption flares (TDF) & $-15..-19$ & 30..350 &            $10^{-6}$\,Mpc$^{-3}$ yr$^{-1}$   & $6,000$        \\
Luminous SNe            & $-19..-23$ & 50..400 &            $10^{-7}$\,Mpc$^{-3}$ yr$^{-1}$   & $20,000$       \\
Orphan afterglows (SHB) & $-14..-18$ & 5..15   & $3\times10^{-7..-9}$\,Mpc$^{-3}$ yr$^{-1}$   & $\sim$10--100 \\
Orphan afterglows (LSB) & $-22..-26$ & 2..15   & $3\times10^{-10..-11}$\,Mpc$^{-3}$ yr$^{-1}$ & $1,000$        \\
On-axis GRB afterglows  &    $..-37$ & 1..15   &            $10^{-11}$\,Mpc$^{-3}$ yr$^{-1}$       & $\sim$50     \\
\hline
\multicolumn{4}{p{0.7\textwidth}}{$^a$Universal rates from \citet{Rau++09}; see references therein.}\\
\multicolumn{4}{p{0.7\textwidth}}{$^b$Time to decay by 2 magnitudes from peak.}
\end{tabular}


\label{tab:rates_distant}
\end{footnotesize}
\end{center}
\end{table*}

In \autoref{fig:survey_rates_distant} we compare the ability of various surveys
to detect cosmological transients. LSST will be the instrument of choice
for finding very rare and faint transients, as well as probing
the distant Universe ($z \sim 2-3$) for the most luminous events.
It will have data collecting power more than 10 times greater than any existing facility,
and will extend the time-volume space available for systematic exploration
by three orders of magnitude. In \autoref{tab:rates_distant} we summarize
the expected event rates of cosmological transients that LSST will find.

The main challenges ahead of massive time-domain surveys are timely recognition
of interesting transients in the torrent of imaging data, and maximizing the utility
of the follow-up observations \citep{2006Natur.442..364T}. For every orphan afterglow present
in the sky there are about 1,000 SNe Ia \citep{2002ApJ...576..120T} and millions
of other variable objects (quasars, flaring stars, microlensing events).
LSST alone is expected to deliver tens of thousands of astrophysical transients every night.
Accurate event classification can be achieved by assimilating on the fly
the required context information: multi-color time-resolved photometry,
galactic latitude, and
possible host galaxy information
from the survey itself, combined with broad-band spectral properties from external catalogs
and alert feeds from other instruments---including gravitational wave and neutrino detectors.
While the combined yield of transient searches in the next decade is likely to saturate the resources
available for a detailed follow-up, it will also create an unprecedented opportunity for discovery.
Much of what we know about rare and ephemeral objects comes from very detailed studies
of the best prototype cases, the ``Rosetta Stone'' events. In addition to the traditional target
of opportunity programs that will continue to play a vital role, over the next few years
we will witness a global proliferation of dedicated rapid follow-up networks of 2-m class imagers
and low resolution spectrographs \citep{2009AN....330....4T,2008AN....329..269H}.
But in order to apply this approach to extremely data intensive sky monitoring surveys
of the next decade, a fundamental change is required
in the way astronomy interacts with information technology \citep{Borne2008b}.

Filtering time-critical actionable information out of $\sim30$ Terabytes of survey data
per night \citep{2008arXiv0805.2366I} is a challenging task \citep{Borne2008a}.
In this regime,
the system must be capable of automatically optimizing the science potential
of the reported alerts and allocating powerful but scarce follow-up instruments.
In order to realize the science goals outlined in previous sections,
the future sky monitoring projects must integrate state of the art
information technology such as computer vision, machine learning,
and networking of the autonomous hardware and software components.
A major investment is required in the development of hierarchical,
distributed decision engines capable of ``understanding'' and refining
information such as partially degenerate event classifications
and time-variable constraints on follow-up assets.
A particularly strong emphasis should be placed on:  1) new classification and anomaly
detection algorithms for time-variable astronomical objects,  2) standards for
real-time communication between heterogeneous hardware and software
agents,  3) new ways of evaluating and reporting the most important
science alerts to humans, and  4) fault-tolerant network topologies
and system architectures that maximize the usability. The need to
delegate increasingly complex tasks to machines is the main driver
behind the emerging standards for remote telescope operation and event messaging
such as RTML (Remote Telescope Markup Language), VOEvent and SkyAlert
\citep{Williams2008}. These innovations
are gradually integrated into working systems, including the GRB Coordinates Network (GCN),
a pioneering effort in rapid alert dissemination in astronomy.
The current trend will continue to accelerate over the next decade.

By the time LSST starts getting data, the field of time domain astronomy will be much
richer in terms of availability of light curves and colors for different types
of objects. Priors, in general, will be available for a good variety of
objects. LSST will add to this on a completely different scale in terms of
cadence, filters, number of epochs, and so on. Virtual Observatory (VO) tools that link new optical
transient data with survey and archival data at other wavelengths are already
proving useful. Newer features being incorporated include semantic linking as
well as follow-up information in the form of a portfolio based on expert
inputs, active automated follow-up in the form of new data from follow-up
telescopes as well as passive
automated follow-up in terms of context-based annotators such as galaxy
proximity, apparent motion, and so on, which help in the classification process.

Since a transient is an object that has not been seen before (by
definition), we are still in the data paucity regime, except
for the possibility that similar objects are known. The approach to
reliable classification involves the following steps: 1) quick
initial classification, involving rejection of several classes and
shortlisting a few likely classes, 2) deciding which possible
follow-up resources are likely to disambiguate the possible classes
best, 3) obtaining the follow-up, 4) reclassification by folding in
the additional data. This schema can then be repeated if
necessary. All these steps can be carried out using Bayesian
formalism.
\begin{figure}[hbt]
\centerline{
\includegraphics[width=0.8\textwidth]{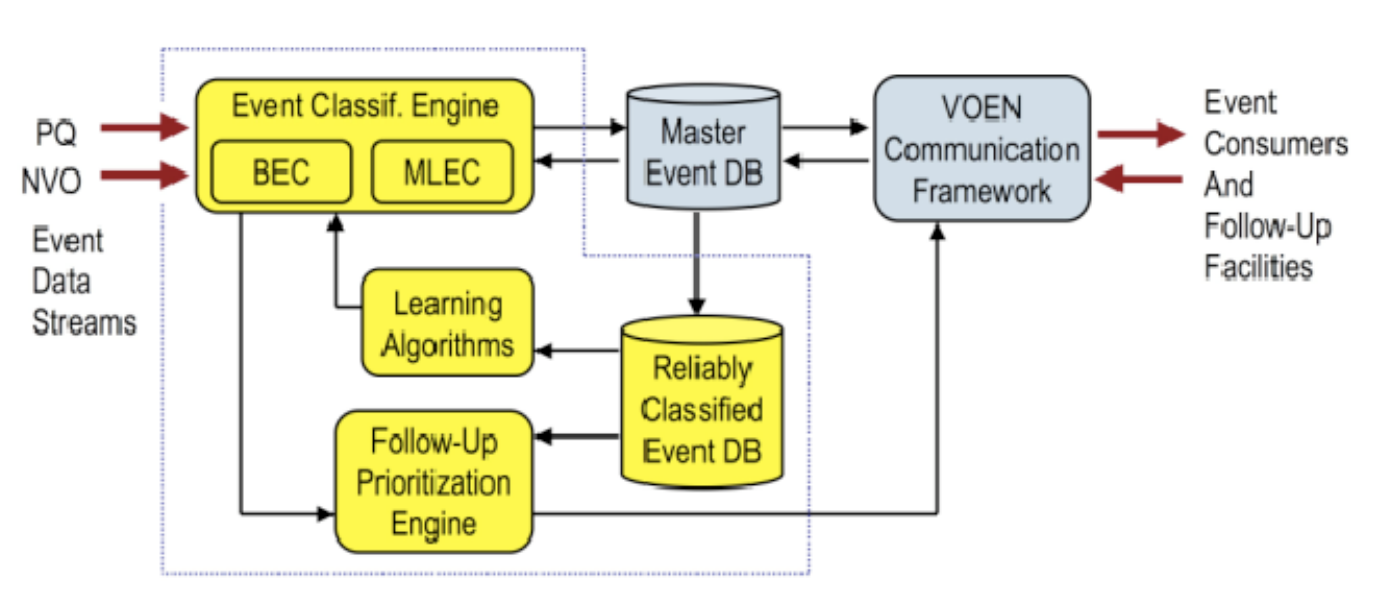}
}
\caption{
A schematic illustration of the desired functionality of the
Bayesian Event Classification (BEC) engine for classifying variables
and transients. The input is generally sparse
discovery data, including brightness in various filters, possibly the rate of
change, position, possible motion, etc., and measurements from available
multi-wavelength archives; and a library of priors giving probabilities for
observing these particular parameters if the event belongs to a class X. The
output is an evolving set of probabilities of belonging to various classes of
interest. Figure reproduced with permission from \citet{HW1}.
}
\label{fig:tr:BEC}
\end{figure}

1) Quick initial classification can be done using a) a Bayesian Network, and/or b)
Gaussian Process Regression. The advantage of the Bayesian Network over some other Machine Learning
applications is that it can operate better when some or most of the input data
are missing. The best approach is to use both the Bayesian and Machine
Learning approaches playing those to their strengths. The inputs from the priors
of different classes are colors, contextual information, light curves, and
spectra. With a subset of these available for the transient, one can get
probabilities of that object belonging to the different classes. \autoref{fig:tr:BEC}
represents such a schematic including both Bayesian and Machine
Learning. 
There is a lot of work that still needs to be done on this topic. Currently the priors
for different classes are very non-uniform in terms of number of examples in
different magnitude ranges, sampling rates, length of time, and so on. Moreover, to
understand transients, we will have to understand variables better. A resource
such as Gaia will be exceptional in this regard.

\begin{figure}
\begin{center}
\includegraphics[width=0.8\textwidth]{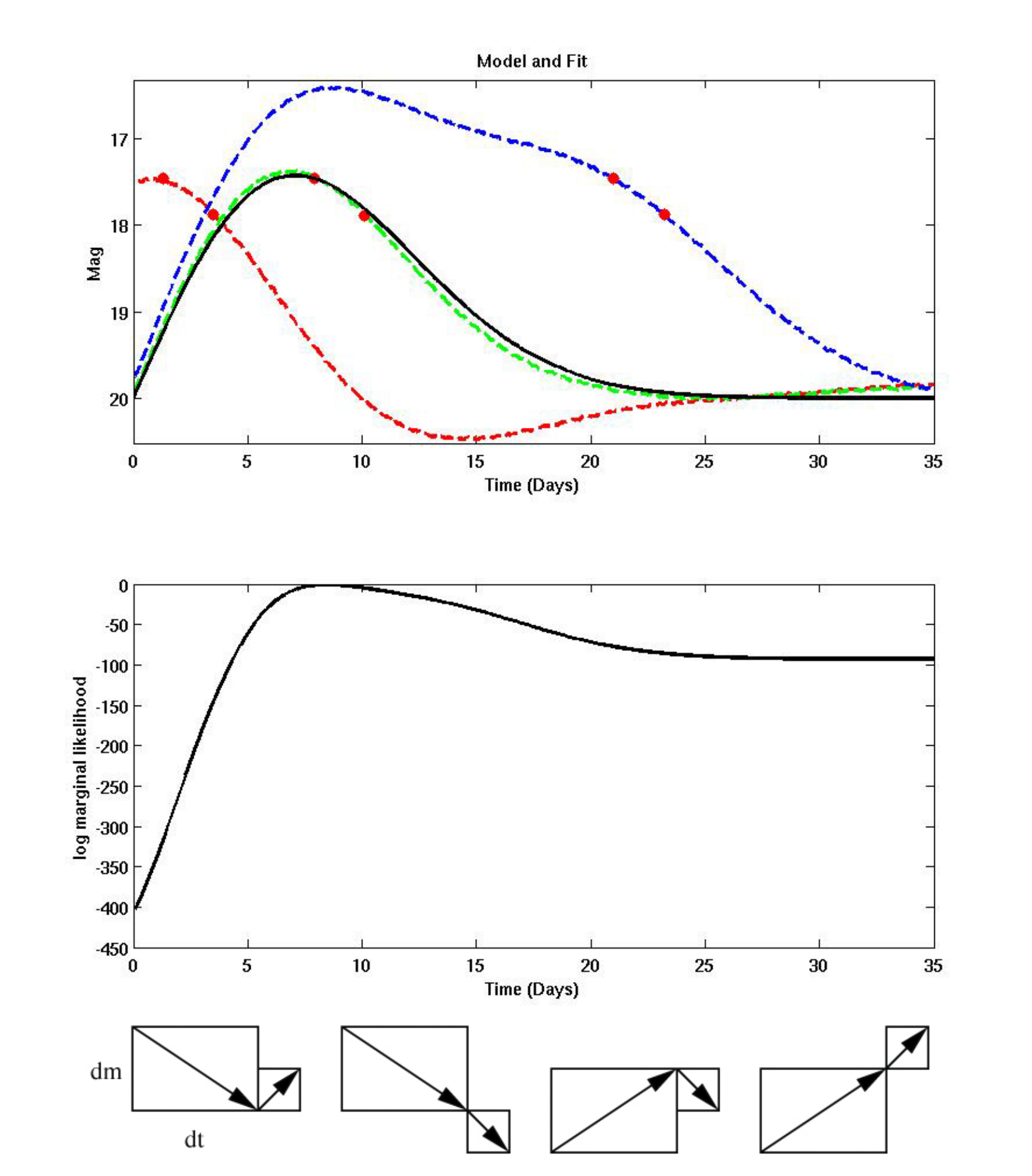}
\caption{\small Illustrated here is the use of the Gaussian Process Regression (GPR) 
technique to determine the likelihood that a newly detected transient is a 
supernova. The solid line in the first panel is a model obtained using GPR. 
The two observed points with given change in magnitude, $dm$, over the 
corresponding time interval, $dt$, allow one to estimate which phase of 
the model they are likely to fit best. Three specific epochs are shown as 
dotted lines. The second panel shows the log marginal likelihood that the 
pair of observed points correspond to the entire model light curve. In order 
to make the best estimate for the class of a given transient, a similar 
likelihood curve has to be obtained for models of different variable types.  
These model curves are obtained using covariance functions, where different 
types of variability require the use of different covariance functions.  
As more observed points become available for comparison, a progressively 
larger number of previously competing hypotheses can be eliminated, thus 
strengthening the classification. The boxes below the second panel show 
the distinct possibilities when three observations are present: in each case, 
the larger box represents the two previously known data points, which are 
decreasing in dm/dt, while the smaller box indicates the direction of the 
light curve based on the new data point. For example, the first of the possibilities 
begins to brighten after initially decreasing in brightness, which is inconsistent 
with the behavior of a supernova light curve.  The three other possibilities, where 
the object continues to dim, dims after an initial brightening, or continues to 
brighten, would all be consistent with different phases in a SN light curve. 
Figure reproduced with permission from \citet[][Figure 2]{class08}.
}
\label{fig:tr:GPR}
\end{center}
\end{figure}

Gaussian Process Regression, illustrated in \autoref{fig:tr:GPR}, is a
technique can be used to build template average light curves if they
are reasonably smooth. One can
use the initial LSST epochs to determine if the transient is likely to belong
to such a class and if so, at what stage of evolution/periodicity it
lies. 

2) Spectroscopic follow-up for all transients is not possible.
Different research groups are inherently interested in different types of objects as
well as different kinds of science.  Some observatories are already beginning
the process of evaluating optimal modes and spectroscopic instruments for
maximal use of LSST transient data.  A well-designed follow-up strategy must
include end-to-end planning and must be in place before first light.

\begin{figure}[hbt]
\centerline{
\includegraphics[width=0.8\textwidth]{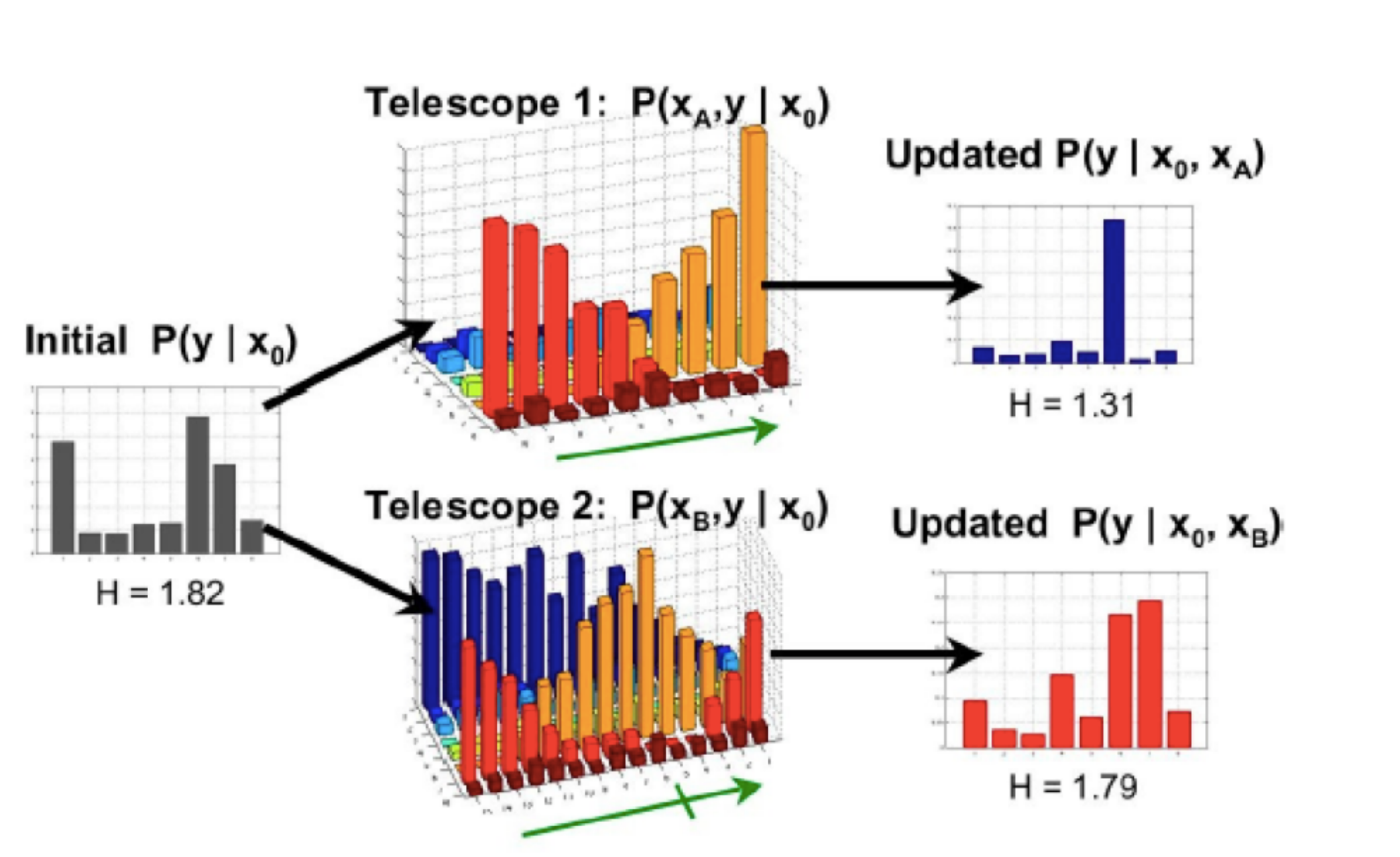}
}
\caption{
A schematic illustration of follow-up observation recommendations:
At left, the initial estimated per-class probabilities for eight object
classes, showing high entropy resulting from ambiguity between the object
classes numbered 1, 6, and 7. Follow-up observations from two telescopes are
possible (center). Their resolving capacity is shown as a function of class
$y$
(left axis) and observed value (right axis parallel to green arrows). In the
diagram, for telescope 1, as observed value, $x_A$, moves up the green arrow, class
6 becomes increasingly preferred. For telescope 2, moderate values (near the
crossbar in the arrow) indicate class 6, and other values indicate class 7.
Finally, at right, are typical updated classifications. The lower-entropy
classification at the top is preferred. Since the particular values used for
refinement $(x_A, x_B)$ are unknown at decision time, appropriate averages of
entropy must be used, as described in the text. Figure reproduced with permission from \citet[][Figure 3]{HW1}.
}
\label{fig:tr:DOBS}
\end{figure}

The very first observations of a transient may not reveal its class
right away, and follow-up photometric observations will be required
for a very large number of objects. Here too there will be a choice
between different bands, available apertures, and sites.  For example,
follow-up with a specific cadence may be necessary for a suspected
eclipsing binary, but with a very different cadence for a suspected
nova.  Follow-up resource prioritization can be done by choosing a
set-up that reduces the classification uncertainty most. One way of
accomplishing this is to use an information/theoretic approach
\citep{Loredo03} by quantifying the classification uncertainty using
the conditional entropy of the posterior for $y$, given all the
available data -- in other words, by quantifying the remaining
uncertainty in $y$ given a set of ``knowns'' (the data). When an
additional observation, $x+$, is taken, the entropy (denoted here as
$H$) decreases from $H(y|x0)$ to $H(y|x0, x+)$. 
This is illustrated in
\autoref{fig:tr:DOBS}. where the original classification, $p(y|x0)$, is ambiguous and may be
refined in one of two ways. The refinement for particular observations, $x_A$
versus $x_B$, is shown.

3) For fast or repeating transients, the LSST deep drilling
sub-survey (\autoref{sec:design:cadence}) will
yield the highest quality data with excellent sampling. 
Much of the transient science
enabled by LSST will rely on additional observations of selected transient
objects on other facilities based on early classification using the LSST data.  Some of the
additional observations will be in follow-up mode, while others will be in a
co-observing mode in which other multi-wavelength facilities monitor the same sky
during LSST operations. For relatively bright transients, smaller robotic
telescopes around the world can be deployed for follow-up.  An example is the
Las Cumbres Observatory Global Telescope Network of 2-m telescopes and
photometric + IFU instruments dedicated to follow-up. 

The information from these follow-up observations needs to be fed back
to the LSST classification. 
VO tools and transient portfolios will allow
LSST and non-LSST observations of the same object to be properly grouped for
the next iteration of classification.

4) Together with any such new data the classification steps are repeated
until a set threshold is reached (secure classification, or $\Delta t$ exceeded
for best classification, or classification entropy cannot be decreased
further with available follow-up resources, etc.) In addition to semantic
linking mentioned earlier, iterated and interleaved citizen science and expert
plus machine learning classification will be heavily used.

\subsection{Prospects for Follow-up and Co-observing}
\label{sec:prospects}

LSST will probe 100 times more volume than current generation
transient searches such as Pan-STARRS1 and PTF. It will also have
somewhat faster cadence and superb color information in six
photometric bands. Much of the science on repeating transients as well
as on explosive one-time events will be accomplished largely from the
LSST database, combined with other multi-wavelength data sets.  The
logarithmic cadence of the primary survey and the deep drilling
sub-survey are designed to optimize the time sampling for a wide range
of variability patterns, both known and predicted from theory.  As the
number of multi-wavelength facilities continuously observing various
areas of the sky continues to grow, ``co-observing" is becoming an
increasingly important avenue to discovery. However, in order to
maximize the science return of LSST, a well designed program of
detailed follow-up observations for a smaller sample of carefully
selected transients will be required.

LSST is expected to deliver data on tens of thousands of
transients every night. By the time of the initial alert, the survey
will have collected detailed information on the presence, morphology,
and photometric properties of the host galaxy, including a photometric
redshift. For a majority of transients, existing catalogs will provide
useful limits on the progenitor across the electromagnetic spectrum,
and for some sources a positive identification can be made. A small
fraction of the best candidates for follow-up will have a high energy
identification and possibly a simultaneous detection by one of the
next-generation all-sky monitors such as EXIST to follow the Swift and
Fermi missions.  With the help of expert systems based on Bayesian
belief and decision networks, the long list of ongoing transients will
be prioritized on science potential. Transients will naturally fall
into two categories: 1) rare bright events and/or well covered
transients with the most complete data and frequently found well
before the peak light and 2) numerous fainter (22-24th mag) objects with
less coverage, but suitable for statistical studies.

Transients in the first category will be relatively rare, 
and efficient follow-up would focus on one object at a
time. New classes of exotic transients can usually be established
based on a few exceptionally well-observed events.  LSST will enable
early detection of prototype cases for a number of theoretically
predicted explosive transients which we've already discussed, including orphan GRB afterglows,
accretion induced collapse events, fall-back and pair-instability
supernovae, and the so called SN 0.Ia.  Several groups are developing
systems for multi-band simultaneous photometry and Integrated Field
Unit (IFU) spectroscopy on rapidly deployed telescopes around the
world that can continuously follow transients brighter than
$\sim$ 22nd mag.  An example is the Las Cumbres Observatory Global
Telescope Network of 2-m telescopes and photometric/IFU instruments
dedicated to follow-up.  

Historically, cutting-edge instruments have not been focussed on
science that can be done with bright objects.  It will be important that 1--4-m
telescopes be instrumented to follow brief transients to their
peak brightnesses, which can get to naked-eye level \citep{Racusin++08}.  
The leaders
in this type of follow-up are observatories that respond to new
discoveries of gravitational microlensing events (e.g., Microlensing
Follow-Up Network ($\mu$-FUN) and Robonet-II). Target-of-opportunity
programs on specialized X-ray, infrared and radio observatories such as
today's Chandra, XMM/Newton, Spitzer, and VLA will continue to provide
broad-band spectra and imaging across the electromagnetic
spectrum. With ALMA and projects such as Constellation-X, IXO, and JWST in the queue, we may
expect higher resolution and more sensitive multi-wavelength follow-up
resources to be available by the time LSST starts operating.

Somewhat less detailed follow-up will be obtained for significant
numbers of fainter transients.  For photometry, LSST itself provides
sparsely time-sampled follow-up on timescales of hours to
days. Spectroscopic and multi-wavelength follow-up is the key to
breaking degeneracies in the classifications and unraveling the
physics, and will necessarily be a world effort.  The amount of large
telescope time required to determine the redshift of optical
afterglows accompanying GRBs localized by Swift is 0.5-2.5 hours, with
a mean response time of 10 hours.  The list of world's large optical
telescopes includes about half dozen instruments in each of the
classes: 9--10 m, 8--9 m, and 5--8 m. As of 2009, there are $\sim20$
optical telescopes with diameter of 3 meters or larger which can
access the Southern Hemisphere at least in part; these are the
facilities which will be well-places to follow up fainter LSST
transients.

Several design
studies for extremely large optical telescopes are in progress
(Euro50, E-ELT, MaxAT, LAMA, GMT, TMT). They will further reduce the
integration time required for spectroscopic follow-up of faint
sources.  Major observatories around the world such as ESO and NOAO,
and the astrophysics community at large are developing optimized
observing modes and evaluating spectroscopic instruments that will
better utilize LSST transient data.  Because we expect many transients
per LSST field of view, efficient spectroscopic follow-up would best
be carried out with multi-slit or multi-IFU systems. BigBOSS is a
newly proposed instrument for the Mayall or Blanco 4-m telescopes, capable of simultaneously measuring 4,000
redshifts over a $3^\circ$ diameter field of view.
Wide field
follow-up would be possible with AAT/AAOmega and Magellan/IMACS
instruments.  Some northern facilities will partially overlap with the
LSST survey: BigBOSS at the Mayall, GTC, Keck MOSFIRE and DEIMOS,
MMT/Hectospec, and LAMOST. Smaller field of view spectroscopic
follow-up in the south can be accomplished with Gemini/GMOS, the
VLTs/FORS1, and SALT/RSS. It is reasonable to expect that new
instruments will be built for these and other spectroscopic facilities
by the time LSST sees first light.

\section{Gravitational Lensing Events} 
\label{sec_microlensing}
{\it Rosanne Di Stefano, Kem H. Cook, Przemek Wozniak, Andrew C. Becker}

Gravitational lensing is simply the deflection of light from a distant
source by an intervening mass. There are several regimes of lensing. 
In strong lensing (\autoref{chp:sl}),
the source is typically a quasar or very distant galaxy, and the lens
is a galaxy or galaxy cluster at an intermediate distance.
Lensed quasars typically have multiple
images, each a distorted and magnified view of the unlensed quasar.
Lensed galaxies may appear as elongated arcs or rings.
{\it Weak lensing} is discussed in \autoref{chp:wl}. Weak lensing is also
a geometrical effect. While no single distant source may exhibit
wildly distorted images, the lensing effect can be measured through 
subtle distortions of many distant sources spread out over a field
behind the lens. In these cases, the main effects of lensing are
detected in the spatial domain.
In this chapter we focus on those cases in which the primary signature
of lensing is in the time domain. That is, we discuss lensing {\it events},
in which the time variability arises because of the relative motion 
of source, lens, and observer.
This is generally referred to as {\it microlensing}. When the lens is
nearby, however, the Einstein ring becomes large enough that
spatial effects can also  be detected. Because of this and other
observing opportunities made possible by the proximity of the lens,
nearby lensing is referred to as {\it mesolensing} \citep{DiStefano2008a,DiStefano2008b}.    
LSST will play a significant role in the discovery and study of both 
microlensing and mesolensing events. 

A lensing event occurs when light from a background source
is deflected by an intervening mass. \citet{Einstein1936}
published the
formula for the brightening expected when the source and lens are
point-like. The magnification is $34\%$ when the
angular separation between source and lens is equal to $\theta_E,$
an angle now referred to as the Einstein angle.
\begin{equation}
\theta_E   =  \Bigg[
   {{4 GM (1-x)}\over{c^2 D_L }}\Bigg]^{{1}\over{2}}
         =  
          0.01''\, \Bigg[{(1-x)\, 
\Big({{M}\over{1.4\, M_\odot}}\Big)\Big({{100\, {\rm pc}}\over{D_L}}\Big)}
\Bigg]^{{1}\over{2}}
\label{eqn:tr:einstein_angle}.
\end{equation}
In this equation, $M$ is the lens mass, $D_L$ is the distance to the lens,
$D_S$ is the distance to the source, and $x=D_L/D_S.$
The time required for the source-lens separation to change by
an Einstein diameter is

\smallskip

\begin{equation}
\tau_E  =  {{2\, \theta_E}\over{\omega}}
      = 
      70\,{\rm days}\, 
         \Big[{{50\, \kms}\over{v_T}}\Big]
         \Bigg[ {{M}\over{1.4\,  M_\odot}}\, 
       {{D_L}\over{100\, {\rm pc}}}\, (1-x)\Bigg]^{{1}\over{2}}.      
\end{equation}

Einstein did not consider the effect to be observable because of the
low probability of such close passages and also because the observer would be
too ``dazzled'' by the nearby star to detect changes in the background star.
\citet{Paczynski1986} answered both of these objections by
noting that low-probability events could be detected because
monitoring of large numbers of stars in dense source fields
had become possible,
and by suggesting lensing as a way to test for the presence of compact
{\it dark} objects.
The linking of the important dark matter problem to lensing,
at just the
time when nightly monitoring of millions of stars had become possible,
sparked ambitious new observing programs designed to discover lensing events.
Given the fact that episodic stellar variability of many types is
$100-1,000$ times more common than microlensing,
success was not assured.
To be certain that events they discovered had actually been caused  by lensing,
the monitoring teams adopted strict selection criteria.

In fact, these early teams and their descendants have been wildly
successful. They have convincingly demonstrated that they
can identify lensing events. More than $4,000$ candidate
events are now known,\footnote{Most of the events discovered so far
were generated by low-flux stellar masses along the direction to the bulge
(see, e.g., \citealt{Udalski03}).},
among them several ``gold standard'' events which
exhibit effects such as parallax and lens binarity.

Perhaps the greatest influence these programs have had is in demonstrating
the power afforded by frequent monitoring of large fields.
In addition to discovering rare events, the ``needles-in
a-haystack'' of other variability, they have also
yielded high returns for a number of other astrophysical investigations,
including stellar structure, variability, and supernova searches.
One may argue that
the feasibility and scientific return of wide field programs such as LSST 
was established by the
lensing monitoring programs.
\begin{figure}
\begin{center}
\includegraphics[width=0.3\textwidth]{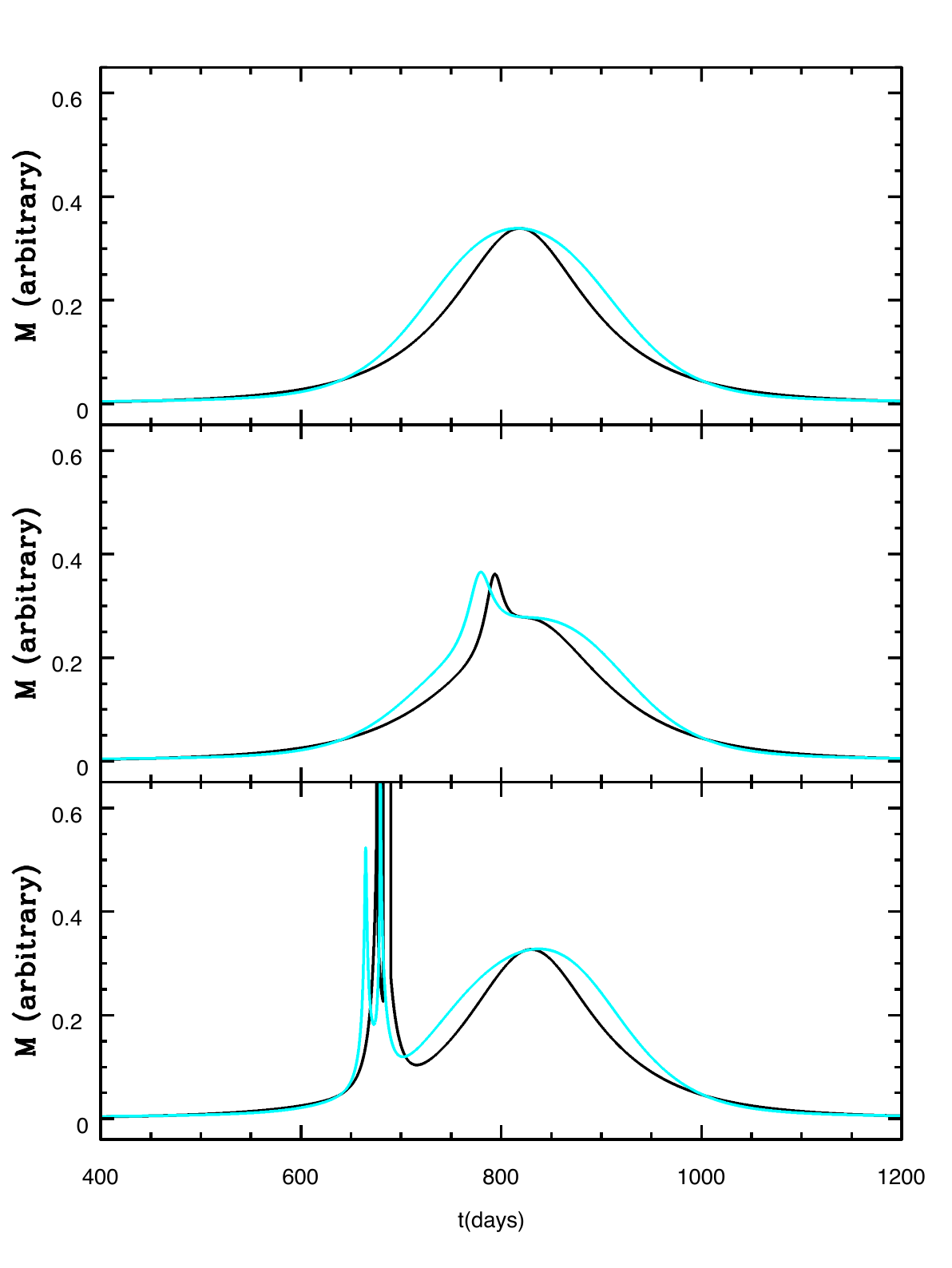}
\caption{\small 
Sample lensing light curves. These particular light curves were 
generated by high-mass lenses (black holes with $M=14\, M_\odot,$
and $D_L = 200$~pc) and the deviations from baseline 
last long enough and evolve slowly enough that LSST can track the event and
provide good model fits. Cyan curves include parallax effects due to
the motion of the Earth around the Sun; black 
curves do not.
{\sl Top:} The lens is an isolated black hole. {\sl Middle and
Bottom:} The lens
has a white dwarf companion with orbital period appropriate
for the end of mass transfer. The orbital phase at the time of peak is
the distinguishing feature between the middle and bottom panels. 
In these cases, data collected by LSST alone could identify the correct 
models. For short duration events or for some planet lenses, 
LSST could discover events and spark alerts to allow more frequent
monitoring.     
\label{fig:tr:microlens}
}
\end{center}
\end{figure}
Every year LSST data will contain the signature of tens of thousands
of lensing events (\autoref{tab:tr:lensrate}).  Many will remain above
baseline for several months.  This means that, with a sampling
frequency of once every few nights, LSST will obtain dozens of
measurements of the magnification as the event progresses.  Meaningful
fits to a point-lens/point-source light curve can be obtained with
fewer than a dozen points above baseline.  LSST will therefore be able
to test the hypothesis that an ongoing event is caused by lensing.

In fact, LSST will also be able to discover if lensing events display
deviations from the point-lens/point-source form.  Such deviations
will be common, because they are caused by ubiquitous astrophysical
phenomena, such as source binarity, lens binarity, and parallax.  The
black light curve in the top panel of \autoref{fig:tr:microlens} shows
a point-lens/point-source light curve, and the blue light curve in the
same panel shows the parallax effects expected if the lens is $200$~pc
away.  In this case, the deviation introduced by parallax is several
percent and lasts for a significant fraction of the event.  The middle
and bottom panels of \autoref{fig:tr:microlens} illustrate that
deviations caused by lens binarity similarly influence the
magnification over extended times in ways that can be well-modeled
\citep[see][for a mathematical treatment]{DiStefano1997}. 

The good photometric sensitivity of LSST will allow it to detect these
deviations. By pinning down the value of the magnification every few
days, LSST will provide enough information to allow detailed model
fits. We have demonstrated that the fits can be derived and refined,
even as the event progresses \citep{DiStefano2007}.  This allows sharp
features (such as those in the bottom panel of
\autoref{fig:tr:microlens}) to be predicted, so that intensive
worldwide monitoring can be triggered.  The LSST transient team will
develop software to classify events in real time to allow it to call
reliable alerts (\autoref{sec_followup}).  While there are some
examples of light curves on which we will want to call alerts (some
planet-lens light curves for example), many special effects will be
adequately fit through LSST monitoring alone.

\autoref{fig:tr:microlens} illustrates another point as well: many
lenses discovered through LSST's wide area coverage will be
nearby. That is, they will be mesolenses.  The black hole in this
example would create a detectable astrometric shift. The size of its
Einstein angle could thereby be measured, while the distance to the
lens could be determined through the parallax effects in the light
curve. \autoref{eqn:tr:einstein_angle} then allows the lens mass to be
determined. Similarly, when nearby low-mass stars serve as
lenses, radiation from the lens provides information that can also
break the degeneracy and measure the lens mass.


\subsection{What Can Lensing Events Teach Us?}

\noindent{\sl 1) Dark Matter:}
It is still controversial whether or not the existing lensing programs have successfully established the
presence or absence of MACHOs in the Galactic halo. 
The upper limit on the fractional component of
MACHOs was found to be approximately $20\%$ by 
\citet{Alcock2000}. 
However, if the experiments on which these estimates are based are 
overestimating their detection efficiencies (perhaps by missing some
lens events that deviate from the point-lens/point-source form, as
suggested by the under-representation of binary lenses and binary
source events in the data; \citealt{Night++08}), 
the true rate
and perhaps the number of MACHOs, would be larger than presently thought.
Additional monitoring can definitively answer the questions
of whether MACHOs exist and, if they
do, whether they comprise a significant component of Galactic dark matter.
To achieve this, we need to develop
improved event identification techniques and 
reliable calculations of the detection efficiencies for 
events of different types.

\begin{figure}[ht]
\begin{center}
\includegraphics[width=0.6\textwidth]{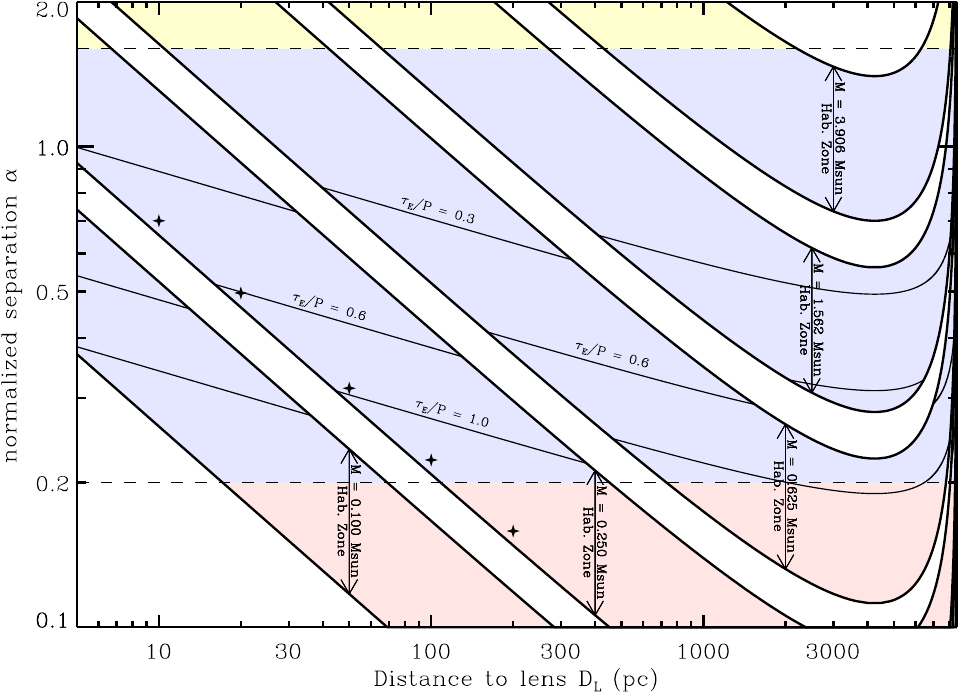}
\caption{ Separation between $\alpha$ vs. $D_L$ for the habitable zones (HZs)
of low-mass stars.
Each colored bar represents a star with a given mass: $M = 0.1\,
{\rm M_\odot}$ on the lower left, increasing by a factor of 2.5 for each
subsequent bar.  The lower (upper) part of each bar corresponds to the
inner (outer) edge of the HZ for a star of that mass.  The
upper horizontal dashed line at $\alpha = 1.6$ marks the approximate
boundary between ``wide'' systems, in which the planet and star act as
independent lenses \citep{DiStefano1999a, DiStefano1999b}, and
``close'' systems in which distinctive non-linear effects, such as
caustic crossings provide evidence of the planet \citep{Mao1991,Gould1992}. All of the planets detected so far have model fits with
$\alpha$ lying between $0.7$ and $1.6.$ In this range, the effects of
caustics are the most pronounced. As $\alpha$ decreases, the effect
of the planet on the lensing light curve becomes more difficult to
discern; the horizontal dashed line at $\alpha = 0.2$ is an estimate
of a lower limit. Contours with constant values of the ratio $\tau_E/P,$
where $P$ is the orbital period, are also shown. This is
because the probability of detecting the planet in
close systems ($\alpha \le 0.5$) is increased by the orbital motion.
For $\alpha>0.2,$ systems with large orbital motion are potentially
detectable by current observations.
\label{fig:tr:habitable}
}
\end{center}
\end{figure}

\noindent{\sl 2) Planets:}
The search for planets is an important ongoing enterprise (\autoref{sec_transits}).
Lensing can contribute
to this search in several important ways. For example, in contrast
to transit and radial velocity methods, lensing is
sensitive to planets in face-on orbits. In addition,
lensing is effective
at discovering both low-mass planets and planets in wide orbits.
Finally, it is ideally suited to discovering planets at large
distances and, therefore, over vast volumes.
In addition, we have recently begun to explore the
opportunities of using lensing to study planets orbiting nearby ($< 1$~kpc)
stars \citep{DiStefano2007,DiStefanoNight2008}.
{Fortuitously, the Einstein ring associated with a nearby M~dwarf
is comparable in size to the semi-major axes of orbits in the M~dwarf's
zone of habitability (\autoref{fig:tr:habitable}).
Events caused by nearby planets can be discovered
by monitoring surveys or through targeted follow-up lensing
observation. 
}

\noindent{\sl 3) Distant stellar populations:}
Lensing can teach us about 
both the star serving as the lens and the source star that was lensed.
Both source and lens can be members of
a distant dense source field, and  
binary-lens effects
and/or binary-source effects should be detectable for a significant fraction
of events. 
When the selection effects are well-understood, the fraction and 
characteristics of binaries in external galaxies can be derived.
It is noteworthy that, unlike eclipse studies, lensing 
is sensitive to binaries with orbits of all orientations. 
In addition, finite-source effects provide information about
surface features of the source star.     

\noindent{\sl 4) The solar neighborhood:}
A significant fraction of all lensing events are generated by
nearby  
masses, most M dwarfs. As noted, lensing is particularly sensitive to 
planets in the habitable zones of nearby M~dwarfs.   
LSST will also discover lensing by white dwarfs, neutron stars,
and black holes (see \autoref{tab:tr:lensrate}). Particularly
because few nearby neutron stars and no 
nearby black holes are known, all-sky monitoring has the potential to make important   
contributions. 

\subsection{What Does LSST Bring to These Studies?}

LSST will sample most parts of the sky every few days. 
Although the telescopes will 
return to some regions more frequently, the cadence is not
well suited to study the rapid changes that can be associated with,
for example, 
caustic crossings. Nevertheless, LSST will become a major player
in the study of lensing. It has several important advantages:

\smallskip

\noindent{\sl 1) All-sky coverage:} 
LSST will be able to find lensing events across most of the sky.
It will probe the Galactic halo in 
many directions, discovering MACHOs or placing tight limits on their 
existence, and exploring the stellar populations of the halo. 
Lensing of stars will be detected in a wide range of external 
galaxies, including Local Group dwarf galaxies and galaxies within
several Mpc. In addition, lensing of stars in our
own Galaxy will also be observed. To illustrate, we
note that limited monitoring has already discovered the lensing
of an A0 star just one kpc away by an unknown intervening mass
\citep{Fukui2007,Gaudi2008}.    

\begin{table*}
\vspace{-.1 true in}
\centering{
\label{tab:tr:lensrate}
\caption{\bf Nearby Microlens Event Rates}
\begin{tabular}{|c||c|c|c|c|}
\hline
 & {\bf Past}  &{\bf Present}& {\bf Future}& {\bf Future}   \\
 & per decade  & per decade  & per decade  & per decade      \\
Lens type&per deg$^2$&per deg$^2$&per deg$^2$&{\bf over 150} deg$^2$\\
\hline
M dwarfs&2.2     &46    &920   &$1.4 \times 10^5$\\
L dwarfs&0.051  & 1.1  &22    &3200\\
T dwarfs&0.36    &7.6   &150   &$2.3\times 10^4$\\
WDs     &0.4     &8.6   &170   &$2.6\times 10^4$\\
NSs     &0.3     &6.1   &122   &$1.8\times 10^4$\\
BHs     &0.018  &0.38  &7.7   &1200
\\

\hline
\end{tabular}
}
\par
\medskip
\begin{minipage}{0.94\linewidth}
\footnotesize
Each predicted rate is valid for the direction toward the Bulge \citep[see][for details]{DiStefano2008a, DiStefano2008b}.
{\it Past:} the observing parameters apply to the first generation of
monitoring programs, including MACHO. {\it Present:} applies to the
present generation, including OGLE~III and MOA.
{\it Future:} applies to upcoming projects such as
Pan-STARRS and LSST. The effective area containing high-density source fields is
$\sim 150$ deg$^2$; this is used in the last column. In fact,
near-field source stars spread across the sky will also be lensed, adding
to the rate of lensing by nearby masses;
the above estimates for lensing by nearby masses
are fairly conservative. 
\end{minipage}
\end{table*}

\noindent{\sl 2) Excellent photometric sensitivity:} When an event deviates
from the point-source/point-lens form, the deviations are 
typically long-lasting, even if
the most dramatic effects occur during a short time interval. 
Sampling the light 
curve with good photometric sensitivity at a modest number of points 
can therefore identify its unique features 
and help determine the physical characteristics of the lens.       

\noindent{\sl 3) The opportunity to develop superior selection criteria:}  
When the lensing monitoring teams first started, they had to prove that it 
is possible to identify lensing events among the much larger background
``noise'' of intrinsic variations exhibited by stellar systems. They,
therefore, used strict criteria designed to identify the point-lens/point-source
light curves first predicted by Einstein. Despite their remarkable 
success, with more than 4,000 lensing candidates identified, many 
of the events that should be associated with common astrophysical
systems (binary sources, binary lenses, etc.) have been found only 
rarely. The detection efficiencies are not well understood,
making it difficult to draw general conclusions based on the
events that have been discovered. 
We have the opportunity to use the years before LSST data acquisition
to develop procedures to identify {\sl all} lensing events with
an efficiency that can be calculated. 

\noindent{\sl 4) The opportunity to predict mesolensing events:}  
LSST will identify and track the motions of many nearby stars,
measuring parallaxes and proper motions. This detailed look at the
local sky will supplement what has been learned from SDSS and
other surveys \citep[see, e.g.,][]{Lepine2008}, and will allow us to predict when nearby stars 
will pass close enough 
to distant objects to generate detectable lensing events.
The ability to predict lensing events, based on LSST data,
will turn lensing into a more flexible tool for astronomical studies.
While the predicted events may be detected with LSST, other
telescopes can learn a good deal by providing frequent
multiwavelength monitoring.  

\noindent{\sl 5) Studies of both the astrometric and photometric effects
for mesolensing events:}  
LSST will make sensitive astrometric as well as sensitive photometric
measurements. Because lensing creates multiple images,
whose positions and intensities change as the event progresses,
astrometric shifts 
are expected \citep[see, e.g.,][]{Dominik2000}. 
For nearby lenses, the shifts can be several milli-arcseconds,
potentially measurable with LSST. Indeed, a unique combination of
astrometric as well as photometric monitoring is possible with LSST and
can be valuable to both discover and study lensing events.

\medskip

The bottom line is that LSST can advance fundamental science through the 
detection, identification, and correct interpretation of lensing
events. In order for this to happen, we will have to devote significant
effort to laying the necessary foundation in theory, event selection,
and analysis.

\section{Identifying Variables Across the H-R Diagram}
\label{sec_variable_stars}
{\it Steve B. Howell, Dante Minniti}

The LSST will address three major science objectives related to variable stars: 1)
production of very large samples of already known variable types, 2)
discovery of theoretically predicted populations of variables not yet
discovered, 
 and 3) discovery of new variable types.  Large samples for
specific types of variable star provide enormous leverage in terms of
the statistical properties attributed to or deduced from
them. For example, small color deviations, unnoticeable in samples of
100-500, may illustrate metallicity effects and other evolutionary
properties for the class. Theoretical models often set boundaries in
\teff{} and $\log g$ space for classes of pulsators. These can be
exquisitely determined from big samples.  Additionally, large samples
obtained in a systematic way with uniform properties and biases enable stronger conclusions from limited samples.  This
type of new knowledge about old, ``well studied" variable star
classes was very well shown by the MACHO observations of RR Lyrae and
Cepheids toward the Galactic bulge. 

\begin{figure}[!hbt]
\centerline{
\includegraphics[width=0.5\textwidth,angle=0]{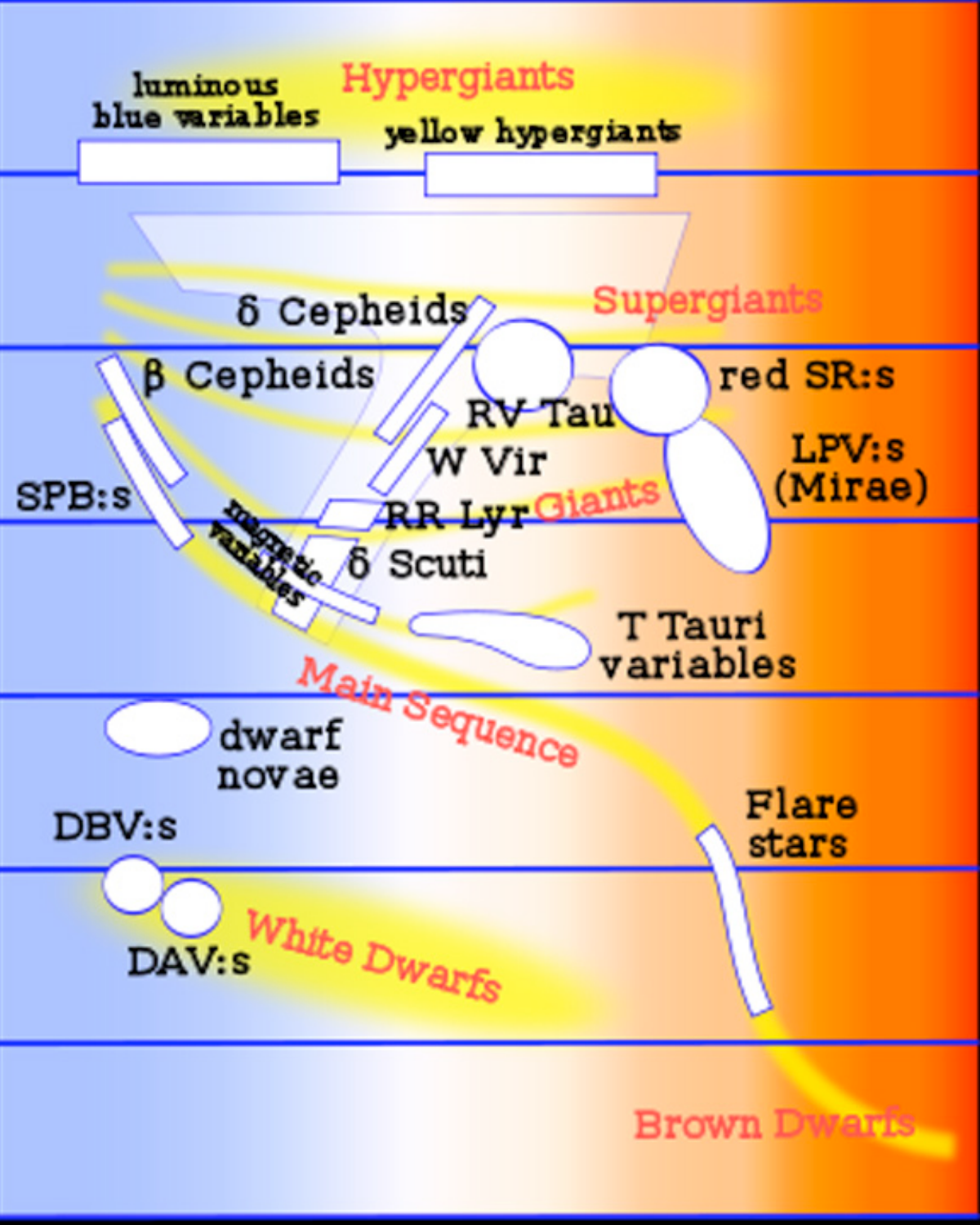}}
\caption[]{
H-R Diagram showing the locations of currently known classes of
variable stars, mostly comprised of pulsating variables: slowly
pulsating B stars (SPBs), red semi-regular variables (red SRs),
pulsating white dwarfs (DAV and DAB), and long period variables
(LPVs). Also shown are cool flare stars and T Tauri stars. Note that
while the absolute luminosity scale covers many orders of magnitude,
the present day set of pulsators is limited to those variables that
are bright. The exception to this is the pulsating white dwarfs due to
their special status and targeted study.  LSST will provide large
uniform samples of pulsating variables and will allow the remainder
of the H-R Diagram to be explored for additional variable types. 
}
\label{fig:tr:hr_diagram}
\end{figure}

Several classes of variable stars are shown in
\autoref{fig:tr:hr_diagram}.  The figure primarily includes pulsating
variables, which are likely to be the focus of many LSST research
programs.  Known types of periodic variables are either quite
luminous, $M_V<+2.5$, or are pulsating variable stars.
\autoref{fig:tr:hr_diagram} also shows non-periodic type low-mass M dwarfs, intrinsically variable objects which produce large amplitude ($>$1 mag)
transient flares, and the T Tauri stars.  Additional periodic
variable star types include eclipsing binaries and solar-like stars,
which show a rotational modulation due to star spots.  

\autoref{fig:tr:hr_diagram} shows that all  known pulsating variables (with the exception of
white dwarfs), have $M_V<+2.5$. The major reason for this ``bright
limit" is observational  selection effects in terms of areal
coverage, limiting magnitude, observed sample size, photometric
precision, and time coverage. LSST will lessen all of these biases by orders of magnitude, which will completely revolutionize
the science of periodic variable stars.

\begin{figure}[hbt]
\centerline{
\includegraphics[width=0.8\textwidth,angle=0]{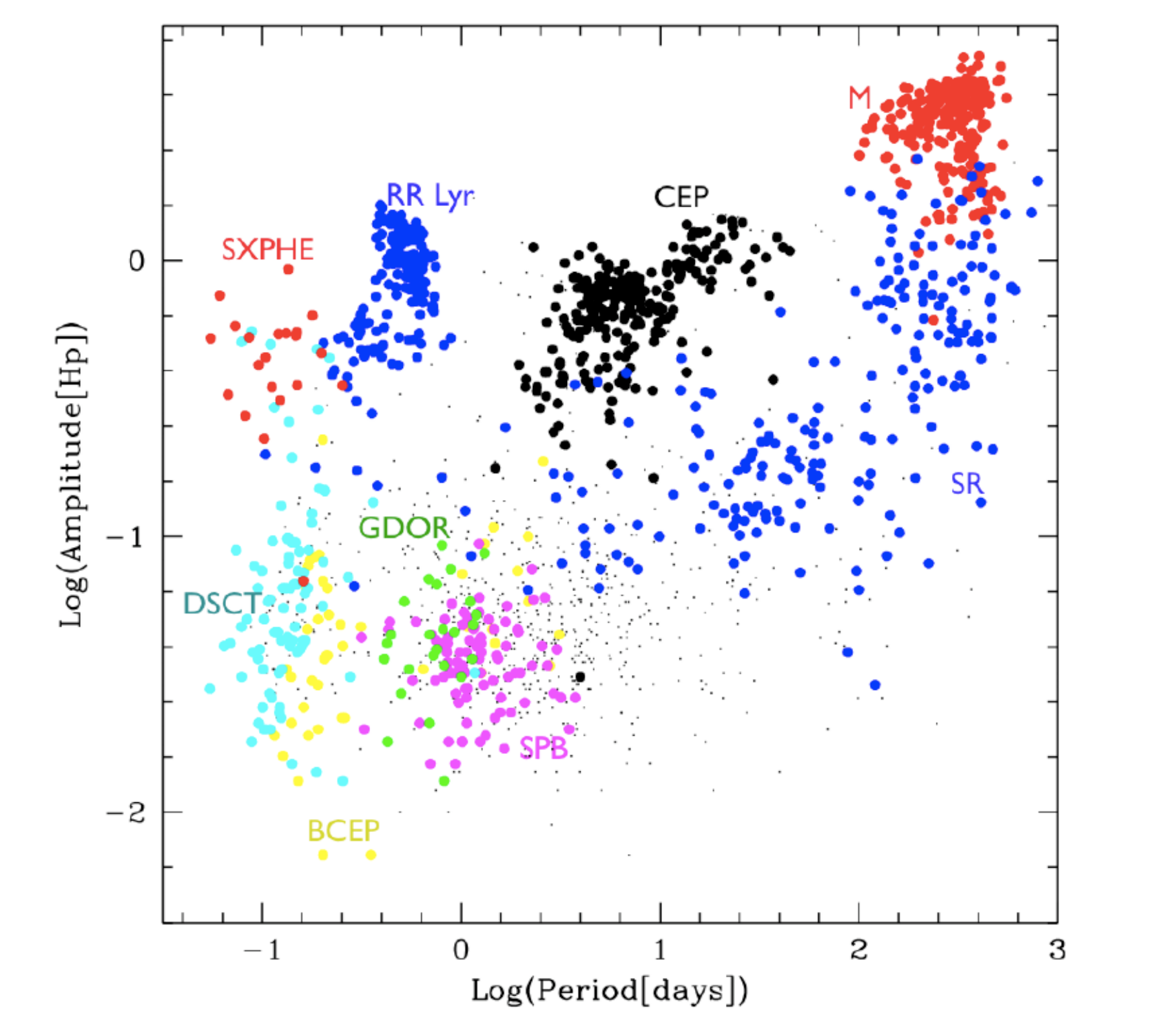}}
\caption[]{
The Period-Amplitude diagram for different classes of pulsating
variables as known today: $\delta$ Scutis (DSCT), SX PHe (SXPHE),
$\gamma$ Dor (GDOR), $\beta$ Cepheid (BCEP), Cepheids (CEP), RR Lyraes
(RRL), semi-regular variables (SR), slowly pulsating B stars (SPB),
and M dwarfs (M). Light curve measurements of pulsating variables
provide two fundamental parameters; the pulsation period and the light
curve amplitude. These measured parameters have a scaling relationship
as they are proxies to energy transport within a stellar
atmosphere. The larger the amplitude of a pulsation, the longer it
takes for energy to be displaced and thus the period of this action is
longer as well. 
The LSST will extend this diagram in the regions of longer periods and
smaller amplitudes as well as identify completely new members to add
to the plot. (Adapted from \citealt{Eyer2008}).
}
\label{fig:tr:period_amplitude}
\end{figure}

A useful overview of stellar variability is presented in
the recent paper by \citet{Eyer2008}. Gathering summary
information from the MACHO, OGLE-II and III, HAT, ASAS, SuperWASP,
HIPPARCOS, and other surveys, Eyer and Mowlavi  attempt to separate
variables based on type (periodic or not) and subtype while providing
summary statistics for each group. While time sampling and photometric
precision vary among the surveys, very useful general trends and
parameters are apparent. Stars are variable throughout the H-R  diagram
but not with the same observed frequency. For example, red giants are
nearly 100\% variable while the A main sequence stars  only show about
a 5\% rate.  For all periodic variables, classification work often
begins with period and amplitude, the two common photometrically
measured parameters, with star color being of additional importance.
The majority of periodic variables are normal pulsators and cover a
large range in  period and amplitude.
\autoref{fig:tr:period_amplitude} gives a schematic view of the
known pulsator types shown in period-amplitude space.  Each point
represents a single well-studied member of the variable star class, and
we can see the general trend for pulsators in that the larger the
pulsation amplitude the longer the pulsation period.  The smallest
amplitude limit is about 0.01 mag and the period limits are $\sim$0.1
day to 1,000 days, values that will be greatly improved upon by the
LSST, thereby likely increasing the discovery space even for normal
pulsators.

\citet{Pietrukowicz2009} recently carried out a deep variability
survey in a field in the Galactic plane using VLT/VIMOS, allowing an
experimental quantification of the numbers and types of variable stars
that the LSST survey may ultimately detect. This work is well suited
for comparison with the LSST due to the large telescope aperture,
similar exposure times, similar limiting magnitudes, and light curve
quality. The survey lasted only four nights, but due to its depth and
improved precision, the total number of variable stars found in this
survey was higher than previous shallower surveys (like MACHO, OGLE,
ASAS, etc.). Over this short time baseline, 0.69\% of the observed
stars had detectable variability.

Extrapolating from their results suggest that LSST will discover of order 
135 million variable stars. Of these,
57 million will be eclipsing/ellipsoidal variables, 59 million will be
pulsating variables, 2.7 million will be flaring stars, and 0.78
million will exhibit variability due to extrasolar planetary transits.

\begin{figure}
\begin{center}
\includegraphics[width=0.6\textwidth]{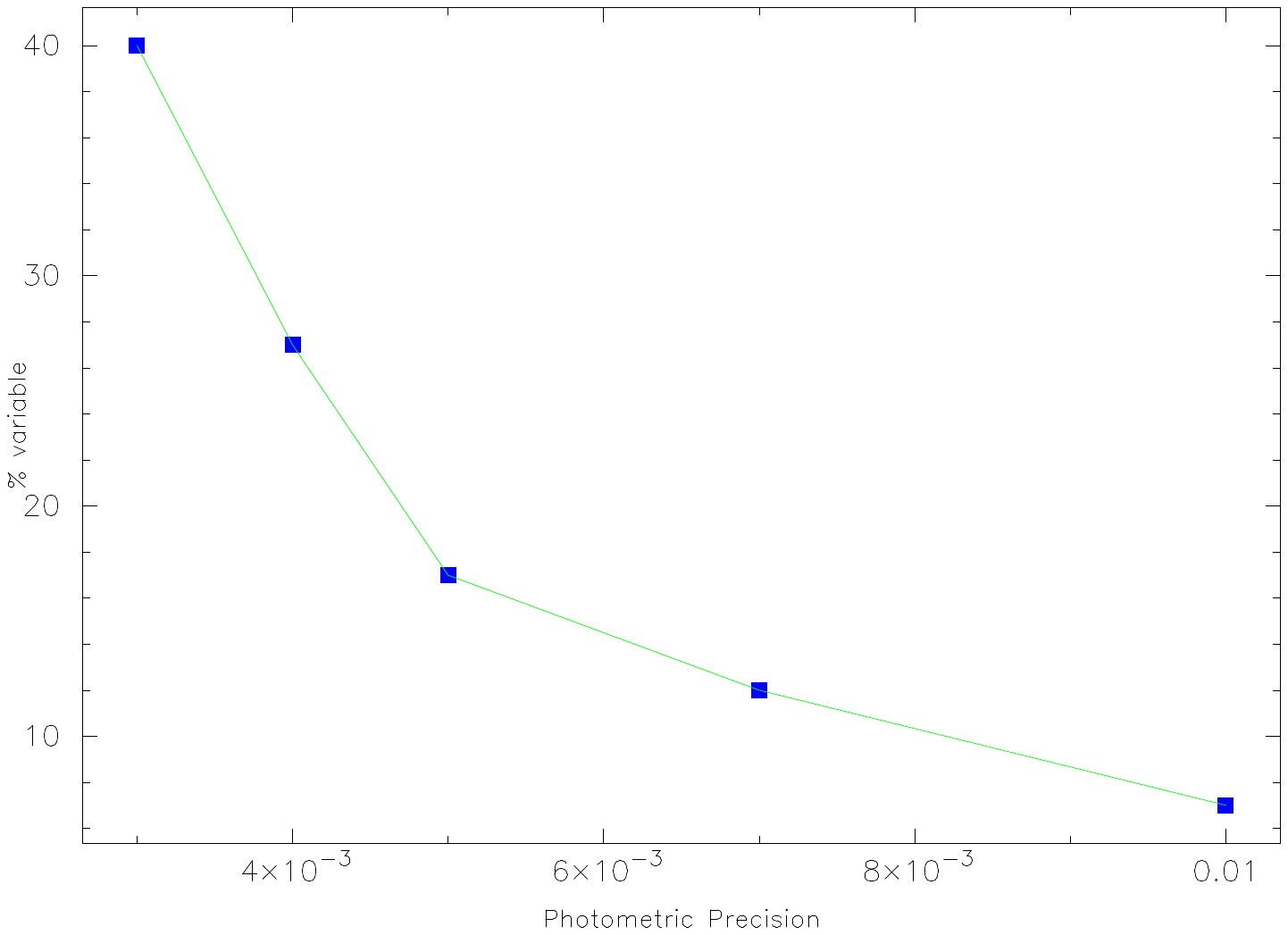}
\caption{Schematic diagram relating the percentage of variable sources a given survey will detect compared with the best photometric precision of that survey. Combining the information gleaned 
from many surveys designed to search for variability, the figure shows
the general trend observed. As the photometric precision of a survey
improves, the number of sources observed to be variable goes up,
steeping considerably below 0.005 mag.}
\label{fig:tr:variable_precision}
\end{center}
\end{figure}

From a sample of four photometric surveys designed to discover
variability, \citet{Howell2008}  discusses the relationship between variable
fraction and the photometric precision of the survey.  The general
finding, shown schematically in \autoref{fig:tr:variable_precision},
illustrates  the exponential increase in variable fraction of the
observed sources as a function of improved photometric precision of
the survey. This plot is averaged over a number  of observational
biases, survey lengths, and other parameters and should be viewed as
an approximate guideline.  However, its predictions for the number of
variables a survey will find at a specific photometric precision are
in fair agreement with the numerical results of the HIPPARCOS, ASAS,
and OGLE surveys.  For the LSST baseline relative photometric
precision per 15 sec exposure at $r$ magnitude of 17-19  ($1\, \sigma \sim
0.005$ mag), it is probable that several tens of percent of the observed
sources will be variable in some manner.  For $r$ magnitudes of 22 to
23, the precision will be of order 0.01 mag, 
suggesting that $\sim$5\% of the sources in this magnitude range will
be variable.

\subsection{Models of Variable Star Light Curves}
\label{tr:opsim}
{\it K. Simon Krughoff, R. Lynne Jones, Andrew C. Becker, Steve B. Howell}

In order to assess the output light curves that the LSST survey will produce for variable stars, we have produced model outputs based on the
LSST light curve interpolation tool, which convolves template
light-curves of objects with the expected cadence of observations
(\autoref{sec:design:opsim}). 
\subsubsection{The Light Curve Interpolation Tool}
\vskip -0.1cm
In brief, the light curve interpolation tool is intended to facilitate
the simulation of observations of time variable objects with LSST
cadences.  Developed in Python by K. Simon Krughoff at the University
of Washington, 
%
the tool operates in three phases.  Phase one is interpolation of the
input time series with optional error estimates. 
One of two fitting methods may be chosen.  A
univariate splining method may be used for smoothly varying idealized
curves.  If the input time series contains noise or has sharp
discontinuities (as is the case with transiting exoplanets, for
example), the polyfit (\url{http://phoebe.fiz.uni-lj.si/?q=node/103})
method is optimal.  In phase two, the user specifies the data
necessary to turn the time series into an observed light curve.  These
parameters include period, position, and version of the Operations
Simulator output.  The tool turns this information into a MySQL query
and sends it to the database.  Exposure time in MJD and the $5\,\sigma$
limiting magnitude for all pointings in the Operations Simulation
(\autoref{sec:design:opsim}) that overlap the specified
position (defined as the 1.75$^\circ$ radius circular aperture) are
returned to the tool.  Phase three is construction of the observed
``light curve" for the input time series.  The time series is sampled
at the times specified by the operations simulation pointings that overlap the
position.  The photometric errors are calculated based on the returned
$5\,\sigma$ limiting magnitude.  Optionally, the interpolated points are
randomly ``jiggled'' by an amount consistent with the computed error,
assuming normal errors.

\subsubsection{Example Observations of Variable Stars}

For a preliminary evaluation of LSST's ability to identify and characterize
different types of variable stars, we have taken ten well known variable stars of various types, 
generated ``template" light curves for each star, and produced ``observations" of each star as would be
seen by LSST using the light curve interpolation tool described above.

The template light curves were created from AAVSO V band data chosen
to cover a representative two-year interval and assigned to the $g$ band. 
We generated input light curves for the
remaining five filters through a simple scaling of the $g$ band light
curve to brighter or fainter magnitudes based on the known colors of
each type of variable star. While many variables actually change color
as they vary, this is a second order effect to our goals in this preliminary
effort. The set of six light curves ($u,g,r,i,z,y$) were then scaled to
represent LSST observed stars with $g$ magnitudes of 18 to 27, in one
magnitude steps, each with properly scaled uncertainties. The input
light curves were smoothed a bit to reduce their sensitivity to
day/night and seasonal variations and to light curve value uncertainty
for a given night. 

The results show that in some regions there is a good chance that a
variable star could be reliably identified after some time period of
data is collected. We have baselined this time period here to be two
years to allow the reader to get a sense of the sampling efficiency
and temporal nature. Using variability time scale, amplitude, and
color information, gross categorization of variable sources from LSST
observations can begin within the first few months of
operations. Not surprisingly, the most complete variability
information comes from the deep drilling fields
(\autoref{sec:design:cadence}) with their rapid candence.
Identification also depends on the average magnitude of the
source itself; sometimes fields are observed
when the star is below the limiting magnitude in the field, and thus
no measurements were simulated with our tool (although the imaging
pipeline will still report a meaningful upper limit for that
position).  Our preliminary study shows that matching each variable
source light curve (time scale, amplitude, color) to well-observed
templates can provide very good to good classification probabilities,
especially as the database grows over the ten years 
of the survey, although further study must be done to expand the range
of templates tested.  

\begin{figure}[htb]
 \begin{centering}
 \includegraphics*[width=0.45\textwidth]{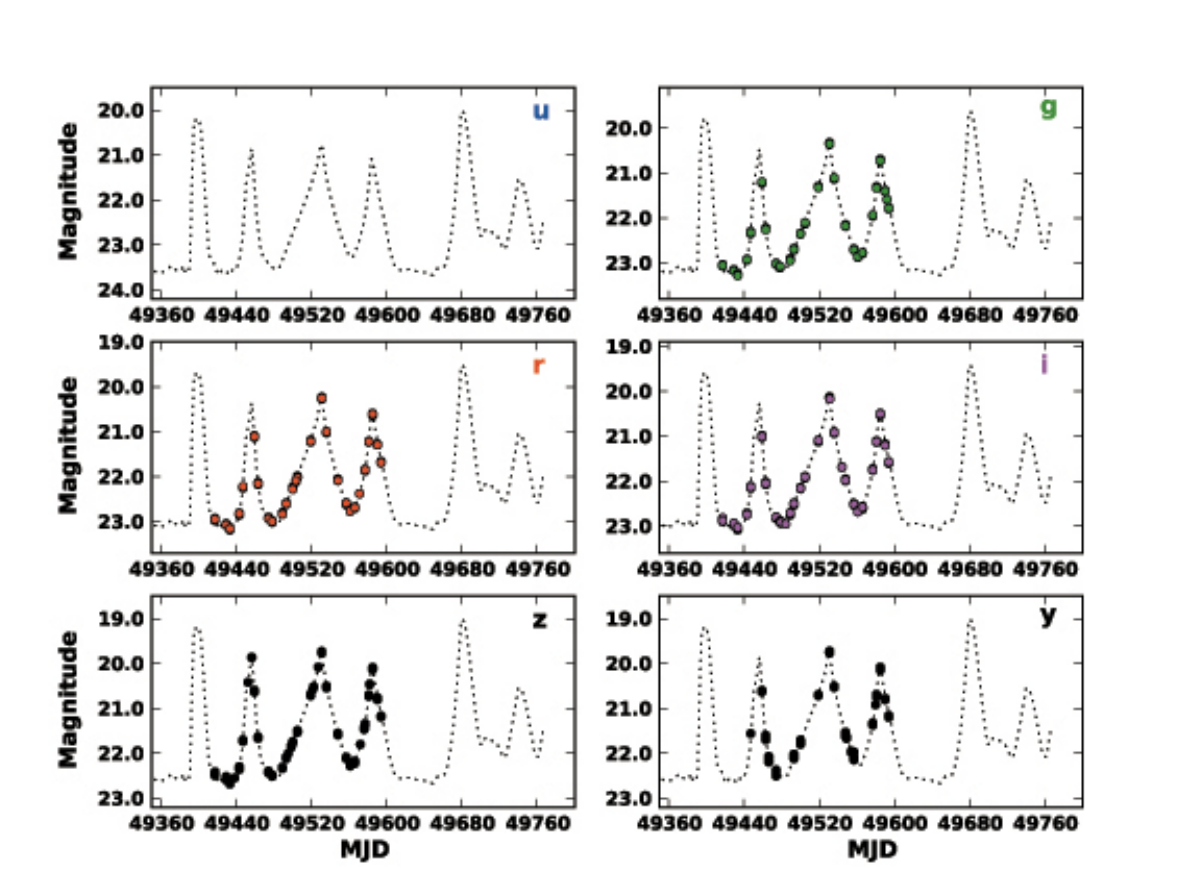}
 \includegraphics*[width=0.45\textwidth]{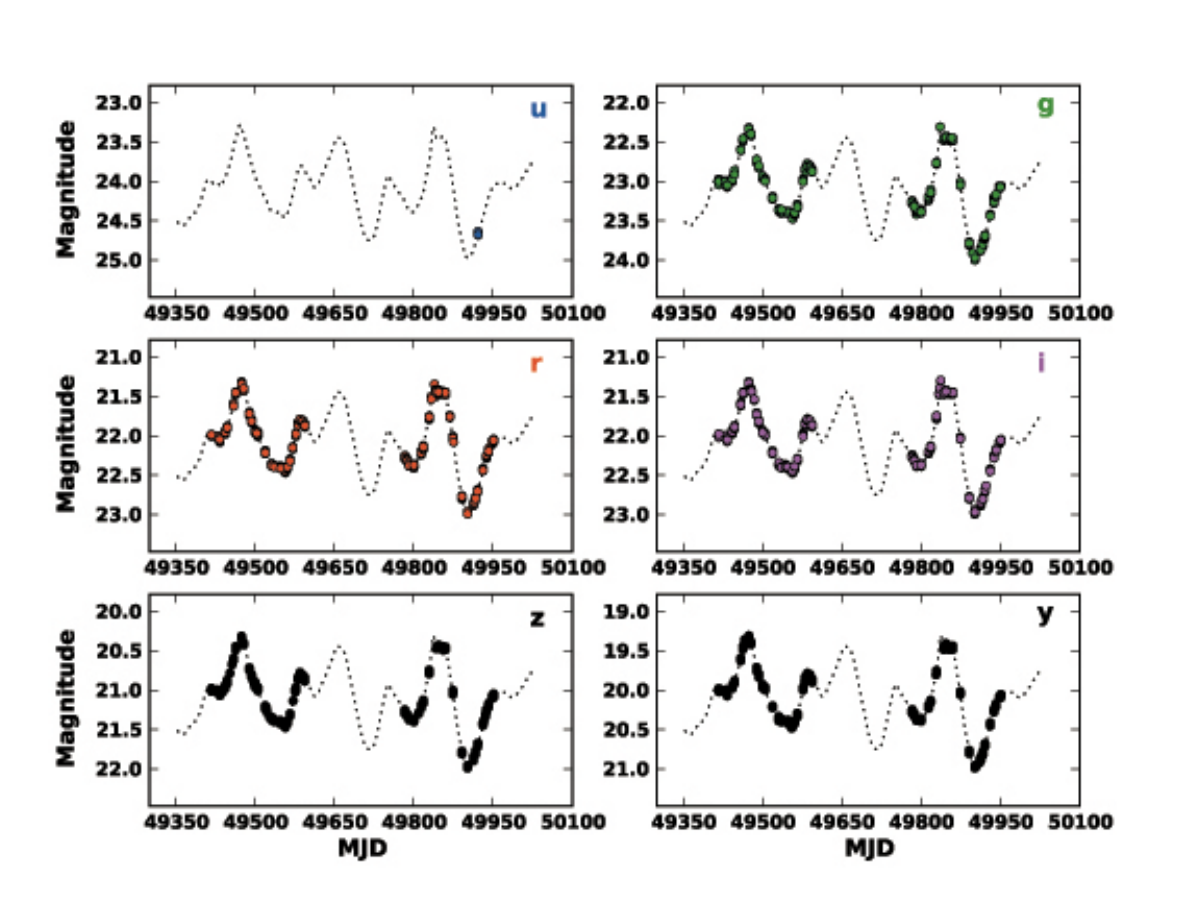}
 \caption{
In these figures we have used the light curve sampling tool to generate
``observations" of two very different variable stars, SS Cyg and Z UMa
(left and right respectively), as they (or stars like them) would be
sampled by LSST for approximately one and two years (respectively), if
the average magnitude of each variable were $\langle g \rangle =22$ or
$ \langle g \rangle=23$
(respectively). The stars were assumed to be placed in the deep
drilling fields.  The dashed line indicates the input variability
light-curve, while the filled circles illustrate the ``observations"
LSST would make in each filter, $u$, $g$, $r$, $i$, $z$, and
$y$. Estimated errors on each datapoint are shown, but are smaller
than the circles. These figures show that template light curve fitting
for LSST should be able to distinguish between variable stars of
different types, as long as those templates are known. These
observations come from what is potentially the best case scenario for
variable stars in LSST's observing cadence -- the deep drilling (or
``supernova") fields.  
}
\label{tr:sscygzuma1}
 \end{centering}
\end{figure}

We show two examples of reasonably well observed variables in
\autoref{tr:sscygzuma1}. The input light
curve templates for these variables are AAVSO observed light curves for
the RV Tau star, Z UMa, and the cataclysmic variable, SS Cyg. These
templates were assembled as described above.  Z~UMa changes its
brightness due to pulsations where a fundamental and fist overtone
period tend to operate simultaneously. SS~Cyg has small amplitude
variability while the larger (2 magnitude) brightenings are due to
semi-regular dwarf nova outbursts. These figures show the
``observations" that LSST would make for each star; with a knowledge of
the typical template light curve, the difference between these variable
stars is measurable.

\subsection{A Study of RR Lyrae Period Recovery}
{\it Hakeem Oluseyi, Julius Allison, Andrew C. Becker, Christopher S. Culliton, Muhammad Furqan, Keri Hoadley}

Here we 
explore
LSST's light curve recovery capability for RR Lyrae stars as a function of stellar
distance and LSST observing cadence. 

Templates for input to the light curve tool (see
\autoref{tr:opsim}) were obtained from \citet{marconi06}. 
The non-linear, non-local time-dependent convective RR Lyrae stellar models used for
this study span a range of metallicity, helium content, stellar mass,
and luminosity for both fundamental and first overtone pulsators. 

The RR Lyrae light curves were tested against a set of many locations
on the sky, distributed between universal cadence overlap regions
(which receive roughly twice the number of observations as the bulk of
the sky) and deep drilling fields. 

The $ugriz$ light-curves of the \citet{marconi06} RR Lyrae model 
were placed in each observation field and
sampled with the LSST simulation tool, which returned realistic
limiting magnitudes and photometric scatter based on historic seeing
and weather data at the LSST site on Cerro Pach\'on, Chile. The LSST
$y$-band data were simulated by using Marconi's $z$-band data. The
$g$-band stellar magnitudes $\langle m_{g} \rangle$ ranged from
$17^{th}$ to $26^{th}$ with $\Delta\langle m_{g} \rangle$ = 0.5 mag,
for survey lengths of 1, 2, 5 and 10 years.

The period of the unequally spaced time-sampled and noised periodic
data was fit using periodograms and least squares estimation methods
\citep{reiman94}. The simulated data was then phased and fit, via a
$\chi^{2}$ minimization, to a Fourier series of the form: 

\begin{equation}
	m_{i}\left(t\right) = \langle m_{i} \rangle~+~ \sum_{k=1}^{5} A_{k} \textup{cos} \left[2\pi k f \left(t - t_{0} \right) + \phi_{k} \right],
\end{equation}

\noindent where $\langle m_{i} \rangle$ represents the mean stellar
magnitude in filter $i$, $A_{k}$ is the amplitude of the $k$-component
of the Fourier series, $f$\,=\,1/$P$ is the frequency (where $P$ is
period of the magnitude variation), and $\phi_{k}$ is the phase of the
$k$-component at $t - t_{0}$. Only the first five Fourier terms were
included in the series, consistent with typical fits described in the
literature. 

\begin{figure}[htb]
  \begin{centering}
  \includegraphics*[width=0.45\textwidth]{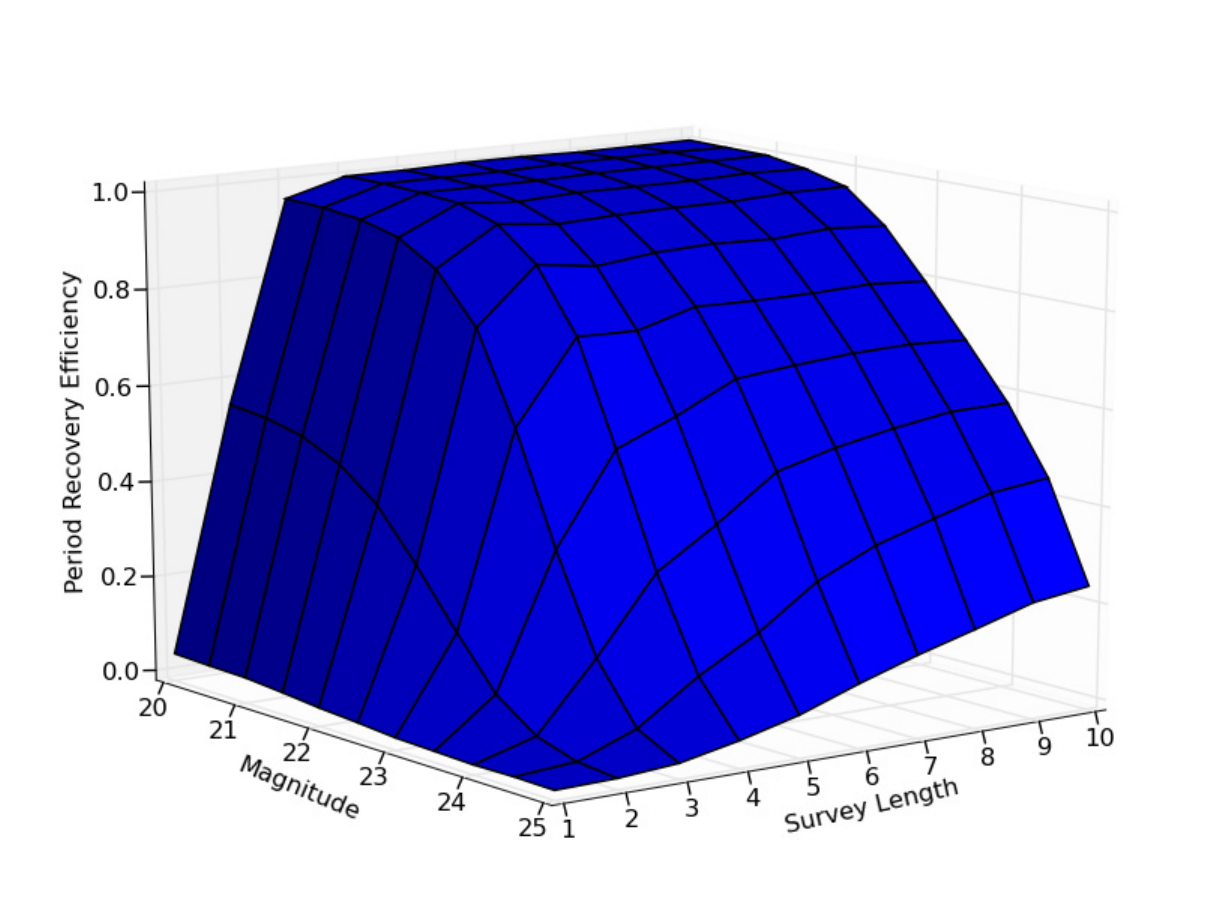}
  \includegraphics*[width=0.45\textwidth]{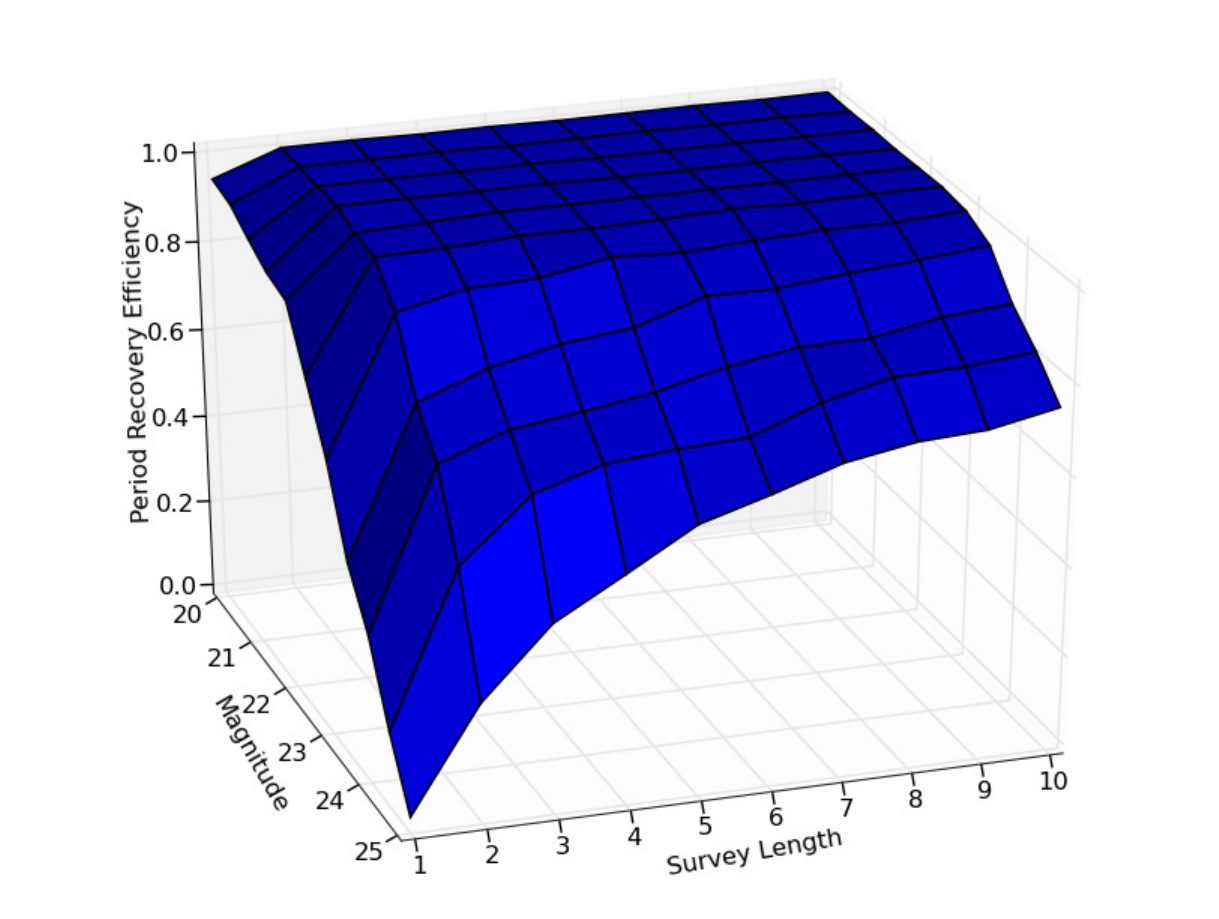}
  \caption{Percentage of the time that the $g$-band light curves of RR
  Lyrae stars placed in the overlap cadence overlap regions (left) or
  deep drilling fields (right) were
  successfully recovered, as a function of $\langle m_{g} \rangle$ and
  survey length. A successful recovery was defined as  determination
  of the light curve's period to within $0.1\%$.} 
  \label{tr:gOverlap}
  \end{centering}
\end{figure}

The calculated period and Fourier parameters were compared with the
input values. A period was considered successfully determined if it
was within 0.1\% of the input value.  
%
\autoref{tr:gOverlap} shows LSST's ability
to successfully recover light curves as a function of stellar magnitude
and survey length, using only $g$ band data. The ability to
successfully recover the pulsational periods and light curve shapes
depended on magnitude, field, and filter. Two years of data were
sufficient to recover $\langle90\%$ of the periods for RR Lyraes
brighter than $g=24$ in the deep drilling fields, while closer to six
years of data were required for the overlap fields covered by the
universal cadence.

\section{Pulsating Variable Stars}

Pulsating stars make up the vast majority of periodic variables on the H-R diagram. The period/luminosity (PL) relations of RR Lyrae, Cepheids and Miras make them useful for calibrating the cosmic distance ladder and tracing Galactic structure. The pulsations of these stars can also shed light on the fundamental physics of stellar atmospheres, e.g., by studying how metallicity variations affect the period/luminosity relationship. Because of their far-reaching utility as a population, these pulsators are discussed in a number of other sections in this book (see, for example, \autoref{sec:sp:rrlyrae}). 

We focus this section
on pulsators {\em not} extensively covered elsewhere: asymptotic giant
branch stars (AGB stars, also discussed briefly in
\autoref{chp:stellarpops}) and pulsating white dwarfs.

\subsection{AGB Stars}
{\it Stephen T. Ridgway, Kem H. Cook, {\v Z}eljko Ivezi{\'c}}

The Asymptotic Giant Branch (AGB) phase of stellar evolution occurs when
core helium is exhausted but the star is not massive enough to ignite
its carbon/oxygen core, so there is helium and hydrogen shell
burning. The helium shell burning is exquisitely temperature sensitive
and thus unstable, resulting in shell ``flashes,'' which can dredge up
carbon.  This mixes CNO products to the surface and creates carbon
stars.  Stars on the AGB are bright, unstable, and
produce prodigious mass loss, returning a large part (up to 50-70\%)
of their mass, including nucleosynthetic products, to the interstellar
medium.  While all low and intermediate mass stars are believed to
pass through the AGB stage, they are rare owing to the brief AGB
lifetime.  Thanks to their ubiquity and their high luminosity, they
are a guide to the evolution of stellar populations.  Accounting for
their numbers and colors is important to modeling color evolution of
galaxies \citep{Maraston2009}.  There are many targets for study.
\citet{Fraser2005} catalog 22,000 AGB variable stars in the Large
Magellanic Cloud.  \citet{Jackson2002} estimate an AGB population of
200,000 for the Milky Way.  LSST will obtain good photometric time
series for AGB stars throughout the Galaxy and the Local Group.
Near-infrared photometry for these stars already exists in the 2MASS
survey.  AGB stars will be easy to identify from a combination of O/IR
colors and variability patterns.  The LSST colors will show an
extremely red star, reddened by a circumstellar shell, with long
period variability.  The infrared colors will show emission from the
photosphere and from the shell. The LSST sampling cadence is well
matched to these slowly varying stars.

The Mira stars are AGB stars in the fundamental mode of pulsation.
For the Miras, distances can be obtained to about 10\% from a PL
relation.  With this information for a large, unbiased set of objects,
it is possible to investigate questions of Galactic structure in the
Milky Way \citep[][see \autoref{chp:mw}]{Feast2000} and in other galaxies \citep{Girardi2007}.
For example, Mira periods select for main sequence mass, and hence
age, and thus periods greater or less than $\sim 300$ days can be used to
distinguish membership in the thick or thin disk \citep{Jura1994}.  By
qualifying this technique in the Local Group, a powerful tool will be
available for population studies far beyond the Local Goup when
future larger apertures with higher spatial resolution are available.

A deep catalog will show throughout the Galaxy where AGB stars are
actively returning nucleosynthesis products to the interstellar
medium.  From AGB distributions, one can infer past rates of star
formation and the current production rates of planetary nebulae and
stellar remnants.  Complementary mid-infrared measurements, such as from
WISE \citep{Mainzer2005}, will stringently constrain the actual mass
loss rates.

Mira stars in the solar neighborhood and the Magellanic Clouds are consistent
with a universal PL relationship \citep{Whitelock2008}, and have been used to
extend Mira-calibrated distances beyond the local group to Cen A, with
$\pm 0.11$
uncertainty in distance modulus \citep{Rejkuba2004}. Parallaxes from
the coming Gaia mission, augmented with LSST astrometry for the most extincted
objects \citep{SahaMonet2005}, can be used to strengthen the PL relation, offering
the opportunity to further refine the usefulness of Miras as a distance
indicator. For a single visit $5\,\sigma$ magnitude of $y=22$, LSST will detect Mira
stars at minimum brightness to a distance of 2.5 Mpc.  With typical
periods of several hundred days, and a $y$ band amplitude $\sim$ 2 magnitudes
\citep{Alvarez1997}, Mira stars are easily recognized from their light
curves, and with 100-200 measurements over a 10-year program, mean
periods will be measured to $\sim$ 1\%. 

AGB stars occasionally show period, amplitude, or mode
shifts. Some of these simply show the complexity of pulsation with multiple
resonances and mixed modes, but some must be associated with changes in the
internal structure involving mixing, shell flashes, or relaxation. A very large
data set will reveal or strongly bound the frequency and character of such
events \citep{Templeton2005}. Mira light curve shapes may in some cases reveal
the action of nucleosynthesis at the base of the convective envelope \citep{Feast2008}.

\subsection{Pulsating White Dwarfs} \label{anjum:pulsWDs}
{\it Anjum S. Mukadam}

Non-interacting white dwarf pulsators are found in distinct instability strips along
the cooling track. Hydrogen atmosphere white dwarf pulsators (DAVs or ZZ~Ceti stars) are observed
to pulsate between 11,000 K and 12,000 K.
Helium atmosphere white dwarf variables (DBVs) pulsate around 25,000 K, while
hot white dwarf pulsators are observed in the broad range of 70,000 K to 140,000 K. All
pulsating white dwarfs exhibit nonradial gravity-mode pulsations. Pulsations probe
up to the inner 99\% of the mass of white dwarf models;  pulsating white dwarfs provide us
with a unique opportunity to probe the stellar interior through seismology.
Each pulsation frequency is an independent constraint
on the structure of the star. A unique model fit to the observed periods of the variable
white dwarf can reveal information about the stellar mass, core
composition, age, rotation rate, magnetic field strength, and
distance. 

The observed pulsation periods of the DAVs and DBVs lie in the range
of about 50-1,400 s with amplitudes in the range of 0.1\% to 10\%
(0.001 to 0.1 mag). Hot white dwarf pulsators show pulsation periods
in the range of a few hundred to a few thousand seconds.  High
amplitude white dwarf pulsators will exhibit a higher photometric
scatter in LSST photometry than non-pulsating
white dwarfs.  Detection probability increases with the number of
measurements irrespective of cadence. White dwarf pulsators of all
amplitudes can be discovered by selecting suitable candidates from a
$u-g$ vs.\ $g-r$ diagram ($0.3\leq u-g \leq 0.6$; $-0.26 \leq g-r \leq
-0.16$; see \autoref{Fig:uggr})
for follow-up photometry with an expected success rate of $\sim$ 25\%
\citep{Mukadamet04}.  Sixty white dwarf pulsators were known in 2002;
discoveries from the SDSS have increased to more than 150. Follow-up
photometry of suitable candidates 
from LSST should increase the known population of white dwarf
pulsators brighter than 20th mag to well over a thousand.

\subsubsection{What Can We Learn from White Dwarf Pulsators?}
The core composition
of a white dwarf is effectively determined by nuclear reaction rates
in the red giant stage. Therefore,
pulsating white dwarfs allow us the opportunity
to study nuclear reaction rates $ ^{12}C(\alpha,\gamma) ^{16}O$
in red giant cores \citep{Metcalfeet01}.
White dwarf models with \teff $\geq $25,000\,K show plasmon neutrinos as a dominant
form of energy loss. Measuring the cooling rates of these stars
can serve as a strong test of electroweak theory \citep{Wingetet04}.
\citet{Montgomery05} fit the observed non-sinusoidal light curves of
large amplitude pulsating white dwarfs to study convection, a fundamental pursuit
widely applicable in several domains of physics and astronomy.

Pulsating low mass ($\log g \leq 7.6$) white dwarfs
are expected to be helium core white dwarfs. Their pulsations
should allow us to probe their currently unknown equation of state with tremendous
implications for fundamental physics. 
\citet{Metcalfeet04} present strong seismological evidence
that the massive ($\log g \geq 8.5$)
cool white dwarf pulsator BPM\,37093 is 90\% crystallized; this directly tests the
theory of crystallization in stellar plasma \citep{Wingetet97}.
Such a study also has implications for
models of neutron stars and pulsars, which are thought to have crystalline crusts.

Measuring the cooling rates of pulsators helps in calibrating the white dwarf cooling curves,
which reduces the uncertainty in using cool white dwarfs at \teff $\leq 4,500$~K as Galactic
chronometers. 
We can also use the cooling rates of \zzc\ pulsators to study exotic particles such as axions \citep{Isernet92,Bischoff-Kimet08}.
Should stable white dwarf pulsators
have an orbiting planet around them, their
reflex motion around the center of mass of the system would
provide a means of detecting the planet.
\citet{Wingetet03} describe the sensitivity of this technique
and \citet{Mullallyet08} find that GD\,66
may harbor a 2\,\mjup\ planet in a 4.5-yr orbit. 
We can use these flickering candles to measure distances that are typically
more accurate than what we determine
from measured parallax (e.g., \citealt{Bradley01}).

\begin{figure}
\begin{center}
\includegraphics[width=0.6\textwidth]{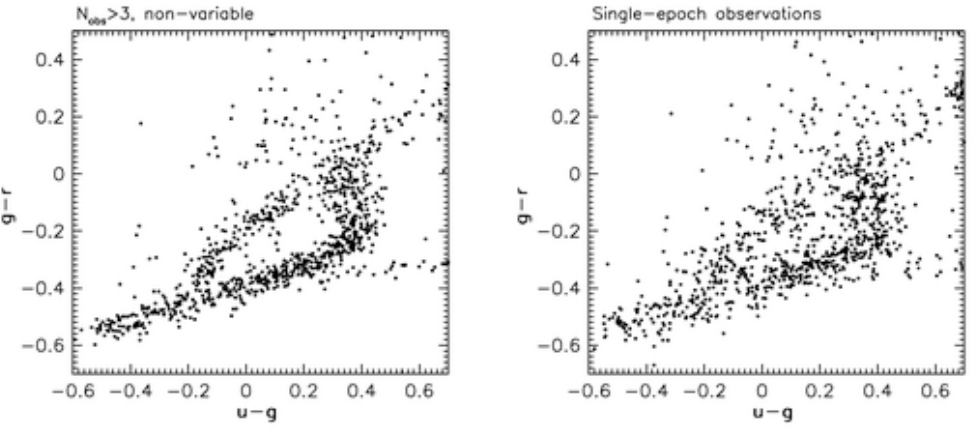}
  \caption{\footnotesize SDSS color--color diagram for objects near
    the white dwarf cooling sequence.  The {\it right} panel shows the
    colors for all sources seen to be non--variable over many epochs,
    but only shows their photometric measurements from one epoch.  The
    {\it left} panel shows the mean colors for these objects over all
    epochs, and resolves cooling sequences much less apparent in the
    single epoch photometry.  Adapted from Figure 24 of
    \citet{Ivezic2007}. A simulated color-color diagram of these two
    white dwarf sequences as observed in LSST is shown in
    \autoref{Fig:uggr}.}
\label{fig:wd}
\end{center}
\end{figure}

An illustrative experiment was carried out by
\citet{Ivezic2007}, using data from SDSS repeat observations.
\autoref{fig:wd} shows the colors of non--variable ($\sigma_{g} <
0.05; \sigma_{r} < 0.05$) objects near the white dwarf cooling
sequences.  The rightmost panel shows single--epoch colors taken from
SDSS DR5.  The left panel shows the averaged colors of the objects
over $\sim 10$ epochs.  With the higher S/N photometry, multiple
sequences are apparent, two of which correspond to 
the cooling curves of H and He white dwarfs
\citep{1995ApJ...443..764B}. These are fundamental tests of degenerate matter that cannot be
replicated in the lab.  While LSST can identify the variability, followup observations in a
blue filter will be needed to pin down the pulsation periods, and
spectra will be needed to determine accurate temperatures.

\section{Interacting Binaries}
{\it Paula Szkody, Scott F. Anderson, Julie Lutz}

As the majority of stars are binaries, it is astrophysically important
to understand the implications of binaries for stellar
evolution. Binaries that form close enough that they will interact at
sometime in their evolutionary lifetime are particularly interesting
for LSST, as the interaction alters the evolution process in many ways
that can result in spectacular transient and variable phenomena. In
addition, the mass transfer and angular momentum losses during the
interaction time have dramatic consequences on the evolution of the
individual stars.  For common low-mass stars, this evolution involves
starting as two normal non-interacting main sequence stars, followed
by the giant stage of the more massive star, which causes a common
envelope resulting in angular momentum loss which brings the stars
much closer together \citep{2005MNRAS.356..753N}.  This stage is followed
by the subsequent evolution of a normal star with a remnant white
dwarf until continued angular momentum losses bring the stars close
enough so the companion fills its Roche lobe and starts mass transfer
\citep{1981A&A...100L...7V}. Variations on this type of scenario can
result in X-ray binaries and symbiotic stars (for more massive stars),
cataclysmic variables (including novae, dwarf novae and novalikes) and
ultimately, the AM CVn systems \citep{1996MNRAS.280.1035T}.

The variability that is known so far for these types of systems is
summarized in \autoref{tab:tr:cv_types} below.  The challenge for LSST
lies in detecting the variability and determining that the object is
one of these types. As discussed in \autoref{tr:opsim},
templates of various light curves have been run 
through the simulators to determine detectability. Determining the
type of object requires both color and variability information. Close
binaries that have 
not yet begun mass transfer will be easy to pick out because the colors of
WD+MS stars are well known from SDSS
\citep{2004ApJ...615L.141S,2006AAS...20916218S,2007AJ....134..741S}.
Other objects such as novae and dwarf novae will be 
selected on the basis of their variability. Objects with disks have a
wide range in color based on the characteristics of the disks but
generally are blue in color because they contain hot sources. This means
that the selection will improve as LSST completes the full color
information of the survey area and as templates of light curves of
various objects are available for match up. After two years of survey operations, we
anticipate both of these will be in place. While individual science
goals (discussed below) vary for each type of object, a common
goal for all close binaries for LSST lies in determining the correct
space density of objects. LSST will reach to much fainter magnitudes
(hence greater distances from Earth),
and be more complete in reaching binaries with lower mass companions
and with lower mass transfer rates than previous surveys.
 The derivation of the correct numbers
will be matched with population and evolution models to determine the
correct scenario for close binary evolution. In all cases, followup of
candidates from the ground will enhance the science output. This
includes determinations of orbital period, mass, and distance from
spectroscopy and time-resolved photometry. Much of this work will
involve the amateur community of observers in conjunction with
professional astronomers.

\subsection{Cataclysmic Variables}

By definition, cataclysmic variables (CVs) are close binaries with
mass transfer from a late type main sequence star to a primary white
dwarf. Depending on the magnetic field strength of the white dwarf,
the mass transfer will result in an accretion disk around the white
dwarf (fields under a MG), an accretion ring with inner area funneled
to the magnetic poles for fields of 1-10 MG (intermediate polars), or
direct transfer to the magnetic poles for fields over 10 MG
(polars). The orbital periods range from 67 min to 2 days, with the
majority of systems having periods under 2 hrs. A comprehensive review
of all CVs can be found in \citet{1995CAS....28.....W}. 

For systems with disks, the mass transfer can lead to a  thermonuclear
runaway on the white dwarf (nova). When the H-rich accreted matter
reaches about $10^{-5}$ M$_\odot$ and 1 km depth, the pressure becomes
large enough to start nuclear fusion, which becomes a runaway due to
the electron degeneracy. The rapid nuclear energy release causes the large
rise in luminosity (7-15 mags) and the ejection of the envelope. These novae
outbursts repeat on timescales of 10 yrs (recurrent
novae) to  10$^4$ yrs. Between nova outbursts, the systems exist as 
dwarf novae or
novalikes. The dwarf nova outbursts are due to a disk instability and
can recur on timescales of days
to decades, with a particular timescale and amplitude associated with
the mass transfer rate. At high rates, the disk is stable with no
outbursts and the systems are termed novalikes. At the lowest rates,
the buildup to an outburst takes decades and the resulting outburst is
very large (8 mags). For unknown reasons, the mass transfer can also
stop completely for months to years, causing a drop in brightness by
4-5 mags (these times are termed low states).  Since polars have no
disks, they do not outburst and only show high and low states of
accretion. 

\begin{table}[ht]
\vskip-12pt

\caption[Summary of Close Binary Timescales and Amplitudes]  {Summary of Close
  Binary Timescales and Amplitudes} 
\label{tab:tr:cv_types}
\begin{tabular*}{\textwidth}{@{\extracolsep{\fill}}lll}
\hline Variability & Typical Timescale &  Amplitude (mag)\cr \hline
Flickering & sec -- min & tenths\cr 
WD pulsation & 4--10 min & 0.01--0.1\cr
AM CVn orbital period & 10--65 min & 0.1--1\cr
WD spin (intermediate polars) & 20--60 min & 0.02--0.4\cr 
CV orbital period & 10 min--10hrs & 0.1--4\cr
Accretion Disks & 2--12 hrs & 0.4\cr 
AM CVn Outbursts & 1--5 days & 2--5\cr
Dwarf novae Outbursts & 4 days--30 yrs & 2--8\cr
Symbiotic Outbursts & weeks--months & 1--3\cr 
Symbiotic orbital period & months--yrs & 0.1--2 \cr
Novalike High-Low states & days--years & 2--5\cr
Recurrent Novae & 10--20 yrs & 6--11\cr
Novae & 1000--10,000 yrs & 7--15 \cr
\hline
\end{tabular*}
\end{table}

The colors of CVs are clues to their accretion rates and their
types. High mass transfer rate CVs are very blue ($u-g<0.0$) because their
light is dominated by the accretion disk or column. Polars can be very red ($i-z\sim1$),
when their emission is primarily from cyclotron harmonics. Very low mass
transfer systems are both blue and red, because the accretion disk or column becomes a minor 
contributor to the optical light and 
the underlying star can be seen.  Because the color range is so
large,  the optimum search for CVs must involve both  variability and
color. 

In some cases, both stars have evolved to white dwarfs and the binary
periods can be even shorter (10 min) while the outbursts will be
hotter due to the presence of a helium rather than a hydrogen accretion disk
(AM CVns, see \autoref{sec:tr:amcvn}). The orbital period can be revealed from
photometry when the inclination is high enough to cause an eclipse,
when there is a prominent hot spot on the disk where the mass transfer
stream intersects the disk, or when the accretion area in a polar
passes by the line of sight. In the latter case, the strong magnetic
field locks the spin of the white dwarf to the orbit so all variation
is at the orbital period. For intermediate polars, the spin of the white dwarf is seen
as a periodic 10-20 min variation in the blue. If the white dwarf is
in a specific temperature range (11,000--16,000K), it may have
non-radial pulsations on the order of minutes. 

If the mass transfer continues onto a white dwarf near the
Chandrasekhar limit, a Type I SN may result
\citep{2004Natur.431.1069R}; see the discussion in \autoref{sec:TheGap}.

Below we summarize two major science drivers for CVs in LSST:

\subsubsection{Determining the Space Density of CVs}

\begin{figure}[hbt]
\centerline{
\includegraphics[width=0.7\textwidth]{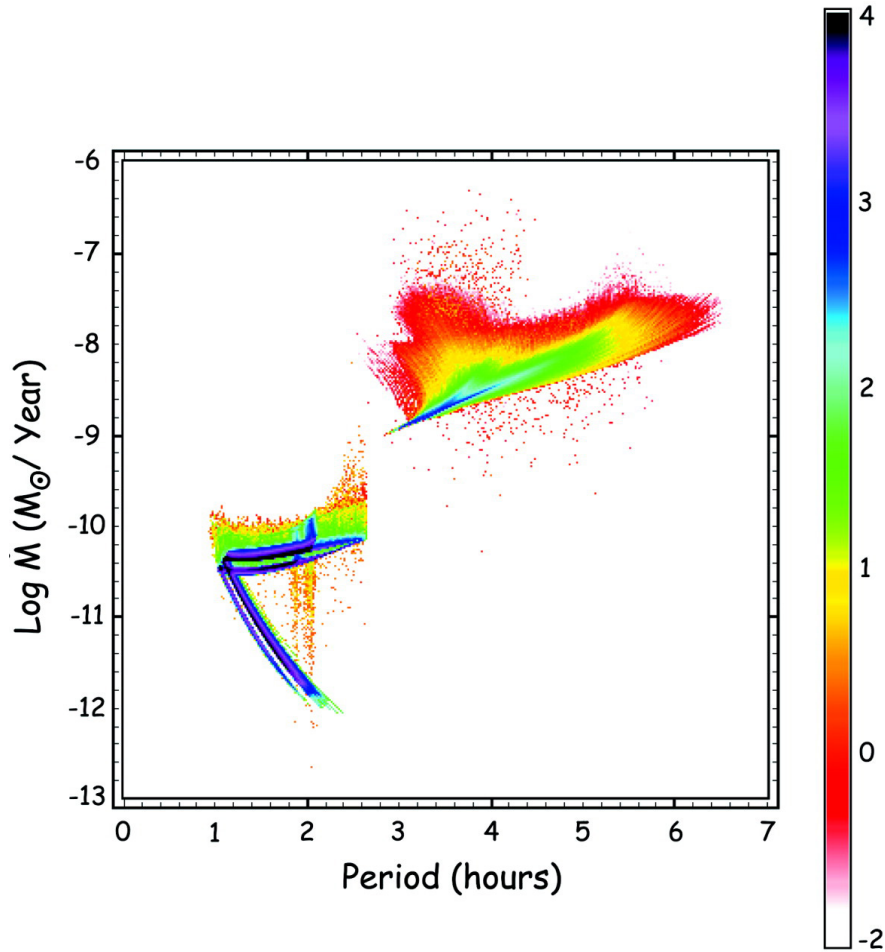}}
\caption{Predicted present-day population of cataclysmic variables in
the Milky Way \citep[from][]{2001ApJ...550..897H}. The models are
presented in the mass transfer ($\dot M$) - orbital period plane and
shown as a density distribution. The scale on the right side gives the
number of CVs per colored dot as a log scale and the
majority of present-day systems are expected to lie at very short
orbital period (less than two hours) and have very little mass transfer, and thus will be intrinsically very faint. The LSST survey will be the first
survey to allow the majority of the systems modeled here to be
discovered.} 
\label{fig:tr:cv_periods}
\end{figure}

Identification of cataclysmic variables with LSST is important
primarily for understanding the long term evolution and ages of close
binary systems. Population models predict almost all close binaries
should have evolved to the period minimum in the lifetime of the
Galaxy (see \autoref{fig:tr:cv_periods} below from \citealt{2001ApJ...550..897H}).
Past surveys have been skewed by selection effects, which find the
brightest and most active systems with outbursts. The true numbers of
systems at various orbital periods are needed to sort out the
evolution for different categories of systems, which is dependent on
the angular momentum losses. The CVs found in SDSS have shown
\citep{Gansicke09} that the evolution model predictions of \citet{2001ApJ...550..897H}
are correct in terms of predicting percentages of systems at various
orbital periods down to 21st mag and in predicting an orbital period
spike near 81 min (critical for confirming the angular momentum
prescription and the white dwarf history).  This model predicts the
majority of CVs (70\%) should be past the period minimum and have
magnitudes of 22--24 \citep{2004ApJ...604..817P}. LSST will be the
first survey to test this endpoint of evolution.  While LSST will
identify the faintest CVs, followup observations (either continuous
photometry to detect orbital variations due to hot spots in the disk
or on the magnetic white dwarf or spectroscopy with large telescopes
to obtain radial velocity curves) will be needed to find the correct
periods.

While no disk CVs are known with periods between two and three hours,
the period distribution of magnetic systems does not show this gap.
This, and the unknown effects of a strong magnetic field on angular
momentum losses, suggests that the evolution of the two types should
be different, which might manifest itself as different number
densities of the two populations. A related question is how much
time systems spend in low states as they evolve. Also, the numbers of
systems that contain pulsating white dwarfs is important in order to
determine the instability zones of accreting pulsating white dwarf vs.
non-interacting ones \citep{2007ApJ...658.1188S}.

SDSS was able to probe deeper than previous surveys and determine a
space density of CVs of 0.03 deg$^{-2}$ down to $r \sim 21$
\citep{2003AJ....126.1499S} and 
$M_{v}=11.6\pm0.7$ at the period minimum \citep{Gansicke09}.  However, SDSS observed
primarily out of the Galactic plane, where the space density is lower.
Estimates of the space density of these objects range from 10$^{-4}$ to 10$^{-6}$ pc$^{-3}$, with a million
objects expected in our Galaxy.  LSST will go almost four mag deeper and
closer to the Galactic plane, the coverage will extend to larger
distances and lower mass transfer rates (both of which contribute to
fainter observed magnitudes). 

\subsubsection{Novae as Probes of Mass, Composition, and Distance}

Novae are the intrinsically brightest CVs during outburst ($M_{v}= -6$ to $-9$)
and can thus
serve as a probe of conditions in our own Galaxy as well as other
galaxies. While novae generally have fast rise times
of a few days, the decline time and shape give important information as
to the mass, distance, and composition.  Due to the mass-radius relation
of white dwarfs, there is a tight correlation of a nova peak luminosity
and time to decline by 2-3 magnitudes \citep{1981ApJ...243..926S}.
 Slow novae are more
common, have
absolute magnitudes fainter than $-7$, show FeII lines in their spectra,
and are located in the bulge of our Galaxy \citep{1998ApJ...506..818D}.
Fast novae occur on more massive white dwarfs, 
are brighter, show He and N in their spectra, forbidden lines of O, Ne, and
Mg in their ejecta \citep{1986ApJ...306L..49G,1992ApJ...391L..71S}, and are
generally found in the disk. 

Since the two types of novae are found in different
locations and are important in the production of CNO isotopes, the
correct rates are needed to understand Galactic chemical evolution
and star formation history (the latter since the rate is dependent on
binary star formation and evolution).
LSST will
be able to find the fainter novae to greater distance in our Galaxy
and provide  improved estimates of the nova rate and type in the Milky
Way to compare to those found in other galaxies.
Estimates of the nova rate in the Milky Way currently 
are on the order of 35 $\pm$ 10/yr, with rates in other galaxies scaling
as the mass \citep{2002AIPC..637..462S}. While LSST will only observe a few novae
from the Milky Way each year, some of these will be close enough to observe
the precursor star within a few days of the actual outburst, thus providing
new information on the outburst process. The numbers of recurrent novae are thought
to be underestimated by a factor of 100 \citep{2009AAS...21349104S} 
 due to missed
outbursts. This number is especially important to pin down because these could
be Type Ia SN progenitors.

\subsection{ AM CVn Systems}
\label{sec:tr:amcvn}

AM CVn binaries are extremely rare relatives of cataclysmic variables
with  ultra-short orbital periods; the most extreme cases have orbital
periods of  tens of minutes, arguably encompassing {\em the} shortest
orbital periods of any  known class of binaries \citep[see review by][]{2005ASPC..330...27N}. AM CVn systems are so compact
that both binary components must be  degenerate (or least partially
so), likely with mass-transfer driven by  gravitational radiation from
a helium-rich degenerate ($\sim 0.02 M_\odot$) onto a
more standard white dwarf. Their optical spectra are  distinct from
typical CVs: membership in the AM CVn class requires a  near absence
of hydrogen, and helium lines are instead prominent. There are about
20 known at the present time.

The unusual nature of the prototype, AM CVn, was recognized some time
ago \citep{1967AcA....17..255S,1967AcA....17..287P}, but the next four
decades yielded less than a dozen additional discoveries. Though
elusive, AM CVns have emerged as objects of renewed interest for
several reasons: their evolutionary link to and possible insights
about an earlier common envelope phase; as possible SN Ia progenitors
\citep[e.g.,][]{1997ase..work..253L}; and perhaps notably on the LSST
timescale, as one of the most common objects likely detectable by
upcoming gravitational wave experiments. For example, some formation
and evolutionary models suggest that up to $\sim10^4$ AM CVns and
related double-degenerates may be detected/resolved in gravity waves
by LISA \citep{2004MNRAS.349..181N}.

Eight new AM CVn systems have been discovered in the past few years
especially from SDSS \citep{2005MNRAS.361..487R,2005AJ....130.2230A},
including the first eclipsing AM CVn. 
The bulk of known AM CVns occupy a relatively small region of
multicolor space \citep[see Figure 2 in][]{2005AJ....130.2230A}, and
so (similar) LSST-filter imaging should provide a basis for multicolor
selection.  However, there are still plenty of other objects in this
region of color space, such as normal white dwarfs, quasars, and
ordinary CVs, and 
only a small
fraction of randomly selected objects with such colors in SDSS (to
$m\le20.5$) are subsequently verified as AM CVns. 
Multicolor selection alone is not efficient.

The additional time domain information available in LSST imaging will
provide an important additional sieve to find these rare objects, as
the variability of AM CVns will distinguish them from normal white
dwarfs. Groups are currently engaged in variability-selection programs
focused especially on AM CVns, and these results can guide LSST
efforts. Optical variability for AM CVn's is typically $\sim$ 0.1 mag
on orbital timescales of tens of minutes, but $\sim$ 1 mag for the one
known eclipsing case (sharp 1-minute eclipses on a 28 min orbital
period); and, often a few tenths of a mag on longer timescales (weeks,
months, years). Strong variations of up to several magnitudes are seen
in the substantial subset of AM CVn's that show outbursts.

The surface density of AM CVns is highly uncertain, due to survey
completeness.  SDSS discoveries suggest a surface density $\ge 0.001 \rm deg^{-2}$ 
\citep{2005AJ....130.2230A}.  But some AM CVn
population models
\citep{2004MNRAS.349..181N} predict a population  of thousands of AM CVn to
a modest depth of $m<22$, easily accessible to  LSST.

The availability of accurate LSST colors, plus LSST light curve
information, should yield an excellent list of AM CVn candidates with
quantifiable selection. Followup confirmation via detailed light curves
and/or spectroscopy will be required in many cases. Candidate lists
from  LSST should, of course, also be cross-correlated with available
all-sky  X-ray surveys (most AM CVn are X-ray sources), and ultimately
(but more  speculatively) future catalogs of sources of gravitational
radiation. As  example science, the AM CVn orbital period distribution
(usually from the  optical), coupled with mass-transfer rates (often
from X-ray measures),  are key ingredients in testing evolutionary
models for AM CVn and related  double degenerates in the presence of
marked gravitational radiation \citep[e.g.,][]{2005ASPC..330...27N}.

\subsection{Symbiotic Stars}

Symbiotic stars (SS) show variability on a variety of timescales and
magnitude ranges. The classic symbiotic system has an M-type star
(often a giant, which is called the primary in these systems) in a
binary system  with a white dwarf (secondary) that is close, but not
so close that the system exhibits the sort of chaotic phenomena
present in cataclysmic variables. Some of the primaries in SS are Mira
variables with periods of hundreds of days and amplitudes of several
magnitudes, while others are semi-regular variables with smaller
amplitudes. Some don't appear to be variable at all. Not all the
primaries are red stars. There is a group of SS known as the ``yellow
symbiotic stars'' which have F, G, or K stars as the primary.  In
addition, a few systems have K stars instead of white dwarf
secondaries.  SS show two distinct groups  in the near-infrared
(S-type symbiotic stars have declining flux  while D-type show the
signature of warm dust accretion disks, presumably around the dimmer
stars). Thus, the range of colors of SS is large. 

Some SS have dramatic sudden non-periodic outbursts of several
magnitudes. The brighter ones are sometimes called  ``slow novae''
because they brighten by a few magnitudes and remain bright for
months.  Others outburst  and decline in weeks. LSST will contribute
greatly to SS research by determining the number, timescales, and
shapes of the outbursts.  LSST alerts will enable  intensive followup
observations as soon as an outburst is reported. Some SS also show
flickering in the $u$ filter (caused by the accretion disk or hot
spot(s)) on timescales of minutes to hours. LSST will determine if
there are long-term changes or periodicities in these systems.  For
many SS, not enough observations have been made to know for sure
whether or not they are variable. They were identified as SS in a
spectroscopic survey (M-type star plus chromospheric emission lines). This is
especially true in the southern hemisphere, where there are many
in the plane of the Galaxy at declinations $< -25^\circ$. Thus, LSST can refine the definition of the SS class 
(e.g., how many SS are really variable and with what timescales and
magnitudes in various wavelength bands?).

\section{Magnetic Activity: Flares and Stellar Cycles}

\subsection{Flaring in Cool Stars}
\label{sec:tr:mdwarf_flare}
 {\it Eric J. Hilton, Adam Kowalski, Suzanne L. Hawley, Lucianne M. Walkowicz, Andrew A. West}

Because low mass stars comprise nearly 70\% of stars in the Galaxy, their
flares represent a major source of transient variability in time domain surveys
such as LSST (\autoref{sec:NewAstronomy}). These flares are manifestations of
internal magnetic field production and the subsequent emergence of
these fields at the stellar surface -- processes which are poorly
understood even in the Sun, but particularly elude physical
description in late-type M  dwarfs. As stars become fully convective
below $\sim 0.3\,M_{\odot}$ \citep{Chabrier1997}, the nature of the
magnetic dynamo changes, which may alter the relationship between {\em
  quiescent} magnetic activity (persistent chromospheric and coronal
emission in the optical, UV, and X-ray) and the large transient
increases in the continuum and line emission caused by {\em flare}
activity.  The flare rate may also be influenced by the effective
temperature of the star, with lower temperatures inhibiting field
emergence in the quiescent state, but promoting field storage and
eruption of huge flares \citep{Mohanty2002}. Stellar flares also have
interesting implications for astrobiology because the cumulative effect of
high energy irradiation by flares may impact the evolution and
eventual habitability of planets \citep{Lammer2007}. Because
low mass, largely convective M dwarfs are the most numerous of
potential planetary hosts,  it is essential to understand how
frequently and powerfully these stars flare. 

LSST will obtain an unprecedented data set of M dwarf flares over a
range in activity level, mass, and age. Since individual flares will
only be observed once or at most twice by LSST, these sparsely sampled
light curves of M dwarfs require sophisticated interpretation. Using
repeat photometric observations of SDSS Stripe 82 ($\sim 250\,\rm
deg^2$) in combination with
a new model of the Galaxy,  we are currently developing the analysis
tools needed for interpreting the LSST flare data. 
Our new Galactic
model includes the most current mass and luminosity functions of low
mass stars \citep{Covey2008, Bochanski2007a, Bochanski2008,
  West2008}. In this section,  we present preliminary estimates of the flares that LSST will observe using this model Galaxy.  
  
\subsubsection{Activity in Low Mass Stars}

Previous large scale surveys have been instrumental in understanding
activity in low mass stars. Observations of over 38,000 M dwarfs in
the SDSS revealed that the fraction of
``active'' stars (those which have H$\alpha$ emission with equivalent
width $>$ 1\AA) increases dramatically from types M0 to M6, peaking
near spectral type M7-M8. The observed active fraction depends on
distance;  stellar activity declines with age, and stars that are
further out of the plane are likely to be older than nearby stars
\citep{West2008}. Although cool stars are
designated as active or inactive by whether they have H$\alpha$ in
emission or absorption, most so-called ``inactive'' stars actually
possess low to moderate levels of magnetic activity. Therefore, while
most flare stars will belong to the ``active'' population, low to
moderate activity stars may also flare. 

Flares have been observed throughout the M spectral class, as well as
on the less massive L dwarfs
\citep[e.g.,][]{Liebert1999,Schmidt2007}. It has been suggested that
activity in ultra-cool dwarfs is confined primarily to large flares,
with little or no ``quiescent'' emission \citep{Rutledge2000,
Linsky++95, Fleming++00}. In contrast, UV spectra
of active M7-M9 dwarfs have shown that quiescent transition region (C
IV) emission is present at levels comparable to those seen in earlier
active M dwarfs \citep{Hawley2003a}. Activity (in the form of
persistent H$\alpha$ emission) has been detected on a number of nearby
T dwarfs, but to date no T dwarf flares have been observed
\citep{Burgasser2002}. 

Flare light curves \citep[see][for examples and
  discussion]{Hawley1991, Eason1992, Hawley2003, Martin2001}
generally consist of a sudden increase in brightness that is most
extreme in the near UV and blue optical, followed by a long tail as
the star gradually returns to its quiescent state. The largest flares
cause brightness enhancements of $\sim$ 5 magnitudes in {\it u} and
persist for over an hour, although small flares of fractions of a
magnitude and a few minutes' duration are much more common.
The magnitude changes associated with flares are dependent on both the
spectral type of the object and the observation filter. In early M
dwarfs, the quiescent flux in the optical is much higher than in the
ultra-cool M and L dwarfs-- therefore the optical contrast between the
quiescent star and the flare emission is higher in later type stars.
While the flare contrast is greatest in {\it u}, most flares will be
visible (although with a smaller increase in brightness) in {\it g},
{\it r}, and to a lesser extent in {\it i}.

\subsubsection{Preparing for LSST: Results from SDSS Repeat Photometry}

The repeat photometry of SDSS Stripe 82 \citep{Ivezic2007} provides a
useful test set of observations for developing analysis tools for
LSST. Although the sky area (250 deg$^2$) is much smaller, the data
less deep and the cadence less
frequent, the Stripe 82 data are qualitatively similar to those that LSST will
produce.  We here summarize the results of the flare analysis of
\citet{Kowalski++09}. 

Considering stars of
spectral types M0--M6 with $u_{quiet}< 22$ on Stripe 82, SDSS detected
270 flares with a $u$-band magnitude change of at least 0.7. Flares as
large as 
$\Delta u \sim 5$ magnitudes were observed in both early and late
type M stars, but flares of $\Delta u < 2$ magnitude dominate the
sample.  Later type stars show a higher flaring rate. Ninety-two
percent of the stars that have spectra and also
show flares in the SDSS photometry have H$\alpha$ emission
during quiescence.  One of 10,000 total SDSS observations
of M dwarfs show flares, but the flare fraction rises to $\sim$ 30 out of 10,000
observations of {\em active} M dwarfs (i.e., those with emission
lines) show flares.  Clearly the stars 
that have quiescent magnetic activity are more likely to flare.

The observed flare rate is very
strongly dependent on the line of sight through the Galaxy, since this
changes both the number of stars per deg$^2$, and the age and
activity of the stars observed. Given the high flare rate for active
stars, it is not surprising that
the flaring fraction decreases sharply with Galactic $Z$ for all
spectral types: $\sim$ 95\% of the flaring 
observations occur on stars that are within 300 pc of the plane, and
the flare rate ranges from 0 to 8 flares$,\rm hr^{-1}\,deg^{-2}$
depending on Galactic latitude. 

\begin{figure}[hbt]
\centerline{
\includegraphics[width=0.6\textwidth]{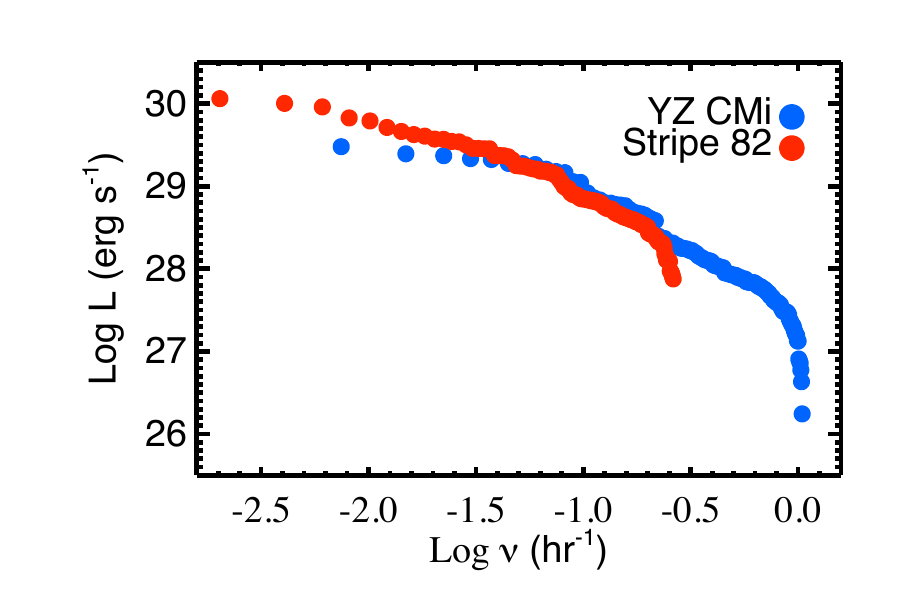}}
\caption{
The flare luminosity distribution of $\sim$ 40 M4-5 stars
      in SDSS Stripe 82 compared to  extensive flare observations of
      the M4.5 flare star YZ CMi (from \citealt{Lacy1976}). We will
      similarly be able to compare sparsely sampled
      photometric observations of millions of M dwarfs in LSST to a
      handful of closely monitored, well-known flare stars in the
      solar neighborhood.}
\label{fig:sdss_lme}
\end{figure}



Analyzing flares at the Stripe 82 cadence (or even that of LSST) is
difficult because any individual flare is only observed once. Without
time resolution, it is impossible to determine when during the flare
the observation occurred, and therefore know the total energy, peak
luminosity, or flare duration. However, we can use the instantaneous
luminosity of the flare to infer its
properties. \autoref{fig:sdss_lme} compares the distribution of flare
luminosities of 
$\sim$ 40 M4-5 dwarfs in Stripe 82 to results from the optical flare rate survey of \citet{Lacy1976} for
the M4.5 dwarf YZ CMi. In the case of the YZ CMi observations, where
the light curve captures the entire flare, we calculate the average
flare luminosity. The close agreement between the flare luminosity
distributions found from the Stripe 82 sparsely sampled light curves
and the very well-sampled light curves of YZ CMi shows that we are
able both to confidently identify flares in the data and to determine
their average luminosity.

\subsubsection{Flares in LSST}

In addition to being astrophysically interesting, stellar flares also
represent a source of confusion for ``true'' transient events, as
stars that are below the detection limit in quiescence may be visible
during flare events (\autoref{sec:NewAstronomy}). The flare energy
cut-off, spectral types, stellar 
ages, and activity status, along with the line of sight, all
contribute to the flaring rate determined for an observed sample of M
dwarfs.  

In order to understand the rate and energy distribution of flares LSST
will observe, we are developing a new model of M dwarf flares in the
Galaxy. Existing observations will be combined with a new flare monitoring
effort (E. Hilton, PhD thesis, in preparation) to empirically determine the
rate and energy distribution of M dwarf flares. We can then use this
flare frequency distribution to construct light curves for each M
dwarf in our model Galaxy. Every star in the model has a position,
distance, and quiescent magnitude in each SDSS/LSST filter, and a light
curve populated with flares, allowing us to ``observe'' stars using
simulated LSST cadences. 

\begin{figure}[!htbp]
    \centering
 \includegraphics*[width=0.6\textwidth]{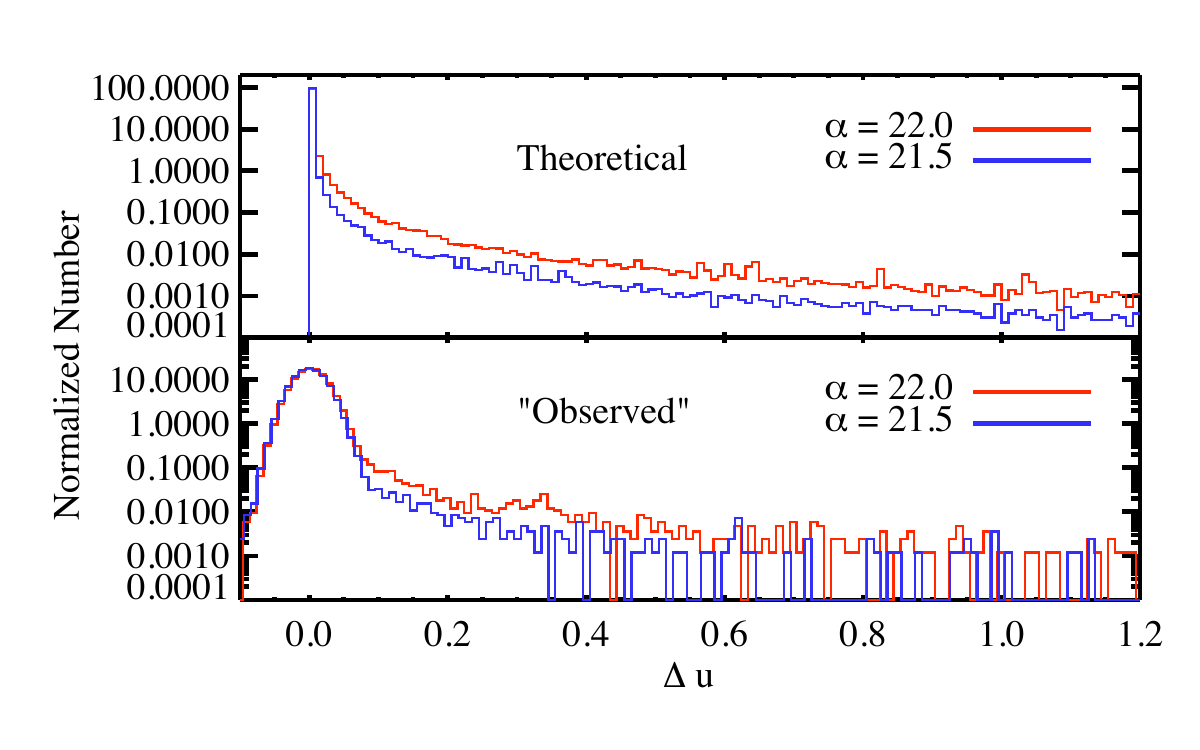}
    \caption{{\it Top panel}: Model flare frequency distributions. {\it Bottom
      panel}: These distributions ``observed'' for 300 identical stars at
      $\it{u}$ = 20 using the operations simulation and simulated seeing effects.}
\label{fig:lsst_observed_compare}
\end{figure}

In \autoref{fig:lsst_observed_compare}, we demonstrate how our
model may be used to interpret flare observations in LSST. The top
panel shows our theoretical prediction for how often a star is
observed at a given increase in brightness ($\Delta${\it u}). Since
stars are in quiescence much of the time, this distribution peaks at
$\Delta${\it u} = 0, and because large flares are much less frequent than
smaller flares, the distribution has a long tail towards larger
$\Delta${\it u}. The red and blue lines represent two flare frequency
distributions, given by  $ \log \,\nu = \alpha + \beta \log \,E_{u} $
, where $\nu$ is the number of flares per hour, $E_{u}$ is the total
flare energy in ergs/sec, $\beta = - 1$,  and $\alpha$ varies from
21.5 to 22.0.

The bottom panel shows the result of sampling this theoretical
distribution using the LSST operations simulation cadences
(\autoref{sec:design:opsim}) for 300 identical
$u = 20$ M dwarfs at eight different telescope pointings. Photon sky
noise  gives the broad peak around
$\Delta u = 0$ magnitudes. For flares  with $\Delta${\it{u}}
$\gtrsim 0.07$ magnitudes, the two distributions are quite distinct. 

\begin{figure}[!htbp]
    \centering
 \includegraphics*[width=0.45\textwidth]{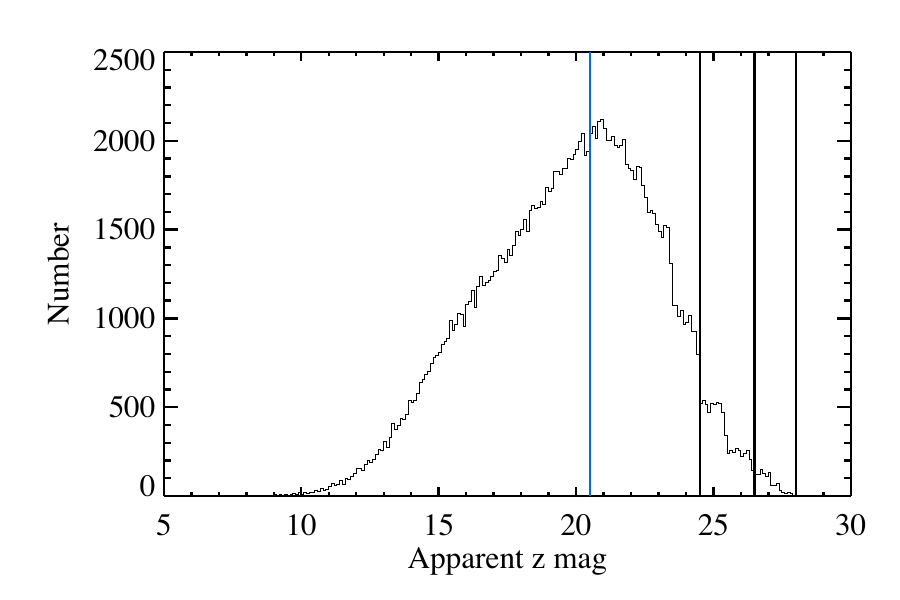}
 \includegraphics*[width=0.45\textwidth]{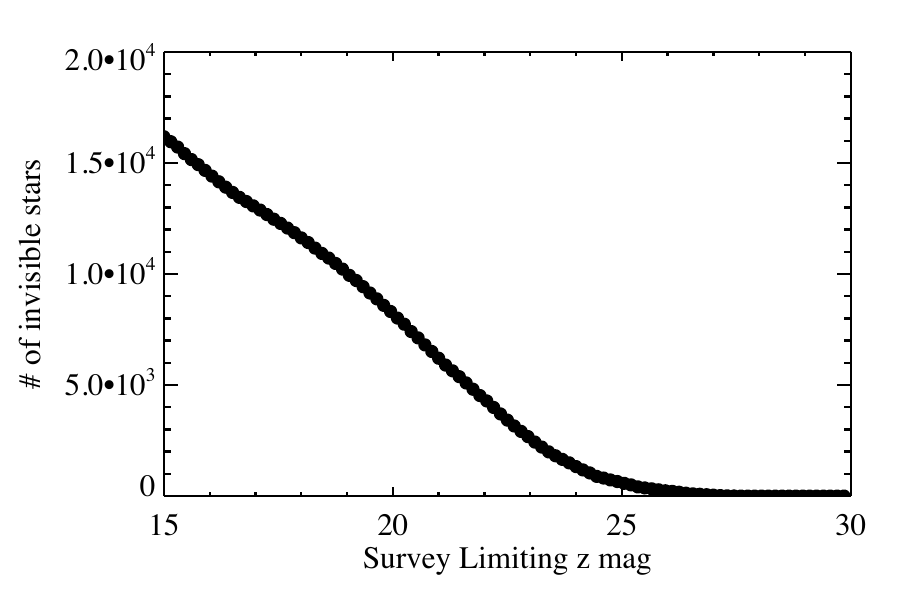}
    \caption{The apparent $z$ magnitude distribution of M dwarfs along a line of sight in a model
representation of the Milky Way. The model includes the most current mass and luminosity functions of low
mass stars \citep{Covey2008, Bochanski2007a, Bochanski2008,
  West2008}. The blue line is the SDSS z-band limiting magnitude. The black lines are LSST
single visit, co-add, and deep drilling limits. Notice that for this sight line, SDSS is unable to detect a large
fraction of the M dwarfs in the Galaxy, while LSST will detect the vast majority.
 Shown on the right is the number of objects
not seen in quiescence as a function of survey limiting depth. As expected, it decreases
monotonically. Note that it becomes completely negligible around $z=27$. 
Flares on objects not seen in quiescence will appear as true optical transients, 
whereas flares on known objects will be easier to categorize. }
\label{fig:stars_vis}
\end{figure}

The model can also be used to calculate the amount of time a flare will appear to be a truly transient event,
 i.e., how often a star not detected in quiescence will brighten above the detection limit. 
In \autoref{fig:stars_vis}, we show the number of stars along a particular line of sight in our model,
 along with the number of stars that are invisible as a function of survey limiting magnitude. 
The superior depth of LSST, particularly in the deep drilling fields, 
means that for certain lines of sight we will likely observe {\it all} M dwarfs. 

Using our model, we can thus predict the number and brightness of
flares that LSST will see before data collection begins. These
predictions will be useful to separate flares from other variable
objects of interest.  Once LSST is gathering data, we will compare our
model predictions to the empirical data to refine the model and
produce  a better description of M dwarf flare frequency distributions
as a function of spectral type. 

The unique power of LSST to open the time domain will allow us to
statistically determine the flare frequency distribution as a function
of stellar type, age, and activity level with unprecedented
accuracy. The flare rates will be interpreted from the data with the
aid of models and tools for finding flares with sparsely sampled light
curves that are already in development. At the excellent photometric
accuracy of LSST, it will be possible to resolve even relatively small
flares, and therefore constrain chromospheric and coronal heating
mechanisms in the outer atmospheres of these stars.

\subsection{Resolving the Stellar Dynamo: Activity Cycles Across the Main Sequence}
{\it Stylani (Stella) Kafka}

The Mount Wilson Observatory HK survey \citep{1982ApJ...257..179W}
produced the first comprehensive sample of long-term stellar light
curves of stars with spectral types F5 to M2. These light curves
indicated that while most stars have activity cycles, which are not
necessarily similar to the solar cycle in amplitude and
duration. Since then, studies of stellar activity have revealed
fundamental properties of magnetic field generation and evolution in
stars of all spectral types, but these revelations have produced a
rather confusing picture.  \citet{Olah.etal(2007)} report variations
of the length of stellar and solar activity cycles and a connection
with stellar rotation in that stars with long rotation periods have
longer cycle periods \citep{Olah(2000)}. The
\citet{Baliunas.etal(1995)} study of 111 dwarfs with spectral types
between G0 and K7 from the Mount Wilson HK sample, indicate that only
1.5$\%$ of the stars display cycles similar to the Sun's. According to
that study, F dwarfs seem to either have non-detectable cycles or very
long ones, while K dwarfs seem to have very pronounced cycles, and
Maunder-minima-like activity levels are detected only in solar-like
stars. Activity cycles in fast rotating, young solar-type stars seem
to range from 2.1 to $\sim$ 10 years \citep{MessinaGuinan(2003)}. At
the same time, the existing theory of magnetic field generation and
stellar activity dictates that an $\alpha\Omega$ type dynamo acting in
the interface between the convection zone and radiative core of stars
(the tachocline) is responsible for the large-scale, solar-like dipole
field. Such a dynamo is in action for {\em all} stars with a tachocline,
and it provides the means for magnetic braking which is the main
angular momentum loss mechanism in stars. However, the diversity in
the cycle behavior -- even within the same stellar spectral types --
suggests that our understanding of the stellar magnetic field
generation and evolution is still not clear.

Current studies of stellar activity use snapshot observations of a
large number of field stars to assess the properties of activity for
stars of various spectral types. However, we still have not been able
to reach a coherent picture of the properties (amplitude, duration,
variability, and so on) of activity cycles for stars across the main
sequence, especially when it comes to the lower mass objects (late K
and M dwarfs). Although the simple model of solar-like magnetic field
generation should not apply in these low mass, fully convective stars,
the level and character of their activity seems to be
indistinguishable from that of earlier spectral type stars. A plethora
of new models have attempted to explain this phenomenon by focusing on
the effects of fast rotation alone on magnetic field generation
(e.g., \citealt{ChabrierKuker(2006)}; \citealt{Baliunas.etal(2006)}) with no
satisfactory results. The lack of evidence that activity changes
character when stars become fully convective and the numerous
alternative proposed mechanisms (none of which can be securely
confirmed or dismissed) demonstrate the need of a large, unbiased
statistical sample to study properties of activity cycles in main
sequence stars.  

An ideal sample would consist of a number of open star cluster members
for which ages and metallicities are easily extracted. Existing data
are restricted to the brighter cluster members, leaving late K and M
stars out of the equation. Long-term monitoring of order thousands 
of K/M dwarfs would provide solid and secure
results freed from small number statistic biases and serendipitous
discoveries.  Spot coverage during activity cycles is a ubiquitous
fingerprint of the underlying dynamo action over long periods of time;
therefore, long-term monitoring is desired in order to reveal the
mechanism in action for stars of different spectral types and in
different environments. Thus far the few existing studies of stellar
cycles resulted from visual observations of bright nearby stars by
amateur astronomers and/or observations with 1-m class automatic
photometric telescopes (APTs) focused on specific objects and/or
limited parts of the sky. Especially for late K/M dwarfs, the length
and intensity of relevant activity cycles has not been systematically
investigated. One of the few examples is Proxima Centauri (M5.5V),
which is found to have a $\sim$ 6 year activity cycle
\citep{Jason.etal2007}\footnote{Although this star is fully convective,
its activity characteristics appear to be solar-like.}. Prior to LSST,
CoRoT and Kepler will define properties of short-term (minutes to
months) secular stellar brightness variations due to starspot
evolution on field stars (over a small portion of their activity
cycle). However, even those observations are time and frequency
limited, and address an inhomogeneous stellar population (field
stars). Long-term, multi-epoch observations, covering the time
and frequency domain for a large number of objects sharing common
properties, are essential to reveal the characteristics of activity
cycles of stars of all spectral types.

For the first time, LSST will provide the large (and faint!) sample
population required to reveal properties of activity cycles in stars
of different spectral types and ages. Using LSST data we will build
the first long-term, multi-frequency light curves of stars past the
mass limit of full convection. This will suffice to reach the end of
the main sequence in a large number of open star clusters, observing
stars of all ages and metallicities -- parameters that are hard to
obtain for field stars. With a limiting magnitude of $r=24.7$ per
visit, LSST will provide $\sim$1,000 points for each star (in each
filter) during the first 5 years of its operation. Although activity
cycles can be longer (5 years is $\sim$ 50$\%$ of the cycle for a solar
twin), a five-year coverage will allow us to detect long-term
modulations in the light curves of thousands of stars. In turn, these
long baseline light curves will allow us to probe the character and
evolution of stellar activity cycles in an unbiased way, deriving
correlations between cycle duration, stellar ages, and spectral type
(or depth of the convective zone) for cluster members. 

The length of activity cycles will provide constraints for existing
dynamo models and identify trends in various stellar cluster
populations. We will be able to determine average timescales and the
amplitude of variations in stellar cycles, answering fundamental
questions: How does the stellar dynamo evolve over a cycle for
different stellar masses? Is there a dependence of the amplitude and
duration of cycles on stellar age and/or metallicity for stars of
specific spectral types? Can we identify changes of the dynamo
properties (in terms of activity cycle characteristics) in M stars at
the mass regime where their interior becomes fully convective? Do
fully convective stars have activity cycles? Do all stars have Maunder
minimum-like characteristics? Furthermore, LSST will resolve how
cycle-related stellar flux variations affect a star's habitable zone,
providing essential information on how common the Earth's environment
is in the Universe.  

\section{Non-Degenerate Eruptive Variables}
\label{sec_eruptive}
\subsection{The Death Rattle of High Mass Stars: Luminous Blue Variables and Cool Supergiants}
{\it Nathan Smith, Lucianne M. Walkowicz}

The scarcity of high mass stars poses a serious challenge to our
understanding of stellar evolution atop the H-R diagram. As O-type
stars evolve off the main sequence, they may become Luminous Blue
Variables (LBVs), red supergiants, yellow hypergiants, blue
supergiants, or they may evolve through several of these
phases sequentially, depending on their initial mass, metallicity, and
rotation \citep[e.g.,][]{Chiosi1986,Meynet1994,Langer2004}. These death throes can sometimes be characterized by
extreme mass loss and explosive outbursts that in some cases are
short-lived and possibly intermittent \citep{Smith2006,Humphreys1994}.  There are only a handful of nearby massive stars
that are caught in this phase at any given time (as in the case of
$\eta$ Carinae), and as a result, they appear unique or exceptional
when considered in context with other stars.  They may nevertheless
represent a very important phase that most massive stars pass through,
but it is difficult to judge how representative they are or how best
to account for their influence in models of stellar evolution.  The lack of extensive data for these stars
makes it very difficult to connect distant explosions (supernovae and
GRBs) to their underlying stellar populations. 

Although LBV outbursts can be seen up to 80 Mpc away, the best-studied LBVs are in the plane of our own Galaxy (predominantly in the southern hemisphere).  Unfortunately, the Galactic LBVs are few,  and so these and cooler outbursting stars have eluded meaningful statistical study to date.  A small number of them have been studied in very nearby galaxies \citep[e.g.,][]{Hubble1953,Tammann1968,Humphreys1994,Drissen1997,Massey2007}. However, the improved breadth and sensitivity of LSST are ideally suited to the study of these intrinsically rare objects.
LSST will make these extremely luminous stars accessible in many galaxies, offering a new opportunity to improve the sample of known evolved massive stars.

Time resolved observations of variability in the new sample provided by LSST will quantify
the statistical distribution of time dependent mass loss rate,
luminosity, radiated energy, total ejected mass, duration of
outbursts, time between outbursts, and connections to the pre-outburst
stars. Outbursts last anywhere from $\sim$ 100 days to a year, so the universal cadence of LSST will revisit their evolving light curves several times in multiple passbands as the outburst progresses.
In the case of red supergiants, variability provides a key
discriminating factor between foreground red dwarfs and extragalactic
massive stars. Time domain observations may also resolve the
evolution in the amplitude and timescale of variability as these stars
expand and cast off their outer layers. 

New observations will inform models of massive star evolution,
providing prescriptions for the time-dependent properties mentioned
above so that they can be included in stellar evolution codes in a
meaningful way. Stellar evolution codes that predict the fates of
massive stars over a range of mass and metallicity
\citep[e.g.,][]{Heger2003} do not currently include the effects of
LBV-like outbursts 
because an empirical assessment of their properties as functions of
initial mass and metallicity does not yet exist.  LBV eruptions are
currently ignored in stellar evolution codes, even though they may
dominate the total mass lost by a massive star in its lifetime and may, therefore, be key to the formation of Wolf-Rayet stars and GRB
progenitors over a range of metallicity \citep{Smith2006}. Many
model grids do not extend to sufficiently cool temperatures at high
masses, leaving the formation of the most luminous red supergiants
largely unexplained by detailed theory.  Most evolution codes also do
not include the effects of pulsations that drive large temperature
variations in red supergiants, in some cases pushing them far cooler than the
Hayashi limit for periods of time \citep[although see][]{Heger1997}.

By providing a new sample of these stars in other galaxies, LSST will also enable studies of
high mass stellar evolution as a function of metallicity. Absorption
by lines of highly ionized metals plays a major role in accelerating
winds in these objects \citep[e.g.,][]{Castor1975}, thereby driving
mass loss and affecting the duration of late stages of evolution
\citep{Chiosi1986}. Current theory holds that metallicity affects
the relative number of blue versus red supergiants by changing the
duration of these end stages \citep{Meynet1994}, but these models 
do not include the effects of pulsation-driven or outburst-driven
mass loss. In a complementary fashion, further study of these stars
will also improve our understanding of their contribution to galactic
feedback and enrichment of the interstellar medium.

LSST will also bring insight to another open question: the true nature of core collapse
supernova (CCSN) progenitors (e.g., \citealt{Smartt++09}). A large sample of
evolved massive stars will propel our understanding of the diversity
of CCSN progenitors.  Among the large sample of luminous stars
monitored in nearby galaxies, some will explode {\it while they are
being monitored}. This will provide not only an estimate of the star's
pre-explosion luminosity and temperature, but also its variability and
potential instability in the final years of its life. For example,
red supergiants that explode as Type II-P SNe may exhibit strong photometric
variability, and this must be accounted for when using pre-explosion
data to infer the star's initial mass and radius.  Additionally, there
is growing evidence that a subset of massive stars suffer violent
precursor outbursts, ejecting large amounts of mass in the decades
leading up to core collapse \citep{Smith2007}.  The resulting SNe
are called Type IIn because of the narrow H lines that arise in the
dense shocked circumstellar gas \citep{Filippenko1997}.  These Type IIn SNe
come in a wide range of luminosity and spectral properties and may
trace a diverse group of progenitors that suffer precursor outbursts.
These pre-SN outbursts are neither predicted by nor explained by
current stellar evolution theory.  With an observed record of pre-SN
variability, we can connect properties of these precursor outbursts
and the resulting SN to determine which outbursts are pre-SN and
which are not.  Furthermore, the enormous outbursts of LBVs themselves
may masquerade as low luminosity Type IIn SNe \citep[e.g.,][]{VanDyk2000}. Time resolved observations of relatively large samples of
potential CCSN progenitors will provide new insight into the nature of
this intriguing population.  With high enough imaging resolution,
stellar population studies of the surrounding field stars can
constrain the local star formation history, and thus constrain the
delay time between star formation and core collapse.

\subsection{Eruptive Variability in Pre-Main Sequence Stars}
{\it Peregrine M. McGehee}

Variability is one of the distinguishing features of pre-main sequence stars
and can result from a diverse collection of physical phenomena including
rotational modulation of large starspots due to kiloGauss magnetic
fields, hot spots formed by the impact of accretion streams onto the
stellar photosphere, variations in the mass accretion rate, thermal
emission from the circumstellar disk, and changes in the line of sight
extinction. These physical processes generate irregular 
variability across the entire LSST wavelength range (320--1040 nm) 
with amplitudes of tenths to several magnitudes on
timescales ranging from minutes to years and will be detectable by LSST.

Due to its sensitivity and anticipated ten-year operations lifetime, LSST
will also address the issue of the eruptive variability 
found in a rare class of young stellar objects - the FUor and EXor stars.
FUor and EXor  
variables are named after the prototypes FU Orionis \citep{hartmann96} and
EX Lupi \citep{herbig01} respectively.
These stars exhibit outburst behavior characterized by an up to 6 magnitude 
increase in optical brightness, with high states persisting from several 
years to many decades.  
Both classes of objects are interpreted as pre-main sequence stars
undergoing significantly increased mass accretion rate possibly due to 
instabilities in the circumstellar accretion disk. The mass accretion
rates during eruption have been observed to increase by 3 to 4 orders of 
magnitude over the $\sim 10^{-9}$ to $10^{-7} \,M_\odot$
per year typical of Classical T Tauri stars.
Whether FUor/EXor eruptions are indeed the signature of an evolutionary
phase in all young stars and whether these outbursts share common mechanisms 
and differ only in scale is still an open issue.

To date only about 10 FUors, whose eruptions last for decades, 
having been observed to transition 
into outburst\citep{aspin09} with the last major outburst
being that of V1057 Cyg \citep{herbig77}. Repeat outbursts of several
EXors have been studied, including those of EX Lupi \citep{herbig01} and 
V1647 Ori \citep{aspin09}, the latter erupting in 1966, 2003, and 2008.
The outbursts of
EXors only persist for several months to roughly a year in
contrast those of FUors, which may last for decades: for example, the 
prototype FU Ori has been in a 
high state for over 70 years.
These eruptions can occur very early in the evolution of a
protostar as shown by the detection of EXor outbursts from a deeply
embedded Class I protostar in the Serpens star formation region
\citep{hodapp96}.
The observed rarity of the FUor/EXor phenomenon may be due to the 
combination of both the relatively brief (less than 1 Myr) duration
of the pre-T Tauri stage and the high line of sight 
extinction to these embedded objects hampering observation at
optical and near-IR wavelengths.

\begin{figure}
\begin{center}
\includegraphics[width=0.9\linewidth,angle=0]{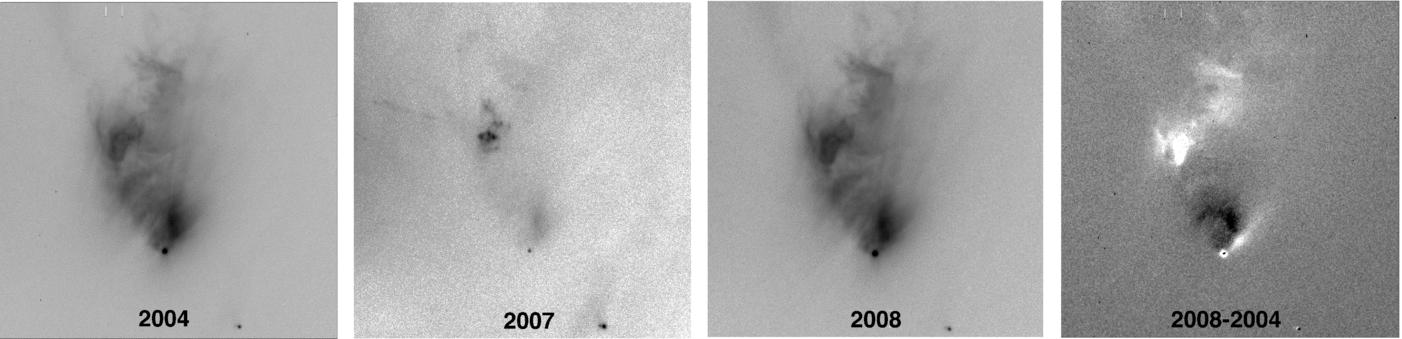}
\caption{
This series of images from \citet{aspin09} shows the region containing 
McNeil's Nebula and the EXor V1647 Orionis (seen at the apex of the nebula).
The observations span ({\it left}) 2004 to ({\it middle right}) 2008
and include two outbursts and an intervening quiescent period. On the right is
the difference image between the 2004 and 2008 outbursts.
\label{fig:tr:v1647ori}}
\end{center}
\end{figure}

V1647 Ori is a well-studied EXor found in
the Orion star formation region ($m-M = 8$) and thus is a suitable case 
study for discussion of LSST observations.  \autoref{fig:tr:v1647ori} shows $r'$ and $R$ imaging of V1647 Ori 
\citep{aspin09} demonstrating the appearance of the protostar and surrounding
nebulosity during the two most recent eruptions and the intervening
low state. 
The inferred extinction is $A_r \sim 11$ magnitudes which coupled with the
observed $r$ range of 23 to nearly 17 during
outburst \citep{mcgehee04} suggest that $M_r$ varies from 4 to $-2$ magnitudes.

The LSST single visit $5\,\sigma$ depth for point sources is $r \sim 24.7$, thus
analogs of V1647 Ori will be detectable in the $r$ band during quiescence
to $(m-M) + A_r = 20.5$ and at maximum light to $(m-M) + A_r = 26.5$.
The corresponding distance limits are 800 pc to 12 kpc assuming $A_r = 11$.
For objects at the distance of Orion the 
extinction limits for LSST $r$-band detections of a V1647 Ori analog are 
$A_r = 12.5$ and $A_r = 17.5$.
These are conservative limits as V1647 Ori was several magnitudes brighter 
at longer wavelengths ($iz$ bands) during both outburst and
quiescence indicating that the LSST observations 
in $izy$ will be even more sensitive to embedded FUor/EXor stars.

LSST will increase the sample size for detailed follow-up
observations due its ability to survey star formations at large heliocentric
distances and to detect variability in embedded and highly extincted 
young objects that would otherwise be missed in shallower surveys. 
During its operations LSST will also
provide statistics on the durations of high states, at least for the 
shorter duration EXor variables.

\section{Identifying Transiting Planets with LSST}
\label{sec_transits}
{\it Mark W. Claire, John J. Bochanski, Joshua Pepper}

Large planets in close orbits (a.k.a ``hot Jupiters") spend
1--5\% of their orbital period transiting their host stars, if
viewed edge-on.  Thus, given optimal geometry, 10-50 of the $\sim$ 1,000 LSST
observations of a given star with a transiting hot Jupiter will occur
in eclipse. Preliminary results from the Operations Simulator
(\autoref{sec:design:opsim}) 
indicate that LSST will dramatically increase the number of known hot
Jupiter systems, expanding their census to greater Galactic depths.

\subsection{What Will We Learn about Transiting Planets from LSST?}
The primary scientific gains will lie in three areas:

\textit{Studying the hot Jupiter frequency distribution at large distances.} By the conclusion of LSST's ten-year science run, the frequency of
nearby hot Jupiters should be relatively well-known as a function of
spectral type and metallicity due to dedicated radial velocity
(eg.,~MARVELS, HARPS) 
and transit (eg.,~COROT, Kepler) surveys.  LSST may detect thousands to
tens of thousands of planetary transit candidates -- numbers which
should remain significant in the mid 2020s.  LSST will thus enable
investigation of how hot Jupiter frequencies derived for the solar
neighborhood extrapolate to thick disk and halo stars. Within the thin
disk, LSST will constrain radial gradients in planetary frequency, and
if these are correlated, with metallicity.

\textit{Providing statistical constraints on planetary migration theory.} Hot Jupiters are not thought to form \textit{in situ}, and hence require migration 
through protoplanetary discs.  Planetary migration theory is still in its infancy 
and cannot yet predict distributions of feasible planetary radii and distances.  
LSST will enable statistically significant constraints of various formation theories 
by revealing how planetary system architecture varies with stellar/planetary masses 
and metallicity.

\textit{Examining the effects of intense stellar irradiation on planetary atmospheres.} The atmospheres of hot Jupiters can be heated enough to drive hydrodynamic escape. By 
identifying the shortest period planets, LSST will help constrain the energy absorption 
limit beyond which a hot Jupiter cannot maintain its atmosphere for the life of the 
stellar system \citep{2007Natur.450..845K}.

Any transiting planet candidates found will require follow-up for full
confirmation. Even in 2025, radial velocity studies with sufficient  
precision will likely be difficult at the distances of most LSST stars. Given that active 
follow-up on potentially tens of thousands of targets may be infeasible, results drawn from 
LSST may be statistical in nature \citep{2006Natur.443..534S}, and care must be taken to 
identify the best candidates for follow-up.

\subsection{How Many Planets will LSST Detect?}


\subsubsection*{Q1) What is LSST's Chance of Detecting a Transiting Planet Around a Given Star?}

A simulation pipeline is initiated by specifying
values of stellar radius ($Rs$), planetary radius ($Rp$), period
($P$), ra ($\alpha$), dec ($\delta$), distance ($d$), and inclination
($i$).   A normalized planetary transit light curve
\citep{2003ApJ...585.1038S} is assigned a random initial phase
($\phi$). Stellar \textit{ugriz} colors are interpolated from
\citet{2007AJ....134.2398C}, and \textit{y} colors are estimated by
the integration of Kurucz model fluxes through LSST filter curves for
warm stars and with template spectra
\citep{2005ApJ...623.1115C,2007AJ....133..531B} for cool stars.  The
specified distance is used to create apparent magnitude light curves
in each filter, which are then realized through the Operations
Simulation via the light curve simulation tool (\autoref{tr:opsim}),
assuming dithering of field locations.

Points with excellent photometry from a six-filter normalized light
curve are scanned for box type periodicity from 0.5 to 40 days in one
second intervals.  Nearly all pipelines runs return either an exact
period/alias or a complete non-detection of a period, with very few
($<$0.1\%) ``false positives'' in which a periodic signal is detected
that was not present in the initial light curve.  Of these false
positives, $\sim$ 60\% are periods of $\simeq $ 1 day, which can be
easily screened.  False positives are reported as negative results in
these estimates, but potentially $\sim$ 1 of every 1,000 planets
``detected'' by this method might be spurious due to inability to cull
false positives.  Further complicating factors such as correlated red
noise and binary contamination may also increase false positives, and
require future attention.

Assuming that detection probabilities $(\Phi)$ from
changing distance, inclination, and position on the sky are
independent of the parameters of the planetary systems enables
computation of  $ \Phi = \Phi_{detect}(Rs, Rp , P, \phi) \times
\Phi_{detect}(\alpha, \delta)\times
\Phi_{detect}(d)\times\Phi_{detect}(i)$.  The effects of initial
phase and instrument properties are averaged over as described below.  

Stars of different spectral types have distance-dependent changes in
their magnitude errors, given differing bright and faint limits in
each filter. In addition, the transit depth signal varies as $(Rp/Rs)^2$. As
$Rs$ is not independent of the other parameters under consideration, $Rs$ 
is fixed and a suite of results for differing spectral types is constructed.
A ``best-case scenario" distance is chosen so 
that a star of that spectral type will be observable at 1\%
photometry in a maximum number of LSST filters. With $Rs$ and $d$
fixed (and with $i=90^\circ$), the remaining independent
variables $(Rp, P, \phi)$ are explored. $\Phi_{detect}(Rs, Rp, P, \phi)$
is reported as the number of positive detections in 50 pipeline runs of varying
initial phase.

\begin{figure}[ht]
\centerline{
\includegraphics[width=0.6\textwidth]{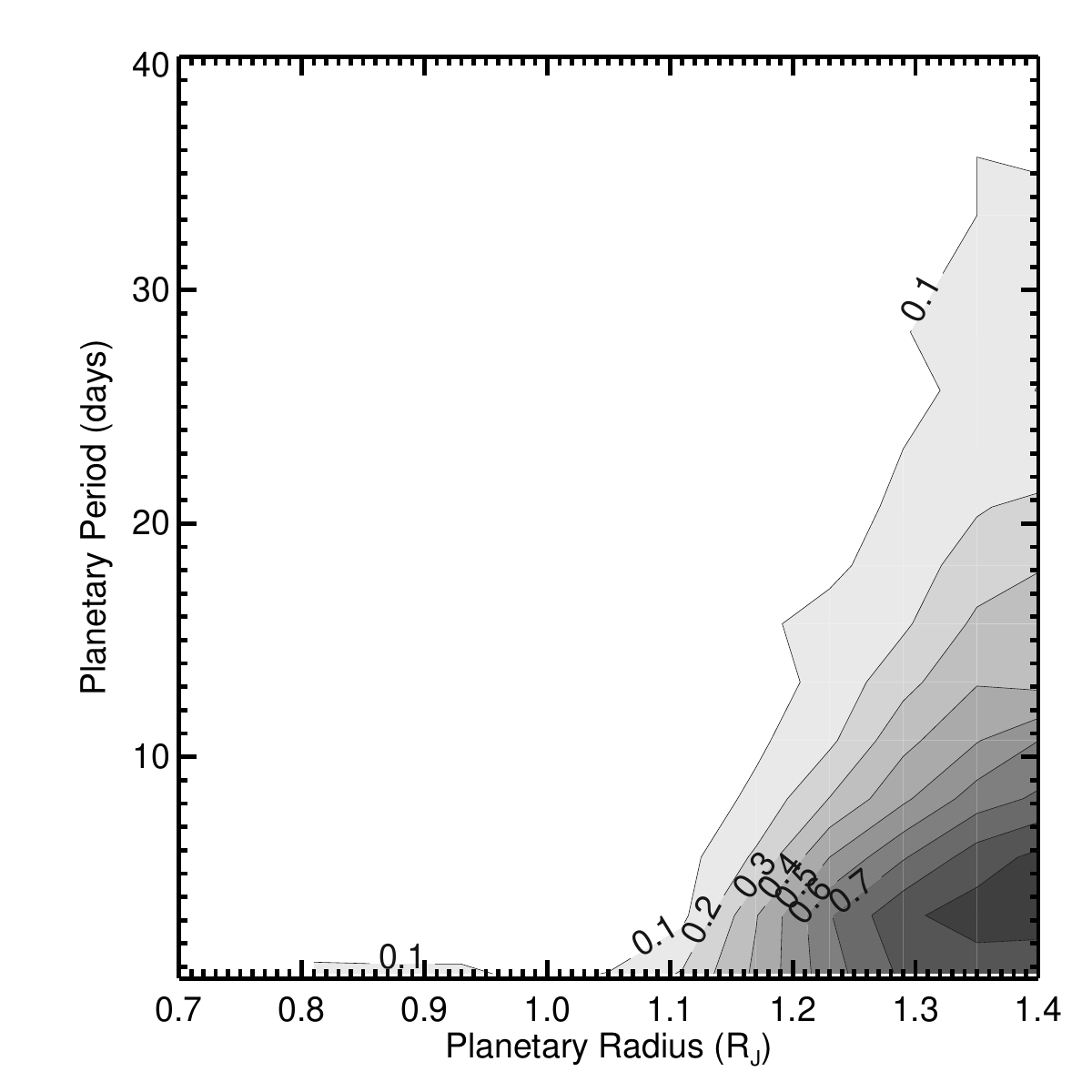}}
\caption[]{
Planetary transit detection probabilities for a 0.7 $R_\odot$ star at 1
kpc, calculated at each grid point as the period recovery percentage for 50 runs at random initial phases. Planetary radii are in units of Jupiter radii.
}
\label{fig:mc}
\end{figure}

\autoref{fig:mc} is a contour plot of $\Phi_{detect}(Rs, Rp,
P, \phi)$  for hot Jupiters around a 0.7 $R_\odot$ star at 1 kpc, calculated for 
a star at ($248^\circ,-30^\circ$). To test the effect of changing sky
position, deviations from a case where $\Phi_{detect}(Rs,
Rp , P, \phi)=100\%$ were computed.  The simulation results (not shown) show
$\sim$10\% deviations in $\Phi_{detect}(\alpha,\delta)$, and generally follow the 
pattern seen in the number of visits per field (\autoref{fig:design:rband}).

\begin{figure}[ht]
\begin{center}
\includegraphics[width=0.6\textwidth]{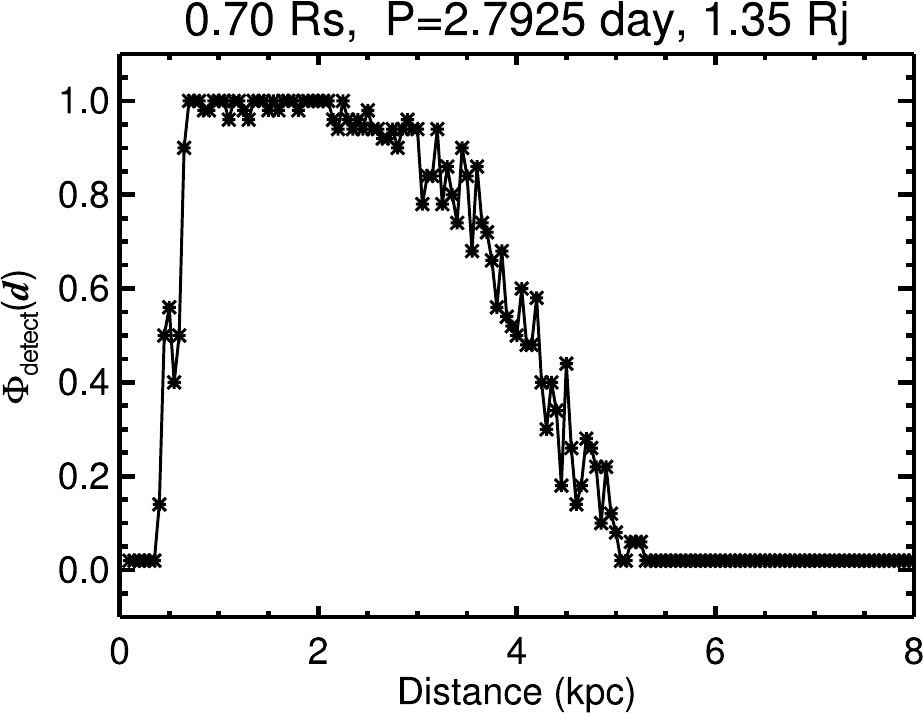}
\caption{Planetary transit detection probability as a function of
distance of the star.  Saturation of the LSST detectors
(\autoref{sec:com:expos}) was taken into account.  
\label{fig:mc2}}
\end{center}
\end{figure}

\autoref{fig:mc2} examines $\Phi_{detect}(d)$ as a 0.7 $R_\odot$ star with a 1.35 Jupiter radii planet in a 2.725 day period observed at ($248^\circ,-30^\circ$) is placed at various distances from the Sun.  

\subsubsection*{Q2) How Many Transiting Hot Jupiters will LSST Detect?}

To predict the number of hot Jupiters that LSST might detect requires an
estimate of the number of observable stars as a function of
spectral type, position, and distance on the sky, making allowances
for the fractions of stars that are non-variable and non-binary, and
those that might have planets in edge-on configurations.  
A Monte Carlo simulation over relevant planetary system
parameters applied to computed detection probabilities will allow 
quantification of the number of detectable planets in that
volume using methodology similar to \citet{2005ApJ...631..581P}.  

A simple analytic calculation predicts that LSST could observe
$\sim$ 20,000 transiting hot Jupiters \citep{2007ASPC..366..273G}, but
cites the need for the more detailed treatment that is underway.
The calculations are too preliminary to provide an answer to
Question 2 at this moment, but the project outlined above will provide
predictions that are more closely tied to the actual observation
conditions.  LSST's planet finding capabilities will be immense, given 
that most of the stars in the sky
have radii smaller than 0.7 $R_\odot$, and thus will have an expanded
phase space in which $\Phi_{detect} = 100\%$.

\section{EPO Opportunities}
{\it Paula Szkody,  Steve B. Howell, Robert R. Gibson }

After the first year or two of operations, LSST 
will have a large collection of well sampled,
multi-color light curves. These can be used by the project to assign a
confidence level to a new ``event'' such as a transient or for
assignment of a sparsely sampled light curve to a specific variable
type (\autoref{sec_followup}). A useful tool for the presentation of
the template observations 
and the additional LSST light curves would be to develop a ``VO
Broker'' that allows a database search ability and can produce a light
curve, a phased light curve, and other variable star tools. This tool
would be highly useful to the project, other scientists, and the
interested public. 

Citizen scientists can play a role in classifying light curves in this
initial archive of several hundred thousand variable stars from the
early science proposal.  By comparing the shape of light curves in the
LSST sample against templates of known sources, initial
classifications can be assigned for further weighting and analysis by
researchers.  This idea could be developed as part of the Light Curve
Zoo Citizen Science Project described in \autoref{chp:epo}. 

Light curves, the graphical representation of changes in brightness
with time, have educational value in several settings.  Learning to
read, construct, and interpret line graphs are critical skills at all
levels of the National Council of Teachers of Mathematics' Standards as is the ability to use
representations to model and interpret physical, social, and
mathematical phenomena \citep{NCTM00}.  Light curves also can be
converted to audio and represented by sound, letting people ``listen to
the light curve'' and hear differences between light curves of
different types of variable objects. A supernova light curve would
sound different from an AGN light curve, which is different from a
Cepheid light curve, and so on.  This learning technique would be useful to
multiple learning styles including the visually impaired, and could be
explored in a web-based tutorial on variability and integrated into a
training module on light curve identification.

Amateur astronomers can provide a valuable service in following up the
brighter sources with their telescopes, enabling identifications in
addition to those completed by professional astronomers. Tens of
thousands of variable object alerts are predicted nightly, a rate
beyond the capability of professional telescopes to monitor.
Effective partnerships between professionals and amateurs can be
developed that capitalize on the 
opportunities offered by LSST alerts and increasingly sophisticated
capabilities of the amateur community.  AAVSO (American Association of
Variable Star Observers), VSNET (Variable Star Network in Japan) and
CBA (Center for Backyard Astrophysics) are prime organizations
with a record of CCD observations of variables and high interest in
their communities.

\bibliographystyle{SciBook} 
\bibliography{transients/transients}

%
%
%
%
%
%
%
%
%
%
%
%
%
%
%
%
%
%
%
%
%
%
%
%
%
%
\chapter[Galaxies]{Galaxies}
\label{chp:galaxies}

\noindent{\it Henry C. Ferguson, Lee Armus, L. Felipe Barrientos, James
  G. Bartlett, Michael R. Blanton, Kirk D. Borne, Carrie R. Bridge, Mark
  Dickinson, Harold Francke, Gaspar Galaz, Eric Gawiser, Kirk Gilmore, Jennifer M. Lotz, R. H. Lupton, Jeffrey A. Newman, Nelson D. Padilla, Brant E. Robertson, Rok Ro\v{s}kar, Adam Stanford, Risa H. Wechsler} 

%
%
%
%
%
%
%
%
%
%
%
%
%
%
%
%
%
%
%
%
%
%
%
%
%
%
\section{Introduction}

The past decade of research has given us confidence that it
is possible to construct a self-consistent model of galaxy evolution
and cosmology based on the paradigm that galaxies form
hierarchically around peaks in the dark matter density distribution.
Within this framework, astronomers have made great progress in
understanding the large-scale clustering of galaxies, as biased
tracers of the underlying dark matter.  We have started to 
understand how baryonic gas within the dark matter
halos cools and collapses to form stars, and how the energy from
star formation can feed back into the surrounding gas and regulate
subsequent star formation. 

However, at a fundamental level we still lack a solid understanding of
the basic physics of galaxy evolution. We do not know for certain that
all galaxies form at peaks in the dark matter density distribution.
There is considerable debate about cold versus hot accretion of gas
onto dark matter halos, and even more debate about which feedback
mechanisms regulate star formation (photo-ionization, supernova winds,
AGN, massive stellar feedback, etc.).  We are reasonably certain that
various feedback processes depend on environment, being different in
rich clusters than in the low-density field, but the mechanisms are
not understood. Some environmental and feedback effects
(e.g., photo-ionization) can have an influence over long distances,
which can in principle affect how stars connected to dark matter halos
in different environments.  We have a long way to go before declaring
victory in our understanding of galaxy formation.

The galaxy evolution process is stochastic.  The hierarchical paradigm
tells us {\it statistically} when dark matter peaks of various masses
and overdensities collapse and virialize. It tells us the statistical
distribution of dark matter halo angular momenta, and it tells us
statistically how dark matter halos grow via successive mergers and
accretion. On top of this dark matter physics, we layer our
understanding of gas cooling, star formation, and feedback. We know
that our current prescriptions for these processes are vastly
oversimplified, but hope to learn about how these processes operate
-- when averaged over entire galaxies -- by comparison to
observations.

Because the overall process is stochastic, some of the most important
tests of the models are against large statistical data sets. These
data sets must be uniform, with known, well-defined selection
functions. SDSS has demonstrated the power of such large-area
surveys, and we can anticipate further progress from SkyMapper, PS1,
DES, and other surveys before LSST comes on line. There will also
undoubtedly be progress in smaller-area deep surveys following on from
the Hubble Deep Fields, GOODS, COSMOS, and the Subaru Deep Fields.

\begin{figure}[ht]
\centerline{\resizebox{4.5in}{!}{\includegraphics{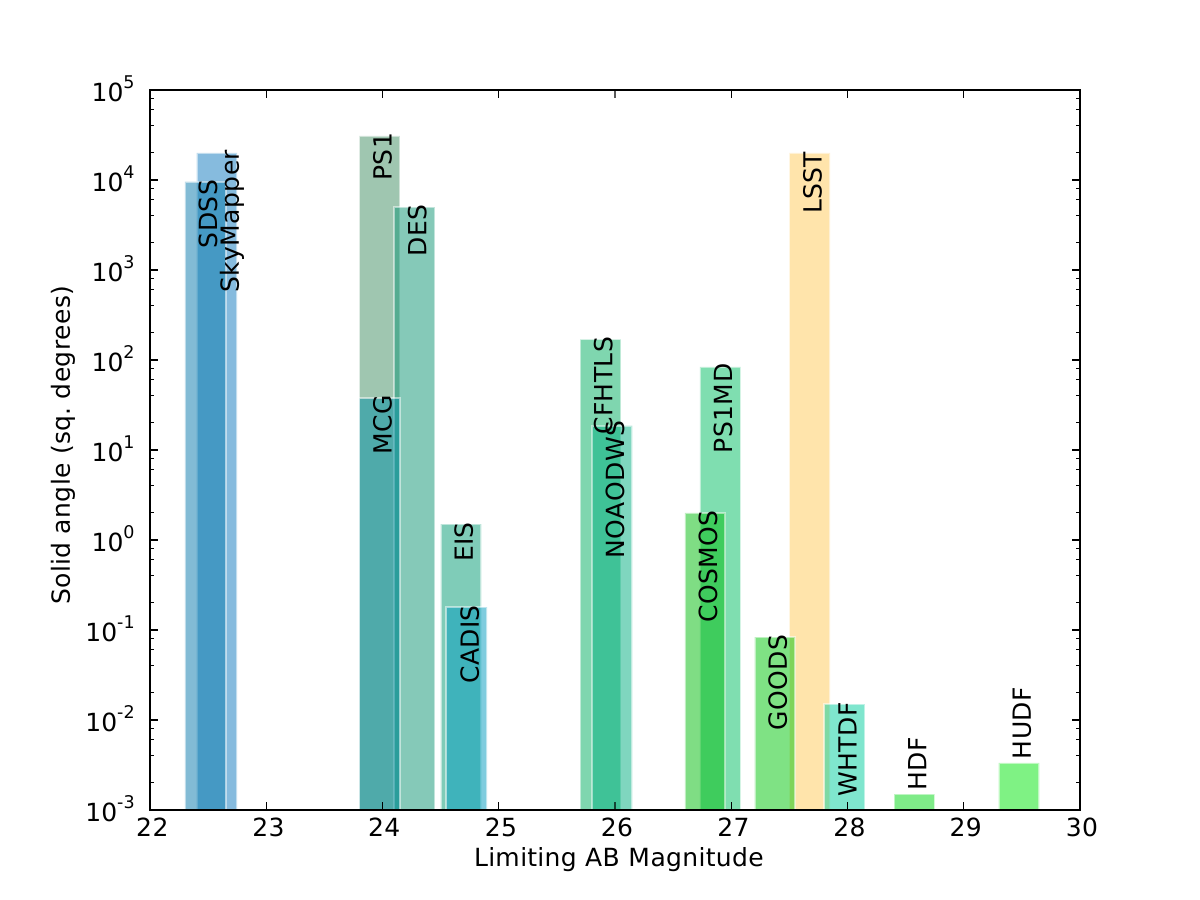}}}
\caption{
Comparison of survey depth and solid angle coverage. 
The height of the bar shows the solid angle covered by the survey. 
The color of the bar is set to indicate a combination of resolution,
area, and depth with $rgb$ values set to $r = V/V({\rm HUDF})$,
$g = (m_{\rm lim}-15/)/16$, and $b = \theta/2^{\prime\prime}$, where
$V$ is the volume within which the survey can detect a typical
$L^*$ galaxy with a Lyman-break spectrum in the r band, $m_{\rm lim}$ is
the limiting magnitude, and $\theta$ is the resolution in arcseconds.
The surveys compared in the figure are as follows:
SDSS: Sloan Digital Sky Survey;  
MGC: Millennium Galaxy Catalog (Isaac Newton Telescope); 
PS1: PanSTARRS-1 wide survey, starting in 2009 in Hawaii; 
DES: Dark Energy Survey (Cerro-Tololo Blanco telescope starting 2009);
EIS: ESO Imaging survey (complete); 
CADIS: Calar Alto Deep Imaging Survey;
CFHTLS: Canada France Hawaii Telescope Legacy Survey;
NOAODWS: NOAO Deep Wide Survey;  
COSMOS: HST 2 deg$^2$ survey with support from many other facilities;
PS1MD: PanSTARRS-1 Medium-Deep Survey covering 84 deg$^2$; 
GOODS: Great Observatories Origins Deep Survey (HST, Spitzer, Chandra, and many other facilities); 
WHTDF: William Herschel Telescope Deep Field; 
HDF, HUDF: Hubble Deep Field and Ultra Deep Field.
}
\label{fig:gal:surveysize}
\end{figure}

\begin{figure}[ht]
\centerline{\resizebox{4.5in}{!}{\includegraphics{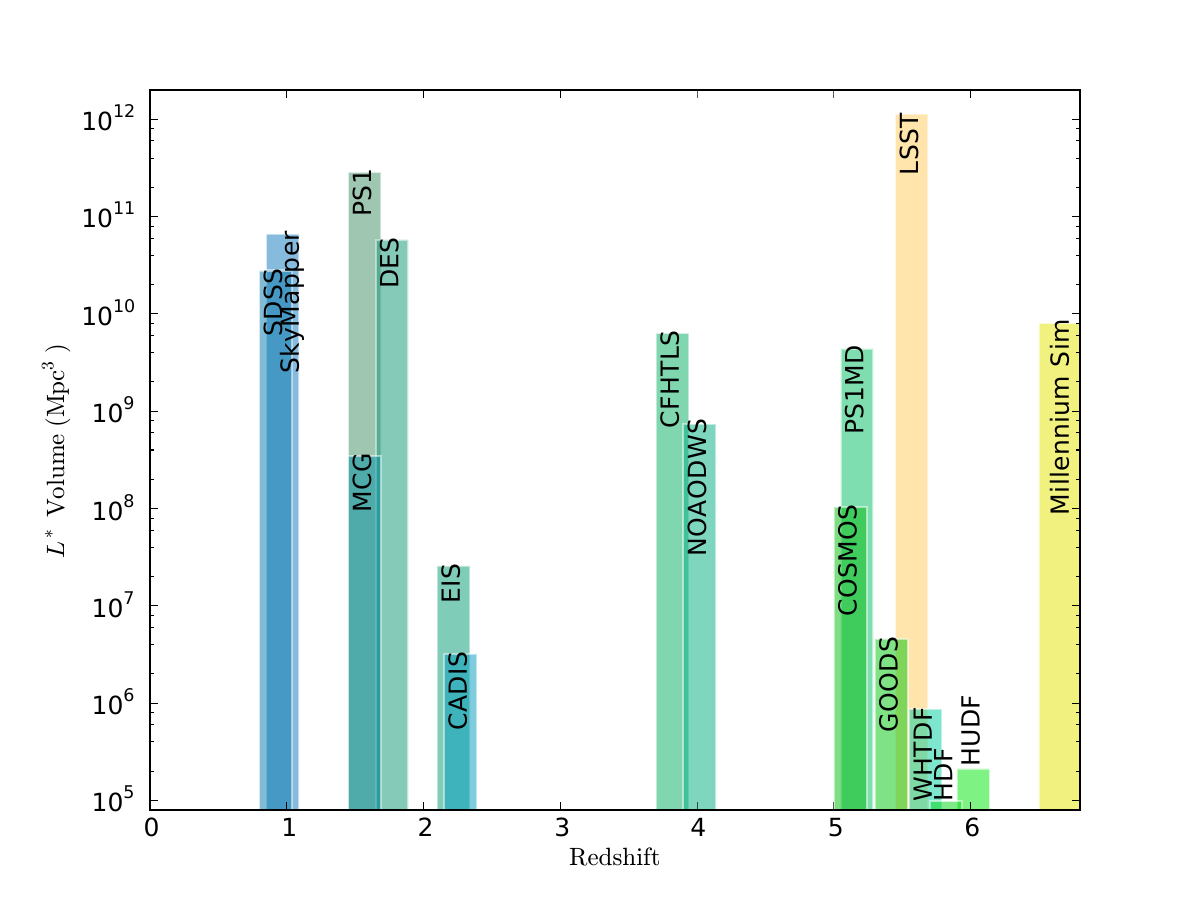}}}
\caption{
Co-moving volume
within which each survey can detect a galaxy with a characteristic
luminosity $L^* \,\,(M_B \sim -21)$ assuming a typical Lyman-break galaxy spectrum.
LSST encompasses about two orders of magnitude more volume than current
or near-future surveys or the latest state-of-the-art numerical
simulations. This figure shows the same surveys as the previous diagram,
with the addition of the Millennium Simulation
\citep{2005Natur.435..629S}.
}
\label{fig:gal:surveyvolume}
\end{figure}

LSST will be a unique tool to study the universe of galaxies. 
The database will provide photometry for $10^{10}$ galaxies, from the local
group to redshifts $z>6$. It will provide useful shape measurements and
six-band photometry for about $4 \times 10^9$
galaxies (\autoref{sec:common:galcounts}). \autoref{fig:gal:surveysize} and
\autoref{fig:gal:surveyvolume} provide indications of the grasp of
LSST relative to  
other existing or planned surveys.

A key to testing our understanding of galaxy formation and evolution
will be to examine the full multi-dimensional distributions of galaxy
properties. Tools in use today include the luminosity function of
galaxies, the color-luminosity relation, size-luminosity relation,
quantitative morphology, and the variation of these distributions with
environment (local density or halo mass). As data sets and techniques
evolve, models will be tested not just by their ability to reproduce
the mean trends but by their ability to reproduce the full
distribution in multiple dimensions. Studies of the tails of these
distributions -- e.g., galaxies of unusual surface brightness
or morphology -- give us the leverage to understand short-lived
phases of galaxy evolution and to probe star formation in a wide
range of environments.

The core science of the Galaxies Science Collaboration will consist of
measuring these distributions and correlations as a function of
redshift and environment. This will make use of the all-sky survey and
the deep fields. Accurate photometric redshifts will be needed, as
well as tools to measure correlation functions, and catalogs of
clusters, groups, overdensities on various scales, and voids, both
from LSST and other sources. 

The layout of this chapter is as follows. We begin
(\autoref{sec:gal:measurements}) by outlining the
measurements and samples that will be provided by LSST. We then
focus on topics that emphasize counting objects as a function of
redshift, proceeding from detection and characterization of objects
to quantitative measurements of evolutionary trends
(\autoref{sec:gal:demo}-\autoref{sec:gal:merger}). In 
\autoref{sec:gal:dm}, we turn to 
environmental studies, beginning with an outline of the different
types of environment and how they can be identified with LSST alone
or in conjunction with other surveys and discussing
measurements that can be carried out on the various environment-selected
galaxy catalogs.  We conclude in \autoref{sec:galaxies:epo} with a
discussion of public involvement in the context of galaxy studies. 

%
%
%
%
%
%
%
%
%
%
%
%
%
%
%
%
%
%
%
%
%
%
%
%
%
%

\section{Measurements}
\label{sec:gal:measurements}

Over the $\sim 12$ billion years of lookback time accessible to LSST, 
we expect galaxies to evolve in luminosity, color, size, and shape.  
LSST will not be the deepest or highest resolution survey in existence. 
However, it will be by far the largest database. 
It will resolve scales of less than $\sim 3$ kpc at any redshift. It
is capable of detecting typical star-forming Lyman-break galaxies $L>L^*$ 
out to $z>5.5$ and passively evolving $L>L*$ galaxies on the red sequence out 
to $z \sim 2$ over 20,000 deg$^2$.  For comparison, the combined area 
of current surveys to this depth available in 2008 is less than 2 deg$^2$. 
In deep drilling fields (\autoref{sec:design:cadence}), the LSST
will go roughly ten times deeper over  tens of square degrees. The
basic data will consist of positions, fluxes, broad-band spectral 
energy distributions, sizes, ellipticities, position angles, and
morphologies for literally billions of galaxies
(\autoref{sec:common:galcounts}).  
Derived quantities include photometric redshifts
(\autoref{sec:common:photo-z}), star-formation rates, internal
extinction, and stellar masses.  

\subsection{Detection and Photometry}

The optimal way to detect an object of a known surface brightness profile
is to filter the image with that surface brightness profile, and apply a $S/N$
threshold to that filtered image. In practice, this is complicated by 
the wide variety of shapes and sizes for galaxies, combined with the fact
that they can overlap with each other and with foreground stars.
The LSST object catalog will be a compromise, intended to enable a broad 
spectrum of scientific programs without returning to the original image data. 

Perhaps the most challenging aspect of constructing a galaxy catalog
is the issue of image segmentation, or deblending.  Galaxies that are
either well resolved, or blended with a physical neighbor or a chance
projection on the line-of-sight can be broken into
sub-components (depending on the $S/N$ and PSF).  Improperly
deblending overlaps can result in objects with unphysical
luminosities, colors, or shapes. Automated deblending
algorithms can be quite tricky, especially when galaxies are irregular
or have real substructure (think of a face-on Sc in a dense stellar
field).  
It will be important
to keep several levels of the deblending hierarchy in the catalog, as
well as have an efficient way to identify close neighbors.  Testing
and refining deblending algorithms is an important aspect of the
near-term preparations for LSST.

%
%
%
%
%
%
%
%
%
%
%
%
%
%
%
%
%
%
%
%
%
%
%
%
%
%
\subsection{Morphology}

\begin{figure}[ht]
\centerline{\resizebox{4.5in}{!}{\includegraphics{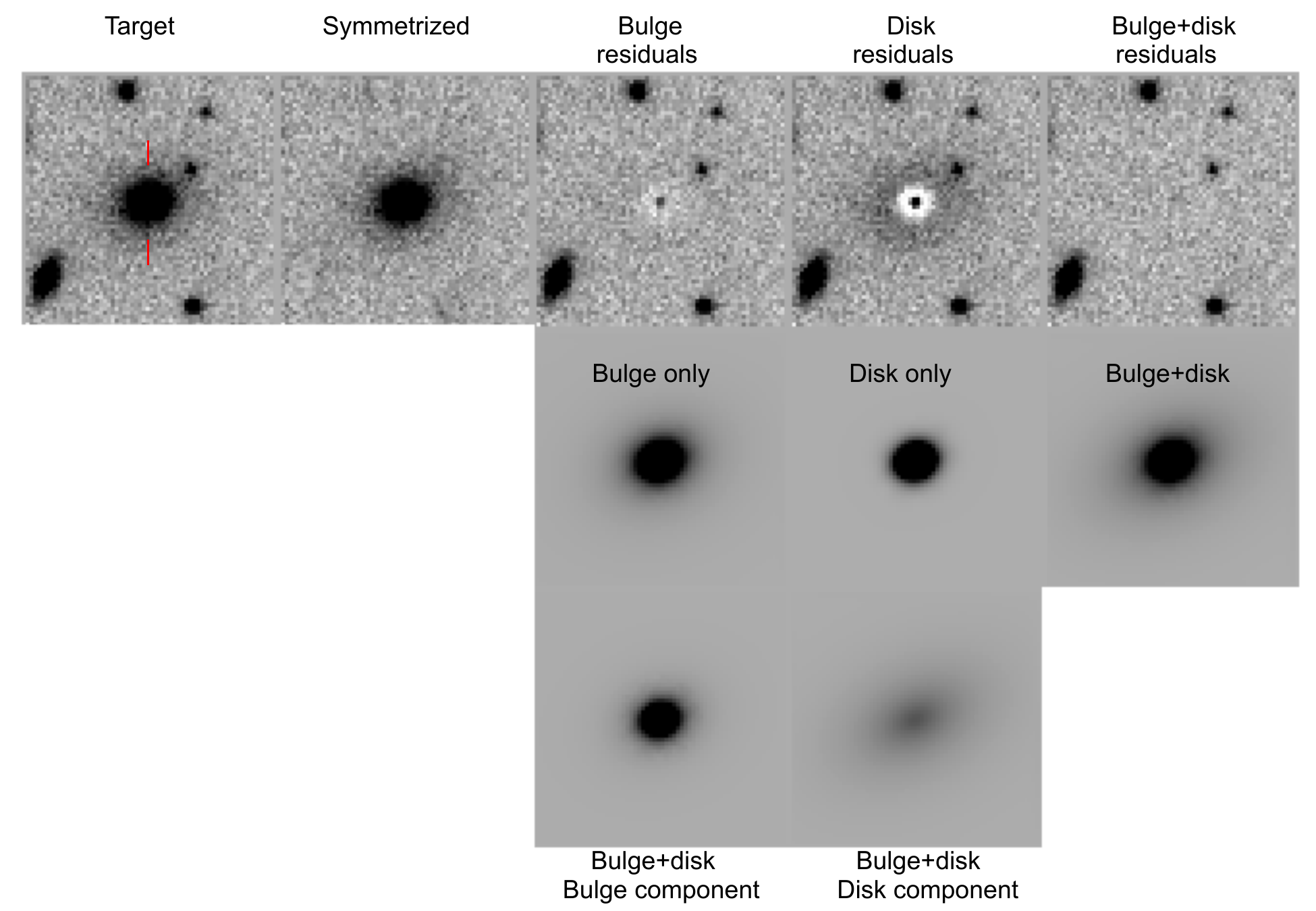}}}
\caption{Example of two-dimensional galaxy light profile fitting. The
  top left panel corresponds to the target galaxy, the next to the
  right is its symmetrized image, the next three show the
  residuals from bulge, disk, and bulge plus disk models respectively. The corresponding models are shown in the lower panel,
  with the bulge and disk (of the bulge plus disk model) components in
  the third row. This galaxy is best fit by an $r^{1/4}-$law or a
  bulge plus disk model.}
\label{2dgalaxymodels} 
\end{figure}

The excellent image quality that LSST will deliver will allow us to
obtain morphological information for all the extended objects with
sufficient signal-to-noise ratio, using parametric model fitting
and non-parametric estimation of various morphology indices.  The
parametric models, when the PSF is properly accounted for, will
produce measurements of the galaxy axial ratio, position angle and 
size. Possible models are a general Sersic model and more classical
bulge and disk decomposition. In the case of an $r^{1/4}-$law, the size
corresponds to the bulge effective radius, while for an
exponential disk, it is the disk scale-length. This process naturally
produces a measurement of the object surface brightness, either
central or median. An example of such a fit is shown in
\autoref{2dgalaxymodels} \citep{barrientos2004}.

The median seeing requirement of $0.7^{\prime\prime}$ corresponds to
$\sim 4$ kpc at $z=0.5$, which is smaller than a typical $L^*$ galaxy
scale-length. Therefore, parametric models will be able to
discriminate between bulge or disk dominated galaxies up to $z \sim
0.5 - 0.6$, and determine their sizes for the brightest ones. 
Non-parametric morphology indicators include concentration, 
asymmetry, and clumpiness (CAS; \citealt{2003ApJS..147....1C}) as
well as measures of the distribution function of galaxy pixel flux
values (the Gini coefficient) and moments of the galaxy image
($M_{20}$; \citealt{2004AJ....128..163L}).  



\section{Demographics of Galaxy Populations}
\label{sec:gal:demo}

It is useful for many purposes to divide galaxies into different
classes based on morphological or physical characteristics.  The
boundaries between these classes are often fuzzy, and part of the
challenge of interpreting data is ensuring that the classes are
defined sensibly so that selection effects do not produce artificial
evolutionary trends. Increasingly realistic simulations can help to
define the selection criteria to avoid such problems. Here we briefly
discuss the detectability of several classes of galaxies of interest
for LSST. We shall discuss the science investigations in more depth
later in this chapter. 

\begin{figure}[ht]
\centerline{\resizebox{4.5in}{!}{\includegraphics{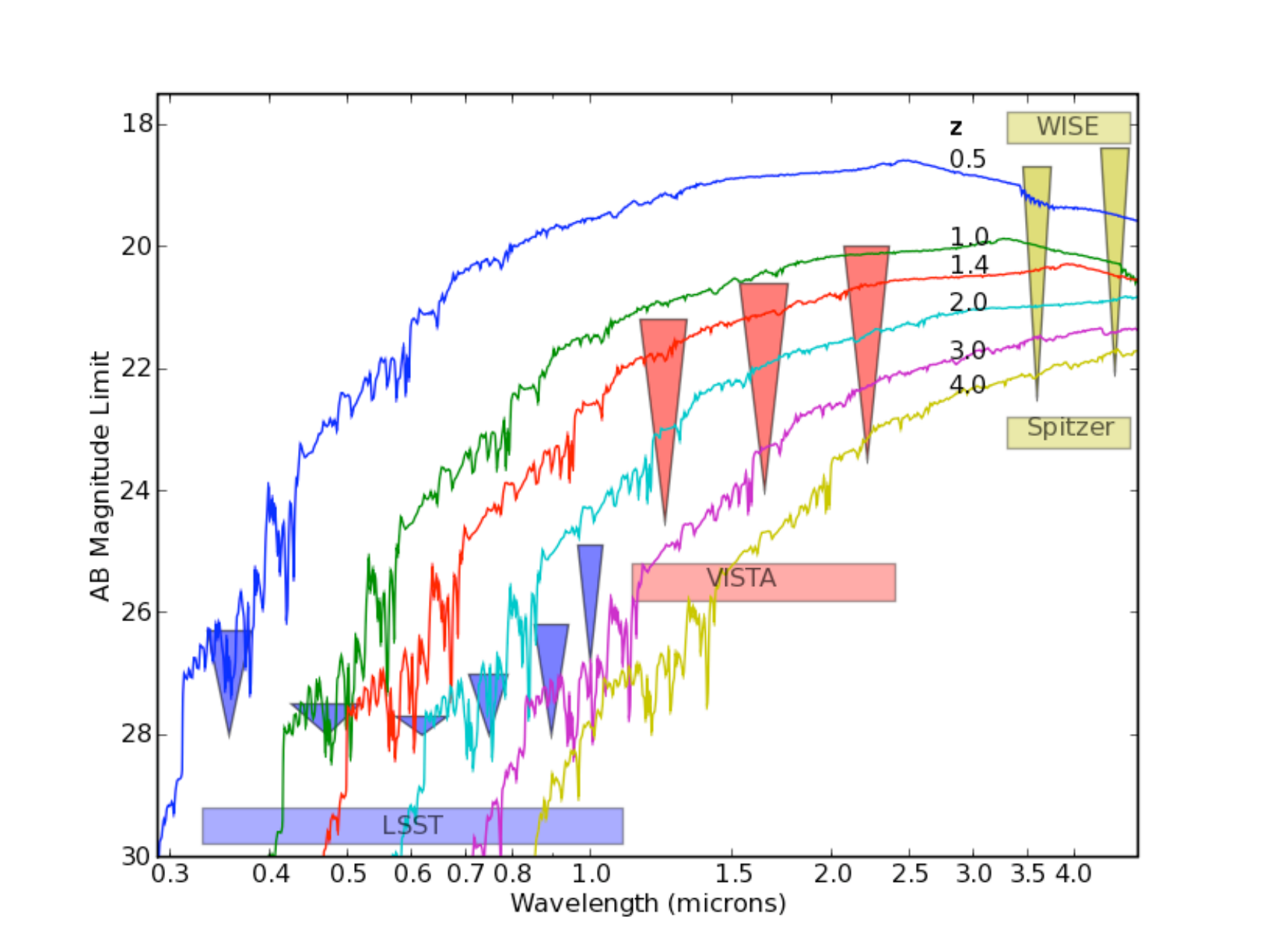}}}
\caption{\label{fig:gal:redsequence}
The spectrum of a fiducial red-sequence galaxy as a function of
redshift. The spectral energy distribution is from a \citet{2005MNRAS.362..799M} 
model, with 
solar metallicity, Salpeter IMF, with a star formation timescale of 0.1 Gyr, 
beginning to form stars at $z=10$, and normalized to an absolute $B_{AB}$ mag of 
$-20.5$ at $z=0$.  
Magnitude limits are indicated as blue triangles in the optical for
LSST, red triangles in the near-IR for VISTA and
yellow triangles in the mid-IR for Spitzer and WISE. The wide top of the triangle
shows the limits corresponding to surveys of roughly 20,000 deg$^2$
(the VISTA Hemisphere Survey, and the WISE all-sky survey). 
The point of the triangle corresponds to depths reached over tens
of square degrees. For LSST we use a strawman for the deep drilling
fields (\autoref{sec:design:cadence})
that corresponds to putting 1\% of the time into each field (i.e., 10\% if there
are 10 separate fields). Apportioning the time in these fields at $9,1,2,9,40,39$\% 
in $ugrizy$ yields $5\sigma$ point source detection depths of
$28.0,28.0,28.0,28.0,28.0,26.8$, which is what is shown.
For VISTA the deep fields correspond to the VIDEO survey; 
for Spitzer they correspond to SWIRE.
}
\end{figure}

{\noindent\it Passively evolving galaxies.} Early-type galaxies, with
little or no star formation, represent roughly one-half of the
present day stellar mass density \citep{2003ApJS..149..289B}. These
galaxies formed their stars earlier and more rapidly than late-type
galaxies. They are more strongly clustered. It is likely that mergers
played an important role in their formation, contributing to their
rapid star formation rates and their kinematically hot structure. It
is also likely that some form of feedback or ``strangulation" prevented
the subsequent accretion and cooling of gas that would have led to
further star formation. With good sensitivity in the $i,z,$ and $y$
bands, LSST will be sensitive to $L^*$ early-type galaxies out to
redshifts $z \sim 2$ for the wide area survey, and, depending on
observing strategy, to $z \sim 3$ for the deep-drilling
fields. \autoref{fig:gal:redsequence} shows the LSST survey limits
compared to a 
passively-evolving $L^*$ early type galaxy.

\begin{figure}[ht]
\centerline{\resizebox{4.5in}{!}{\includegraphics{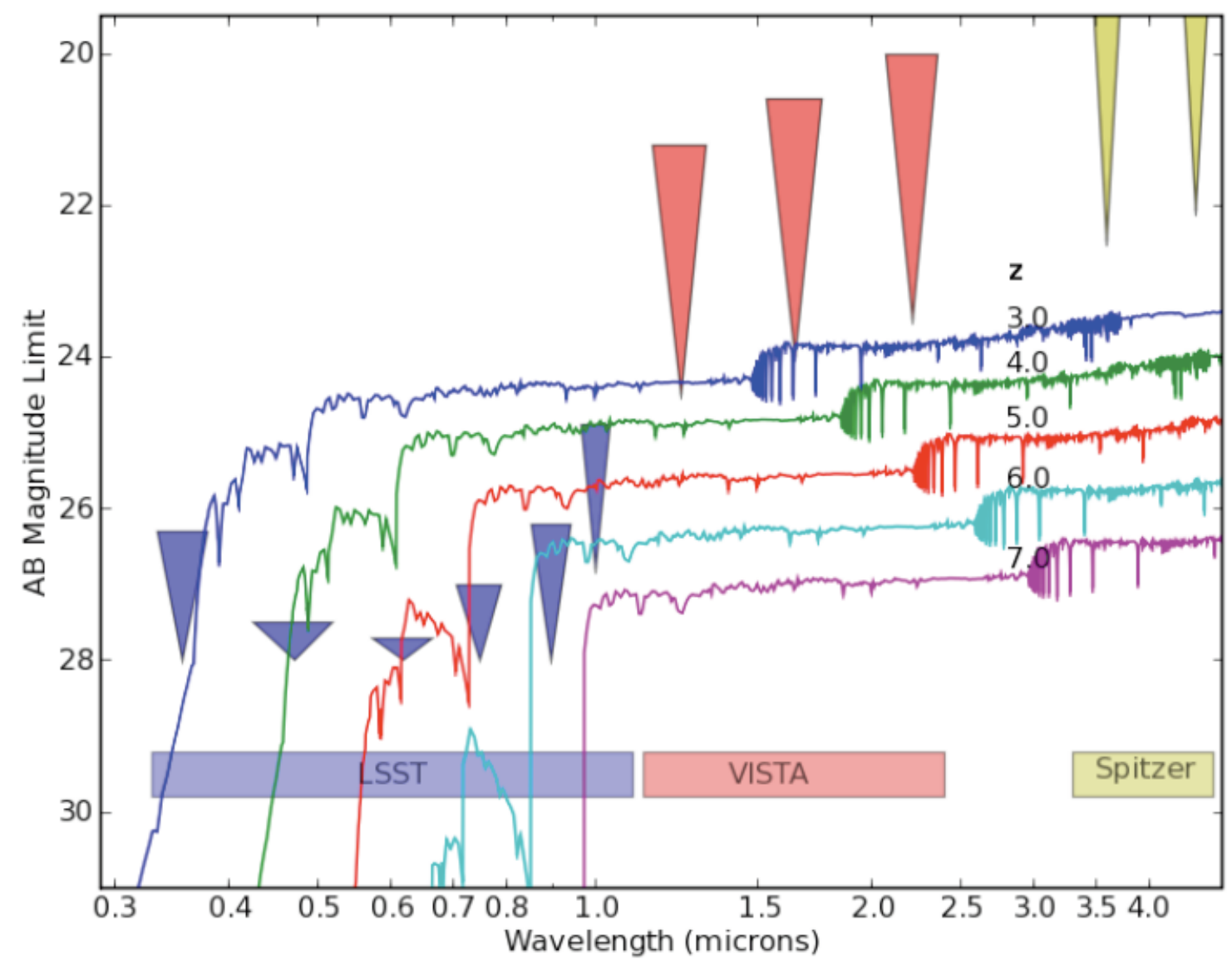}}}
\caption{\label{fig:gal:lbg}
Fiducial Lyman-break galaxy as a function of redshift. The spectral energy
distribution is a \citet{2003MNRAS.344.1000B} model, with 
solar metallicity, a Salpeter IMF, an age of 0.2 Gyr and a constant star formation rate,
viewed through a \citet{2000ApJ...533..682C} extinction screen with
$E(B-V) = 0.14$ \citep{2008ApJS..175...48R}. 
This is normalized to an absolute ${AB}$ mag at 1600 {\AA} of $-20.97,
-20.98, -20.64, -20.24$, and $-19.8$ 
at $z = 3,4,5,6,$ and 7 respectively \citep{2009ApJ...692..778R,2008ApJ...686..230B}.
Magnitude limits are the same as those shown in \autoref{fig:gal:redsequence}.
}
\end{figure}

{\noindent\it High-redshift star forming galaxies.} In the past decade
or so, deep surveys from the ground and space have yielded a wealth of
data on galaxies at redshifts $z>2$.  Photometric sample sizes have
grown to $>10^4$ galaxies at $z \sim 3$ and $>10^3$ galaxies at
$z>5$. (Spectroscopic samples are roughly an order of magnitude
smaller.)  However, we still have only a rudimentary understanding of
how star formation progresses in these galaxies, we do not know how
important mergers are, we and have only very rough estimates of the
relations between galaxy properties and halo mass. LSST will provide
data for roughly $10^9$ galaxies at $z>2$, of which $\sim 10^7$ will
be at $z>4.5$. Detection limits for LSST compared to a fiducial
evolving $L^*$ Lyman break galaxy are shown in \autoref{fig:gal:lbg}.

\begin{figure}[ht]
\begin{minipage}[b]{0.5\linewidth}
\centering
\includegraphics[width=7cm]{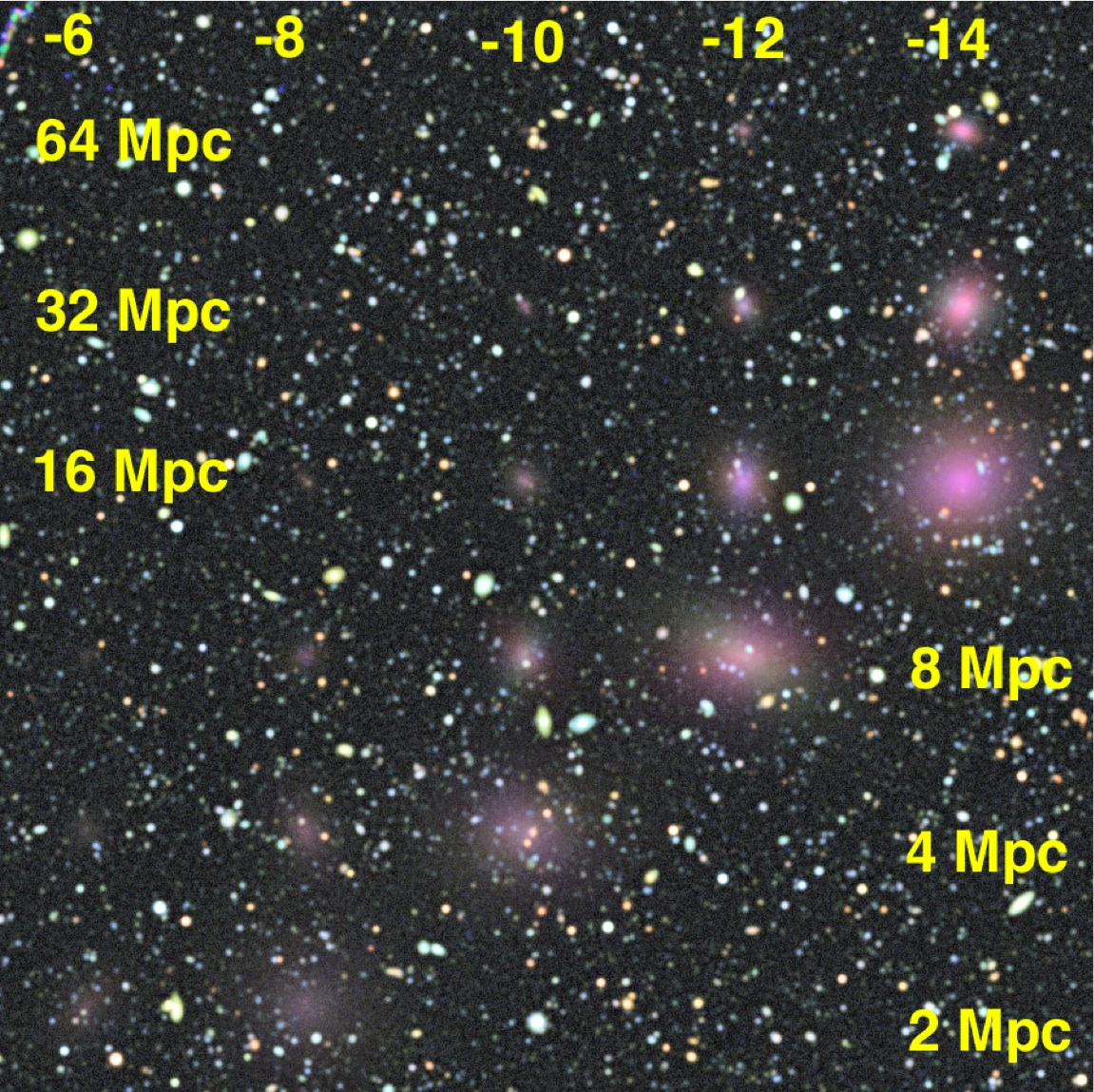}
\caption{
Dwarf spheroidal galaxy visibility. Dwarfs of various distances and
absolute magnitudes have been inserted into a simulated LSST image.
The simulation is for 50 visits (1500s)  each in dark time with
$g,r,i$. The background image is from the GOODS program
\citep{2004ApJ...600L..93G}, 
convolved with a $0.7^{\prime\prime}$ PSF with appropriate noise added. Sizes and colors for
the dwarfs are computed from the size-magnitude and mass-metallicity
relations of \citet{2008MNRAS.390.1453W} assuming a 10-Gyr old population.
}\label{fig:gal:dwarfs_image}
\end{minipage}
\hspace{0.2cm}
\begin{minipage}[b]{0.5\linewidth}
\centering
\includegraphics[width=8.3cm]{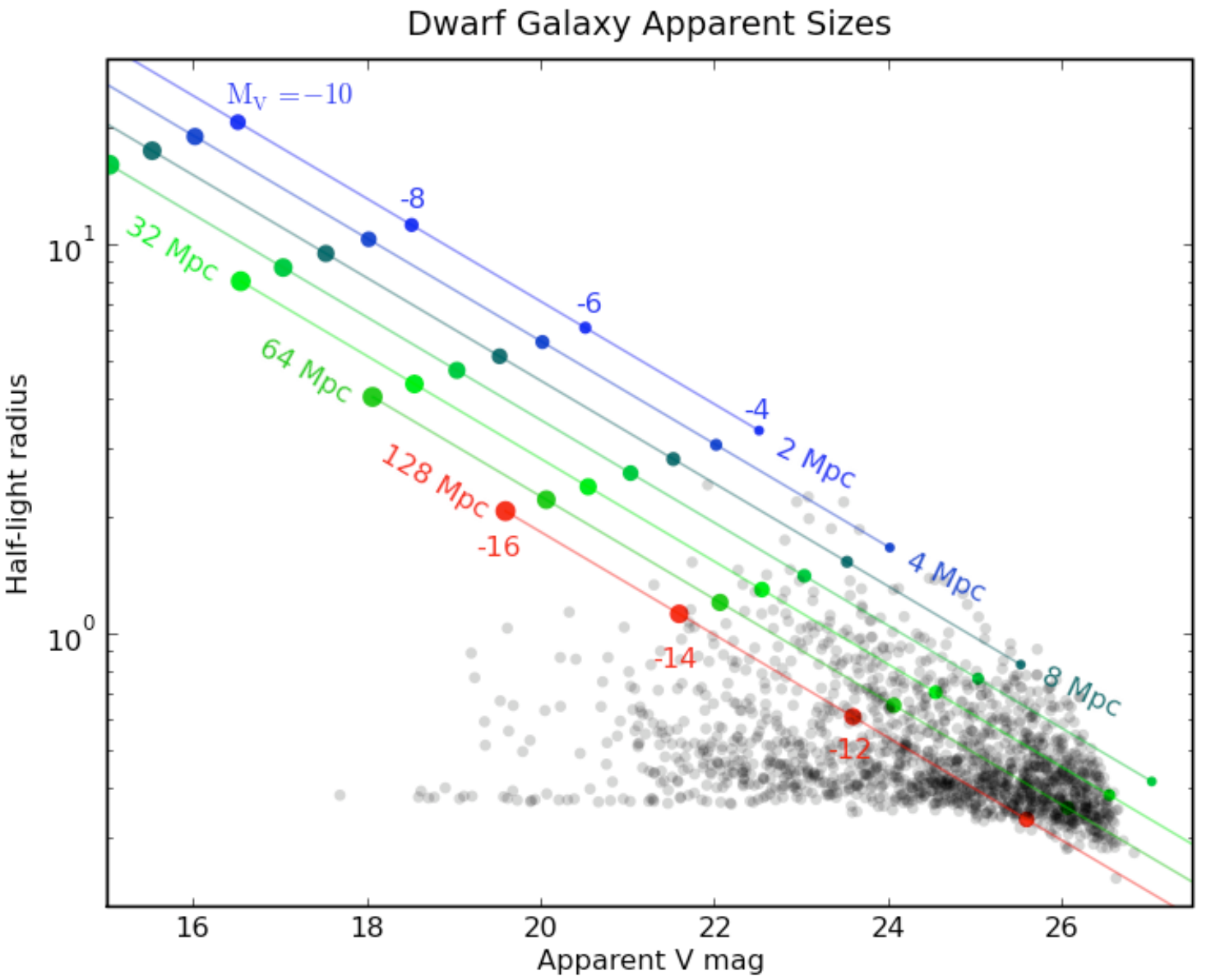}
\caption{
The colored points and lines show the half-light radii in arcsec for dwarf
galaxies as a function of magnitude for distances ranging from 2
to 128 Mpc computed from the scaling relation of 
\citet{2008MNRAS.390.1453W}. 
The gray points show the sizes of typical background galaxies
measured from the simulation in \autoref{fig:gal:dwarfs_image}.  
A dwarf galaxy with $M_V= -4$ should be visible and distinguishable from 
the background out to $\sim 4$ Mpc; a dwarf with $M_V= -14$ at 128 Mpc will be 
larger than most of the background galaxies of the same apparent magnitude.
}\label{fig:gal:dwarfs_magradius}
\end{minipage}
\end{figure}
\nocite{2004ApJ...600L..93G}
\nocite{2008MNRAS.390.1453W}

{\noindent\it Dwarf galaxies.} LSST will be very useful for studies of
low luminosity galaxies in the nearby Universe. Blind H I surveys and
slitless emission-line-galaxy surveys have given us reasonable
constraints on the luminosity function and spatial distribution of
gas-rich, star forming galaxies. However, most of the dwarfs in the
local group lack H I or emission lines. Such dE or dSph galaxies tend
to have low surface brightnesses (\autoref{sec:gal:lsb}) and are difficult to find in shallow
surveys like the SDSS or 2MASS, and are also difficult targets for
spectroscopy.  Our census of the local Universe is highly incomplete
for such galaxies.  \autoref{fig:gal:dwarfs_image} shows some typical
example morphologies.  \autoref{fig:gal:dwarfs_magradius} shows the
magnitude--radius relation for dwarf galaxies at a variety of
distances.  Nearby dwarf galaxies 
within a few Mpc and distant faint galaxies are well-separated in this
space; their low photometric redshifts will further help to
distinguish them.  An important question will be the extent to which
systematic effects in the images (scattered light, sky subtraction
issues, deblending, flat-fielding) will limit our ability to select
these low surface brightness galaxies. 

{\noindent\it Mergers and interactions.} The evolution of the galaxy
merger rate with time is poorly constrained, with conflicting results
in the literature. LSST will provide an enormous data set not only for
counting mergers as a function of redshift, but also for quantifying
such trends as changes in color with morphology or incidence of AGN
versus merger parameters. LSST is comparable to the CFHTLS-Deep survey
in depth, wavelength coverage, seeing, and plate scale --- but covers
an area 5000$\times$ larger. Scaling from CFHTLS, we expect on the order of 15 million galaxies will have detectable signs of strong tidal
interactions.  At low redshift, LSST will be useful for detecting
large-scale, low surface brightness streams, which are remnants of
disrupted dwarf galaxies (\autoref{sec:gal:merger}).  

\section{Distribution Functions and Scaling Relations}
\label{sec:galaxies:distfunct}

One of the key goals of the Galaxy Science Collaboration
is to measure the
multivariate properties of the galaxy population including trends with
redshift and environment.  This includes observed galaxy properties,
including luminosities, colors, sizes, and morphologies, as well as
derived galaxy properties, including stellar masses, ages, and star
formation rates, and how the joint distribution of these galaxy
properties depends on redshift and environment as measured on a wide
range of scales.

Galaxy formation is inherently stochastic, but is fundamentally
governed (if our theories are correct) by the statistical properties
of the underlying dark matter density field.  Determining how the
multivariate galaxy properties and scaling relations depend on this
density field, and on the underlying distribution and evolution of dark
matter halos, is the key step in connecting the results of large
surveys to theoretical models of structure formation and galaxy
formation.  We describe this dark matter context further in the
following section. 

A complete theory of galaxy formation should reproduce the fundamental
scaling relations of galaxies and their scatter as a function of
redshift and environment, in the high dimensional space of observed
galaxy properties.  Unexplained scatter, or discrepancies in the
scaling relations, signals missing physics or flaws in the model.  We
need to be able to subdivide by galaxy properties and redshift with
small enough errors to quantify evolution at a level compatible with
the predictive capability of the next generation of simulations.
By going both deep and wide, LSST is unique in its ability to quantify the
global evolution of the multivariate distribution galaxy properties.

Indeed, the consistency of these properties (e.g., the
luminosity function in different redshift slices) across the full
survey may well be an important cross-check of calibration and
photometric redshift accuracy. The massive statistics may reveal
subtle features in these distributions, which in turn could lead to
insight into the physics that governs galaxy evolution.

\subsection{Luminosity and Size Evolution}

The tremendous statistics available from billions of galaxies will
allow the traditional measures of galaxy demographics and their
evolution to be determined with unprecedented precision.  The
luminosity function, $N(L)\,dL$, gives the number density of galaxies
with luminosity in the interval $[L,L+dL]$.  It is typically
parametrized by a \citet{Schechter76} function.
LSST data will enable us to measure the luminosity functions of all
galaxy types at all redshifts, with the observed bands corresponding
to rest-frame ultraviolet-through-near-infrared at low redshift
($z<0.3$), rest-frame ultraviolet-to-optical at moderate redshift
($0.3<z<1.5$), and rest-frame ultraviolet at high redshift ($z>1.5$).
We will also determine the color distribution of galaxies in various
redshift bins, where color is typically measured as the difference in
magnitudes in two filters e.g., $g-r$, and this has a direct
correspondence to the effective power law index of a galaxy's spectrum
in the rest-frame optical.  Color reveals a combination of the age of
a galaxy's stellar population and the amount of reddening caused by
dust extinction, and we will use the great depth of LSST data to
expand studies of the galaxy color-morphology relation to higher
redshift and lower luminosity.  Image quality of $0.7''$ in deep
$r$-band images will allow us to measure the sizes of galaxies
(typically parametrized by their half-light or effective radii) out
to $z \sim 0.5$ and beyond. Size studies at higher redshift are
hampered by the nearly unresolved nature of galaxies caused by the
gradual decrease in galaxy sizes with redshift and the increase in
angular diameter distance until its plateau at $z>1$. Nevertheless,
LSST will provide unique data on the incidence of large galaxies at
high redshifts, which may simply be too rare to have appeared in any
great quantity in existing surveys.

\subsection{Relations Between Observables}

Broadly speaking, galaxies fall into two populations,
depending on their mass and their current star formation
rate \citep{kauffmann2003,bell2004}.  Massive galaxies
generally contain old, passively evolving stellar populations, while
galaxies with ongoing star formation are less massive. This bimodality
is clearly expressed in color-magnitude  diagrams.  Luminous
galaxies populate a tight red sequence, and star forming ones inhabit
a wider and fainter blue cloud, a landscape that is observed at all
times and in all environments from the present epoch out to at least
$z \sim 1$.

The origin of this bimodality, and particularly of the red sequence,
dominates much of the present discussion of galaxy formation.  The
central questions include:  1) what path in the color-magnitude
diagram do galaxies trace over their evolutionary history?  2) what
physical mechanisms are responsible for the necessary ``quenching'' of
star formation which may allow galaxies to move from the blue cloud to
the red sequence?  and 3) in what kind of environment do the relevant
mechanisms operate during the passage of a typical galaxy from the
field into groups and clusters?  Answering the first question would
tell us whether galaxies are first quenched, and then grow in mass
{\it along} the red sequence (e.g., by dry mergers, in which there are
no associated bursts of star formation), or grow primarily through
star formation and then quench directly onto their final position on
the red sequence (see e.g., \citealt{faber2007}).  The latter two
issues relate more specifically to galaxy clusters and dense
environments in general.  We know that galaxies move more quickly to
the red sequence in denser environments -- the red sequence is already
in place in galaxy clusters by $z \sim 1.5$ when it is just starting
to form in the field.  So by studying the full range of environments
we should be able to make significant advances in answering the
central questions.

Because a small amount of star formation is enough to remove 
a galaxy from the red sequence, it is of great interest to quantify
the distribution of galaxy colors near the red sequence, in multiple
bands, and in multiple environments. This should allow us to make
great progress in distinguishing bursty and episodic star formation
from star formation that is being slowly quenched. 

\citet{2003ApJ...594..186B,2005ApJ...629..143B} 
have looked at the paired relations between photometric quantities in the
SDSS, and such relations have provided insights into the successes
and failures of the current generation of galaxy evolution models 
\citep{2008arXiv0812.4399G}. LSST will push these relations to lower
luminosities and surface brightnesses and reveal trends as a function
of redshift with high levels of statistical precision. 

The physical properties of galaxies can be more tightly constrained
when LSST data are used in conjunction with data from other
facilities.  For high-redshift galaxies, the rest-frame ultraviolet
luminosity measured by LSST reveals a combination of the
``instantaneous'' star formation rate (averaged over the past $\sim
10$ Myr) and the dust extinction; the degeneracy is broken by
determining the dust extinction from the full
rest-ultraviolet-through-near-infrared spectral energy distribution
(SED) and/or by revealing the re-radiation of energy absorbed by dust
at far-infrared-to-millimeter wavelengths.  At lower redshifts, LSST
probes the stellar mass, stellar age, and dust extinction at
rest-frame optical wavelengths, and degeneracies can be mitigated
using the full rest-ultraviolet-through-near-infrared SED to measure
stellar mass.  Luminosities in additional wavebands such as $L_X,
L_{\rm NIR}, L_{\rm MIR}, L_{\rm FIR}, L_{\rm mm}$, and $L_{\rm
  radio}$ can be added to the distribution function, revealing
additional fundamental quantities including the AGN accretion rate,
dust mass, and dust temperature.  Because most surveys that are deep
enough to complement LSST will cover much smaller area, coordinating
the locations of the LSST deep fields to maximize the overlap with
other facilities will be important.

\subsection{Quantifying the Biases and Uncertainties}

Because much of the power of LSST for galaxies will come from the
above-mentioned statistical distributions, it will be crucial
to quantify the observational uncertainties, biases, and
incompleteness of these distributions.  This will be done through
extensive simulations (such as those in
\autoref{fig:gal:dwarfs_image}), analyzed with the same pipelines
and algorithms 
that are applied to real data. The results of these simulations can be
used to construct transfer functions, which simultaneously
capture uncertainties, bias and incompleteness as a function of the
input model properties of the galaxies. 
A given galaxy image 
will suffer from different noise and blending
issues depending on where it falls in which images with which
PSFs. 
With thousands of realizations sampling the observational parameter
space of galaxies, one can build a smooth representation of the
probability distribution of recovered values with respect to the input
parameters, and thus quantify errors and biases. The deep drilling
fields can be used to validate these transfer functions for the
wide-deep survey.  These probability distributions can then be used
when trying to derive true scaling relations from the LSST data or to
compare LSST data to theoretical predictions.  

%
%
%
%
%
%
%
%
%
%
%
%
%
%
%
%
%
%
%
%
%
%
%
%
%
%
\section{Galaxies in their Dark-Matter Context}

\label{sec:gal:dm}


In the modern galaxy formation paradigm, set in the context of
$\Lambda$CDM, structure forms hierarchically from small to large
scales.  Galaxies are understood to form at the densest peaks in this
hierarchical structure within bound dark matter structures (halos and
subhalos).  The properties of galaxies themselves are determined by
the physics of gas within the very local overdensity that forms the
galaxy (which depends on density, metallicity, and angular momentum),
and on the interactions between that specific overdensity and the
nearby overdensities (mergers, tides, and later incorporation into a
larger halo).  For instance, the dark matter dominates the potential
well depth and hence the virial temperature of a halo, which sets the
equilibrium gas temperature, determining whether infalling material
can cool efficiently and form stars \citep{1977ApJ...211..638S,
  1977MNRAS.179..541R, 1977ApJ...215..483B, 1978MNRAS.183..341W,
  2009MNRAS.395..160K, 2009Natur.457..451D}.  Similarly, galaxy
properties can change radically during mergers; and the merger history
of a galaxy is also intimately related to the merger history of its
underlying halo, which can be very different in halos of different
masses (e.g., \citealt{1993mnras.262..627L},
\citealt{2002ApJ...568...52W}).


Connecting galaxies to their underlying dark matter halos allows one
to understand their cosmological context, including, in a statistical
sense, their detailed merging and formation histories.  These
relationships are not merely theoretical; the distribution of galaxy
properties changes radically from the low-mass, high star formation
rate galaxies near cosmic voids, where halo masses are low, to the
quiescent, massive early-type galaxies found in the richest clusters,
where dark matter halo masses are very high.


\subsection{Measuring Galaxy Environments with LSST}

One way of exploring these relationships is to measure the variation
of galaxy properties as a function of the environment in which a galaxy is
found (e.g., cluster vs. void), using the local overdensity of
galaxies as a proxy for the local dark matter density.  However,
environment measures for individual galaxies are noisy even with
spectroscopic samples, due to sparse sampling and the increase of
peculiar velocities in dense environments; a solution is to measure
the average overdensity of galaxies as a function of their properties
\citep{2003ApJ...585L...5H}, a formulation which minimizes errors and
bin-to-bin correlations.  The situation is much worse for photometric
samples, as simulations demonstrate that local overdensity is very
poorly determined if only photometric redshifts are available
\citep{2005ApJ...634..833C}.  The problem is straightforward to see:
the characteristic size of clusters is $\sim 1 h^{-1}$ Mpc co-moving,
and the characteristic clustering scale length of galaxies is $\sim 4
h^{-1}$ Mpc co-moving for typical populations of interest, but even a
photometric redshift error as small as 0.01 in $z$ at $z=1$
corresponds to an 18 $h^{-1}$ Mpc error in co-moving distance.

As a consequence, with photometric redshifts alone it is {\it
  impossible} to determine whether an {\it individual object} is
inside or outside a {\it particular structure}.
%
Hence, just as with spectroscopic surveys, it is far more robust to
measure the {\it average overdensity as a function of galaxy
  properties}, rather than {\it galaxy properties as a function of
  overdensity}.  This may seem like an odd thing to do - after all, we
tend to think of mechanisms that affect a galaxy associated with a
particular sort of environment -- but in fact, a measurement of
average overdensity is equivalent to a measurement of the {\it
  relative large-scale structure bias} of a population -- a familiar
way of studying the relationship between galaxies and the underlying
dark matter distribution.

There are a variety of methods we will apply for this study.  Simply
counting the average number of neighbors a galaxy has (within some
radius and $\Delta z$) as a function of the central galaxy's
properties, will provide a straightforward measure of overdensity
analogous to environment measures used for spectroscopic surveys
\citep{2003ApJ...585L...5H, 2006MNRAS.370..198C}.  This idea is
basically equivalent to measuring the cross-correlations between
galaxies (as a function of their properties) and some tracer
population, a technique that can yield strong constraints on the
relationship between galaxies and the underlying dark matter
(\autoref{sec:galaxies:clustering}).

These cross-correlation techniques can even be applied to study galaxy
properties as a function of environment, rather than the reverse:
given samples of clusters (or voids) in LSST, we can determine their
typical galaxy populations by searching for excess neighbors around
them with a particular set of galaxy properties.  Such techniques
(analogous to stacking the galaxy populations around a set of
clusters) have a long history
(e.g., \citealt{1974ApJ...194....1O,1980ApJ...236..351D}; more recently
\citealt{2009ApJ...699.1333H}), but will have unprecedented power in
  LSST data thanks to the accurate photometric redshifts (reducing
  contamination), great depth (allowing studies deep into the
  luminosity function), and the richness of galaxy properties to be
  measured.


\subsection{The Galaxy--Halo Connection}
\label{sec:gal:gal-halo}

Given our knowledge of the background cosmology (e.g., at the pre-LSST
level, assuming that standard dark energy models are applicable,
\autoref{sec:com:cos}), we can calculate the distribution of dark
matter halo masses as a function of redshift
\citep{2008ApJ...688..709T}, the clustering of those halos
\citep{2001MNRAS.323....1S}, and the range of assembly histories of a
halo of a given mass \citep{2002ApJ...568...52W}.  Both semi-analytic
methods and N-body simulations can predict these quantities, with
excellent agreement.  Uncertainties in the processes controlling
galaxy evolution are currently much greater than the uncertainties in
the modeling of dark matter.

What is much less constrained now, and almost certainly will still be
unknown in many details in the LSST era, is how {\it visible} matter
relates to this underlying network of dark matter.  It is in general
impossible to do this in an object-by-object manner (except in cases
of strong gravitational lensing, \autoref{chp:sl}), but in recent
years there has been considerable success in determining how many
galaxies of given properties will be found in a halo of a given dark
matter mass \citep[e.g.,][]{2002MNRAS.329..246B, 2004ApJ...608...16Z,
  2007MNRAS.376..841V, 2008ApJ...676..248Y, 2007ApJ...667..760Z, 2009ApJ...696..620C}.


The connection between a population of galaxies and dark matter halos
can be specified by its halo occupation distribution (or HOD)
\citep[][] {2002ApJ...575..587B}, which specifies the probability
distribution of the number of objects of a given type (e.g.,
luminosity, stellar mass, color, or star formation rate) and their
radial distribution given the properties of the halo, such as its mass
(and/or formation history).  The HOD and the halo model have provided
a powerful theoretical framework for quantifying the connection
between galaxies and dark matter halos.  They represent a great
advance over the linear biasing models used in the past, which assume
that the clustering properties of some population of interest will
simply be stronger than the clustering of dark matter by a constant
factor at all scales.

In the simplest HOD model, the multiplicity function $P(N|M)$ (the
probability distribution of the number of subhalos found within halos
of mass $M$) is set by the dark matter \citep{2004ApJ...609...35K}, and
the details of galaxy star formation histories map this multiplicity
function to a conditional luminosity function, $P(L|M)$.  This
deceptively simply prescription appears to be an excellent description
of the data \citep{2003MNRAS.340..771V, 2005ApJ...633..791Z,
  2006MNRAS.365..842C, 2007MNRAS.376..841V}.  A variety of recent
studies have also found that this approach can be greatly simplified
with a technique called abundance matching, in which the most massive
galaxies are assigned to the most massive halos monotonically (or
with a modest amount of scatter).  This technique has also been shown to
accurately reproduce a variety of observational results including
various measures of the redshift-- and scale--dependent spatial
clustering of galaxies \citep{1999ApJ...523...32C, 2004ApJ...609...35K,
  2006ApJ...647..201C, 2006MNRAS.371.1173V}.

There are several outstanding issues.  In the HOD approach, it is
unclear whether the galaxy distribution can be described solely by
properties of the halo mass, or whether there are other relevant halo
or environmental properties that determine the galaxy populations.
Most studies to date have considered just one galaxy property
(e.g., luminosity or stellar mass).  With better observations, the HOD
approach can be generalized to encompass the full range of
observed properties of galaxies.  Instead of the conditional
luminosity function $P(L|M)$ at a single epoch, we need to be
considering multi-dimensional distributions that capture the galaxy
properties we would like to explain and the halo properties that we
believe are relevant: $P(L,a,b,c,... | M,\alpha,\beta,\gamma,...)$,
where $a,b,c,...$ are parameters such as age, star formation rate,
galaxy type, etc., and $\alpha,\beta,\gamma,...$ are parameters of the
dark-matter density field such as overdensity on larger scales or
shape.

Measuring the distributions of galaxy properties
(\autoref{sec:galaxies:distfunct}) and their relationship to
environment (i.e., average overdensity) and clustering
(\autoref{sec:galaxies:clustering}) measurements on scales ranging
from tens of kiloparsecs to hundreds of Megaparsecs will allow us to
place strong constraints on this function, determining the
relationship between galaxy properties and dark matter, which will be
key for testing theories of galaxy evolution and for placing galaxies
within a cosmological context.  In addition to providing an
unprecedentedly large sample, yielding high precision constraints,
LSST will be unique in its ability to determine the dark matter host
properties for even extremely rare populations of galaxies.
We describe several of the techniques we can
apply to LSST data to study the relationship between galaxies and dark
matter in the remainder of this section.  In general, they are
applicable to almost any galaxy property that can be measured for a
sample of LSST galaxies, and thus together they give extremely
powerful constraints on galaxy formation and structure formation
models.



%
%
%
%
%
%
%
%
%
%
%
%
%
%
%
%
%
%
%
%
%
%
%
%
%
%
\subsection{Clusters and Cluster Galaxy Evolution}

The large area and uniform deep imaging of LSST will allow us to find
an unprecedented number of galaxy clusters.  These will primarily be
out to $z \sim 1.3$ where the LSST optical bands are most useful,
although additional information can be obtained using shear-selected
peaks out to substantially higher redshift \citep{2009ApJ...702..603A}.
This new sample of clusters will be an excellent resource for galaxy
evolution studies over a wide redshift range. LSST will allow studies
of the galaxy populations within hundreds of $\sim 1\times10^{15}
M_{\odot}$ clusters as well as hundreds of thousands of
intermediate mass clusters at $z > 1$.   For a discussion of
cluster-finding algorithms, the use of clusters as a
cosmological probe and estimates of sample sizes, see
\autoref{sec:lss:cluster} and \autoref{sec:sl:clusmf}.

\begin{figure}[ht]
\centerline{\resizebox{3.5in}{!}{\includegraphics{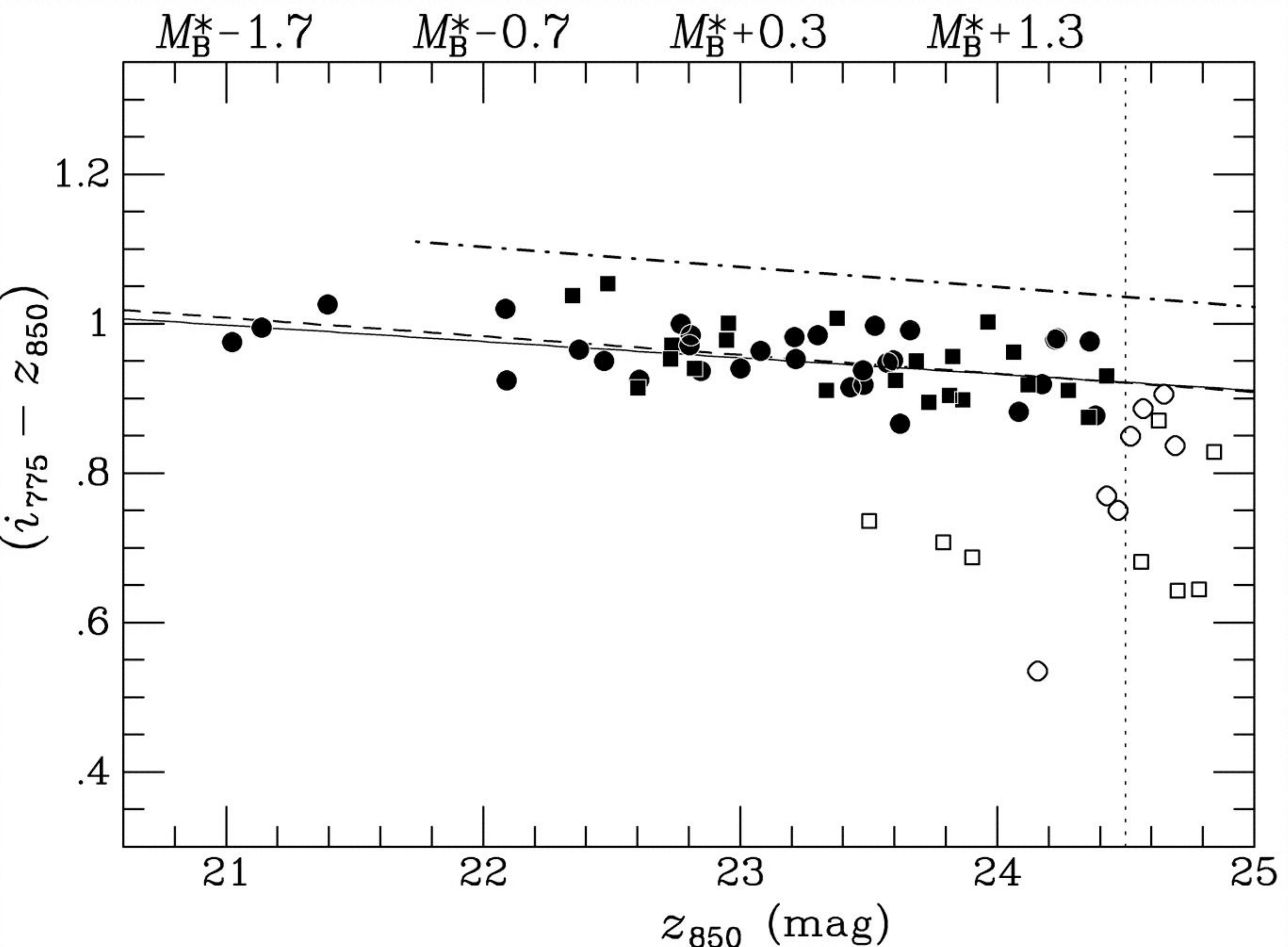}}}
\caption{Color-magnitude diagram for the cluster RDCS1252-2927 at
$z=1.24$ \citep{blakeslee2003}. A color-magnitude diagram of this
quality will be achieved with LSST in a single visit in two bands. 
}
\label{cmd_rdcs1252}
\end{figure}

Of particular interest will be the study of the red sequence populated
by early type galaxies, which is present in essentially all rich
clusters today.  This red sequence appears to be in place in at least
some individual clusters up to $z\sim 1.5$.  As an example,
\autoref{cmd_rdcs1252} shows that the red sequence is already well
defined in the cluster RDCS1252.9-2927 at $z=1.24$. The homogeneity in
the colors for this galaxy population indicates a high degree of
coordination in the star formation histories for the galaxies in this
cluster.  In general however, the role of the galaxy cluster
environment in the evolution of its member galaxies is not yet well
understood.  Issues include on what timescale and with what mechanism
the cluster environment quenches star formation and turns galaxies
red, and how this relation evolves with cosmic time.

Galaxy populations in a photometric survey can be studied using the
cross-correlation of galaxies with clusters, allowing a full
statistical characterization of the galaxy population as a function of
cluster mass and cluster-centric radius, and avoiding many of the issues
with characterizing galaxy environment from photometric redshift surveys. This
has been applied with much success to the photometric sample of the
SDSS \citep{2009ApJ...699.1333H}, where excellent statistics have
allowed selecting samples, which share many common properties (e.g., by
color, position, and whether central or satellite) to isolate
different contributions to galaxy evolution.  Although such studies
will be substantially improved by pre-LSST work e.g., from DES, LSST
will be unique in a few respects: 1) studies of the galaxy population
for the most massive clusters above $z \sim 1$; 2) studies of faint
galaxies in massive clusters for a well-defined sample from $z \sim
1.3$ to the present; and 3) studies of the impact of large scale
environment on the galaxy population.  At lower redshifts or with the aid of follow-up high-resolution imaging, LSST will also allow these
studies to be extended to include galaxy morphology information, so
that the morphological Butcher-Oemler effect \citep{goto2004} can be studied
as a function of redshift for large samples over a wide luminosity
range.

We are only beginning to systematically examine the outskirts of
clusters and their infalling groups at moderate redshifts, and this
effort will most likely be ongoing when LSST begins operations.  LSST
will be able to produce significant gains over the  state of
cluster research in the middle of the next decade by focusing on the
interface regions between cluster cores, groups, and superclusters.
These areas are particularly hard to study currently because the
galaxy densities are too low for targeted spectroscopic follow-up, and
large area spectroscopic studies do not cover enough volume at
moderate redshifts to effectively sample the relatively rare
supercluster type environments.

\subsection{Probing Galaxy Evolution with Clustering Measurements}
\label{sec:galaxies:clustering}

The parameters of halo occupation models of galaxy properties (described in \autoref{sec:gal:gal-halo}) may be established by
measuring the clustering of the population of galaxies of interest.
This may be a sample of galaxies selected from LSST alone or in
concert with photometry at other wavelengths.  The principal measure
of clustering used is the two-point correlation function, $\xi(r)$:
the excess probability over the expectation for a random, unclustered distribution that one object of a given class
will be a distance $r$ away from another object in that class.  This
function is generally close to a power law for observed populations of
galaxies, $\xi(r)=(r/r_0)^{-\gamma}$, for some scale length $r_0$,
typically $\sim 3-5 h^{-1}$ Mpc co-moving for galaxy populations of
interest, and slope $\gamma$, typically in the range 1.6--2.  However,
there is generally a weak break in the correlation function
corresponding to the transition between small scales (where the
clustering of multiple galaxies embedded within the same dark matter
halos is observed, the so-called ``one-halo regime") to large scales
(where the clustering between galaxies in different dark matter halos
dominates, the ``two-halo regime").  The more clustering properties are
measured (e.g., higher-order correlation functions or redshift-space
distortions, in addition to projected two-point statistics), the more
precisely the parameters of the relevant halo model may be determined
\citep{2004ApJ...608...16Z, 2008ApJ...688..709T}.

\begin{figure}[ht]
\label{fig:halomodel}
\centerline{\resizebox{6.0in}{!}{
\includegraphics{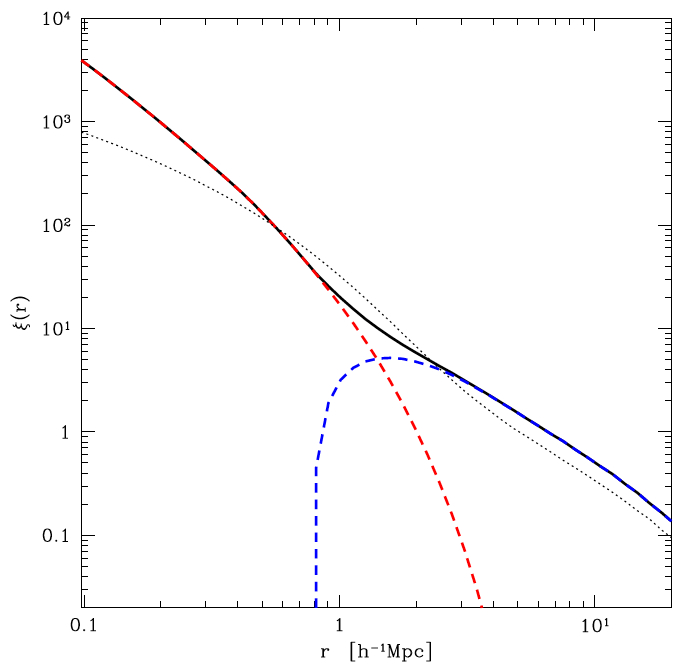}
\includegraphics{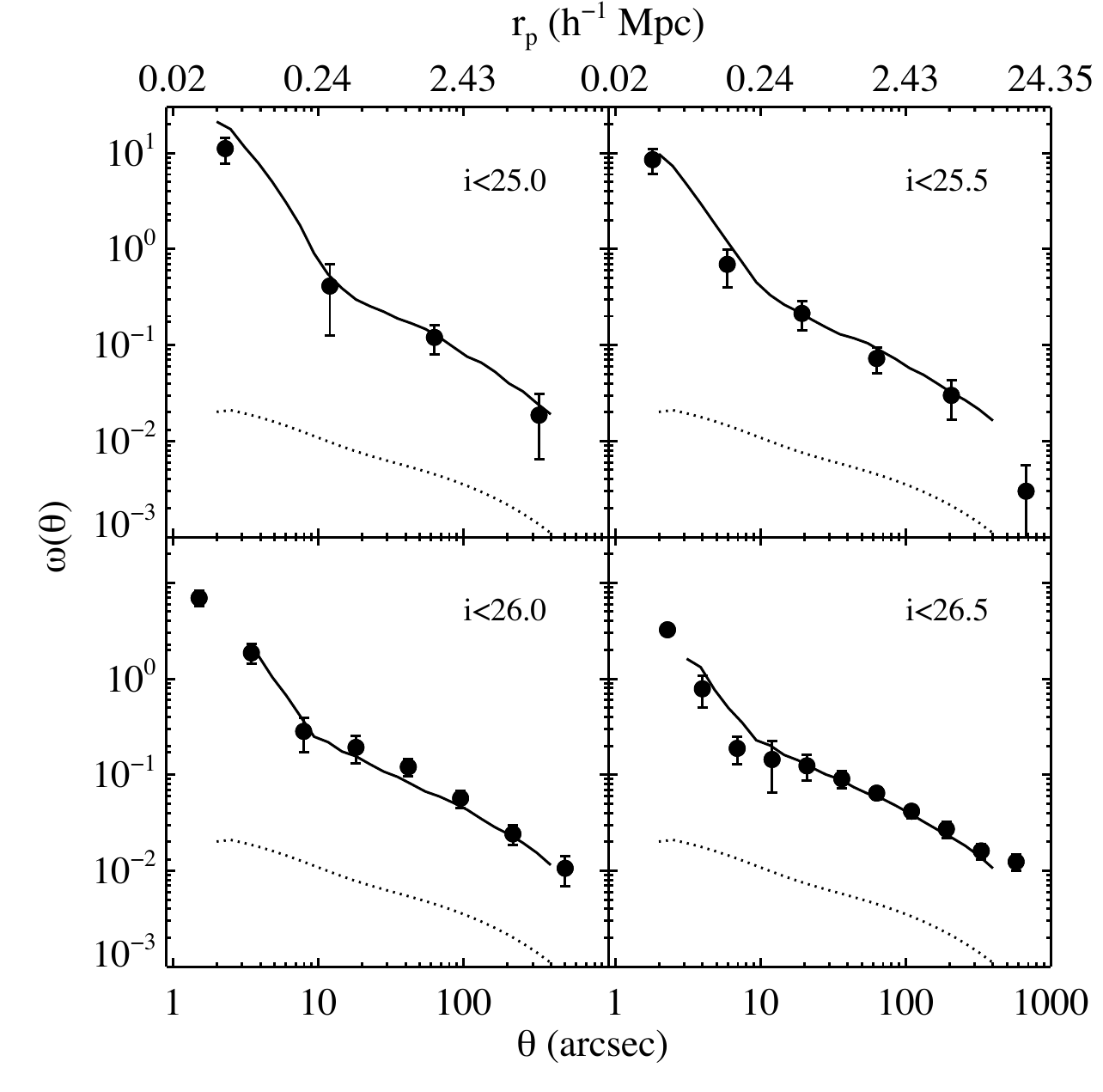}}}
\caption{
{\it Left:} A prediction for the correlation function $\xi(r)$ from the halo
model.  The dotted line shows the two-point real-space correlation
function for dark matter in a consensus cosmology
(\autoref{sec:com:cos}), while the solid 
black curve shows the predicted correlation function for a sample of
local galaxies with $r$-band absolute magnitude $M_r < -21.7$.  This
is the sum of two contributions.  The first is from galaxies within
the same dark matter halo (the ``one-halo term''), reflecting the
radial distribution of galaxies within halos, and shown by the red
dashed curve.  The second contribution comes from the clustering
between galaxies in different dark matter halos, reflecting the
clustering of the underlying halos; this ``two-halo term'' will be
greater for populations of galaxies found in more massive (and hence
more highly biased) dark matter halos.  Figure provided
by A. Zentner.
{\it Right:} Projected correlation function for galaxies at $z\sim4$
from the GOODS survey, compared to a model based on abundance
matching with dark matter halos and subhalos.  The scale of the typical
halo hosting the galaxies is clearly seen even in the projected correlations.
Figure from \protect{\cite{2006ApJ...647..201C}}.
}
\end{figure}


\subsubsection{Measuring Angular Correlations with LSST}

While models predict the real-space correlation function, $\xi(r)$,
for a given sample, we are limited to photometric redshifts, and thus
we we will measure the angular
two-point auto-correlation function $w(\theta)$: the excess probability
over random of finding a second object of some class (e.g., selected
in a slice in photometric redshift) an angle
$\theta$ away from the first one.  Given modest assumptions, the
value of $w(\theta)$ can be determined from knowledge of $\xi(r)$
through Limber's Equation \citep{1953ApJ...117..134L, 1980lssu.book.....P}:

\begin{equation}
w(\theta)= \frac{\int d\bar{z}  \left (\frac{dN}{d\bar{z}} \right)^2 \int
dl\, \xi(r(\theta,l),\bar{z})}{\left(  \int \frac{dN}{dz}\, dz
\right)^2}, 
\end{equation}

where $dN/dz$ is the redshift distribution of the sample (which may
have been selected, e.g., by photometric redshifts, to cover a
relatively narrow range), $l$ is the
co-moving separation of two objects along the line of sight, and
$\bar{z}$ is their mean redshift.  The quantity $r\approx (D_c^2 \theta^2 +
l^2)^{1/2}$, where $D_c$ is the co-moving angular size distance.  The
amplitude of $w$ will increase proportionally as $\xi$ grows greater,
but decrease as the redshift distribution $\frac{dN}{dz}$ grows wider,
as the angular correlations are diluted by projection effects.  

If we assume that $\xi(r)$ evolves only slowly with redshift, then for
a sample of galaxies with redshift distribution given by a Gaussian
centered at $z_0$ with RMS $\sigma_z$ (e.g., due to photometric
redshift errors or other sample selection effects), the amplitude of
$w$ will be proportional to $\sigma_z^{-1}$.  For the sparsest LSST
samples, the error in measuring $w(\theta)$ within some bin will be
dominated by Poisson or ``shot'' noise, leading to an uncertainty
$\sigma(w)=(1+w(\theta)) N_{p}(\theta)^{-1/2}$, where $N_{p}$ is the
number of pairs of objects in the class whose separations would fall
in that bin if they were randomly distributed across the sky.  This is
$N_p={1 \over 2} N_{gal}
\Sigma_{gal} (2\pi\theta \Delta\theta)$, where $N_{gal}$ is the number
of objects in the class of interest, $\Sigma_{gal}$ is the surface
density of that sample on the sky, and $\Delta\theta$ is the width of
the bin in $\theta$; since $N_p$ scales as $\Sigma_{gal}^2$,
$\sigma(w)$ decreases proportional to $\Sigma_{gal}^{-1}$.  

For large samples or at large scales, the dominant contribution to errors when measuring correlations with standard techniques is associated with the variance of the integral constraint \citep{1980lssu.book.....P}.  This variance is related
to the ``cosmic'' or sample variance due to the finite size of a field
-- i.e., it is a consequence of the variation of the mean density from
one subvolume of the Universe to another -- and is roughly equal to
the integral of $w(\theta)$ as measured between all 
possible pairs of locations within the survey \citep{Bernstein94}.  This
error is independent of separation and is highly covariant amongst all
angular scales.  However, it may be mitigated or eliminated via a suitable choice of correlation estimator (e.g., \citealt{2007MNRAS.376.1702P}).  For a 20,000 deg$^2$ survey with square
geometry (the most pessimistic scenario), for a sample with
correlation length, $r_0$, correlation slope, $\gamma=1.8$, and redshift
distribution described by a uniform distribution about $z=1$ with
spread, $\Delta z$, the amplitude of $w(\theta)$ will
be 
\begin{equation} 
w = 0.359 \left(\frac{\theta}{1~ {\rm arcmin}}\right)^{1-\gamma} 
\left(\frac{r_0}{4\, h^{-1} \rm Mpc}\right)^{\gamma} 
\left(\frac{\Delta z}{0.1}\right)^{-1}, 
\end{equation}
while the contribution to errors from the variance of the integral constraint will be approximately (Newman \&
Matthews, in preparation): 
\begin{equation} 
\sigma_{w,ic} \approx 5.8\times 10^{-4} \left(\frac{r_0}{4\, h^{-1}\rm
Mpc}\right)^{\gamma} \left(\frac{\Delta z}{0.1}\right)^{-1}. 
\end{equation}

For
sparse samples at modest angles, where Poisson noise dominates (or equivalently if we can mitigate the integral constraint variance) if we
assume the sample has a surface density of $\Sigma_{gal}$ objects
deg$^{-2}$ over the whole survey, the signal-to-noise ratio for a 
measurement of the angular correlation function in a bin in angle with
width 10\% of its mean separation will be 
\begin{equation} 
S/N = 47.4 \left(\frac{r_0}{4 \,h^{-1} {\rm Mpc}}\right)^{\gamma}
\left(\frac{\Delta z}{0.1}\right)^{-1} \left(\frac{\Sigma_{gal}}{100
~\rm deg^{-2}}\right)
\left(\frac{\theta}{1 ~{\rm arcmin}}\right)^{2-\gamma}.
\label{eqn:wtheta_sn}
\end{equation}

In contrast, for larger samples (i.e., higher $\Sigma_{gal}$), for which the variance in the integral constraint dominates, the
S/N in measuring $w$ will be nearly independent of sample properties,  $\sim 600(\theta/1 ~{\rm arcmin})^{1-\gamma}$.  Thus, even if the variance in the integral constraint is not mitigated, $w$ should be measured with
S/N of 25 or better at separations up to $\sim 0.9^\circ$, and with
S/N of 5 or better at separations up to $\sim 7^\circ$.  The
effectiveness of LSST at measuring correlation functions changes
slowly with redshift: the prefactor in \autoref{eqn:wtheta_sn} is
91, 55, 50, or 71 for $z=0.2$, 0.5, 1.5, or 3. 

As a consequence, even for samples of relatively rare objects -- for
instance, quasars (see \autoref{sec:agn:clustering} and
\autoref{fig:qso_correlation}), supernovae, or massive clusters of
galaxies -- LSST will be able to measure angular correlation functions
with exceptional fidelity, thanks to the large area of sky covered and
the precision of its photometric redshifts.  This will allow detailed
investigations of the relationship between dark matter halos and
galaxies of all types: the one-halo--two-halo transition
(cf. \autoref{fig:halomodel}) will cause $\sim 10$\% deviations of
$w(\theta)$ from a power law at $\sim$ Mpc scales in correlation
functions for samples spanning $\Delta z\sim 0.1$
\citep{2008MNRAS.385.1257B}, which will be detectable at $\sim
5\sigma$ even with highly selected subsamples containing $<0.1\%$ of all
galaxies from LSST.  The ensemble of halo models (or
parameter-dependent halo models) resulting from measurements of
correlation functions for subsets of the LSST sample split by all the
different properties described in \autoref{sec:gal:measurements} will
allow us to determine the relationship between the nature of galaxies
and their environments in unprecedented detail.  In the next few
years, we plan to develop and test techniques for measuring halo model
parameters from angular correlations using simulated LSST data sets, so
that we may more precisely predict what can and cannot be measured in
this manner.

\begin{figure}[ht]
\label{fig:biasz_lsst}
\begin{center}
\includegraphics[width=0.6\linewidth]{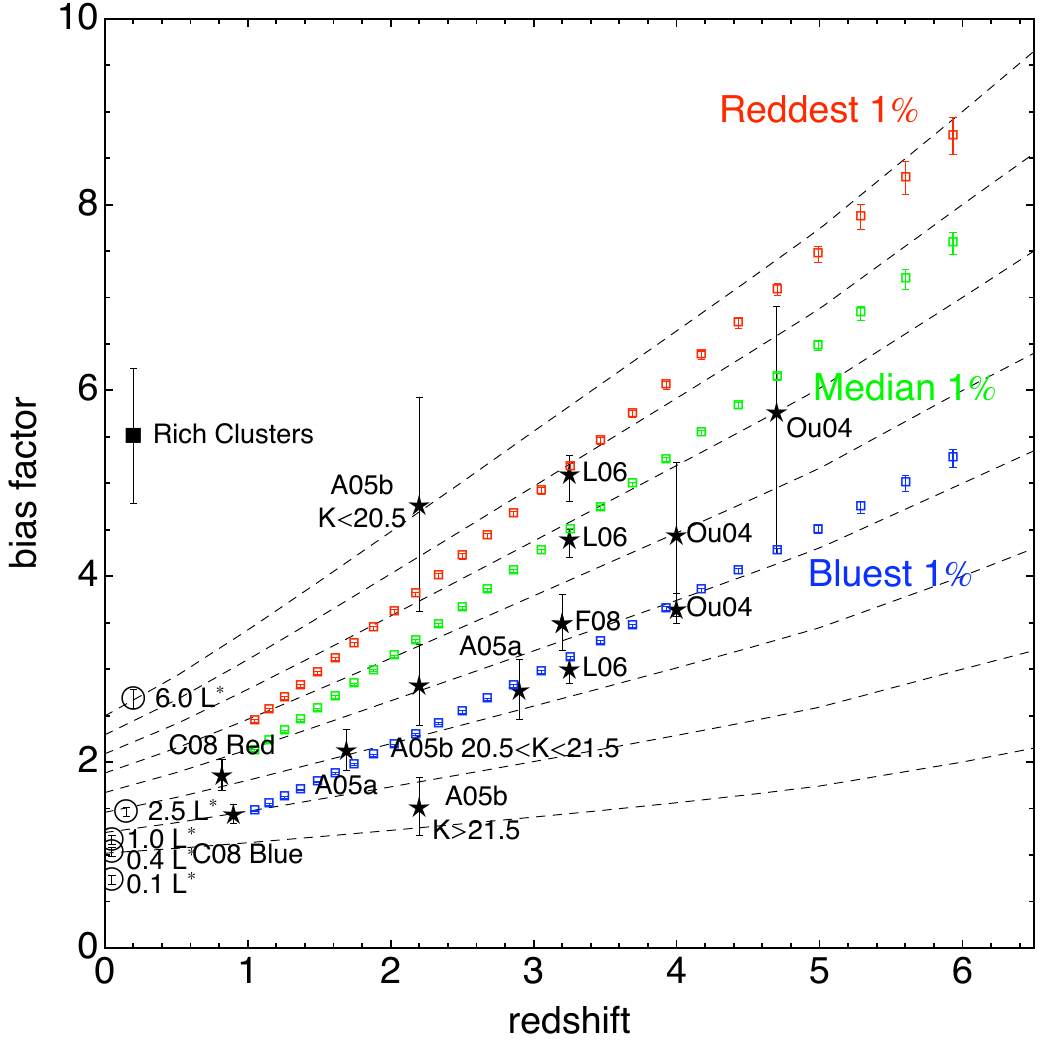}
\caption{
Evolution of galaxy bias versus redshift for three LSST galaxy samples at $1<z<6$.  The three samples are selected to be 
the 1\% of all 
LSST galaxies at each redshift 
that is the bluest/median/reddest in rest-ultraviolet color.
The dashed evolutionary tracks show the
evolution in bias factor versus redshift based on the Sheth-Tormen conditional mass function.
Points with error bars show a compilation of literature bias values for 
$z=1.7$ color-selected galaxies  
(A05a, \citealt{adelbergeretal05a}), 
$z=2.1$ color selected galaxies 
(A05b, \citealt{adelbergeretal05b}), 
$z\sim3$ Lyman break galaxies 
(A05a; F08, \citealt{franckeetal08}; 
L06, \citealt{leeetal06}), 
and $z>4$ Lyman break galaxies  
(Ou04, \citealt{ouchietal04}).  
Also shown are $z\sim1$ galaxies separated by color (C08, \citealt{coiletal08}), 
$z\sim0$ galaxies labeled by their optical luminosity, from \citet{zehavietal05}
and
rich galaxy clusters from \citet{bahcalletal03}.
}
\end{center}
\end{figure}

Measuring the spatial clustering of the dark matter halos hosting galaxies 
over a wide range of cosmic time will 
allow us to trace the evolution of galaxy populations 
from one epoch to another by identifying progenitor/descendant relationships. 
\autoref{fig:biasz_lsst} shows an example of what LSST will reveal about the clustering of galaxy populations as
a function of redshift.  Here, ``bias'' refers to the average fluctuation in number density of
a given type of galaxies divided by that of the dark matter particles.
The redshift bins were chosen to have width of  
$\delta_z=0.05 \times (1+z)$, i.e., somewhat broader than the
expected LSST photometric redshift uncertainties.
In this illustrative model, high-redshift galaxies discovered by LSST are broken into 100 subsets, with three of those subsets corresponding to the bluest, median, and reddest rest-ultraviolet color plotted.  Those three subsets 
were assumed to have a correlation length evolving as $(1+z)^{0.1}$.
Uncertainties were generated by extrapolating results from the 0.25
deg$^2$ field of \citet{franckeetal08},  
assuming Poisson statistics and a constant observed galaxy number density over $1<z<4$ that falls by a factor of ten by $z=6$ due to
the combination of intrinsic luminosity evolution and the LSST imaging depths.
In this particular model, the bluest galaxies at $z=6$ evolve into typical galaxies at $z\sim2$,
and typical galaxies at $z=6$ evolve into the reddest galaxies at $z\sim3$. The breakdown into 100 galaxy subsets based on color, luminosity, size, etc.
with such high precision represents
a tremendous improvement over current observations;  the 
figure illustrates the large error bars that result when breaking current 
samples into 2--3 bins of color (points labeled C08,A05b) or luminosity (points labelled L06,Ou04).

\subsubsection{Higher-order Correlation Functions}

Measuring higher-order correlation statistics (such as the three-point
function -- the excess probability of finding three objects with
specified separations from each other -- or the bispectrum, its
Fourier counterpart) provides additional constraints on the
relationship between galaxies and dark matter not available from
two-point statistics alone \citep[e.g.,][see also \autoref{sec:lss:bisp}]{Verde++02}.  Whereas the
techniques for measuring two-point statistics 
\citep{Martinez+Saar02} are quite mature, measuring and interpreting
higher-order correlation functions is an active field which will
evolve both before and during LSST observations.  Therefore, although
these higher-order correlation functions will be used to constrain the
relationship of galaxies to dark matter for broad galaxy samples
(e.g., linear, non-linear, or stochastic biasing models, or HOD-based
models), we expect that most of the effort in this field in the LSST
context will be in the large-scale-structure context, as described in
\autoref{sec:lss:bisp}, rather than focused specifically on galaxy evolution. 
The ultimate result of this research will be a calibration of the
large-scale structure bias of samples of galaxies observed by LSST, putting
relative bias measurements coming from two-point functions on an
absolute scale and improving all halo modeling.

%
%
%
%
%
%
%
%
%
%
%
%
%
%
%
%
%
%
%
%
%
%
%
%
%
%
\subsubsection{Cross-correlations}


\label{sec:galaxies:cross}

As described above, the auto-correlation function -- which measures
the clustering of objects in some class with other objects of the same
type -- can provide information about the relationship of those
objects to the underlying hierarchy of large-scale structure.  A
related quantity, the angular two-point, cross-correlation function
(the excess probability over Poisson of finding an object of one type
near an object of a second type, measured as a function of separation)
is a sensitive probe of the underlying relationships between any two
different classes of extragalactic objects.  

As an example, the
clustering of galaxies of some type (e.g., blue, star-forming galaxies)
around cluster centers provides a measurement of both the fraction of
those galaxies that are associated with clusters, and their average
radial distribution within a cluster.  Hence, even though with
photometric redshifts we {\em cannot} establish whether any {\em
individual} galaxy belongs to one particular cluster, we {\em can}
determine with high precision the {\em average} galactic populations
of clusters of a given sort (mass, richness, and so on).   

Cross-correlation functions are particularly valuable for studying
rare populations of objects for which they may be measured with much
higher S/N than auto-correlations.  The amplitude of the
angular correlation function between two classes, $A$ and $B$, with
redshift distributions, $d N_A/dz$ and $d N_B/dz$, is:

\begin{equation}
w_{AB}(\theta)= \frac{\int dz \, \frac{dN_A}{dz} \int dz'\, \xi_{AB}(r\theta,z')
\,\frac{dN_B}{dz'}}{\left( \int dz'\, \frac{dN_A}{dz'} \right) \left(
\int dz\, \frac{dN_B}{dz} \right)}.
\end{equation}

In the weak-clustering regime, which will generally be applicable for
LSST samples at small scales, the error in $w_{AB}(\theta)$ will be
$(1+w_{AA})^{1/2} (1+w_{BB})^{1/2} N_{AB}(\theta)^{-1/2}$, where
$w_{AA}$ is the auto-correlation of sample A, $w_{BB}$ is the
auto-correlation of sample B, and $N_{AB}$ is the number of pairs of
objects in each class separated by $\theta$, if the samples were
randomly distributed across the sky.

In the limit that the redshift distributions of samples A and B are
identical (e.g., because photometric redshift errors are comparable for
each sample), the auto-correlations and cross-correlations of samples A
and B have the familiar power law scalings, 
the S/N for measuring $w_{AB}$ 
will be larger than that of $w_{AA}$ on large scales by a factor of $(r_{0,AB}/r_{0,AA})^\gamma
(\Sigma_B/\Sigma_A)^{1/2}$, where $\Sigma_A$ and $\Sigma_B$ are the
surface densities of samples A and B on the sky and $r_{0,AB}$ and
$r_{0,AA}$ are the scale lengths for the cross-correlation and
auto-correlation functions.  

\subsubsection{Cross-correlations as a Tool for Studying Galaxy Environments}

\label{sec:galaxies:crossenv}

It would be particularly desirable to measure the clustering of
galaxies of a given type with the underlying network of dark matter.
One way of addressing this is measuring the lensing of background
galaxies by objects in the class of interest (\autoref{wl-sec-gglensing}); 
this is not possible for rare objects, however.  An
alternative is to determine the cross-correlation between objects in
some class of interest with all galaxies at a given redshift
(``tracers'').  This function, integrated to some maximum separation
$r_{max}$, will be proportional to the average overdensity of galaxies
within that separation of a randomly selected object.  For linear
biasing, this quantity is equal to the bias of the tracer galaxy
sample times the overdensity of dark matter, so it is trivial to
calculate the underlying overdensity.  The mapping is more complicated
if biasing is not linear; however, the exquisitely sensitive
correlation function measurements that LSST will provide will permit halo
modeling of nonlinear bias allowing accurate reconstruction. 

\begin{figure}[ht]
\centerline{\resizebox{5in}{!}{
\includegraphics{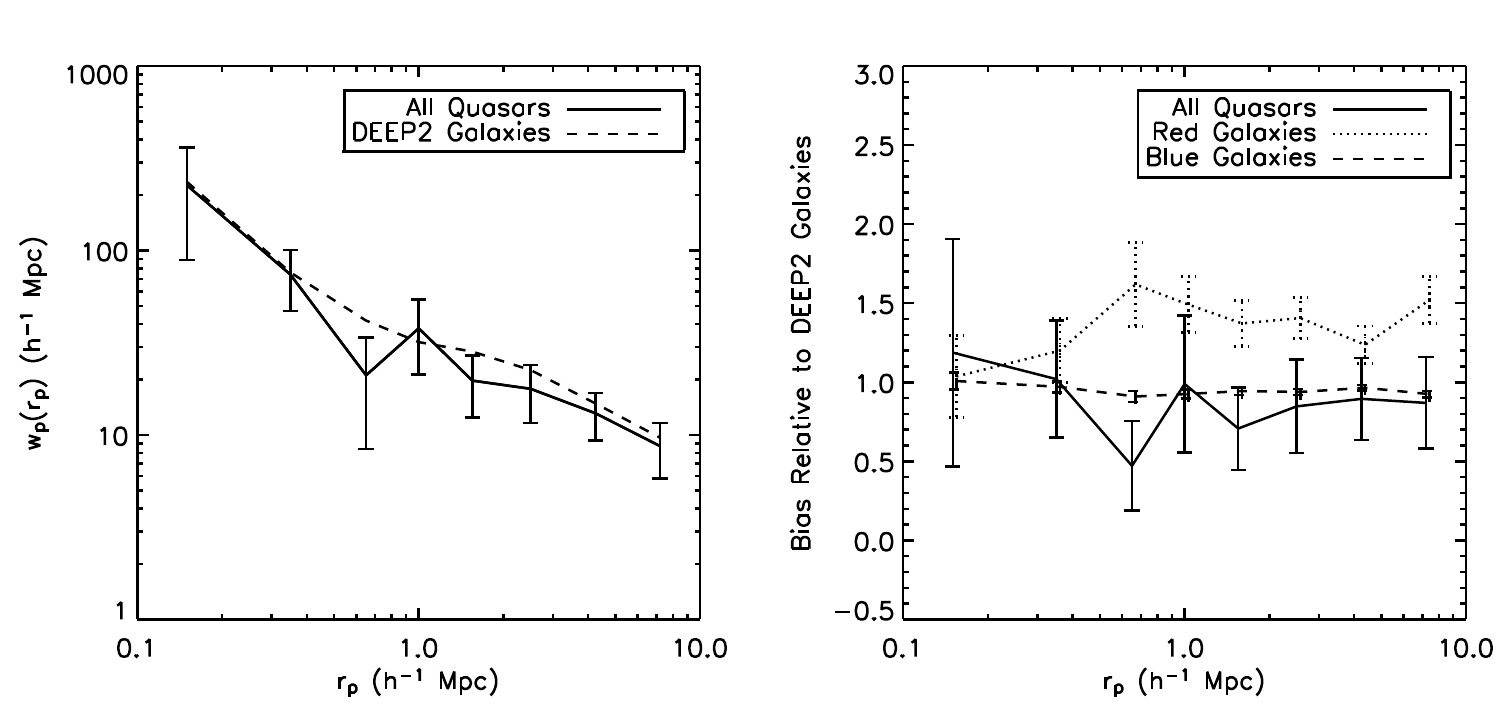}}}
\caption{\label{fig:gal:xcorr}
A demonstration of the power of cross-correlation techniques for rare samples, from
\citet{2007ApJ...654..115C}.  The left panel shows the projected two-point cross-correlation between
a sample of only 52 quasars at $0.7 < z < 1.4$ identified using
spectroscopy from the SDSS or the DEEP2
Galaxy Redshift Survey, and a comparison sample of $\sim 5000$ DEEP2
galaxies.  The dashed curve indicates the auto-correlation of the
comparison galaxy sample.  From these measurements, Coil et al. determined
the relative bias of quasars to the DEEP2 galaxy sample, and with similar
techniques measure the relative bias of blue or red galaxies within
DEEP2 to the overall sample, as shown in the right panel.  
See \autoref{fig:qso_correlation} for predicted errors for LSST data.}
\end{figure}

This measurement is essentially equivalent to the average
overdensity measured from large-scale galaxy environment studies
\citep{2007ApJ...664..791B, 2005ApJ...629..143B, 2008MNRAS.390..245C};
an advantage is that clustering measurements can straightforwardly
probe these correlations as a function of scale.  With LSST, such
comparisons will be possible for even small samples, establishing the
relationship between a galaxy's multivariate properties and the
large-scale structure environment where it is found; see
\autoref{fig:gal:xcorr} for an example of the utility of
cross-correlation techniques.

The cross-correlation of two samples is
related to their auto-correlation functions by factors involving both
their relative bias and the stochastic term \citep{Dekel+Lahav01},
thus one can learn something about the extent that linear 
deterministic bias holds for the two samples \citep{Swanson++08}. 
As another example, associating blue,
star forming galaxies with individual galaxy clusters will be fraught
with difficulties given photometric redshift errors, but the
cluster-blue galaxy cross-correlation function will determine both the
fraction of blue galaxies that are associated with clusters and also
their average radial distribution within their host clusters
(see \citealt{Coil++06} for an application with spectroscopic samples).  
This will
allow us to explore critical questions such as what has caused the
strong decrease in galaxies' star formation rates since $z\sim 1$,
what mechanism suppresses star formation in early-type galaxies, and
so on.  

AGN may provide one critical piece of this puzzle;
feedback from AGN can influence the cooling of gas both on the scale
of galaxies and within clusters 
\citep{2006MNRAS.365...11C, 2008ApJS..175..356H}
and the black-hole mass/bulge-mass correlation strongly suggests that
black hole growth and galaxy growth go hand-in-hand.  By measuring the
cross-correlation of AGN (e.g., selected by variability) with galaxies (as a
function of their star formation rate, for instance) and with
clusters, we can test detailed scenarios for these processes.  See the
discussion in \autoref{sec:agn:clustering}.  
The
evolution of low-mass galaxies within larger halos could also be
influenced by tides, mergers, gas heating and ionization from nearby
galaxies, and other effects;
mapping out the types of
galaxies found as a function of cluster mass and clustocentric
distance can constrain which of these phenomena is most important.

Cross-correlation against LSST samples will also boost the utility of a
variety of future, complementary multi-wavelength data sets.  Even unidentified classes of
objects found at other wavelengths (e.g., sub-millimeter sources,
sources with extreme X-ray to optical brightness ratios,
etc.) may be localized in redshift and their dark matter context
identified by measuring their correlation with galaxies or structures
of different types and at different redshifts; cross-correlations will
be strong only when objects of similar redshift and halo mass are used
in the correlation.  In this way, LSST data will be a vital tool for
understanding data sets which may be obtained long after the survey's
completion.


\subsubsection{Cross-correlations as a Tool for Studying Galaxy Dust}

\label{sec:galaxies:crossdust}

Another application of cross-correlation techniques is to measure properties of the dust 
content of dark matter halos and the intergalactic medium.  For a given
redshift slice of galaxies, the light from galaxies behind the sample
has to travel through the dust associated with the foreground galaxies.
\citet{2009arXiv0902.4240M} show that the dust halos surrounding
field galaxies in the SDSS generates a detectable reddening in the
colors of background quasars.  By cross-correlating quasar colors (rather than the positions of quasars) with
foreground galaxy density, \citet{2009arXiv0902.4240M} were able to
detect dust halos extending well beyond 100$h^{-1}$kpc for typical
0.5$L^*$ galaxies.  This, in turn, leads to an opacity of the Universe
which is a potential source of systematic bias for planned supernova
surveys \citep{2009arXiv0903.4199M}.

With LSST, we will be able to extend these measurements in a number
of ways.  Of particular interest is looking at the evolution of
these dust halos as a function of redshift.  With the relatively
shallow depth of the SDSS data (and the need for high foreground and
background object density on the sky to detect the signal),
measurements with current data will be limited to redshifts below $z
\sim 0.5$.  With the much greater depth available in LSST, these
limits should be doubled at the least, perhaps even taken as high as
$z \sim 2$, depending on the efficiency of finding $r$ and $i$ band
drop-out galaxies.  Going to higher redshifts will mean a stronger
signal as the rest-frame wavelength of the background sample light
shifts to the ultraviolet where extinction should be stronger.  More
importantly, however, this shift into the UV will break a number of
degeneracies in the current measurements, which are unable to
distinguish between Milky Way or LMC-like extinction curves.  This, in
turn, would tell us if the bulk of the dust was more silica or
graphite-based \citep{1984ApJ...285...89D} and offer clues as to how
these extended dust halos may have formed.

%
%
%
%
%
%
%
%
%
%
%
%
%
%
%
%
%
%
%
%
%
%
%
%
%
%

\section{Galaxies at Extremely Low Surface Brightness}
\label{sec:gal:lsb}

As the deepest wide-field optical survey currently planned, LSST will
push observations of galaxies to lower surface brightness than has
ever been available over such a large field. This capability will
allow a better understanding of the outskirts of galaxies, of the
merger history of galaxies, of the role of tidal stripping in groups
and clusters, and of the lowest surface brightness dwarfs and their
evolution.  In \autoref{sec:MW:UFs}, we discussed the
discovery of nearby examples of extremely faint galaxies in resolved
stars; here we do so in diffuse light. To push LSST data to its
faintest limits will 
require a dedicated analysis effort; as found in SDSS, detection, deblending, and photometry at low surface brightness
levels requires a different analysis than that necessary for stellar
photometry. For example, while the formal signal-to-noise ratio of the
data will be sufficient to detect signal at less than 1/1000 the sky
level on scales of many arcseconds, clearly to really achieve that
precision requires an exquisite understanding of scattered light and
other systematics, to distinguish true galaxies with, for example, ghosts
from bright stars, variations in the background sky, and other artifacts.  


\subsection{Spiral Galaxies with Low Surface Brightness Disks}

Low surface brightness (LSB) spirals are diffuse galaxies with disk
central surface brightness fainter than 22.5 mag arcsec$^{-2}$ in the
$B$ band. They are generally of quite low metallicity, and thus
exhibit little dust or molecular gas, but have quite large neutral
hydrogen content \citep{O'Neil++00a, O'Neil++00b, O'Neil++03,
  Galaz++02, Galaz++06, Galaz++08} and star formation rates lower
than $1\, M_\odot$ yr$^{-1}$ 
\citep{Vallenari++05}. Rotation curves of LSBs extend to large radii
\citep{deBlok2002}, and, therefore, their dynamics are dark matter
dominated. Several studies have shown that LSBs dominate the volume
density of galaxies in the Universe (e.g., \citealt{Dalcanton++97}), and
thus it is of prime importance to understand them in the context of
the formation of spiral galaxies.

Given the depth and scattered light control that LSST will have
(\autoref{sec:common:scattered}), it should be sensitive to galaxies
with central surface brightness as low as 27 mag arcsec$^{-2}$ in $r$
in the ten-year stack -- compared with SDSS, where the faintest
galaxies measured have $\mu_r \sim 24.5$ mag arcsec$^{-2}$
\citep{Zhong++08}. Scaling from the estimates of LSB surface density
from \citet{Dalcanton++97}, we conservatively estimate that LSST will
discover $10^5$ objects with $\mu_0 > 23$ mag arcsec$^{-2}$.  Indeed,
this estimate is quite uncertain given our lack of knowledge of the
LSB population demographics.  LSST's combination of depth and sky
coverage will allow us to
settle at last the contribution of very low surface brightness
galaxies to the volume density of galaxies in the Universe. 

LSST will also discover large numbers of giant LSB spirals, of which
only a few, such as Malin 1 \citep{Malin1}, are known, and tie down
the population of red spiral LSBs.  ALMA will be ideal
for studying the molecular content and star formation of these
objects.

\subsection{Dwarf Galaxies}
The other prominent members of the LSB world are dwarf galaxies.  
Low luminosity galaxies are the most numerous galaxies in 
the Universe, and are interesting objects for several 
reasons. They tend to have had the least star formation per 
unit mass of any systems, making them interestingly pristine 
tests of small-scale cosmology. For the same reason, they 
are important testbeds for galaxy formation: Why is their 
star formation so inefficient? Does the molecular cloud 
model of star formation break down in these systems? Do 
outflows get driven from such galaxies? Does reionization 
photo-evaporate gas in the smallest dwarfs? 
However, dwarf galaxies also tend to be the galaxies of lowest surface 
brightness. For this reason, discovery of the faintest known 
galaxies has been limited to the Local Group, where they can 
be detected in resolved stellar counts (\autoref{sec:MW:UFs}).  Here
we discuss the discovery of such objects in diffuse light at larger
distances.   

We know that for larger galaxies, the effects of environment are
substantial --- red galaxies are preferentially foud in dense
environments. Thus, we need to study dwarfs in environments beyond the
Local Group. 
Questions about the
importance of reionization relative to ram pressure and tidal
stripping hinge crucially on the field dwarf population --- and
extremely deep, wide-field surveys are the only way to find these
galaxies, especially if reionization has removed their gas.

Based on the early-type galaxy luminosity function of
\citet{2005MNRAS.356.1155C}, with a faint-end slope $\alpha=-0.65$, we
can expect $\sim 2 \times 10^5$ early-type dwarfs brighter than $M_V =
-14$ within 64 Mpc. \autoref{fig:gal:dwarfs_image} and
\autoref{fig:gal:dwarfs_magradius} suggest that such galaxies will be
relatively easy to find within this distance. Pushing to lower
luminosities, the same luminosity function predicts $8 \times 10^3$
dwarf spheroidal galaxies at $D < 10$ Mpc brighter than $M_V =
-10$. However, we have no business extrapolating this luminosity
function to such low luminosities. Using the same $M^*$ and $\phi^*$,
but changing the slope to $\alpha = -1$, changes the prediction to
$2.5 \times 10^5$ galaxies. Clearly, probing to such low luminosities
over large areas of the sky will provide a lot of leverage for
determining the true faint end slope and its dependence on
environment.

\begin{figure}[ht]
\centerline{\resizebox{4.0in}{!}{\includegraphics{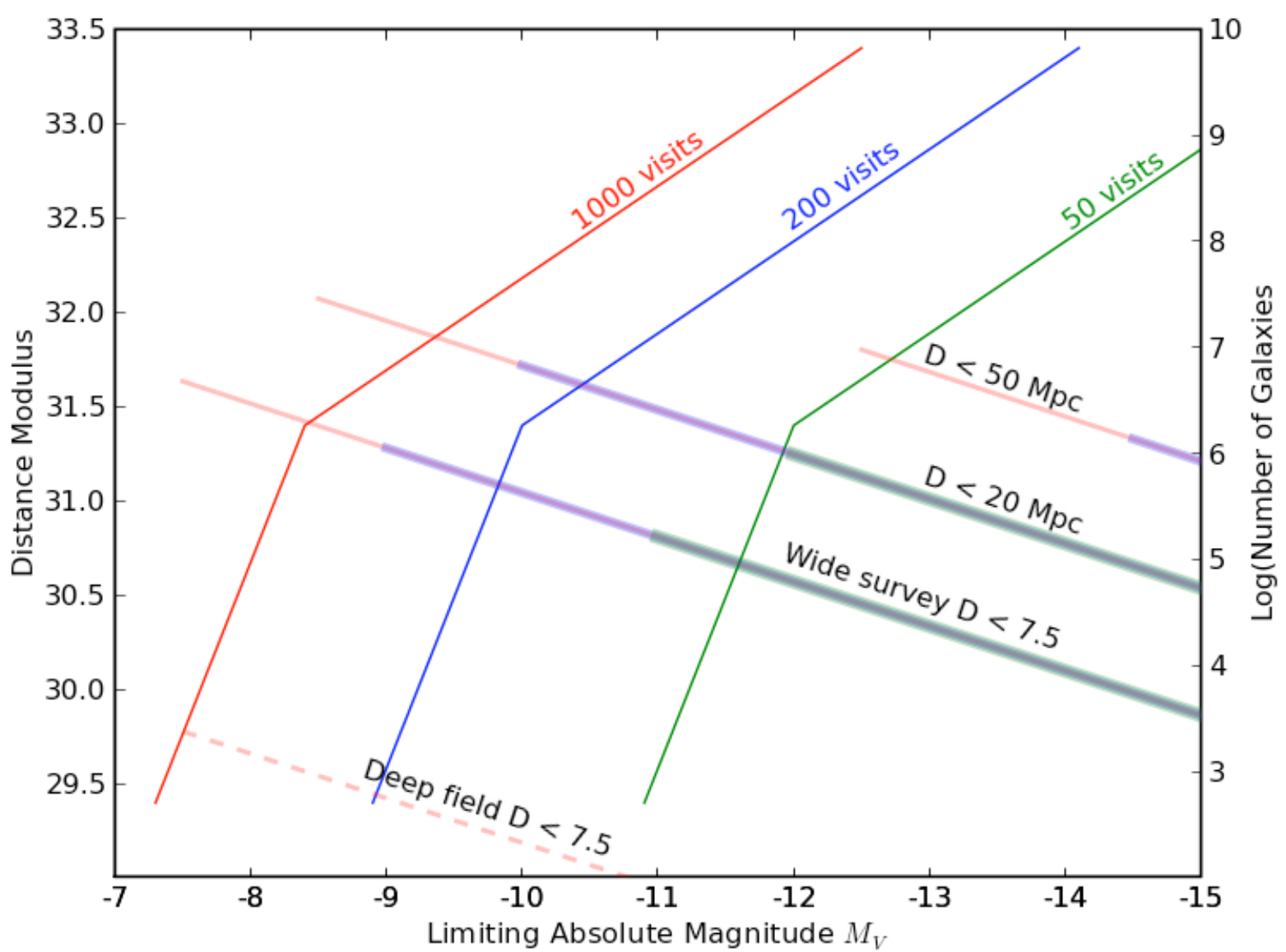}}}
\caption{\label{fig:gal:sbf} LSST surface brightness fluctuations,
  whereby mottling of the galaxy image due to the finite number of
  stars in each pixel is a measure of the distance to the galaxy. The
  curves moving upwards to the right show distance modulus vs. absolute magnitude for distance modulus determination to a precision
  of 0.5 mag for 50, 200, and 1,000 $r$-band visits (the latter
  appropriate to the deep drilling fields). This is derived by
  scaling from the realistic image simulations of
  \citet{2003A&A...403...43M}, which
include the
effects of photon statistics, resolution, and image size. The curves
moving upwards to the left show the expected number of galaxies in
a 20,000 deg$^2$ survey (solid lines) or a 10 deg$^2$
deep-drilling field with 1,000 visits (dashed line near the bottom).
Numbers are based on the luminosity function of
\citet{2005MNRAS.356.1155C}. 
}
\end{figure}

Spectroscopy may not be the most efficient way to confirm that these
are actually nearby dwarf galaxies
(\autoref{fig:gal:dwarfs_magradius}). At $M_V = -10$, the surface
brightnesses are generally too low for most spectrographs. However,
many will be well enough resolved to measure surface-brightness
fluctuations (\autoref{fig:gal:sbf}). Followup observations with HST,
JWST, or JDEM can resolve the nearby galaxies into individual stars,
confirming their identification and measuring distances from the tip
of the red-giant branch.

\subsection{Tidal Tails and Streams}

One of the major recent advances in astronomy has been the discovery
of ubiquitous tidal streams of disrupted dwarf galaxies surrounding
the Milky Way and other nearby galaxies
(\autoref{sec:MW:streams}). The existence of such streams fits well
into the hierarchical picture of galaxy formation, and has caused a
re-assessment of traditional views about the formation and evolution
of the halo, bulge, and disk of our Galaxy.

\begin{figure}[ht]
\centerline{\resizebox{3.5in}{!}{\includegraphics{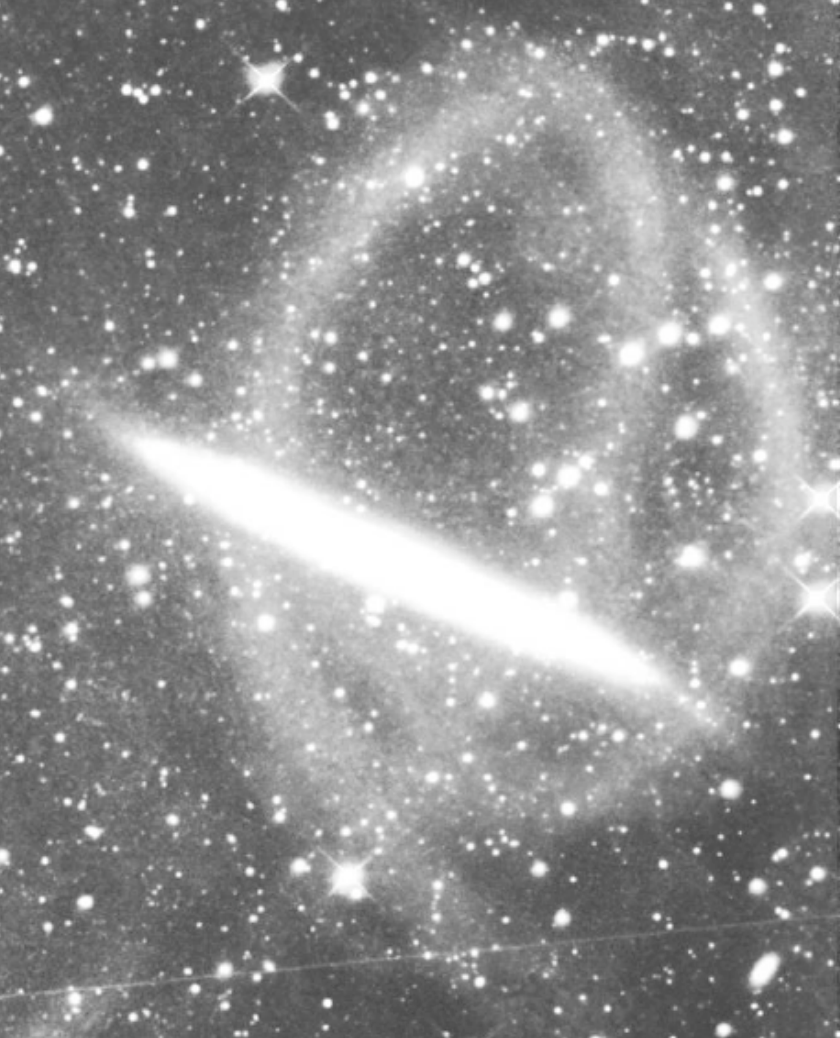}}}
\caption{
Low surface brightness tidal streams surrounding NGC5907.  This is a
$>10$ hr exposure taken on a 0.5-meter telescope; the faintest
features apparent have a surface brightness below 28 mag arcsec$^{-2}$
in $r$.  
From \citet{2008ApJ...689..184M}.
\label{fig:gal:ngc5907}}
\end{figure}

\begin{figure}[ht]
\centerline{\resizebox{4.5in}{!}{
\includegraphics{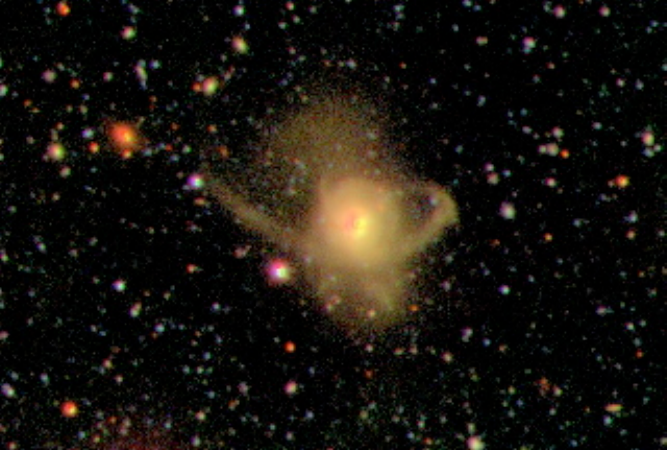}
\includegraphics{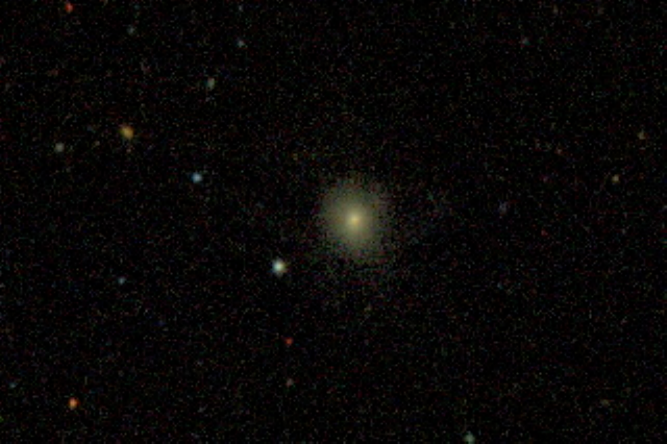}}}
\caption{
{\it Left:}
MUSYC $UVR$ image of a $z=0.1$ galaxy with red low surface brightness features 
revealing a recent interaction without active star formation 
\citep{vanDokkum05}.  This image reaches a  $1\sigma$ surface brightness 
limit of mag 29.5 arcsec$^{-2}$, a good match to the expected LSST depth.  
{\it Right:} 
SDSS $gri$ image of the same $3' \times 2'$ piece of sky, showing that these features 
are not accessible to the current generation of full-sky surveys beyond the 
very local Universe.
\label{fig:gal:lsb_musyc_sdss}}
\end{figure}

The streams can be studied in detail through resolved stars, but only
a few galaxies are close enough to be studied in this way.  Studies of
more distant galaxies in diffuse light will be important for
understanding the demographics of streams in general.  Such studies
have a bearing on a variety of interesting issues.  The streams are
heated by interaction with dark matter sub-halos within the larger
galaxy halo. Statistical studies of the widths of tidal streams may
thus provide some constraints on the clumpiness of dark matter
halos. This is important because $\Lambda$CDM models predict hundreds
of dwarf galaxy mass halos in Milky Way size galaxies, whereas we
only know of a few dozen such galaxies. This could be telling us that
the dark matter power spectrum cuts off at dwarf galaxy scales, or it
could be signaling that star formation is suppressed in low-mass
halos. The shapes of tidal streams also provide constraints on the
shapes of galaxy halos. This can be studied statistically using large
samples of tidal streams revealed by deep images
(e.g. \autoref{fig:gal:ngc5907}, \autoref{fig:gal:lsb_musyc_sdss}).  
By the time LSST begins observing, we
expect that hundreds of individual galaxies will have been targeted
for deep study with other facilities. LSST, however, will allow us to create a deep, unbiased statistical
survey of thousands more galaxies.

\nocite{2008ApJ...689..184M}

\nocite{2008ApJ...689..184M}


At $z=0.1$, a semi-major axis of 50 kpc corresponds to 
$27 ^{\prime\prime}$. A dwarf galaxy of absolute magnitude 
$M_{R,AB} = -16$ stretched uniformly around a circular stream of radius
$27''$ with a half-light width of 2 kpc will have a 
mean surface brightness of $\mu_{R,AB} = 29.2$, 0.07\%
the mean dark-sky brightness. The ability to detect and measure
the parameters of such streams depends critically on the 
flatness of the LSST sky background or the ability to model it.  

\subsection{Intracluster Light}

Moving from individual galaxies to groups and clusters, we expect
the tidal streams that existed during the early stages of galaxy formation
to have been smoothed out into a diffuse stellar halo interspersed between
the galaxies. \citet{2007ApJ...666...20P} calculate that the fraction of the
total stellar mass in this intra-halo population should range from $\sim 8$\%
to $\sim 20$\% for halo masses ranging from $10^{13}$ to $10^{15}
M_\odot$; these numbers are roughly confirmed in the deep imaging
study of the Virgo cluster by \citet{Mihos++05}. 
The uniformity of the LSST data should enable careful measurements of 
this diffuse light with very large samples of nearby groups and clusters,
to probe both the trend with group mass and the trends with other properties
of the groups. Stacking large numbers of groups after masking the galaxies
will enable the mean halo light profile to be traced to very low surface 
brightness.

Novae will provide a unique way to probe diffuse light.
\citet{2006AJ....131.2980S} estimates that LSST will obtain good light
curves and hence distance estimates for $\sim 50$ {\it tramp} novae per year
within 40 Mpc if the diffuse stellar mass is 10\% of the stellar mass
in galaxies.  We might consider putting one of the LSST deep drilling
fields (\autoref{sec:design:cadence}) on a nearby cluster of
galaxies such as Fornax.  If 10\% of the total stellar mass of Fornax
($\sim 2.3 \times 10^{11} M_\odot$) in intracluster light, and we
observed it 9 months of every year, we would discover 
roughly 170 intra-cluster novae. 

%
%
%
%
%
%
%
%
%
%
%
%
%
%
%
%
%
%
%
%
%
%
%
%
%
%

\section{Wide Area, Multiband Searches for High-Redshift Galaxies}
\label{sec:gal:highz}

Deep, narrow surveys with space-borne telescopes have identified new 
populations of high-redshift galaxies at redshifts $z>5$ through 
photometric dropout techniques.  While these observational efforts have 
revolutionized our view of the high-redshift Universe, the small fields 
of such surveys severely limit their constraining power for understanding 
the bright end of the high-redshift galaxy luminosity function
and for identifying other rare objects, including the most massive, oldest, 
and dustiest galaxies at each epoch.
By combining the power of multi-band photometry for dropout selection and the 
unprecedented combination of wide area and deep imaging, LSST will 
uncover the rarest high-redshift galaxies (\autoref{fig:gal:lbg}).  
The discovery and characterization of the most massive galaxies at 
high redshift will provide new constraints on early hierarchical structure 
formation and will 
reveal the galaxy formation process associated with high-redshift 
quasars (\autoref{sec_cens_select}).  

\begin{figure}
\begin{center}
\includegraphics[width=0.5\linewidth]{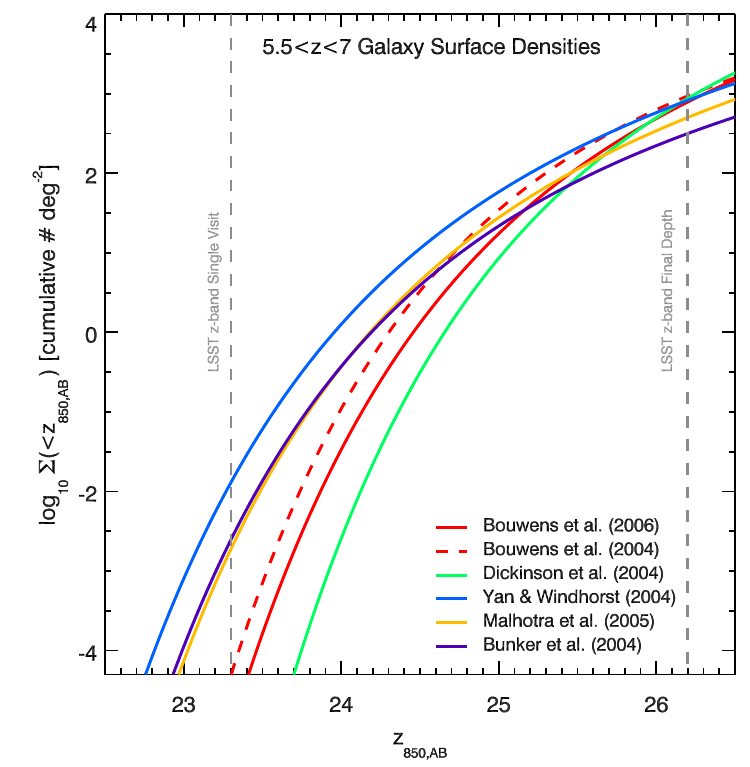}
\caption{\label{fig:gal:sd.z55}
 Fits to measured galaxy source count surface densities for 
$5.5\ltsim z \ltsim 7$ galaxies in the $z$-band, as measured by
 different surveys.  Note the tremendous variation, especially at the
 bright end, caused by the small areas that these surveys cover.
}
\end{center}
\end{figure}

Observations of $i$-dropout and 
$z$-dropout galaxies in the Hubble Ultra Deep and GOODS fields have enabled a 
determination of the rest-UV luminosity function of $z\sim6$ galaxies
\citep[e.g.,][]{Dickinson++04,Bunker++04,Yan+Windhorst04, 
Malhotra++05, Yan++06, Bouwens++04, Bouwens++06}. 
There is a scatter of 1-2 orders of magnitude in determinations of
the bright end of the LF.
\autoref{fig:gal:sd.z55} 
 shows the galaxy source count surface densities for 
$5.5\ltsim z \ltsim 7$ galaxies in the $z$-band calculated from a range of 
$z \sim 6$ rest-frame 
UV luminosity functions taken from the literature. 
LSST will increase the counts of galaxy candidates at $z>5.5$ by 
$\sim 5$ orders of magnitude. 
The LSST survey will probe almost the 
entire luminosity range in this figure and should find hundreds of 
$z_{850} \sim 23-24$ galaxies at $z \sim 6$. 
The resulting uncertainty on their abundance will be a few percent,
2--3 orders of magnitude better than currently
available. 

Observatories such as JWST will reach
extremely deep sensitivities, but it cannot survey large areas of sky;
for example, the Deep Wide Survey discussed in
\citet{2006SSRv..123..485G} will be only $\sim$100 arcmin$^2$. For
extremely rare objects such as luminous high-redshift galaxies, JWST
will rely on wide-area survey telescopes such as LSST for follow-up
observations.  
Wide-area surveys of the far infrared, submillimeter, and
millimeter sky may also be capable of finding rare, massive
galaxies at high-redshift through dust emission powered by
star formation or AGN activity.  
The Herschel-ATLAS survey\footnote{\url{http://h-atlas.astro.cf.ac.uk/science/h-atlas\_final\_proposal.pdf}}
will 
survey 550 deg$^2$ at $110-500\mu$m down to sensitivities
of $<100$ mJy.  The SCUBA-2 ``All-Sky'' Survey\footnote{\url{http://www.jach.hawaii.edu/JCMT/surveys/sassy/}}
will map the entire
$850\mu$m sky available to the James Clerk Maxwell Telescope to $30$ mJy/beam.
The South Pole Telescope \citep{carlstrom2009a} will detect dusty galaxies 
over 4000 deg$^2$ at $90-270$ GHz to a $1\,\sigma$ sensitivity of 1 mJy at 150 GHz.
Surveys such as these will complement
the LSST wide area optical survey by providing star formation rate
and bolometric luminosity estimates for rare, high-redshift galaxies.

\begin{figure}
\begin{center}
\includegraphics[width=0.75\linewidth]{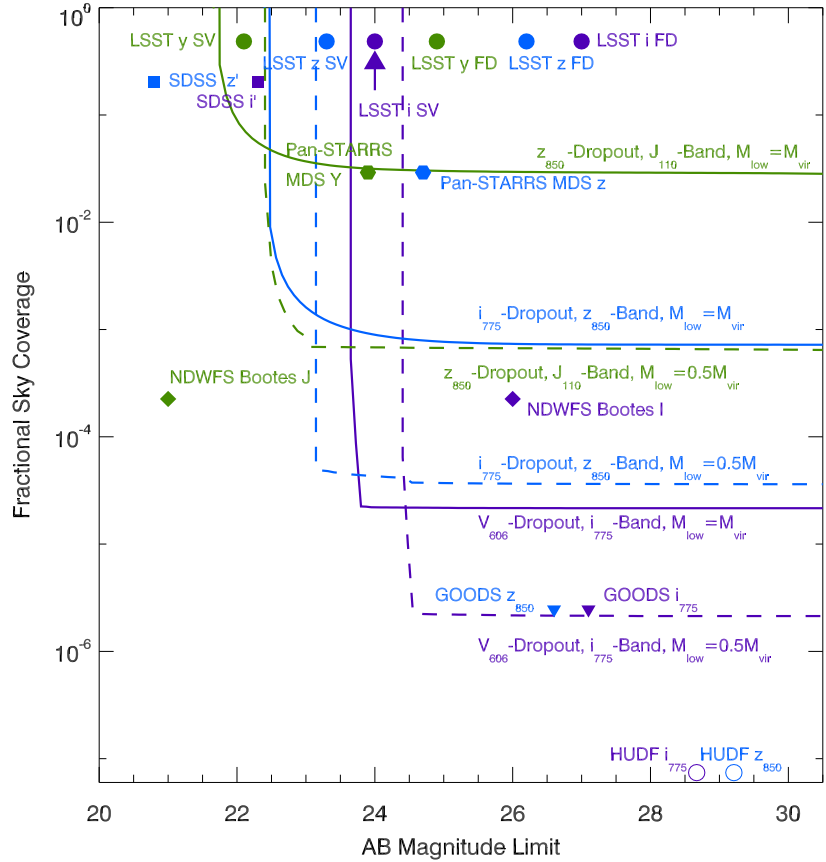}
\caption{\label{fig:gal:dropout_survey}
Estimated survey parameters required to find $z\gtrsim7$ quasar
progenitors and quasar descendants at redshifts
$7\gtrsim z\gtrsim4$.
Shown is the fractional sky coverage and
AB magnitude limit needed to build $V_{606}$-dropout (purple),
$i_{775}$-dropout (blue), and $z_{850}$-dropout (green)
samples that include
a galaxy more massive than the virial mass $M_{\mathrm{vir}}$ (solid
line) or $0.5M_{\mathrm{vir}}$ (dashed line) of the simulated $z\sim6$
quasar host from \citet{2007ApJ...665..187L}.  As the dropout
selection moves to redder bands and higher redshifts, the co-moving
volume and redshift interval over which massive galaxies satisfy
the selection criteria decreases.  The co-moving number density of massive
galaxies, calculated using the \citet{1999MNRAS.308..119S} mass
function, also declines rapidly at high redshifts.  The combination
of these effects requires large fractional sky coverage to find
starbursting quasar progenitors at $z\gtrsim7$.
The circles show the parameters of the existing Hubble UDF ($i_{775}$-
and $z_{850}$-bands, open circles), 
GOODS ($i_{775}$- and $z_{850}$-bands, triangles),
SDSS ($i$- and $z$-bands, squares), and NOAO Wide Deep Field Survey
($I$- and $J$-bands, diamonds)
observations.  
Future wide area surveys with red sensitivity, such as 
LSST ($i$, $z$, and $y$ Single Visit and Final Depths, solid circles),
or possibly the Pan-STARRS Medium Deep Survey
($z$- and $Y$-band, hexagons),
could find quasar
progenitors at $z\gtrsim6$ if their two reddest bands reach
$\gtrsim22$ AB magnitude sensitivity. 
Adapted from \citet{2007ApJ...667...60R}.
}
\end{center}
\end{figure}

A primary goal for studying rare, high-redshift galaxies will be to
understand the galaxy formation process associated with luminous
quasars at $z= 5-6$ with supermassive black holes (SMBHs) of mass
$>10^9\, M_\odot$.  This problem, first popularized by
\citet{Efstathiou+Rees88}, involves finding a robust way of growing
SMBHs quickly in the limited time available before $z\sim6$.  Recent
work simulating the formation of a $z\sim6$ quasar with a SMBH mass of
$\sim 10^9$ M$_\odot$ \citep{2007ApJ...665..187L} has provided a
theoretical argument that high-redshift quasars can be explained
naturally in the context of the formation of galaxies in rare density
peaks in the $\Lambda$CDM cosmology.  A clear test of this picture is the
predicted population of very rare, massive starburst galaxies at
redshifts $z>5$.  \citet{2007ApJ...667...60R} performed a detailed
characterization of the observable ramifications of this scenario, the
foremost being the possible detection of the starbursting progenitors
of $z\sim 4-6$ quasars with massive stellar populations ($M_* \sim
10^{11-12}\,M_\odot$) at higher redshifts in wide area, Lyman-break
dropout samples.  Such objects should be very strongly clustered, as
is found for high-redshift quasars (\citealt{She++07}; see the
discussion in \autoref{sec:agn:clustering}).  

\autoref{fig:gal:dropout_survey} shows the area and depth required for
a photometric survey to identify the high-redshift progenitors of
$z\sim 4-6$ quasars.  To find a single such galaxy in $i$-dropout and
$z$-dropout samples at $z>5$ and $z>6$, a survey must cover ${}>5$\% of
the sky with a depth of $z_{AB}\sim 23-24$ and $y\sim 21.7-22.5$,
respectively. These requirements are remarkably well-matched to the
single visit LSST limiting magnitudes ($z_{AB}\sim 23.3$, $y \sim
22.1$, see \autoref{fig:gal:lbg}). Given that these requirements are
unlikely to be realized by surveys before LSST, the identification of
rare high-redshift galaxies could provide an exciting early LSST
science discovery. With the substantially deeper co-added depth of the
repeated LSST visits, the sample of rare, high-redshift galaxies would
increase rapidly.

\section{ Deep Drilling Fields}
\label{sec:gal:deep_drilling}

The currently planned LSST cadence (\autoref{sec:design:cadence})
involves $\sim$10 pointings on the  
sky that will be observed more frequently, with a cadence and filter 
distribution that can be optimized for finding e.g., supernovae. 
Significantly enhanced science can be achieved by spending proportionally 
more time in $uzy$ than $gri$ in order to achieve more equal depth in
the six filters.   By switching to a fractional observing 
time distribution of 9, 1, 2, 9, 40, 39 \% in $ugrizy$ respectively,
and 1\% of total the LSST observing time on each drilling field, we
would achieve 
$5\,\sigma$ point source detection depths of 
28.0, 28.0, 28.0, 28.0, 28.0, and 26.8 respectively\footnote{The
  detection limit for  
  resolved galaxies will be brighter than that for point sources.}.  
This is shown by the triangles in 
\autoref{fig:gal:lbg}. This would also avoid hitting the confusion
depth at $g$ and $r$ of $\sim 29$ mag.

These deep drilling fields present a number of opportunities for
coordinated deep multiwavelength imaging to select targets for
narrow field follow-up with JWST, ALMA, and other facilities.  Deep
infrared coverage is critical for photometric redshifts.  Ideal field
locations for extragalactic work are those at high Galactic latitude
with minimal dust extinction.  There are several existing fields with
wide-deep multiwavelength coverage that represent likely locations for
LSST Deep Fields, e.g.,  the Extended Chandra Deep Field-South,
COSMOS/Ultravista, the equatorial complex of Subaru-XMM Deep
Survey/Deep2 Field 4/VVDS 0226-0430/CFHT LS D1/XMM-LSS/NDWFS
Cetus/UKIDDS UDS/SpUDS, VISTA VIDEO fields, and the Akari Deep
Field-South.  None of these yet covers a full LSST field of view at
the desired depth for complementary wavelengths, but once the LSST
Deep Field locations are declared, the international astronomy
community will be encouraged to conduct wide-deep surveys with
near-infrared (VISTA, NEWFIRM), mid-infrared (warm Spitzer),
far-infrared (Herschel), ultraviolet (GALEX), sub-millimeter (APEX,
ASTE, LMT), and radio (EVLA, SKA) telescopes on these locations.

These multiwavelength concentrations will be natural locations for
extensive spectroscopic follow-up, yielding three-dimensional probes
of large-scale structure and allowing the calibration of LSST-only
photometric redshifts for use elsewhere on the sky.  They will enable
a nearly complete census of baryonic matter out to $z\sim 7$ traced via
the star formation rate density (rest-ultraviolet plus far-infrared to
get the total energy from stars), stellar mass density, and gas mass
density as a function of redshift.  Thus LSST will move us towards a
complete picture of galaxy formation and evolution.


%
%
%
%
%
%
%
%
%
%
%
%
%
%
%
%
%
%
%
%
%
%
%
%
%
%

\section{Galaxy Mergers and Merger Rates}
\label{sec:gal:merger}

Galaxies must grow with time through both discrete galaxy mergers and
smooth gas accretion.  When and how this growth occurs remains an
outstanding observational question.  The smooth accretion of gas and
dark matter onto distant galaxies is extremely challenging to observe,
and complex baryonic physics makes it difficult to infer a galaxy's
past assembly history.  In contrast, counting galaxy mergers is
relatively straightforward.  By comparing the frequency of galaxy
mergers to the mass growth in galaxies, one can place robust
constraints on the importance of discrete galaxy mergers in galaxy
assembly throughout cosmic time.  The mass accretion rate via mergers
is likely to be be a strong function of galaxy mass, merger ratio,
environment, and redshift \citep{2009ApJ...702.1005S}; these dependencies
can test both the cosmological model and the galaxy-halo connection.

In addition to contributing to the overall buildup of galaxy mass,
the violent processes associated with mergers are expected to
significantly influence the star formation histories, structures,
and central black hole growth of galaxies.  However, other physical
mechanisms may influence galaxy evolution in similar ways, so direct
observations of galaxy mergers are needed to answer the following
questions:

$\bullet$ What fraction of the global star formation density is
driven by mergers and interactions?  Is the frequency of galaxy
mergers consistent with the ``tightness" of the star formation
per unit mass vs. stellar mass relation throughout cosmic time?

$\bullet$ Are typical red spheroids and bulges formed by major
mergers, or by secular evolution?  Do $z>1$ compact galaxies grow in
size by (minor) mergers?

$\bullet$ Are today's most massive ellipticals formed via
dissipationless mergers?  If so, when?

$\bullet$ Do gas-rich mergers fuel active galactic nuclei? Which forms
first, the bulge or super massive black hole?           

In a $\Lambda$CDM model, the rate at which dark matter halos merge is
one of the fundamental processes in structure formation.  Numerical
simulations predict that this rate evolves with redshift as
$(1+z)^{m}$, with $1.0<m<3.5$ \citep{2001ApJ...546..223G,
  2006ApJ...652...56B, 2008MNRAS.386..577F, 2009ApJ...702.1005S}.
It is difficult to directly compare the predicted dark matter halo merger rate
with the observed galaxy merger rate due to the uncertainty in the halo
occupation number. However, if this comparison is done
self-consistently, measuring the merger frequency as a function of
cosmic epoch can place powerful constraints on models of structure
formation in the Universe.

Numerous observational studies over the past two decades have focused
on measuring the galaxy merger rate, yielding highly discrepant values of
$m$, ranging from no evolution ($m \sim 0$) to strong evolution ($m
\sim 5$)
\citep{1989ApJ...337...34Z, 1994ApJ...435..540C, 2000ApJ...532L...1C,
  2000ApJ...536..153P, 2002ApJ...565..208P, 2004ApJ...601L.123B,
  2004ApJ...617L...9L, 2007AAS...21112605B, 2008ApJ...672..177L}.  As
a consequence, the importance of galaxy mergers to galaxy assembly,
star formation, bulge formation, and super-massive black hole growth
is strongly debated.  These observational discrepancies may stem from
small sample sizes, improperly accounting for the timescales over
which different techniques are sensitive, and the difficulty in tying
together surveys at high and low redshift with different selection
biases.

The galaxy merger rate is traditionally estimated by measuring the
frequency of galaxies residing in close pairs, or those with
morphological distortions associated with interactions (e.g., double
nuclei, tidal tails, stellar bridges).  The detection of distortions
is done either by visual analysis and classification
(\citealt{2000MNRAS.311..565L}; Bridge et al. 2009 in preparation) or
through the use of quantitative measures \citep{1996ApJS..107....1A,
  2003ApJS..147....1C, Lotz++08}.
A key uncertainty in calculating the galaxy merger rate is the
timescale associated with identifying a galaxy merger.  The merger of
two comparable mass galaxies may take 1--2 Gyr to complete, but the
appearance of the merger changes with the merger stage, thus a given
merger indicator (i.e., a close companion or double nucleus) may only
be apparent for a fraction of this time \citep{Lotz++08}.
Galaxies at $z<1$ with clear signatures of merger activity are
relatively rare ($<$10-15\% of $\sim$ L$^*$ galaxies at $z < 1$, $<$5\%
at $z=0$), although the fraction of galaxies which could be considered
to be `merging' may be significantly higher.

No single study conducted so far has been able to uniformly map the
galaxy merger rate from $z=0$ to $z\geq 2$, as current studies must
trade off between depth and volume.  An additional limiting factor is
the observed wavelength range, as galaxy morphology and pair
luminosity ratios are often a strong function of rest-frame
wavelength.  Very few merger studies have been done with SDSS
(optimized for the $z< 0.2$ Universe).  SDSS fiber collisions and low
precision photometric redshifts prevent accurate pair studies, while
the the relatively shallow imaging and moderate ($\sim 1.4''$) seeing
reduce the sensitivity to morphological distortions.  Deeper
spectroscopic and imaging studies probe the $z \sim 0.2 - 1$
Universe, but do not have the volume to also constrain the low
redshift Universe or the depth at near-infrared (rest-frame optical)
wavelengths to constrain the $z>1$ Universe. Ultra-deep Hubble Space
Telescope studies (GOODS, UDF) can detect $L^*$ mergers at $z > 1$,
but have very small volumes and are subject to strong cosmic variance
effects.  The CFHTLS-Deep survey is well matched to the proposed LSST
depths, wavelengths, and spatial resolution, but, with an area 5000
times smaller, is also subject to cosmic variance (Bridge et al.,
2009, in preparation).

\begin{figure}[ht]
\centerline{\resizebox{6.5in}{!}{\includegraphics{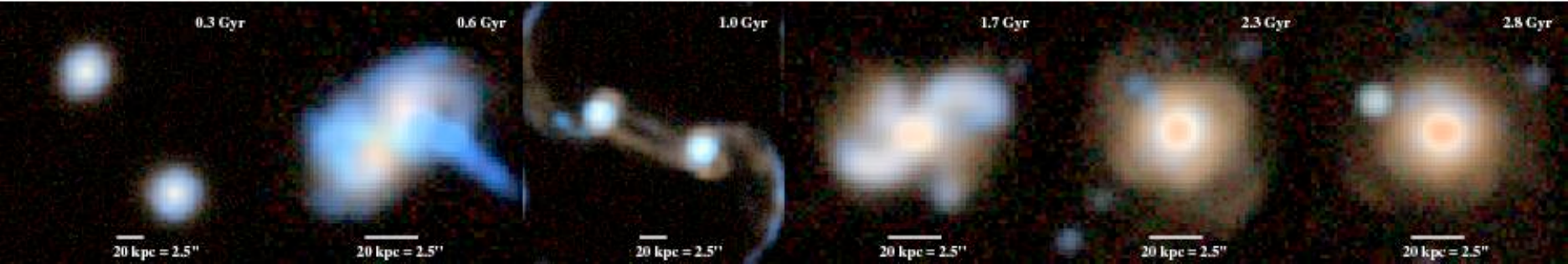}}}
\caption{\label{fig:gal:merger_sim}
At full LSST depth, strong asymmetries, tidal tails, and post-merger
fine structures will be observable for gas-rich major mergers at
$z \gtrsim 1$.  These images show the progression of a  gas-rich equal
mass disk merger, as it would appear at $z=1$ in $r-i-z$ color. 
These are from a hydrodynamical simulation which includes gas, star 
formation, and dusty radiative transfer \citep{Lotz++08}. 
During the initial encounter at $t=0.6$ Gyr and the final merger at $t=1.7$
Gyr, strong blue distortions are visible on scales of a few arc-seconds.
After the first pass at $t=1$ Gyr,  tidal tails of $\sim 5-10''$ are
detectable at $\mu_i  < 27$ mag/arcsec$^2$.  Faint shells, 
tidal features, and blue tidal dwarfs will be apparent at full LSST depth
for up to a Gyr after the final merger ($t=2.3-2.8$ Gyr), and are observed
in deep HST and CFHTLS images.  } 
\end{figure}

Unlike the current studies, LSST has the depth,
volume, and wavelength coverage needed to perform a uniform study of
$L^*$ mergers out to $z \sim 2$, and a statistical study of bright galaxy
mergers out to $z \sim 5$.  The wide area coverage of LSST will be critical 
for addressing the effects of cosmic variance on measures of the merger rate, 
which can vary by a factor of two or more even on projected scales of a square 
degree (Bridge et al., 2009, in preparation). 
A variety of approaches will be used to identify mergers in the LSST data:
\begin{itemize} 
\item Short-lived strong morphological disturbances, such as strong
asymmetries and double nuclei, which occur during the close encounter
and final merger stages and are  apparent for only a few 100 Myr.
These will be most easily found in $z \ltsim 0.2$ galaxies,
where the LSST $0.7^{\prime\prime}$ spatial resolution corresponds to 1--2
kpc.  Lopsidedness in galaxy surface brightness profiles
can provide statistical constraints on minor mergers and requires 
similar spatial resolution.
\item Longer-lived but lower surface brightness extended tidal tails,
which occur for $\sim 0.5$ Gyr after the initial encounter and for up
to 1 Gyr after the merger event.  These tails are the longest-lived
merger signatures for disk-galaxy major mergers, and should be easily
detected at $z \leq 1$ in the full depth LSST $r\,i\,z$ images
(\autoref{fig:gal:merger_sim} and \autoref{fig:gal:lsb_musyc_sdss}).  Scaling from the CFHTLS-Deep survey,
LSST should detect on the order of 15 million galaxies undergoing a
strong tidal interaction.
\item Residual fine structures (faint asymmetries, shells, and dust
features) detected in smooth-model subtracted images.  These
post-merger residual structures are visible for both gas-rich
and gas-poor merger remnants, and contribute $<1$--5\% of the 
total galaxy light, with surface brightnesses $\sim 28$ mag arcsec$^{-2}$. 
\item The statistical excess of galaxy pairs with projected separations
small enough to give a high probability for merging within a few hundred
Myr.  With $\sim 0.7^{\prime\prime}$ seeing, galaxies with projected
separations $> 10$ kpc will be detectable to $z \sim 5$.  LSST's
six-band photometry will result in photometric redshift accuracies of
about $0.03(1+z)$ (\autoref{sec:common:photo-z}).  This is comparable
to or better than those used in other studies for the identification
of close galaxy pairs, and will allow for the selection of merging
galaxies with a wide range in color. With LSST's high quality
photometric redshifts and large number statistics, it will be possible
to accurately measure the galaxy pair fraction to high precision
(although the identification of any given pair will be uncertain).
Current surveys detect only about 50-70 red galaxy pairs per square
degree for $0.1 < z < 1.0$. With LSST, we should be able to observe
more than a million ``dry'' mergers out to $z \sim 1.0$.
\end{itemize}

One of the advantages of the LSST survey for studying the evolving
merger rate is the dense sampling of parameter space.  A large number
of merger parameters --- galaxy masses and mass ratio, gas fractions,
environment --- are important for understanding the complex role of
mergers in galaxy evolution.  For example, mergers between gas-poor,
early-type galaxies in rich environments have been invoked to explain
the stellar mass build-up of today's most massive ellipticals (e.g.,
\citealt{vanDokkum05, Bell++07}.  Each of the approaches given above
will yield independent estimates of the galaxy merger rate as a
function of redshift, stellar mass, color, and environment.  However,
each technique probes different stages of the merger process, and is
sensitive to different merger parameters (i.e., gas fraction, mass
ratio). Therefore, the comparison of the large merger samples selected
in different ways can constrain how the merger sequence and parameter
spaces are populated.

Finally, the cadence of the LSST observations will open several
exciting new avenues.  It will be possible to identify optically
variable AGN (\autoref{sec:agn:var}) in mergers and constrain the SMBH
growth as a function of merger stage, mass, and redshift.  With
millions of galaxy mergers with high star formation rates, we will
detect a significant number of supernovae over the ten-year LSST
survey.  We will be able to determine the rate of SN I and II
(\autoref{chp:sne}) in mergers, and obtain independent constraints on
the merger star formation rates and initial stellar mass functions.


%
%
%
%
%
%
%
%
%
%
%
%
%
%
%
%
%
%
%
%
%
%
%
%
%
%
\section{Special Populations of Galaxies}



There are a variety of approaches to classifying galaxies and
searching for outliers. In broad terms, one attempts to define a 
manifold of galaxies through the multi-dimensional space of the
measured parameters. Scientific discoveries come both from
defining this manifold --- which is equivalent to measuring 
galaxy scaling relations, their linearity, and their scatter ---
and trying to understand the outliers. 

Approaches to defining the manifold include training sets
and neural networks, principal component analysis, decision
trees, self-organizing maps, and a variety of others. 
Training sets of millions of objects in each redshift interval
will reveal subtle variations within known astrophysical 
phenomena. For example, for well-resolved 
galaxies at $z \lessim 0.2$, LSST can characterize lopsidedness --
as a function of color and environment -- at a
much higher $S/N$ than any previous survey. Phenomena
that were either overlooked or ascribed to cosmic variance in
smaller samples will be revealed and quantified using the LSST
images and database. 

With millions of high $S/N$ training examples to define the locus of
``normal'' galaxies, we can expect a wealth of scientific discoveries
in the outlier population. The outlier population will include rare
kinds of strong gravitational lenses (\autoref{chp:sl}), which may
have slipped through automated lens finders. It will include unusual
galaxy interactions --- e.g., ring galaxies, polar ring galaxies, or
three and four-body interactions --- follow-up studies of which may
yield insights into the shapes of dark-matter halos
\citep{2003ApJ...585..730I} or how star formation is triggered in
merger events \citep{2008A&A...492...31D}.  It will include rare
projections of galaxies that can be used to probe dust within spiral
arms \citep{2007AJ....134.2385H}.

%
%
%
%
%
%
%
%
%
%
%
%
%
%
%
%
%
%
%
%
%
%
%
%
%
%
\section{Public Involvement}
\label{sec:galaxies:epo}


We have already described in \autoref{sec:epo:Citizens} the very
successful Galaxy Zoo project, whereby hundreds of thousands of
citizen scientists have made a real contribution to scientific
research by classifying the SDSS images of galaxies by eye.  This
motivates a new generation of ``Zoos'', and one that would provide
equally remarkable science value will be Merger Zoo. In fact, the SDSS
Galaxy Zoo team has specifically indicated that this is needed.
Galaxy Zoo-Classic provides morphological classifications of spirals
and ellipticals and ``mergers.'' All oddball galaxies that cannot be
placed into one of the other two classes are classified as mergers. When
LSST generates deeper images of larger numbers of mergers, at
increasing look-back times, then something must be done to
classify these mergers (\autoref{sec:gal:merger}).  Over the past
three decades, collision/merger modelers have succeeded in deriving
reasonable (unique?) models for of order 100--200 merging
systems. This is a very small number.  It is exceedingly hard because there
are one to two dozen input parameters to even the simplest models,
there are unknown 
viewing angles, and there is an unknown age for each system. It is
virtually impossible to train a computer model to emulate the human
pattern recognition capabilities that our eyes and our brains
provide. This has been tried with genetic algorithms with limited
success. Fortunately, as  Galaxy Zoo has demonstrated, we can enlist
the aid of hundreds of thousands of pairs of eyes to look at merger
models, to compare with images, and to decide which model matches a
given observation. Plans for Merger Zoo are now under way, involving
the original Galaxy Zoo team, plus merger scientists at George Mason
University and outreach specialists at Adler Planetarium and Johns
Hopkins University. It will be deployed to work with SDSS mergers, so
that Merger Zoo (or more likely, its descendant) will be ready for the
flood of galaxy data from LSST in the future.



\bibliographystyle{SciBook}
\bibliography{galaxies/galaxies}

%
%
%
%
%
%
%
%
%
%
%
%
%
%
%
%
%
%

\chapter[Active Galactic Nuclei]{Active Galactic Nuclei}
\label{chp:agn}

{\it W. N. Brandt, 
Scott F. Anderson, 
D. R. Ballantyne,
Aaron J. Barth, 
Robert J. Brunner,
George Chartas,
Willem H. de Vries,
Michael Eracleous,
Xiaohui Fan, 
Robert R. Gibson,
Richard F. Green, 
Mark Lacy,
Paulina Lira,
Jeffrey A. Newman,
Gordon T. Richards, 
Donald P. Schneider, 
Ohad Shemmer,
Howard A. Smith,
Michael A. Strauss,
Daniel Vanden Berk}



Although the numbers of known quasars and active galactic nuclei (AGN)
have grown considerably in the past decade, a vast amount of
discovery space remains to be explored with much larger and deeper
samples.  LSST will revolutionize our
understanding of the growth of supermassive black holes with cosmic
time, AGN fueling mechanisms, the detailed physics of
accretion disks, the contribution of AGN feedback to galaxy evolution,
the cosmic dark ages, and gravitational lensing. 
The evolution of galaxies
is intimately tied with the growth and energy output from the
supermassive black holes which lie in the centers of galaxies.  The
observed correlation between black hole masses and the velocity
dispersion and stellar mass of galaxy bulges seen at low redshift \citep{Tremaine++02}, and
the theoretical modeling that suggests that feedback from AGN
regulates star formation, tell us that AGN play a key role in galaxy
evolution.  

The goal of AGN statistical studies is to define the changing
demographics and accretion history of supermassive black holes (SMBHs)
with cosmic time, and to relate these to the formation and evolution
of galaxies.  These results are tightly coupled to the evolution of
radiation backgrounds, particularly the ultraviolet ionizing
background and extra-Galactic X-ray background, and the co-evolution
of SMBHs and their host galaxies.  The LSST AGN sample
(\autoref{sec:agn:census}) will be used by
itself and in conjunction with surveys from other energy bands to
produce a measurement of the AGN luminosity function and its evolution
with cosmic time (\autoref{sec:agn:LF}) and the evolution of the
bolometric accretion luminosity density.  LSST will break the
luminosity-redshift degeneracy inherent to most 
flux-limited samples and will do so over a wide area, allowing 
detailed explorations of the physical processes probed by the 
luminosity function.  Indeed, the AGN sample will span a luminosity
range of more than a
factor of one thousand at a given redshift, and will
allow detection of AGN out to redshifts of approximately seven,
spanning $\sim95$\% of the age of the Universe.  

AGN clustering is a reflection of the dark matter halos in which these
objects are embedded.  LSST's enormous dynamic range in
luminosity and redshift will place important constraints on models for
the relationship between AGN and the dark matter distribution, as
described in \autoref{sec:agn:clustering}.  LSST will significantly
increase the number of high-redshift quasars, where the average co-moving
separation of currently known luminous quasars is as high as 150 $h^{-1}$ Mpc
(at $z\sim4$) --- so sparse as to severely limit the kinds of clustering
analyses that be can done, hindering our ability to distinguish
between different prescriptions for AGN feedback.

AGN are an inherently broad-band phenomenon with emission from the
highest-energy gamma-rays to long-wavelength radio probing different
aspects of the physics of the central engine.  LSST will overlap
surveys carried out in a broad range of wavelengths, allowing studies
of a large number of multi-wavelength phenomena
(\autoref{sec:agn:multilambda}).  LSST's multiwavelength power comes
from the ability to compare with both wide-area and ``pencil-beam''
surveys at other wavelengths.  The former are important for
investigations of ``rare'' objects, while the latter probe
intrinsically more numerous, but undersampled populations.

In all, the LSST AGN survey will produce a high-purity sample of at
least ten million well-defined, optically-selected AGNs
(\autoref{sec:agn:census}).  Utilizing the large sky coverage, depth,
the six filters extending to 1$\mu$m, and the valuable temporal
information of LSST, this AGN survey will dwarf the largest current
AGN samples by more than an order of magnitude.  Each region of the
LSST sky will receive roughly 1000 visits over the decade-long survey,
about 200 in each band, allowing variability to be explored on
timescales from minutes to a decade, and enabling unique explorations
of central engine physics (\autoref{sec:agn:var}).

The enormous LSST AGN sample will enable the discovery of
extremely rare events, such as transient fueling events from stars
tidally disrupted in the gravitational field of the central SMBH
(\autoref{sec:agn:transient}) and large numbers of multiply-lensed AGN
(\autoref{sec:agn:lens}).  Lastly, the giant
black holes that power AGN 
inspire strong interest among students and the general public,
providing natural avenues for education and public outreach
(\autoref{sec:agn:epo}).

\section{AGN Selection and Census}
\label{sec:agn:census}

\noindent{\it 
Scott F. Anderson,
Xiaohui Fan, 
Richard F. Green,
Gordon T. Richards,
Donald P. Schneider, 
Ohad Shemmer,
Michael A. Strauss,
Daniel Vanden Berk
}

%
%
%
%
%
%
%
%
%
%
%
%
%
%
%
%
%
%
%
%
%
%
%
%
%




%
%


\subsection{AGN Selection}
\label{sec_cens_select}

There are three principal ways in which AGN will be identified in LSST
data: from their colors in the LSST six-band filter system, from their
variability, and from matches with data at other wavelengths.  

\subsubsection{Color Selection}
\label{sec_cens_color}


Unobscured AGN with a broad range of redshifts can be isolated in
well-defined regions of optical--near-IR multicolor space
\citep{Fan99,Ric++01}.  At low redshifts ($z \lesssim 2.5$), quasars
are blue in $u-g$ and $g-r$ (these are the ultraviolet excess sources
of \citealt{San65} and \citealt{Schmidt83}), and are well-separated from stars
in color-color space.
At higher redshift, the Ly-$\alpha$ forest (starting at 1216\AA) and
the Lyman limit (at 912\AA) march to ever longer wavelengths, making
objects successively redder.

\begin{figure}
\begin{center}
  \includegraphics[angle=0,scale=0.4]{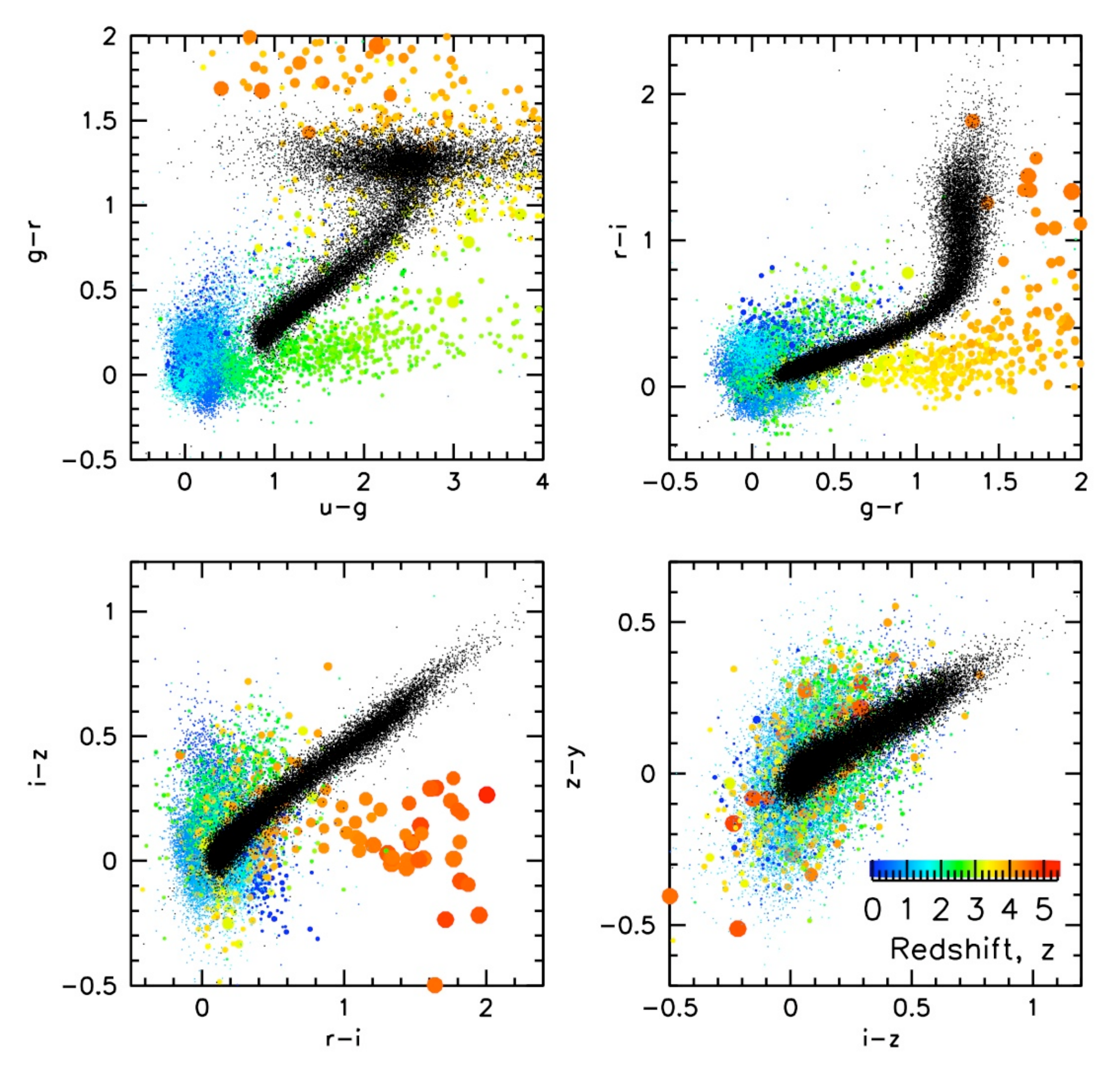}
  \caption{Color-color plots of known quasars from SDSS (colored dots)
  and stars 
    (black dots) in the LSST photometric system. The quasars are color
    coded by redshift according to the color key, and for clarity, the
    dot size is inversely proportional to the expected surface density
    as a function of redshift. Since there is no $y$ filter in the
    SDSS system, a random Gaussian color offset has been added to the
    $z-y$ color according to the width of the stellar locus in the
    $i-z$ color.  The quasar colors become degenerate with those of F
    stars at redshifts between about 2.5 and 3. See
    \autoref{fig:agn:z7quasars_cm} and \autoref{fig:agn:z7quasars} for redshifts above 5.}
\label{fig:agn:color_select}
\end{center}
\end{figure}

\autoref{fig:agn:color_select} shows the colors of quasars and stars
as convolved with the LSST filters, with data taken from the SDSS. 
The $u$-band data are crucial
for selection of low-redshift ($z < 2$) AGN; observations in this
filter allow one to distinguish between AGN and stars (in particular
white dwarfs and A and B stars). High-redshift AGN will be easily
distinguished; as we discuss in more detail below, the $y$ filter
should allow quasars with redshifts of 7.5 to be selected (compare SDSS, whose filter set ends with the $z$ band;
it has discovered quasars with redshift up to 6.4, \citealt{Fan06}).
As with SDSS, most of the sample contamination is in the range
$2.5<z<3$, where quasar colors overlap the stellar locus in most
projections.  It is also difficult to select quasars at $z \sim 3.5$,
where Lyman-limit systems cause quasars to be invisible in $u$ and $g$
but quasars have similar colors to hot stars at longer wavelengths.
However, lack of proper motion and variability will allow quasars to
be efficiently separated from stars in these redshift ranges, as we
describe below.

AGN color selection will proceed in much the same manner as for the
SDSS; however, LSST's increased depth and novel observing strategy
require consideration of the following issues:
\begin{itemize} 
\item LSST (unlike SDSS) will not measure a given area of sky
  through the various filters simultaneously.  Because of variability,
  the colors measured will, therefore, not exactly reflect the colors of
  the object at a given moment in time.  However, the large number of
  epochs in each bandpass mean that the {\em average} colors of objects
  will be exactly what they would have been if the observations in
  each bandpass were simultaneous.
\item For low-luminosity systems, the colors of AGN will be
  contaminated by the colors of their hosts.  Simulated LSST
  images (e.g., \autoref{sec:design:imsim}) will help to characterize
  this effect.  Variability will allow objects with appreciable
  host-galaxy contribution to
  be selected, as will photometric measurements of unresolved point
  sources in the centers of galaxies.
\item The majority of quasars used in \autoref{fig:agn:color_select}
  are not significantly reddened.  However, there is great interest in
  the reddened population \citep{Ric++03,Maddox08}. While the most
  heavily obscured (``type 2'') quasars will not be recognized as AGNs
  using LSST alone, LSST will detect millions of type 2 quasars via emission from their host galaxies and narrow-line 
  regions.  These objects can be recognized as AGNs by their infrared, 
  radio, or X-ray emission, as we describe below.

\end{itemize}


\subsubsection{Selection by Lack of Proper Motion}
\label{sec_cens_prop}

Lack of proper motion will further distinguish faint quasars from
stars.  The 3$\sigma$ upper limit on
proper motion for the full 10 years of the LSST survey is intended to
be 3 milli-arcsec at $r \sim 24$, and five times better at $r =
21$. The stringent upper limit on proper motions 
will essentially eliminate the relatively nearby L and T dwarfs as
contaminants of the very high-redshift quasar candidate lists, and
will also remove many of the white dwarfs and subdwarfs.  

It is illustrative to consider the case of contamination of the color
selection by white dwarfs, which can overlap as ultraviolet excess
objects at low redshift, and (for cooler white dwarfs) as objects with
similar colors to $3.2 < z < 4.0$ quasars.  
For each quasar redshift, we use the white dwarf
color-absolute magnitude diagram to estimate the white dwarf
properties most closely matched to the quasar energy distribution as \citet{Holberg+Bergeron06}.  The typical distances of
these objects at $r = 24$ place these contaminants in the thick
disk population.

\begin{table}[ht]
\caption{Elimination of White Dwarf
Contaminants}
\smallskip
\begin{center}
\begin{tabular*}{\textwidth}{@{\extracolsep{\fill}}lccccc}
\hline
Quasar $z$& WD $M_V$ for& WD $T_{\rm eff}$& Distance (pc)&$3\sigma$ limit $v_{tan}$&Fraction\cr
& quasars $(V-I)$&& at $r=24$&  $\kms$&excluded\cr
\hline
3.2&13.7&6500&1260&17.6&77\%\cr
3.6&15.7&4500&660&9.4&88\%\cr
4.0&16.5&3500&500&7.1&92\%\cr
\hline
\end{tabular*}
\end{center}
\end{table}

The width of the distribution of the thick disk component of the
velocity dispersion is $\sim60\, \kms$ \citep{Beers+Sommer-Larsen95}.  
With a Gaussian form for the velocity dispersion and the $3\sigma$
upper limits for proper motion quoted above, we compute the fraction
of the thick disk white dwarfs excluded.

If we consider a halo subdwarf at the main sequence turn-off detected
at $r=24$, the distance is some 50 kpc. Even then, the proper
motion upper limit rejects half of the tangential velocity
distribution of the outer halo, with its dispersion of $\sim130\, \kms$.

The conclusion is, therefore, that moderate to low-temperature white
dwarfs will be effectively screened.  The space distributions of
hotter white dwarfs and main sequence stars earlier than K spectral
type place the vast majority of them in the brighter magnitude range
typical of the current SDSS samples.  They would, therefore, not be
expected to be significant contaminants at these faint magnitudes.  An
increasing fraction of the halo subdwarfs will remain as contaminants
as the LSST survey limits are approached. The surface
density of very distant halo main sequence stars is lower, which will
minimize the contamination due to the poorer proper motion
measurements at the faintest survey magnitudes.

\subsubsection{Selection by Variability}
\label{sec_cens_var}

Variability will add a powerful dimension to AGN selection by 
LSST, since AGN vary in brightness at optical and ultraviolet
wavelengths with a red-noise power spectrum. It is expected that the
efficiency of AGN selection by variability may be comparable to the
color selection efficiency \citep{Sesar07}.
The amplitude of AGN variability depends upon rest-frame variability
timescale, wavelength, luminosity, and possibly redshift \citep[e.g.,][]
{DVB04}. We use the parametrized description of AGN variability
from the SDSS \citep{Ivezic04}, extrapolated to fainter apparent
magnitudes, to estimate the fraction of AGN in the LSST that may be
detected as significantly variable.
Given the depth of individual LSST exposures, we calculate the
magnitude difference at which only 1\% of the non-variable stars will
be flagged as variable candidates due to measurement uncertainty, first assuming
two measurement epochs separated by a month, and also assuming 12 measurement
epochs spanning a year in total.

\begin{figure}
\begin{center}
\includegraphics[width=0.95\linewidth]{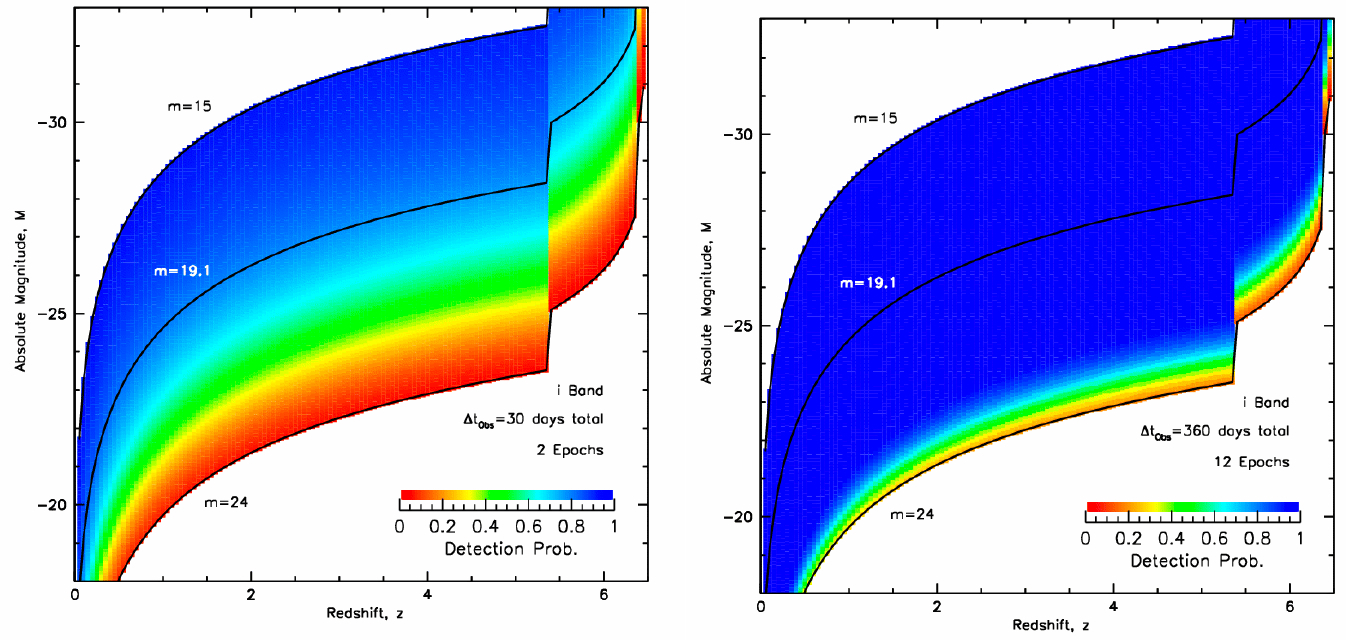}
  \caption{The probability of detecting an AGN as variable as a
    function of redshift and absolute magnitude. {\it Left:} two
    epochs separated by 30 days. {\it Right:} 12 epochs spanning a
    total of 360 days. Nearly all of the AGN between the limiting
    apparent magnitudes would be detected as variable after one
    year. }
\label{fig:agn:var_select}
\end{center}
\end{figure}

The probability that the single-band rms magnitude difference of an
AGN will exceed this value, and will, therefore, be flagged as a
variable candidate, depends upon redshift (as it determines rest
wavelength and rest-frame variability timescale), luminosity, observed
temporal baseline, and the number of observing epochs.  Here we follow
the model of \citet{Ivezic04}, and show the
results in \autoref{fig:agn:var_select}.  

Even with only two epochs separated by 30~days, a large fraction of
AGN will be detected as variable objects. The fraction of AGN detected
depends strongly on absolute magnitude at each redshift; intervening
Lyman series absorption shortward of the 1216\AA\ Lyman~$\alpha$
emission line also affects the detection probability. After 12 epochs
with a total temporal baseline of 360 days, nearly all of the AGN to a
limiting apparent magnitude of 24 will be detected as variable. The
detection fraction will increase as the number of epochs increases,
and the use of all six bands will improve the detection fraction even
further. Ultimately, LSST will provide $\sim200$ epochs for each AGN
candidate in each band, thus increasing the detection fraction as well
as increasing the limiting magnitude.

The LSST temporal information will be especially useful for selecting
low-luminosity AGN which would otherwise be swamped by their hosts, as
well as radio-loud AGN, which have larger variability
amplitudes and shorter variability timescales
\citep[e.g.,][]{Giveon99}. Variability will also allow selection of
AGN which are confused with 
stars of similar color, particularly at $z\sim2.7$ where SDSS is
highly incomplete.  Variability timescales, coupled with LSST's 6-band
photometry will allow clean separation of AGNs from variable stars.
For example, RR Lyrae stars have similar colors to AGNs 
but have very different variability timescales \citep{Ivezic++01}, and can thus 
efficiently be identified as contaminants.

\subsubsection{Selection by Combination with Multiwavelength Surveys}
\label{sec_cens_multiwav}

Cross-correlation of LSST imaging with multi-mission, multiwavelength
surveys will also contribute to the AGN census by allowing selection
of sources, such as optically obscured quasars,  that cannot easily be identified as AGN by color selection,
lack of proper motion, or variability.  LSST's ``deep-wide'' nature will allow it to be combined
both with shallower all-sky surveys at other wavelengths in addition
to having both the areal coverage and depth to be paired with the
growing number of multi-wavelength pencil beam surveys.
For example, cross-correlations of LSST images with Chandra or
XMM-Newton observations can reveal obscured AGN that are not
easily identifiable via standard optical techniques; X-ray sources
that have no LSST counterparts in any band may be candidates of $z>7.5$
quasars. Many high-redshift AGN may also be detected by matching LSST
images with a growing array of future multi-wavelength surveys; see
\autoref{sec:agn:multilambda} for further discussion.
 


\subsection{Photometric Redshifts}
\label{sec:cens:photoz}

As LSST is a purely photometric survey and AGN science generally
requires having accurate redshifts, photometric redshift
determinations are a crucial part of the project.  To zeroth order,
the continuum of an unobscured quasar longward of Ly$\alpha$ is a
power-law, and thus its colors are independent of redshift.  However,
the broad strong emission lines of high equivalent width modulate the
colors as a function of redshift, allowing photometric redshifts to be
determined with surprising fidelity \citep{Wei++04,Richards09},
especially once the Ly$\alpha$ forest enters the filter set.  SDSS was
able to determine photometric redshifts for quasars to $\Delta z = \pm 0.3$ 
for 80\% of SDSS quasars ($i\sim19$; \citealt{Richards09}).  Even
without the $y$-band LSST will do at least that well to $i\sim24$.

\begin{figure}
\begin{center}
    \includegraphics[scale=0.35]{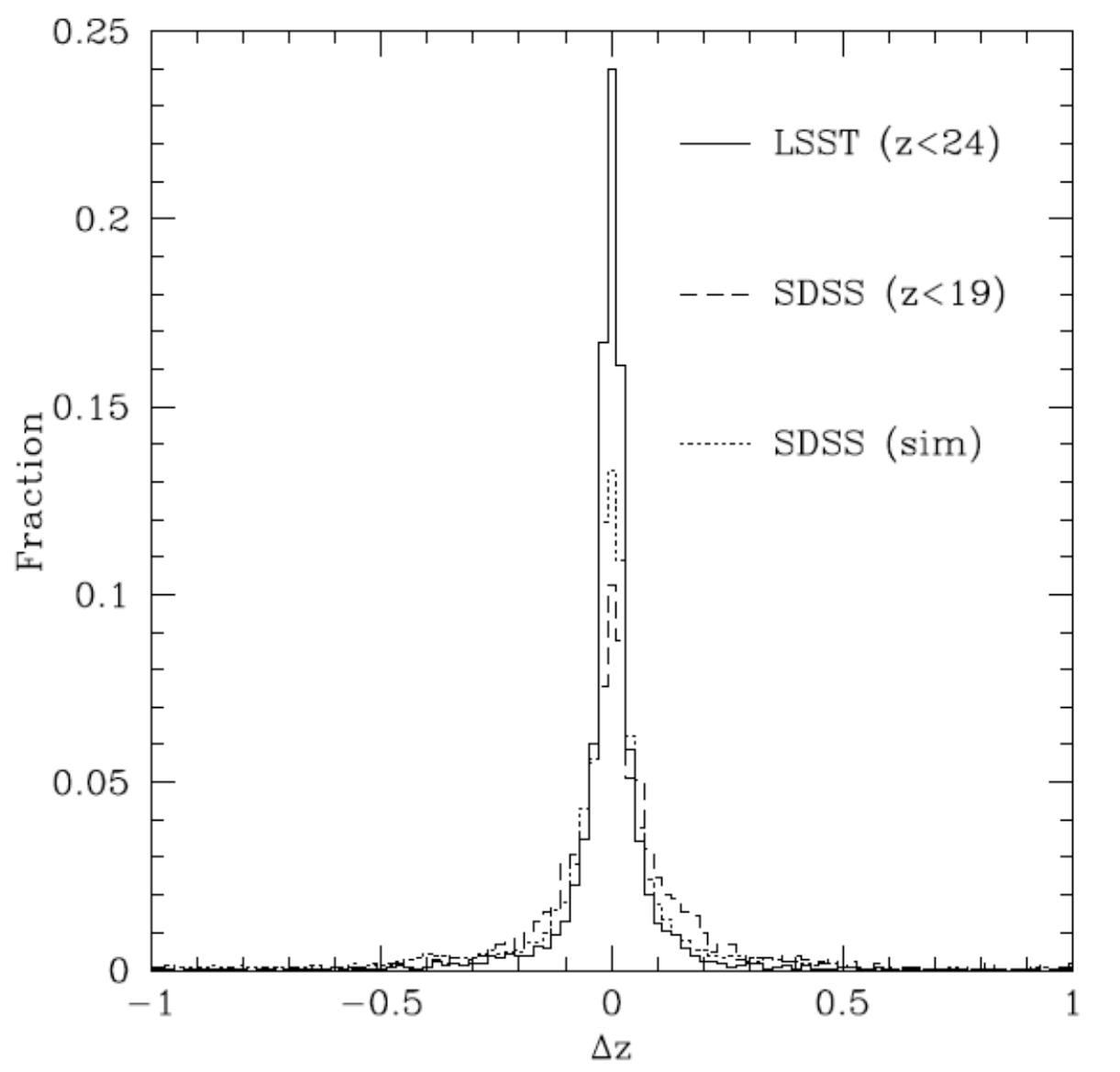}
  \caption{Distribution of the difference between photometric and
  spectroscopic redshifts ($\Delta z = z_{\rm spec} - z_{\rm phot}$)
    for UV-excess ($z\le2.2$) quasars (at higher redshifts, the sharp
    discontinuity in the spectral energy distribution (SED) caused by the onset of the Ly$\alpha$
    forest is measureable by the LSST filter set, making photometric
    redshifts quite accurate).  Results for known SDSS quasars to $i =
    19.1$ ({\em dashed}) are compared with expected results from LSST
    to $r\sim 26.5$ ({\em
      solid}).  LSST results assume the full 10-year co-added
      photometry. The simulated LSST
    quasar colors follow \citet{Fan99} (see also
    \citealt{Richards06}), using a distribution of power-law colors
    modulated by broad emission line features and inter-galactic
    hydrogen absorption.  Photometric redshifts were determined using
    the algorithm of \citet{Wei++04}. The dotted line shows the
    results from {\em simulated} SDSS quasars; they do better than the
    real data because of the limitations of our simulations.  
LSST's deeper imaging will allow accurate photometric redshifts
    to much fainter magnitudes than SDSS, while the addition of the
    $y$ bandpass reduces the overall scatter.}
\label{fig:agn:photoz}
\end{center}
\end{figure}

\autoref{fig:agn:photoz} shows that for UV-excess quasars, LSST will
produce considerably more precise photometric redshifts, with more than
80\% of quasars having photometric redshifts accurate to $\Delta z = \pm
0.1$ (SDSS did this well for only 60\% of quasars).  With LSST's
exquisite astrometry (\autoref{sec:Gaia}), the subtle effects of
emission lines on chromatic aberration will be measureable, allowing
an independent estimate of redshift \citep{Kacz09}.  When combined
with photometric redshifts, we estimate that the fraction of quasars
with redshifts correct to $\Delta z = \pm 0.1$ will be of order 90\%. 

\subsection{Expected Number of AGN}
\label{sec_cens_numbers}

The growth in the benchmark sizes of individual quasar samples is 
impressive over the past several decades, starting at $N\sim10^{0-1}$ 
\citep[e.g.,][]{Schmidt63, Schmidt68}, but growing rapidly to $10^2$ 
\citep[e.g.,][]{Braccesi70,Schmidt83}, and then to $N\sim10^3$ 
\citep[e.g.,][]{Hewett93} by the 1990s. The most recent decade has seen 
the continuation of an exponential expansion in the number of quasars 
identified in homogeneously selected samples, extending to moderate 
depth: 25,000 color-selected quasars to $b_J<20.85$ are 
included in the final 2dF QSO Redshift Survey catalog \citep{Croom04},
and  SDSS is approaching $N\sim10^5$ spectroscopic quasars (mostly
with $i<19.1$) \citep{Schneider07}, and $N\sim10^6$
photometrically-selected quasars to  $i<21.3$ \citep{Richards09}. LSST
will provide a major leap forward in  
quasar sample size, plausibly identifying over $10^7$ quasars to 
beyond $m\sim24$ through the variety of selection approaches we've
just outlined. 

An estimate of LSST's coverage of the quasar redshift-magnitude plane
is given in \autoref{tab:agn:agn_counts}.  The numbers of quasars in
the various bins were calculated using the quasar luminosity
function of \citet{Hopkins07}, extrapolated to low luminosities.  
\citet{Hopkins07} combines the most recent measurements of the
luminosity function from optical, IR, and X-ray data to provide the
most robust determination available to date of the {\em bolometric}
luminosity function over the redshift and luminosity ranges that LSST
will survey.  These results are in good agreement 
with those of the 2dF SDSS Luminous Red Galaxy and Quasar Survey data
\citep{Croom09b}, which is restricted to lower redshift and lower
luminosity than LSST will probe.  In all, LSST will detect over 10
million type 1 AGN with $M_i \le -20$, $i \le 24.5$ and redshifts
below 6.5; this number rises to as many as 16~million for $i \le
26.25$.  

\begin{figure}
\begin{center}
    \includegraphics[scale=0.5]{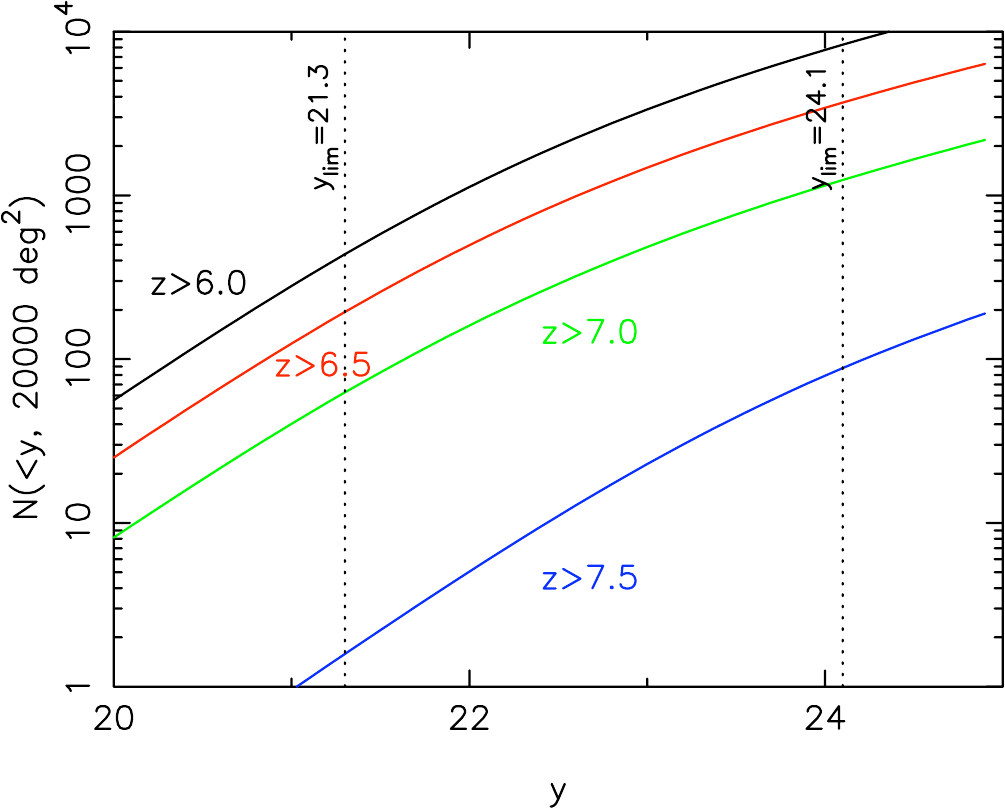}
  \caption{
Number of high-redshift ($z>6$) quasars expected to be discovered in a
20,000 deg$^{2}$ area as a function of redshift and limiting
magnitude.  We use the luminosity function (LF) at
$z\sim6$ measured by \citet{Jiang09}.
We assume that the density of quasars declines with redshift as measured in
\citet{Fan01,Fan06} and continues to $z>6$, with the same LF shape.
Two vertical dashed lines indicate the 10-$\sigma$ detection limit for
LSST for a single visit and for the final coadd.  
}
\label{fig:agn:highz}
\end{center}
\end{figure}

At very high redshift ($z>6$) and faint luminosities, a better estimate
is provided by the \citet{Fan06b} and \citet{Jiang09} samples.
Predictions for $z>6$ quasars from these studies are shown in
\autoref{fig:agn:highz}.  LSST can detect significant number of
quasars up to $z\sim7.5$, after which quasars become $y$ drop-outs.
Indeed, one of the most important discoveries of LSST is expected to
be the detection of many AGN at the end of the cosmic ``Dark Ages."
\autoref{fig:agn:z7quasars} shows that the $y$-band filter will permit
selection of quasars out to $z\sim7.5$ and down to moderate AGN
luminosities ($\approx10^{45}$~ergs~s$^{-1}$) in impressively high
numbers due to the steepness of the luminosity function at high
redshifts.  
Such quasars should be detected as $z$-band dropouts and will be
followed up spectroscopically from the ground and with 
  JWST. This will exceed the current number of the most distant SDSS
quasars at $5.7<z<6.4$ by an order of magnitude
\citep[e.g.,][]{Fan06}.  The LSST census of $z\sim7$ quasars
will place tight constraints on the cosmic environment at the end of
the reionization epoch and on the SMBH accretion history in the
Universe.



\begin{table}
\vbox{
\halign{\hskip 3pt
\hfil # \hfil\tabskip=1em plus 1em minus1em&
\hfil # &
\hfil # &
\hfil # &
\hfil # &
\hfil # &
\hfil # &
\hfil # &
\hfil # \cr 
\noalign{\medskip}
\noalign{\medskip\hrule\smallskip\hrule\medskip}
\hfil $i$ \hfil
& \hfil 0.5 \hfil
& \hfil 1.5 \hfil
& \hfil 2.5 \hfil
& \hfil 3.5 \hfil
& \hfil 4.5 \hfil
& \hfil 5.5 \hfil
& \hfil 6.5 \hfil & \hfil Total \hfil \cr
\noalign{\medskip\hrule\bigskip}
16 & 	666 & 	597 & 	254 & 	36 & 	0 & 	0 & 	0 & 	1550 \cr 
17 & 	4140 & 	4630 & 	1850 & 	400 & 	54 & 	0 & 	0 & 	11100 \cr 
18 & 	19600 & 	28600 & 	10700 & 	1980 & 	321 & 	19 & 	0 & 	61200 \cr 
19 & 	68200 & 	131000 & 	53600 & 	8760 & 	1230 & 	115 & 	0 & 	263000 \cr 
20 & 	162000 & 	372000 & 	194000 & 	35000 & 	4290 & 	441 & 	1 & 	767000 \cr 
21 & 	275000 & 	693000 & 	453000 & 	113000 & 	14000 & 	1380 & 	34 & 	1550000 \cr 
22 & 	336000 & 	1040000 & 	756000 & 	269000 & 	41200 & 	3990 & 	157 & 	2450000 \cr 
23 & 	193000 & 	1440000 & 	1060000 & 	476000 & 	103000 & 	10900 & 	527 & 	3280000 \cr 
24 & 	0 & 	1370000 & 	1360000 & 	687000 & 	205000 & 	27400 & 	1520 & 	3660000 \cr 
25 & 	0 & 	314000 & 	1540000 & 	888000 & 	331000 & 	60800 & 	4100 & 	3140000 \cr 
26 & 	0 & 	0 & 	279000 & 	760000 & 	358000 & 	86800 & 	7460 & 	1490000 \cr 
\noalign{\smallskip}
Total & 	1060000 & 	5390000 & 	5720000 & 	3240000 & 	1060000 & 	192000 & 	13800 & 	16700000 \cr 
\noalign{\medskip\hrule\smallskip\hrule}
}}
\caption{Predicted Number of AGN in 20,000 deg$^2$ over $15.7<i<26.3$
and $0.3<z<6.7$  with $M_i \le -20$.  The ranges in each bin are
$\Delta i=1$ and $\Delta  z_{em}=1$, except in the first and last bins
where they are 0.8 and 0.7, respectively.} 
\label{tab:agn:agn_counts}
\end{table}

The Chandra Deep Fields show a surface density of order 7000 AGN per
deg$^2$ \citep[e.g.,][]{Bauer++04, Brandt05}, which, when extrapolated to the 20,000
deg$^2$ of LSST, implies a total count of over $10^8$ AGN, an order of
magnitude larger than the optical AGN luminosity function would
predict.  This may be thought of as a reasonable upper limit to the
number of AGN that LSST might find, as it includes optically obscured
objects and may include objects of
intrinsically lower luminosity than we have assumed, and may also point
to errors in our 
extrapolation of the measured
luminosity function.  Indeed, this gives us motivation to {\em
measure} the luminosity function, as we discuss below in
\autoref{sec:agn:LF}.

\begin{figure}
\begin{center}
 \includegraphics[angle=0,scale=0.3]{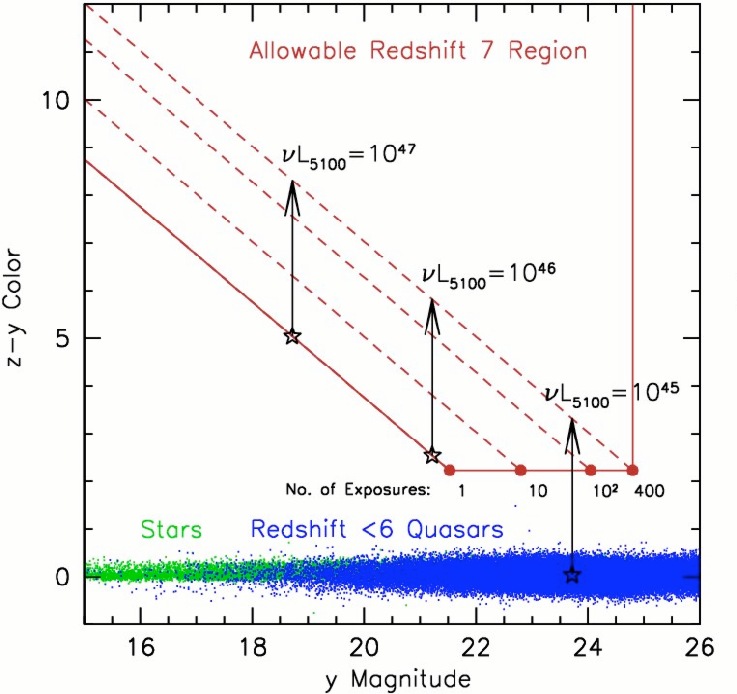}
  \caption {LSST $z-y$ vs. $y$ color-magnitude diagram,
    showing the expected region in which $z\sim7$ quasar
    candidates will lie. The region limits are defined by the $5~\sigma$
    $y$-magnitude detection limits, and $2~\sigma$ $z$-magnitude
    detection limits, as a function of the number of co-added $15$~s
    exposures. An object is considered a $z\sim7$ candidate if it is
    detected at the $>5~\sigma$ level in the $y$-band, and does not
    exceed the $z$ band $2~\sigma$ detection limit.  
The
    $y$-magnitudes and $z-y$ color limits are shown for simulated
    $z=7$ quasars at three different $5100$~\AA\ continuum
    luminosities. The open stars show the minimum $z-y$ color limits
    required for a single $15$~s exposure, and the ends of the arrows
    show the limits for $400$ exposures (the full extent of
    the 10-year survey).  
The
    expected $z-y$ colors of stars (green) and $z<6$ quasars (blue),
    based on results from the SDSS, are shown for comparison.  }
\label{fig:agn:z7quasars_cm}
\end{center}
\end{figure}

\begin{figure}
\begin{center}
 \includegraphics[width=0.5\linewidth]{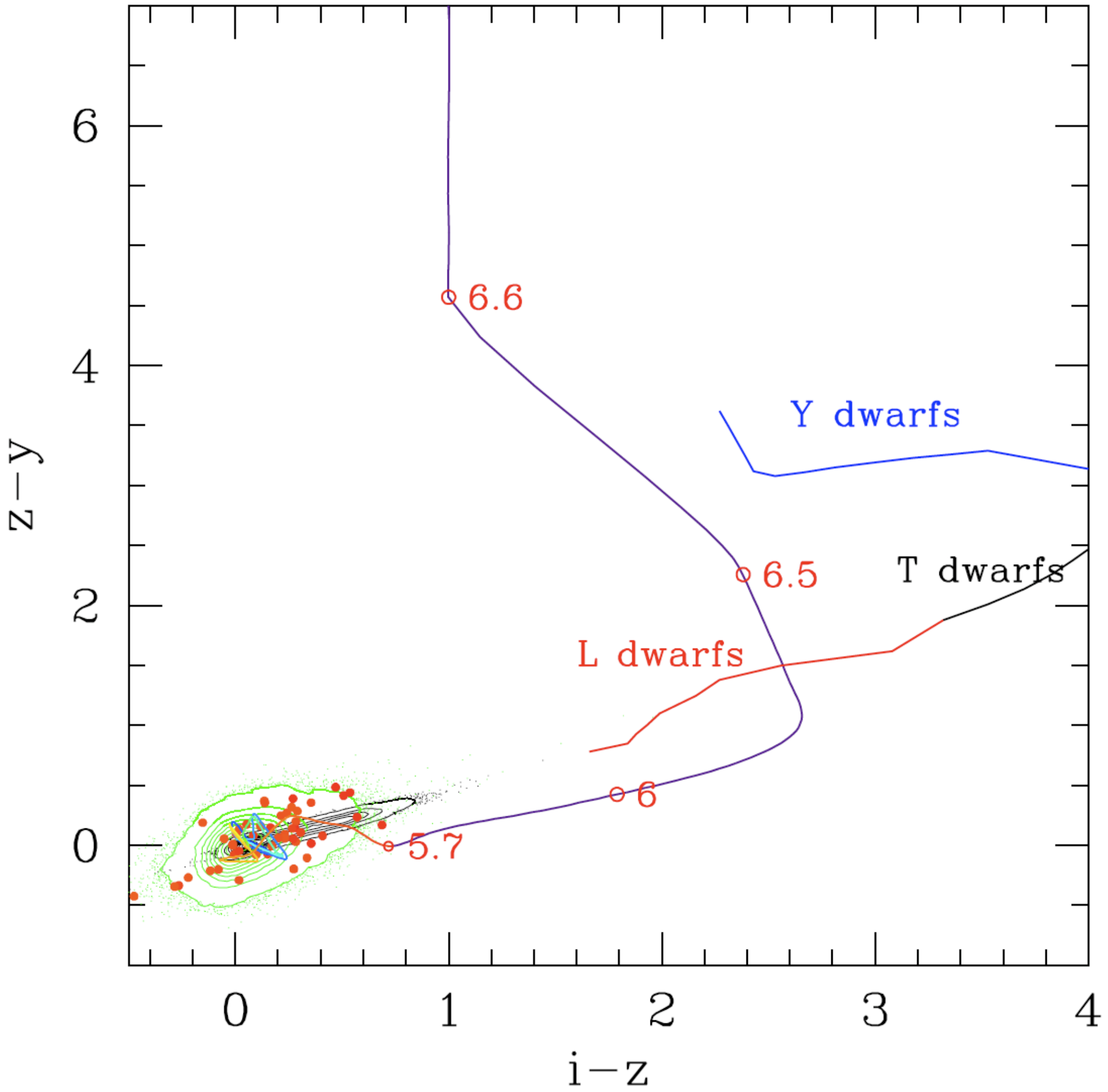}
  \caption{Distribution of objects in $i-z,z-y$ color space.  Ordinary
  stars are shown as the black contours, while low-redshift quasars
  are the green contours; those quasars with redshift above 5 are red
  dots.  The loci show the expected tracks of higher-redshift quasars
  (with redshifts labeled), and brown dwarfs of type L, T, and Y, as
  labelled.}
\label{fig:agn:z7quasars}
\end{center}
\end{figure}

\section{AGN Luminosity Function}
\label{sec:agn:LF}

\noindent{\it 
Scott F. Anderson,
Richard F. Green,
Gordon T. Richards,
Donald P. Schneider,
Ohad Shemmer}

%
%
%
%
%
%
%
%
%
%
%
%
%
%
%
%
%
%
%
%
%
%
%
%
%
%



The census of AGN through cosmic time, tracing the evolution of
supermassive black holes, may be quantified via the AGN luminosity
function (hereafter, LF), as well as closely related empirical
measures such as the $\log N - \log S$ curve.  The LF impacts studies
of the ionizing background radiation, the X-ray background, and quasar
lensing, and constrains a variety of parameters in physical models for
the evolution of AGNs, including black hole masses, accretion rates
and Eddington ratios, the fraction of massive galaxies (perhaps most)
that undergo an AGN phase and the lifetime of this phase, and cosmic
down-sizing \citep[e.g.,][]{K+H00,Wyithe02,Hopkins07}.  The
variety and sensitivity of LSST-enabled, AGN-selection metrics will
result in a high-quality, representative AGN sample required for
detailed LF studies.  In particular, the large area, depth, and
dynamic range of LSST form a superb basis to study the populous faint
end of the LF at moderate to high redshifts.

While there has been exponential growth in quasar survey samples in
the last few decades (\autoref{sec_cens_numbers}), there has been far
less progress at ``ultrafaint" ($m > 22.5$) magnitudes that sample the
low-luminosity end of the LF.
For example, the pioneering photographic studies of 
Koo, Kron, and collaborators \citep[e.g.,][]{Koo86}
which extend to $B<22.6$ over 0.3 deg$^2$, are still often quoted in current studies as 
among the handful of reliable points in LF studies of the 
faint AGN population. Significant expansions in areal coverage from 
a few $m\sim22.5$ modern CCD-based surveys are underway \citep[e.g., 
3.9~deg$^2$ in the SDSS faint quasar survey of][]{Jiang06}.

Yet there are strong motivations in LF studies to explore much fainter
than the break in the number counts distribution, and this sparsely
sampled
ultrafaint regime is one where LSST is poised to have significant
impact on LF studies.  Only ultrafaint ($m>22.5$) surveys can probe
the populous, faint end of the AGN LF, especially at moderate to
high redshifts. For example, an AGN with absolute magnitude $M=-23$,
i.e., a high space density object from the faint end of the luminosity
function, will have apparent magnitude $m>22.5$ at $z>2.1$.
%
%

\begin{figure}[!htb]
\begin{center}
  \includegraphics[angle=0,scale=0.5]{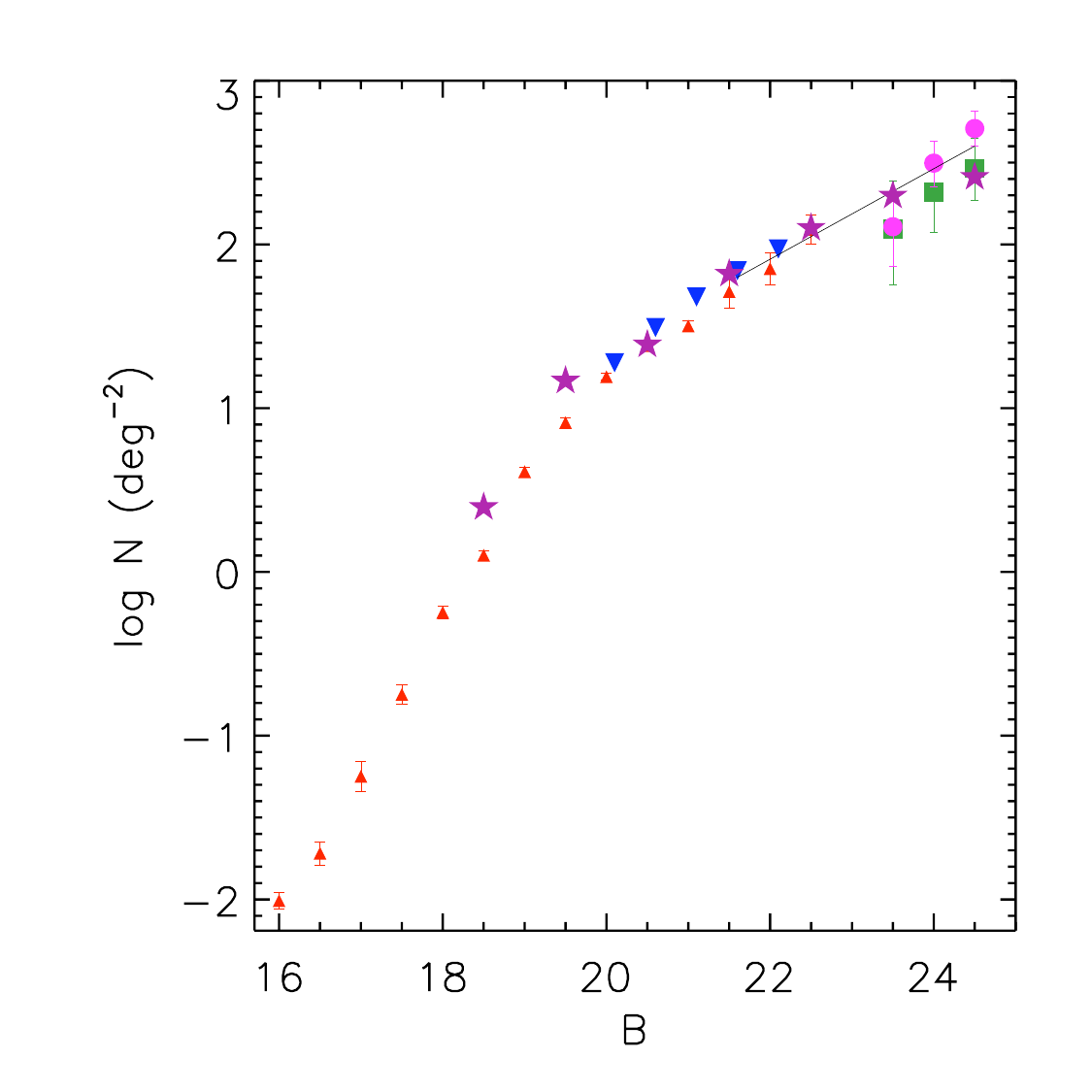}
  \caption{A summary of our current understanding of the numbers of
  AGNs per square degree of sky brighter than a given apparent
  magnitude, adapted from \citet{Beck07}.  The
    ultrafaint points are from the COMBO-17 survey 
    \citep[purple stars;][]{Wolf03} and HST based surveys 
    \citep[pink circles and green squares;][]{Beck07}. Shown for broad comparison
    are: brighter 2SLAQ points \citep[blue, upside-down triangles;][]{Richards05}; 
    a simple extrapolation of 2SLAQ points to ultrafaint
    magnitudes (solid line); and the \citet{Hartwick90} compilation
    (small, red triangles), which incorporates many earlier quasar
    surveys.  The data show $\sim500$ AGNs~deg$^{-2}$ to 
    $m<24.5$ and $z<2.1$. The LSST AGN surveys will extend both
    fainter and across a much wider redshift range, suggesting a
    sample of at least $\sim10^{7}$ AGNs.}
\label{fig:agn:number_counts}
\end{center}
\end{figure}

\autoref{fig:agn:number_counts} shows our current understanding of the 
optical AGN counts as a function of magnitude.  Among the most
reliable points well beyond $m>22.5$ are those of the 
\citet{Wolf03} COMBO-17 survey, although there are a handful of 
other smaller area optical surveys using a variety of selection
criteria that give similar 
results at least for AGNs out to $z<2.1$.  This figure suggests
that LSST will discover on the order of 500 photometric  
AGNs/deg$^2$ to $m<24.5$ and $z<2.1$, in rough agreement with the
numbers we found above.  To the extent that we can identify AGNs
from the co-added data below the single-visit limits, we should be able
to find appreciably more objects.  

Given the very large numbers of AGNs that LSST will find, a bin
of a few tenths in redshift covering a decade in luminosity will 
include thousands of AGNs over much of the redshift range, allowing
statistical errors to be negligible, and systematic errors (due to
errors in photometric redshifts, bolometric corrections, or selection
efficiency) will dominate our measurements.  
Of course the efficacy of any AGN census for establishing the LF is 
not measured merely by the numbers of objects sampled. Survey depth, 
sky coverage, dynamic range, completeness, contamination, redshift 
range, and wavelength selection biases/limitations, are all additional 
key elements. As an example, a recent survey embodying many of these as 
attributes is the 2SLAQ survey of 8700 AGNs over 190 deg$^2$, which extends to $g<21.85$ 
\citep{Croom09}.  But the dynamic range, redshift range, depth, and sky coverage of the LSST AGN sample will be much more impressive.

The impact of LSST depth and dynamic 
range in magnitude and redshift for ultrafaint AGN LF studies may be 
seen in the context of current LF models. One popular form for the LF 
considered in many recent studies is a double power law with 
characteristic break at luminosity $L_{*}$. The LF shape might evolve 
with redshift in either luminosity, density, or both 
\citep[e.g.,][]{Schmidt83}. For several decades, studies tended to 
favor pure luminosity evolution models, but some recent studies from
various wavebands \citep[some extending quite deep in small areas,
such as the X-ray studies of][]{Ueda03,Hasinger05} have found
markedly disparate evolutionary rates, depending on their energy
selection wavebands. Preliminary indications are that the slope of the
AGN luminosity function varies considerably from $z = 2$ to $z = 6$
\citep{Richards++06,Jiang08}.  In reconciling multiple survey results from various wavebands, there has been a recent resurgence in combined
luminosity/density evolution
models \citep[e.g.,][]{Schmidt83,Hasinger05,Croom09}, which
incorporate ``cosmic downsizing'' \citep{Cowie96} scenarios for the
LF.  These are well represented by the bolometric LF studies
of \citet{Hopkins07}, who argue that the peak of the AGN space density
occurs at increasing redshifts for more luminous AGNs (see also \citealt{Croom09b}). 

In the currently popular merger plus feedback model
of \citet{Hopkins06}, 
the faint-end slope of the luminosity function is a measure of how much time
quasars spend accreting at sub-Eddington rates (either before or after
a maximally accreting state).  The bright-end slope, on the other
hand, tells us about the intrinsic properties of quasar hosts (such as
merger rates).  If these two slopes are fixed with cosmic time,
then the space density of AGN will peak at the same redshift at all
luminosities --- contrary to recent results demonstrating downsizing,
whereby less luminous AGNs peak at lower redshift as the average mass
of accreting supermassive black holes moves to lower scales with
cosmic time.  Thus, understanding the evolution of the bright- and
faint-end quasar LF slopes is central to understanding cosmic downsizing.

\begin{figure}[!htb]
\begin{center}
  \includegraphics[angle=0,scale=0.5]{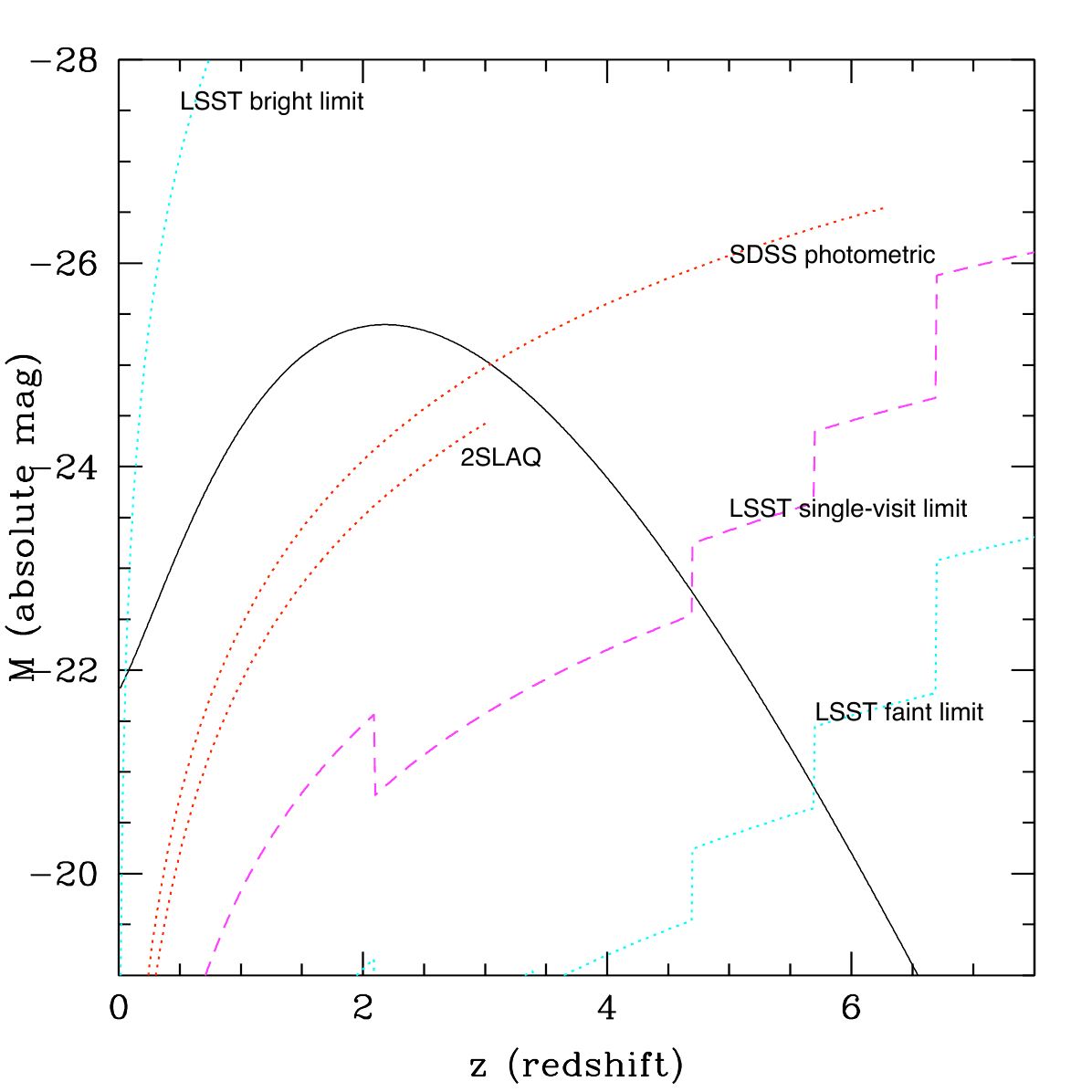}
  \caption{Depth and redshift coverage of large, optical surveys,
    compared to a representation of the LF of \citet{Hopkins07}.
    The evolution of the break luminosity $L_{*}$ with
    redshift is shown by the solid black curve (adapted from Hopkins
    et al.). The corresponding sensitivity of two current large
    quasar surveys is depicted by the dotted red curves 
    \citep[2SLAQ and SDSS photometric surveys;][]{Richards05,Richards09}.
    The depth of the
    LSST AGN survey will permit a much more sensitive measure of
    the break luminosity evolution at intermediate to high redshifts,
    encompassing (in a single sample) $0<z<4.5$ (magenta curve
    reflecting LSST single-visit depth), and perhaps $0<z<5.5$ (lower
    cyan curve reflecting final, stacked LSST depth). The cyan curve
    to the upper left reflects the bright limit of LSST (in a single
    visit).}
\label{fig:agn:lf_break}
\end{center}
\end{figure}

\autoref{fig:agn:lf_break} (adapted from figure 8 of 
\citealt{Hopkins07}) shows a realization of one of these 
downsizing models: it adopts the usual double power-law shape, but
allows for a break luminosity $L_{*}$ that evolves with redshift, as
shown by the solid line.  Superposed are dotted red curves
representative of the faint limits of the 2SLAQ and the SDSS
photometric surveys \citep{Richards05,Richards09}.  These surveys,
however, don't probe significantly beyond the break luminosity for redshifts
much larger than 2.
%
The bright limit is indicated by the cyan curve, and the faint limit
in a single visit probes to the break luminosity to $z = 4.5$, and to
$z = 5.5$ in the co-added images, even in this model in which the break
luminosity decreases rapidly at high redshift.  Thus the
LSST-determined quasar LF will provide crucial insights to our understanding
of AGN feedback in the early Universe and how it influences the
evolution of massive galaxies.

With the large number of objects in the sample, the 
dominant
uncertainties in LF studies will be systematics, such as the
contamination of the sample by non-AGNs, completeness, and
uncertainties associated with photometric redshifts.  Internal
comparison of LSST color-, variability-, and proper motion-selected
AGN surveys will limit contamination and enhance completeness
(\autoref{sec_cens_select}), while comparison with deep Chandra X-ray and
Spitzer mid-IR data will allow the selection effects to be
quantified. 
There is clearly a need as well for spectroscopic follow-up
of a modest subset of the full LSST sample to further quantify the
contamination of the sample from non-AGN. 

\section{The Clustering of Active Galactic Nuclei}
\label{sec:agn:clustering}

\noindent{\it 
Michael A. Strauss, 
Robert J. Brunner,
Jeffrey A. Newman}

%
%
%
%
%
%
%
%
%
%
%
%
%
%
%
%
%
%
%
%
%
%
%
%
%
%
One way that we can get a handle on the physical nature of
the galaxies that host AGN and the conditions that cause infall and
growth of the black hole is to study the spatial clustering of the
AGN.  The relationship between AGN clustering and that of ``ordinary''
galaxies can give important clues about how the two are physically
related.  

  The luminous parts of galaxies of course represent only a small
  fraction of the clustered mass density of the Universe, and there is
  no guarantee that the clustering apparent from the matter that we
  see matches that of the underlying dark matter perfectly
  (\autoref{sec:gal:dm}).  A common 
  hypothesis, which is predicted, e.g., in so-called {\em threshold bias}
  models in which galaxies form only in regions of high density
  contrast in the dark matter, is that the fractional density contrast 
$
\delta({\bf r}) \equiv \frac{\rho({\bf r}) - \langle \rho\rangle}
   { \langle \rho\rangle}
$
as measured for galaxies is proportional to that of the dark matter:
$$\delta_{galaxies} = b\,\delta_{dark\ matter}.$$ 
Here where the bias factor
$b$ may be a function of the smoothing scale on which $\rho$ is
measured.  This simple relation is often referred to as a linear bias
model (as opposed to models which include higher-order terms or
scatter around this simple deterministic relation; see also the
discussion of halo occupation distribution models in
\autoref{sec:galaxies:distfunct} and
\autoref{sec:galaxies:clustering}).  

  In threshold bias models, the bias factor $b$ is directly related to
  the value of the threshold.  Thus one can
  determine the characteristic mass of the dark matter halos associated
  with a given sample of galaxies directly from a measurement of their
  clustering.   The higher the halo mass
  associated with the galaxy population in question, the higher the
  bias, and, therefore, the stronger the
  expected clustering.  

  In practice, clustering is quantified by measuring the correlation
  function $\xi(r)$ (or
  its Fourier Transform, the power spectrum) of the galaxy sample, as
  described in \autoref{sec:galaxies:clustering}, 
and comparing it with that of the underlying dark matter as predicted
from linear theory (on large scales) or N-body simulations (on
smaller, non-linear scales).  The linear bias model states
\begin{equation} 
\xi_{galaxies} = b^2 \xi_{dark\ matter},
\end{equation}
where again $b$ may be a function of scale. Our current cosmological
model is precise enough to allow a detailed prediction for $\xi_{dark\
  matter}$ to be made. 

  The galaxy correlation function at low redshift has been measured
  precisely, using samples of hundreds of thousands of galaxies (from
  the redshift survey of the SDSS; see,
  e.g., \citealt{Zeh++05,Eis++05}), allowing quite accurate
  determination of  
  the bias as a function of scale for various subsets of galaxies.
  However, AGN are rarer, and the measurements are not as accurate
  (for example, the mean separation between $z \sim 3$ quasars in the
  SDSS is of order 150 co-moving Mpc).
  The enormous AGN samples selectable from LSST data
  (\autoref{sec:agn:census}) will cover a very large range of
  luminosity at each redshift, allowing the clustering, and thus bias
  and host galaxy halo mass, to be determined over a large range of
  cosmic epoch and black hole accretion rate. 


  While gravitational instability causes the contrast and, therefore,
  the clustering of dark matter to grow monotonically with time,
  observations of galaxies as a function of redshift shows their
  clustering strength measured in co-moving units to be essentially
  independent of redshift (albeit with increasingly larger error bars
  at higher redshift).  This is roughly as expected, if (for a given
  population of galaxies) the characteristic halo mass is independent
  of redshift.  As one goes to higher redshift, and therefore further
  back in time, the amplitude of the underlying dark matter clustering
  decreases, meaning that this characteristic halo mass represents an
  ever-larger outlier from the density contrast distribution, and is
  therefore ever more biased.  Quantifying this relation allows one to
  measure the characteristic halo mass of galaxies as a function of
  redshift \citep{Ouchi++05}. 

  We would like to do the same for AGN, to determine the masses of
  those halos that host them.  The observed correlation function
  of luminous quasars at all redshifts below $z \sim 3$ is very
  similar to that of luminous red ellipticals, suggesting that they
  live in similar mass halos, and perhaps that these quasars are
  hosted by these elliptical galaxies \citep[e.g.,][and references
  therein]{Ross++09}. 

  How does the clustering depend on AGN luminosity?  The AGN
  luminosity depends on the mass of the central black hole, and the
  Eddington ratio. 
It has been suggested that the mass of
  the central black hole is correlated with that of its host halo at
  low redshift
  \citep{Fer02}; after all, these black holes are correlated with
  the mass of the spheroidal components of galaxies, and the masses of
  these spheroids are plausibly correlated with the mass of the halo,
  as modern Halo Occupation Distribution (HOD) models would
  suggest.  Thus if most AGN are accreting at close to the Eddington
  limit \citep{Kol++06,She++08}, one might imagine a
  fairly significant correlation of clustering strength with
  luminosity.  If, on the other hand, luminosity is driven more by a
  range of Eddington ratios, such luminosity dependence becomes quite
  weak \citep{Lidz++06}.  Models of black hole growth differ largely
  on questions of the duration of the accretion and the level and
  constancy of the 
  Eddington ratio, thus measurements of the luminosity dependence of
  the clustering strength become particularly important.

  Current samples, however, simply do not have the dynamic range in
  luminosity at any given redshift to allow this test to be done
  robustly.  For example, the SDSS quasar sample \citep{Richards++06}
  has a range of only 
  about two magnitudes (a factor of less than 10 in luminosity) over
  most of its redshift range.  Samples going deeper do exist over
  small areas of sky, but do not probe the large scales where the
  linear clustering is best measured.  The current measurements of the
  luminosity dependence are poor: the data are consistent with no
  luminosity dependence at all (although there is a hint of an upturn
  for the highest luminosity decile, \citealt{Shen++08}), but the
  error bars are large, the range of luminosities tested is small, and
  redshift and luminosity evolution are difficult to separate out.

  LSST will increase the dynamic range enormously over existing
  samples.  At most redshifts, we will be able to select AGN with
  absolute magnitudes ranging from $-29$ to $-20$
  (\autoref{tab:agn:agn_counts}), a factor of several 
  thousand in luminosity, and the numbers of objects in moderate
  luminosity bins will certainly be large enough to measure the
  correlations with high significance.  There must be a luminosity
  dependence to the clustering at some level if black hole masses are
  at all correlated with halo masses; this may only become apparent
  with samples of such large dynamic range.  

\begin{figure}
\begin{center}
\includegraphics[width=12cm]{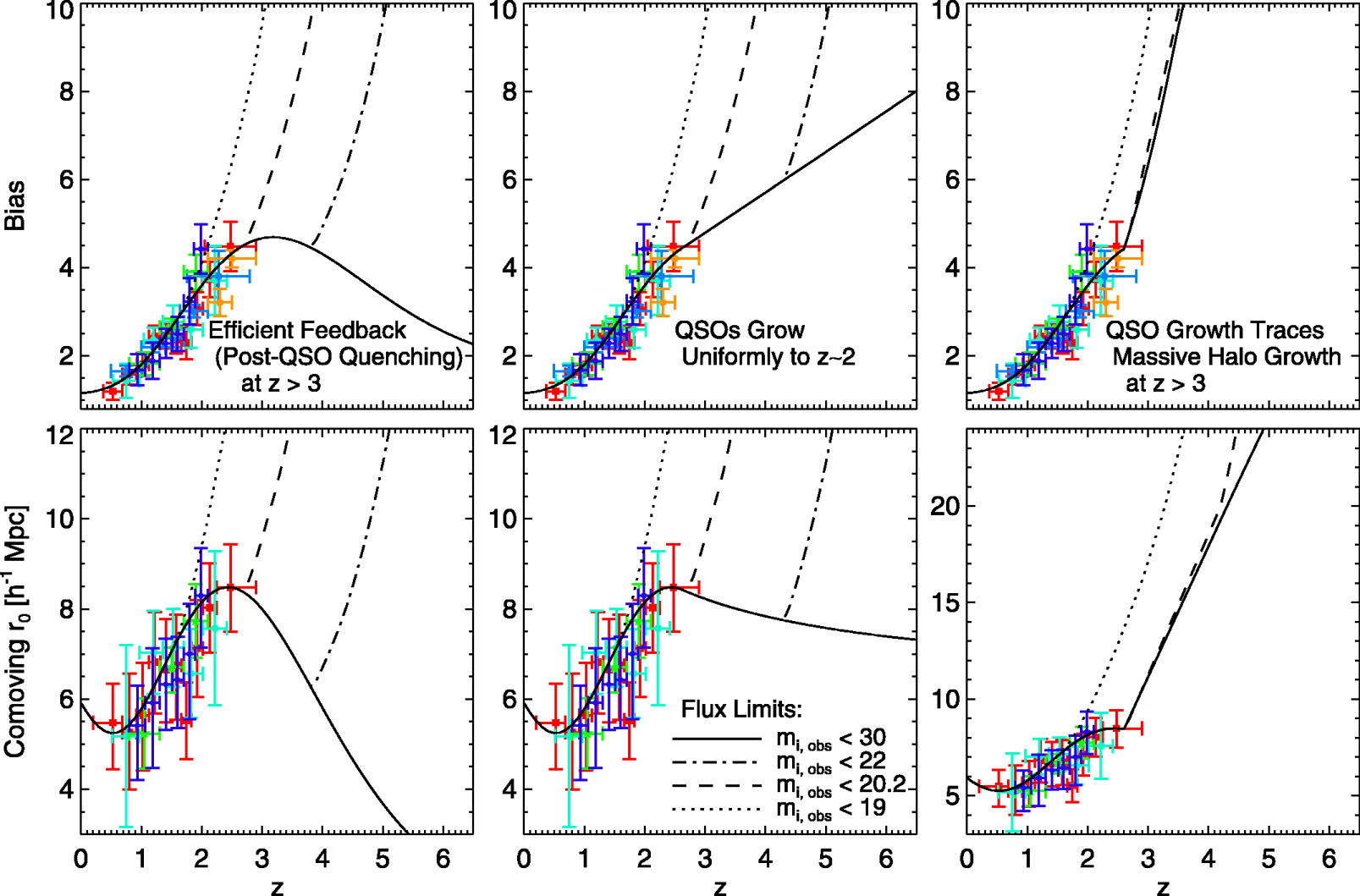}
\caption{The bias (top panels) and comoving clustering length (lower panels) of
quasars in three models of quasar growth, for samples of various
limiting magnitude.  LSST will be able to probe to limiting
magnitudes of $m \sim 26$ reliably.  Measured data points, entirely
limited to $z < 2.5$, are shown as colored points with error bars.
Note that the models are essentially entirely degenerate, with no
luminosity dependence, in this redshift range; all the action is at $z
> 3$.  Even at $z > 3$, one needs to go appreciably fainter than the
SDSS magnitude limit to break the degeneracy.  The three models are (left to right): an efficient feedback
model (in which infall to the SMBH halts immediately after a quasar
episode); a model in which SMBHs grow smoothly to $z=2$; and a model
in which black hole growth is tied to that of the dark matter halo to
$z = 2$.  Figure from \citet{2007ApJ...662..110H}, with permission.}
\label{fig:agn:luminosity_clustering}
\end{center}
\end{figure}

\begin{figure}
\begin{center}
\includegraphics[width=6cm]{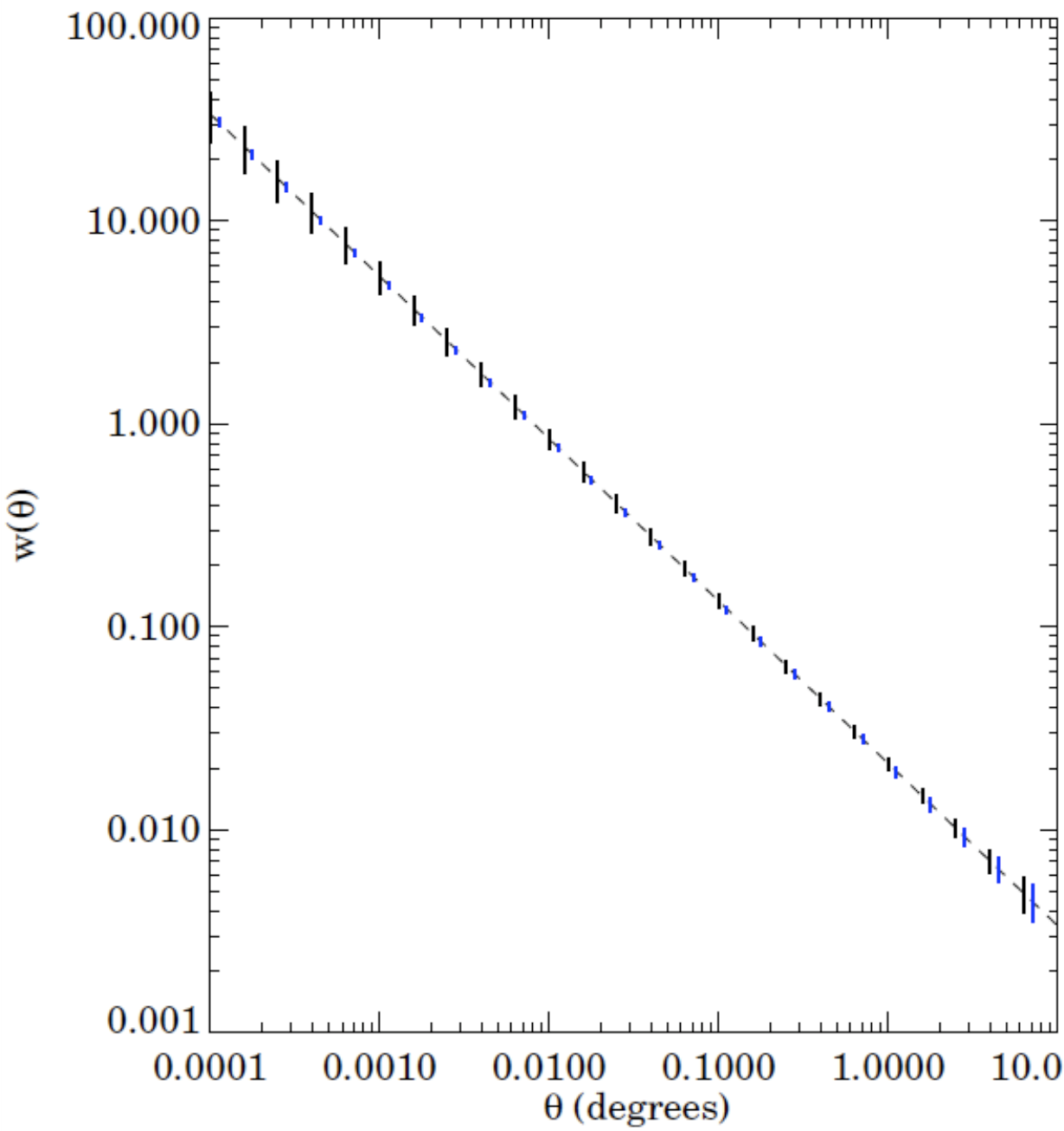}
\caption{The predicted angular auto-correlation of quasars (black) and
cross-correlation between quasars and galaxies (blue), for a sample of
250,000 quasars with $g < 22.5$ with redshifts between 2.75 and 3.25.
The underlying three-dimensional auto- and cross-correlation functions
are assumed for the purposes of the figure to be the same pure
powerlaw, $\xi(r) = (r/10\,h^{-1}{\rm Mpc})^{-1.8}$ in co-moving
coordinates.  The galaxy sample extends to $i < 25$.  The error bars
are calculated using the formalism of \citet{Bernstein94}.  Even
$10\times$ smaller quasar samples will yield useful clustering
measurements via cross-correlation techniques.}
\label{fig:qso_correlation}
\end{center}
\end{figure}

At higher redshifts, \citet{She++07} have
found that the clustering length grows with redshift: $17\pm
2\,h^{-1}$Mpc at $z \sim 3.2$, and $23 \pm 3\,h^{-1}$Mpc at $z \sim 
4$; \citep{She++07}.  This suggests both 
that the most luminous objects at these redshifts are accreting at
close to the Eddington limit (and, therefore, their luminosities reflect
their black hole masses), {\em and} the black hole masses are tightly
coupled to their halo masses \citep{2008MNRAS.390.1179W}.  Exploring
these connections at lower luminosities is crucial, as has been
emphasized by \citet{2007ApJ...662..110H}, where different models for
AGN feeding can be distinguished by the luminosity dependence of
clustering at $z > 3$.  This is illustrated in
\autoref{fig:agn:luminosity_clustering}, which shows the substantial
dependence of the quasar bias and co-moving clustering length on
redshift and luminosity as predicted in various models.  Most of the
luminosity dependence, and the distinction between models, becomes
apparent at $z > 3$, where existing data are very limited.
\autoref{fig:qso_correlation} shows both the angular quasar
auto-correlation, and the quasar-galaxy cross-correlation, that we
might expect for a sample of 250,000 quasars with $2.75 < z < 3.25$
with $g < 22.5$ (i.e., easily visible in a single visit).  The error
bars are calculated using the formalism of \citet{Bernstein94}.  In
fact, our photometric redshifts will be good enough to explore
clustering in substantially finer redshift bins, strengthening the
clustering signal (\autoref{sec:cens:photoz}). Even with broad
redshift bins, correlation function errors are small enough that we
can divide the sample into many bins in luminosity, color, or other
properties, allowing us to explore both the redshift and luminosity
dependence of the clustering strength.  For cross-correlation studies,
errors in the clustering measurements at small scales depend only
weakly on sample size (as $1/sqrt{N}$), allowing $S/N>10$ measurements
even for samples 10 times smaller than the one shown here.
 
From the measurement of quasar clustering, we get an estimate of the
minimum mass of the halos hosting them.  Given a
cosmological model, the number density of halos of that mass can be
predicted, and the ratio to the observed number density of quasars
allows inference of the duty cycles of quasars.  With existing data
\citep{She++07}, this test gives uncertainties of an order of
magnitude; with LSST, this can be done much more precisely 
and explored as a function of luminosity, thereby
further constraining models of AGN growth.  

The small-scale clustering of AGN can be studied in great detail with
LSST; the co-added photometry will go deep enough to see host clusters,
for example, to at least $z = 4$.  This gives an independent test of
bias relations as a function of redshift; given that the
highest-redshift quasars are so strongly biased, they live in
particularly massive halos and, therefore, are likely to lie in regions
in galaxy overdensity.  These data will allow us to explore how
quasars fit into the Halo Occupation Distribution picture as a
function of luminosity and redshift
(\autoref{sec:galaxies:distfunct}).  Indeed, the quasar-galaxy
cross-correlation function can be measured to much higher precision
than the auto-correlation function, simply because there are so many
more galaxies in the sample (see the discussion in
\autoref{sec:galaxies:clustering}).  As \citet{Padmanabhan++08} describe,
the cross-correlation of quasars with either the general galaxy
population or specific galaxy subsamples can be directly compared to
the auto-correlation of that galaxy sample to place constraints on the
quasar bias, its evolution with redshift and luminosity, and the
quasar host halo mass at different cosmic epochs \citep{Coil++07}.

LSST will also be able to resolve close companion galaxies to quasars,
allowing us explore how mergers drive quasar activity.  Finally, the
stacked images will go to low enough surface brightness and have
enough dynamic range to separate out quasar host galaxy light; an
important exercise for the future is to quantify to what extent this
will be doable as a function of luminosity and redshift.

Finally, LSST will explore the
nature of quasar pairs and the quasar correlation function on small
($< 1\,$Mpc) scales.  It is known that quasars show an excess of pairs
over what is expected given an extrapolation of the power-law from
larger scales \citep{Djo91,Hen++06,Mye++06}.  Is this excess due to
triggering of quasar activity in dense environments?  This will be explored with exquisite statistics and over a wide
range of luminosity and redshift with LSST.  Even projected pairs are tremendously
useful; follow-up spectroscopy allows the environments (IGM, companion
galaxies) and isotropy of the emission of the foreground object to be probed
from their signature in the absorption spectra of the background
object \citep{2007ApJ...655..735H}.

\section{Multi-wavelength AGN Physics}
\label{sec:agn:multilambda}

\noindent{\it 
Scott F. Anderson,
D.R. Ballantyne,
Robert R. Gibson,
Gordon T. Richards, 
Ohad Shemmer}

%
%
%
%
%
%
%
%
%
%
%
%
%
%
%
%
%
%
%
%
%
%
%
%
%
%
AGN emit strongly across a very broad energy range, typically with 
prodigious luminosity spanning at least from the infrared through the 
X-ray, and sometimes extending to radio and/or gamma-ray energies as well.  
Although a power-law is often used to describe the broad underlying 
spectral energy distribution (SED) of AGNs, such a characterization is a 
marked oversimplification: quasars display a rich diversity of radiation 
emission and absorption features that add complexity to their SEDs, but 
also enable a more detailed understanding of their complicated, 
multi-region structure. In many cases, a specific structure --- such as an 
accretion disk, a disk corona, a jet, or an outflowing wind --- is 
primarily associated with emission or absorption in a particular energy 
band.  A multi-wavelength view
of AGN is needed to understand the total
(bolometric) energy output of AGN, and also to study particular
structures that may differ dramatically among AGN subclasses.
For example, dust-obscured quasars are more 
readily found and studied in the infrared, and some quasar central engines 
enshrouded by moderately thick columns of intrinsic absorbing gas are 
best studied via hard X-rays.

Observations at other wavelengths are thus essential companions to LSST 
optical studies to obtain a reliable physical understanding of the structure 
of AGN, and to count and classify the wide range of observed 
multi-wavelength AGN phenomena with minimal bias. Moreover, the 
combination of multi-epoch LSST optical photometry with overlapping 
contemporaneous multiwavelength observations will provide unprecedented, 
time-dependent coverage of the AGN SED. Because LSST is repeatedly 
scanning a large portion of the sky, it will be possible to match LSST 
optical observations to any overlapping fields or individual AGN sources 
targeted contemporaneously by other missions, providing a 
near-simultaneous, multi-wavelength ``snapshot'' of the SED, as well as a 
description of the history and evolution of the source in LSST wavebands. 
Such time-dependent data will, for example, expand our knowledge of the 
co-evolution of accretion structures \citep[e.g.][]{DVB04},
and provide a unique view of remarkable sources 
such as blazars, highly-absorbed quasars, and perhaps new types of AGN 
that LSST will discover. LSST AGN studies will benefit from data taken 
with other telescopes or observatories of sources in the LSST sky,
including upcoming or ongoing 
wide surveys such as VISTA \citep{VISTA09}, { WISE} 
\citep{WISE09}, { EXIST} \citep{EXIST09}, { JANUS} \citep{JANUS09},
and { Fermi} \citep{Michelson03},
and existing wide surveys such as NVSS 
\citep{Condon++98}, SUMSS \citep{Bock++99},
2MASS \citep{Skrutskie++06}, COSMOS 
\citep{Scoville+07}, { GALEX} \citep{Martin++05}, { ROSAT} 
\citep{Truemper82}, and { XMM-Newton} \citep{Jansen++01}.

\subsection{Multi-wavelength AGN Classification}
\label{sec_multiwav_class}

\begin{figure}[!htb]
\begin{center}
  \includegraphics[angle=0,scale=0.3]{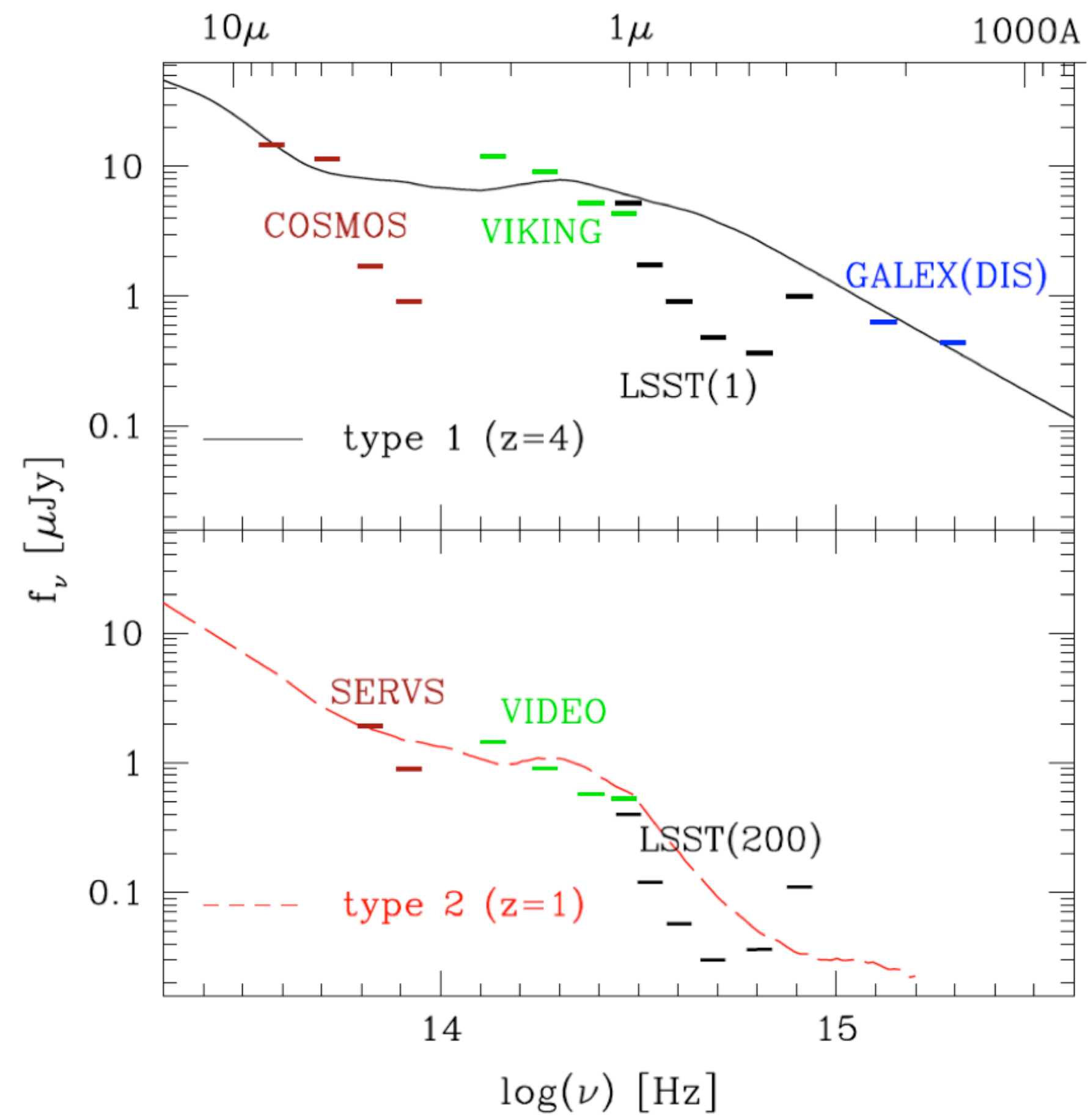} \caption{{\em
  top:} Type 1 quasar SED \citep{Richards06} at $z=4$ compared to
  the overlapping depth of the COSMOS \citep{SCOSMOS07} data in the
  mid-IR, the proposed VISTA-VIKING \citep{VISTA07} survey in the
  near-IR, the LSST single epoch data in the optical, and the
  GALEX \citep{Martin++05} Deep Imaging Survey (DIS) in the UV.  {\em
  bottom:} Type 2 quasar SED \citep{Zak++03} at $z=1$ compared to the
  overlapping depth of the SERVS \citep{SERVS09},
  VISTA-VIDEO \citep{VISTA07}, and multi-epoch LSST surveys.}
\label{fig:agn:sed}
\end{center}
\end{figure}

The depth and sky coverage provided by LSST are essential for
characterizing and classifying optically faint AGN that are prominent
in other wavebands, but that cannot be studied with shallower optical
surveys such as the SDSS \citep{York00}. Any sky areas---whether by
design or by serendipity---in which past, present, or future deep
multiwavelength surveys overlap with LSST sky coverage, will be
promoted by LSST investigations to ``optical plus multiwavelength
Selected Areas."  \autoref{fig:agn:sed} demonstrates that AGN SEDs are
well probed across a broad range of wavelengths both in terms of depth
and areal coverage.  LSST AGN with multiwavelength data available will
have less selection bias than AGN selected by LSST optical colors
alone (\autoref{sec:agn:census}), allowing large samples to be
constructed that are representative of the overall AGN population.
Combining multi-wavelength data sets with the LSST optical catalogs will 
also reveal new views of the wide range of AGN phenomena.

For example, overlapping \mbox{X-ray} observations will be a valuable 
component of source-classification algorithms for LSST AGN; 
\mbox{X-ray-to-optical} flux ratios of AGN are roughly $\sim0.1-10$ 
\citep[e.g.,][]{Schmidt98,Barger03,Bauer++04}. The { ROSAT} All-Sky Survey 
\citep{Voges99} and the XMM-Newton Slew Survey \citep{Saxton08} will 
overlap the LSST survey region, giving at least shallow to moderate-depth 
\mbox{X-ray} coverage to nearly all LSST AGN. There are already $\sim 
10^2\,\rm deg^2$ of sky covered with {Chandra} to a depth sufficient to 
detect $10^2\,\rm AGN\,deg^{-2}$ \citep[e.g.,][]{Green09}, and of course 
this area will continue to expand. LSST imaging of the Chandra Deep 
Field South region in the ``deep drilling'' LSST mode 
(\autoref{sec:design:cadence}) will enable detailed studies of heavily 
obscured AGN. Although such deep Chandra data are concentrated in a 
few pencil-beam fields, they yield very impressive AGN surface densities 
of $\sim7000$ deg$^{-2}$, and the obscured sources comprise a significant 
fraction of the AGN population that are missed by shallower optical 
surveys \citep[e.g.,][]{Brandt05}. \mbox{X-ray} observations can also 
reveal important AGN characteristics that can be compared to 
optically-derived measures of spectral shape, luminosity, and temporal 
evolution.  For example, the X-ray SED slope (represented by the ratio 
between hard and soft X-ray flux) is an indicator of \mbox{X-ray} 
absorption, and can be used to classify Type~2 AGN 
\citep[e.g.,][]{Mainieri02}.

Heavily obscured LSST AGN may also be identified by combining LSST optical 
colors with sub-millimeter surveys \citep[e.g.,][]{Alexander05}, or mid-IR 
photometry from Spitzer \citep[e.g.,][]{Polletta06}.  There are of 
order $10^2\,\rm deg^2$ of deep mid-IR imaging data from surveys like 
SWIRE \citep{Lonsdale++03}; these surveys have AGN surface densities 
approaching $10^3\,\rm deg^{-2}$. Combining LSST data with these surveys 
and X-ray data may even be used to identify Compton thick AGNs, and 
mid-infrared photometry can also improve photometric redshift estimates 
over purely optical estimates. Cross-correlating mid--far-IR data (e.g., 
from Spitzer and Herschel) with LSST AGN will also improve our 
understanding of the starburst-AGN connection across cosmic time.

Radio survey data of LSST AGN will allow us to distinguish between 
radio-loud and radio-quiet AGN, test the dependence of radio power on 
luminosity and redshift, and probe unification models.  The combination of 
\mbox{X-ray}, radio, and LSST photometry may identify new blazars from 
their unusual location in \mbox{X-ray}-radio-optical 
multi-band diagrams \citep[e.g.,][]{Perlman01}. Additional 
gamma-ray information from the {Fermi} Gamma-ray Space Telescope will 
improve our understanding of how accretion processes accelerate immense 
jets of material to nearly the speed of light. Although only early
{Fermi} results are available at the time of writing, more than  
$\sim10^{3}$ gamma-ray blazars may be selected and monitored at high 
energies \citep[e.g.,][]{Abdo09}.
Moreover, LSST may contribute significantly to {Fermi} blazar 
identifications: for example, LSST may discover transient/variable optical 
objects coincident with radio sources and inside { Fermi} persistent 
gamma-ray error circles, or transients/variables may be caught flaring 
contemporaneously in both LSST and {Fermi}.  Blazars display dramatic 
SED changes, which are associated with the jet acceleration mechanism.  
LSST will provide optical light curve information on few day (or better) 
timescales for $10^{2-3}$ {Fermi} blazars (with $m>17$) in the LSST sky 
region; { Fermi}'s lifetime will plausibly suffice to provide 
extraordinary contemporaneous blazar gamma-ray lightcurves extending down 
to intra-day time resolution, for high-energy comparison to corresponding 
LSST optical lightcurves of the full ensemble.

Multiwavelength data for the LSST AGN census will produce the 
largest inventory of AGN SEDs over a very wide wavelength range, allowing 
better constraints on typical accretion and reprocessing mechanisms.
A combination of multiwavelength data from optical, radio, 
infrared, ultraviolet, and \mbox{X-ray} bands is also essential to avoid 
missing ``drop-outs'' from sources that are atypically faint in some 
wavebands, including such interesting classes as high-redshift AGN, 
obscured AGN \citep[e.g.,][]{Brandt05}, ``\mbox{X-ray} bright, 
optically-normal galaxies'' \citep[XBONGs; e.g.,][]{Comastri++02}, or 
intrinsically \mbox{X-ray} weak AGN \citep[e.g.,][]{Just07,Leighly07}. 
LSST will likely also discover interesting AGN that cannot be 
straightforwardly classified based on their multiwavelength properties, 
highlighting the most promising subset for follow-up. Such sources may 
include remarkable outliers, ``borderline'' sources in classification 
schemes, as well as interesting classes of AGN that are strongly 
distinguished by their unusual radio, infrared, optical-UV, and/or 
\mbox{X-ray} colors.

\subsection{Time-Dependent SEDs}
\label{sec_multiwav_time}

Augmenting LSST photometry with multiwavelength data will also enable 
unprecedented temporal investigations. For example, AGN that flare or 
exhibit other unusual temporal behavior in LSST will trigger alerts for 
multiwavelength follow-up in other relevant energy bands.  In principle, 
these alert criteria could be quite complex, identifying the onset of 
strong absorption or a ``state change'' in the variation properties 
(\autoref{sec:agn:var}). In such cases, it will be particularly 
interesting to obtain multiwavelength observations to determine how the 
inner AGN regions (e.g., the jet base or disk corona) are affected.

LSST will, over its lifetime, measure optical variation amplitudes and 
colors for AGN over a wide range of luminosities and redshifts. Hundreds 
of repeat LSST observations in each band will reveal the extent 
to which the scatter in measured SED shapes \citep[e.g.,][]{Steffen06, 
Gibson08a} can be attributed to emission or absorption variability over 
observed time scales of days to a few years.  Additionally, any AGN in the 
LSST sky coverage targeted contemporaneously (purposefully or 
serendipitously) in another energy band by future ground- or space-based 
observations can be matched to the most recent LSST optical photometry in 
order to derive statistical inferences about the shape of the AGN SED and 
its evolution on shorter time scales.

SED variation will be particularly interesting for strongly-absorbed AGN 
in order to constrain the size scales, evolution, lifetimes, and 
large-scale impact of absorbing outflows. As one example, Broad absorption 
line (BAL) outflows, and their AGN hosts, have been studied in the radio, 
infrared, optical-UV, and \mbox{X-rays} \citep[e.g.,][and references 
therein]{Gibson++09}.  Their SEDs reveal information about the structure 
and evolution of UV and \mbox{X-ray} absorbers in the central region of an 
AGN. LSST will monitor the light curves and colors of the $\sim$ 5000 BAL 
quasars identified in current catalogs \citep{Gibson++09}, and (at least) thousands 
more will be identified in the LSST fields by SDSS--III and future 
surveys. LSST monitoring will enable other observatories to trigger 
follow-up observations based on dramatic changes in the absorption of 
these sources, and will provide detailed light curves useful for studies 
of absorber photoionization.  Multi-wavelength follow-up observations will 
examine connections among the various structures that absorb radiation in 
the different wavebands.

In some cases, AGN have demonstrated coordinated variability across 
multiple wavebands that is presumably driven by physical relations among 
the structures responsible for emission in each waveband 
\citep[e.g.,][]{UttleyMcHardy04}.  Coordinated campaigns to monitor AGN in 
other wavebands could, in principle, generate multi-wavelength light 
curves for large numbers of AGN (or for interesting classes of AGN) 
sampled on rest-frame time scales of days or shorter. Because SED 
wavebands are associated with different physical processes, the 
correlations (including lead or lag times) between wavebands can reveal 
relationships among emitting structures such as the accretion disk and its 
corona.


\section{AGN Variability}
\label{sec:agn:var}

\noindent{\it 
W. N. Brandt,
Willem H. de Vries,
Paulina Lira,
Howard A. Smith}

%
%
%
%
%
%
%
%
%
%
%
%
%
%
%
%
%
%
%
%
%
%
%
%
%
%
%
%

One of the key characteristics of AGNs is that their emission is variable 
over time. In addition to aiding effective AGN selection (see \autoref{sec:agn:census}), 
this time dependence offers a probe of the physics associated with the
accretion process. While there is no model capable of explaining all
aspects of AGN variability in a compelling manner, 
accretion-disk instabilities, changes in accretion rate, the evolution 
of relativistic jets, and line-of-sight absorption changes have all been 
invoked to model the observed variability. 

The characteristics of AGN variability are frequently used to 
constrain the origin of AGN emission 
\citep[e.g.,][]{Kawaguchi98,Trevese01,Kelly09}. 
AGN variability is observed to depend upon
luminosity, wavelength, time scale, and the presence of strong radio 
jets. However, despite considerable efforts over last few decades, 
conflicting claims about correlations with physical properties
exist. This is at least in part due to the fact that many early 
studies included at most only \hbox{50--300} objects and had a limited 
number of observation epochs 
\citep[see][]{Giveon99,Helfand01}. 

Significant progress in the description of AGN variability has 
recently been made by employing SDSS data 
\citep[][]{devries03,devries05,DVB04,Ivezic04,
Wilhite05,Wilhite06,Sesar06}. 
\citet[][]{DVB04} used two-epoch photometry for 25,000 
spectroscopically confirmed quasars to constrain how 
quasar variability in the optical/UV regime depends upon rest-frame 
time scale (up to $\sim 2$ years in the observed frame), luminosity, rest wavelength, 
redshift, and other properties. They found that accretion-disk 
instabilities are the most likely mechanism causing the majority of 
observed variability. 
\citet[][]{devries05} and 
\citet[][]{Sesar06} 
utilized SDSS and Palomar 
Observatory Sky Survey (POSS) measurements for 40,000 quasars 
spectroscopically confirmed by SDSS, and constrained quasar continuum 
variability on time scales of \hbox{10--50}~yr in the observer's frame. 
In the context of a shot-noise light-curve model, 
\citet[][]{devries05} 
found evidence for multiple variability timescales in long-term 
variability measurements.
Using SDSS repeat spectroscopic observations obtained more than 50 
days apart for 315 quasars which showed significant variations, 
\citet[][]{Wilhite05}
demonstrated that the difference spectra are 
bluer than the ensemble quasar spectrum for rest-frame wavelengths 
shorter than 2500~\AA\ with very little emission-line variability. 
The difference spectra in the rest-frame wavelength range 
\hbox{1300--6000}~\AA\ could be fit by a standard thermal 
accretion-disk model with a variable accretion 
rate \citep[][]{Pereyra06}. 

However, the above efforts were limited in what they could study, given
that each object in their sample was observed only twice.  The LSST 
variability survey will be unrivaled in its combination of size
(millions of AGNs), number of observation 
epochs, range of timescales probed (rest-frame 
minutes-to-years), multi-color coverage, and photometric accuracy. 
Relations between AGN variability properties and luminosity, redshift, 
rest-frame wavelength, time scale, color, radio-jet emission, and other 
properties will be defined with overwhelming statistics over a 
wide range of parameter space. Degeneracies between the potential 
controlling parameters of variability will thereby be broken, enabling 
reliable determination of which parameters are truly fundamental. 
With appropriate spectroscopic follow-up, it will also be possible 
to relate AGN variability to emission-line and absorption-line 
properties, as well as physical parameters including black-hole mass 
and Eddington-normalized luminosity 
\citep[e.g.,][]{Oneill05}.  
Both the observed luminosity and spectral variability 
of the optical/UV AGN continuum will used to test accretion 
and jet models. 

The LSST AGN variability survey will also greatly improve our 
categorization of the range and kinds of AGN variability. Rare
but physically revealing events, for example, will be detected in 
sufficient numbers for useful modeling. These are expected to 
include transient optical/UV obscuration events due to gas and
dust moving temporarily into the line of sight 
\citep[e.g.,][]{Goodrich95,Lundgren07,Gibson08b}, 
strong intranight variability events 
\citep[e.g.,][]{Stalin05,Czerny08}, 
and perhaps quasi-periodic oscillations. Notable events discovered 
by LSST will trigger rapid follow-up with other facilities, and 
LSST photometry will automatically synergize with many AGN monitoring 
efforts (e.g., wide-field X-ray and gamma-ray monitors; 
reverberation-mapping projects). AGN lifetimes, or at least
the timescales over which they make accretion-state transitions, 
will also be constrained directly by looking for objects that
either rise or drop strongly in flux 
\citep[e.g.,][]{Martini03}.


\section{Transient Fueling Events: Temporary AGNs and Cataclysmic AGN Outbursts}

\label{sec:agn:transient}
\noindent{\it 
Aaron J. Barth, 
W. N. Brandt, 
Michael Eracleous,
Mark Lacy}  

%
%
%
%
%
%
%
%
%
%
%
%
%
%
%
%
%
%
%
%
%
%
%
%
%
%

Strong transient outbursts from galactic nuclei can occur when a star, 
planet, or gas cloud is tidally disrupted and partially accreted by a 
central SMBH. The tidal field of a SMBH is sufficient
to disrupt solar-type stars that approach within $\sim5M_7^{-2/3}$ Schwarzschild
radii, where $M_{SMBH} = M_7 \times 10^7 M_{\odot}$
\citep{1975Natur.254..295H}.
An optical flare lasting several months is expected when the star
disintegrates outside the event horizon, i.e., for $M_7 < 20$.  %
Transient variability  may also arise during the inspiral 
and merger phases of binary SMBHs. LSST will be a premier facility for 
discovering and monitoring such transient SMBH phenomena, enabling and 
aiding studies across the electromagnetic spectrum as well as
detections with gravitational waves. 

\subsection{Tidal Disruption Events by Supermassive Black Holes}

Models of tidal disruptions predict optical
emission from a hot optically thick accretion disk dominating the
continuum and enhanced by line emission from unbound ejecta
\citep{Roos92,1999ApJ...514..180U}.  The peak brightness can reach
$M_R = -14$ to $-19$ mag approaching that of a supernova.
The expected full sky rate of events down to a 24 mag threshold
(z $\sim$ 0.3) is $10^4 M_7^{3/2}$ yr$^{-1}$. Multi-epoch X-ray and UV
observations have discovered about eight  
candidates for tidal-disruption events in the form of large-amplitude 
nuclear outbursts \citep[e.g.,][]{Don++02,Kom++04,VEW04,Gez++06,Gez++08}. 
These 
events have large peak luminosities of 
\hbox{$\sim 10^{43}$--$10^{45}$~erg~s$^{-1}$}, optical-to-X-ray 
spectral properties broadly consistent with those expected from tidal 
disruptions, and decay timescales of months. The inferred event rate 
per galaxy is \hbox{$10^{-5}$--$10^{-4}$~yr$^{-1}$} 
\citep{Don++02,Gez++08,Luo++08}, roughly consistent 
with the predicted rate for stellar tidal disruptions \citep[e.g.,][]{W+M04}. 
These X-ray and UV outbursts are theoretically expected and in some cases
observed \citep{BPF95,Gru++95,Gez++08}
to induce accompanying optical nuclear variability that will be detectable 
by LSST. 

LSST will dramatically enlarge the sample of detected tidal-disruption 
events, thereby providing by far the best determination of their
rate.   \citet{Gez++08} and \citet{Gez++09} have used
the currently known UV/optical events to estimate rates, and predict
that LSST should detect at least 130 tidal disruptions per
year. With such a large sample, it will be possible to  
measure outburst rates as a function of redshift, host-galaxy type, 
and level of nuclear activity. This will allow assessment of the role 
that tidal disruptions play in setting the luminosity function of 
moderate-luminosity active galaxies 
\citep[\eg][]{MMH06}. 

An interesting subset of tidal-disruption events involves the
disruption of a white dwarf by a black hole of mass
$<10^5$~M$_{\odot}$ \citep[e.g.,][]{Rosswog++08,Ses++08}. Such events are
interesting for the following reasons. First, the debris released from
the disruption of a white dwarf is virtually devoid of hydrogen,
giving rise to a unique spectroscopic signature. Second,
since white dwarfs are tightly bound objects, their tidal disruption
radius is smaller than the Schwarzschild radius of a black hole for
black hole masses greater than $2\times 10^5$~M$_{\odot}$. In other
words, black holes more massive than this limit will swallow white
dwarfs whole without disrupting them. Third, unlike main sequence
stars, the strong tidal compression during the disruption of a white
dwarf triggers thermonuclear reactions which release more energy than
the gravitational binding energy of the white dwarf
\citep{Rosswog++08}. Thus, such an event could resemble a supernova,
albeit with a different light curve and a different spectral
evolution.  Fourth, the disruption of a white dwarf in an initially
bound orbit around a black hole is accompanied by a strong
gravitational wave signal, detectable by {LISA\/}, considerably
stronger than that of a main sequence star.

Detection of the prompt optical flash of such a white dwarf disruption 
event with the LSST would allow rapid follow-up spectroscopy to confirm 
the nature of the event through 
the composition of the debris and the shape of the light
curve. Such events are of particular interest because they can reveal
the presence of moderately massive black holes in the nuclei of
(presumably dwarf) galaxies. Black holes in this mass range are
``pristine'' examples of the seeds that grow to form the most massive
black holes known today \citep[see][and references therein]{Vol08}.
As such they provide useful constraints on models of hierarchical
galaxy assembly and growth of their central black holes.

The tidal disruption events that have been discovered to date were
mostly identified 
after they were largely over.  However, LSST data processing will provide near-instant
identification of transient events in general and new tidal
disruptions in particular (\autoref{sec:design:dm}), so that intensive optical spectroscopic and 
multiwavelength follow-up studies will be possible while the events are in 
their early stages.  Prompt and time-resolved optical spectroscopy, 
for example, will allow the gas motions from the tidally disrupted object to 
be traced and compared with computational simulations of such events
\citep[e.g.,][]{Bog++04}. Joint observations with LSST and X-ray 
missions such as the {Black Hole Finder Probe\/} \citep[e.g.,][]{Gri05}, 
{JANUS\/}, and {eROSITA\/} will allow the accreting gas to to be 
studied over the broadest possible range of temperatures and will also 
constrain nonthermal processes such as Compton upscattering and shocks. LSST 
identifications of tidal disruptions will also complement {LISA\/}
detections as these events are expected to create gravitational-wave 
outbursts \citep[e.g.,][]{Kob++04}. 

\subsection{Inspirals of Binary Supermassive Black Holes}

SMBH mergers are an expected component of models of galaxy evolution
and SMBH growth. The correlation of the masses of the central SMBHs in 
galaxies today and the velocity dispersions of their bulges suggests a
close link between the build-up of mass in galaxies and in their
central SMBHs, perhaps driven by mergers, as many models suggest
\citep[e.g.,][]{K+H00,diM++08}. 

Several dual SMBH systems have already been found in the form of quasar 
pairs, but most have relatively wide ($\sim 10$~kpc) separations
(\citealt{Hen++06}, \citealt{Comerford++08}).
At lower redshift, there are now several examples of dual AGN with
$\sim$~kpc separation in merging galaxies, the best-known case being
NGC~6240 \citep[e.g.,][]{Kom++03}. True binary systems, in which
the two SMBHs are tightly gravitationally bound to each other,
have proved more difficult to find, and the single nearby example is 
a binary with 7~pc separation discovered in the radio with VLBI
\citep{Rod++06}. Theory indicates that dynamical friction
will cause the SMBHs in galaxy merger events to sink to the 
bottom of the common potential well formed at the end of the
merger on a timescale of $\sim 10^7$~yr. There they form a SMBH
binary system with pc-scale separation, primarily by ejecting stars
from the core of the galaxy \citep[e.g.,][]{BBR80}.  
These binary systems may be, however, resistant to further decay
\citep{Yu09} until 
the separation reaches less than about $10^{-3}$~pc, when gravitational 
radiation becomes an effective mechanism for angular momentum loss 
(the ``inspiral'' phase).

The solution to the stalling of the binary separation at the parsec scale
probably lies in gas. In the most-likely case of an unequal mass merger, 
an accretion disk around the primary SMBH can exert a torque on the 
secondary component, reducing its angular momentum over a period of
$\sim 10^7$~yr \citep[e.g.,][]{A+N02}. Furthermore, in this 
scenario, a spike in the accretion rate will occur during the inspiral 
phase as gas trapped between the two SMBHs is accreted (over a period 
of $\sim 10^3$~yr). More detailed predictions of the accretion rate 
as a function of time during the binary phase were performed by 
\citet{Cua++09}. They argue that the accretion rate onto both 
SMBHs will vary on timescales corresponding to the binary
period. For example, a $\sim 0.01$~pc separation of two 
$\sim 3\times 10^6\,~M_{\odot}$ SMBHs leads to a variability period 
of $\sim 1$ month, well suited for detection within the enormous
sample of LSST AGN with high-quality photometric monitoring. 

Another prominent observational signature of sub-pc binaries can
come about from the interaction of one of the two black holes with the
accretion disk surrounding the other. Such an interaction (and the
resulting signal) is likely to be periodic, but with periods on the order of decades to centuries. Thus, we are likely to 
observe individual events and perceive them to be isolated
flares. Some initial theoretical work attempting to predict the
observational signature of such an interaction has been carried out by
\citet{Bog++08}. Candidates for such systems have also been
found. The best known example is OJ~287 where more than a dozen pairs
of outbursts have been observed with a recurrence time between pairs of
\hbox{10--12} years 
\citep[e.g.,][and references therein]{Val07,Val++08}. 
Less persuasive claims for recurring outbursts have 
also been made for 3C~390.3 and PKS~0735+178 \citep{Q+T04,Tao++08}.
The role of the LSST in identifying similar outbursts 
will be extremely important. After the initial identification, candidates
can be studied further with continued long-term photometry and
spectroscopy, in order to verify the nature of the system and derive 
its properties. 

\subsection{Mergers of Binary Supermassive Black Holes}

The proposed gravitational wave telescope {LISA\/} will have the
capability to detect gravitational waves from SMBH mergers out to
\hbox{$z\sim 10$} or higher. In favorable cases, {LISA\/} will be able
to localize a source to within a few arc-minutes to a few degrees on
the sky. Furthermore, the gravitational-wave signal from binary SMBH
coalescence serves as a ``standard siren'' that gives the luminosity
distance to the event (limited by uncertainties in gravitational
lensing along the line-of-sight), so {LISA\/} can provide a
three-dimensional localization for a detected event.  Determination of
the luminosity distance is possible because the shape of the
gravitational waveform (i.e., the variation of the frequency as a
function of time) depends on the {\it chirp} mass of the binary
(${\cal M} \equiv [(M_1 M_2)^3/(M_1+M_2)]^{1/5}$, where $M_1$ and
$M_2$ are the masses of the two components), while the amplitude of
the wave depends on the ratio of the chirp mass to the luminosity
distance \citep{Hughes09}. Therefore, fitting the waveform yields the
chirp mass, which can then be combined with the measured amplitude to
yield the luminosity distance. The uncertainty in the luminosity
distance is ultimately set by the signal-to-noise ratio of the
gravitational-wave amplitude \citep[see][]{Finn+Chern93}.
Identification of the electromagnetic counterparts to such events will
be of great importance, both for studying the physics of accretion
during SMBH mergers \citep[e.g.,][]{M+P05} and for measurement of the
redshift. The redshift can be combined with the luminosity distance
measured by {LISA\/} to provide new constraints on cosmological
parameters \citep[\eg][]{H+H05}.

The LSST data stream has the potential to be one of the most important
resources for identifying the electromagnetic counterparts to SMBH
mergers. During the final month before SMBH coalescence, there may be
a periodic signature in the accretion luminosity due to the binary
orbit, with a period of minutes to hours. The electromagnetic
afterglow following the coalescence may be primarily luminous in
X-rays \citep{M+P05}, but reprocessing or ionization of emission-line
gas could make the source detectable in the optical and near-infrared.
And once the coalescence takes place, LSST will be able to localize
the host object \citep{Koc++06}.  Indeed, the LISA error volume in
angle and distance may be
small enough to identify the object uniquely, given LSST's photometric
redshifts and AGN identification.  

\section{Gravitationally Lensed AGNs}
\label{sec:agn:lens}

\noindent{\it 
W. N. Brandt,
George Chartas}  

%
%
%
%
%
%
%
%
%
%
%
%
%
%
%
%
%
%
%
%
%
%
%
%
%
%

As discussed in the strong lens chapter (\autoref{sec:sl:yield}), we
estimate that in its single-visit images, LSST will discover $\sim
4000$ luminous AGN that are gravitationally lensed into multiple images 
(\autoref{sec:sl:agn}).
This more than ten-fold increase in the number of known 
gravitationally-lensed quasar systems, combined with the 
high cadence of observations of these 
systems will allow a variety of studies of these systems.  We discuss
the lensing-specific issues in \autoref{sec:sl:agn}, while here we
focus on what we can learn about the AGN themselves.  


\subsection{Microlensing as a Probe of AGN Emission Regions}

Resolving the emission regions of distant quasars is beyond the capabilities
of  present-day telescopes, and thus indirect methods have been applied to
explore these regions. Such methods include  reverberation mapping of the
broad line region \citep[e.g.,][]{Pet93,N+P97}, measurements
of occultations of the central X-ray source  by absorbing clouds
\citep{Ris++07}, and  microlensing of the continuum and emission-line regions
\citep[e.g.,][]{GKR88,GKS91,SEF92,G+G97,A+K99,M+Y99,Yon++99,Cha++02,Pop++03,
Bla++06,Koc++07,Poo++06,Poo++07,Jov++08,Mor++08}.

Since LSST will be monitoring the fluxes of $\simeq 4000$ gravitationally
lensed AGN, it is ideally suited to tracking microlensing events in these
systems. These events are  produced by the lensing effect of a star or group
of stars in the lensing galaxy. As the caustic network produced by the stars
traverses the AGN accretion disk and other emission sources, regions near the
caustics will be magnified. This causes uncorrelated variability in the
brightnesses of the images of a lensed quasar,  where the amplitude of the
variability is determined by the ratio of the emission regions to the
Einstein radius \citep[e.g.,][]{Lew++98,P+C05}. The
largest components, such as the radio and optical  emission-line regions,
should show little or no microlensing variability.   The thermal continuum
emission from the disk should show greater variability  at shorter
wavelengths, corresponding to smaller disk radii and higher  temperatures.
This wavelength dependent variability has been observed by  \citet{Ang++08}
and \citet{PMK08}, and LSST should enable  its study for
large numbers of gravitationally lensed AGN. 

The timescale of a microlensing event will depend in general on the
size of the source, the relative transverse velocity of the caustic
with respect to the source, and the angular diameter distances $D_{\rm
os}$ and $D_{\rm ol}$, where the subscripts o, s, and l refer to the
observer, source, and lens respectively.  The caustic crossing time
can be expressed as $t_{\rm cross} = (1+z_{\rm lens}) (R_{\rm
source}/v) (D_{\rm ol}/D_{\rm os})$, where $z_{\rm lens}$ is the
redshift of the lens, $R_{\rm source}$ is the size of the emitting
region, and $v$ is the relative transverse velocity of the caustic with
respect to the source.  Thus, for AGNs with redshifts in the range of
\hbox{1--4} we expect typical timescales for a caustic to cross the optical
emission region of the disk to be of the order of a few weeks. The
cadence of LSST is, therefore, well suited to map out microlensing
light curves of AGNs.


\subsection{LSST Microlensing Constraints on Accretion Disks}

The first step in large scale LSST microlensing studies will necessarily be
the identification of the lensed AGN. Good candidates for lensed AGN will
be identified using photometric redshift information for objects with
small angular separations. These candidates may then be confirmed either 
via follow-up spectroscopic observations or via LSST studies of intrinsic 
variability. In the latter case, one will be searching for similar light 
curves from the putative lensed AGN images that are temporally shifted due 
to the different light-travel times associated with each image. The detection 
in deep LSST images of a foreground galaxy or cluster that could act as the 
lens will also aid the identification process and allow lenses to be
distinguished from binary quasars. 

Once the light-travel time delay is determined via a cross-correlation
analysis from a given lens light curve, the data can be searched for
evidence of microlensing. 
The LSST cadence will 
be sufficient for many microlensing analyses. However, to obtain even
better temporal sampling (e.g., for rare, high-magnification events
that have relatively short duration), it will make sense to 
target identified microlensing events with additional telescopes. 
Ultraviolet and X-ray observations using facilities with 
sufficient angular resolution, such as Chandra in the X-ray 
band, will also be pursued as appropriate. 

The large number of lensed quasars from \hbox{$z\approx 1$--6}
will allow a search for evolution of AGN structure across this redshift 
range and a large range of luminosity and Eddington ratio. For
example, a change in the mode of accretion from the  
standard thin accretion-disk solution may be revealed by changes
in the scalings between wavelength, emission radius, and SMBH
mass.  
Microlensing analyses will help to determine whether the observed ``downsizing''
in the luminosity function (\autoref{sec:agn:LF}) is accompanied by
downsizing in accretion-disk size and SMBH mass.

\section{Public Involvement with Active Galaxies and Supermassive Black Holes}
\label{sec:agn:epo}

\noindent{\it 
W. N. Brandt,
Ohad Shemmer}  

%
%
%
%
%
%
%
%
%
%
%
%
%
%
%
%
%
%
%
%
%
%
%
%
%
%

Active galaxies and the supermassive black holes that power them
are of strong interest to the public. LSST will greatly advance  
understanding of both the demography and physics of active galaxies, 
and thus there are numerous approaches that can be used to involve 
non-astronomers in LSST active-galaxy discoveries. Effective themes 
for engaging the public include understanding the engines of the 
most powerful sources in the Universe, using active galaxies to 
trace large-scale structures, and finding the most distant cosmic 
objects.

Advanced high-school students, college students, and science teachers
(at the elementary school through high school levels) can learn about 
the methods by which LSST finds active galaxies by working with 
multi-color and multi-epoch LSST images. These students will 
categorize the various types of cosmic objects LSST detects using their color and variability properties, and then isolate 
the ranges of these properties corresponding to active galactic
nuclei (for example, strong blue emission and significant variability). 
This can lead to discussions of why active galaxies have the 
colors they do (i.e., accretion disks around supermassive black 
holes), the sizes and structures of their central engines, and 
extreme strong-gravity conditions. Special rare cases of color and 
variability behaviors will be used to explore remarkable objects. 
For example, the highest redshift quasars found by LSST that
probe the cosmic dark ages will appear in only the reddest 
filter. Similarly, remarkable variability can be used to identify 
transient fueling events of supermassive black holes by stellar 
tidal disruptions. 

We will include computer-based modules on the LSST World Wide
Web site that will illustrate basic LSST active galaxy concepts to 
the general public. These will include a tool that shows the 
connection between an active galaxy's spectrum and its multi-band 
LSST images (with connections to photometric redshifts for advanced 
learners) and interactive three-dimensional movies of the Universe 
as traced by the LSST AGN population. These modules will also
include elementary school activities such as building, with basic
materials (e.g., beads and string), a patch of the Universe based 
on LSST active-galaxy large-scale structure data. 

Dedicated amateur astronomers acting as ``Citizen Scientists,'' 
including faculty and students at small colleges and high schools, 
can play a valuable role by spectroscopically investigating remarkable 
bright LSST active galaxy phenomena. For example, amateur groups 
with access to telescopes of $32^{\prime\prime}$ or more using modern 
spectrographs can study the optical spectra of many $m_{\rm r}<19$ 
AGN that LSST will detect during its 10-year mission. Such studies 
may also include real-time spectroscopy of active-galaxy flares 
and spectroscopic variability monitoring aimed at revealing 
active-galaxy structure. This will allow the Citizen Scientists 
to complement the research of professional astronomers by making 
many instruments available at a particular moment of interest with 
the advantage of increased flexibility and shortening of 
observational response times.
Citizen Scientists will also be employed to classify the 
morphologies of nearby active galaxies via 
an ``Active Galaxy Zoo'' (see \autoref{sec:epo:Citizens} and 
\autoref{sec:galaxies:epo}).



\bibliographystyle{SciBook}
\bibliography{agn/agn}

\newcommand{\sne}{SNe\xspace}
\newcommand{\snia}{SN~Ia\xspace}
\newcommand{\sneia}{SNe~Ia\xspace}
\newcommand{\snii}{SN~II\xspace}
\newcommand{\sneii}{SNe~II\xspace}
\newcommand{\sniin}{SN~IIn\xspace}
\newcommand{\sneiin}{SNe~IIn\xspace}
\newcommand{\sniip}{SN~IIp\xspace}
\newcommand{\sneiip}{SNe~IIp\xspace}
\newcommand{\sniil}{SN~IIl\xspace}
\newcommand{\sneiil}{SNe~IIl\xspace}
\newcommand{\snibc}{SN~Ib/c\xspace}
\newcommand{\sneibc}{SNe~Ib/c\xspace}
\newcommand{\snib}{SN~Ib\xspace}
\newcommand{\sneib}{SNe~Ib\xspace}
\newcommand{\snic}{SN~Ic\xspace}
\newcommand{\sneic}{SNe~Ic\xspace}

\newcommand{\mlcs}{{\tt MLCS2k2}\xspace}
\newcommand{\SALTII}{{\tt SALT-II}\xspace}
\newcommand{\snana}{{\tt SNANA}\xspace}
\newcommand{\simlib}{{\tt SIMLIB}\xspace}
\newcommand{\Trest}{T_{\rm rest}\xspace}

\newcommand{\effrat}{{\cal R}_{\rm eff}}

\newcommand{\NFILTSYM}{N_{\rm filt}}
\newcommand{\SNRMAXSYM}{{\rm S/N}_{\rm max}}

\newcommand{\lamobs}{\bar{\lambda}_{\rm obs}}

\newcommand{\CCDEFF}{0.95\xspace}  
\newcommand{\OVPEFF}{0.81\xspace}  

\newcommand{\NFLDDEEP}{7\xspace}  
\newcommand{\OMEGADEEPVAL}{0.01945\xspace}         
\newcommand{\OMEGADEEPSYM}{\Omega_{\rm DEEP}} 

\newcommand{\NFLDMAIN}{3173\xspace}  
\newcommand{\OMEGAMAINVAL}{7.139\xspace}         
\newcommand{\OMEGAMAINSYM}{\Omega_{\rm MAIN}\xspace} 

\newcommand{\NFLDCUTSMAIN}{2686\xspace}  

\newcommand{\OPSIMVERSION}{OPSIM1.29\xspace}
\newcommand{\ZPTSYM}{{\rm ZPT}\xspace}

\newcommand{\wwwSNANA}{\url{http://www.sdss.org/supernova/SNANA.html}}
\newcommand{\wwwOPSIM}{\url{http://opsimcvs.tuc.noao.edu/index.html}}

\chapter[Supernovae]{Supernovae}
\label{chp:sne}


{\it W. Michael Wood-Vasey, David Arnett, S. J. Asztalos, Stephen Bailey,
Joseph P. Bernstein, Rahul Biswas, David Cinabro, Kem H. Cook, Jeff Cooke, Willem H. de Vries, Benjamin Dilday, 
Brian D. Fields, Josh Frieman, Peter Garnavich, Mario Hamuy, Saurabh W. Jha, Richard
Kessler, Stephen Kuhlman, Amy Lien, Sergei Nikolaev, Masamune Oguri,
Scot S. Olivier, Philip A. Pinto, Jeonghee Rho, Evan Scannapieco,
Benjamin D. Wandelt, Yun Wang,
Patrick Young, Hu Zhan }

\section{Introduction}
\label{sec:sn:intro}
\noindent{\it Josh Frieman, W. Michael Wood-Vasey} 


In 1998, measurements of the Hubble diagram of Type Ia supernovae (SNe Ia) provided the 
first direct evidence for cosmic acceleration \citep{riess98,
  perlmutter99}. 
This discovery rested on observations of several tens of supernovae at low and 
high redshift. In the intervening decade, several dedicated \snia surveys have 
together measured light curves for over a thousand \sneia, confirming and 
sharpening the evidence for accelerated expansion. 

Despite these advances (or perhaps because of them), a number of concerns have 
arisen about the robustness of current \snia cosmology constraints. The \snia Hubble 
diagram is constructed from combining low- and high-redshift \snia samples that 
have been observed with a variety of telescopes, instruments, and photometric 
passbands. Photometric offsets between these samples are degenerate with 
changes in cosmological parameters. In addition, the low-redshift \snia measurements 
that are used both to anchor the Hubble diagram and to train \snia distance estimators 
were themselves compiled from combinations of several surveys using different 
telescopes and selection criteria. When these effects are combined with uncertainties in intrinsic 
\snia color variations and in the effects of dust extinction, the result is that the current 
constraints are largely dominated by systematic as opposed to statistical errors. 

Roughly $10^3$ supernovae have been discovered in the history of
astronomy.  In comparison, LSST will discover over ten million supernovae during its ten-year survey, spanning a 
very broad range in redshift and with precise, uniform photometric calibration.
This will enable a dramatic step forward in supernova studies, using the power of 
unprecedented large statistics to control systematic errors and thereby lead to major advances in 
the precision of supernova cosmology. 
This overwhelming compendium of stellar explosions will also allow for novel
techniques and insights to be brought to bear in the study of 
large-scale structure, the explosion physics of supernovae of all types, and star formation and
evolution. 
The LSST sample will include hundreds of thousands of well-measured Type Ia
supernovae (\sneia), replicating the current generation of \snia cosmology
experiments several hundred times over in different directions
and regions across the sky, providing a stringent test of homogeneity and isotropy.
Subsamples as a function of redshift, galaxy type, environment, and supernova
properties will allow for detailed investigations of supernova evolution and of 
the relationship between supernovae and the processing of baryons in galaxies,
one of the keys to understanding galaxy formation.  Such large samples
of supernovae, supplemented by follow-up observations, will reveal
details of supernova explosions both from the large statistics of typical 
supernovae and the extra leverage and perspective of the outliers of the supernova population.  The
skewness of the brightness distribution of \sneia as a function of redshift will
encode the lensing structure of the massive systems traversed by the \snia 
light on the way to us on Earth.  There will even be enough \sneia to construct
a \snia-only baryon acoustic oscillation measurement that will provide
independent checks on the more precise galaxy-based method as well as allowing
for a \snia-only constraint on \Om.  This will allow \sneia alone to constrain $\sigma_8$ and \Om, as well as 
the properties of dark energy over the past 10 billion years of cosmic
history. Finally, millions of supernovae will allow for population
investigations using supernovae that were once reserved for large
galaxy surveys \citep{wood-vasey09}. 

Supernovae are dynamic events that occur on time scales of hours to months, but they allow us to probe billions of years into the past.  From studying nearby progenitors of these violent deaths of stars, to studying early massive stellar explosions and connecting with the star formation in between, to probing properties of host galaxies, clusters and determining the arrangement of the galaxies themselves, supernovae probe the scales of the Universe from an AU to a Hubble radius.

The standard LSST cadence of revisits every several days carried out
over years, in addition to specialized ``deep-drilling'' fields to
monitor for variations on the time scale of hours
(\autoref{sec:design:cadence}), offers the perfect laboratory to study
supernovae.  In this chapter we consider the science possible with two
complementary cadences for observations with the LSST, one based on
the standard cadence of 3--4 days (the ``main'' survey in what
follows) and another survey over a smaller area of sky using a more
rapid cadence, going substantially deeper in a single epoch (the
``deep'' survey).  As described in \autoref{sec:design:cadence}, the
detailed plans for the deep-drilling fields are still under
discussion, but they have two potential benefits: allowing us to get
improved light curve coverage for supernovae at intermediate redshifts
($z \sim 0.5$), and pushing photometry deep enough to allow \sneia to $z
\sim 1$ to be observed.  With this in mind, the ideal cadence would be
repeat observations of a single field on a given night, totalling
10-20 minutes in any given band.  We would return to the field each
night with a different filter, thus cycling through the filters every
five or six nights.  Because the \sneia discovered in this mode will be in
a small number of such deep fields, it is plausible to imagine
carrying out a follow-up survey of their host galaxies
with a wide-field, multi-object spectrograph to obtain spectroscopic
redshifts.  


The outline of this chapter is as follows.  \autoref{sec:sn:snia_rate_opsim} describes detailed simulations of \snia light
curves for both the main and deep LSST fields and the resulting expected
numbers of measured SNe Ia as a function of redshift for different selection
criteria, while \autoref{sec:sn:cc_contamination} quantifies the
contamination of the \snia sample by core-collapse supernovae.  
In \autoref{sec:sn:photoz} we discuss photometric redshift estimation with \snia 
light curves, the expected precision of such redshift estimates, and their
suitability for use in cosmological studies. 
\autoref{sec:sn:darkenergy} describes the constraints on dark energy that will
result from such a large \snia sample. 
The large sky coverage of the LSST \snia sample will enable a novel probe of
large-scale homogeneity and isotropy, as described in \autoref{sec:sn:isotropy}. 
\autoref{sec:sn:evolution}, presents the issue of \snia evolution, a
potential systematic for \snia cosmology studies. 
In \autoref{sec:sn:snia_rates} we discuss \snia rate models and the use of LSST to probe
them. 
\autoref{sec:sn:bao} describes how \snia measurements can be used to measure the
baryon acoustic oscillation feature in a manner complementary to that using the
galaxy distribution. 
The effects of weak lensing on the distribution of \snia brightness and its
possible use as a cosmological probe are described in
\autoref{sec:sn:wl}.  We then turn to core-collapse SN, discussing
their rates in 
\autoref{sec:sn:sneii} and their use for distance
measurements and cosmology in \autoref{sec:sn:sneii}. 
\autoref{sec:sn:light_echoes} discusses the prospects for measuring SN light
echoes with LSST.
Pair-production SNe, the hypothetical endpoints of the evolution of
supermassive stars, are the subject of \autoref{sec:sn:ppsne}. 
Finally, we conclude with a discussion of the opportunities for education and
outreach with LSST supernovae in \autoref{sec:sn:epo}.

\section{Simulations of \snia Light Curves and Event Rates}
\label{sec:sn:snia_rate_opsim}
\noindent{\it Stephen Bailey, Joseph P. Bernstein, David Cinabro, Richard Kessler, Steve Kuhlman} 

We begin with estimates for the anticipated 
\snia sample that LSST will observe. We put emphasis on those objects
with good enough photometry that detailed light curves can be fit to
them, as these are the objects that can be used in cosmological
studies as described below.  

\sneia are simulated assuming a volumetric rate of
\begin{equation}
    r_V(z) = 2.6 \times 10^{-5}(1+z)^{1.5}\ 
    {\rm SNe~}h_{70}^3~{\rm Mpc}^{-3}~{\rm yr}^{-1}
\end{equation}
based on the rate analysis in \citet{Dilday_08a}.  See
\autoref{sec:sn:snia_rates} for how LSST can improve the measurement of
\sneia rates.  At redshifts beyond  
about $z\simeq 1$, the above rate is likely an overestimate since it  
does not account for a correlated decrease with star formation rates  
at higher redshifts.  We simulated one year of LSST data; thus the
total numbers of SNe we present below should be multiplied by ten for
the full survey.  

For each observation the simulation determines the (source) flux  
and sky noise in photoelectrons measured by the telescope CCDs, and  
then estimates the total noise, assuming PSF fitting, that would be  
determined from an image subtraction algorithm that subtracts the host 
galaxy and sky background.  For this initial study we have ignored  
additional sky noise from the deep co-added template used for the  
image subtraction, as well as noise from the host galaxy; we expect  
these effects to be small.

We use the outputs of the Operations Simulator
(\autoref{sec:design:opsim}) to generate \snia light curves with a
realistic cadence, including the effects of weather.  The light curves were made by \snana
\citep{kessler09a}, a publicly available\footnote{\wwwSNANA} simulation and light curve fitter.  We carry out
simulations separately for the universal cadence, and for seven deep
drilling fields, which are visited in each filter on a five-day
cadence (see the discussion in \autoref{sec:design:cadence}).  


We explore the contamination of the \snia sample by other
types of \sne in \autoref{sec:sn:cc_contamination}.

  \subsection{Light Curve Selection Criteria}
  \label{sec:sn:cuts}

A \snia light curve depends on a number of parameters: a stretch
(shape) parameter, redshift, extinction, intrinsic color, and so on.
Fitting for all these parameters requires high-quality photometric
data.  Here, we apply a series of cuts on the data to identify those supernovae
with good enough repeat photometry of high signal-to-noise ratio (S/N)
in multiple bands to allow a high-quality light curve to be fit to it.
Defining $\Trest$ as the rest-frame epoch where
$\Trest=0$ at peak brightness, the selection requirements are
\begin{itemize}
 \item  at least one epoch with $\Trest < -5$ days (no S/N cut);
 \item  at least one epoch with $\Trest > +30$ days (no S/N cut);
 \item  at least seven different nights with one or more observations,
        $-20 < \Trest < +60$ (no S/N cut);
 \item  largest ``near-peak'' gap in the coverage (no S/N cut) is $<15$ rest-frame days, 
        where ``near-peak'' means that the gap overlaps the 
        $-5$ to $+30$ day region;
 \item  at least $\NFILTSYM$ passbands that have a measurement 
        with $\SNRMAXSYM>\{10,15,20\}$ (we explore variations in $\SNRMAXSYM$ below).
 \item  All observations used in the cuts above satisfy 
        $3000 < \lamobs/(1+z) < 9000$~\AA,
        where $\lamobs$ is the mean wavelength
        of the observer-frame filter.
\end{itemize}

We applied no S/N requirement on the light curve sampling
requirements; the S/N requirement is made only on the highest S/N value in a 
given number of passbands. 
The large temporal coverage ($-5$ to $+30$ days) is motivated
by the necessity to reject non-Ia SNe based solely on
photometric identification (see \autoref{sec:sn:cc_contamination}).  


  \subsection{Rate Results}
  \label{sec:sn:rateresults}

\begin{figure}
\begin{center}
\includegraphics[width=0.9\linewidth]{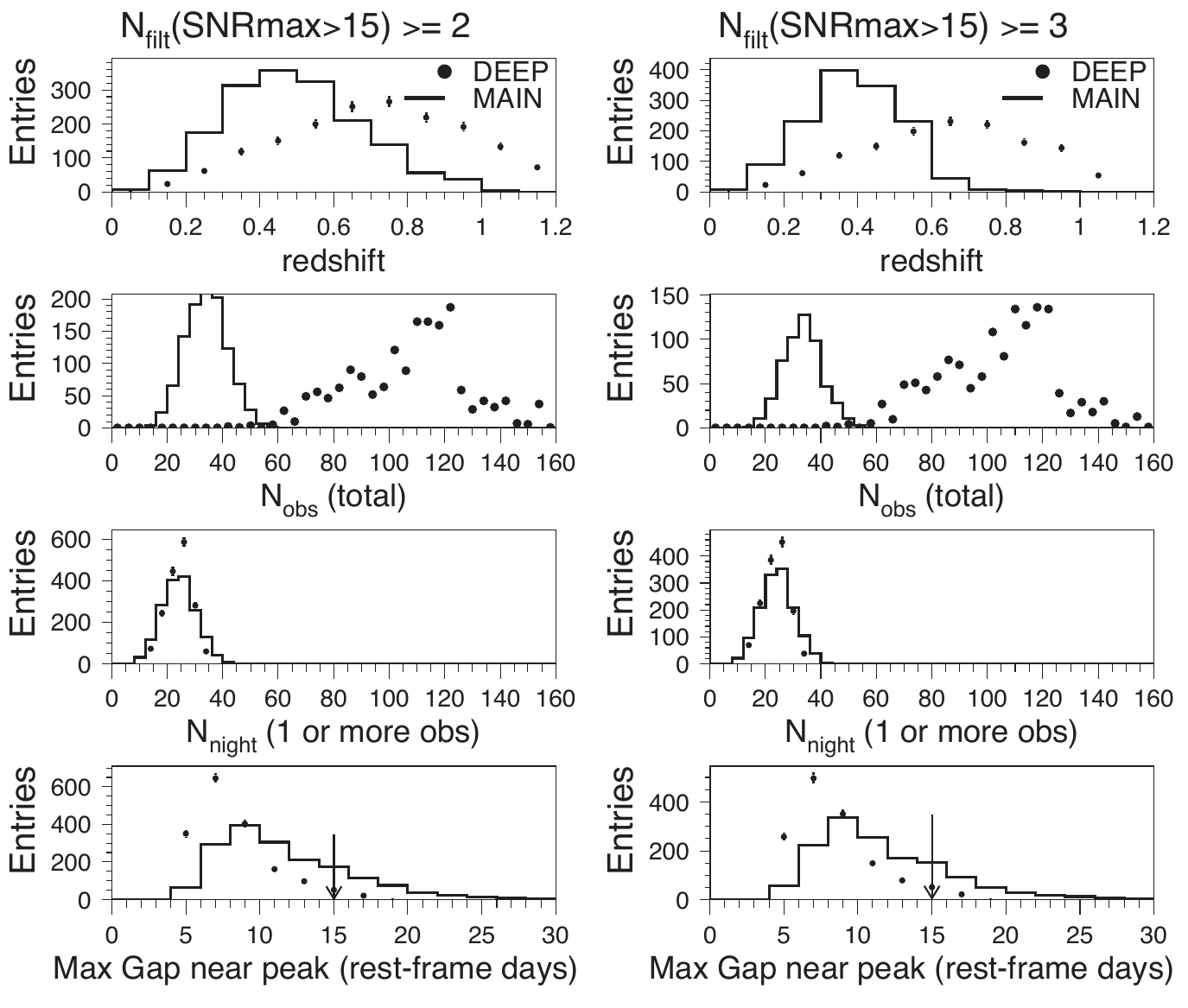}
  \caption[\snia distributions for the main and deep drilling surveys]{\small
        Distributions of \snia lightcurve characteristics for our
	Monte Carlo simulations of one year of the deep and main surveys,
        where the main field histograms have been scaled to have
        the same statistics as for the deep fields.  The quantities
	shown are the redshift distribution of \snia (top), the total
	number of obervations of each supernova (summed over filters,
	second row), the total number of nights each object has been
	observed (third row; thus the ratio of $N_{obs}$ to
	$N_{night}$ is a measure of the number of observations of the
	supernova per night), and the largest gap in days of the
	coverage of the supernova.  
        The S/N-related cut (see text) is indicated at the top of each column.
        All selection requirements have been applied, 
        except for the cut on the maximum gap in the light curve in the bottom plot.
        The vertical arrow at 15~days shows this nominal cut.
      }
  \label{fig:sn:sniarate}
\end{center}
\end{figure}

\autoref{fig:sn:sniarate} compares distributions of various
observables obtained
from the deep drilling and main fields.  As expected, the deep survey
has several times more total observations per light curve than the
main survey, and the deep fields probe higher redshifts. 
However, the number of nights with an observation (third panel down) 
is comparable for the deep and main surveys.  
The distribution of the largest time with no photometric data points
is slightly narrower for the deep survey.

\begin{figure}[ht]
\begin{center}
\includegraphics[width=0.6\linewidth]{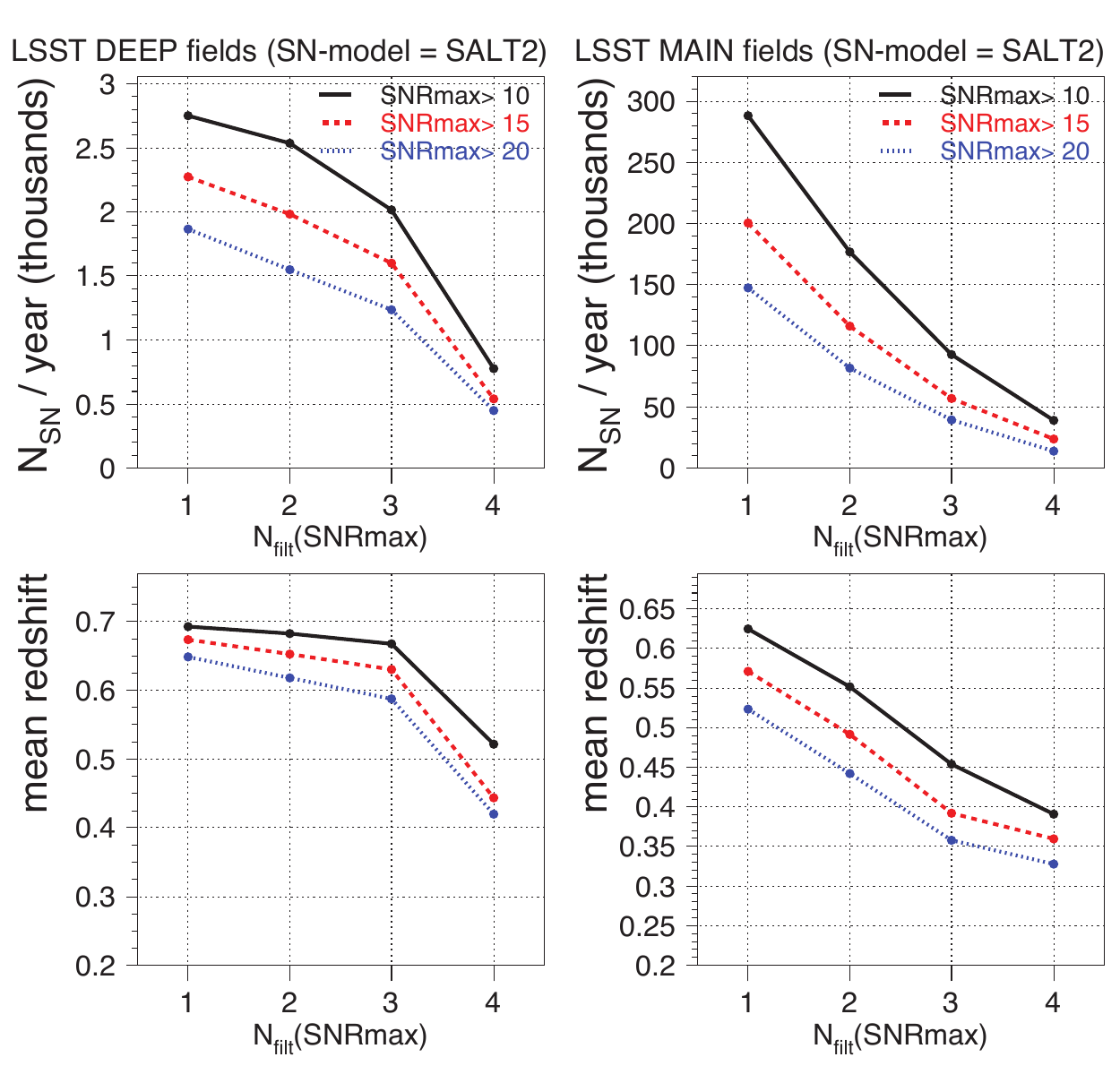}
  \caption[Number of \sneia/yr with good light curves]{\small
        The upper panels show the number of \sneia detected in a
	single year in the deep drilling fields (left) and universal
	cadence fields (right) which have good
	enough photometry to allow fitting of a high-quality light
	curve, using the criteria outlined in the text, for various
	values of $\SNRMAXSYM$.  The more filters in which
	high-quality photometry is available, the better the resulting
	constraint on supernova parameters; the number of filters is
	shown along the x-axis.  The lower panel shows the mean
	redshift of the resulting \sneia samples.  These simulations
	use the \SALTII model \citep{guy07} to generate light curves;
	we find similar results using the MLCS method \citep{jha07}.
}
\label{fig:sn:rate_salt2}
\end{center}
\end{figure}

A total of $10^4$ and $1.4\times 10^5$
SNe~Ia were generated for the deep and main surveys, respectively in a
single year. 
Each deep field is sampled more than 1,000 times,
and each main survey field is sampled about 40 times in that year. 
The number of \sneia per season after selection requirements 
is shown in \autoref{fig:sn:rate_salt2}. 
The SN sample size is plotted as a function of
$\NFILTSYM$ for $\SNRMAXSYM$ values of 10, 15, and 20.
Even with the strictest requirements of four filters with
$\SNRMAXSYM>20$, we recover 500 \sneia per season in the deep fields and
roughly 10,000 in the main survey.  

  \subsection{ Visual Examination of SN Light Curves}
  \label{sec:sn:lc}

\begin{figure}
\begin{center}
\includegraphics[width=0.6\linewidth]{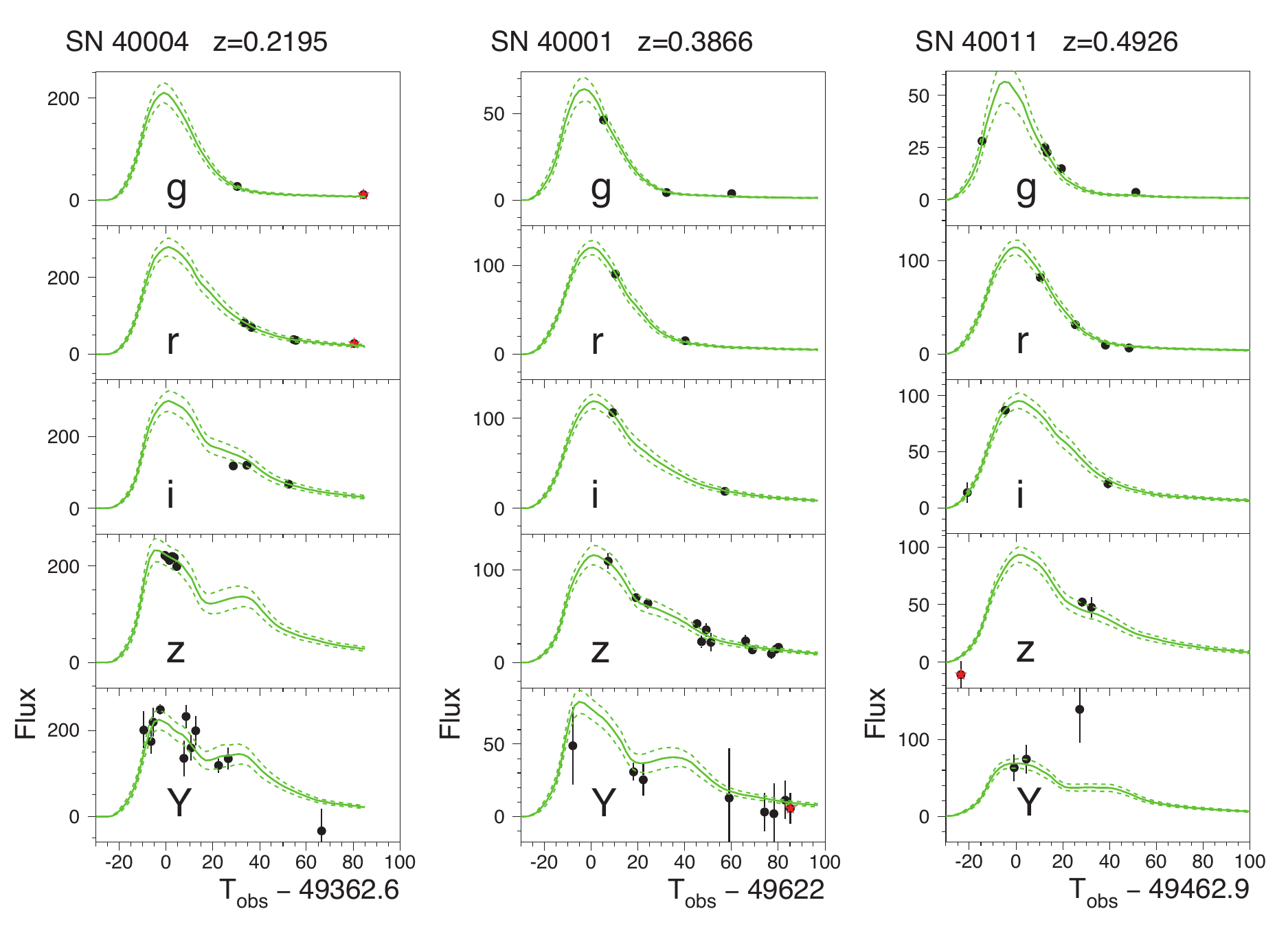}
\caption[Main field \snana Light Curves with $\NFILTSYM(\SNRMAXSYM>15) \ge 3$]
{\small
Simulated \sneia light curves (dots),
along with the best fit model (green curve),
for three representative \sneia from the main field that satisfy the 
selection requirements (\autoref{sec:sn:cuts}) with
three or more passbands having a measurement with S/N$>15$.
The redshift is shown above each panel.
The dashed green curve represents the model error. 
The red stars are measurements excluded from the fit because
$\Trest < -15$ days or $\Trest > +60$ days.
}
\label{fig:sn:main_snana_sneia}
\end{center}
\end{figure}

\begin{figure}
\begin{center}
\includegraphics[width=0.6\linewidth]{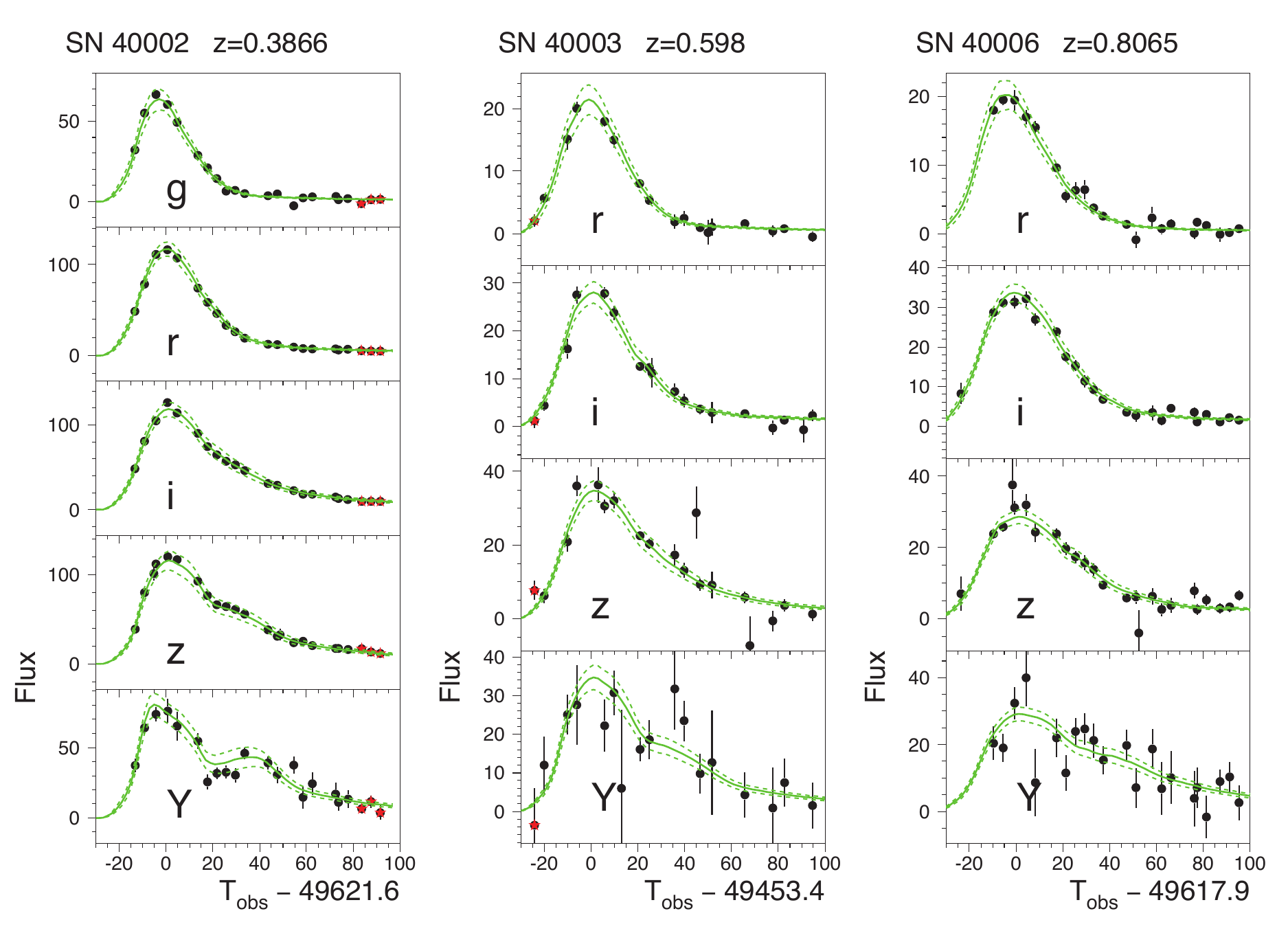} 
\caption[Deep field \snana Light Curves with $\NFILTSYM(\SNRMAXSYM>15) \ge 3$]
{\small
Same as for \autoref{fig:sn:main_snana_sneia}, but now for the
deep drilling survey.  Note how much better sampled the light curves are than for the main survey fields.
}
\label{fig:sn:deep_snana_sneia}
\end{center}
\end{figure}

Not surprisingly, the main survey light curves are considerably
sparser than for the deep survey, as shown in simulated data for the
two in \autoref{fig:sn:main_snana_sneia} and
\autoref{fig:sn:deep_snana_sneia}.    All light curves satisfy the
requirement $\NFILTSYM(\SNRMAXSYM>15) \ge 3$ so that we have
at least two well-measured colors.


For the deep fields, the excellent sampling in all passbands
results in measured colors that do not require any interpolation
between data points.  
In contrast, the main fields typically have poor sampling
in any one passband, even though the combined sampling
passes the selection requirements.
SN~40001 (\autoref{fig:sn:main_snana_sneia}, center panel), for example,
has just one observation before the peak in the $y$-band, and hence
no pre-max color measurements. 
SN~40004  (left panel) has a decent $y-z$ color measured
near peak, but the $g,r,i$ measurements are so far past
peak that there is essentially no second color measurement
near peak. 
Although one can introduce more ad-hoc selection requirements
to ensure visually better sampling in multiple passbands,
it would be better to define cuts based on how the sampling 
quality is related to the precision of the cosmological parameters.

Future work should include running light curve fits on large simulated
SN samples to extract both a distance modulus and redshift (i.e., a
photometric redshift fit; \autoref{sec:sn:photoz}), and then determining
the cosmological constraints, 
biases, and contamination.  These results can then be used to optimize the light curve sampling
requirements per passband.  

\section{Simulations of Core-Collapse Supernova Light Curves and Event Rates}
\label{sec:sn:cc_contamination}
\noindent{\it Joseph P. Bernstein, David Cinabro, Richard Kessler, Stephen Kuhlman} 


Because the vast majority of \sne that LSST will observe have no spectroscopic
follow-up, the scientific return from LSST \sne will be strongly
dependent on the ability to use photometric typing to classify and determine redshifts
for these events. 
In this section we consider the contamination rate of the
cosmological sample of \sneia by core-collapse SNe.  The simulations
described here assume that we have spectroscopic redshifts for either
the \sne or, more likely, the SN host galaxy.  This is certainly not going to be the case for LSST, and
these investigations need to be repeated in the context of photometric
redshifts.  

Light curves of core-collapse SNe (henceforth ``SNcc''; i.e., \sneibc
and \sneii) are less standard and less comprehensively studied
than are the detailed models of normal \sneia we used above.  
Therefore, we take a template approach 
to modeling SNcc. We utilized the following spectral templates constructed 
by Peter Nugent\footnote{\url{http://supernova.lbl.gov/~nugent/nugent_templates.html}}:
\begin{itemize}
\item \sneibc based on SN1999ex, which lies in the middle of the range
  defined by the three SNe discussed in \citet{ham02};
\item \sneiil from \citet{gil99}; 
\item \sneiin based on SN1999el as discussed in \citet{dic02} (note
  however that this supernova is almost 2 magnitudes 
dimmer than a typical \sniin; we have corrected this back to
``normal'' luminosity, which we correct for),
\item composite \sneiip based on \citet{bar04},
\end{itemize} 
As discussed in \citet{nug02}, one should use caution when applying
the above templates to for example, making K-corrections for determination
of rates or cosmology.  

These templates do not include intrinsic magnitude or color fluctuations. We added 
intrinsic magnitude fluctuations, coherent in all passbands, based on \citet{ric02}.  

In order to simulate SNcc with \snana, one must define the input supernova rate. \citet{Dilday_08a} 
found the \snia rate from SDSS to be of the form $\alpha(1+z)^{\beta}$ 
with $\alpha_{Ia} = 2.6\times10^{-5}\,\rm SNe\,Mpc^{-3}\,yr^{-1}$ and $\beta_{Ia} = 1.5$. For SNcc, we 
take $\beta_{cc} = 3.6$ to match the observed star formation rate. Various studies, the most
recent being the SuperNova Legacy Survey (SNLS) \citep{baz09}, have shown this assumption to be valid, albeit with low 
statistics and limited redshift range.  We normalize the SNcc rate
using the observed ratio of cc/Ia from the SNLS survey of 4.5 at $z <
0.4$, giving $\alpha_{cc} = 6.8\times10^{-5}\,\rm
SNe\, Mpc^{-3}\,yr^{-1}$.  Further discussion may be found in
\autoref{sec:sn:cc_rate}. The relative numbers of different types of
core-collapse supernovae are poorly known.  Our guesses, which we used
in the simulations, are shown in \autoref{tab:sn:ccfrac}; they are
based on \citet{man05} and \citet{cap99} for the Ib/c's, \citet{cap99}
for \sneiin, and private communications from Peter Nugent for
\sneiil.  

\begin{figure}
\begin{center}
\includegraphics[width=0.6\linewidth]{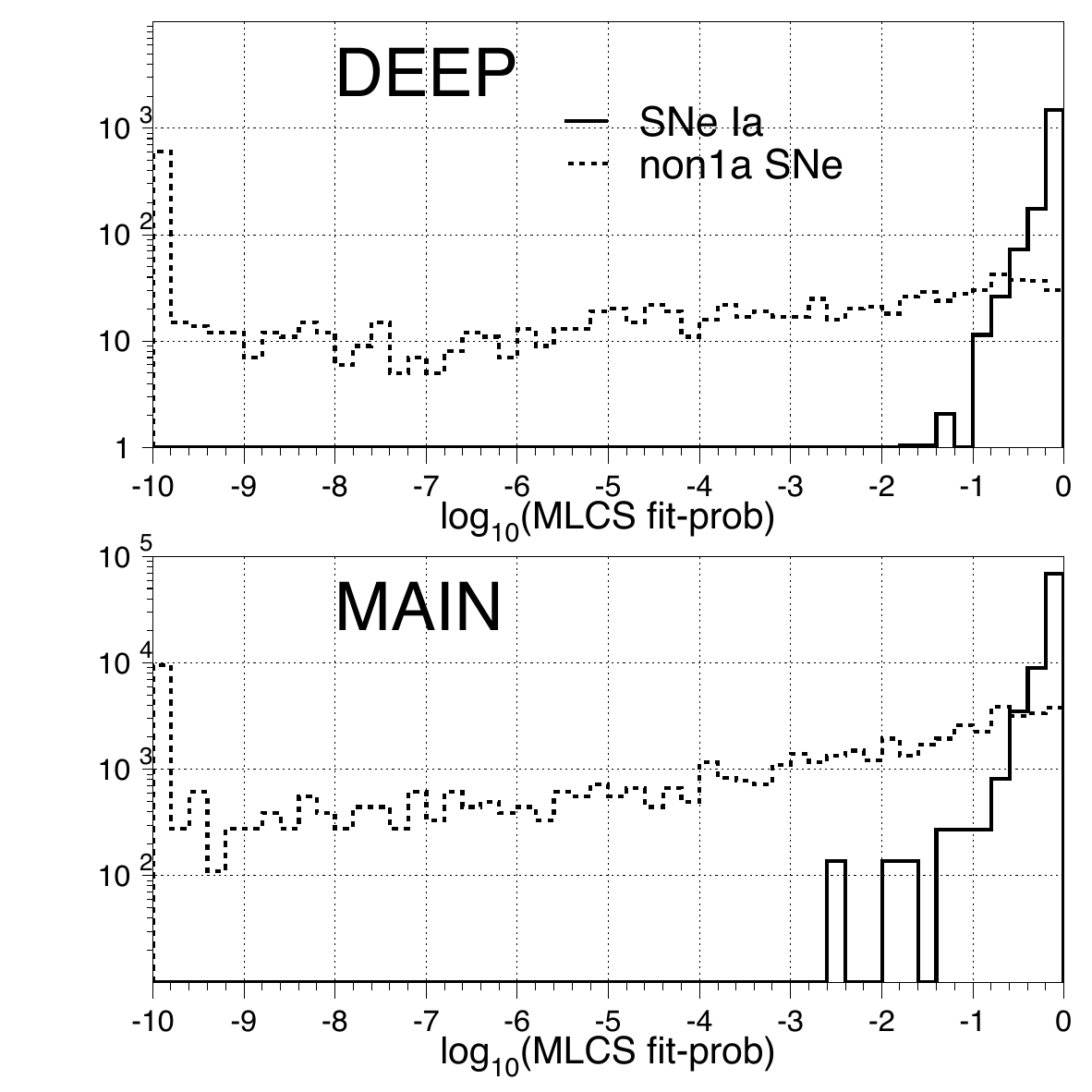}
\caption[Distinguishing \sneia and core-collapse \sne from light curve data alone.]{\small
Demonstrating our ability to distinguish \sneia and
  core-collapse supernovae from light curve data alone.  We fit each
  simulated \snia and SNcc light curve with an \snia
  model, and from the $\chi^2$ of the fit, tabulated $P_{\chi^2}$, the
  probability of getting a value of $\chi^2$ larger than that value.
  Plotted is the distribution of values (note the logarithmic y-axis)
  for those objects that are true core-collapse SN (dashed line) and
  true \snia (solid line), for simulations of the deep drilling fields
  and the main survey fields.  Cutting on high $P_{\chi^2}$ gives a
  clean \snia sample with little contamination.}
\label{fig:sn:fitprob}
\end{center}
\end{figure}


\begin{table}
\centering%
\begin{tabular}
[c]{c|c}\hline
SN Type & Relative Fraction of Core-Collapse SNe\\
\hline
IIP & 0.70 \\
Ib/c & 0.15 \\
IIL & 0.10 \\
IIn & 0.05 \\
\hline
\end{tabular}
\caption[Assumed Relative Core-Collapse SN Fraction]{Assumed
  Distribution of Different Types of Core-Collapse Supernovae}
\label{tab:sn:ccfrac}
\end{table}

\begin{figure}
\begin{center}
\includegraphics[width=3in]{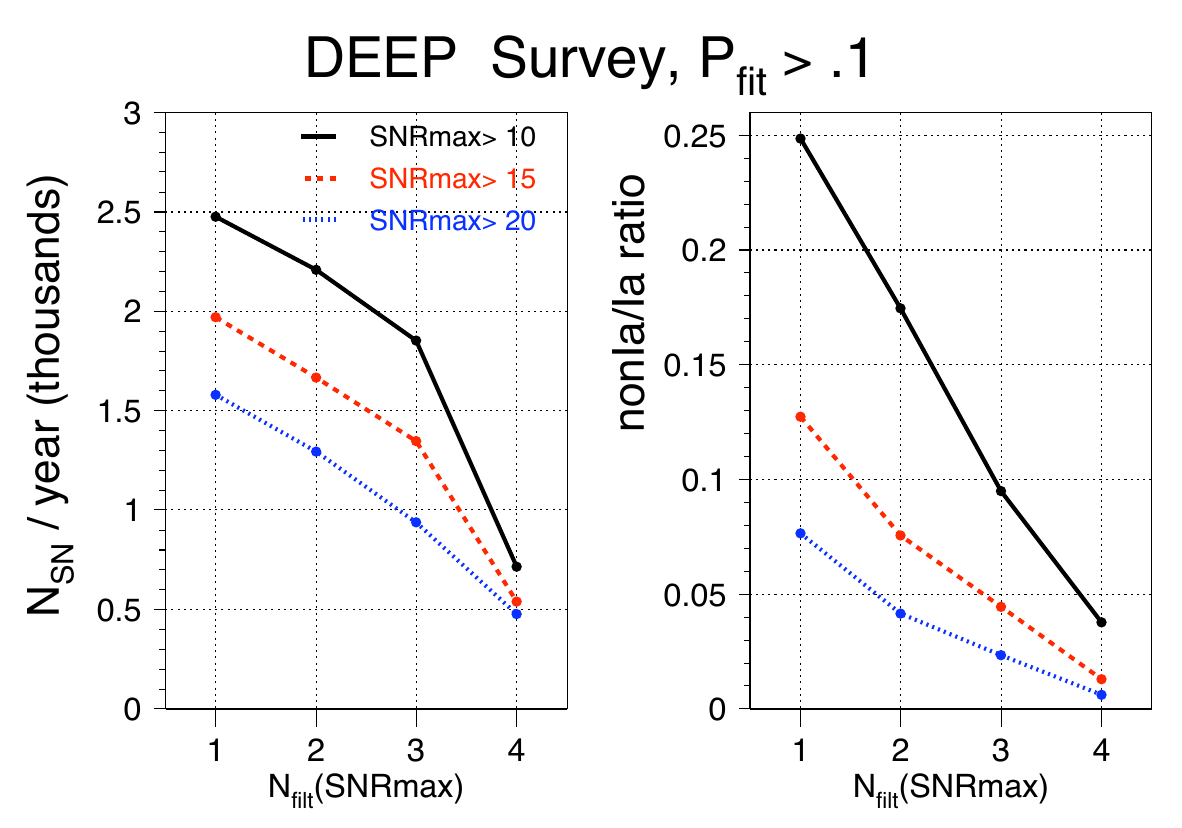}
\includegraphics[width=3in]{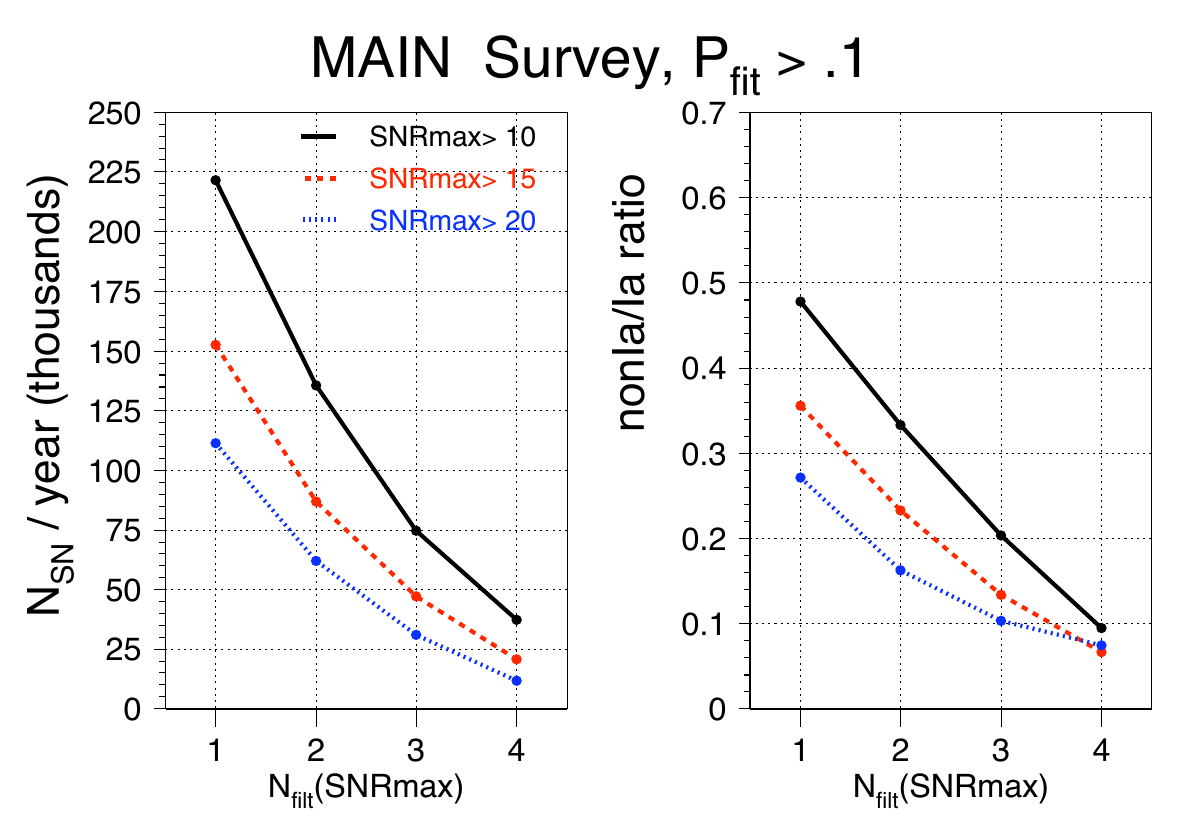}
\caption[Rates of \sneia and non-\sneia as a function of number of filters and $P_{\chi^2}>0.1$.]{\textit{Left}: Rates of Type Ia and Non-Ia supernovae are shown as a function of the number of 
filters passing S/N cuts plus $P_{\chi^2}$ $>$0.1 for the deep
sample. \textit{Right}: Same but for the main
 sample. 
Note that both samples use an assumed host galaxy spectroscopic redshift.}
\label{fig:sn:cc_rate_1}
\end{center}
\end{figure}

\begin{figure}
\includegraphics[width=3in]{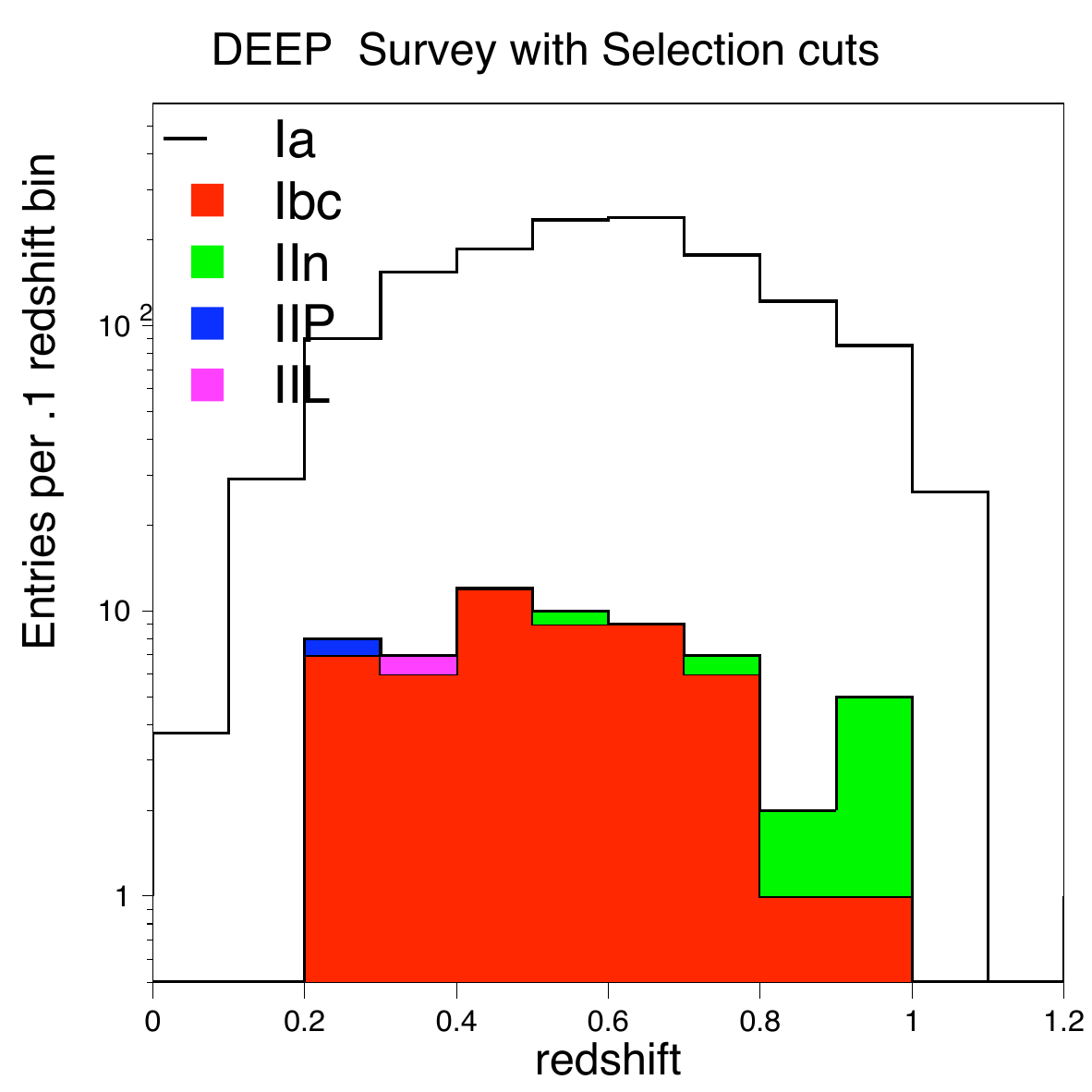}
\includegraphics[width=3in]{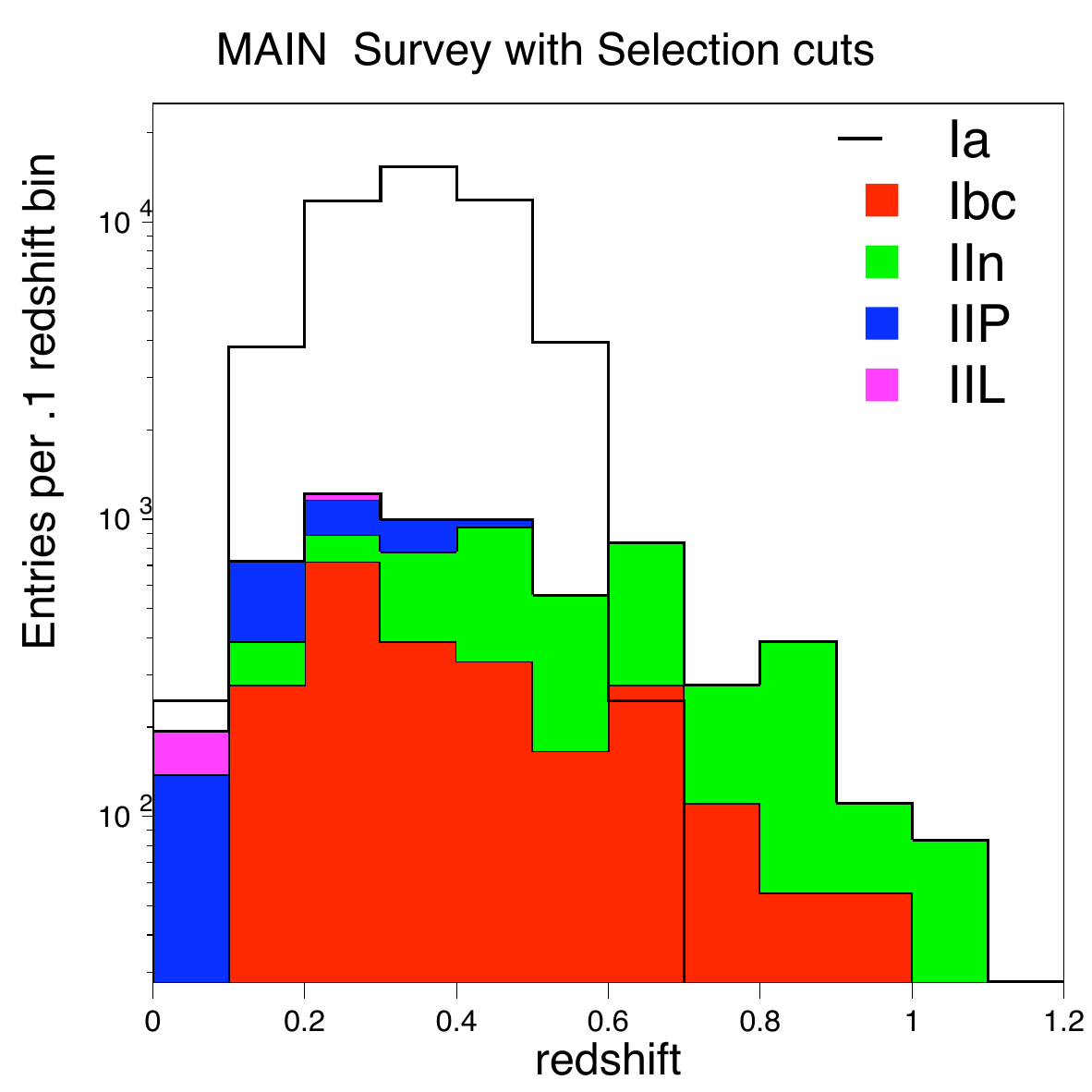}
\caption[Rates of \sneia and non-\sneia as a function fo redshift]{\textit{Left}: Rates of Type Ia and Non-Ia supernovae are shown as a function of the redshift
for the deep sample. \textit{Right}: Same but for the main sample. 
Note that both samples use an assumed host galaxy spectroscopic redshift.}
\label{fig:sn:cc_rate_3}
\end{figure}

Given these assumptions, we simulated a population of core-collapse
objects using the \snana code described in
\autoref{sec:sn:snia_rate_opsim}.  We then fit the simulated
photometry of each object in the combined \sneia and
core-collapse sample to templates using the \sneia MLCS2k2 model of
\citet{jha07}.  Our hope is that the goodness of fit (as quantified in
terms of the probability of observing a value of $\chi^2$ greater than
the measured value, $P_{\chi^2}$) would be a clean
way to distinguish the two classes of objects.  This was borne out, as
shown in \autoref{fig:sn:fitprob}.   The distributions of $P_{\chi^2}$
peaked sharply near $P_{\chi^2}=1$ for
\snia, and are very flat, extending to low probabilities for core collapse supernovae.  
\autoref{fig:sn:cc_rate_1} is analogous to
\autoref{fig:sn:rate_salt2}, now showing the number of \sneia with an
additional cut of $P_{\chi^2}>0.1$, and the contamination rate.  The
main sample has 2--3 times more contamination due to the sparser light
curves.  \autoref{fig:sn:cc_rate_3} shows the redshift distributions
for the deep and main samples, for the
\sneia and different types of core collapse supernovae. The contamination with
our assumptions is dominated by \sneibc, with the bright \sneiin contributing
at large redshift.


Improvements in our knowledge of the fraction of Type Ib/c supernovae
would help us better understand the overall contamination of the LSST
\snia sample.  The LSST \snia supernova cosmology samples will have
some level of core collapse contamination, and minimizing and
understanding that contamination is important for precision cosmology.

\section{\snia Photometric Redshifts}
\label{sec:sn:photoz}
\noindent{\it Scot S. Olivier, Sergei Nikolaev, Willem H. de Vries, Kem H. Cook, S. J. Asztalos} 
%

Supernova cosmology is currently based on measurements of two
observable \snia quantities: the brightness at several epochs (light
curve) observed in one or more bands and the redshift of
features in the spectrum of the supernova (or the host galaxy).


Only a small fraction of the \sneia discovered by LSST will have spectroscopic measurements of redshift.
Instead the redshift determination for most of these \sneia will be
based on photometric measurements of the broad-band colors from either
the host galaxy or the SN.  Redshift determinations
of galaxies, for which spectroscopy was unavailable, have been
estimated based on their colors (\autoref{sec:common:photo-z}).  This type of
``photometric redshift'' can also be used for \sneia
\citep{barris04,wang07a,wang07b}.   

\begin{figure}
\begin{center}
\includegraphics[width=0.6\linewidth]{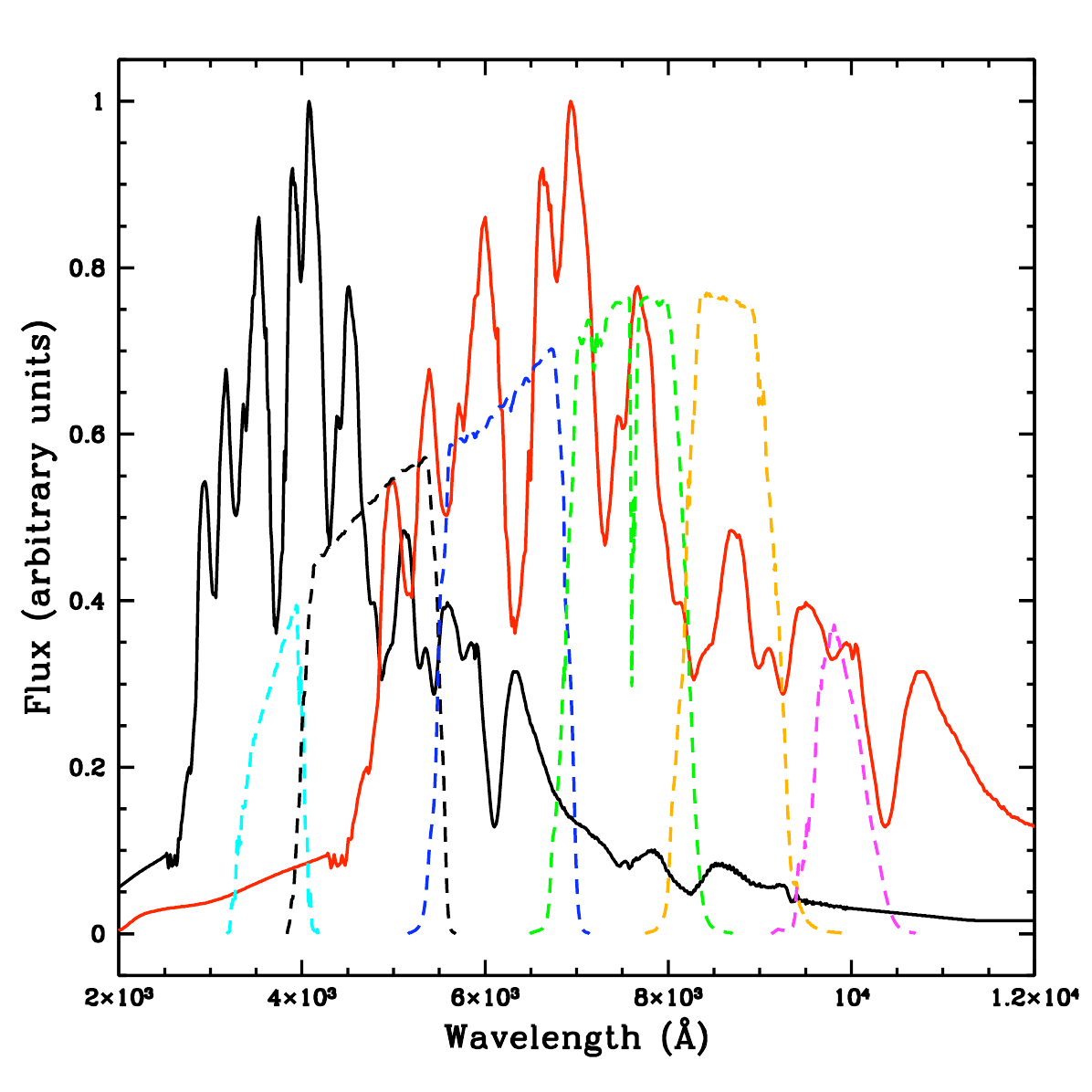}
\caption
[\snia light curves at different redshifts]
{
High quality \snia spectra at $z=0$ (black) and $z=0.7$ (red) overlaid
with the LSST filter bands.  The observed relative intensity in each
filter (i.e., the colors) can be used to estimate the redshift. 
}
\label{fig:sn:snia_spectrum_filters}
\end{center}
\end{figure}
\snia spectra are characterized by broad emission and absorption
features.  \autoref{fig:sn:snia_spectrum_filters} shows a nominal
\snia spectrum at two redshifts along with the wavelength bands for
the LSST filters.  If the intrinsic spectrum is known a priori, the colors
have the potential to accurately determine the redshift.  Errors in
this redshift 
estimate will be introduced by noise in the photometric measurements, variability in the intrinsic spectrum, degeneracies in the redshift
determination, and uncorrected extinction due to dust
along the line of sight.  Since \snia spectra evolve in time, errors
in the redshift estimate are also introduced by imprecise knowledge of
the epoch of observation relative to the time of the explosion.
However, this time evolution also provides an opportunity to use
observations at multiple epochs in order to improve the redshift
estimation. 

\begin{figure}
\includegraphics[width=\columnwidth]{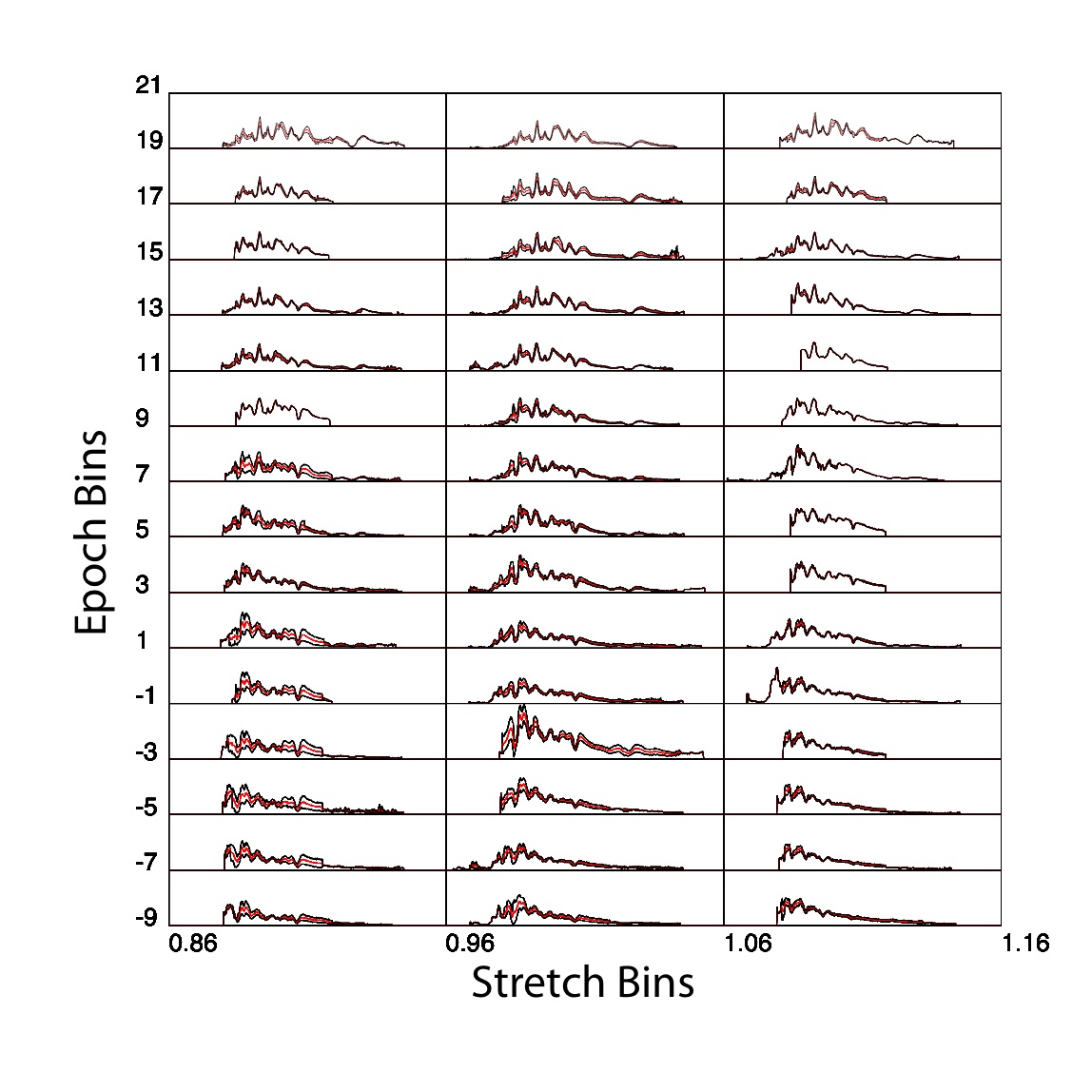}
\caption[\snia spectra and their variability limits in ``stretch-rest
  frame epoch'' space]{\snia spectra and their variability limits in
  ``stretch-rest frame epoch'' space.  Each bin shows the average
  rest-frame spectrum in relative units (red) and $\pm1\sigma$ limits
  (black).  The x and y labels indicate the bin boundaries in stretch and
  rest-frame epoch (in days); the wavelength scale within each bin is from
  $1,000$\AA\ to $15,000$\AA.  A stretch factor of unity corresponds
  to a supernova whose light curve drops by 1.0 mag in 15 days in the
  B band \citep{jha06a}.} 
\label{fig:sn:snia_spectra_stretch}
\end{figure}

\autoref{fig:sn:snia_spectra_stretch} shows the evolution of \snia
spectra as well as their variability limits as a function of stretch
and rest-frame epoch.  For each stretch-epoch bin, the average
spectrum was derived by co-adding \snia ``normal'' spectra from the
literature \citep[e.g.,][]{jha06a,matheson08,foley08a}.  The spectra
can be calibrated to absolute flux units by comparing to the actual
measured magnitudes of Type Ia (after the latter have been corrected
for reddening and standardized based on the observed
stretch). Focusing on the variability of spectra derived from real
supernovae allows for more realistic simulations of supernovae
observed by LSST. 

The LSST SN Science Team is currently studying the accuracy with which
the redshifts of \sneia can be estimated using the broadband
photometry provided by the LSST observations.  The approach is based
on simulating the supernovae using the methods described in
\autoref{sec:sn:snia_rate_opsim}.  

The redshift of the supernova is then another free parameter in a
model which includes the stretch and intrinsic color of the SN light
curve and the dust extinction \citep{devries09}. 
This
procedure does not use the apparent brightness as an
indicator of redshift, but uses the shape and color of the light curve as it
evolves over time and across different passbands.  Therefore, it is
sensitive to the $(1+z)$ time dilation effect which, like a
spectroscopic redshift, provides a distance-modulus independent
measurement. 

\begin{figure}
\begin{center}
\includegraphics[width=0.5\columnwidth]{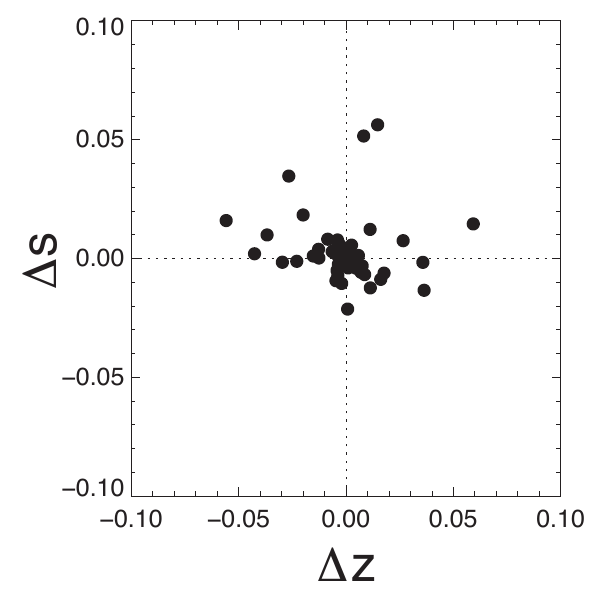}
\caption[Recovery of spectroscopic redshifts and stretch using only photometric data.]
{Recovery of spectroscopic redshifts and stretch using
  photometric information alone for 100 \sneia simulated from $0<z<1$,
  randomly populated across the LSST-visible sky, and propagated
  through the LSST Operations Simulator cadence for those fields.  Milky Way
  reddening (with $R_V = 3.1$) is assumed here; the redshift
  determination is likely to be worse if this assumption is not valid.}
\label{fig:sn:dzds}
\end{center}
\end{figure}

Fitting the light curve of each model supernova gives the estimate of
the photometric redshift and stretch, which can be compared to the
original values used in generating the supernova.
\autoref{fig:sn:dzds} shows the accuracy of the photometric redshift
and stretch determination based on simulation of 100 \sneia.  
These \sneia were selected uniformly from $z=[0,1]$, stretch
$s=[0.86,1.16]$, and randomly across the LSST sky.  Each \snia light
curve was then propagated through the results of the Operations
Simulator (\autoref{sec:design:opsim}) and photometric uncertainties
were applied according to the LSST exposure time calculator (\autoref{sec:com:expos}).  The
light curves were then fit with the photometric redshift fitting
code. 

In
the figure, we turned off the effects of reddening and the variability
in the Type Ia spectra.  Reddening in particular will have an
important impact of our ability to recover photometric redshift.  If
$R_V=3.1$ as in the Milky Way, we will be able to calibrate out
the effect of reddening on redshift determination to an accuracy of a
percent in $(1+z)$, but if the reddening law is unconstrained, one
might get biases in estimated $(1+z)$ at the few percent
level.  

\begin{figure}
\begin{center}
\includegraphics[width=8cm,angle=270]{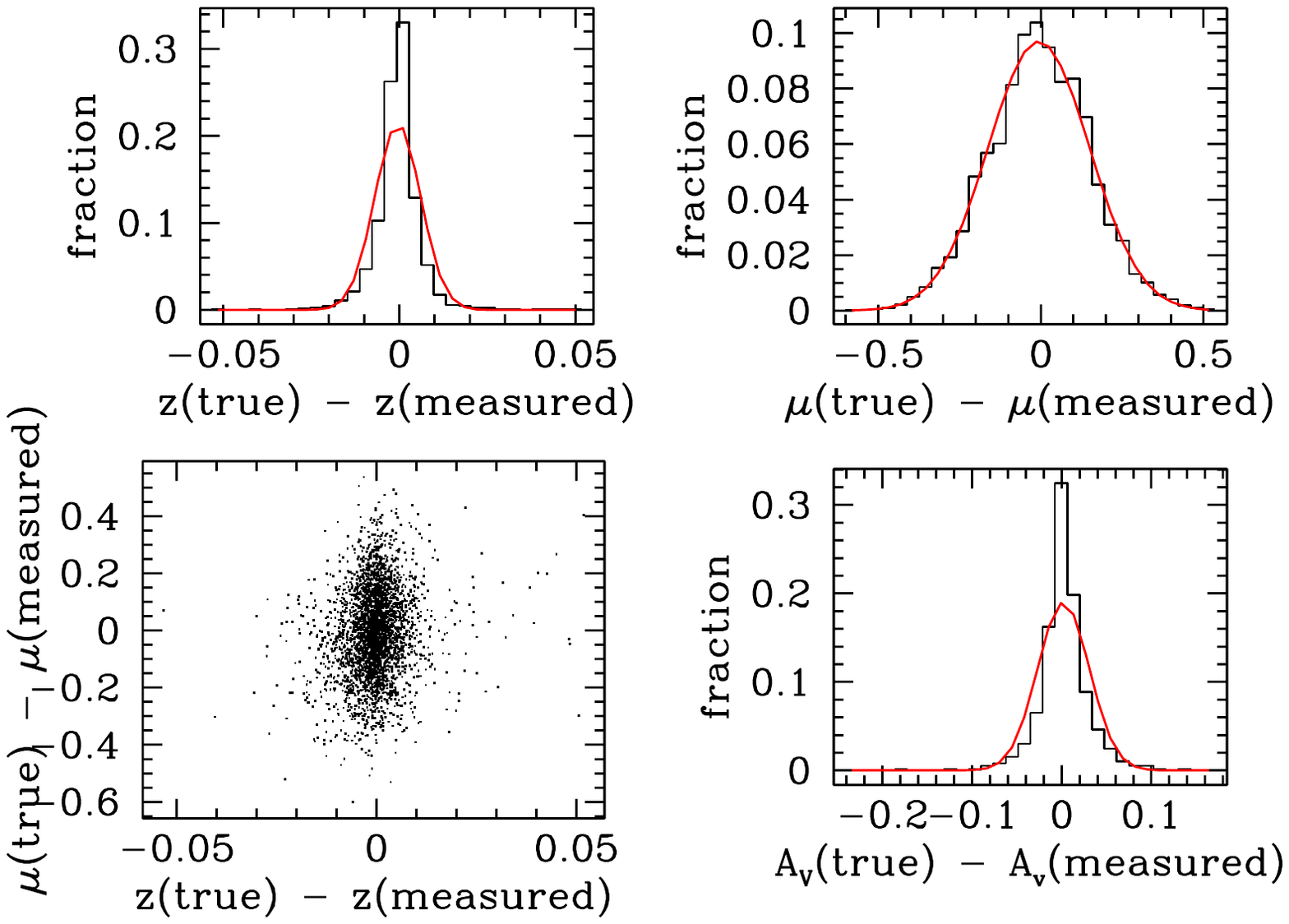}
\caption[Deep Drilling \snia Light Curve Parameters]{
Simulations showing the errors in recovering $z < 1$ SN properties from light
curves obtained in the deep-drilling mode. 
(clockwise from top left): 
1) redshift errors; mean = $-$0.0004, $\sigma_z=0.007$. 
2) distance modulus $\mu$ error; mean = $-$0.006, $\sigma_\mu=0.16$. 
3) reddening error; mean = 0.002, $\sigma_{A_V} = 0.03$.
4) correlation between redshift and distance modulus errors. The
covariance is weak.
}
\label{fig:sn:sn4plot}
\end{center}
\end{figure}

This is explored in a bit more detail in \autoref{fig:sn:sn4plot},
which focuses on the deep drilling fields; the ability to recover
photometric redshift, distance modulus, and reddening (again assuming
a fiducial value for $R_V$) from our light curves is shown.  

Although initial results quantifying the accuracy of photometric
redshifts for \sneia are promising, more work is needed in order to
fully understand the effect of photometric errors and intrinsic
spectral variability, as well as inaccuracies in the spectral
templates and incompleteness in the light curve sampling.

\section{Constraining the Dark Energy Equation of State}
\label{sec:sn:darkenergy}
\noindent{\it Philip A. Pinto, Peter Garnavich, W. Michael Wood-Vasey,
Rahul Biswas, Benjamin D. Wandelt} 

Type Ia supernovae (\sneia) are the best standard candles at large
distances \citep{gibson00,parodi00}.  Supernovae provided
the first of the triad of observational constraints on which the
now-standard dark energy-dominated model of cosmology is based 
\citep{riess98,perlmutter99}.  
The challenge of the next decade of
supernova research is to explore the physics of supernovae
themselves, their relationship with their environments,
and the nature of the redshift-luminosity relation for \sneia.
Massive samples of supernovae at all redshifts with superb data are
required for these goals. As we have seen in
\autoref{sec:sn:snia_rate_opsim}, the LSST data will produce on the order
of 50,000 \sneia per year with photometry good enough for accurate
light curve fitting.  The sample will have a mean redshift of $z \sim
0.45$, stretching up to $z \sim 0.8$.  

Supernova color statistics and good light curves, combined with a relatively
small number ($\sim1$\%) of sample spectra, will reveal any
dependence of the supernova standard candle relation on parameters
other than light curve shape and extinction, shedding light on any
systematic errors in the \snia technique.  

%

\begin{figure}
\begin{center}
\includegraphics[width =0.4\textwidth]{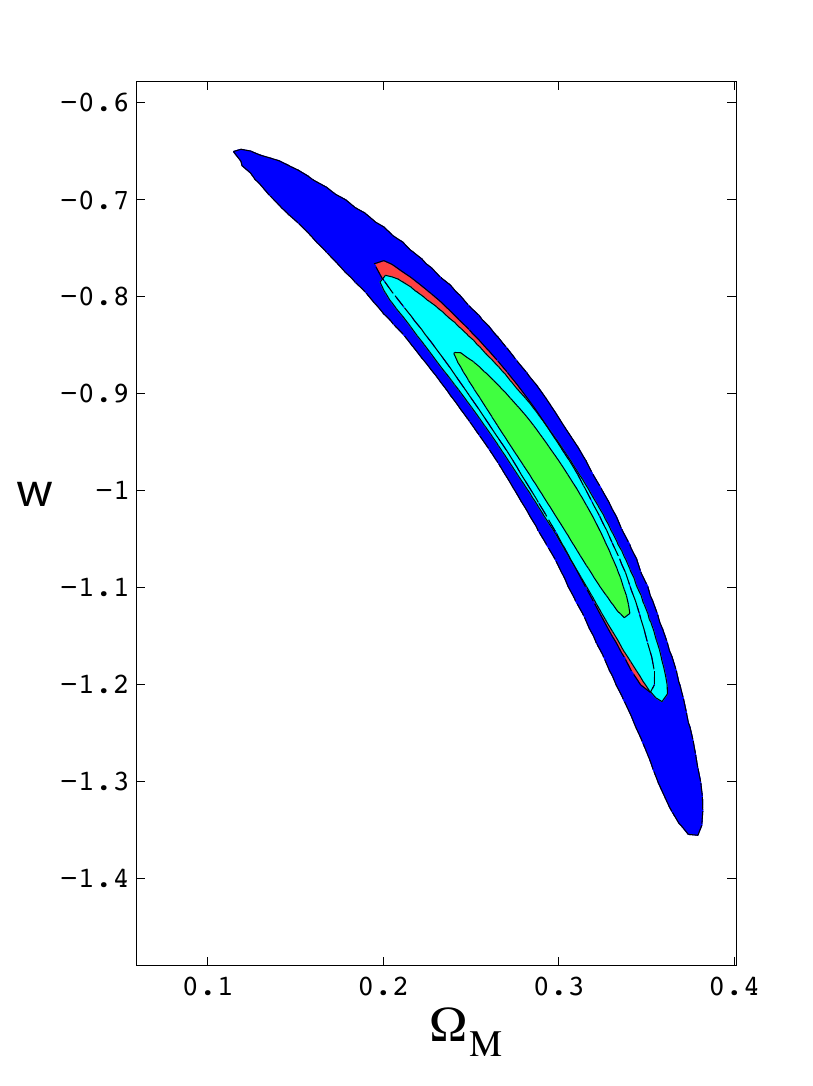}
\includegraphics[width =0.4\textwidth]{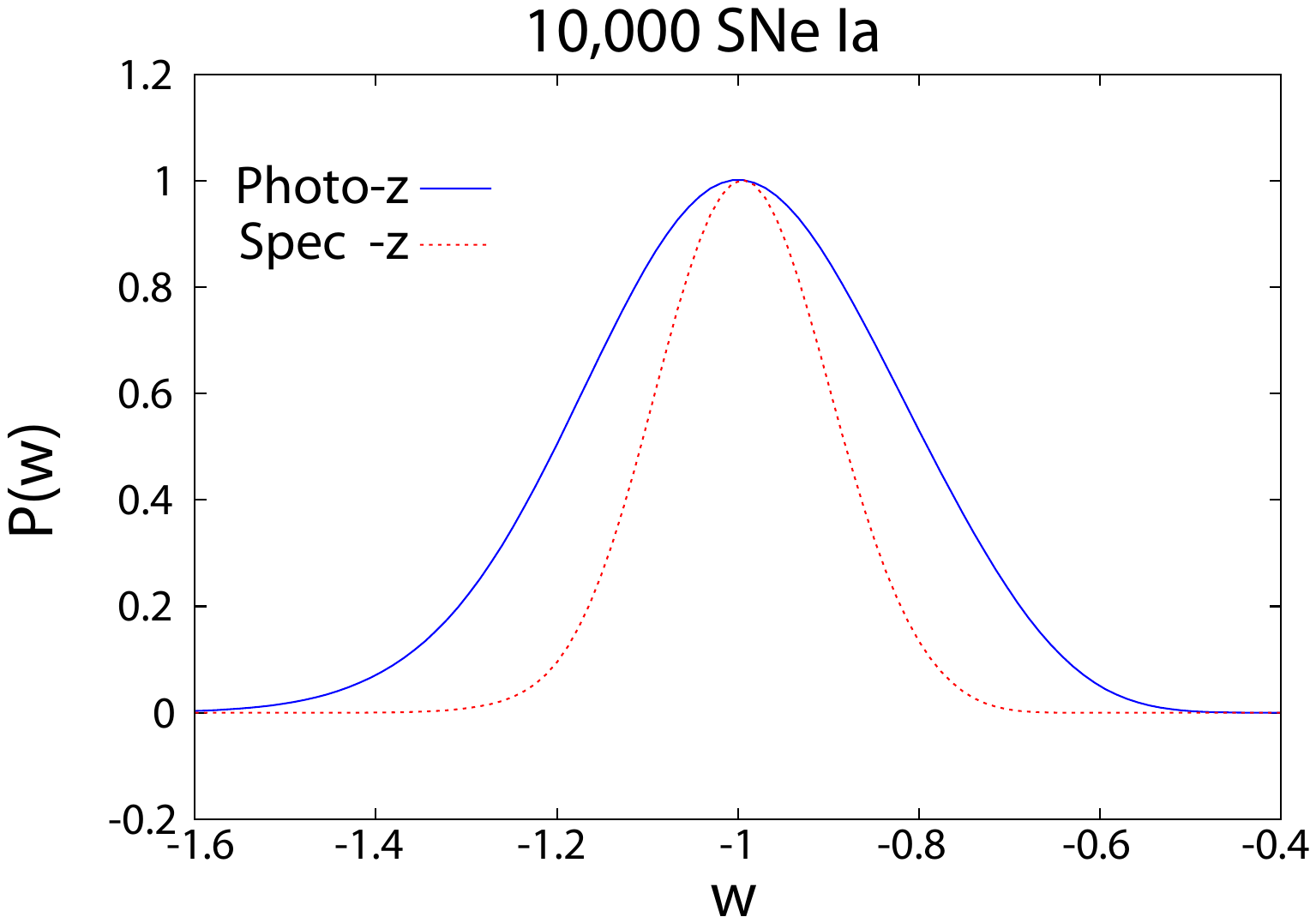}
\caption[$\Omega_M-w_0$ constraints from 10,000 \sneia]{Forecasted
  constraints on a constant equation of state $w$ in a flat cosmology
  from about 10,000 supernovae. The left panel shows the joint
  posterior distribution on $\Omega_m$ and $w$, assuming an intrinsic
  distribution in the distance indicator of 0.12 mag. The green and
  cyan contours show the 68\% and 95\% constraints when photometric
  redshifts are used, while the red and blue contours show the same
  constraints with spectroscopic redshifts. The right panel shows the
constraints on $w$ marginalized over $\Omega_M$. Results are shown
separately assuming that we have spectroscopic redshifts for all
supernova hosts, and the more realistic case of photometric redshifts
only.  No other priors were used.} 
\label{fig:sn:w_omegam_10000sneia}
\end{center}
\end{figure}

The large number of supernovae detected allows for a number of 
different approaches. Rough models show that even $\sim 10,000$~\sneia with intrinsic
scatter in the distance indicator of 0.12 mag can constrain a constant equation of state $w$ for a flat 
cosmology to better than 10\% with no additional priors
(\autoref{fig:sn:w_omegam_10000sneia}).  
Over 10 years one should thus have $\sim50$ independent measurements
of $w$, each to 10\%.    
This of course assumes that there is no systematic floor in
the supernova distance determination. 
With such a large sample, we could imagine deriving $w$ independently
for subsamples of supernovae with identical properties (e.g., light
curve decay times, host galaxy types, etc.), to look for such systematics.  Each subclass will
provide an independent estimate of $w$, and consistency will indicate
lack of serious systematic effects such as supernova evolution (\autoref{sec:sn:evolution}). 

\begin{figure}
\begin{center}
\includegraphics[width=0.6\textwidth]{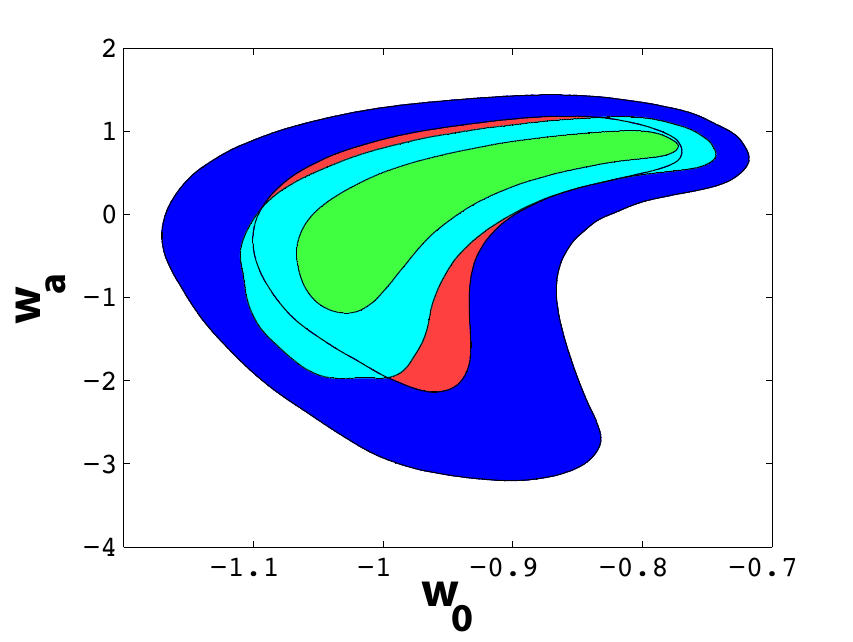}
\caption[$w_0-w_a$ constraints from 50,000 \sneia]{Forecast for joint
  posterior distribution on the parameters of a time evolving equation
  of state parametrized by $w(a) = w_0 + w_a (1-a)$ in a flat
  cosmology from 50,000 supernovae (i.e., one year of the LSST
  survey). The green and cyan contours show the 68\% and 95\%
  constraints including photometric errors on redshift as a Gaussian
  with an error $\sigma_z=0.01(1+z)$, while the red and blue contours
  ignore photometric errors and only include an intrinsic dispersion
  of $0.12$ mag in the distance indicator.}
\label{fig:sn:w0_wa_50000sneia}
\end{center}
\end{figure}

With a sample of 50,000~\sneia (i.e., about one
tenth of the full sample LSST will gather in ten years), one can put
constraints better than 5\% on a constant dark energy equation 
of state $w$ (\autoref{sec:com:cos}).  With a redshift-dependent
equation of state, these data will constrain $w_0$ to 0.05, and $w_a$
to accuracy of order unity (\autoref{fig:sn:w0_wa_50000sneia}).  

With weak lensing (\autoref{chp:wl}), constraining cosmological
parameters requires a model for the growth of structure with epoch.
By contrast, \snia luminosity distances constrain cosmology by directly measuring the
redshift-distance relation, and therefore the metric itself.   If dark
energy is a manifestation of something radically new 
in space-time gravity, a comparison of the two approaches will reveal
discrepancies, which will give us clues about this new physics.

\section{Probing Isotropy and Homogeneity with \sneia}
\label{sec:sn:isotropy}
\noindent{\it W. Michael Wood-Vasey} 


The most basic cosmological question about dark energy is whether it
is constant in space, time, and local gravitational potential (\autoref{sec:cp:newphys}).  One of
the most powerful properties of SNe Ia as cosmological probes is that
even a single object provides useful constraints.

For general cosmological investigations it will be possible to minimize various
systematic difficulties inherent in the analysis of \sneia by calibrating the LSST
\snia sample vs.~other cosmological probes. 
But one can then start to determine whether the dark energy is uniform across
space and time. 

The all-sky nature of the sample of 500,000~\sneia that the LSST will
identify in its 10 years of operations will enable searches for an
angular dependence in the redshift-distance relation, thus determining
whether the dark energy equation of state as characterized by $w$, and
possibly even $w_a$, are directionally dependent.  Any such signature
would surely be an indication of fundamental new physics.

\begin{figure}
\begin{center}
\includegraphics[width=.6\columnwidth]{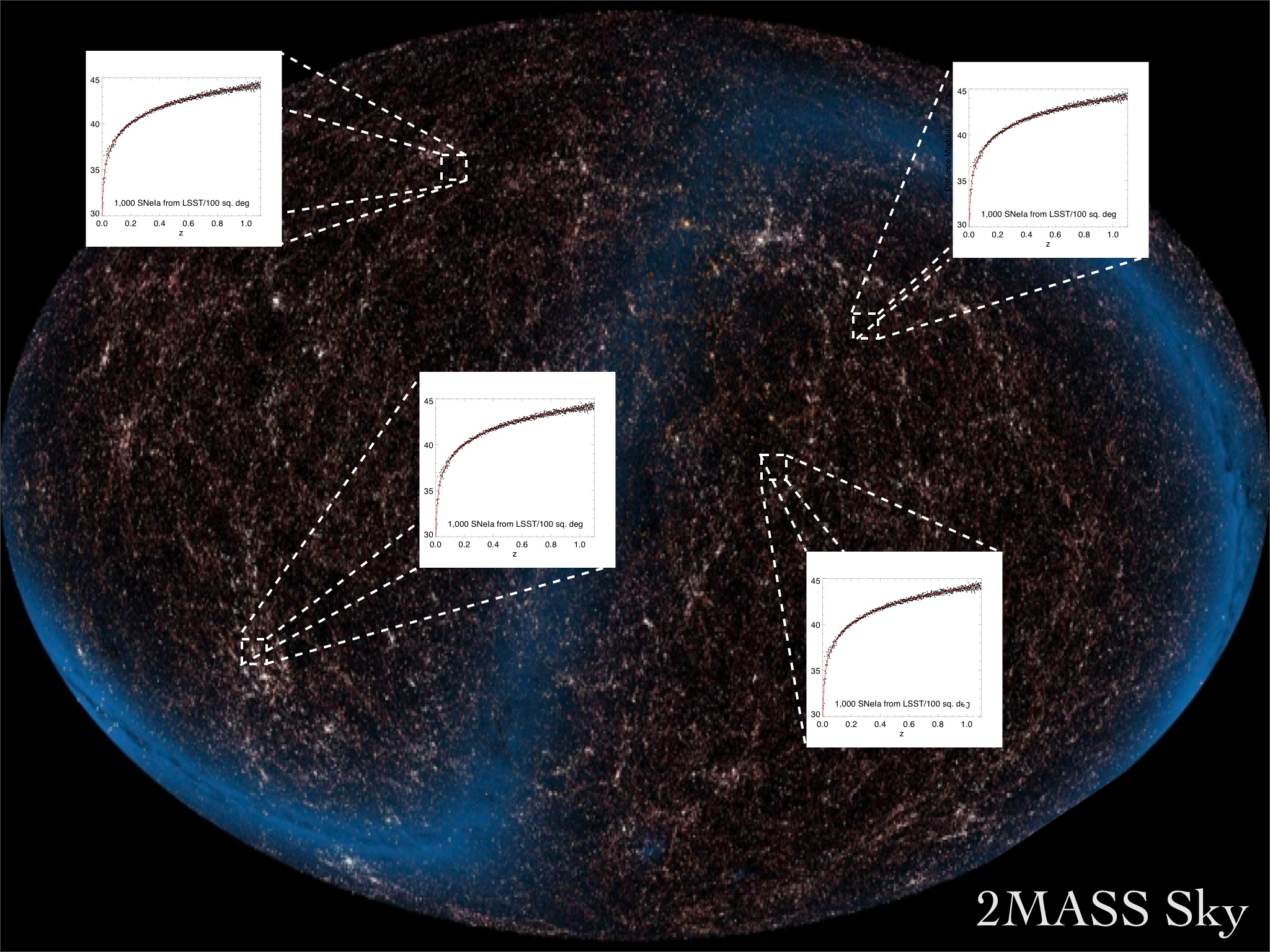}
\caption[Is dark energy isotropic?]{Testing the isotropy of dark
energy by obtaining \snia luminosity-redshift measurements in each of
500 pixels on the sky. Each such pixel of 40 deg$^2$ will have on the
order of 1000 \sneia, and cosmological parameters can be estimated
from each of these independently.  The background sky image is from
the 2MASS survey, and shows the distribution of nearby galaxies.} 
\label{fig:sn:pixel_cosmology}
\end{center}
\end{figure}

To investigate isotropy, one can divide up the LSST \sneia into 500
40~deg$^2$ pixels on 
the sky, each containing one thousand \sneia
(\autoref{fig:sn:pixel_cosmology}).  Given the Hubble diagram for each
pixel, one obtains 500 independent measures of $w$ to look for
variations which violate assumptions of isotropy and homogeneity: What
is the rms for $w$ over all the 
pixels? If there is variation in $w$, is it smoothly varying
(i.e., correlated), and is the variance correlated with large-scale
structure?  These isotropy tests  will be directly comparable to the
large-scale structure maps that will come from galaxy photometric
redshift surveys, weak lensing, and strong-lensing from clusters.
Further explorations can be made by investigating shells of redshift
and the correlation of \snia vs.~local environment, both in galaxy
properties and gravitational potential.   See \autoref{sec:cp:newphys}
for further probes of the isotropy of dark energy.

\section{\snia Evolution}
\label{sec:sn:evolution}
\noindent{\it David Cinabro, Saurabh W. Jha } 

One reason for a focus on \sneia is their observational homogeneity.
When a small subset of peculiar \sneia are removed, easily identified
by unusual colors at the peak of the light curve, the dispersion of
\snia peak luminosities at rest-frame blue bands is about 20\%,
corresponding to 10\% dispersion in distance.  This is perfectly
adequate for measurements of cosmological parameters with sample sizes
of a few hundred.  But without understanding the physical origins of
this dispersion, and whether it has systematic effects that depend on
redshift, we will not be able to use the full statistical power of the
tens to hundreds of thousands of \sneia that LSST will find. Already
the latest measurements of cosmology parameters with \sneia
\citep{hicken09,kessler09b} have uncertainties with significant
contributions from systematic effects.

With such a large sample size it becomes possible to test for the
underlying causes of the intrinsic dispersion of \snia peak
luminosity.  Dependence on cosmic time, or redshift, would most likely
indicate evolution of the \snia progenitor population.  There is
already good evidence for a difference in the properties of \sneia as
a function of host galaxy type.  Higher luminosity, slowly
brightening, and slowly declining \sneia are preferentially associated
with star-forming galaxies, while dimmer, fast brightening, and
rapidly declining \sneia are preferentially in passive galaxies
\citep{sullivan06}.  At the present level of measurement, the two
populations of supernovae follow the same correlation between light
curve stretch and peak luminosity, but it is suggestive that more
subtle effects may be associated 
with the \snia environment such as the host galaxy's historical
development and properties such as metallicity.  Indeed there is good
evidence for a correlation between age, or metallicity, and \snia peak
luminosities.  In a study of 29 galaxy spectra that hosted \sneia,
\citet{gallagher08} show that \snia peak luminosities are correlated
with stellar population age and therefore metallicity as well.  
They also note a suggestive correlation between age, or metallicity, and 
residual on the Hubble diagram.   

The effect is large, of the same order as the 20\% intrinsic scatter,
and validates theoretical models of the effect of population age, or
metallicity, on \snia progenitor composition
\citep{hoeflich98,umeda99,timmes03}.
Given that stellar populations will naturally be younger in
higher redshift supernovae, this could lead to a systematic effect as
a function of redshift.  
Measurement of \snia evolution, or lack thereof, will provide valuable
constraints on these models.  LSST will allow correlations of
supernovae properties with those of their host galaxies.  For example, 
\citet{howell09} correlated \snia luminosities with host
metallicities estimated from the host photometric colors, and
\citet{hicken09} found that host morphology correlates with
\snia extinction and scatter on the Hubble diagram.  
The large LSST sample would allow one to subdivide by galaxy
properties, measuring cosmological parameters for each subsample
separately. 

Our lack of understanding of what \sneia progenitors really are
(\autoref{sec:sn:snia_rates}) limits
our confidence in using them for cosmology at a precision level.  We
would like to understand the progenitors and explosion models that
lead to both the intrinsic scatter of normal \snia properties and give
rise to the peculiar population.  Understanding the physical causes of
peculiar \sneia, such as SN~2005hk
\citep{sahu08,stanishev07,phillips07,chornock06} and 2005gj
\citep{hughes07,prieto07,aldering06}, has the promise to constrain
models for normal \sneia and illuminate the underlying reason for the
diversity of normal \sneia.  We now have only a handful of truly
peculiar objects, i.e., objects not well described by the family of
light curves with one free parameter, because they are only a few
percent of all \sneia. The LSST sample of \sneia will yield a larger
sample of peculiar objects that can be targeted for further study.

\section{\snia Rates}
\label{sec:sn:snia_rates}
\noindent{\it Evan Scannapieco, Benjamin Dilday, Saurabh~W.~Jha} 

There is now broad consensus that a Type Ia supernova is the
thermonuclear explosion of a carbon-oxygen white dwarf star that
accretes mass from a binary companion until it approaches the
Chandrasekhar mass limit \citep[e.g.,][]{Branch_95}. However, much
remains to be learned about the physics of \sneia , and there is
active debate about both the nature of the progenitor systems and the
details of the explosion mechanism.  For example, the binary companion
may be a main sequence, giant, or sub-giant star (the
single-degenerate scenario), or a second white dwarf (the
double-degenerate scenario).  The type of the companion star
determines in part the predicted time delay between the formation of
the binary system and the SN event~\citep{Greggio_05}. The time delay
can be constrained observationally by comparing the \snia rate as a
function of redshift to the star formation history
\citep{Strolger_04,Cappellaro_07,Pritchet_08}.

In order to test such a model for the evolution of the \snia rate,
improved measurements of the rate as a function of redshift and of
host galaxy properties are needed.  The LSST is well suited to provide
improved measurements of the \snia rate, with unprecedented
statistical precision, and over a wide range of redshifts. Rate
studies are much less sensitive to photometric redshift uncertainties
than \snia cosmology measurements (e.g., \citealt{Dilday_09}), 
and the vast photometric LSST \snia sample will be
directly applicable to this problem.

The insight into the nature of the progenitor systems that \snia rate
measurements provide can also potentially strengthen the utility of
\sneia as cosmological distance indicators (\autoref{sec:sn:evolution}). Although the strong
correlation between \snia peak luminosity and light curve decline rate
was found purely empirically \citep{Psk_77,Phillips_93}, the physics
underlying this relation has been extensively studied
\citep{Hoeflich_95,Hoeflich_96,Kasen_07}.  
There is hope that improved
physical understanding and modeling of SNe Ia explosions, coupled with
larger high-quality observational data sets, will lead to improved
distance estimates from \sneia.  As part of this program, deeper
understanding of the nature of the progenitor systems can help narrow
the range of initial conditions that need to be explored in carrying
out the costly simulations of SNe Ia explosions that in principle
predict their photometric and spectroscopic properties.

Measurement of the \snia rate may also have a more direct impact on
the determination of systematic errors in \snia distance estimates.
The empirical correlation discussed in \autoref{sec:sn:evolution}
between stretch parameter and stellar populations in the host 
\citep{Hamuy_96,Howell_01,vandenbergh_05,jha07} suggest 
connection between the age of the stellar population and the \snia
rate, e.g., a ``prompt'' channel from progenitors found in
star-forming regions and a ``delayed'' channel which depends perhaps
only on the integrated star formation history of the host.

\citet{Mannucci_05}, \citet{ScanBildsten}, \citet{Neill_06}, 
\citet{sullivan06}, and \citet{aubourg2008} have argued that a two-component model of the SN
Ia rate, in which a prompt SN component follows the star formation
rate and a second component follows the total stellar mass, is
strongly favored over a single \snia channel.  In this picture, since
the cosmological star formation rate increases sharply with lookback
time, the prompt component is expected to dominate the total \snia
rate at high redshift.  \citet{Mannucci_05} and \citet{Howell_07}
pointed out that this evolution with redshift can be a potential
source of systematic error in \snia distance estimates, if the two
populations have different properties and are not properly disentangled.

To model the contributions of each of these two types of progenitors,
\citet{ScanBildsten} write the total \snia rate as
\begin{equation}
{\rm Rate}_{\rm Ia} (t) = A M_{\star}(t) + B \dot{M}_{\star} (t)
\end{equation}
Here the A-component or delayed component is proportional to the total
stellar mass of the host, and the B-component or prompt component is
proportional to the instantaneous star formation
rate (as measured from model fits to the broad-band spectra energy distribution (SED) of the
host). \citet{sullivan06} later measured these proportionality 
constants to be $A = 5.3 \pm 1.1 \times 10^{-14} \, M_{\odot}^{-1} \,
{\rm yr}^{-1}$ and $B = 3.9 \pm 0.7 \times 10^{-4} \, M_{\odot}^{-1}$,
and the large sample of \sneia from the LSST will clearly improve
these error bars.  However, pinning the prompt component to the
instantaneous star formation rate is theoretically unpleasing since
the formation time of a white dwarf is no less than 40 Myr. It is far
more likely that the ``prompt'' component exhibits a characteristic
short delay $\tau$, namely,
\begin{equation}
{\rm Rate}_{\rm Ia} (t) = AM_{\star}(t) + B \dot{M}_{\star} (t - \tau)
\end{equation}
\citet{Mannucci_05}
proposed modeling the B-component as a Gaussian centered at 50 Myr,
but the true value of this delay remains extremely uncertain.
Recently, \citet{Fruchter06} developed an observational method
that constrains the properties of core-collapse SNe and gamma-ray bursts (GRBs). The
method involves observing the spatial locations of SNe in their host
galaxies and calculating the fraction of the total host galaxy light
contained in pixels fainter than these locations. Transients
associated with recent star-formation are systematically located in
the brightest pixels, while transients arising from older stellar
populations are anti-correlated with the brightest regions.  

\citet{Raskin_08} carried out detailed modeling to interpret such data
and proposed a modified pixel method that can be used to constrain the
properties of \sneia progenitors. What is needed is a procedure that
correlates \sneia with the properties of their immediate environment in
the galaxy, rather than
with the host as a whole. In a spiral host, the ideal method for
constraining \sneia progenitors would be to measure the relative
brightness of pixels within annuli. In this case, as the density wave
of star formation moves around the annulus, \sneia would appear behind
it at a characteristic surface brightness determined by the level to
which a stellar population fades away before \sneia appear. However,
observations are never ideal, and observing a single annulus of a
spiral host is subject to complications such as spurs, knots, and
gaps.  The solution to this problem is the doughnut method, which
builds directly on the method described in \citet{Fruchter06}. The
idea is to expand an annulus radially by some small but appreciable
radius, so as to encompass enough of the host's morphological
peculiarities to have a good representative sample, yet narrow enough
to represent local variations in the host light. 
In \citet{Raskin_09}, this method was applied
to SDSS images using a sample of 50 local \sneia, finding clear
evidence that the delay time $\tau$ associated with the B component
exceeds 200 Myr. In the LSST data set with $0.7''$ resolution, the
method can be applied to \sneia hosts out to $\sim$250 Mpc. This will
give over 1,000 \sneia from the main LSST survey per year, which will
be sufficient to map out the short-time distribution of \sneia in
exquisite detail, providing strong constraints on the relationship
between star formation and \snia production.

\section{\snia BAO} \label{sec:sn:bao}
\noindent{\it Hu Zhan } 

Since \sneia explode in galaxies, they can, in 
principle, be used as the same tracer of the large-scale structure as 
their hosts to measure baryon acoustic oscillations (BAOs, see 
\autoref{sec:lss:bao} for details) in the power spectrum of their 
spatial distribution. Considerations for measuring BAO with \sneia
are as follows.
\begin{itemize}
\item As described elsewhere in this chapter, \sneia have rich and 
time-varying spectral features, so that their \phz{}s can be 
determined more accurately than galaxy \phz{}s 
\autoref{sec:sn:photoz}, which compensates 
for the sparsity of the \snia sample for constraining dark energy. 
Indeed, within the same redshift range of $z < 0.8$, measuring BAO
with the LSST \snia sample will place
slightly tighter constraints on $w_0$ and $\wa$ than the LSST galaxy BAO. 

\item The \snia sample has a very different selection 
function from conventional galaxy samples that are selected by 
luminosity or color. Hence, \snia BAO can provide a weak consistency 
check for galaxy BAO.

\item \sneia are standardizable candles. The narrow range of the 
standardized \snia intrinsic luminosity reduces the effect of 
Malmquist-like biases and luminosity evolution, as seen in galaxy 
surveys. 

\item Although \snia BAO only places weak constraints on
$w$, it can significantly improve the constraints from 
\snia luminosity distances of the same data, leading to results 
comparable to those of LSST WL two-point shear tomography. In other
words, the extra information from \snia BAO can make LSST \sneia
more competitive with other LSST probes.
\end{itemize}

\begin{figure}[ht]
\begin{center}
\includegraphics[width=0.9\linewidth]{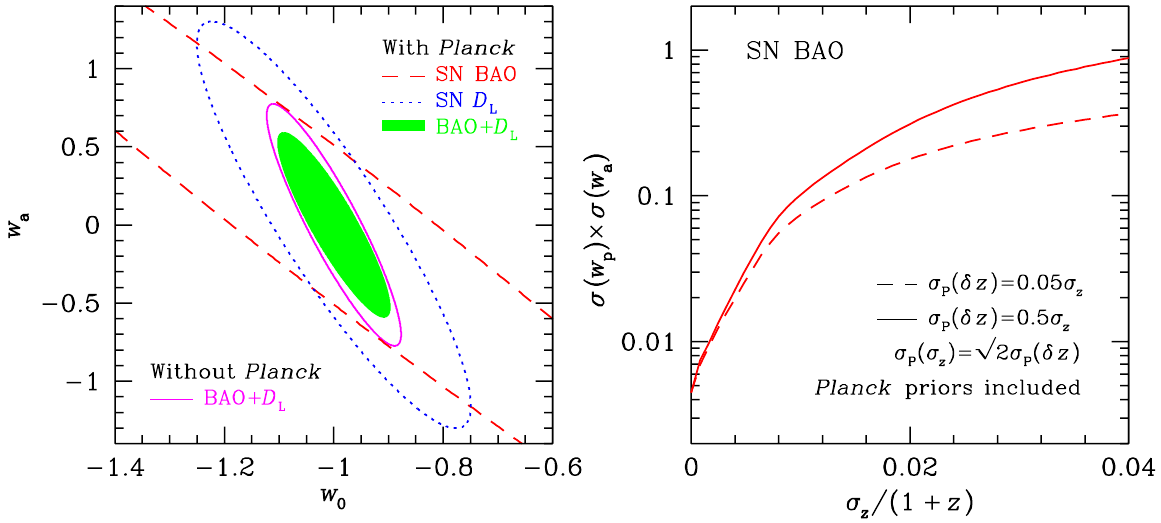}
\caption[Dark energy constraints from \snia BAO]
{
\emph{Left panel}: Marginalized $1\,\sigma$ error contours of the 
dark energy EOS parameters $w_0$ and $w_{\rm a}$ from \snia BAO with Planck (dashed line), luminosity distances with Planck (dotted line), and the two combined with 
(shaded area) and without (solid line) Planck. 
The mean curvature parameter, $\Ok$, is allowed to float. Even 
though results of the \snia BAO and \snia $D_{\rm L}$ techniques are 
sensitive to CMB priors individually, the combined result is much 
less so. \emph{Right panel}: The error product 
$\sigma(w_{\rm p}) \times \sigma(w_{\rm a})$ from LSST \snia BAO
as a function of the rms \phz ~error $\sigma_z$. The error 
$\sigma(w_{\rm p})$ equals the error on $w_0$ when $\wa$ is 
held fixed. The priors on the \phz ~biases are 
taken to be $0.5 \sigma_z$ (solid lines) and $0.05 \sigma_z$ (dashed
lines), which correspond to calibrations with four and four hundred spectra 
per redshift bin, respectively, in the Gaussian case.
To reduce the dimensions, we peg the prior on the \phz ~rms 
to that on the \phz ~bias: $\sigma_{\rm P}(\sigma_z) = 
\sqrt{2} \sigma_{\rm P}(\delta z)$. 
For comparison, LSST weak lensing, galaxy BAOs, and the two combined 
will achieve error products of $\sim 0.01$, $0.02$, and $0.002$, 
respectively \citep{zhan06d}. The behavior of the \snia BAO error 
product as a function of the \phz ~rms is not specific to \sneia
and is generally applicable to any \phz ~BAO survey.
Figure from \citet{zhan08}, with permission.
}
\label{fig:sn:bao}
\end{center}
\end{figure}

For the BAO technique to be useful, one must survey a large volume 
uniformly at a sufficient sampling density.
Although \snia events are rare, the spatial density of \sneia
accumulated by LSST over several years will be 
comparable to the densities targeted for 
future spectroscopic galaxy BAO surveys. 
In its wide survey mode, LSST will obtain half a million \sneia over 
20,000 deg$^2$ to redshift 
$z = 0.8$ (\autoref{sec:sn:snia_rate_opsim}). Such a sample is capable of measuring the baryon signature 
in the \snia spatial power spectrum. The significance of detection, 
however, depends on the assumptions about cosmology. For example,
the baryon signature has been detected at the $\sim 3\,\sigma$ level
(constraining $\om$ to 10\%)
from SDSS
Luminous Red Galaxies, both spectroscopically
\citep{eisenstein05} and photometrically \citep{blake07,padmanabhan07}.
These detections assume a flat Universe with a cosmological constant
and a fixed scalar spectral index $n_{\rm s}$. 
Under the same assumptions, LSST 20,000 deg$^2$ \snia BAO can constrain 
$\om$ to 8\% and $\ob$ to 15\% \citep{zhan08}. 
If $\omega_{\rm b}$ is fixed as well
\citep[as in][]{eisenstein05}, the same \snia BAO data can achieve
$\sigma(\ln\omega_{\rm m})=1.5\%$.

Dark energy constraints from \snia BAOs are much weaker than those 
from luminosity distances of the same \sneia, 
but these two techniques are highly complementary to each other. 
The left panel of \autoref{fig:sn:bao} illustrates that the 
combination of the two techniques improves the dark energy constraints
significantly over those of luminosity distances, and the results
are no longer sensitively dependent on CMB priors. 
The right panel of \autoref{fig:sn:bao} shows the degradation to
dark energy constraints from LSST \snia BAO as a function of the 
\phz ~rms and priors on the \phz ~parameters. The slope of the 
error product changes around $\sigma_z \sim 0.01(1+z)$, because 
radial BAO information becomes available when \phz ~errors are 
small enough.


\section{\snia Weak Lensing}
\label{sec:sn:wl}
\noindent{\it Yun Wang} 

\begin{figure} 
\begin{center}
\includegraphics[width=3in]{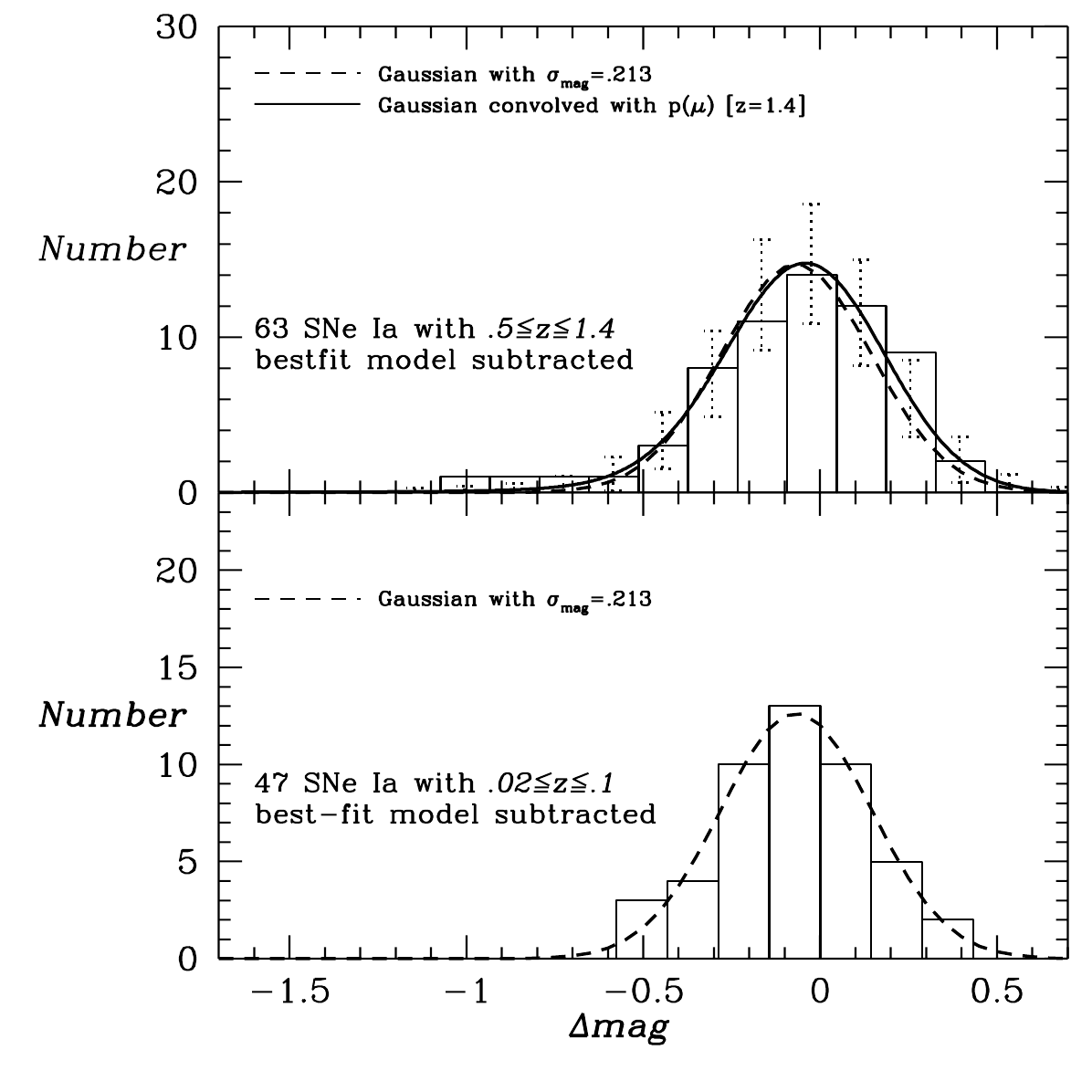}
\caption[Lensing with \sneia]{The distributions of fractional differences between
  the peak flux and that predicted by the best-fit model
  \citep{wang04} of 63 SNe~Ia
with $0.5 \leq z \leq 1.4$ (top panel) and 47 SNe~Ia with $0.02 \leq z \leq 0.1$ 
(bottom panel). 
The weak lensing predictions are the solid lines in the top panels 
(with the error bars indicating the Poisson noise of these predictions), 
and depend on the assumption that the \snia intrinsic peak 
brightness distribution is Gaussian in flux. 
Figure used with permission from \citet{wang05}.
}
\label{fig:sn:wl}
\end{center}
\end{figure}

The effect of weak lensing adds an additional uncertainty
to using SNe~Ia as cosmological standard candles at high redshift. 
Fluctuations
in the matter distribution in our Universe deflect the
light from SNe~Ia, causing either demagnification or 
magnification \citep[see, e.g.,][]{kantowski95,branch95,frieman97,wambsganss97,holz98b,metcalf99,wang99,valageas00,munshi00,barber00,premadi01}.
The weak lensing effect of SNe~Ia can be analytically 
modeled by a universal probability distribution function (UPDF) derived from 
the matter power spectrum \citep{wang99,wang02}.
\citet{wang05} derived the observational signatures of weak lensing
by convolving the intrinsic distribution in SN~Ia peak luminosity, $p(L_{SN})$,
with magnification distributions of point sources derived
from the UPDF, $p(\mu)$. 
\autoref{fig:sn:wl} shows the difference between peak brightness and that
predicted by the best-fit cosmological model for 63 SNe~Ia
with $0.5 \leq z \leq 1.4$ (top panels) and
47 SNe~Ia with $0.02 \leq z \leq 0.1$ (bottom
panels), taken from the data of \citet{riess04b}. 
The distribution of residuals of the low-$z$ SNe~Ia from current data is
consistent with a Gaussian (in both flux and magnitude), while the high-$z$
SNe~Ia seem to show both signatures of weak lensing
(high magnification tail and demagnification shift of the 
peak to smaller flux).

With hundreds of thousands of low- to medium-redshift 
SNe~Ia from the LSST, and the thousands of high-redshift SNe~Ia
from the Joint Dark Energy Mission (JDEM), the error bars in the measured \snia peak brightness
distributions at low, medium, and high redshifts will shrink by 
1-2 orders of magnitude compared to the current data.
This will enable us to rigorously study weak lensing effects
on the SN~Ia peak brightness distribution and derive
parameters that characterize $p(L_{SN})$ and $p(\mu)$.
Since SNe Ia are lensed by the foreground matter distribution, 
the large scale structure traced by galaxies in the foreground
can be used to predict $p(\mu)$ directly, allowing us to
cross-correlate
with the $p(\mu)$ derived from the measured \snia brightness 
distributions.

\begin{figure}
\begin{center}
\includegraphics[width=3in]{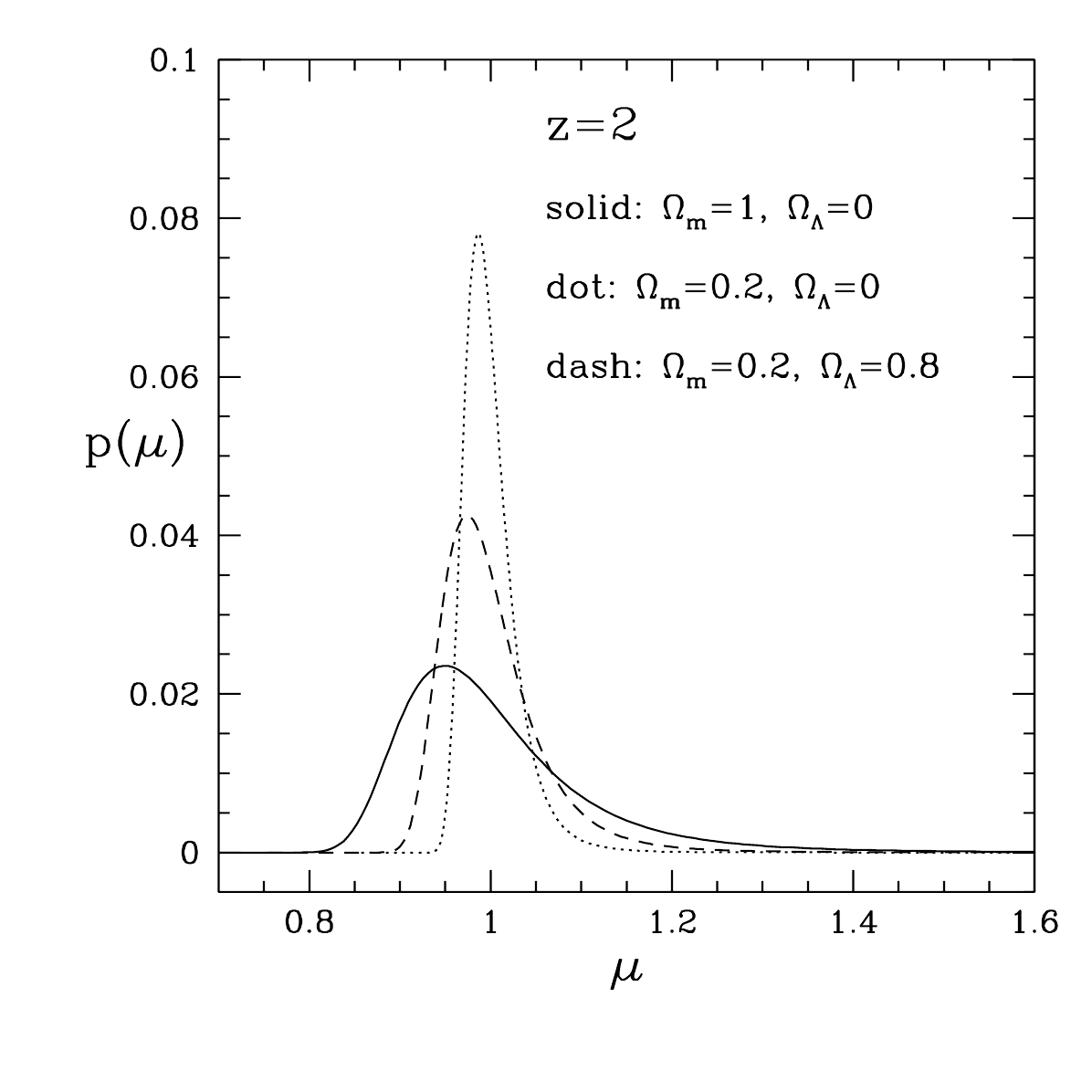}
\caption[Magnification distribution of point sources]{Magnification distributions of point sources for three
different cosmological models at $z=2$. Used with permission from
\citet{wang99}.
}
\label{fig:sn:pmu}
\end{center}
\end{figure}

The measured $p(\mu)$ is a probe of cosmology, since it
is sensitive to the cosmological parameters 
(\autoref{fig:sn:pmu}) \citep{wang99}.
Thus the weak lensing of SNe Ia can be used to tighten constraints on 
cosmological parameters, and cross check the dark energy constraints 
from other LSST data \citep{wang99,cooray06,dodelson06}.

%
%
%

%

\section{Core-Collapse Supernovae}
\label{sec:sn:cc_rate}
\noindent{\it Amy Lien, Brian D. Fields} 


The LSST will discover nearly as many Type II supernovae as Type Ia
supernovae (\autoref{sec:sn:cc_contamination})
and will similarly obtain finely-sampled light curves in many colors. 
Core-collapse and Type Ia supernovae share very similar
observational properties (light curve histories,
maximum brightness), and thus the LSST strategies for
optimizing Type Ia discovery will automatically discover
an enormous number of core-collapse events.
Indeed, the LSST will harvest core-collapse supernovae
in numbers orders of magnitude greater than have ever been observed to date.

Most \sneii can be distinguished from other types of SN by the
duration and color evolution of their light curves. The supernova
rates themselves, together with photometric redshifts which the LSST
will obtain of their host galaxies (\autoref{sec:common:photo-z}), will be a direct measure of the
star formation history of the Universe. Late-time light curves will
provide a direct measure of type II supernova $^{56}$Ni (and hence
iron) yields.  The amount of iron which is released in the supernova
explosion depends sensitively on the fraction of the total produced by
explosive burning in the silicon shell that falls back into the
compact object at the center.  The watershed mass coordinate dividing
what falls back from what escapes (the so-called ``mass cut'') can be
measured from the $^{56}$Ni yield, and is crucial for our
understanding of cosmic chemical evolution of iron-group elements and
the mass function of compact remnants.

The survey will map out the cosmic core-collapse supernova redshift
distribution via direct {\em counting}, with very small statistical
uncertainties out to a redshift depth that is a strong function of the
survey limiting magnitude (\autoref{sec:sn:cc_contamination}; see also
\citealt{lien09}).  
Over all redshifts, the total annual
harvest of core-collapse supernovae with one or more photometric
points is predicted by \citet{lien09} to be $\sim 3 \times 10^5$
events to $r = 23$.   

The core-collapse supernova redshift history encodes rich information about
cosmology, star formation, and supernova astrophysics and phenomenology;
the large statistics of the supernova sample will be crucial to
disentangle possible degeneracies among these issues.
For example, the cosmic supernova rate
can be measured to high precision 
out to $z \sim 0.5$ for all core-collapse types,
and out to redshift $z \sim 1$ for Type IIn events
if their intrinsic properties remain the same
as those measured locally. \citet{lien09} showed that in a single year
of observation, LSST will determine the cosmic core-collapse supernova
rate to an accuracy of 10\% to $z \sim 0.9$. 

A precise knowledge of the cosmic supernova rate would remove the
cosmological uncertainties in the study of the wealth of observable
properties of the cosmic supernova populations and their evolution
with environment and redshift.  Because of the tight link between
supernovae and star formation, synoptic sky surveys will also provide
precision measurements of the normalization and $z \leq 1$ history of
cosmic star-formation rate in a manner independent of, and
complementary to, current data based on UV and other proxies for
massive star formation.

Furthermore, Type II supernovae can serve as distance
indicators and would independently cross-check
Type Ia distances measured in the same surveys (\autoref{sec:sn:sneii}).
Arguably the largest and least-controlled uncertainty in all of these
efforts comes from the poorly understood evolution of
dust obscuration of supernovae in their host galaxies;
\citet{lien09} outline a strategy to determine empirically the obscuration
properties by leveraging large supernova samples over
a broad range of redshift.

\section{Measuring Distances to Type IIP Supernovae}
\label{sec:sn:sneii}

\noindent{\it Mario Hamuy} 


The subclass of
Type II plateau supernovae can be used as distance
indicators in a manner complementary to \sneia, although to
smaller redshifts due their fainter intrinsic luminosities. The method
is called the Expanding Photosphere Method (EPM)
\citep{schmidt94a,hamuy01,jones09}, and relies on the fact that the
velocity of expansion of the photosphere (as measured from emission
lines in the supernova spectrum) determines the size, and thus
luminosity, of the photosphere.  This technique
needs at least two photometric observations over the first 50 days since
discovery, using two optical filters (optimally in the $g-i$
range), as well as at least two spectroscopic
observations contemporaneous to the photometric data in
order to determine the photospheric expansion velocity as a function of
time.  Thus such work will require extensive access to 8-10-meter class
telescopes with spectrographs. The calibration of the EPM is based on
theoretical atmosphere models. Systematic differences in the two model
sets available to date \citep{eastman96,dessart05} lead to 50\%
differences in the EPM distances. However, once corrected to a common
zero-point, both models produce relative distances with a 12\% scatter
\citep{jones09}, which reflects the internal precision of this
technique. This is somewhat higher than the 7-10\% internal precision that
characterizes the techniques based on Type Ia supernovae. Since only
relative distances are required for the determination of cosmological
parameters, the thousands of Type II supernovae that LSST will discover will enable a completely independent determination of cosmological
parameters.

Type II plateau supernovae can also be used for the determination of
distances using the SCM (Standardized Candle Method), which does not
require a theoretical calibration
\citep{hamuy02,nugent06,olivares07,poznanski09}. This technique, which
relies on an empirical correlation between expansion velocity and peak
luminosity, 
requires observations through two filters (e.g., $r$ and $i$), at least on
two epochs toward the end of the plateau phase \citep{olivares07}.  Because 
this technique is based on an empirical luminosity-velocity relation, a
minimum of one spectroscopic observation is needed (preferentially
contemporaneous to the photometric data), although two spectra would be
desirable. This method yields relative distances with a precision of
10-15\% \citep{olivares07,poznanski09}, thus offering a independent
route to cosmological parameters.

While EPM employs early-time data and SCM requires late-time
observations, the two techniques are independent of one another.  Thus
two independent and complementary Hubble diagrams will be produced
from the same data set of Type II plateau events. Since these objects
are very different from \sneia in their explosion physics and
progenitors, these data will provide a valuable assessment of the
potential systematic errors that may affect the distances obtained
from \sneia.

Optical observations covering the first 100 days of evolution of the Type
II plateau events will provide information to estimate their bolometric luminosities
\citep{bersten09}, plateau lengths, and luminosity function. Through a
comparison with hydrodynamic models \citep{litvinova85,utrobin07},
these observables can be converted into physical parameters
such as explosion energy and progenitor mass. This information will
provide important advances in our understanding of the progenitor stars
that produce these supernovae and their explosion mechanisms.


\section{Probing the History of SN Light using Light Echoes}
\label{sec:sn:light_echoes}

\noindent{\it Jeonghee Rho} 

The light from supernovae can be visible as scattered-light echoes centuries
after the explosion, whereby light from the supernova (in our own
Milky Way or nearby galaxies) scatters off
interstellar dust. These are identified in wide-field difference
imaging on timescales over which the supernova evolves (weeks).  This
has been used to identify the type of supernovae associated with
supernova remnants.
There are a few light echo measurements that have been carried out of
LMC supernova remnants \citep{rest05}, and light curve constructions using light
echoes have been done for a few objects in nearby galaxies
\citep{rest08}. Searching for light echoes from historic Galactic SNe
has been challenging because of the need for repeated deep wide-angle
imaging. Echoes of Galactic supernova remnants were first found in the infrared in
Cas A \citep{krause05} with the Spitzer Space Telescope.  Here the
infrared ``echo'' is the result of dust absorbing the SN outburst
light, being heated and then re-radiating at longer wavelengths.
Optical follow-up observations revealed the directly scattered light
echo of Cas A \citep{rest08}.  

LSST offers excellent opportunities to find the structures and
evolution of light echoes of supernova remnants both in the Milky Way
and in nearby galaxies. The structures of echoes change on timescales
of days, months, and years, allowing one to construct accurate light
curves and to constrain the properties of the progenitors.

Constraints on the light curves and accurate masses of progenitors of young
supernova remnants are important for understanding nucleosynthesis  and dust
formation in SNe.  Many species of nucleosynthetic yields and dust emission are
more easily observable in supernova remnants than in supernovae, because after
the reverse shock encountered by the ejecta, both the ejecta and dust are 
sufficiently heated to emit in both optical and infrared wavelengths.


%

\section{Pair-Production SNe}
\label{sec:sn:ppsne}
\noindent{\it Evan Scannapieco, David Arnett} 

Pair-production supernovae (PPSNe; \autoref{sec:tr:pair_sn}) are the uniquely calculable result
of non-rotating stars that end their lives in  the $140$--$260$~M$_{\odot}$ mass
range \citep{heger02}.
Their collapse and explosion result from
an instability that generally occurs whenever the central temperature
and density of a star moves within a well-defined regime  
\citep{barkat67}.
While this instability arises irrespective of
the metallicity of the progenitor star, PPSNe are expected only in 
primordial environments. 
In the present metal-rich Universe, it appears that stars
this massive are never assembled, as supported by a wide range of
observations 
\citep[e.g.][]{figer05, oey05b}.  However, 
molecular hydrogen is a relatively inefficient coolant, so under
primordial conditions the fragmentation of primordial molecular clouds
was likely to have been  biased towards the  formation of stars with very high
masses 
\citep{nakamura99,abel00,schneider02,tan04}.
Indeed, because very massive stars
are only loosely bound and they exhibit large line-driven winds which
scale with metallicity as $Z^{1/2}$ or faster  
\citep{vink01,kudritzki02},
$140$--$260$~M$_{\odot}$ mass stars would quickly shed a large fraction
of  their gas unless they were extremely metal poor.

\begin{figure}
\begin{center}
\includegraphics[width=\columnwidth]{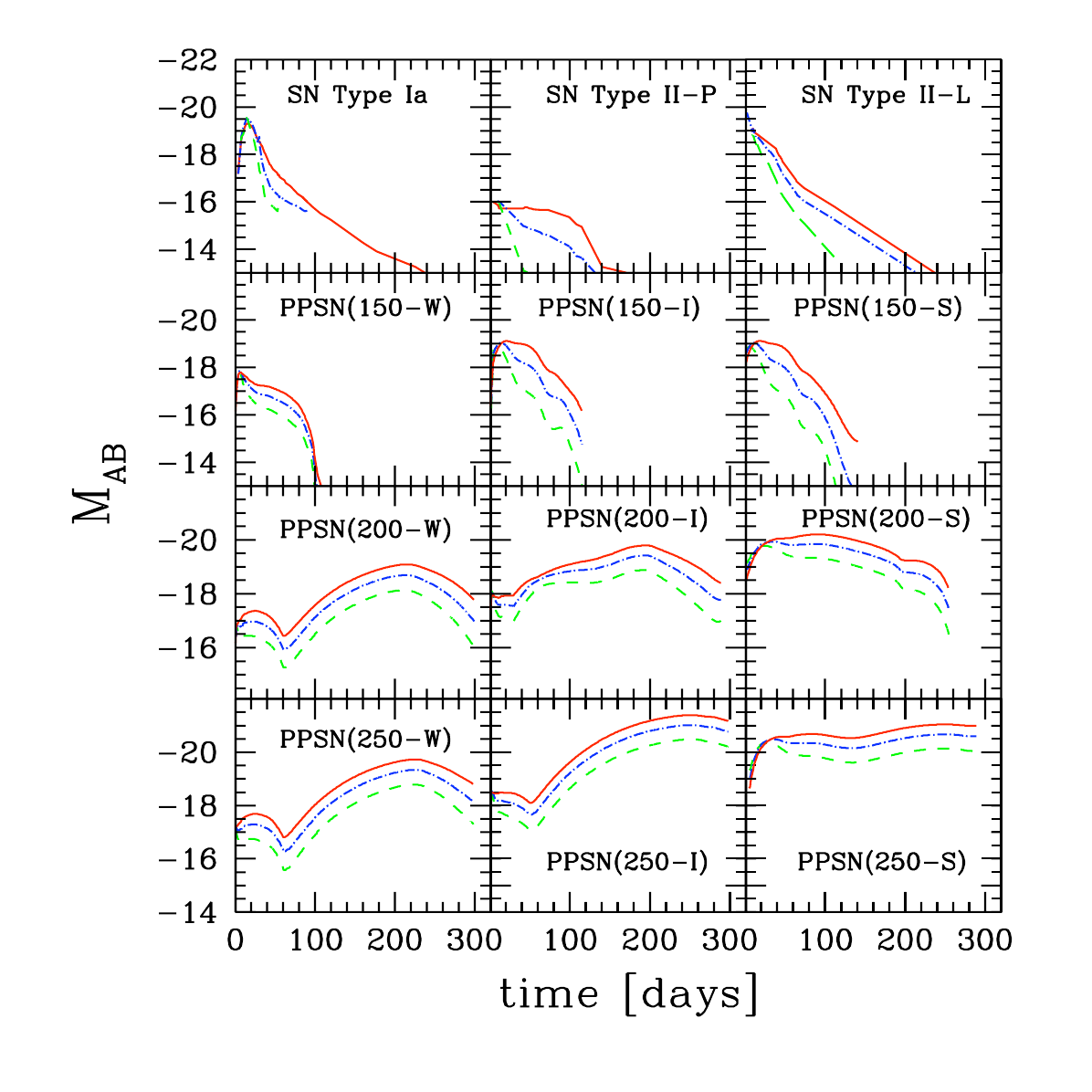}
\caption[Light curve comparison of Type Ia, IIp, IIl, and Pair-Production SNe]
{Comparison of light curves of a \snia, a \sniip, a bright \sniil, and
  PPSNe models with varying progenitor masses and levels of dredge-up.
  The models are labeled by the level of mixing from the core into
  the envelope (W-weak; I-Intermediate; S-Strong) and the mass of the
  progenitor star (150, 200, and 250 M$_\odot$).     In all cases  the
  solid lines are absolute V-band AB  magnitudes, the dot-dashed lines
  are the absolute B-band AB magnitudes, and  the dashed lines are the
  absolute U-band AB magnitudes.   In  general, less mixing leads to
  more $^{56}$Ni production, which makes  the SNe brighter at late
  times, while more mixing expands the envelope, which makes the SNe
  brighter at early times.  Peak brightness also increases strongly
  with progenitor mass.  Figure from \citet{scannapieco05b}, with
  permission.} 
\label{fig:sn:ppsn_lcs}
\end{center}
\end{figure}

\citet{scannapieco05b} calculated approximate PPSNe light curves,
varying parameters to blanket the range of theoretical uncertainties
and possible progenitor masses.  These are shown in \autoref{fig:sn:ppsn_lcs}, in
which they are compared with \snia and core-collapse SN light curves.
Despite enormous kinetic energies of  $\sim 50 \times
10^{51}$ ergs,  the peak optical luminosities of PPSNe are similar to
those of other SNe, even falling below the luminosities of \sneia and \sneii in many cases.
This is because the higher ejecta mass produces a large optical depth
and most of the internal energy of the gas is converted into kinetic
energy by adiabatic expansion \citep[see, e.g.,][]{arnett82}.  The colors of the  PPSN
curves are also similar those to more usual SNe.

\begin{figure}
\begin{center}
\includegraphics[width=0.85\columnwidth]{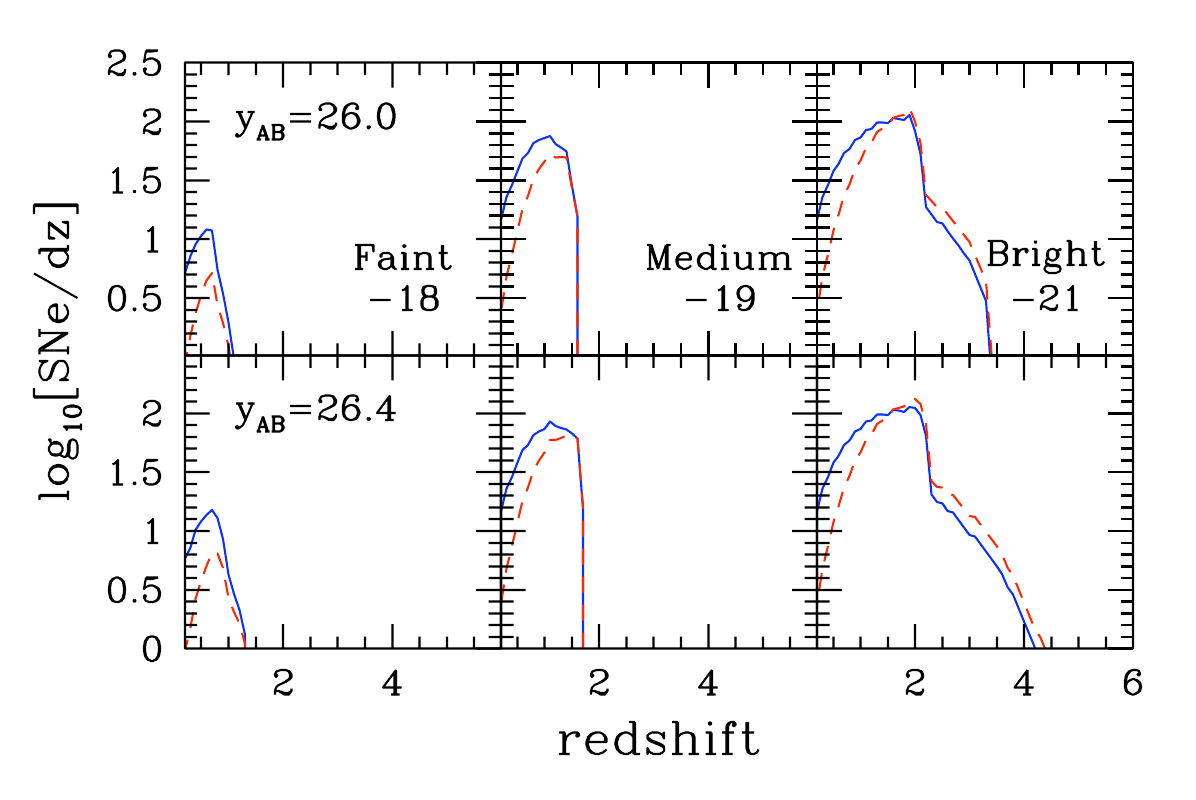}
\includegraphics[width=0.85\columnwidth]{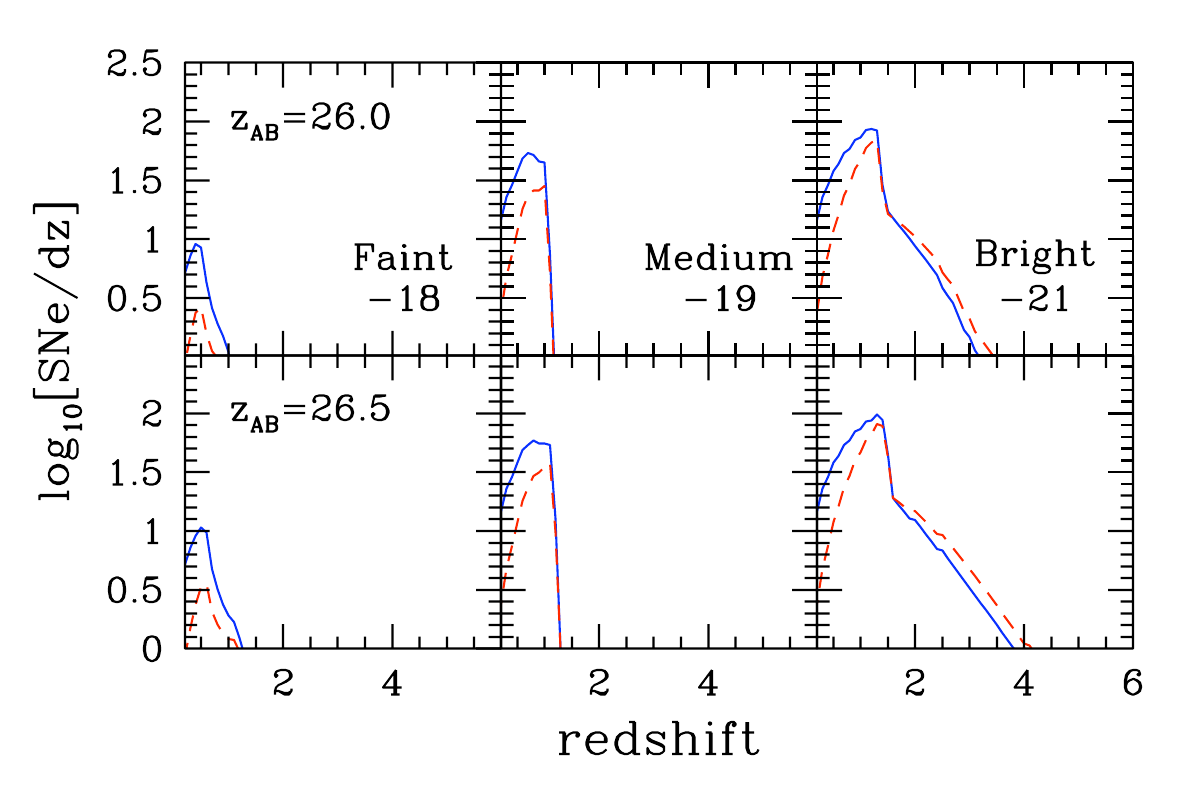}
\caption[Visibility of pair-production SNe in LSST]{ Number of PPSNe per unit redshift observable by  
LSST in $y$ (top) and $z$ (bottom) per unit redshift in a single 9.6 deg$^2$ field. Lines show models  
in which metal-free star formation occurs at a rate of 1\% of the  
overall star-formation rate (red lines) or at a fixed rate of 0.001  
$M_\odot\,\rm yr^{-1}\,Mpc^{-3}$ (blue lines), assuming one PPSN per
$1000\,M_\odot$ of metal-free stars formed.}
\label{fig:sn:ppsn_visibility}
\end{center}
\end{figure}

Thus distinguishing PPSNe from other SNe will require multiple
observations that constrain the time evolution of these objects. In
particular, there are two key features that are uniquely
characteristic to PPSNe. The first is a dramatically extended
intrinsic decay time, which is especially noticeable in the models
with the strongest enrichment of CNO in the envelope. This is due to
the long adiabatic cooling times of supergiant progenitors, whose
radii are $\sim 20$ AU, but whose expansion velocities are similar
to, or even less than, those of other SNe.  Second, PPSNe are the only
objects that show an extremely late rise at times $\geq 100$ days.
This is due to energy released by the decay of $^{56}$Co, which unlike
in the \snia case, takes months to dominate over the internal energy
imparted by the initial shock.

Such constraints will require an extremely long cadence, roughly 100
days in the rest frame, or $\sim1$ year for SNe at theoretically
interesting redshifts $\geq 1.$ While the very faintest PPSNe, such as
the 150 $M_\odot$ models in \autoref{fig:sn:ppsn_lcs} cannot be
meaningfully constrained by LSST, co-adding the $\sim16$ images
taken of each patch of sky each year in the $z$ band will place
exquisite constraints on 200 $M_\odot$ and 250 $M_\odot$ progenitor
models.  Indeed observations down to $z=26.0$ covering 16,000 deg$^2$
will be able to detect thousands of $200-250 M_\odot$ PPSN if very
massive metal-free stars make up even 0.01\% of the stars formed at a
redshift of 2, well within the range of theoretical uncertainties
\citep{scannapieco03,jimenez06,tornatore07}.
Even if very massive metal-free star formation does not occur below $z
= 2,$ hundreds of $z=2-4$ PPSNe will be detected by LSST
(\autoref{fig:sn:ppsn_visibility}).  At the same time such
long-cadence studies will turn up large numbers of long duration SNe, such as the extremely bright SN~2006gy \citep{sn2006gy}, which, while not
likely to be of primordial origin, nevertheless will provide unique
probes into extreme events in stellar evolution (\autoref{sec:TheGap}).


%



\section{Education and Public Outreach with Supernovae}
\label{sec:sn:epo}
\noindent{\it W. Michael Wood-Vasey} 

Supernovae have always fascinated and engaged the public. The great wealth of supernovae that will come from a decade of LSST are an excellent opportunity to share the discoveries and science of LSST with the world.  The hexa-color LSST movie of the sky leads to natural learning opportunities from elementary students through college and life-long learners.  See \autoref{chp:epo} for a general discussion of EPO activities in the context of LSST.  Here we focus on the unique engagement and educational opportunities related to supernova science.

Students can search for and study supernovae in the LSST data.  From
simple exercises in visual comparison, school children will learn that
supernovae rise and fall in brightness and that they are associated
with galaxies.  This level of understanding is the perfect time to
talk about brightness, cooling due to expansion, and radioactivity
(the decay $^{56}$Ni is the dominant source of energy after a week or
two in a supernova).  More advanced college students can learn about
the image differencing, the expansion of the Universe, the life cycle
of stars, and the surface brightness of expanding explosions.  By
measuring the light curve of a supernova, they will learn
about measurement uncertainties and fitting data to empirical and
analytic curves.  Using the brightness of \sneia to measure the
expansion of the Universe has already become a standard lab in
astronomy courses.  With LSST, each student could take their own patch
of the Universe and compare with their classmates to learn about
systematic errors, methods, techniques, and ``global'' measurements.
More advanced opportunities to identify the type of supernovae based
on their light curve properties could be effectively done either as
individual labs or as a Supernova Zoo-type Citizen Science Project
(c.f., the Palomar Transient Factory, \url{http://www.astro.caltech.edu/ptf/} or Galaxy Zoo collaboration, 
\url{http://www.galaxyzoo.org/}) to
benchmark and test the automated transient classification of LSST
while teaching participants about redshift and time-dilation, color,
luminosity, and every astronomer's favorite topic, extinction due to
dust.

The basic scientific investigations that will be one of the important
science topics for supernova science with LSST are extremely
accessible.  Do supernovae come from big galaxies or small?  Are they
close in to the center of galaxies or are they found in intra-cluster
spaces?  These topics will benefit from visual inspection and will
teach basic concepts of sizes, projected distance, angles, as well as
more advanced topics of cosmological distances, galaxy evolution, and
metallicity. 

The participation of the amateur/semi-professional astronomical
community has always been a key aspect of time-domain astronomy.
These opportunities will multiply a thousand-fold in the LSST era, and
integration with robotic and individual telescopes and observing
programs around the world will both share the LSST science with the
world and significantly contribute to a number of the main LSST
supernova science topics.  By thoroughly examining the patterns of
supernovae across the sky, students and the public can learn how
supernovae match the distribution of galaxies and about the structure of
the cosmos.  At the most basic level, the general public and students
will learn how the dramatic deaths of stars throughout the cosmos tells
us about the fundamental nature of our Universe and the elements that
make life possible.

\bibliographystyle{SciBook}
\bibliography{supernovae/supernovae}

\chapter[Strong Lenses]{Strong Gravitational Lenses}
\label{chp:sl}

\noindent {\it 
Phil~Marshall, 
Maru\v{s}a~Brada\v{c}, 
George~Chartas, 
Gregory~Dobler, 
\a'Ard\a'\i s~El\'iasd\'ottir, 
Emilio~Falco, 
Chris~Fassnacht,  
James Jee, 
Charles~Keeton,  
Masamune~Oguri, 
Anthony~Tyson}


%
%
%
%
%
%
%
%
%
%

LSST will contain more strong gravitational lensing events than any
other survey preceding it, and will monitor them all at a cadence of a
few days to a few weeks. Concurrent space-based optical or perhaps
ground-based surveys may provide higher resolution imaging: the biggest
advances in strong lensing science made with LSST will be in those areas
that benefit most from the large volume and the high accuracy,
multi-filter time series. In this chapter we propose an array of science
projects that fit this bill. 

We first provide a brief introduction to the basic physics of gravitational
lensing, focusing on the formation of multiple images: the strong lensing
regime. Further description of lensing phenomena will be provided as they
arise throughout the chapter. We then make some predictions for the properties
of samples of lenses of various kinds we can expect to discover with LSST:
their numbers and distributions in redshift, image separation, and so on. This
is important, since the principal step forward provided by LSST will be one of
lens  sample size, and the extent to which new lensing science projects will
be enabled depends very much on the samples generated. From 
\autoref{sec:sl:galev} onwards we introduce the proposed LSST 
science projects. This is by no means an exhaustive list, but should serve as
a good starting point for investigators looking to exploit the strong lensing
phenomenon with LSST.


\section{Basic Formalism}
\label{sec:sl:basics}

\noindent{\it\a'Ard\a'\i s El\'iasd\'ottir, Christopher D. Fassnacht} 

The phenomenon of strong gravitational lensing, whereby multiple
images of a distant object are produced by a massive foreground object
(hereafter the ``lens''), provides a powerful tool for investigations
of cosmology and galaxy structure.  The 2006 Saas Fee lectures provide an
excellent introduction to the physics of gravitational lenses 
\citep{Sch06,Koc06}; we provide here a short summary of the basics.
An extension of the discussion here with details more relevant to
weak lensing can be found in \autoref{chp:wl}.

\subsection{The Lens Equation}
\begin{figure}
\centering\epsfig{file=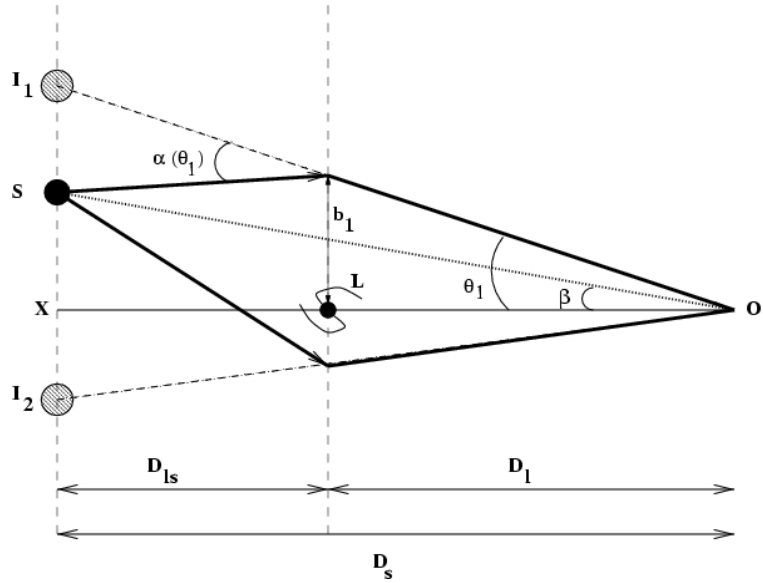,width=0.6\linewidth}
\caption{Cartoon showing lens configuration.  The light coming from the source,
S, is deflected due to the potential of the lensing object, L.
$D_l$ is the angular diameter distance from the observer to the lens,
$D_s$ is the angular diameter distance from the observer to the source
and $D_{ls}$ is the angular diameter distance from the lens to the
source (note that $D_{ls}\ne D_s-D_l$).  Due to the lensing effect, an
observer O sees two images I$_1$ and I$_2$ of the original background
source.  The images are viewed at an angle $\theta$ which differs from
the original angle $\beta$ by the reduced deflection angle
$\hat{\alpha}(\theta) = (\Dds / \Ds ) \alpha(\theta)$.  The magnitude
of the deflection angle $\alpha$ is determined by its impact parameter
($b$) and the distribution of mass in the lensing object.  Figure from
\citet{Fas99}.}
\label{fig:sl:lenscartoon}
\end{figure}

The geometrical configuration of the lensing setup is most simply expressed in
terms of angular diametric distances, which are defined so that  ``normal"
Euclidean distance-angle relationships hold.  From
\autoref{fig:sl:lenscartoon} one sees that  for small angles
\begin{equation}
\alpha\left(\theta\right) \Dds + \beta \Ds = \theta \Ds
\end{equation}
i.e.,
\begin{equation}
 \beta = \theta  - \hat{\alpha}\left(\theta\right)
 \label{eq:lenseq}
\end{equation}
where $\hat{\alpha}\left(\theta\right) \equiv \frac{\Dds}{\Ds}  \alpha\left(\theta\right)$ is the reduced deflection angle.  
\autoref{eq:lenseq} is referred to as the lens equation, and its solutions 
$\theta$ give the angular position of the source as seen by the
observer.  It is in general a non-linear equation and can have
multiple possible solutions of $\theta$ for a given source position
$\beta$.

For a point mass $M$ and perfect alignment between the observer, lens and
source (i.e. ${\bf \beta=0}$), the solution to the lens equation is given by
\begin{equation}
\thetaE=\sqrt{ \frac{\Dds}{\Dd \Ds} \frac{4 G M}{c^2}},
\end{equation}
which defines a ring centered on the lens with angular radius, $\thetaE$,
called the Einstein radius. The Einstein radius defines the angular scale for
a lensing setup, \ie, the typical separation of multiple images for a multiply
imaged background source.  
For an extended mass distribution that has
circular symmetry in its projected surface mass density, the Einstein
ring radius is given by
\begin{equation}
\thetaE=\sqrt{ \frac{\Dds}{\Dd \Ds} \frac{4 G M(\thetaE)}{c^2}},
\end{equation}
where $M(\thetaE)$ is the projected mass contained within a cylinder
of radius $R = \Dd \thetaE$.  For a galaxy, the typical Einstein
radius is of the order of $1\arcsec$, while for galaxy clusters the
Einstein radii are of the order of $10\arcsec$.   This direct
relationship between the Einstein ring radius and the mass of the lensing
object provides a powerful method for making precise measurements
of the masses of distant objects.

Various relations and quantities in lensing can be simplified by expressing
them in terms of the projected gravitational 
potential.  
This potential, $\psi(\theta)$, is just the three-dimensional Newtonian
gravitational potential of the lensing object, $\Phi(r)$, projected onto
the plane of the sky and scaled:
\begin{equation}
\label{chapter1:eq:lenspot}
\psi(\vec{\theta}) = \frac{\Dds}{\Dd \Ds} \frac{2}{c^2} 
 \int \Phi(\Dd \vec{\theta}, z) dz,
\end{equation} 
where $z$ represents the coordinate along the line of sight.
The scaled potential can also be written in terms of the mass surface
density of the lensing object:
\begin{equation}
\label{chapter1:eq:lenspot2}
\psi(\theta)= \frac{1}{\pi} \int \kappa(\theta^{'}) \ln \left| \theta-\theta^{'}\right| \mathrm{d}^2 \theta^{'},
\end{equation} 
where $\kappa$ is the dimensionless surface mass density (or convergence)
\begin{equation}
\label{chapter1:eq:kappa}
\kappa\equiv\Sigma/\Sigma_{\mathrm{crit}},
\end{equation}
and $\Sigma_{\mathrm{crit}}$ is the ``critical surface mass density'' 
defined as 
\begin{equation}
\Sigma_{\mathrm{crit}}=\frac{c^2}{4 \pi G}\frac{\Ds}{\Dd \Dds}.
\label{chapter1:eq:critical}
\end{equation}


The distortion and magnification of the lensed images is given by the
magnification tensor
\begin{equation}
M(\vec{\theta}) = A^{-1}(\vec{\theta}),
\end{equation}
where
\begin{equation}
A(\vec{\theta}) = \frac{\partial \beta}{\partial \vec{\theta} } = \left(\delta_{ij} - \frac{\partial^2\psi(\vec{\theta})}{\partial\theta_i\partial\theta_j}\right)=\left( \begin{array}{cc} 
  		 1-\kappa-\gamma_1 & -\gamma_2 \\
		 -\gamma_2             & 1-\kappa+\gamma_1
                     \end{array}\right)
\end{equation}
is the Jacobian matrix and $\gamma\equiv\gamma_1+\mbox{\bf i}\gamma_2$ is the shear (see also \autoref{chp:wl}).  The shape distortion of the lensed images is described by the shear while the magnification depends on both $\kappa$ and $\gamma$.

\subsection{The Fermat Potential and Time Delays}

Another useful quantity is the Fermat potential, $\tau(\theta;\beta)$,
defined as
\begin{equation}
\tau(\theta;\beta) = \frac{1}{2}\left(\theta-\beta\right)^2 - \psi(\theta),
\label{chapter1:eq:ferpot}
\end{equation}
which is a function of $\theta$ with $\beta$ acting as a parameter.  
The lensed images form at the extrema of $\tau(\theta)$, namely at
values of $\theta$ that satisfy
the lens equation, \autoref{eq:lenseq},  with
the deflection angle $\alpha = \nabla_\theta \psi(\theta)$.

The travel time of the light ray is also affected by gravitational lensing. 
The delay compared to a light ray traveling on a direct path in empty space
is given in terms of the Fermat potential (\autoref{chapter1:eq:ferpot})
and equals
\begin{equation}
\Delta t = \frac{\Dd \Ds}{c \Dds}\left(1+\zd\right)\tau(\theta;\beta) + constant,
\end{equation}
where $\zd$ is the redshift of the lens plane and the constant arises
from the integration of the potential along the travel path.  The
indeterminate constant term means that the time delay for a particular
single image cannot be calculated.  However, the {\em difference} in the
travel time between two images, $A$ and $B$, for a multiple imaged
source can be measured, and is given by
\begin{equation}
\Delta t_{A,B}= \frac{\Dd \Ds}{c \Dds}\left(1+\zd\right)\left(\tau(\theta_A;\beta) - \tau(\theta_B;\beta)\right).
\label{eqn:sl:preamble:timedelay}
\end{equation}
The time delay equation provides a direct link between the
distribution of mass in the lens, which determines $\psi$, and the
time delay, scaled by the factor outside the brackets on the right
hand side of the equation, which is inversely proportional to $H_0$
through the ratio of angular diameter distances:
\begin{equation}
\frac{\Dd \Ds}{\Dds} = \frac{1}{H_0}\ f(\zd,\zs,\Om,\OL).
\end{equation}
The dependence of this ratio on the cosmological world model
($\Om,\OL$) is rather weak, changing by $\sim$10\% over a wide range
of parameter choices.  Thus, if $\Delta t_{A,B}$ can be measured and
observations can constrain $\psi$ for a given lens, the result is a
determination of $H_0$ modulo the choice of ($\Om,\OL$).  Conversely,
if $H_0$ is known independently and $\Delta t_{A,B}$ is measured, the
lens data provide clear information on the mass distribution of the
lensing galaxy.  This simple and elegant approach, developed by
\citet{Ref64} long before the discovery of the first strong lens
system, relies on a time variable background object such as an active
galactic nucleus. LSST opens up the time domain in a way no previous
optical telescope has: many of the most exciting LSST strong lenses
will have variable sources.

\subsection{Effects of the Environment}
\label{sec:sl:basics:env}

Frequently galaxy-scale lenses reside in dense environments and, therefore, it
is necessary to consider not only the lensing effects of the main lensing
object but also that of the environment.  In modeling, these are referred to
as ``external convergence'' and ``external shear.''  

To first order the external convergence can often be taken to be  a constant
over the relevant  area (approximately $1\arcsec$ radius around the main
lensing object for a galaxy-scale lens)  and, therefore, cannot be separated
from the convergence of the main lensing object.  This is due to an effect
called the ``mass-sheet degeneracy,'' which simply states that if $\kappa(
\vec{\theta})$ is a solution to the lensing constraints (image positions) then
$\kappa'( \vec{\theta})=\lambda+(1-\lambda)\kappa( \vec{\theta})$ is  also a
solution (where $\lambda$ is a real number).  The first term is equivalent to
adding a constant convergence, hence the name ``mass-sheet.''
Additional information about the projected mass associated with the lens
environment or density profile must be supplied in order  to break this
degeneracy: in practice, this can come from stellar dynamics, the image time
delays plus an assumed Hubble constant, weak lensing measurements of the
surrounding field, and so on. The external shear does have a non-degenerate
effect on the image configuration, and is frequently needed in lens models to
achieve satisfactory fits to the observational constraints.

%


%
%
%
%
%
%
%
%
%
%
%
%

\section{ Strong Gravitational Lenses in the LSST Survey }
\label{sec:sl:yield}

\noindent{\it Masamune~Oguri, Phil~Marshall}   

\begin{figure*}[!p]
\centering\epsfig{file=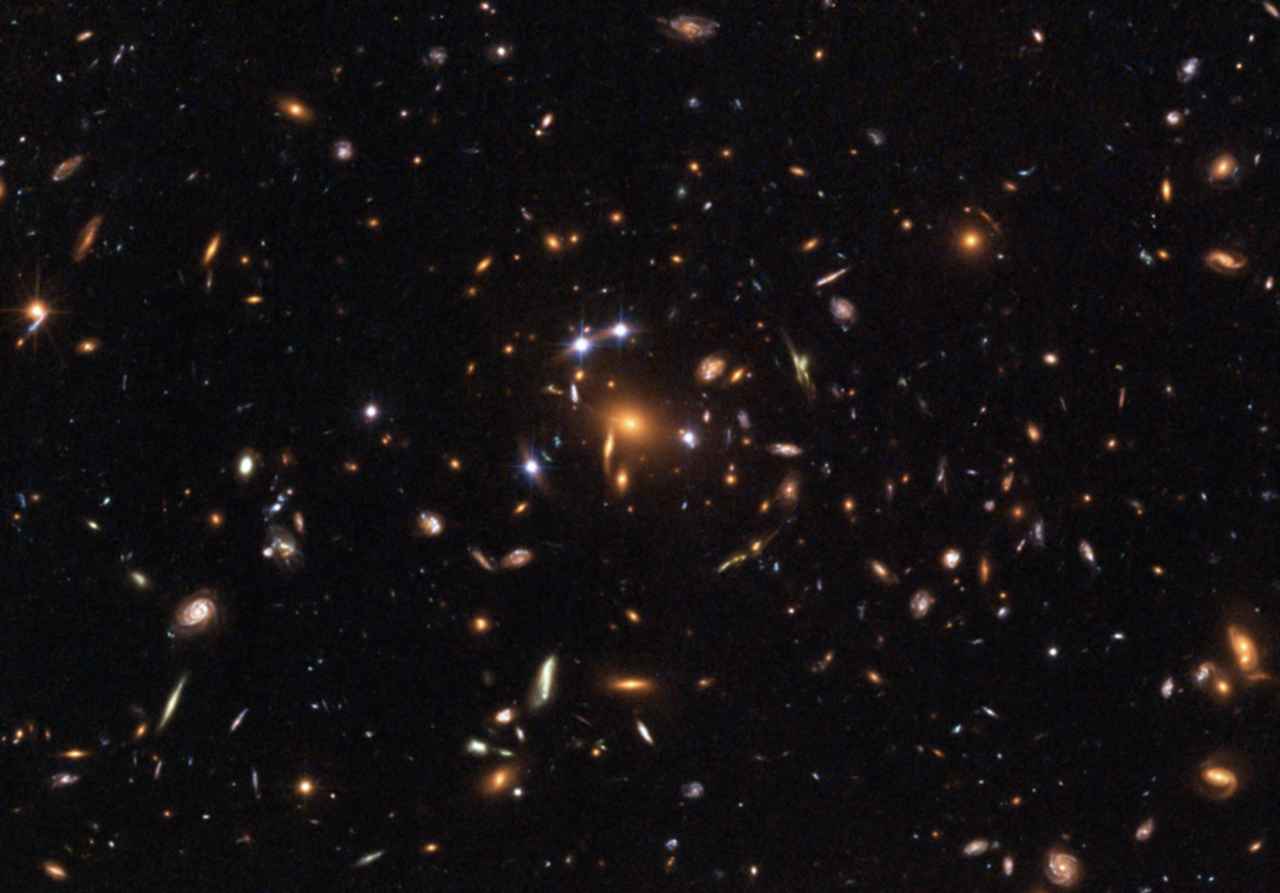,width=0.8\linewidth}
\caption{Lensed quasar SDSS J1004+4112 \citep{Ina++03}. Shown is a
  color-composite HST image from
  \url{http://hubblesite.org/newscenter/archive/releases/2006/23/}. 
  The lens is a cluster of galaxies, giving rise to five images with
  a maximum separation of $15''$. LSST will act as a {\it finder} for
  exotic objects such as this. 
}
\label{stronglens:fig:SDSSJ1004}
\end{figure*}

\begin{figure*}[!p]
\begin{center}
\epsfig{file=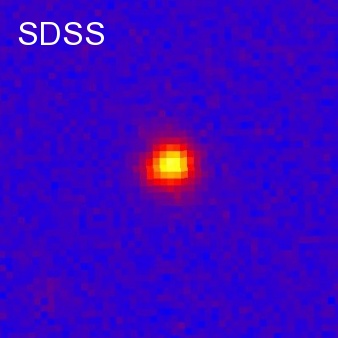,width=0.4\linewidth}
\epsfig{file=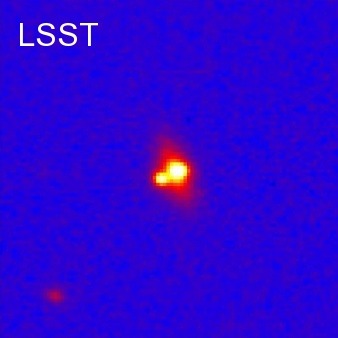,width=0.4\linewidth}
\end{center}
\caption{
A comparison of the images of gravitationally lensed quasars. The 
left panel
shows the image of SDSS J1332+0347 \citep{Mor++07} (a double lens with
a separation of $1.14''$) obtained by the
SDSS (median seeing of $1.4''$), while the right panel shows an image
of the same object taken with Suprime-Cam on Subaru, with seeing of
$0.7''$, comparable to that of LSST. The drastic difference of appearance 
between these two images demonstrates the importance of high spatial
resolution for strong lens searches.  
}
\label{stronglens:fig:sdss1332}
\end{figure*}

LSST will discover an enormous number of strong gravitational lenses,
allowing statistical studies and exploration of rare classes of
lenses not at present possible with the small samples currently known.
In 2009, 
when the first version of this book was published,
strong lenses were still considered to be rather rare objects -- 
in the LSST era
this will no longer be the case. The large number of strong lenses expected to
be found
with LSST suggests that it will also be effective in locating rare,
exotic strong lensing events (\autoref{stronglens:fig:SDSSJ1004}). A
big advantage of LSST will be its excellent image quality. The high
spatial resolution is crucial for strong 
lens searches, as the typical angular scales of strong lensing are quite
comparable to the seeing sizes of ground-based observations (see
\autoref{stronglens:fig:sdss1332}).

In this section we give a brief overview of the samples of strong lenses
we expect to find in the LSST database; these calculations are used in
the individual project sections.  We
organize our projected inventory in order of increasing lens mass,
dividing the galaxy-scale lenses by source type before moving on to
groups and clusters of galaxies. 

\subsection{Galaxy-scale Strong Lenses}
\label{sec:sl:yield:gal}

\begin{figure*}[!t]
\centering\epsfig{file=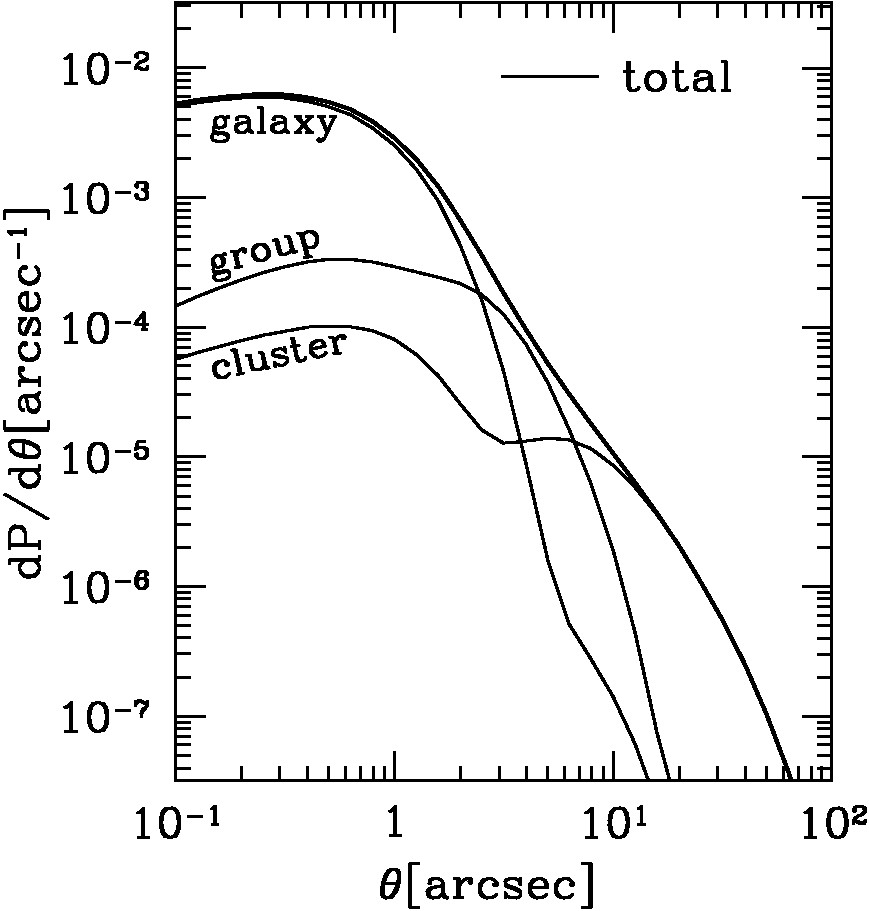,width=0.5\linewidth}
\caption{The distribution of lens image separations (approximately twice the
  Einstein radius) for three different scales: galaxy, group, and
  cluster-scales, predicted by a halo model 
  \citep[figure from][]{Ogu06}. The total 
  distribution is shown by the thick line.}
\label{stronglens:fig:prob_dif}
\end{figure*}

Most of the cross-section for strong gravitational lensing in the
Universe is provided by massive elliptical galaxies \citep{TOG84}.  A
typical object's lensing 
cross-section is a strong function of its mass;
the cross-section of an object is (roughly) proportional to  its central
velocity dispersion to the fourth power. However, the mass function is
steep, and galaxies are far more numerous than the more massive groups
and clusters -- the total lensing optical depth peaks at around
$220\, \kms$. The larger the cross-section of an object, the larger its
Einstein radius; the predicted distribution of strong lens Einstein
radii is shown in \autoref{stronglens:fig:prob_dif}.

A good approximation for computing the lensing rate at galaxy scales is
then to focus on the massive galaxies. 
The observed SDSS velocity 
dispersion function \citep[e.g.,][]{CPV07} gives a measure of the
number density of these objects at low redshift ($z\sim0.1$ or so).
We must expect this mass function to have evolved since redshift 1, but
perhaps not by much -- attempts to use the observed numbers of lenses as
a way of measuring cosmic volume (and hence, primarily, $\OL$), get
answers for the cosmological parameters in agreement with other
cosmographic probes without having to make any such evolution
corrections \citep[e.g., see][]{Mit++05,Ogu++08}.   

The simplest realistic model for a galaxy mass distribution is the
elliptical extension of the singular isothermal sphere $\rho(r)\propto
r^{-2}$, namely the singular isothermal ellipsoid \citep[\eg][]{KSB94}:
\begin{align}
\kappa(\theta) &=
\frac{\thetaE}{2}\frac{1}{\sqrt{(1-e)x^2+(1-e)^{-1}y^2}},\\
\thetaE &=  4\pi\left(\frac{\sigma}{c}\right)^2\frac{D_{\rm
    ls}}{D_{\rm s}}.
\end{align}
This turns out to be remarkably
accurate for massive galaxies that {\it are} acting as strong lenses
\citep[see e.g.,][]{RK05,Koo++06}.
For our model lenses, 
the ellipticity of the lenses is assumed to be distributed as a Gaussian
with mean of $0.3$ and scatter of $0.16$. We also include external
shear, with median of $0.05$ and scatter of 0.2~dex, which is the
level expected in ray-tracing simulations \citep[e.g.,][]{Dal05}. 
The orientation of
the external shear is taken to be random. 

Given the lens ellipticity and external shear distribution, together
with a suitable distribution of background sources, we can now calculate the
expected galaxy-scale lens abundance. 
%
We consider three types of sources:
faint blue galaxies, quasars and AGN, and supernovae. Of course the
latter two also have host galaxies -- but these may be difficult to
detect in the presence of a bright point source. As we will see, in a
ground-based imaging survey, time-variable point-like sources are the
easiest to detect. 


\subsubsection{Galaxy-Galaxy Strong Lenses}
\label{sec:sl:yield:gal:gg}

\begin{figure*}[!t]
\centering\epsfig{file=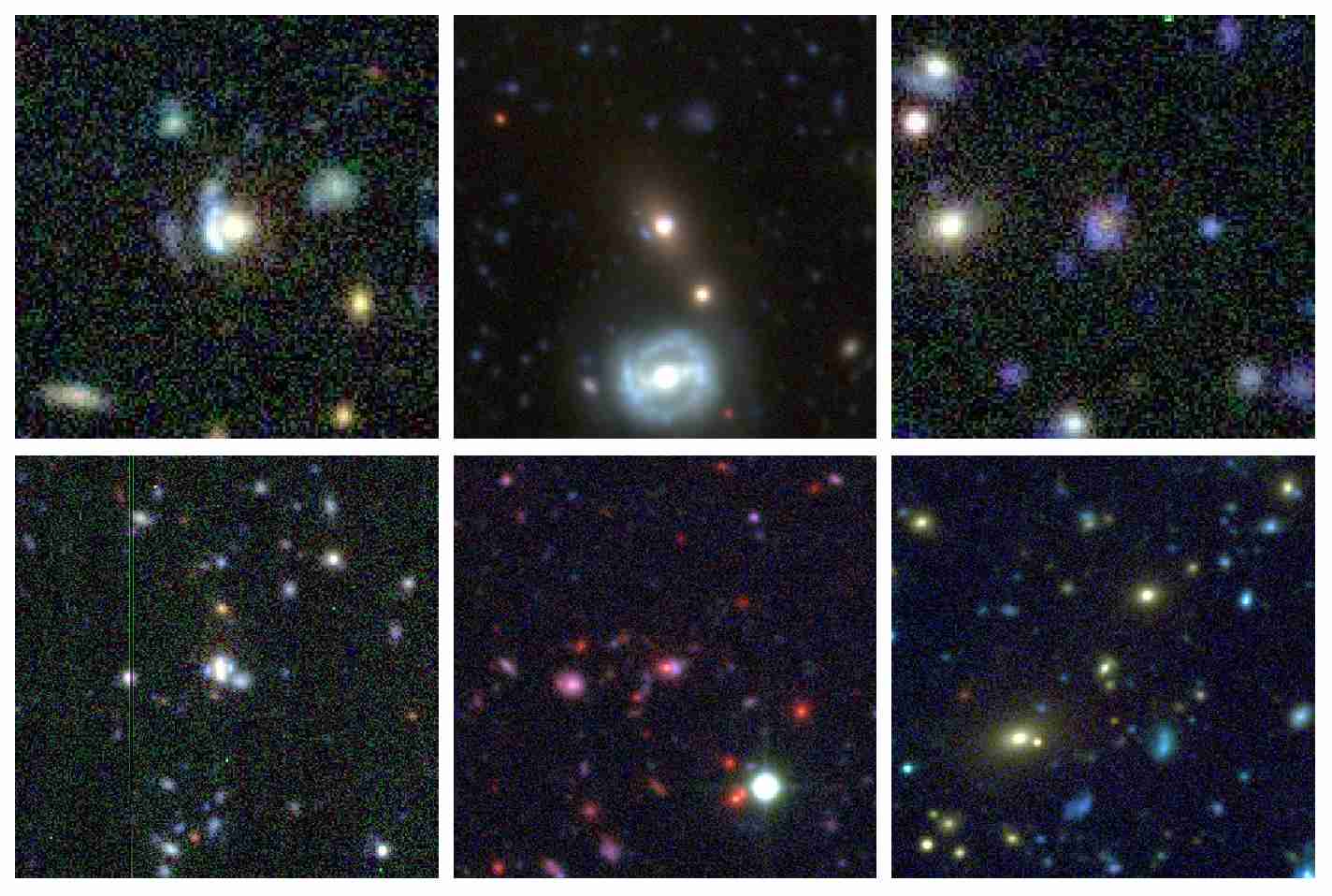,width=0.96\linewidth}
\caption{Galaxy-scale strong lenses detected in the CFHTLS images,
from the sample compiled by Gavazzi et al. (in prep.). All were
confirmed by high resolution imaging with HST. 
Color images kindly
provided by R.\ Cabanac and the SL2S collaboration.}
\label{fig:sl:yield:gglensgallery}
\end{figure*}

We expect the galaxy-scale lens population to be dominated by 
massive elliptical galaxies at redshift 0.5--1.0, whose background
light sources are the ubiquitous faint blue galaxies. The typical
gravitational lens, therefore, looks like a bright red galaxy, with some
residual blue flux around it. The detection of such systems depends on
our ability to distinguish lens light from source light -- this
inevitably means selecting against late-type lens galaxies, whose blue
disks provide considerable confusion. An exception might be edge-on
spirals: the high projected masses make for efficient lenses and 
the resulting cusp-configuration arcs are easily
recognizable.

The SLACS survey has provided the
largest sample of galaxy-scale lenses to date, with almost 100 lenses detected
and measured \citep{Bol++08}: the sources are indeed faint blue galaxies,
selected by their emission lines appearing in the (lower redshift)
SDSS luminous red galaxy spectra. Due to their selection for spectroscopic
observation, the lens galaxies tend to be luminous elliptical galaxies at
around redshift 0.2.
Extending this spectroscopic search to the
SDSS-III ``BOSS'' survey should increase this sample by a factor of two or
more (A.~Bolton, private communication), to cover lens galaxies at somewhat higher
redshift. Optical imaging surveys are beginning to catch up, with various 
HST surveys beginning to provide samples of several tens of lenses
\citep[e.g.,][Marshall et al. in preparation]{Mou++07,Fau++08}. From the ground, the
SL2S survey is finding similar numbers of galaxy-scale lenses in the 
CFHTLS survey area \citep{Cab++07}.

\autoref{fig:sl:yield:gglensgallery} shows a gallery
of galaxy-galaxy lenses detected in the CFHT legacy survey images by
the SL2S project team \citep[][Gavazzi et al. in preparation]{Cab++07}. 
This survey is very well matched to what LSST will provide: the
4 deg$^2$ field is comparable in depth to the LSST 10-year stack,
while the 170 deg$^2$ wide survey is not much deeper than a
single LSST visit.\footnote{The service-mode CFHTLS is quite
uniform, having image quality around $0.9''$ with little scatter. The
``best-seeing'' stack has image quality closer to $0.65''$. With LSST
we expect median seeing of better than $0.7''$
(\autoref{fig:design:seeing}), 
but a broader distribution of PSF
widths.} 

\begin{figure*}[!t]
\centering\epsfig{file=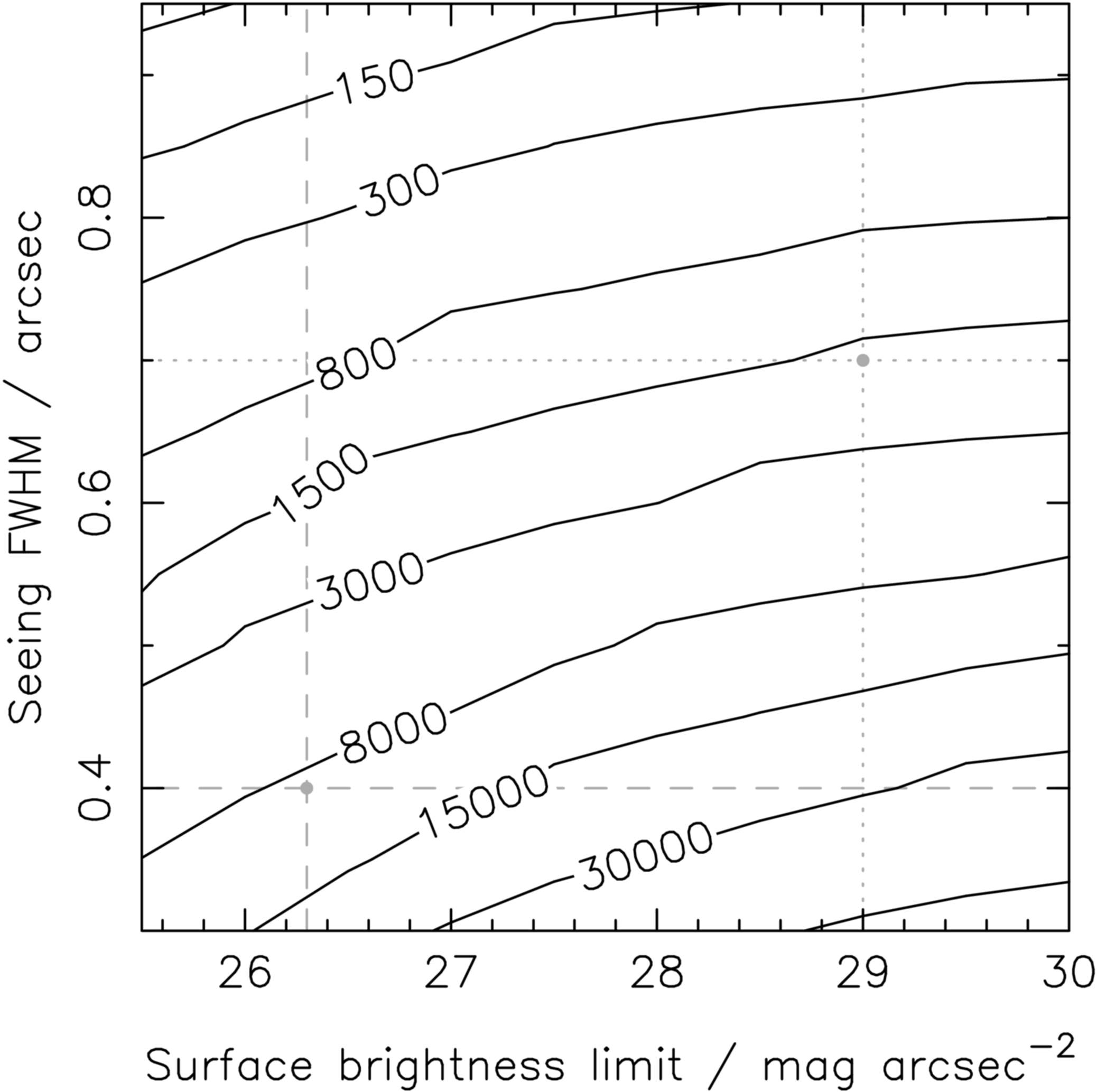,width=0.55\linewidth}
\caption{Number of galaxy-galaxy strong lenses expected in the 20,000
  deg$^2$ LSST survey, as a function of seeing FWHM and surface
  brightness limit. The dashed lines show the approximate expected
  surface brightness limit for one visit and the approximate seeing
  FWHM in the best visit's image. The dotted lines mark the median
  seeing for the survey and the approximate  surface brightness limit
  of the 10-year stacked image. 
}
\label{fig:sl:yield:ggabundance}
\end{figure*}

The SL2S galaxy-scale lens sample contains about 15 confirmed
gravitational lenses or about 0.1 deg$^{-2}$.  Higher
resolution imaging (from HST) was used to confirm the lensing nature
of these objects, at a success rate of about 50\%. The number of
cleanly detected CFHT-only lenses is rather lower, perhaps just a
handful of cases in the whole 170 deg$^2$ survey. This is in
broad agreement with a calculation like that described above, once we
factor in the need to detect the lens features above the sky
background. The results of this calculation are shown in   
\autoref{fig:sl:yield:ggabundance}, which shows clearly how the
detection rate of galaxy-scale strong lenses is a strong function of
image quality. The right-hand dotted cross-hair shows the expected
approximate image quality and surface brightness limit of the 10-year
stack, or equivalently, the Deep CFHTLS fields, and suggests a lens
detection rate of 0.075 per square degree. The left-hand cross-hair
shows the surface brightness limit of a single LSST visit, and the
best expected seeing. By optimizing the image analysis (to capitalize
on both the resolution and the depth) we can
expect to discover $\sim 10^4$ galaxy-galaxy strong lenses in the
10-year 20,000 deg$^2$ LSST survey. The challenge is to make the first cut
efficient: fitting simple models to galaxy images for photometric and
morphological studies will leave residuals that contain information allowing
lensing to be detected, but these residuals will need to be both available and
well-characterized. This information is also required by, for example, those
searching for galaxy mergers (\autoref{sec:gal:merger}).
Given its survey depth, the Dark Energy Survey (DES) should yield a number density of
lenses somewhere in between that of the CFHTLS Wide and Deep fields, and so 
in
its 4000 deg$^2$ survey area DES will discover at least a few hundred
strong galaxy-galaxy
lenses. Again, this number would increase with improved image quality and
analysis.

In \autoref{sec:sl:galev} below we describe the properties of these
lenses, and their application in galaxy evolution science.


\subsubsection{Galaxy-scale Lensed Quasars}
\label{sec:sl:yield:gal:qsos}

Galaxy-scale lensed quasars were the first type of strong
lensing to be discovered \citep{WCW79};
the state of the art in lensed quasar searching is the SDSS quasar lens survey
\citep{Ogu++06}, which has (so far) found 32 new lensed quasars and
rediscovered 13 more. 
The bright and compact
nature of quasars makes it relatively easy to locate and characterize such
strong lens systems (see \autoref{stronglens:fig:sdss1332} for an
example). An advantage of lensed quasars is that the sources are very often
variable: measurable time
delays between images provide unique information on both the lens
potential and cosmology (see \autoref{sec:sl:H0} and
\autoref{sec:sl:H0stat}). We expect that lensed quasars in LSST will be
most readily detected using their time variability \citep{Koc++06}.  
In \autoref{sec:sl:agn} we also discuss using strong lenses to provide a
magnified view of AGN and their host galaxies.

\begin{figure}[!t]
\begin{center}
\includegraphics[width=0.48\linewidth]{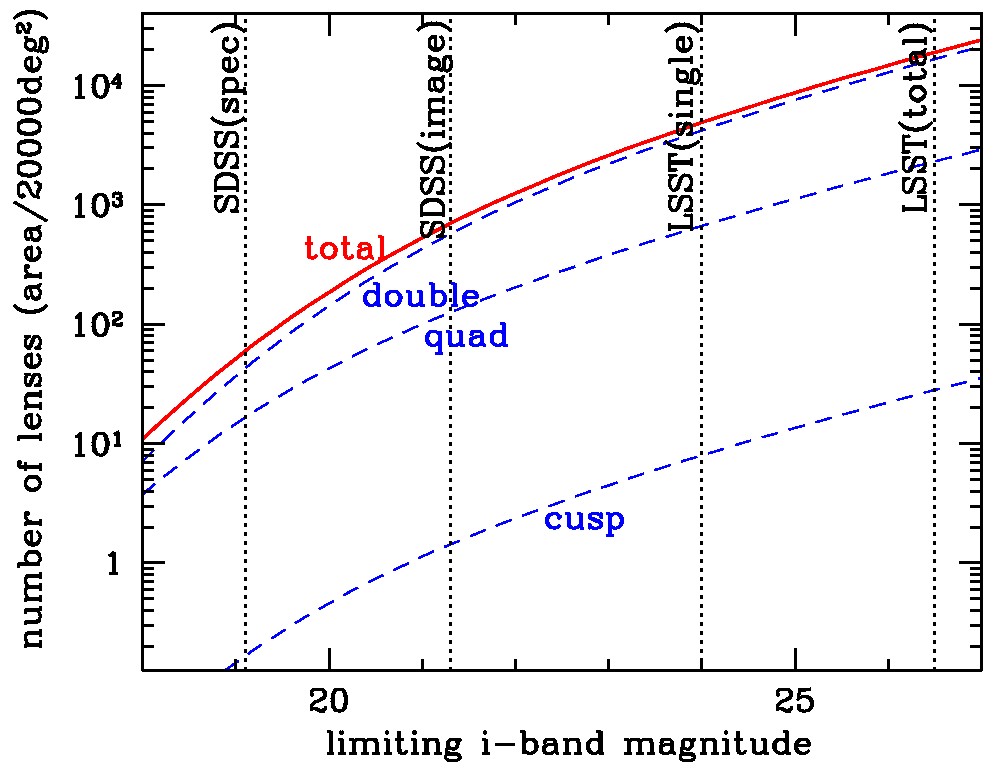}
\includegraphics[width=0.48\linewidth]{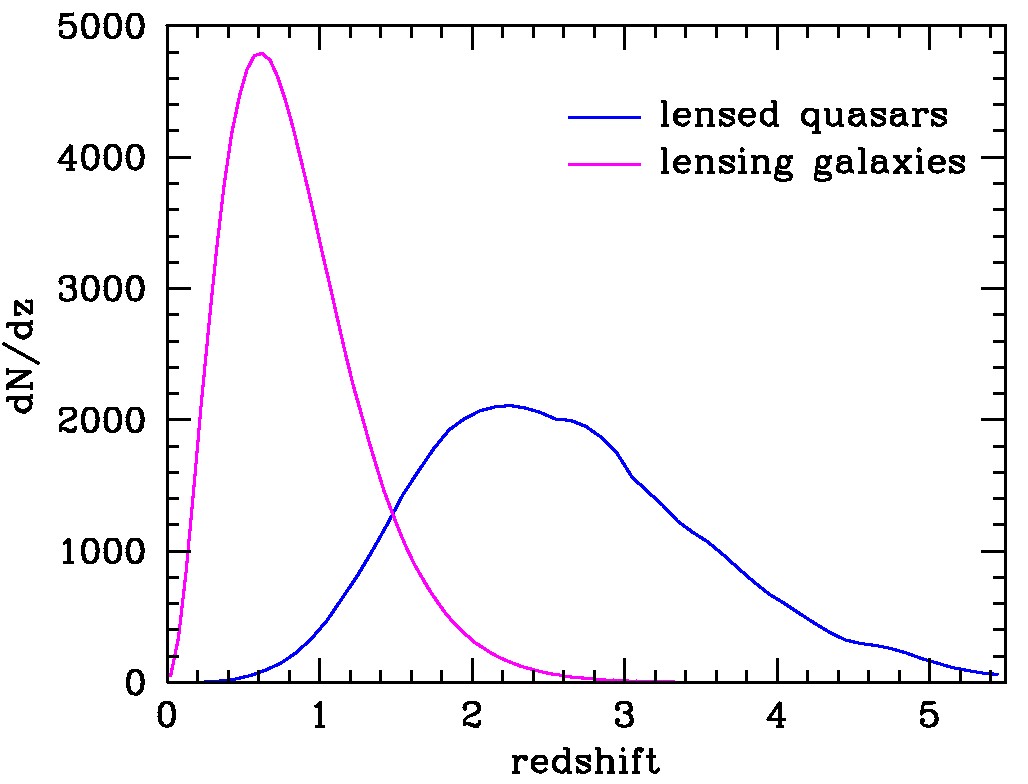}
\end{center}
\caption{The number of lensed quasars expected in LSST. {\it Left:}
  The number of lenses in 20,000~deg$^2$ region as a function of
  $i$-band limiting magnitude. Dashed lines are numbers of lenses for
  each image multiplicity (double, quadruple, and three-image naked cusp
  lenses). The total number is indicated by the red solid
  line. Limiting magnitudes in SDSS and LSST are shown by vertical
  dotted lines. {\it Right:} The redshift distributions of lensed
  quasars ({\it blue}) and lensing galaxies ({\it magenta}) adopting
  $i_{\rm lim}=24$.}
\label{fig:sl:yield:qso}
\end{figure}

We compute the expected number of lensed quasars in LSST as follows. 
We first construct a model quasar luminosity function of
double-power-law form, fit to the
SDSS results of \citet{Ric++06}, assuming the form of
the luminosity evolution proposed by \citet{MHR99}. 
To take LSST observable limits into account, we reject lenses with
image separation $\theta < 0.5''$, and only include those lenses whose
fainter images  
(the third brightest images for quads) have 
$i<24.0$.  Thus these objects will be detectable in each visit, and
thus recognizable by their variability.  This will also allow us to
measure time delays in these objects. 

\autoref{fig:sl:yield:qso} shows the number of lensed quasars expected in
LSST as a function of limiting magnitude.  We expect to find $\sim
2600$ well-measured lensed quasars.  Thus the 
LSST lensed quasar sample will be nearly two orders of magnitude larger than
the current largest survey of lensed quasars.
There are expected to be as many as
$\sim10^3$ lensed quasars detectable in the PS1 $3\pi$ survey, but these will
have only sparsely-sampled light curves (six epochs per filter in
three years). The 4000 deg$^2$ DES should also contain $\sim500$
lensed quasars, but with 
no time variability information to aid detection or to provide
image time delay information.

The calculation above also predicts the distribution of 
image multiplicity. In general, the number of
quadruple lenses decreases with increasing limiting magnitude, because
the magnification bias becomes smaller for fainter quasars. For the LSST
quasar lens sample, the fraction of quadruple lenses is predicted to
be $\sim 14$\%. The lensed quasars are typically located at $z\sim
2-3$, where the space density of luminous quasars also peaks
\citep[e.g.,][]{Ric++06}. The lensing galaxies are typically at $z\sim
0.6$, but a significant fraction of lensing is produced by galaxies at
$z>1$. We discuss this in the context of galaxy evolution studies in 
\autoref{sec:sl:galev}.


\subsubsection{Galaxy-scale Lensed Supernovae}
\label{sec:sl:yield:gal:sne}

Strongly lensed supernovae (SNe) 
will  provide
accurate estimates of time delays between images, because we have an a
priori understanding of their light curves. Furthermore, the SNe fade, 
allowing us to study the structure of the lensing galaxies in great
detail. 

\begin{figure}[t]
\begin{center}
\includegraphics[width=0.48\linewidth]{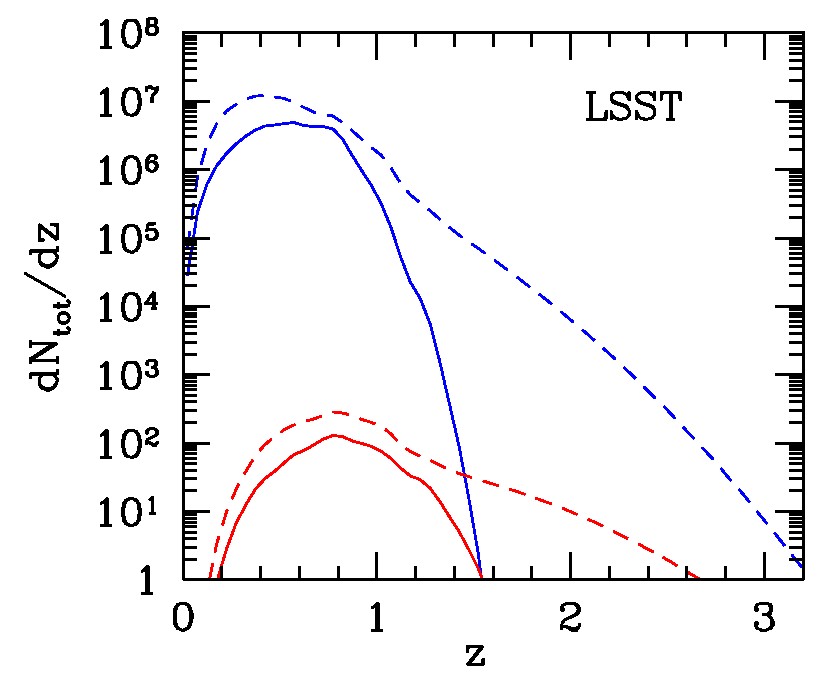}
\includegraphics[width=0.48\linewidth]{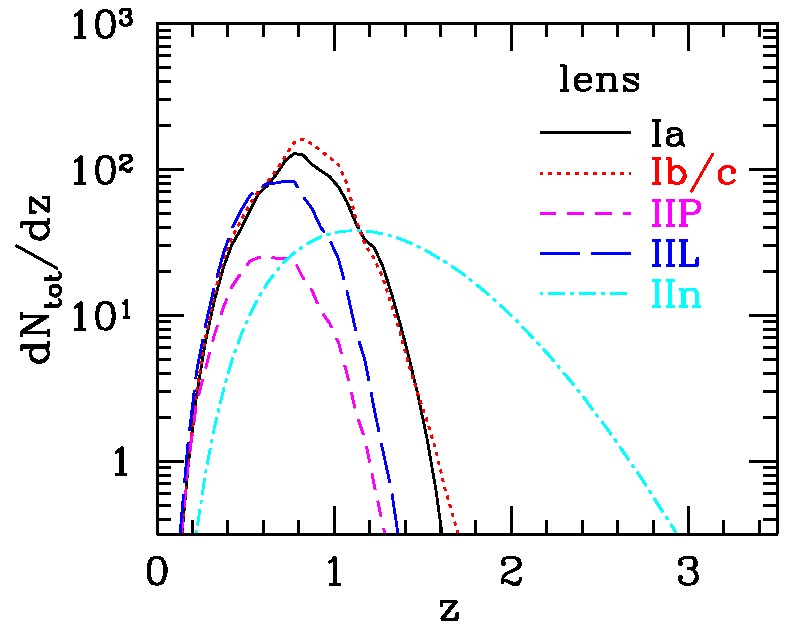}
\end{center}
\caption{The abundance of expected lensed SNe observed with the 10-year
  LSST survey as a function of the redshift. The left panel shows
  the unlensed population in blue, for comparison. Dashed curves show
  core-collapse SNe, solid curves type Ia SNe. In the right panel, the
  contribution of each SN type to the number of lensed SNe is shown.} 
\label{fig:sl:yield:sne}
\end{figure}

We calculate the number of lensed SNe as follows. First we adopt the
star formation rate from \citet[][assuming the initial mass function
  of \citealt{B+G03}]{H+B06}. The Ia rate is then 
computed from the sum of ``prompt'' and ``delay'' components,
following \citet[][see the discussion in
  \autoref{sec:sn:snia_rates}]{Sul++06}. The core-collapse supernova (SN) rate is
assumed to be 
simply proportional to the star-formation rate \citep{H+B06}. For the
relative rate of the different types of core-collapse supernovae
(Ib/c, IIP, IIL, IIn), we use the compilation in \citet{O+T05}. The 
luminosity functions of these SNe are assumed to be Gaussians (in
magnitude) with different means and scatters, which we take from
\citet{O+T05}. For lensed SNe to be detected by LSST, we assume that
the $i$-band peak magnitude of the fainter image must be brighter than
$i=23.3$, which is a conservative approximation of the $10\,\sigma$ point
source detection limit for a single visit.  We insist the image
separation has to be larger than $0.5''$ for a clean identification,
arguing that the centroiding of the lens galaxy, and first image, will
be good enough that the appearance of the second image will be a
significantly strong trigger to justify confirmation follow-up of some
sort. We return to the issues of detecting and following-up lensed SNe
in \autoref{sec:sl:H0stat}.  We assume that each patch of the sky is
well sampled for three months during each year; thus for a 10-year LSST
survey the effective total monitoring duration of the SN search is 2.5
years. 
The right-hand panel of \autoref{fig:sl:yield:sne} shows the expected
total number of strongly lensed SNe in the 10-year LSST survey, as a
function of redshift, compared to their parent SN distribution.
It is predicted that $330$ lensed SNe will be discovered in total, $90$
of which are type Ia and $240$ are core-collapse SNe. Redshifts of
lensed SNe are typically $\sim 0.8$, while the lenses will primarily be
massive elliptical galaxies at $z \sim 0.2$. Similar distributions may be
expected for DES and PS1 prior to LSST -- but the numbers will be far smaller
due to the lower resolution and depth (PS1) or lack of cadence (DES). While we
may expect the first discovery of a strongly lensed supernova to occur prior
to LSST, they will be studied on an industrial scale with LSST. 


\subsection{Strong Lensing by Groups}
\label{sec:sl:yield:groups}

\begin{figure}[!t]
\centering\epsfig{file=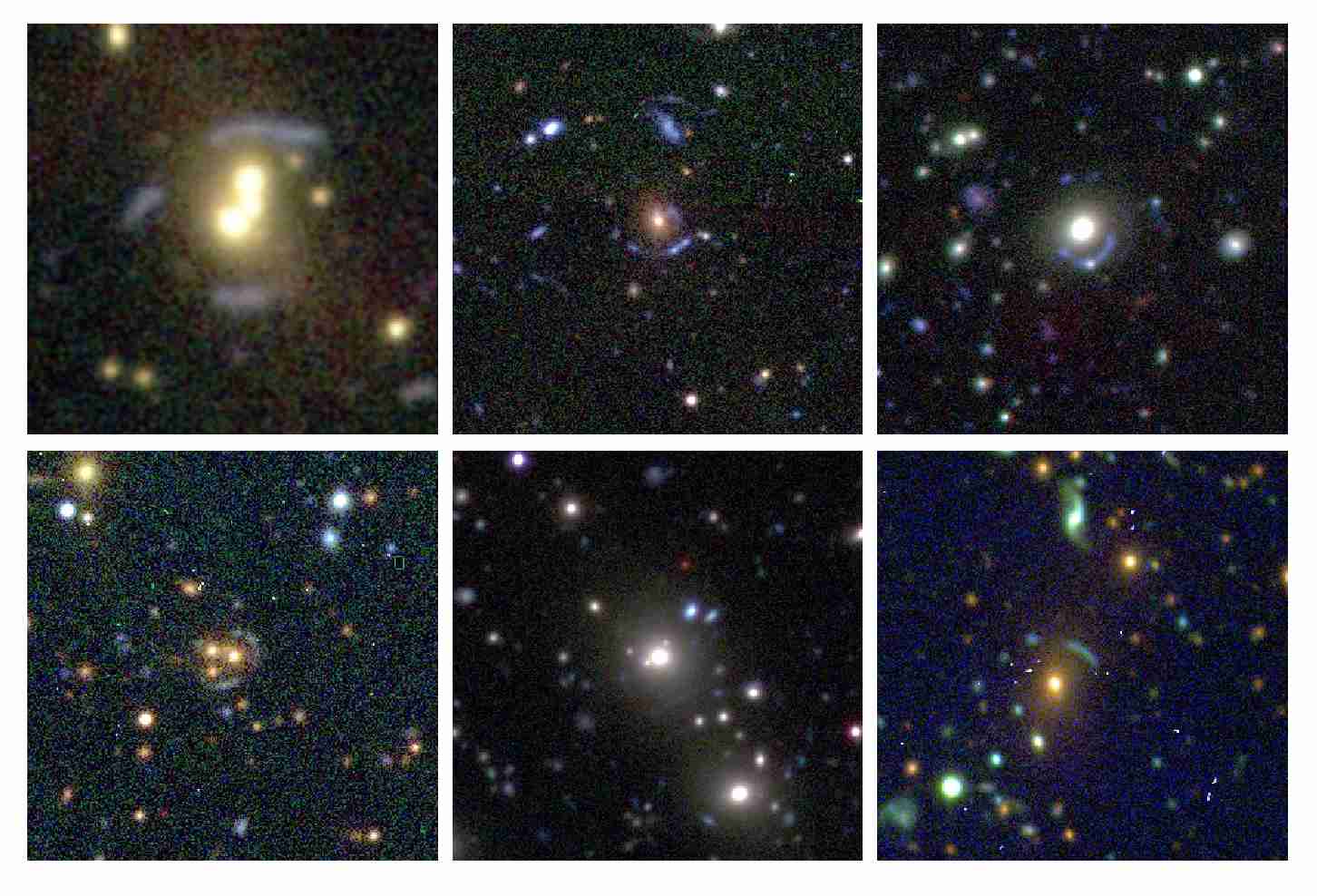,width=0.9\linewidth}
\caption{Group-scale strong lenses detected in the CFHTLS images, from the
sample compiled by \citet{Lim++09}. Color images kindly provided by
R.\ Cabanac and the SL2S collaboration.}
\label{fig:sl:yield:grouplensgallery}
\end{figure}

Strong lensing by galaxy groups has not been studied very much to date,
because finding group-scale lensing requires a very wide field survey. 
Galaxy groups represent the transition in mass between galaxies
and clusters, and are crucial to understand the formation and evolution of
massive galaxies. \citet{Lim++09} presented a sample
of 13 group-scale strong lensing from the SL2S (see
\autoref{fig:sl:yield:grouplensgallery}), and used it to explore
the distribution of mass and light in galaxy groups. By extrapolating
the SL2S result (see also \autoref{stronglens:fig:prob_dif}), we
can expect to discover $\sim 10^3$ group-scale strong lenses in  the
10-year LSST survey. Strong lensing by groups is often quite
complicated, and thus is a promising site to look for exotic lensing
events such as higher-order catastrophes \citep{O+M09}. This is an area where
DES and PS1 are more competitive, at least for the bright, easily followed-up
arcs. LSST, like CFHTLS, will probe to fainter and more numerous arcs.


\subsection{Cluster Strong Lenses}
\label{sec:sl:yield:arcs}

\begin{figure}[!t]
\begin{center}
\includegraphics[width=0.48\linewidth]{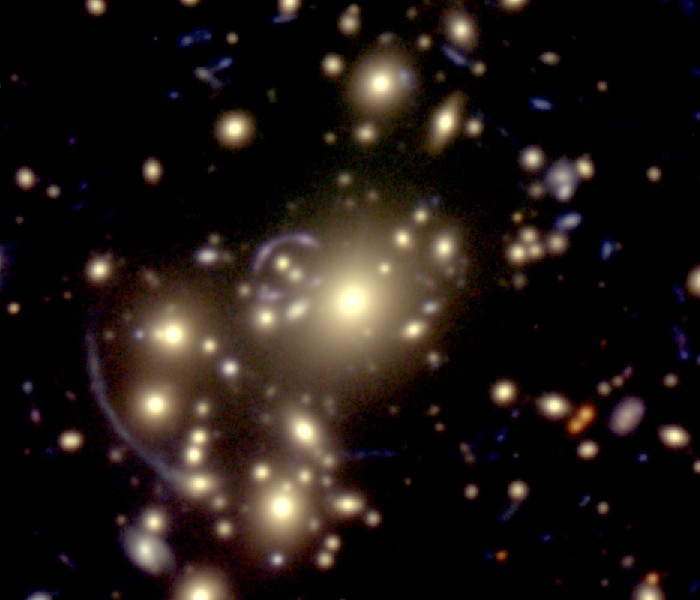}
\includegraphics[width=0.48\linewidth]{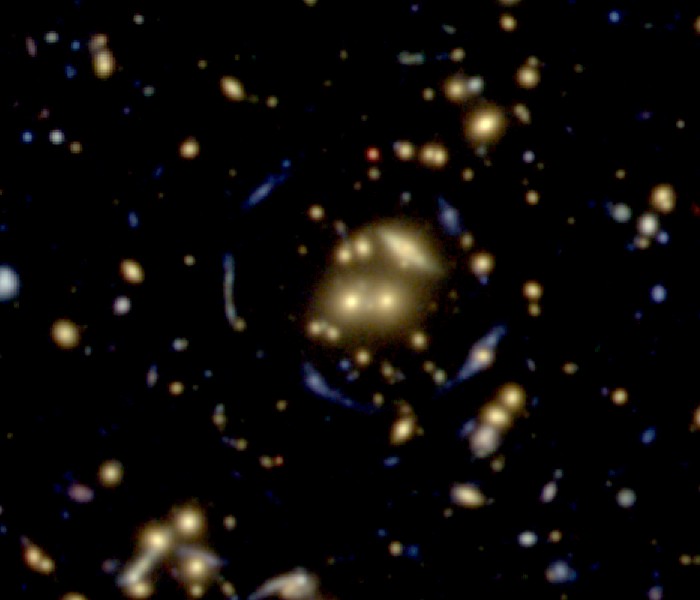}
\end{center}
\caption{Example of giant arcs in massive clusters. Color composite
  Subaru Suprime-cam images of clusters Abell 1703 and SDSS J1446+3032
  are shown \citep{OHG09}.}
\label{fig:sl:yield:clustersgallery}
\end{figure}

Since the first discovery of a giant arc in cluster Abell~370
\citep{LP86,Sou++87}, many lensed arcs have been discovered in  clusters
(\autoref{fig:sl:yield:clustersgallery}). The number of lensed arcs
in a cluster is a strong function of the cluster mass, such that the
majority of the massive clusters ($>10^{15}M_{\odot}$) exhibit strongly
lensed background galaxies when observed to 
the depth achievable in the LSST survey
\citep[e.g.,][]{Bro++05}. Being able to identify systems of multiple images via their colors and morphologies, requires high resolution imaging. It is
here that LSST will again have an advantage over precursor surveys like DES
and PS1.

%

\begin{figure}[ht]
\begin{center}
\includegraphics[width=0.45\linewidth]{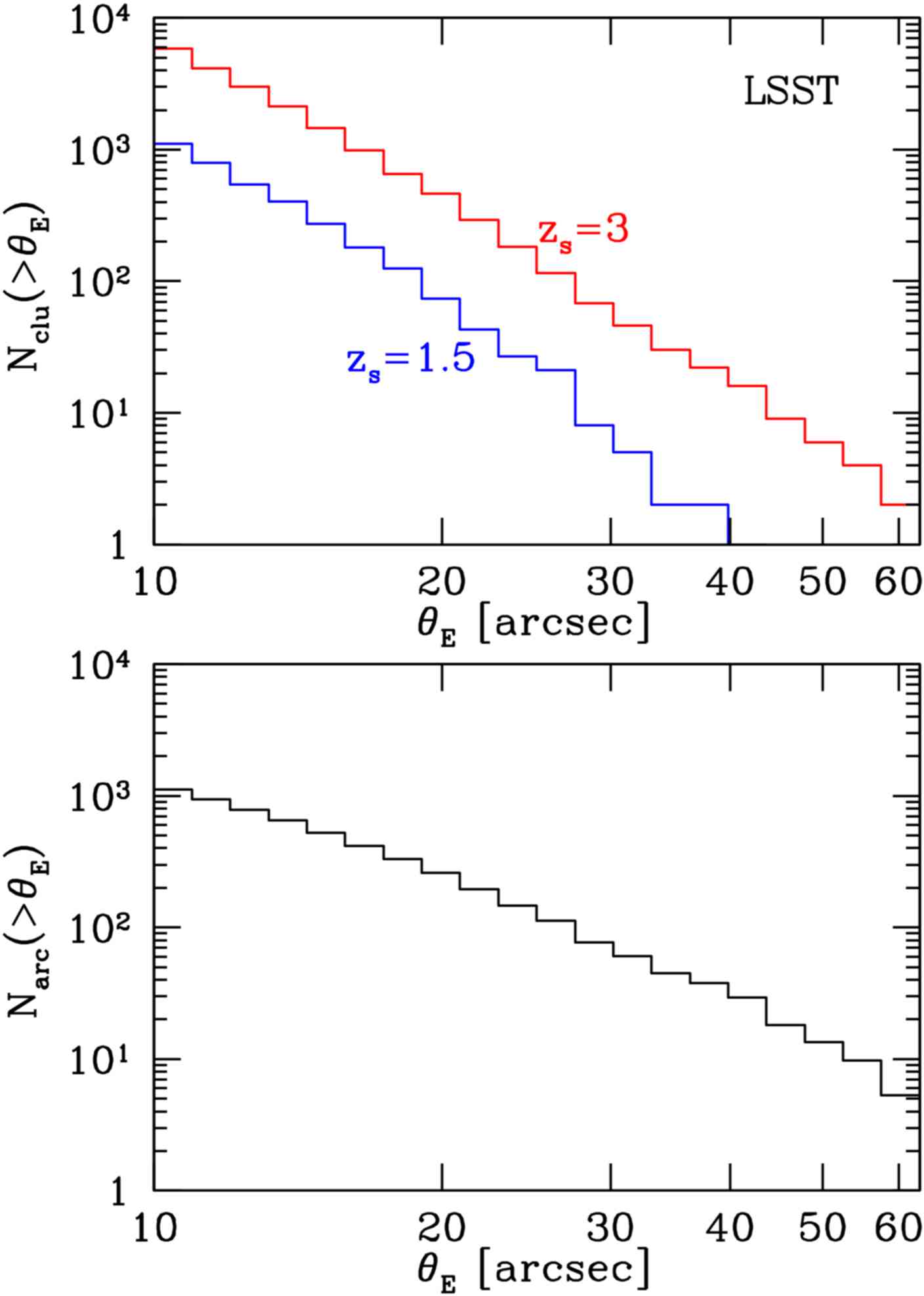}
\end{center}
\caption{Estimated numbers of clusters with larger Einstein radii
(top) and cluster multiple image systems (bottom) in the 
  $i=27$ LSST survey,
  based on the semi-analytic model of \citet{O+B09}, which assumes
  smooth triaxial halos as lensing clusters. Here we consider
  only clusters with $M>10^{14}M_\odot$. The background galaxy number
  density is adopted from \citet{Zhan++09}. Although our model does
  not include central galaxies, the effect of the baryonic
  concentration is not very important for our sample of lensing
  clusters with relatively large Einstein radius, $>10''$. The
  magnification bias is not included, which provides quite
  conservative estimates of multiple image systems. The number of
  multiple image systems available is of order $10^3$, with these
  systems roughly evenly divided over a similar number of clusters.} 
\label{fig:sl:yield:arcstats}
\end{figure}

In \autoref{fig:sl:yield:arcstats} we plot estimates for the number of
multiple-image systems produced by massive clusters. As can be seen in
this plot, we can expect to detect several thousand massive clusters
in the stacked image set whose Einstein radii are $10''$ or
greater. Not all will show strong lensing: the number of multiple 
image systems detectable with LSST is likely to be $\sim 1000$, but
with the more massive clusters being more likely to host many multiple
image systems. Very roughly, we expect that clusters with Einstein
radius greater than $\sim 30''$ should host more than one strong lens system 
detectable by
LSST: there will be
$\sim 50$ such massive systems in the cluster sample.
Given the relative scarcity of these most massive clusters, we
consider the sample of LSST strong lensing clusters to number 1000 or
so, with the majority displaying a single multiple image system (arcs and
counter arcs). This number is quite uncertain: the detectability and
identifiability of strongly lensed features is a strong function of
source size, image quality, and the detailed properties of cluster
mass distributions: detailed simulations will be required to understand
the properties of the expected sample in more detail, and indeed to
reveal the information likely to be obtained on cluster physics and
cosmology based on arc statistics. 

Strong lensing by clusters is enormously useful in exploring the
mass distribution in clusters. For instance, merging clusters of
galaxies serve as one of the best astronomical sites to explore
properties of dark matter (\autoref{sec:sl:dmclus}). Strong
lensing provides robust measurements of cluster core masses and,
therefore, by combining them with weak lensing measurements, one can
study the density profile of clusters over a wide range in radii,
which provides another test of structure formation models
(\autoref{sec:sl:clusmf}). Once the mass distribution is
understood, we can then use strong lensing clusters to find and
measure distant faint sources by making use of these high
magnification and low background ``cosmic telescopes''
(\autoref{sec:sl:telescopes}).

\section{ Massive Galaxy Structure and Evolution }
\label{sec:sl:galev}

\noindent{\it Phil~Marshall, Christopher D. Fassnacht, Charles R. Keeton}   

The largest samples of strong lenses discovered and measured with LSST will be
galaxy-scale objects (\autoref{sec:sl:yield}), which (among other
things) will allow us to measure lens galaxy mass.  In this
section we describe several approaches towards measuring the gross mass
structure of massive galaxies, allowing us to trace their evolution
since they were formed. 

%


\subsection{Science with the LSST Data Alone}

Optimal combination of the survey images should permit:
\begin{itemize}
\item accurate
measurements of the image positions ($\pm0.05''$), fluxes,
and time delays ($\pm$few days) for several thousand
(\autoref{sec:sl:yield}) lensed quasars,
AGN, and supernovae and
\item detection and associated modeling, of $\simeq10^4$ lensed
galaxies.
\end{itemize}
The multi-band imaging will yield photometric
redshift estimates for the lens and source.  The most robust output
from all these data will be the mass of the lens galaxy enclosed within the
Einstein radius.  When combined with the photometry, this provides
 an accurate aperture mass-to-light ratio for each strong lens galaxy
regardless of its redshift.  \citet[][and references therein]{RK05}
illustrate a method to use strong lens ensembles to probe the mean
density profile and luminosity evolution of early-type galaxies.
The current standard is the SLACS survey 
\citep[][and subsequent papers]{Bol++06,Bol++08}: with 70
spectroscopically-selected low redshift (median 0.2), luminous 
lenses observed with
HST, the SLACS team
was able to place robust constraints on the mean 
logarithmic slope of the density profiles (combining the lensing image separations with the SDSS
stellar velocity dispersions). 
However, the first thing we can do with LSST lenses is increase the ensemble
size from tens to thousands, pushing out to higher redshifts and lower lens
masses. 

Note the distinction between \emph{ensemble}
studies that do not require statistical completeness and
\emph{statistical} studies that do.  LSST will vastly expand both
types of samples. Statistical studies will be more easily carried out
with the lensed quasar sample, where the selection function is more
readily characterized. The larger lensed galaxy sample will require more
work to render its selection function.

The first thing we can do with a large, statistically complete sample
from LSST is measure the mass function of lens galaxies.  
Since we know
the weighting from the lensing cross section, we will be able to probe
early-type (and, with fewer numbers, other types!) galaxy mass evolution
over a wide range of redshift, up to and including the era of elliptical
galaxy formation ($\zd \simeq 1-2$). From the mock catalogue of well-measured
lensed quasars
described in
\autoref{sec:sl:yield:gal}, we expect about 25\% ($\simeq 600$) 
of the lenses to lie at
$z_d > 1$, and 5\% ($\simeq 140$) to lie at $z_d > 1.5$, if the assumption of
a non-evolving velocity function is valid to these redshifts. While it seems
to be a reasonable model at lower redshifts \citep{Ogu++08}, it may not be at
such high redshifts: the high-$z$ lenses are sensitive probes of the evolving
mass function.

The time delays contain information about the density profiles of the
lensing galaxies although it is combined with the Hubble parameter
\citep{CSKtdel}.  
\citet{CSK0435} fixed $H_0$ and then used the time
delays in the lens HE~0435$-$1223 to infer that the lens galaxy has a
density profile that is shallower than the mean, quasi-isothermal
profile of lens galaxies. With the LSST ensemble we can consider
simultaneously fitting for $H_0$ and the mass density profile
parameters of the galaxy population (see \autoref{sec:sl:H0} and
\autoref{sec:sl:H0stat} for more discussion of strong lens
cosmography).

The lensed quasar sample has a further appealing property: it will be
selected by the properties of the sources, not the lenses. 
When
searching for lensed {\em galaxies} in ground-based imaging data, the
confusion between blue arcs and spiral arms is severe enough that one
is often forced to focus on elliptical galaxy  lens candidates. 
In a source-selected lensed {\em quasar}, however, the lens galaxies
need not be elliptical, or indeed regular in any way. This means we
can aspire to compile samples of massive lens galaxies at high
redshift that are to first order selected by their mass.

Understanding the distribution of galaxy density profiles
out to $z \sim 1$ will strongly constrain models of galaxy formation,
including both the hierarchical formation picture (what range of
density profiles are expected if ellipticals form from spiral
mergers?) and environmental effects such as tidal stripping (how
do galaxy density profiles vary with environment?).
In this way, we anticipate LSST providing the best
assessment of the distribution of galaxy mass density profiles out to
$z \sim 1$.




\subsection{Science Enabled by Follow-up Data}

While the LSST data will provide a wealth of new lensing measurements, 
we summarize very briefly some of the additional opportunities provided by
various follow-up campaigns:

\begin{itemize}
\item {\bf Spectroscopic redshifts.} While the LSST photometric redshifts will
be accurate to $0.04(1+\zd)$ for the lens galaxies
(\autoref{sec:common:phz:cal}), the source redshifts will be somewhat more
uncertain. High accuracy mass density profiles will require spectroscopic
redshifts. Some prioritization of the sample may be required: we can imagine,
for example, selecting the most informative image configuration 
lenses in redshift bins for spectroscopic follow-up.
\item {\bf Combining lensing and stellar dynamics.} Stellar velocity
dispersions provide valuable additional information on galaxy mass profiles,
to first order  providing an additional aperture mass estimate at a different
radius to the Einstein radius
\citep[see e.g.,][]{T+K04,Koo++06,Tro++08}. 
More subtly the stellar dynamics probe the three-dimensional potential,
while the lensing is sensitive to mass in projection, meaning that some
degeneracies between bulge, disk, and halo can be broken. 
Again, since these measurements are
expensive, we can imagine focusing on a particular well-selected 
subset of LSST lenses.
\item {\bf High resolution imaging.} The host galaxies of lensed quasars may
appear too faint in the survey images -- but distorted into Einstein rings,
they provide valuable information on the lens mass distribution. They are also
of interest to those interested in the physical properties of quasar host
galaxies, since the lensing effect magnifies the galaxy, making it much more
easily studied that it otherwise would be.
\item  {\bf Infrared imaging.} There is an obvious 
synergy with concurrent near-infrared surveys such as VISTA and SASIR.
Infrared photometry will enable the study of 
lens galaxy stellar populations; one more handle on the mass distributions of
massive galaxies.   
\end{itemize}

Note that the study of massive galaxies and their evolution with detailed
strong lensing measurements can be done simultaneously with the cosmographic
study discussed in \autoref{sec:sl:H0}: there will be some considerable
overlap between the cosmographic lens sample and that defined for galaxy
evolution studies. The optimal redshift distribution for each ensemble is a
topic for research in the coming years.

\section{ Cosmography from Modeling of Time Delay Lenses and Their Environments}
\label{sec:sl:H0}

\noindent{\it Christopher D. Fassnacht, Phil~Marshall, Charles R. Keeton, Gregory Dobler, Masamune Oguri}  


\subsection{Introduction}
\label{sec:sl:H0:intro}

Although gravitational lenses provide information on many cosmological
parameters, historically the most common application has been the use
of strong lens time delays to make measurements of the distance scale
of the Universe, via the Hubble Constant, $H_0$
(\autoref{eqn:sl:preamble:timedelay}).  The current sample of lenses
with robust time delay measurements is small, $\sim20$ systems or
fewer, so that the full power of statistical analyses cannot be
applied.  In fact, the sample suffers from further problems in that
many of the lens systems have special features (two lensing galaxies,
the lens galaxy sitting in a cluster potential, and so on) that complicate
the lens modeling and would conceivably lead to their being rejected
from larger samples.  The large sample of LSST time delay lenses will
enable the selection of subsamples that avoid these problems.  These
subsamples may be those showing promising signs of an observable point
source host galaxy distorted into an Einstein ring, those with a
particularly well-understood lens environment, those with
especially simple lens galaxy morphology. While all current time delay lenses
have AGN sources, a significant fraction of the LSST sample will be lensed
supernovae (\autoref{sec:sl:yield:gal}).  These will be especially
useful if the lensed supernova is a Type Ia, where
it may be possible to directly determine the magnification factors of
the individual images.  

In \autoref{sec:sl:yield} we showed that the expected
sizes of the LSST samples of well-measured
lensed quasars and lensed supernovae are some 2600 and 330 respectively; 90 of
the lensed SNe are expected to be type Ia. Assuming the estimated quad
fraction of~14\%, we can expect to have 
some 400 quadruply-imaged variable sources to work with. Cuts in
environment complexity and lens morphology will reduce this further -- a
reasonable goal would be to construct a sample of 100 or more high quality
time delay lenses for cosmographic study. From simple counting statistics this
represents an order of magnitude increase in {\it precision} over the current
sample.

We can imagine studying this
cosmographic sample in some detail: with additional information on the
lens mass distribution coming from the extended images observed at
higher resolution (with JWST or ground-based adaptive optics imaging)
and from spectroscopic velocity dispersion measurements, and with
spectroscopically-measured lens and source redshifts, we can break
some of the modeling degeneracies and obtain quite tight constraints
on $H_0$ given an assumed cosmology \citep[e.g.,][]{Koo++03}.  With a
larger sample, we can imagine relaxing this assumption and providing
an independent cosmological probe competitive with those from weak
lensing (\autoref{chp:wl}), supernovae (\autoref{chp:sne}), and BAO
(\autoref{chp:lss}).  Note that the approach described here is
complementary to the statistical method of \autoref{sec:sl:H0stat},
which aims to use much larger numbers of individually less-informative
(often double-image) lenses.


\subsection{$H_0$ and the Practicalities of Time Delay Measurements}

As shown in \autoref{eqn:sl:preamble:timedelay}, the
measurement of $H_0$ using a strong lens system requires that the time
delay(s) in the system be measured.  In the past, this has been a
challenging exercise.  The measurement of time delays relies on
regular monitoring of the lens system in question.  Depending on the
image configuration and the mass of the lens, time delays can range
from hours to over a year.  The monitoring should return robust
estimates of the image fluxes for each epoch.  This can be easily
achieved in the monitoring of radio-loud lenses
\citep[e.g.,][]{Fas++02}, where contamination from the lensing galaxy is
typically not a concern.  With optical monitoring, however, the fluxes
of the lensed AGN images must be cleanly disentangled from the
emission from the lensing galaxy itself.  The standard difference imaging
pipeline may not provide accurate enough light curves, and in most cases we
anticipate needing to use the high resolution exposures to inform the
photometry in the poorer image quality exposures:
to achieve this, fitting
techniques such as those developed by \citet{Cou++98} or
\citet{Bur++01} must be employed.  These techniques are straightforwardly
extended to incorporate even 
higher-resolution follow-up 
imaging (\eg, from HST or adaptive optics) of the system as a basis
for the deconvolution of the imaging.

\begin{figure}
\begin{center}
\epsfig{file=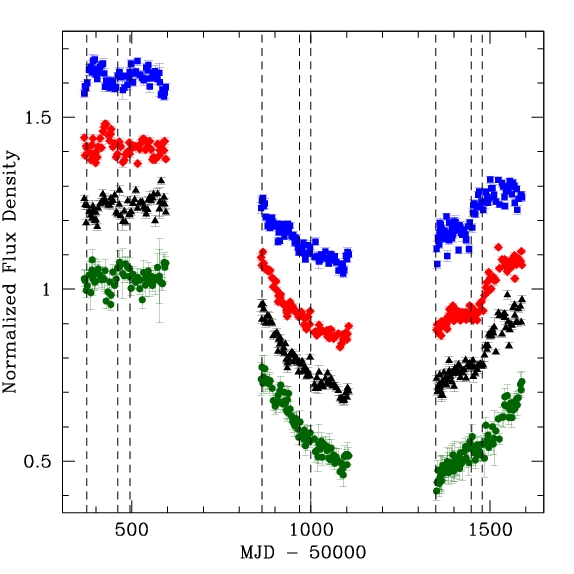,width=0.4\linewidth}
\epsfig{file=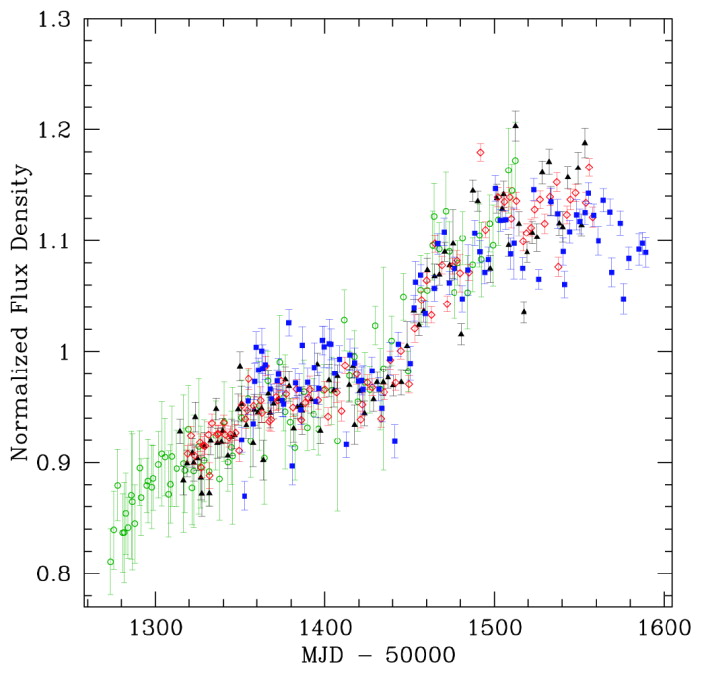,width=0.4\linewidth}
\end{center}
\caption{Example of time-delay measurement.  {\it Left:} light
curves of the four lensed images of B1608+656, showing three ``seasons''
of monitoring.  The seasons are of roughly eight months in duration and
measurements were obtained on average every 3 to 3.5 days.  Each
light curve has been normalized by its mean flux and then shifted
by an arbitrary vertical amount for clarity.  {\it Right:} the
B1608+656 season 3 data, where the four images' 
light curves have been shifted by
the appropriate time delays and relative magnifications, and 
then overlaid (the symbols match those in the left panel). 
Figures from \citet{Fas++02}.}
\label{fig:sl:delays_B1608}
\end{figure}

Once obtained, the light curves must be evaluated to determine the
best-fit time delays between the lensed components.  The statistics of
time delays has a rich history
\citep[e.g.,][]{PRH1,PRH2,Pelt++94,Pelt++96}, driven in part by the
difficulty in obtaining a clean delay from Q0957+561 (the first lensed
quasar discovered) until a sharp
feature was finally seen in the light curves \citep{Kun++95,Kun++97}.
Most lens monitoring campaigns do not obtain regular sampling; dealing
properly with unevenly sampled data is a crucial part of a successful
light curve analysis.  Two successful approaches are the
``dispersion'' method of \citet{Pelt++94,Pelt++96}, which does not
require any data interpolation, and fitting of smooth functions to the
data \citep[e.g., Legendre polynomials;][]{CSK0435}.  With these
approaches and others, time delays have now been successfully
measured in $\sim15$ lens systems
\citep[e.g.,][]{Barkana++97,Biggs++0218,Lovell++1830,Koop++1600,Bur++1600,Bur++1520,Bur++2149,Fas++02,Hjorth++02,CSK0435,cosmograilV,cosmograilVII}.
\autoref{fig:sl:delays_B1608} shows an example of light curves from a monitoring program
that led to time delay measurements in a four-image lens system. 

It is yet to be seen what time delay precision the LSST cadence will allow:
experiments with simulated image data need to be performed based on
the operations simulator output. We note that the proposed main survey
cadence (\autoref{sec:design:cadence}) leads to an exposure in some filter being taken every week (or
less) for an observing season of three months or so.  This cadence is
not very different from the typical optical monitoring campaigns
referred to above.  However, the length of the monitoring season is
significantly shorter, as lens monitoring campaigns are typically conducted
for the entire period that the system is visible in the night sky,
pushing to much higher airmass than the LSST will use.
That being said, typical time delays for four-image galaxy-scale
lens systems range from a few days to several tens of days 
\citep[e.g.,][]{Fas++02}, so a three-month observing season will be adequate
for measuring delays if the lensed AGN or supernova varies such that
the leading image variation occurs near to the beginning of the
season.  Ascertaining the precise 
effect of the season length on the number of well-measured lens
systems will require some detailed simulations. This program could also be
used to assess the gain in time delay precision in the 10-20\% of
systems that lie in the
field overlap regions, and hence get observed at double cadence.

A further systematic process that can affect optical lens monitoring
programs is microlensing, whereby individual stars in the lensing
galaxy can change the magnification of individual lensed images.
Both gradual changes, due to slow changes in the magnification pattern
as stars in the lensing galaxy move, and short-scale variability,
due to caustic crossings, have been observed 
\citep[e.g.,][]{Bur++1520,C+S03}.

\subsection{Moving Beyond $H_0$}

Fundamentally a time delay in a given strong lens system measures
the ``time delay distance,''
\begin{equation} \label{eq:sl:tdeldist}
  D \equiv \frac{\Dd \Ds}{\Dds} = \frac{\Delta t_{\rm obs}}{f_{\rm mod}}\ ,
\end{equation}
where $\Dd$ and $\Ds$ are angular diameter distances to the lens
(or deflector) and the source, $\Dds$ is the angular diameter
distance between the lens and source, and $f_{\rm mod}$ is a factor
that must be inferred from a lens model.  In individual lens systems
people have used the fact that $D \propto H_0^{-1}$ to measure the
Hubble constant.  With a large ensemble, however, we can reinterpret
the analysis as a measurement of distance (the time delay distance)
versus redshift (actually both the lens and source redshifts),
which opens the door to doing cosmography in direct analogy with
supernovae, BAO, and the CMB.

\begin{figure*}[!t]
\centering\epsfig{file=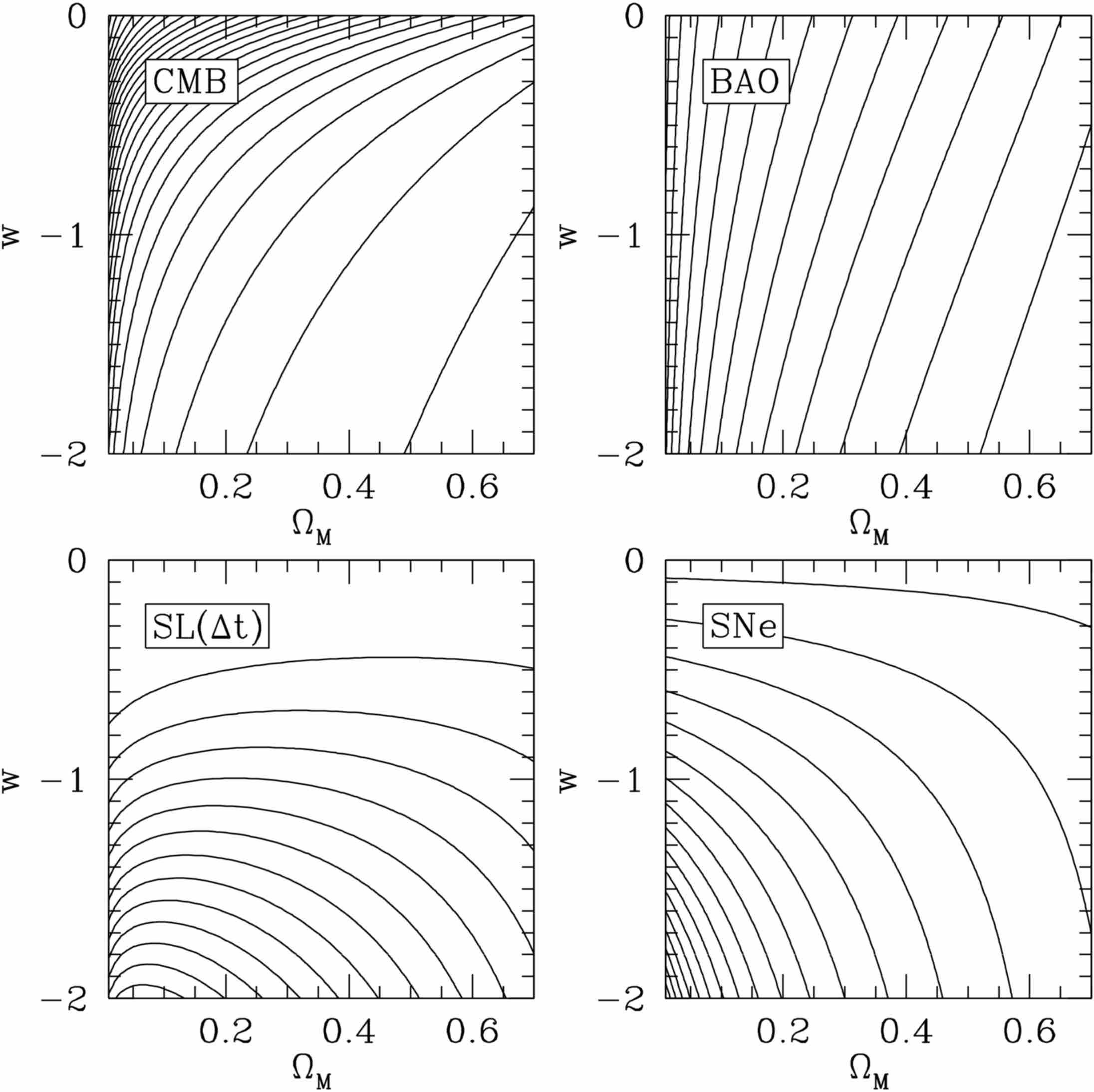,width=0.9\linewidth}
\caption{
Contours of key cosmological quantities that are constrained from CMB ({\it upper left}),
BAO ({\it upper right}), lensing time delay ({\it lower left}), and Supernovae Ia 
({\it lower
right}), which indicate the degeneracy direction from each
observation. For CMB and BAO, we plot contours of
$D_A(z_{\rm CMB}=1090)\sqrt{\Omega_Mh^2}$ and $D_A(z_{\rm
  BAO}=0.35)\sqrt{\Omega_Mh^2}$, respectively which are measures of the angular
scale of the acoustic peak at two redshifts.  For time delays, $\Delta t$, we show
contours of $D \equiv \Dd \Ds/\Dds$, where we adopted
$z_l=0.5$ and $z_s=1.8$. Contours of SNe are simply constant
luminosity distance,  $D_L(z_{\rm SNe}=0.8)$. All the contours are
shown on the $\Omega_M$-$w$ plane, assuming a flat Universe. 
}
\label{sl:fig:distances}
\end{figure*}

In principle, strong lensing may be able to make a valuable contribution
to cosmography because of its independence from and 
complementarity to these other probes.
\autoref{sl:fig:distances} illustrates this point by comparing
degeneracy directions of cosmological constraints from lens time
delays with those from CMB, BAO, and SNe. The lensing constraints
look quite different from the others, with the notable feature that
the contours are approximately horizontal --- and thus particularly
sensitive to the dark energy equation of state parameter $w$ ---
in much of the region of interest.  The value of strong lensing
complementarity is preserved even when generalized to a
time-dependent dark energy equation of state \citep{Lin04}.

In practice, of course, the challenge for strong lensing cosmography
is dealing with uncertainties in the $f_{\rm mod}$ factor in
\autoref{eq:sl:tdeldist}.  In this section we discuss an approach
that involves modeling individual lenses as carefully as possible,
while in \autoref{sec:sl:H0stat} we discuss a complementary statistical
approach.

One way to minimize uncertainties in $f_{\rm mod}$ is to maximize
the amount and quality of lens data.  We will require not just
image positions and time delays but also reliable (ideally
spectroscopic) lens and source redshifts; we would like to have
additional model constraints in the form of arc or ring images
of the host galaxy surrounding the variable point source, or some
other background galaxy; and we would make good use of dynamical
data for the lens galaxy if available.  All of this calls for
follow-up observations, likely with JWST or laser guide star adaptive
optics on 10-m class 
or larger telescopes.  We expect
to use the full sample of time delay lenses to select good
sub-samples for such follow-up, as described in \autoref{sec:sl:H0:intro}.

There are systematic errors associated not just with the lens
galaxy but also with the influence of mass close to the line of
sight, either in the lens plane or otherwise
\citep[e.g.,][]{KZ04,Fas++06,Mom++06}. 
Some constraints on this ``external convergence'' 
(\autoref{sec:sl:basics:env}) can be placed by
modeling all the galaxies in the field using the multi-filter
photometry, and perhaps the weak lensing signal \citep[e.g.,][]{Nak++09}.
This degeneracy between lens and environment can also be broken if
we know the magnification factor itself, which is approximately true
if the source is a type Ia supernova, whose light curves may be
guessed a priori.  
The hundred or so multiply-imaged SNeIa  (\autoref{fig:sl:yield:sne}) will
be especially valuable for this time delay lens
cosmography~\citep{O+K03}, and may be expected to make up a significant
proportion of the cosmographic strong lens sample.

\section{ Statistical Approaches to Cosmography from Lens Time Delays }
\label{sec:sl:H0stat}

\noindent{\it Masamune Oguri, Charles R. Keeton, Phil~Marshall}   


The large number of strong lenses discovered by LSST will permit
statistical approaches to cosmography -- the measurement of the distance
scale of the Universe, and the fundamental parameters associated with it
-- that complement the detailed modeling
of individual lenses.  Statistical methods will be particularly
powerful for two-image lenses, which often have too few lensing
observables to yield strong modeling constraints, but will be so
abundant in the LSST sample (see \autoref{sec:sl:yield}) that we
can leverage them into valuable tools for cosmography. Note that to
first order we do
not need a complete sample of lenses for this measurement: we can work
with any ensemble of lenses, provided we understand the form of the 
distributions of its members' structural parameters.



The basic idea is to construct a statistical model for the likelihood
function $\Pr(\datav|\mathbf{q},\mathbf{p})$, where the data
$\datav$ concisely characterize the image configurations and time
delays of all detected lenses, while $\mathbf{q}$ represents
parameters related to the lens model (the density profile and
shape, evolution, mass substructure, lens environment, and so on),
and $\mathbf{p}$ denotes the cosmological parameters of
interest.  We can then use Bayesian statistics to infer posterior
probability distributions for cosmological parameters, marginalizing
over the lens model parameters $\mathbf{q}$ (which are nuisance
parameters from the standpoint of cosmography) with appropriate
priors.

Here we illustrate the prospects for cosmography from statistical
analysis of a few thousand time delay lenses that LSST is likely to
discover.
We use the statistical methods introduced by \citet{Ogu07}, working
with the ``reduced time delay'' between images $i$ and $j$,
$\Xi_{ij} \equiv 2c\Delta t_{ij}\Dds/[\Dd \Ds(1+\zd)(r_i^2-r_j^2)]$,
which allows us to explore how time delays depend on the ``nuisance
parameters'' related to the lens model.  Here $r_i$ and $r_j$ are
the distances of the two images from the center of the lens galaxy.
Following \citet{Ogu07}, we conservatively model
systematics associated with the lens model using a log-normal
distribution for $\Xi$ with dispersion $\sigma_{\log\Xi}=0.08$.  
We can then combine this statistical model for $\Xi$ with observed
image positions and time delays to infer posterior probability
distributions for cosmological parameters (which enter via the
distances $\Dd$, $\Ds$, and $\Dds$). See \citet{C+M09} for detailed
discussions of how $\Xi$ depends on various cosmological parameters.



From the mock catalog of
lensed quasars in LSST (see \autoref{sec:sl:yield}), 
we choose two-image lenses
because they are particularly suitable for statistical analysis.
We only use systems whose image separations are in the range
$1''<\Delta\theta<3''$ and whose configurations are asymmetric with
respect to the lens galaxy: specifically, we require that the asymmetry parameter
$R_{ij} \equiv (r_i-r_j)/(r_i+r_j)$ be in the range
$0.15<R_{ij}<0.8$, where $\Xi$ is less sensitive to the complexity of
lens potentials. This yields a sample of $\sim$ 2600 lens systems.  
We assume positional uncertainties of $0.01''$ and time delay
uncertainties of 2~days. We assume no errors associated with lens and
source redshifts. 

\begin{figure*}[!t]
\centering\epsfig{file=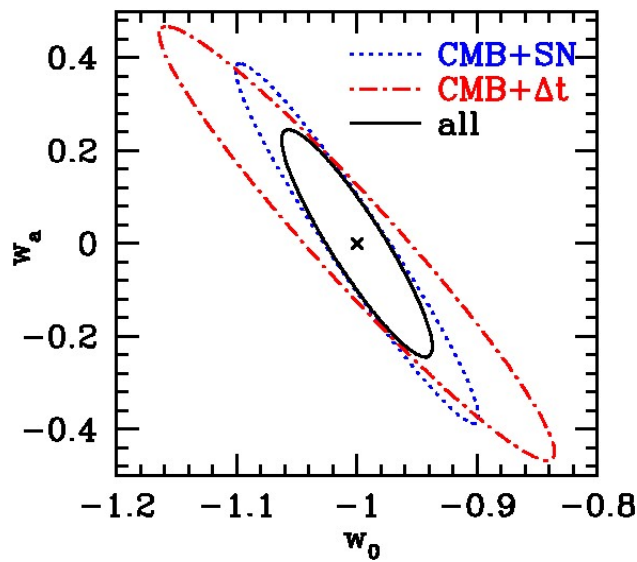,width=0.5\linewidth}
\caption{
Forecast constraints on cosmological parameters in the $w_0$-$w_a$
plane, assuming a flat Universe. The CMB prior from Planck is adopted
for all cases.  The SN constraint represents expected constraints from
a future SNAP-like JDEM mission supernova survey. 
Constraints from time delays in LSST are denoted by $\Delta t$.}
\label{sl:fig:cont-lsst_vs_snap}
\end{figure*}

\autoref{sl:fig:cont-lsst_vs_snap} shows the 
corresponding
constraints on the dark energy equation of state, $w_0$ and $w_a$,
from a Fisher matrix analysis  (\autoref{sec:app:stats:numtech:fisher}) 
of the combination of CMB data (expected from Planck), supernovae
(measured by a SNAP-like JDEM mission),  and the strong lens time
delays measured by LSST. In all cases, a flat Universe is
assumed. This figure indicates that the constraint from 
time delays can be competitive with that from supernovae. By combining
both constraints, we can achieve higher accuracy on the dark energy
equation of state parameters.



One important source of systematics in this analysis is related to the 
(effective) slope
of lens galaxy density profiles.  While the mean slope only affects
the derived Hubble constant, any evolution of the slope with
redshift would affect other cosmological parameters as well.
We anticipate that the distribution of density slopes (including
possible evolution) may be calibrated by other strong lensing
data (see \autoref{sec:sl:galev}). In practice, this combination will have 
to be done carefully to avoid using the same data twice. Another
important systematic error comes from uncertainties in the lens and
source redshifts. The effect may be negligible if we measure all the redshifts
spectroscopically as we assumed above, but this would require considerable
amounts of spectroscopic follow-up observations. Instead we can use
photometric redshifts for lenses and/or sources; in this case, errors
on the redshifts will degrade the cosmological use of time delays
\citep[see][]{C+M09}.

\section{ Group-scale Mass Distributions, and their Evolution }
\label{sec:sl:groups}

\noindent{\it Christopher D. Fassnacht}   

Galaxy groups are the most common galaxy environment in the local
Universe \citep[e.g.,][]{TG76,GH83,Eke++04}.
They may be responsible for driving much of the evolution in galaxy
morphologies and star formation rates between $z \sim 1$
and the present \citep[e.g.,][]{AF80,Barnes85,Merritt85}, and their mass
distributions represent a transition 
between the dark-matter dominated NFW profiles seen on cluster scales
and galaxy-sized halos that are strongly affected by baryon cooling
\citep[e.g.,][]{Ogu06}. LSST will excel in finding galaxy groups beyond the local
Universe and measuring the evolution in the mass function. 


Groups have been very well studied at low redshift
\citep[e.g.,][]{ZM98,MZ98,OPGEMS04} but very little is known about
moderate-redshift ($0.3<z<1$) groups.  This is because unlike
clusters they are difficult to discover beyond the local Universe.
At optical wavelengths, their modest galaxy overdensities make these
systems difficult to pick out against the distribution of field
galaxies, while their low X-ray luminosities and cosmological dimming
have confounded most X-ray searches.  This situation is beginning to
change with the advent of sensitive X-ray observatories such as
Chandra and XMM
\citep[e.g.,][]{Willis05,Mulchaey06,Jel++06,Jel++07}.
However, the long exposure times required to make high-SNR detections
of the groups have kept the sample sizes small and biased detections
toward the most massive groups, i.e., those that could be classified
as poor clusters.
Large spectroscopic surveys are also producing samples of group
candidates, although many of the candidates are selected based on only
3--5 redshifts and, thus, the numbers of false positives in the
samples are large.  Here, too, the sample sizes are limited by the
need for intensive spectroscopic followup in order to confirm the
groups and to measure their properties \citep[e.g.,][]{Wil++05}.


Both X-ray and spectroscopic data can, in principle, be used to
measure group masses.  However, these mass estimates are based on
assumptions about, for example, the virialization of the group, and may be
highly biased.  Furthermore, velocity dispersions derived from only a
few redshifts of member galaxies may be poor estimators of the true
dispersions \citep[e.g.,][]{ZM98,Gal++08}, further biasing dynamical mass
estimates.  In the case of the X-ray measurements, even deep exposures
($\sim$100~ksec) may not yield high enough signal-to-noise ratios to
measure spectra and, thus, determine the temperature of the intragroup
gas \citep[e.g.,][]{Fas++08}.


In \autoref{sec:sl:yield}, we estimated that $\sim 10^3$ galaxy
groups will be detected by their strong lensing alone.  This will
greatly advance the state of group investigations.  The
number of known groups beyond the local Universe will be increased
by a factor of 10 or more.  These groups should fall in a broad
range of redshifts, including higher redshifts
than those probed by X-ray--selected samples, which are typically
limited to $z \leq 0.4$.  More importantly, however, these
lens-selected groups will all have highly precise mass measurements
that are not reliant on assumptions about hydrostatic equilibrium
or other conditions.  Strong lensing provides the most precise
method of measuring object masses beyond the local Universe,
with typical uncertainties of $\sim$5\% or less \citep{PW89}.
The combination of a wide redshift range, a large sample size,
and robust mass measurements will enable unprecedented explorations
of the evolution of structure in this elusive mass range.

\section{ Dark Matter (Sub)structure in Lens Galaxies }
\label{sec:sl:cdmsub}

\noindent{\it Gregory~Dobler, Charles R. Keeton, Phil~Marshall}  



Gravitationally lensed images of distant quasars 
(\autoref{fig:sl:cdmsub:0924,2045}) contain a wealth of information
about small-scale structure in both the foreground lens galaxy and the
background source.  Key scales in the lens galaxy are:
\begin{itemize}
\item Macrolensing ($\sim 1$ arcsec) by the global mass distribution sets
the overall positions, flux ratios, and time delays of the images;
\item Millilensing ($\sim 1$ mas) by dark matter substructure perturbs the
fluxes by tens of percent or more, the positions by several to tens of
milli-arcseconds and the time delays by hours to days; and
\item Microlensing ($\sim 1$ $\mu$as) by stars sweeping across the images
causes the fluxes to vary on scales of months to years
(\autoref{fig:sl:cdmsub:2237}).
\end{itemize}
Additional scales are set by the size of the source.  Broad-band optical
observations measure light from the quasar accretion disk, which can be
comparable in size to the Einstein radius of a star in the lens galaxy.
Differences in the size of the source at different wavelengths can make microlensing
chromatic (see \autoref{fig:sl:cdmsub:doblermulens} and \autoref{sec:sl:agn}).

\begin{figure}[t]
\begin{center}
\includegraphics[width=0.5\linewidth,angle=0]{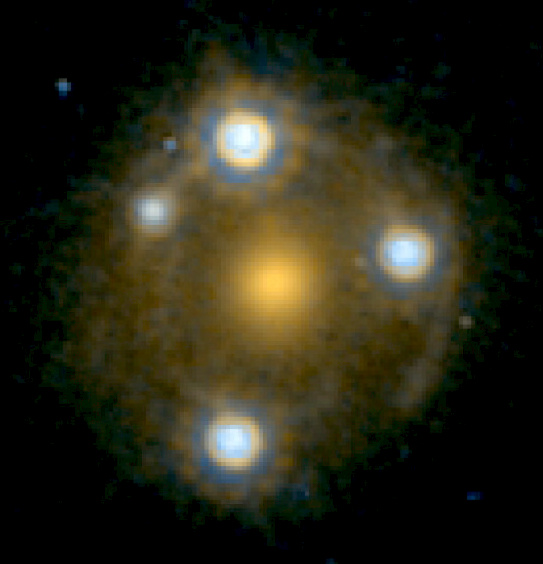}
\includegraphics[width=0.36\linewidth,angle=0]{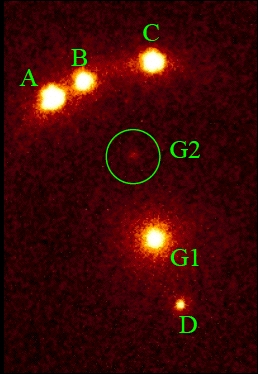}
\end{center}
\caption{
Two of the most extreme ``flux ratio anomalies'' in four-image lenses.
{\it Left:} HST image of SDSS~J0924+0219 \citep[from][]{Kee++06}.
A quasar at redshift $\zs=1.52$ is lensed by a galaxy at $\zd=0.39$
into four images, which lie about $0.9''$ from the center of the galaxy.
Microlensing demagnifies the image at the top left by a factor $> 10$,
relative to a smooth mass distribution.
{\it Right:} Keck adaptive optics image of B2045+265
\citep[from][]{McK++07} showing the lensing galaxy (G1) and four lensed
images of the background AGN (A--D).  Smooth models predict that image
B should be the brightest of the three close lensed images, but instead
it is the faintest, suggesting the presence of a small-scale perturbing
mass.  The adaptive optics imaging reveals the presence of a small
satellite galaxy (G2) that may be responsible for the anomaly.
}
\label{fig:sl:cdmsub:0924,2045}
\end{figure}

\begin{figure}[t]
\begin{center}
\includegraphics[width=0.9\linewidth,angle=0]{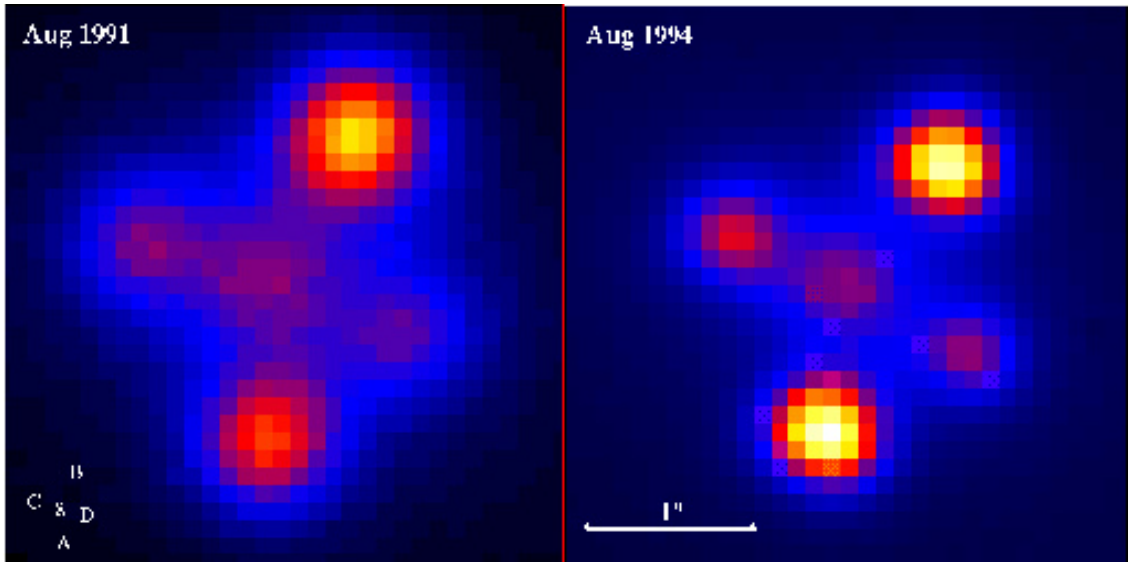}
\end{center}
\caption{Microlensing-induced variability in the 
``Einstein Cross'' lens Q2237$+$0305, from
Lewis, Irwin, et al.,
\url{http://apod.nasa.gov/apod/ap961215.html}.  The relative fluxes
in the four images are noticeably different in images taken three
years apart. }
\label{fig:sl:cdmsub:2237}
\end{figure}

\begin{figure}[t]
\begin{center}
\begin{minipage}{0.9\linewidth}
\centering\includegraphics[width=\linewidth,angle=0]{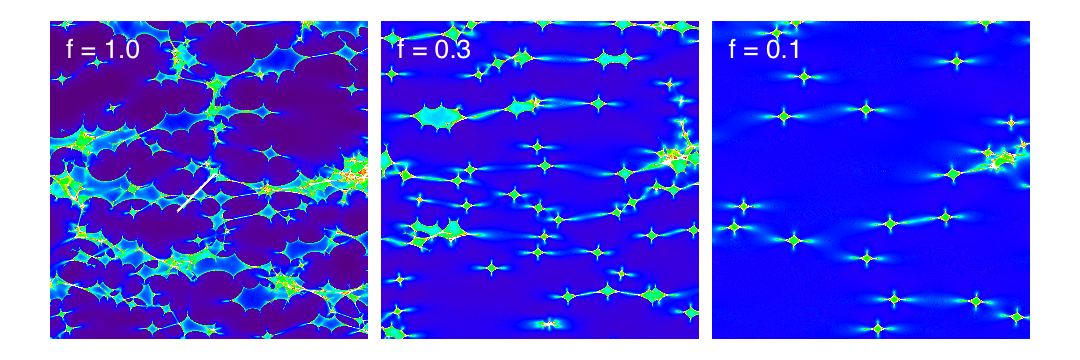}
\end{minipage}\hfill
\begin{minipage}{0.9\linewidth}
\centering\includegraphics[width=\linewidth,angle=0]{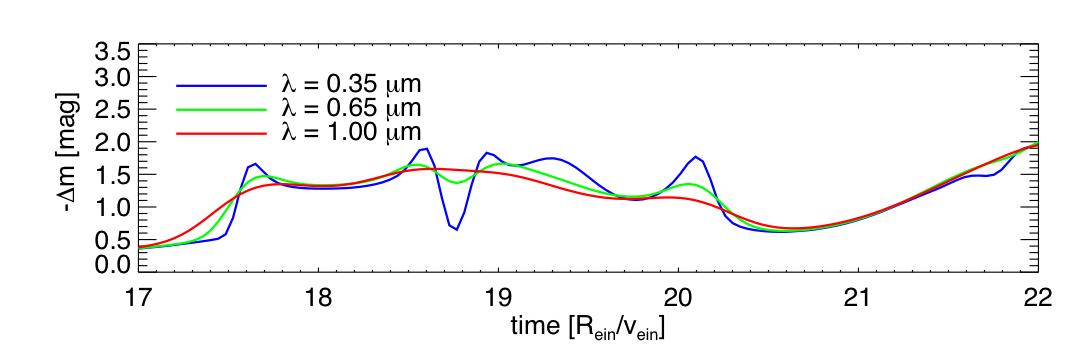}
\end{minipage}\hfill
\end{center}
\caption{
The maps in the top panels show examples of the lensing magnification for
a very small patch of the source plane just $30 R_E$ cm on a side, where
$R_E \sim 5\times10^{16}$ cm is the Einstein radius of a single star in
the lens galaxy.  The panels have the same total surface mass density
but a different fraction $f^{*}$ of mass in stars (the remaining mass
is smoothly distributed).  The magnification varies over micro-arcsecond
scales due to light bending by individual stars.  As the background
source moves relative to the stars (as shown by the white line in the
upper left panel), it feels the changing magnification leading to
variability in the light curves as shown in the bottom panel.  The
variability amplitude depends on wavelength because AGN have different
effective sizes at different wavelengths (see \autoref{sec:sl:agn}).
From \citet{Kee++09}.
}
\label{fig:sl:cdmsub:doblermulens}
\end{figure} 

The various phenomena have distinct observational signatures that
allow them to be disentangled.  Time delays cause given features in
the intrinsic light curve of the source to appear in all the lensed
images but offset in time; so they can be found by cross-correlating
image light curves.  Microlensing causes uncorrelated, chromatic
variations in the images; so it is revealed by residuals in the
delay-corrected light curves.  Millilensing leads to image fluxes,
positions, and time delays that cannot be explained by smooth lens
models; it can always be identified via lens modeling, and in certain
four-image configurations the detection can be made model-independent
\citep{Kee++03,Kee++05}.

The key to all this work is having the well-sampled, six-band light
curves provided by LSST.  The planned cadences should make it
possible to determine the time delays of most two-image lenses, and
multiple time delays in many four-image lenses.  The multicolor light
curves will have more than enough coverage to extract microlensing
light curves, which will not only enable their own science but also
reveal the microlensing-corrected flux ratios that can be used to
search for CDM substructure (\autoref{sec:sl:millilensing}).

In this section we discuss using milli- and microlensing to probe
the distribution of dark matter in lens galaxies on sub-galactic
scales.  In \autoref{sec:sl:agn} and \autoref{sec:agn:lens} we discuss
using microlensing to probe the structure of the accretion disks in
the source AGN.


\subsection{Millilensing and CDM Substructure}
\label{sec:sl:millilensing}

Standard lens models often fail to reproduce the fluxes of multiply-imaged
point sources, sometimes by factors of order unity or more (see
\autoref{fig:sl:cdmsub:0924,2045}).  In many cases these ``flux ratio
anomalies'' are believed to be caused by subhalos in the lens galaxy
with masses in the range $\sim 10^6$--$10^{10} M_\odot$.  CDM simulations
predict that galaxy dark matter halos should contain many such subhalos
that are nearly or completely dark
\citep[for some recent examples, see][]{Die++08,Spr++08}.  Strong lensing
provides a unique opportunity to detect mass clumps and thus test CDM
predictions, probe galaxy formation on small scales, and obtain
astrophysical evidence about the nature of dark matter
\citep[e.g.,][]{M+M01,Chiba02}.

The LSST lens sample should be large enough to allow us to probe the
mass fraction contained in substructure, its evolution with redshift,
the mass function and spatial distribution of subhalos within parent
halos, and the internal density profiles of the subhalos.  LSST
monitoring of the lenses will be essential to remove flux perturbations
from smaller objects in the lens galaxy, namely microlensing by stars.
This will enable us to expand millilensing studies from the handful of
four-image radio lenses available today to a sample of well-monitored
optical lenses that is some two orders of magnitude larger.

\subsubsection{Flux Ratio Statistics}

Millilensing by CDM substructure is detected not through variability
(the time scales are too long) but rather through observations of
image flux ratios, positions, and time delays that cannot be produced
by any reasonable smooth mass distribution.  Flux ratio anomalies
consistent with CDM substructure have already been observed in a
small sample of four-image lenses; they provide the only existing
measurement of the amount of substructure in galaxies outside the
Local Group \citep{D+K02}. 

Presently, constraints on substructure in distant galaxies are
limited by sample size because the analysis has been restricted to
four-image radio lenses.  Four-image lenses have been the main focus
for millilensing studies because they provide many more constraints
than two-image lenses, and that will continue to be the case with LSST.
Radio flux ratios have been required to date because optical flux
ratios are too contaminated by stellar microlensing (only at radio
wavelengths is the source large enough to be insensitive to stars).
The breakthrough with LSST will come from exploiting the time domain
information to measure microlensing well enough to remove its effects
and uncover the corrected flux ratios.  In this way LSST will finally
make it possible to use optical flux ratios to study CDM substructure.
The data volume will increase the sample of four-image lenses available
for millilensing by some two orders of magnitude.

The statistics of flux ratio anomalies reveal the overall abundance
of substructure (traditionally quoted as the fraction of the projected
surface mass density bound in subhalos), followed by the internal
density profile of the subhalos \citep{S+E07}.  These lensing
measurements are unique because most of the subhalos are probably
too faint to image directly.  With the large sample provided by LSST,
it will be possible to search for evolution in CDM substructure with
redshift (see below).

\subsubsection{Time Delay Perturbations}

\begin{figure}[t]
\begin{center}
\begin{minipage}{0.9\linewidth}
\begin{minipage}{0.48\linewidth}
\centering\includegraphics[width=\linewidth,angle=0]{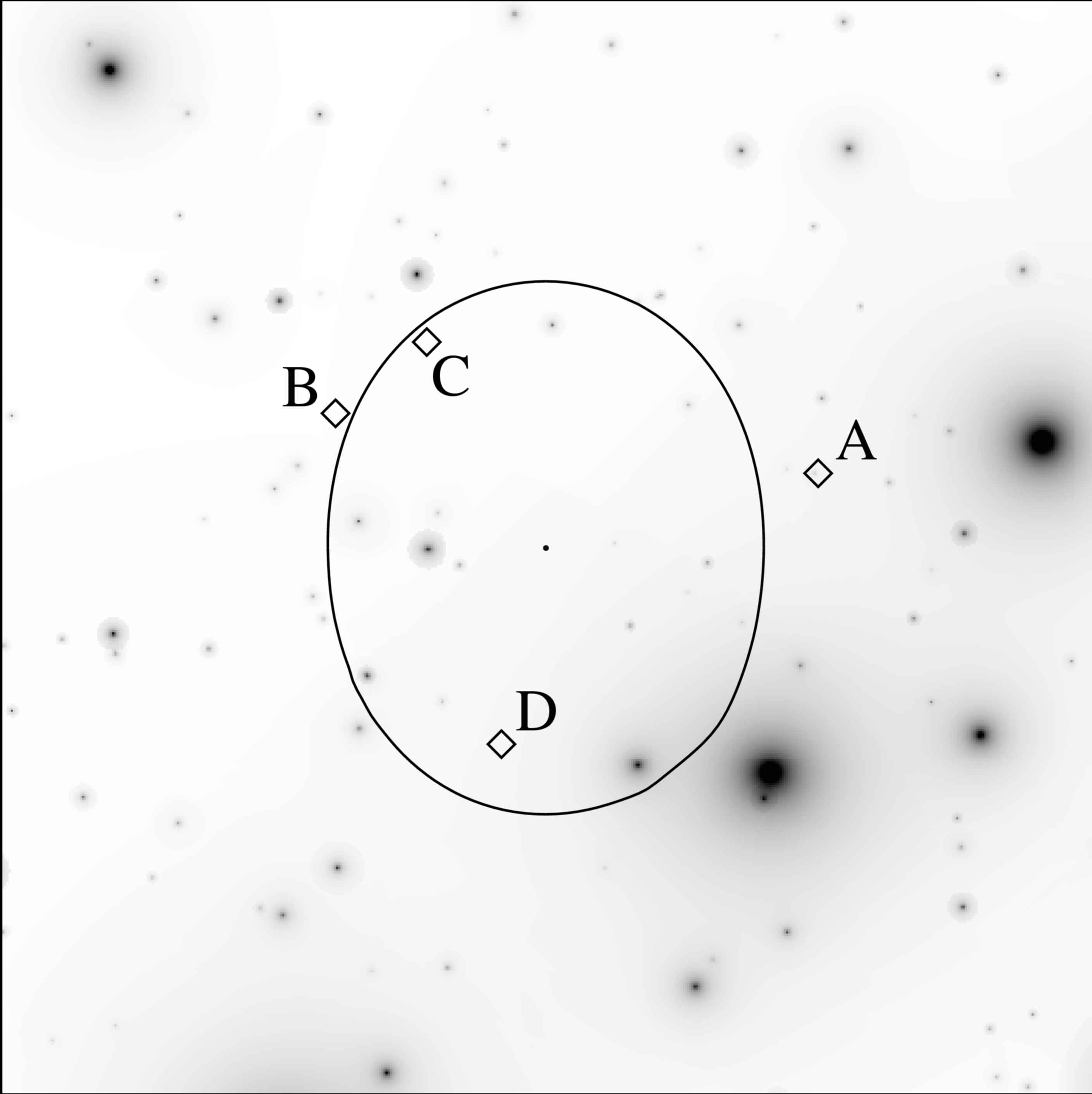}
\end{minipage}\hfill
\begin{minipage}{0.48\linewidth}
\centering\includegraphics[width=\linewidth,angle=0]{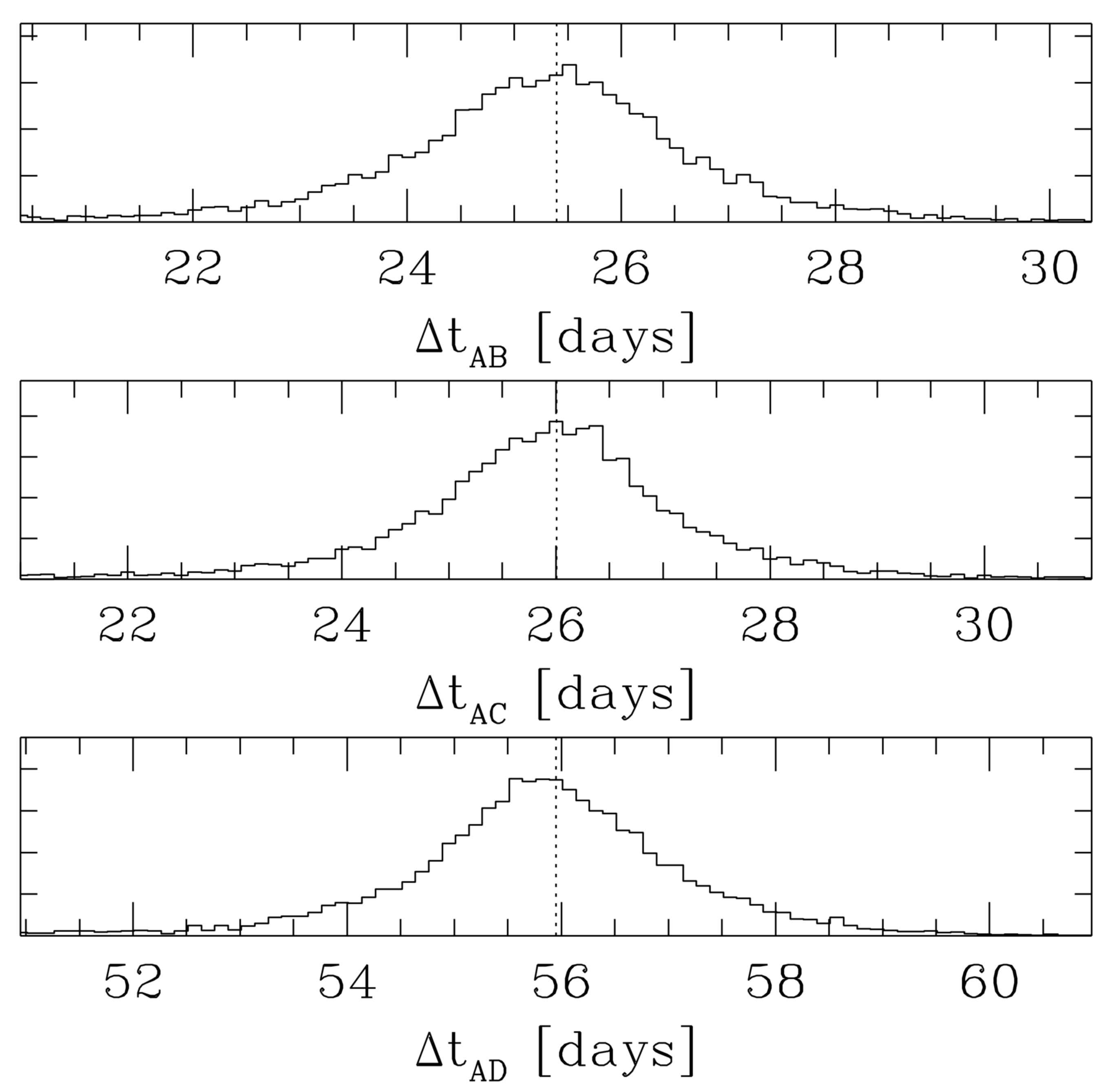}
\end{minipage}
\end{minipage}
\end{center}
\caption{{\it Left:} sample mass map of CDM substructure (after
  subtracting away the smooth halo) from
semi-analytic models by \citet{Z+B03}. 
The points indicate example lensed image positions, and the critical
curve is shown.  
{\it Right:} histograms of the time delays between the images, for
$10^4$ Monte Carlo simulations of the substructure (random positions
and masses).  The dotted lines show what the time delays would be
if all the mass were smoothly distributed.  \citep[Figures from][]{Kee09}
}
\label{fig:sl:cdmsub:cdmsim}
\end{figure} 

In addition to providing flux ratios, LSST will open the door to using
time delays as a new probe of CDM substructure.  
\autoref{fig:sl:cdmsub:cdmsim} shows an example of how time delays are
perturbed by CDM subhalos.  The perturbations could be detected either
as residuals from smooth model fits \citep{K+M09} or as inconsistencies
with broad families of smooth models \citep{congdon-tdel}.  Time delays
are complementary to flux ratios because they probe a different moment
of the mass function of CDM subhalos, which is sensitive to the physical
properties of the dark matter particle.  Also, time delay perturbations
are sensitive to the entire population of subhalos in a galaxy, whereas
flux ratios are mainly sensitive to subhalos projected in the vicinity
of the lensed images \citep{K-stochastic}.

The number of LSST lenses with time delays accurate enough 
($\lesssim 1$ day) to constrain CDM substructure will depend on the
cadence distribution and remains to be determined.  It is clear, though,
that LSST will provide the first large sample of time delays, which will
enable qualitatively new substructure constraints that provide indirect
but important astrophysical evidence about the nature of dark matter.

\subsubsection{The Evolution of CDM Substructure}

The lens galaxies and source quasars LSST discovers will span a
wide range of redshift, and hence cosmic time (\autoref{sec:sl:yield}).  
The sample will be large enough that we can search for any change in
the amount of CDM substructure with redshift/time.  Determining whether
the amount of CDM substructure increases or decreases with time will
reveal whether the accretion of new subhalos or the tidal disruption
of old subhalos drives the abundance of substructure.  Also, the cosmic
evolution of substructure is a key prediction of dark matter theories
that is not tested any other way.


\subsection{Microlensing Densitometry}
\label{sec:sl:densitometry}

It is not surprising that the amplitude and frequency of microlensing
fluctuations are sensitive to the density of stars in the vicinity
of the lensed images.  What may be less obvious is that microlensing
is sensitive to the density of smoothly distributed (i.e., dark)
matter as well.  The reason is twofold: first, the global properties
of the lens basically fix the \emph{total} surface mass density
at the image positions, so decreasing the surface density in stars
must be compensated by increasing the surface density in dark matter;
second, there are nonlinearities in microlensing such that the smooth
matter can actually enhance the effects of the stars \citep{S+W02}.
These effects are illustrated in \autoref{fig:sl:cdmsub:doblermulens}.

The upshot is that measuring microlensing fluctuations can reveal
the relative densities of stars and dark matter at the positions of
the images.  This makes microlensing a unique tool for measuring
\emph{local densities} (as opposed to integrated masses) of dark
matter in distant galaxies.  The large LSST sample of microlensing light
curves will make it possible to measure stellar
and dark matter densities as a function of both galactic radius and
redshift.

\section{ Accretion Disk Structure from 4000 Microlensed AGN }
\label{sec:sl:agn}

\noindent{\it George~Chartas, Charles R. Keeton, Gregory~Dobler}  


Microlensing by stars in the lens galaxy creates independent variability
in the different lensed images.  With a long, high-precision monitoring
campaign, the microlensing variations can be disentangled from intrinsic
variations of the source.  While light bending is intrinsically
achromatic, color effects can enter if the effective source size varies
with wavelength (see \autoref{fig:sl:cdmsub:doblermulens}).  Chromatic
variability is indeed observed in lensed AGN, indicating that the
effective size of the emission region -- the accretion disk -- varies
with wavelength.  This effect can be used to probe the temperature
profile of distant accretion disks on micro-arcsecond scales.  The
LSST sample will be two orders of magnitude larger than the lensed
systems currently known, so we can
study accretion disk structure as function of AGN luminosity, black
hole mass, and host galaxy properties.

This is a joint project with the AGN science collaboration.
\autoref{sec:agn:lens} contains the AGN science case; here we discuss
very briefly the microlensing physics.



\subsubsection{Quasar Accretion Disks under a Gravitational Microscope}
 

Individual stars in a lens galaxy cause the lensing magnification to
vary across micro-arcsecond scales.  As the quasar and stars move, the
image of the accretion disk responds to the changing magnification, leading to variability
that typically spans months to years but can be more rapid when the
source crosses a lensing caustic. We show typical 
microlensing source-plane magnification maps in \autoref{fig:sl:cdmsub:doblermulens};
the caustics are the bright bands.  The variability amplitude depends
on the quasar size relative to the Einstein radius of a star (projected
into the source plane), which is
\begin{equation}
  R_{\rm E} \sim 5\times10^{16} {\rm cm} \times \left( \frac{m}{M_{\odot}} \right)^{1/2}
\end{equation}
for typical redshifts.  According to thin accretion disk theory, the
effective size of the thermal emission region at wavelength $\lambda$ is
\begin{equation}
  R_{\lambda} \simeq 9.7\times10^{15} {\rm cm} \times 
     \left( \frac{\lambda}{\mu{\rm m}} \right)^{4/3}
     \left( \frac{M_{\rm BH}}{10^9 M_{\odot}} \right)^{2/3}
     \left( \frac{L}{\eta L_{\rm Edd}} \right)^{1/3},
\end{equation}
where $\eta$ is the accretion efficiency.  By comparing the variability
amplitudes at different wavelengths, we can determine the relative
source sizes and test the predicted wavelength scaling.  With black
hole masses estimated independently from emission line widths, we can
also test the mass scaling.  These methods are in use today
\citep[e.g.,][]{Koc++06}, but the expense of dedicated monitoring has
limited sample sizes to a few.  

\section{ The Dust Content of Lens Galaxies }
\label{sec:sl:dust}

\noindent{\it \a'Ard\a'\i s~El\'iasd\'ottir, Emilio E. Falco} 



The interstellar medium (ISM) in galaxies
causes the extinction of light passing through it, with the dust particles
scattering and absorbing the incoming light and re-radiating them as thermal
emission.  The resulting extinction curve, i.e., the amount of dust extinction
as a function of wavelength, is dependent on the composition, amount, and grain
size
distribution of the interstellar dust.   
Therefore, extinction curves
provide important insight into the dust
properties of galaxies. 

Probing dust extinction at high redshift is a challenging task: the
traditional method of comparing lines of sight to two standard stars is not
applicable, since individual stars cannot be resolved in distant galaxies.  
Various methods have been proposed to measure extinction at high redshift,
including analysis of SNe curves 
\citep{1998AJ....116.1009R,1997ApJ...483..565P,1996ApJ...473..588R,
2000ApJ...539..658K, 2006ApJ...645..488W}, gamma-ray burst light curves
\citep{2004A&A...427..785J, 2008arXiv0810.2897E}, comparing reddened (i.e.,
dusty) quasars to standard quasars \citep[see e.g.,][]{1991ApJ...378....6P, 
2004MNRAS.354L..31M, Hopkins++04, 2005AJ....130.1345E, 2006MNRAS.367..945Y,
1997ApJ...488L.101M},  and lensed quasars.

Gravitationally lensed multiply imaged background sources provide two or four
sight-lines through the deflecting galaxy (see \autoref{fig:sl:lenscartoon}), 
allowing the
differential extinction curve of the intervening galaxy to be deduced.  
This
method has already been successfully applied to the current, rather small, 
sample of 
multiply imaged quasars 
\citep[see
e.g.,][]{1999ApJ...523..617F,2003A&A...405..445W,
2004ApJ...605..614M,2004AN....325..135W,2005MNRAS.360L..60G,
2006ApJS..166..443E}.  

As discussed in \autoref{sec:sl:yield}, it is expected that the LSST
will discover several thousand gravitationally lensed quasars with lens
galaxy redshifts ranging from $0$--$2$.
In addition, around 300 gravitationally lensed SNe
are expected to be found.  This combined
sample will allow us to conduct statistical
studies of the extinction properties of high redshift galaxies and the
evolution of those dust distributions with redshift.  
%
Furthermore, 
it will be possible to determine the differences in the extinction
properties as a function of galaxy type.  Although lensing galaxies are
predominantly early-type, we expect that $20$--$30$\% may be
late-type.  The study of dust extinction properties of spiral galaxies
will be especially relevant for SNe studies probing dark energy, as
dust correction could be one of the major sources of systematic error in
the analysis.  The lensing sample will provide an independent and
complementary estimate of the dust extinction for use in these
surveys.
%
All these studies will be possible
as a function of radius
from the center of the lensing galaxy, since the images of 
lensed quasars 
typically probe
lines of sight at different radii.

One of the major
strengths of the LSST survey is that it will monitor all lens systems
over a long period, making it possible to correct for contamination
from both microlensing and intrinsic variation in the background
object.  In addition, for the lensed SNe, once the background source
has faded, it will be possible to do a followup study of the dust
emission in the lensing galaxy.  This will make it possible for the
first time to do a comparative study of dust extinction and dust
emission in galaxies outside the Local Group.


\subsection{Lens Galaxy Differential Extinction Curves with the LSST}

The LSST sample
will yield extinction curves from the ultraviolet to the infrared;
different regions of the extinction curve 
will be sampled in different redshift bins.  The
infrared slope of extinction curves will be best constrained by the
lower-redshift systems, whereas the UV slope will be best constrained by
the higher-redshift systems.  One of the most prominent features of
the Milky Way extinction curve, a ``bump'' at about $2200$~\AA\
of excess extinction, possibly due to  
polycyclic aromatic hydrocarbons (PAHs), will be probed by any system at a
redshift of $\zd \gtrsim 0.4$.

%

It is important to keep in mind that the derived extinction curves will be 
{\it
differential} extinction curves  and not absolute ones, except in the limit
where one line of sight is negligibly extinguished compared to the other. 
Therefore, the derived total amount of extinction (e.g., scaled to the V-band,
$A_{\rm V}$), is always going to be a lower bound on the absolute extinction 
for the
more extinguished line of sight.

\begin{figure}
\centering\includegraphics[width=12cm]{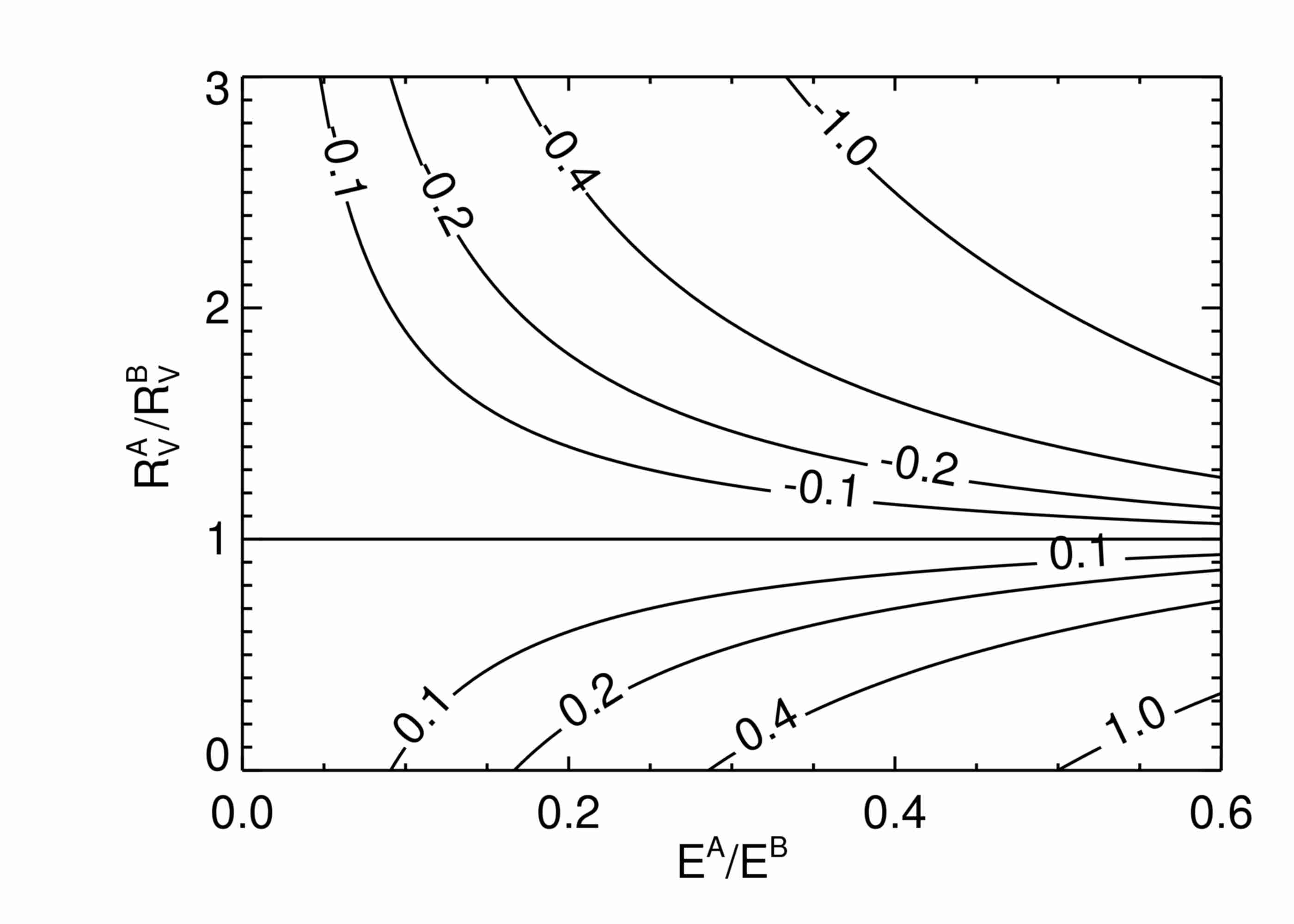}
\caption{A contour plot showing the bias in the derived value of 
$R_{\rm V}^{\rm diff}$
due to extinction along both lines of sight.   The bias is negligible if the
extinction along one line of sight dominates the other
or if the $R_{\rm V}$'s along
the two lines of sight are similar. Figure taken from
\citet{2006ApJS..166..443E}.\label{fig:eta}}
\end{figure}
The differential extinction law is of the same type for both linear
(Small Magellanic Cloud, SMC) and Milky Way (MW)  
extinction laws.  
Defining $R_{\rm V}\equiv A_{\rm V}/E$, where the reddening~$E = E(B-V)$
between the B and V bands is given by
$E \equiv A_{\rm B}-A_{\rm V}$,
we can write the differential extinction between images A and B 
for MW dust as \citep[for details, see
][]{2006ApJS..166..443E}:
\begin{eqnarray}
\frac{R_{\rm V}^{\rm diff}}{R_{\rm V}^{\rm B}} & = & \frac{E^{\rm B} - E^{\rm A} R_{\rm V}^{\rm A}/R_{\rm V}^{\rm B}}
                                         { E^{\rm B} - E^{\rm A}} \\
   & = & 1 + \frac{E^{\rm A}}{E^{\rm B}-E^{\rm A}}\left( 1 -
   \frac{R_{\rm V}^{\rm A}}{R_{\rm V}^{\rm B}}\right) \nonumber \\
   & \equiv & 1 + \eta \nonumber.
\label{eq:rdiff}
\end{eqnarray}
This says that the error due to the non-zero extinction of image A 
that is introduced when using $R_{\rm V}^{\rm diff}$ 
as an estimator for $R_{\rm V}$ is
\begin{equation}
\eta = \frac{E^{\rm A}/E^{\rm B}}{1-E^{\rm A}/E^{\rm B}}\left( 1 - \frac{R_{\rm V}^{\rm A}}{R_{\rm V}^{\rm B}}\right).
\label{eq:eta}
\end{equation}

A biased estimate will only arise if either the amount of
extinction along the two lines of sight is very similar, or if the type of
extinction along the two lines of sight is very different (see
\autoref{fig:eta}).   In the first case, the measured differential
extinction will, however, be close to zero, so these systems can be
automatically
excluded from the sample.  In the second case, we must consider 
how much the extinction law might vary within a given galaxy.  
For SMC, LMC
and MW dust most lines of sight have very similar values of $R_{\rm V}$, 
but outliers do
exist.  For example, if we take the most extreme values of the Milky Way
we need $E^{\rm A}/(E^{\rm B}-E^{\rm A})\le0.05$ for reaching a $10\%$ accuracy in $R_{\rm V}$.


\subsection{Dealing with Microlensing and Intrinsic Variations}

While lensing is in general achromatic,
microlensing can resolve color gradients in accretion disks
(\autoref{sec:sl:agn}), therefore can cause a color 
variation which may mimic dust extinction.
Microlensing typically causes changes in
brightness $\lesssim
1$~mag, and the chromatic variation is typically $\lesssim 10$\%, so for
strongly extinguished systems this effect is expected to be minor
\citep[see e.g., ][]{2008arXiv0810.1626M}.  
However, as LSST will monitor the lenses over long time spans,
independent estimates of the potential microlensing bias
will be obtained, and a correction can be made if necessary.
%
Likewise, the intrinsic (and chromatic) 
variability of the AGN can be corrected for, provided
the lens time delay between images is known. The long-term monitoring will be
essential for this: how well the time delayed intrinsic variations can be
removed with three-month monitoring seasons is to be demonstrated.


\subsection{Follow-up Observations}

It would be desirable to obtain X-ray measurements for at least a subsample of
the gravitationally lensed systems studied in this survey \citep[see e.g.,
][]{2008arXiv0803.1679D}.  
In the case where dust extinction is detected, an
X-ray measurement can provide the neutral hydrogen column density along
the two (or four) image sight-lines.  
This can be used to check the reliability that one line
of sight is significantly less extinguished than the other 
and  to
estimate the dust-to-gas ratio.  
Also, one can get an estimate
of how many of the ``zero differential extinction" systems contain dust 
and how many of them are
truly dust free.  
An X-ray telescope with a resolution of the order of $\sim 1\arcsec$ will be
required for such a followup to resolve the different lines of sight.  Chandra
could easily be used if still in operation, while the planned resolution for
the International X-ray Observatory (IXO) would limit the study to large separation lenses.

In the case of the multiply imaged SNe, a deep study of the lensing galaxies
can be obtained once the SNe have faded.  The PAHs thought to give
rise to the bump in the MW extinction curve 
have characteristic emission lines in the infrared which can
be measured to give an estimate on the PAH abundance 
\citep[see e.g., ][]{2007ApJ...663..866D}.
With the lensing information about the 
type of dust extinction, it would be possible
to correlate the strength of the bump to the PAH abundance.  If the bump is
due to the PAHs, a strong correlation should be seen.


\subsection{Technical Feasibility}

LSST data are expected to have ample sensitivity and dynamic range to yield a
sample of lensed quasars with measured extinction curves up to two orders of
magnitude larger than the existing one.  Clear requirements for extracting
high-quality extinction curves are broad wavelength coverage, seeing
consistently below 1 arcsec, and stable and consistent PSFs.  The high
resolution and broad wavelength coverage make the LSST data set ideal for such
a dust study while the frequent sampling with a consistent set of filters
will  be crucial to address the systematic effects of microlensing and time
delays.  While PS1 and DES are expected to yield similar data sets, they lack 
the frequent sampling required to deal with these systematic effects.  

A reasonably accurate redshift for the lensing galaxy is 
also required to
calibrate the extinction curves and to reliably search for the excess
extinction around 2200~\AA.  
The photometric redshifts from the LSST should be sufficient:
an uncertainty of $0.04(1+z)$ per galaxy
(\autoref{sec:common:phz:cal}) in the photometric redshift translates to an
$\sim150$~\AA\ uncertainty in the center of the bump for a lensing galaxy at
$z\sim0.8$, which is smaller than the typical width of the bump.  

The PAH abundance study will require additional follow-up observations of
selected systems in the infrared.  The PAH emission features range from
$3$--$11$~$\mu$m in the rest-frame which will correspond to a range of
$7$--$23$~$\mu$m for redshifts of $z=1.0$.  The planned MIRI spectrograph on
the JWST has imaging capabilities and a spectrograph capable of covering this
range with a sensitivity far superior to that of current facilities.  

\section{Dark Matter Properties from Merging Cluster Lenses}
\label{sec:sl:dmclus}

\def\bulletcluster{1E0657$-$56}
\def\babybullet{MACS~J0025.4$-$1222}
\def\bulletii{MACSJ2243.3\nobreakdash--0935}
\def\rxj{RX~J1347$-$1145}
\noindent{\it Maru\v{s}a~Brada\v{c}, Phil~Marshall, Anthony~Tyson}  


Clusters of galaxies are composed of large amounts of dark matter. These clusters are unique in their power to directly probe and place limits on the
self-interaction cross-section of dark matter \citep{clowe06, bradac06,
bradac08b}.  Furthermore, in clusters we can probe the
spatial distribution of dark matter and its interplay with the baryonic
mass component \citep[e.g., as shown with RXJ1345$-$1145;][]{bradac08}, and
thereby study effectively the formation and evolution of clusters, one
of the more robust predictions of currently favored $\Lambda$CDM
cosmologies.


\subsection{Merging Galaxy Clusters as Dark Matter Laboratories}

The standard Cold Dark Matter model makes definite predictions for the
characteristics of dark matter halos, including their inner slopes and
concentrations.  These predictions have not been tested accurately
yet.  Massive clusters are the best places we know for doing so; the
dark matter distribution in clusters even has the potential of
constraining the interaction cross-sections of the dark matter
particles themselves.

\begin{figure*}[!t]
\centering
\begin{minipage}{0.55\linewidth}
\centerline{\includegraphics[width=0.9\linewidth]{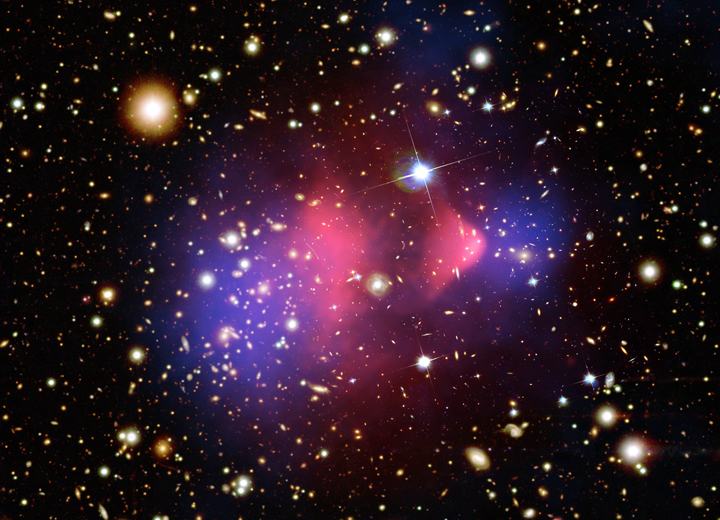}}
\end{minipage}\hfill
\begin{minipage}{0.4\linewidth}
\centerline{\includegraphics[width=0.9\linewidth]{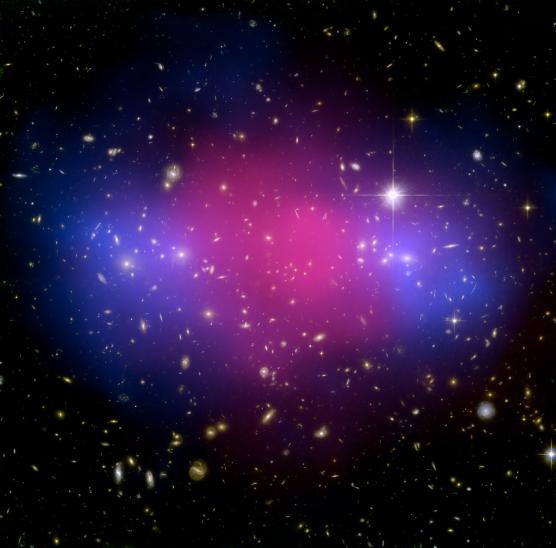}}
\end{minipage}
\caption{The color composite of the Bullet cluster \protect{\bulletcluster}
 ({\it left}) and  {\babybullet} ({\it right}). Overlaid in {\it blue} shade is the 
 surface mass density map from the weak lensing mass reconstruction. 
 The X-ray emitting plasma is shown in {\it red}. 
 Both images subtend $\sim10$~arcmin on the vertical axis.
 Credit (left): X-ray NASA/CXC/CfA Optical: NASA/STScI;
  Magellan/U.Arizona; \protect{\citet{clowe06,bradac06}} 
  (right) X-ray (NASA/CXC/Stanford/S.Allen); 
  Optical/Lensing (NASA/STScI/UCSB/M.Bradac); \protect{\citet{bradac08}}.}
\label{fig:bullet}
\end{figure*}

The most striking examples of such investigations to date have been the
Bullet cluster {\bulletcluster} \citep{clowe06,bradac06} and
{\babybullet} \citep{bradac08b}.  These are examples of merging or
colliding clusters, where the interaction has happened in the plane of
the sky; the (almost) collisionless dark matter and galaxies have
ended up with different, more widely-separated distributions than the
collisional X-ray emitting gas, which remains closer to the
interaction point.  In these systems, the positions of the
gravitational potential wells and the dominant baryonic component are
well separated, leading us to infer the clear presence and domination
of a dark matter mass component (see \autoref{fig:bullet}). A union of
the strong lensing data (information from highly distorted arcs) and
weak lensing data (weakly distorted background galaxies) for the
cluster mass reconstruction has been demonstrated to be very
successful in providing a high-fidelity, high signal-to-noise ratio mass
reconstruction over a large area \citep{bradac06, bradac08b}. It was
this gravitational lensing analysis that first allowed the presence of
dark matter to be confirmed, and then limits on the self-interaction
cross-section of dark matter particles to be estimated. The latter are
currently at $\sigma/m < 0.7 \,\mbox{cm}^2\mbox{g}^{-1}$
\citep{randall08}. LSST can act as a finder for more massive, merging systems,
and allow accurate lensing 
measurements of the relative positions of the stellar and
dark matter distributions.


\subsection{Breaking Degeneracies with Multiple Data Sets}

Merging clusters are, however, not the only places where dark matter
can be studied. It was first proposed by \citet{navarro97} that the
dark matter halos on a variety of scales should follow a universal
profile (the so-called Navarro Frenk and White or NFW profile) within
the currently accepted $\Lambda$CDM paradigm. The three-dimensional density
distribution of dark matter should follow $\rho_{\rm DM} \propto
r^{-1}$ within a scale radius $r_{\rm s}$ and fall off more steeply
at radii beyond that ($\rho_{\rm DM} \propto r^{-3}$). In practice,
however, the density profiles of simulated clusters appear to be less
concentrated and often at odds with gravitational lensing observations
\citep{limousin07,Bro++05,sand08}. A central baryon enhancement
that could explain these discrepancies is not observed, leaving us
with a puzzle.  However, now for the first time supercomputers and
simulations have become powerful enough to give clear predictions of
not only the distribution and amounts of dark matter, but also its
interplay with the baryons and the effects the baryons have on the
formation of dark matter halos (e.g., through adiabatic contraction,
\citealp{gnedin04,nagai07}).

 It is, therefore, high time that the predictions of simulations are
 paired with state-of-the-art observations of evolving clusters. A
 high-resolution, absolutely-calibrated mass map of galaxy clusters in
 various stages of evolution at all radii will allow us to measure the
 slopes of dark matter and baryonic profiles, which are a critical
 test of cosmology and a key to understanding the complicated baryonic
 physics in galaxy clusters. Several works have previously studied
 mass distributions in number of clusters using combined
 strong (information from highly distorted arcs) and weak (weakly
 distorted background galaxies) lensing reconstruction \citep[see
   e.g.,][]{bradac06,bradac08,natarajan96,kneib03, marshall02, diego05, Jee07,
   limousin07} and combined strong lensing and stellar kinematics data
 of the dominating central galaxy \citep{sand08}. These approaches
 offer valuable constraints for determining the mass distributions. At
 present, results range from consistent to inconsistent with
 $\Lambda\mbox{CDM}$ (\citealp{bradac08,sand08,medezinski09}, see also
 \autoref{fig:rxj}). However, these studies all lack either a
 sophisticated treatment of baryons or a self-consistent combination
 of data on either large (around the virial radius) or small scales
 (around the core radius). It is, therefore, crucial to use a
 combination of weak and strong lensing data, matched with the method
 allowing mass reconstruction in the full desired range, from the
 inner core ($\sim100\mbox { kpc}$) to the outskirts
 ($\gtrsim1000\mbox { kpc}$), with accuracy in total mass estimates of
 $\lesssim 10\%$, and proper account of baryons.


\subsection{A Kilo-cluster Sample with LSST}

LSST will be well-placed to support this project, since it combines
multi-color high-resolution imaging over a large field of view. This
will allow us to detect clusters in two different ways: by optical and
photometric redshift overdensity and by weak lensing shear strength.
We can expect the LSST massive cluster sample to number in the thousands
(\autoref{sec:sl:yield}, \autoref{sec:wl:shselcl}), 
with the fraction of clusters showing strong lensing
effects increasing to unity at the high mass end.

Once found, the high quality cluster images will permit the
identification of large numbers of 
strongly lensed multiple imaged systems, 
as well as allow us to perform
weak lensing measurements all the way out to the virial radius
($\gtrsim1000\mbox{ kpc}$). One significant technical challenge will be to
automate the multiple image system identification procedure: currently, each
individual cluster is analyzed in great detail, with a mass model being 
gradually built up as more and more systems are visually identified.

The statistical
uncertainties in the measured photometric redshifts of these systems will
be more than adequate for this project. The ``catastrophic outliers''
do not pose a significant problem, since the geometry of lensing is
redshift-dependent, allowing us to distinguish a low from
high-redshift solution. The spectroscopic follow-up is therefore not
crucial, however it eases the analysis considerably, increases the
precision, and can be achieved with a moderate/high investment (300
nights on a 10-m class telescope for a total of 1000 clusters). 

To
study the influence of baryons, auxiliary X-ray data to study the gas
distribution in the clusters will be needed. Many clusters already have such data
available (typically one needs $20\mbox { ks}$ exposures with Chandra
for a $z\sim0.3$ cluster), and we expect many more to be covered with
current and future X-ray surveys. To probe dark matter properties with
merging clusters, an additional requirement will be a deeper X-ray
follow up. These follow-ups are currently already possible with
Chandra and since the sample for this part of the study is much
smaller (10-100), we expect to be able to use future X-ray missions to
achieve this goal.

With a sample of $\gtrsim 1000$ clusters capable of strong lensing
paired with accurate mass reconstruction from the very center to the
outskirts, LSST will be able to achieve a number of interesting science
goals:
\begin{enumerate}
\item Clusters of galaxies are unique in their power to directly probe
  and place limits on the self-interaction cross-section of dark
  matter. With a subsample of clusters that are merging clusters ($10-100$ out of $\sim 1000$ capable of strong lensing), these
  limits can be significantly improved and systematic errors inherent
  to studies of single clusters (1E0657$-$56, MACSJ0025-1222) can be
  reduced to negligible amounts.
\item Studying the distribution of dark matter in
  $\gtrsim 1000$ clusters of galaxies will allow
  us to follow the growth of dark matter 
  structure through cosmic time, including its
  interplay with the baryonic mass component, thereby allowing us to
  effectively study cluster formation and evolution and test a scenario
  which is one of the more robust predictions of currently favored
  $\Lambda$CDM cosmology. Examples of CDM-predicted quantities 
  that can be probed are the profile concentration (and its relation to halo
  mass and redshift), halo ellipticity, and substructure mass function.
\item Well-calibrated clusters can also be used as cosmic telescopes
  (see \autoref{sec:sl:telescopes}), thereby enabling the study of
  intrinsically lower luminosity galaxies than would otherwise be
  observable with even the largest telescopes.
\end{enumerate} 

\section{LSST's Giant Array of Cosmic Telescopes}
\label{sec:sl:telescopes}

\noindent{\it Maru\v{s}a~Brada\v{c}, Phil~Marshall}  

What are the sources responsible for cosmic reionization?  The most
efficient way to study galaxy populations shortly after the reionization
epoch is to use clusters of galaxies as gravitational telescopes. With a
cluster-scale gravitational lens one can gain several magnitudes of
magnification, enabling the study of intrinsically lower luminosity
galaxies than would otherwise be observable with even the largest
telescopes \citep[e.g.,][]{ellis01,richard06,hu02}. With a sample of
well-chosen clusters (to achieve the best efficiency and to beat the
cosmic variance) the properties of these first galaxies can be
determined, enabling us to address the question of whether these objects
were responsible for reionizing the Universe. Furthermore, gravitational
lensing is very efficient in rejecting possible interlopers (cool stars,
$z \approx 2$ old galaxies) that plague such surveys in the field
environment.


\subsection{Galaxy Clusters as Tools to Explore Reionization}

Theoretical studies suggest that the Universe
underwent a transition from highly neutral to a highly ionized
state in a relatively short period (``reionization'') at $z>6$
\citep{dunkley08}. It is thought that $z>6$ 
proto-galaxies were responsible for this process. However, the
luminosity function of $z\gtrsim 7$ objects is quite uncertain (e.g.,
\citealp{stanway08,henry07,stark06,bouwens08}), as is their role in
reionization.  
If these objects did indeed reionize the Universe,
non-standard properties (such as unusually high abundance of faint
sources, large stellar masses, and/or very low metallicities) 
may need to be invoked. These results are based on tiny samples ($\lesssim 10$)
and so need to be confirmed with larger samples across different
patches of sky in order to beat the sample variance. 
Finding more sources at the
highest redshifts is therefore crucial.  Observations at these high
redshifts are extremely challenging, not only due to the large
luminosity distance to these objects, but also due to their intrinsically low
luminosity
(stemming from their presumably lower stellar masses
compared to moderate redshifts).


\subsection{Observational Issues}

One can find high redshift
galaxies by searching for the redshifted Lyman break using broad-band
photometry (\autoref{sec:gal:highz}. $z\simeq 6$ objects will not be
detected at $i$ and blueward, and $z\simeq 7$\nobreakdash--$8$
will be $z$-band dropouts (e.g., \citealp{henry07,stanway08, bouwens08}). The
main limitations of experiments to look for such objects to date have been 
the small fields examined and the difficulty of spectroscopic
confirmation. 

\begin{figure}
\centering
\begin{tabular}{cc}
\begin{minipage}{0.45\textwidth}
\centerline{\includegraphics[width=0.9\textwidth]{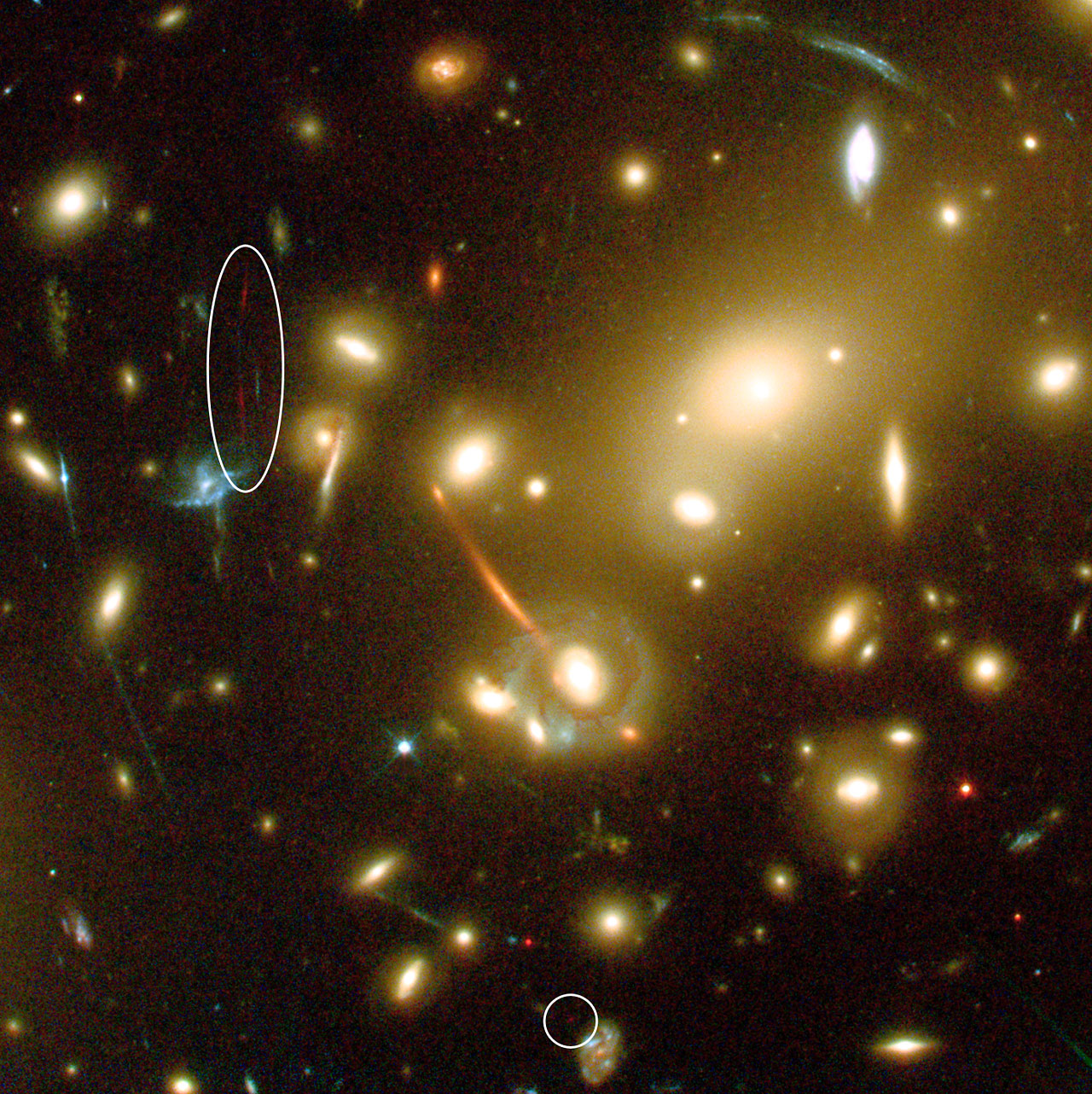}} 
\end{minipage} &
\begin{minipage}{0.45\textwidth}
\centerline{\includegraphics[width=0.9\textwidth]{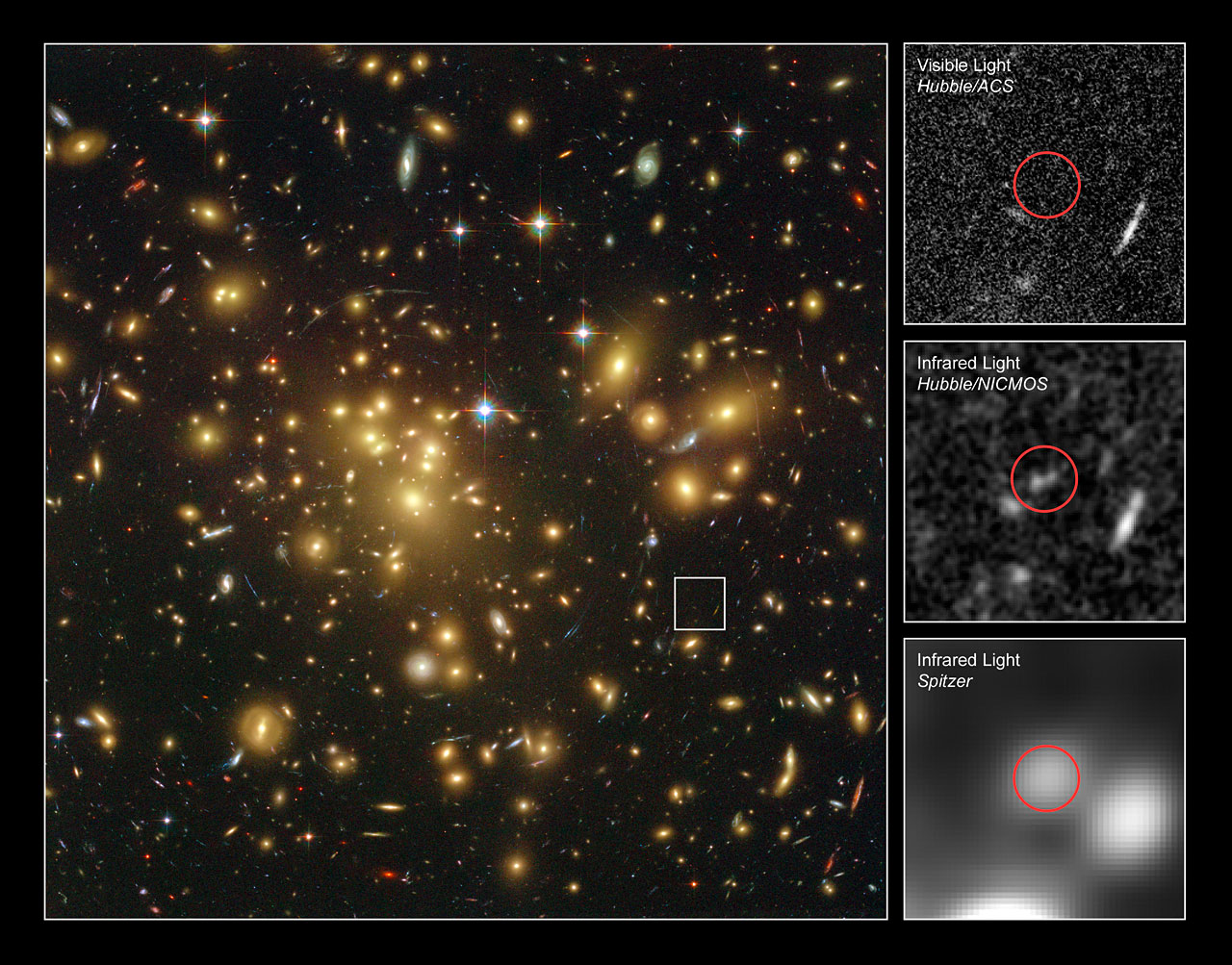}}
\end{minipage}\\[-0.2cm]
(a) & (b) \\
\begin{minipage}{0.45\textwidth}
\centerline{\includegraphics[width=0.9\textwidth]{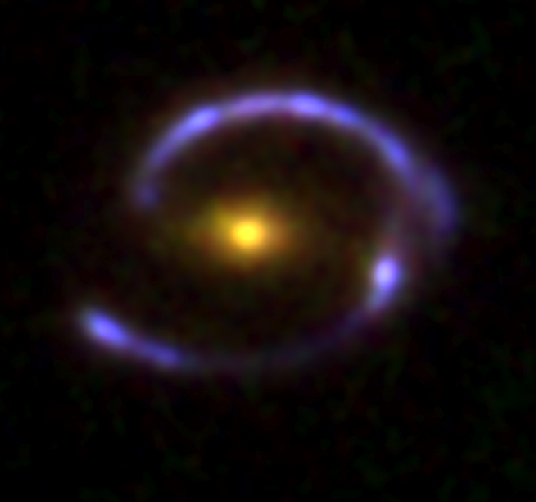}} 
\end{minipage} &
\begin{minipage}{0.45\textwidth}
\centerline{\includegraphics[width=0.9\textwidth]{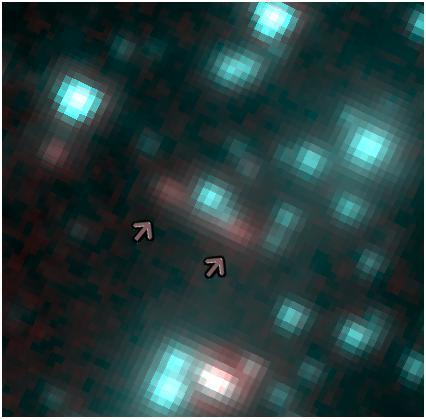}}
\end{minipage}\\[-0.2cm]
(c) & (d) \\
\multicolumn{2}{c}{%
\begin{minipage}{\textwidth}
  \caption{{\it (a)} This close-up of the galaxy cluster Abell 2218  shows a
  $z\sim 7$ galaxy that was magnified by the cluster acting as a  gravitational
  telescope \protect{\citep{kneib04}}. {\it (b)} $z \sim
  7.6$ galaxy behind A1689 \protect{\citep{bradley08}}. {\it (c)} LBG J2135-0102
  (also known as the ``Cosmic Eye'') is a typical star-forming galaxy at $z=3.07$.
  Resolved spectroscopy was made possible for this high redshift, regular
  star-forming galaxy because of the magnifying power of the foreground galaxy
  \protect{\citep{smail07}}.  {\it (d)} Bright IRAC (shown) and sub-mm
  source at a redshift of $z=2.7$, lensed by the Bullet Cluster \protect{\citep{gonzalez09}}. \label{fig:ctel}}
\end{minipage}%
}
\end{tabular}
\end{figure}

Both these observational issues can, however, be addressed when using
galaxy clusters as {\it gravitational telescopes}. This technique was
proposed shortly after the discovery of the first arcs in galaxy
clusters \citep{soucail90} and has been consistently delivering record
holders in the quest for the search for high redshift galaxies
(\citealp{kneib04,bradley08}, see also \autoref{fig:ctel}a,b): the
brightest objects of a given class are often, if not always lensed.
This is of course also the case at ``lower'' ($z\sim3$) redshifts,
some of the examples are the ``Cosmic Eye'' \citep{smail07} and the
bright sub-mm galaxy behind the Bullet Cluster (\citealp{gonzalez09,
  wilson08}, see also \autoref{fig:ctel}c,d). The 
magnification effect provided by the deep gravitational potential well
of a massive cluster allows detections of objects more than a
magnitude fainter than the observation limit. Hence, clusters of
galaxies offer the best opportunity to study the faintest,
smallest and most distant galaxies in the Universe.  Due to this
magnification, the solid angle of the search area effectively
decreases --- but since the luminosity function is practically
exponential at the magnitudes one needs to probe, we can make
substantial gains using the lensing magnification. In addition, these
sources are observed with increased spatial resolution. As a result, we
can resolve smaller physical scales than would otherwise be possible,
and begin to actually measure the properties of $z\gtrsim7$ galaxies
on an individual basis.

Searches for $z\gtrsim 7$
objects in the field are plagued by the fact that, based on optical and
IR colors alone, it is very difficult to distinguish between 
$z\gtrsim 7$ objects, old and dusty elliptical galaxies (the 4000$\,$\AA  break at
$z\approx2$ can potentially be mistaken for a Lyman break at $z>7$),
and cool T and L-dwarf stars (see e.g., \citealp{stanway08}). As shown
by these authors, Spitzer/IRAC data can help to exclude some, but not
all, of the interlopers. Gravitational lensing alleviates all these
problems if the objects observed are multiply imaged. Since the
geometry of the multiple images is redshift dependent, we can
not only remove stars as main contaminants, but also remove
contaminant galaxies at redshifts $\lesssim 3$ by using the constrained lens
model to essentially measure a ``geometric redshift,'' and rule out low
redshift false positives.  
In addition, even if the sources are not multiply
imaged, they will likely be highly distorted, allowing one to
discriminate them from stellar objects and 
unlensed lower redshift galaxies.


\subsection{The Need for a Well-calibrated Telescope}

LSST will help us measuring the luminosity function of $z\gtrsim 7$
galaxies by using $\sim 1000$ galaxy clusters as cosmic
telescopes. One will also need $J$ and $H$-band follow-up imaging to a
depth of at least $H_{\rm AB} = 27.5$, which will require future
space-based missions such as JWST.  The resulting sample of $\gtsim2000$
sources at $z\gtrsim7$ will allow us to measure the full shape of the
luminosity function at $z\gtrsim 7$. Comparing 
the results with simulations (see e.g., \citealp{choudhury07}) will
allow us to answer the question of whether this population was
responsible for reionization.

At $z \sim 3-4$, LSST will study the lensed galaxy population to a depth
far beyond the luminosities reached by the deepest field
surveys (\autoref{sec:gal:highz}), albeit surveying a substantially
smaller solid angle. In combination, these surveys
will provide a very accurate luminosity function of Lyman break galaxies (LBGs) from the
bright to the faint end.


%



%
%
%
%
%
%
%
%
%
%
%
%
%
%
%
%
%
%
%
%
%
%
%
%
%
%

\section{Calibrating the LSST Cluster Mass Function using Strong and Weak Lensing}
\label{sec:sl:clusmf}
\noindent{\it M. James~Jee, Maru\v{s}a~Brada\v{c}, Phil Marshall}  


\subsection{Introduction}

\begin{figure}
\begin{center}
\includegraphics[width=0.49\textwidth]{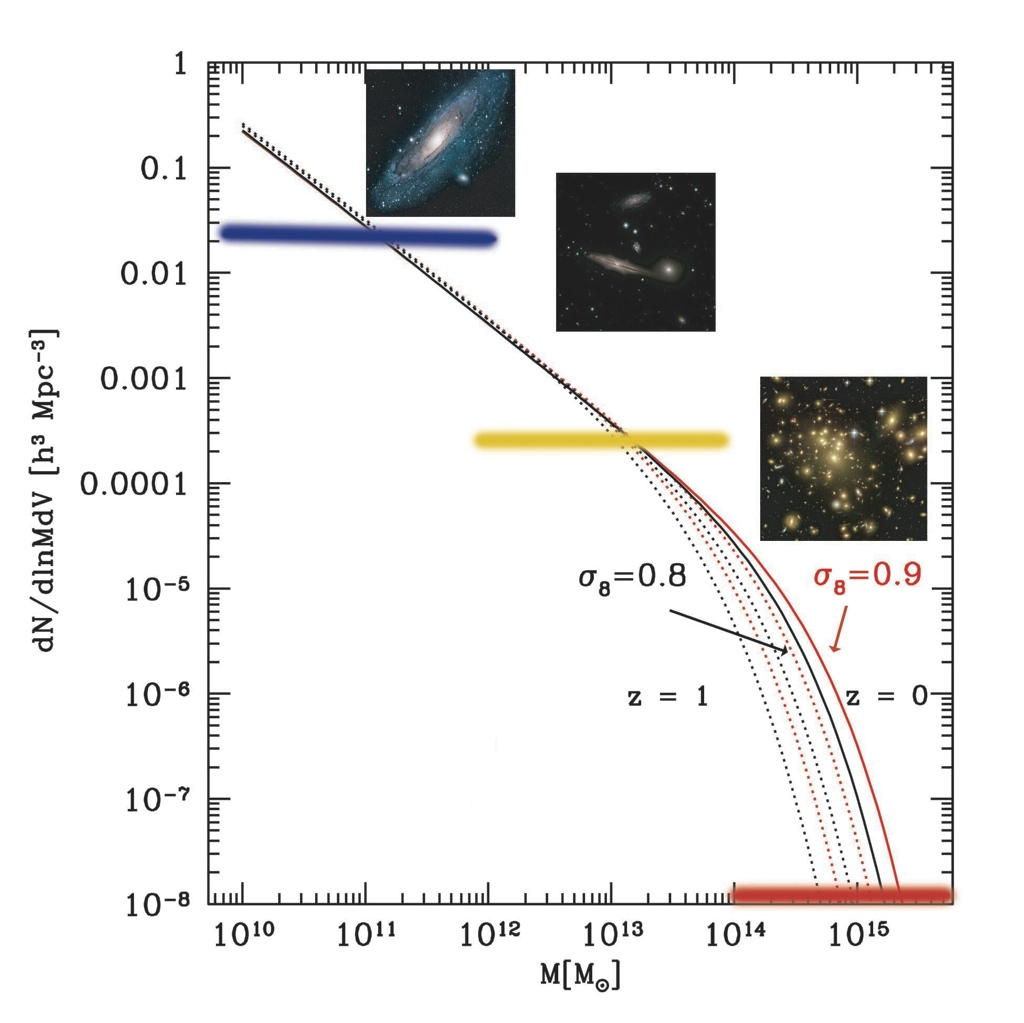}
\includegraphics[width=0.46\textwidth]{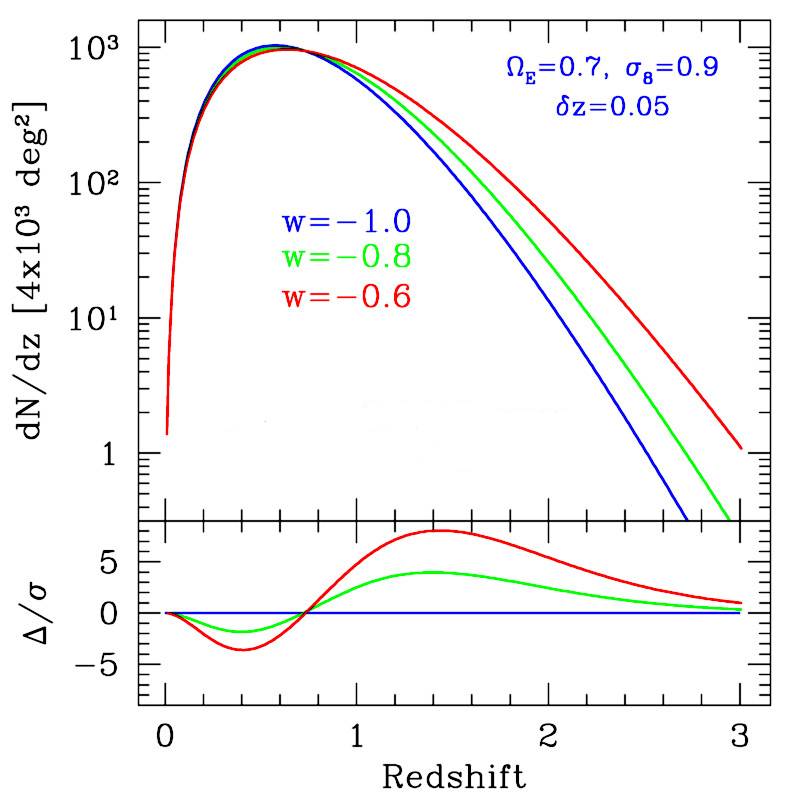}
\end{center}
\vspace*{-0.5cm}
\caption{{\it Left:} schematic representation of the dependence of mass
function on cosmological parameter $\sigma_8$.  The
sensitivity to cosmological parameters is highest at the high-mass end. {\it Right:} the
number of
clusters above an observable mass threshold 
as a function of redshift for different values of dark energy equation-of-state
parameter $w$ \citep{Mohr2005}. 
The volume effect is dominant at $z\sim0.5$ whereas
the  growth rate sensitivity determines the cluster numbers at $z\sim1.5$.
\label{fig:massfct}}
\end{figure}

The cluster mass function $dn(M)/dz$ is one of the four most promising dark
energy probes that the Dark Energy Task Force (DETF) recommends 
\citep{albrecht06}. Both growth and expansion rates due to the presence of
dark energy sensitively affect the abundance of collapsed structures,  
and the sensitivity increases toward the high mass end
(\autoref{fig:massfct}).  The cluster counting cosmology experiment is
introduced in \autoref{sec:lss:cluster}, which we recommend be read
for context for the present discussion. In this section we discuss the
contribution that strong lensing can make towards a better calibrated
cluster mass function.

Most of the galaxy
clusters in the high mass regime have 
core surface densities high enough
to create conspicuous strong lensing features such as multiple
images, arcs, and arclets. A critical curve, whose
location depends on the source redshift, defines the aperture
inside which the total projected mass becomes unity\footnote{
Here the mass is defined in units of the critical density
$\Sigma_{crit}$ defined in \autoref{chapter1:eq:critical}.}
Therefore, more than one multiple
image system at significantly different redshifts enables us
to obtain an absolutely calibrated mass profile.
The superb
image resolution of LSST, as well as the deep six-band
data, will facilitate 
the identification of multiple systems, which can be
automated by robotic searches \citep[e.g.,][]{Marshall09}.
Still, not all clusters will show strong lensing features; we can expect a
sub-set of all weak lensing or optically selected clusters to be suitable for
strong plus weak lensing mass measurement. How this ``calibration set'' should
be optimally combined with the larger sample is a topic for research.


%

An absolutely calibrated mass profile is useful for revealing systematic
errors in weak lensing such as shear calibration, source redshift estimation,
PSF modeling, and so on. 
Even when the cluster field does not reveal more than one
multiple image system at different redshifts, 
the ellipticity of the strongly
distorted galaxies can put  strong constraints on the mass slope, again
because there are many such arclets sampling the gravitational field at many
radii.
Once properly calibrated through the above method,  the signal strength is
often comparable to that from strong lensing. Therefore, the slope  ambiguity
arising from using only a single multiple image system can be effectively
lifted by including the more
numerous arclets, which sample the location of the
critical curves at different redshifts. Currently, one of the major obstacles
in this approach is the lack of reliable redshift information for the
individual sources: the well-calibrated photometric redshifts needed for LSST
weak lensing studies should enable robust statistical cluster analyses as well.

\begin{figure}
\begin{center}
\includegraphics[width=0.505\linewidth]{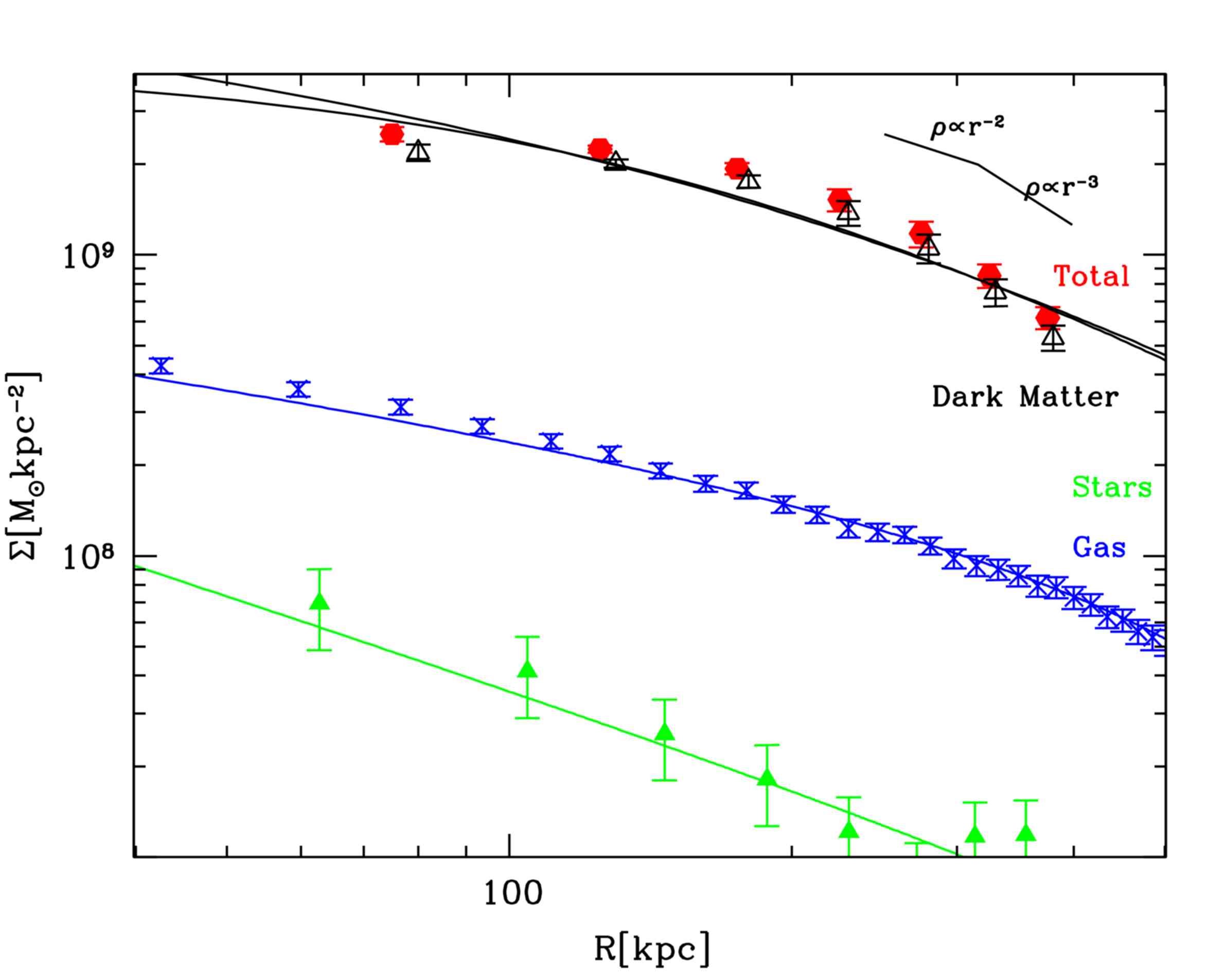}
\includegraphics[width=0.395\linewidth]{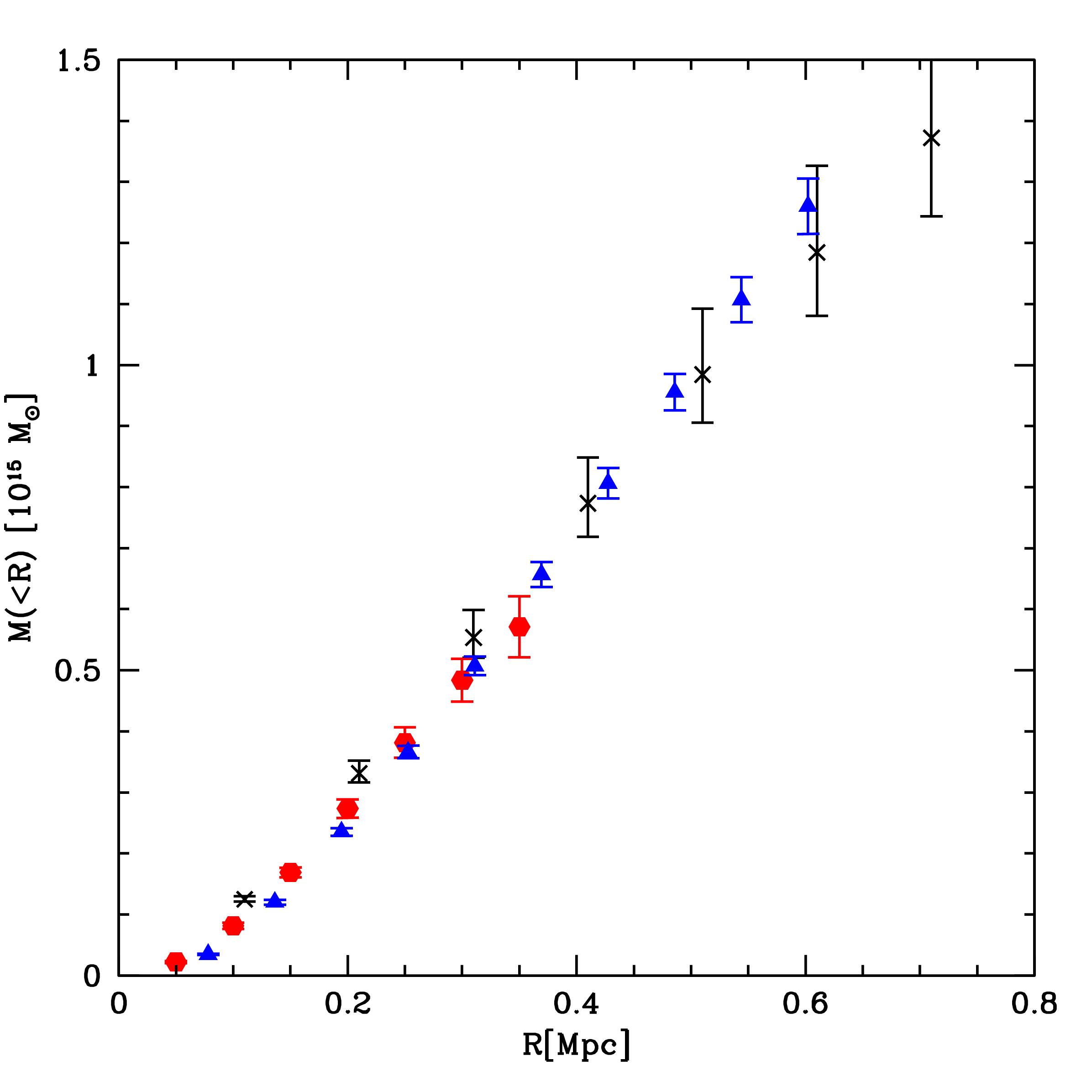}
\end{center}
\vspace*{-0.5cm}
\caption{{\it Left: } the projected mass density
  $\Sigma$ profile of {\protect \rxjthirteenfortyseven } showing separately the stellar
  mass profile (filled triangles -- {\it green}), the gas profile (stars -
  {\it blue}), dark matter profile (open triangles -- {\it black}) and the total
  profile from strong and weak lensing (hexagons -- {\it red}). The dark
  matter profile has been fitted using the generalized NFW model -
  shown is the best fit model (inner slope $\beta = 0$, scale radius
  $r_{\rm s} = 160\mbox{kpc}$), and also a best fit NFW model ($\beta
  = 1$, $r_{\rm s} = 350\mbox{kpc}$). We we combine with constraints on
  larger scale (from weak lensing) we will be able to break these
  degeneracies. {\it Right:} Comparison of the projected total mass
  profiles for {\protect \rxjthirteenfortyseven } 
  determined from the strong-plus-weak
  lensing analysis ({\it red} hexagons), the Chandra X-ray data ({\it black}
  crosses), and strong lensing analysis where a profile slope has been
  assumed ({\it blue} triangles). The presence of significant non-thermal
  pressure support would cause the X-ray measurements to
  underestimate the total mass. Data
  from \protect{\citealp{bradac08}}.}
\label{fig:rxj}
\end{figure}

The most significant merit of LSST in cluster mass estimation lies in its
ability to measure shears even a few Mpcs away from the cluster center 
(see \autoref{chp:wl})
thanks
to its unprecedentedly large field of view. Because the mass-sheet degeneracy
($\kappa \longrightarrow \lambda \kappa + (1 - \lambda)$) 
is small in the low $\kappa$ ($\ll 1$) region, weak lensing cluster 
masses from LSST 
(estimated with e.g., aperture mass densitometry \citep{Fahlman94}
or some variation \citealt[e.g.,][]{Clowe98})
should be relatively free of this effect.
In any case, 
if the weak lensing data are combined
with strong lensing constraints, the mass-sheet degeneracy can be broken in
the intermediate regime (where $\kappa$ is not close to zero) because the
redshift  distribution of the source galaxies serves as an alternative source
plane different from the one defined by the strong lensing galaxies (see
\autoref{fig:rxj} for an example).

Of course, the tangential  shears around the
cluster at a few Mpcs are non-negligibly contaminated by cosmic shear and
thus without correcting the effect, it is not feasible to reduce the total halo
mass
uncertainty to smaller than $\sim10$\% \citep{Hoekstra03}; 
extending the field
beyond the $10-15$ arcmin radius does not decrease the uncertainty. However,
considering the ever improving photometric redshift techniques
(\autoref{sec:common:photo-z})
and the LSST wavelength coverage
from ultraviolet to near-infrared with significant depth, we anticipate
that
future tomographic weak lensing analysis \citep{Hu02b} will provide a high quality,
three-dimensional map of the target field and enable us to separate the cluster mass from
the background structure. How well we will be able to do this in practice can
be determined from ray tracing simulations: such a program will yield a
quantified cluster detection function, essential for accurate cosmology
studies. 

The LSST weak lensing survey will enable the detection of tens of
thousands of clusters over a wide mass range (\autoref{sec:wl:shselcl}), the
number showing detectable strong lensing signatures will be lower. While the 
highest  mass clusters will contain many arcs, clusters of all masses will be
represented in the strong lens sample.
LSST will detect $\gtsim 1000$ clusters with giant arcs 
(\autoref{sec:sl:yield}), and so
provide efficient, unbiased probes of these 
cluster masses from the cluster core
($10-100$ kpc) to well beyond the virial radius ($\gg 1$ Mpc). 
This will offer
a unique opportunity to study the cluster dynamics in unprecedented detail, 
to construct a well-calibrated mass function, and thus to quantify the 
effects of
dark energy.


\subsection{Science with LSST Data Alone}

Most of the calibration methods above can be implemented without follow-up
data. We can summarize the cluster mass function science program in the
following steps:

\begin{itemize}
\item Cluster search with red-sequence and/or weak lensing analysis. This
should provide a well-understood sample of clusters, selected by
mass and/or richness (see \autoref{sec:lss:cluster} and 
\autoref{sec:wl:shselcl} for discussion of these selection techniques).
\item Multiple image identification with morphology and color based on initial
mass model. With current data, this step is carried out by hand: with such a
large sample of clusters, we can either a) enlist many more lensing analysts
to do the image identification (see \autoref{sec:sl:epo}) or b) attempt to
automate the process.
\item Shear calibration from the comparison with the strong-lensing
constraints in the non-linear regime.
This requires a good strong lensing mass model, based on the
identified multiple-images, which are the end-product of the previous
step. 
\item Mass distribution estimation. This can be done with the 
parametrized models built up during the image identification process, or 
by moving to more flexible grid-based methods. Since the shear calibration was
carried out using the strong plus weak lensing data, the major
remaining
source of systematic error is in the choice of mass model parametrization, and
thus it makes sense to use several approaches to probe this systematics.  
\item Construction of cluster mass function, based on the different
mass modeling methods. Accurate 
uncertainties on the cluster masses will be important,
especially in clusters lying close to the mass threshold.
\item The mass functions obtained in the previous step at different
redshifts are used to quantify dark energy via the evolution of the cluster mass
function (\autoref{sec:lss:cluster}). Statistical inference of the mean
density profile, concentration-mass relation, and so on can be carried out
simultaneously (see \autoref{sec:sl:dmclus} for more discussion of this
project).
\end{itemize}


\subsection{Science Enabled by Follow-up Observations}

Although the above science can be carried out with LSST alone, follow-up
observations provide important cross-checks on systematics:

\begin{itemize}
\item Spectroscopic observations will confirm the multiple image
identification, as well as improve upon
the photometric redshift obtained from the LSST photometry. Large optical
telescopes with multi-object spectrographs will be required: exploring
synergies with spectroscopic surveys may be worthwhile, since observing more
than a few tens of clusters may be impractical.
\item X-ray analysis of the clusters detected by LSST will determine the bias
factor present in the X-ray mass estimation method, and suggest an improved
approach to convert the X-ray observables into cluster masses. Detailed
comparison of gas density and
temperature structure with the mass maps per se provides a
crucial opportunity to learn about cluster physics and perhaps the properties of dark matter.
\item Near-infrared follow-up would 
improve the accuracy of the arc photometric redshifts, and thus
enhance our ability to break the mass-sheet/slope ambiguity. Imaging at high
resolution with JWST should provide better arc astrometry and morphology (to
confirm the multiple image identification). More lensed features should also
come into view, as small, faint, high redshift objects come into view.
\end{itemize}



\subsection{Technical Feasibility}

The idea of utilizing both strong and weak lensing data simultaneously in a
single mass reconstruction is not new \citep[e.g.,][]{Abdelsalam98}, and the
technique has improved substantially in the past decade to the point where
reconstructing cluster mass distributions on adaptive  pixelized grids is now
possible \citep{bradac06,Jee07}. Mass distribution
parameters (surface density, or lens potential, pixel values)
are inferred from the data via a joint likelihood, consisting of 
the product of a weak lensing term with a strong lensing term.
Although the exploration and optimization 
of the thousands of parameters (e.g., a mass/potential $50
\times 50$ grid
has 2,500 parameters to be constrained)
involved is certainly 
a CPU-intensive operation,
future parallelization of the mass reconstruction algorithm will overcome this
obstacle and allow us to significantly 
extend the grid size
limits. Automatic searches for multiple images
have been implemented for galaxy-scale lenses \citep{Marshall09}. Although a
significant fraction of the machine-identified candidates will need to be
individually confirmed by human eyes, it still dramatically exceeds the purely
manual identification rate. Adapting this to cluster scale will require an
iterative scheme, mimicking the current human approach of 
trial-and-error cluster modeling. The central principle will remain; however, confirmation of a lensing event requires a successful mass model.

%


%
%

%
%
%
%
%
%
%
%
%
%
%
%
%
%

\section{Education and Public Outreach}
\label{sec:sl:epo}

\noindent{\it Phil~Marshall}  

In this final section we give brief outlines of two possible EPO
projects connected to strong lensing science. Both are based around the
``Galaxy Zoo'' concept described in the Galaxies and EPO chapters 
(\autoref{sec:galaxies:epo} and \autoref{sec:epo:Citizens} respectively).


\subsection{Finding Complex Lenses}

Simple galaxy-scale lenses and giant arcs can be found using automated
detection routines. However, more exotic lenses are more difficult to find.
Group-scale lenses may contain a lot of confusing lens-plane structure, while
arguably the most interesting lenses are the ones least likely to be found by
robots trained on more typical lenses. To date, many complex lenses have been
found by eyeball inspection of images -- this process can be continued into
the LSST era provided we increase the number of eyeballs accordingly. The
Citizen Science website, Galaxy Zoo, has, at the time of writing, 
a community of over 200,000 people enthusiastically inspecting images of galaxies and
classifying their morphology. Systematic lens-finding will be a feature of the
second generation ``Zoo2'' site, from which we will learn much about how the
process of lens detection and identification can be ``crowd-sourced.''
Starting from the simple tutorial and decision tree of Zoo2, we can imagine
moving on from  simple lens configurations and building up to the truly exotic
lenses in time for LSST. Interestingly, the Galaxy Zoo forum thread, ``Are
these gravitational lenses,'' is after a promising beginning already overflowing with low-probability lens candidates, illustrating the
need for well-designed tutorials and sources of more information. This is the
motivation for the second Galaxy Zoo strong lensing EPO project.

Eyeball classification works on color
JPG images made from cutout ``postage stamp'' images. 
Ideally, these images will represent the optimal resolution
and signal-to-noise ratio available. We might consider providing deconvolved
images (where the deconvolution is performed in a stable, inferential
way) as a way of increasing the resolution over the basic stacked
images, but this will incur a significant image processing overhead.
Even at 1 sec per cutout, deconvolving the $10^8$ bright, 
massive galaxies detected by LSST would take $3.2$ CPU-years (12 days on
a 100-CPU farm). 


\subsection{Modeling Gravitational Lenses by Hand}

A key feature of the previous project is that it touches a particular nerve of
the Citizen Science community: the strong desire to be the first to discover a
new and exciting celestial object. Still, the bulk of the Galaxy Zoo
classifications will be done (it seems) by infrequent or low attention span
users, whose drive for discovery wanes after $\sim100$ galaxy inspections.  
However, there is a particular class of Zooites who actively want to spend
time investigating a smaller number of systems in some detail, and spend
significant amounts of time and energy learning about new things that are
posted to the discussion forums (often from each other). 
We can think of reaching this
community not just by providing more data, but better tools with which to
investigate the interesting subset of data they have themselves distilled
from the survey. We propose to have this team perform the necessary
``expert'' human classification of lens candidates generated in the first
project above.  

The only difference between professional gravitational lens astrophysicists
and the amateur astronomers of the Galaxy Zoo community is that the former are
able, through experience and physical intuition, to quickly assess a lens
candidate's status: they do this by essentially modeling the system as a lens
in their heads, and rejecting objects  that do not fit. The Zooites will be
able to do this just as well, if they are provided with a {\it tool for
modeling gravitational lenses}. We can imagine a web interface where the model
parameters can be dialed up and down, and the corresponding  predicted image
displayed and compared with the LSST object postage stamp in real time.

This modeling process will not only yield a much purer sample of new complex
lenses, it will also provide an excellent platform for teaching scientific
data analysis and inference in the classroom. It will introduce the key
concept of fitting a model to data in a very clear and, one hopes, satisfying
way. One can imagine basing a high school lesson series or an undergraduate
laboratory exercise on this tool.

\bibliographystyle{SciBook}
\bibliography{stronglens/stronglens}


%
%
%
%
%
%
%
%
%
%
%
%
%
%
%
%
%
%
%
%
%
%
%
\chapter[Large-Scale Structure]
{Large-Scale Structure and Baryon Oscillations} \label{chp:lss}
{\it Hu Zhan, 
Wayne A. Barkhouse, James G. Bartlett, S\'ebastien Fromenteau, 
Eric Gawiser, Alan F. Heavens, Leopoldo Infante, Suzanne H. Jacoby, Christopher J.  Miller, 
Jeffrey A. Newman, Ryan Scranton, Anthony Tyson, Licia Verde}

\section{Introduction}

The six-band (\emph{ugrizy}) LSST survey will yield a sample of ten 
billion galaxies (\autoref{sec:common:galcounts}) over a huge
volume. It will be the largest  
photometric galaxy sample of its time for studies of the large-scale
structure of the Universe, and will characterize the 
distribution and evolution of matter on extragalactic scales
through observations of baryonic matter at a broad range of wavelengths.
In standard cosmology, structures on scales from galaxies to superclusters grow 
gravitationally from primordial adiabatic fluctuations that were 
modified by radiation and baryons between the Big Bang and 
recombination. Therefore, the large-scale 
structure encodes crucial information about 
the contents of the Universe, the origin of the 
fluctuations, and the cosmic expansion background in which the 
structures evolve.

In this Chapter, we focus on the potential of LSST to
constrain cosmology with a subset of techniques that utilize 
various galaxy spatial correlations, counts of galaxy clusters, 
and cross correlation between galaxy overdensities and the cosmic 
microwave background (CMB) as described below.  \autoref{chp:wl}
describes the measurement of weak lensing with LSST, and in
\autoref{chapter-cosmology}, we combine these with each other and
other cosmological probes to break degeneracies and put the tightest
possible constraints on cosmological models.  

The shape of the galaxy two-point correlation function (or
power spectrum in Fourier space) depends on that of the primordial 
fluctuations and imprints of radiation and baryons, which are well 
described by a small set of cosmological parameters  
(\autoref{sec:lss:galps}). Hence, one can constrain
these parameters with the galaxy power spectrum after accounting for
the galaxy clustering bias relative to the underlying dark matter. Of particular
interest is the imprint on galaxy clustering of baryon acoustic oscillations (BAOs), 
which reflects the phases of acoustic waves at recombination as 
determined by their wavelengths and the sound horizon then
 ($\sim 100\, \hmpc$). 
The BAO scales are sufficiently small that it is possible to 
measure them precisely with a large volume survey; yet, they are 
large enough that nonlinear evolution does not alter the scales 
appreciably. The BAO features can be used as a 
standard ruler to measure distances and constrain dark energy 
(\autoref{sec:lss:bao}). One can measure not only the 
auto-correlation of a galaxy sample but also cross-correlations 
between different samples (\autoref{sec:galaxies:clustering}). The latter can provide valuable 
information about the redshift distribution of galaxies (see also 
\autoref{sec:common:photo-z}). The matter power spectrum on very 
large scales has not been modified significantly by radiation or 
baryons, so it is one of the
few handles we have on primordial fluctuations and inflation 
(\autoref{sec:lss:lscale}). The bispectrum will arise from nonlinear
evolution even if the perturbations are Gaussian initially. It contains
cosmological information and can be used to constrain the galaxy bias
(\autoref{sec:lss:bisp}). The abundance of clusters as a 
function of mass and redshift is sensitive to cosmological 
parameters, and the required knowledge of the mass--observable relation
and its dispersion may be achieved through multi-wavelength 
observations (\autoref{sec:lss:cluster}, see also 
\autoref{sec:sl:clusmf}).
Finally, CMB photons traveling through a decaying potential well, such
as an overdense region in the $\Lambda$CDM universe, 
will gain energy. In other words, the large-scale structure causes 
secondary anisotropies in the observed CMB. This effect can be 
measured from the correlation between galaxy overdensities and 
CMB temperature fluctuations, and it provides direct evidence for the
existence of 
dark energy (\autoref{sec:lss:isw}).

%
%
%
%
%
%
%
%
%
%
%
%
%
%
%
%
%
%
%
%
%
%
%
%
%
%
\section{Galaxy Power Spectra: Broadband Shape on Large Scales } \label{sec:lss:galps}
{\it Christopher J. Miller, Hu Zhan}

The overall shape of the matter power spectrum is determined by the
physical matter density, $\om$, the physical baryon
density $\ob$, the primordial spectral index, $n_s$, 
the running of the index, $\alpha_s$, and
neutrino mass, $m_\nu$. For example, the prominent turnover at 
$k \sim 0.02\, \hmpci$
is related to the size of the particle horizon at matter--radiation
equality and hence is determined by $\omega_m$ (and $T_\mathrm{CMB}$,
which is precisely measured). The most significant
component to the broadband shape of the power spectrum is this
turnover, which has yet to be robustly detected in any galaxy survey. LSST provides the
best opportunity to confirm this turnover and probe
structure at the largest scales.

\subsection{The Large Scale Structure Power Spectrum}
\label{subsec:intro}

For discrete Fourier modes, we define the three-dimensional power spectrum to be 
$P(k) \propto \langle |\hat{\delta}_k|^2 \rangle$, where
$\hat{\delta}_k$ is the Fourier transform of the density
perturbation field, i.e., the overdensity, in a finite volume
(see also \autoref{eq:app:pk} -- \autoref{eq:app:gps}).
The choice of the underlying basis
used to estimate the spatial power spectrum can be tailored to best suit the shape of
the data; however, most work is expressed in the Fourier basis where the
power spectrum is the Fourier inverse of the two-point spatial correlation function
\citep{vogeley92, fisher93,park94,dacosta94,retzlaff98,miller01, percival01, percival07b}.
In the case of the CMB,
we are accustomed to measuring the two-dimensional power spectrum, $C(\ell)$ of temperature
fluctuations on the sky via spherical harmonics \citep[e.g.,][]{bond98}.
The LSST survey will be studied
using two-dimensional projections of the galaxy data as well (see
\autoref{sec:lss:bao}), but in this section
we focus on the three-dimensional Fourier power spectrum.

Theoretical models for the shape of the power spectrum, which governs structure
formation, start from the primordial $P_{pr}(k)$ spectrum. These primordial fluctuations
are thought to have been generated during an inflationary period where
the Universe's expansion 
was driven by a potential-dominated scalar field \citep[e.g.,][]{liddle2000}. If this
scalar field has a smooth potential, then $P_{pr}(k) \propto k^{n_{\rm
    s}}$ with $n_{\rm s} \sim 1$. However, it is possible that the
scalar field was based on multiple or non-smooth potentials.  Thus, determining the
primordial power spectrum remains one of the most important challenges for cosmology.
In the specific case of a galaxy survey we measure the spectrum
of galaxy density fluctuations, and relate this to the matter density
fluctuations through a bias model, as described, e.g., in \autoref{sec:galaxies:distfunct}.

Over time, the shape of the primordial power spectrum is modified by
the effects of self-gravitation, pressure, dissipative processes, and
the physical processes that determine the expansion rate; these
effects take place on scales smaller than the horizon size at any
given epoch. 
To account for
these effects, the power spectrum is usually written as: $P(k,z) =
T^2(k,z)P_{pr}(k)$, where $T(k,z)$ is the matter transfer
function. 

Calculating the transfer function involves solving the Boltzmann
equation for all the constituents that play a role. For adiabatic
models, the value of $T(k)$ approaches unity on large scales and
decreases towards smaller scales. The degree of damping depends on the
type of particles and processes. For instance, pure Cold Dark Matter
(CDM) models produce less damping than Hot Dark Matter 
universes. Regardless of the species of matter, a turnover in the
power spectrum will occur at the scale where pressure can effectively
oppose gravity in the radiation-dominated era, 
halting the growth of perturbations on small
scales. This bend in the shape of the power spectrum depends 
solely on the density of matter, $\om$, in the Universe and will be
imprinted in the observed power spectrum.

Non-cosmological effects also change the shape of the observed power spectrum. The bias
parameter between the dark matter and the galaxies will change the amplitude.
While it is often assumed that bias is scale-independent
\citep{Scherrer+Weinberg98}, there are observational
\citep[e.g.,][]{percival07b} and theoretical
(\autoref{sec:lss:nongauss}) 
reasons to think otherwise.
An unknown scale-dependent and stochastic bias will limit our ability
to determine the matter power spectrum. The power spectrum of SDSS
galaxies \citep{tbs04} is consistent with a bias independent of scale over 
$0.02\,\mpci{} < k < 0.1 \,\mpci{}$, but the data are not terribly
constraining, and there is room for subtle effects, which
will become more apparent with better data.  
Extensive studies with the halo model, weak lensing,
and simulations (\citealt{hoekstra02, hj04, weinberg04, smm05}) will help us
better understand the limits of galaxy bias.  In this section, we assume that the
bias is known and scale independent on scales of interest; see further discussion in
\autoref{sec:lss:bisp}.  

Any interactions between dark
matter particles and baryonic matter will dampen the power at small
scales \citep{miller02}, and 
photometric redshift errors will even suppress power on fairly large
scales. In particular, in the presence of these errors, 
the observed power spectrum will become 
$P^\prime(k) \propto P(k)\exp[-(k_{||} \sigma_z)^2]$, where $k_{||}$ are
the modes parallel to the observer's line of sight and $\sigma_z$
are the errors on the photometric redshifts \citep{seo03}. This damping
will be large for scales smaller than 150 Mpc.

\begin{figure}
\centering
\includegraphics[height=3.1in]{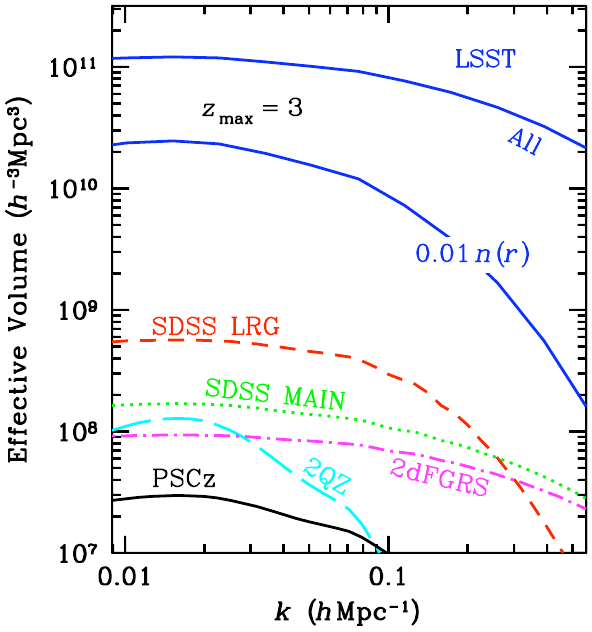}
\caption{Effective survey volumes (see \autoref{eq:lss:Veff}). 
The survey data, except 
that of LSST, are from \citet{eisenstein05}. LSST survey parameters
are for the fiducial 20,000 deg$^2$ survey. The curve labeled with
$0.01 n(r)$ approximates the case where subvolumes or subsamples of
LSST data are selected to explore systematic effects.}
\label{fig:lss:effv}
\end{figure}

Thus while LSST's galaxy photometric redshift survey will be both very
wide and very deep, the errors on 
these redshifts will greatly reduce its statistical power compared to
spectroscopic surveys of the same size \citep[e.g.,][]{blake03}. 
Regardless of the
challenges to measuring the power spectrum described here, LSST
will provide the community with the largest view of our Universe in
terms of the effective survey volume (see \autoref{fig:lss:effv}),
defined as
\begin{equation} \label{eq:lss:Veff} 
V_{\rm eff}(k) = \int \left [ 
\frac{{n_g}(\bit{r})P_g^\prime(k)}
{{n_g}(\bit{r})P_g^\prime(k)+1}
\right ]^2 d^3r,
\end{equation}
where $n_g(\bit{r})$ is the galaxy number density, and $P_g^\prime(k)$
is the galaxy power spectrum in \phz{} space.

\subsection{Measuring the Turnover in the Power Spectrum for Photometric Redshift Surveys}
In \autoref{subsec:intro}, we discussed the various ways in which the observed
power spectrum of galaxies differs from the true underlying dark
matter power spectrum. 
In past and current galaxy surveys, the size and shape of the survey volume limited
the largest scale for which power could be accurately measured.
There is a minimum 
survey volume that is required to detect the turnover in the power spectrum,
and LSST will be the first experiment to go well beyond that required size. 

\begin{figure}
\centering
\includegraphics[width=6in]{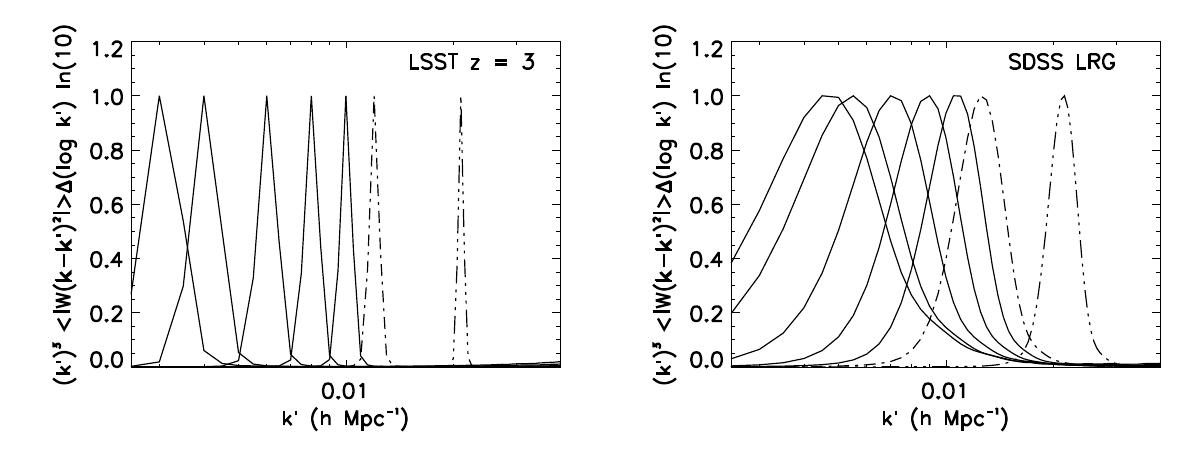} 
\caption{Measurement of the power spectrum at a given wavenumber
  invariably involves a window function, quantifying the range of
  values of $k$ contributing \citep{lin96}.  In a survey of infinite volume, the windows
would be a series of $\delta$ functions.  Shown are effective
  survey volume window functions for the LSST survey at $z = 3$ and the equivalent
  for the SDSS LRG sample. This is the integrand of \autoref{eq:window_convolv}
multiplied by $k'ln10\Delta(log_{10}k')$ in this log-linear plot.  
The width and shape of the $k$-mode windows depend on the shape and size of the 
volume used for the Fourier analysis. The LSST windows are compact
compared to the SDSS LRG windows. The LSST volume is large enough
for the windows to provide uncorrelated measurements of the power spectrum to
very large scales beyond the turnover (i.e., $k \sim 0.006\,\hmpci$).  The
window heights have been renormalized to emphasize the shapes and widths of
adjacent windows.}
\label{fig:lss:windows}
\end{figure}

The shape and size of the survey volume affects the power spectrum by
damping and smearing the power. Ideally, the survey volume would allow
one to measure power in
Fourier modes that are independent and uncorrelated between adjacent
nodes.  In practice, the size and shape of the volume puts limits on
which nodes can be used in any analysis before severe window
convolution and aliasing effects destroy the signal. In practice, the
true underlying power spectrum, $P_{true}(k)$, is convolved with the survey window, $W(k)$:
\begin{equation}
\label{eq:window_convolv}
\langle P_{windowed}(k) \rangle  \propto \int |\hat{W}({\bf k} - {\bf k'})|^2 P_{true}({\bf k'}) k'^2 dk'.
\end{equation}
 \autoref{fig:lss:windows} graphically describes these window effects for
various modes within the LSST volume via the integrand of \autoref{eq:window_convolv},
where we assume that the ratio of the observed power to the true power is constant over the
small $k$-range of each window (see \citealt{lin96}).
The narrower the window, the
cleaner the measurement.  Indeed, the modes are essentially
independent of one another and the survey volume window function biases the
measurement little for $0.006\,\hmpci \le k <
0.02\,\hmpci$, and for $k >
0.002\,\hmpci$, there is little or no ``leakage'' or aliasing of power
into nearby bins.  Compare with the SDSS LRG sample \citep{Tegmark++06}, 
where modes $k \le 0.01\,\hmpci$ are strongly
overlapping\footnote{Note this situation can be mitigated by 
  using Karhunen-Lo{\`e}ve eigenfunctions to find statistically independent modes
  \citep{szalay03,vogeley96,tbs04}.}.

For shallow galaxy redshift surveys at low $z$, measuring $P(k)$ is straightforward. First,
galaxy redshifts are converted to distances using the Hubble Law. The comoving density
can then be determined. The power spectrum is then directly measured in Fourier space.
The most significant challenge here is dealing with the window function, which describes
the angular geometry of the survey. If the geometric basis used to estimate the power spectrum
is not entirely orthogonal to the survey geometry, an effective window will distort the shape of the power
spectrum and can even smear out sharp features like the baryon acoustic
oscillations \citep[e.g.,][]{miller02}.
From a statistical perspective, the problem here is that the power measured in any
given wave-band becomes correlated with other wave-bands. This is not a significant issue for
inferring the cosmological parameters, as one simply convolves the model power spectra with
the same window during the analysis \citep[e.g.,][]{miller01,tegmark98}.
This convolution comes at the expense of lost statistical
power in the determination of the inferred parameters.

For deep photometric redshift surveys at high $z$, measuring $P(k)$ becomes more
challenging. Arguably, if the window of a photometric survey like LSST is large
and contiguous, then the window
effects seen in many of the shallow low-redshift data sets (e.g., SDSS, the 2dF Survey) will not be
significant.
However, the Hubble Law no longer suffices as a distance estimator. In addition,
the evolution of the bias parameter
(how light traces mass) and its dependence on scale also become issues.
Last but not least, the photometric errors and their distributions
affect the shape of the measured power spectrum.

\citet{blake03} and \citet{seo03} compare the power of redshift and
photometric surveys to measure the power spectrum. They find that photometric
surveys require 10 -- 20 times more sky coverage than
redshift surveys; there are fewer useful radial Fourier modes in
photometric surveys because of the large errors in the photometric redshifts.
On the other hand, \citet{blake05} show how the
tangential modes, which are not affected by photometric redshift
errors, can be used to provide good constraints on the large-scale 
shape of the large-scale structure power spectrum. 

However, on the largest scales (i.e., comparable to the turnover
scale), all Fourier 
modes can be used in the analysis. As an example from \citet{blake05},
for a survey with $\sigma_0 = \sigma_z/(1 +z) \sim  0.03$
(\autoref{sec:common:photo-z}) with a survey volume effective depth
corresponding to $z = 0.5$, all modes with $k < 0.02\,\hmpci$ will
survive the damping caused by the use of photometric redshifts. They find the
turnover at good statistical confidence (99.5\%) for a photometric
survey of 10,000 deg$^2$ to a limiting magnitude of $r = 24$. The LSST
large-scale structure 
galaxy survey will have a much larger effective volume (due to its larger area and deeper magnitude
limit) with \phz{} errors comparable to $\sigma_0 \sim 0.03$ as used in \citet{blake05}.


\subsection{Other Systematics}
In addition to the major systematics caused by the photometric redshifts,
the large-scale structure power spectrum is also sensitive to
effects such as star/galaxy misclassifications, dust extinction fluctuations on the sky, seeing
fluctuations, color errors, which bias the \phz{}s, and so on.  In
this section, we discuss how large these effects might be. 

On very large scales, the variance of density fluctuations in
logarithmic $k$ bins is very small,
e.g., $\Delta^2(k) \equiv k^3P(k)/2\pi^2 \sim 10^{-3}$ at $k = 0.01\,\mpci{}$.
An uncorrected-for varying Galactic extinction over the wide survey area can
cause fluctuations in galaxy counts that may swamp the signal.
If the logarithmic slope of galaxy counts $\bar{n}_{\rm g}(<m)$
as a function of magnitude is
$s = {\rm d}\log \bar{n}_{\rm g}/{\rm d} m$, then the fractional
error in galaxy counts is
\begin{equation}
\frac{\delta n_{\rm g}}{n_{\rm g}} = \ln 10\, s\, \delta m
= 2.5\, s \frac{\delta f}{f},
\end{equation}
where $\delta f/f$ is the fractional error in flux caused by, e.g.,
extinction correction residuals. Observationally, 
$s$ varies from $0.6$ at blue wavelengths to $0.3$ in the red
\citep[e.g.,][]{t88,pmz98,yfn01}, and tends to be smaller for fainter
galaxies \citep{msc01,lld03}. To keep this systematic angular
fluctuation well below $\Delta(k)$, one has to reduce the flux error
to 1\% or better over the whole survey area (thus motivating our
requirements on photometric uniformity; see
\autoref{sec:introduction:reqs}). This is a very 
conservative estimate, because the power spectrum receives
contributions from not only angular clustering but also radial
clustering of galaxies on large scales, which is much less affected
by extinction or photometry errors.

To understand how these effects might affect our measurements of
galaxy fluctuations, we introduce some formalism from
\citet{zhan06b}. 
The average number of galaxies within an angular window
$\Theta(\hat{\mathbf{r}})$ is
\begin{equation}
\bar{N}_{\rm g} = \int \bar{n}_{\rm g}(r) \Theta(\hat{\mathbf{r}})
{\rm d}^3 r,
\end{equation}
and the variance is
\begin{equation} \label{eq:varN}
\sigma_{N_{\rm g}}^2 = \sum_{lm} \int P(k) |n_l(k) \Theta_{lm}|^2
k^2 {\rm d} k,
\end{equation}
where $\bar{n}_g$ is the mean number density, 
$P(k)$ is the matter power spectrum at $z = 0$,
\begin{eqnarray} \nonumber
n_l(k) &=& \sqrt{\frac{2}{\pi}}\int \bar{n}_{\rm g}(z)b(z)g(z) j_l(kr)
              {\rm d} z, \\ \nonumber
\Theta_{lm} &=& \int \Theta(\hat{\mathbf{r}})
Y_{lm}^*(\hat{\mathbf{r}}) {\rm d} \hat{\mathbf{r}},
\end{eqnarray}
$b(z)$ is the linear galaxy bias, and $g(z)$ is the linear growth 
function of the large-scale structure. Since the scales of interest
are very large, the linear approximation for the galaxy power spectrum
is sufficient. 

\begin{figure}
\centering
\includegraphics[width=0.6\linewidth]{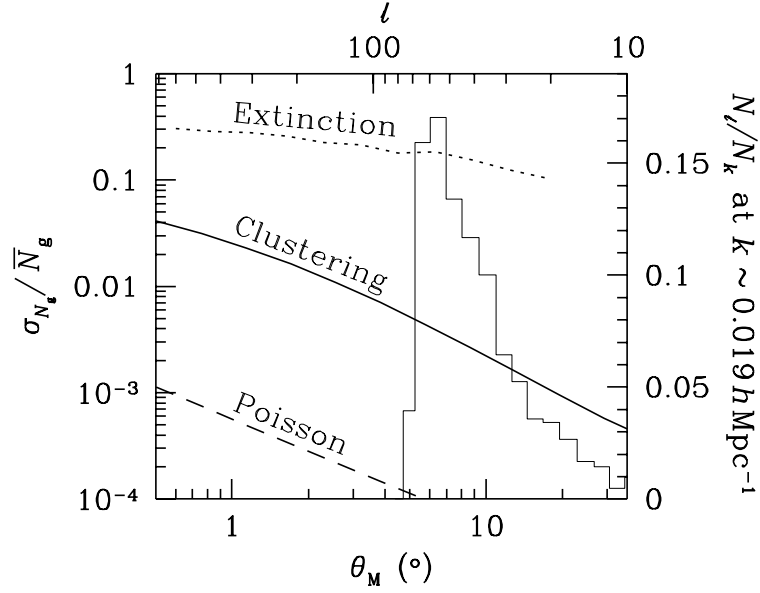}
\caption{Rms fluctuations of galaxy counts on the sky due to the
extinction (dotted line), galaxy clustering (solid line), and
Poisson noise $1/\sqrt{\bar{N}_{\rm g}}$ (dashed line) within
an angular window of size $\theta_{\rm M}$.
The histogram shows the contribution to the number of independent
modes, $N_k$ ($k \sim 0.019\,\mpci{}$ with $\Delta k = 0.16k$), from
each band of multipoles in the spherical harmonic analysis. 
The scale of the histogram is marked on the
right axis. For lower wavebands, the distribution moves
to lower multipoles. The multipole number, $l$, is related to
$\theta_{\rm M}$ by $l \sim 360^{\rm o} / \theta_{\rm M}$.
Figure from \citet{zhan06b}, with permission.
\label{fig:lss:varN}}
\end{figure}

\citet{zhan06b} use a Gaussian window function to demonstrate the 
effects:
\[
\Theta(\theta, \phi) \equiv \Theta(\theta) = 
e^{- \theta^2/2\theta_{\rm M}^2},
\]
where $\theta$ and $\phi$ are respectively the polar and azimuthal
angles. For Galactic extinction, the \citet{sfd98} map is used,
and the Galactic latitude is restricted to 
$|b_{\rm c}| > 20^o + 1.5 \theta_{\rm M}$.
The rms fluctuation of $g$-band galaxy counts within the window
function due to reddening alone is calculated with the conversion
$\delta n_{\rm g}/n_{\rm g} \sim \delta A_B = 4.3 \delta E(B-V)$.

The results are shown in \autoref{fig:lss:varN}. The Galactic
extinction (if it were not corrected for!) would dominate over galaxy
clustering.  However, we can 1) correct for the 
extinction to fairly high accuracy, using maps like that of
\citet{sfd98}, and 2) use photometric redshifts to measure the
clustering, thus greatly reducing the projection effects and
increasing the true clustering signal. Analysis based on SDSS
data demonstrates that the error caused by extinction (and 
photometry calibration) is an order of magnitude lower than the
signal, the angular galaxy power spectrum, at multipoles of a few 
hundred \citep{tegmark02}. With better photometric calibrations
(\autoref{sec:design:calsim}) and additional mapping of Galactic 
dust using the stellar locus (\autoref{sec:MW:dust}), LSST will be
likely to reduce the error even further.

The rms fluctuations in projected galaxy number density 
due to large-scale structure is well under 1\% on scales above several
degrees, suggesting that the galaxy counts can be used to constrain
photometric calibration drifts on these large scales. 
%
Further improvement on the relative flux error is possible by combining
galaxy counts with multi-band galaxy photometry
\citep[e.g.,][]{bwm05}, HI and CO surveys, and stellar locus analyses
(\autoref{sec:MW:dust}). An advantage of counting galaxies is that it 
does not rely on color information and, hence, is sensitive to gray 
dust. Since the Poisson noise in the galaxy counts is an order of 
magnitude lower than that caused by galaxy clustering, one can also 
divide the galaxies into groups of similar properties and compare them 
in one field with those in another to better determine the differential
extinction.  


%
%
%
%
%
%
%
%
%
%
%
%
%
%
%
%
%
%
%
%
%
%
%
%
%
%
\section{Baryon Acoustic Oscillations} \label{sec:lss:bao}
{\it Hu Zhan}

\subsection{Introduction}

Before the Universe became neutral for the first time at recombination, 
the cosmic plasma was tightly coupled with photons. Perturbations
(acoustic waves) in the relativistic fluid  
propagated at the 
speed of sound ($v_s \sim c/\sqrt{3}$ for a relativistic fluid) 
but stopped after recombination
when the fluid lost pressure support by the photons. The primary CMB 
temperature anisotropy is a snapshot of these acoustic waves at the 
last scattering surface, which can be characterized by the sound horizon, 
$r_s$ ($\sim 100 \hmpc$ co-moving), at that time
\citep{peebles70,bond84,holtzman89}. 

The low-redshift signature of the acoustic waves before recombination
is a slight enhancement of the correlation between density fluctuations
separated by a distance, $r_s$, and it is named after the source of the
effect -- Baryon Acoustic Oscillations (BAOs). Because there is only
a single scale in effect, the acoustic peak can be easily identified in 
the two-point correlation function in configuration space 
\citep{eisenstein05}. In Fourier space, the 
imprint becomes a series of oscillations in the power spectrum at 
$k \sim 0.1 \hmpci$. 
Since galaxies trace matter fairly well on large scales, and since there
is no known astrophysical process that can produce similar oscillatory 
features on the same scale, the BAO features must exist in the galaxy
distribution as well. Indeed, they have been detected from SDSS and
2dF galaxy surveys, both spectroscopically and photometrically
\citep{eisenstein05,cole05,padmanabhan06,blake06b,percival07,percival2009}.

The scale of the BAO features shifts only slightly after recombination 
due to nonlinear evolution \citep[e.g.,][]{seo05,huff07,crocce08},
which can be quantified 
by cosmological simulations \citep[e.g.,][]{seo08}. Hence, the 
BAO scale can be used as a CMB-calibrated standard
ruler for measuring the angular diameter distance and for constraining 
cosmological parameters (\citealt*{eisenstein98}; \citealt{cooray01b};
\citealt{hu03b}; \citealt{blake03}; \citealt{linder03b}; 
\citealt{seo03}; \citealt{wang06}; \citealt{zhan06d}; 
\citealt*{zhan09a}). Of particular interest are
the dark energy equation of state and the 
mean curvature of the Universe.

LSST will observe $\sim 10^{10}$ galaxies over 20,000 deg$^2$ with 
redshifts estimated from its six-band photometry data. As we discussed
in the previous section, the errors in 
the photometric redshifts (\phz{}s) severely suppress the radial 
BAO information \citep{seo03, blake05}. 
Therefore, LSST will consider angular BAO only in redshift (or
``tomographic'') shells. 
With their superior photometric redshifts, supernovae should allow a
full three-dimensional analysis; see the discussion in
\autoref{sec:sn:bao}.  

Errors in photometric redshift cause bins in redshift to overlap in
reality, giving rise to correlations between fluctuations of galaxy 
number density in adjacent bins. Such cross-bin correlations can
be fairly strong when there is a significant overlap between two 
galaxy distributions, and, hence, provide useful information 
about the \phz{} error distribution (see
\autoref{sec:common:photo-z}). This is a crucial advantage of 
jointly analyzing galaxy and shear power spectra (see 
\autoref{sec:cp:wlbao}).

For the science case presented here, we assume that by the time LSST is 
in full operation, we will have essentially perfect knowledge of the 
matter power spectrum at least in 
the quasi-linear regime (see \autoref{sec:cp:cos_sim}), 
so that we can sufficiently account for the 
slight evolution of the BAO features to achieve percent-level 
measurements of distances. We also assume that practical issues
such as masks and angularly varying selection functions will be 
handled in such a way as to give systematic effects much smaller than the statistical errors 
of measured quantities. Finally, even though we adopt a very simple
\phz{} error model, we expect the results to be valid for a
more realistic \phz{} error distribution \emph{if} the distribution 
can be modeled by a small number of parameters with high fidelity. 

The rest of this section is organized as follows. 
\autoref{sec:lss:bao:gaps} describes the galaxy auto- and cross-power
spectra in multipole space that will be measured by LSST.  
It also elaborates the assumed survey data and configuration 
for the forecasts. 
Treatment of \phz{} errors is provided in \autoref{sec:lss:bao:phz}.
Our estimates of constraints on distance, dark energy equation of state,
and curvature are given in \autoref{sec:lss:bao:conc}. 
We demonstrate the constraining power of galaxy cross power spectra 
on the \phz{} error distribution in \autoref{sec:lss:bao:conz}. 
We discuss limitations of the results, further work, and computational needs in \autoref{sec:lss:bao:dis}.

\subsection{Galaxy Angular Power Spectra} \label{sec:lss:bao:gaps}
As discussed in \autoref{sec:lss:galps} and 
\autoref{sec:app:stats:numtech:fisher},
the power spectrum $P(k)$ completely characterizes the statistics of a
Gaussian random  
field, which is a reasonable approximation for the cosmic density 
field on large scales, if primordial non-Gaussianity is negligible. 
With the Limber approximation \citep{limber54,kaiser92}, we project
the three-dimensional matter power spectrum $\Delta_\delta^2 (k)$ into angular power spectra
$P(\ell)$ in multipole space
\begin{eqnarray} \label{eq:lss:gaps}
P_{ij}(\ell) &=& \frac{2\pi^2}{c\ell^3} \int_0^\infty {\rm d} z\, H(z) 
D_{\rm A}(z) W_i(z) W_j(z) \Delta^2_\delta(k; z)  + 
\delta_{ij}^{\rm K} \frac{1}{\bar{n}_i} \\
W_i(z) &=& b(z) n_i(z)/\bar{n}_i, \nonumber
\end{eqnarray}
where subscripts correspond to different photometric redshift bins, $H(z)$
is the Hubble parameter, $D_{\rm A}(z)$ is the comoving angular 
diameter distance, $\Delta_\delta(k;z) = k^3P(k;z)/2\pi^2$, 
$k = \ell/D_{\rm A}(z)$, the mean 
surface density $\bar{n}_i$ is the total number of galaxies per
steradian in bin $i$, $\delta_{ij}^{\rm K}$ is the Kronecker delta 
function, and $b(z)$ is the linear galaxy clustering bias.
(See \autoref{sec:galaxies:clustering} for calculations of the
two-point correlation function in configuration space.)
The true redshift distribution of galaxies in the $i$th tomographic 
bin, $n_i(z)$, is an average of the underlying three-dimensional 
galaxy distribution over angles. It is sampled from an overall galaxy 
redshift distribution, $n(z)$, with a given \phz{} model, as described
in the next subsection.
The last term in \autoref{eq:lss:gaps} is the shot noise due to 
discrete sampling of the continuous density field with galaxies.
The covariance between the power spectra $P_{ij}(\ell)$ 
and $P_{mn}(\ell)$ per angular mode is
\begin{equation} \label{eq:lss:covgps}
C_{ij,mn}(\ell) = P_{im}(\ell) P_{jn}(\ell) + P_{in}(\ell) P_{jm}(\ell),
\end{equation}
and the 1$\sigma$ statistical error of $P_{ij}(\ell)$ is then
\begin{equation} \label{eq:lss:gpserr}
\sigma[P_{ij}(\ell)] = \left[\frac{P_{ii}(\ell)P_{jj}(\ell) + 
P_{ij}^2(\ell)}{f_{\rm sky}(2\ell + 1)}\right]^{1/2},
\end{equation}
where $f_{\rm sky} = 0.485$, corresponding to the sky coverage 
of 20,000 deg$^2$.

For the forecasts in this section, we only include 
galaxy power spectra on largely linear scales, 
so that we can map the matter power spectrum to galaxy power
spectrum with a scale-independent but time-evolving linear 
galaxy bias \citep{verde02,tbs04}. Specifically, we require that 
the dimensionless power spectrum $\Delta_\delta^2(k;z) < 0.4$ in each 
tomographic bin. In addition, only multipoles in the range 
$40 \le \ell \le 3000$ are used. Very large-scale information is excluded
here, but see \autoref{sec:lss:lscale}, \autoref{sec:lss:isw}, 
and \autoref{sec:cp:newphys} for its applications in testing 
non-standard models, such as primordial non-Gaussianity and 
dark energy clustering.

\begin{figure}
\centering
\includegraphics[width=5in]{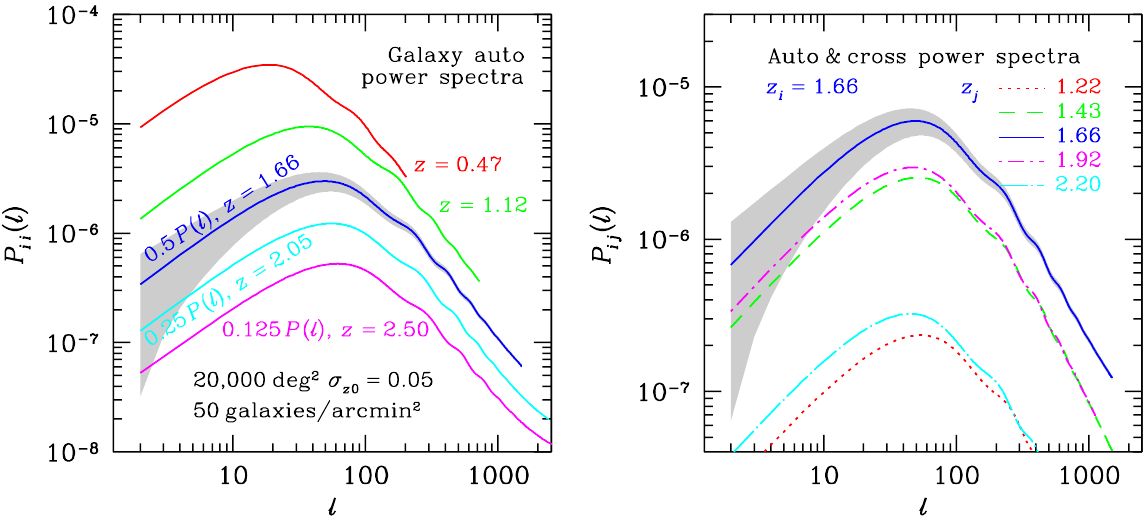}
\caption{{\it Left panel}: Galaxy angular auto power spectra in five
redshift bins (shifted for clarity). The central \phz{} of each bin 
is as labeled, and the bin width is proportional to $1+z$, 
increasing from 0.07 to 0.16 for the bins shown. We assume \phz{}
errors with rms
$\sigma_z = 0.05(1+z)$.
The BAO features are prominent at multipole $\ell$ of 
several hundred. The gray area indicates the statistical error 
(cosmic variance and shot noise) \emph{per multipole} for the bin 
centered at z = 1.66. Each power spectrum is shown to a value of
$\ell$ beyond which nonlinear evolution would significantly
contaminate our analysis. The 
flattening of the power spectra at $\ell \gtrsim 1000$, visible for
the high-redshift curves, is due to shot noise. 
{\it Right panel}: Cross power spectra $P_{ij}(\ell)$ between bin 
$i$ centered at $z = 1.66$ and bin $j$ centered at 
$z = 1.22$ (4th neighbor, dotted line), $1.43$ (2nd 
neighbor, dashed line), $1.66$ (solid line), $1.92$ (2nd neighbor,
dash-dotted line), and $2.20$ (4th neighbor, long-dash-dotted line).
These quantify the effect of overlap between these bins, and can be
used to quantify the photometric redshift error distribution. 
\label{fig:lss:bcl}}
\end{figure}

The linear galaxy bias assumes a fiducial model of $b(z) = 1+0.84z$, 
which is estimated from the simulation results in \citet{weinberg04}.
The exact value of $b(z)$ is not important for our purpose, though
a higher bias does produce stronger signals (galaxy power spectra) 
and hence tighter parameter constraints. The dependence of the dark 
energy equation of state ($w$) error on the power spectrum amplitude can be found in 
\citet{zhan06d}. We use {\sc cmbfast} version 4.5.1 \citep{zaldarriaga00} 
to calculate the matter transfer function at $z = 0$, and then apply 
the linear growth function and \citet{peacock96} fitting formula to 
obtain the nonlinear matter power 
spectrum at any redshift. A direct application 
of the fitting formula to the CDM power spectrum would cause a large
shift of the BAO features. In addition, it has difficulty processing 
power spectra that have an oscillating logarithmic slope 
\citep{zhan06d}. Thus, we calculate the multiplicative nonlinear correction to a
linear matter power spectrum with no BAO features
that otherwise matches the CDM power spectrum \citep{eisenstein99a},
and apply this ratio
to the linear CDM power spectrum with BAO features \citep[see also][]{eisenstein05}.

We assign LSST galaxies to 30 bins from \phz{} of $0.15$ to $3.5$ with
the bin width proportional to $1+z$ in order to match the \phz{} rms,
$\sigma_z = 0.05(1+z)$. The left panel of \autoref{fig:lss:bcl} shows
five auto power spectra labeled with their central \phz{}. One can 
clearly identify the BAO features at multipole $\ell \gtrsim 100$ 
despite the radial averaging over the bin width. 
Note that the broadband turnover in \autoref{fig:lss:bcl} between 
$\ell = 10$ and 100 does not directly correspond to the broadband 
turnover in the three-dimensional matter power spectrum $P(k)$. In full calculations 
without the Limber approximation, the angular power spectrum becomes 
flat on large scales \citep[see, e.g.,][]{loverde08b}. Since we 
exclude modes $\ell < 40$ and since smaller scale modes carry more 
statistical power, the errors of the Limber approximation on
large scales have little impact on our results. The flattening of 
the $z=2.05$ and $2.50$ power spectra at $\ell \gtrsim 1000$ is due 
to the shot noise. However, this is not relevant, because the shot 
noise depends on binning (hence, $\bar{n}_i$); what is relevant is 
the amount of information that can be extracted with a particular 
binning scheme \citep[see][]{zhan06d}.

The right panel of \autoref{fig:lss:bcl} shows four cross power spectra 
between the bin centered on $z = 1.66$ and its 
neighbors. The auto spectrum at $z = 1.66$ is included for reference.
The amplitude of the cross power spectrum is largely determined by the overlap
between the two bins in true redshift space, so it decreases rapidly
with the bin separation (given our Gaussian \phz{} model). The
cross-bin power spectra can be used to self-calibrate the \phz{} error
distribution (see also \autoref{sec:photoz:cross} and
\autoref{sec:galaxies:clustering}).

\subsection{Photometric Redshift Treatment} \label{sec:lss:bao:phz}

We assume that the \phz{} error distribution follows a truncated 
Gaussian:
\begin{equation} \label{eq:lss:pzpz}
\mathcal{P}(\zp;z) \propto \left\{ \begin{array}{ll}
\exp\left[-\frac{(\zp - z - \delta z)^2}{2\sigma_z^2}\right] & 
\zp \ge 0 \\ 0 & \zp < 0, \end{array} \right .
\end{equation}
where the subscript p signifies \phz{}s, $\delta z$ is the \phz{} bias,
and $\sigma_z$ is the \phz{} rms error. Since any \phz{} bias known a
priori can be taken out, one can set the fiducial value $\delta z = 0$ and
allow it to float. For the rms, we adopt the fiducial model 
$\sigma_z = \sigma_{z0} (1+z)$ with $\sigma_{z0}=0.05$.
The truncation in \autoref{eq:lss:pzpz} implies that 
galaxies with negative \phz{}s have been discarded from the sample,
which is not essential to our analysis.
The \phz{} bias and rms error at an arbitrary redshift are linearly 
interpolated from 30 \phz{} bias, $\delta z_i$, and rms, $\sigma_{zi}$, 
parameters evenly spaced between $z = 0$ and 4 \citep*{ma06,zhan06d}; 
they are linearly extrapolated from the last two rms and bias 
parameters between $z=4$ and 5, beyond which we assume practically 
no galaxy in the sample. We describe in \autoref{sec:lss:bao:conz} how we 
might constrain these quantities.  
Note that the \phz{} parameters are assigned in true-redshift space
independent of galaxy bins, which are specified in \phz{} space. 

The underlying galaxy redshift distribution can be characterized by
\citep{wittman00}
\begin{equation} \label{eq:lss:nz}
n(z) \propto z^\alpha \exp\left[-(z/z^*)^\beta\right]
\end{equation}
with $\alpha = 2$, $z^*=0.5$, $\beta=1$, and a 
projected galaxy number density of $n_{\rm tot} = 50$ per square 
arc-minute for LSST (see \autoref{sec:common:galcounts}).
This distribution peaks at $z=1$ with 
approximately $10\%$ of the galaxies at $z > 2.5$. 
The galaxy distribution $n_i(z)$ in the $i$th bin 
is sampled from $n(z)$ by \citep{ma06,zhan06d}
\begin{equation} \label{eq:lss:niz}
n_i(z) = n(z) \mathcal{P}(z_{{p},i}^{\rm B}, z_{{p},i}^{\rm E}; z),
\end{equation}
where $z_{{p},i}^{\rm B}$ and $z_{{p},i}^{\rm E}$ define the 
extent of bin $i$, and $\mathcal{P}(a,b;z)$ is the probability of 
assigning a galaxy that is at true redshift $z$ to 
the \phz{} bin between $z_{p} = a$ and $b$.
With \autoref{eq:lss:pzpz}, the probability becomes
\begin{eqnarray} \nonumber
\mathcal{P}(z_{{p},i}^{\rm B}, z_{{p},i}^{\rm E}; z)
&=& I(z_{{p},i}^{\rm B}, z_{{p},i}^{\rm E}; z)
 / I(0,\infty; z), \\ \nonumber
I(a, b; z) &=& \frac{1}{\sqrt{2\pi}\,\sigma_z} \int_a^b d z_{p} 
\,\exp\left[-\frac{(z_{p} - z - \delta z)^2}{2\sigma_z^2}\right].
\end{eqnarray} 
We have discarded the possibility of negative \phz{}s here
by normalizing the probability with $I(0,\infty;z)$.
It is worth mentioning that even though the probability distribution 
of \phz{}s at a given true redshift is assumed Gaussian, the reverse is
not true. In other words, the Gaussian assumption is flexible enough
to allow for modeling of more complex galaxy distributions in
tomographic bins \citep{ma06}.

\subsection{Constraints on Distance, Dark Energy, and Curvature}
\label{sec:lss:bao:conc}

We apply the Fisher matrix analysis \citep[e.g.,][see 
\autoref{sec:app:stats:numtech:fisher} 
for details]{tegmark97b} to estimate the 
precisions LSST BAO can achieve on distance, dark energy, and 
curvature parameters. This involves two separate calculations: 1)
estimating the constraints on distance (and growth) parameters with
a set of cosmological and nuisance parameters that are modified to
have no effect on distance (or growth of the large-scale structure),
and 2) estimating the constraints on the set of cosmological and 
nuisance parameters specified in \autoref{sec:com:cos} and previous
subsections. The latter can be done by a projection of the results of
the former. We give a brief account here; a full discussion of 
the subtle details is given in \citet{zhan09a}.

We assign 14 co-moving distance parameters $D_i$ ($i = 1 \ldots 14$) 
at redshifts evenly spaced in $\log(1+z)$ from $z_1 = 0.14$ to 
$z_{14}=5$\footnote{The actual calculation is done with 15 Hubble 
parameters $H_i$ ($i = 0, \ldots, 14$) and then projected into $D_i$
($i = 1, \ldots, 14$ and $H_0$ unchanged)
for reasons stated in \citet{zhan09a}. We include 15 growth 
parameters as well, but the growth measurements contribute little 
to the cosmological constraints in this section.}.  For the BAO
measurement, we'll need the standard angular diameter distance.  But
for the weak lensing analysis (\autoref{chp:wl}), we will find it
useful to define the more general co-moving angular diameter distance
$D_{A}(z,z')$ of $z'$ as viewed from $z$. This quantity is related to the co-moving
radial distance $D(z,z')$ between $z'$ and $z$ via
\begin{equation}  \label{eq:lss:DA-Ok}
D_{A}(z,z') = \left\{ \begin{array}{ll}
K^{-1/2} \sin[D(z, z') K^{1/2}] & \quad K > 0 \\
D(z, z') & \quad K = 0 \\
|K|^{-1/2} \sinh[D(z, z') |K|^{1/2}] & \quad K < 0 \end{array} \right.,
\end{equation}
where the curvature $K=-\Ok'(H_0/c)^2$. Since the co-moving distance is
interpolated from the distance parameters, the curvature parameter has
no effect  on distance except through \autoref{eq:lss:DA-Ok}. Hence,
we label it  as $\Ok'$ to distinguish from the real curvature
parameter $\Ok$.  

Constraints on $\Ok'$ will hold for any model that preserves the 
Friedmann-Robertson-Walker metric and (the form of) 
\autoref{eq:lss:gaps}, whereas curvature constraints from exploiting 
the full functional dependence of the angular diameter distance or 
luminosity distance on $\Ok$ (e.g., \citealt*{knox06c}; 
\citealt{spergel07}) are
valid only for a particular cosmological model. For this reason, 
measurements of $\Ok'$ are considered pure metric tests for curvature 
\citep{bernstein06}. However, because BAO (or weak lensing) alone 
does not constrain $\Ok'$ \citep{bernstein06,zhan09a}, we defer further 
discussion until \autoref{sec:cp:wlbao} where joint analyses of 
multiple techniques are presented.

\begin{figure}
\centering
\includegraphics[width=5in]{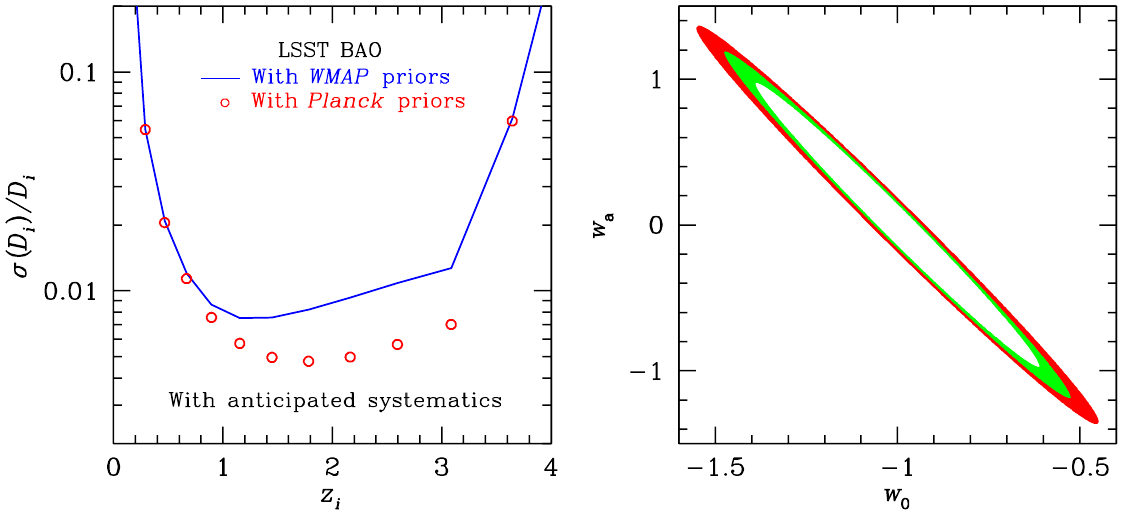}
\caption{{\it Left panel}: Marginalized 1-$\sigma$ errors on the 
co-moving distance from LSST angular BAO measurements. We have assumed
that the \phz{} bias $\delta z_i$ is known within $\pm0.2\sigma_{z,i}$, 
or $\sigma_P(\delta z_i) = 0.01(1+z_i)$, per redshift interval of 
$\sim 0.13$ 
from independent sources. For Gaussian \phz{} errors, this prior on 
$\delta z_i$ would mean a calibration sample of 188 galaxy spectra per 
unit redshift. Figure from \citet{zhan09a}, with permission.
{\it Right panel}:  Marginalized 1-$\sigma$ error contours of the 
dark energy equation of state parameters $w_0$ and $w_a$. The innermost
contour assumes that the linear galaxy clustering bias $b_i$ is known 
within $15\%$ and that $\sigma_P(\delta z_i) = 0.05\sigma_{z,i}$, i.e.,
3000 spectra per unit redshift for calibration in the Gaussian case. 
The outer contour of the green shaded area corresponds to no prior on
the galaxy bias [for numerical reasons, we take 
$\sigma_P(\ln b_i) = 1000$]
and $\sigma_P(\delta z_i) = 0.05\sigma_{z,i}$. The outermost
contour further relaxes $\sigma_P(\delta z_i)$ to  $0.4\sigma_{z,i}$. 
To reduce the number of parameters that are varied, we 
peg the uncertainty in the \phz{} rms to 
$\sigma_P(\sigma_{z,i}) = \sqrt{2} \sigma_P(\delta z_i)$ in both panels.
\label{fig:lss:bdwes}}
\end{figure}

The left panel of \autoref{fig:lss:bdwes} demonstrates that LSST BAO
can achieve percent level precision on nine co-moving distances between 
$z = 0.29$ and $3.1$ with WMAP five-year priors
(\citealt{komatsu2009}, solid line). 
Stronger priors from Planck will further reduce the errors 
to $\sim 0.5\%$ (open circles). The results include an additive noise
power of $10^{-8}$ per galaxy bin. The galaxy bias and growth 
parameters are allowed to float freely. 
We have applied fairly weak priors
to the \phz{} parameters described at the beginning of the previous
subsection: $\sigma_P(\delta z) = 2^{-1/2}  
\sigma_P(\sigma_z) = 0.2\sigma_z$, which would take only 25 galaxy 
spectra for calibration around each \phz{} bias parameter (and given
the spacing of these parameters, this corresponds to 188 spectra per
unit redshift) in the 
Gaussian case. 

The right panel of \autoref{fig:lss:bdwes} shows error contours
of the dark energy equation of state parameters, $\w0$ and $\wa$, 
with different priors on the galaxy bias $b$ and \phz{} parameters.
The innermost contour assumes $\sigma_P(b)/b = 15\%$, which is 
aggressive but comparable to current determination for low redshift 
galaxies \citep{hoekstra02,verde02,seljak05}, and 
$\sigma_P(\delta z) = 2^{-1/2} \sigma_P(\sigma_z) = 0.05\sigma_z$, 
i.e., 400 spectra for calibration around each \phz{} bias 
parameter in the Gaussian case. 
The intermediate contour allows the galaxy bias parameters
to float freely while keeping the same \phz{} priors. The outermost
contour also allows $b$ to float freely but relaxes the \phz{} priors
to $\sigma_P(\delta z) = 2^{-1/2} \sigma_P(\sigma_z) = 0.4\sigma_z$.
It is not surprising that the LSST BAO constraints on $\w0$ and $\wa$
change only mildly with wild variations in the priors, because the 
distances are determined from the BAO features in the galaxy angular 
power spectra not from the amplitudes and because the cross-bin power 
spectra can self-calibrate the \phz{} error distribution.

The curvature constraint depends on the parametrization of the dark 
energy equation of state. With the equation of state parametrized 
as $w(a) = w_0 + w_a(1-a)$, 
LSST BAO can achieve $\sigma(\Ok)\sim 10^{-3}$ \citep{zhan06d,knox06c}.
This is an order of magnitude improvement over the current result 
with the assumption of a constant equation of state \citep[e.g.,][]{spergel07}. 

\begin{figure}
\centering
\includegraphics[width=6in]{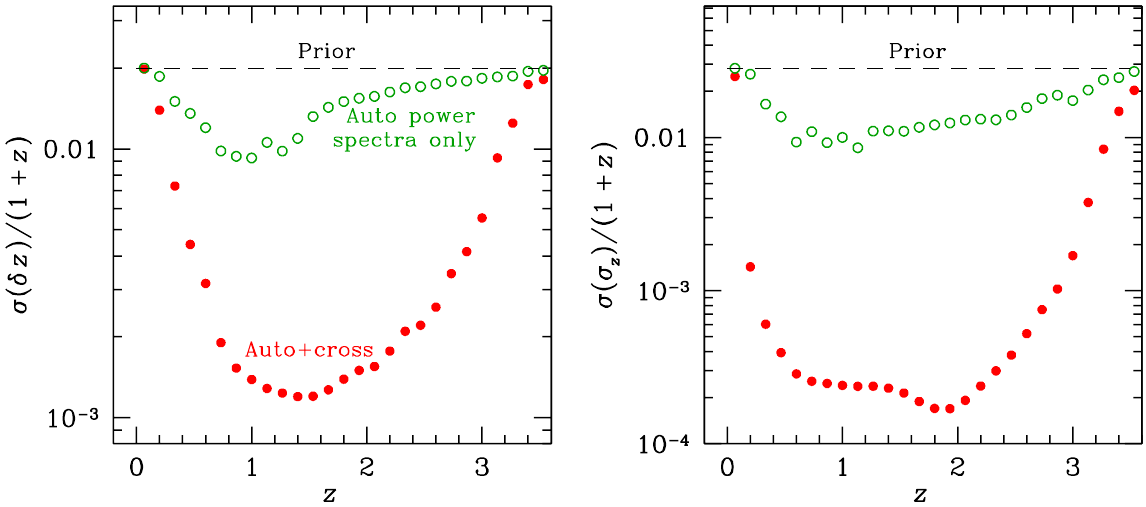}
\caption{{\it Left panel}: Marginalized $1\sigma$ constraints on 
the \phz{} bias parameters from the galaxy auto power spectra 
(open circles) and
full set of galaxy auto and cross power spectra (filed circles).
The thin dashed line marks the imposed weak prior 
$\sigma_P(\delta z_i)=0.4\sigma_{z,i}=0.02(1+z_i)$. The cross power 
spectra can self-calibrate the \phz{} bias to $10^{-3}$ level, which 
is very useful for weak lensing.
{\it Right panel}: Same as the left panel but for the \phz{} rms
parameters. 
\label{fig:lss:bzcon}}
\end{figure}

\subsection{Constraining Photometric Redshift Parameters}
\label{sec:lss:bao:conz}

\autoref{sec:common:photo-z} discusses direct methods for
determining galaxy \phz{}s, including 
calibration of the \phz{} error distribution using cross-correlations
between the \phz{} sample and a spatially overlapping spectroscopic 
sample \citep{newman08}.  Here we present another method to calibrate
the \phz{} error distribution using cross-power spectra between \phz{}
samples themselves \citep{zhan06d,schneider06}.  Even bins that
don't overlap in \phz\ will overlap in true redshift and, therefore,
include galaxies that are physically correlated with one another.  
\autoref{fig:lss:bzcon} shows marginalized $1\sigma$ constraints
on the \phz{} bias (left panel) and rms\footnote{The smallest errors
of the rms occur at $z\sim 1.9$ where the galaxy bin widths 
match the \phz{} parameter spacing of $\Delta z=0.13$. We have
replaced the uniform sampling of the galaxy distribution $n_i(z)$ in 
\citet{zhan06d} by an adaptive sampling to improve the accuracy of
the tails of $n_i(z)$. This, in turn, leads to tighter constraints on 
$\sigma_z$ in \autoref{fig:lss:bzcon} than those in \citet{zhan06d},
adjusted for different number of parameters and number of galaxy 
bins.} (right panel) parameters
corresponding to the outermost contour in the right panel of 
\autoref{fig:lss:bdwes}. Results of using only the auto power spectra
are shown in open circles (which are statistically incorrect 
because the correlations between the bins -- the cross power spectra 
-- have been neglected), and those from the full set of power spectra
are in filled circles. The thin dashed line in each panel represents
the priors. It is remarkable that the cross-bin power spectra can
place such tight constraints on the \phz{} parameters. This explains
why the dark energy constraints in \autoref{fig:lss:bdwes} are not
very sensitive to the \phz{} priors. Moreover, as we discuss in 
\autoref{sec:cp:wlbao}, the capability of self-calibrating the 
\phz{} error distribution with galaxy power spectra is a 
crucial advantage for combining BAO with weak lensing over the 
same data. However, we emphasize as well that 
the BAO self-calibration of the \phz{} parameters cannot replace 
spectroscopic calibrations, because without knowing how to 
faithfully parametrize the \phz{} error distribution, 
the self-calibration will be less informative. 

\subsection{Discussion} \label{sec:lss:bao:dis}
\Phz{} errors are one of the most critical systematics for an imaging 
survey, as redshift errors directly affect the interpretation 
of the distance--redshift and growth--redshift relations, from which
constraints on dark energy and other cosmological parameters are
derived. Even though the galaxy cross power spectra
can self-calibrate the parameters of a Gaussian \phz{} error model, 
such capability must be quantified for realistic \phz{} errors. 
Another method of calibrating the \phz{} error distribution is to
cross-correlate the \phz{} sample with a spatially overlapping 
spectroscopic sample \citep[see \autoref{sec:common:photo-z}]{newman08}, 
which does not have to be as 
deep as the \phz{} sample. These indirect methods hold promise for 
application to future surveys, though it is also noted that lensing
by foreground galaxies can produce spurious cross-correlations and
contaminate the results \citep*{loverde08a,bernstein09}.

The Limber approximation is accurate only when the width of the 
redshift bin is much larger than the linear size corresponding to 
the angular scale of interest \citep{limber54,kaiser92}. In other
words, the angular power spectra calculated using
\autoref{eq:lss:gaps} are not accurate on large scales (low $\ell$s)
\citep[e.g.,][]{loverde08b}. 
Since we do not use multipoles $\ell < 40$, the impact on our results 
is small. Nevertheless, the inaccuracy of the Limber approximation is 
not necessarily a loss of information, but one should do the full 
calculation without the approximation if low multipoles are included
in the parameter estimation. Similarly, one should model the 
correlations induced by lensing \citep{loverde08a} based on the 
foreground galaxy distribution. 

The two-point correlations in configuration space are
calculated by counting pairs and hence scale with $N^2$, where $N$ is
the number of objects. With a hierarchical algorithm, the computational
cost can be reduced to $N\log N$. In Fourier space, the power spectrum
calculation scales as $N\log N$ with Fast Fourier Transforms. The 
advantage of working in Fourier space is that the errors of the modes
are independent of each other, but one has to deconvolve 
unavoidable masks and anisotropic selection function to obtain the 
true power spectra, which can give rise to correlations between the 
modes.


%
%
%
%
%
%
%
%
%
%
%
%
%
%
%
%
%
%
%
%
%
%
%
%
%
%
\section{Primordial Fluctuations and Constraints on Inflation} \label{sec:lss:lscale}

{\it Licia Verde, Hu Zhan}

Very large-scale fluctuations in the matter distribution entered the
horizon after the epoch of matter--radiation equality and grew primarily 
under gravity since then. Therefore, these fluctuations preserve 
the imprint of primordial quantum perturbations. 
This provides a handle on models of inflation. 
More specifically, the overall shape of
the primordial matter power spectrum, as described by the primordial 
spectral index and its running, is controlled by inflationary slow-roll 
parameters, which  also determine the shape of the inflation potential.
Furthermore, any departure from the approximately featureless power 
law or non-Gaussianity detected on very large scales will require some
detailed modeling in the context of inflation. Of course, other
causes such as dark energy clustering are equally interesting to
explore.

\subsection{Features in the Inflation}

\begin{figure}
\centering
\includegraphics[width=3.5in]{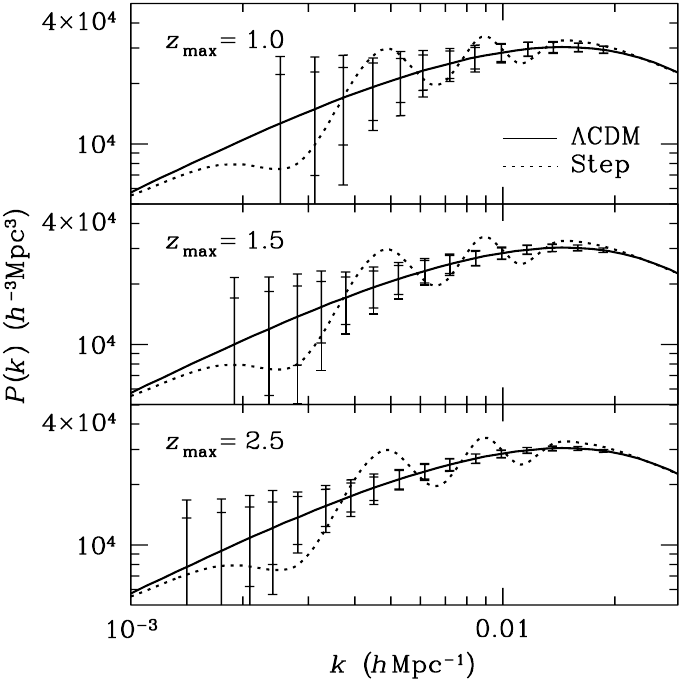}
\caption{
Effect of depth ($z_{max}$) on error forecasts for measurements of the
matter power spectrum with LSST \citep{zhan06b}. 
A 20,000 deg$^2$
photometric galaxy redshift survey is assumed. The solid line is the
fiducial model power spectrum, while the dotted line is the power
spectrum generated by the step inflation potential \citep{peiris03}. 
The error bars are 1$\sigma$ statistical errors of the power
spectrum measured in non-overlapping logarithmic bins with bin width
$\Delta k \sim 0.16 k$. The inner error bars are based on simple
mode-counting in a cubic volume, while the outer ones count spherical
harmonic modes. All the power spectra are scaled to $z = 0$. 
Figure from \citet{zhan06b}, with permission.
\label{fig:lss:varp}}
\end{figure}

Features in the inflation can generate features in the primordial
spectrum of perturbations that make it deviate from a simple power
law. For example, it can induce features such as a step or a bump,
which should then be detectable in the CMB power spectrum and in the
galaxy power spectrum. \autoref{fig:lss:varp} illustrates that a step
inflation potential consistent with \emph{WMAP} three-year 
data \citep{peiris03} induces oscillations in $P(k)$ that can be detected
by the full ten-year LSST survey 
\citep{zhan06b}.

\begin{figure}
\centering
\includegraphics[width=3.5in]{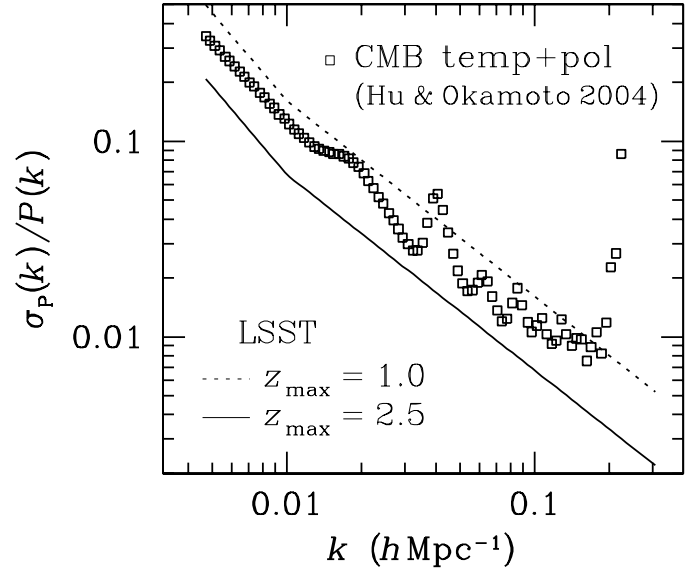}
\caption{
Forecasts of sample variance errors on the primordial matter power
spectrum \citep{zhan06b}. 
For the LSST, we set $z_{\rm max} = 1.0$ (dotted line) and 2.5
(solid line). The forecast for the CMB (open squares) includes both
temperature and polarization information, and it is taken from 
\citet{hu04a}. Both forecasts assume a binning of $\Delta k = 0.05 k$.
Figure from \citet{zhan06b}, with permission.
\label{fig:lss:cmpv}}
\end{figure}

On scales larger than $k \sim 0.05\,\mpci$, CMB data are
already cosmic variance dominated. A deep and wide \phz{} galaxy
survey such as LSST can provide measurements on these scales with
comparable errors. Moreover, those features should be more pronounced
in the three-dimensional matter power spectrum (even with photometric
redshift errors) than in the
projected two-dimensional CMB temperature power spectrum. 
\autoref{fig:lss:cmpv} demonstrates that the estimated statistical 
errors of the primordial power spectrum from LSST \citep{zhan06b}
are competitive with those from the CMB \citep{hu04a}. Hence, the
addition of large-scale structure data will significantly improve 
our knowledge about the primordial fluctuations. 

\subsection{Non-Gaussianity from Halo Bias}
\label{sec:lss:nongauss}

\begin{figure}
\centering
\includegraphics[width=3.5in]{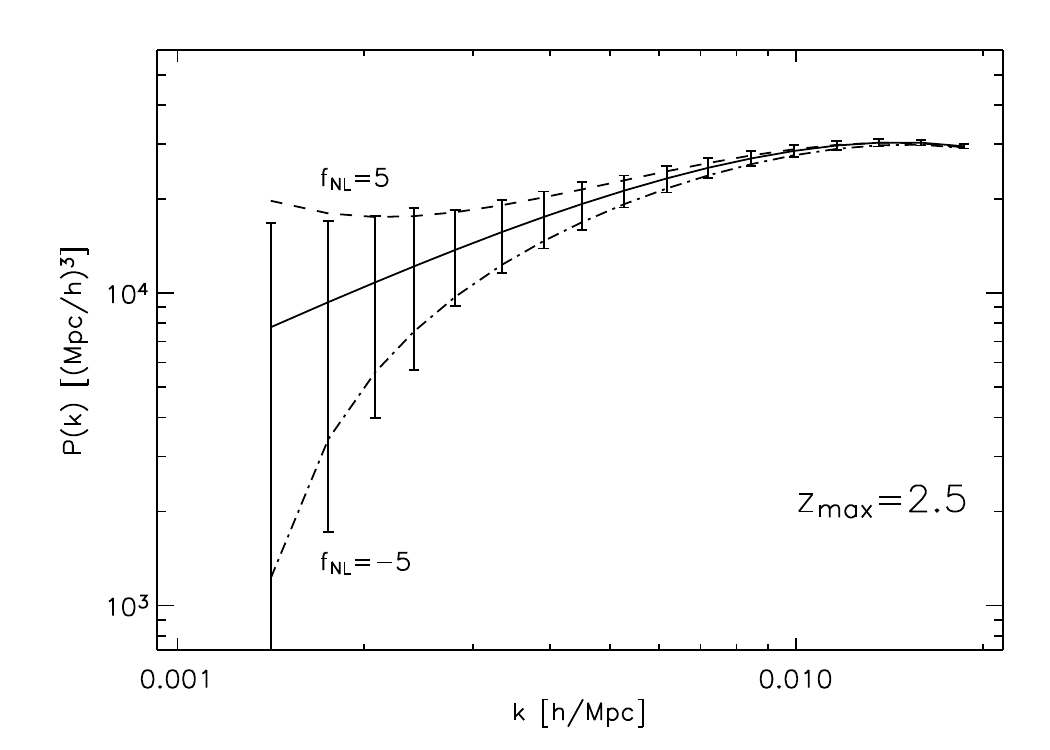}
\caption{
Effect on the large scale observed galaxy power spectrum of a 
primordial non-Gaussianity of the local type described by an 
$f_{NL}$ parameter of the values $\pm 5$.  Such departure from Gaussian 
initial conditions can be detected at the level of several $\sigma$ 
by LSST, while the CMB Planck  experiment is expected to 
have an error bound of $\sigma(f_{NL}) = 5$. 
While the simplest, single field, slow-roll inflation models predict 
$f_{NL}<1$, several models (multi-field models, non-slow roll models)
 yield much larger deviations from Gaussianity, which would be detectable 
with LSST. We have adopted the same conventions as in 
\autoref{fig:lss:varp}.
\label{fig:lss:pkNG}}
\end{figure}

\citet{dalal08} and \citet{matarrese08} have shown that
primordial non-Gaussianity affects  
the clustering of dark matter halos, inducing a scale-dependent
bias, arising even for Gaussian
initial conditions. The workhorse non-Gaussian model is the so-called
local model: $\Phi=\phi+f_{NL}(\phi^2-\langle\phi^2 \rangle)$ where
$\phi$ denotes a Gaussian random field, $\Phi$ denotes the Bardeen
potential (which on sub-horizon scales reduces to the negative of the
gravitational potential), and $f_{NL}$ is the non-Gaussian
parameter. Local non-Gaussianity arises in inflationary models where
the density perturbations are created outside the horizon and can have
a large $f_{NL}$ in models like the curvation or in multi-field
inflation. In this case, the non-Gaussian correction $\Delta
b^{F_{NL}}$, to the standard halo bias increases as $\sim 1/k^2$ at large
scales and roughly as $(1+z)$ as function of redshift. 
Galaxy surveys can be
used to detect this effect, which would appear as a difference between the shape
of the observed power spectrum on large scales, and that expected for
the dark matter. The signature of
non-Gaussianity is a smooth feature, thus photometric surveys are
well suited to study this effect. \citet{carbone08}
estimate that LSST would yield a 1-$\sigma$ error on $f_{NL}$
$\lesssim 1$. \autoref{fig:lss:pkNG} illustrates the effect of the 
large-scale non-Gaussian bias. 
This error could be in principle reduced further if cosmic
variance could be reduced \citep[see][]{seljak09,slosar09}. In any case
this limit of $\sigma(f_{NL}) \lesssim 1$ is particularly interesting for
two reasons: 1) it is comparable to, if not better than, the limit
achievable from an ideal CMB experiment, making this approach highly
complementary to the CMB approach and 2) inflationary models such as
curvation or multi-fields can yield $f_{NL}$ as large as $\sim
10$, while $f_{NL}$ from standard slow-roll inflation is expected to
be $\ll 1$; a constraint of  $\sigma(f_{NL}) \lesssim 1$ from LSST 
can be a useful test for these inflationary models.



%
%
%
%
%
%
%
%
%
%
%
%
%
%
%
%
%
%
%
%
%
%
%
%
%
\section{Galaxy Bispectrum: Non-Gaussianity, Nonlinear Evolution, and Galaxy Bias} \label{sec:lss:bisp}
{\it Licia Verde, Alan F. Heavens}

\begin{figure}
\centering
\resizebox{1.0\columnwidth}{!}{
\includegraphics{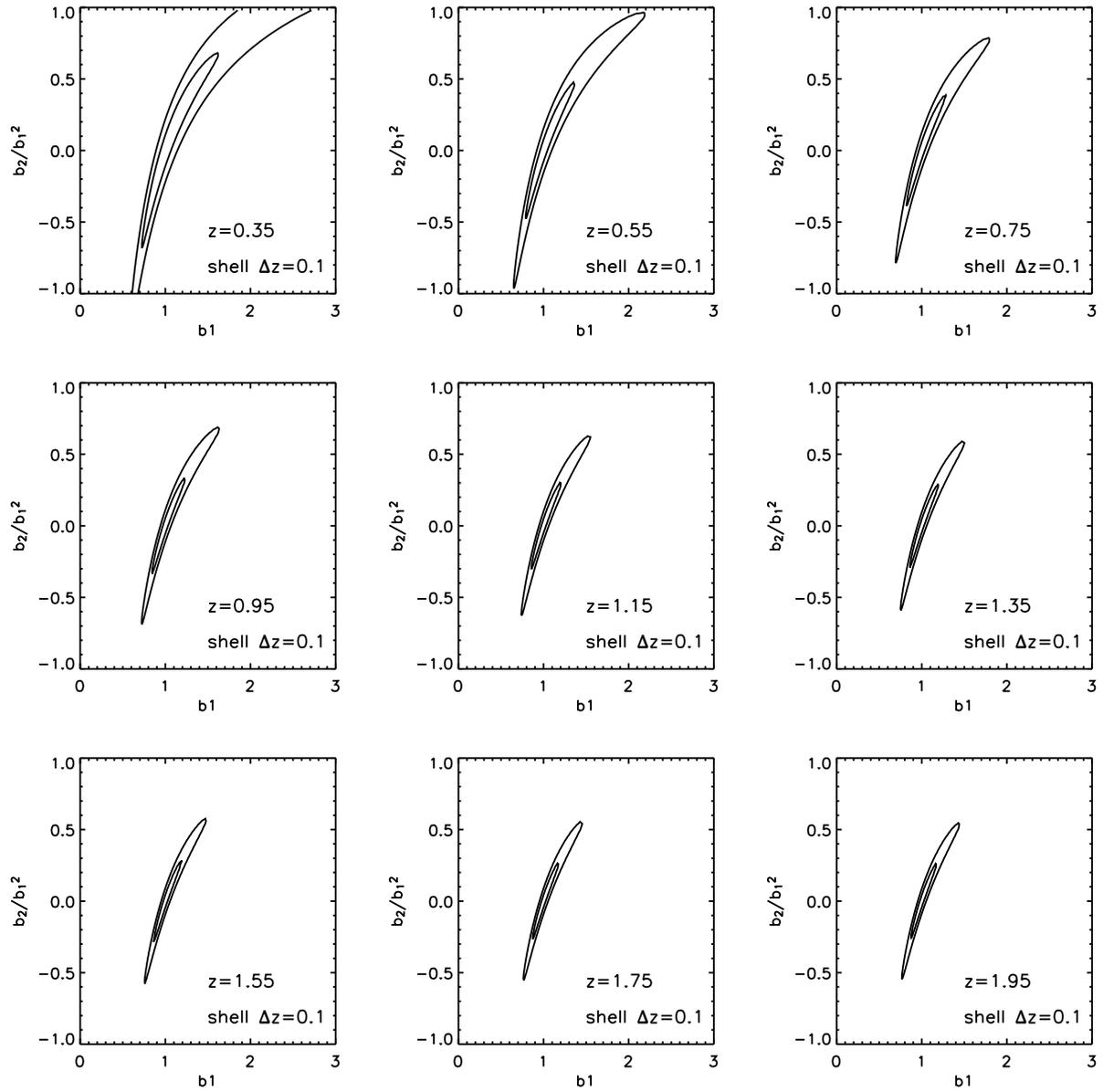}}
\caption{Predicted $1\,\sigma$ and $2\,\sigma$ uncertainties on the first- and
  second-order bias parameter as measured from the bispectrum of
  galaxies in redshift shells.  Only alternate redshift  bins are shown to  
illustrate the scaling of the errors with redshift. }
\label{fig:lss:gbz}
\end{figure}

The classic paper of  \citet{Kaiser84} suggested that galaxies form at
high peaks of the dark matter distribution, and are thus 
{\it biased} tracers of the mass distribution
(\autoref{sec:galaxies:distfunct}).  There 
are many theoretical models for galaxy bias, and observations have
shown that it depends on galaxy type, redshift, and possibly scale
\citep[e.g.,][]{SDSSswanson,SDSSBlanton,SDSSZehavi,
  Moetal97,CresswellPercival09,Norberg01}. However to a good
approximation (and ignoring the effects of {\em primordial}
non-Gaussianity described above), on large scales the effect of bias
can be summarized as  
\begin{equation}
P_g=b^2P_{DM},
\end{equation}
where $P_g$ denotes the galaxy power spectrum, $P_{DM}$ denotes the underlying
dark matter power spectrum, and $b$ denotes the bias parameter.  The
{\em relative} bias of galaxies is relatively straightforward to
measure using the power spectrum or two-point function of galaxy
clustering split by type (\autoref{sec:galaxies:clustering}), but the
absolute bias is more difficult to 
establish.  It can be measured from the observed galaxy power spectrum, given
predictions for the underlying clustering of dark matter given our
concordance cosmological model (e.g., \citealt{Lahav02}).  However, if
we had an independent measurement of the bias factor, we could combine
CMB and galaxy clustering measurements to make more precise
measurements of cosmological parameters and the growth rate of
large-scale structure under gravity. 
In linear theory one cannot use measurements of large-scale structure
to distinguish between bias and the growth rate of structure.
However, to second order, the degeneracy is lifted \citep{fry94}.  The
second-order corrections depend on the gravitational clustering of dark matter, and one can determine the bias factor by measuring
the shape dependence of the three-point correlation function (or its
Fourier analogue, the bispectrum).   Indeed, going to second order in
perturbation theory suggests going to second order in the bias model,
and we parametrize bias as:
\begin{equation} 
\delta({\bf r})_{galaxies} = b_1 \delta({\bf r})_{DM} +
b_2\left(\delta^2({\bf r})_{DM} -
\left\langle\delta_{DM}^2\right\rangle\right). 
\label{eq:nonlinbias}
\end{equation}
The \citet{fry94} approach allows both $b_1$ and $b_2$ to be
determined. 
While primordial perturbations are
expected to be Gaussian (\autoref{sec:lss:lscale}), the observed
galaxy distribution is 
non-Gaussian for two reasons: 1) non-linear gravitational instability
introduces a skewness in the density distribution and thus
non-Gaussianity and 2) the non-linear bias of \autoref{eq:nonlinbias}
also skews the distribution. 

The bispectrum approach has been applied
successfully to spectroscopic surveys to measure bias (e.g.,
\citealt{verde02}). Photometric redshift errors will not allow us to
use any radial clustering information in the LSST data on the mildly
nonlinear scales of relevance.  \cite{Verdeprojbisp} computed the
expected errors on the bias parameters for a photometric survey with
the depth of APM ($r \sim 20$).  We scale these results to LSST as
follows: 1) we can use photometric redshifts to divide the LSST sample in
shells of width $\Delta z=0.1$ 
yielding negligible cross-correlation between shells; 2) in each of
these shells we compute how many volumes of the \cite{Verdeprojbisp}
set-up will fit within the volume, and rescale the errors accordingly;
3) we assume that the shot 
noise level is similar to that of \cite{Verdeprojbisp}; 4)
we conservatively assume that in all shells perturbation theory breaks
down 
at the same scales as it does at $z < 0.1$; and 5) we assume the effects of
bias evolution and growth factor with redshift cancel. 
The resulting predicted uncertainties in $b_1$ and $b_2$ are shown in 
\autoref{fig:lss:gbz} for a few redshift slices. In the figure
$b_1=b_{linear}\times G(z)$, where $G(z)$ is the linear growth factor.

We can do better if we assume a functional form for the evolution of
bias.   For example, for a one-parameter toy model where
$b_{linear}(z)=b_1/G(z)$, the constraint on $b_1$ (marginalized over
$b_2$), we estimate $1\,\sigma$ and $2\,\sigma$ errors on $b_1$ of 
$0.045$ and $0.1$ 
respectively.




%
%
%
%
%
%
%
%
%
%
%
%
%
%
%
%
%
%
%
%
%
%
%
%
%
%
\section{The LSST Cluster Sample} \label{sec:lss:cluster}

{\it James G. Bartlett, Wayne A. Barkhouse,   S\'ebastien Fromenteau,
  Licia Verde, Jeffrey A. Newman}


The number density of clusters as a function of redshift depends on
the rate at which cosmic structures grow; it also depends on the cosmic
volume element as a function of redshift. 
It, therefore, probes both dynamical and geometrical aspects of the cosmological 
model as a function of redshift.  This is a powerful combination for discovering
the nature of dark energy and any deviations from standard gravity
(see also \autoref{sec:sl:clusmf}), as emphasized, 
for example, by the Dark Energy Task Force \citep{DETF,FoMSWG}.
  
The method relies on our ability to accurately predict cluster
abundance and its evolution as a function of observable cluster
properties and of the cosmological parameters.  The feasibility of
this, in turn, rests on connecting observable cluster properties to
those of the host dark matter halos.  N-body simulations then robustly
provide the halo abundance function, often fit by simple
analytical forms \citep{PS74,Shethetal99,Jenkinsetal01}.

The critical link is the relationship between observable quantities of
clusters, and the properties of the halos in which clusters live.
Fortunately, clusters obey a number of simple {\em scaling relations}
between cluster observables themselves, on the one hand, and halo mass
and redshift on the other.  In other words, there is a one--to--one
relation (albeit with scatter) between a cluster and its host halo
governed by well--defined correlations.  This differs significantly
from the case of a typical galaxy, which does not on average identify
with an individual dark matter halo.

We can, therefore, view clusters as dark matter halos ``tagged'' with 
different observational signatures: a grouping of galaxies in 
physical and color space; X--ray emission from the hot intracluster 
medium; the Sunyaev--Zel'dovich (SZ) signal in the Cosmic Microwave
Background caused by the same hot gas; and peaks in gravitational
shear maps (\autoref{sec:wl:shearpeaks}). 
  The existence of many different observational signatures
plays an important role in identifying and controlling cluster modeling
and corresponding systematic effects in cosmological analyses.  

As we discuss in this section, LSST will produce a large catalog
of clusters detected through their member galaxy population out 
to and beyond
redshift unity.  This catalog will on its own enable an
important cosmological study of dark energy and gravity.
Moreover, a number of wide--area cluster surveys in other wavebands 
will also have produced catalogs in the LSST era, including the SPT 
SZ survey \citep{SPT,Stanetal08} over several thousand deg$^2$, the
Planck all--sky SZ  
\footnote{\url{http://sci.esa.int/science-e/www/area/index.cfm?fareaid=17}} survey and 
an all--sky X--ray survey from eROSITA\footnote{\url{http://www.mpe.mpg.de/projects.html\#erosita}}.   

For all of these, LSST will identify optical counterparts and provide
deep optical--band imaging.  The resulting host of multi--band 
catalogs will be highly valuable for two reasons:  Firstly, the
comparison of catalogs in different wavebands will allow us to ferret
out systematics 
related to cluster detection and will tighten the modeling of 
survey selection functions.  Secondly, the deep imaging will
allow us to calibrate the cluster observables--mass distribution
function through gravitational lensing measures of cluster mass
(\autoref{sec:sl:clusmf}, where the use of clusters to constrain
cosmological parameters is discussed).  

\subsection{The Method}

In this section we describe implementation of the cluster counting method in detail.  
We keep the discussion general to serve as a reference not only to analysis of
LSST's own cluster catalog, but also to the improvements LSST will bring in
its application to catalogs at other wavebands.

We may usefully break the method down into three main steps:
\begin{enumerate}
\item Catalog construction
\item Mass determination
\item Cosmological analysis
\end{enumerate}

At the heart of catalog construction is the cluster detection algorithm,
which determines the catalog selection function ({\em completeness}) and
{\em contamination rate} by false detections.  In the case of LSST, for example,
we detect clusters via the observation of their member galaxies in the six 
LSST bands; X--ray satellites and SZ observations, on the other hand,
detect clusters through their hot intracluster medium.  
In the following, we refer to the completeness function as $\Pi\left(M,z,\Theta_N\right)$ and treat it as a function of 
halo mass $M$, redshift $z$ and a set of parameters $\Theta_N$.
The latter depend on the nature of the cluster detection algorithm and
describe both observational effects as well as astrophysical 
effects tied to cluster physics.  

In addition to position and redshift, we characterize our clusters with a 
set of measurable observables, $\vec{O}$, such as member galaxy count 
(richness), X-ray flux or SZ signal.  Via the cluster scaling relations, 
we use these to construct an estimate of cluster mass, 
referred to in the following as the cluster {\em observable mass}, $M_O$.  
The key quantity then is the distribution between the observable 
mass and true cluster halo mass, $M$: $P\left(M_O|M,z,\Theta_N\right)$.
Specifically, this is the probability distribution of $M_O$ given the true mass $M$ and redshift; it also depends on a number of 
parameters, most notably astrophysical parameters describing 
cluster physics, which we include among the nuisance parameters, $\Theta_N$.  

The objective is to relate the observed
cluster distribution to the theory through the cosmological parameters,
$\Theta_C$:
\begin{eqnarray}
\nonumber
\frac{dN}{dzdM_O} & = &\frac{dV}{dz}\left(z,\Theta_C\right) \int d\ln M\;
P\left(M_O|M,z,\Theta_N\right) \Pi\left(M,z,\Theta_N\right) \frac{dn}{d\ln M}
\left(M,z,\Theta_C\right)\\ 
&& + \left.\frac{dN}{dzdM_O}\right|_{false}
\end{eqnarray}
Here, the last term accounts for catalog contamination and 
the quantity $dn/d\ln M$ is the mass function of dark matter halos
giving their co-moving number density as 
a function of mass, redshift and cosmological parameters.
This function can be written as
\begin{equation}
\frac{dn}{d\ln M} = \frac{\bar{\rho}}{M}F(M,z,\Theta_C)
\end{equation}
where $\bar{\rho}$ as the co-moving mass density 
and the function, $F$, as the multiplicity function -- often
simply referred to as the mass function itself.  
Numerical N--body simulations confirm the theoretical
expectation \citep{Jenkinsetal01} for a universal function, $F$, dependent 
only on the amplitude of the matter power spectrum at 
each redshift: 
$\sigma(M,z,\Theta_C)=g(z,\Theta_C)\sigma(M,z=0,\Theta_C)$,
where $g$ is the linear growth factor (defined
so that $g=1$ at $z=0$). In Gaussian theories, the mass
function is, in fact, an exponential function of this 
amplitude, giving the method strong leverage on 
cosmological parameters.  

The cosmological parameters are constrained 
by fitting the above equation to the observed distribution, 
$dN/dzdM_O$, and marginalizing over the nuisance parameters, $\Theta_N$, 
incorporating as much prior information as possible on the latter.  
The nuisance parameters account for a host of systematic effects in the
procedure, and their proper definition is crucial to an 
unbiased cosmological analysis including the selection effects of the
sample.  
Properly defined nuisance parameters 
allow us to
incorporate what are often strong prior constraints on their
values when marginalizing in the final analysis.

Mock catalogs of a given survey and catalog construction algorithm
guide the choice of parameters describing the selection function.  We
also empirically control cluster selection functions 
by comparing different kinds of surveys, e.g., optical versus 
X-ray versus SZ surveys, all of which will be available in the
LSST era.  The LSST survey will, in fact, find so many clusters
that we will be able to use comparison of different catalog construction 
methods and different selection cuts on the survey data itself
as a powerful control of the selection function. 

Parameters of the observable mass-distribution are 
primarily related to cluster physics, a subject of great 
interest in its own right.  Cluster masses can be determined directly through application of the virial theorem 
to member galaxy dynamics, through application of hydrostatic equilibrium
to observations (X--ray, SZ) of the intracluster medium, and 
through gravitational lensing experiments.  The latter 
two methods have been particularly powerful in the 
establishment of cluster scaling relations in recent years.
With LSST we will use gravitational lensing to constrain the
observable--mass relation; for example, by stacking objects it is
possible to calibrate the mean relation down to very low 
masses (e.g., \citealt{Johnstonetal07b}).  Furthermore, LSST lensing
measurements 
will help calibrate observable mass relations for other wavebands,
such as the X--ray and millimeter (SZ catalogs).

\subsection{The LSST Cluster Catalog}
Numerous methods exist and have successfully been used for finding
clusters in large multi-band imaging surveys.  They are distinguished
by their emphasis on different aspects of the cluster galaxy
population.  This is a strength, because it will be important to
implement a variety of cluster detection methods to best understand
the selection criteria defining the final catalog.  All methods
provide, in the end, a list of cluster positions, photometric
redshifts, and observable properties, such as richness, total
luminosity, and so on.  We here describe one such method, based on Voronoi
tessellation of galaxies on the red sequence.  Many other approaches,
including matched filters, will also be considered. 

\subsubsection{The Cluster Red Galaxy Population}

Most detection methods rely on the presence of the characteristic red, early-type galaxy 
population in clusters, which displays a well--defined color-magnitude relation
known as the ``red sequence'' (RS).  The detectability of this population as a function
of cluster mass and redshift is therefore a central issue for most detection algorithms.

\begin{figure}
\centering
\includegraphics[width=3.45in]{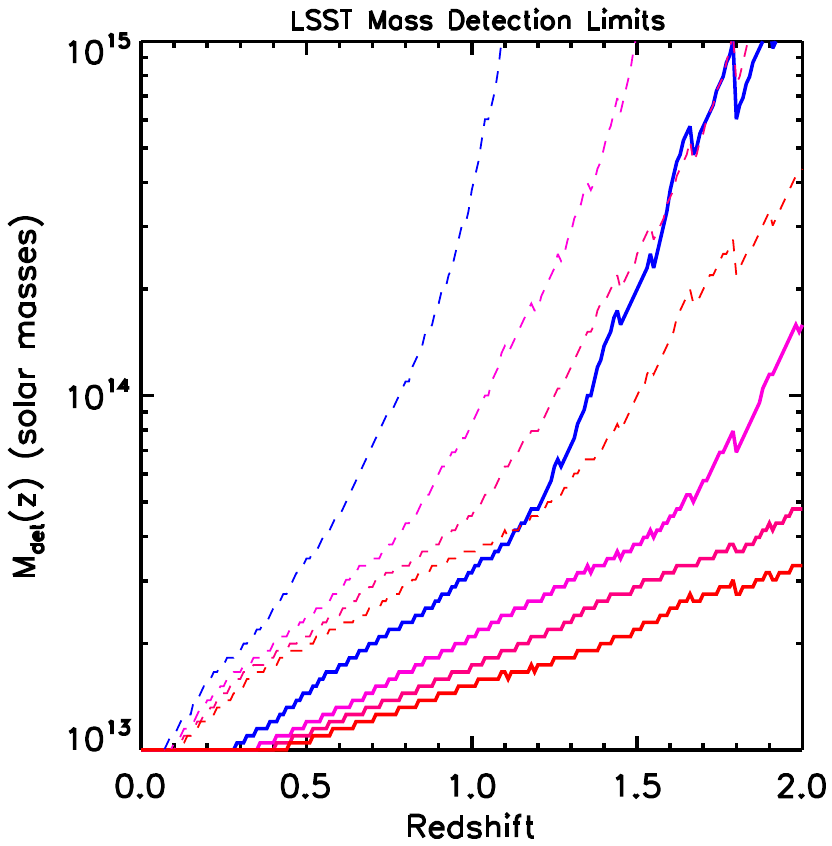}
\caption{Minimum detectable cluster mass as a function of redshift for
  $r, i, z$ and $y$ bands (blue to red curves).  Cluster detection
  requires at least ten red--sequence galaxies detected in--band at
  $10\,\sigma$ and with $L_R>0.4L_*$ (Fromenteau et al., in
  preparation).  The dashed 
  lines correspond to single-visit images and the solid lines to the
  complete ten--year survey.} 
\label{fig:lss:clus:MdetSummary_RIZY}
\end{figure}

\begin{figure}
\centering
\includegraphics[width=3.45in]{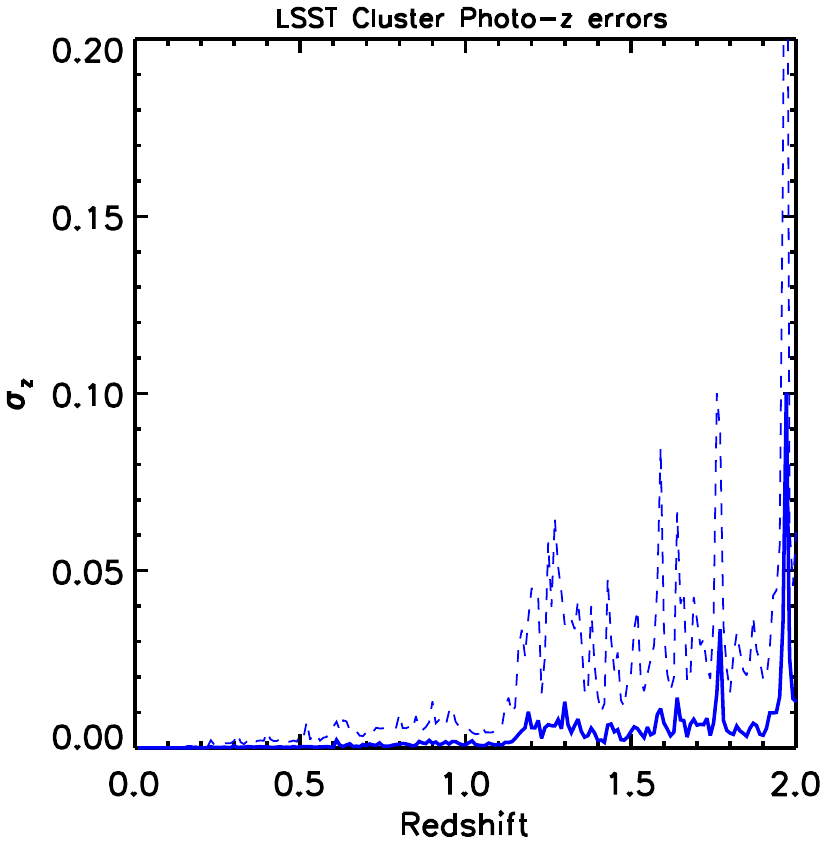}
\caption{Estimated cluster photometric redshift errors for
  single--visit (dashed) and full ten-year (solid) survey images.  They are underestimated
because the model uses a single red--galaxy template, but 
give an idea of the expected errors and their 
variation with redshift.}
\label{fig:lss:clus:photoz}
\end{figure}

To address this point, we show the estimated mass detection threshold
as a function of redshift, $M_{det}(z)$, individually for the LSST $r,
i, z$ and $y$ bands in \autoref{fig:lss:clus:MdetSummary_RIZY}.  We
consider as detected those clusters for which ten red--sequence
galaxies with $L_R>0.4L*$ are seen at
$>10\sigma$ in the band in question.  For our cluster model we use
red--galaxy conditional luminosity functions from SDSS
\citep{Yangetal08} evolved with passive stellar evolution to higher
redshift (Fromenteau et al., in preparation).  Fromenteau et
al. examine the appropriateness of passive evolution by comparing the
cluster red galaxy luminosity functions from this model to luminosity
functions extracted from the halo occupation distribution
(\autoref{sec:galaxies:distfunct}) constrained in the NOAO deep wide field
by \cite{Brownetal08}.  

That comparison supports the idea that the colors of red sequence cluster populations are well described by
passive evolution out to $z\sim 1$.  Less certain is what fraction of
clusters will {\it possess} a red sequence, however, and how many
members that red sequence will have.  A number of studies have found
that the abundance of red galaxies with luminosities of $L*$ and below has increased by a
factor of 2-4 since $z\sim 1$ \citep{Brownetal08,2006ApJ...647..853W},
so we know the overall red sequence population must have grown since
$z\sim 1$.  Detailed investigations of galaxy populations in groups
and low-mass clusters \citep{2007MNRAS.376.1425G} and of the
dependence of galaxy color on environment \citep{2007MNRAS.376.1445C}
to $z\sim 1.4$ have found that the fraction of red galaxies in
clusters of modest mass ($\gtsim10^{13}\,M_\odot$) is
indistinguishable from the fraction in the field at $z\sim 1.35$, but
grows steadily at lower redshifts.  Since it takes $\sim 1$ Gyr for a
galaxy's color to turn red after star formation ends, this requires
that star formation began to be strangled in clusters with a mass of
$\gtsim 10^{13}\,M_\odot$ around redshift 2, with the process ongoing
to $z=1$ or beyond.  However, at least some massive systems {\it do}
contain a well-defined red sequence at $z >1$; e.g., the $z=1.24$
system studied by \autoref{cmd_rdcs1252}. 

As expected in hierarchical structure formation models, it appears
that 
galaxy evolution proceeds fastest in the most massive clusters, which
also will host the most massive galaxies.  In those systems, a red
sequence may be apparent by $z\sim 1.5-2$, while in lower-mass
clusters or groups it may appear only after $z\sim 1$.  This is
predicted in models where the near total quenching of star formation
necessary to produce a red sequence galaxy requires the presence of a
dark matter halo above the threshold mass where cooling becomes
inefficient (e.g., \citealt{1977ApJ...211..638S, 1977MNRAS.179..541R,
  1977ApJ...215..483B, 1978MNRAS.183..341W, 2006MNRAS.365...11C}).  In
such a scenario, massive clusters pass that threshold mass at $z\sim
3$, while a typical weak cluster or group will pass it at $z\sim 2$ or
later, consistent with observations.  Our uncertainty in the evolution
of galaxies within clusters to high redshift is one reason it will be
important to compare cluster samples selected via different means. 

Assuming the the red sequence galaxy fraction does not
change with redshift, LSST will be able to detect clusters well down
into the group range, in both single visit and complete survey
images (\autoref{fig:lss:clus:MdetSummary_RIZY}).  The mass threshold
decreases as we move redward because the 
RS galaxies are dominated by red light, and inflects upwards in a
given band when the 4000\AA\ break moves through that band.
Single visit $r, i$ and $z$ images will be comparable in depth to the
Dark Energy Survey (DES) survey.  The LSST $y$ band and the deeper
imaging of the complete survey will allow us to go to appreciably
higher redshifts at a given mass threshold than is possible for DES.

In \autoref{fig:lss:clus:photoz} we give an estimate of the expected
photometric redshift errors for a cluster of $10^{14}\,M_\odot$ as a
function of redshift (Fromenteau et al., in preparation).  The errors are
slightly underestimated, especially at low redshift, because we have only
employed a single galaxy template for this estimate.  Nevertheless, we
see that cluster photometric redshifts should be very good out to
redshift unity, after which they suffer some degradation for
single visit images; the degradation, however, is not severe. The
redshift precision remains very good out to $z=2$ with the deeper
complete survey images.


\subsubsection{Galaxy Cluster Finder: Red-Sequence Voronoi Tessellation 
and Percolation Method}

\begin{figure}
\centering
\includegraphics[width=4in]{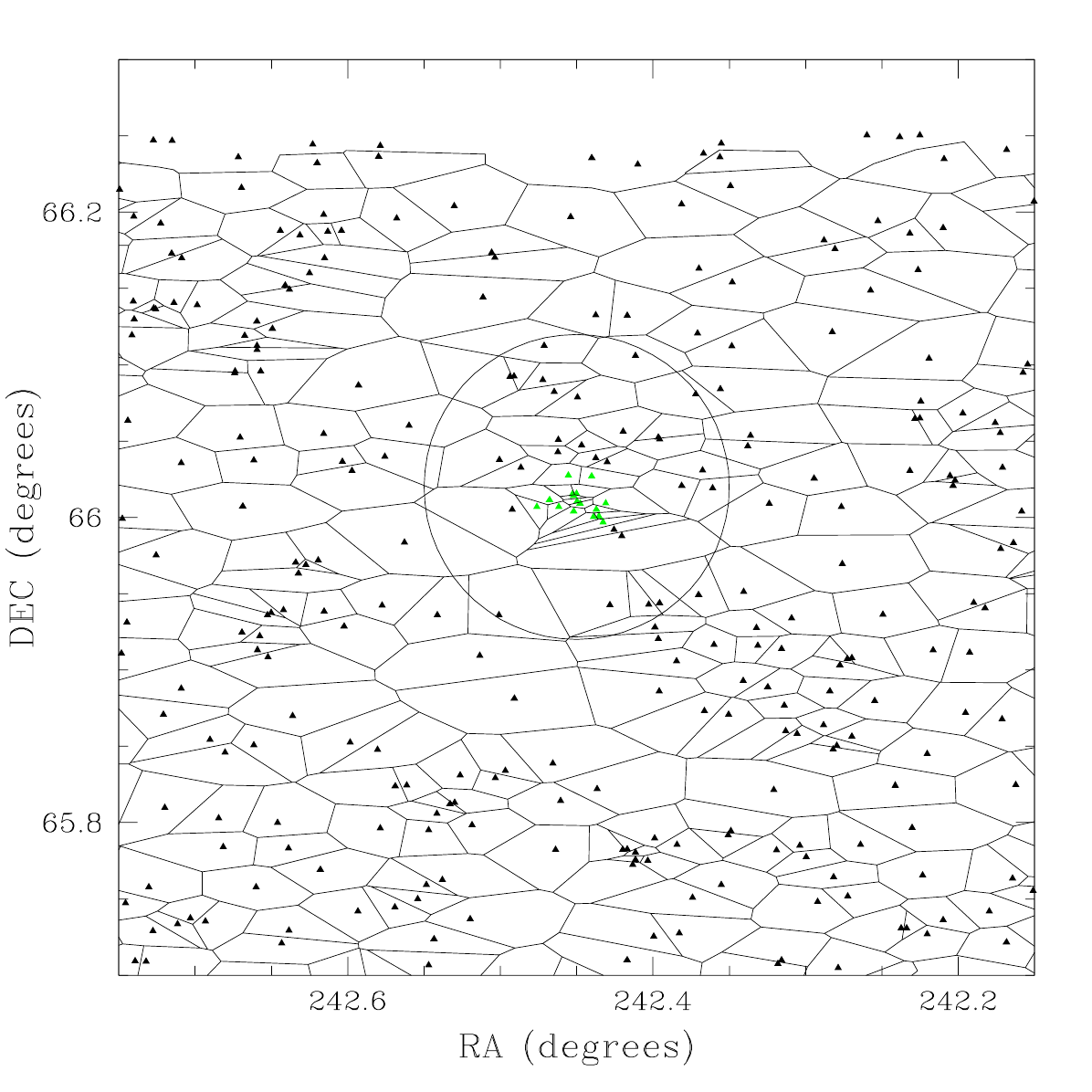}
\caption{The Voronoi tessellation on the galaxy distribution (from the
  CTIO and KPNO 4-m telescopes) in the
  field of an extended X-ray source observed by Chandra. Only galaxies satisfying 
the $r-i$ color cut expected for a cluster red sequence at $z=0.475$ are
depicted.  The imaging data used here do not go as deep as a single
LSST visit.  Figure from \citet{Barkhouseetal06}, with permission. 
\label{fig:lss:clus:voronoi}}
\end{figure}

\begin{figure}
\centering
\includegraphics[width=4in]{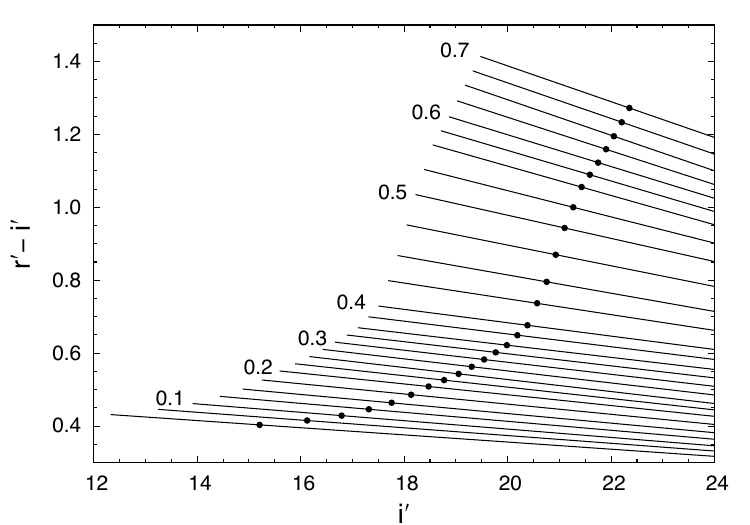}
\caption{Selected red-sequence model color slices used to choose galaxies for a range of redshifts (indicated to the left 
of the lines) for the VTP cluster detection method
\citep{Barkhouseetal06}. The solid circles indicate positions of the brightest 
cluster galaxies. 
Figure from \citet{Barkhouseetal06}, with permission. 
\label{fig:lss:clus:colorslices}}
\end{figure}

The detection of galaxy clusters based on the red-sequence Voronoi
tessellation and percolation method (VTP) utilizes the red-sequence
property of early-type cluster galaxies. The method works by dividing
the galaxy spatial plane into polyhedral cells, each containing a
unique galaxy (see \autoref{fig:lss:clus:voronoi}). The cells are then
grouped together using a percolation method, and galaxy clusters are
detected as over densities of cells from that expected for a random
distribution. In order to maximize the signal-to-noise ratio of a given
cluster above the background field population, the galaxy plane is
first divided into multiple overlapping red-sequence slices based on
the expected color of early-type cluster galaxies for a range in
redshifts (\autoref{fig:lss:clus:colorslices}). Galaxy clusters are
then detected using individual color slices, with overlapping
detections in adjacent slices merged based on the significance of the
detected clusters as output by the VTP algorithm. The non-parametric
VTP method employs no assumption
regarding cluster shape; thus the VTP technique is
sensitive to symmetric as well as irregular clusters.  However, with
this and other optical cluster-finding algorithms, it will be
necessary to determine the relationship of the identified systems to
the individual virialized halos counted in cosmological tests;
e.g., these techniques will often identify the combined population of
groups along the line of sight as belonging to a single cluster
(e.g., \citealt{2007ApJ...655..128G}).   

\subsubsection{Relation to Other Cluster Catalogs (X--ray, SZ, IR, Shear)}
{\bf SZ Cluster Scaling Relations}\\

\cite{SealfonWL} showed how to place constraints on cluster physics by
stacking the weak lensing signal from multiple clusters found through
the SZ effect. The next generation of
SZ surveys will provide a catalog of thousands
of clusters \citep{Carlstrometal02}.  The SZ flux is proportional to
the integral of the product of the density and the temperature of the
hot gas in 
the cluster \citep{SZ80}, and the flux limit of the catalogs will
approximately translate into a mass limit. However, the actual value
of the cluster's mass for a given SZ flux depends on the details of
the SZ-mass relation, which in turn is governed by cluster physics.
Cluster scaling relations involving SZ have been extensively studied
as a tool to investigate cluster physics (e.g., \citealt{daSilva03,
  Bensonetal, McCarthy03,VerdeHS}).

Recent work has explored the SZ-mass relation using numerical
simulations (e.g., \citealt{Oh03,daSilva03,Motl,Nagai}) and analytical
approximations (e.g., \citealt{DosSantos01,ReidSpergel, Roy05,
  Ostriker05}).  Numerical models find that the SZ-mass relation is
expected to be very tight, implying that cluster masses can be
directly read out from the SZ observations once the SZ-mass relation
has been calibrated.

One possibility is to calibrate this relation on cosmological
simulations, but this requires understanding and modeling all the
relevant baryonic physics. 
Alternatively, we would like a robust method to estimate cluster
masses directly from observations, independently of intra-cluster
medium (ICM) modeling, and without relying on numerical simulations.
Gravitational lensing provides the most direct way to measure the mass of
clusters, and we discuss this in detail in
\autoref{sec:sl:clusmf} and \autoref{sec:wl:shearpeaks}.  

With ground-based experiments, a direct mass determination is possible
only for fairly massive clusters, $ \approx 10^{15} \, M_{\odot}$
(e.g., \citealt{Marian}).  However \cite{SealfonWL} argue that by
stacking the weak lensing signals from multiple clusters
with roughly the same SZ luminosity, otherwise undetectable shear
signal can be amplified, allowing one to determine an average mass in
bins of SZ luminosity. 

While a non-parametric technique to reconstruct the average mass
profile from measurements of clusters-shear correlation function has
been presented in \cite{Johnstonetal07a, Johnstonetal07b}, for this
signal-to-noise ratio calculation we will take a complementary approach
followed by \cite{SealfonWL}: we assume a cluster profile (a Navarro-Frenk-White (NFW)
radial cluster profile; \citealt{NFW}) and we will recover the average
mass.

Both the total SZ luminosity and the central Compton
parameter $y_0$ are expected to depend on cluster
mass approximately as power laws, with some intrinsic scatter. 
It is customary to parametrize these
relations over a range of masses as power laws with two free
parameters (an amplitude and a slope of a linear fit in log-log space)
and study how these parameters are expected to change for different
assumptions about cluster physics. 

\autoref{lss:tab:scalingrelations} shows the constraints on these
parameters from a 5000 deg$^2$ SZ survey overlapping the LSST
footprint.  Given these errors, we will be able to 
distinguish at more than the $3\,\sigma$ level between:
\begin{table}
\begin{center}
\begin{tabular}{l|cc|cc}
$\Delta z=0.2$   & $M$-$L_{\rm SZ}$&       & $M$-$y_0$ &      \\
bin&  Amplitude      & Slope & Amplitude & Slope\\
\hline
$z\sim 0.1$  &     2.8\%         & 2.4\%   &    2.4\%    &  3.6\%    \\
$z\sim 0.5$   &     1.6  \%       & 1.2\%   &    1.6\%    &  2.4\%   \\
$z\sim 1$  &     9 \%       & 9\%  &    5.3\%    &  12\%    \\
\end{tabular}
\caption{Forecasted fractional errors on the slope and amplitude of
  the scaling relations between mass and SZ luminosity and Compton
  parameter $y_0$.  Despite having
  larger  scatter, the $M$-$y_0$ relation is flatter, yielding
  fractional errors on the 
  amplitude comparable (or better) than those for the $M$-$L_{\rm SZ}$
  relation.}
\label{lss:tab:scalingrelations}
\end{center}
\end{table}

\begin{itemize} 
\item A  self-similar model with only 
gravitational physics (i.e. no heating nor cooling), where the gas temperature is given solely by the 
dark matter virial temperature;
\item a model with a pre-heating  and a cooled gas fraction of $f_{cool}=0.42$;
\item a model with an accretion pressure decreased by a factor of 3.5
  from self-similar spherical collapse; 
\item a model with an accretion pressure increased by a factor of 3.5 from 
self-similar spherical collapse;
\item a purely adiabatic cluster model from one that includes cooling 
and a star formation model;
\item models in which the exponent of the radial entropy profile are
  1.1 and 1.5, respectively; and
\item models in which the entropy profile normalization differ by a
  factor of 1.5. 
\end{itemize}

These small error bars suggest that we may be able to 
constrain how the mass-SZ scaling evolves with redshift. A mass
measurement from weak lensing and a SZ measurement in different
redshift bins can constrain the evolution of hot gas as a function of
redshift, which in turn would enable one to constrain feedback
evolution.

The approach presented here can, of course, be applied also to other analytical models and to predictions from numerical work.
In the cases where the two-parameter fit yields a S/N $> 5$, one could
add parameters to the fit to test, e.g., if deviations from a power 
law in the scaling relation can yield additional information about cluster physics.
\section{Cross-Correlations with the Cosmic Microwave Background}
\label{sec:lss:isw}
{\it Ryan Scranton, Leopoldo Infante}

Cross-correlations between a large galaxy survey like the LSST and a map of
the cosmic microwave background radiation from either WMAP or Planck can
provide a number of useful measurements of various physical processes.  The
most studied of these is the detection of the late Integrated Sachs-Wolfe
(ISW) effect \citep{SachsWolfe1967}, a positive cross-correlation between foreground galaxies and
background CMB temperature induced by dark energy.  In the first year of
operation, LSST should measure the ISW effect to greater precision than
current efforts involving combinations of galaxy catalogs from multiple
sources.   Over the longer term, measuring the ISW with LSST has the potential
to provide unique insight into the nature of dark energy by placing
constraints on the smoothness of the dark energy potential at the 3-5\% level
on scales around 1 Gpc.  The richness of the LSST galaxy sample will also
allow us to greatly expand upon related measurements involving galaxy-CMB
cross-correlations.  This will include magnification-induced ISW signal at
high redshifts, ISW detection using superclusters and voids and cluster
peculiar velocity measurements through the kinetic Sunyaev-Zel'dovich effect.

\subsection{Dark Energy and Structure Formation}

As the Universe expands, light travels from the surface of last scattering
through the intervening large scale structure to observers here on Earth.  In
doing so, it passes through any number of local gravitational
potentials (i.e., regions in which structure is forming), experiencing
a gravitational blueshift as they fall into the potential and a
redshift as they exit it. During the matter-dominated phase of the
Universe's evolution (when $\Omega_{matter} \approx 1$), the rate of
structure growth matches the rate of 
universal expansion to first order, so the extra energy lost by photons climbing out of
potential wells that have been growing during the photon's traversal 
matches the general expansion of the Universe.  Thus, the photon energy is
practically identical to one that  avoided the potential altogether.  As the Universe
transitions to a dark energy dominated phase, the universal expansion begins
to accelerate, outpacing the growth of structure.  Hence, photons passing
through intervening potentials will not lose all the energy they
gained when they entered, as the potential wells are shallower as they
leave. This induces a positive correlation between the background CMB
temperature and the projected matter (or galaxy) density.

The first measurements of the ISW effect were done using CMB maps from
the COBE DMR mission cross-correlated with the NVSS (radio) and HEAO-1
(X-ray) galaxy 
surveys.  With the release of higher resolution CMB maps from WMAP, a
second wave of analysis was done cross-correlating the CMB against
galaxy surveys from 2MASS \citep{Afshordi2004}, SDSS
\citep{Scranton2003, Fosalba2003}, NVSS \citep{Raccanelli08}, and the
photometric quasar survey from the SDSS \citep{Giannantonio2006}.
Each of these measurements yielded detections of the expected signal
in the 2-3 $\sigma$ range.  The current state of the art comes from
combining the various measurements from the various surveys into a
single detection \citep{Giannantonio2008, Ho2008}.  This allows for a
redshift coverage from $0 < z < 2$, albeit with varying sky coverage
over the course of that range.  These assemblies yield detections in
the 3.5-4.5 $\sigma$ range.  While the ISW
signal itself is not currently capable of constraining cosmological
parameters to the extent of other probes, its likelihood contours are
very complementary to those from baryon acoustic oscillations, CMB,
supernovae, and weak lensing.

\subsection{ISW Formalism}

The ISW effect dominates the cross-correlation signal at an angular scale of
$\theta \gtrsim 1^\circ$ (at this scale, the choice of WMAP or Planck for
the CMB map should be irrelevant), while at smaller angular scales, the signal
is dominated by the thermal SZ effect. The cross-correlation function of fields
$A$ and $B$, $\omega_{AB}(\theta)$ can be written in terms of the
angular power spectrum multipoles ($C_{AB}(l)$) by expanding it with Legendre
polynomials,
\begin{equation}
  \omega_{AB} (\theta)\ =\ \sum_{l=0}^{\infty} \frac{(2l+1)}{4 \pi} C_{AB}(l)
  P_l(\cos{\theta}) ,
\end{equation}
where $P_l$ is the Legendre polynomial of order $l$. It is possible to show
that for small angles, or large $l$, \citep{Afshordi2004, Cooray02b}
\begin{equation}
  C_{AB}(l)\ =\ \int_0^{\infty} \frac{dr}{r^2}\ P(k)\ W^A(k,r)\ W^B(k,r) ,
\end{equation}
where $P(k)$ is the initial power spectrum of matter,
$k = \frac{l + 1/2}{r} ,$ and $W^X (k,r)$ is the window function of the
field $X$. This approximation holds up to a good degree of accuracy for
$l \ge 2$. 

The window function of the anisotropy field in the CMB map generated by the
ISW effect can be written as
\begin{equation}
  W^{ISW}(r,k)\ =\ -3 T_0\ \frac{\Omega_m}{k^2} \frac{H_0^2}{c^2}
  \frac{\partial G(z) (1+z)}{\partial z} , \label{eq:isw_window}
\end{equation}
where $T_0$ is the mean temperature of the CMB,  $\Omega_m$ is the matter
density of the Universe in units of the critical density, $H_0$ is Hubble's
constant, $c$ is the speed of light, and $G(z)$ is the growth factor of the
gravitational potential.

For the galaxy side, the window function is given by
\begin{equation}
W^g\ =\ b_g\ \frac{H(z)}{c}\ G(z)\ n(z) , \label{eq:galaxy_window}
\end{equation}
where $b_g$ is the bias factor, $H(z)$ is the Hubble parameter as a function of
redshift, and $n(z)$ is the galaxy density distribution, which will
depend directly on the characteristics of the observations.  This model of
bias is very simple, obviously, but we are only interested in the signal on
large scale and, as will be seen shortly, the signal-to-noise ratio is
independent of our choice.

The variance ($\sigma^2_{C_{gT}}$) on each
multipole of the angular cross-power spectrum, ($C_{gT}$), is given by
\begin{equation}
  \sigma^2_{c_{gT}}(l) \ =\ \frac{1}{f_{sky}(2l+1)}\ \left\{C^2_{gT}(l) + C_{TT}(l)
    \left[C_{gg}(l)+\frac{1}{\bar{N}}\right]\right\},
\end{equation}
where $C_{TT}$ and $C_{gg}$ are the CMB and galaxy angular power spectra,
respectively.  $\bar{N}$ is the mean number of galaxies per steradian on the
survey (the ``shot-noise'' term in the galaxy map) and $f_{sky}$ is the fraction
of the sky used for the cross-correlation.  This can be propagated to the
cross-correlation function, such that
\begin{equation}
  \sigma^2_{\omega_{gT}}\ =\ \sum_l \frac{(2l+1)}{f_{sky} (4\pi)^2}\
  P_l^2(\cos{\theta})\ \left\{C^2_{gT}(l) + C_{TT}(l)
    \left[C_{gg}(l)+\frac{1}{\bar{N}}\right]\right\} .
\end{equation}

\subsection{Cross-Correlating the CMB and the Stacked LSST Galaxy Sample}

With the full LSST survey, we will be able to cross-correlate the CMB
fluctuations with different subsamples of galaxies selected by redshift or type.  
This will allow us to measure
how the ISW signal changes over the course of the history of the Universe, how
different populations experience multiple effects that contribute to their
cross-correlation with the CMB, and how the local over-(under-)densities contribute
to the correlation.  

\subsubsection{Dark Energy Clustering}

As mentioned above, the ISW constraints on the cosmological parameters
for the fiducial model (\autoref{sec:com:cos}) are relatively weak.
However, for more exotic models of dark energy this is not true.  In
particular, if the dark energy field can cluster, ISW measurements
become our best means to detect this effect, which would be a key
indicator for physical dark energy models.

\citet{Hu2004} consider a dark energy model where the clustering is
parametrized by the sound speed ($c_s$) of the dark energy field
(which we here assume to be independent of $w$).  Given this speed,
one can determine a scale $\eta$ at which perturbations in the fluid
enter the horizon and begin to gravitationally collapse.  At this
point, dark energy would begin to fall into gravitational potentials
on the largest scales, and the freezing out process that began as dark
energy became the driver for the expansion of the Universe would
reverse itself.  Since this happens on the largest scales and proceeds
inward, the effect would be first detectable in the ISW effect, as
well as presenting the longest baseline for measuring dark energy
clustering.

For the calculations presented in \citet{Hu2004}, the fiducial model used has
$w = -0.8, c_S = 0.1\,c$.  The survey depth considered is 70 galaxies
arcmin$^{-2}$, which is approximately what would be possible with LSST
if we only 
require $10\,\sigma$ photometry in the three deepest bands
(\autoref{sec:common:galcounts}).  With these model 
parameters, they estimate that the smoothness of the dark energy potential
could be constrained at 3\% on scales of 1 Gpc.  While $c_s$ remains largely
unconstrained by current measurements, the combined constraints from
CMB, baryon oscillations, and supernovae put the value of $w$ used in those calculations outside of the 95\%
contours for a constant $w$ model.  

\subsubsection{Magnification}

The peak in expected ISW $S/N$
for a $\Lambda$CDM cosmology happens for galaxies around $z \sim 0.5$.  The window function
for the CMB side of the cross-correlation peaks at $z = 0$, but the competing
volume effects push the peak in $S/N$ to somewhat higher redshift.  For
$z > 1$, the effects of a cosmological constant are generally too small to
generate a deviation from simple CDM growth measurable via the ISW effect.
However, as pointed out by \citet{LoVerde07}, this ignores the effects of
magnification.

For higher redshift samples, we need to include the effect of lensing by
foreground structure.  While the dark matter potentials will not have
experienced significant decay at those redshifts, the galaxies inside those
potentials will be lensed by foreground structures where the ISW effect is
considerably stronger.  This replaces the galaxy window function from
\autoref{eq:galaxy_window} with
\begin{equation}
W^{\mu,i}\ =\ 3\Omega_m\frac{H_0^2}{c^2} (2.5s_i - 1) \ G(z)\ (1 + z) g(z, z_i) ,
\label{eq:lensing_window}
\end{equation}
where $s_i$ is the power-law slope of a given photometric redshift bin's galaxy
number counts.  $g(z, z_i)$ is the lensing weight function for that redshift
bin,
\begin{equation}
g(z, z_i) = \chi(z) \int dz^\prime \frac{\chi(z^\prime) - \chi(z)}{\chi(z^\prime)}
n_i(z^\prime),
\end{equation}
where $\chi$ is the comoving distance and $n_i(z)$ is the redshift distribution
for the bin.  For a given set of galaxies around $\chi$, $g$ will peak at
roughly $\chi/2$.  This, in turn, implies that even a set of galaxies at
$z > 2$ can experience significant cross-correlation with the CMB due to
lensing by galaxies at much lower redshift.  This signal is, of course,
dependent on the value of $s$ for that sample, although as pointed out by
\citet{Menard02}, one can apply an optimal estimator where each galaxy is
weighted by $2.5s - 1$, which only yields a null result for $s = 0.4$.

Using two samples which are roughly equivalent to the first year and full LSST
data set, \citet{LoVerde07} show that the contribution of magnification for
a given high redshift bin ($z \sim 3.5$) can result in nearly an order of
magnitude increase in the expected cross-correlation.  This in turn leads to
an increase in the $S/N$ at those redshifts, although the aggregate $S/N$ as a
function of redshift does not improve significantly due to strong correlations
between the signals in each redshift bins induced by the magnification.  More
importantly, including the effects of magnification improves the constraints
on $w$ at higher redshift from ISW by roughly a factor of 2, bringing them
in line with the constraints at lower redshifts ($\delta w \sim 0.2$ at
$z \sim 1$).  With the galaxy color information available in LSST, this could
be further enhanced by looking at galaxy sub-populations like luminous red
galaxies or Lyman-break galaxies which have especially steep number
count relations ($2.5s - 1 \gtrsim$ 2).

\subsubsection{Superclusters and Voids}

The standard detection of the ISW effect is through cross-correlation
between the local large scale structure and CMB temperature fluctuations.
These detections have been at levels below 3$\sigma$.  In coming years through
large area surveys (DES, VST, and so on), the $S/N$ for ISW detection will be
increased by perhaps $\sim$ 50\% and by a factor of 4 with the early LSST
20,000 deg$^2$ data. A recent paper by \citet{Granett08} claims a $S/N > 4$
detection of the ISW effect. To trace the highest and lowest
density peaks, which presumably trace the highest and lowest mass structures,
they identified 50 candidate superclusters and 50 potential
supervoids at redshifts $\sim 0.5$ from the SDSS data. They stacked the WMAP five-year temperature
pixels corresponding to these regions, and found an increase 
in temperature towards the potential superclusters and a decrease towards the
potential supervoids, detecting the ISW effect at above 4 $\sigma$. 

To carry out this experiment with LSST, we will need to clearly
identify massive supercluster-scale structures on scales up to 150 Mpc. LSST will provide
photometric redshifts as good as $0.02(1+z)$ for luminous red galaxies (LRGs)
(\autoref{sec:common:photo-z}).  We will be able to identify coherent
structures to $z\sim 1.5$, where $\Delta z \sim 0.04$
corresponds to $\Delta(D_{com})\sim 93$ Mpc or $\theta\sim3^\circ$. 

Extrapolating from the \citet{Granett08} results and scaling by the
larger solid angle of the LSST analysis, we estimate that we
will be able to detect the ISW effect in this way to $7\,\sigma$ total
using perhaps half a dozen redshift shells to $z \sim 2.$

\section{Education and Public Outreach}

{\it Eric Gawiser, Suzanne H. Jacoby}

The Large-Scale Structure Science Collaboration explores ``the big
picture'' of how the Universe is organized on a grand scale and over
grand expanses of time.  Characterizing the evolution of the
distribution of matter on extragalactic scales is a primary science
goal.  The large-scale structure of the Universe encodes crucial
information about its contents, how it is organized and how this
organization has evolved with time.   This way of viewing the Universe
is well aligned with several ``big ideas'' in science education reform
as described in Project 2061:  Science for All
Americans\footnote{\url{http://www.project2061.org/publications/sfaa/default.htm}}
and the 
National Science Education Standards \citep{NRC96}.  These ideas, called
Unifying Concepts, include concepts such as systems, order, and
organization; patterns of change, evolution, and scale.  Unifying
Concepts can serve as a focus for instruction at any grade level; they
provide a framework within which science can be learned and a context
for fostering an understanding of the nature of science.  The
Large-Scale Structure team will support the EPO group in providing
this framework of Unifying Concepts in materials developed for
classroom learning experiences. 

One thread of the LSST Education and Public Outreach program
emphasizes visualization of LSST data in science centers and on
computer screens of all sizes. Each LSST public data release can be
viewed using two- and three-dimensional visualization programs (e.g.,
Google Sky, WWT, the Digital Universe).  Large-Scale Structure Science
Collaboration team members will assist informal science centers in
incorporating the LSST data into their visualization platforms and
conveying their meaning to the public.  This will enable LSST
discoveries to be featured in weekly, live planetarium shows and to
actively involve our audience in LSST's mission of mapping the
structure of all matter in the Universe. 
The use of LSST data in an informal museum or science center setting
represents an ideal opportunity to expose large numbers of people to
the magnificence of the vast LSST data set.  Image browsers such as
Google Sky, WWT, and the Digital Universe will broaden LSST's
availability to everyone with a home computer or laptop (or PDA or
cellphone, given the rapid growth of technology), to truly enable
visualization on ``computer screens of all sizes.''


\bibliographystyle{SciBook}
\bibliography{lss/lss}

%
%
%
%
%
%
%
%
%
%
%
%
%
%
%
%
%
%
%
%
%
%
%
%
%
%
\chapter[Weak Lensing]
{Weak Gravitational Lensing}
\label{chp:wl}

{\it David Wittman, Bhuvnesh Jain, Douglas Clowe, Ian P. Dell'Antonio,
  Rachel Mandelbaum, Morgan May, Masahiro Takada, Anthony Tyson, Sheng
  Wang, Andrew Zentner}  

%
%
%
%
%
%
%
%
%
%
%
%
%
%
%
%
%
%
%
%
%
%
%
%
%
%
Weak lensing (WL) is the most direct probe of the mass distribution in the
Universe. It has been applied successfully on many different scales,
from galaxy halos to large-scale structure. These measurements in turn
allow us to constrain models of dark matter, dark energy, and
cosmology. The primary limitation to date has been statistical:
Lensing causes a small perturbation to the initially random
orientations of background galaxies, so large numbers of background
galaxies are required for high signal-to-noise ratio measurements. The
LSST survey, encompassing billions of galaxies, will dramatically
improve the statistical power of weak lensing observations. At large
scales, cosmic variance is the limiting factor, and the extremely wide
footprint of the LSST survey will bring this limit down as well. At
the same time, the greatly increased statistical power means that
systematic errors must be carefully examined and controlled. 

The key observables to be extracted from the LSST data set are shear
from galaxy shapes and source redshifts from photometric
estimates. These must be derived for as many galaxies as possible,
over as wide a range in redshift as possible. Subsequent analysis can
be in terms of two- or three-point correlation functions, or shear
profiles or mass maps depending on the specific project, but nearly
all analyses rest on these two fundamental quantities. The projects
are listed below in increasing order of angular scale. In combination
(and especially when combined with LSST baryon acoustic oscillation
data normalized by Planck data; \autoref{sec:cp:wlbao}), these WL
measurements will provide powerful constraints on dark energy and
modified gravity, the mass power spectrum, and on the distribution and
mass profiles of galaxy and cluster halos.  Due to its precision and
wide-area coverage, the LSST WL survey data will uniquely probe the
physics of dark matter and cosmological issues of large-scale
isotropy.  Below, after reviewing weak lensing basics (\autoref{weaklens-basics}), we discuss lensing by galaxies (\autoref{sec:wl:gg}), lensing by clusters
of galaxies (\autoref{wl-sec-cluster}), lensing
by large scale structure (\autoref{sec:wl:lss}), and finally systematic issues which touch on
all these areas (\autoref{wl-sec-sys}).



\section{Weak Lensing Basics}
\label{weaklens-basics}
%
%
%
%
%
%
%
%
%
%
%
%
%
%
%
%
%
%
%
%
%
%
%
%
%
%

Massive structures along the line of sight deflect photons originating
from distant galaxies.  \autoref{fig:sl:lenscartoon} shows the
geometry in the general case where multiple photon paths from source
to observer are possible.  Outside the densest lines of sight, only
one, slightly deflected, path is possible, and this is the domain of
weak lensing.  If the source is small compared to the scales on which
the deflection angle varies, the effect is a (re)mapping of $f^{\rm
s}$, the source's surface brightness distribution (see
\citealt{Bartelmann99} for more details):

\begin{equation}
f^{\rm obs}(\theta_i)=f^{\rm s}({\cal A}_{ij}\theta_j),
\end{equation}

\noindent where ${\cal A}$ is the distortion matrix (the Jacobian
of the transformation)

\begin{equation}
{\cal A}=\frac{\partial(\delta\theta_i)}{\partial \theta_j}
=(\delta_{ij}-\Psi_{,ij})=
\left(
    \begin{array}{cc}
        1-\kappa-\gamma_1 & -\gamma_2 \\
        -\gamma_2        & 1-\kappa+\gamma_1 \\
    \end{array}
\right).
\label{distort}
\end{equation}

Here $\Psi$ is the two-dimensional lensing
potential introduced in \autoref{chapter1:eq:lenspot}, and 
$\Psi_{,ij}\equiv\partial^2 \Psi/{\partial
\theta_i\partial\theta_j}$. The lensing convergence $\kappa$, defined in \autoref{chapter1:eq:kappa},
is a scalar quantity which can also be defined as a weighted
projection of the mass density fluctuation field:
\begin{equation}
\kappa(\bm{\theta})=\frac{1}{2}\nabla^2\Psi(\bm{\theta})
=\int\!\!d\chi W(\chi) 
\delta[\chi, \chi\bm{\theta}],
\label{eqn:kappa}
\end{equation}
where the Laplacian operator $\nabla^2\equiv
\partial^2/{\partial\bm{\theta}^2}$ is defined using the flat sky
approximation, $\delta$ is the fractional deviation of the density
field from uniformity, and $\chi$ is the co-moving
distance (we have assumed a spatially flat Universe).  Note that $\chi$
is related to redshift $z$ via the relation $d\chi=dz/H(z)$, where $H(z)$ is
the Hubble parameter at epoch $z$.  
The lensing efficiency function $W$ is given by
\begin{equation}
W(\chi)=\frac{3}{2}\Omega_{m0}H_0^2 a^{-1}(\chi)
\chi \int\!\!d\chi_s~ 
n_s(\chi_s) \frac{\chi_{\rm s}-\chi}{\chi_s},
\label{eqn:weightgl}
\end{equation}
where $n_s(\chi_s)$ is the redshift selection function of source galaxies
and $H_0$ is the Hubble constant today. 
If all source galaxies are at a single redshift
$z_s$, then $n_s(\chi)=\delta_D(\chi-\chi_s)$.  

In \autoref{distort} we introduced the components of
the complex shear $\bm{\gamma}\equiv\gamma_1+i\gamma_2$, which can
also be written as $\bm{\gamma}= \gamma\ {\rm exp}(2i\alpha)$, where $\alpha$
is the orientation angle of the shear. The Cartesian components of the
shear field are related to the lensing potential through
\begin{equation}
\gamma_{1}=\frac{1}{2}(\Psi_{,11}-\Psi_{,22})\hspace{1em}{\rm and}\hspace{1em}
\gamma_2=\Psi_{,12}.
\label{sheardef}
\end{equation}

In the weak lensing regime, the convergence $\kappa$ gives the magnification
(increase in size) of an image and the shear $\gamma$ gives the ellipticity
induced on an initially circular image (see
\autoref{fig-wlfeelgood} for an illustration). Under the assumption
that galaxies are randomly oriented in the absence of lensing, the
strength of the tidal gravitational field can be inferred from the
measured ellipticities of an ensemble of sources
(see \autoref{weaklens-intrinsic} for a discussion of intrinsic
alignments). In the absence of observational distortions, the observed
ellipticity, $e^{\rm obs}$, is related to its unlensed value, $e^{\rm
int}$, through
\citep{Seitz97,Bartelmann99}:

\begin{figure}
\centerline{\includegraphics[width=4.5in]{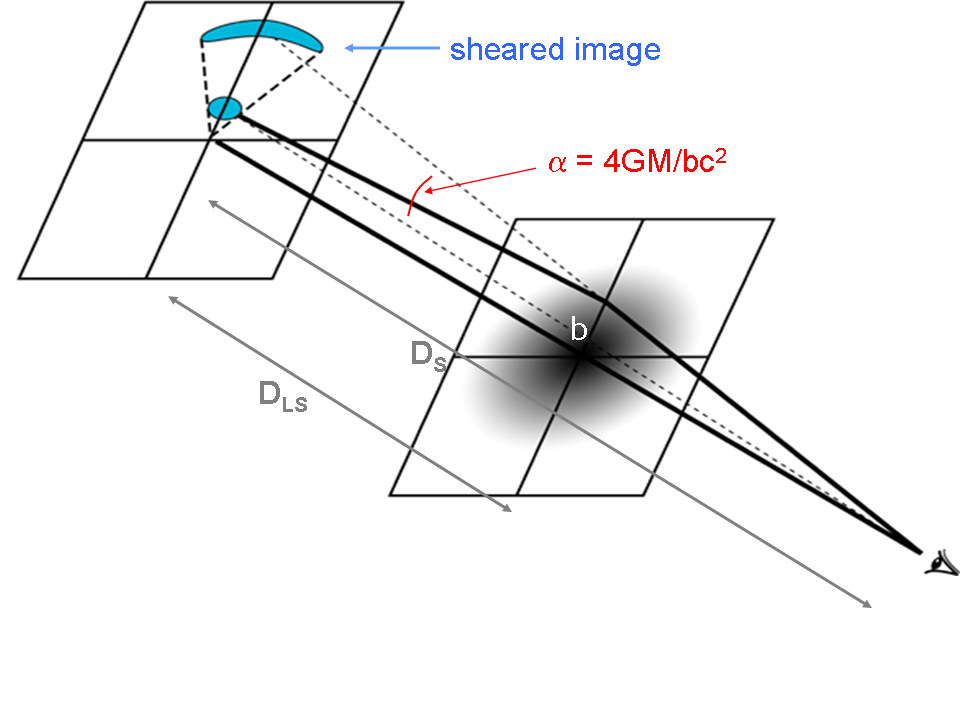}}
\caption{\label{fig-wlfeelgood} 
The primary observable effect of weak lensing is to impose an apparent tangential ellipticity on background galaxies which are truly randomly aligned.
}
\end{figure}

\begin{equation}
e^{\rm obs}=\frac{e^{\rm int}+ \gamma}{1+\gamma^* e^{\rm int}},
\label{ave}
\end{equation}
where $e \simeq \left[(1-b/a)/(1+b/a)\right]{\rm exp}(2i\alpha)$ for
an ellipse with major and minor axes, $a$ and $b$, respectively and
orientation angle $\alpha$. $\gamma^*$ is the complex conjugate of the
lensing shear. The average value of $e^{\rm obs}\approx \gamma$ in the
weak lensing regime. To be more precise, the observable is the reduced
shear $\gamma/(1-\kappa)$. Hence, the unbiased measurement of the
shapes of background galaxies (which constitute the small, faint end
of the galaxy sample) lies at the heart of any weak lensing analysis.
Some of the difficulties inherent in this task are discussed
in \autoref{wl-sec-sys}. 

Shear and magnification are both observable.  Shear is inferred from
the distribution of galaxy {\it shapes}; it is assumed that galaxy
shapes are randomly distributed in the absence of lensing (but
see \autoref{weaklens-intrinsic} for violations of that assumption and
how they may be dealt with).  More concretely, if galaxies are
approximated as ellipses with ellipticity, $\epsilon_i$, and position
angles, $\phi_i$, then the ellipticity components,
$\epsilon_{1,i} \equiv \epsilon_i \cos(2\phi_i)$, and
$\epsilon_{2,i} \equiv \epsilon_i \sin(2\phi_i)$ are randomly
distributed with zero mean.  Shifts away from zero mean are
proportional to the shear (about a 1\% effect in a typical
area of sky), but the constant of proportionality depends on various
factors, thus making {\it shear calibration} a potential source of
systematic error (discussed in \autoref{subsec-shearcalib}).  In
addition, the intrinsic spread presents a source of noise, {\it shape
noise}, which can only be reduced by averaging together more galaxies
per unit area of sky.  We can write the shear noise in a one-arcminute
patch of sky as $\gamma_{\rm rms}/\sqrt{N_{\rm eff}}$, where
$\gamma_{\rm rms}$ encodes the shape noise, which we take to be
$0.4/\sqrt{2}=0.28$ per component, and $N_{\rm eff}$ is the effective
number of galaxies.  The effective number of galaxies is defined to
account for the fact that many galaxies are imaged at low S/N and
therefore do not fully contribute to beating down the 
shape noise.  The effective number is the number of perfectly measured
galaxies which would be required to yield the same shear noise as the
actual set of measured galaxies:
\begin{equation} 
\displaystyle N_{\rm eff} = \sum_{i=1}^{n_{gals}}
{\sigma_{SN}^2\over \sigma_{SN}^2 + \sigma_{meas,i}^2},
\end{equation}
are themselves imperfectly
measured where
$\sigma_{meas,i}$ is the measurement error on the shape of the $i$th
galaxy and $\sigma_{SN}$ is the rms shape noise (distinct from
$\gamma_{\rm rms}$ only because this calculation is in shape rather than
shear space).

Contributions to $\sigma_{meas,i}$ come from photon noise and from
finite angular resolution.  Photon noise determines the uncertainty in
measuring the shape as observed after convolution with the PSF.  This
uncertainty is then amplified when extrapolating to the true pre-PSF
shape, by a factor which depends on the relative sizes of the galaxy
and the PSF, but which is never less than unity.  Therefore, to
maximize the effective number of galaxies, a survey must go deep and
have good angular resolution.  LSST will have both, and will maximize
angular resolution by taking data in $r$ and $i$ (the most sensitive
bands and the ones to be used for lensing) only when the seeing is
better than $0.7''$.

How many galaxies per square arcminute will LSST effectively measure,
and what shear noise level will it reach?  Figure A.1
of \cite{CloweEDisCS06} shows the shear noise level in 45-minute and
2-hour exposures taken with the 8.2-m VLT in various seeing
conditions.  These data show that for LSST resolution and depth in $r$
{\it or} $i$ band (slightly better than $0.7''$, maintained
over 200 30-second visits) the shear noise is 0.05 arcmin$^{-2}$; this
corresponds to $N_{\rm eff}=31$ galaxies arcmin$^{-2}$ 
for our fiducial $\gamma_{\rm rms}$.  $N_{\rm eff}$ will be increased
by combining information from multiple filters, boosting more
galaxies above the useful S/N threshold.  \cite{JarvisJain08} showed
that the number density scales as the square root of the number of
filters, assuming the filters are roughly equally deep.  Counting only
$r$ and $i$ for LSST, this yields $N_{\rm eff}=44\,\rm arcmin^{-2}$.  The actual gain
is likely to be somewhat lower because the additional galaxies will be
the least well resolved and the most affected by crowding, background
estimation errors, etc, and so we adopt $N_{\rm eff}=40\,\rm arcmin^{-2}$ in the
sections below.  We emphasize that LSST will detect many more
galaxies, as estimated in \autoref{sec:common:galcounts}; this is the
effective number usable for weak lensing.

We are now beginning to analyze high-fidelity simulations (\autoref{sec:design:imsim}), which will
give us more concrete numbers for LSST data (including the redshift
distribution, not just the overall number, of usable galaxies).  The
statistical errors over 20,000 deg$^2$, even including cosmic
variance, are very small and naturally raise the question of the
systematic error floor.  This is a primary concern of the weak lensing
collaboration, and we devote \autoref{wl-sec-sys} to it.

Measuring magnification is more difficult than shear because
the unlensed distribution of galaxy fluxes is roughly a power law,
making it difficult to measure the small departures induced by
lensing.  It has been observed and can be used in some cases to break
degeneracies involved in using shear alone, but by and large weak
lensing projects focus almost exclusively on shear.  Magnification
will find some uses in the cluster context, but as far as large-scale
statistics are concerned, magnification can only be measured as a
cross-correlation with foreground galaxies (and so far with lower
signal-to-noise ratio than the shear). Whether this will prove to be a
useful complementary measure of lensing for LSST remains to be
determined.

\section{Galaxy-Galaxy Lensing\label{wl-sec-gglensing}}
\label{sec:wl:gg}
%
%
%
%
%
%
%
%
%
%
%
%
%
%
%
%
%
%
%
%
%
%
%
%
%
%
{\it Rachel Mandelbaum}

\subsection{Motivation and Basic Concepts}

Weak lensing around galaxies (or galaxy-galaxy lensing, hereafter, g-g
lensing) provides a {\it direct} probe of the dark matter that
surrounds galaxies (for a review, see \citet{2001PhR...340..291B}; see
also \autoref{sec:gal:dm} for a galaxy-based view of some of the
issues examined here).  Gravitational lensing induces tangential shear
distortions of background galaxies around foreground galaxies,
allowing direct measurement of the galaxy-mass correlation function
around galaxies.  The distortions induced by individual galaxies are
small (on the order of 0.1\%), but by averaging over all foreground galaxies
within a given subsample, we obtain high signal-to-noise ratio in the shear
as a function of angular separation from the galaxy.  If we know the
lens redshifts or some approximation thereof (e.g., photometric redshift
estimates), the shear signal can be related to the projected mass
density as a function of proper distance from the galaxy.  Thus we can
observe the averaged dark matter distribution around any given galaxy sample.

Mathematically, g-g lensing probes  the
connection  between galaxies  and matter  via  their cross-correlation
function
\begin{equation}
\xi_{gm}(\vec{r}) = \langle \delta_g (\vec{x})
\delta_{m}(\vec{x}+\vec{r})\rangle, 
\end{equation}
where $\delta_g$  and $\delta_{m}$  are overdensities of  galaxies and
matter respectively.  This cross-correlation  can be related  to the  
projected surface density,
\begin{equation}\label{E:sigmar}
\Sigma(R) = \overline{\rho} \int \left[1+\xi_{gm}\left(\sqrt{R^2 + \chi^2}\right)\right] d\chi
\end{equation}
(for $r^2=R^2+\chi^2$), where we ignore the radial window, which is
much broader than the typical extent of the lens.  This surface
density is then related  to the observable
quantity for lensing,
\begin{equation}\label{E:ds}
\Delta\Sigma(R) = \gamma_t(R) \Sigma_c= \overline{\Sigma}(<R) - \Sigma(R), 
\end{equation}
where the second relation is  true only for a matter distribution that
is axisymmetric along the line of sight.  This observable quantity can
be  expressed  as the  product  of  two  factors, a  tangential shear,
$\gamma_t$, and a geometric factor, the critical surface density 
\begin{equation}\label{E:sigmacrit}
\Sigma_c = \frac{c^2}{4\pi G} \frac{D_S}{D_L D_{LS}},
\end{equation}
where $D_L$ and $D_S$ are angular diameter distances to the lens and
source, $D_{LS}$ is the angular diameter distance between the lens and
source.  Note that the $\kappa$ defined in \autoref{weaklens-basics} is
equal to ${\Sigma_{ }\over\Sigma_c}$.

Typical practice is to measure the g-g weak lensing  signal around a stacked
sample of  lenses to obtain the average $\Delta\Sigma(R)$ for  the whole
sample, as the signal from a individual galaxy is too weak to be
detected over the shape noise.  For shallow surveys it may be necessary to stack
tens of thousands of galaxy-mass lenses to obtain reasonable $S/N$
\citep{2006MNRAS.368..715M}; clearly, for the much deeper 
LSST, an equal $S/N$ can be achieved by stacking far fewer lenses,
thus allowing finer divisions in galaxy properties and more
information to be extracted.  This stacked lensing signal  can be
understood in terms of what information is available 
on  different scales.  
The lensing signal on $\ltsim 0.3
h^{-1}$Mpc\ 
scales  tells  us about the  dark matter  halo  in  which the  galaxy
resides; the  signal from $\sim  0.3$ -- $1h^{-1}$Mpc  reveals the
local environment (e.g., group/cluster membership) of the galaxy; and
the signal on larger scales 
indicates the large-scale correlations of the galaxy sample, similar
to the information present in the galaxy-galaxy autocorrelations. 

The shear systematics requirements for g-g lensing are less
rigorous than for cosmic shear because g-g lensing is a
cross-correlation function.  As a result, if the shape measurements of
galaxies used to compute the shear have some multiplicative and
additive bias, the additive bias term can be entirely removed through
cross-correlation with a random lens sample, and the multiplicative
bias enters only once. 

\subsection{Applications with LSST}


Galaxy-galaxy lensing on its own can be used to explore many
properties of galaxies, to relate them to the underlying host dark
matter halos and, therefore, to constrain galaxy formation and
evolution.  Below are two examples of applications of g-g lensing.

\subsubsection{Galaxy Host Halo Mass as a Function of Stellar Mass}

After estimating the stellar mass of galaxies (using luminosities and
colors) and binning the galaxies by stellar mass, it is possible to
study the lensing properties of the galaxies as a function of stellar
mass.  The results would provide important information about the
connection between the visible (stellar) component of the galaxy, and
its underlying dark matter halo.  Thus, they would be very useful for
constraining theories of galaxy formation and evolution. 

\begin{figure}
\begin{center}
\includegraphics[width=0.45\linewidth]{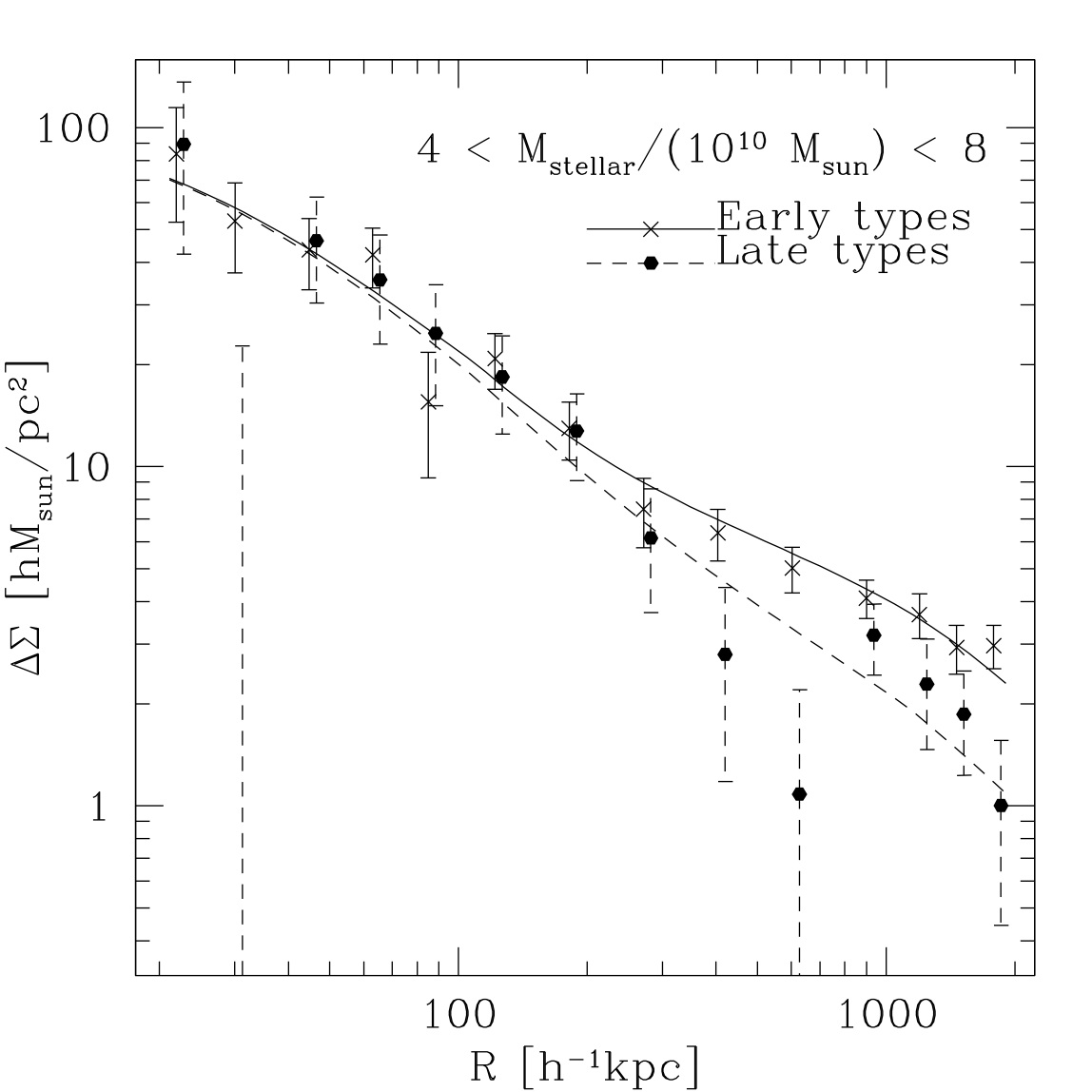}
\includegraphics[width=0.45\linewidth]{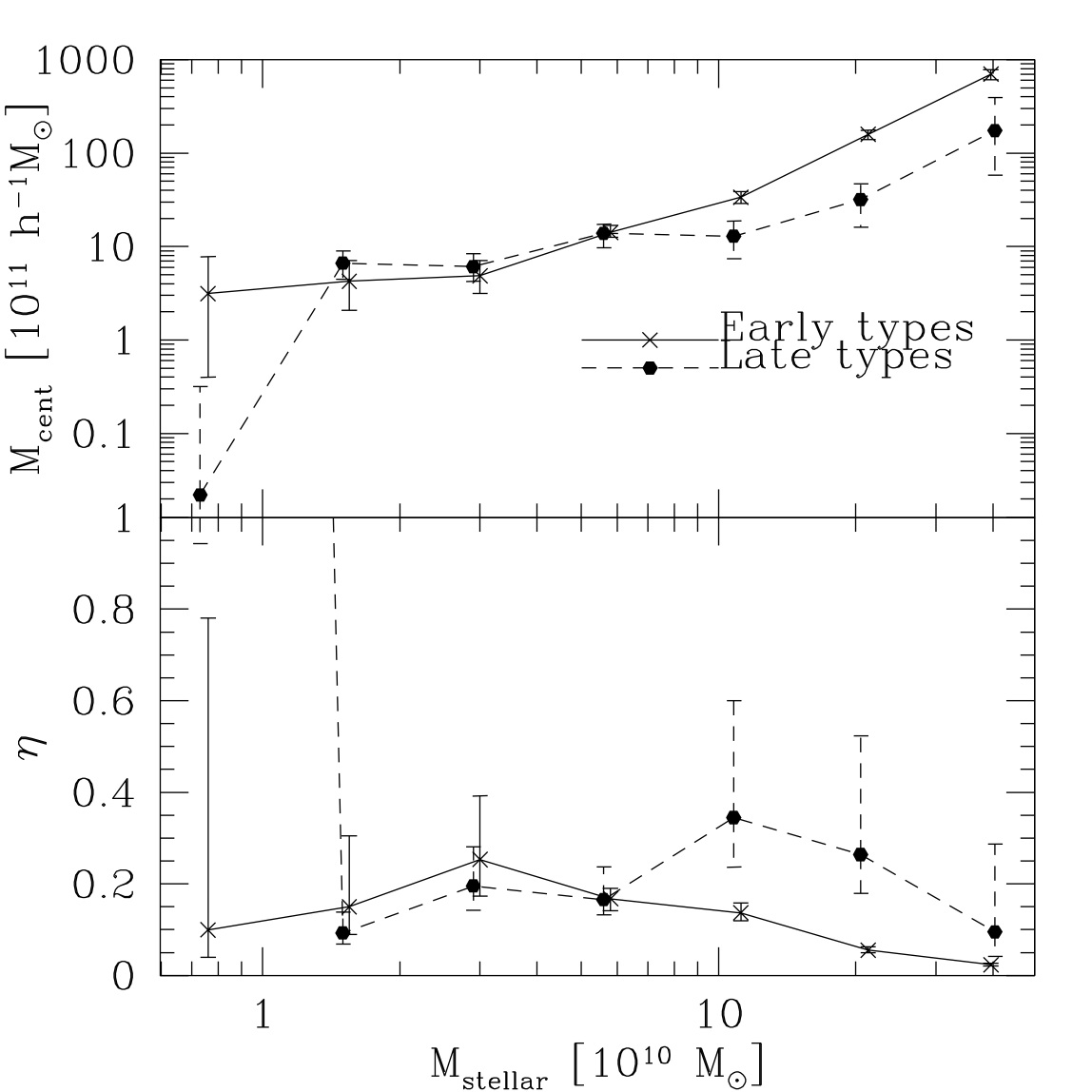}
\caption{\label{F:sm}At left, measured lensing signal from SDSS for a
  stellar mass bin containing Milky Way-type galaxies, with a model
  shown as the smooth lines.  Such fits were used as input for the
  figure at the right, which shows central halo mass (top) and
  conversion efficiency in central galaxy $\eta =
  (M_*/M_{halo})/(\Omega_b/\Omega_m)$ (bottom) as a function of
  stellar mass for early-type and late-type galaxies.  Constraints are
  extremely weak for galaxies with low stellar mass and for the
  relatively rare high stellar mass late-type galaxies.  Figure uses
  data originally published in \protect\cite{2006MNRAS.368..715M}.}
\end{center}
\end{figure}

This procedure has been done in several surveys at lower
redshift, such as SDSS (see \autoref{F:sm}, \citealt{2006MNRAS.368..715M}) and RCS
\citep{2005ApJ...635...73H}, and as high as 
$z=0.8$ but with relatively low $S/N$ using GEMS
\citep{2006MNRAS.371L..60H}.  LSST will 
have the power to vastly improve the precision of these constraints,
which would be particularly interesting at the low stellar mass end,
below $L\sim L_*$, 
where these previous surveys lack the statistical power to make any
interesting weak lensing constraints (and where strong lensing
constraints are unlikely because lower mass galaxies are typically not
strong lenses).  Furthermore, LSST will enable studies to at least a
lens redshift $z\sim 1$, thus giving a measure of galaxy mass assembly and the relation between
stellar and halo mass over the second half of the lifetime of the
universe.  Finally, LSST will be able to extend all these measurements
out to much larger angular scales, up to $10h^{-1}$ Mpc, with high $S/N$.


\subsubsection{Halo Ellipticity}

Dark matter halo ellipticity, a robust prediction of $\Lambda$CDM
according to N-body and hydrodynamic simulations, can in principle be
detected using galaxy-galaxy weak lensing.  Ellipticity of dark matter halo profiles has
been predicted in CDM N-body simulations (e.g.,
\citealt{1991ApJ...378..496D}), and observed with
non-lensing methods on scales $<20$ kpc (for a review, see
\citealt{1999ASPC..182..393S}). 

Given the need to stack multiple objects, a measurement of dark matter
halo ellipticity naturally depends on 
alignment between the
ellipticity of the light distribution and of the matter distribution.
Current measurements have relatively poor signal-to-noise ratio
\citep{2004ApJ...606...67H,2006MNRAS.370.1008M,2007ApJ...669...21P}
though there is a suggestion of halo ellipticity in CFHTLS data and
for very luminous red galaxies in SDSS.  A robust measurement of halo
ellipticity would both confirm the predictions of simulations, and
serve as a measurement of the alignment between the light and mass
distributions in galaxies, which is in itself an important clue to
environmental effects on galaxy shapes during their formation and
evolution.   Again, due to the depth of LSST, it should be possible to
explore the ellipticity of dark matter halos as a function of galaxy
type and of redshift to unprecedented precision relative to what is
possible now.  These results may also be a stringent test of the theory of
gravity particularly for isolated halos
\citep{1983ApJ...270..365M,1986MNRAS.223..539S}.  

\subsection{Combining Galaxy-Galaxy Lensing with Galaxy Clustering\label{sec:ggl_bias}}


Galaxy-galaxy lensing is a powerful probe of cosmological parameters
when combined with measurements of the clustering of galaxies.  The
former measures the galaxy-mass correlations, $\xi_{gm}$, whereas the
latter measures the galaxy autocorrelations, $\xi_{gg}$.  On large
scales, where the mass and galaxy fluctuations are highly correlated
(though with possible multiplicative offsets in amplitude, known as
``galaxy bias''), the two probes together can be used to estimate the
matter correlation function, $\xi_{mm}$.  Furthermore, the galaxy bias
estimated in this way is not limited by cosmic variance, because the
same exact density fluctuations are used for both measurements.
Because the systematics in g-g lensing are different than in cosmic
shear, this measurement is a useful cross-check on the matter power
spectrum reconstructed from cosmic shear.

\section{Galaxy Clusters}
\label{wl-sec-cluster}
%
%
%
%
%
%
%
%
%
%
%
%
%
%
%
%
%
%
%
%
%
%
%
%
%
%
Clusters of galaxies are the largest virialized structures in the
Universe and are useful both as cosmological probes and as
astrophysical laboratories in their own right.  The two aspects are
heavily intertwined, because we must understand the physical processes
in clusters to have confidence that they can be used as accurate
cosmological probes.  Readers interested in the astrophysical
laboratory aspect are referred to \autoref{sec:lss:cluster} and \autoref{sec:gal:dm}
respectively.   

Clusters represent the most extreme overdensities, and thus probe the
growth of structure.  An accurate census of clusters by mass and
redshift can thus be used to infer cosmological parameters.  Weak
lensing can play a critical role here because it can measure total
cluster mass without regard to baryon content, star formation history,
or dynamical state.  Thus, even if clusters are discovered by optical,
X-ray, or Sunyaev-Zel'dovich (SZ) surveys, lensing is crucial for
calibrating the mass-observable relation.  \autoref{sec:lss:cluster} explores
in more detail searches for clusters in the galaxy distribution and the synergies with 
future surveys at other wavelengths. 
Weak lensing studies of clusters also naturally 
complement the studies of cluster-scale strong lensing (\autoref{sec:sl:clusmf}).  
Strong lensing reconstructions of mass distributions in the cluster 
cores can eliminate mass sheet degeneracies in the weak lensing 
measurements \citep{broadhurst95}, allowing more accurate total masses to be calculated. 
At the same time, weak lensing measurements are essential to refining models for 
the mass distribution and substructure in cluster cores \citep{Bradac06}.

We focus here on what weak lensing measurements of clusters of galaxies with
LSST will mean for the study of the properties of clusters themselves.
First, we examine mass measurements of individual clusters.  Then, we
briefly discuss {\it searching} for clusters with lensing, and finally
we generalize this to a search for peaks in the shear field, regardless of whether
they are truly clusters of galaxies.  We show that the shear peak
distribution revealed by LSST will provide a powerful probe of
cosmology.

\subsection{Cluster Mass Distributions}
%
%
%
%
%
%
%
%
%

%
%
%
%
%
%
%
%
%
%
%
%
%
%
%
%
{\it Douglas Clowe, Ian P. Dell'Antonio}

LSST will generate the largest and most uniform sample to date of galaxy
clusters with gravitational lensing measurements.  This sample
of clusters will be important for cosmological measurements and for
the study of the assembly history of galaxy clusters.  

At their most basic level, the cluster weak lensing measurements will
consist of measurements of shapes, magnitudes, and colors (and hence
photometric redshift estimates) for $\sim40$ galaxies arcmin$^{-2}$
behind the clusters (for low redshift clusters at least; 
some of these galaxies will be in front of higher redshift clusters, and so
would not be distorted).  These measurements can be combined many
different ways to constrain the mass profiles of clusters and the mass
distribution in clusters.  For the regions of sky that comprise the
LSST deep fields (\autoref{sec:design:cadence}), we expect almost 70\%
more resolved galaxies, with much more precise SED measurements for
them. These deep fields will be very useful as calibrators of the
contamination at the faint end of the LSST weak lensing mass
distribution. 

\subsection {One-dimensional Cluster Lensing:  Radial Cluster Profile Fitting}

The mean radial density profile of clusters can be calculated from the
azimuthally averaged tangential reduced shears of the background
galaxies (e.g., \citealt{tvw_1990}).  The deflection of a galaxy at an
observed radius, $r$, from the cluster center is a function of the
enclosed mass.  Because the distortion of a galaxy is a function of
the deflection and its gradient, the tangential shear as a function of
radius encodes information about the slope of the profile.  

In these one-dimensional measurements (and also in two-dimensional
lensing maps), the quantity that is calculated depends on the
line-of-sight surface density (the integral of the mass density along
the line of sight) modulated by the changes in $\Sigma_{\mathrm{crit}}$
with the redshift of the mass.  As a result, all of these techniques
are sensitive to projection effects whereby unrelated (and often quite
distant) structures along the line-of-sight contribute to the signal.
Some progress can be made in disentangling the signal along the line
of sight using photometric redshifts, but in general there is always
contamination because the redshift dependence of the lensing kernel is
weak, and because large-scale structure at very similar redshift to
the target clusters cannot be resolved by photometric redshifts.
Nevertheless, one-dimensional techniques can provide powerful (and
comparatively high S/N) measures of cluster properties.

Two complementary approaches have been introduced to recover the mass
distribution from the shear profile.  \cite{fahlman94} introduced a
non-parametric technique of ``aperture densitometry," based on the
realization that the difference between the mean surface mass density
inside a radius, $R$, and the mean surface mass density out to some
fiducial outer annulus or radius, $R_m$, is given by an integral over
the (seeing-corrected) tangential shear between $R$ and $R_m$:
\begin{equation}
\label{eq:densitometry}
\zeta = \bar{\kappa}(r<R) - \bar{\kappa}(R<r<R_m) = {2\over(1-R^2/R_m^2)} \int_R^{R_m} \langle\gamma_T\rangle d\ln r.
\end{equation}
The $\langle \rangle$ notation
indicates an azimuthally averaged value around a chosen cluster center.
A variant on this statistic which uses a fixed outer aperture instead
of one which extends all the way to $R$ is:
\begin{equation}
\label{eq:densitometry2}
\zeta_c = \bar{\kappa}(r<R_1) - \bar{\kappa}(R_2<r<R_m) = 2 \int_{R_1}^{R_2} \langle\gamma_T\rangle d\ln r + {2\over (1 - R_2^2/R_m^2)}\int_{R_1}^{R_m} \langle\gamma_T\rangle d\ln r,
\end{equation}
where $R_1$ is the radius inside which one is measuring the mean
surface density and $R_2$ is the radius of the inner edge of the fixed
outer aperture \citep{clowe00}.  In the case where $R_m \gg R_1,R_2$,
as can be the case for LSST studies, these two statistics become
identical and the error on both is simply $\sigma_{\zeta}
= \sigma_{\gamma}/\sqrt{\pi R_1^2 n}$, where $n$ is the number density
of background galaxies in the survey.  From this we get an estimate
for LSST that $\sigma_{\zeta} = 0.028/R_1$ with $R_1$ measured in
arcseconds.  As with $\kappa$, one can convert this to a surface
density by multiplying by $\Sigma_{\mathrm{crit}}$.

The aperture densitometry technique has the advantage of being
non-parametric.  However, it is limited by the extent to which the
outer annulus can be measured (which has been a limiting factor up to
now but will be a non-issue given LSST's sky coverage).  Furthermore, a
radial profile made from $\zeta$ has points that are strongly
correlated, and it provides only a measurement of the integrated
surface density with no way to convert to a three-dimensional mass
density profile. Finally, the above analysis assumes that one is
measuring the shear $\gamma$ when in fact one measures the reduced
shear $g = \gamma/(1-\kappa)$.  This latter effect leads to severe
overestimation of the surface density of clusters in their cores.

A related statistic which can be used to compute $\kappa$ as a
function of radius is:
\begin{equation}
\label{eq:densitometry3}
\kappa (R) = 1 - {1\over 1-g(R)} \exp \left( -\int\limits^{R_m}_{R} {2 \langle g\rangle\over r (1 - \langle g\rangle)} dr - {2 \langle g(R_m)\rangle\over \alpha}\right),
\end{equation}
which assumes that at $r>R_m$ the reduced shear profile behaves as a
power law, $\langle g(r)\rangle \propto r^{-\alpha}$ \citep{clowe01}.
This assumption is usually the limiting factor in the
accuracy of this statistic outside of the cluster core.  For the LSST
survey, however, $R_m$ can be made so large that $\langle
g(R_m)\rangle \to 0$ and, therefore, the effect of the power-law assumption is
negligible.  While this statistic properly uses $g$ instead of
$\gamma$ and is, therefore, valid in massive cluster cores, it is still
a measurement of the surface density, the individual radial $\kappa$
points are still strongly correlated, and one is still sensitive to
superposition of unrelated structures along the line-of-sight.  A
final statistic in this family is the aperture mass statistic
\citep{Schneider98}, which sums up $\langle g\rangle$ convolved with
a given radial kernel over a given aperture, and is mostly used to
detect mass peaks and measure their significance of detection in a
non-parametric method.

Alternatively, one can use parametric fitting techniques in which a
parametrized radial profile family (usually as an NFW profile) is
assumed, and the measured tangential reduced shear is compared to the
model to fit the parameters of the model (e.g., \citealt{clowe01}).
Although the choice of model family can affect the results (because
the model is ``forced" to follow the profile even in the presence of
other signals such as extra substructure), this technique allows one
to convert the two-dimensional lensing measurement to a
three-dimensional density measurement assuming the chosen model family
is valid.  It also allows for an easier calculation of the
significance of measurement, as one can easily compute a $\Delta
\chi^2$ between the best fit model and a zero mass model or similar
statistic.  The primary uncertainties with this technique, aside from
the noise in the shear measurement and the question about whether the
assumed mass profile family is valid, are the triaxiality in the
cluster population (the model families are usually spherically
symmetric) and the presence of significant mass substructure.  Despite
the biases, one-dimensional measurements will provide the highest
significance detections of the lowest mass systems.  With LSST, we
should be able to measure the masses of clusters with mass
$M_{200}>1.5\times 10^{14}\,M_\odot$ to 10\% statistical accuracy. 
Given the area coverage of LSST, this level of accuracy will be
reached for approximately 20,000 clusters of galaxies \citep{hty04}.
To measure the mean properties of even less massive systems, we can
combine the tangential shear around large samples of systems selected
through optical, X-ray, or SZ techniques.  This ``stacking''
\citep{Sheldon01,sheldon07,johnston07} of clusters sorted according to an
observable will allow the precise calibration of the relationship
between mass and observable (\autoref{sec:sl:clusmf}).

Systematic errors arise in individual clusters due to projection
effects, both non-local (multiple clusters or groups widely separated
along the line-of-sight) and local (clusters embedded in local large-scale structure). The availability of accurate photometric redshift information
will allow the calibration of the first of the two projection effects,
but the second effect will remain an irreducible source of error in
the mass measurements.  Other sources of error include intrinsic
alignments (\citealt{Mackey02}; see \autoref{weaklens-intrinsic}),
although these can be minimized with the 
photometric redshift information \citep{heymans04} and effects from
the triaxiality and substructure within individual clusters.  Because
a one-dimensional treatment necessarily averages over the actual
distribution of matter, it will produce a biased measurement for
systems that depart greatly from spherical symmetry.  While the stacking
or shear cross-correlation methods mentioned above can remove these
effects for mean measurements of large samples, they are not applicable 
to the study of individual systems, which will be subject to bias due 
to departures from spherical symmetry.  To deal with this bias, 
two-dimensional weak lensing techniques will need to be employed.  
Nevertheless, because one-dimensional lensing can be used to detect 
lower mass systems, one-dimensional measures will be essential for 
constructing the cluster mass function both in a fixed redshift range and 
as a function of redshift (which provides an independent test of hierarchical
clustering).

\subsection{Two-dimensional Cluster Lensing Maps}

Given a catalog of galaxy positions, sizes and reduced shears, one can
construct a two-dimensional map of the convergence either directly
through a convolution of the reduced shears with a window function
\citep{ks, ksb, tyson_fischer95} or indirectly by modeling
the gravitational potential and matching the predicted shear to the
observations \citep{Seitz96,seitz_schneider98,bradac05,khiab08}.  The
resulting map can be used to detect samples of clusters and the
cluster mass function for cosmological purposes
(\autoref{sec:sl:clusmf}), but also to study the distribution of mass in
individual clusters.  In particular, LSST will be sensitive (at roughly 4$\sigma$)
to individual clusters with $\sigma>500$ \kms\ or $M_{200} > 0.5\times
10^{14} M_\odot$.  Furthermore, the weak lensing data will lead to
detection of substructures with $M/M_{cl}$ greater than 10\% for
approximately $10^4$ clusters of galaxies with $0.05<z<0.7$.  This
large sample will be an extremely important comparison set to match
with the gravitational clustering simulations \citep{Millenium} to
constrain the growth of structure, and to compare with the galaxy
spatial and star formation distribution to study the evolution
properties of galaxies as a function of {\it mass} environment.  Just
as fundamentally, the LSST measurements of these clusters will
determine mass centroids for these clusters accurate to better than 1
arcminute.  These centroids can be compared with the optical and x-ray
centroids of the cluster to determine whether there are offsets
between the various components.  The presence of mass offsets in
merging clusters provides a sensitive test of modified gravity
theories (\citealt{clowe06}; see also \autoref{sec:sl:cdmsub}).

\subsection{Magnification by Clusters}

LSST will also study the magnification induced by cluster lenses.
Unlike shear, magnification is not affected by
the mass-sheet degeneracy (\autoref{sec:sl:basics}).  This means that
magnification measurements 
can be combined with shear measurements to calibrate the masses of
clusters derived from shear
lensing, even in the absence of any strong lensing features.  Although
there have been  
attempts to use magnification to determine the properties 
of clusters (cf. \citealt{broadhurst95}), it is difficult to measure for two reasons.  First, the competing 
effects of the dilution of objects due to the magnification of the 
background area and the increase in detected objects due to the 
magnification means that for different classes of objects, the magnification 
can either enhance or decrease the number counts at a given flux level.
More seriously unlike shear lensing, there is not an a priori determination of what the unmagnified population should look like.
However, LSST will probe a significant fraction of the
observable Universe at these redshifts, allowing the mean number
density of galaxies to be measured to great accuracy.  Stacking a
large number of clusters, 
LSST will
be able to measure the mean magnification for the ensemble of clusters
$M_{vir} > 5\times 10^{14}$ $M_\odot$ and $0.1<z<0.6$ (of which LSST should
detect $\sim 3000$) with a statistical error smaller than 0.2\% (of course, the 
systematic errors in field-to-field photometric calibration will likely limit the
overall mass scale calibration to $\sim 1$\%.)  

One of the systematic sources of uncertainty in magnification measurements is ignorance about the number counts of objects in the magnitude range immediately below the magnitude limit of the survey.  Here, the LSST deep fields will prove exceptionally valuable -- they will cover enough area to reach a representative sample of the faint Universe, and will allow accurate estimates of the population of galaxies up to 1 magnitude deeper than the main survey. This should reduce the contribution of the unknown slope of the number counts to the magnification calculation below the 1\% level. 

\subsection{Three-dimensional Cluster Lensing}

We can study the variation of the shear signal from clusters as a
function of photometric redshift (\autoref{sec:common:photo-z}), which
allows fundamentally new science.  
First, it provides an independent technique for measuring the angular
diameter distance to clusters. \citet{wittman03} have demonstrated that
one can estimate the redshift of a galaxy cluster to an accuracy of
roughly 0.1 simply based on the variation of the shear of the
background galaxies with redshift.  Second, the exact shape of the
shear versus redshift profile is a function of the cosmological
parameters.  This has important consequences for cosmology.
\citet{jt03} showed that for a large enough sample of clusters, one could
use the variation of shear with redshift as a tomographic measurement.
For a given foreground mass distribution, the measured shear at a
fixed angular position $(\theta,\phi)$ is only a function of the
angular diameter distance ratio, $D_L D_{LS}/D_S$.  Thus, the shear
tomography provides a geometry-only measurement of the dark energy
equation of state; a measurement that is complementary to the
cosmological measurement derived from shear correlation functions.

Tomographic measurements such as these can also be used to test
photometric redshift
estimates for large samples of galaxies that are too faint to be
measured spectroscopically.  Using the ensemble of weak lensing detected 
clusters of galaxies, comparison of the shear vs. photometric redshift
for galaxies both bright enough for spectroscopic confirmation and too
faint for spectroscopy should allow the tomographic profile of the
former to calibrate the redshift normalization for the latter,
effectively extending the magnitude limit for verifying photometric redshift
estimates by at least one magnitude.

\subsection{Panoramic Mass Maps}
The techniques for surface mass density
reconstruction we've discussed will be applied to the entire 20,000
deg$^2$ LSST 
survey area, yielding a 2.5-dimensional atlas of ``where mass is'' that will
simultaneously be useful to professional astronomers and fascinating
for the public.  Astronomers will use these maps to examine galaxy
properties as a function of environment.  
``Environment'' in this context is usually defined in terms of
neighboring galaxies, 
but mass maps covering half the sky will enable a new way of looking
at the concept of environment.  With the LSST sample it will be possible to map
galaxy bias (\autoref{sec:ggl_bias}) over the sky, cross-correlate with
the CMB to map CMB lensing, and cross-correlate with supernovae to map
(and possibly reduce the scatter due to) supernova lensing
(\autoref{sec:sn:wl}).

Mass distributions are also important for testing alternative theories
of gravity (\autoref{sec:cp:grav}).  Unambiguous examples of
discrepancies on large scales between the mass and galaxy distribution
will not only constrain
these theories, but will also derive astrophysical constraints on
dark matter interaction cross sections, as was done for the visually
stunning example of the Bullet Cluster (\citealt{2004ApJ...606..819M},
\autoref{sec:sl:dmclus}).  

At the same time, these maps are a prime opportunity to bring science
to the public.  Everyone understands maps, and the public will be able
to explore mass (and other) maps over half the sky.  The maps will
also have some coarse redshift resolution, enabling them to gain a
sense of cosmic time and cosmic history.

\subsection{Shear-selected Clusters}
\label{sec:wl:shselcl}

The mass maps lead naturally to the idea of searching for clusters
with weak lensing.  Weak lensing has traditionally been used to
provide mass measurements of already known clusters, but fields of
view are now large enough(2--20 deg$^2$) to allow blind surveys for mass
overdensities 
\citep{2006ApJ...643..128W,
  2007A&A...470..821D,2007A&A...462..459G,2007ApJ...669..714M,2007Natur.445..286M}.
Based on these surveys, a conservative estimate
is that LSST will reveal two shear-selected clusters deg$^{-2}$ with
good signal-to-noise ratio, or 40,000 over the full survey area.
Results to date suggest that many of these will not be strong X-ray
sources, and many strong X-ray sources will not be selected by shear.
This is an exciting opportunity to select a large sample of clusters
based on mass only, rather than emitted light, but this field is
currently in its infancy.  Understanding selection effects is critical
for using cluster counts as a cosmological tool (see
\autoref{fig:massfct} and \autoref{sec:lss:cluster}) because mass, not
light, clustering is the predictable quantity in cosmological models;
simulations of structure formation in these models
(\autoref{sec:cp:cos_sim}) will be necessary to interpret the data.
Shear selection provides a unique view of these selection effects, and
LSST will greatly expand this view.  

Because shear selection uses background galaxies rather than cluster
members, it is difficult to detect clusters beyond $z\sim 0.7$.  Hence
the proposed deep LSST fields will be very useful in accumulating a
higher redshift, shear-selected sample in a smaller (but still $\sim$
100 deg$^2$!) area.  This will be critical in comparing with X-ray,
optically, and SZ selected samples, which all go to higher redshift.

Shear selection has the property that it selects on {\it projected}
mass density.  Therefore, shear peaks may not be true three-dimensional
density peaks.  From a cluster expert's point of view, this results in
``false positives,'' which must be eliminated to get a true
shear-selected cluster sample.  Using source redshift information to
constrain the structure along the line of sight (tomography) helps
somewhat but does not eliminate these ``false positives'' \citep{hs05},
because the lensing kernel is quite broad.  We will see in the next
section how to turn this around and use shear peaks as a function of
{\it source} redshift (rather than lens redshift) to constrain
cosmological parameters.

\subsection{Cosmology with Shear Peaks \label{sec:wl:shearpeaks}} 
%
%
%
%
%
%
%
%
%
%
%
%
%
%
%
%
%
%
%
%
%
%
%
%
%
%
{\it Sheng Wang, Morgan May}

Structure formation is a hierarchical process in which gravity is
constantly drawing matter together to form increasingly larger
structures.  Clusters of galaxies currently sit atop this hierarchy as
the most massive objects that have had time to collapse under the influence
of their own gravity.  Their appearance on the cosmic scene is
relatively recent, but they also serve as markers for those locations
of the highest density fluctuations in the early
Universe \citep{bbks}, which make them unique tracers of cosmic
evolution \citep{hmh01}.  Analytic predictions exist for the mass
function of these rare events per unit co-moving volume per unit
cluster mass \citep[][see also \autoref{fig:massfct}
and \autoref{sec:lss:cluster}]{ps,bcek}.  Gravitational $N$-body
simulations can produce even more precise predictions of the mass
function \citep[][see also \autoref{sec:cp:cos_sim}]{st99,jen01}.

The abundance of clusters on the sky is sensitive to dark energy in
two ways: first, the co-moving volume element depends on dark energy,
so cluster counts depend upon the cosmological expansion history;
second, the mass function itself is sensitive to the amplitude of
density fluctuations (in fact it is exponentially sensitive to the
growth function at fixed mass); see \autoref{fig:massfct}
and \autoref{sec:lss:cluster}.  The LSST weak lensing selected lensing
sample of mass selected clusters will be ideal for dark energy
diagnostics. 

Common
two-point statistics, such as the cosmic shear power spectra
(\autoref{sec:wl:lss}), do not contain all the statistical information
of the WL field, and clusters are a manifestation of the non-Gaussian
nature of the field. 
Recent studies \citep{fh07,tb07} have shown that
by combining the power spectrum with the redshift evolution of the
cluster abundance, the constraints on dark energy parameters from
shear power spectra can be tightened by roughly a factor of two.

However, even with photometric redshifts, one has little radial
information \citep{wvm02,hty04,hs05}. 
All matter along the line-of-sight to the distant source galaxies
contributes to the lensing signal.  Consequently false detections of
overdensities arise,
due mostly to this projection effect: small mass objects along the same
line of sight but physically separated in redshift would be attributed
to a single object.  \citet{us08} suggest a simple alternative
observable to cluster abundance, the fractional area of
high significance hot spots in WL
mass maps to determine
background cosmological parameters.  A similar idea is to use the
projected WL peaks \citep{jv00,msb08} regardless of whether they
correspond to real galaxy clusters.

Here we focus on the fractional area of the high convergence regions,
which has the advantage that it takes into
account projection effects by construction. Analogous to the
\citet{ps} formalism, it is determined by the
high convergence tail of the probability distribution function (PDF). 
Previous works have shown that the one-point PDF is a simple yet
powerful tool to probe non-Gaussian features
\citep{rkjs99,jsw00,ks00,vmb05}.  Since the non-Gaussianity in the
convergence field is induced by the growth of structure, it holds
cosmological information mainly from the nonlinear regime and
complements the well-established statistics in the linear regime.

\cite{us08} used a Fisher matrix approach to forecast LSST cosmological 
parameter constraints with this method, dividing the galaxies into
three bins with mean redshift $z_s=0.6$, 1.1, and 1.9 so that each bin
contains the same number of galaxies.  They took a Gaussian smoothing
scale of $\theta_G=1$ arcmin, and considered seven different $S/N$
thresholds, from $\nu=2.0$ to $5.0$ in increments of $\Delta\nu=0.5$,
in each redshift bin to utilize the information contained in the shape
of the PDF.  They assumed a 20,000 deg$^2$ survey with an rms shear
noise of 0.047 per square arcminute, just slightly lower than the 0.05
cited in the introduction to this chapter, and a fiducial 
$\sigma_8$ of 0.9 (the assumed $\sigma_8$ has a strong effect on peak
statistics).  They also examined ``pessimistic'' and ``optimistic''
scenarios for systematic errors, the former being 1\% priors on
additive and multiplicative shear errors, and the latter being 0.01\%
and 0.05\% priors respectively.  (Note that neither of these accounts
for photometric redshift errors.)  They found, for example, a $w_0$
precision ranging from 0.55 for the pessimistic scenario to 0.16 for
the optimistic scenario for LSST alone, decreasing to 0.12 and 0.043
respectively for LSST plus Planck priors.  

Further theoretical work is needed to exploit this statistic.  
The one-point PDF has not been tested extensively in simulation for
different cosmologies.  For instance, \citet{us08} adopt the expression
given by \citet{do06}, which has been checked for only one flat $\Lambda$CDM
cosmology with $\Omega_m=0.3$.  In order not to dominate the
observational errors, the 
theoretical prediction for the PDF has to be accurate at the $\sim 1\%$
level, a level at which it has not been tested as a function of
cosmological model.  However, the one-point PDF is such a simple statistic, and
its derivation adds almost no extra computational cost, once WL
simulations are made.  One expects to obtain accurately calibrated
formulas for the PDF from currently ongoing or planned large WL
simulations (\autoref{sec:cp:cos_sim}).  

%
%

\section{Weak Lensing by Large-scale Structure}
\label{sec:wl:lss}
%
%
%
%
%
%
%
%
%
%
%
%
%
%
%
%
%
%
%
%
%
%
%
%
%
%
Lensing by large-scale structure (cosmic shear) is best characterized
by two- and three-point correlation functions, or equivalently their
Fourier transforms: power spectra.  Much of the information contained
in shear correlations measured by LSST will lie in the nonlinear
regime of structure formation.  Analytical models such as the halo
model, N-body simulations, and hydrodynamical simulations
(\autoref{sec:cp:cos_sim}) are all
required to obtain accurate predictions from arcminute to degree
scales.

LSST will measure these correlation functions in source redshift
shells, as well as cross-correlations between redshift shells.  These
correlations are sensitive to both the growth of structure and the
expansion history of the Universe, making this a particularly powerful
cosmological probe. The growth of structure can be separated from the
expansion history in combination with
expansion-history probes such as Type Ia Supernovae and 
baryon acoustic oscillations
(\autoref{sec:cp:wlbao}), thus providing stringent tests of dark energy
and modified gravity models.

\subsection{Two-point Shear Correlations and Tomography}
\label{sec:wl:twopoint}
%
%
%
%
%
%
%
%
%
%
%
%
%
%
%
%
%
%
%
%
%
%
%
%
%
%
{\it Bhuvnesh Jain, Anthony Tyson}

To quantify the lensing signal, we measure the shear
correlation functions from galaxy shape catalogs. The
two-point correlation function of the shear, for source galaxies in
the $i^{th}$ and $j^{th}$ redshift bin, is defined as
\begin{equation}
\xi_{\gamma_i\gamma_j}(\theta)=\langle\bm{\gamma_i}(\bm{\theta}_1)
\cdot\bm{\gamma_j}^\ast(\bm{\theta}_2)\rangle,
\label{eqn:shearcorrelation}
\end{equation}
with $\theta = |\bm{\theta}_1 - \bm{\theta}_2|$. 
Note that the two-point function of the convergence is identical to
that of the shear. It is useful to
separate $\xi_\gamma$ into two separate correlation functions by 
using the $+/\times$ decomposition: 
the $+$ component is defined parallel or
perpendicular to the line connecting the two points
taken, while the $\times$ component is defined along $45^\circ$.  
This allows us to define the rotationally invariant two-point
correlations of the shear field:
$\xi_{+}(\theta)=\langle\gamma_{i+}(\bm{\theta}_1)
\gamma_{j+}(\bm{\theta}_2)\rangle$, and 
$\xi_{\times}(\theta)=\langle\gamma_{i\times}(\bm{\theta}_1)
\gamma_{j\times}(\bm{\theta}_2)\rangle.
$
The correlation function of \autoref{eqn:shearcorrelation} is
simply given by $\xi_{\gamma_i\gamma_j} = \xi_+ + \xi_-$. 

There is information beyond tangential shear. The lensing signal is 
caused by a scalar potential in the lens and, therefore, should be curl-free. 
We can decompose each correlation function into one that measures the 
divergence (E-mode) and one that measures the curl (B mode).
The measured two-point correlations can be expressed as contributions
from E- and B-modes (given by linear superpositions of integrals over
$\theta$ of $\xi_+$ and $\xi_-$). Since lensing is essentially derived
from a scalar potential, it contributes (to a very good approximation) 
only to the E-mode.  
A more direct way to perform the E/B decomposition is
through the mass aperture variance, $M_{\rm ap}^2(\theta)$, which is a
weighted second moment of the tangential shear measured in
apertures. This provides a very useful test of systematics in the
measurements; we will not use it here, but refer the reader
to \cite{Schneider02}. All two-point statistics such as $M_{\rm
ap}^2(\theta)$ can be expressed in terms of the shear correlation
functions defined above.

The shear power spectrum at angular wavenumber $\ell$ 
is the Fourier transform of $\xi_{\gamma_i\gamma_j}(\theta)$. 
It is identical to the power spectrum of the 
convergence and can be expressed as a
projection of the mass density power spectrum, $P_\delta$. 
For source galaxies in the $i^{th}$ and $j^{th}$ redshift bin, it is
\citep{Kaiser92,Hu99}: 
\begin{equation}
C_{\gamma_i\gamma_j}(\ell) = 
\int_0^{\infty} dz \,{W_i(z)\,W_j(z) \over \chi(z)^2\,H(z)}\,
 P_\delta\! \left ({\ell\over \chi(z)}, z\right ),
\label{eqn:pkappa}
\end{equation} 
where the indices $i$ and $j$ cover all the redshift bins. The
geometric factors $W_i$ and $W_j$ are defined in \autoref{eqn:weightgl}.
The redshift binning is assumed to be provided by photometric
redshifts (\autoref{sec:common:photo-z}).  The redshift binning 
is key to obtaining dark energy information from weak lensing. 
A wealth of cosmological information can be extracted by using 
shear-shear correlations, galaxy-shear cross-correlations (\autoref{sec:wl:gg})
and galaxy clustering in multiple redshift bins. As discussed 
in \autoref{wl-sec-sys}, 
this must be done in the presence of several systematic errors, which 
have the potential to degrade errors on cosmological parameters. 
Note that if both source galaxy
bins are taken at redshift $z_s$, then the integral is dominated by
the mass fluctuations at a distance about half-way to the source
galaxies. This is a useful guide in estimating the lens redshift best
probed by a set of source galaxies. 

\begin{figure}[ht!]
\begin{center}
\includegraphics[width=8cm,angle=0]{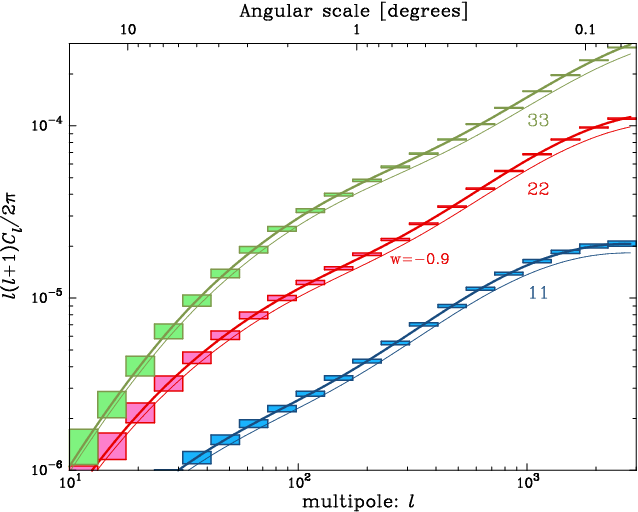}
\caption{\footnotesize 
The lensing power spectra constructed from galaxies split into three
broad redshift bins: $z < 0.7, 0.7 < z < 1.2$, and $1.2 < z < 3$. 
The solid curves are predictions for the fiducial $\Lambda$CDM
model and include nonlinear evolution. The boxes show the expected
measurement error due to the sample variance and intrinsic ellipticity
errors (see text for details). 
The thin curves are the predictions for a dark energy model
with $w = -0.9$. Clearly such a model can be distinguished at very high significance using information from all bins in $\ell$ and
$z$.  Note that many more redshift bins are expected from
LSST than shown here, leading to over a hundred measured auto- and 
cross-power spectra. 
\label{fig:pkappa}}
\end{center}
\end{figure}

\autoref{fig:pkappa} shows the predicted power spectra from the
ten-year LSST stack for
galaxies split into three redshift bins: $z<0.7,\  
0.7<z<1.2, \ 1.2<z<3$. The fiducial $\Lambda$CDM model is used
for the predictions and the 
error bars indicate experimental uncertainty due to sample variance
(which dominates at low $\ell$) and shape noise (which dominates at
high $\ell$). The thin curves show the predictions for a $w=-0.9$
model. The figure shows that for much of the range in $\ell$ and $z$
accessible to LSST, such a model can be distinguished using a single
bin alone (based on statistical errors); by combining the information 
in all bins, it can be distinguished at very high significance.  
Note that the actual measured power spectra include
contributions from systematic errors, not included here but discussed
in \autoref{wl-sec-sys}. We also discuss below the covariance between
spectra in different wavenumber bins due to nonlinear effects which
degrade the errors at high $\ell$. We
have plotted just three spectra in \autoref{fig:pkappa} for
illustrative purposes. With LSST we expect to use many more bins, and also
utilize cross-correlations between redshifts.  
About ten redshift bins exhaust the available weak lensing information 
in principle, but in practice up to twenty bins are likely to be used  
to carry out tests related to photometric redshifts 
and intrinsic alignments and to study questions such as the relation
of luminous tracers to dark matter. 

\autoref{eqn:pkappa} shows how the observable shear-shear power
spectra are sensitive both to the geometric factors given by $W_i(z)$
and $W_j(z)$, and to the growth of structure contained in the
mass density power spectrum, $P_\delta$. Both are sensitive to dark energy
and its possible evolution, which determine the relative amplitudes of
the auto- and cross-spectra shown in \autoref{fig:pkappa}. $P_\delta$ also 
contains information about the primordial power
spectrum and other parameters such as neutrino masses. 
In modified gravity theories, the shape and time evolution of the density
power spectrum can differ from that of a dark energy model, even one
that has the same expansion history. Lensing is a powerful means of
testing for modifications of gravity as well 
\citep{Knox06,Amendola07,Jain07,Heavens07,Huterer07}. The
complementarity with other probes of each application of lensing is
critical,  especially with the CMB and with measurements of the
distance-redshift relation using Type Ia Supernovae and baryonic
acoustic oscillations in the galaxy power spectrum. 

The mass power spectrum is simply related to the linear 
growth factor $D(z)$ on large scales (low $\ell$):
$P_\delta \propto D^2(z)$. However, 
for source galaxies at redshifts of about 1, observable
scales $\ell \gtsim 200$ 
receive significant contributions from nonlinear gravitational
clustering. So we must go beyond the linear regime 
using simulations or analytical
fitting formulae to describe the nonlinear mass power spectrum
\citep{Jain97,jsw00,White04,Francis07}. To
the extent that only gravity describes structures on scales larger
than the sizes of galaxy clusters, this can be done with high
accuracy. There is ongoing work to determine what this scale is
precisely and how to model the effect of baryonic gas on smaller
scales \citep{zentner_etal08}, as discussed in more detail in
\autoref{sec:wl:baryons}. 

\subsection{Higher-order Correlations}
\label{sec:wl:threepoint}
%
%
%
%
%
%
%
%
%
%
%
%
%
%
%
%
%
%
%
%
%
%
%
%
%
%
{\it Masahiro Takada}

\subsubsection{Weak Lensing Covariances}

It is important to understand the statistical precision of cosmic shear
observables and error propagation in the determination of cosmological
parameters.  Since cosmic shear probes the projected mass distribution,
the statistical properties of the cosmic shear field reflect those of
the mass distribution. The statistical precision of the cosmic shear
power spectrum is determined by the covariance, which contains three
terms: shot noise contamination due to
intrinsic ellipticities, and Gaussian and non-Gaussian sample
variance caused by imperfect sampling of the fluctuations
\citep{SZH99, C+H01, T+J09}. The
non-Gaussian sample variance arises from the projection of the mass
trispectrum weighted with the lensing efficiency kernel. In fact, most
of the useful cosmological information contained in the lensing power
spectrum 
lies on small angular scales, which are affected by nonlinear clustering.
Therefore, the non-Gaussian errors can be significant in weak lensing
measurements, and cannot be ignored in the precision measurements
delivered by LSST. 

The non-Gaussian errors cause two additional uncertainties in measuring
the cosmic shear power spectrum. First, they degrade accuracies in
measuring band powers of the spectrum at each multipole bin via the
trispectrum contribution to the power spectrum covariance. Second,
nonlinearities in the mass distribution cause correlations between the
band powers at different multipoles, decreasing the effective number
of independent degrees of freedom of multipoles measured from LSST. 

\citet{T+J09}
investigated the impact of the non-Gaussian errors on the cosmic shear
power spectrum measurement using a dark matter halo approach. In the
$\Lambda$CDM scenario, the cumulative 
signal-to-noise ratio for measuring the power spectrum over a
range of angular scales, from degrees down to a few arcminutes scale,
can be degraded by up to a factor of two compared to the Gaussian error
case. Adding the tomographic redshift information slightly mitigates the
degradation, but the non-Gaussian errors remain 
significant.

Given the LSST measurements, accuracies of estimating cosmological
parameters can be obtained by propagating the statistical
uncertainties of power spectrum measurements into parameter space. The
marginalized errors on individual parameters would be degraded by less
than 10-20\% after the proper analyses. The smaller degradation is
primarily because: 1) individual parameters in a high-dimensional parameter
space are degraded much less than the volume of the full Fisher
ellipsoid in a multi-dimensional parameter space; 2) lensing involves
projections along the line-of-sight, which reduce the non-Gaussian
effect; and 3) some of the cosmological information comes from geometric
factors which are not degraded at all. These are promising prospects;
 a large number of ray-tracing simulations will be needed to
calibrate the impact of the non-Gaussian covariances on the
parameter estimations, taking into account the
effects of survey geometry and masking (\autoref{sec:cp:cos_sim}).

\subsubsection{Three-point Correlation Functions of Cosmic Shear}

The non-Gaussian
signatures measured via the higher-order moments themselves carry additional
information that cannot be extracted with the power spectrum. In fact,
the higher-order moments are complementary to
the power spectrum because they depend differently on the lensing efficiency
function, which in turn is sensitive to the geometry of the Universe.
\citet{T+J04} showed that 
combining the three-point correlation of cosmic shear or the Fourier
transform, bispectrum, with the power spectrum can improve the
cosmological constraints by up to a factor of three. 

However a more realistic forecast is needed that takes into account
the non-Gaussian covariances of the bispectrum, which requires
knowledge to the six-point correlation functions to calculate. The
preliminary result is that bispectrum tomography is more degraded
by non-Gaussian errors than is the power spectrum, but combining
the two- and three-point correlation information can improve the dark
energy constraints. This result indicates that most of the
cosmological information inherent in cosmic shear can be extracted by
using the two- and three-point correlation functions in CDM-dominated
structure formation (we need not go to the four-point
correlation!). This is because nonlinear clustering is physically
driven by the quadratures of density and velocity perturbations in the
continuity and Euler equations, which derive originally from Gaussian seed
fluctuations. That is, the power spectrum gives us most of the
Gaussian information in the original density 
field.  The bispectrum encodes most of the
non-Gaussian signatures that arise from nonlinear
mode-coupling of the Gaussian field and the quadrature fields.
The most important systematic in our understanding of the non-linear
part of the mass spectrum is the effect of baryons
(\autoref{sec:wl:baryons}), which will have to modeled statistically.

The complementary sensitivities of the two- and three-point
correlations to cosmological parameters offer a useful way of
discriminating cosmological information from systematic errors in
cosmic shear (such as shape measurement, photometric redshifts).  The two- and
three-point correlations are affected by these systematics
in different ways.  If we can model the systematic errors by
physically motivated models with a small number of nuisance
parameters, combining the two spectra allows determination of
cosmological parameters and the nuisance parameters simultaneously,
protecting the cosmological information against systematic errors
-- the self-calibration regime \citep{Hut++06}. This method will
be a powerful tool for controlling systematics and achieving the
desired accuracy for constraining dark energy parameters with LSST.


%
%


\section{Systematics and Observational Issues\label{wl-sec-sys}}

LSST weak lensing using a sample of several billion galaxies
provides wonderful statistical precision. Realizing that potential
involves significant effort using multiple cross checks to detect and then 
control systematic errors.    Here, we
address observational systematics, including photometric redshift
errors (\autoref{sec:wl:zphot}), shear calibration and additive shear
errors (\autoref{subsec-shearcalib}), and intrinsic 
alignments of galaxies (\autoref{weaklens-intrinsic}).  We also
discuss our limited theoretical understanding of the effects of
baryons on the small-scale lensing signal in
\autoref{sec:wl:baryons}.  

\subsection{Photometric Redshift Systematics}
\label{sec:wl:zphot}
%
%
%
%
%
%
%
%
%
%
%
%
%
%
%
%
%
%
%
%
%
%
%
%
%
%
{\it David Wittman}

Accurate inference of source galaxy redshifts is a fundamental
requirement for weak 
lensing.  The photometric redshift performance of the LSST survey is
discussed in \autoref{sec:common:photo-z}.  Here we discuss how
photometric redshift errors relate to weak lensing science, and how
systematics can be controlled to the required level.

Unlike some other science areas, in weak lensing and in BAO, the accuracy and 
precision of the photometric redshifts, $z_p$, matter less than how well we
{\it know} the distribution function of photometric redshift errors in
any photometric redshift bin.  
Because the weak lensing kernel is broad 
in redshift, wide photometric redshift bins may be used.
The science analysis proceeds by integrating over the
distribution and is biased only to the extent that the assumed distribution is
incorrect.  In the simplified case of a Gaussian
distribution, \citet{Hut++06} found that the mean and scatter of the
distribution should be
known to about 0.003 in each redshift bin.  \citet{Ma08} extended this
to arbitrary distributions, which can be represented as a sum of
Gaussians, but this more sophisticated analysis did not qualitatively
change the result.  \citet{Hut++06} also investigated the effect of
$z_p$ systematics on combined two-point and three-point statistics, 
and found that requirements are much reduced: with only a 20-30\% dark
energy constraint degradation, $z_p$ errors could be self-calibrated
from the data (\autoref{sec:wl:threepoint}).  However, this assumed a
very simple $z_p$ error model, 
and so may be overly optimistic.  It may not be necessary 
to reach a precision of 0.003 per bin for precision cosmology: combining
WL and BAO data can significantly reduce the required photometric 
redshift precision (\autoref{sec:cp:wlbao}).  

It is traditional to ``measure'' the $z_p$ bias and scatter by
obtaining spectroscopic redshifts of a subset of galaxies.  If this
were true, the spectroscopic sample requirement is simply to amass a
large enough sample to beat down the noise in each bin to less than
0.003. In this context, \citet{Ma08} estimate that $10^5-10^6$
spectroscopic redshifts would be sufficient for LSST.  However,
estimates of bias and scatter from the $z_p-z_s$ relation contain
systematics because the spectroscopic sample is never completely
representative of the photometric sample, especially for deep
photometric surveys where spectroscopy cannot go as faint as the
photometry.  Even for less deep surveys ($r \sim 24$), the
spectroscopic incompleteness rate is a function of type, magnitude, color,
and redshift \citep[e.g.,][]{Cannon06}.

We are currently modeling photometric redshifts with the LSST filter
set and depth to better understand the range of possible systematic
errors.  We have already found one way to dramatically reduce
systematic errors, namely using a full probability distribution, $p(z)$,
rather than a single point estimate for each
galaxy \citep{Wittman09}.  \citet{Mandelbaum07} also found this in the
context of galaxy-galaxy lensing.  We plan to store a full $p(z)$, or a 
compressed version of it, for each galaxy.

Systematic errors in the $z_p$ distribution will be a function of
type, color, magnitude and redshift.  We are planning a
``super-photometric redshift'' field observed in a large number of
filters by the 
time of LSST commissioning to calibrate these quantities;
see \autoref{sec:common:phz:cal}.  We are also developing a new method
of calibrating redshift distributions using angular cross-correlations
with a spectroscopic sample \citep{Newman08};
see \autoref{sec:photoz:cross} for more details.  
One potential
difficulty relates to lensing: magnification induces angular
cross-correlations between foreground structures and background galaxy
populations \citep{BernsteinHuterer09}.  However, the deep fields can
be used to characterize the fainter population available for
magnification into the gold sample, and thus correct for this effect.


\subsection{Shear Systematics}
\label{subsec-shearcalib}
\subsubsection{Shear Calibration}
%
%
%
%
%
%
%
%
%
%
%
%
%
%
%
%
%
%
%
%
%
%
%
%
%
%
{\it David Wittman, Anthony Tyson}

The primary systematic of concern in weak lensing is the smearing of
galaxy shapes due to the telescope point-spread function (PSF).  LSST
delivered image quality will be good, with a nominal $0.7''$ FWHM cutoff
for weak lensing observations, but it still will be a challenge to
accurately infer the true shapes of galaxies, which are often smaller
than this.

There are many contributions to the PSF: atmospheric
turbulence or ``seeing,'' optics and perturbations on the
optics, non-flatness of the focal plane and the CCDs,
charge diffusion in the detector, etc.  Simulations that
include these effects are discussed in detail in
\autoref{sec:design:imsim}.  Here we discuss these issues as they
relate to weak lensing specifically and describe how systematics will be
controlled to the required level.

If the PSF were perfectly round (isotropic), it would change galaxy
shapes by making them appear more round, thus diluting the lensing
signal.  Recovering the true amplitude of the shear is a problem
referred to as {\it shear calibration}, or reducing multiplicative
errors.  Real PSFs are themselves anisotropic, and thus may imprint
{\it additive} shear systematics onto the galaxies, as we discuss in
the next subsection.  In any case, the
observed galaxy shape is the true galaxy shape convolved with the
PSF. Thus, for barely resolved galaxies, this effect is very
large and must be removed to high precision.  \citet{Hut++06} found
that for LSST to maintain precise dark energy constraints, shear must
be calibrated to 0.1\%.  And as with photometric redshifts, they found
that combining the two- and three-point functions substantially
reduced this requirement.

The current state of the art is 1\% calibration accuracy, based on
blind analysis of simulated data \citep{STeP1,STep2}.  The lensing
community is now working with the statistics community, bringing in
new ideas through blind analysis challenges \citep{GREAT08}.
Meanwhile, the LSST Weak Lensing Science Collaboration is developing
LSST-specific ways to reduce this problem. Among them are:

\begin{itemize}

\item MultiFit as described in detail in
\autoref{sec:design:algorithms}, whereby properties of each galaxy are
determined not by a co-addition of all the images covering a given field, but by
a simultaneous fit to each image.  This uses the PSF information of
each exposure in an optimal way, and thus should allow maximal control
of PSF-related systematics. 

\item \citet{Jain06} suggested a method of canceling out many PSF-related
errors: when computing correlation functions, {\em cross}-correlate shapes from
two separate sets of images of a given field.  The PSFs are
independent, so the correlation functions should be free of
PSF-related systematics, again barring a very global systematic.  This
idea can be combined with the previous idea by cross-correlating
shapes produced by MultiFit from two separate sets of 100 exposures
each, for example.

\item High-fidelity image simulations as described in
\autoref{sec:design:imsim} are 
  underway to test the LSST pipeline's ability to model the PSF.
  Subtle effects such as color dependence of the PSF from differential
  chromatic refraction will be included.

\item \cite{Jarvis04} demonstrated the effectiveness of principal component
analysis (PCA) for accurately modeling PSF variations in data sets where many
images were taken with the same instrument.  LSST is the ultimate in
this regard, with several {\it million} images to be taken by a single
camera.  The PCA can also be augmented with a physically motivated optical model
\citep{Jarvis08}, which takes advantage of the extensive modeling done for LSST.

\end{itemize}

Simulations including everything from all relevant cosmological
effects to PSF distortions due to the atmosphere and telescope
optics, will be invaluable for testing this machinery.
Typically one checks for systematics by measuring the B-mode, which
should be very near zero in the absence of systematics.  However,
multiplicative errors can change the E-mode more than they change the
B-mode, by up to an order of magnitude \citep{Guzik05}.  The LSST
image simulator (\autoref{sec:design:imsim}) will allow these tests to be
carried out.  The LSST deep fields (\autoref{sec:design:cadence}) will
also be valuable for assessing the 
accuracy of shear measurements using the faintest galaxies, by
comparing shear measurements using only the main imaging data
and those using the deep data.

\subsubsection{Additive Shear Errors}
%
%
%
%
%
%
%
%
%
%
%
%
%
%
%
%
%
%
%
%
%
%
%
%
%
%
{\it David Wittman}

Even if shear calibration were perfect, spurious shear can be added by
an anisotropic PSF.  In this section we highlight several tests that
demonstrate that this effect can be controlled to high precision.
There are three main sources of PSF anisotropy: focal plane, optics, and atmosphere.

In long-exposure astronomical images, PSF anisotropy induced by
atmospheric turbulence is negligible because the turbulence flows
across the aperture and averages out after several seconds.  LSST's
exposure time of 15 seconds is just short enough that atmospheric
anisotropy must be considered.  \citet{Wittman05} examined a set of 10
and 30 second exposures from the Subaru Telescope, which is a good
match to LSST, 
with roughly the same aperture and a wide field camera (Suprime-Cam).  These
exposures contained a high density (8 arcmin$^{-2}$) of well-measured
stars; a randomly selected subset representing typical survey
densities ($\sim 1$ arcmin$^{-2}$) were used to build a PSF model and
the remaining $\sim$7 stars arcmin$^{-2}$ used to assess residuals.
On the 0.3-3 arcminute scales examined, the residual ``shear''
correlations were 1--2 orders of magnitude below the cosmological
signal in a single exposure, and were further averaged down by
multiple exposures.  After five exposures, the residuals were less
than the projected LSST statistical error on scales 3 arcmin and
larger.  On larger scales, the effects of the atmosphere falls off
quite rapidly, and 
other systematics are more worthy of attention as described below.  On
smaller scales, it seems likely that this systematic will be well
controlled with hundreds of LSST exposures, but large data sets or
simulations will be required to prove this assertion.

\begin{figure}
\begin{center}
\includegraphics[width=0.5\linewidth]{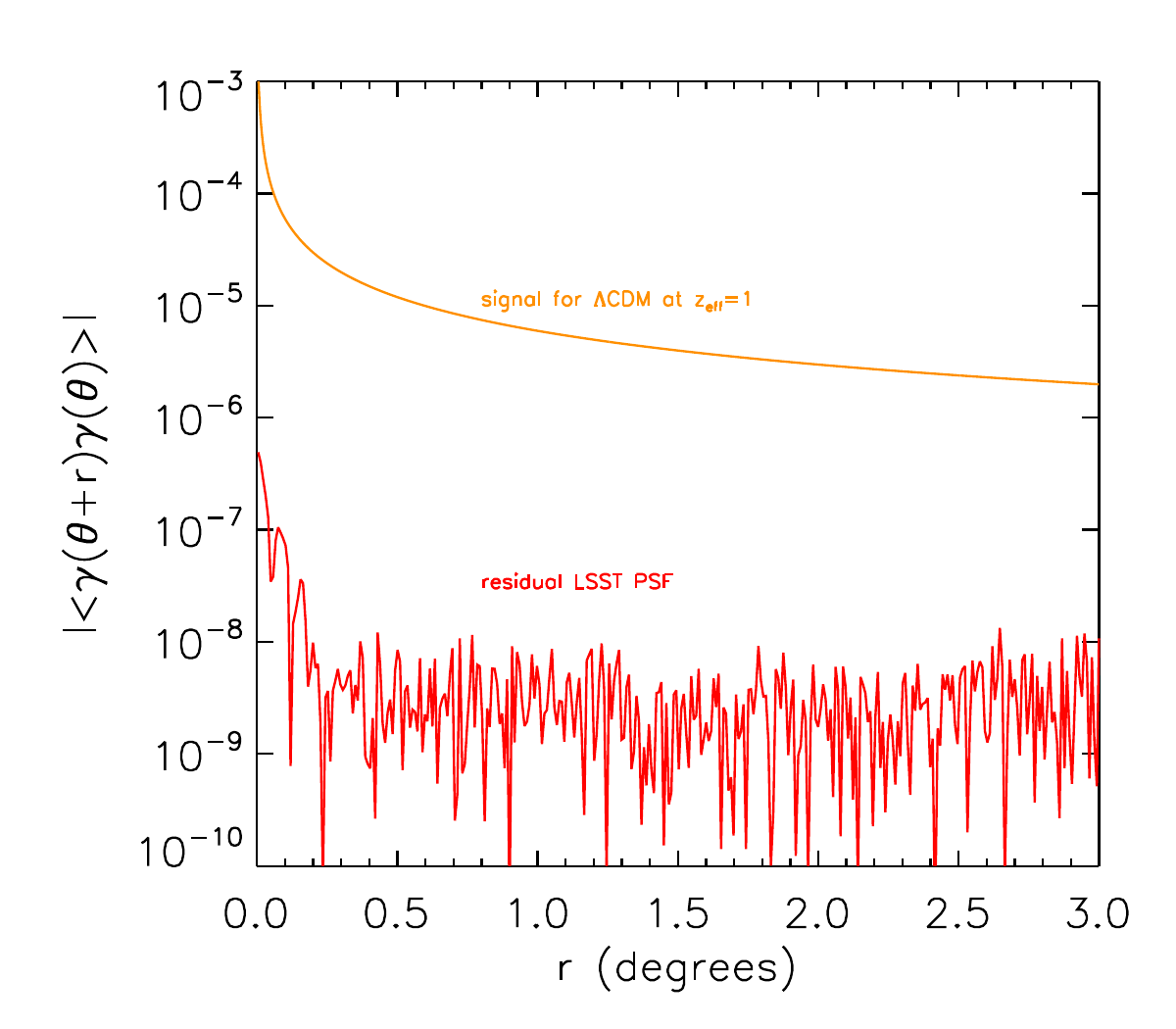}
\caption{Absolute value of shear residuals as a function of angular separation,
for a simulated exposure.  As described in the text and
in \citet{Jee07b}, the simulated image includes atmosphere, perturbed
LSST optics, and focal plane non-flatness.  The simulation included
zero true lensing signal, thus assessing additive errors only.  About
one high-S/N 
star arcmin$^{-2}$ (close to the density that the LSST data will have)
was used to model the spatial variation of the PSF,
while many additional point sources were used to assess
the residuals (a luxury we will not have with the real data!).  LSST
{\em statistical} errors on the shear will be roughly  
1\% of the lensing signal for the scales shown here.  Note that
because the absolute value is shown, residuals are actually consistent
with zero beyond 0.25$^\circ$.\label{fig-jee}}
\end{center}
\end{figure}

\citet{Jee07b} extended this type of analysis to 3$^\circ$ scales
using simulated star fields (\autoref{sec:design:imsim}) imaged
through a simulated atmosphere, a 
model of the LSST optics with realistic perturbations, and simulated
focal plane departures from flatness, including piston errors and
potato chip curvature of the sensors.
\autoref{fig-jee} shows the
residual correlations in a single image full of high-S/N stars.  The
residuals are three orders of magnitude less than the lensing signal
out to 0.25$^\circ$ scales, beyond which they are consistent with zero
(\autoref{fig-jee} shows the absolute value of the residuals; in fact,
the sign fluctuates at large scales).  For comparison, LSST
statistical errors are typically about two orders of magnitude smaller
than the signal over the range of scales shown.  Note that this
discussion applies to additive errors only, as the simulations lacked
input shear.  More sophisticated simulations are now being assembled
to assess the ability to recover a given input shear.

\subsubsection{The Power of Many Exposures}

The large number of exposures in the LSST survey provides built-in
advantages for reducing shear errors. Hundreds of dithered exposures 
per filter at various pupil rotations and image rotations will help 
isolate and reduce systematics. In addition to the MultiFit,
PCA, and cross-image correlation methods described above, the LSST
data set will enable many windows into systematics which are not
available today.  For example, one could divide the data set into
seeing bins and examine the trends of cosmic shear with seeing.
Furthermore, the LSST scheduler will be able to {\it control} some
important aspects such as instrument rotation, so that each field will
be seen by many different rotations and dither positions.  This 
will allow us to  {\it examine} the systematics in many different
subsets and use this information to further control systematics.

\subsection{Intrinsic Alignments of Galaxies}
\label{weaklens-intrinsic}
%
%
%
%
%
%
%
%
%
%
%
%
%
%
%
%
%
%
%
%
%
%
%
%
%
%
{\it Rachel Mandelbaum}

Weak lensing analyses begin with the assumption that galaxy shapes are
uncorrelated in the absence of gravitational shears.  Intrinsic
alignments of galaxy shapes violate this
assumption, and are typically due to correlations with local tidal fields
and/or the larger scale cosmic web.  $\Lambda$CDM $N$-body simulations
and analytic models show that such alignments of dark matter halos are
expected on large scales
\citep{2000ApJ...545..561C,2000MNRAS.319..649H,2001MNRAS.320L...7C,2001ApJ...559..552C,2002MNRAS.335L..89J,2005ApJ...618....1H},
but the question of whether observed galaxy shapes, i.e., the
shapes of the baryonic components, also show such alignments cannot be
answered with $N$-body simulations.  While many observational studies
have found intrinsic alignments within structures such as galaxy
groups and clusters \citep[e.g.,
][]{1982A&A...107..338B,1999ApJ...519...22F,2000ApJ...543L..27W,2005ApJ...627L..21P,2006ApJ...644L..25A,2007ApJ...662L..71F},
of greater concern are alignments that persist on the large scales used
for cosmological lensing studies.

Alhough this section focuses on intrinsic alignments as a contaminant to
lensing surveys, these alignments are also an interesting problem in
galaxy formation.  In brief, observations of intrinsic alignments can
help us learn 1) how the shape of a galaxy is affected by tidal
fields as it is formed; 2) how that shape evolves with time given the
(also-evolving) tidal field of large-scale structure; and finally, 3)
how galaxy interactions and mergers erase information about the
original projected shapes of the galaxies.  Because these questions
depend on the galaxy formation history, we expect the intrinsic
alignment signal to depend on morphology, luminosity, and possibly
environment (and thus redshift).

To determine the effects of intrinsic alignments on the measured
lensing signals, we express the shape of a galaxy as the sum of
three shear terms,
\begin{equation}
\gamma = \gamma_{\rm rand} + \gamma_{I} + \gamma_{G}.
\end{equation}

The first term is simply the random shape and orientation of the
galaxy on the sky, the second is the tidal shear that causes intrinsic
alignment with local or large-scale structures, and the third is the
gravitational lens shear from foreground mass.  In the absence of $\gamma_{I}$,
shear-shear autocorrelations (cosmic shear) simply measure $\langle
\gamma_G \gamma_G\rangle$.  However, if $\gamma_I\ne 0$, the
shear-shear autocorrelation signal acquires two additional terms,
$\langle \gamma_I \gamma_I\rangle$ and $\langle \gamma_G
\gamma_I\rangle$.

The $\langle \gamma_I \gamma_I\rangle$ term
\citep[e.g.,][]{2000ApJ...545..561C},
often known as the II 
alignment, is due to the fact that galaxies that experience the same
local tidal field 
become aligned with that field and, therefore, with each other.
Consequently, this term is important only for two galaxies at the same
redshift.  The $\langle \gamma_G \gamma_I\rangle$ term \citep{2004PhRvD..70f3526H},
often known as the GI 
alignment, arises when a tidal field at some redshift causes intrinsic
alignment of a galaxy at that redshift, and also gravitationally
shears a more distant galaxy, anticorrelating their shapes.  Similarly, one can
define intrinsic alignment terms for any other
lensing statistic determined using galaxy shape measurements (e.g.,
three-point functions have GGI, GII, and III intrinsic alignment
terms).  

Because intrinsic alignments are difficult to predict theoretically,
observational measurements have attempted to determine at what level
these alignments will contaminate the lensing signal for surveys such
as LSST.  Observations of II alignments at low redshift
\citep{2006MNRAS.367..611M,2008arXiv0809.3790O} suggest that while II
alignments may be present at a low level particularly for luminous red
galaxies, these alignments should affect the cosmic shear signal for
an LSST-like survey at the several percent level (at most) if not
removed.  The GI term, which has been robustly detected for red galaxies to
tens of Mpc, may be more important than the II term, possibly
contaminating the estimated $\sigma_8$ inferred from the cosmic shear
signal for an $r<24$ survey at the $-2$\% level, or most
pessimistically $-10$\%
\citep{2006MNRAS.367..611M,2007MNRAS.381.1197H}.  Our
current understanding will likely be supplemented in the coming years 
with  constraints for 
fainter and bluer galaxies, and at higher redshift.  It is
clear, however, that an 
LSST-like survey must estimate and remove this GI alignment
term.

The use of photometric redshifts to avoid correlating 
galaxies at the same redshift can eliminate II alignments
\citep{2002A&A...396..411K,2003MNRAS.339..711H,2003A&A...398...23K,2004ApJ...601L...1T}.
Unfortunately, this scheme will actually exacerbate the GI
alignments, since the GI alignment depends on the correlation with a
gravitational shear that increases as the pair separation in redshift
space increases.  However,
\cite{2008A&A...488..829J,2009arXiv0905.0393J} have explored a method
that works perfectly in the presence of spectroscopic redshifts to
remove both the II and GI terms, based on their known dependence on
the redshifts of the galaxy pairs.  This geometric scheme results in
the loss of some statistical power that depends on the number of
redshift bins, as quantified in \cite{2008A&A...488..829J}, but
typically expands the areas of confidence regions in parameter space
by several tens of percent for a large number of bins.  
\cite{2009arXiv0905.0393J} show that the need to remove intrinsic
alignments in this way places stringent requirements on the photometric redshift
quality.

Other schemes, such as that proposed by \cite{2005A&A...441...47K},
project out the II and GI terms using some dependence on galaxy type,
redshift, transverse separation on the sky, and luminosity.  In that
case, observations that already have constrained these dependencies
serve as model inputs, so that one can avoid losing as much
information as when using the \cite{2008A&A...488..829J} method.
\cite{2007NJPh....9..444B} find that intrinsic alignments can double
the number of tomographic bins required to recover 80\% of the
available information from the lensing analysis, while requiring three
times smaller photometric redshift errors than in the case of no
intrinsic alignment contamination.  However, 10\%-level priors on the
intrinsic alignment power spectrum, which should be achievable given
data that will be available at the start of LSST, are very helpful in
minimizing information loss due to intrinsic alignments.

Other useful diagnostics that may help minimize the contamination of the
LSST cosmic shear measurement by intrinsic alignments include the
intrinsic alignments three-point functions
\citep{2008MNRAS.388..991S}, the galaxy-galaxy lensing signal, and the
$B$-mode signal \citep{2004PhRvD..70f3526H,2006MNRAS.371..750H,2007NJPh....9..444B}.
Since most of the weak lens source galaxies are faint and blue, and no significant 
alignment for that population has yet been found, it is possible that the 
intrinsic alignment corrections will be small.

\subsection{Theory Systematics: The Effect of Baryons}
\label{sec:wl:baryons}
%
%
%
%
%
%
%
%
%
%
%
%
%
%
%
%
%
%
%
%
%
%
%
%
%
%
{\it  Andrew Zentner}

Utilizing shear correlations to constrain dark energy puts demands 
on theorists to make accurate predictions for
these quantities.  In particular, the matter power spectrum must 
be computed on scales $k \sim \mathrm{a}\ \mathrm{few}\ \mathrm{Mpc}^{-1}$ 
to better than a percent to render biases in dark energy parameters
negligible \citep{huterer_takada05}.  This goal may be achievable with 
dissipationless $N$-body simulations of
cosmological structure growth in the near future.  However, current $N$-body
calculations amount to treating all matter as dark matter and neglecting
the non-gravitational interactions of the baryonic component of the
Universe during structure growth and the gravitational response of the
dark matter to the redistribution of baryons
(\autoref{sec:cp:cos_sim}).  The process of galaxy 
formation is not well understood in its detail, and this introduces a
potentially important theoretical uncertainty in predictions of
lensing observables.  However, it is likely that this challenge can be
addressed adequately and weak lensing observables from LSST 
used to constrain the physics of both dark energy and galaxy formation.

Following earlier analytic studies \citep{White04,zhan_knox04}, recent
simulations have addressed the influences of baryonic processes on 
lensing observables.  Alhough results differ in their details, all 
studies indicate that baryonic effects can modify lensing statistics 
relative to dissipationless $N$-body predictions by amounts that are large
compared to the statistical limits of LSST 
\citep{jing_etal06,rudd_etal08}.  If unaccounted for, these 
offsets translate into biases in dark energy parameters comparable to 
or larger than their statistical uncertainties \citep{zentner_etal08}.  
One could apply a nulling procedure to mitigate the contamination of small-scale
processes, but the cost is a factor of $\sim 3$ increase in dark 
energy equation-of-state constraints and a corresponding 
factor of $\sim 10$ decrease in the Dark Energy Task Force 
\citep{albrecht_etal06} figure of merit \citep{zentner_etal08}.

It should be feasible to model such effects in the near future 
and salvage much of the information contained in the high-multipole 
range of the power spectrum.  
The power spectra of the \cite{rudd_etal08} simulations 
differ from $N$-body results in a regular fashion.  
These authors showed that the bulk of the difference 
is due to markedly different mass distributions within dark matter halos 
in baryonic simulations compared to the same halos in $N$-body simulations.  
To the degree that this is valid, standard $N$-body-based techniques can 
be corrected for baryonic effects, eliminating biases on scales $\ell < 5000$, 
by modifying the internal structures of halos.  Scales of $\ell < 2000$ are 
important for cosmic shear.

%

It is necessary to re-assess cosmological constraints in models that 
contain the additional parameter freedom of a baryonic correction.  
\cite{zentner_etal08} studied the influence of the
additional parameter freedom on dark energy constraints from shear
power spectra under various assumptions on the relationship between
halo concentration and mass. 
With reliable tomographic binning, the additional freedom is not
strongly degenerate with dark 
energy and the degradation in the constraints on $w_0$ and
$w_a$ is less than 20\% in one specific (and reasonable) model for the
concentration.  
Alternatively, theoretical or observational prior knowledge of the concentrations of halos near 
$M \sim 10^{14}h^{-1}\mathrm{M}_{\odot}$ of 
better than $30\%$ significantly reduce
the influence of the degradation due to baryonic effects on halo
structure \citep{zentner_etal08}.  Contemporary constraints from 
galaxy-galaxy lensing are already approaching this level 
\citep{mandelbaum_etal08}. This is also constrained by requiring 
consistency with galaxy-galaxy lensing. Finally, strong+weak lensing 
precision studies of several thousand clusters, taken together with 
optical and X-ray data, will constrain baryon-mass models.

A comprehensive simulation program is underway 
to assess baryonic effects, understand them, and 
account for them (\autoref{sec:cp:cos_sim}).  This program will result in 
a robust treatment of lensing observables measured with 
LSST and will be a boon for galaxy 
modelers as well.  Internal calibration of halo structure in an 
analysis of shear spectra will translate into valuable information 
about galaxy formation.  The program has two aspects.  
One is to simulate several self-consistent models of 
galaxy formation in cosmologically large
volumes.  Such simulations are limited due to computational expense.  
The second aspect is to treat baryonic processes with effective models 
that are not self-consistent, but aim 
to capture the large-scale dynamical influences of baryonic 
condensation and galaxy formation in a manner that is 
computationally inexpensive.  This second set of simulations 
allows for some exploration of the parameter space of
cosmology and baryonic physics.

\subsection{Systematics Summary}

We have examined both multiplicative and additive shear
systematics. Multiplicative systematics arise from the convolution of
a galaxy image with a finite PSF.  These systematics are a function of
galaxy size relative to the PSF and the limiting surface brightness of
the image.  Additive shear systematics arise from anisotropic PSFs and
also depend on galaxy size. We have described above how the telescope
design, survey strategy, and algorithmic advances incorporated into the
LSST image analysis will enable us to control and mitigate shear
systematics.

Our methods need to be thoroughly tested
through simulations, such as the ones we discussed in this chapter,
\autoref{sec:design:imsim}, and \autoref{sec:cp:cos_sim}. 
One of the
most challenging areas for systematics will be at large scales.  Many
of the tests we have cited apply mostly to small scales of a focal
plane or less.  The current record for largest scale cosmic shear
detection is 4$^\circ$, which is just larger than an LSST focal plane.
The cosmic shear signal is small at very large scales and experience
is limited.  However, large scales do have some advantages, such as
the large number of PSF stars.  More work must be done to assure the
control of systematics on these scales including simulations of
different dither patterns.

Beyond shear systematics, the three primary sources of systematics
we addressed relate to photometric redshifts, intrinsic alignments, and
theoretical uncertainties. While advances in methodology and modeling
will help mitigate these, empirical information will provide the
surest check. Thus for both photometric redshift errors and intrinsic alignments,
spectroscopic data at high redshifts will enable us to calibrate and
marginalize over the systematic error contributions. For the
uncertainty in theoretical predictions at small scales, a powerful
consistency check (within the halo model framework) will be provided
by high precision measurements of the shear profiles of clusters of
different masses. Thus the LSST data set will itself provide the most
reliable test of the theoretical model.






\bibliographystyle{SciBook}
\bibliography{weaklens/weaklens}


%
%
%
%
%
%
%
%
%
%
%
%
%
%
%
%
%
%
%
%
%
%
%
%
%
%
\chapter{Cosmological Physics\label{chapter-cosmology}}
{\it Hu Zhan, Asantha Cooray, Salman Habib, Alan F. Heavens, Anthony Tyson, Licia Verde, Risa H. Wechsler}

The ultimate goal of astronomical surveys is to deepen our 
fundamental understanding of the Universe. One specific question to be
addressed is the cosmological framework within which we interpret the
observations. Because it is not feasible to physically perturb the 
Universe for investigation, cross checks and confirmations by multiple
lines of evidence are extremely important in cosmology. 
The acceptance of dark matter by the community nearly 
50 years after the seminal proposal of \citet{zwicky33, zwicky37} is a 
perfect example of how observations of
galaxy rotational curves \citep*[e.g.,][]{rubin78}, dynamics of galaxy
groups and clusters \citep[for a review, see][]{faber79}, and
galaxy X-ray emission \citep*[e.g.,][]{fabricant80}
from various surveys, working in unison, eventually shifted our 
paradigm of the Universe.

The accelerated cosmic expansion, inferred from luminosity 
distances of type Ia supernovae \citep{Riess,perlmutter99a} 
and reinforced by large-scale structure and CMB observations
\citep{spergel03} has led to yet another puzzle -- dark energy.
It poses a challenge of such magnitude that, as 
stated by the Dark Energy Task Force (DETF), 
``nothing short of a revolution in our understanding of 
fundamental physics will be required 
to achieve a full understanding of the cosmic acceleration'' 
\citep{albrecht06b}. 

The lack of multiple complementary precision observations is a major 
obstacle in developing lines of attack for dark energy theory. 
This lack is precisely what LSST will address via the powerful 
techniques of weak lensing (WL, \autoref{chp:wl}) 
and baryon acoustic oscillations (BAO, \autoref{chp:lss}) 
-- galaxy correlations more generally -- 
in addition to SNe (\autoref{chp:sne}), cluster counts
(\autoref{sec:lss:cluster}), and other probes of geometry and 
 growth of structure (such as $H_0$ from strong lensing time delay
measurements in \autoref{sec:sl:H0}). 
The ability to produce large, uniform
data sets with high quality for multiple techniques is a crucial
advantage of LSST. It enables not only cross checks of the result from 
each technique but also detections of unknown systematics and 
cross-calibrations of known systematics. Consequently, one can achieve
far more robust and tighter constraints on cosmological parameters and 
confidently explore the physics of the Universe beyond what we know now.

Because the observables of these probes are extracted from the same
cosmic volume, correlations between different techniques can be 
significant. New observables can also emerge, e.g., galaxy-shear 
correlations. A joint analysis of all the techniques with LSST must
involve careful investigations of the cross-correlations between these
techniques. 

In this Chapter, we describe several cosmological investigations 
enabled by the combination of various LSST and external data sets. 
\autoref{sec:cp:wlbao} mainly demonstrates the complementarity 
between BAO (galaxy angular power spectra, \autoref{sec:lss:bao}) and WL (shear power spectra, \autoref{sec:wl:lss})
techniques in constraining the dark energy equation 
of state (EOS), especially in the presence of systematic uncertainties.
Results from LSST SNeIa and cluster counts are shown at the end of this
section. Density fluctuations measured by BAO and WL are sensitive to
the sum of the neutrino masses. We estimate in \autoref{sec:cp:mnu} 
that LSST, in combination with Planck, can constrain the neutrino 
mass to $\Delta m_\nu < 0.1$ eV and determine the mass hierarchy.
Conventional dark energy affects the growth of structure indirectly
through the expansion background, i.e., the Hubble ``drag,'' assuming 
that dark energy clusters occur only on very large scales and with
small amplitude. In contrast, gravity that deviates from General
Relativity (``modified gravity'') can have a direct
impact on clustering at all scales. Because the probes mentioned above 
and in previous chapters are sensitive to both the expansion history of 
the Universe and the growth of structures, LSST can place useful 
constraints on gravity theories as well. Several examples are given in 
\autoref{sec:cp:grav}. \autoref{sec:cp:newphys} shows that
LSST can take the advantage of being a very wide and deep survey to 
test the isotropy of distance measurements across the sky and 
constrain anisotropic dark energy models. We finish with a discussion
of requirements on the cosmological simulations needed for carrying
out the analyses described in this and previous chapters.  The details of the
statistical analyses are given in \autoref{chp:analysis}.

\section[Joint Analysis of BAO and WL]
{Joint Analysis of Baryon Oscillations and Weak Lensing\footnote{Weak lensing in this section refers to
two-point shear tomography (\autoref{sec:wl:twopoint}) only. In the
joint analysis of baryon oscillations and weak lensing, 
galaxy--shear cross power spectra (\autoref{wl-sec-gglensing}) are also included.}}
\label{sec:cp:wlbao}
{\it Hu Zhan}

\subsection{Introduction}
BAO (galaxy power spectra, \autoref{sec:lss:bao}) and WL (shear 
power spectra, \autoref{sec:wl:lss}) techniques each have their 
own systematics and parameter degeneracies. When the
shear and galaxy distribution are analyzed jointly, one gains from the
extra information in the galaxy--shear cross power spectra, which is
not captured in either technique alone. Moreover, the two
techniques can mutually calibrate some of their systematics and
greatly strengthen parameter constraints. 

The WL technique extracts cosmological information from the
distribution of minute distortions (shear) to background galaxies
caused by foreground mass (see \autoref{weaklens-basics}). 
It has the advantage that it measures the
effect of all matter, luminous or not, so that the shear statistics
reflect the clustering properties of the bulk of the matter -- dark
matter, for which gravity alone can provide fairly robust predictions
via linear theory on large scales and $N$-body simulations on small
scales. To achieve its power, however, the WL technique requires
unprecedented control of various systematic effects. One example is
the \phz{} error distribution. Because the kernel of the WL shear
power spectrum peaks broadly between the observer and the source, the
shear power spectrum is not very sensitive to the redshift
distribution of source galaxies. For a redshift bin at $z = 2$, a shift
of the bin by $\Delta z = 0.1$ causes very little change in the shear
power spectrum. However, the inference for cosmological parameters can
change more markedly with the redshift difference $\Delta z = 0.1$.
In other words, one must know the true-redshift distribution of 
galaxies in each \phz{} bin accurately in order to interpret the WL
data correctly.

Galaxies provide a proxy for mass. One may relate fluctuations in the
galaxy number density to those in all matter by the galaxy clustering
bias, which evolves with time and is assumed to be scale independent
only on large scales. Although this does not severely impact the BAO
technique, which utilizes the small oscillatory features in the galaxy
power spectrum to measure distances (see \autoref{sec:lss:bao}), 
knowing the galaxy bias
accurately to the percent level does help improve cosmological
constraints from BAO. Because the kernel of the galaxy (angular) power
spectrum is determined by the true-redshift distribution of galaxies
in a \phz{} bin, galaxy power spectra can be sensitive to the \phz{}
error distribution. For instance, with the Limber approximation, the
cross power spectrum between two redshift bins is given by the overlap
between the two bins in true-redshift space. A small shift to one of
the redshift bins can change the amplitude of the cross power spectrum
significantly, suggesting that the galaxy power spectra can help
calibrate the \phz{} error distribution (\autoref{sec:wl:zphot}).

\subsection{Galaxy and Shear Power Spectra}
We extend the definition of galaxy power spectrum in 
\autoref{sec:lss:bao} and shear power spectrum in
\autoref{sec:wl:lss} to include the galaxy--shear power 
spectrum \citep{hu04b, zhan06d}
\begin{equation} \label{eq:cp:aps}
P_{ij}^{XY}(\ell) = \frac{2\pi^2\ell}{c} \int_0^\infty {\rm d} z\,  
H(z) D_{\rm A}(z) W_i^X(z) W_j^Y(z) \Delta^2_\phi(k; z),
\end{equation}
where lower case subscripts correspond to the tomographic bins, upper 
case superscripts label the observables, i.e., $X=$ g for galaxies 
or $\gamma$ for shear; $H(z)$
is the Hubble parameter, $D_{\rm A}(z)$ is the co-moving angular diameter 
distance, $\Delta^2_\phi(k;z)$ is the dimensionless power spectrum of 
the potential fluctuations of the density field, 
and $k = \ell/D_{\rm A}(z)$. 
BAO and WL do not necessarily use the same binning. In other words, 
the bin number is  defined for each technique separately. The window 
function is
\begin{equation} \label{eq:cp:Wi}
W_i^X(z) = \left\{ \begin{array}{ll}\frac{n_i(z)}{\bar{n}_i} 
\frac{2 a\,b(z)}{3 \Omega_{\rm m} H_0^2 D_{\rm A}^2(z)}
& X = \mbox{g} \\ \frac{1}{c\,H(z) D_{\rm A}(z)}
\int_z^\infty \!{\rm d} z'\, \frac{n_i(z')}{\bar{n}_i}
\frac{D_{\rm A}(z,z')}{D_{\rm A}(z')} & X = \gamma, 
\end{array} \right .
\end{equation}
where $b(z)$ is the linear galaxy clustering bias, and 
$\Omega_{\rm m}$ and $H_0$ are, respectively, the matter 
fraction at $z = 0$ and Hubble constant. 
The galaxy redshift distribution, $n_i(z)$, in the
$i$th tomographic bin is an average of the underlying 
three-dimensional galaxy distribution over angles, and the mean
surface density, $\bar{n}_i$, is the total number of galaxies per
steradian in bin $i$. The distribution, $n_i(z)$, is broader 
than the nominal width of the tomographic bin (defined in \phz{} 
space) because of \phz{} errors.

We only include galaxy power spectra on largely linear scales, 
e.g., the scales of BAOs, in our analysis, 
so that we can map the matter power spectrum to galaxy power
spectrum with a scale-independent but time-evolving linear 
galaxy bias \citep{verde02,tegmark04a}. 
One may extend the analysis to smaller scales with a halo model 
to describe the scale dependency of the galaxy bias and, in fact, 
can still constrain the scale-dependent galaxy bias to $1\%$ 
level \citep{hu04b}. 

The observed power spectra have contributions from
galaxy shot (shape) noise $\bar{n}_i^{-1}$ 
($\gamma_{\rm rms}^2 \bar{n}_i^{-1}$), multiplicative
errors $f_i^X$, and additive errors $A_i^X$:
\begin{equation} \label{eq:cp:totps}
(C_\ell^{XY})_{ij} = (1 + \delta_{X\gamma}^{\rm K} f_i^X + 
\delta_{Y\gamma}^{\rm K} f_j^Y) P_{ij}^{XY}(\ell) +
\delta_{XY}^{\rm K} \left[\delta_{ij}^{\rm K} 
\frac{X_{\rm rms}^2}{\bar{n}_i} + \rho_{ij}^X A_i^X A_j^Y
\left(\frac{\ell}{\ell_*^X}\right)^{\eta^X}\right], 
\end{equation}
where $\delta_{XY}^{\rm K}$ and $\delta_{ij}^{\rm K}$ are 
Kronecker delta functions, $\rho_{ij}^X$ determines how strongly the 
additive errors of two different bins are correlated, 
and $\eta^X$ and $\ell_*^X$ account for the scale dependence of 
the additive errors.  Again, $X$ and $Y$ refer to galaxies g or shear
$\gamma$.  For galaxies, ${\rm g_{rms}} \equiv 1$, 
and, for the shear, $\gamma_{\rm rms}\sim 0.2$ is due to the 
intrinsic shape of galaxies and measurement errors.
Note that the multiplicative error of galaxy number density is 
degenerate with the galaxy clustering bias and is hence absorbed
by $b_i$. Below the levels of systematics future surveys aim to 
achieve, the most important aspect of the (shear) additive error is
its amplitude \citep{huterer06}, so we simply fix $\rho^X = 1$
and $\eta^X = 0$. For more comprehensive accounts of the above 
systematic uncertainties, see 
\citet{huterer06}, \citet{jain06}, \citet{ma06}, and \citet{zhan06d}.

We forecast LSST performace with a Fisher matrix analysis.
See \autoref{sec:app:stats:numtech:fisher} for a detailed description 
of the Fisher matrix calculation. The BAO aspect of the calculations,
which includes the galaxy binning, galaxy bias, and \phz{} treatment,
is the same as that in \autoref{sec:lss:bao}. 
WL results in this section are based on the two-point shear 
tomography described in \autoref{sec:wl:lss}, 
and the joint BAO and WL results include the galaxy--shear power 
spectra (\autoref{eq:cp:totps}) as well. We use 10 shear bins evenly
spaced between $z = 0.001$ and 3.5. 

\subsection{Complementarity Between BAO and WL}

BAO and WL have their unique strengths and are very complementary 
to each other. A joint analysis benefits from the strength of each 
technique: BAO can
help self-calibrate the \phz{} error distribution, while WL can help
constrain the galaxy bias as the different power spectra have
different dependence on the galaxy bias. 

\Phz{} errors are one of the most critical systematics for WL
tomography. Redshift errors directly affect the interpretation 
of the distance--redshift and growth--redshift relations.
Because of its broad lensing kernel, WL cannot self-calibrate the 
\phz{} error distribution, but, as shown in \autoref{sec:lss:bao}, 
the cross-bin galaxy power spectra can calibrate the \phz{} rms
and bias parameters to $\sim 10^{-3}(1+z)$, which is sufficient for 
WL \citep{ma06,zhan06d}.

\begin{figure}
\centering
\includegraphics[width=5.1in]{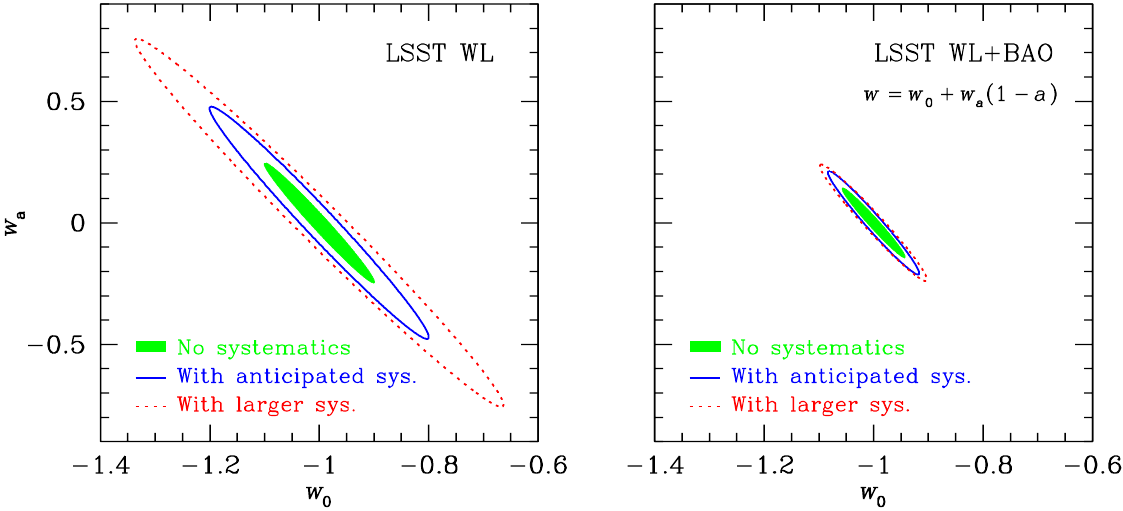}
\caption{1--$\sigma$ error contours of the dark energy EOS parameters
$w_0$ and $w_a$ from LSST WL shear power spectra (left panel) and 
joint LSST WL and BAO (right panel). 
The shaded areas represent the results with 
statistical errors only. The solid contours correspond to those
with the anticipated level of systematic errors, which include 
the uncertainty in the \phz{} error distribution and additive and 
multiplicative errors in the power spectra (see, e.g., 
\autoref{sec:common:photo-z} and \autoref{wl-sec-sys}). 
The assumed \phz{}
systematics would require a redshift calibration sample of 3000 
spectra per unit redshift interval if the \phz{} 
error distribution were Gaussian. The dotted contours 
 relax the requirement to 188 spectra per unit redshift. A much larger 
sample will be needed for realistic \phz{}s. The joint WL and BAO 
results are less affected by the systematics 
because of the ability to self-calibrate the systematics. 
See text for details of the calculations.
\label{fig:cp:wsys}}
\end{figure}

\autoref{fig:cp:wsys} 
demonstrates that while the WL constraints on the dark energy 
EOS parameters, $w_0$ and $\wa$, are sensitive to systematic 
uncertainties in the \phz{} error distribution, the joint 
BAO and WL results remain fairly immune to these systematics. 
The dramatic improvement of the BAO+WL results
over the WL-alone results is due to the cross-calibration 
of galaxy bias and \phz{} uncertainties 
and is independent of the dark energy EOS parametrization.

The statistics-only results in \autoref{fig:cp:wsys} are 
marginalized over the other nine cosmological parameters listed in
\autoref{tab:com:cosp} and 30 galaxy clustering bias 
parameters. We impose no prior on the galaxy bias (for numerical 
reasons, we take $\sigma_P(\ln b_i)=1000$). The \phz{} parameters
are fixed, and the power spectra errors in \autoref{eq:cp:totps}
are not included. 

For anticipated systematics, we assume that 
$\sigma_P(\delta z_i) = 2^{-1/2} \sigma_P(\sigma_{z,i}) = 
0.05\sigma_{z,i} = 0.0025(1+z)$, $\sigma_P(f_i) = 0.005$,
$A^\mathrm{g}_i = 10^{-4}$, and $A^\gamma_i=10^{-5}$. 
For larger systematics, we relax the \phz{} priors to
$\sigma_P(\delta z_i) = 2^{-1/2} \sigma_P(\sigma_{z,i}) = 
0.2\sigma_{z,i} = 0.01(1+z)$ and $A^\gamma_i=10^{-4.5}$. 
See \autoref{sec:common:photo-z} and \autoref{wl-sec-sys}
for discussions about the systematics.

The linear galaxy clustering bias, $b$, is degenerate
with the linear growth function, $G$, for the angular BAO technique. 
Therefore, one cannot extract
much useful information from the growth of the large-scale structure
with \phz{} BAO. One can break this degeneracy by jointly analyzing the 
galaxy and shear power spectra, because the galaxy--galaxy, 
galaxy--shear, and shear--shear power spectra depend on different
powers of the linear galaxy bias ($b^2$, $b^1$, and $b^0$ respectively). 
CMB data help as well by providing an accurate 
normalization of the matter power spectrum. The resulting constraints
on the linear galaxy bias parameters can reach the percent level 
\citep*{hu04b,zhan06d,zhan09a}, so that growth information 
can be recovered from galaxy power spectra as well.

\subsection{Precision Measurements of Distance and Growth Factor
\label{sec:cp:dgc}}

Dark energy properties are derived from variants of the 
distance--redshift and growth--redshift relations. Different dark
energy models feature different parameters, and various 
phenomenological parametrizations may be used for the same  
quantity such as the EOS. In contrast, distance and growth 
measurements are model-independent, as long as dark energy does 
not alter the matter power spectrum directly. Hence, it is 
desirable for future surveys to provide results of the distance and 
growth of structure, so that different theoretical models can 
be easily and uniformly confronted with the data.

\begin{figure}
\centering
\includegraphics[width=3in]{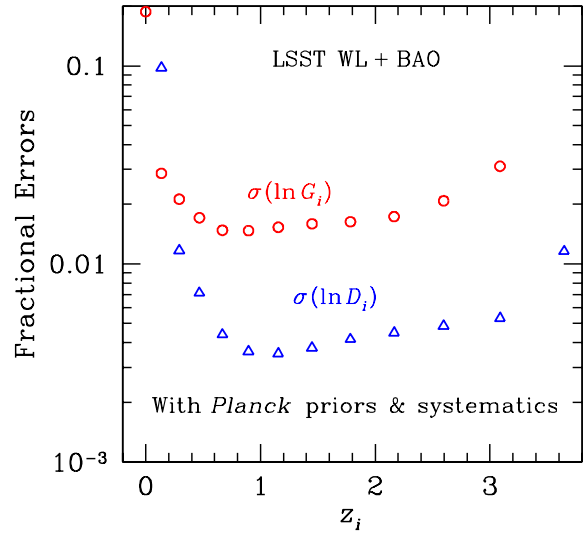}
\caption{Marginalized $1\sigma$ errors on the co-moving distance 
(open triangles) and growth factor (open circles) parameters from 
the joint analysis of LSST BAO and WL (galaxy--galaxy, galaxy--shear,
and shear--shear power spectra) with a 
conservative level of systematic uncertainties in the photometric redshift error 
distribution and additive and multiplicative errors in the shear and 
galaxy power spectra. The maximum multiple used for WL is 
2000, and that for BAO is 3000 (with the additional requirement
$\Delta_\delta^2(\ell/D_{A};z) < 0.4$).
The growth parameters, $G_0 \ldots D_{14}$, are evenly spaced in 
$\log(1+z)$ between $z = 0$ and 5, and the distance parameters, 
$D_1 \ldots D_{14}$, start at $z_1 = 0.14$ (see text for details).
The error of each distance (growth) parameter is marginalized 
over all the other parameters including growth (distance) parameters
and other distance (growth) parameters. The joint constraints on 
distance are relatively insensitive to the assumed systematics. 
Figure from \citet{zhan09a}, with permission.
\label{fig:cp:dges}}
\end{figure}

\autoref{fig:cp:dges} demonstrates that 
joint LSST BAO and WL can achieve 
$\sim 0.5\%$ precision on the distance and $\sim 2\%$ on the growth 
factor from $z = 0.5$ to $3$ in each interval of $\Delta z \sim 0.3$ 
\citep{zhan09a}. Such measurements can test the consistency of dark 
energy or modified gravity models 
\citep[e.g.,][]{knox06a,Heavens07}.

\subsection{Constraining the Mean Curvature}

The mean curvature of the Universe has a significant impact on dark 
energy measurements. For example, the curvature parameter $\Ok$ is 
completely degenerate with a $w=-1/3$ dark energy if dark energy 
clustering (\autoref{sec:lss:isw}) is neglected. In the 
concordance \lcdm{} model ($w=-1$), allowing $\Ok$ to float 
greatly weakens the ability of supernovae at $z<1.7$ to constrain 
$\wa$ \citep{linder05b,knox06a}. LSST BAO and WL can 
determine $\Ok$ to $\sim 10^{-3}$ separately and $< 10^{-3}$ jointly, 
and their results on $w_0$ and $\wa$ are not affected in practice 
by the freedom of $\Ok$ \citep{zhan06d,knox06a}. The reason is that
low-redshift growth factors, which can be measured well by WL, and 
high-redshift distances, which can be measured well by BAO, are 
very effective for measuring $\Ok$ and, hence, lifting the 
degeneracy between $\Ok$ and $w_a$ \citep{zhan06e}. 
Given its large area, LSST 
can place a tight upper limit on curvature 
fluctuations, which are expected to be small ($\sim 10^{-5}$) at 
the horizon scale in standard inflation models. 

The aforementioned results are obtained either
with the assumption of matter dominance at $z \gtrsim 2$ and 
precise independent distance measurements
at $z\gtrsim 2$ and at recombination \citep{knox06a} or with a specific 
dark energy EOS: $w(z) = w_0 + w_az(1+z)^{-1}$ \citep{knox06c,zhan06d}. 
However, if one assumes only the Robertson-Walker metric without 
invoking the dependence of the co-moving distance on cosmology, then 
the pure metric constraint on curvature from a simple combination of
BAO and WL becomes much weaker: $\sigma(\Omega_k^\prime) \simeq 0.04
f_{\rm sky}^{-1/2}(\sigma_{z0}/0.04)^{1/2}$ 
\citep{bernstein06}\footnote{$\Omega_k$ affects both the co-moving 
distance and the mapping between the co-moving distance and 
the angular diameter distance, while $\Omega_k^\prime$ affects
only the latter. See \autoref{eq:lss:DA-Ok}.}.

Our result for $\Omega_k^\prime$ from LSST WL or BAO alone is not 
meaningful, in agreement with \citet{bernstein06}. However,
because WL and BAO measure very different combinations of 
distances \citep[see, e.g., Figure 6 of][]{zhan09a}, breaking the
degeneracy between $\Omega_k^\prime$ and other parameters,
the joint analysis of the two leads to
$\sigma(\Omega_k^\prime)=0.017$, including 
anticipated systematics in \phz{}s and power spectra for LSST.
This result is better than the forecast derived from the shear 
power spectra and galaxy power spectra in \citet{bernstein06} because
we include in our analysis more information: the
galaxy--shear power spectra.

\begin{figure}
\centering
\includegraphics[width=3in]{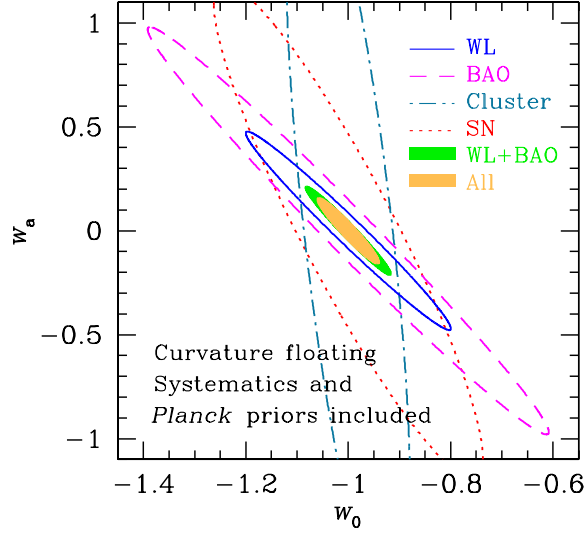}
\caption{Joint $w_0$--$\wa$ constraints from 
LSST BAO (dashed line), cluster counting (dash-dotted line), 
supernovae (dotted line), WL (solid line), joint BAO and WL 
(green shaded area), and all combined (yellow shaded area).
The BAO and WL results are based on galaxy--galaxy, galaxy--shear,
and shear--shear power spectra only. 
Adding other probes such as strong lensing time delay
(\autoref{sec:sl:H0}), ISW effect (\autoref{sec:lss:isw}), and 
higher-order galaxy and shear statistics (\autoref{sec:lss:bisp}
and \autoref{sec:wl:lss}) will further improve the constraints.
\label{fig:cp:cswb}}
\end{figure}

\subsection{Results of Combining BAO, Cluster Counting, Supernovae, 
and WL}

We show in \autoref{fig:cp:cswb} $w_0$--$\wa$ constraints 
combining four LSST probes 
of dark energy: BAO, cluster counting, supernovae, and WL.
The cluster counting result is from \citet{fang07} and the 
supernova result is based on \citet{zhan08}. Because each probe has
its own parameter degeneracies, the combination of any two of them
can improve the result significantly. As mentioned above, BAO and
WL are highly complementary to each other. Much of the 
complementarity is actually in parameter space (such as \phz{}s
and galaxy bias) that has been marginalized over. 
In \autoref{fig:cp:cswb}, we see that cluster counting is 
quite effective in constraining $w_0$ and that it is directly 
complementary to WL and BAO in the $w_0$--$w_a$ plane. 
When all the four probes are combined, the error ellipse 
area decreases by $\sim 30\%$ over the joint BAO and WL result. 

\begin{figure}
\centering
\includegraphics[width=5.5in]{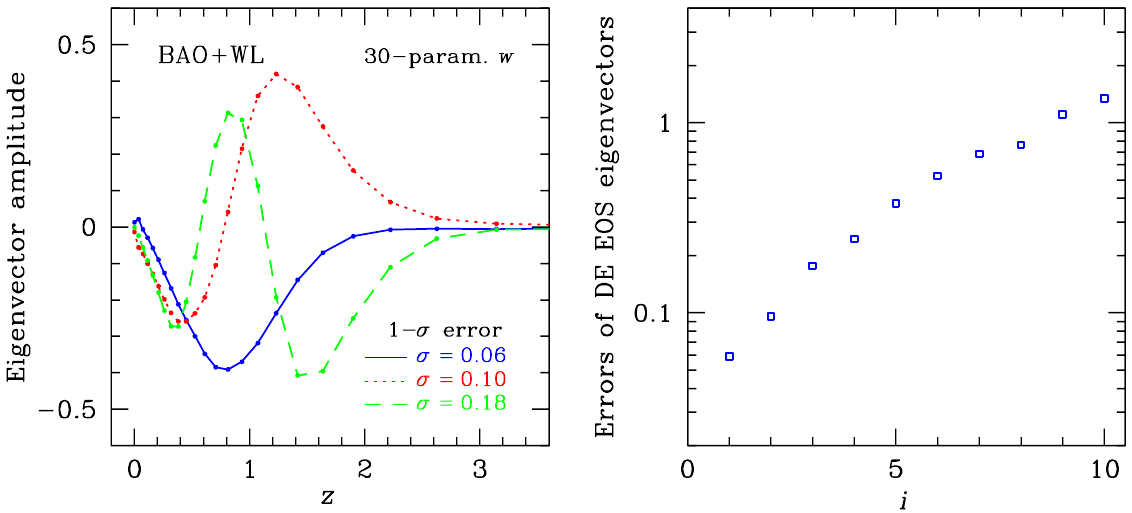}
\caption{Eigensystem analysis of joint LSST BAO and WL 
(galaxy--galaxy, galaxy--shear, and shear--shear power spectra) 
constraints
on a 30-dimensional dark energy EOS model. The dark energy EOS is 
spline-interpolated between 30 EOS parameters evenly spaced between 
$a = 0$ and 1 with a fiducial model of $w=-1$. \emph{Left panel}: 
The first three best-determined dark energy EOS eigenmodes 
(departure from $w=-1$) with LSST BAO+WL. 
\emph{Right panel}: 1-$\sigma$ errors of the EOS eigenmodes.
The errors will be proportional to the square root of the dimension
of the dark energy EOS for sufficiently large dimensions.
\label{fig:cp:w30}}
\end{figure}

The $w_0$--$\wa$ parametrization in \autoref{fig:cp:cswb} 
 does not capture the complexity of all dark energy models. 
It also significantly underestimates the full capabilities of Stage 4
surveys \citep{albrecht07}, such as that of the LSST. More generally,
one may allow the EOS to vary independently at different redshifts
and let the data determine the EOS eigenmodes and their errors, 
which can then be used to constrain dark energy models.
\autoref{fig:cp:w30} presents the best determined dark energy EOS 
eigenmodes and their errors from LSST BAO+WL for a 30-dimensional 
EOS model \citep{albrecht07,albrecht09}. It is seen that the best
determined mode is sensitive to the dark energy EOS at $z\sim 0.8$.
The eigensystem analysis gives the expected noise of the eigenmodes,
and one can then project dark energy models into the orthogonal 
eigenmode space to constrain the models. Detailed calculations show 
that LSST can eliminate a large space of quintessence models 
\citep[e.g.,][]{barnard08}.


%
%
%
%
%
%
%
%
%
%
%
%
%
%
%
%
%
%
%
%
%
%
%
%
%
%
\section{Measurement of the Sum of the Neutrino Mass\label{sec:cp:mnu}} 

{\em Licia Verde, Alan F. Heavens}

Current limits on the neutrino mass from cosmology come most robustly
from the CMB, $\sum m_\nu < 1.3$ eV at 95\% confidence
\citep{Dunkley}, and with the addition of large-scale structure and
supernova data to $<0.6$ eV \citep{Komatsu}.  With the addition of
Lyman $\alpha$ data, the limits may be pushed as low as 0.17 eV
\citep{Seljak2006}, with model-dependent assumptions.  For a summary
of experimental limits on neutrino masses, see \citet{Fogli}.  In the
future, the primary robust tools for constraining massive neutrinos
are the CMB combined with a large scale structure survey and
measurement of the three-dimensional cosmic shear. In the
three-dimensional cosmic shear technique \citep{heavens03, Castro,
  Heavens06, Kitching07} the
full three-dimensional shear field is used without redshift binning, maximizing the
information extracted.  Massive neutrinos suppress the growth of the
matter power spectrum in a scale-dependent way, and it is from this
signature that cosmic shear measurements can constrain neutrino
properties. Inevitably, there is a degeneracy with dark energy
parameters, as dark energy also affects the growth of perturbations
\citep{Kiakotou}.  Following \citet{Kitching08nu} we explore the
constraints on neutrino properties obtained with a Fisher matrix
approach for a survey with the characteristics of LSST assuming a Planck prior. Unless otherwise stated, the reported constraints are
obtained allowing for running of the spectral index of the primordial
power spectrum, non-zero curvature and for a dark energy component
with equation of state parametrization given by $w_0,w_{a}$; all
results on individual parameters are fully marginalized over all other
cosmological parameters.

\begin{figure}
\centering
\includegraphics[width=4in]{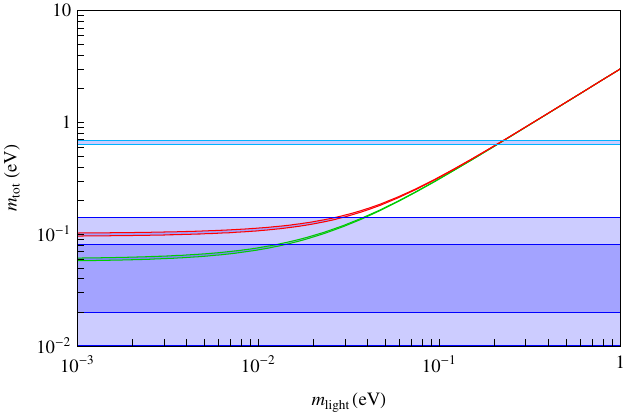}
\caption{ Forecasted constraints in the context of what is known to
  date from neutrino oscillations experiments. The x-axis is
  the mass of the lightest neutrino in eV, and the y-axis is the sum
  of neutrino masses (also in eV). The narrow green band represents
  the normal hierarchy  and the narrow red band the inverted one: for
  light neutrinos the two hierarchies are in principle
  distinguishable. The light blue (horizontal shaded) regions
  represent the $1-\sigma$ constraints for the combination Planck+LSST three-dimensional lensing, for two fiducial models of massive and
  nearly massless lightest neutrino, consistent with the normal hierarchy.
  The lighter regions are $1-\sigma$ constraints for a  general
  cosmological model with massive neutrinos  (see text for
  details). The darker horizontal  band shows the  forecasted
  $1-\sigma$ constraint obtained in the context of  a power-law
  $P(k)$, $\Lambda CDM$ + massive neutrinos model.  These constraints
  offer the possibility in principle to distinguish between the normal
  and inverted hierarchies. Figure courtesy of
  E. Fernandez-Martinez. } 
\label{fig:cp:numass}
\end{figure}

\begin{figure}
\centering
\includegraphics[width=3.6in]{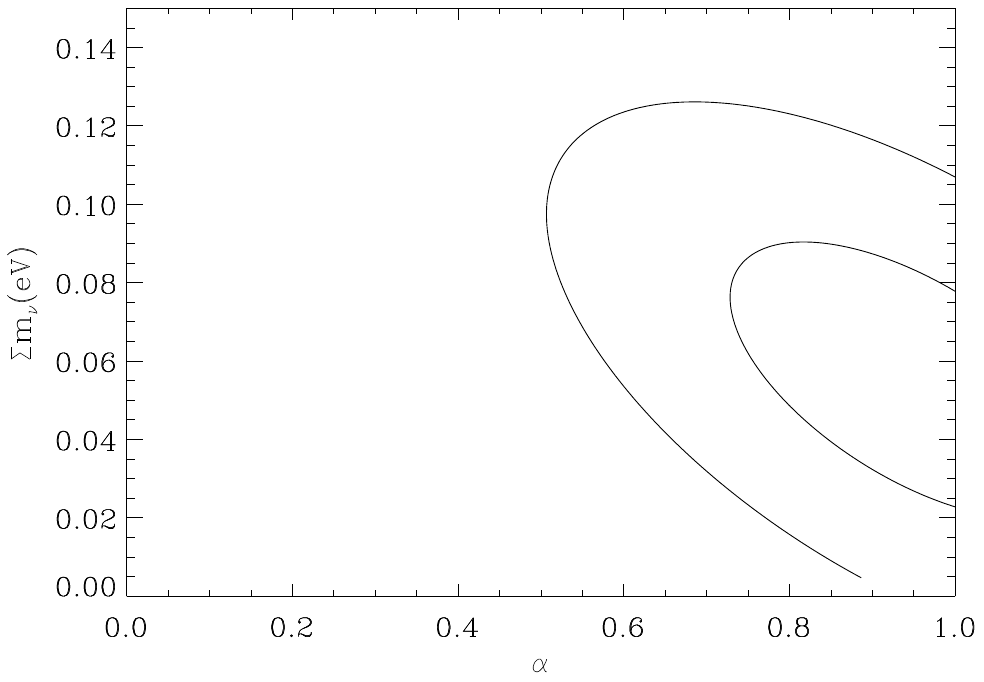}
\caption{The 68\% and 95\% confidence intervals for the sum of
  neutrino masses and the fraction of mass in the heaviest neutrino,
  using Planck and LSST's WL survey.  marginalized over other
  parameters in a 10-parameter cosmological model. Figure from
  \citet{deBernardis09}, with permission. }
\label{fig:cp:nuhierarchy}
\end{figure}

By combining three-dimensional cosmic shear constraints achievable
with a survey like LSST with constraints achievable with Planck's data, the massive neutrino (fiducial values $m_{\nu}=0.66$
eV ; $N_{\nu}=3$) parameters could be measured with marginal errors of
$\Delta m_{\nu}\sim 0.03$ eV and $\Delta N_{\nu}\sim 0.08$, a factor
of 4 improvement over Planck alone.  If neutrinos are massless
or have a very small mass (fiducial model $m_{\nu}=0$ eV ;
$N_{\nu}=3$) the marginal errors on these parameters degrade ($\Delta
m_{\nu}\sim 0.07$ eV and $\Delta N_{\nu}\sim 0.1$), but there is still
a similar improvement over Planck alone.  This degradation in the
marginal error occurs because the effect of massive neutrinos on the
matter power spectrum and hence on three-dimensional weak lensing is
non-linear. These findings are in good agreement with an independent
analysis \citep{Hannestad06}.
Alternatively, the constraints could improve by as much as a factor of
2 if complementary data sets (such as direct measurements of the
expansion history from BAO or supernovae) were used to lift the degeneracies
between $m_{\nu}$ and the running of the spectral index, $w_a$ and
$w_0$ (Kitching, private communication).  As
discussed in \citet{Kitching08} and \citet{Kitching08nu}, a
degradation in errors by a factor of $\sim \sqrt{2}$ is expected due
to systematics.

\autoref{fig:cp:numass} shows these constraints in the context of what
is known currently from neutrino oscillations experiments.  Particle
physics shows that neutrinos come in three flavors: muon, tau, and
electron neutrinos and that they oscillate i.e., as they propagate they
can change flavor; the neutrino flavor eigenstates are not the same as
the neutrino mass eigenstates (simply called 1 2 3). In the standard
model for particle physics, the existence of flavor oscillations
implies a non-zero neutrino mass because the amount of mixing between
the flavors depend on their mass differences. The properties of the
mixing are described by a ``mixing matrix" which is like a rotation
matrix specified by the mixing angles $\theta_{12}$, $\theta_{13}$,
etc. Oscillation experiments have so far determined absolute values of
neutrino mass differences, one mass difference being much smaller than
the other one.  However neither the sign of the mass difference 
nor the absolute mass scale are known. There are, therefore, two possibilities: a)
the ``normal hierarchy,'' two neutrinos are much lighter than the
third or b) an ``inverted hierarchy,'' in which one
neutrino is much lighter than the other two.  Cosmology, being sensitive
to the sum of the neutrino masses, can offer complementary information
to particle physics experiments in two ways: a) a determination of the
total neutrino mass will give an absolute mass scale and b) since in
the normal hierarchy the sum of neutrino masses is lower (by up to a
factor of 2, depending on the absolute mass scale) than in the
inverted hierarchy, a determination of the total neutrino mass with an
error $\ll 0.1$ eV may select the neutrino hierarchy. This can be
appreciated in the green and red narrow bands of
\autoref{fig:cp:numass}.

Particle physics experiments that will be completed by the time of
LSST do not guarantee a determination of the neutrino mass $\Sigma m_{\nu}$
if it lies below $0.2$ eV.  Neutrino-less double beta decay experiments
will be able to constrain neutrino masses only if the hierarchy is
inverted and neutrinos are Majorana particles (i.e., their own
anti-particle).  On the other hand, oscillation experiments will
determine the hierarchy only if the composition of electron flavor in
all the neutrino mass states is large (i.e., if the mixing angle
$\theta_{13}$ is large).  Cosmological observations are principally
sensitive to the sum of neutrino masses.  However, there is some
sensitivity to individual masses, due to features in the power
spectrum arising from the different redshifts when the neutrinos
become non-relativistic.  The effects are weak \citep{Slosar}, but a
large, deep weak lensing survey in combination with Planck, could
exploit this signal and tighten the above constraints further.  Thus,
the LSST survey, together with CMB observations, could offer valuable
constraints on neutrino properties highly complementary to particle
physics parameters.

\citet{deBernardis09}  parametrized neutrino masses  using $\alpha$,
where $m_3 \equiv \alpha \sum m_\nu$ under the weak assumption that
$m_1=m_2$. $\alpha \sim 1$ represents the normal hierarchy  for very
low  mass of the lightest neutrino, and $\alpha \simeq 0$ represents
the inverted hierarchy.  They compute the expected marginal error on
$\alpha$ for a fiducial model consistent with the direct hierarchy:
this is shown in \autoref{fig:cp:nuhierarchy}: distinguishing the
hierarchy is within reach of a large-scale weak lensing survey such as
could be undertaken with LSST, with an expected marginal error on
$\alpha$ of 0.22 (for normal hierarchy).  \citet{deBernardis09} also
point out that assumption of the wrong hierarchy can bias other
cosmological estimates by as much as $1-2\sigma$. 

\begin{figure}
\centering
\includegraphics[width=3in]{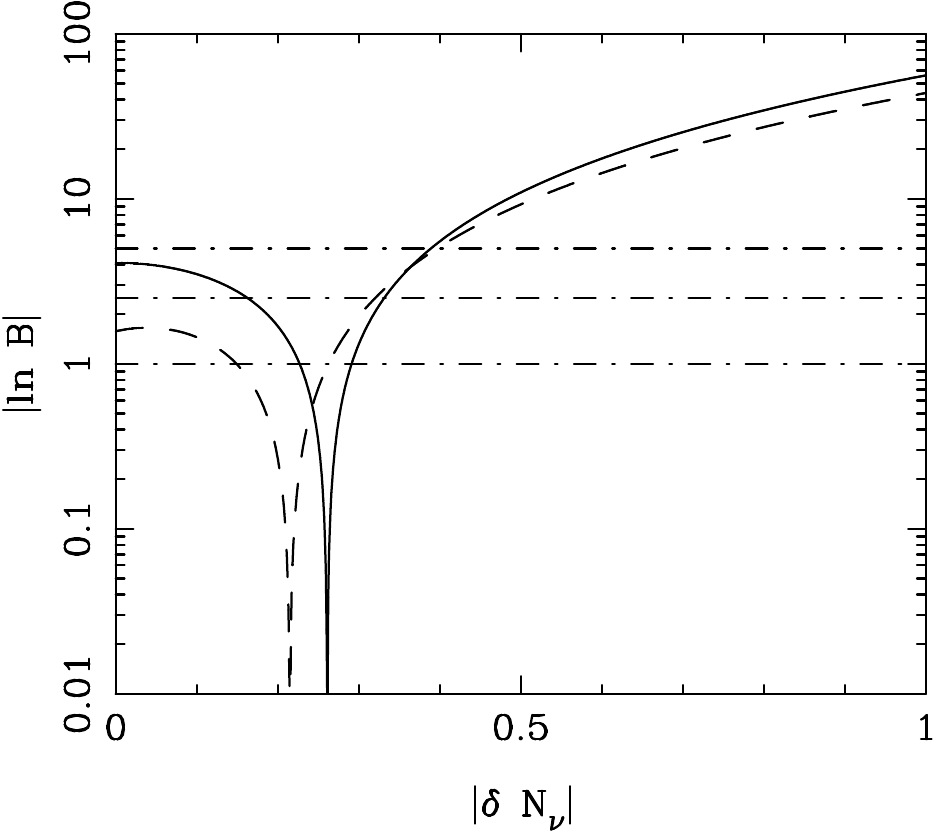}
\caption{The predicted evidence for the number $N_{\nu}$ of 
  neutrinos   individually for three-dimensional cosmic shear using the fiducial survey combined with a Planck prior.
  The solid line shows the conditional evidence assuming that the 
  other parameter is fixed at its fiducial value, the dashed line shows the 
  marginal expected 
  evidence when the possible values of the hidden parameter are taken into account. The dot-dashed lines show the defining evidence limits on 
  the Jeffreys scale where $\ln B<1$ is 
  ``inconclusive," $1<\ln B<2.5$ is ``substantial" evidence in favor of a model, 
  $2.5<\ln B< 5$ is ``strong," and $\ln B>5$ is ``decisive."
Reprinted figure with permission from \citet{Kitching08nu}, \url{http://link.aps.org/abstract/PRD/v77/e103008}.  
}
\label{fig:cp:figure1D}
\end{figure}

These constraints have also been considered in the framework of
Bayesian evidence \citep{Kitching08nu}. The Bayes factor is a tool for
model selection, and can be used to quantify an experiment's ability
to distinguish one model from another
(\autoref{sec:app:stats:evidence}).  The standard model predicts three
neutrino species; corrections to account for quantum electrodynamics (QED) effects and for
neutrinos being not completely decoupled during electron-positron
annihilation imply $N_{eff}$ = 3.04.  Any light particle that does not
couple to electrons, ions, and photons will act as an additional
relativistic species.  Departures from the standard model, which are
described by a deviation $N_{eff} \ne 3.04$, can arise from the decay
of dark matter particles, quintessence, exotic models, and additional
hypothetical relativistic particles such as a light majoron or a
sterile neutrino.  Such hypothetical particles are constrained from
standard big bang nucleosynthesis (BBN), where the allowed extra
relativistic degrees of freedom are $N^{BBN}_ {eff} = 3.1_{-1.2}^{
  +1.4}$ \citep[see e.g.,][]{Mangano}. BBN constraints rely on
different physics and the energy density in relativistic species may
easily change from the time of BBN to the later Universe.

\autoref{fig:cp:figure1D} shows the predicted evidence for the number
$N_{\nu}$ of neutrinos from the analysis described above.  The solid
line shows the conditional evidence assuming that the other parameter
is fixed at its fiducial value, the dashed line shows the marginal
expected evidence when the possible values of the hidden parameter are
taken into account.  We see that only if $N_{eff}>3.4$ will the
evidence against the standard model be decisive.

To summarize: the errors on the sum of the neutrino masses from a weak lensing analysis of a
fiducial LSST survey are impressively small, and there is some sensitivity to individual neutrino masses, enough in principle to
distinguish between the normal and inverted hierarchies.  Improvement in
constraints on the effective number of neutrinos is also possible, but the
constraints are not expected to be particularly tight.  


%
%
%
%
%
%
%
%
%
%
%
%
%
%
%
%
%
%
%
%
%
%
%
%
%
%
\section{Testing Gravity}
\label{sec:cp:grav}


{\em Alan F. Heavens, Licia Verde}

\subsection{Introduction}
The acceleration of the Universe is such an unexpected feature that it has
spawned a number of explanations, many of which are very far-reaching in their
consequences.   The simplest solution is found in Einstein's General Theory
of Relativity (GR), in the form of the infamous cosmological constant.  In a more
modern guise, this term is placed on the opposite side of Einstein's field
equation, as a source term, and is interpreted as a vacuum energy density. 
This opens up a wealth of more general possibilities that the source is not
actually vacuum, but a slowly rolling dark energy field, which may evolve. 
Since this field would need to account for about 75\% of the Universe's energy
budget, determining its properties and nature is essential for a full
understanding of the Universe.  In addition to this possibility, there is an
even more radical solution.  As a cosmological constant, Einstein's term
represents a modification of the gravity law itself, rather than an unusual source of
gravity.  Thus it is compelling to raise the question of whether the
acceleration is driven by a new, beyond-Einstein theory of gravity.  Although
no compelling theory currently exists, suggestions include modifications to GR
arising from extra dimensions, as might be expected from string theory
braneworld models.  There are potentially measurable effects of such exotic
gravity models that LSST could probe 
\citep[e.g.,][]{lue04,song05,Ishak,Knox2006,zhang07}, 
and finding evidence for extra
dimensions would of course signal a radical departure from our conventional
view of the Universe.

In this section we focus on measurements that might be made to distinguish GR from modified gravity models.  We will restrict the discussion to scalar perturbations, and how they are related to observation.   The interval in the conformal Newtonian gauge may be written in terms of two scalar perturbations, $\psi$ being the potential fluctuation and $\phi$ the curvature perturbation, as follows
\begin{equation}
ds^2 = a^2(\eta)\left[(1+2\psi)d\eta^2 - (1-2\phi)d\vec x^2\right],
\end{equation}
where we assume a flat background Universe for simplicity.  This assumption is easily relaxed.  $a(\eta)$ is the scale factor as a function of conformal time.

 In GR, and in the absence of anisotropic stresses (a good
 approximation for epochs when photon and neutrino streaming are
 unimportant),  $\phi=\psi$.  In essence, the information is all
 contained in these potentials and how they evolve, and these will
 depend on the gravity model.  In modified gravity, one expects that the
 Poisson law is modified, changing the laws for $\psi$ and $\phi$.
WL and spectroscopic galaxy surveys together can provide consistency 
tests for the metric perturbations, density fluctuations, and velocity 
field in the GR framework \citep[e.g.,][]{zhang07,Song}. Furthermore,
\citep{zhang07} show that the ratio of the Laplacian of the 
Newtonian potentials to the peculiar velocity divergence can be a clean 
probe of modified gravity -- independent of galaxy-mass bias and the scale 
of mass perturbations.

 The difference between $\phi$ and $\psi$ can be characterized \citep{Daniel} by the 
 {\em slip}, $\varpi$.  Since this may be scale- and time-dependent,
 we define in Fourier space 
 \begin{equation}
 \psi(k,a) = \left[1+\varpi(k,a)\right]\phi(k,a) ,
\end{equation}
where $\varpi \equiv 0$ in GR.  
We may also write the modified Poisson equation in terms of the matter perturbation $\delta_m$ and density $\rho_m$ as \citep{Amendola}
\begin{equation}
-k^2\phi = 4\pi G a^2 \rho_m \delta_m Q(k,a),
\end{equation}
which defines $Q$ as an effective change in the Gravitational Constant
$G$.

Different observables are sensitive to $\psi$ and $\phi$ in different
ways \citep{JainZhang}.  For example, the ISW effect depends on
$\dot\psi + \dot\phi$, but the effect is confined to large scales, and
cosmic variance precludes accurate use for testing modified
gravity. Peculiar velocities are sourced by $\psi$, and LSST may be a
useful source catalog for later spectroscopic surveys to probe this.
Lensing is sensitive to $\psi+\phi$, and this is the most promising
route for LSST to probe beyond-Einstein gravity.  The Poisson-like
equation for $\psi+\phi$ is
\begin{equation}
-k^2(\psi+\phi) = 2\Sigma\frac{3H_0^2 \Omega_m}{2a} \delta_m
\label{eq:modified_poisson},
\end{equation}
where $\Sigma \equiv Q(1+\varpi/2)$.   For GR, $Q=1$, $\Sigma=1$, and
$\varpi=0$.  
The Dvali-Gabadadze-Porrati (DGP) braneworld
model \citep{DGP} has $\Sigma=1$, so mass perturbations deflect light
in the same way as GR, but the growth rate of the fluctuations
differs.  Thus we have a number of possible observational tests of
these models, including probing the expansion history, the growth rate
of fluctuations, and the mass density-light bending relation.

%

Some methods, such as study of the luminosity distance of Type Ia
supernovae \citep[SNe; e.g.,][]{Riess}, baryonic acoustic oscillations
\citep[BAO; e.g.,][]{EisensteinHu97}, or geometric weak lensing methods
\cite[e.g.,][]{TKBH06} probe only the expansion history, whereas
others such as three-dimensional cosmic shear weak lensing or cluster
counts can probe both.  Models with modified gravity laws predict
Universe expansion histories which can also be explained with standard
GR and dark energy with a suitable equation of
state parameter $w(z)$.  In general, however, the growth history of
cosmological structures will be different in the two cases, allowing
the degeneracy to be broken (e.g.,
\citealt{Knox2006}; \citealt{HutererLinder06}; but see
\citealt{Kunz}).

\subsection{Growth Rate}
The growth rate of perturbations in the matter density
$\rho_m$, $\delta_m \equiv \delta \rho_m/\rho_m$, is parametrized as a function of scale factor $a(t)$ by
\begin{equation}
\frac{\delta_m}{a} \equiv g(a) =
\exp\left\{\int_0^a\,\frac{da'}{a'}\left[\Omega_m(a')^\gamma-1\right]\right\},
\end{equation}
where $\Omega_m(a)$ is the density parameter of the matter. 
In the standard GR cosmological model, $\gamma\simeq 0.55$,  whereas in modified gravity 
theories it deviates from this value \citep{Linder05}.  As a strawman example, 
the flat DGP braneworld model \citep{DGP} has $\gamma\simeq 0.68$
on scales much smaller than those where cosmological 
acceleration is apparent \citep{LinderCahn07}. 
While this is not the most general model (in principle $\gamma$ might,
for example,  depend on scale) this   offers a convenient Minimal
Modified Gravity parametrization \citep{LinderCahn07,HutererLinder06}.

\begin{figure}
\centering
\includegraphics[width=3in]{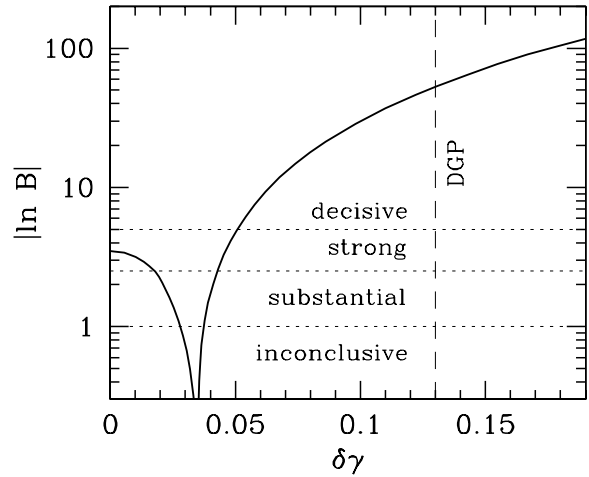}
\caption{Bayesian evidence $B$ for GR as a function of the true
deviation of the growth index from GR, 
$\delta \gamma = \gamma - 0.55$, for a Stage 4 WL survey comparable to 
our fiducial survey in combination with Planck.
The larger the $B$ value, the greater the statistical power of this 
survey to distinguish the models. If modified gravity is the true
model, GR will be favored by the data to the left of the cusp
($B>1$), and increasingly disfavored to the right ($B<1$). 
The Jeffreys scale of evidence \citep{Jeffreys61} is as labeled.
Joint BAO and WL will place stronger constraints.
Figure from \citet{Heavens07}, with permission.
\label{fig:cp:dgamma}}
\end{figure}

Measurements of the growth factor can be used to determine the growth
index $\gamma$. It becomes a very interesting question to ask whether
a given method or observational set up could distinguish between the
dark energy and modified gravity scenarios.  In contrast to {\em
  parameter estimation}, this is an issue of {\em model selection},
which has been the subject of recent attention in cosmology.
That is, one might 
compare a dark energy
model that has a fixed GR value for $\gamma$ with a modified gravity
model whose $\gamma$ is determined by the data and ask 
``do the data require the additional parameter and therefore 
signal the presence of new physics?'' This question may 
be answered with the Bayesian evidence, $B$, which is proportional to
the ratio of 
probabilities of two or more models, given some data
(see \autoref{sec:app:stats:evidence} 
for more details).
To quantify how LSST will help in addressing the issue of testing
gravity using the growth of structures, we follow  \cite{Heavens07}.  

\autoref{fig:cp:dgamma} shows how the Bayesian evidence for GR changes
with increasing true deviation of $\gamma$ from its GR value for a
combination of a Stage 4 WL survey (comparable to our fiducial LSST survey)
and Planck \citep{Heavens07}.  The expected evidence ratio
changes with progressively greater differences from the GR growth
rate.  The combination of WL and Planck could strongly
distinguish between GR and minimally-modified gravity models whose
growth index deviates from the GR value by as little as $\delta \gamma
= 0.048$.  Even with the WL data alone, one should be able to
decisively distinguish GR from the DGP model at $\ln B \simeq 11.8$,
or, in the frequentist view, $5.4\sigma$ \citep{Heavens07}.  The
combination of WL+Planck+BAO+SN should be able to distinguish
$\delta \gamma = 0.041$ at $3.41\,\sigma$. A vacuum energy General Relativity
model will be distinguishable from a DGP modified-gravity model with
log evidence ratio $\ln B \simeq 50$ with LSST + Planck; the three-dimensional lensing data alone
should still yield a ``decisive" $\ln B \simeq 11.8$.  An alternative
approach is to explore whether the expansion history and growth rate
are consistent assuming GR \citep{lue04,song05,Ishak,Knox2006}.

\begin{figure}
\centering
\includegraphics[width=3in]{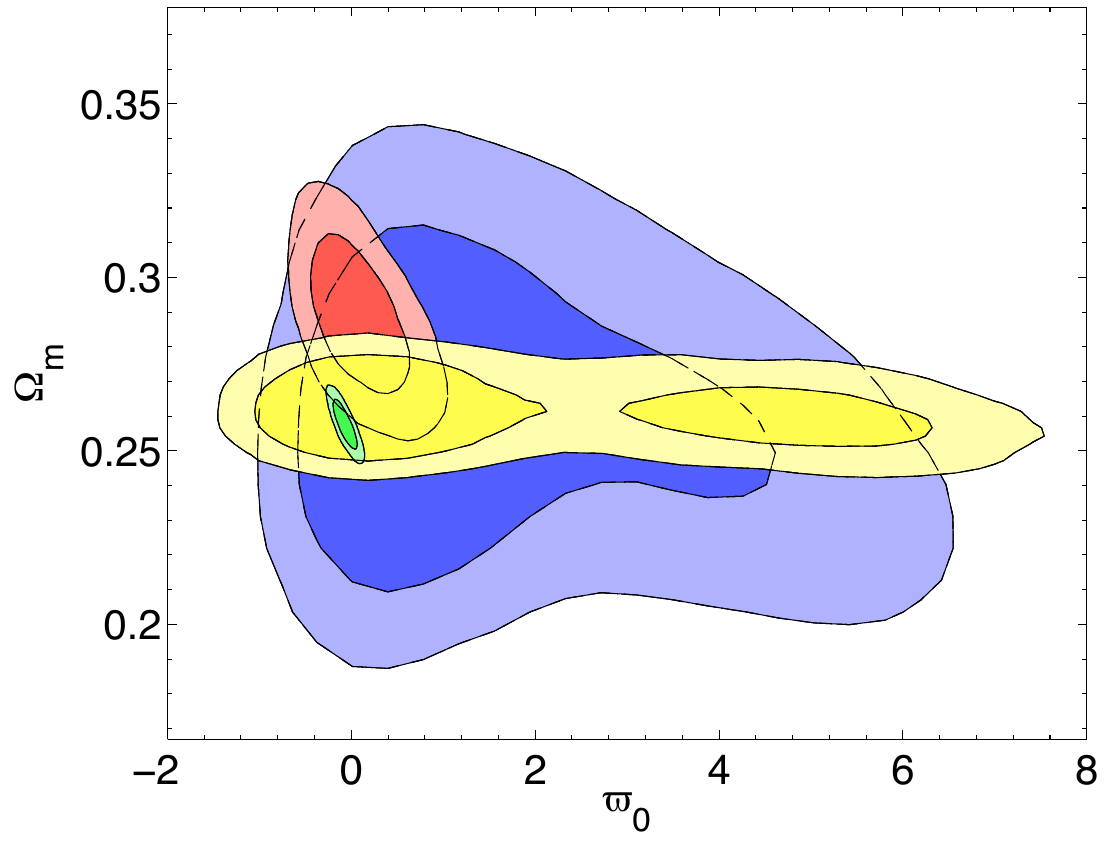}
\caption{Projected 68\% and 95\% likelihood contours of the matter 
fraction $\Om$ and gravitational slip $\varpi_0$ for WMAP 5-year data
(blue), adding current weak lensing and ISW data (red).  Yellow is
mock Planck CMB data, and green adds weak lensing from a 20,000 deg$^2$
survey. 
Figure from \citet{Daniel}, with permission. 
\label{fig:cp:Daniel}}
\end{figure}

Next we turn to constraints on the slip.  Considering a simplified model \citep{Daniel}, where $\varpi=\varpi_0(1+z)^{-3}$ is a specific function of scale factor only, the expected errors on $\varpi_0$, after marginalizing over other cosmological parameters, are shown in \autoref{fig:cp:Daniel}. We see that LSST could improve vastly on what is currently possible.

\begin{figure}
  \begin{center}
  \includegraphics[width=3.5in]{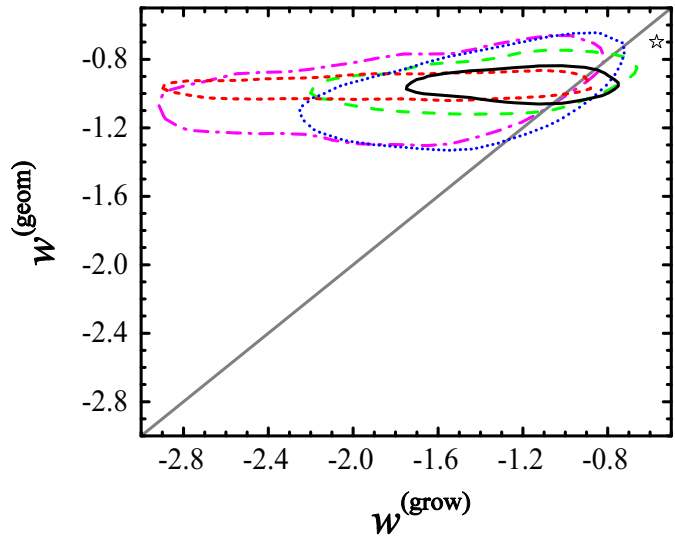}
  \caption{\label{fig:cp:QCDM}Joint constraints on $w({\rm geom})$ and
    $w({\rm grow})$ in a model in which the two EOS parameters are
    allowed to be different.  The contours show the $68\%$
    confidence limits from the marginalized distributions.  The thick
    gray line shows $w({\rm geom})=w({\rm grow})$.  Different contours
   and curves represent constraints from different combinations of the
    current data sets (CMB, SNe, galaxies and WL).  The smallest contour and the
    most narrow curve (black solid line) represent constraints from all
    the data.  No significant difference is found and deviations are
    constrained to
    $w({\rm geom})-w({\rm grow})=0.37^{+0.37+1.09}_{-0.36-0.53}$ ($68\%$
    and $95\%$ C.L.).  The star-shaped symbol corresponds to the
    effective $w({\rm geom})$ and $w({\rm grow})$, which approximately
    match the expansion history and the growth history, respectively, of
    a flat DGP model with our best-fit $\Omega_m$. Adapted from \citet{us07}.}
  \end{center}
\end{figure}

An alternative approach is to look for inconsistencies in the $w$
derived from the growth rate and that derived from the
distance-redshift relation.  Given a dark energy parameter, such as
the energy density $\Omega_\Lambda$ or equation of state $w$, we split
it into two parameters with one controlling geometrical distances, and
the other controlling the growth of structure.  Observational data are
then fitted without requiring the two parameters to be equal.
Recently, \citet{us07} applied this {\em parameter-splitting}
technique \citep{simpson05,zhs05,zhan09a} to the current data sets,
and \autoref{fig:cp:QCDM} shows the main result.  It reveals no
evidence of a discrepancy between the two split meta-parameters.  The
difference is consistent with zero at the $2\sigma$ level for the
quintessence(Q)-CDM model.  The existing data sets already pose tight
constraints on $w$ derived from geometry.  However, the constraint is
much weaker for $w$ derived from growth, because currently galaxy data
are limited by the uncertainty of the bias factor and WL data are
restricted by both the width and depth of the survey.  LSST will open
these windows dramatically.

The parameter-splitting technique can also check for 
{\it internal} inconsistency within any {\it one} data set, such as
shear tomography, that is sensitive to both geometry and growth.
It can be thought of as a crude way to parametrize
the space of these theories.  As such, the constraints can be viewed
as putting restrictions on modified gravity theories, with the caveat
that the precise constraints on any particular theory must be worked
out on a case by case basis.  

\begin{figure}
\centering
\includegraphics[width=3in]{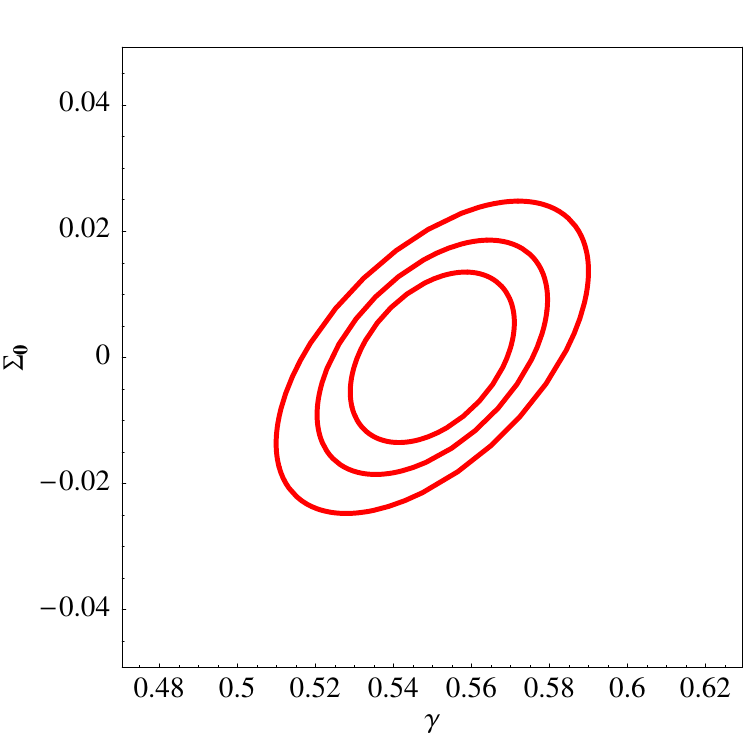}
\caption{Projected 68\% likelihood contours of $\Sigma$, the parameter
describing the effective modification to the lensing potential, and 
the growth index $\gamma$
for weak lensing surveys from a full sky survey with median $z=0.9$, 
and surface densities of sources of 35, 50, and 75 galaxies per 
arcminute.  LSST is likely to achieve a surface density near the 
bottom of this range.  Errors scale inversely with the square root 
of the sky fraction. Figure from \citet{Amendola}, with permission.
\label{fig:cp:Sigma0Gamma}}
\end{figure}

Finally, we show in \autoref{fig:cp:Sigma0Gamma} how accurately LSST
could measure evolution of $\Sigma$, which describes the modification
to the lensing potential (\autoref{eq:modified_poisson}). Assuming
$\Sigma=1+\Sigma_0 a$, $\Sigma_0$ may be determined to an accuracy of
a few hundredths \citep{Amendola}.  One caveat on all of these
conclusions is that WL requires knowledge of the nonlinear regime of
galaxy clustering, and this is reasonably well-understood for GR, but
for other models, further theoretical work is required. This has
already started \citep{Schmidt2008}. 




%
%
%
%
%
%
%
%
%
%
%
%
%
%
%
%
%
%
%
%
%
%
%
%
%
%



\section{Anisotropic Dark Energy and Other Large-scale Measurements}
\label{sec:cp:newphys}
{\it Anthony Tyson, Hu Zhan}

By providing measurements of WL shear, BAO, and other observables in
different directions on the sky covering $\sim 100^\circ$ scales, LSST
will address specific questions related to clustering on the largest
scales. These range from the clustering of dark energy to exotic
models that require horizon scale tests (see also 
\autoref{sec:lss:lscale} and \autoref{sec:lss:isw}). 
Because of its wide and deep
coverage, the 20,000 deg$^2$ LSST survey of billions of galaxies has
the power to test isotropy to percent level precision.


There is compelling evidence that the mean expansion of the Universe
is accelerating. At this time there are no plausible theoretical
models for the dark energy.  We are far from understanding the
nature of this phenomenon.  In some respects this is similar to the
earliest days of the CMB observations. We should, therefore, examine the
consequences of an anisotropic dark energy in cosmological models and
estimate their observability.

While there are even fewer plausible theories of anisotropic dark
energy, there are several logical possibilities that can be checked
through direct observation. It is in principle possible to have
anisotropy in an otherwise homogeneous Universe described by a
cosmological constant. If dark energy is something other than a
cosmological constant, it will in general have anisotropic stresses at
some level. This is also a generic prediction of modified gravity
theories.  Because covariance implies that a time-varying field is
equivalently spatially-varying, dynamical dark energy is necessarily
inhomogeneous. Inhomogeneities in the surrounding radiation and matter
fields can drive fluctuations in dynamical dark energy. Spatial
variations in the expansion rate should accompany fluctuations in dark
energy. Distortions of the expansion rate and luminosity distance may
also arise if the observed cosmic acceleration is due to gravitational
effects in a strongly inhomogeneous Universe.

In some models the small Jeans scale of the effective dark energy
forms small wavelength perturbations which can be
probed via weak lensing \citep{Amendola}. In
general, it may be possible to distinguish between cosmological
constant, dynamical dark energy, and modified gravity. 
\citet{jimenez07} studied the consequences of a homogeneous
dark energy fluid having a non-vanishing velocity with respect to the
matter and radiation large-scale rest frames. They found that in
scaling models, the contributions to the quadrupole can be
non-negligible for a wide range of initial conditions. Anisotropies
have been considered as potentially observable consequences of vector
theories of dark energy \citep{armendariz-picon04}.

\subsection{Possible Relation to CMB Large Scale Anisotropy}

The CMB exhibits less power on large scales than predicted \citep{spergel03}.  Perhaps
this is a statistical fluke, due to cosmic variance in an ensemble of
possible universes.  However, we should also explore alternative
explanations. Perturbations whose wavelengths enter the horizon
concurrent with dark energy domination distort the CMB during the late
time acceleration. Is the cosmological principle more fundamental than general
relativity? It is worthwhile to observationally study this assumption
of perfect homogeneity on large scales. We know that large
imhomogeneities exist on smaller scales. Large scale CMB anisotropies
develop for off-center observers in a spherically symmetric
inhomogeneous universe \citep{alnes06b,enqvist07}.
\citet{koivisto08} investigated
cosmologies where the accelerated expansion of the Universe is driven
by a field with an anisotropic pressure. In the case of an anisotropic
cosmological constant, they find that in the current data the tightest
bounds are set by the CMB quadrupole. 

\subsection{Matter Inhomogeneities}

One might ask if we can tell the difference between anisotropic
matter cosmic variance and dark energy cosmic variance. Depending on the
redshift, that could be settled by looking at the galaxy number count
quadrupole, etc. over a cosmological volume. What effects might be
present in LSST data if there were an anisotropic distribution of
matter on Gpc scales \citep[viz.][]{caldwell08}?  Our current
framework for cosmology will be violated if the anisotropy is at a
level well above the cosmic variance. If the cosmological principle
does not hold, we cannot assume that physics here applies to other
places in the Universe. Hence, there is a logical inconsistency in
predicting observables for a truly anisotropic Universe (i.e., more
anisotropic than allowed by cosmic variance) based on the Friedman-Robertson-Walker (FRW)
model. Anisotropic matter is in some sense worse than anisotropic dark
energy, only because we know more about matter fluctuations and hence
their cosmic variance. Our ignorance about the appropriate cosmic
variance for dark energy gives us some room for anisotropy.

\subsection{Observations Enabled by LSST}

\begin{figure}
\centering
\includegraphics[width=3in]{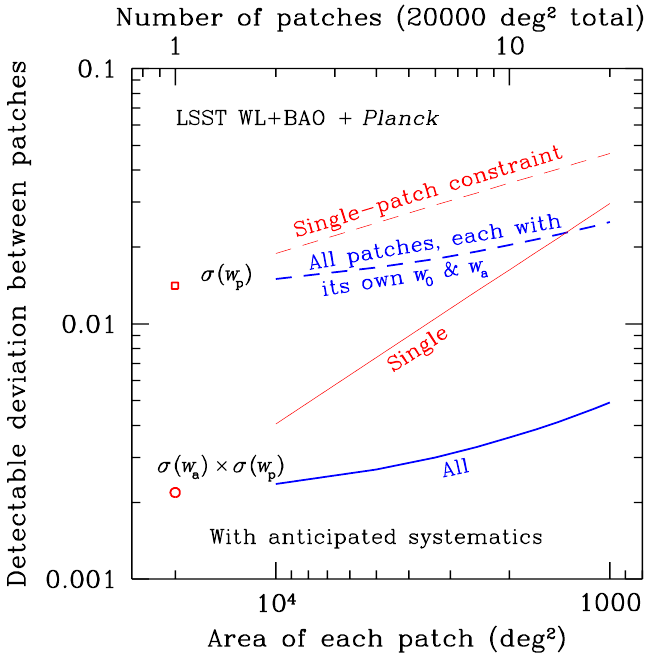}
  \caption{Detectable deviation between LSST measurements of dark
    energy parameter, $w_p$, and error product as a function of the
    number of patches. The constraints are derived from the joint
BAO and WL (galaxy--galaxy, galaxy--shear, and shear--shear power 
spectra) analysis in \autoref{sec:cp:wlbao}. 
Note that there is little degradation in
    sensitivity as one goes from quadrupole to higher moments if the
    other marginalized cosmological parameters are shared: a 16 patch
    survey degrades the error product per patch by less than a factor
    of two.  The red curves labeled ``single'' are for independent
    determination of all parameters for each patch. Estimated LSST
    systematics are included.}
\label{fig:cp:warea}
\end{figure}

Anisotropic dark energy can be probed via distance and
growth measurements over the sky with all the dark energy probes.
LSST is particularly suited for testing the anisotropy of dark energy. 
Its wide survey area enables one to measure dark energy properties in
many patches and to potentially detect their variations across
the sky; its deep imaging not only results in more usable 
galaxies for more accurate measurements of the distances and 
growth function, but also allows one to probe the differences in the 
evolution of dark energy properties across the patches.

The results of dark energy anisotropy tests using the joint LSST
BAO (galaxy power spectra) and WL (shear power spectra) analysis 
are shown in \autoref{fig:cp:warea}. See \autoref{sec:cp:wlbao} 
for  details of the calculation such as the anticipated 
systematics and priors.
The only difference from \autoref{sec:cp:wlbao} is that we let
each patch of sky have its own $w_0$ and $\wa$ in this 
section. The rest of the parameters are assumed not to vary from 
patch to patch and, thus, are shared among the patches.

In \autoref{fig:cp:warea} we find only a small degradation in 
sensitivity to the dark energy EOS parameters that are allowed to 
vary independently in up to $\sim 10$ different patches of the LSST
survey area, if the other marginalized
cosmological parameters are shared between patches. Each patch 
achieves nearly the same precision of measurement of the
dark energy parameters as the full 20,000 deg$^2$ survey.  
For comparison, the
red curves labeled ``single'' are for independent determination of all
parameters for each patch. The single patch $\sigma(w_p)$ line (and
the single patch error product) is what one would get from just doing 
the LSST survey over that smaller patch area.

In the single patch case, the Fisher matrix (see 
\autoref{sec:app:stats:numtech:fisher}) is proportional to the area
of the survey\footnote{CMB observations are reduced to the same 
area.}, so that the estimated error on a single parameter scales
as square root of the number of patches, $N_\mathrm{pat}$, and the 
error product $w_p\times\wa \propto N_\mathrm{pat}$.
Because $w_0$ and $w_p$ consist only a small subset of many 
cosmological and nuisance parameters that BAO and WL can constrain, 
increasing the degrees of freedom of the dark energy EOS to 
$2N_\mathrm{pat}$ does not inflate individual parameter errors by 
a factor of $N_\mathrm{pat}^{1/2}$. Therefore, the ``all'' results 
in \autoref{fig:cp:warea} degrades more slowly than the ``single'' results. 

Adding SNeIa would have little impact in \autoref{fig:cp:warea}, 
just as they do over the whole 
sky (see \autoref{fig:cp:cswb}). A SN-only survey will 
behave very much like the single patches in \autoref{fig:cp:warea},
because the only shared parameter among
all patches on which SNeIa have some marginal constraint is curvature;
in particular, the evolution parameters are not constrained.  
In other words, the gain of the ``all'' results over 
the ``single'' results is due to sharing these parameters across the
sky.

A separate investigation using the angular power spectrum of SNeIa
luminosity by \citet{Cooray++08} finds that an LSST-like survey of 
one million SNeIa at $z \lessim 1$ can limit rms dark energy 
fluctuations at the horizon scale down to a fractional energy 
density of $\sim 10^{-4}$. This limit is much higher than the 
$1.92\times10^{-5}$ horizon-scale matter fluctuations measured from COBE \citep{bunn97},
but as we have demonstrated above with the dark energy EOS, one can 
combine more LSST probes as well as external data sets to improve
the constraints on dark energy density fluctuations.




%
%
%
%
%
%
%
%
%
%
%
%
%
%
%
%
%
%
%
%
%
%
%
%
%
%
\section{Cosmological Simulations}
\label{sec:cp:cos_sim}

{\em Salman Habib, Risa H. Wechsler}

An enormous amount of science can be extracted from a synoptic survey
as deep, wide, and rich as LSST. In order to maximize the quantity and
quality of the science, the underlying theory must be developed -- as
far as possible -- to a point where analysis and interpretation of the
data are not systematics-limited. In contrast to the analysis of the
CMB temperature anisotropy, where, for the most part, only linear
physics need be considered, predictions for LSST observations and
analysis of the data will overwhelmingly involve nonlinear aspects of
structure and galaxy formation. The nonlinear aspects include gravity,
hydrodynamics, and a variety of astrophysical feedback mechanisms.
Cosmological numerical simulations thus play a critical role in combining
these into a precision theoretical framework.  Meeting the demands of
LSST requires a numerical simulation campaign at the cutting edge of
present and future computational astrophysics and cosmology.

\subsection{The Simulation Context}
\label{sim_context}

In the context of a deep, wide survey such as LSST, which is exploring
new regimes in parameter space, there are four general uses for
cosmological simulations.  The first is an accurate calculation of the
underlying dark matter framework for a given cosmological model.
Because dark matter outweighs baryons by roughly a factor of five,
gravity-only N-body simulations provide the bedrock on which all other
techniques rest.  The second use is a detailed prediction of the
observed galaxy population and other observables in the survey,
including all relevant baryonic effects.  Third, simulation based
mock-catalogs provide an essential tool for the planning of the LSST
survey strategy and for testing its technical aspects.  The final
critical use of simulations is to provide a framework for interpreting
the LSST data in the context of models, both cosmological models and
models of baryonic physics.

These different uses require a wide range of input physics into the
simulation.  Ideally, we would like to be able to simulate the full
structure formation problem, including galaxy and star formation, in a
box the size of the LSST survey volume.  However, it will not be
possible now, or in the foreseeable future, to develop an accurate
first principles simulation capability that can address the needs of
all of the observational probes enabled by LSST; the dynamic range is
too vast and the underlying physics too complex.  On the largest
scales where gravity from dark matter dominates, gravity-only
simulations are sufficient for predictions of structure.  However, on
scales smaller than a few Mpc, complex baryonic physics also enters
requiring the modeling of (magneto-)hydrodynamic, thermal, chemical,
and radiative processes.  The computational challenges are immense,
even in the dark matter only case, and become prohibitive when solving
the full baryonic problem.

To overcome the computational hurdle, several strategies exist.  To
begin with, we will need to produce a large, LSST-volume simulation of
the dark matter distribution at modest resolution with as many
simulation particles as possible.  This will provide the backbone of
the simulation analysis.  What is required is a collection of robust
phenomenological approaches for combining the observations with
simulation results, appropriately tuned for each of the observational
probes. We will refer to these approaches collectively as examples of
``self-calibration."  One technique is to incorporate results from
hydrodynamic simulations done over small yet representative volumes by
developing semi-analytic models that can be used in large-volume, gravity-only simulations. Statistical methods can also be used to add
galaxies to the N-body simulation.  These should reproduce the
observed distribution of galaxies with as much fidelity as possible
based on existing data, including upcoming stage III experiments such
as DES and PanSTARRS as well as the results from deeper smaller volume
surveys from the ground and from space.

A more computationally intensive approach is the use of large-scale
hydrodynamical simulations with sub-grid models developed using
smaller runs. As simulation capabilities continue their explosive
growth (supercomputer performance has increased by a factor of 10,000
over the last 15 years), we can expect a major improvement in the
sophistication of self-calibration techniques over what is available
now. Indeed, many of the phenomenological modeling ideas currently in
use will be obsolete by the time the LSST data stream is in full
flow. The LSST simulation campaign must have the flexibility to change
as this situation evolves.

Finally, LSST is developing a detailed image simulation pipeline,
described in \autoref{sec:design:imsim}, which uses the properties of
galaxies in cosmological simulations as a key input.  These
simulations are essential for developing the data reduction,
management, and analysis pipelines and also provide critical input to
the various science teams in planning and running their analysis.

It is also important to keep in mind that, because LSST will exist on
the same timescale as other important large scale structure probes
(JDEM, eROSITA, SKA), there will be substantial overlap in the
simulation requirements. The power of combining various data sets will
significantly lower the systematics floor and impose additional
demands on simulations. For example, the simple dark energy
parametrizations currently in use may have to be abandoned in favor
of a non-parametric approach embedded within the simulation
framework. Additionally, an advanced framework of simulations will be
necessary in guiding how we combine the multiple data sets from
different surveys.  This will become easier as computer power grows
enabling ``end-to-end'' style modeling.

\subsection{LSST Simulations I: Main Requirements}
\label{sim_lsst_req}
The large-scale distribution of matter (on scales greater than several
Mpc) is determined primarily by gravity.  Therefore, the first set of
simulation desiderata are determined by what is required of N-body
simulations. For LSST, the largest scales of interest cover studies of
Baryon Acoustic Oscillations (BAO), which occur on scales of 150
$h^{-1}$Mpc.  LSST BAO studies will require 
$0.1\%$ accuracy for the matter power spectrum to scales $k\sim
0.3$~$h$Mpc$^{-1}$ over a range of redshifts $0.1<z<5$ (for a recent
discussion of BAO simulation errors, see \citealt{Nis08}).  In
the high redshift range, perturbation theory may suffice, but for
$z<1$, perturbative results rapidly lose accuracy in ways that are
difficult to predict \citep{Car++09}. In any case, the mock
catalogs required to develop BAO analysis as well as to understand the
dependence on galaxy properties can only be carried out with
simulations.

Additionally, the cluster-scale halo mass distribution and its
dependence on cosmology can be calibrated with N-body simulations. 
It is already known to be non-universal in a way that can influence
precision cosmological analyses \citep{Ree++07g, Luk++07, Tin++08,
  Luk++09, Bha++09}.
Because significant constraints come from the high end of the mass
function, where the number density drops exponentially, simulation of
large volumes are also needed to more accurately characterize the mass
function in this range.

In general, the N-body simulation task for galaxy clustering and
cluster counts is targeted at precision studies of the dark matter
halo and subhalo distribution, construction of sufficiently
fine-grained merger trees for improved empirical and semi-analytic
modeling of the galaxy distribution, and running a large number of
simulations to understand errors and their covariances.  The
fundamental science requirements are: 1) sufficient mass resolution
to resolve dark matter halos and subhalos hosting the target galaxies,
2) sufficient volume to control sample variance uncertainties, and
3) enough information to model galaxy bias. For LSST, this
translates into simulations with multi-Gpc box-sizes and particle
counts in the $10^{11-12}$ range (for mass resolutions of order
$10^9$~$M_{\odot}$), all with $\sim$kpc force resolution. These
simulations will probe the scales of 
superclusters, voids, and large-scale velocity fields, and are
well-matched to petascale computing resources 
\citep{Hab++09}.

Full-up simulations of this kind cannot be performed with petascale
resources but require going to the exascale: resolving subhalos in a
Hubble volume demands $10^{13-14}$ particles, in turn requiring
$10^{24}$ flop of computation. By 2015, assuming current trends hold,
computing performance will be at the 100~Petaflop level. Therefore,
provided that N-body code development keeps pace with architecture
developments, simulations at the scale demanded by LSST will be
available by the time the data makes its appearance.

While such simulations will provide a significant amount of power for
understanding the LSST data, they need additional refinement.  In
particular, we will need targeted simulations which investigate higher
resolution to calibrate the impact of the smaller scale clustering
and simulations including gas physics to understand the impact of
baryons on the dark matter distribution.  Finally, we will need models
for putting LSST observables into the N-body simulations.  These
models can take the form of direct hydrodynamical simulations of small
scales, semi-analytic modeling of high-resolution merger trees that
incorporate sub-grid physics calibrated to the higher resolution
simulations, and empirically constrained statistical models for
incorporating the observational galaxy distribution into the
simulation.

Simulations with smaller box sizes (linear scales of Gpc and less) are
needed to deal with further challenges posed by LSST weak lensing
observations, requiring absolute error control at the
sub-percent/percent level up to scales of $k\sim 10$~$h$Mpc$^{-1}$
\citep{H+T05}. Currently, the best demonstrated N-body
error control is at the $1\%$ level out to $k\sim 1$~$h$Mpc$^{-1}$
\citep{Hei++08b}, and this can be extended and improved to $k\sim
10$~$h$Mpc$^{-1}$ with a petascale computing campaign. 

Recent studies have shown that the evolution of the baryonic component
of the Universe imprints itself even on the largest scales, including
a roughly 5\% impact on the cluster mass function and the BAO peak
\citep{SRE09}. 
This is also true on larger scales
than may have been expected for the weak lensing power spectrum;
baryonic effects become important beyond scales of $k\sim
1$~$h$Mpc$^{-1}$ \citep{Whi04b, Z+K04, Jin++06f, RZK08}.  This creates a systematic error present in any
attempt to extract cosmological parameters calibrated using N-body
only simulations.  Unfortunately, the computational power necessary
for even a rough treatment of the baryons on the LSST scales is not
present.  It will, therefore, be necessary to complement our above large
volume simulations with a series of much smaller, higher-resolution
simulations that include accurate treatment of the gas physics.
Although a full treatment will require AMR and SPH simulations with
absolute error controls and physics treatments beyond the current
state of the art, there are encouraging signs that self-calibration
approaches can be used successfully.  In particular, \citet[][see
  also \autoref{sec:wl:baryons}]{ZRH08} 
showed that the impact of baryonic physics on the weak
lensing power spectrum can be measured from hydrodynamic simulations and
added as a correction to dark matter only simulations with high
accuracy.

Similar self-calibration approaches can be applied to modeling cluster
counts, which for LSST will be in the hundreds of thousands. The
critical requirement for simulations is to properly model the form of
various mass--observable relations; the parameters specifying this form
can them be self-calibrated with the data itself (e.g., \citealt{Roz09}).  
Still, it is essential to determine the sensitivity of these
relations to baryonic treatment and the full range cosmological
parameter space.  Although this is a very large computational task,
petascale capabilities are sufficient to accomplish it within the next
few years.

The associated simulations are an order of magnitude more expensive
than the underlying pure N-body runs, and extensive numerical
exploration is required to understand the effects of parametric
variations in sub-grid models. Additionally, it must be kept in mind
that modeling of various sources of systematic error, such as
intrinsic galaxy alignments (\autoref{weaklens-intrinsic}) and the
photometric redshift distribution, is actually 
a bigger concern for weak lensing measurements, and will require
substantial observational input.

Finally, resolving the smallest length scales using a series of nested
boxes will be necessary for investigating the physics and dark matter
and baryonic structure of individual galaxies and galaxy
clusters. These simulations will be useful for providing an additional
source of sub-grid inputs into the large-scale codes and will also be
essential for weak and strong lensing studies of cluster masses
especially the influence of substructure. Detailed simulations for
investigations of the properties of dark matter-dominated dwarf
satellite galaxies of the Milky Way are needed, incorporating
modifications of CDM (e.g., warm and interacting dark
matter). Detecting these galaxies is a prime target for LSST 
as described in Chapters \ref{chp:mw} and \ref{chp:galaxies}.

Given the computational challenges of hydrodynamical simulations, as
well as the fact that they still involve significant unresolved
sub-grid physics that impacts the observations, it is essential to
develop in parallel empirical and semi-analytic models for connecting
the well-modeled dark matter distribution with the observed galaxy
distribution.  Current approaches range from those which require
resolved substructure \citep{CWK06}, to halo occupation
approaches \citep{B+W02} that only require resolving
central halos, to algorithms which are designed specifically for
modeling larger volumes with poor resolution (ADDGALS; \citealt{Wec++09}). 
Generically, these models use available data to constrain a
statistical relation between the observed galaxy and properties of the
dark matter distributions.  Such models let us build mock galaxy
catalogs that are designed to reproduce particular observed galaxy
properties, such as clustering, bias, and mass-richness relations.
These models can be run on top of the lightcone outputs of N-body
simulations to create mock catalogs that will help us bridge this gap
between dark matter simulations and observable properties.  Indeed,
with the level of precision that modern cosmology has reached,
understanding of the dark matter-observable relation will likely be a
dominant source of error when extracting cosmological information from
LSST data.

The requirements of such mock catalogs for LSST are quite stringent.
In order to test the full analysis pipeline, these catalogs should
produce galaxies with realistic photometric and morphological
properties down to $r=28$ and cover as much volume as possible, and
must be produced for a range of cosmological models (described more
fully in the following section). Additionally, mocks should include the
correlation between galaxy properties and, for example, AGN properties and
supernovae, and should include the full range of source light curves is
needed to model the types of variability that will be seen by LSST.

\subsection{LSST Simulations II: New Directions}
\label{sim_lsst_new}
Aside from investigating the role of dark energy as a background
phenomenon, LSST observations can put constraints on dark energy
anisotropy and clustering, enable tests of gravity on cosmological and
astrophysical scales, investigate the primordial density fluctuation
power spectrum including the existence of features and running of the
spectral index, and study primordial non-Gaussianities
(\autoref{sec:lss:nongauss}). A suite of 
simulations must be developed to address all of these questions. Most
of the simulation capability would be based around that discussed
above, but several new directions will be explored.

Fully self-consistent simulations for dynamical dark energy models,
with initial conditions set by transfer functions incorporating dark
energy fluctuations, will be necessary to make realistic predictions
for LSST observations. Because dark energy clustering occurs on very
large scales, the basic simulation requirements are not too different
from those for galaxy clustering. The added wrinkle will be the need
for a PDE-solver for the quintessence field, but the dynamic range for
this is limited and will not be a significant overhead.

Investigations of the primordial power spectrum essentially involve
running the standard simulations but with a modified initial
condition. Although large scales may be in the linear or quasi-linear
regime, simulations are important in understanding systematics issues,
such as the role of scale-dependent bias and modeling of galaxy
evolution. Explorations outside the simple assumptions underlying
current approaches (adiabatic Gaussian fluctuations, power-law
primordial spectra) are essential to establish the robustness of
inferences made from the observations, especially since -- the
inflationary paradigm notwithstanding -- there is no firmly
established theory of initial conditions. An example of this sort of
exploration is scale-dependent halo bias induced by non-Gaussianity
\citep[][see also \autoref{sec:lss:nongauss}]{Dal++08c, M+V08} in the initial
conditions. Primordial non-Gaussianity has been traditionally
parametrized by the parameter, $f_{NL}$, which LSST can constrain
extremely tightly (to $f_{NL}\sim 1$, \citealt{CVM08}), however,
this is only a particular case, and other aspects of non-Gaussianity
should be investigated (for example, by using the Edgeworth expansion
to set up initial conditions).

Simulations aiming to study the effects of modified gravity are based
either on the use of specific models or on the so-called parametrized
post-Friedmann approach, which uses one post-Newtonian parameter,
$\gamma$, to signify departures from general relativity (e.g.,
\citealt{B+Z08}). Simulations such as \citet{S+J06},
\citet{L+B08}, \citet{Oya08}, and others that include
long-range dark matter interactions \citep{NGP05}, will be
useful foils for the main line of the numerical effort, important to
clarify the precise nature of LSST observational results (in terms
of acceptance or rejection of alternative models).

\subsection{Calibration Frameworks}
\label{sim_cal}

The simulation requirements for next-generation cosmological surveys
such as LSST are very demanding and cover not only scanning over
cosmological and physical modeling parameters, but in the case of
end-to-end modeling, a slew of observational and instrumental
variables. The final number of parameters can range from tens to
thousands, depending on the particular application. While
post-processing results from expensive simulations can sometimes be
used to incorporate more parameters, this approach is far from being
universal. The basic difficulty that must be faced is that the number
of complex simulations that can be performed will always be finite --
based on this finite set of results, one has to effectively
interpolate across a high-dimensional space of parameters.

The most direct way to approach this problem is through a brute-force
comparison of simulated output with the most recent observational
data.  A number of upcoming surveys are expected to be in the final
stages of taking data by the time LSST comes online in 2015.  In
particular, the Dark Energy Survey (DES) and Pan-STARRS are expected
to map out thousands of square degrees down to 24th magnitude, roughly
the depth of a single LSST exposure.  

Incorporating observational information such as color distributions
and clustering from these surveys into, for example, galaxy halo occupation distribution (HOD) modeling, as
well as comparing with predictions from hydro simulations and
semi-analytic modeling, will be necessary to begin immediate analysis
of the LSST data as it becomes available.  In particular, this will
allow us to create a stronger understanding of our mass-observable
relations, and, with the help of a realistic transfer function for
creating mock sky images, can help in getting an initial handle on
systematics.  However, these comparisons will only help to understand
how to accurately represent the galaxy distribution of a given
(simulated) cosmology and set of model parameters, making it very
difficult to scan across a wide range of cosmological models, not all
of which will have been directly simulated.

Recent advances in applications of Bayesian and other statistical
techniques to modeling simulation results has resulted in the
development of the cosmic calibration framework \citep{Hei++06, 
Hab++07, Sch++08}, an approach targeted
precisely to the problem identified above. Given a smooth enough
response surface in the high-dimensional parameter space, this
methodology has been shown to give excellent results, in particular,
percent level predictions for the nonlinear matter power spectrum
\citep{Hei++09}. The procedure consists of four interlocking
steps: 1) defining the simulation design, which determines at what
parameter settings to generate the training sets, 2) generation of
the emulator -- using PCA-based Gaussian process models -- which
replaces the simulator as a predictor of results away from the points
that were used to generate the training set, 3) an uncertainty and
sensitivity analysis associated with the emulator, and 4) the
(self-) calibration against data via MCMC methods to determine
parameter constraints. The calibration methodology can be adapted to
multiple tasks within the LSST science campaign and, significantly,
can be used similarly for cosmological parameters, the parameters
specifying the empirical or semi-analytic models, as well as
uncertainties in the instrumental response.  In each case unknown
parameters can be determined in the final MCMC calibration against
observational results.

\bibliographystyle{SciBook}
\bibliography{cosmological_physics/cosmophys}



\appendix
\setboolean{appendix}{true}

\renewcommand\thechapter{\Alph{chapter}}

\renewcommand{\chaptermark}[1]{\markboth{Appendix \thechapter: #1}{}}
\renewcommand{\sectionmark}[1]{\markright{\thesection\ #1}{}}


\chapter[Assumed Cosmology]{Assumed Cosmology}
\label{sec:com:cos}

{\it Hu Zhan}

One of the most important scientific goals of the LSST is to refine
and rigorously test our current ``standard model'' of cosmology.  In
predictions of LSST's performance, however, we must agree on a
fiducial cosmology, which we describe here.  This book will describe
our predictions for LSST's ability to tighten our constraints on these
parameters, test for consistency among a variety of cosmological
probes, and test some of the basic assumptions of the model, from the
Cosmological Principle, to the clustering and isotropy of dark
energy. 

Our fiducial model is a cold dark matter (CDM) universe with a large
fraction of its energy density in the form of dark energy that has an
equation of state $w=p/\rho$ (\wcdm).  This model is characterized by
the 11 parameters listed in \autoref{tab:com:cosp}, which are taken
from the WMAP five-year data analysis \citep{dunkley08}. 
We use the WMAP-only results to avoid dealing with the complex 
correlations between LSST probes and other probes incorporated in 
\citet{dunkley08}. Slight changes to the fiducial model do not 
affect our assessment of the LSST performance.
Since the cosmic microwave background (CMB) alone cannot constrain 
all 11 parameters, we center the fiducial model on the 
concordance \lcdm{} model (i.e., $\w0=-1$, $\wa=0$, and $\Ok=0$) and 
allow all the 11 parameters to float in the forecasts. 

We adopt a phenomenological parametrization for the dark energy
equation of state used by the report of the Dark Energy Task Force
\citep{albrecht06}: $w(a) = \w0 + \wa (1-a)$, where $a$ is the
expansion factor.  The rest of the parameters are chosen to be
convenient for techniques such as baryon acoustic oscillations (BAO)
and weak lensing and for combining LSST constraints with 
CMB results. For example, the lensing potential
scales with the physical matter density $\om$, not by the matter
fraction $\Om$ alone ($\om=\Om h^2$ and $h$ is the reduced Hubble
constant).  Likewise, the BAO features are determined by $\om$ and the
physical baryon density $\ob=\Ob h^2$, where $\Ob$ is the baryon
fraction.

\begin{table}  
\begin{center}
\caption{Cosmological parameters from WMAP five-year 
results$^\dagger$ \label{tab:com:cosp}}
\begin{tabular}{@{}lll}
\hline Symbol & Value & Remarks \\ \hline

$w_0$ & $-1$ & dark energy equation of state at $z=0$ \\

$w_a$ & 0 & rate-of-change of the dark energy EOS as in 
$w(a) = w_0 + w_a (1-a)$ \\

$\omega_{\rm m}$ & 0.133 & physical matter density  
  $\omega_{\rm m} = \Omega_{\rm m} h^2$, $\Omega_{\rm m}=0.258$ \\

$\omega_{\rm b}$ & 0.0227 & physical baryon density 
  $\omega_{\rm b} = \Omega_{\rm b} h^2$, $\Omega_{\rm b}=0.0441$ \\

$\theta_{s}$ & 0.596$^\circ$ & angular size of the sound horizon at 
  the last scattering surface \\

$\Omega_{\rm k}$ & 0 &  curvature parameter \\

$\tau$ & 0.087 & optical depth to scattering by electrons in the 
  reionized \\  & & intergalactic medium \\

$Y_{p}$ & 0.24 & primordial helium mass fraction \\

$n_{s}$ & 0.963 & spectral index of the primordial 
scalar perturbation power spectrum\\

$\alpha_{s}$ & 0 & running of the primordial 
scalar perturbation power spectrum\\

$\Delta_R^2$ & $2.13 \times 10^{-9}$ & normalization of the 
primordial curvature power spectrum at \\
& & $k^*=0.05\,\mbox{Mpc}^{-1}$ ($\sigma_8=0.796$ or 
$\Delta_R^2 = 2.41 \times 10^{-9}$ at $k^*=0.002\,\mbox{Mpc}^{-1}$ )\\
\hline
\end{tabular}
\end{center}
{\small
$^\dagger$ The reduced Hubble constant $h = 0.719$ and 
the present equivalent matter fraction of dark energy 
$\Omega_{X} = 0.742$ are implicit in this parametrization, 
meaning that either one of them can replace $\theta_{s}$ or
any parameter that affects $\theta_{s}$.\\
}
\end{table}

In addition to \autoref{tab:com:cosp}, we also make standard 
assumptions about other parameters and processes, e.g., adiabatic
initial condition, standard recombination history, three effective 
number of neutrino species, etc. We fix the neutrino mass to zero in 
all but \autoref{sec:cp:mnu} where we estimate the upper limit that 
can be placed by LSST shear and galaxy clustering data. 
The actual values of the neutrino masses have little impact on most 
forecasts, as long as they are held fixed. 

\bibliographystyle{SciBook}
\bibliography{appendix/assumed_cosmology}

%
%
%
%
%
%
%
%
%
%
%
%
%
%
%
%
%
%
%
%
%
%
%
%
%

\chapter[Analysis methods]{Analysis Methods}
\label{chp:analysis}
{\it Phil Marshall, Licia Verde, Hu Zhan}


This chapter describes the statistical analysis methods that are 
used in previous chapters either to forecast LSST performance
or as suggested to analyze LSST data. We start with an introductory 
review before moving on to some practical examples.


\section{Basic Parameter Estimation}
\label{sec:app:stats:basics}

Very readable introductions to probabilistic data analysis are given
by~\citet{Sivia}, \citet{MacKay} and~\citet{Jaynes}; an introduction 
to the basics is given in this section.  
A single piece of experimental data
is often presented in the form
$x=x_0\pm\sigma_0$, with $x_0$ being the result of the measurement and 
$\sigma_0$ the estimate of its uncertainty. This is shorthand for
something like the
statement ``I believe that the quantity I am trying to measure, $x$, 
is most likely from my experiments to be
$x_0$, but I could also believe that it was actually less than
or greater than this by roughly $\sigma_0$.'' 
That is, the relation $x=x_0\pm\sigma_0$
is a compressed version of the probability distribution (or probability
density function, PDF) $\pr(x_0|x,H)$,
to be read as the probability
of getting $x_0$ given assumptions about $x$ and~$H$. When written as a
function of the model parameters, this PDF is referred to 
as the likelihood. 
Since our observed data come in probability distribution form, any
conclusions we draw from them will necessarily be probabilistic in
nature as well. 

Traditionally there are two interpretations of probability: ``Frequentist'' 
and ``Bayesian.'' For frequentists, probabilities are just frequencies of
occurrence: ${\cal P}=n/N$ where $n$ denotes the number of successes and $N$
the total number of trials. Probability is then defined as the limit for the
number of independent trials going to infinity. In the example above if one
were to repeat the experiment an infinite number of times, then $x$ will fall
in the range $[x_0-\sigma_0, x_0+\sigma_0]$, say, 68\% of the time. 
Bayesians instead interpret
probability as a degree of belief in a hypothesis -- a quantified version of
the original statement above.
%
%

In cosmology, statistical analysis tends to be carried out in the Bayesian
framework. It is easy to understand why: cosmic variance makes cosmologists
only too aware of the limited information available to them. Only one
realization of the CMB anisotropy and the large scale structure is  accessible
to our telescopes, and so while this is not a technical barrier to our happily
simulating large numbers of fictitious universes in order  to either compute
or interpret our uncertainties, it is perhaps something of a psychological
one, promoting the acceptance of the  Bayesian notion of probability.

Bayesian cosmologists, seeking a steady point for their lever, assume that
the observable universe is just one particular realization of a true
underlying stochastic  model of the Universe: the cosmological  parameters of
this model can be inferred from this one realization via the rules of
probability.  Only if we could average all possible (unobservable)
realizations of the underlying model could we recover the true values of the
parameters with certainty -- but since we can only observe one of the infinite
possible realizations of it, we have to settle for probability distributions
for the parameters of the underlying model instead.

This mental approach has the distinct advantage that it keeps cosmologists
honest about the assumptions they are making, not only  about the underlying
world model, but also every other aspect of the data set they are attempting to
model: systematic errors should, in principle, be already at the forefront of
her mind! The catch is that to interpret probability as a degree of belief in
an hypothesis, Bayesian cosmologists have to assume a probability distribution
for the hypothesis itself.  This step can be somewhat  arbitrary and thus
subjective: this is the age-old point of friction between frequentists and
Bayesians. In this appendix we will use the Bayesian framework, in keeping
with the tradition of cosmology. We will nevertheless try to point out where
the ``subjectivity'' of being Bayesian is introduced (and where it is not).

After this aside, let us return to practicalities.
The precise functional form of the likelihood
is always unknown, and so an assumption must be made 
about it before any interpretation of the data can occur.
This assumption forms part of a model for the data, 
which we denote by~$H$, whilst
$x$ is a variable parameter of this model. More often than not, the
physical nature of the object being studied is best understood in terms
of some different parameter, $\theta$, rather than $x$: 
in this case the model still allows the datum
$x_0$ to be predicted, and describes how its probability is distributed
through~$\pr(x_0|\theta,H)$.
If more than one datum
is available, and they came from independent attempted measurements of $x$, 
we can write the joint likelihood as  
\begin{equation}
\pr(x_0,x_1,x_2,\ldots|\theta,H) =
\pr(x_0|\theta,H)\pr(x_1|\theta,H)\pr(x_2|\theta,H)\ldots,
\label{eq:lhood1}
\end{equation}
the product rule for combining independent probabilities. This makes
clearer the distinction between the parameter~$\theta$ and the 
data (which can be conveniently packaged into the vector~$\datav$ having
components~$x_i$). Indeed, a more complicated model for the data would
make use of more than one parameter when predicting the data; these can
be described by the parameter vector~$\parsv$. The generalization of
\autoref{eq:lhood1} to $N_d$ independent data sets, 
$\{\datav_j\}$, is then:
\begin{equation}
\pr(\datav|\parsv,H) = \prod_{j=1}^{N_d} \pr(\datav_j|\parsv,H).
\label{eq:lhood2}
\end{equation}

Within a given model then, the 
likelihood~$\pr(\datav|\parsv,H)$ 
can be calculated for any values of the model parameters~$\parsv$.
However, as outlined above, cosmologists want
statistical inferences, i.e., we want to learn more about our model and
its parameters
from the data, by calculating the posterior 
distribution~$\pr(\parsv|\datav,H)$. 
This distribution contains all the information about the model
supplied
by the data, as well as all the information we had about the model other than
that provided by the data: this can be seen by applying the product 
rule of 
conditional
probability to give Bayes' theorem,
\begin{equation}
\pr(\parsv|\datav,H) = \frac{\pr(\datav|\parsv,H)\,\pr(\parsv|H)}{\pr(\datav|H)}. 
\label{eq:bayes}
\end{equation}
The prior~$\pr(\parsv|H)$ encodes the additional 
information 
(this is where the subjectivity of the Bayesian approach comes in), and
is a PDF normalized over the parameter space.
The likelihood is also a frequentist quantity 
(without dependence on the prior) 
while the posterior is a Bayesian construct.
In practical applications of Bayesian parameter inference it is 
good practice therefore to check how much the reported result 
depend on the choice of prior: reliable results depend very weakly on the 
prior chosen. This is in fact a form of model comparison: for Bayesians, a
complete data model consists of a parameter set {\it and} the prior PDF for 
those parameters: some priors are more appropriate than others. We discuss
quantitative model comparison below: in this context it
provides a way of recovering some objectivity in Bayesian analysis.


\section{Assigning and Interpreting PDFs}
\label{sec:app:stats:pdfs}

As \autoref{eq:bayes} shows, computing the probability distribution for a
parameter (and hence measuring it) necessarily involves the assignment of a
prior PDF for that parameter. There are two types of prior we can assign:
\begin{itemize}
  \item{\bf Uninformative priors,} such as uniform distributions in
  the parameter or its logarithm (the Jeffreys prior) 
  are often assumed. Sometimes we genuinely know very little, and so
  minimizing the average information content of a prior PDF 
  (or maximizing its entropy) makes sense. In other situations we do know
  something about a model parameter, but choose to assign an uninformative
  prior in order to investigate cleanly
  the information content of the data. Sometimes the reason given is to ``give
  an unbiased result.'' This makes less sense, given that Bayesian
  inferences are biased by design -- biased towards what is already known
  about the system.  
  \item{\bf Informative priors:} it is very rare to know {\it nothing} 
  about a model
  and its parameters. An experiment has usually been carried out before, with
  different data! The best kind of prior PDF is the posterior PDF of a
  previous experiment -- this is exactly equivalent to combining data sets in 
  a joint analysis (\autoref{eq:lhood2} above).
\end{itemize}

Given suitably assigned priors and likelihoods then, the posterior
distribution gives the probability of the parameter vector lying between
$\parsv$ and $(\parsv + d\parsv)$.  This is the answer to the problem, the
complete inference within the framework of the model.  However, we typically
need to present some compressed version of the posterior PDF: what should we
do? 

The probability distribution for a single parameter $\theta_{N}$ is
given by marginalization over the other parameters, 
\begin{equation}
\pr(\param_{N}|\datav,H) = \int \pr(\parsv|\datav,H)\,d^{N-1}\parsv. 
\label{eq:bayes:marg}
\end{equation}
This is the addition rule for probabilities, extended to the 
continuous variable 
case\footnote{Sometimes \autoref{eq:bayes:marg} is used with  the posterior
$\pr(\parsv|\datav,H)$ substituted by the likelihood. Even in this case a
Bayesian step has been taken: a uniform prior is ``hidden'' in the parameter 
space ``measure'' $d^{N-1}\parsv$. }. 
This single parameter, one-dimensional marginalized
posterior is most useful when the parameter is the only one of
interest. In other cases we need to represent the posterior PDF and the
parameter constraints that it describes in higher dimensions -- although
beyond two dimensions the posterior PDF becomes very difficult to plot. 

The one-dimensional marginalized
posterior PDFs can be further compressed into their means, or medians, or
confidence intervals containing some fraction of the total probability --
confidence intervals need to be defined carefully as the integrals can be
performed a number of different ways.
However, note that the
set of one-dimensional marginalized posterior means (or medians, etc.) 
need not itself represent a model that is a good
fit to the data.
The ``best-fit'' point is the position in parameter space where the likelihood
function has a global maximum. This point is of most interest when the prior
PDF is uninformative: in the case of uniform prior PDFs on all parameters, the
peak of the likelihood coincides with the peak of the posterior PDF, but in
general it does not. Moreover, the maximum likelihood model 
is necessarily the model most affected by the noise in the data -- if any
model ``over-fits'' the data, it is that one! Graphical displays of
marginalized posterior PDFs remain the most complete way to present
inferences; propagating the full posterior PDF 
provides the most robust estimates of individual parameters. 

One class of parameters that are invariably marginalized over in the final
analysis are the so-called nuisance parameters. The model $H$ is a model for
the data, not just the physical system of interest: often there are aspects of
the experimental setup that are poorly understood, and so best included in the
model as functions with free parameters with estimated prior PDFs. This
procedure allows the uncertainty to be propagated into the posterior PDF for
the interesting parameters. Examples of nuisance parameters might include:
calibration factors, unknown noise or background levels, point spread function
widths, window function shapes, supernova dust extinctions, weak lensing mean
source redshifts, and so on. If a systematic error in an experiment is
identified, parametrized and then that nuisance parameter marginalized over,
then it can be said to have been translated into a statistical error (seen as
a posterior PDF width): a not unreasonable goal is to translate all systematic
errors into statistical ones.


\section{Model Selection}
\label{sec:app:stats:evidence}


While the goal of parameter estimation is to determine the  posterior PDF for
a model's parameters, perhaps characterized simply by the most probable or
best-fit  values and their errors, model selection seeks to distinguish
between different models, which in general will have different sets of
parameters.  Simplest is the case of {\em nested} models,  where the more
complicated model has additional parameters, in addition to those in the
simpler model. The simpler model may be interpreted as a particular case for
the more complex model, where the additional parameters are kept fixed at some
fiducial values. The additional parameters may be an indication of new
physics, thus the question one may ask is: ``would the experiment provide data
with enough statistical power to require additional parameters and therefore
to signal the presence of new physics if the new physics is actually the true
underlying model?'' Examples of this type of question are:  ``do the
observations require a modification to general relativity as well as a
universe dominated by dark energy?''(\autoref{sec:cp:grav}), or, ``do the
observations require a new species of neutrino?''  (\autoref{sec:cp:mnu}).
However, completely disparate models, with entirely different parameter sets
can also be compared using the Evidence ratio. In this case, in is even more
important to assign realistic and meaningful prior PDFs to all parameters.

These questions may be answered in a Bayesian context by considering
the Bayesian Evidence ratio, or Bayes factor, ~$B$: 
\begin{equation}
B = \frac{\pr(\datav|H_1)}{\pr(\datav|H_2)}, 
\label{eq:bayesratio}
\end{equation}
where $H_1$ and $H_2$ represent the two models being compared.
The Bayes factor is related to the perhaps more desirable posterior ratio
\begin{equation}
\frac{\pr(H_1|\datav)}{\pr(H_2|\datav)} =
\frac{\pr(\datav|H_1)}{\pr(\datav|H_2)}\,\frac{\pr(H_1)}{\pr(H_2)}. 
\label{eq:bayesratio2}
\end{equation}
by the ratio of model prior probabilities. The latter is 
not, in general straightforward to
assign with differences of opinion between analysts common.
However, the
Bayes factor itself can be calculated objectively once $H_1$ and $H_2$ have
been defined, and so is the more useful quantity to present. 

If there is no 
reason to prefer one hypothesis over another other than
that provided by the data, the ratio of the 
probabilities of each of the two models being true
is just given by the ratio of evidences. Another way of
interpreting a value of $B$ greater than unity 
is as follows: if models $H_1$ and $H_2$ are still to be 
presented as
equally probable after the experiment has been performed, then 
proponents of the lower-evidence model $H_2$ must be willing to offer odds of
$B$ to one against $H_1$. In practice, the Bayesian Evidence ratio can be
used simply to say that ``the probability of getting the data would be
$B$ times higher if model $H_2$ were true than if $H_2$ were.'' 
Indeed, \citet{Jeffreys61} proposed that $1<\ln B<2.5$ be considered as
``substantial'' evidence in favor of a model, $2.5<\ln B< 5$ as
``strong,'' and $\ln B>5$ as ``decisive.'' 
Other authors have introduced different terminology
\citep[e.g.,][]{Trotta05}.

The evidence $\pr(\datav|H)$ is the normalization of the posterior PDF 
for the parameters, and so is given by integrating the product of the
likelihood and the prior over all $N$~parameters:
\begin{equation}
\pr(\datav|H) = \int \pr(\datav|\parsv,H)\, \pr(\parsv|H)\, d^N\parsv. 
\label{eq:evidence}
\end{equation}
There is ample literature on 
applications of Bayesian Evidence ratios in
cosmology 
\citep[e.g.][]{Jaffe96, Hobson, Saini, 
LMPW, MRS06, PML06, Mukherjeeetal06, Pahud06, Szydlowski06a, Szydlowski06b, 
Trotta07, Pahud07}.  
The evidence 
calculation typically involves  computationally expensive integration
\cite{Skilling04, Trotta05, Beltranetal05, 
Mukherjeeetal06, MPL06, PML06}; however, it can often be approximated just as the model
parameter posterior PDF can. 
For example, \citet{Heavens07} shows how, by making 
simplifying assumptions in the same  spirit of Fisher's analysis 
\citep{Fisher},  one can compute the expected evidence for a given
experiment, in advance of taking any data, and forecast the extent
to which an experiment may be able to distinguish between different
models. 
We implement this in \autoref{sec:cp:mnu} and \autoref{sec:cp:grav}. 
In \autoref{sec:cp:mnu} we consider the  issue of deviations from the
standard number of three neutrino species. The simplest model has
three neutrino species, but effectively this number can be changed by
the existence of a light particle that does not couple to electrons,
ions or photons, or by the decay of dark matter particles, or indeed
any additional
relativistic particle. Given the observables and errors achievable from
a survey  with given specifications, we use the evidence in order to address
the issue of how much different from the standard value the
number of neutrino species should be such that the alternative model
should be favored over the reference model. 

In  \autoref{sec:cp:grav} we also employ the Bayesian evidence: this
time  the reference model is  a cold dark matter + dark energy model,
where gravity is described by General Relativity (GR). In the
alternative model, GR is modified so that the growth of cosmological
structure is different. Again, given the observables and errors achievable
from a survey  with given specifications, we use the evidence to
quantify how different from the GR
prediction the growth of structure would have to be such 
that the alternative model should be preferred.

%

\section{PDF Characterization} 
\label{sec:app:stats:numtech}

The conceptually most straightforward way to carry out parameter
inference (and model selection) is to tabulate the posterior PDF 
$\pr(\parsv|\datav,H)$ on a suitable grid, and normalize it via simple
numerical integration. This approach is unlikely to be practical unless the 
number of parameters is very small and the PDF is very smooth. The number of
function evaluations required increases exponentially with the dimensionality
of the parameter space; moreover, the following marginalization 
integrals will all be correspondingly time-consuming. In this section we
consider two more convenient ways to characterize the posterior PDF ---
the multivariate Gaussian (or Laplace) approximation, and Markov Chain Monte
Carlo sampling. 


\subsection{The Laplace Approximation}
\label{sec:app:stats:numtech:gaussian}

By the central limit theorem, the product of a set of 
convex functions tends to
the Gaussian functional form in the limit of large set
size~\citep{Jaynes}; the
posterior probability distribution of \autoref{eq:bayes}
often fits this bill, suggesting that the 
Gaussian distribution is likely to be a good approximation to the 
posterior density. 
Approximating probability distributions with
Gaussians is often referred to as the Laplace approximation~\citep[see
\eg][]{Sivia,MacKay}. 

In one dimension, a suitable Gaussian can be found by Taylor
expansion about the peak 
position $\theta_0$ of the logarithm of the posterior PDF~$P(\theta)$ 
(where the
conditioning on the data and the model have been dropped for clarity):
\begin{equation}
\log{P(\theta)} \approx \log{P(\theta_0)} 
+ \frac{1}{2}(\theta-\theta_0)^2 \frac{d^2 P}{d
\theta^2}{\bigg\vert}_{\theta_0}.
\end{equation}
Exponentiating this expression gives the   
Gaussian approximation to the function, 
\begin{equation}
g(\theta) \approx P(\theta_0) \exp{\left[ -
\frac{(\theta-\theta_0)^2}{2 \sigma^2}\right]}.
\end{equation}
The width $\sigma$ of this Gaussian 
satisfies the following relation:
\begin{equation}
\frac{d^2 \log{P} }{d \theta^2}{\bigg\vert}_{\theta_0} = -\frac{1}{\sigma^2}. 
\end{equation}

The extension of this procedure to multivariate distributions is
straightforward: instead of a single variance $\sigma^2$, an 
$N \times N$
covariance matrix~$\mathsf{C}$ is required, such that the posterior
approximation is
\begin{equation}
g(\parsv) = P(\parsv_0) \exp{\left[ -\frac{1}{2}(\parsv-\parsv_0)^{\rm
T}\mathsf{C}^{-1}(\parsv-\parsv_0) \right] },
\label{eq:bayes:mvgauss}
\end{equation}
and the covariance matrix has components
\begin{equation}
\left(\mathsf{C}^{-1}\right)_{ij} = - \frac{\partial^2 \log{P}}
{\partial \theta_i \partial \theta_j }{\bigg\vert}_{\parsv_0}.
\label{eq:laplacecovmat}
\end{equation}
(This matrix is very unlikely to be diagonal -- correlations, or degeneracies,
between parameters are encoded in its off-diagonal elements.)
The problem is now reduced to finding (numerically) 
the peak of the log-posterior, and its second derivatives at that 
point. 
When the data quality is good, one may expect the individual datum likelihoods
to be already quite convex, giving a very peaky unimodal posterior:  in this
case the Gaussian approximation is likely to be both accurate, and more
quickly and easily located. More commonly, techniques such as simulated
annealing may be necessary when finding the maximum of  complex functions of
many parameters; in this case a Gaussian may not be such a good approximation
anyway.


\subsection{Fisher Matrices}
\label{sec:app:stats:numtech:fisher}

The Fisher information matrix \citep{Fisher} is widely used for 
forecasting survey performance and for identifying dominant 
systematic effects \citep[see e.g.,][]{albrecht09}. 
The Fisher matrix formalism 
is very closely related to the 
Laplace approximation to the parameter posterior described above.
The  discussion that follows may seem unconventional to those
more familiar with its frequentist origins and presentation: our aim is to
show how the formalism has been adapted to modern Bayesian cosmology.  


The Fisher matrix was originally defined to be
\begin{equation}
F_{\alpha\beta}=-\left\langle\frac{\partial^2\ln P(\bit{x}|\bit{q})}
{\partial q_\alpha\partial q_\beta}\right\rangle,
\end{equation}
where
$\bit{x}$ is a data vector, $\bit{q}$ is the vector of model parameters,
and $\langle\ldots\rangle$ denotes an ensemble average. If the prior PDFs are
non-uniform, we must replace the likelihood $P(\bit{x}|\bit{q})$ by the
posterior $P(\bit{q}|\bit{x})$. In any case, we recognize the Laplace
approximation and identify
(by comparison with \autoref{eq:laplacecovmat}) 
the ensemble average
covariance matrix of the inferred parameters as $\bit{F}^{-1}$. 

This estimate of the forecast parameter uncertainties
really corresponds to the best 
case scenario, as dictated by the Cramer-Rao theorem. More 
specifically, the estimated error of the parameter $q_\alpha$ is 
$\sigma(q_\alpha) \ge (F_{\alpha\alpha})^{-1/2}$ if all other 
parameters are known precisely, or 
$\sigma(q_\alpha) \ge [(\bit{F}^{-1})_{\alpha\alpha}]^{1/2}$ if all 
the parameters are estimated from the data.

Cosmological applications of the Fisher matrix were introduced by
\citet{jungman96,vogeley96,tegmark97a,tegmark97b}. 
The key is to identify the
correct likelihood function $P(\bit{x}|\bit{q})$ (which is anyway 
crucial for all inference techniques).  However, the Fisher matrix 
analysis has a further limitation due to the Gaussian approximation of 
$P(\bit{q}|\bit{x})$ with respect to $\bit{q}$: the likelihood, priors and
indeed choice of parametrization need to be such that this approximation is a
good one. Usual practice is to approximate the likelihood function as
Gaussian, and assert either Gaussian or uniform priors (both of which
guarantee 
the Gaussianity of the posterior PDF).

If we approximate the likelihood function
by a Gaussian distribution then, 
\begin{equation}
P(\bit{x}|\bit{q}) = \frac{1}{(2\pi)^{N/2} \det [\bit{C}(\bit{q})]} 
\exp \left\{\left[\bit{x}-\bar{\bit{x}}(\bit{q})\right]^{\rm T}
\frac{\bit{C}^{-1}(\bit{q})}{2}\left[\bit{x}-\bar{\bit{x}}(\bit{q})
\right]\right\},
\end{equation}
where $N$ is the dimension of the observables \bit{x}, 
$\bar{\bit{x}}(\bit{q})$ is the ensemble average of \bit{x}, 
$\bit{C}(\bit{q}) = \langle (\bit{x}-\bar{\bit{x}})
(\bit{x}-\bar{\bit{x}})^{\rm T} \rangle$ is the covariance of \bit{x}.
The Fisher matrix is then \citep{tegmark97a}
\begin{equation} \label{eq:app:gfsh}
F_{\alpha\beta}=\frac{1}{2} \mbox{Tr} \left(\bit{C}^{-1}
\frac{\partial \bit{C}}{\partial q_\alpha} \bit{C}^{-1}
\frac{\partial \bit{C}}{\partial q_\beta} \right)+ 
\frac{\partial \bar{\bit{x}}}{\partial q_\alpha} \bit{C}^{-1}
\frac{\partial \bar{\bit{x}}}{\partial q_\beta},
\end{equation}
where we have dropped the variables \bit{q} in \bit{C} and 
$\bar{\bit{x}}$ for clarity. To avoid confusion, we note that
\bit{C} is the covariance matrix of the data \bit{x}, whereas 
$\bit{F}^{-1}$ is the covariance matrix of the parameters \bit{q}.

In the Gaussian approximation, marginalization, and moment-calculating
integrals are analytic.
Independent Fisher matrices are additive; a Gaussian prior 
on $q_\alpha$, $\sigma_{\rm P}(q_\alpha)$, 
can be introduced via $F_{\alpha\alpha}^{\rm new} = F_{\alpha\alpha}
+\sigma_{\rm P}^{-2}(q_\alpha)$.
A Fisher matrix of the parameters
$\bit{q}$ can be projected onto a new set of parameters 
$\bit{p}$ via
\begin{equation} \label{eq:app:projfsh}
F_{\mu\nu}^{\rm new}= \sum_{\alpha,\beta}
\frac{\partial q_\alpha}{\partial{p_\mu}} F_{\alpha\beta}
\frac{\partial q_\beta}{\partial{p_\nu}}.
\end{equation}
Fixing a parameter is equivalent to striking out its corresponding row
and column in the Fisher matrix. To obtain a new Fisher matrix after
marginalizing over a parameter, one can strike out the parameter's
corresponding row and column in the original covariance matrix 
(i.e., the inverse of the original Fisher matrix) and then invert the
resulting covariance matrix\footnote{For a better numerical
treatment, see \citet{albrecht09}.}.



\subsection{Examples}

At this point, we give two worked examples from observational cosmology,
describing the data model and Fisher matrix forecasts of parameter
uncertainties.

\noindent{\bf Example 1: Type Ia Supernovae}\\
For SNe, the observables are their peak magnitudes in a certain band
\begin{equation}
m_i = \bar{m}(\bit{q},z_i) + n_i,
\end{equation}
where the subscript $i$ labels each SN, $\bar{m}$ is the mean 
value of the SN peak magnitude at redshift $z_i$, the 
parameters \bit{q} include both cosmological and nuisance parameters, 
and $n_i$ represents the observational noise 
and intrinsic scatter of the peak 
magnitude. The mean peak magnitude is given by
\begin{equation}
\bar{m}_i = M + 5\log\left[D_{\rm L}(w_0,\wa,\Om,\Ok, h,\ldots,z)\right]
+ \mbox{evolution terms} + \mbox{const},
\end{equation}
where $M$\footnote{$M$ is degenerate with the Hubble constant.} 
is the mean absolute peak magnitude at $z = 0$, 
$D_{\rm L}$ is the luminosity distance, and the evolution terms 
account for a possible drift of the mean absolute peak magnitude 
with time. In a number of forecasts, the evolution terms are simply 
model with a quadratic function $a z + b z^2$ 
\citep[e.g.,][]{albrecht06b, knox06c, zhan08}. 

We assume that the scatter $n_i$ 1) does not 
depend on cosmology or redshift, 2) is uncorrelated with each
other, and 3) is normally distributed, i.e., 
\begin{eqnarray}
\left\langle (m_i - \bar{m}_i)(m_j-\bar{m}_j)\right\rangle &=& 
\sigma_m^2 \delta^{\rm K}_{ij}, \\
P(\bit{m}|\bit{q}) = \Pi_i P(m_i|\bit{q},\sigma_m) &=&
\Pi_i \frac{1}{\sqrt{2\pi}\sigma_m}
\exp\left[-\frac{(m_i-\bar{m}_i)^2}{2\sigma_m^2}\right],
\end{eqnarray}
where $\delta^{\rm K}_{ij}$ is the Kronecker delta function, and
$\sigma_m\sim 0.15$, then Fisher matrix reduces to
\begin{equation}
F_{\alpha\beta}=\Sigma_i\frac{\partial \bar{m}_i}{\partial q_\alpha}
\frac{1}{\sigma_m^2}\frac{\partial \bar{m}_i}{\partial q_\beta}.
\end{equation}
With a \phz{} SN sample, the Fisher matrix has to be integrated over
the \phz{} error distribution \citep{zhan08}:
\begin{eqnarray}
F_{\alpha\beta} &=& \frac{1}{\sigma_{\rm m}^2}\int 
n_{\rm p}(z_{\rm p}) 
\frac{\partial \bar{m}_{\rm p}(z_{\rm p})}{\partial q_\alpha}
\frac{\partial \bar{m}_{\rm p}(z_{\rm p})}{\partial q_\beta} 
dz_{\rm p} \\ \nonumber
\bar{m}_{\rm p} &=& \int 
\left[5 \log D_{\rm L}(w_0,\wa,\Om,\Ok, h,\ldots,z) + M + 
\mbox{evol. terms} + \mbox{const.}\right] 
p(z | z_{\rm p}) dz,
\end{eqnarray}
the subscript p signifies \phz{} space, $n_{\rm p}(z_{\rm p})$ is
the SN distribution in \phz{} space, and 
$p(z|z_{\rm p})$ is the probability density 
of a SN at $z$ given its \phz{} $z_{\rm p}$. 

\vspace{\baselineskip}
\noindent{\bf Example 2: Gaussian Random Fields}\\
The power spectrum is the covariance of the Fourier modes of the field.
For an isotropic field, the modes are uncorrelated, i.e.,
\begin{equation} \label{eq:app:pk}
\langle \hat{\delta}(\bit{k})\hat{\delta}^*(\bit{k}')\rangle
= P(k)(2\pi)^3\delta^{\rm D}(\bit{k}-\bit{k}'),
\end{equation}
where $P(k)$ is the power spectrum, and 
$\delta^{\rm D}(\bit{k}-\bit{k}')$ is the Dirac delta function. 
By definition, the modes $\hat{\delta}(\bit{k})$ have zero mean.
Since surveys are limited by volume, the wavenumbers must be
discrete. For a cubic survey of volume $V=L^3$, we have
\begin{equation} \label{eq:app:dpk}
\langle \hat{\delta}(\bit{k})\hat{\delta}^*(\bit{k}')\rangle
= P(k)V\delta^{\rm K}_{\bit{n},\bit{n}'},
\end{equation}
where $\bit{k} = 2\pi \bit{n}/L$, and $\bit{n} = (n_1,n_2,n_3)$ 
with integer $n_i$s running from $-\infty$ to $\infty$. If the 
density field is discretized, e.g., on a grid, then $n_i$s are 
limited by the Nyquist frequency. For 
convenience, we use \bit{k} and \bit{n} interchangeably, with
the understanding that \bit{k} is discrete. 
If the power spectrum is calculated based on discrete objects, 
e.g., galaxies, then we have
\begin{equation} \label{eq:app:gps}
P_{\rm g}(k) = P(k) + n_{\rm g}^{-1},
\end{equation}
where $n_{\rm g}$ is the galaxy number density.

For a Gaussian random field sampled by galaxies, the modes are 
normally distributed and are completely characterized by the 
power spectrum $P_{\rm g}(k)$. Using
the Fourier modes (rather than the power spectrum) as observables,
we obtain the Fisher matrix using \autoref{eq:app:gfsh} and 
\autoref{eq:app:dpk}
\begin{equation}  \label{eq:app:sumfsh}
F_{\alpha\beta}=\frac{1}{2} \Sigma_{\bit{n}}
\frac{\partial \ln P_{\rm g}(n)}{\partial q_\alpha} 
\frac{\partial \ln P_{\rm g}(n)}{\partial q_\beta},
\end{equation}
where the summation runs over all available modes. When the 
survey volume is sufficiently large, one can replace the summation
in \autoref{eq:app:sumfsh} with an integral \citep{tegmark97b}
\begin{equation} \label{eq:app:intfsh}
F_{\alpha\beta} = \frac{V}{2} \int
\frac{\partial \ln P_{\rm g}(k)}{\partial q_\alpha} 
\frac{\partial \ln P_{\rm g}(k)}{\partial q_\beta}
\frac{k^2 dk}{4 \pi^2}.
\end{equation}
In terms of angular power spectra, the Fisher matrix becomes
\begin{equation} \label{eq:app:trfm}
F_{\alpha\beta} = f_{\rm sky} \sum_\ell \frac{2\ell + 1}{2} 
\mbox{Tr} \left[ \bit{P}^{-1}(\ell)
\frac{\partial \bit{P}(\ell)}{\partial q_\alpha} \bit{P}^{-1}(\ell)
\frac{\partial \bit{P}(\ell)}{\partial q_\beta} \right],
\end{equation}
where $f_{\rm sky}$ is the fraction of sky covered by the survey, 
and for each multipole $\ell$, $\bit{P}(\ell)$ is a matrix of power 
spectra between pairs of redshift bins (see e.g, 
\autoref{eq:lss:gaps} and \autoref{eq:cp:totps}).


\subsection{Sampling Methods}
\label{sec:app:stats:numtech:mcmc}

Probability distributions calculated on
high-dimensional regular 
grids are rather unwieldy. A Gaussian approximation allows integrals over the
posterior to be performed analytically -- but may not provide sufficient
accuracy especially if the PDF is not unimodal.

A far more useful characterization of the posterior PDF is a list of samples
drawn from the distribution. By definition, the number density of these
samples is proportional to the probability density, such that (given enough
samples) a smoothed histogram is a good representation of the probability
density function.  Each histogram bin  value is an integral over this PDF, as
are all other inferences.  By sampling from the distribution,  these integrals
are calculated by Monte Carlo integration (as opposed to the simple summation
of a gridding algorithm). Marginal distributions are trivial to calculate --
the histogram needs only to be constructed in the required dimensions, usually
just one or two. Samples are also useful for the simple reason that they
represent acceptable fits: visualization of the model corresponding to each
sample can provide much insight into the information content of the data.

The problem is now how to draw samples from a general PDF $P(\parsv)$.
The Metropolis-Hastings algorithm (and various derivatives)
provides a flexible solution
to this 
problem: see the books by \eg ~\citet[][]{MCMC,Ruanaidh,Neal,MacKay} for good
introductions. 
This is the basic Markov chain Monte Carlo method, and works as
follows. 
A Markov chain is a series of random variables (specifically representing points in a parameter space) whose values are each 
determined only by the previous point in the series. 
Generation of a Markov chain proceeds as follows: a candidate
sample point is drawn from a suitably chosen ``proposal
density''~$S(\parsv',\parsv)$, and then accepted with 
probability~$A(\parsv',\parsv)$ -- if not accepted, the current sample
is repeated to preserve the invariance of the target distribution.
In the Metropolis-Hastings algorithm, the acceptance probability is given by 
\begin{equation}
A(\parsv',\parsv) = \textrm{min} \left[ 1,
\frac{P(\parsv)}{P(\parsv')}\right],
\label{eq:metro}
\end{equation}
provided the proposal distribution~$S$
is symmetric about the previous sample point.

In other words, we accept the new sample if the probability density at that point is
higher, and otherwise accept it with probability equal to the ratio  of new to
old probability densities. Note that since the sampling procedure depends only
on a probability ratio; the normalization of the PDF need not be known: this
is just the situation we find ourselves in when analyzing data,  able only to
calculate the unnormalized product of likelihood and prior.  

As seen in the previous paragraphs, the basic MCMC algorithm is very simple;
whilst it guarantees that the output list of sample points will have been
drawn from the target density~$P$, that is not the same as fully sampling
the distribution in a finite time.
Consequently, the computational challenge lies in the choice of 
proposal density~$S$. If~$S$ is too compact, the chains take too
long to explore the parameter space; too broad, and the sample rejection
rate becomes very high as too much time is spent testing regions of low
likelihood. 

\citet{L+B02} provide a useful primer to the use of MCMC in
cosmological parameter estimation, and in particular CMB analysis.
In the next sections we show some example sampled PDFs, and then outline some
common problems encountered when sampling.

\subsubsection{Example: WMAP5}

\begin{figure*}[!ht]
\centering
\begin{minipage}{0.9\linewidth}
\begin{minipage}{0.48\linewidth}
\centering\includegraphics[width=\linewidth]{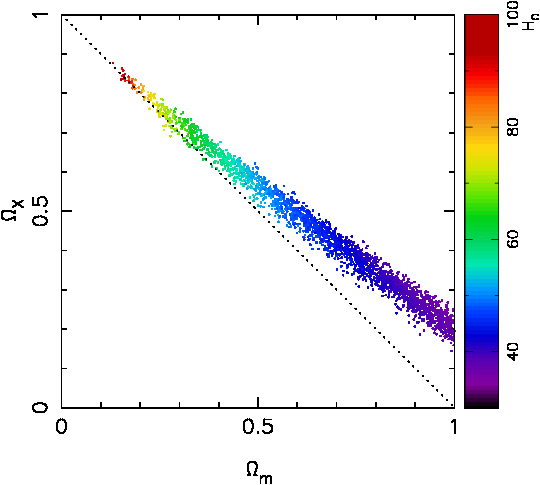}
\end{minipage}\hfill
\begin{minipage}{0.48\linewidth}
\centering\includegraphics[width=\linewidth]{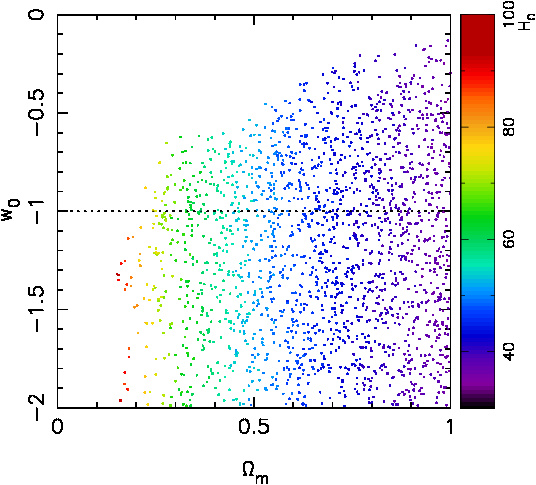}
\end{minipage}
\begin{minipage}{0.48\linewidth}
\centering\includegraphics[width=\linewidth]{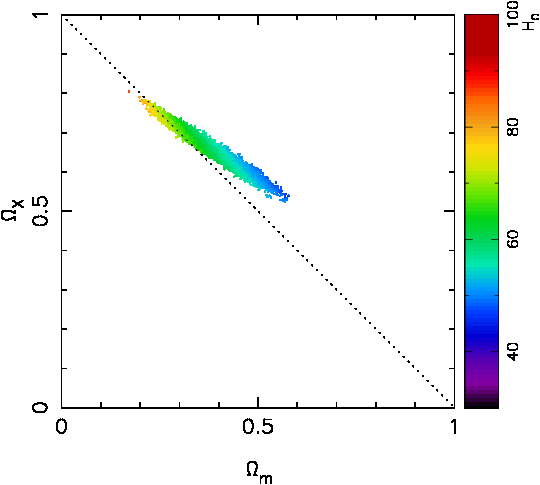}
\end{minipage}\hfill
\begin{minipage}{0.48\linewidth}
\centering\includegraphics[width=\linewidth]{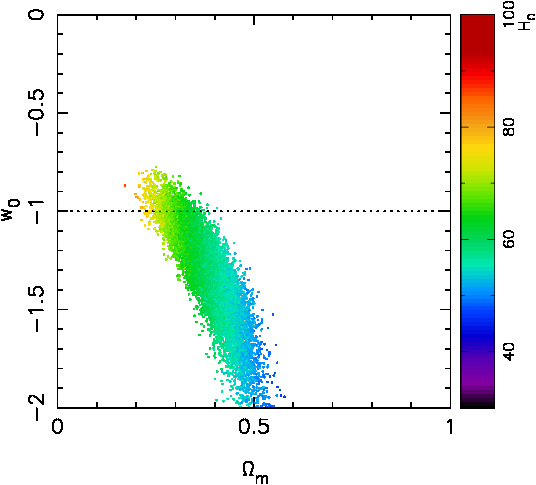}
\end{minipage}
\end{minipage}
\caption{Marginalized posterior PDFs for cosmological parameters given the
WMAP 5 year data set, represented by MCMC sample density.
{\it Top}: $\Pr(\Om,\Omega_X|{\rm WMAP5\, only})$ (left) 
and $\Pr(\Om,w_0|{\rm WMAP5\, only})$ (right). {\it Bottom}: 
$\Pr(\Om,\Omega_X|{\rm WMAP5+SNe})$ (left) 
and $\Pr(\Om,w_0|{\rm WMAP5+SNe})$ (right). In each row,
uniform priors were assumed for $\Om$, $\Omega_X$ (the dark energy
density parameter), $w_0$ (the non-evolving dark energy equation of state) 
and $H_0$. Samples are color-coded by $H_0$ to allow a third dimension to be
visualized. The dashed lines are loci representing universes with flat geometry
amd a cosmological constant. The Markov chains in the two rows contain 
different numbers of samples.}
\label{fig:WMAP5mcmc}
\end{figure*}

The WMAP team provide their cosmological parameter inferences in Markov Chain
form, downloadable from their
website\footnote{\url{http://lambda.gsfc.nasa.gov/}}  \citep[see the papers
by][for details]{Dunkley,Komatsu}.  In \autoref{fig:WMAP5mcmc} we display
posterior PDFs in a 4-dimensional cosmological parameter space 
by plotting the WMAP team's MCMC samples, marginalizing by
projecting the samples 
onto two different planes, and color-coding them by the samples' Hubble
constant values in order to
visualize this third dimension. 
In the top row, the likelihood function is for the WMAP 5th
year data set alone, while in the second row the likelihood for the ``SN all''
combined supernova type Ia data set has been multiplied in. 

\subsubsection{MCMC Sampling Issues}

MCMC sampling is often the preferred way to approximate a posterior PDF: the 
CPU time taken scales (in principle) only linearly with the number of
parameter space dimensions (as opposed to exponentially in the case of
brute-force gridding), it provides (in principle) accurate 
statistical uncertainties that take into account the various (often
non-linear) parameter
degeneracies, and avoids (in principle) the false maxima in the PDF that can
cause Laplace approximation maximizers either to need restarting in multiple
locations, or worse, to give misleading results.

However, MCMC sampling can be affected by a number of problems. We give a very
brief overview here and refer the reader to the cited textbooks for more
information. 
As the number of dimensions increases, finding isolated sharp peaks in the
very large volumes involved is a particularly difficult problem.  ``Cooling''
the process (starting by sampling  from the prior, and gradually increasing
the weight of the likelihood during a ``burn-in'' phase)  can help alleviate
this -- this also allows evidence estimation via ``thermodynamic integration''
\citep{Ruanaidh}.  These burn-in samples are to be discarded. 

It also gets
progressively more difficult to  move away from a false maximum:  proposal
distributions that are too broad  can lead to very high sample rejection rates
and low chain mobility.  Similarly very narrow degeneracies also lead to high
sample rejection rates. In these cases, the design of the proposal
distribution is key! One partial solution is to re-parametrize such that the
degeneracies are not so pronounced: a simple example is in CMB analysis, where
working with $\omega_b = \Omega_b h^2$ instead of $\Omega_b$ removes a
particularly pronounced ``banana'' degeneracy. However, the prior PDFs need
especially careful attention in this case, since a uniform prior in A is never
a  uniform prior in B(A).

To increase efficiency,
updating the covariance matrix of the proposal distribution as sampling
proceeds has some considerable appeal (arising from the intuition that the
best proposal distribution must be close to the target PDF itself),  but the
updating must be done carefully to preserve detailed balance in the chains.
Finally, how do you know when you are finished?  Various convergence tests on
the chains have been proposed \citep[\eg,][]{Gelman-Rubin}; Dunkley et al. look at the
chain  power spectrum to check for unwanted correlations. In general, it is
usually found that  running multiple parallel Markov chains allow more tests
and provide greater confidence.

\subsubsection{Importance Sampling}

We can incorporate new information into an MCMC inference 
by importance-sampling
the posterior distribution \citep[see e.g.,\ ][for an
introduction]{L+B02}. This procedure allows us to calculate integrals (such as means and
confidence limits) over
the updated posterior PDF, $P_2$, by re-weighting the samples drawn from the
original  PDF,
$P_1$. For example, the posterior mean value of a parameter $x$:
\begin{align}
\langle x \rangle_2 &= \int x \cdot P_2(x)\, dx, \notag \\
                    &= \int x \frac{P_2(x)}{P_1(x)} \cdot P_1(x)\, dx.
\label{eq:impsamp}
\end{align}
By weighting the samples from $P_1$ by the ratio $P_2(x)/P_1(x)$, we can
emulate a set of samples drawn directly from $P_2$.  It works most
efficiently  when $P_1$ and $P_2$ are quite similar, and fails if $P_1$ is
zero-valued over some of the range of $P_2$ or if the sampling of $P_1$ is
too sparse. In many cases though, it provides an efficient way to explore the
effects of both additional likelihoods and alternative priors.

\begin{figure*}[!ht]
\centering
\begin{minipage}{0.9\linewidth}
\begin{minipage}{0.48\linewidth}
\centering\includegraphics[width=\linewidth]{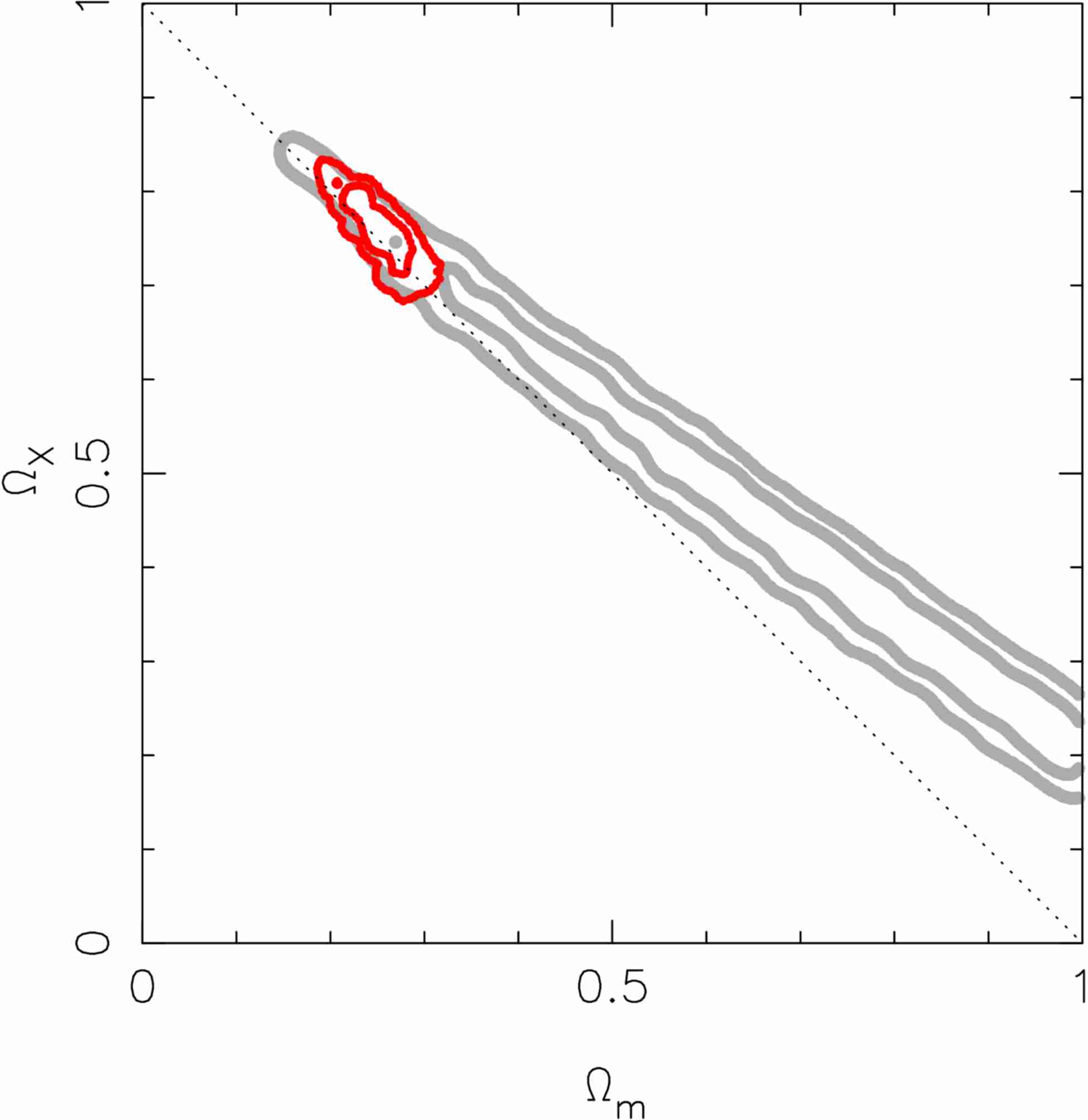}
\end{minipage}\hfill
\begin{minipage}{0.48\linewidth}
\centering\includegraphics[width=\linewidth]{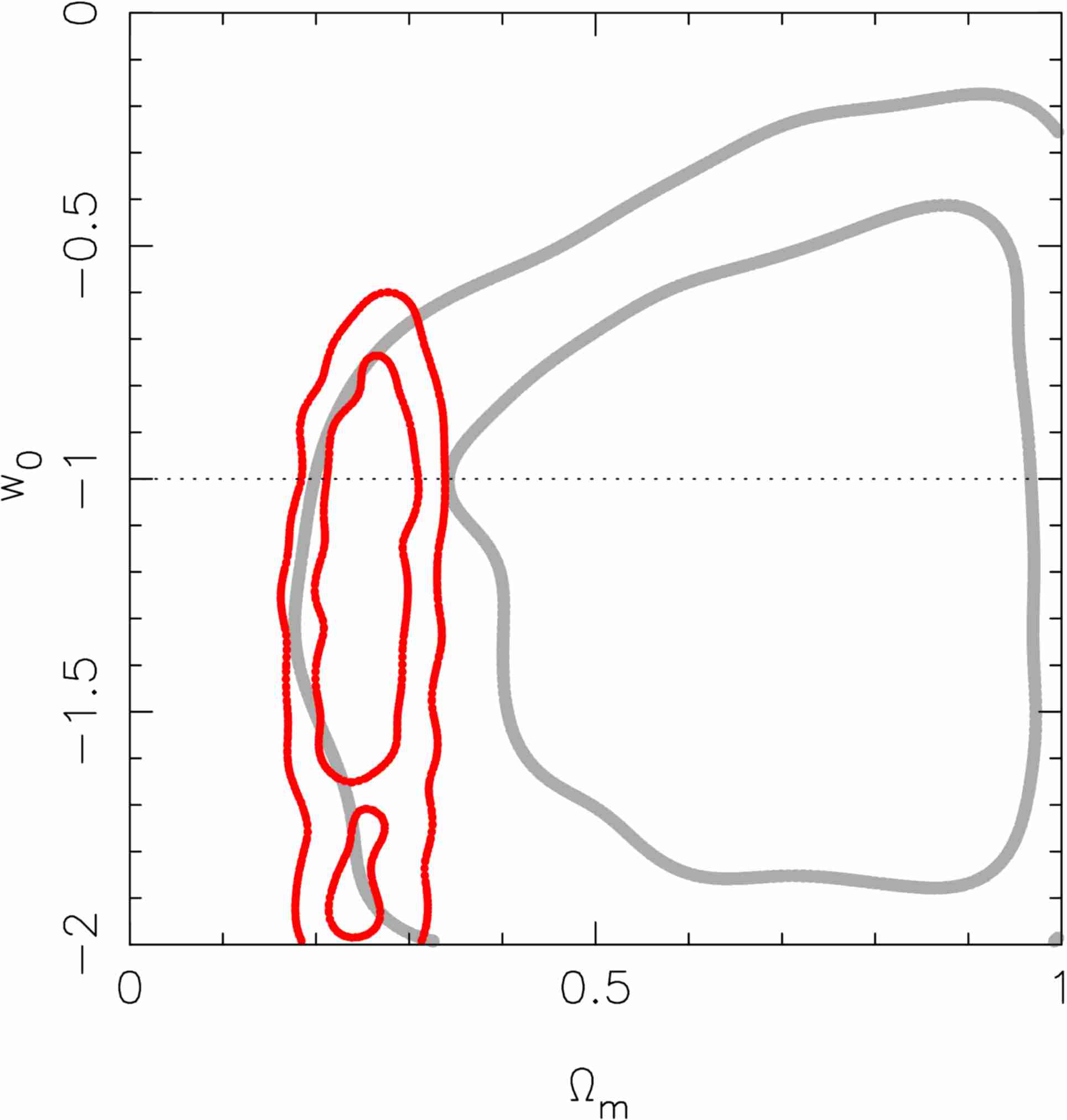}
\end{minipage}
\end{minipage}
\caption{Marginalized posterior PDFs for cosmological parameters given the
WMAP 5 year data set and the \citet{Riess++2009} Hubble constant measurement, 
obtained by importance-sampling (red curves). 
{\it Left}: $\Pr(\Om,\Omega_X|{\rm WMAP5,\, Riess})$
and {\it right}: $\Pr(\Om,w_0|{\rm WMAP5,\, Riess})$. The contours shown 
contain 68\% and
95\% of the integrated posterior probability: the gray contours in the
background show the PDFs
without the Riess et al Hubble constant constraint.}
\label{fig:WMAP5impsamp}
\end{figure*}

As an example, \autoref{fig:WMAP5impsamp} shows the same marginalized
posterior PDF from the WMAP 5-year data set as in \autoref{fig:WMAP5mcmc}, but
after importance sampling using the Hubble constant measurement of 
$H_0 = 74.2 \pm 3.6 {\rm km s}^{-1} {\rm Mpc}^{-1}$
by \citet{Riess++2009}. To make this plot, we interpret the Riess et al. measurement as the Gaussian  
PDF $\Pr(H_0|{\rm Riess})$, and then write the updated posterior PDF 
arising from
a joint analysis of the WMAP 5-year data
and the Riess et al. data as follows:
\begin{align}
\Pr(H_0,\boldsymbol{q}|{\rm WMAP5,\, Riess}) 
  &\propto \Pr({\rm WMAP5}|H_0,\boldsymbol{q}) \cdot 
            \Pr(H_0|{\rm Riess})\Pr(\boldsymbol{q}) \\
  &= \Pr(H_0,\boldsymbol{q}|{\rm WMAP5\, only}) \cdot \frac{\Pr(H_0|{\rm Riess})}{\Pr(H_0)}.
\end{align}
Here, for clarity, the cosmological parameters other than $H_0$ 
are denoted by the ``vector''
$\boldsymbol{q} = \{Om,\Omega_X,w_0,\ldots\}$. 

\noindent In the second line, we have
substituted the original posterior 
PDF, $\Pr(H_0,\boldsymbol{q}|{\rm WMAP5\, only})$: from this equation 
it is clear that the
weight, $P_2(x)/P_1(x)$, from \autoref{eq:impsamp} is just given by the value of
the PDF, $\Pr(H_0|{\rm Riess})$, if (as was the case) the original prior 
on $H_0$ was uniform. To make the plots, each sample was added to a
two-dimensional histogram according to its weight; the histogram was then
minimally smoothed and contours computed.

\clearpage

%
%
%
%
%





\bibliographystyle{SciBook}
\bibliography{appendix/analysis/analysis}

%


\chapter[Common Abbreviations and Acronyms]{Common Abbreviations and Acronyms}
\label{chp:glossary}
{\it Michael A. Strauss}

This glossary of abbreviations used in this book is not complete, but
includes most of the more common terms, especially if they are used
multiple times in the book.  We split the list into
general astrophysics terms and the names of various telescopic
facilities and surveys.   

\begin{center}
\Large
{\bf General Astrophysical and Scientific Terms}
\end{center}
\begin{itemize}
\item {\bf $\Lambda$CDM:} The current ``standard'' model of cosmology,
  in which Cold Dark Matter (CDM) and a cosmological constant
  (designated ``$\Lambda$''), together with a small amount of ordinary
  baryonic matter, together give a mass-energy density sufficient to
  make space flat.  Sometimes written as ``LCDM''.  
\item {\bf AAAS:} American Association for the Advancement of Science.
\item {\bf AAVSO:} American Association of Variable Star Observers (\url{http://www.aavso.org}).
\item {\bf ADDGALS:} An algorithm for assigning observable properties to
  simulated galaxies in an N-body simulation. 
\item {\bf AGB:} Asymptotic Giant Branch, referring to red giant stars with
  helium burning to carbon and oxygen in a shell. 
\item {\bf AGN:} Active Galactic Nucleus.
\item {\bf AM CVn:} An AM Canum Venaticorum star, a cataclysmic variable
  star with a particularly short period. 
\item {\bf AMR:} Adaptive Mesh Refinement, referring to a method of gaining
  dynamic range in the resolution of an N-body simulation. 
\item {\bf APT:} Automatic Photometric Telescope, referring to small
  telescopes appropriate for photometric follow-up of unusual
  variables discovered by LSST. 
\item {\bf BAL:} Broad Absorption Line, a feature seen in quasar spectra. 
\item {\bf BAO:} Baryon Acoustic Oscillation, a feature seen in the power spectra of
  the galaxy distribution and in the fluctuations of the Cosmic
  Microwave Background. 
\item {\bf BBN:} Big-Bang Nucleosynthesis, whereby deuterium, helium, and
  trace amounts of lithium were synthesized in the first minutes after
  the Big Bang. 
\item {\bf BD:} Brown Dwarf star. 
\item {\bf BH:} Black Hole.
\item {\bf CCD:} Charge-Coupled Device, the sensors to be used in the LSST
  camera. 
\item {\bf CCSN:} Core-Collapse Supernovae, due to the implosion of a
  massive star.  As opposed to Type Ia supernovae, in which a white
  dwarf exceeds the Chandrasekhar limit. 
\item {\bf CDM:} Cold Dark Matter. See $\Lambda$CDM.  
\item {\bf CMB:} Cosmic Microwave Background.
\item {\bf CMD:} Color-Magnitude Diagram, relating the brightnesses and
  colors of stars or galaxies. 
\item {\bf CME:} Coronal Mass Ejection, from the Sun. 
\item {\bf CMR:} Color-Magnitude Relation; see CMD. 
\item {\bf CPU:} Central Processing Unit, referring to processing power in a
  computer. 
\item {\bf CV:} Cataclysmic Variable, a close binary star consisting
  of a white dwarf with mass transfer from a secondary. 
\item {\bf DAV:} Pulsating white dwarf stars with hydrogen atmospheres. 
\item {\bf DBV:} Pulsating white dwarf stars with helium atmospheres. 
\item {\bf DETF:} Dark Energy Task Force, which produced an influential
  report \citep{albrecht06} outlining future experimental probes of dark energy. 
\item {\bf DWF:} ``Deep-Wide-Fast'', characterizing LSST's ability to survey
  the sky. 
\item {\bf EOS:} Equation of State, referring to the relationship between
  density and pressure of dark energy.  
\item {\bf EPM:} Expanding Photosphere Method, a method of measuring
  distances to Type II supernovae. 
\item {\bf EPO:} Education and Public Outreach. 
\item {\bf ETC:} Exposure Time Calculator, which returns estimates of
  survey depth and S/N under different assumptions. 
\item {\bf EXors:} EX Lupi-type stars, a type of T Tauri star that undergoes
  irregular outbursts. See also FUOr. 
\item {\bf FMF:} First Mass Function, referring to the mass distribution of
  the first generation of stars. 
\item {\bf FOV:} Field of View (of a telescope). 
\item {\bf FUOr:} FU Orionis Stars, which differ from EXors in that their
  outbursts last for a longer time. 
\item {\bf FWHM:} Full Width at Half Maximum.
\item {\bf GCVS:} General Catalogue of Variable Stars.
\item {\bf GRB:} Gamma-Ray Burst.
\item {\bf GZK:} Greisen-Zatsepin-Kuzmin effect, whereby 
  photons and cosmic rays with energies above $\sim 10^{19}$ eV
  scatter off CMB photons inelastically to produce pions.  This causes
  the Universe to be opaque to 
  such high-energy particles at distances above roughly 100 Mpc.  
\item {\bf HOD:} Halo Occupation Distribution, referring to the distribution
  of galaxies within dark matter halos. 
\item {\bf HFC:} Halley Family Comet. 
\item {\bf HR {\rm or} H-R diagram:} The Hertzsprung-Russell diagram,
  plotting the luminosity (or absolute magnitude) of stars versus
  their surface temperature (or color). 
\item {\bf HVS:} Hypervelocity stars, i.e., stars in the Milky Way
  travelling above the escape speed of the Galaxy.  
\item {\bf ICM:} Intercluster Medium, i.e., the hot gas within clusters of
  galaxies.  
\item {\bf ICME:} Interplanetary Coronal Mass Ejections, i.e., the gas
  associated with the CME at large heliocentric distances. 
\item {\bf ICRF:} International Celestial Reference Frame, with respect to
  which astrometric calibration will be done. 
\item {\bf ISM:} Interstellar Medium, the gas and dust between the
  stars. 
\item {\bf ISW:} Integrated Sachs-Wolfe Effect, an imprint in the CMB
  fluctuations due to the propagation of photons through the potential
  field of matter in an $\Omega_m \ne 1$ Universe. 
\item {\bf IFU:} Integrated Field Unit, an instrument which can obtain spatially
  resolved spectra of objects. 
\item {\bf JFC:} Jupiter Family Comet. 
\item {\bf KBO:} Kuiper Belt Object, an asteroid with an orbit beyond that
  of Neptune. 
\item {\bf LBV:} Luminous Blue Variable star. 
\item {\bf LGRB:} Long duration Gamma-Ray Burst
\item {\bf LGSAO:} Laser Guide Star Adaptive Optics, a technique to correct
  for atmospheric turbulence in imaging from ground-based telescopes. 
\item {\bf LMC:} Large Magellanic Cloud, a companion galaxy to the Milky
  Way. 
\item {\bf LPC:} Long Period Comet. 
\item {\bf LPV:} Long Period Variable star. 
\item {\bf LRG:} Luminous Red Galaxy. 
\item {\bf LRN:} Luminous Red Novae. 
\item {\bf LSB:} Long Soft gamma-ray Burst, or Low Surface Brightness
  galaxy. 
\item {\bf LV:} Local Volume, referring to the galaxies with 10-20 Mpc
 of the Milky Way.
\item {\bf MBA:} Main Belt Asteroid, i.e., with an orbit between that of
  Mars and Jupiter. 
\item {\bf MBC:} Main Belt Comet, i.e., with an orbit between that of
  Mars and Jupiter. 
\item {\bf MCMC:} Markov-Chain Monte-Carlo, a technique for exploring
  likelihood surfaces in multiparameter space. 
\item {\bf MDF:} Metallicity Distribution Function of stars. 
\item {\bf ML:} Machine Learning
\item {\bf MMR:} Mean-Motion Resonance, referring to resonance between
  orbits of different Solar System bodies. 
\item {\bf MMRD:} Maximum Magnitude Rate of Decline, referring to the rate
  at which transient objects, such as novae, decline in brightness. 
\item {\bf MN:} Macronova, a subrelativistic stellar explosion with
  sub-supernova energies and emission powered by radioactive decay. 
\item {\bf MOID:} Minimum Orbital Intersection Distance of asteroids. 
\item {\bf MOPS:} The Moving Object Processing System, the software package
   LSST will use to determine orbits of asteroids. 
\item {\bf MS:} Main Sequence (of stars on the Hertzsprung-Russell
  Diagram). 
\item {\bf MySQL:} My Structured Query Language, a relational database
  management system. 
\item {\bf NAS:} National Academy of Sciences.
\item {\bf NASA:} National Aeronautics and Space Administration. 
\item {\bf NEA:} Near-Earth Asteroid, i.e., an asteroid whose orbit takes it
  within 1.3 AU of the Sun. 
\item {\bf NEO:} Near-Earth Object, including Near-Earth Asteroids and
  Near-Earth Comets. 
\item {\bf NFW:} Navarro-Frenk-White, referring to a standard density
  profile for dark matter halos \citep{NFW}. 
\item {\bf NIR:} Near Infrared, typically referring to the wavelength
  range from 1 to 2.5 microns (although some will call wavelengths as
  short as 0.7 microns, and as long as 8 microns, as part of the NIR
  range). 
\item {\bf NRC:} National Research Council.
\item {\bf NS} Neutron Star.
\item {\bf OC:} Oort Cloud, where most long-period comets are thought to
  reside. 
\item {\bf PAH:} Polycyclic Aromatic Hydrocarbon, complex organic molecules
  that are found in the interstellar medium. 
\item {\bf PDE:} Partial Differential Equation. 
\item {\bf PDF:} Probability Distribution Function (also {\bf UPDF},
  the Universal Probability Distribution Function).  
\item {\bf PHA:} Potentially Hazardous Asteroid, the subset of NEAs that
  pass within 0.05 AU of the Earth's orbit. 
\item {\bf PMS:} Pre-Main Sequence star, i.e., one which is still
  gravitationally collapsing before coming into hydrostatic
  equilibrium on the main sequence. 
\item {\bf PPSN:} Pair-Production Supernova (sometimes called Pair
  Instability Supernova), the result of collapse of stars
  in the 140-260$\,M_\odot$ mass range, where the energy density in
  the center is large enough to create electron-positron pairs.  This
  reduces the internal thermal pressure, leading to contraction,
  further heating, and catastrophic collapse. 
\item {\bf PSF:} Point Spread Function, referring to the response of the
  telescope plus camera to a point source of light, such as a star. 
\item {\bf QED:} Quantum Electrodynamics. 
\item {\bf QSO:} Quasi-Stellar Object, synonymous with Quasar. 
\item {\bf RGB:} Red Giant Branch.  See also TRGB. 
\item {\bf RS:} The Red Sequence, a narrow locus of red elliptical
  galaxies on a color-magnitude diagram.  
\item {\bf RTML:} Remote Telescope Markup Language.
\item {\bf SBF:} Surface Brightness Fluctuation, the mottling of the images
  of nearby elliptical galaxies due to the finite number of stars in
  each pixel. 
\item {\bf SCM:} Standardized Candle Method, a method of determining the
  distances to core-collapse supernovae. 
\item {\bf SDO:} Scattering Disk Objects, a subclass of Trans-Neptunian
  Objects in orbits that gravitationally interact with Neptune. 
\item {\bf SED:} Spectral Energy Distribution, i.e., the spectrum of an
  astronomical object over a broad range of wavelengths. 
\item {\bf SETI:} Search for Extraterrestrial Intelligence.
\item {\bf SFD:} The Size-Frequency Distribution of asteroids.  The acronym
  can also refer to the paper by \citet{SFD} giving Galactic
  extinction maps. 
\item {\bf SFH:} Star Formation History.
\item {\bf SFR:} Star Formation Rate.
\item {\bf SGRB:} Short duration Gamma-Ray Burst.
\item {\bf SHB:} Short Hard gamma-ray Burst.
\item {\bf SMBH:} Super-Massive Black Hole.
\item {\bf SMC:} Small Magellanic Cloud, a companion galaxy to the Milky
  Way. 
\item {\bf SNIa:} Type Ia Supernova, i.e., one caused by the collapse of a
  white dwarf star pushed over the Chandrasekhar limit. 
\item {\bf SNR:} Signal-to-Noise Ratio, or Supernova Remnant
\item {\bf SPC:} Short-Period Comet. 
\item {\bf SPH:} Smooth-Particle Hydrodynamics, a computational technique
  for including hydrodynamical effects into N-body simulations. 
\item {\bf SZ:} Sunyaev-Zel'dovich effect, whereby Cosmic
  Microwave Background photons are Compton-scattered to higher energy
  in interactions with  electrons in the hot gas in clusters of
  galaxies. 
\item {\bf TDF:} Tidal Disruption Flare, the disruption of a star by a
super-massive Black Hole.
\item {\bf TNO:} Trans-Neptunian Object, an asteroid with orbit beyond that
  of Neptune. 
\item {\bf TO:} Turn-off (of stars from the main sequence on the Hertzsprung-Russell
  Diagram). 
\item {\bf TRGB:} Tip of the Red Giant Branch, referring to the fact
  that red giant branch stars in a given stellar
  population have a well-defined upper limit of luminosity, making
  this a useful distance indicator. 
\item {\bf VHE:} Very High Energy Photon, i.e., one with energies of at
  least TeV. 
\item {\bf VLM:} Very Low Mass (stars), i.e., M stars and later. 
\item {\bf VLBI:} Very Long Baseline Interferometry, a technique for
  high resolution imaging at radio wavelengths. 
\item {\bf VO:} Virtual Observatory.
\item {\bf VOEvent:} A National Virtual Observatory standard for
exchanging information on astronomical transients.
\item {\bf VTP:} Voronoi Tessellation and Percolation method, a way to look
  for clusters in point data.  
\item {\bf WCS:} World Coordinate System, a transformation between
  coordinates on the focal plane and those on the sky (such as right
  ascension and declination).  
\item {\bf WD:} White Dwarf star. 
\item {\bf WDLF:} White Dwarf Luminosity Function. 
\item {\bf WL:} Weak (gravitational) Lensing, the subtle distortion of
  galaxy images by the gravitational field of foreground
  overdensities. 
\item {\bf YORP:} The Yarkovsky-Radzievskii-O'Keefe-Paddock effect, whereby
  the spin state of an asteroid is systematically changed by
  anisotropic thermal emission of its surface. 
\item {\bf YSO:} Young Stellar Object.
\end{itemize}
\bigskip

\begin{center}
\Large
{\bf Past, Present, and Future Astronomical Facilities, Surveys, and Organizations}
\end{center}

\begin{itemize} 
\item {\bf 2dF:} Two-Degree Field, referring to a wide-field multi-object
  spectrograph on the Anglo-Australian Telescope. 
\item {\bf 2dFGRS:} The Two-Degree Field Galaxy Redshift Survey, which
  obtained redshifts for almost 250,000 galaxies. \url{http://www2.aao.gov.au/2dFGRS/}.
\item {\bf 2MASS:} Two-Micron All-Sky Survey, which surveyed the entire sky
  in $J$, $H$, and $K$.  \url{http://www.ipac.caltech.edu/2mass/}.
\item {\bf 2QZ:} The 2dF Quasar Redshift Survey, which obtained redshifts of
  over 23,000 quasars.  \url{http://www.2dfquasar.org/}. 
\item {\bf 2SLAQ:} The 2dF-SDSS LRG and QSO survey, which obtained spectra
  of LRGs (op.cit.) and QSOs from catalogs selected from SDSS imaging
  data.  \url{http://www.2slaq.info/}. 
\item {\bf AAOmega:}  A multiobject/integral-field spectrograph at the
Anglo-Australian Telescope. \url{
http://www.aao.gov.au/local/www/aaomega/}. 
\item {\bf AAT:} The Anglo-Australian Telescope, 3.9 m in
  diameter. \url{http://www.aao.gov.au/}. 
\item {\bf AGES:} The AGN and Galaxy Evolution Survey, a redshift survey in
  the NOAO Deep Wide Field. \url{
    http://cmb.as.arizona.edu/~eisenste/AGES/}. 
\item {\bf AKARI:} A Japanese satellite that mapped the sky at infrared
  wavelengths.  \\
\url{http://www.ir.isas.jaxa.jp/ASTRO-F/Outreach/index_e.html}. 
\item {\bf ALMA:} Atacama Large Millimeter Array, operating between
  0.3 and 9.6 mm. \\
\url{http://www.alma.nrao.edu}. 
\item {\bf A-LIGO:} Advanced LIGO; the proposed next generation of the Laser
  Interferometer Wave Observatory, \url{
    http://www.ligo.caltech.edu/advLIGO/scripts/summary.shtml}. 
\item {\bf APEX:} The Atacama Pathfinder Experiment, a 12-meter
  sub-millimeter telescope placed in the Chilean Andes. \url{
    http://www.mpifr-bonn.mpg.de/div/mm/apex.html}. 
\item {\bf APM:} Automated Plate Measuring facility, which digitized
  photographic sky survey plates to make one of the premier galaxy
  catalogs in the 1990s.  \citet{Maddox++90}. 
\item {\bf ASAS:} All-Sky Automated Survey, which repeatedly images the
  entire sky to about 14th magnitude to look for variable stars. \url{http://www.astrouw.edu.pl/asas/}
\item {\bf ASTE:} The Atacama Submillimeter Telescope Experiment,  a 10-meter
  sub-millimeter telescope placed in the Chilean Andes. \url{
    http://www.ioa.s.u-tokyo.ac.jp/~kkohno/ASTE/}. 
\item {\bf ATCA:} The Australia Telescope Compact Array is a six-dish radio
  interferometer. \\
 \url{http://www.narrabri.atnf.csiro.au/}. 
\item {\bf AURA:} Association of Universities for Research in
  Astronomy, the parent organization that operates the Gemini
  Observatory and NOAO, among others. \url{
    http://www.aura-astronomy.org/}. 
\item {\bf BigBOSS:} A proposed multi-object spectrograph to study
  baryon acoustic oscillations. \url{http://bigboss.lbl.gov/}.
\item {\bf Black Hole Finder Probe:} A proposed NASA satellite to study
  accretion onto black holes, as part of their ``Beyond Einstein''
  mission concept.  EXIST (see below) is one possible
  implementation of this concept.  
\item {\bf BOSS:} The Baryon Oscillation Spectroscopic Survey, a redshift
  survey over 10,000 deg$^2$ of galaxies to $z = 0.7$ and quasars to
  $z \sim 3$ to study baryon oscillations.  One of the components of
  the SDSS-III. \url{http://www.sdss3.org}.  
\item {\bf BTC:} Big Throughput Camera, a wide-field imaging camera on the
  4-meter Blanco Telescope at CTIO.
\item {\bf CADIS:} Calar Alto Deep Imaging Survey, covering 0.3 deg$^2$ in
  three broad bands and 13 medium bands, \citet{CADIS}. 
\item {\bf CANGAROO:} Collaboration of Australia and Nippon for a GAmma Ray
  Observatory in the Outback, an array of imaging Cherenkov telescopes
  to search for very high-energy gamma-rays.  \url{
    http://icrhp9.icrr.u-tokyo.ac.jp/}. 
\item {\bf Catalina Sky Survey:} Uses telescopes in the US and
  Australia to look for asteroids and comets.  \url{
    http://www.lpl.arizona.edu/css/}. 
\item {\bf CBA:} Center for Backyard Astrophysics, a global network of small
  telescopes dedicated to the photometry of cataclysmic variables.
  \url{http://cbastro.org/}. 
\item {\bf CFHT:} The Canada-France-Hawaii Telescope, a 3.6-meter optical
  telescope on Mauna Kea in Hawaii. \url{
    http://www.cfht.hawaii.edu/}. 
\item {\bf CFHTLS:} The Canada France Hawaii Telescope Legacy Survey, which
  is surveying up to 400 deg$^2$ in optical bands.  \url{
    http://www.cfht.hawaii.edu/Science/CFHLS/}. 
\item {\bf Chandra X-ray Observatory:} One of NASA's Great
  Observatories. \\
  \url{http://chandra.harvard.edu/}. 
\item {\bf COBE:} Cosmic Background Explorer, which made the first detection
  of fluctuations in the CMB in an all-sky map.  \url{
    http://lambda.gsfc.nasa.gov/product/cobe/}. 
\item {\bf COMBO-17:} Classifying Objects by Medium-Band Observations, an
  imaging survey carried out on the Calar Alto 3.5m telescope of one deg$^2$
  through 17 medium bands. \\ \url{
    http://www.mpia-hd.mpg.de/COMBO/combo_index.html}. 
\item {\bf COROT:} COnvection, ROtation, and planetary Transits, a
  European space mission to look for transiting planets. \url{
    http://smsc.cnes.fr/COROT/}.
\item {\bf COSMOS:} Cosmological Evolution Survey over 2 deg$^2$ with the
  Hubble Space Telescope, together with follow-up with many other
  facilities. \\
 \url{http://cosmos.astro.caltech.edu/index.html}. 
\item {\bf COVET:} A repeat imaging survey of nearby clusters of galaxies to
  search for transients. 
\item {\bf CTIO:} Cerro Tololo Inter-American Observatory.  Cerro Tololo is
  adjacent to Cerro Pach\'on, where LSST will be sited. 
\item {\bf DEEP and DEEP2:} The Deimos Extragalactic Probe, a redshift
  survey of roughly 50,000 galaxies with redshifts of order unity,
  carried out on the Keck Telescopes. \\
\url{
    http://deep.berkeley.edu/}. 
\item {\bf DEIMOS:} DEep Imaging Multi-Object Spectrograph on the Keck-II
telescope. 
\item {\bf DENIS:} The Deep Near Infrared Survey of the Southern Sky,
  similar in scope to 2MASS. \url{
    http://www-denis.iap.fr/denis.html}. 
\item {\bf DES:} Dark Energy Survey, a wide-angle survey to be carried out
  on the 4-meter Blanco Telescope at CTIO. \url{
    https://www.darkenergysurvey.org/}. 
\item {\bf DLS:} The Deep Lens Survey, a deep four-band survey of 20
  deg$^2$ carried out with on the CTIO 4-meter Blanco Telescope. \url{
    http://dls.physics.ucdavis.edu}.  
\item {\bf DMT:} Dark Matter Telescope, an early incarnation of the LSST
  concept. 
\item {\bf E-ELT:}  The European Extremely Large Telescope, a proposed
  telescope with a mirror 42 meters in diameter. \url{
    http://www.eso.org/sci/facilities/eelt/}. 
\item {\bf E-LIGO:} Enhanced LIGO; intermediate between LIGO and A-LIGO.
\item {\bf EIS:} The ESO Imaging survey, covering several deg$^2$ in optical
  and near-IR bands. \\
\url{
    http://www.eso.org/sci/activities/projects/eis/}.
\item {\bf eROSITA:} The extended ROentgen Survey with an Imaging Telescope
  Array, a planned medium-energy X-ray survey of the sky.  \url{
    http://www.mpe.mpg.de/projects.html\#erosita}. 
\item {\bf Euro50:} A proposed 50-meter optical and infrared
telescope \url{http://www.astro.lu.se/~torben/euro50/}.
\item {\bf ESA:} The European Space Agency. 
\item {\bf ESO:} The European Southern Observatory, which operates
  telescopes at La Silla and Cerro Paranal in the Chilean Andes.  
\item {\bf ESSENCE:} ``Equation of State: SupErNovae trace Cosmic
  Expansion'', an imaging survey to find supernovae on the CTIO
  4-meter Blanco Telescope. \\
\url{http://www.ctio.noao.edu/essence/}.
\item {\bf EUCLID:}, Proposed European near-infrared wide-angle sky
  survey from space to explore dark energy. \url{
    http://sci.esa.int/science-e/www/area/index.cfm?fareaid=102}.  
\item {\bf EVLA:} The Expanded Very Large Array, an extension of the premier
  radio interferometer facility in the world. \url{http://www.aoc.nrao.edu/evla/}.
\item {\bf EXIST:} Energetic X-ray Imaging Survey Telescope, a proposed hard
  X-ray survey of the sky.  \url{http://exist.gsfc.nasa.gov/}
\item {\bf FASTSOUND:} Fiber Multi-Object Spectrograph Ankoku Shind\={o} Tansa
  Subaru Observation Understanding Nature of Dark Energy, a proposed
  spectroscopic survey on the Subaru Telescope to study baryon
  oscillations.  
\item {\bf Fermi Gamma-Ray Space Telescope:} Formerly known as GLAST, this
  is surveying the sky at 10 keV to 300 GeV. \url{
    http://fermi.gsfc.nasa.gov/}. 
\item {\bf FIRST:} Faint Images of the Radio Sky at Twenty Centimeters, a
  VLA survey at 1.4 GHz.  \url{http://sundog.stsci.edu}.  
\item {\bf FORS:} Visual and near UV FOcal Reducer and low dispersion
Spectrograph for the Very Large Telescope.  \url{
  http://www.eso.org/instruments/fors1/}. 
\item {\bf Gaia:} A planned European satellite for precision
  astrometry. \url{http://www.rssd.esa.int/Gaia}.
\item {\bf Galaxy Zoo:} A project to visually classify over a million
  galaxies from the Sloan Digital Sky Survey. \url{http://www.galaxyzoo.org/}
\item {\bf GALEX:} The Galaxy Evolution Explorer, which is surveying the sky in
  the ultraviolet. \\
\url{http://www.galex.caltech.edu/}. 
\item {\bf GCN:} Gamma-Ray Burst (GRB) Coordinates Network.
\item {\bf GTC:} The Gran Telescopio Canarias, a 10.4-meter telescope
  on the Canary Islands. \\
\url{http://www.gtc.iac.es/en/}.
\item {\bf Gemini:} A pair of 8-meter telescopes, one on Mauna Kea
  (Hawaii), and the other on Cerro Pach\'on (Chile).  \url{
  http://www.gemini.edu}. 
\item {\bf GEMS:} Galaxy Evolution From Morphology and SEDs, a wide-field
  imaging survey with the Hubble Space Telescope. \url{
    http://www.mpia-hd.mpg.de/GEMS/home0.htm}. 
\item {\bf GMOS:} Gemini Multi-Object Spectrographs (one one each of the
two telescopes).  \\
\url{http://www.gemini.edu/node/10625}.
\item {\bf GMT:}  The Giant Magellan Telescope, a proposed telescope
  with an effective aperture of 24.5 meters. \url{http://www.gmto.org/}
\item {\bf GOODS:} The Great Observatories Origins Deep Survey, a wide-field
  imaging survey with the Hubble Space Telescope and other
  facilities.  \url{http://www.stsci.edu/science/goods/}.
\item {\bf Google Sky:} A display of wide-field imaging data for the
  public. \url{http://www.google.com/sky/}. 
\item {\bf GSMT:} The Giant Segmented Mirror Telescope, a generic name
  for a future US 20-30 meter telescope. 
\item {\bf HAT:} The Hungarian-made Automated Telescope, a network of small
  wide-field telescopes to survey the sky. \url{
    http://www.cfa.harvard.edu/~gbakos/HAT/}. 
\item {\bf HDF, HUDF:} Hubble Deep Field and Ultra Deep Field, extremely
  deep exposures of the sky with the Hubble Space Telescope. 
\item {\bf HEAO-1:} The High-Energy Astrophysics Observatory, one of the first
  X-ray surveys of the sky in the 1970s.  \url{
    http://heasarc.gsfc.nasa.gov/docs/heao1/heao1.html}. 
\item {\bf Hectospec:} A moderate-resolution, multi-object optical
spectrograph fed by 300 optical fibers, on the Multiple Mirror
Telescope. \url{http://www.cfa.harvard.edu/mmti/}. 
\item {\bf HESS:} The High Energy Stereoscopic System, a system of Imaging
  Atmospheric Cherenkov Telescopes that investigates cosmic gamma rays
  in the 100 GeV to 100 TeV energy range.  \url{
    http://www.mpi-hd.mpg.de/hfm/HESS/}. 
\item {\bf HIPPARCOS:} High Precision Parallax Collecting Satellite, which
  did accurate astrometry of bright stars over the entire sky. \\
\url{
    http://www.rssd.esa.int/index.php?project=HIPPARCOS}.
\item {\bf HSC:} Hyper-SuprimeCam, a planned wide-field imager for the Subaru
  Telescope. 
\item {\bf HST:} The Hubble Space Telescope. 
\item {\bf IceCube:} A telescope in Antartica which uses Cherenkov
  light in deep ice from secondary particles due to collisions from
  high-energy neutrinos.  \url{http://icecube.lbl.gov/}.  
\item {\bf IMF:} Initial Mass Function, the distribution of masses
  of stars when they are first born. 
\item {\bf IMACS:} The Inamori Magellan Areal Camera and Spectrograph,
  for the Magellan Telescope at Las Campanas, Chile. 
\item {\bf IRAC:} The Infrared Array Camera on the Spitzer Space
  Telescope, with filters at 3.6, 4.5, 5.8, and 8 microns.  \url{
    http://ssc.spitzer.caltech.edu/irac/}. 
\item {\bf ISO:} Infrared Space Observatory, a European space-based mission
  of the 1990s. \\
\url{http://iso.esac.esa.int/}.  
\item {\bf IXO:} International X-ray Observatory, a proposed
  facility with superior collecting area and spectral resolution. \url{
    http://ixo.gsfc.nasa.gov/}.  
\item {\bf JANUS:} A proposed near-infrared low-resolution spectroscopic
  survey of the sky, designed to find high-redshift quasars and
  gamma-ray bursts. 
\item {\bf JDEM:} The Joint Dark Energy Mission, the generic name for the
  proposed NASA satellite mission to study dark energy. 
\item {\bf JWST:} James Webb Space Telescope, a 6.4-meter telescope
  sensitive from 0.6 to 25$\mu$m, which 
  NASA will launch in 2014.  \url{http://www.jwst.nasa.gov/}. 
\item {\bf KAIT:} The Katzman Automatic Imaging Telescope, which is
  surveying nearby galaxies to search for supernovae.  \url{
    http://astro.berkeley.edu/~bait/kait.html}. 
\item {\bf Kepler:} A NASA mission doing photometry of stars to look
  for transiting planets. \\
\url{http://kepler.nasa.gov}. 
\item {\bf KPNO:} Kitt Peak National Observatory. \url{
  http://www.noao.edu/kpno}. 
\item {\bf LAMA:} Large-Aperture Mirror Array, a proposed array of
  10-meter liquid-mirror telescopes. \url{http://www.astro.ubc.ca/lmt/lama/}.
\item {\bf LAMOST:}  Large Sky Area Multi-Object Fibre Spectroscopic
Telescope, a Chinese 4-meter telescope devoted to spectroscopic
surveys. \url{http://www.lamost.org/en/}. 
\item {\bf LBT:} The Large Binocular Telescope, a pair of 8.4-meter telescopes
  on a common mount.  \url{http://medusa.as.arizona.edu/lbto/}.
\item {\bf LCOGTN:} Las Cumbres Observatory Global Telescope Network,
  dedicated to study of transient and variable objects. \url{
    http://lcogt.net/}. 
\item {\bf LIGO:} The Laser Interferometer Gravitational Wave Observatory, now
  in operation. \\
 \url{
  http://www.ligo.caltech.edu/}.
\item {\bf LISA:} The Laser Interferometer Space Antenna, a proposed space-based
  gravitational-wave detector. \url{http://lisa.nasa.gov}
\item {\bf MACHO:} Massive Compact Halo Object, which can cause
  gravitational microlensing of background objects.  Also, a survey
  carried out with the Anglo-Australian Telescope to find such
  objects, \citet{MACHO}.
\item {\bf MAGIC:} Major Atmospheric Gamma Imaging Cherenkov. A pair of
  telescopes looking for Cherenkov radiation from high-energy cosmic
  rays. \url{http://magic.mppmu.mpg.de/}. 
\item {\bf MaxAT:} the Maximum Aperture Telescope, a generic name for
  a future 30-50 meter telescope.  See also GSMT. 
\item {\bf MGC:} Millennium Galaxy Catalog, a 37.5 deg$^2$ imaging survey
  carried out on the Isaac Newton Telescope to a depth of $B \sim
  24$. \url{http://www.eso.org/~jliske/mgc/}. 
\item {\bf Micro-FUN:} Microlensing Follow-Up Network, which uses
  small telescopes to get high time resolution on microlensing
  stars. \url{http://www.astronomy.ohio-state.edu/~microfun/}. 
\item {\bf MIPS:} The Multiband Imaging Photometer for SIRTF, on the Spitzer
  Space Telescope, with filters at 24, 70, and 160
  microns. \url{http://ssc.spitzer.caltech.edu/mips/}. 
\item {\bf MIRI:} Mid-Infrared Instrument, for the JWST. \url{
  http://ircamera.as.arizona.edu/MIRI/}. 
\item {\bf MMT:} Multiple Mirror Telescope, a 6.5-meter single-mirror
  telescope, despite its name. \\
\url{http://www.mmto.org/}.
\item {\bf MOA:} Microlensing Observations in Astrophysics, a 0.6-meter
  telescope in New Zealand used for studies of gravitational
  microlensing.  \url{http://www.phys.canterbury.ac.nz/moa/}. 
\item {\bf MOSAIC:} A wide-field imaging camera used on the 4-meter
  telescopes at CTIO and KPNO. 
\item {\bf MOSFIRE:} Multi-Object Spectrometer for Infra-Red
  Exploration, being built for the Keck Telescopes. \url{
    http://irlab.astro.ucla.edu/mosfire/}.
\item {\bf NANTEN:} A 4-meter submillimeter telescope in Chile.  \\
\url{
  http://www.astro.uni-koeln.de/nanten2/}. 
\item {\bf NCSA:} National Center for Supercomputing Applications at
  the University of Illinois.  b\url{http://www.ncsa.illinois.edu/}. 
\item {\bf NDWFS:} NOAO Deep Wide-Field Survey, an optical/near-IR imaging
  survey of 9 deg$^2$.  \url{http://www.noao.edu/noao/noaodeep/}.  
\item {\bf NEWFIRM:} The NOAO Extremely Wide-Field Infrared Imager, a
  imaging camera with a field of view of 1/4 deg$^2$ for the KPNO
  4-meter Mayall Telescope. \url{http://www.noao.edu/ets/newfirm/}. 
\item {\bf NOAO:} National Optical Astronomical Observatory, the parent
  organization of CTIO and KPNO.  \url{http://www.noao.edu}. 
\item {\bf NVSS:} The NRAO VLA Sky Survey, which covered the entire Northern
  sky at 1.4 GHz.  \url{http://www.cv.nrao.edu/nvss/}. 
\item {\bf ODI:} One Degree Imager on the WIYN 3.5-m telescope at Kitt
  Peak. \\
\url{  http://www.noao.edu/wiyn/ODI/}. 
\item {\bf OGLE:} Optical Gravitational Lensing Experiment, which carries
  out repeat imaging of the sky. \url{
    http://www.astrouw.edu.pl/~ftp/ogle/}.  
\item {\bf P60-FasTING:} Palomar 60-inch Fast Transients in Nearby Galaxies
  carries out repeat imaging of nearby galaxies.  \url{
    http://www.astro.caltech.edu/ptf/}.  
\item {\bf Pan-STARRS:} Panoramic Survey Telescope and Rapid Response
  System.  A dedicated survey telescope based at the University of
  Hawaii.  Pan-STARRS1 consists of a single 1.8-meter telescope with
  a $3^\circ$ field of view, and has seen first light.  Pan-STARRS4
  will consist of four such telescopes on a common mount.  \url{
    http://pan-starrs.ifa.hawaii.edu/public/}.  
\item {\bf Planck:} A recently launched satellite which is mapping
  fluctuations in the Cosmic Microwave Background.  \url{
    http://www.rssd.esa.int/index.php?project=planck}.  
\item {\bf POSS:} Palomar Observatory Sky Survey, a photographic survey of
  the sky started in the 1950s.  \url{
    http://www.astro.caltech.edu/~wws/poss2.html}. 
\item {\bf PS1, PS4:} Abbreviations for Pan-STARRS1, Pan-STARRS4. See
  above. 
\item {\bf PSCz:} Point-Source Catalog Redshift Survey, of galaxies detected
  by the Infrared Astronomical Satellite at 60 microns, \citet{PSCz}. 
\item {\bf PTF:} Palomar Transit Survey.  See P60-FasTING above. 
\item {\bf ROSAT:} The R\"ontgen Satellite, which carried out an X-ray
  survey of the sky. \\
 \url{http://www.mpe.mpg.de/xray/wave/rosat/}.  
\item {\bf RSS:} The Robert Stobie Spectrograph on the SALT
  telescope. \\
\url{
    http://www.salt.ac.za/telescope/instrumentation/rss/}. 
\item {\bf SAGE:} Surveying the Agents of Galaxy Evolution, a Spitzer
  imaging study of the Magellanic Clouds. \url{
    http://sage.stsci.edu}. 
\item {\bf SALT:} Southern African Large Telescope, with a primary
  mirror 11 meters across. \\
\url{http://www.salt.ac.za/}. 
\item {\bf SASIR:} The Synoptic All-Sky Infrared Survey, a dedicated
  6.5-meter telescope which will go appreciably deeper than 2MASS.
  \url{http://sasir.org}.  
\item {\bf SDSS:} The Sloan Digital Sky Survey, an imaging and spectroscopic
  survey of the Northern Sky.  \url{http://www.sdss.org}.  
\item {\bf SEGUE:} Sloan Extension for Galactic Understanding and
  Exploration, a component of the SDSS focussed on the structure of
  the Milky Way. \url{http://ww.sdss.org/segue}.  
\item {\bf SERVS:}, The Spitzer Extragalactic Representative Volume
  Survey, an imaging survey of 18 deg$^2$ of high-latitude sky at 3.6
  and 4.5 microns. \\
 \url{
    http://www.its.caltech.edu/~mlacy/servs.html}. 
\item {\bf SINGS:} Spitzer Infrared Nearby Galaxies Survey, a comprehensive
  survey of 75 nearby galaxies in the infrared. \url{
    http://sings.stsci.edu}.  
\item {\bf SKA:} The Square Kilometre Array, a proposed enormous radio
  survey telescope.  \\
\url{http://www.skatelescope.org/}
\item {\bf SkyMapper:} A 1.3-meter telescope at Siding Spring Observatory,
  which is imaging the Southern skies. It will produce the SkyMapper
  Southern Sky Survey (SSSS).  \\
\url{
    http://msowww.anu.edu.au/skymapper/}.  
\item {\bf SMEI:} Solar Mass Ejection Imager, which is flying on the US Air
  Force's Coriolis spacecraft.  \url{http://smei.ucsd.edu/}.  
\item {\bf SNLS:} The SuperNova Legacy Survey, which was carried out as part
  of the CFHTLS (see above).  \url{http://cfht.hawaii.edu/SNLS/}.  
\item {\bf SOAR:} Southern Astrophysical Research Telescope, a 4.1-m
  telescope located on Cerro Pach\'on in Chile. \url{
    http://www.soartelescope.org/}. 
\item {\bf SOFIA:} The Stratospheric Observatory for Infrared
  Astronomy, a mid-infrared 2.5-meter telescope in development, which
  flies on a specially modified airplane.  \\
\url{http://www.sofia.usra.edu/}. 
\item {\bf SkyAlert:} A website that collects and distributes astronomical
events (such as transients) over the Internet in near-real time. \url{
  http://www.skyalert.org/}. 
\item {\bf Spacewatch:} This project uses telescopes at Kitt Peak to
  search for asteroids. \\
\url{http://spacewatch.lpl.arizona.edu/}. 
\item {\bf Spitzer Space Telescope:} One of NASA's Great Observatories, it
  is sensitive from 3 to 160 microns.  \url{
    http://www.spitzer.caltech.edu/}.  
\item {\bf SPT:} South Pole Telescope, a 10-meter millimeter telescope
  designed to measure fluctuations in the CMB.  \url{
    http://pole.uchicago.edu}.  
\item {\bf STEREO:} The Solar Terrestrial Relations Observatory, a pair of
  spacecraft which monitor the Sun. \url{
    http://stereo.gsfc.nasa.gov/}. 
\item {\bf Subaru Telescope:} A 8.2-meter wide-field optical telescope
  operated by the Japanese astronomical community. \url{
    http://www.naoj.org}.  
\item {\bf SUMSS:} The Sydney University Molonglo Sky Survey, which covered
  the Southern sky at 843 MHz. \url{
    http://www.physics.usyd.edu.au/sifa/Main/SUMSS}. 
\item {\bf SuperCOSMOS:} Digitized scans of photographic survey
  plates. \\
  \url{http://www-wfau.roe.ac.uk/sss/}. 
\item {\bf SUPERMACHO:} A survey for gravitational microlenses and
  supernovae using the MOSAIC imager on the CTIO 4-meter Blanco
  Telescope.  \url{http://www.ctio.noao.edu/supermacho/}.  
\item {\bf SuperWASP:} Wide-Angle Search for Planets, which consists of two
  imaging telescopes looking for planetary transit events. \url{
    http://www.superwasp.org}.  
\item {\bf Swift:} A gamma-ray burst satellite. \url{
  http://swift.gsfc.nasa.gov/}. 
\item {\bf SWIRE:} The Spitzer Wide-area InfraRed Extragalactic survey, a
  legacy mapping program covering 50 deg$^2$ with the Spitzer Space
  Telescope.  \\
\url{http://swire.ipac.caltech.edu/swire/swire.html}.  
\item {\bf THINGS:} The HI Nearby Galaxy Survey of 34 galaxies observed with
  the Very Large Array at 21 cm. \citet{THINGS}. 
\item {\bf TMT:}  Thirty Meter Telescope, a proposed telescope whose
  name says it all. \url{http://www.tmt.org}.  
\item {\bf TSS:} The Texas Supernova Search, carried out with a telescope
  from the ROTSE collaboration.  \url{
    http://grad40.as.utexas.edu/~quimby/tss/index.html}.  
\item {\bf UKIDSS:} The UKIRT (United Kingdom InfraRed Telescope)
  Infrared Deep Sky Survey, covering 7500 deg$^2$.  \url{http://www.ukidss.org/}.
\item {\bf USNO-B:} The United States Naval Observatory astrometric
  calibration and catalog of the POSS plates.  \url{
    http://www.nofs.navy.mil/data/fchpix/}.  
\item {\bf Veritas:} Very Energy Radiation Imaging Telescope System, an
  array of atmospheric Cherenkov telescopes at Mount Hopkins.  \url{
    http://veritas.sao.arizona.edu/}.  
\item {\bf VISTA:} The Visible and Infrared Survey Telescope for Astronomy, a
  dedicated 4-meter survey telescope operated by the European Southern
  Observatory.  \url{http://www.vista.ac.uk/}. 
\item {\bf VLT:} The Very Large Telescope, a set of four 8-meter
  telescopes at Cerro Paranal in Chile, operated by the European
  Southern Observatory. \url{http://www.eso.org/projects/vlt/}. 
\item {\bf VST:} The VLT Survey Telescope, a 2.6-meter telescope with a 1
  deg$^2$ Field of View at Cerro Paranal in Chile. \url{
    http://vstportal.oacn.inaf.it/}. 
\item {\bf VVDS:} The VIRMOS-VLT Deep Survey, a spectroscopic redshift
  survey of 150,000 faint galaxies over 16 deg$^2$.  \url{
    http://www.oamp.fr/virmos/vvds.htm}.  
\item {\bf WEBDA:} A database of open star clusters, \url{
  http://www.univie.ac.at/webda/}. 
\item {\bf WHTDF:} William Herschel Telescope Deep Field, a $7' \times 7'$
  survey imaged to close to LSST co-added depths.  \url{
    http://astro.dur.ac.uk/~nm/pubhtml/herschel/herschel.php}.  
\item {\bf Wigglez:} A spectroscopic survey of galaxies to $z \sim 1$ on the
  Anglo-Australian Telescope to study baryon acoustic oscillations.  \url{
    http://wigglez.swin.edu.au/}. 
\item {\bf WISE:} Wide-field Infrared Survey Explorer, a NASA satellite that
  will survey the sky from 3.3 to 23 microns.  \url{
    http://www.astro.ucla.edu/~wright/WISE/}. 
\item {\bf WMAP:} The Wilkinson Microwave Anisotropy Probe, a satellite that
  has made full-sky maps of fluctuations in the Cosmic Microwave
  Background.  \url{http://map.gsfc.nasa.gov/}. 
\item {\bf WWT:} The WorldWide Telescope, a web resource for exploring
  images of the sky. \\
\url{
    http://www.worldwidetelescope.org/Home.aspx}.  
\item {\bf XMM-Newton:} The X-ray Multi-Mirror Mission is an X-ray telescope
  with particularly high throughput.  \url{http://xmm.esac.esa.int/}. 
\item {\bf zCOSMOS:} A spectroscopic follow-up survey of the COSMOS field.  \\
 \url{http://cosmos.astro.caltech.edu/index.html}. 
\end{itemize}

\bibliographystyle{SciBook}
\bibliography{appendix/acronym/acronyms}

%
%
%
%
%
%
%
%
%
%
%
%
%
%
%
%
%
%
%
%
%
%
%
%
%
%

\chapter[Authors]{List of Contributing Authors}
\label{chp:ap:authors}

An asterisk indicates those responsible for the overall editing of
this book. 

\begin {tabbing}							
\= \hspace{50 mm}		\=Scott Anderson	\=\hspace{50 mm}		\kill
\>	Paul A. Abell		\>	Planetary Science Institute/NASA Johnson Space Center \\
\>	Julius Allison		\>	Alabama A\&M University \\
\>	Scott	 F. Anderson	\>	University of Washington	\\
\>	John R. Andrew	\>	National Optical Astronomy Observatory (NOAO) \\
\>	J. Roger P. Angel	\>	University of Arizona \\
\>	Lee	Armus		\>	Spitzer Science Center	\\
\>	David Arnett		\>	University of Arizona\\
\>	S. J. Asztalos  		\>	Lawrence Berkeley National Laboratory \\
\>	Tim	S. Axelrod		\>	LSST Corporation 	\\
\>	Stephen	Bailey	\>	Laboratoire de Physique Nucleaire et des Hautes Energies, CNRS/IN2P3	\\
\>	D. R.  Ballantyne 	\>	Georgia Institute of Technology\\
\>	Justin R. Bankert	\>	Purdue University	\\
\>	Wayne A. Barkhouse	\>	University of North Dakota
% ...
\\
\>     	Alex Szalay 		\> 	Johns Hopkins University \\
\>	Paula Szkody		\>	University of Washington	\\
\>	Masahiro	Takada	\>	The Institute for the Physics and Mathematics of the Universe (IPMU)\\
\>      Paul Thorman 		\>   	University of California, Davis\\
\>	David E. Trilling      	\> 	Northern Arizona University \\
\>	Virginia Trimble	\>	University of California, Irvine	\\
\>	Anthony Tyson$^*$	\>	University of California, Davis	\\
\>	Richard Van Berg	\>	University of Pennsylvania \\
\>	Daniel Vanden Berk	\>	St.~Vincent College \\
\>      Jake VanderPlas 	\> 	University of Washington\\
\>	Licia Verde		\>	ICREA, Barcelona, Spain	\\
\>      Bojan Vr\v{s}nak	\>	Hvar Observatory, Zagreb, Croatia \\
\>	Lucianne	M. Walkowicz	\>	University of California, Berkeley	\\
\>      Benjamin D. Wandelt  	\>      University of Illinois at Urbana-Champaign \\
\>	Sheng Wang		\>	University of Chicago	\\
\>	Yun	Wang		\>	University of Oklahoma	\\
\>	Michael Warner	\>	NOAO/Cerro Tololo Inter-American Observatory\\
\>	Risa	H. Wechsler	\>	Stanford University	\\
\>	Andrew A. West	\>	Boston University	\\
\>	Oliver Wiecha		\>	National Optical Astronomy Observatory \\
\>	Benjamin	F. Williams	\>	University of Washington	\\
\>	Beth	Willman		\>	Haverford College	\\
\>	David Wittman		\>	University of California, Davis	\\
\>	Sidney C. Wolff$^*$	\>	LSST Corporation	\\
\>	W. Michael Wood-Vasey	\>	University of Pittsburgh	\\
\>	Przemek Wozniak	\>	Los Alamos National Laboratory	\\
\>	Patrick Young		\>	University of Arizona	\\
\>	Andrew Zentner	\>	University of Pittsburgh	\\
\>	Hu Zhan			\>	National Astronomical Observatories of China	\\
\end{tabbing}

\end{document}